\documentclass[a4paper,11pt]{article}
\pdfoutput=1

\usepackage{longtable}
\usepackage[utf8]{inputenc}
\usepackage[T1]{fontenc}
\usepackage{textcomp}
\usepackage{microtype}
\usepackage{bm}
\usepackage{physics}
\usepackage{siunitx}
\usepackage{graphicx}
\usepackage{booktabs}
\usepackage{enumitem}
\usepackage{afterpage}

\usepackage{cancel}

\usepackage[normalem]{ulem}
\usepackage{xcolor}

\usepackage[english]{babel}
\usepackage[autostyle, english = american]{csquotes}
\MakeOuterQuote{"}

\usepackage{amsmath}
\usepackage{amssymb}
\usepackage{mathtools}
\usepackage{upgreek}
\usepackage{slashed}

\usepackage{authblk}
\usepackage{multirow}
\usepackage{rotating}
\usepackage{mathrsfs}
\usepackage{nicefrac}

\usepackage{subcaption}
\usepackage{booktabs}
\usepackage{multirow}
\usepackage{graphicx}
\usepackage{adjustbox}
\usepackage{xcolor}
\usepackage{wrapfig}
\usepackage{xspace}

\usepackage{bm}
\usepackage{soul}
\usepackage{amsmath}
\usepackage{fontawesome}
\usepackage{yfonts}
\usepackage{makecell}
\usepackage{xspace}
\usepackage{url}
\usepackage{verbatim}
\usepackage{upgreek}
\usepackage{amstext}
\usepackage{times}
\usepackage{tabulary}
\usepackage{tabularx}
\usepackage{etoolbox}
\usepackage[version=4]{mhchem}
\usepackage[utf8]{inputenc}

\usepackage{tcolorbox}

\usepackage{rotating}
\usepackage{lineno}
\usepackage{geometry}
\usepackage{float}
\usepackage{caption}

\pdfoutput=1
\usepackage{amsfonts}
\usepackage{color}
\usepackage{hyperref}
\hypersetup{colorlinks=true,linkcolor=blue}
\usepackage{listings}
\usepackage[capitalise]{cleveref}

\definecolor{MH}{rgb}{0.0,0.9,0}

\newcommand\ee{e^+e^-} %Crivelli
\newcommand\aee{A'\to e^+e^-} %Crivelli
\newcommand\xee{X\to e^+e^-} %Crivelli
\newcommand{\gev}{\ensuremath{\text{Ge\!V}}}
\newcommand{\tev}{\ensuremath{\text{Te\!V}}}
\newcommand{\ifb}{\ensuremath{\text{fb}^{-1}}}
\newcommand{\iab}{\ensuremath{\text{ab}^{-1}}}
\begin{document}

%==================
\begin{titlepage}
%==================

\pagenumbering{roman}

% Header ---------------------------------------------------
\vspace*{-2.5cm}
%\centerline{\large EUROPEAN ORGANIZATION FOR NUCLEAR RESEARCH (CERN)}
\vspace*{0.2cm}
\hspace*{-15mm}
\begin{tabular*}{16.5cm}{lc@{\extracolsep{\fill}}r}
  \vspace*{-25mm}
    \includegraphics[width=0.15\linewidth]{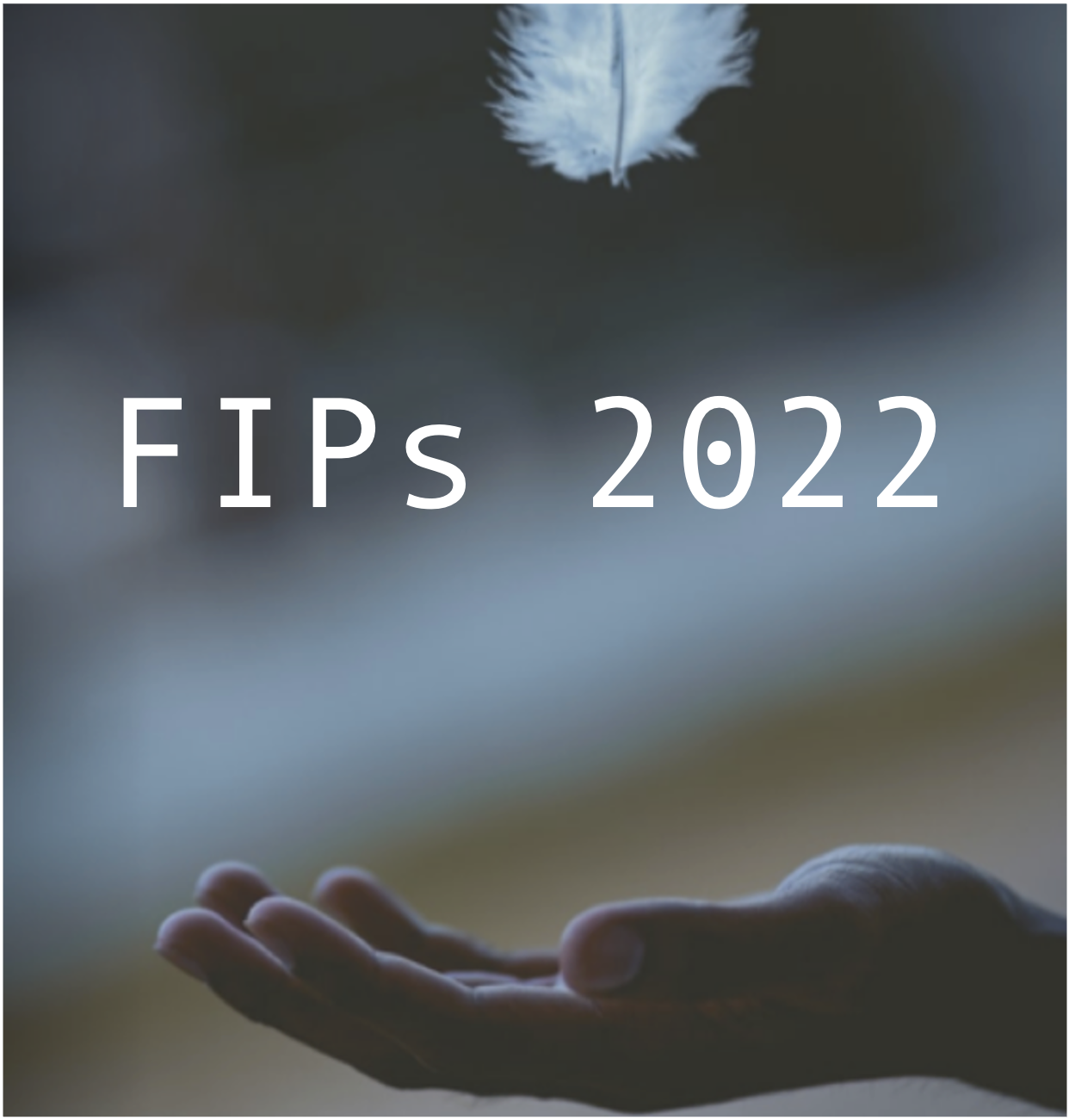} & & \\
%& & CERN-PBC-REPORT-2023-xxx \\
& & {CERN-TH-2023-061} \\
& & DESY-23-050 \\
& & FERMILAB-PUB-23-149-PPD \\
& & INFN-23-14-LNF \\
& & JLAB-PHY-23-3789 \\
& & LA-UR-23-21432 \\
& & MITP-23-015 \\
\hline
\end{tabular*}

\vspace*{1.0cm}

% Title --------------------------------------------------
{\bf\boldmath\LARGE
  \begin{center}
    Feebly-Interacting Particles: \\ 
    FIPs 2022 Workshop Report
\end{center}
}

% Authors -------------------------------------------------
\vspace*{0.5cm}
\begin{center}

\vskip 2mm
C. Antel$^{1}$,
M.~Battaglieri$^{2}$,
J.~Beacham$^{3,a}$,
C.~Boehm$^{4}$,
O.~Buchm\"uller$^{5}$,
F.~Calore$^{6}$,
P.~Carenza$^{7}$,
B.~Chauhan$^{8}$,
P.~Cladè$^{9}$,
P.~Coloma$^{10}$,
P.~Crivelli$^{11}$,
V.~Dandoy$^{12}$,
L.~Darm\'e$^{13}$,
B.~Dey$^{14}$,
F.~F.~Deppisch$^{15}$,
A.~De~Roeck$^{16,a}$,
M.~Drewes$^{17,a}$,
B.~Echenard$^{18,a}$,
V.~V.~Flambaum$^{19}$,
P.~Foldenauer$^{10}$,
C.~Gatti$^{20}$,
M.~Giannotti$^{21,a,*}$,
A.~Golutvin$^{5}$
M.~C.~Gonzalez-Garcia$^{22,23,24}$,
S.~Gori$^{25}$,
E.~Goudzovski$^{26}$,
A.~Granelli$^{27,28}$,
H.~Grote$^{29}$,
S.~Guellati-Khelifa$^{9,30}$,
J.~Hajer$^{31}$,
P.~Harris$^{32}$,
C.~Hearty$^{33}$,
D.~Heuchel$^{34}$,
M.~Hostert$^{35,36,37}$,
S.~Junius$^{38}$,
F.~Kahlhoefer$^{39}$,
J.~Klaric$^{17,a}$,
F.~Kling$^{34}$,
P.~Klose$^{40}$,
J.~Knolle$^{41}$,
J.~Kopp$^{16,42}$,
O.~Kwon$^{43}$,
O.~Lantwin$^{44}$,
G.~Lanfranchi$^{20,a,*}$,
L.~Li$^{45}$,
A.~Lindner$^{34}$,
J.~Lopez-Pavon$^{46,a}$,
J.~Marocco$^{47}$,
J.~W.~Martin$^{48}$,
S.~Middleton$^{18}$,
S.~Milstead$^{49}$,
I.~Oceano$^{34}$,
C.~A.~J.~O'Hare$^{4}$,
A.~Paoloni$^{20}$,
S.~Pascoli$^{27,28,a}$
S.~T.~Petcov$^{50,51,52}$,
M.~Pospelov$^{35,36,37}$,
R.~P\"ottgen$^{53}$,
M.~Raggi$^{54}$,
G.~Ripellino$^{55}$
I.~B.~Samsonov$^{19}$,
S.~Sandner$^{46}$
S.~S\"oldner-Rembold$^{56}$,
J.~Shelton$^{57}$,
N.~Song$^{58}$,
C.~Sun$^{59}$,
Y.~V.~Stadnik$^{4,a,*}$,
J.-L.~Tastet$^{60}$,
N.~Toro$^{61}$,
N.~Tran$^{62}$,
N.~Trevisani$^{63}$,
S.~Ulmer$^{64,65}$,
S.~Urrea$^{46}$,
B.~Velghe$^{66}$,
B.~Wallisch$^{67,68}$,
Y.~Y.~Y.~Wong$^{19}$,
C.~Zorbilmez$^{69}$,
K.~Zurek$^{18}$

\end{center}

\vspace*{0.5cm}
% Abstract -----------------------------------------------
\begin{abstract}
  \noindent
  {\bf Abstract:} 
  Particle physics today faces the challenge of
explaining the mystery of dark matter, the origin of matter over anti-matter in the Universe, the origin of the neutrino masses, the apparent fine-tuning of the electro-weak scale, and many other aspects of fundamental physics. Perhaps the most striking frontier to emerge in the search for answers involves new physics at mass scales comparable to familiar matter, below the GeV-scale, or even radically below, down to sub-eV scales, and with very feeble interaction strength. 
New theoretical ideas to address dark matter and other fundamental questions predict such feebly interacting particles (FIPs) at these scales, and indeed, existing data 
%may even provide 
provide numerous hints for such possibility.  
A vibrant experimental program to discover such
physics is under way, guided by a systematic theoretical approach firmly grounded on the underlying principles of the Standard Model. This document represents the report of the FIPs 2022 workshop, held at CERN between the 17 and 21 October 2022 and aims to give an overview of these efforts, their motivations, and the decadal goals that animate the community involved in the search for FIPs.
\end{abstract}

\vskip 1cm
    {\footnotesize
      \noindent
$^{*}$ Corresponding authors: Maurizio Giannotti {\it (MGiannotti@barry.edu)}, Gaia Lanfranchi {\it (Gaia.Lanfranchi@lnf.infn.it)}, Yevgeny Stadnik {\it (yevgenystadnik@gmail.com)}\\
$^{a}$ Editorial Team \\
}
\end{titlepage}

\clearpage
\pagestyle{empty}  % no page number for the title 
\bigskip
    {\it \footnotesize
      \noindent
      $^{1}$ D\'epartement de Physique Nucl\'eaire et Corpusculaire, Universit\'e de Gen\`eve, Gen\`eve; Switzerland \\
      $^{2}$ Istituto Nazionale di Fisica Nucleare, Sezione di Genova, Genova, 16146, Italy \\
     $^{3}$ Department of Physics, Duke University, Durham NC, USA \\
     $^{4}$ The University of Sydney, Physics, A28 Physics Road, NSW 2006 Camperdown, Australia \\
      $^{5}$ Blackett Laboratory, Imperial College London, Prince Consort Road, London, SW7 2AZ, UK\\
      $^{6}$ Laboratoire d’Annecy-le-Vieux de Physique Théorique (LAPTh), CNRS, F-74000 Annecy, France \\
      $^{7}$ The Oskar Klein Centre, Department of Physics, Stockholm University, Stockholm 106 91, Sweden\\
      $^{8}$ Department of Physics and Astronomy, University of Iowa, Iowa City, IA 52242, USA \\
      $^{9}$ Laboratoire Kastler Brossel, Sorbonne Universit\'e, CNRS, ENS-PSL, Coll\`ege de France, 4 place Jussieu, 75005 Paris, France.\\ 
      $^{10}$ Instituto de F\'isica Te\'orica UAM/CSIC, calle de
        Nicol\'as Cabrera 13--15, Universidad Aut\'onoma de Madrid, Cantoblanco, E-28049 Madrid, Spain.\\
      $^{11}$ ETH Zurich, Institute for Particle Physics and Astrophysics, Auguste-Piccard-Hof 1, 8093 Zurich.\\
      $^{12}$ Institut f\"ur Astroteilchenphysik, Karlsruhe Institute of Technology (KIT), Hermann-von-Helmholtz-Platz 1, 76344 Eggenstein-Leopoldshafen, Germany.\\
      $^{13}$ Institut de Physique des 2 Infinis de Lyon (IP2I), UMR5822, CNRS/IN2P3, F-69622 Villeurbanne Cedex, France.\\
      $^{14}$ Eotvos Lorand University, Budapest, Hungary \\
      $^{15}$ Department of Physics and Astronomy, University College London, Gower Street, London WC1E 6BT, United Kingdom.\\
      $^{16}$ CERN, 1 Esplanade des Particules, 1211 Geneva 23, Switzerland\\ 
      $^{17}$ Centre for Cosmology, Particle Physics and Phenomenology, Université catholique de Louvain, Louvain-la-Neuve B-1348, Belgium\\
      $^{18}$ California Institute of Technology, Pasadena, California 91125, USA\\ 
    $^{19}$ School of Physics, University of New South Wales, Sydney 2052, Australia \\
    $^{20}$ Laboratori Nazionali di Frascati, via Enrico Fermi 54, 00044 Frascati (Rome, Italy). \\
    $^{21}$ Department of Chemistry and Physics, Barry University,  Miami Shores, USA.\\
    $^{22}$ Departament  de  Fisica  Quantica  i  Astrofisica
    and  Institut  de  Ciencies  del  Cosmos,  Universitat de Barcelona, Diagonal 647, E-08028 Barcelona, Spain.\\
    $^{23}$ Instituci\'o Catalana de Recerca i Estudis Avancats, Barcelona, Pg. Lluis  Companys  23,  08010 Barcelona, Spain.\\
    $^{24}$ C.N. Yang Institute for Theoretical Physics, Stony Brook University, Stony Brook NY11794-3849,  USA.\\
    $^{25}$ Department of Physics, University of California Santa Cruz, 1156 High St., Santa Cruz, CA 95064, USA\\
and Santa Cruz Institute for Particle Physics, 1156 High St., Santa Cruz, CA 95064, USA\\
$^{26}$ School of Physics and Astronomy, University of Birmingham, Edgbaston, Birmingham, B15 2TT, United Kingdom \\
$^{27}$ Dipartimento di Fisica e Astronomia, Universit\`a di Bologna, via Irnerio 46, 40126 Bologna, Italy \\
$^{28}$ INFN, Sezione di Bologna, viale Berti Pichat 6/2, 40127 Bologna, Italy\\
$^{29}$ School of Physics and Astronomy,
Cardiff University,
CF243AA,
United Kingdom \\
    $^{30}$ Conservatoire National des Arts et Métiers, Paris, France \\
     $^{31}$ Centro de Física Teorica de Particulas (CFTP), Instituto Superior
Tecnico (IST), Universidade de Lisboa, 1049-001 Lisboa, Portugal \\
 $^{32}$ Laboratory for Nuclear Science, Massachusetts Institute of Technology, Cambridge, MA 02139, USA\\
     $^{33}$ Department of Physics and Astronomy, University of British Columbia, Vancouver, BC, Canada{~~and} Institute of Particle Physics Victoria, BC, Canada \\
     $^{34}$ Deutsches Elektronen-Synchrotron DESY, Notkestr. 85, 22607 Hamburg, Germany \\
     $^{35}$ {Perimeter Institute for Theoretical Physics, Waterloo, ON N2J 2W9, Canada}
$^{36}${School of Physics and Astronomy, University of Minnesota, Minneapolis, MN 55455, USA}
$^{37}${William I. Fine Theoretical Physics Institute, School of Physics and Astronomy, University of
Minnesota, Minneapolis, MN 55455, USA}\\
     $^{38}$ Service de Physique Th\'eorique, Universit\'e Libre de Bruxelles, C.P. 225, B-1050 Brussels, Belgium.\\
     $^{39}$ Institute for Theoretical Particle Physics (TTP), Karlsruhe Institute of Technology (KIT), 76128 Karlsruhe, Germany \\
     $^{40}$ Institut f\"ur theoretische Physik, Universit\"at Bern.\\
     $^{41}$ Ghent University, Sint-Pietersnieuwstraat 33, 9000 Gent, Belgium\\
     $^{42}$ PRISMA Cluster of Excellence and Mainz Institute for Theoretical Physics,
Johannes Gutenberg University, Staudingerweg 7, 55099 Mainz, Germany\\
$^{43}$  Center for Axion and Precision Physics Research, Institute for Basic Science, Daejeon 34051, Republic of Korea \\
    $^{44}$ Affiliated with an institute covered by a cooperation agreement with CERN \\
     $^{45}$ Department of Physics, Brown University, Providence, RI 02912, USA \\
    $^{46}$ Instituto de F\'isica Corpuscular, Universidad de Valencia and CSIC, Edificio Institutos de Investigaci\'on, Calle Catedr\'atico Jos\'e Beltr\'an 2, 46980 Paterna, Spain.\\
    $^{47}$ Physics Division, Lawrence Berkeley National Laboratory, Berkeley, CA, USA \\  
    $^{48}$ Department of Physics The University of Winnipeg
Winnipeg, MB R3B 2E9 CANADA\\
    $^{49}$ Fysikum, Stockholms Universitet, 10691 Stockholm, Sweden \\
    $^{50}$ SISSA, via Bonomea 265, 34136 Trieste, Italy\\
    $^{51}$ INFN, Sezione di Trieste, via Valerio 2, 34127 Trieste, Italy\\
    $^{52}$ Kavli IPMU (WPI), UTIAS, The University of Tokyo, Kashiwa, Chiba 277-8583, Japan\\
     $^{53}$ Dep. of Physics, Lund University, Professorsgatan 1, 22363 Lund, Sweden \\
    $^{54}$ Universit\'a La Sapienza e INFN, sezione di Roma 1 \\
  $^{55}$ Uppsala University, Uppsala, Sweden \\
    $^{56}$ The University of Manchester, Manchester M13 9PL, United Kingdom\\
    $^{57}$ Illinois Center for Advanced Studies of the Universe and Department of Physics,
    University of Illinois Urbana-Champaign, Urbana, IL 61801 \\
    $^{58}$ Department of Mathematical Sciences, University of Liverpool, Liverpool, L69 7ZL, United Kingdom\\
    $^{59}$ Theoretical Division, Los Alamos National Laboratory, Los Alamos, NM 87545, U.S.A. \\
    $^{60}$ Departamento de F\'isica Te\'orica and Instituto de F\'isica Te\'orica UAM/CSIC,
        Universidad Aut\'onoma de Madrid, Cantoblanco, 28049, Madrid, Spain \\
    $^{61}$ SLAC National Accelerator Laboratory,
2575 Sand Hill Road, Menlo Park, CA 94025, USA. \\    
    $^{62}$ {Fermi National Accelerator Laboratory, Batavia, IL 60563, USA}\\
    $^{63}$ Karlsruhe Institute of Technology (KIT), 76128 Karlsruhe, Germany \\
    $^{64}$ Institut f\"ur Experimentalphysik, Heinrich Heine Universit\"at, 40225 D\"usseldorf, Germany \\ 
    $^{65}$ RIKEN, Fundamental Symmetries Laboratory, Wako, Saitama, Japan\\
    $^{66}$ TRIUMF, 4004 Wesbrook Mall, Vancouver V6T 2A3 Canada\\
    $^{67}$ Oskar Klein Centre, Department of Physics, Stockholm University, 10691~Stockholm, Sweden \\
    $^{68}$ Texas Center for Cosmology and Astroparticle Physics, Weinberg Institute for Theoretical Physics, Department of Physics, University of Texas, Austin, TX~78712, USA \\
    $^{69}$ Cukurova University, Physics Department, Science and Art Faculty, Adana, Turkey \\     
}

%\title{FIPs2020 Workshop Report}
%\author{The authors}
%\date{December 2020}

%\maketitle

\clearpage
\setcounter{tocdepth}{2}
\pagestyle{empty}  % no page number for the title 
\tableofcontents

\cleardoublepage
\pagestyle{plain} % restore page numbers for the main text
\setcounter{page}{1}
\pagenumbering{arabic}

%\linenumbers

\clearpage
%----------------------------------------
\noindent{\large \bf Executive Summary}
%---------------------------------------
%This should contain an overview of the content of the Report to guide the reader through the document and indicate the reasoning behind.

\vskip 2mm

\noindent
The origin of neutrino masses and oscillations, the nature of dark matter and dark energy, the mechanism generating the matter-antimatter asymmetry in our universe or the origin of the hierarchy of scales are some of the deepest mysteries in modern particle physics. 
So far the vast majority of experimental efforts to answer these questions have been driven by theoretical arguments favoring searches for new particles with sizeable couplings to the Standard Model (SM) and masses commensurate with the Higgs boson. Exploring this paradigm is one of the main goals of the Large Hadron Collider (LHC) at CERN. 

\vskip 2mm
An alternative framework to explain new phenomena introduces new light particles interacting only feebly with Standard Model fields. Feebly-interacting particles (FIPs) are currently one of the most debated and discussed topics in fundamental physics, recognized by the recent European Strategy for Particle Physics as one of the compelling fields to explore in the next decade. The workshop FIPs 2022, held at CERN between the 17th and 21st of October 2022, has played a central role in this worldwide discussion.  This event was the second edition of a series of workshops fully dedicated to the physics of feebly-interacting particles, gathering experts from particle physics, nuclear physics, astroparticle physics and cosmology. About 320 participants from many countries, including representatives from all major particle physics laboratories, participated in the discussions. 

\vskip 2mm
The workshop was organized around three main themes, that represent the backbone of the present Report: i) ultra-light FIPs and their connection with cosmology and astrophysics; ii) light Dark Matter in particle, astro-particle and cosmology;  iii) Heavy Neutral Leptons and their connection with active neutrino physics and neutrino-less double beta decays experiments. In addition, dedicated sessions on {\it New Ideas} featured 20 talks given by brilliant young researchers, offering them an opportunity to discuss their work in front of renowned researchers while attending a series of presentation on a wide range of topics.

\vskip 2mm
The objective of FIPs 2022 was to set a common vision for a multi-disciplinary and interconnected approach to fundamental questions in modern particle physics.  Their breadth and deep interconnection requires more than ever a diversified research portfolio with a large set of experimental approaches and techniques, together with strong and focused theoretical involvement. No single experiment or laboratory can  cover the large parameter space in terms of masses and couplings that FIPs models can suggest, calling for a broad, diverse, but interconnected network of experimental projects. 

\vskip 2mm
FIPs 2022 %strongly linked with the current effort pursued at CERN within the Physics Beyond Colliders initiative, 
aims to shape the scientific programme in Europe for the physics of feebly-interacting particles. Such a programme would extend far beyond the traditional searches performed at colliders and extracted beam lines, including elements from star evolution, cosmological models, indirect dark matter detection, neutrino physics, gravitational waves physics and AMO (atomic-molecular-optical) physics. In that context, synergies between experimental facilities are paramount, calling for deeper collaboration across main laboratories in the world. 

\vskip 2mm
In a geo-political situation with unprecedented economical, political and climate-oriented challenges, we believe that a network of interconnected laboratories can become a sustainable, flexible, and efficient way of addressing the particle physics questions in this century.

\clearpage
%%=====================
%% INTRODUCTORY TALKS
%%=====================

%%============================
\section{Introductory talks}
\label{sec:introduction}
%%============================

%%-----------------------------------------
\subsection{The FIP/Hidden Sector/Valley paradigm: opportunities and challenges -- {\it K.~Zurek}}
\label{ssec:zurek}
%%-----------------------------------------
{\it Author: Kathryn Zurek, <kzurek@caltech.edu>}  \\
%\subsection{Introduction}

\begin{figure}[hbt]
\centering
\includegraphics[scale=0.4]{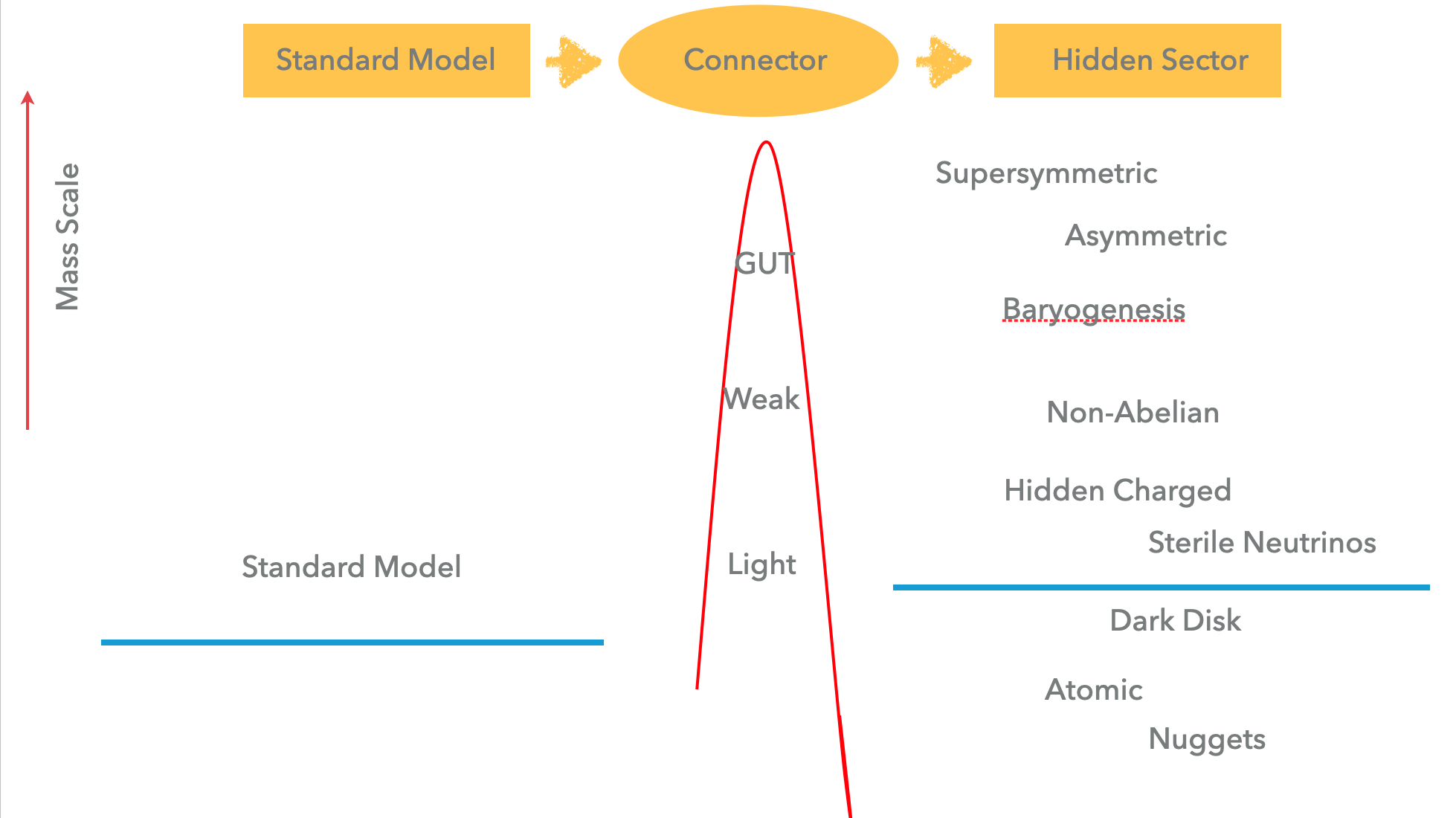}
\caption{A schematic of low mass Hidden Valleys / Hidden Sectors / FIPs.  The hidden sector is separated from the visible sector by a barrier, which corresponds to feeble interactions mediated by either a heavy state (labeled here by the GUT scale), an intermediate scale like the weak scale, or some light state.  Within the hidden sector(s), there are many types of states which can comprise the matter and force content, and we have named a few (far from complete) examples.  Hidden Valleys are motivated by many UV-complete theories, such as string theory and Grand Unification, both of which tend to give rise to a wide variety of states at low energy. }
\label{ref:fig1}
\end{figure}

Traditionally, high energy physics has focused on chasing particles at colliders at ever higher energies.  However, light states may reside in a hidden sector, at mass scales well below the weak scale, and remain hidden from the visible sector on account of a high barrier, corresponding to sub-weak interactions between the sectors.  This situation, known as the Hidden Valley (HV) / Sector / Feebly Interacting Particle (FIP) framework encompasses a broad class of theories, and is shown schematically shown in Fig.~\ref{ref:fig1}.  A proposal to systematically search for light hidden sectors, from a diverse class of hidden sectors, at colliders was made in Ref.~\cite{Strassler:2006im}, with the class of low-mass Hidden Sectors called Hidden Valleys.  A field of research around hidden sector theories and their phenomenological signatures has grown up, which we attempt to overview here~\cite{Alexander:2016aln,Battaglieri:2017aum,Beacham:2019nyx,Agrawal:2021dbo}. 

The examination of HVs or FIP sectors has gained momentum along with a shift away from a single-minded focus on states at ever higher energies.  What has emerged is a complex tapestry of new physics phenomena that occur at lower energies but are still very weakly coupled to the Standard Model (SM).  Further, as emphasized in Ref.~\cite{Strassler:2006im}, hidden sectors appear generically in many UV complete theories, such as string theory.  Such a Hidden Sector need not solve the SM's problems, but it does naturally provide a dark matter candidate.  One only requires some symmetry to stabilize at least one of the states in the hidden sector.

The structure of the hidden sector can take many forms.  It can be strongly or weakly coupled, with dark photons or Higgses, a dark $SU(N)$ with $N_f$ flavors.  It can have sterile neutrinos or asymmetric dark matter, and have a wide range of dynamics associated with it.  The highly malleable nature of low mass hidden sectors, which couple only weakly to the SM, has led, in the last 10-15 years, to a wide range of model building efforts, from light hidden supersymmetric sectors, composite dark matter (from asymmetric and flavored dark matter  in many realizations such as nugget and atomic dark matter), to hidden valleys that do the work of baryogensis.  The list of theories would be too broad to give justice to here, so we refer the reader to community reports~\cite{Alexander:2016aln,Battaglieri:2017aum,Beacham:2019nyx} (and references therein) to grasp the scope of the efforts.  There has been accordingly a renewal in the search for mechanisms to set the relic abundance from Asymmetric Dark Matter, freeze-in, $3 \rightarrow 2$ mechanisms, dark Affleck-Dine, and WIMPless miracles, to name a few.

\begin{figure}[hbt]
\centering
\includegraphics[scale=0.45]{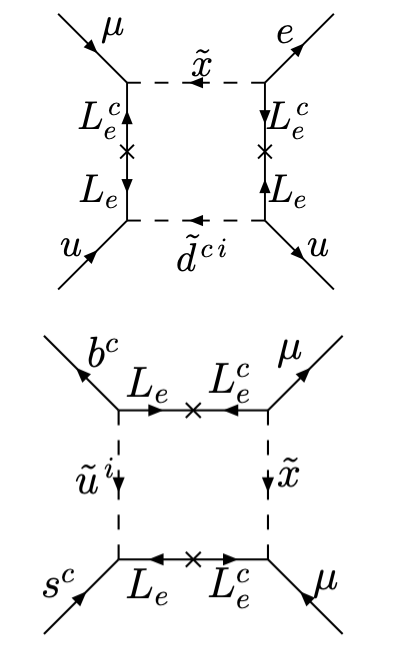}
\includegraphics[scale=0.45]{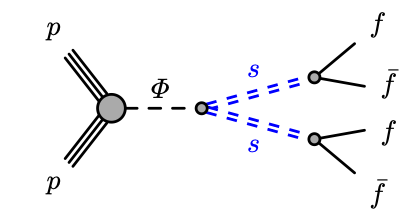}
\includegraphics[scale=0.45]{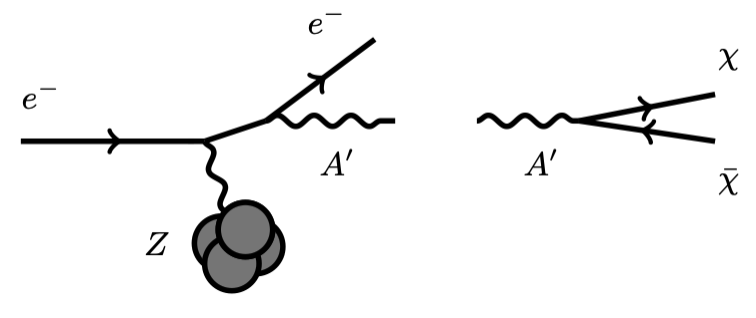}
\caption{Examples of three types of signatures that occur via the three types of mediators shown in Fig.~\ref{ref:fig1}.  In the left figure, we show flavor-violating interactions mediated by a heavy state, as discussed in Ref.~\cite{Kim:2013ivd}.  In the center figure, we show displaced decays to a light hidden sector through a weak scale state.  And in the right figure, intensity production of a light vector state which subsequently decays to dark matter \cite{Izaguirre:2013uxa}. }
\label{ref:signatures}
\end{figure}

While the model building activity has been inspiring, the challenge is to detect such sectors through their interactions with the SM, extending the symmetry structure of the SM.  The types of FIP signatures depend on the nature of the mediator, as shown schematically in Fig.~\ref{ref:signatures}.  Weak scale mediators can be produced resonantly at high energy colliders.  The mediators then decay to a high multiplicity of states in the hidden sector.  Those light states may then decay back to the SM through the heavy mediators, in some cases with long lifetimes.  This is the classic Hidden Valley set-up.  If the mediator is much lighter than the weak scale, low energy intensity experiments can produce the FIP.  A broad range of proposals have been made in the last decade, from the Heavy Photon Search, DarkLight and APEX (see Ref.~\cite{Beacham:2019nyx} for details).  Finally, heavy mediators between the hidden and visible sectors can mediate rare decays or flavor violating processes, as frequently occurs in models of Asymmetric or Flavored DM \cite{Kim:2013ivd}.  In short, the Hidden Valley / Sector / FIP revolution has dramatically reshaped the theoretical and experimental particle physics landscape, and the corresponding search for Physics Beyond the Standard Model (BSM).

Traditionally, the search for physics BSM has focused on new states at the weak scale, as solutions to the hierarchy problem.  If we relax the motivation to search for new physics only as solutions to the SM's puzzles, where do we focus on attention? One powerful motivation is dark matter, and its relic density, being careful to satisfy cosmological constraints on dark sectors.  The old paradigm is weak scale dark matter, with the relic abundance fixed by freeze-out.  Fixing the dark matter abundance to be the observed abundance natural places the freeze-out cross-section in the electroweak range.  This paradigm is highly testable, either through annihilation byproducts with the predicted cross-section, $\langle \sigma v \rangle \simeq 3 \times 10^{-26} \mbox{ cm}^3/\mbox{s}$, or through Higgs-interacting dark matter in direct detection experiments.

\begin{figure}[hbt]
\centering
\includegraphics[scale=0.4]{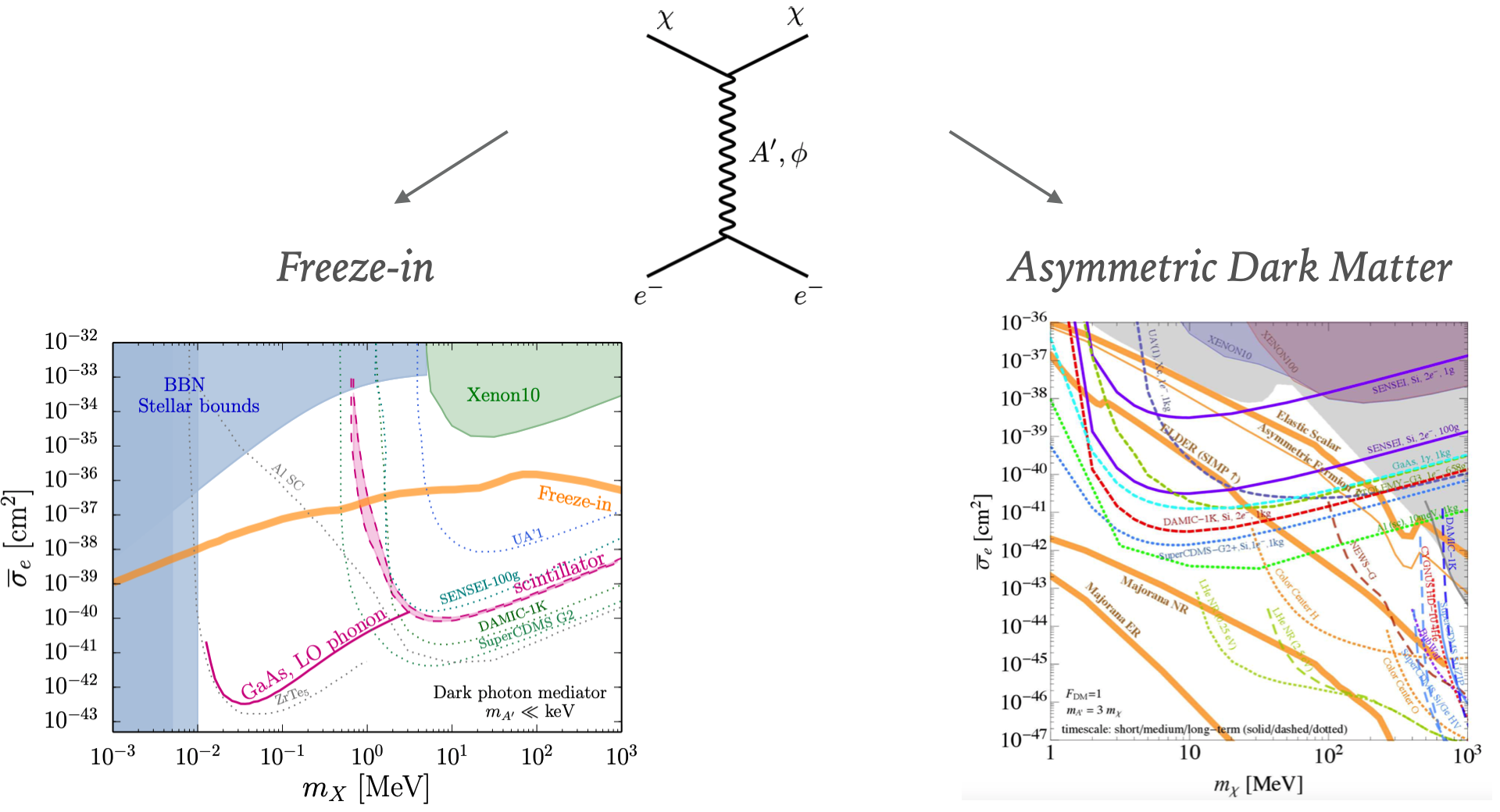}
\caption{Dark matter relic density can be a guide for where to look for light dark matter.  The same process that sets the relic abundance (or relic asymmetry) can also mediate a scattering process in a direct detection experiment, shown here as dark matter production through $e^+e^-$ annihilation (freeze-in) or Asymmetric freeze-out.  The left and right figures (taken from Refs.~\cite{Knapen:2017ekk,Battaglieri:2017aum} show the predicted interaction cross-section as a shaded band.}
\label{fig:relic}
\end{figure}

However, even in a hidden sector, the relic abundance may still be set by interactions with the SM.  These same interactions give rise to predictive signatures in terrestrial experiments, via crossing symmetry.  For example, rare $e^+e^-$ annihilations to produce DM through freeze-in predict a scattering cross-section in direct detection.  In addition, Asymmetric Dark Matter has a {\em minimum} annihilation cross-section to be consistent with observations in the Cosmic Microwave Background, which also gives rise to a predictive direct detection cross-section.  These points are demonstrated in Fig.~\ref{fig:relic}.  More generally, light mediators yield large scattering cross-sections in direct detection experiments \cite{Knapen:2017xzo,Lin:2011gj,Battaglieri:2017aum}, and give rise to the possibility of direct DM production in intensity experiments, such as SHIP, FASER, Codex and MATHUSLA (see Ref.~\cite{Beacham:2019nyx} for a review).  In some cases, the relic abundance requirement also gives rise to predictive production rates in intensity experiments such as LDMX, as emphasized in Ref.~\cite{Izaguirre:2015yja}.

As there has been an effort to build and modify low intensity experiments to search for FIPs, there has also been developed in the last decade a suite of experiments to search for sub-GEV light dark matter in direct detection experiments.  These efforts were summarized in the community document Ref.~\cite{Beacham:2019nyx}.  The standard nuclear recoil experiments, which are kinematically well-matched to weak scale dark matter, do not operate well on sub-nuclear mass scales.  Instead, research has turned to the development of electronic excitations, and collective modes such as phonons and magnons, which have smaller gaps (eV for electronic excitation and meV for collective excitation) opening the way for detection of light dark matter that deposits less energy.

After this discussion, one has the impression that there are a broad range of signatures in a broad range of models.  Some may give up hope that any one of them will lead us to dark matter.  While we are offered no guarantees let me close by offering an example of a concrete model of dark matter which gives rise to a broad range of signals from a single, UV complete model.  We have argued that such a broad range of signals are the rule, rather than the exception, in Hidden Valley models.  To be concrete, we consider Asymmetric Dark Matter as proposed in Ref.~\cite{Kaplan:2009ag}.  Asymmetric Dark Matter requires interactions that, from the SM perspective, violate baryon or lepton number (but conserve these global symmetries when the hidden dark matter sector is included).  These operators are naturally higher dimension and of the form
\begin{equation}
{\cal L} = \frac{{\cal O}_{B-L} {\cal O}_X}{M^{d-4}}, 
\end{equation}
where $M$ is the mass scale of the baryon- or lepton-number violating interaction, which will generally be above the weak scale.  Such operators naturally induce flavor-violating interactions, as discussed at length in Ref.~\cite{Kim:2013ivd}.

The ADM interaction generates a particle-anti-particle asymmetry, which must then be annihilated to allow the small asymmetry to shine through.  In the SM this occurs naturally through the presence of forces, {\em e.g.} $e^+ e^- \rightarrow \gamma \gamma$.  This most naturally suggests that ADM exists in a hidden sector with dark forces that mediate such interactions to leave us with only the asymmetric component.  Both the light and heavy forces that mediate interactions in such an ADM model give rise to observable signatures that we have discussed, such a flavor violation and displaced decays at colliders, as well as signatures at low energy colliders through kinetic mixing of the dark photon.

\begin{figure}[htb]
\centering
\includegraphics[scale=0.4]{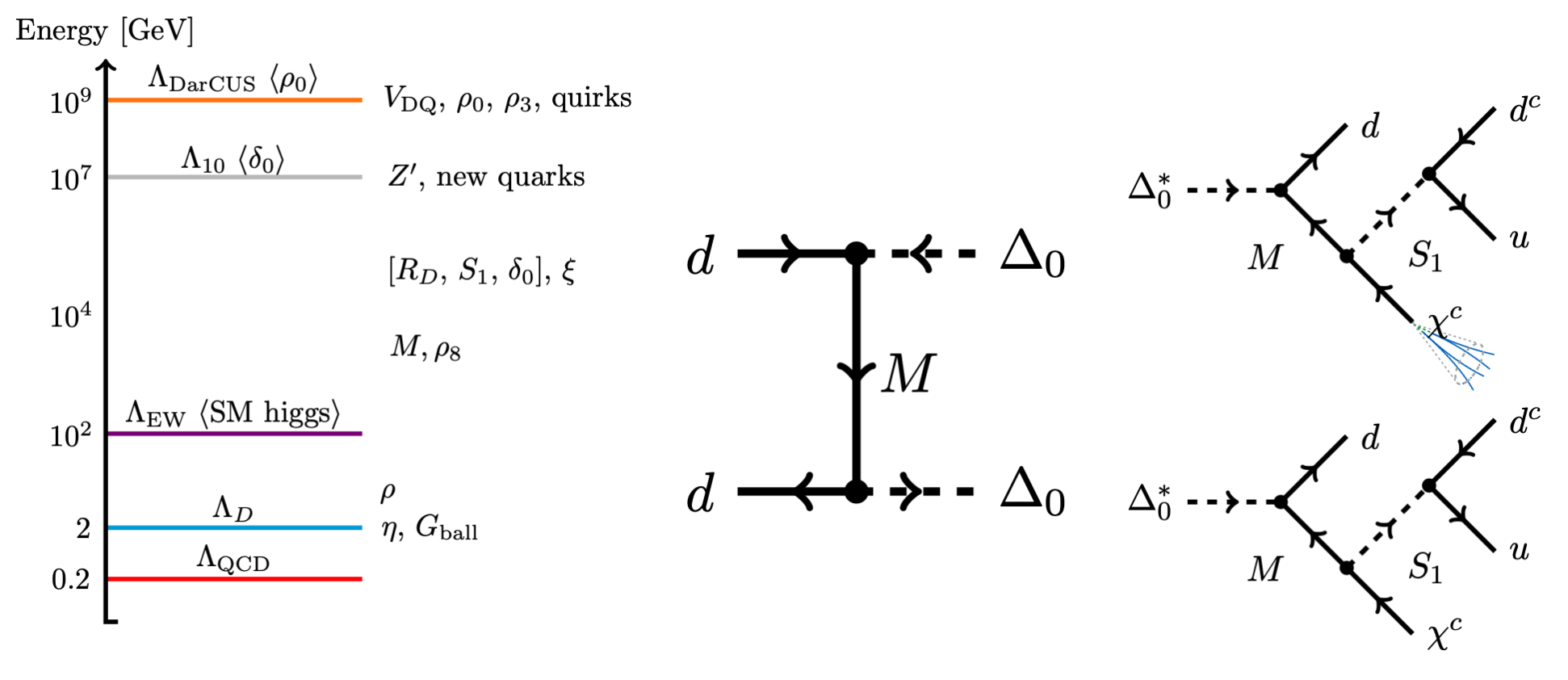}
\caption{A broad array of states, and observational signatures, is predicted in a UV-complete theory of dark matter.  Here we show schematically a case developed in Ref.~\cite{Murgui:2021eqf}, where the Unified structure of the theory gives rise to many states (shown in the left side of the figure), as well as a broad range of signatures at colliders (shown in the right panel.  This includes the displaced decay of connector particles to a high multiplicity of final states, including the possibility of dark showers and semi-visible jets.}
\label{fig:DarkUnified}
\end{figure}

While the considerations just discussed appear very generically from a bottom-up, they appear in a UV complete theory of ADM.  To be concrete, we utilize the model proposed recently in Ref.~\cite{Murgui:2021eqf}, where the mass scale of ADM was generated by unification of the dark gauge group with SM SU(3), where the confinement of both dark and visible sectors is triggered by a common origin.  A schematic is shown in Fig.~\ref{fig:DarkUnified}.  Hidden Valleys were motivated from the top-down ({\em e.g.} in string constructions or grand unification), but engineered from the bottom-up.  In Dark Unification, we generate the whole panoply of signatures at a collider, from Connector particles that decay promptly to missing energy plus jets, to pairs of displaced vertices, as well as semi-visible jets.

To summarize, there is no {\em single} way to search for signatures of hidden sectors.  In general, UV complete models feature multiple signatures, and will require multiple different experiments working together to identify the nature of the dark sector.  This include light dark matter detection, prompt and displaced decays of light states at a collider, searches for light states at intensity machines, flavor, and even gravitational wave signatures from a hidden sector.  Fortunately, there is now a community of people dedicated to probing light dark sectors from many facet.

\clearpage
%===============
% Ultralight FIPs
%===============

%==========================================
\section{Ultra-light FIPs: theory and experiments}
\label{sec:ultralight} 
%==========================================

%-------------------------------------------
\subsection{Introduction}
\label{ssec:ULF_intro}
%-------------------------------------------

The term \emph{ultralight FIP} is normally used to indicate the lowest mass range of feebly interacting particles. 
There is not a clear and crisp definition of the ultralight mass region, but it is common to adopt the $\sim$keV threshold (see, e.g., Sec.~\ref{ssec:pospelov}).
Even lighter FIPs, in the sub-eV mass region, are normally identified as WISPs, which stands for weakly interacting sub-eV (or slim) particles (see, e.g., Ref.~\cite{Jaeckel:2010ni}). 
WISPs may play an important role in cosmology, as discussed in the contribution in Sec.~\ref{ssec:wallish} below.

Ultralight FIPs have been studied extensively over the past several decades. 
Arguably, the most well known and studied example is the QCD axion. 
Originally proposed as a solution of the strong CP problem~\cite{Peccei:1977hh,Peccei:1977ur,Weinberg:1977ma,Wilczek:1977pj}, axions were soon recognized as one of the primary dark matter candidates~\cite{Abbott:1982af,Dine:1982ah,Preskill:1982cy} (see Refs.~\cite{Marsh:2015xka,Adams:2022pbo} for recent accounts).
Furthermore, as discussed below (see, in particular, Sec.~\ref{ssec:carenza}), they may play significant roles in astrophysics. 
Besides QCD axions, the ultralight FIP panorama includes other well-studied candidates such as axion-like particles (ALPs), dark photons and light scalars. 
On purely theoretical grounds, it is possible to construct several models of ultralight FIPs by considering different sets of low-dimension operators. 
The discussion of appropriate criteria, such as \emph{simplicity} or \emph{technical naturalness},  to classify the different models is among the aims of the FIP Physics Center (see Sec.~\ref{ssec:pospelov}).

As discussed in the following contributions, the phenomenological motivations for ultralight FIPs are numerous, and span from particle physics to astrophysics and cosmology. 
A broad range of probes is available to study FIPs, including laboratory, astrophysical and cosmological probes. 
The types of signatures can be further categorised depending on whether the ultralight FIPs contribute to the dark matter, dark radiation or dark energy (i.e., are of a cosmological nature), are produced in stars (i.e., are of an astrophysical nature), or are produced in the laboratory (e.g., via FIP-photon interconversion or as virtual mediators of new forces). 
Searches for FIPs of these different natures generally involve different sets of assumptions, providing complementarity between these different types of searches. 

\begin{figure*}[ht]
\centering
\includegraphics[width=10cm]{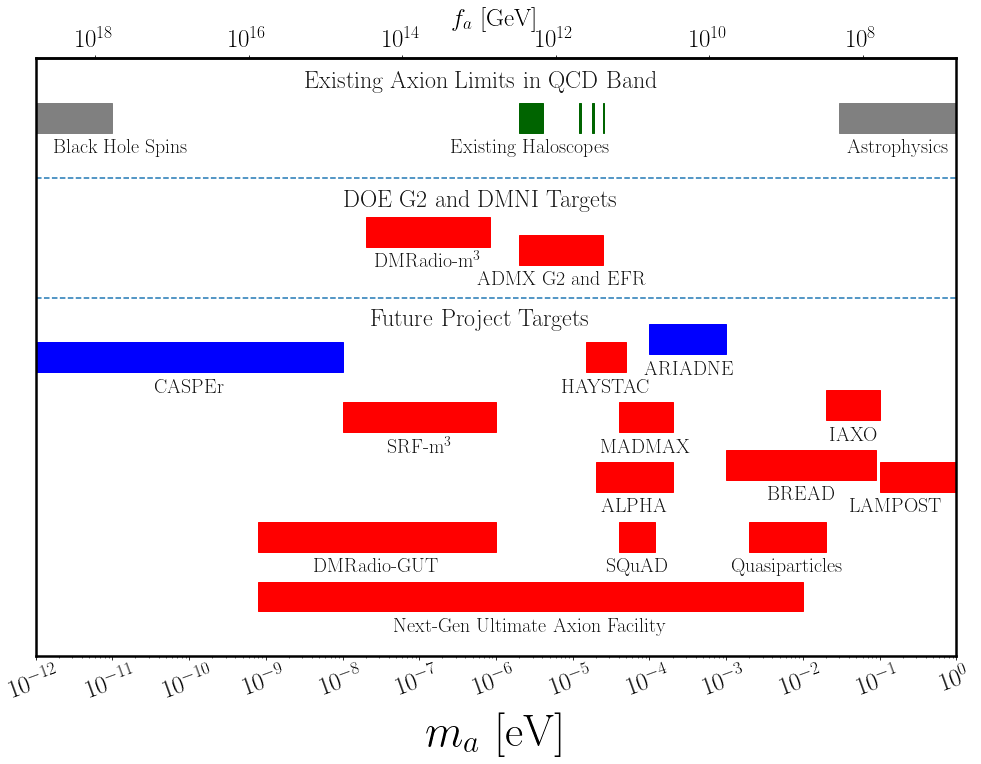}
\caption{ %\normalsize 
Landscape overview of experiments available to probe ultralight axion dark matter, in the context of canonical QCD axion models. 
Green regions denote mass ranges already probed with laboratory dark matter searches for the axion-photon coupling. 
Red and blue regions denote mass ranges to be probed with ongoing and future laboratory dark matter searches for the axion-photon and axion-nucleon couplings, respectively. 
Grey regions indicate astrophysical bounds on the axion-photon coupling. 
Figure from Ref.~\cite{Adams:2022pbo}. 
}
\label{Fig:ULF_Landscapes_Axion}
\end{figure*}

\begin{figure*}[ht]
\centering
\includegraphics[width=10cm]{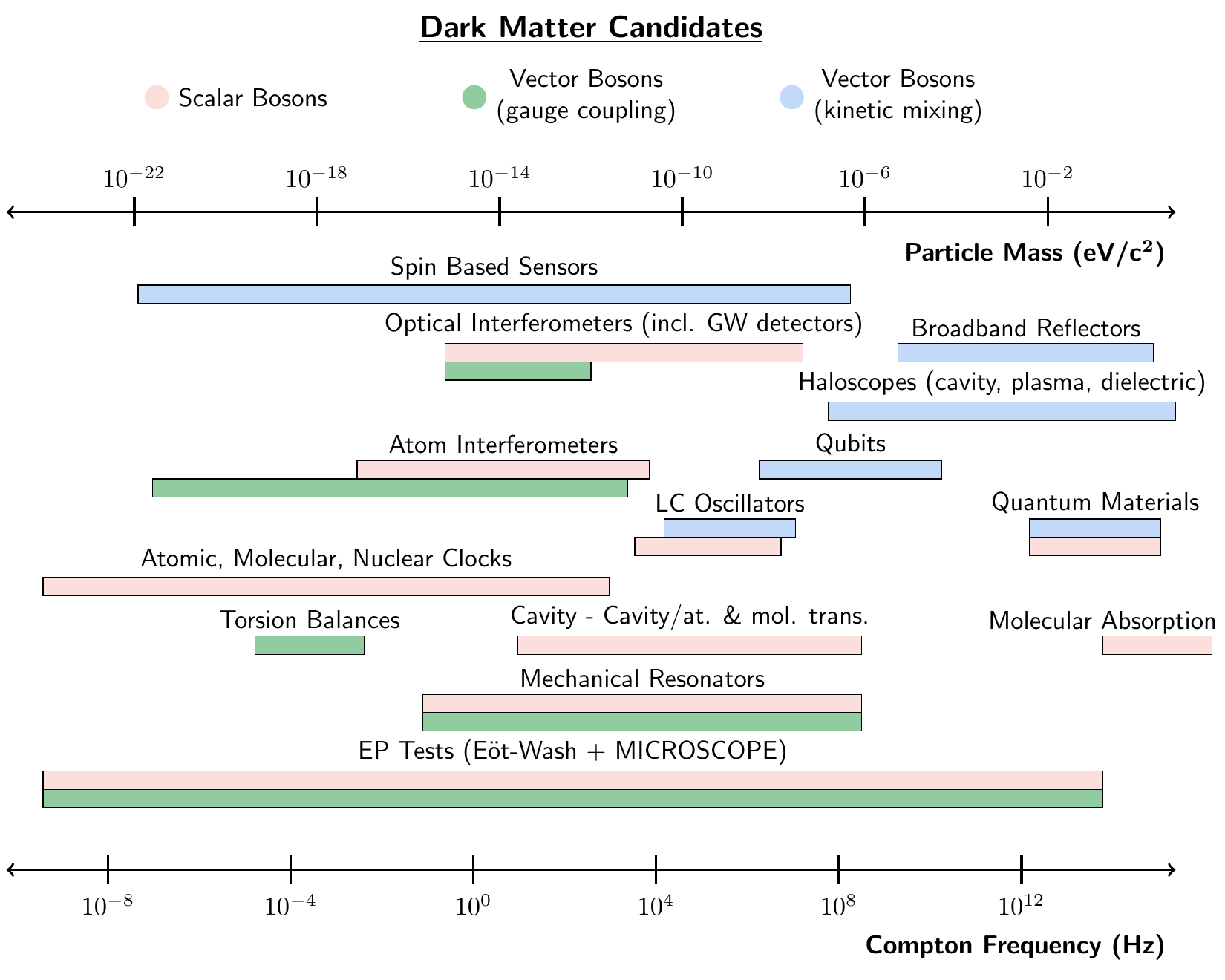}
\caption{ %\normalsize 
Landscape overview of experiments available to probe ultralight scalar and vector dark matter. 
Pink regions denote searches for scalar boson dark matter, green regions denote searches for vector boson dark matter via gauge couplings, and blue regions denote searches for vector boson dark matter via the kinetic mixing portal. 
Figure from Ref.~\cite{Antypas:2022asj}. 
}
\label{Fig:ULF_Landscapes_Scalar+Vector}
\end{figure*}

A general feature of ultralight FIP models is the enormous potential of the astrophysical and cosmological signatures they can yield. 
Their feeble couplings can allow them to escape dense environments (such as stellar cores), providing a useful telescope into regions that are inaccessible to photons (see Sec.~\ref{ssec:carenza}). 
Furthermore, being very light and feebly interacting, such FIPs are often stable (or have extremely long lifetimes), and are thus good dark matter candidates. 
Several non-thermal production mechanisms allow an efficient cosmological production of a non-relativistic population even in the case of extremely light (sub-eV) FIPs. 
%Several cosmological signatures of FIPs are presented in Sec.~\ref{ssec:wallish}. 
In this sense, the sub-eV region is particularly interesting and well-motivated. 
If extremely light particles are to account for the totality (or a large fraction) of the dark matter in the universe, they must have huge occupation numbers and thus they would exhibit wavelike behavior. 
This observation turns out to have rich consequences. 
In particular, the detection of wavelike dark matter requires specific techniques which have advanced impressively in the past decade, in particular precision measurements and probes (see Figs.~\ref{Fig:ULF_Landscapes_Axion} and \ref{Fig:ULF_Landscapes_Scalar+Vector} for landscape overviews). 
An increasing number of these probes adopt the use of quantum techniques and technologies. 
Numerous contributions in these proceedings present the progress in specific detection techniques as well as near-future expectations. 
In particular, we report on the status of searches in some of the major laboratories in the world (see Secs.~\ref{ssec:lindner}, \ref{ssec:gatti} and \ref{ssec:Kwon}).
%

%%%%

%%%%

The relevance of experimental studies of FIPs in the ultralight region and the progress in specific detection techniques have been long recognized by the international community and have been discussed in recent strategy reports.
The US Report on Basic Research Needs for Dark-Matter Small Projects New Initiatives~\cite{osti_1659757} acknowledged that 
the ``Discovery of dark matter waves would provide a glimpse into the earliest moments in the origin of the universe and the laws of nature at ultrahigh energies, beyond what can be probed in colliders”. 
Several strategic directions have been presented in numerous contributions to the Snowmass 2021 reports (see Figs.~\ref{Fig:ULF_Landscapes_Axion} and \ref{Fig:ULF_Landscapes_Scalar+Vector})~\cite{Antypas:2022asj,Adams:2022pbo,Jaeckel:2022kwg,Chou:2022luk,Dvorkin:2022jyg}.

In this section, we provide an overview of the main theoretical and experimental results and, in particular, provide an update on the most recent searches and results, as well as future perspectives. 
%We open with a presentation, in Sec.~\ref{ssec:pospelov}, of the model building panorama effective operators associated with ultralight FIPs.
In Sec.~\ref{ssec:pospelov}, we open with an account of the theoretical panorama, presenting some general criteria to guide the theoretical developments of ultralight FIP models. 
This defines the FIP Physics Centre approach to classifying and studying ultralight FIPs. 
Thereafter follow a survey of cosmological (Sec.~\ref{ssec:wallish}) and astrophysical (Sec.~\ref{ssec:carenza}) probes. 
We note that these probes are not necessarily limited to the sub-eV mass range (although this is the predominant mass region) --- 
for some astrophysical probes, the same methods adopted for the sub-eV mass range also apply for FIP masses up to several keV or even (in the case of supernovae) a few 100 MeV. 
The following three contributions present the status and perspectives from DESY (Sec.~\ref{ssec:lindner}), the INFN laboratories of Frascati (LNF) and Legnaro (Sec.~\ref{ssec:gatti}), and IBS-CAPP (Sec.~\ref{ssec:Kwon}). 
Following these general reports, Secs.~\ref{ssec:martin} and \ref{ssec:flambaum} present updates on EDM searches, Sec.~\ref{ssec:pierre_clade} discusses precision measurements of the fine-structure constant, while Sec.~\ref{ssec:stefan_ulmer} discusses tests of the fundamental symmetries with matter-antimatter conjugate systems. 
Sections \ref{ssec:buchmuller} and \ref{ssec:grote} discuss the use of gravitational-wave detectors to search for FIPs. 
Sections \ref{Song} to \ref{Li} present some new theoretical and experimental ideas in the realm of ultralight FIPs. 
Finally, in Sec.~\ref{ssec:ULF_conclusions}, we summarise and discuss the outlook of the field.

%-------------------------------------------
\subsection{Ultra-light FIPs: the FIP Physics Centre approach - {\it M.~Pospelov}}
\label{ssec:pospelov}

{\it Author: Maxim Pospelov, <pospelov@umn.edu> -- Joint Session with PSI 2022 }
%-------------------------------------------

%{\bf Abstract: } 
Benchmark models formulated by the PBC have played a catalyzing role for the experimental and theoretical studies of MeV-to-GeV scale FIPs. At the moment, there is some interest in expanding this activity towards ultra-light field models, in hope of a similar effect on the field. 
Here we present a possible approach to such a classification, and formulate some models that can serve as benchmarks.

\subsubsection{Principles behind the classification} 

Ultra-light field (ULF) models is a particular example of FIPs where the masses of particles can be below keV, eV, 
and sometimes much below. For example quintessence field models have mass parameters in the potential as small as the Hubble constant today $\sim 10^{-33}$ eV. 
In recent years, there has been an intensification of precision measurements/probes of this type of new physics. Many models exist on the market, but a comprehensive list of them has not been created yet.

In devising the models, one could adhere to the following general philosophy: 

\begin{enumerate}

\item {\em Relative simplicity} is still a criterion, as we want a manageable parameter space, and there is no "clear winner" among any of the complicated models. 

\item  {\em Renormalizability is not a criterion} like it was for the majority of the PBC benchmark models before. (The reason is that these models are often in the same class as gravity that by itself is not renormalizable.) That does not prevent us to treat all models within the effective field theory (EFT) framework. 

\item {\em Technical naturalness}, or stability of the chosen model and all its parameters against radiative corrections without the necessity of extreme fine tuning or unnaturally small cutoff,  is the hardest criterion to address. In the past, particle physics literature would tend to adhere to technical naturalness verbatim, and ignore models that tend to contradict this principle. On the other hand, some other communities (AMO, and some cosmologists) tended to ignore this entirely, arguing that more creative UV completion may avoid the problem (that sometimes proves to be correct). 

Here we will give mild priority to models with explicit technical naturalness, but would not entirely discard unnatural models if they give promising phenomenology. 

\end{enumerate}

Lastly, we also want some relevance of formulated models for the on-going experimental program of precision 
physics for BSM. If there is no connection, there is little impetus for FCP/PBC to ``maintain" such models. This is a practical consideration, rather than a criterion.

\subsubsection{Possible benchmark cases} 

The set of models we are going to consider is going to be mostly bosonic, not least because bosons can have very large occupation numbers, and this can partially offset the smallness of their coupling. More importantly, light bosons can saturate dark matter. Many models would use $X \sim \sqrt{2\rho_{\rm DM}/m_X^2}$ ansatz for the amplitude of field $X$ when it saturates the local dark matter energy density $\rho_{\rm DM}$. 

Below I give an account of some reasonable/popular models of ULFs. In the ``A list" I will include models that to a large extent conform the idea of technical naturalness. 

{\bf A category}. Technically natural. 

{\bf M0.} ``Model 0" is a non-interacting ULF model with some scalar field $\phi$ and potential $V(\phi)$ 
that gives it a slow roll and modifies equation of state for dark energy, $w$ \cite{Peebles:2002gy,Linder:2002et}. These models have been extensively studied in cosmology, but unfortunately they have no other signatures. (These models can be technically natural if there is no couplings to SM, and because it is not clear what the cutoff in these theories is.) 

{\bf M1.} Axion-like particles. They are technically natural on account of the shift symmetry, and the shift symmetry is broken softly by the mass. Possible space of mass \& couplings: $\{m_a, g_{a\gamma\gamma}, g_{aee}, g_{app}, g_{ann} \}$ or any combination of those.
These are couplings to $F\tilde F$, and the spins of electrons and nucleons. These couplings can be relatively easily mapped to the models/couplings featured in the PBC set \cite{Beacham:2019nyx}. $m_a$ can be anything including 0. ALPs may or may not saturate DM, and may contribute to the cosmic history \cite{Marsh:2015xka}. There is a large range of experimental and astro searches of these models.

 {\bf M2.}  ULF vector fields. One can have very light vector fields coupled to conserved portals, such as EM current (dark photon), or $B-L$ current. Standard coupling vs mass parameter space, and possibility to saturate the DM abundance with these vector fields. Potentially many interesting applications with broad-band and resonant searches
 with a variety of methods depending on the mass \cite{Chaudhuri:2014dla,An:2014twa,Baryakhtar:2018doz}. 
 
 {\bf M3.} Light thermal freeze-in dark matter, with possibly sub-eV light mediators (e.g. dark photons). If the mass of mediators is equal to zero, the model starts being equivalent to millicharge. This type of models has many astrophysical application, direct detection applications, and novel schemes of detecting light millicharged particles (see review \cite{Knapen:2017xzo} for light dark matter).  

{\bf B category}. These are models that could be natural under certain exotic conditions (see e.g. \cite{Hook:2018jle,Brzeminski:2020uhm}). 

{\bf M4}. Scalar field models (oscillating scalar, or smoothly evolving scalar as in M0) + linear non-derivative couplings 4a to Higgs portal (relaxion) \cite{Graham:2015cka}, 4b to trace of stress -energy tensor, 4c to spins, 4d to $F_{\alpha\beta}^2$ of electromagnetism \cite{Bekenstein:1982eu} etc.
The reference list here is rather elaborate.  Saturating dark matter in such models is an option. This type of models have large AMO opportunities as they generally predict apparent evolution of coupling constants
and masses in time, or an abnormal spin precession.

{\bf M5.} ULF scalars derivatively coupled to matter fields. (Some variants are called ``disformal couplings" in the cosmological
literature \cite{Bettoni:2013diz}). The couplings are given by 
 $T_{\alpha\beta} \partial_\beta \phi \partial_\alpha\phi/\Lambda^4$, $H^\dagger H  \partial_\alpha \phi \partial_\alpha\phi/\Lambda^2$ \cite{Weinberg:2013kea}, where $T_{\alpha\beta}$ is the stress-energy tensor, and $H$ is the SM Higgs field. There is not much signatures here for precision physics, other than cosmology studies of $N_{eff}$ and also collider studies of missing energy. (Also, the model can be put in the ``A list"  if $\Lambda$ is at the weak scale and above.)

{\bf C category}. Technically unnatural, but with interesting phenomenology

{\bf M6.}  Chameleon-type models. (Mass increases inside matter \cite{Khoury:2003aq}). 
Couplings needed in the simplest case are 
$T_{\alpha\alpha} \phi^2/\Lambda^2$, $H^\dagger H\phi^2$ \cite{Olive:2007aj}. If scalar field $\phi$ contributes to DM, the chameleonic mass can keep it outside of over-dense regions, creating space-varying properties of matter \cite{Olive:2007aj,Hees:2018fpg}. 

{\bf M7.}  ULF models with nontrivial clustering properties due to self-interaction. For example,  
DM out of lumps of light fields (Q-balls). Coupling to matter may introduce transient effects, when e.g. atomic properties temporarily change when a lump passes through \cite{Pospelov:2012mt,Derevianko:2013oaa}.

{\bf M8.}  ALP couplings to EDMs. This is inspired by the QCD axion, but the dipole coupling is taken to be much larger than the QCD-derived value \cite{Budker:2013hfa}. Relevant for some experimental proposals.

\subsubsection{ Suggested parameter space} 

{\bf M0.} ${\cal L}  =  \frac12 (\partial \phi)^2 -V(\phi)$. 
It is reasonable to try several  generic possibilities, such as models with ``late motion of field" when there is a 
constant linear forcing, and the value of the field is $\phi(z=0) = 0$ by construction. some  massive field with some initial condition at early times, 
\begin{eqnarray}
A: ~ V(\phi) = V_0 +V'\phi ,~~ \phi(z=0) = 0,\\
B: ~V(\phi)  = V_0 +\frac12 m_0^2 ( \phi)^2,~~ \phi({\rm large} ~z) = \phi_0\\
C: ~ ~V(\phi)  = V_0 +V_1 \cos ( \phi/f),~~ \phi({\rm large} ~z) \sim f\\
\end{eqnarray} 
First model has parameter space $\{V_0,V'\}$, and is about dark energy. 
Second model has parameter space $\{V_0,m_0,\phi_0\}$ and has dark energy and dark matter associated with it. 

{\bf M1.} This is more or less the same parameter space as in ALPs models. Since we are talking about low energies, 
it is reasonable, of course, to switch to the language of nucleons rather than quarks and gluons, 
and therefore we have:
\begin{equation}
{\cal L} =  \frac12 (\partial a)^2 - \frac12 m_a^2 a^2 -\frac{a}{4f_\gamma}F_{\mu\nu}\tilde F_{\mu\nu} 
-\sum_{i=e,p,n} \frac{\partial_\mu a}{f_i}\bar\psi_i \gamma_\mu\gamma_5 \psi_i.
\end{equation}
Parameter space is in principle a multi-dimensional $\{m_a, f_i\}$, and one could take different $f_i$ slices of it. 
Initial position of $a$ field, $a_0$ at earlier times, is also important and a unique $a_0$, and/or randomly distributed $a_0$ are possible. 
Alternatively, all models that aspire to have some contribution to DM, we may quantify their late-time contribution to 
the DM energy density by a parameter $\kappa_i \equiv \Omega_i/\Omega_{DM}$, $i= a,\phi,V$ etc. 

{\bf M2.}  Ultra-light vector fields can have the following Lagrangian:
\begin{eqnarray}
A:~~{\cal L}= -\frac{1}{4} V_{\mu\nu}^2 -\frac{\epsilon}{2} V_{\mu\nu}F_{\mu\nu} +\frac12 m_V^2 
V_\mu^2  \\
B:~~ {\cal L}= -\frac{1}{4} V_{\mu\nu}^2  +\frac12 m_V^2 +{\cal L}_{SM}(D_\mu \to D_\mu - i Q_{B-L} g_{B-L} V_\mu).
\end{eqnarray}
The parameter space is evidently $\{m_V, \epsilon \}$ and $\{m_V, g_{B-L} \}$. (If $m_V$ is tiny, the model may prefer Dirac SM neutrinos). 

{\bf M3.}  The model has the same parameter set as Benchmark model 2 (PBC set). Specifically, one takes small $m_{A'}$, and $m_\chi > O({\rm keV})$:
\begin{equation}
{\cal L} = -\frac{1}{4} V_{\mu\nu}^2 -\frac{\epsilon}{2} V_{\mu\nu}F_{\mu\nu} +\frac12 m_V^2+\bar \chi(i\gamma_\mu D_\mu-m_\chi)\chi,
\end{equation} 
and the covariant derivative contains $g_{d}$ dark gauge coupling. The model parameter space is 
$\{  m_V, \epsilon, m_\chi, g_d \}$. The difference with PBC is a smaller mass range, and 
freeze-in conditions for creating DM. 

{\bf M4}.  In this class of models, to the scalar models encountered before (M0) we add
\begin{eqnarray}
A:~~ {\cal L}_{int} =A\phi (H^\dagger H -  \langle H^\dagger H\rangle)\\
B:~~ {\cal L}_{int} =\frac{\phi}{M_T} \times T_\mu^\mu(SM)\\
C:~~ {\cal L}_{int} =\frac{\phi}{4M_\gamma} \times (F^{EM}_{\mu\nu})^2 + \frac{\phi}{4M_e} \times 
\bar e  e + \frac{\phi}{4M_N} \bar NN \\
D:~~ {\cal L}_{int} = - \sum_{i=e,p,n} \frac{\partial_\mu \phi}{M_i}\bar\psi_i \gamma_\mu\gamma_5 \psi_i\\
E:~~ {\rm Same~structures~with}~~ \phi\to \phi^2. 
\end{eqnarray}
Comments: model A is a more restrictive realization of model C. 
Model D is basically the same as the M1. Simultaneous presence of {\em e.g.} C and D will lead to mass-spin coupling, also searched in experiment. 

{\bf M5.} The parameter space is the mass of ``disformal" scalar and its coupling, $\{m_\phi, \Lambda\}$. 
(Representative couplings are given by $T_{\alpha\beta} \partial_\beta \phi \partial_\alpha\phi/\Lambda^4$ or $H^\dagger H  \partial_\alpha \phi \partial_\alpha\phi/\Lambda^2$). On theoretical grounds, we expect that coupling $\Lambda$ is larger than the EW scale. See the latest paper Ref. \cite{Bauer:2022rwf}. 

{\bf M6.}  Chameleon or ``symmetron"-type models have a large variety. 
For example, one can consider 
\begin{equation}
{\cal L} = \frac12(\partial \phi)^2 - \frac12 m_\phi^2 \phi^2 - \lambda_\phi \phi^4 + \lambda_{H\phi} 
(H^\dagger H -  \langle H^\dagger H\rangle)\phi^2.
\end{equation}
In-medium value of $H^\dagger H -  \langle H^\dagger H\rangle$ is non-zero, and if $m_\phi^2$ is small, it has consequences for spatial distribution of $\phi$, especially if it is dark matter. If $m_\phi^2$ is negative, $\phi$ field will have a nonzero v.e.v. in vacuum, and matter effects can restore symmetry.

{\bf M7.}  Parameter space of models with extended DM objects are difficult to describe in a few numbers. 
Let us approximate DM field profile inside a ``defect'' by some Gaussian field:
\begin{equation}
\phi(r) = \phi_0 \times \exp( -r^2/(2R^2) )
\end{equation}
$\phi_0$ and $R$ will describe the amplitude and the extent of the field configuration, that will have a mass $\propto \phi_0^2R$, so that number density of these objects $n$ should obey $ \sim n \phi_0^2R \leq \rho_{DM}$. The interaction
of such an object with matter can be described by Eqs. (7)-(11).

{\bf M8.} Finally, an ULF coupling to an EDM can be described as 
\begin{equation}
{\cal L_{\rm EDM\phi}} = \sum_{i = e,n,p} \phi  \frac{d_i}{f} \psi_i \sigma_{\mu\nu} \tilde F_{\mu\nu} \psi_i
\end{equation} 
The parameter space of the model is then $\{m_\phi, d_i/f, \Omega_\phi \}$.

{\bf Conclusion}: This is the first, and incomplete step towards the classification 
of ultra-light new physics. A lot more efforts are required in analyzing these models, and assembling constraints on their parameter space.

%-------------------------------------------
\subsection{Observational Searches for Ultra-Light FIPs with Cosmological Surveys -- {\it B.~Wallisch}}
\label{ssec:wallish}
{\it Author: Benjamin Wallisch, <benjamin.wallisch@fysik.su.se>} \\
%-------------------------------------------
% \documentclass[11pt,a4paper]{article}
% \pdfoutput=1
% \usepackage{jcapmod}

% \usepackage{amsmath, amsfonts, amssymb}
% \usepackage[english]{babel}
% \usepackage{subcaption}
% \usepackage[range-units=brackets, range-phrase={,}, per-mode=reciprocal, mode=math]{siunitx}[=v2]
% \usepackage[babel=true]{microtype}

% \setlength{\textwidth}{480pt}
% \setlength{\topmargin}{-1.3cm} \setlength{\textheight}{730pt} \setlength{\oddsidemargin}{-10pt} \setlength{\footskip}{37pt} \linespread{1.1}
% \setlength{\parindent}{0.2in}

% \numberwithin{equation}{section}
% \allowdisplaybreaks[1]

% \begin{document}
% 	\begin{center}
% %%		{\fontsize{20.74}{24}\selectfont \sffamily \bfseries Observational Searches for Ultra-Light FIPs\\[8pt]with Cosmological Surveys}
% 		{\fontsize{20.74}{24}\selectfont \sffamily \bfseries Searches for Light FIPs with Cosmological Surveys}
% 	\end{center}
	
% 	\vspace{-0.1cm}
% 	\begin{center}
% 		{\fontsize{12}{30}\selectfont Benjamin Wallisch\footnote{\href{mailto:benjamin.wallisch@fysik.su.se}{benjamin.wallisch@fysik.su.se}}}
% 	\end{center}
	
% 	\begin{center}
% 		\textsl{Oskar Klein Centre, Department of Physics, Stockholm University, 10691~Stockholm, Sweden}
		
% 		\vskip8pt
% 		\textsl{Center for Cosmology and Astroparticle Physics, Weinberg Institute for Theoretical Physics,\\Department of Physics, The University of Texas at Austin, Austin, TX~78751, USA}
% 	\end{center}

\subsubsection{Introduction}

Cosmological observations are a sensitive probe of particle physics and have become precise enough to start complementing terrestrial experiments. For example, they have the potential to shed new light on the properties of neutrinos and discover particles that are even more weakly coupled to the rest of the Standard Model~(SM). In particular measurements of the cosmic microwave background~(CMB) and the large-scale structure~(LSS) of the universe are remarkably sensitive to light, feebly-interacting particles~(FIPs). Importantly, these observations are and will be providing interesting bounds, both leading and complementary to those from astrophysics, colliders and laboratories, on both hot~(thermal) and cold~(non-thermal) FIP~populations.\medskip

Light thermal relics of the hot big bang are one of the primary targets of current and especially future cosmological surveys. Apart from new massive particles, an interesting consequence of many proposals for physics beyond the Standard Model~(BSM) are extra light species~\cite{Essig:2013lka, Asadi:2022njl, Dvorkin:2022jyg}, such as axions~(we will refer to both QCD~axions and axion-like particles as axions for simplicity)~\cite{Peccei:1977hh, Weinberg:1977ma, Wilczek:1977pj, Arvanitaki:2009fg, Hook:2018dlk}, dark photons~\cite{Holdom:1985ag, Galison:1983pa} and light sterile neutrinos~\cite{Abazajian:2012ys}. While searching for these particles is one of the main objectives of particle physics, their detection could be difficult in terrestrial experiments because their couplings might be too small or their masses too large. Interestingly, the temperature in the early universe was probably high enough to make the production of weakly-coupled and/or massive particles efficient. Their gravitational influence could then be detected if the energy density carried by these particles was significant. This will be the case for light relics which were in thermal equilibrium with the Standard Model at early times and subsequently decoupled from the SM~degrees of freedom. In addition, even the absence of a detection would result in new insights by providing constraints on the SM~couplings of these new particles. This sensitivity to extremely weakly interacting particles is a unique advantage of cosmological probes of BSM~physics.\medskip

Classes of models with non-thermally produced, light particles are another interesting target for BSM~searches for which cosmological surveys can uniquely contribute important information over a large range of scales. The very long wavelengths associated with the classical wave-like nature of ultra-light cold bosons means that their main signatures can be found in astrophysical and especially cosmological environments. As for thermally produced, light~FIPs, a cold population of such particles can be found in many SM~extensions, including the so-called axiverse in string theory which contains many ultra-light axions~\cite{Arvanitaki:2009fg}, for instance. There also exist several different production mechanisms which can leave distinct observational signatures.\medskip

While cosmological probes of~FIPs are a broad and growing field of study, we will focus on light thermal~FIPs and briefly discuss (ultra-)light non-thermal axions at the end. (In the cosmological context, we usually refer to sub-eV particles as being light.) We will first generally describe advantages and the physical origin of the constraining power of cosmology~(and astrophysics) to light thermal relics. Then, we provide explicit examples of bounds on the SM~couplings of these elusive particles, in particular axions. We finally give an introduction to ultra-light axions before we conclude with a summary. We note that we will mainly focus on the particle physics aspects of these searches and refer to~\cite{Hou:2011ec, Baumann:2015rya, Marsh:2015xka, Abazajian:2016yjj, Baumann:2017gkg, Wallisch:2018zrj, Green:2019glg, Grin:2019mub,  DePorzio:2020wcz, Xu:2021rwg, Dvorkin:2022bsc, Dvorkin:2022jyg, Ge:2022qws, Green:2022bre} and references therein for details on the underlying cosmological probes and more general reviews~(see also the related literature on sub-MeV dark matter, such as~\cite{Green:2017ybv, Knapen:2017xzo}).

\subsubsection{The Power of Cosmology (and Astrophysics)}

The detection of new light species is difficult since their couplings to the Standard Model are constrained to be small. Given the resulting small scattering cross sections, terrestrial experiments in the intensity or energy frontier of particle physics are challenging~(see other contributions in these proceedings). The study of astrophysical and cosmological systems, however, provide us access to high-density environments and/or the ability to follow the evolution over long time scales which can overcome the small cross sections and allow a significant production of these extra particles. This then allows us to put some of the best constraints on~FIPs.

It is illustrative in this context~\cite{Wallisch:2018rzj}  to consider the fractional change in the number densities of the particles involved in the production process, which is schematically given by the interaction rate $\Gamma \sim n\sigma$, with number density~$n$ and thermally-averaged cross section~$\sigma$, times the interaction time~$\Delta t$:
\begin{equation}
	\frac{\Delta n}{n} \sim n\sigma \times \Delta t\, .	\label{eq:schematicBoltzmannEquation}
\end{equation}
This highlights how small cross sections can be compensated for by high densities, e.g.\ $n \sim (1\,\mathrm{keV})^3$ and $(10\,\mathrm{MeV})^3$ in stellar interiors and supernova explosions, and long time scales, such as the very long lifetime of stars, $\Delta t \sim O(10^8)\,\mathrm{yrs}$, or $\Delta t \sim 10\,\mathrm{s}$ for supernovae. We therefore find significant changes in the evolution of these astrophysical systems, $\Delta n/n \gtrsim 1$, if $\sigma \gtrsim (n \Delta t)^{-1} \sim \left(10^{10}\,\mathrm{GeV}\right)^{\!-2}$. A similar argument can be applied to the early universe, which was dominated by radiation. The high densities of the early universe, $n \sim T^3 \gg (1\,\mathrm{MeV})^3$ before neutrino decoupling, allow light particles to have been in thermal equilibrium with the~SM~(and therefore efficiently produced) for time scales of $\Delta t < 1\,\mathrm{s}$. They can therefore make a significant contribution to the radiation density, and be detected in CMB~and LSS~observations. Based on this rough estimate, cosmological constraints should improve over astrophysical bounds for temperatures above~$10^4\,\mathrm{GeV}$.

Another advantage of cosmological observations is that they can provide broad constraints on phenomenological descriptions. On the other hand, particle physics searches can be blind to unknown or incompletely specified forms of new physics. This means that terrestrial experiments may give strong constraints on specific scenarios, while cosmological measurements are less sensitive to the details of the models and can compress large classes of BSM~physics into broad categories. For example, cosmology can constrain all couplings of new particles to the Standard Model, while astrophysical systems and laboratory experiments are often only sensitive to a subset of these interactions, such as the coupling to photons. This universality of cosmological constraints is one of the reasons why the search for light thermal relics has been adopted as one of the main science targets of future cosmological surveys~(see e.g.~\cite{Annis:2022xgg, Chang:2022lrw}), such as the next-generation CMB~and LSS~experiments \mbox{CMB-S4}~\cite{Abazajian:2016yjj, Abazajian:2019eic, CMB-S4:2022ght} and Spec-S5~\cite{DESI:2022lza, Schlegel:2022vrv}, respectively. In the next section, we will provide a few explicit examples for the constraining power of cosmology to light thermal relics.

\subsubsection{Cosmological Constraints on Relic Axions}

From both the particle physics and cosmology perspective, it is useful to theoretically describe light thermal relics within an effective field theory~(EFT) framework organized according to their spin~\cite{Brust:2013ova} and observationally search for their contribution to the radiation density in the early universe as parametrized by the effective number of relativistic species,~$N_\mathrm{eff}$. We will therefore briefly review these topics before discussing bounds in the decoupling and rethermalization scenarios.\medskip

Given the universal sensitivity of cosmology, it is more efficient to study the interactions between the new species with the SM~degrees of freedom within the framework of effective field theory~\cite{Brust:2013ova} and thereby capture their main phenomenology instead of working through BSM~models one by one. Generally speaking, this means parametrizing these interactions as
\begin{equation}
	\mathcal{L} \supset \frac{g}{\Lambda^n}\, \mathcal{O}_X \mathcal{O}_\mathrm{SM}\, ,	\label{eq:schematicEFT}
\end{equation}
where~$g \sim O(1)$ is a dimensionless coupling constant, $\Lambda$~is the associated energy scale, $\mathcal{O}_X$~and $\mathcal{O}_\mathrm{SM}$ are operators of dark sector and SM~fields of dimension~$\Delta_X$ and~$\Delta_\mathrm{SM}$, respectively, and $n=\Delta_X+\Delta_\mathrm{SM}-4$. To prevent large quantum corrections of the small masses of~$X$, we employ (approximate)~symmetries which restrict the allowed couplings in~\eqref{eq:schematicEFT}. This naturally separates the~EFT according to the spin of the new particle(s): shift symmetries for scalars, chiral and axial symmetries for spin-$\frac{1}{2}$ particles and gauge symmetries for vector bosons. By dimensional analysis, the interaction rate~$\Gamma_X$ grows with temperature~$T$ for $n > \frac{1}{2}$, i.e.\ the new species is potentially in thermal equilibrium at high temperatures and decouples at a lower freeze-out temperature~$T_F$:
\begin{equation}
	\Gamma_X(T) \sim \frac{T^{2n+1}}{\Lambda^{2n}} \qquad \xrightarrow{\ H(T_F)\,\sim\,\Gamma_X(T_F)\ } \qquad T_F \sim \left(\frac{\Lambda^{2n}}{M_\mathrm{pl}}\right)^{\!1/(2n-1)}\, ,	\label{eq:interactionRateFreezeOut}
\end{equation}
where we used the Hubble parameter during radiation domination in the early universe, $H(T) \sim T^2/M_\mathrm{pl}$. For $\Lambda \ll M_\mathrm{pl}$, decoupling therefore happens at $T_F \ll \Lambda$, i.e.\ in the regime of validity of the~EFT, provided that~$n$ is not too large. If the decoupling temperature is smaller than the reheating temperature of the universe, $T_F < T_R$, the dark sector thermalizes with the Standard Model producing a detectable relic abundance~(see below). Conversely, excluding such a relic abundance would lead to a bound on the interaction strength, $\Lambda \gtrsim \left(T_R^{2n-1} M_\mathrm{pl}\right)^{1/(2n)}$, and severely constrain the currently available parameter space, in particular for high-scale reheating~\cite{Baumann:2016wac}. More generally, we can constrain the interaction terms~\eqref{eq:schematicEFT} by measuring the relic density since it is governed by the decoupling temperature~$T_F(\Lambda)$.\medskip

\begin{figure}[t]
	\centering
	\includegraphics{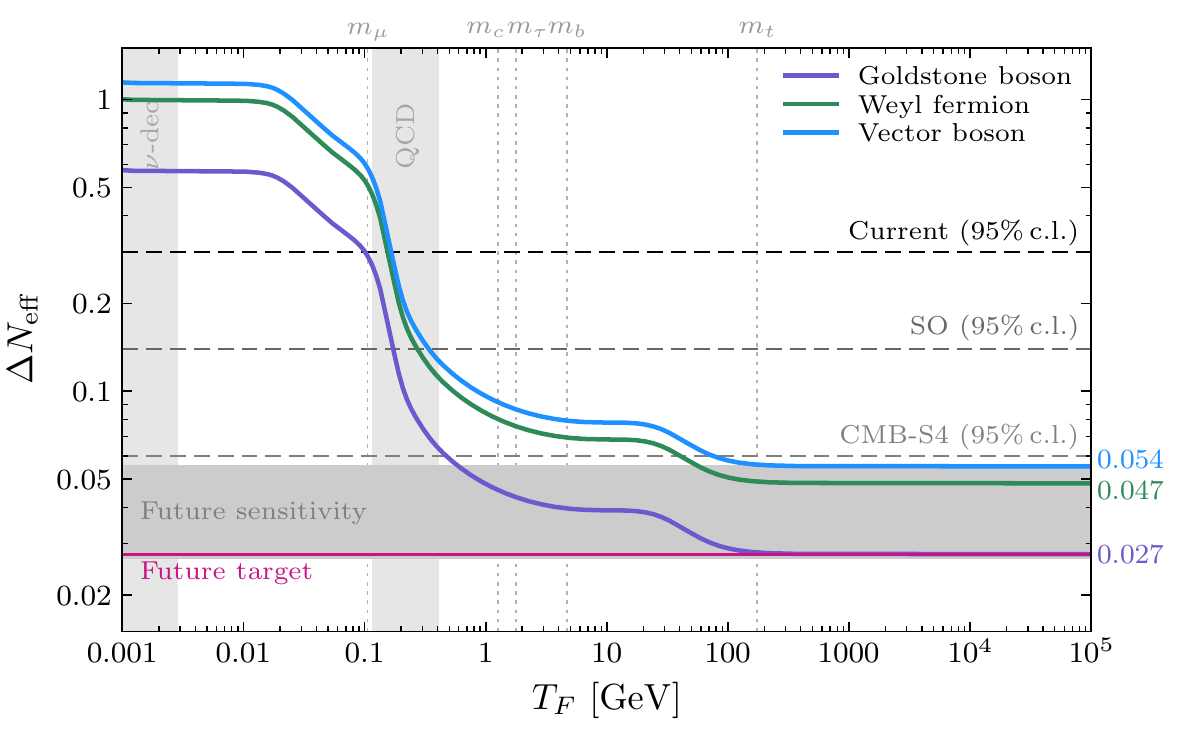}
	\caption{Contributions of a single massless particle, which decoupled from the Standard Model at a decoupling temperature~$T_F$, to the effective number of relativistic species, $N_\mathrm{eff} = N_\mathrm{eff}^\mathrm{SM} + \Delta N_\mathrm{eff}$, with the SM~expectation $N_\mathrm{eff}^\mathrm{SM} = 3.044$ from neutrinos~(from~\cite{Wallisch:2018rzj, Green:2019glg, Dvorkin:2022jyg} where we also refer to for additional details). The purple, green and blue lines show the contributions for a real scalar, Weyl fermion and vector boson, respectively. The two vertical gray bands indicate neutrino decoupling and the QCD~phase transition. The current limit on~$\Delta N_\mathrm{eff}$ at 95\%~c.l.\ from Planck~2018 and BAO~data~\cite{Planck:2018vyg}, and the anticipated sensitivity of the Simons Observatory~(SO)~\cite{SimonsObservatory:2018koc} and \mbox{CMB-S4}~\cite{Abazajian:2019eic, CMB-S4:2022ght} as examples for upcoming and next-generation CMB experiments are included with dashed lines. The horizontal gray band illustrates the future sensitivity that might potentially be achieved with a combination of cosmological surveys of the~CMB and~LSS, such as CMB-HD~\cite{CMB-HD:2022bsz}, Spec-S5~\cite{Schlegel:2022vrv} and PUMA~\cite{PUMA:2019jwd}, cf.~\cite{Baumann:2017gkg, Sailer:2021yzm, MoradinezhadDizgah:2021upg}. The displayed values on the right are the observational thresholds for particles with different spins and arbitrarily large decoupling temperature in minimal scenarios.}
	\label{fig:deltaNeff_freezeout}
\end{figure}

We usually infer the relic density~$\rho_X$ in terms of~$N_\mathrm{eff}$, which parametrizes the radiation density~$\rho_r$ of the universe as
\begin{equation}
	\rho_r = \rho_\gamma + \rho_\nu + \rho_X = \left[ 1 + \frac{7}{8}\left(\frac{4}{11}\right)^{\!4/3}\left(N_\mathrm{eff}^\mathrm{SM} + \Delta N_\mathrm{eff}\right)\right] \rho_\gamma\, .
\end{equation}
Here, we used the well-measured photon energy density~$\rho_\gamma$ and the neutrino energy density~$\rho_\nu$, given by the SM~value of $N_\mathrm{eff}^\mathrm{SM} = 3.044$~\cite{Akita:2020szl, Froustey:2020mcq, Bennett:2020zkv}, which has now been theoretically calculated with an uncertainty in the fourth decimal place. The contribution to~$N_\mathrm{eff}$ from any extra light relic is given by
\begin{equation}
	\Delta N_\mathrm{eff} = \frac{4}{7}\hskip1pt g_{*,X} \left(\frac{43/4}{g_*(T_F)}\right)^{4/3} \approx 0.027 g_{*,X} \left(\frac{g_{*,\mathrm{SM}}}{g_*(T_F)}\right)^{\!4/3} ,	\label{eq:DeltaNeff}
\end{equation}
where~$g_{*,X} = 1, 7/4, 2$ is the effective number of degrees of freedom for a real scalar, Weyl fermion and massless vector boson, respectively, $g_*(T)$ is the effective number of total relativistic degrees of freedom in thermal equilibrium at temperature~$T$, and $g_{*,\mathrm{SM}} = 106.75$ is the maximum SM~value at temperatures above the scale of electroweak symmetry breaking. We present this contribution to~$N_\mathrm{eff}$ from a single thermally-decoupled species as a function of the decoupling temperature~$T_F$ and the spin of the particle in Fig.~\ref{fig:deltaNeff_freezeout}.%
Here, we assumed that there is no significant entropy production in the universe for $T < T_F$ which would otherwise decrease~$\Delta N_\mathrm{eff}$. If we additionally consider a minimal extension of the Standard Model, $g_*(T \gg m_t) \approx g_{*,\mathrm{SM}}$, with top mass~$m_t$, we get $\Delta N_\mathrm{eff} \geq 0.027 g_{*,X}$. (We refer to~\cite{Wallisch:2018rzj} for a detailed discussion of these assumptions and their implications.) This in particular implies that cosmological surveys which are sensitive to $\Delta N_\mathrm{eff} = 0.027$ will either detect the existence of any light particle that was ever thermalized in the history of the universe or provide strong constraints on their potential SM~interactions. Interestingly, this threshold is within reach of future observations.\medskip

We can exemplify the potential sensitivity of cosmological $N_\mathrm{eff}$~measurements to constrain couplings of SM~particles to BSM~pseudo-Nambu-Goldstone bosons~(pNGBs), which are scalar particles that arise as a consequence of the breaking of (approximate) global symmetries, such as axions~(shift symmetry)~\cite{Peccei:1977hh, Weinberg:1977ma, Wilczek:1977pj, Arvanitaki:2009fg, Hook:2018dlk}, familons (flavor symmetry)~\cite{Davidson:1981zd, Wilczek:1982rv, Reiss:1982sq, Feng:1997tn} or majorons~(neutrino masses)~\cite{Chikashige:1980ui, Chacko:2003dt}. For the dimension-5 interaction of photons and gluons with axions~$\phi$, $\mathcal{L} \supset -\phi/(4\Lambda_{\gamma,g}) X_{\mu\nu} \tilde{X}^{\mu\nu}$, with respective field-strength tensor~$X_{\mu\nu} = F_{\mu\nu}, G_{\mu\nu}^a$ and its dual~$\tilde{X}_{\mu\nu}$, we have~\eqref{eq:interactionRateFreezeOut} with $n = 1$. Observationally excluding a contribution of $\Delta N_\mathrm{eff} = 0.027$ to the radiation density therefore implies reheating-temperature-dependent constraints on the interaction strengths of~\cite{Baumann:2016wac}
\begin{equation}
	\Lambda_\gamma \gtrsim 1.4\times10^{13}\,\mathrm{GeV} \left(\frac{T_R}{10^{10}\,\mathrm{GeV}}\right)^{\!1/2}, \qquad \Lambda_g \gtrsim 5.4\times10^{13}\,\mathrm{GeV} \left(\frac{T_R}{10^{10}\,\mathrm{GeV}}\right)^{\!1/2}\, .
\end{equation}
This limits would be stronger than current~(astrophysical) constraints for reheating~(or similar entropy production eras above the electroweak symmetry breaking scale) temperatures of about~$10^4\,\mathrm{GeV}$ for the axion-photon coupling and generally for the interaction with gluons. Similar bounds can also be inferred for couplings to SM~fermions and neutrinos~\cite{Baumann:2016wac}~(see also~\cite{Hannestad:2005ex, Archidiacono:2013dua, Darme:2020ral, DEramo:2021lgb, DEramo:2021psx, DEramo:2021usm}, for instance).\medskip

Another scenario occurs for effective operators with $n < 1/2$ in~\eqref{eq:schematicEFT} for which the new light particles are out of equilibrium at high temperatures and thermalize at lower temperatures, cf.~\eqref{eq:interactionRateFreezeOut}. This is in particular the case for dimension-5 pNGB~couplings to SM~fermions~$\psi$, such as the SM~axial vector current,~$\partial_\mu \phi\, \bar{\psi} \gamma^\mu \gamma^5 \psi$, which become effective dimension-4 operators after electroweak symmetry breaking leading to $\Gamma_\phi \propto m_i/\Lambda_i\, T$, where~$m_i$ is the respective SM~fermion mass. Once these particles thermalize with the~SM at $T \gtrsim m_i$, they will contribute to~$N_\mathrm{eff}$ at an observable level, but at a negligible level for $T \lesssim m_i$ since the fermion number density becomes Boltzmann suppressed. In fact, this would be easier to detect since the contribution to~$N_\mathrm{eff}$ in this rethermalization scenario is larger than their equivalent freeze-out contribution.

\begin{figure}[t]
	\centering
	\includegraphics{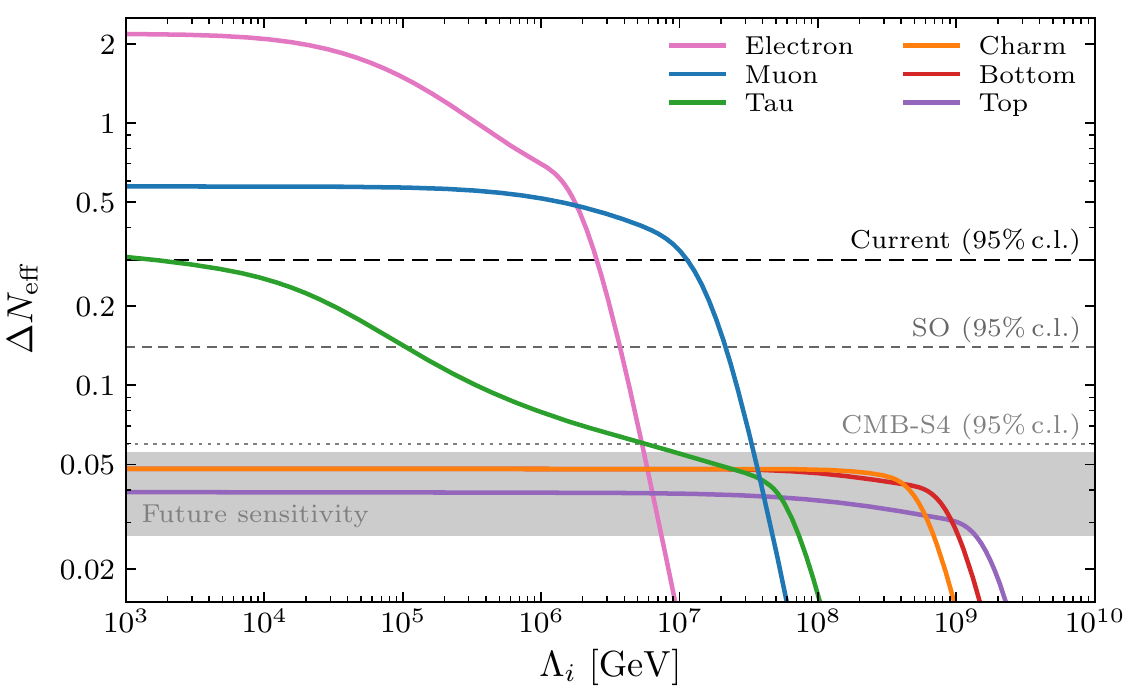}
	\caption{Contribution of a rethermalized light axion to~$\Delta N_\mathrm{eff}$ as a function of its coupling~$\Lambda_i$ to different SM~fermions~$\psi_i$ for the dimension-5 derivative coupling to the SM~axial vector current~$-\Lambda_i^{-1} \partial_\mu \phi\, \bar{\psi}_i \gamma^\mu \gamma^5 \psi_i$~(from~\cite{Green:2021hjh} where we also refer to for a detailed discussion). The displayed values for the bottom and charm couplings are conservative and may be~(significantly) larger. The horizontal dashed lines and gray band are the same as in Fig.~\ref{fig:deltaNeff_freezeout}. Given the displayed results, current and future constraints on~$N_\mathrm{eff}$ can be directly translated into the equivalent bounds on~$\Lambda_i$.}
	\label{fig:deltaNeff_axionMatterCouplings}
\end{figure}

At the same time, the absence of a detection would allow us to place direct constraints on the interaction strength between the light particles and the SM~fermions: we require that the recoupling temperature is smaller than the effective decoupling temperature so that the interaction rate is already Boltzmann suppressed when the~pNGBs would rethermalize~\cite{Baumann:2016wac}. While these constraints are usually weaker than the freeze-out bounds, they do not make any assumptions about the reheating temperature~(as long as $T_R > m_i$). As illustrated in Figure~\ref{fig:deltaNeff_axionMatterCouplings},%
both current constraints on~$N_\mathrm{eff}$ from Planck and, in particular, future measurements at the projected sensitivity of~CMB-S4 set interesting constraints on the couplings of some SM~fermions to axions and other~pNGBs~\cite{Baumann:2016wac, Ferreira:2018vjj, DEramo:2018vss, Arias-Aragon:2020shv, Ghosh:2020vti, Ferreira:2020bpb, Dror:2021nyr, Green:2021hjh}. At the moment, the cosmological constraints on axion-muon and axion-tau couplings are complementary to astrophysical constraints from supernova~1987A~(cf.~\cite{Bollig:2020xdr, Croon:2020lrf, Caputo:2021rux}) and white dwarf cooling, respectively, but will be competitive or even supersede these bounds with future $N_\mathrm{eff}$~measurements~\cite{Green:2021hjh}. Upcoming CMB~and LSS~surveys will also be sensitive to the couplings of axions to heavy quarks even though the derivation of precise constraints depends on the strong-coupling regime of quantum chromodynamics. More generally, the environment, required modeling and physical understanding in the early universe is quite different to the interiors of stars, i.e.\ cosmological probes can be an important complementary test of these axion-matter couplings~(see e.g.~\cite{DeRocco:2020xdt}).

\subsubsection{Cosmological Constraints on Ultra-Light Axions}

So far, we have considered a potential thermal population of light species, but there are also several ways of non-thermally producing light BSM~particles which may also constitute~(part of) the dark matter. One such example are (ultra-)light axions which have an interesting phenomenology and distinct observational signatures, with cosmological observations setting the current bounds in a sizable part of parameter space~(see~\cite{Marsh:2015xka, Grin:2019mub, Ferreira:2020fam, Hui:2021tkt, Dvorkin:2022bsc, Adams:2022pbo} for recent reviews).\medskip

Non-thermally produced bosons with masses in the range $10^{-33}\,\mathrm{eV} \lesssim m_\phi \lesssim 10^{-2}\,\mathrm{eV}$ are particularly interesting since they may contribute the observed dark matter and dark energy densities. These particles generally behave like dark energy with an equation of state $w = -1$ when the Hubble parameter is smaller than their mass, $H \lesssim m_\phi$. For larger values of~$H$ at later times, the axion field oscillates rapidly, behaving like (dark)~matter with $w = 0$ on average. This implies that these~FIPs either contribute to dark matter or dark energy throughout the history of the universe. The behavior of an axion with $m_\phi \lesssim 10^{-33}\,\mathrm{eV}$ cannot be distinguished from a cosmological constant as dark energy today since its Compton wavelength exceeds the cosmic horizon. If its mass is smaller and larger than about~$10^{-27}\,\mathrm{eV}$, respectively, it behaves like early dark energy and dark matter. Through this behavior, the wave-like nature with macroscopic de Broglie wavelengths and broader phenomenology, these particles also have the potential to address some open questions in astrophysics and cosmology.

Ultra-light axions have various observable implications that affect several cosmological probes~(and other astrophysical measurements and terrestrial experiments). For instance, the slowly rolling axion field in the dark energy regime in particular impacts the largest angular scales and the wave-like nature of these~FIPs suppresses density fluctuations on smaller scales below the (comoving)~Jeans scale, $\lambda_J = 0.1\,\mathrm{Mpc}\, (m_\phi/10^{-22}\,\mathrm{eV})^{-1/2}\, (1+z)^{1/4}$. These are signals that are especially imprinted in both the primary and secondary CMB~anisotropies, and in LSS~observations of galaxy clustering, weak gravitational lensing and the Lyman-$\alpha$ forest. Current CMB~and LSS~surveys constrain the ultra-light axion energy density to $\Omega_\phi \lesssim 0.01$ in the mass range $10^{-32}\,\mathrm{eV} \lesssim m_\phi \lesssim 10^{-26}\,\mathrm{eV}$, with Lyman-$\alpha$ observations adding additional constraints for $10^{-23}\,\mathrm{eV} \lesssim m_\phi \lesssim 10^{-20}\,\mathrm{eV}$. Future surveys, especially CMB-S4, are forecasted to close this gap and improve the current bounds by one to two orders of magnitude~(see Fig.~4 of~\cite{Dvorkin:2022bsc} for a recent compilation of current and future CMB~and LSS~constraints~\cite{Bozek:2014uqa, Hlozek:2014lca, Hlozek:2016lzm, Irsic:2017yje, Kobayashi:2017jcf, Poulin:2018dzj, Rogers:2020ltq, Rogers:2020cup, Lague:2021frh, Farren:2021jcd, Dentler:2021zij}). An additional powerful signature of ultra-light axions is a spectrum of isocurvature fluctuations, which depends on whether the underlying $U(1)$~symmetry of these~FIPs is broken before or after inflation and may provide further insights together with CMB~B~modes in the former case~(see e.g.~\cite{Turner:1983sj, Axenides:1983hj, Hertzberg:2008wr, Hlozek:2017zzf, Feix:2020txt}). Somewhat more model-dependent constraints may also be inferred through CMB~birefringence observations~(e.g.~\cite{Sigl:2018fba, Fedderke:2019ajk, Jain:2021shf, Obata:2021nql, Komatsu:2022nvu}), for instance, if these BSM~particles couple to photons.

\subsubsection{Summary}

The wealth and precision of current CMB~and LSS~data does not only allow us to probe the universe on large scales, but we can also use it to search for BSM~particles. These cosmological observations are particularly well suited to study light~FIPs, such as axions, which arise in many well-motivated extensions of the Standard Model. Current surveys already provide interesting bounds on many scenarios through broad, model-independent data analyses which in particular capture the gravitational influence of new species on the evolution of the universe. While the main parameters of the effective number of relativistic species~$N_\mathrm{eff}$ and the axion energy density~$\Omega_\phi$ for a thermal and non-thermal population of such particles, respectively, parametrize their energy density, we also note that these cosmological measurements can shed light on additional properties, such as their (non-)free-streaming nature and, therefore, their self-interactions or dark couplings~(cf.~\cite{Bashinsky:2003tk, Friedland:2007vv, Follin:2015hya, Baumann:2015rya, Baumann:2017lmt, Brust:2017nmv, Choi:2018gho, Baumann:2019keh, Blinov:2020hmc, Abazajian:2022ofy}). Upcoming and future CMB~and LSS~surveys are designed to improve current bounds on~$N_\mathrm{eff}$ and~$\Omega_\phi$ by at least one order of magnitude which will translate into many orders of magnitude in coupling strengths and large parts of currently viable parameter spaces. Cosmological probes of light~FIPs therefore have a bright future by themselves, and through their combination and complementarity with astrophysical and terrestrial experiments.

%\vskip10pt
%\paragraph{Acknowledgments}
%B.\,W.\ acknowledges support from the Swedish Research Council under Grant~\mbox{638-2013-8993}).
%\vskip10pt

% \phantomsection
% \addcontentsline{toc}{section}{References}
% \bibliographystyle{utphys}
% \bibliography{references}

% \end{document}

%-------------------------------------------
\subsection{Ultra-light FIPs: What we know from stars/supernovae/neutron stars/white dwarfs/etc -- {\it P.~Carenza}}
\label{ssec:carenza}
{\it Author: Pierluca Carenza, <pierluca.carenza@fysik.su.se>} \\
%-------------------------------------------
% \documentclass[twocolumn,superscriptaddress,floatfix,preprintnumbers, nofootinbib,hyperref]{revtex4-2} 
% \pdfoutput=1
% \usepackage[colorlinks=true,breaklinks=true]{hyperref}
% \usepackage[normalem]{ulem}
% %\usepackage{ulem,cancel}
% \usepackage[utf8]{inputenc}
% \hypersetup{allcolors=[rgb]{0.0 0.0 0.6},linkcolor=[rgb]{0.75 0.05 0.05}}
% \usepackage{amsmath,amssymb}
% \usepackage{epsfig}  
% \usepackage{graphicx}   
% \usepackage{slashed}       
% \usepackage{tikz}
% \usepackage{tikz-feynman}
% \usepackage{url}
% \usepackage{color}
% \usepackage{multirow}
% %\usepackage{placeins}
% \usepackage{comment}
% \usepackage{amssymb}

% %\clubpenalty=1000
% %\widowpenalty=10000

% \allowdisplaybreaks

% \setlength{\bibsep}{0cm}
% \bibpunct{[}{]}{,}{n}{}{,}

% \begin{document}

% \title{Ultra-light FIPs: What we know from stars/supernovae/neutron stars/white dwarfs/etc}

% \author{Pierluca Carenza}\email{pierluca.carenza@fysik.su.se}
% \affiliation{The Oskar Klein Centre, Department of Physics, Stockholm University, Stockholm 106 91, Sweden}

% %\date{\today}
% \smallskip

% \maketitle

Feebly Interacting Particles (FIPs) play a major role in a plethora of astrophysical phenomena. 
An extremely important phenomenological aspect is the impact that FIPs have on stellar evolution. 
Indeed, exotic particle with feeble interactions with ordinary matter, once produced in stars, escape draining energy from the stellar core.
Thus, stars are efficient FIP factories given their extreme density and temperature conditions.

Axions and Axion-Like Particles (ALPs), collectively called `axions', are among the most studied FIPs since their intriguing connection to the strong CP problem and the possibility of explaining the nature of Dark Matter (see, e.g., Refs.\cite{Giannotti:2022euq,Adams:2022pbo,Baryakhtar:2022hbu,DiLuzio:2020wdo,DiVecchia:2019ejf} for recent reviews). 
A very generic property of these particles is a coupling to photons through the following interaction Lagrangian
\begin{equation}
    \mathcal{L}=-\frac{g_{a\gamma}}{4}a\,F_{\mu\nu}\tilde{F}^{\mu\nu}\,,
    \label{eq:agg}
\end{equation}
which enables the axion production in external electromagnetic field and the possibility of a radiative decay for massive axions.
At low-energy, another phenomenologically relevant coupling is the one with fermions 
\begin{equation}
    \mathcal{L}=\frac{g_{af}}{2m_{f}}\bar{\Psi}\gamma^{\mu}\gamma^{5}\Psi\,\partial_{\mu}a\,,
    \label{eq:fermion}
\end{equation}
where $\Psi$ is any fermion field (electrons, nucleons, ecc...) with mass $m_{f}$.
This interaction is important in environments with a high density of electrons or nucleons, like in stars. Here, axions can be produced by various processes involving fermions and their consequences will be extensively discussed.

Also neutrinos can be listed as FIPs, and their properties are efficiently probed by astrophysics. In particular, their electromagnetic properties, as the magnetic moment, are of great interest to unveil their nature. A neutrino with a non-vanishing magnetic moment interacts with photons through the following Lagrangian
\begin{equation}
    \mathcal{L}=-\frac{\mu^{ij}}{2}\bar{\Psi}_{i}\sigma_{\mu\nu}\Psi_{j}F^{\mu\nu}\,,
\end{equation}
where the indices $i,j$ indicate the neutrino flavors, and $F^{\mu\nu}$ is the electromagnetic tensor. This interaction makes possible for right-handed (sterile) neutrinos to be produced by oscillations of a left-handed neutrino in an external electromagnetic field. 

The last FIP we consider is an exotic gauge boson, usually called hidden photon or Dark Photon (DP). 
In general, DPs are kinetically mixed with ordinary photons by means of the following interaction
\begin{equation}
    \mathcal{L}=-\frac{\epsilon}{2}X_{\mu\nu}F^{\mu\nu}\,,
    \label{eq:DP}
\end{equation}
where $\epsilon$ is the mixing angle and $X_{\mu\nu}$ is the field tensor associated with DPs. 
This vertex allows DPs to convert into photons independently of external fields, unlike axions. 

In the following we explore the phenomenology associated with FIPs produced in stars and their consequences on the stellar evolution.

\subsubsection{Stellar bounds on FIPs}

Stellar astrophysics is a well-developed branch of astrophysics that, while far from being a closed subject, provides an accurate and consistent picture of the stellar evolution, with detailed information about the stellar interior. 
The properties of stars are strongly affected by the microphysics of their constituents. 
Therefore, stellar evolution is sensitive to particle physics and the extreme conditions reigning in stellar interiors make stars ideal laboratories for particle physics. Every evolutionary stage has unique properties that can be used to test various FIPs. Typically stars are grouped depending on their visible properties, as luminosity and surface temperature. In the following we identify different stars and discuss how they can be used to probe FIPs.
Details on the global analyses of stellar bounds on FIPs, especially axions, can be found in~\cite{DiLuzio:2021ysg,DiLuzio:2020jjp,Giannotti:2017hny,Giannotti:2015kwo,Raffelt:2006cw,Raffelt:1996wa,Raffelt:1990yz}. 
Here we provide only a brief summary with updated results.

{\bf  Sun - } The Sun is the most studied and well-known star. Its interior is composed by ionized hydrogen, that is converted into helium to produce energy and balance the gravitational collapse. The detailed internal structure is revealed by helioseismology, the study of pressure waves propagating to the solar surface. This study allows to reconstruct the sound-speed profile in the solar interior, which turns out to be a sensitive probe of density and temperature inside the star. 

Axions coupled to photons through Eq.~\eqref{eq:agg} are produced in the Sun via Primakoff conversion of thermal photons in the electrostatic field of ions, \mbox{$\gamma+Ze\to a\to Ze$}. The constant flux of energy subtracted from axions to the Sun during its evolution would change the equilibrium structure compared to a case without axions. This imprint would affect the sound-speed profile inside the Sun and also the surface helium abundance~\cite{Schlattl:1998fz}.

Solar neutrinos are a precious messenger of the innermost structure of the Sun. The flux of neutrinos produced in nuclear reactions is extremely sensitive to the solar temperature, since production rates have a steep temperature dependence. 
A sizable amount of energy-loss through FIP productions might results into a higher temperature and higher neutrino fluxes. 
Solar neutrino data (for the $\,^{8}{\rm B}$ neutrino flux) collected from the Sudbury Neutrino Observatory (SNO) were used to set an upper limit on exotic losses equal to $L_{x}\lesssim 0.1 L_{\odot}$, where $L_{\odot}=3.84\times10^{33}~{\rm erg}{\rm s}^{-1}$ is the solar photon luminosity~\cite{Gondolo:2008dd}. This probe is more sensitive to exotic losses than helioseismology, since $L_{x}\lesssim 0.2 L_{\odot}$ is the constrained obtained by helioseismological studies~\cite{Schlattl:1998fz}.

A joint analysis of both solar neutrinos and helioseismology leads to a constraint on the axion-photon coupling equal to $g_{a\gamma}<4.1\times10^{-10}~{\rm GeV}^{-1}$ at $99\%$ CL for $m_{a}\lesssim 1~{\rm keV}$~\cite{Vinyoles:2015aba}. 
With a similar approach it is possible to constrain DPs, that are produced by conversion of thermal photons. Precisely, in a plasma there are two photon modes (transversal and longitudinal) with different dispersion relations. Both modes contribute to the DP production, especially in resonant conditions, when the DP mass (for resonant conversion of transverse modes) or its energy (for resonant conversion of longitudinal modes) match the plasma frequency~\cite{Redondo:2013lna,An:2013yfc}.
Analyses of solar properties constrain DP mixing and mass to be $\epsilon\, m_{\chi} < 1.8 \times10^{-12}~{\rm eV}$ at $99\%$ CL~\cite{Vinyoles:2015aba}.
In analogy, also electromagnetic neutrino properties, as a milli-charge or a large magnetic moment, are constrained by solar physics~\cite{Raffelt:1999gv} (see also Ref.~ \cite{Hardy:2016kme} for other FIP models).

Until now the discussion was focused on how FIPs affect solar observables indirectly. However, the Sun is the closest star and potentially the brightest source of sub-keV FIPs. Therefore, it is possible to constrain FIP properties by direct detection experiments pointing to the Sun. For example, light axions (with masses $m_{a}\lesssim{\rm keV}$) coupled to photons have a flux at Earth equal to $\phi_{a}\simeq(g_{a\gamma}/10^{-10}~{\rm GeV}^{-1})^{2}\,4\times10^{11}~{\rm cm}^{-2}{\rm s}^{-1}$. This large axion flux is potentially detectable through conversion of axions into X-rays in intense magnetic fields: this is the idea behind the helioscope design, of which the CERN Axion Solar Telescope (CAST) is the most currently developed example~\cite{CAST:2017uph}. 
CAST set the most stringent experimental constraint on the axion-photon coupling in a wide mass range, $g_{a\gamma}<0.66\times10^{-10}~{\rm GeV}^{-1}$ at $95\%$ CL for $m_{a}\lesssim20~{\rm meV}$~\cite{CAST:2017uph,Giannotti:2017law}.
An upgraded version of CAST is the planned helioscope International Axion Observatory (IAXO), expected to probe axions with a photon coupling down to $g_{a\gamma}\sim10^{-12}~{\rm GeV}^{-1}$~\cite{Armengaud:2014gea,IAXO:2019mpb}. 
The first stage in the IAXO development, BabyIAXO, is already expected to probe unexplored parameter space about a factor of 3 below the CAST bound~\cite{IAXO:2020wwp}.

The astonishing sensitivity of these searches allows to probe other FIP properties. For example, the axion-electron coupling in Eq.~\eqref{eq:fermion} opens many channels for the axion production in the Sun: axion-Compton scattering $\gamma e^{-}\to a e^{-}$, electron Bremsstrahlung on electrons or ions $e^{-}+Ze\to a+e^{-}+Ze$, atomic axio-deexcitation $I^{*}\to I\,a$ and axio-recombination $e^{-}+I\to a+I^{-}$, where $I$ represents any atomic species~\cite{Redondo:2013wwa}. 
The Bremsstrahlung is the main component of the solar axion flux at low energies, with a peak at $\sim 1~{\rm keV}$. At higher energies, above $\sim 5~{\rm keV}$ the axion-Compton scattering  is the most important contribution and atomic processes introduce peculiar spectral lines in the whole energy range. 
Constraints on the solar structure set a bound $g_{ae}\lesssim2.3\times10^{-11}$~\cite{Redondo:2013wwa,Gondolo:2008dd} that is weaker than other stellar constraints. However, CAST is able to probe axions coupled to both electrons, important in the production, and photons, for the detection. 
The bound obtained by CAST is $g_{a\gamma}g_{ae}<8.1\times10^{-23}~{\rm GeV}^{-1}$ at $95\%$ CL for $m_{a}\lesssim10~{\rm meV}$~\cite{Barth:2013sma}.
Since IAXO will probe a motivated region of the axion parameter space, some studies discussed, in case of a discovery, the possibility of IAXO to distinguish between axion models~\cite{Jaeckel:2018mbn}, determining its mass~\cite{Dafni:2018tvj} and also probe solar properties~\cite{Jaeckel:2019xpa} (see Ref.~\cite{Hoof:2021mld} for a comprehensive study). 

Solar axion observations can also be used to probe the axion-nucleon coupling, a fundamental property of the most motivated axion models. 
The idea of axions produced in nuclear deexcitation processes is old~\cite{Moriyama:1995bz,Krcmar:1998xn} and CAST data was already used to set constraints on axions coupled to both photons and nucleons~\cite{CAST:2009jdc,CAST:2009klq}.
An updated analysis of the next generation helioscope sensitivity was recently presented in Ref.~\cite{DiLuzio:2021qct}. 
In addition, axions produced in the $p + d \rightarrow\rm{^3He}+ a(5.49 \,\  MeV)$ reaction have been probed, through photon and electron interactions, by experiments designed for neutrinos, like Borexino~\cite{Borexino:2012guz}, and the perspectives for the future Jiangmen Underground Neutrino Observatory (JUNO) are bright~\cite{Lucente:2022esm}.

This discussion motivates why the Sun is an excellent axion source. However, not all the FIPs produced in the Sun manage to escape its gravitational field. A small portion of these is gravitationally trapped around the star, forming a basin~\cite{VanTilburg:2020jvl}. Thus, we expect that the density of non-relativistic FIPs is locally higher, enhancing the perspectives for these searches.

{\bf  Red giant stars - }
A star like the Sun, which burns hydrogen in the core, is called main-sequence star. 
Most of the stars in the Universe are in this phase. 
When the hydrogen in the core is exhausted, the nuclear source of energy to contrast the gravitational collapse disappears and stars with a mass $M\lesssim2M_{\odot}$ develop an inert helium core, surrounded by a burning shell of hydrogen. 
In this phase, the external layers of the star expand, increasing the stellar luminosity even though the surface temperature drops. This is the Red Giant (RG) phase.
The helium produced in the hydrogen shell continues to fall on the core, that becomes very dense and electrons are degenerate. In order to contrast the gravitational collapse, the degenerate core shrinks to increase the degeneracy pressure. In this process the gravitational attraction at the edge of the core, where the burning hydrogen shell lies, increases and heats up this layer. The luminosity increases as the temperature of the hydrogen shell grows and this process is mostly determined by the core mass. Together with the hydrogen layer, also the temperature of the core increases until, suddenly, the temperature is high enough to ignite helium. This process is very fast, given the steep temperature dependence of the helium burning rate. For this reason, this process is called helium-flash. The properties of the RG when the helium-flash happens determine the location of the so-called RGB tip in the color-magnitude diagram. 

The RGB tip is extremely sensitive on possible exotic losses.
The additional loss delays the He ignition giving time to the core to grow further and thus making the star at the RGB tip brighter. 
This observable provides a very effective way to constrain the axion-electron coupling. 
Indeed, axions are produced very efficinetly in RGs via Bremsstrahlung.
The latest analyses constrain $g_{ae}\lesssim1.48\times10^{-13}$ at $95\%$ CL for light (sub-keV) axions~\cite{Straniero:2020iyi,Capozzi:2020cbu}.
The standard energy-loss in a RG core is dominated by plasmon decay into neutrinos $\gamma^{*}\to \nu\bar{\nu}$. This neutrino production rate might be increased in presence of a large NMM. The constraint evaluate in Ref.~\cite{Capozzi:2020cbu} is $\mu<1.2\times10^{-12}~\mu_{\rm B}$, where $\mu_{\rm B}$ is the Bohr magneton. These are the most stringent bounds on FIP properties, showing the power of astrophysical studies compared to laboratory experiments. Just like in the Sun, DPs can be copiously produced also in RGs by photon-DP oscillations in the DP mass range $3~{\rm keV}\lesssim m_{\chi}\lesssim 30~{\rm keV}$. A simple criterion, requiring that the emissivity in the core is less than $10$~${\rm erg}{\rm g}^{-1}{\rm s}^{-1}$, excludes $\epsilon\gtrsim10^{-15}$~\cite{An:2014twa}.

{\bf  Horizontal Branch stars - }
As explained in the previous section, a relatively low mass star ignites helium in its core after the RG phase. 
The stable configuration with a core composed by helium, surrounded by a burning hydrogen shell, is called Horizontal Branch (HB) star. 
The high temperature reached during this stage ($\sim10$~keV) makes electrons in the core non-degenerate. 
Thus, HB stars are efficient in producing axions coupled to photons via the Primakoff process. 
The energy subtracted by axions speeds up the HB phase, compared to the RG phase that is unaffected by axion production~\cite{Raffelt:1987yu}. 
Indeed, the high electron degeneracy suppresses the axion production via Primakoff conversion. 
The duration of these phases is measurable by counting stars in a particular phase: the longer the duration, the more stars are found in a given phase. 
The most relevant observable in this respect is the $R$ parameter, defined as the ratio of the number of stars in the HB and in the RG phase, $R = N_{\rm HB}/N_{\rm RGB}$. 
This quantity is typically measured by counting stars in Globular Clusters (GCs), gravitationally bound systems of coeval stars differing only in their initial mass. 
In Ref.~\cite{Ayala:2014pea} it was obtained that $R=1.39\pm0.03$ from the analysis of 39 GCs. 
Exotic losses would reduce this parameter, since the HB phases is shortened compared to the RG one.
Precisely, in absence of exotic particles the duration of the HB phase is $\tau_{\rm HB}\simeq88.4$~Myr and the uncertainty on the $R$ parameter reflects into a maximal reduction of the $\sim 15\%$ within $2\sigma$. 

\begin{figure}[t!]
    \centering
    \includegraphics[width=0.95\columnwidth]{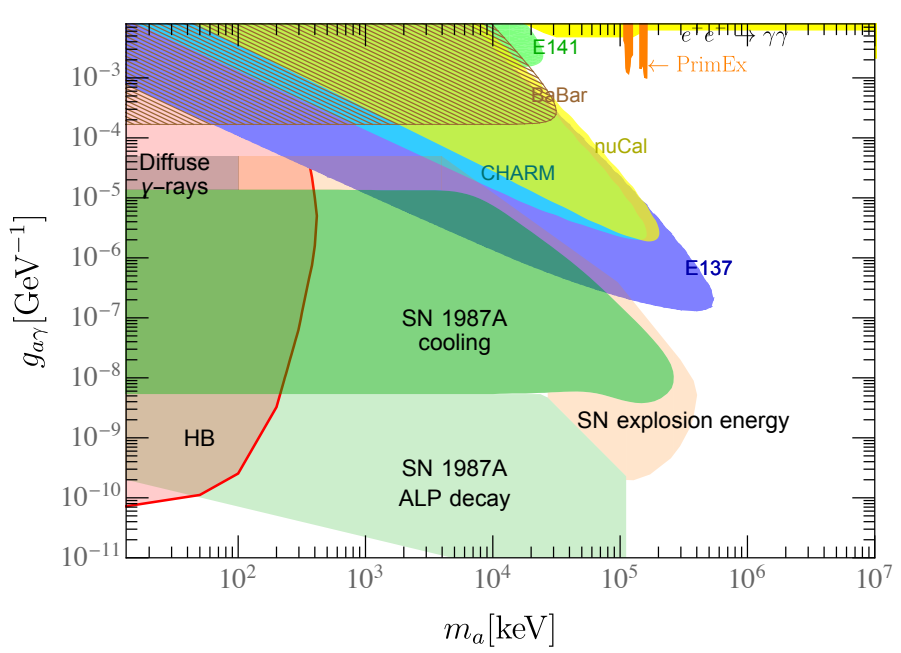}
        \caption{Bounds on the axion-photon coupling $g_{a\gamma}$ for massive axions.	The red region represents the HB bound~\cite{Carenza:2020zil,Lucente:2022wai}.
		The SN 1987A bound (green)~\cite{Lucente:2020whw} and the experimental limits from Refs.~\cite{Dolan:2017osp,Dobrich:2019dxc}, are also shown.
		The light green bound refers to the constraint obtained in Ref.~\cite{Jaeckel:2017tud}. The orange constraint is obtained by considering the energy-loss criterion applied to low-energy SNe~\cite{Caputo:2022mah}.}
    \label{fig:massive}
\end{figure}

Applying this criterion to axions coupled to photons, it is obtained that $g_{a\gamma}<0.65\times10^{-10}~{\rm GeV}^{-1}$ at $95\%$ CL~\cite{Ayala:2014pea,Straniero:2015nvc} for massless axions. 
This constraint was later generalized to masses up to $m_{a}\sim 300~{\rm keV}$~\cite{Carenza:2020zil} and carefully taking into account that radiative axion decays constitute a new energy transfer channel~\cite{Lucente:2022wai}. 
The HB constraint is shown in Fig.~\ref{fig:massive} and compared to other bounds on heavy axions.

The DP emission also causes a shortening of the HB phase, which is significant if the DP luminosity exceeds the $10\%$ of standard HB luminosity (assumed to be $60L_{\odot}$). This criterion leads to a bound that reaches down to $\epsilon\lesssim 10^{-15}$ for $m_{\chi}\sim 2.6~{\rm keV}$~\cite{An:2014twa}.

Eventually, a HB star burns all the helium in the core, and it is left with an inert carbon and oxygen core surrounded by a shell of burning helium. A star in this phase is in the Asymptotic Giant Branch. 
This phase can also be used to test FIPs since these stars are very bright and their properties are well-known. 
In addition to reveal exotic energy-losses, this phase gives a snapshot of how the nucleosynthesis works and its possible interplay with FIPs. 
The case of axions is discussed in this context~\cite{Dominguez:1999gg}, concluding that energy-losses would lower the mass of the remnant core, a White Dwarf (WD), and affect the temperature of the burning helium shell, affecting the chemical composition of the star.

{\bf  White Dwarfs - }
A WD is the last evolutionary stage for stars with a mass $\lesssim8\,M_{\odot}$. 
The chemical composition of WDs depends on the mass of their core: for masses smaller than $\sim0.4M_{\odot}$ they are composed by helium; for masses up to $\sim 1.05M_{\odot}$, carbon and oxygen accumulate in the core and more massive WDs are made of oxygen and neon.
In a WD the gravitational collapse is balanced by the pressure of degenerate electrons, making the core isothermal given their high thermal conductivity. 
The long evolution of WDs, on the scale of Gyr, is a simple cooling. 
A young WD cools mostly through neutrinos produced in plasmon decay in the core.
Older WDs lose energy via surface photon emission. 
FIPs would change this picture by accelerating the cooling process.

Axions coupled to electrons can be produced in WDs via electron Bremsstrahlung~\cite{Isern:1992gia} (see Ref.~\cite{Carenza:2021osu} for an extensive discussion on this process).
The axion emission, for sufficiently high axion-electron coupling, might be competitive with standard losses soon after the neutrino emission ceases to be dominant.
The WD cooling rate is measured with asteroseismology techniques, but monitoring the evolution of a single WD requires an accurate understanding of the internal structure and composition of the star. By constrast, it is possible to reconstruct global properties of WDs via the WD Luminosity Function (WDLF), a relation between WD mass and luminosity. However, the WDLF is sensitive to properties of the stellar population as the Initial Mass Function and Star Formation Rates. This second method based on the WDLF was used to set a constraint $g_{ae}\lesssim 1.4\times10^{-13}$ on the axion-electron coupling~\cite{MillerBertolami:2014rka}.
In a similar fashion to the RG case, a large NMM would enhance the neutrino production via plasmon decay. Observations exclude a NMM larger than $\mu\gtrsim5\times10^{-12}\,\mu_{\rm B}$~\cite{MillerBertolami:2014oki}. 

Another important observable is related with a peculiar class of WDs: the pulsating WDs. The pulsation in a WD is caused by a competition between the cooling rate, that increases the core degeneracy pressure, and the gravitational contraction. Pulsations are especially important for young WDs (see Ref.~\cite{Corsico:2019nmr} for a comprehensive review).
Since the pulsation is related to cooling processes, measurements of the pulsation period are able to probe FIPs~\cite{Corsico:2012sh,Isern:2019nrg}. The analysis of this observable reveals a hint for a DFSZ axion with $g_{ae}\sim5\times10^{-13}$~\cite{Corsico:2012ki} and a constraint $g_{ae}\lesssim7\times10^{-13}$~\cite{Corsico:2016okh}. Regarding non-standard neutrino properties, a large NMM can be similarly constrained giving $\mu\lesssim5\times10^{-12}\,\mu_{\rm B}$~\cite{Corsico:2014mpa}.

Recently, it was pointed out that the WD Initial-Final Mass Relation (IFMR) is sensitive to exotic physics~\cite{Dolan:2021rya}. The IFMR maps the initial mass of a main sequence star to the final mass of the WD into which it evolves. This is a completely new approach for WD constraints on FIPs. When applied to the axion case, this idea leads to a constraint competitive with the HB one, especially for heavy ($m_{a}\sim300$~keV) axions~\cite{Dolan:2021rya}.

{\bf  Bounds from Massive Stars - }
The use of massive stars in constrained FIPs have been considerably more limited. 
In recent years, however, some works have attempted to extract information on the FIP-matter couplings through the observations of these stars. 
In Ref.~\cite{Heger:2008er}, it was shown that even a relatively small neutrino magnetic moment, $(2-4) * 10^{-11} \mu_B$ (below the experimental bound but above the WD and RG constraints) can cause observable changes to the evolution of a massive ($10-20M_{\odot}$) star.
Specifically, it would cause a shifts in the threshold masses for creating core-collapse supernovae and oxygen-neon-magnesium white dwarfs, and the appearance of a new type of supernova in which a partial carbon-oxygen core explodes within a massive star. 

More recently, massive stars in the $8-12M_{\odot}$ range were employed to extract bounds on the axion-photon coupling~\cite{Friedland:2012hj}. 
During the He-burning stage, these stars experience a journey in the color magnitude diagram when their surface temperature increases and then decreases again, maintaining roughly a constant luminosity. 
This short period of the stellar evolution is called blue loop and its existence is assessed by observation of the Cepheid stars. 
In the presence of exotic losses this stage may disappear, leading to a bound on the axion-photon coupling \mbox{$g_{a\gamma}\lesssim0.8\times10^{-10}~{\rm GeV}^{-1}$}~\cite{Friedland:2012hj,Carosi:2013rla}.

{\bf  Supernovae - } 
As we discussed, stars burn lighter elements to produce energy and, as byproduct, heavier elements. For stars more massive than $8\,M_{\odot}$, it is possible to reach core temperatures high enough to make carbon burning possible. 
Later in the evolution, the core temperature will increase and permit the fusion of heavier and heavier elements.
This will result in an onion-like structure, with the heaviest elements in the most inner shells. 
This procedure ends with the creation of an iron core. 
Iron, the most stable nucleus, cannot undergo fusion to produce energy. 
As the core mass increases, it becomes progressively more difficult to contrast the gravitational collapse with the electron degeneracy pressure. When the core reaches the Chandrasekhar mass, $1.4 M_{\odot}$, the collapse shrinks the core up to nuclear densities, $\rho\sim 10^{14}{g\rm cm}^{-3}$, and matter will stop compressing. 
This is the origin of a bounce of external layers onto the degenerate core, thus the implosion becomes an explosion triggering a shock wave that drives the Supernova (SN) explosion. The core is a Proto-Neutron Star with a radius $r\sim 10{\rm km}$ and a density comparable to the nuclear one.
%The only Galactic SN detected with modern technologies is SN 1987A. 
SNe are expected to produce a very large number of neutrinos, whose detection could reveal a plethora of information about the SN physics and the explosion mechanism.  
%The most important SN messengers are neutrinos.
So far, it was possible to detect SN neutrinos only in the case of SN 1987A, over 30 years ago.
% when they were detected by three different detectors in coincidence with SN 1987A. 
The small sample of neutrino events collected by Kamiokande II, Baksan and Irvine–Michigan–Brookhaven detector was studied in details suggesting that the observed signal is compatible with the standard neutrino-driven explosion~\cite{Loredo:2001rx}.

The SN neutrino burst can reveal also FIP footprints. 
An exotic emission would constitute an additional energy-loss channel, stealing energy from the SN core and reducing the neutrino burst duration. 
A shorter neutrino burst corresponds to fewer events from the observed SN 1987A.
Thus, a very efficient FIP cooling would be incompatible with the measurements associated with SN 1987A~\cite{Raffelt:2006cw}. 
This energy-loss argument was originally applied to axions coupled with nucleons and produced by Nucleon-Nucleon (NN) bremsstrahlung, $NN\rightarrow NNa$. 
The most recent analysis in Ref.~\cite{Carenza:2019pxu} highlights a reduction of the OPE emissivity up to more than one order of magnitude when accurate nuclear interactions and high density corrections are accounted for. 
A significant advancement followed from the realization that even the small ($\sim1\%$) fraction of pions in a SN core~\cite{Fore:2019wib} is sufficient for the process $\pi^{-}p\to n a$ to dominate the axion emission~\cite{Carenza:2020cis}. 
These processes were included in SN simulations, showing a good agreement with theoretical expectations~\cite{Fischer:2021jfm}.
Recently, the SN axion emission was revisited to include the contact interaction, a four particle vertex nucleons-axion-pion, and extend the calculation to non-vanishing axion masses~\cite{Lella:2022uwi}. 
Furthermore, it was noted that the pion-conversion rate is significantly enhanced when including the intermediate $\Delta$ resonance in the pion-nucleon scattering~\cite{Ho:2022oaw}. 

\begin{figure}[t]
    \centering
    \includegraphics[width=0.95\columnwidth]{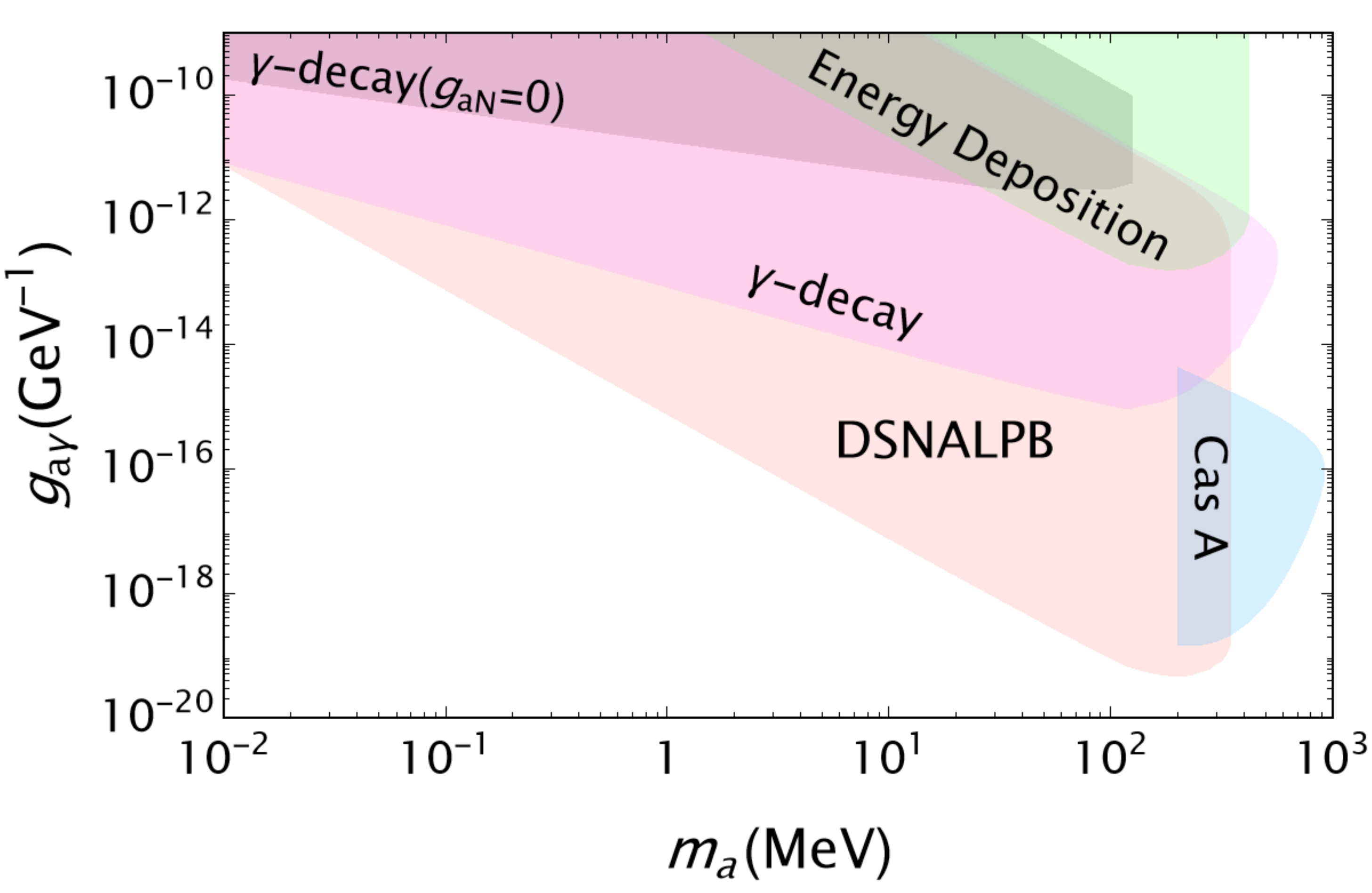}
    \caption{\label{fig:gagamma}Bounds on the axion-photon coupling $g_{a\gamma}$ for massive axions from SNe coupled to protons, $g_{ap}=4.8\times10^{-10}$ and $g_{an}=0$. The green area is referred to the energy deposition mechanism in SN 1987A, the magenta region is associated to radiative decays of axions from SN 1987A, the pink area is the excluded region by means of of the Diffuse SN axion background, while the light blue one refers to the decay of axions gravitationally trapped around the SN remnant Cas A. Finally, the gray region is the excluded region discussed in Ref.~\cite{Jaeckel:2017tud}, considering  radiative decays of massive axions produced by the Primakoff process. Figure taken from Ref.~\cite{Lella:2022uwi}.}
\end{figure}

A whole set of additional SN considerations apply to very massive axions (several 10 to 100 MeV), coupled to nucleons and to photons.
Such particles can still be efficiently produced in the SN core, $T\sim 30$ MeV, through nuclear bremsstrahlung and pion processes, and decay into photons outside the SN envelop, producing 
high energetic and potentially detectable $\gamma$ rays. 
We are not going to discuss these cases in details here but provide a summary of the current constraints in Fig.~\ref{fig:gagamma}.
The updated results are thoroughly discussed in Ref.~\cite{Lella:2022uwi}.
%The recent progress shows that this field is still subject of active and lively research. 
%

Various other axion properties can be probed by SNe. For example, axions coupled to photons would be produced in SNe via Primakoff conversion and, for massive axions, photon coalescence $\gamma\gamma\to a$~\cite{Lucente:2020whw}. The SN bound in this case is weaker than the HB bound for light axions, but it is an important constraint for heavy axions, up to $m_{a}\sim 300$~MeV as shown in Fig.~\ref{fig:massive} together with HB star and laboratory constraints on heavy axions.

Similarly also axions coupled to electrons and produced by Bremsstrahlung and electron-positron fusion $e^{+}e^{-}\to a$, are constrained by a similar argument~\cite{Lucente:2021hbp}. Once more, the SN bound allows to probe heavy axions in a region of the parameter space that is not in the reach of laboratory experiments. The original contribution of Ref.~\cite{Ferreira:2022xlw} discussed the phenomenology of axions coupled at tree level with electrons, acquiring a loop-induced photon coupling. 

Several other works applied the SN cooling argument to constrain various FIPs as DPs~\cite{Hardy:2016kme,Chang:2016ntp}, sterile neutrinos~\cite{Mastrototaro:2019vug} and neutrinos with a large NMM~\cite{Raffelt:1999gv}.

%Until now we discussed the simplest and most general effect of FIPs on a SN explosion. 
In addition to cooling effects, the production of exotic particles in SNe may also affect other observables. 
Recent analyses discussed, for example, the effects of axions on nucleosynthesis occurring in massive stars before the SN phase~\cite{Aoyama:2015asa}. 
Other investigations focused on the impact that light FIPs have on the threshold mass for which a star ends in a SN explosion.
Ref.~\cite{Dominguez:2017mia} found that a shift of $\sim 2M_{\odot}$ would be in contrast with observations, paving the road for the development of a new FIP probe.
More intriguingly, Ref.~\cite{Straniero:2019dtm} found that the relation between initial SN progenitor mass and final SN luminosity shows a slight preference for exotic losses, otherwise many SN events appear dimmer than expected.

{\bf  Neutrons Stars -}
The remnant left after a SN explosion is a Neutron Star (NS), a very dense star composed mostly of degenerate neutrons. 
A NS cools, for the first hundred thousands of years, via neutrino emission through direct $nn\to pn e^{-}\bar{\nu}_{e}$, $np\to pp e^{-}\bar{\nu}_{e}$, inverse Urca processes $pne^{-}\to nn \nu_{e}$ and $pp\,e^{-}\to np \nu_{e}$, and Bremsstrahlung $NN\to NN\nu\bar{\nu}$. 
In addition, when the nuclear matter in the NS cools enough to become superfluid, formation and breaking of Cooper pairs of nucleons are other neutrino production mechanisms. In the final NS stage, the cooling is dominated by photon emission from the surface.

Axions coupled to nucleons can be produced by the same processes that produce neutrinos and their emission would accelerate the NS cooling. The analysis of the NS in the SN remnant HESS J1731-347 gives a constraint on the axion-neutron coupling $g_{an}<2.8\times10^{-10}$ at $90\%$ CL for $m_{a}\lesssim 10$~keV~\cite{Beznogov:2018fda}. The advantage of this study is that the young NS analyzed is hotter than NSs in a more advanced phase, therefore more axions are produced. The first direct detection of the NS cooling was performed in Ref.~\cite{Heinke:2010cr} for Cassiopeia A (Cas A), supporting the hypothesis of a superfluid state in which the Cooper pair processes are efficient~\cite{Page:2010aw,Shternin:2010qi}. The time behavior of the NS average temperature was used to constrain exotic losses associated with axions~\cite{Leinson:2014ioa,Sedrakian:2015krq}. These studies focus on simulating the period when the superfluid transition occurs, but more comprehensive studies, simulating the whole NS life were also performed obtaining an axion constraint comparable with the SN one~\cite{Hamaguchi:2018oqw}. A recent study considered NSs with an age around a few hundred thousand years, a moment of the NS cooling in which the axion losses should dominate, concluding that the constraint is also in this case at the level of SN 1987A~\cite{Buschmann:2021juv}.

\subsubsection{Conclusions}
Stars are exceptional laboratory for FIPs, as well as factories producing possibly detectable FIPs fluxes. 
The great majority of stellar studies refer to axions, though several considerations have been extended to other FIPs as well. 
The enormous progress in the last couple of years, since the publication of the previous FIP overview~\cite{Agrawal:2021dbo}, shows that this is a quite active and lively research field.

%\subsubsection{Acknowledgements}
%The work of P.C. is supported by the European Research Council under Grant No.~742104 and by the Swedish Research Council (VR) under grants  2018-03641 and 2019-02337. 

% \bibliographystyle{bibi}
% \bibliography{biblio.bib}

% \end{document}

%-------------------------------------------
\subsection{Review of axion programme at DESY -- {\it A.~Lindner}} 
\label{ssec:lindner}
{\it Authors: D. Heuchel, A. Lindner, I. Oceano Contact: <axel.lindner@desy.de>} \\
%------------------------------------------- 
% \documentclass{article}
% \usepackage{blindtext}
% \usepackage{multicol}
% \usepackage{geometry}
% \usepackage{epsfig}
% \usepackage{amssymb}
% \setlength{\columnsep}{1cm}
% \usepackage{caption}
% \usepackage{graphicx}
% \usepackage{subfig}
% \usepackage{siunitx}
 
% \title{The DESY axion search program}
% \author{D. Heuchel, A. Lindner, I. Oceano \\
% Deutsches Elektronen-Synchrotron DESY, Notkestr. 85, 22607 Hamburg, Germany}
% \date{\today}
% \geometry{
%  a4paper,
%  total={170mm,257mm},
%  left=20mm,
%  top=20mm,
%  }
%   \begin{document}
% \maketitle

% %\begin{multicols}{2}

\subsubsection{Introduction}
Feebly Interacting Particles (FIPs) might offer the solution to (some of) the open questions beyond the Standard Models of particle physics and cosmology.   
At DESY in Hamburg, three non-accelerator-based experiments will search for FIPs as dark matter candidates (ALPS~II, BabyIAXO) or constituting the dark matter in our home galaxy (MADMAX). Such experiments have to strive for sensitivities many orders beyond the reach of collider or beam-dump experiments. 
Among FIPs, the axion as motivated by the lack of any observed CP violation in Quantum Chromodynamics (QCD) \cite{Abel:2020pzs, Peccei:1977hh,Weinberg:1977ma,Wilczek:1977pj}, is frequently being used as a benchmark to compare the sensitivities of  experimental efforts.
Axions result from a new global Peccei-Quinn symmetry $U(1)$ that spontaneously breaks at the scale $f_a$.
%While the weak interactions are known to violate CP, the strong interactions also contain a CP-violating term in the Lagrangian,  $\frac{\theta}{32 \pi^2} G_{\mu\nu}\widetilde{G}{\mu\nu}$, where $G_{\mu\nu}$ is the gluon field strength. For non-zero quark masses, this term leads to (unobserved \cite{Abel:2020pzs}) CP-violating effects of the strong interactions. This so-called ``strong CP problem" is often exemplified by the lack of observation of a neutron electric dipole moment down to a present experimental upper limit $10$ orders of magnitude smaller than what is expected from a CP-violating QCD. 
\begin{figure}[!ht]
\begin{center}
\includegraphics[scale=0.5]{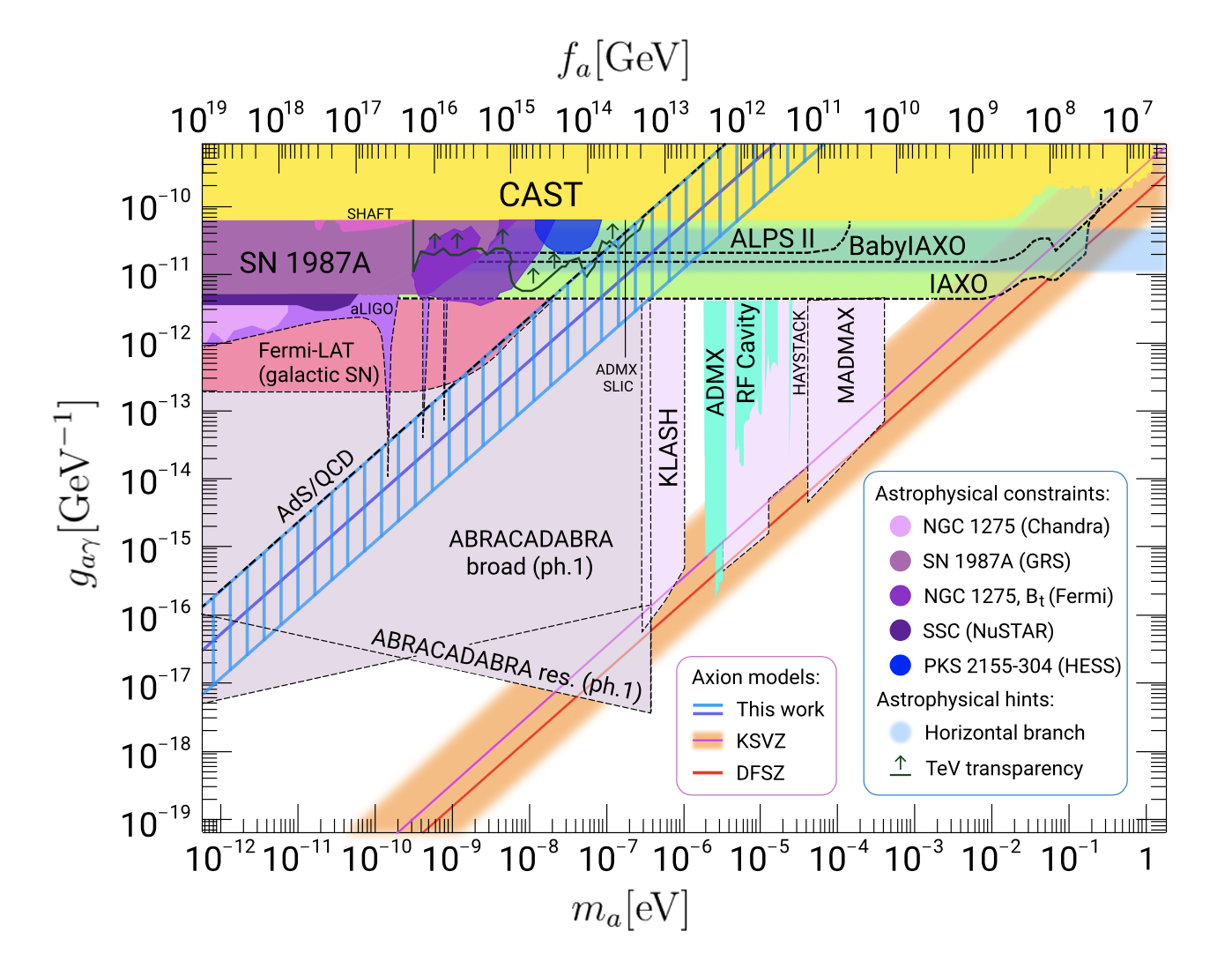}
\caption{Axion-photon coupling as a function of axion mass and decay constant for various axion models. The plot shows the existing and predicted constraints (dotted lines) from experiments together with those derived from astrophysical data and theoretical models. We thank the authors of \cite{Sokolov:2021ydn} for the figure.\\
This article reports on the statuses of the experiments ALPS~II, BabyIAXO and MADMAX.}
\label{Fig:ax-phCoupling}
\end{center}
\end{figure}
For the detection of axions, all three experiments rely on the axion-photon mixing in a background magnetic field (the Sikivie effect \cite{Sikivie:1983ip}) given by the following Lagrangian term:
\begin{equation}
    L_{a\gamma} =  g_{a \gamma } \phi _a \vec{E}\cdot \vec{B}
\end{equation}
where $g_{a \gamma }$ is the axion-photon coupling strength proportional to $1/f_a$ times a model dependent constant, $\phi_a$ is the axion field, $\vec{E}$ is the oscillating electric field of the photon and $\vec{B}$ represents the static magnetic field of the experiments.
Figure \ref{Fig:ax-phCoupling} shows the axion-photon coupling as a function of axion mass, together with existing and predicted constraints from various experiments and astrophysical observations. For reference, the axion-photon couplings in the KSVZ models \cite{Kim:1979if,Shifman:1979if} and in the DFSZ model \cite{Dine:1981rt,Zhitnitsky:1980tq} are compared to a more recent model \cite{Sokolov:2021ydn} indicating the large range of theoretical predictions.

Additional motivation to search for axions and axion-like particles (ALPs, which are not related to the CP conservation in QCD) comes from astrophysical riddles related to the evolution of stars and the propagation of high energy photons in the universe. 
They have been interpreted in various ways, however, a global analysis of all the data indicates a preference for axions and ALPs \cite{Pallathadka:2020vwu,DiLuzio:2021ysg}. 
Interestingly, just one axion with a mass $m_a \approx \SI{e-7}{\electronvolt}$ and $g_{a \gamma } \approx \SI{e-11}{\giga \electronvolt^{-1}}$, as predicted by \cite{Sokolov:2021ydn}, could very well explain the above mentioned riddles, constitute the dark matter and explain the CP conservation in QCD. 
This axion might even be in reach of ALPS~II and BabyIAXO.
In the remainder of this text, we will not strictly differentiate between axions and ALPs anymore.

%The most common experimental searches for axions rely on the EM interaction mediating the axion-photon coupling.
Basically, axions and other FIPs are searched for by:
\begin{itemize}
    \item Haloscopes looking for axions constituting the cold dark matter halo of our home galaxy;
    \item Helioscopes relying on the Sun as a source of relativistic axions;
    \item Purely laboratory-based experiments not requiring cosmological or astrophysical assumptions.
\end{itemize}

 At DESY, all three techniques are exploited by MADMAX, IAXO and ALPS~II, respectively.
 All are reusing the infrastructure of the former HERA collider\footnote{https://www.desy.de/sites2009/site\_www-desy/content/e409/e69110/e4948/e5101/e5142/e5144/infoboxContent6626/\\
 HERA\_en\_eng.pdf}.
Short summaries of these experiments are given in the following section.
 
%\end{multicols}

\subsubsection{Axion searches at DESY}
{\bf MADMAX - }

The MAgnetized Disks and Mirror Axion eXperiment (MADMAX) collaboration \cite{MADMAX:2019pub} lead by the Max-Planck Institute for Physics (Munich, Germany) is developing a new approach to search for axion dark matter in a mass region around $\rm{100\,\mu eV}$ currently not accessible by more traditional approaches based on microwave cavities. 
This mass region is promising as it corresponds to the mass range predicted by post-inflation models \cite{Borsanyi:2016ksw} and the high mass region of pre-inflation models.
%\footnote{The Peccei-Quinn symmetry breaking in the early universe could happen before or after an inflation phase.}.

MADMAX relies on the conversion of dark matter axions into microwave photons, where the photon energy is given by the axion mass plus an order $\rm{10^{-6}}$ correction due to the dark matter velocity distribution.
The main idea of MADMAX is to exploit the constructive interference of electromagnetic radiation emitted by different surfaces to resonantly enhance the conversion of axions to photons. This is achieved through a series of parallel dielectric disks with a mirror on one side, all within a magnetic field $B$ parallel to the disk surfaces, creating a so-called dielectric haloscope.
\begin{figure}[!ht]
\begin{center}
\includegraphics[scale=0.4]{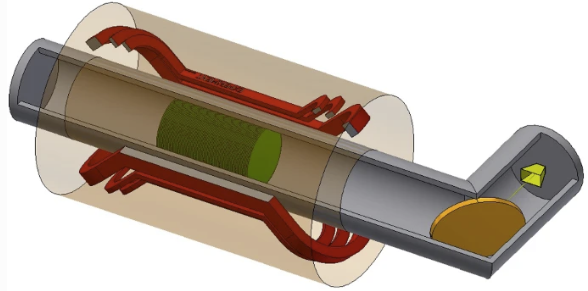}
\caption{Preliminary baseline design of the MADMAX approach (copied from \cite{MADMAX:2019pub}). In the Figure, the three parts of the experiment are shown: (1) magnet (red racetracks); (2) booster, consisting of the mirror (copper disk at the far left) and the dielectric disks (green); (3) the receiver, consisting of the horn antenna (yellow) and the cold preamplifier inside a separated cryostat. The focusing mirror is shown as an orange disk at the right .}
\label{Fig:MADMAX}
\end{center}
\end{figure}
The conceptual sketch of the MADMAX experiment is shown in Figure \ref{Fig:MADMAX}. 
The magnet provides a dipole field of \SI{\sim 9}{\tesla} and an opening of \SI{1.35}{\meter}: dielectric discs with a diameter of more than \SI{1}{\meter} and a mirror reflecting the signal towards the receiver system. 
The distance between the discs can be changed to tune the resonance frequency, which is required to probe different axion masses. 
The booster enhances the axion-induced power by up to four orders of magnitude.
The receiver shall enable the detection of signal photons in the frequency range of \SIrange{10}{100}{\giga \hertz} with a sensitivity of \SI{\sim e-22}{\watt}.

MADMAX is planned to be located in HERA's North Hall \SI{\sim 25}{\meter} below surface. Here, the so-called Cryoplatform will provide the magnet with liquid helium. Furthermore, the iron yoke of the former HERA-experiment H1 will care for the shielding of the magnetic fringe fields and even provide a field strength enhancement at the booster position by about \SI{10}{\percent}.
Initial RF background measurements have demonstrated the suitability of the location.

In the last year, the collaboration has achieved very substantial progress regarding all components of the experiment.
\begin{itemize}
    \item The design of the huge dipole magnet is based on a superconducting cable-in-conduit (CIC). To protect the magnet, a reliable quench-protection system is required in case superconductivity breaks down anywhere in the magnet. In a complex test campaign, CEA/Saclay has clearly demonstrated that the quench propagation velocity within the CIC is fast enough to facilitate the quench detection. Thus, the magnet development has mitigated a crucial potential show-stopper.     
    \item Handling, mounting and positioning of dielectric discs made out of sapphire could be demonstrated. Again, one of the potential show-stoppers could be fully mitigated: a dedicated piezo-motor jointly developed with the company JPE was tested in a strong magnetic field of \SI{5.3}{\tesla} at cryogenic temperatures in vacuum without any issues.
    \item A series of prototype booster tests took place at MPI Munich and CERN and are planned for the future. At CERN the collaboration is using the MORPURGO magnet with a dipole field of up to \SI{1.6}{\tesla} when test beams are shut down. In the year 2022, it was successfully demonstrated that the CERN test beam area allows for physically interesting and competitive axion-like particle dark matter searches. At Munich, a reliable and stably working calibration procedure for a so-called closed-booster-system has been developed. This was extremely important on the path towards a full understanding of the response of the final MADMAX experiment to axion dark matter. 
\end{itemize}

The critical path in the future schedule of MADMAX is mainly given by the availability of funds for the large dipole magnet. Data taking with the final set-up may start in 2030, but already earlier a new prototype magnet could allow for very competitive direct dark matter searches at DESY.

{\bf BabyIAXO and IAXO - }
The International Axion Observatory (IAXO) is a next generation axion helioscope searching for solar axions and ALPs with unprecedented sensitivity. X-ray photons produced via the Sikivie effect (typically in the range of \SIrange{1}{10}{\kilo \electronvolt}) in the magnet bores are focused by high precision X-ray telescopes down to small focal spots at ultra-low background X-ray detectors \cite{IAXO:2019mpb}. The experiment's main goal is to improve the axion-photon coupling sensitivity $g_{a\gamma}$ by more than one order of magnitude with respect to its predecessor experiment CAST \cite{CAST:2007jps}, as illustrated by Figure \ref{Fig:ax-phCoupling}.
IAXO will probe also for the astrophysical hints and search for axions in the eV mass range, which is not accessible in any other axion experiment. 
Furthermore, the physics program is complemented with the possibility to probe different axion production models in the Sun by investigating the axion-electron $g_{ae}$ and the axion-nucleon $g_{an}$ couplings  \cite{Jaeckel:2018mbn,DiLuzio:2021qct}.
At a later point in time, the IAXO magnet could be used to search for dark matter halo axions by the accommodation of additional equipment, like microwave antennas and cavities similar to the RADES project at CERN \cite{IAXO:2019mpb}. 

\begin{figure}[!ht]
\begin{center}
\includegraphics[scale=0.4]{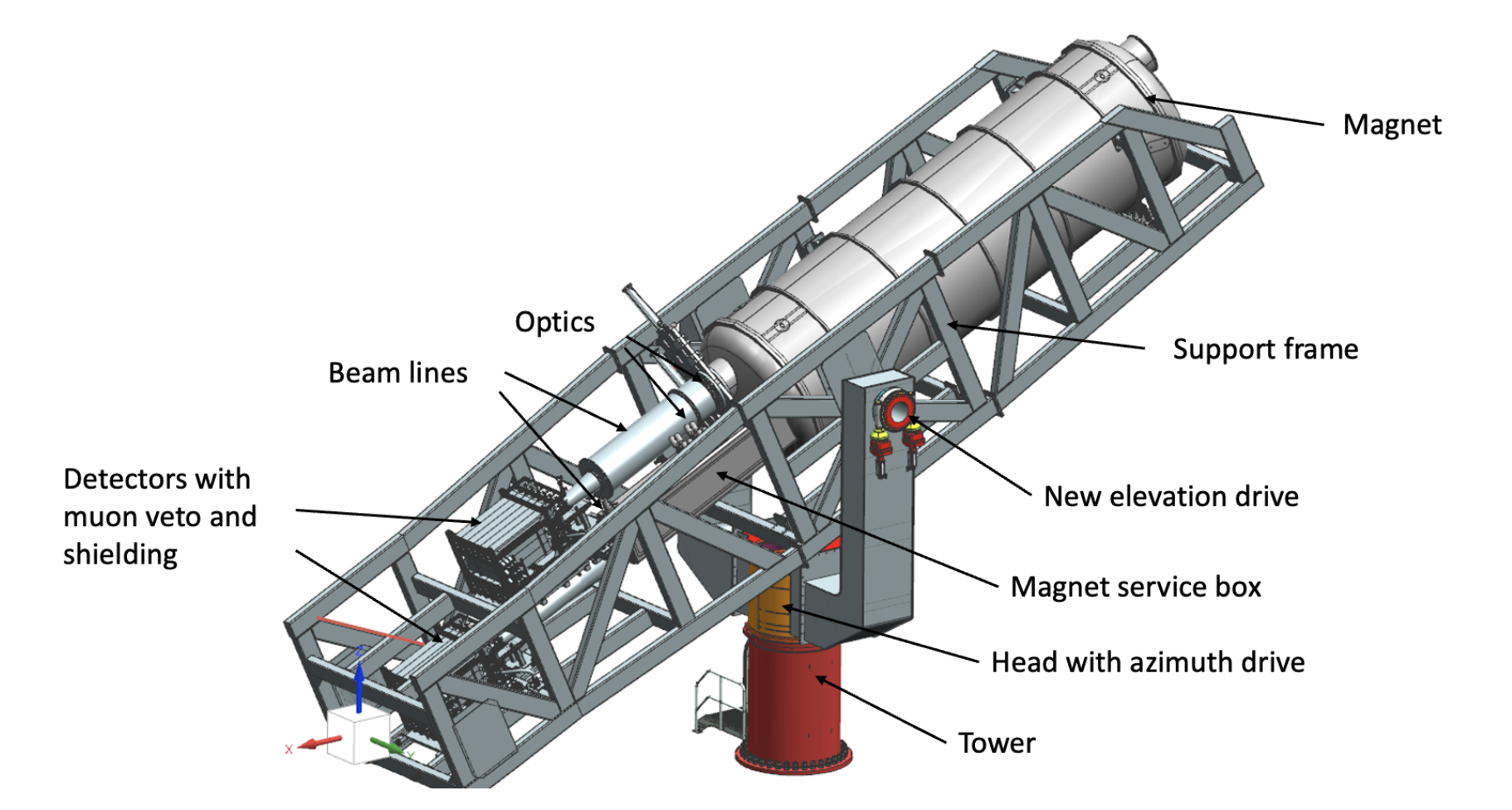}
\caption{Overview of the BabyIAXO experiment. The system features a total length of \SI{\sim 21}{\meter} \cite{IAXO:2020wwp}.}
\label{Fig:BabyIAXO_CDR}
\end{center}
\end{figure}

The first step towards the full IAXO will be the BabyIAXO experiment.
It is conceived with two main objectives: First, to serve as a full technological prototype for all subsystems of IAXO to prove full system integration and hence mitigate risks. Second, to operate as a fully fledged helioscope with own potential for discovery, e.g. by exceeding CAST sensitivity on $g_{a\gamma}$ by a factor $\sim 4$ in the same axion mass range. BabyIAXO will be based on a \SI{10}{\meter} long superconducting magnet (\SI{\sim 2}{\tesla}) with two bores, each with a diameter of \SI{70}{\centi \meter}. As depicted by Figure \ref{Fig:BabyIAXO_CDR}, the two equipped individual detection lines will feature an X-ray optics and an ultra-low background X-ray detector each, with comparable parameters and dimensions as foreseen for IAXO \cite{IAXO:2020wwp}. 

The BabyIAXO magnet relies on a common coil design based on two flat racetrack-coils with counter-flowing currents providing a dipole field in both bores. Superconducting Al-stabilised Rutherford cables will allow for a challenging dry detector magnet concept achieved by cryocoolers and He-gas circulators. However, since the Russian invasion of Ukraine in February 2022, collaboration with Russian institutes is frozen and the collaboration cannot get the magnet cable from Russia as planned before.  
No vendor for such cables, produced with co-extrusion technology, is available in non-Russian industry anymore. Due to the fact that this problem concerns many other particle physics experiments as well, a global effort to find sophisticated alternatives has been initiated and coordinated by CERN, resulting in a dedicated workshop in September 2022.
Different possible solutions have been identified and are being followed up. 
It is hoped that by summer 2023,   updated time schedules and cost estimates for the BabyIAXO magnet can be provided.

With respect to the X-ray optics, BabyIAXO will use one XMM-Newton flight spare module on loan by ESA in one magnetic bore, while the other bore will be equipped with a custom-made and newly built X-ray optics module. To detect the X-rays with high sensitivity, a variety of detector technologies are considered, mainly divided into \textit{discovery} and \textit{energy resolving} detectors \cite{IAXO:2020wwp}. The baseline option relies on small gas chambers (typically \SI{3}{\centi \meter} thick and \SI{6}{\centi \meter} wide) read out by a finely segmented micro-mesh gas structures (Micromegas).
Similar detectors were already successfully operated in CAST. While the other detector technologies are in the R\&D phase, the BabyIAXO Micromegas detector prototypes are currently undergoing final tests in the low background environments of the laboratories at Canfranc. In addition, special efforts and studies are conducted with respect to radio-pure materials, high efficiency veto systems and dedicated shielding with a special focus on neutron backgrounds. Lastly, the main design of the structure and drive system, which re-uses a modified tower and positioner of a Cherenkov Telescope Array (CTA) prototype (from DESY in Zeuthen), is close to being finished and ready for production.

After the approval of the experiment to be hosted at DESY, the collaboration has already taken first steps towards its construction. Currently, the first data taking with the full BabyIAXO experiment is foreseen for 2028. However, the future schedule is mainly driven by the status and availability of the superconducting cable for the magnet.

{\bf ALPS~II - }
The Any Light Particle Search (ALPS) number 2 is a light-shining-through-a-wall (LSW) experiment \cite{Bahre:2013ywa}. 
It will improve sensitivity on the axion-photon coupling by a factor of $10^3$ compared to its predecessors. 
This jump in sensitivity will be achieved by a long string of superconducting dipole magnets and two mode-matched 
optical cavities before and after the light-tight wall as proposed for the first time more than 30 years ago \cite{Hoogeveen:1990vq}. The presence of these cavities will resonantly increase the probability of the production of axions and the probability of their re-conversion into photons.
For the first time, ALPS~II will allow probing for axions in a model-independent fashion beyond present-day limits from astrophysics. 
\begin{figure}[!ht]
\begin{center}
\includegraphics[scale=0.35]{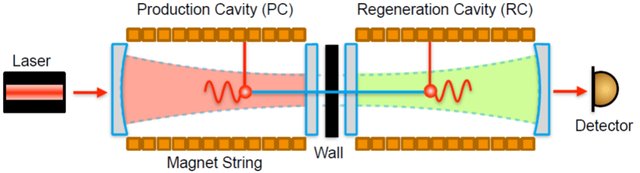}
\includegraphics[scale=0.3]{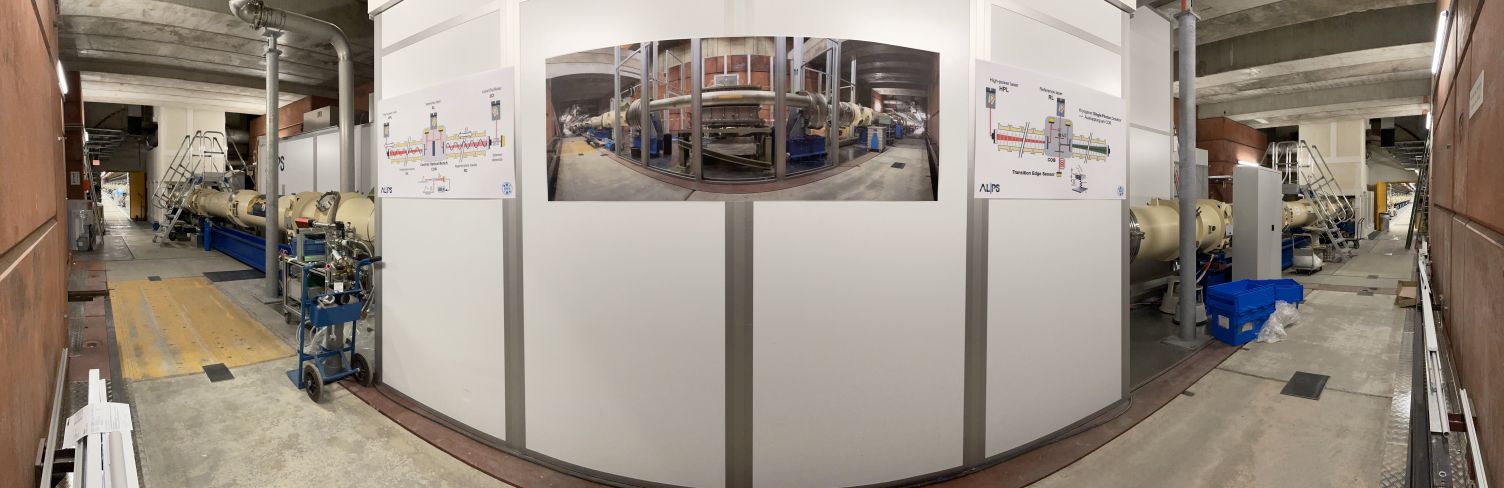}
\caption{Schematic layout of the ALPS~II experiment \cite{Albrecht:2020ntd}, left, and a panoramic picture of the installation in the straight HERA tunnel section around the North hall (right).}
\label{Fig:ALPSII}
\end{center}
\end{figure}

ALPS~II, see Figure \ref{Fig:ALPSII}, consists of two \SI{122}{\meter} long high-finesse optical cavities \cite{Ortiz:2020tgs} inside two strings of 12 superconducting HERA dipole magnets each \cite{Albrecht:2020ntd}. 
The probability for light converting to axions (which easily pass any barrier) and axions converting back to light is given by (for axion masses below \SI{0.1}{\milli \electronvolt})
\begin{equation}
    P_{\gamma \to \, a \, \to \gamma} = 
    \frac{1}{16} \beta_{PC} \beta_{RC} (g_{a\gamma \gamma} B L)^4 = 
    6 \times 10^{-38 } \beta{F}_{PC} \beta{F}_{RC} \times 
    \left(\frac{g_{a\gamma \gamma}}{10^{-10} \rm{~GeV^{-1}}} \frac{B}{1 \rm{~T}} \frac{L}{10 \rm{~m}}\right)^4
\end{equation}
resulting in $10^{-25}$ for the ALPS~II parameters $\beta_{PC}=5000$, $\beta_{RC}= 40000$, $B = \SI{5.3}{\tesla}$, $L = \SI{105.6}{\meter}$ and $g_{a\gamma \gamma} = \SI{2e-11}{\giga \electronvolt^{-1}}$ (motivated by astrophysics).
Thus, with \SI{30}{\watt} of \SI{1064}{\nano \meter} photons injected in the PC, about $2$ photons/day behind the wall are expected.
%The regenerated number of photons is \cite{Arias:2010bh}:
%\begin{equation}
%    n_{reg} = \frac{\eta}{16} (g_{a\gamma \gamma} F(qL_B) B_0 L_B )^4 \frac{P_{PC}}{h \nu} \beta _{PC}\beta _{RC}
%\end{equation}
%where $\eta$ is the power coupling efficiency between eigenmodes of both cavities, $L_B = 106 \rm{~m}$ is the length of each magnet string, $\beta_{PC}$ is the power build-up of the production cavity, $\beta_{RC}$ is the power build-up of the regeneration cavity and $|F(qL_B)|$ is the form factor $|F(qL_B)| \approx |\frac{2}{q L_B} \sin(\frac{q L_b}{2})|$ with $q= m_a^2/ 2 h \nu$.
%This is a typical phase-matching condition which accounts for the possible mass $m_a$, of the relativistic axion-like particles. 
%For masses $m_a<0.1 \rm{~meV}$, the form factor is essentially unity in ALPS II.  The coupling efficiency $\eta$ between the relativistic axion field and the RC eigenmode takes into account all transverse and spectral misalignments between the axion mode, which is identical to the PC eigenmode, and the RC eigenmode. Here $\eta$ is given in terms of the coupling between axions and photons and thus the coupling between the axionic field and the electromagnetic field would be given by $\sqrt{\eta}$. It will be possible to verify $\eta$ before and after measurements by opening a shutter in the light barrier and allowing the field transmitted by the PC to couple to the RC. 

ALPS~II will exploit two independent signal detection systems, each with very different systematic uncertainties. 
This approach will help to increase confidence that any signals observed with the same intensity in both detectors are indeed the result of a photon-axion conversion-re-conversion process. Since the two detectors require different optical systems to operate, they cannot be used in parallel. 
The first scheme to be implemented will be a heterodyne detection (HET) scheme \cite{Hallal:2020ibe,Spector:2023nap}, which measures the interference beat note between a laser, called a local oscillator (LO), and the regenerated photon field. 
The second detection scheme will use a Transition Edge Sensor (TES) \cite{Shah:2021wsp,Dreyling-Eschweiler:2015pja} operated at about \SI{100}{\milli \kelvin}. It allows for counting individual \SI{1064}{\nano \meter} photons with an energy resolution of \SI{\sim 7}{\percent}.

The installation of ALPS~II began in 2019.
In March 2022 the magnet string was successfully tested and in September 2022 the optics installation was completed for the initial science run.
The experiment is now close to start operation. 

This run based on the HET detection scheme will happen in the first months of 2023. 
It will not include the production cavity before the wall to optimise for the study of stray light, but already go  by a factor of 100 in the axion-photon coupling beyond earlier LSW experiments.
The full optical system will be used in the second half of 2023 and a HET science run with upgraded optics is planned for 2024. The further scheduling depends on the outcome of the HET science run, results of ongoing R\&D, resources and worldwide science advancements. The future program might include a TES-based science run, vacuum magnetic birefringence measurements, FIP searches with optimized optics and/or extension of the ALP mass reach and a dedicated search for high-frequency gravitational waves.

\subsubsection{Conclusion}
DESY in Hamburg is planning for three larger scale axion  experiments exploiting the LSW technique, solar axion and axion dark matter searches, all strongly pushed for by international collaborations.
The first one, ALPS~II, will start its science program soon. 
BabyIAXO is ready to launch construction when a new road-map for realizing the magnet exists and the funding is clarified; 
MADMAX is in the prototyping phase. 
\begin{figure}[!bp]
  \centering
  {\includegraphics[width=0.52\textwidth]{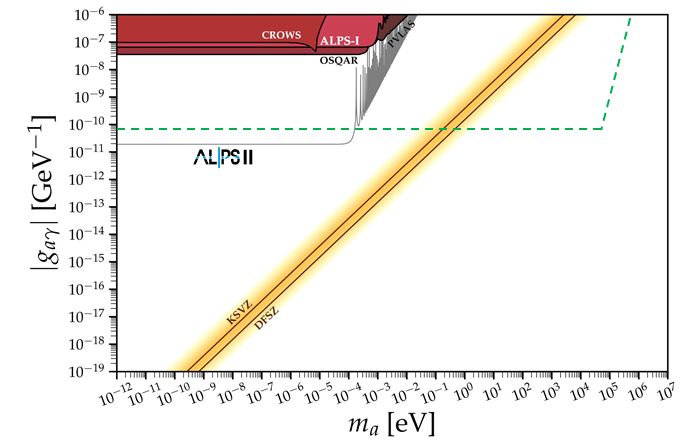}\label{Fig:ALPSIIlim}}
  %\hfill
  {\includegraphics[width=0.48\textwidth]{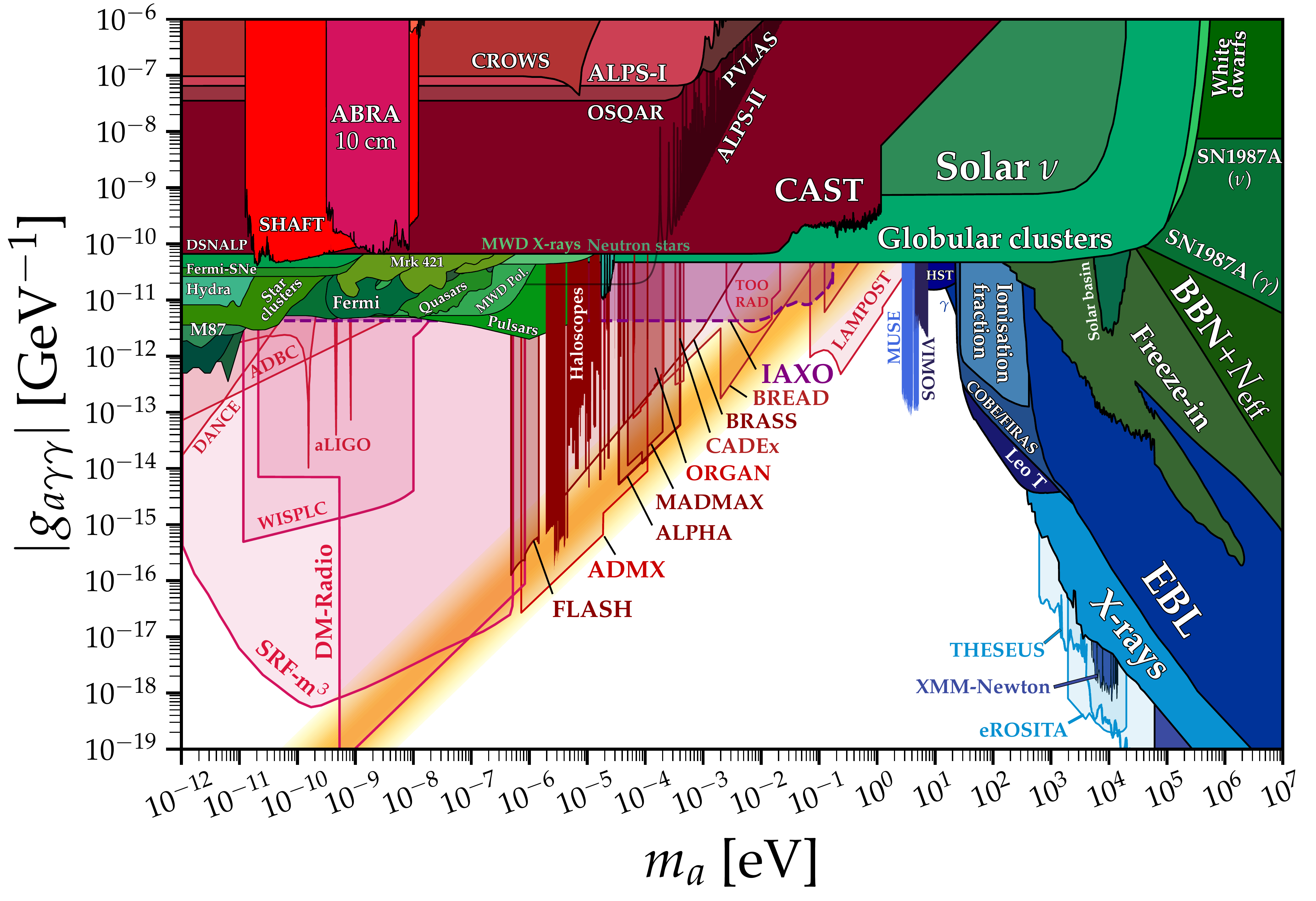}\label{Fig:ALPSIIlim2}}
  \caption{(a) ALP parameter space for model independent searches experiments \cite{AxionLimits}, the dashed line indicates limits from astrophysics. (b) Overview of results, prospects, and hints in the axion/ALP parameter space \cite{AxionLimits}.}
  \label{Fig:Lim_}
\end{figure}
Figure \ref{Fig:Lim_} shows the projection of the sensitivities of all three experiments together with existing limits.
It is very obvious that this unique set of axion experiments in the old HERA premises at DESY offers a likewise unique discovery potential to solve major particle physics, astrophysics and cosmological questions.

% %\bibliographystyle{plain}
% \bibliographystyle{unsrt}
% \bibliography{FIP2022/bib_ALPS.bib}
% \end{document}

%-------------------------------------------
\subsection{Review of axion programme at LNF and Legnaro -- {\it C.~Gatti}}
\label{ssec:gatti}
{\it Author: Claudio Gatti , <claudio.gatti@lnf.infn.it>} 
%-------------------------------------------

% %
% \documentclass[12pt,aps,prd,onecolumn,groupedaddress,superscriptaddress]{revtex4-2}

% \usepackage{lipsum}
% \usepackage{graphicx}  % needed for figures
% \usepackage{dcolumn}   % needed for some tables
% \usepackage{bm}        % for math
% \usepackage{amssymb}   % for math
% \usepackage{amsmath}
% \usepackage{xcolor}
% \usepackage{hyperref}
% \usepackage{amsmath}
% \usepackage{amsfonts}

% \hypersetup{colorlinks=true, linkcolor=blue, citecolor=blue, urlcolor=blue}
% %
% %
% \begin{document}
% %
% \title{
%   Review of the axion program at LNF and LNL
% }
% \author{C.~Gatti for the QUAX Collaboration}
% \affiliation{INFN, Laboratori Nazionali di Frascati, Frascati, Roma, Italy}

% \maketitle
%
%\baselineskip=14pt
%
\subsubsection{Introduction}

The Low-Energy Frontier of Particle Physics~\cite{Jaeckel:2010ni} provided a well motivated case for physics at the subelectronvolt scale that inspired the design and realization of several experiments within the reach of small and medium laboratories. In Italy, the National Laboratories of INFN in Legnaro (LNL) hosted the PVLAS (Polarizzazione del Vuoto con LASer) experiment~\cite{EJLLI20201} aiming at the measurement of the vacuum magnetic birefringence and at the observation of effects due to axion-like or milli-charged particles. Today, the experimental activity continues with a $5^{th}$-force experiment~\cite{PhysRevD.105.022007} looking for monopole-dipole axion interaction and the ferromagnetic and Sikivie haloscopes of the QUAX experiment~\cite{alesini2019galactic,alesini2021search,QUAX2022,barbieri2017searching,crescini2018operation,crescini2020axion}.
At the National Laboratories of Frascati~\cite{universe7070236} (LNF), the interest in the {\em dark sector} first started at the KLOE experiment~\cite{ANASTASI2015633} with the search of light vector-mediators produced in $e^{+}e^{-}$ collisions at 1~GeV, and continues today with the PADME experiment~\cite{instruments6040046}, that looks for dark photons with a positron-beam dump experiment, a second Sikivie haloscope of the QUAX experiment, and a proposal for a large haloscope~\cite{gatti2018klash,alesini2019klash} for searches of axions, dark photons and high-frequency gravitational waves.

Here, we report the results and prospects of the QUAX experiment and the results of ongoing R\&D on high quality-factor cavities and signal amplification. The prospects for a large haloscope at LNF are also discussed.

\subsubsection{QUAX}

\begin{figure}[!ht]
  \centering
      \includegraphics[width=0.47\textwidth]{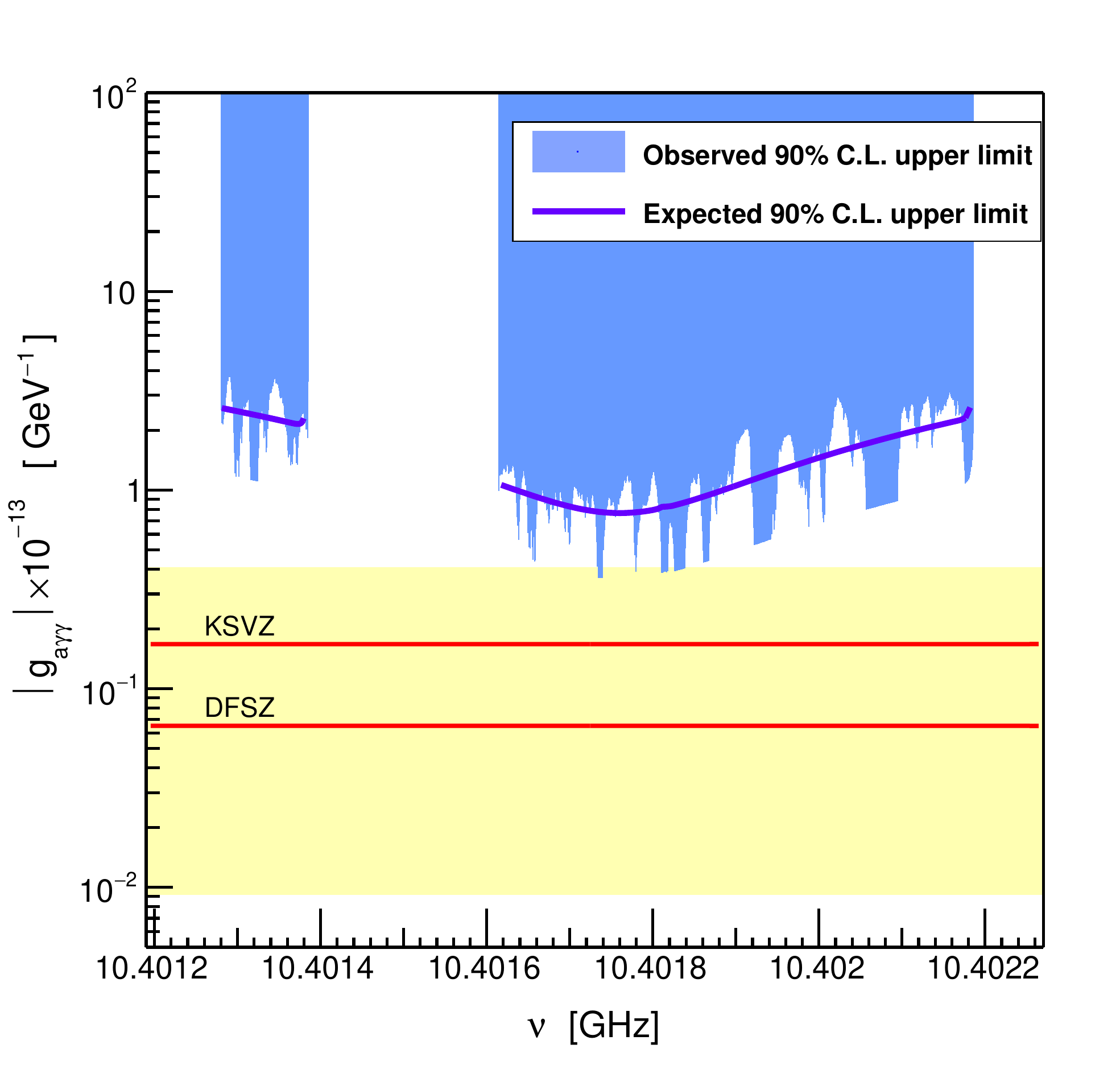}
\caption{\small 90\% single-sided C.L. upper limit for the axion coupling constant $g_{a \gamma \gamma}$. Each point corresponds to a test axion mass in the analysis window. The solid curve represents the expected limit in case of no signal. The yellow region indicates the QCD axions model band. We assume $\rho_a\sim0.45$~GeV/cm$^3$.}
\label{fig:Bin90limit}
\end{figure}

The haloscope, proposed by Sikivie~\cite{sikivie1983experimental,PhysRevD.32.2988}, is based on the immersion of a resonant cavity in a strong magnetic field in order to stimulate the inverse Primakoff effect, converting an axion in a cavity mode excitation. The QUAX experiment is looking for axion dark-matter with a mass around 40~$\mu$eV within the range of masses predicted by post-inflationary scenarios~\cite{Buschmann2022}. The frequency of operation, about 10~GHz, is experimentally very challenging since it involves a small volume of the resonant cavity (fraction of liter) and large quantum-fluctuations limiting linear amplifiers.

In 2021 QUAX reached the sensitivity to QCD-axions (Fig.~\ref{fig:Bin90limit})~\cite{alesini2021search}. The haloscope, assembled at LNL, was composed by a cylindrical OFHC-Cu cavity, with inner radius of 11.05~mm and length 210~mm, inserted inside the 150~mm diameter bore of an 8~T superconducting (SC) magnet of length 450~mm. The whole system was hosted in a dilution refrigerator with base temperature of 90~mK. The first amplification stage was done with a JPA~\cite{Roch} with noise temperature at the quantum-limit of about 0.5~K.

\begin{figure}[!ht]
  \centering
     \includegraphics[width=0.6\textwidth]{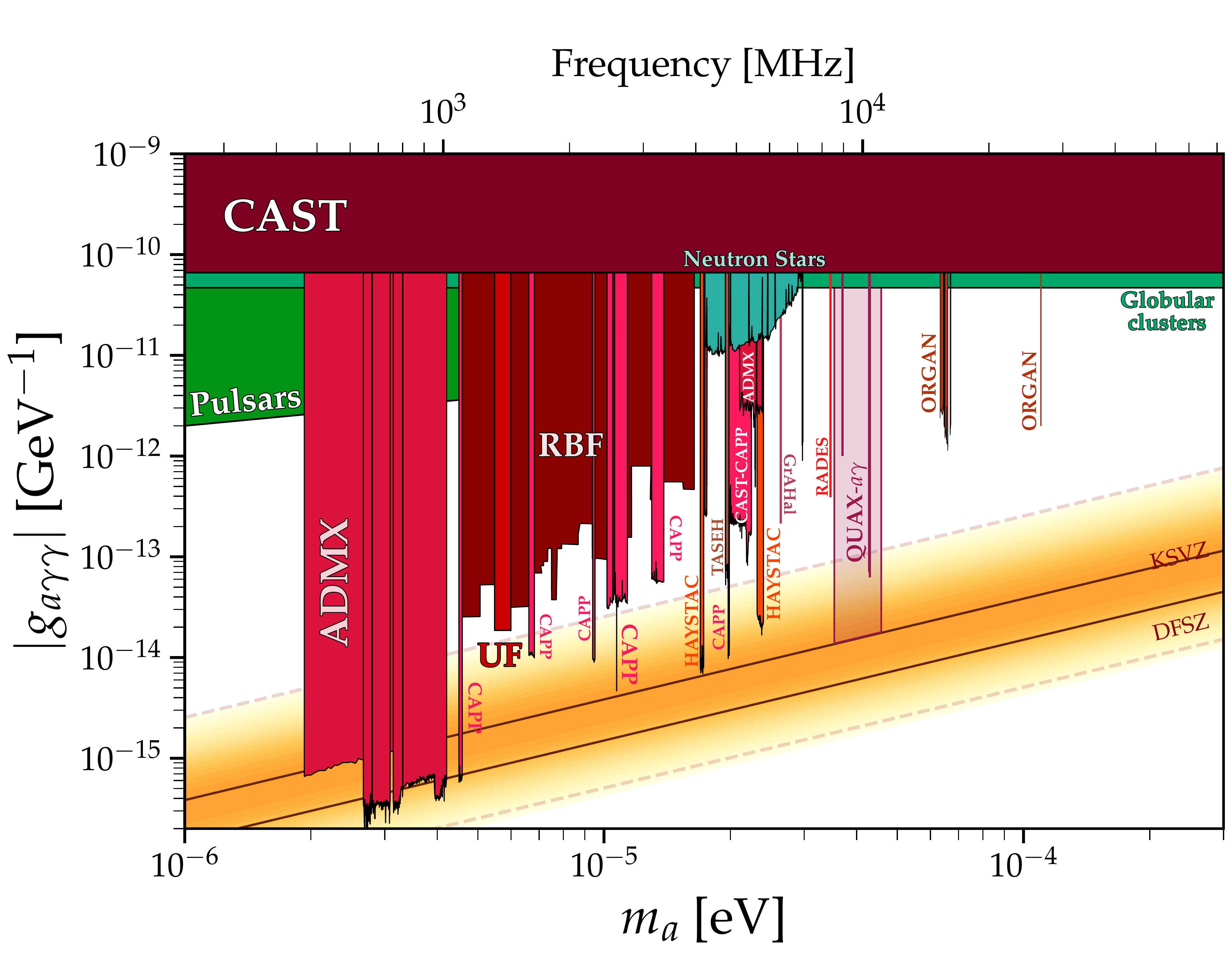}
\caption{\small Projections for axion searches by the QUAX haloscopes in the next few years are shown by the shaded area. The image is realized with~\cite{AxionLimits}}
\label{fig:QUAXProjections}
\end{figure}

In the next few years, the QUAX collaboration will probe the frequency region between 8 and 11 GHz with two complementary Sikivie haloscopes, one located at LNL and the other at LNF. The two haloscopes will be working at two different frequency ranges, implementing two different types of microwave cavities, dielectric and superconducting, and Travelling Wave Josephson Parametric Amplifiers (TWJPA)~\cite{BraggioTWPA,RETTAROLI2023167679}. High quality-factor cavities able to operate in a strong magnetic field have been already fabricated and tested by the QUAX collaboration~\cite{di2019microwave,alesini2020high,alesini2021realization} and used in axion searches~\cite{alesini2019galactic,QUAX2022}. Different geometries, including a multicavity approach, are forseen. The expected limits within 2025 are shown in Fig.~\ref{fig:QUAXProjections}.

\begin{figure}[htbp]
  \begin{center}
    \includegraphics[totalheight=6cm]{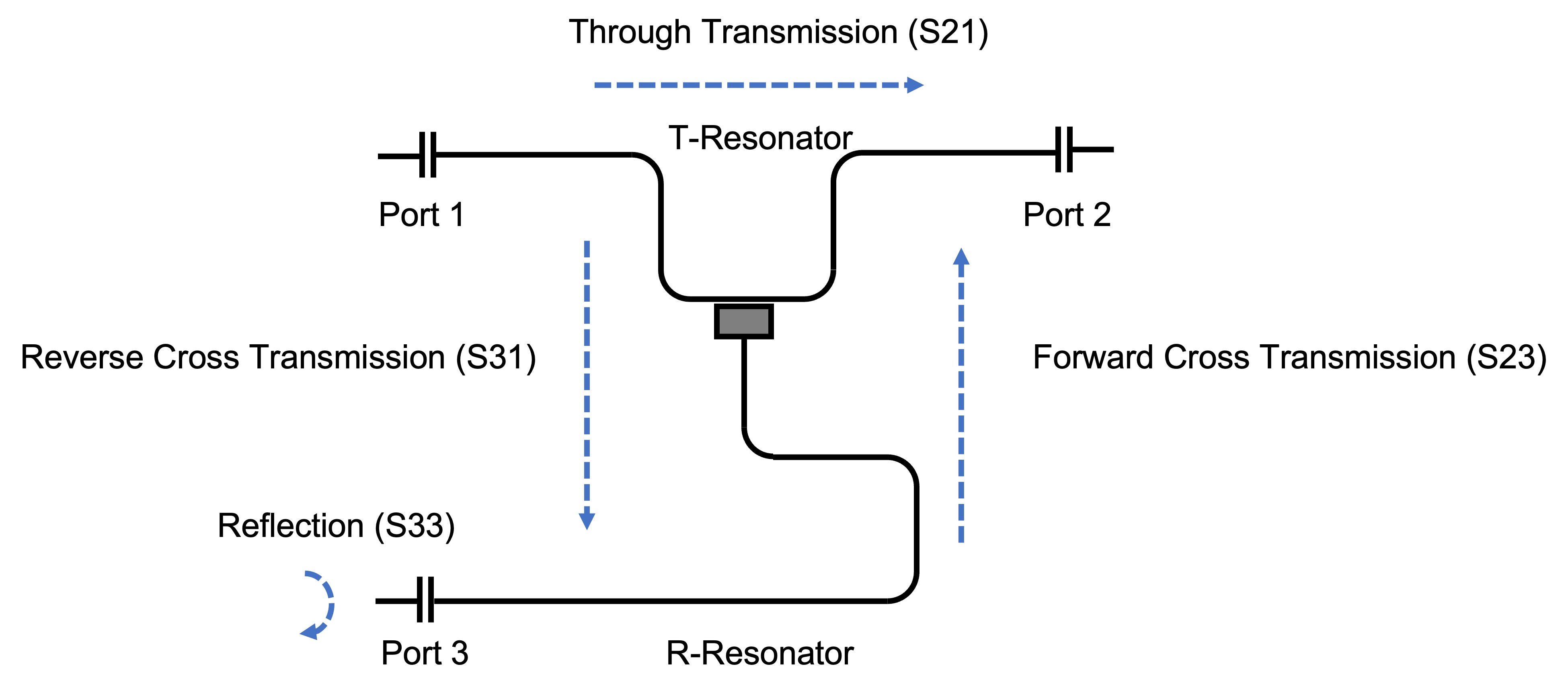}
    \caption{Scheme of the T-type device composed of two resonators {\em T} and {\em R} coupled by a SQN (gray box).}
    \label{fig:Tscheme}
  \end{center}
\end{figure}

Due to the large quantum-fluctuations at this frequency, linear amplifiers are not suited to reach the sensitivity to axions predicted by the DFSZ model, and new counters sensitive to single microwave-photons with low dark-counts must be used~\cite{NatureFlurin,kuzmin2018single,DElia2022}. In particular, a Superconducting Qubit Network (SQN) could be used to enhance the detection sensitivity to single microwave-photons. Recently, a device, arranged in a transistor-like geometry as in Fig.~\ref{fig:Tscheme}, was tested at LNF within the Supergalax project: an SQN working as a coupling element between two perpendicular resonators such that the transmission properties of the device are modified by the presence of few microwave photons~\cite{DElia2022}. The advantage of using a SQN over a single qubit is that of a predicted scaling of the signal-to-noise ratio as $N$ instead of $\sqrt{N}$, where $N$ is the number of qubits in the network~\cite{PhysRevB.105.104516,PhysRevB.103.064503}.

\begin{figure}[htbp]
  \begin{center}
    \includegraphics[totalheight=6cm]{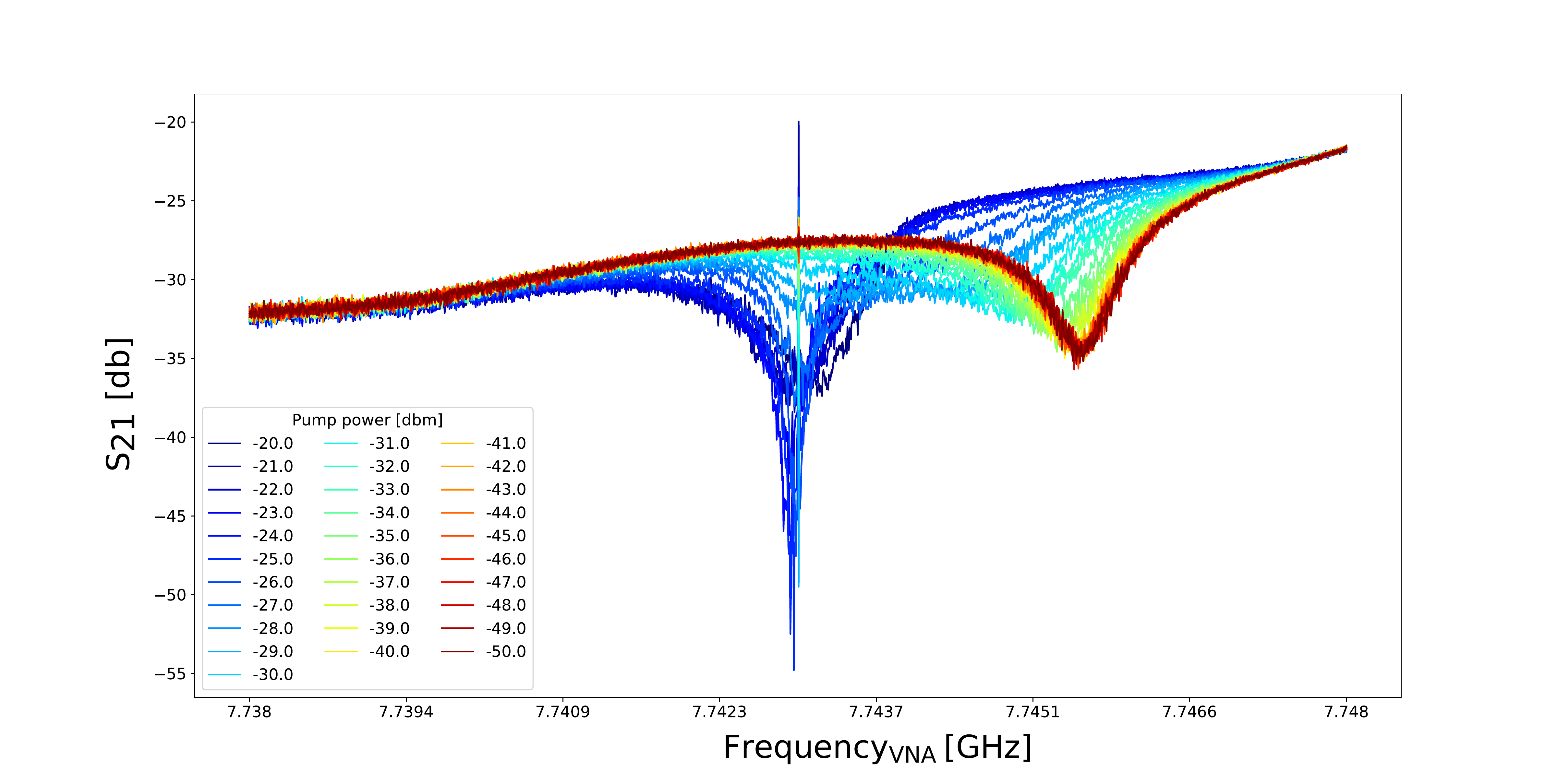}
    \caption{ Two-tone spectra measurements at frequency 7.74~GHz.  First-tone through-transmission (S21) vs VNA-frequency dependencies recorded at different powers of the second-tone signal of frequency of 7.743~GHz applied to the Port 3. }
    \label{fig:S21_vs_frequency}
  \end{center}
\end{figure}

The device was tested at LNF in a Leiden Cryogenics CF-CS110-1000 dilution refrigerator at a temperature of 15~mK.
The third-harmonic absorption-peak of the R-resonator at 7.74~GHz was considered. The VNA output-power was set to
-40~dBm, corresponding to about -100~dBm at the device, and the through transmission (S21) was measured. At the same time, a single tone of frequency 7.743~GHz was sent to the R-resonator with the Rohde{\&}Schwarz SMA100B connected to the Port 3, and the output power of the generator varied from -40 to -20~dBm (Fig.~\ref{fig:S21_vs_frequency}). By increasing the power sent to Port 3 a variation of the resonant-drop frequency in the through transmission-spectrum (S21) was clearly observed, confirming the feasibility of the device, but further optimization and engineering is needed to reach the single photon sensitivity.

\subsubsection{FLASH}

\begin{figure}[!ht]
  \centering
     \includegraphics[width=0.6\textwidth]{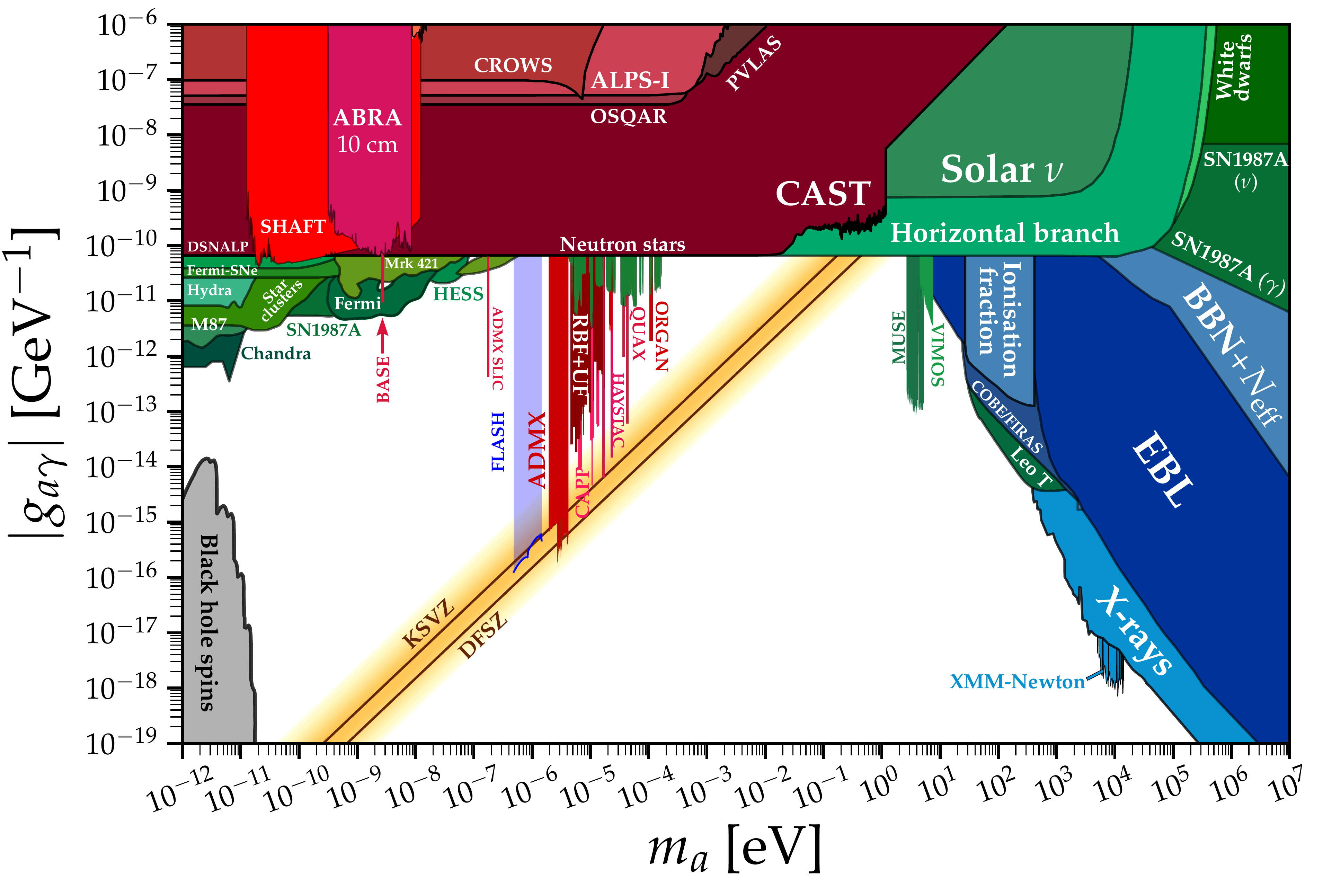}
\caption{\small Projection for axion searches with the FLASH haloscope is shown by the shaded area. Image realized with~\cite{AxionLimits}}
\label{fig:FLASHProjections}
\end{figure}

The FLASH experiment (FLASH, Finuda magnet for Light Axion SearcH), previously called KLASH~\cite{alesini2019klash}, proposes the realization of a haloscope devoted to the detection for sub-$\mu$eV axion by recycling the no-longer used FINUDA~\cite{bertani1999finuda,Modena:1997tz} magnet and the DA$\Phi$NE cryogenic-plant, at LNF. The FINUDA magnet is an iron-shielded solenoid-coil made from an aluminium-stabilised niobium-titanium superconductor providing an homogeneous axial field of 1.1\,T with very large size bore, able to accommodate a cryrogenic resonant cavity with a diameter of up to 2.1~m. The FLASH haloscope will be composed of such a large resonant cavity made of OFHC copper, inserted in a cryostat cooled down to 4.5\,K hosted inside the FINUDA magnet. The operation frequency will be tuned by three metallic movable-rods and the signal will be amplified by a Microstrip SQUID cooled down to 300 mK, for a total temperature noise of 4.9~K.
With this setup, it would be possible to search for KSVZ-axions scanning in the frequency range 117-360~MHz, corresponding to the well motivated mass region between 0.5 and 1.5 $\mu$eV (Fig.~\ref{fig:FLASHProjections}) in a total integrated time of about two years.

FLASH will be sensitive also to dark matter composed of dark photons with kinetic-mixing parameter down to few $10^{-17}$, and to high-frequency gravitational-waves with strain of about few $10^{-22}$.

In a possible second phase of the experiment, the cavity will be cooled to 100~mK. This will increase the axion sensitivity of about 1 order of magnitude, down to $g_{a\gamma\gamma}\sim 2\times 10^{-17}~\mbox{GeV}^{-1}$, allowing us to probe the existence of DFSZ-axions in this mass range.

%\subsubsection*{Acknowledgement}

%Partially supported by EU through FET Open SUPERGALAX project, grant agreement N.863313

%
% \bibliography{biblioFile}
%
% \end{document}

%-------------------------------------------
\subsection{High sensitivity axion dark matter experiments at IBS-CAPP in the 1-8 GHz range and plans for the 8-25 GHz range -- {\it O.~Kwon}}
\label{ssec:Kwon}
{\it Author:  Ohjoon Kwon, <o1tough@gmail.com>} 
%-------------------------------------------

\subsubsection{abstract}
Dark Matter is undeniably one of the most crucial questions in particle physics. Its coupling is exceedingly weak, demanding cutting-edge technology and the employment of high-risk, high-potential physics ingenuity. Center for Axion and Precision Physics Research (CAPP), founded in 2013, has successfully acquired top-of-the-line equipment and developed the necessary know-how, technology, and infrastructure to effectively probe the 1$-$8\,GHz range with the sensitivity required by the Dine-Fischler-Srednicki-Zhitnitsky (DFSZ) model within the next five years

\subsubsection{Introduction}

For the last decade, research on axions has greatly increased because it is not only a strong dark matter candidate but also a most believable solution to the strong $CP$ problem \cite{Peccei:1977hh,Weinberg:1977ma,Wilczek:1977pj,Kim:1979if}. Although the extremely weak conversion into detectable physical quantities limits the experimental proof of its existence, advances in technology such as superconducting cavities in strong magnetic fields, quantum-limited noise amplifiers, and dilution refrigerators etc. enable to search axions with meaningful sensitivity \cite{Kim:1979if,Shifman:1979if,Kim:1986ax,Zhitnitsky:1980tq,Dine:1981rt}. \\
\indent Established in South Korea in 2013 as part of the Korean government's basic science investment plan, the center for Axion and Precision Physics research (CAPP) at the Institute for Basic Science (IBS) aims to explore the axion frequency range above 1\,GHz with DFSZ sensitivity by utilizing state-of-the-art technologies. 
CAPP utilizes sikivie's haloscope method \cite{Sikivie:1983ip}, which is the most sensitive among currently feasible axion detection methods.
A large-capacity microwave cavity of more than 30 liters, which can be tuned in a frequency range of 1\,GHz or more, is installed in the center of the low temperature superconducting magnet of 12\,T magnetic field with a diameter of 32\,cm bore, enabling a high level of axion photon conversion output. 
In addition, the world's lowest level of system noise is achieved, less than 2\,quanta, by use of powerful dilution refrigerator, advanced thermal engineering, and the Josephson parametric amplifier (JPA) from the collaborator - Nakamura group in Tokyo university. 
CAPP has searched axions around 1.1\,GHz with $\>$1\,MHz/day of scan rate with DFSZ sensitivity for the case of 100$\%$ axions in the dark matter halo \cite{Yi:2022fmn} and even for the case of 20$\%$ axions, CAPP will be soon available to search axions with more than a megahertz per day of scanning speed with DFSZ sensitivity with the potential for even faster scanning speeds if high temperature superconducting cavity technology is implemented.

\begin{figure}
\centering
\includegraphics[width=.8\textwidth]{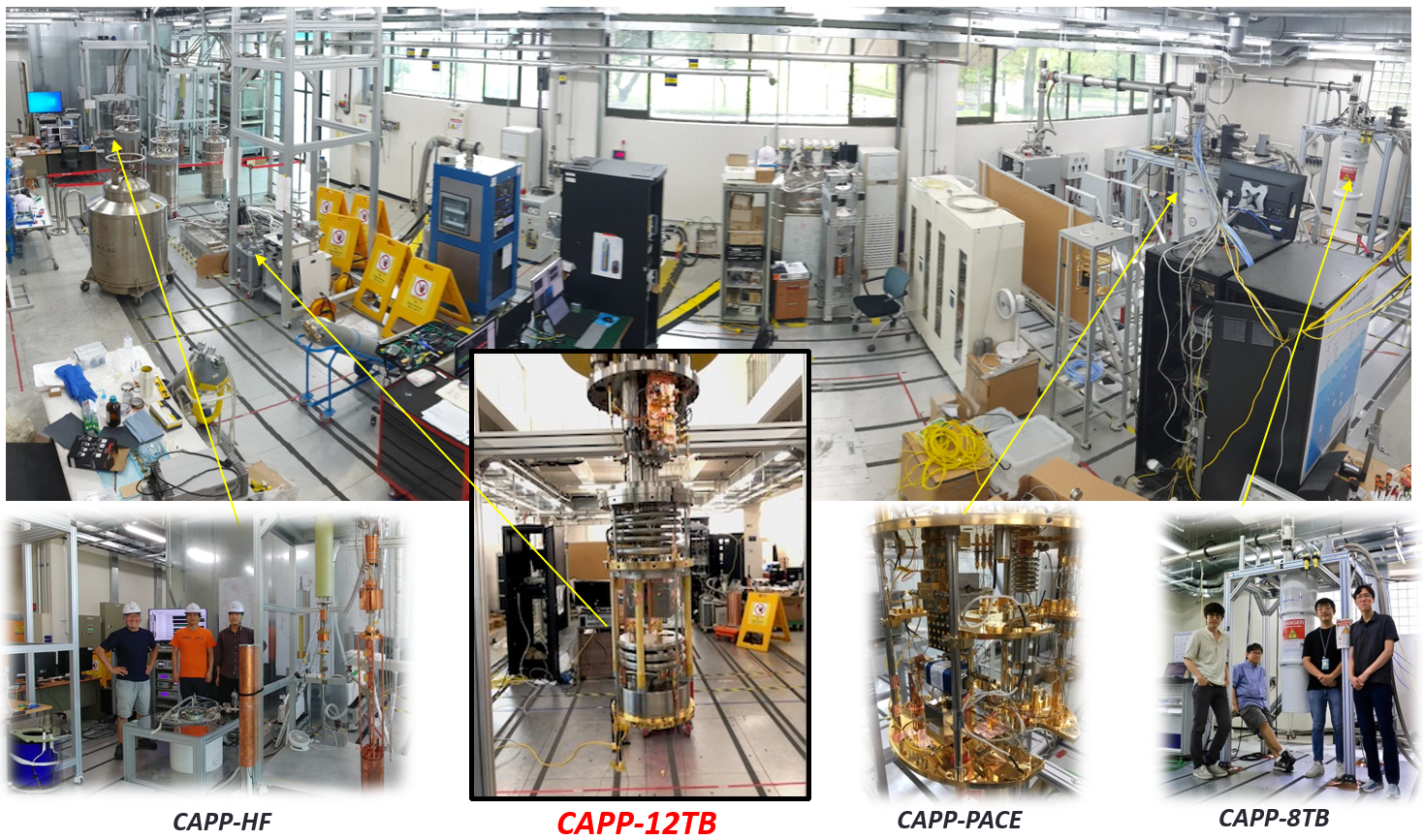}
\caption{\label{fig:call_hall} CAPP experimental hall, top view.}
\end{figure}

\subsubsection{Dark matter and axion}
The origins of dark matter can be traced back to Isaac Newton, who discovered gravity and connected the motions of celestial bodies in the universe. His work gave scientists the ability to detect what they cannot see through visible light. This was exemplified by the discovery of Neptune, which was first detected through deviations from the expected orbits of other planets. Similarly, the discovery of dark matter was first proposed when Vera Rubin observed that the rotational speed of galaxies did not decrease as expected, implying the presence of an unknown heavy mass \cite{Rubin:1980zd}. Numerous experiments and observations have since confirmed the existence of dark matter \cite{Sofue:2000jx}, which is estimated to make up around 25$\%$ of the total mass of the universe \cite{Spergel:2015noa}. Scientists have proposed many different models to explain dark matter, with the axion model being a particularly promising candidate. 

One key feature of the axion is its long de Broglie wavelength, which is estimated to be $\lambda \approx 300\, \text{m} \times \left( \frac{1 \, \mu eV} {m_{\text{a}}}\right) $, significantly larger than the size of the cavity. In the frequency range relevant to CAPP, multiple axion detectors can be positioned within a space that is smaller than the de Broglie wavelength to maximize the speed of axion search through phase matching. This implies that the search speed for axions can be dramatically increased if sufficient infrastructure, such as refrigerators, magnets, and quantum noise-limited amplifiers, is provided. This approach coincided with the strategy of the Institute for Basic Science (IBS) in Korea, which led to the establishment of CAPP. Over the past eight years, CAPP has become the only laboratory in the world capable of searching for axions at several megahertz units per day with DFSZ sensitivity in Korea, where there was no previous axion experimental research.

\subsubsection{Axion searching experiment at IBS-CAPP}
{\bf CAPP's infrastructure -}

\begin{table}[b]
{\small
\centering
\begin{tabular}{|c|r|c|c|r|r|l|c|} 
\hline
\multicolumn{3}{|c|}{\textbf{Refrigerator}}                                  & \multicolumn{3}{c|}{\textbf{Magnet}}                                                    & \multicolumn{2}{c|}{\textbf{Experiment}  }                                                                  \\ 
\hline
\textbf{Vendor}   & \multicolumn{1}{c|}{\textbf{Base T}}   & \multicolumn{1}{c|}{\textbf{Type}} & \textbf{Vendor}         & \multicolumn{1}{c|}{\textbf{B-field}} & \multicolumn{1}{c|}{\textbf{Bore}} & \multicolumn{1}{c|}{\textbf{Exp.} } & \textbf{Freq[GHz]}                                                               \\ 
\hline
Leiden   & 10\,mK                         & wet                       & Oxford Instr.        & 12\,T                         & 32\,cm                          & CAPP-12TB                 & 1.0$-$1.5                                                                 \\ 
\hline
Bluefors & 10\,mK                         & dry                       & AMI            & 12\,T                         & 10\,cm                          & CAPP-HF                   & $>$5                                                                       \\ 
\hline
Bluefors & 10\,mK                         & dry                       & AMI            & 12\,T                         & 10\,cm                          & CAPP-QNA                  & $>$5                                                                       \\ 
\hline
Bluefors & 10\,mK                         & dry                       & AMI            & 8\,T                          & 12\,cm                          & CAPP-PACE                 & \begin{tabular}[c]{@{}c@{}}2.0$-$4.0\\(5.8$-$6.0)\end{tabular}              \\ 
\hline
Bluefors & 10\,mK                         & dry                       & AMI            & 8\,T                          & 16.5\,cm                        & CAPP-8TB                  & \begin{tabular}[c]{@{}c@{}}1.5$-$2.0\\(5.8$-$6.0)\end{tabular}              \\ 
\hline
Janis    & 300\,mK                        & wet                       & \begin{tabular}[c]{@{}c@{}}Cryo-\\Magnetics\end{tabular} & 9\,T                          & 12\,cm                          & CAPP-HF                   & \begin{tabular}[c]{@{}c@{}}3.5$-$4.5\\($\sim$10)\end{tabular}  \\ 
\hline
Oxford Instr.   & 30\,mK & wet                       & SuNAM          & 18\,T                         & 7\,cm                           & CAPP-18T                  & 4.5$-$6.0                                                                 \\
\hline
\end{tabular}
\caption{\label{tab:capp_infra} CAPP's equipments and relevant experiments}
}
\end{table}

CAPP built a state-of-the-art axion search laboratory from scratch. The facility includes seven low-vibration pads, powerful refrigerators, and various magnets, allowing multiple axion experiments to be conducted simultaneously, as shown in Fig. \ref{fig:call_hall}. 
CAPP's flagship experiment, CAPP-12TB, which aims to achieve DFSZ sensitivity in the 1$-$8 GHz range within the next five years, utilizes a Leiden dilution refrigerator with a cooling power of over 1\,mW at 120\,mK \footnote{Leiden Cryogenics, https://leidencryogenics.nl} and a 12\,T, 32\,cm bore Oxford Instruments magnet \footnote{Oxford Instruments, https://www.oxinst.com}.
Two dry type Bluefors dilution refrigerators \footnote{Bluefors Oy, https://bluefors.com}, each integrated with an 8\,T, 12\,cm bore and an 8\,T, 16.5\,cm bore magnet from AMI \footnote{American Magnetics Inc., http://www.americanmagnetics.com}, are used for smaller experiments (CAPP-PACE, CAPP-8TB) as well as for cavity R$\&$D. 
Two other Bluefors dilution refrigerators are currently being used as test benches for quantum noise limited amplifiers and will later be used to search for high mass axions. Details are described in Table \ref{tab:capp_infra}. As a result, CAPP can simultaneously run six (seven in the future) independent axion search experiments.

{\bf CAPP's R$\&$D - }
\subsubsection{Superconducting cavity in a strong magnetic field}
\begin{figure}
\centering
\includegraphics[width=.9\textwidth]{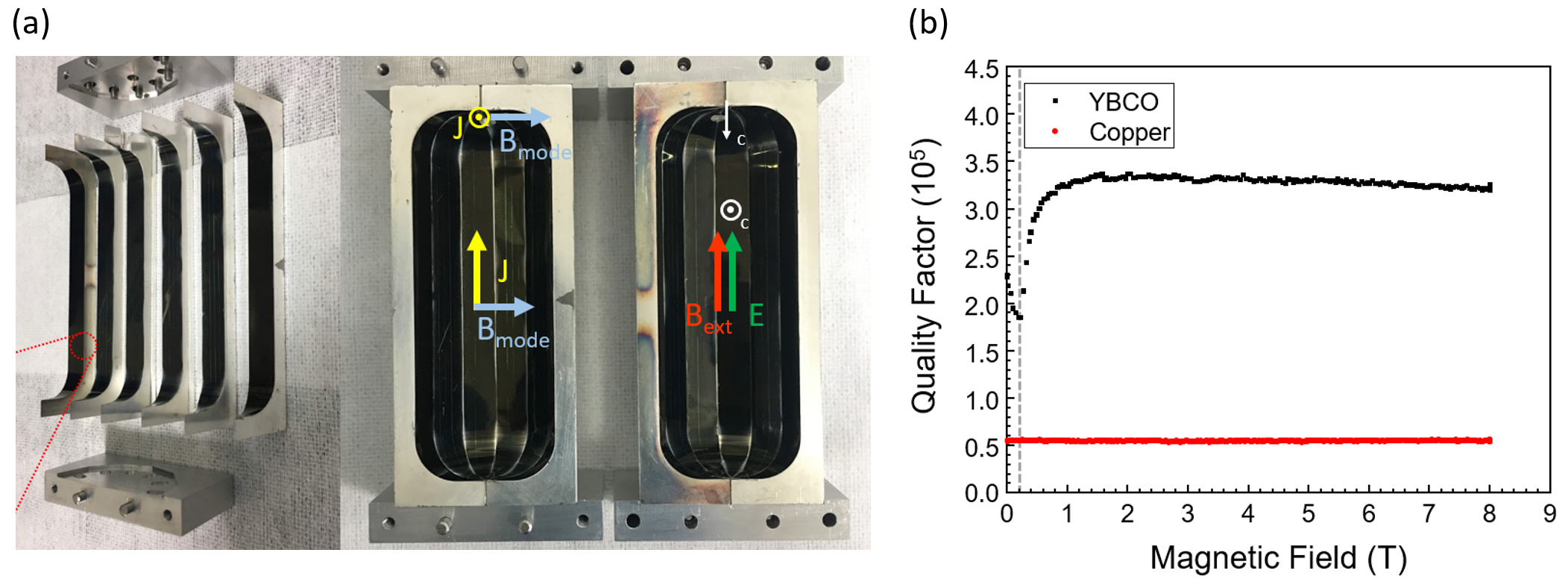}
\caption{\label{fig:capp_ybcoresult} (a) Photo of CAPP's prototype superconducting cavity (b) $Q$-factor measurements of CAPP's 1st generation prototype superconducting cavity. These are imported from figure 1, 4 of \cite{Ahn:2021fgb} }
\end{figure}

CAPP has conducted extensive research and development aimed at achieving high sensitivity in the search for axions across a broad range of frequencies. One of the key areas of focus has been the development of superconducting cavities capable of maintaining a high quality factor ($Q$-factor) even in the presence of a strong magnetic field. 
With the high-$Q$ cavity, the axion signal can remain in the cavity for a longer duration, resulting in a higher signal-to-noise ratio and a significant improvement in the speed of axion searches.
Superconductivity, which is in general essential to make high $Q$-factor cavity, however, usually breaks when a magnetic field exceeds a certain level, and even in cases where superconductivity is well maintained in a strong magnetic field, such as high-temperature superconductors (HTS), the use of superconducting cavities in actual axion research is not practical, as it faces the limitations of 3-dimensional cavity fabrication technology \cite{norton:1996ybco,Li:2001ybco,Reade:2002ybco}.
In 2019, CAPP has solved the problem by attaching the well-grown 2-dimensional REBCO film directly to the inner wall of a 3-dimensional cavity.
\begin{table}[b]
{\small
\centering
\begin{tabular}{|c|c|c|c|c|}
\hline
\textbf{Generation}      & \textbf{HTS type} & \textbf{Volume {[}L{]}} & \textbf{Frequency {[}GHz{]}} & \textbf{$Q$-factor} \\ \hline
1st Gen                  & YBCO          & 0.3                     & 6.9                          & 330,000           \\ \hline
2nd Gen                  & ReBCO         & 1.5                     & 2.3                          & $\sim$500,000     \\ \hline
\multirow{2}{*}{3rd Gen} &  \multirow{2}{*}{Y-based BCO}      & 1.5                     & 2.3                          & $\sim$4,500,000   \\ \cline{3-5} 
                         &       & 0.2                     & 5.5                          & $\sim$13,000,000  \\ \hline
\end{tabular}
\caption{\label{tab:capp_sccavity} HTS cavities developed by CAPP}
}
\end{table}
CAPP developed a prototype cavity in which the TM$_{010}$ mode resonated at 6.9\,GHz using HTS tape and achieved a $Q$-factor six times higher than that of copper resonators at 4\,K and 8\,T magnetic field \cite{Ahn:2021fgb}, as shown in Fig \ref{fig:capp_ybcoresult}.
 A second-generation cavity with a larger volume and a sapphire tuning system was then created to achieve a $Q$-factor of half a million and used in the physics run of axion search in 2021. Finally, CAPP's third-generation superconducting cavity development has achieved a $Q$-factor of order of $10^7$, which is 10 times the axion $Q$-factor, even in an 8\,T magnetic field.

\subsubsection{Josephson parametric amplifiers (JPA)}
\begin{figure}
\centering
\includegraphics[width=.8\textwidth]{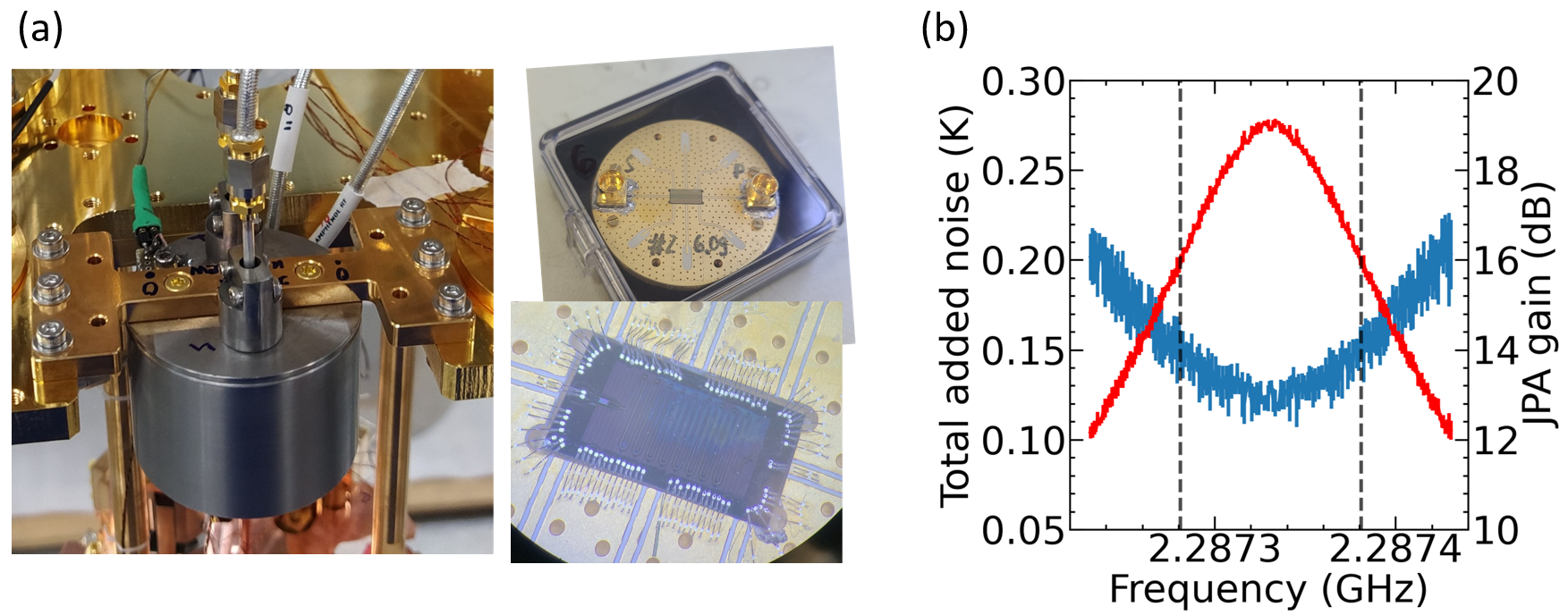}
\caption{\label{fig:capp_jpa} (a) Implementing Nakamura group's JPA into axion haloscope system. Right upper: A JPA chip integrated with PCB board. Right bottom: JPA chip magnified version of right upper one. Left: Fullly pacakaged JPA installed on the mixing plate of dilution refrigerator. (b) Total added noise measurement after implementing a JPA. This is imported from figure 2 of \cite{kim:2022capp} }
\end{figure}

One important focus of R$\&$D in the axion search experiment is the development of quantum noise limited amplifiers (QNAs). Typically, most of the RF chain noise originates from the physical temperature of the device under test (DUT), i.e., the microwave cavity temperature in the haloscope, and the preamplifier noise, with the latter usually being dominant. By developing QNAs, the dominant preamplifier noise can be suppressed down to the quantum limit, significantly enhancing the SNR.
In collaboration with Prof. Nakamura's group at Tokyo University, CAPP was able to implement quantum noise limited JPAs into the axion search experiments. About 200 JPA chips, each with a working range of 3$-$$5\%$ for the JPA resonant frequency, were delivered to CAPP, where they were then packaged and characterized. JPAs for 1\,GHz, 2\,GHz, and 6\,GHz were implemented in the axion haloscope system, and all of them demonstrated noise levels close to the standard quantum limit. Figure \ref{fig:capp_jpa} shows the JPA used in the axion search experiment in 2020 and the corresponding noise. In the JPA resonance region, the measured noise was approximately 120\,mK. Taking into account the pre-JPA attenuation and the downstream noise introduced by the JPA backstage, we can assume that the noise generated by the JPA is an idler noise, which is essentially a copy of the noise injected into the JPA.

\subsubsection{High frequency cavities}
\begin{figure}
\centering
\includegraphics[width=.8\textwidth]{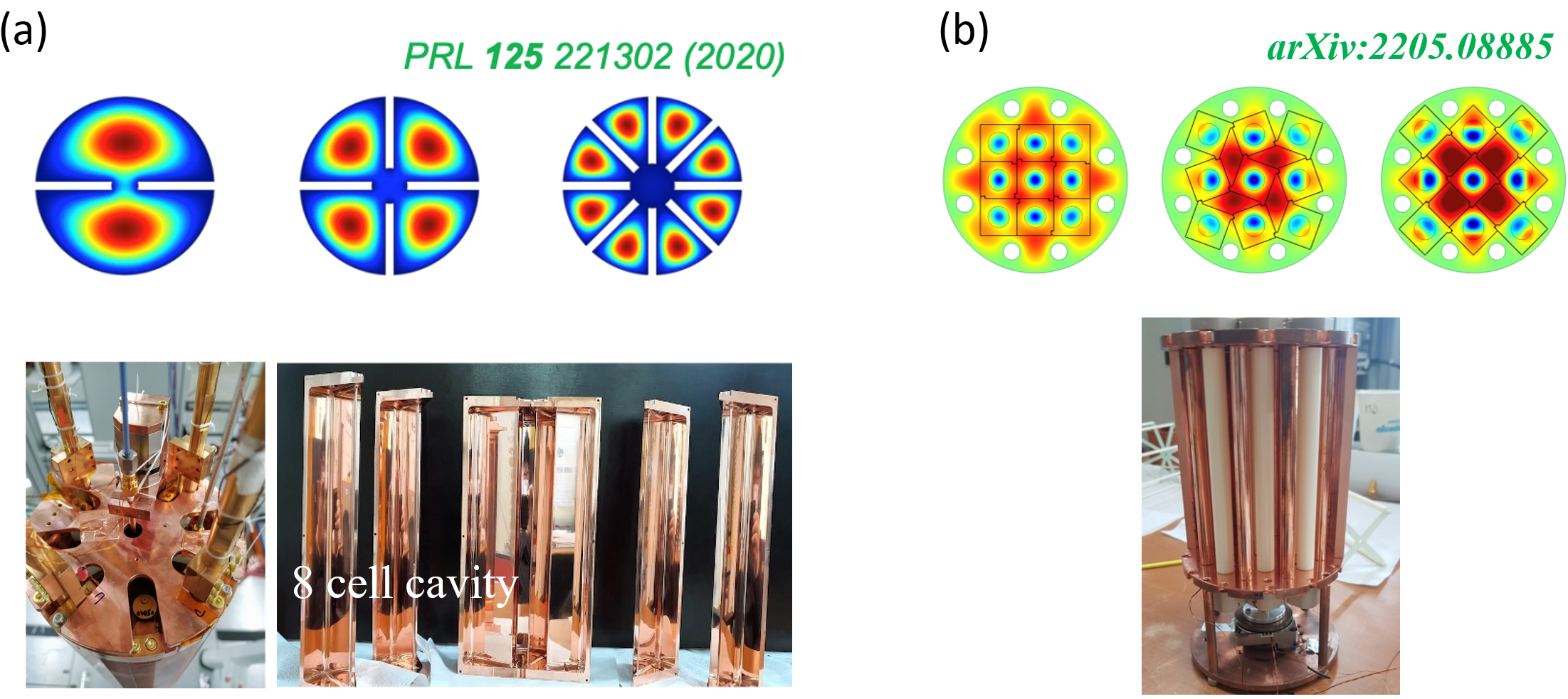}
\caption{\label{fig:multicell} (a) Multiple-cell cavity. Top figures are the top-view of an electric field (E-field) profile of 2-, 4-, and 8-cell cavity. Bottom photos are the fabricated 8 cell cavity. The field profile is imported from figure 1 of \cite{multicell} (b) E-field profile and a photo of the photonic crystal haloscope cavity. This is imported from figure 6 of \cite{bae:2022pc} }
\end{figure}

CAPP has developed several high-frequency cavities for broadband axion searches. One of these designs is the multiple-cell cavity \cite{Jeong:2020cwz}, which comprises several cells vertically separated by walls that do not cross the cavity center, as depicted in Fig. \ref{fig:multicell} (a). The benefits of this design include increasing frequency with more cells, minimal volume loss since only thin walls are added, maintaining the cylinder shape, and extracting all cell energy using a single antenna at the center. CAPP has produced a double-cell and 8-cell cavity, successfully adjusting the frequency in the 3.14$-$3.36\,GHz and around 5.8\,GHz regions, respectively, by placing a sapphire rod in each cell of the multi-cell cavity and rotating and moving it from the center of the cell to the wall separating the cells.

Furthermore, CAPP developed a photonic crystal (PC) haloscope cavity to explore higher frequencies \cite{bae:2022pc}. This cavity contains periodically arranged dielectric rods with low volume loss and high permittivity that act as independent cavities, forming a TM$_\text{010}$-like mode, as illustrated in Fig. \ref{fig:multicell} (b). 
This design offers a higher resonant frequency as the number of distributed dielectrics increases. While the haloscope form factor is relatively low due to the concentration of the electric field in the high-permittivity dielectric volume, the high $Q$-factor offsets this due to far fewer metal walls being required to achieve the same frequency. CAPP has successfully adjusted the frequency of the PC cavity using the auxetic behavior of rotating rigid squares to change the distance between the dielectric rods simultaneously and uniformly. The PC cavity will be utilized to search for axions in high frequency bands around and above 10\,GHz.

\subsubsection{Ultra-light cavity (ULC) for 12TB experiment}
Producing a cavity for the CAPP-12TB experiment posed several challenges due to the enormous weight of a copper cavity with a diameter of 27\,cm and a height of 64\,cm or more, which requires significant resources for installation, movement, and modification work. Furthermore, copper with high mass has a large heat capacity and high hydrogen content, making it difficult to lower its temperature below 100\,mK. Additionally, the tuning rod used in developing a large axion haloscope cavity also presents a major challenge as it grows with the cavity's size, making it difficult to move and minimize heat generation at low temperatures. The physical contact between the tuning rod and the cavity body is also weak, making it harder to cool the rod as much as the cavity.

To address these issues, the CAPP research team roll-banded 0.5\,mm thick Oxygen-Free High Thermal Conductivity (OFHC) copper foil to create a 27\,cm diameter cylinder and combined it with two OFHC copper discs, resulting in an ultra-light cylindrical cavity (ULC) weighing approximately 5\,kg. The tuning rod was also made from an OFHC pipe with a diameter of 30\,mm and wall thickness of 0.7\,mm, and shaped into a closed cylinder to weigh only about 200\,g. The ULC was installed in the Leiden refrigerator (See Table \ref{tab:capp_infra}.) and achieved a temperature of 30\,mK, which is similar to or lower than the cavity temperatures of other axion experiments. Additionally, by applying a 12\,T magnetic field at the temperature, it was possible to adjust the frequency using only one attocube piezo rotator at 1.025$-$1.185\,GHz \footnote{attocube systems AG, https://www.attocube.com}. The $Q$-factor was 90k when the tuning rod was farthest from the center, over 120k when positioned at the center, and maintained more than 100k throughout most of the tunable frequency range.

\begin{figure}
\centering
\includegraphics[width=.8\textwidth]{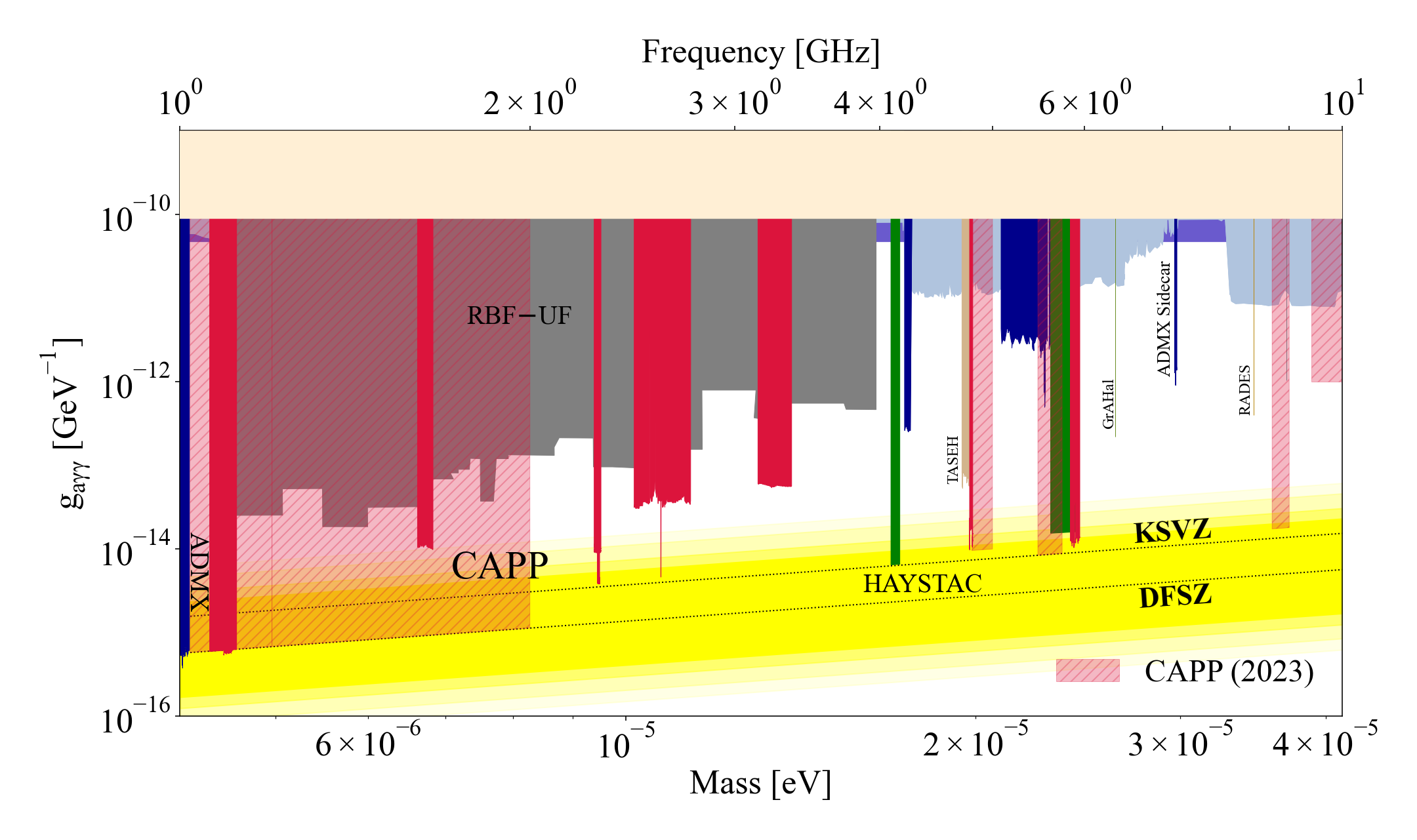}
\caption{\label{fig:exclusion}Exclusion limit. The red area has been excluded by CAPP, and the red shaded area is scheduled to be excluded in 2023.}
\end{figure}

{\bf Experiments at CAPP - }
CAPP has successfully conducted numerous axion experiments thanks to its cutting-edge infrastructure and extensive R$\&$D efforts. The experiments are conducted in conjunction with a superconducting magnet and refrigerator, which are listed in Table \ref{tab:capp_infra}. More information on the specific experiments conducted can be found in Table \ref{tab:capp_exp}.

\begin{table}[b]
{\small
\begin{tabular}{|@{\,}c@{\,}|c|@{\,}c@{\,}|@{\,}c@{\,}|@{\,}c@{\,}|@{\,}c@{\,}|@{\,}c@{\,}|@{\,}c@{\,}|@{\,}c@{\,}|}
\hline
\textbf{Experiment}                                                          & \textbf{year} & \textbf{B {[}T{]}} & \textbf{m$_\text{a}$ {[}GHz{]}} & \textbf{$\Delta$m$_\text{a}$ {[}MHz{]}} & \textbf{Sensitivity}                                    & \textbf{T$_\text{phy}$ {[}K{]}} & \textbf{T$_\text{sys}$ {[}K{]}} & \textbf{Publication} \\ \hline
\textbf{CAPP-PACE}                                                           & 2018          & 8                  & $\sim$2.5             & 250                    & \begin{tabular}[c]{@{}c@{}}10*KSVZ\\ +KSVZ\end{tabular} & \textless 0.05        & $\sim$1               & PRL                  \\ \hline
\textbf{CAPP-8TB}                                                            & 2019          & 8                  & $\sim$1.6             & 200                    & 4*KSVZ                                                  & \textless 0.05        & $\sim$1               & PRL                  \\ \hline
\textbf{CAPP-MC}                                                             & 2019          & 9                  & $\sim$4.0             & 250                    & 10*KSVZ                                                 & $\sim$2               & $\sim$2               & PRL                  \\ \hline
\textbf{\begin{tabular}[c]{@{}c@{}}CAPP-PACE-\\      JPA\end{tabular}}       & 2020          & 8                  & $\sim$2.3             & 30                     & 2*KSVZ                                                  & $\sim$0.05            & $\sim$0.2             & PRL                \\ \hline
\textbf{\begin{tabular}[c]{@{}c@{}}CAPP-PACE-\\      JPA-6cell\end{tabular}} & 2021          & 8                  & $\sim$5.6             & 80                     & 3*KSVZ                                                  & $\sim$0.05            & \textless{}0.3        & -                    \\ \hline
\textbf{\begin{tabular}[c]{@{}c@{}}CAPP-8TB-\\      JPA-8cell\end{tabular}}  & 2022          & 8                  & $\sim$5.8             & \textgreater{}100      & KSVZ                                                    & $\sim$0.03            & \textless{}0.3        & -                    \\ \hline
\textbf{\begin{tabular}[c]{@{}c@{}}CAPP-PACE-\\      JPA-SC\end{tabular}}    & 2022          & 8                  & $\sim$2.3             & 30                     & KSVZ                                                    & $\sim$0.04            & \textless{}0.2        & -                    \\ \hline
\textbf{CAPP-12TB}                                                           & 2022          & 12                 & $\sim$1.1             & 20 or 60                     & DFSZ                                                    & $<$0.05            & \textless{}0.3        & PRL                \\ \hline
\end{tabular}
}
\caption{\label{tab:capp_exp} Physics runs performed by CAPP}

\end{table}

\begin{enumerate}
\item CAPP-12TB is the primary experiment conducted by CAPP, which employs a 12\,T/32\,cm bore magnet and Leiden refrigerator to scan the 1$-$8\,GHz frequency range over the next five years with DFSZ sensitivity. The installation of the Oxford Instr. magnet was delayed by nearly a year due to the Covid-19 pandemic, but the commissioning run of CAPP-12TB was eventually carried out successfully in 2022, eight years after the inception of CAPP. During the experiment, a ULC operating in the 1.025$-$1.185\,GHz range and a JPA optimized at 1.06$-$1.12\,GHz were utilized, and axion data was received within the 1.09$-$1.11\,GHz range \cite{Yi:2022fmn}. The system noise was kept at 200\,mK, the lowest among all axion experiments, resulting in a scanning rate of 1.4\,MHz per day with DFSZ sensitivity. Following the commissioning run, the system was further optimized, achieving a total system noise of 170\,mK, and extending the entire JPA operating range (1.06$-$1.12\,GHz) resulted in axion data being obtained at a faster rate of 2\,MHz/day. Furthermore, CAPP aims to explore axions in the 1$-$2\,GHz range with DFSZ sensitivity by the end of 2023.
\item CAPP's pilot axion cavity experiment (CAPP-PACE), initially developed as a R$\&$D machine for CAPP, utilizing an 8\,T/12\,cm bore LTS magnet a Bluefors dilution refrigerator, has grown into an independent experiment serving as a forefront base for applying many of CAPP's R$\&$D achievements to real axion experiments. 
In 2018, CAPP conducted its first highly sensitive haloscope experiment, achieving a cavity temperature of less than 50\,mK, the first of its kind in the world, by optimizing the axion detection chain in a dilution refrigerator (CAPP-PACE in Table \ref{tab:capp_exp}) \cite{CAPP:2020utb}. 
In 2020, the experiment successfully implemented a JPA, a quantum noise amplifier, achieving the world's lowest system noise of less than 200\,mK (CAPP-PACE-JPA) \cite{Kim:2022hmg}. 
Subsequently, in 2021, a 6-cell cavity (CAPP-PACE-JPA-6cell) and, in 2022, finally, a practically usable HTS cavity (CAPP-PACE-JPA-SC), were successfully incorporated into the axion experiments in series.
\item CAPP-8TB shares almost the same configuration as CAPP-PACE, but features a larger magnet with a 16.5 cm bore, which results in relatively higher sensitivity for axion experiments. In 2019, CAPP-8TB acquired its first axion search data in the frequency range of 1.60$-$1.65 GHz \cite{Lee:2020cfj}, utilizing a 3.47 L OFHC cavity, which was sensitive to axion photon coupling $g_{a\gamma \gamma}$, down to the QCD axion band \cite{Cheng:1995fd}. In 2022, an 8-cell cavity was introduced in the CAPP-8TB system, operating around 5.8 GHz with a volume of over 3.5 L, much larger than the capable cavity with a conventional design. Moreover, a JPA working in the corresponding range was also incorporated in the system, forming CAPP-8TB-JPA-8cell. This new setup managed to establish new exclusion limits, possessing near-Kim-Shifman-Vainstein-Zakharov (KSVZ) axion coupling sensitivity.
\item CAPP's multiple-cavity experiment (CAPP-MC) aimed at high-frequency axion searching, utilizing the Janis cryostat \footnote{Janis in Lakeshore Cryotronics, https://www.lakeshore.com/products/product-detail/janis} and the 9\,T/12\,cm bore magnet by Cryo-Magnetics \footnote{Cryomagnetics, Inc., https://cryomagnetics.com}. The multiple-cell cavity concept introduced in section 3.2.2 was successfully adapted in axion search experiments CAPP-MC \cite{Jeong:2020cwz} and CAPP-8TB-JPA-8cell, as seen in Table \ref{tab:capp_exp}.
Recently, CAPP launched CAPP-HF, a new initiative to explore high-frequency axions in a more diverse manner by incorporating a new 12\,T/8\,cm magnet into a dilution fridge. CAPP-HF aims to search for axions above 10\,GHz by utilizing not only the existing technology of multiple-cell cavities but also the newly announced photonic crystal cavity \cite{bae:2022pc} and single photon detection, which is currently being developed using the expertise available in Korea. CAPP-HF is expected to commence exploration of sub-10\,GHz axions by 2023.
\end{enumerate}

\subsubsection{Conclusion and future plan}
IBS-CAPP has opened up the possibility of overcoming the feeble coupling of the axion in the past 8 years since its establishment. Based on the Oxford Instruments 12\,T/32\,cm magnet and a Leiden dilution refrigerator system, they used the JPA with the world's lowest noise and a ULC weighing less than 5\,kg with a volume of 37 liters, achieving an axion search speed of 1.4\,MHz/day with DFSZ sensitivity. Various R$\&$D achievements, represented by the successful development of a HTS cavity that maintains a high $Q$-factor even in a strong magnetic field, will further accelerate the search for axions. CAPP's objective is to find or exclude the axion from a wide frequency range of 1$-$8 GHz within the next five years and then further extend the probing region to up to 25\,GHz in the next ten years. Finding the axion as dark matter will provide us with invaluable information on the universe at its earliest moments of creation, as well as insight into the strong $CP$ problem of QCD and the structure of matter, the secret of building our world.

%-------------------------------------------
\subsection{Neutron EDM Searches -- {\it J.~Martin}}
\label{ssec:martin}
{\it Author:Jeffery Martin, - Joint Session with PSI 2022}
%-------------------------------------------

%New measurements of the neutron electric dipole moment (nEDM) will
%place even tighter constraints on theories involving new sources of CP
%violation beyond the standard model. It is believed these are required
%in order to explain the predominance of matter over antimatter (baryon
%asymmetry) in the universe. A new measurement also could impact a
%longstanding mystery in the standard model, involving the apparent
%lack of CP violation arising from the strong sector. A nonzero nEDM
%would be a major discovery impacting a broad range of theories.  The
%most precise nEDM experiments involve new intense sources of ultracold
%neutrons (UCN). UCN are slow moving neutrons that can be stored in
%material, magnetic, and gravitational bottles. The ability to trap the
%UCN makes them ideal for nEDM measurements. The most precise
%experiment to date was conducted at PSI and determined an upper bound
%on the nEDM of $|d_n|<1.8\times 10^{-26}~e$cm (90\% C.L.). The next
%generation of neutron EDM experiments aim to improve the uncertainty
%to the $1\times 10^{-27}~e$cm level. In this paper, I will review the
%current status of nEDM experiments worldwide, recent experimental
%progress, and plans for each project. This will include some
%additional focus on the TUCAN EDM experiment, in which I am involved.
%\end{abstract}

\subsubsection{Introduction and Theoretical Overview}

The neutron electric dipole moment (nEDM) is an experimental
observable of considerable interest in fundamental physics.  The nEDM
violates time-reversal symmetry, and hence measurements of the nEDM
are regarded as testing CP symmetry.  To date, all experiments have
demonstrated that the nEDM is zero.  Improving the experimental
precision of the measurements places tighter and tighter constraints
on new sources of CP violation beyond the standard model.

Measurements of the electric dipole moment of the neutron are
complementary to those conducted in other nuclear, atomic, and
molecular systems
\cite{bib:pospelov2005,bib:engel,bib:chupp,bib:flambaum}.  The most
precise experiments can be divided into those using paramagnetic
atoms, diamagnetic atoms, and bare nucleons.  For the paramagnetic
case, the most precise recent experiment used molecules of
ThO~\cite{bib:acme} and can be interpreted as placing an upper bound
on the electric dipole moment of the electron of $|d_e|<1.1\times
10^{-29}~e$cm at 90\% confidence level.  In the diamagnetic case, the
most precise experiment has used
$^{199}$Hg~\cite{bib:graner,bib:graner2} finding the atomic EDM to be
$|d_{\rm Hg}|<7.4\times 10^{-30}~e$cm (95\% C.L.).  Using nuclear and
atomic theory, this result implies a constraint on the nEDM of
$|d_n|<1.6\times 10^{-26}~e$cm.  The free neutron EDM was more
recently constrained by a measurement done using ultracold neutrons at
the Paul Scherrer Institute (PSI) which determined $|d_n|<1.8\times
10^{-26}~e$cm (90\% C.L.)~\cite{bib:psi2020}.  This work is noteworthy
in that it is the first nEDM measurement conducted using a new kind of
UCN source based on superthermal production using a spallation-driven
neutron source.  The experiment used an improved version of the nEDM
apparatus that was used previously at the Institute Laue-Langevin
(ILL) reactor~\cite{bib:baker,bib:afachnew}, but connected to this new
source of UCN.

Recent theoretical work addressing the physics impact of a new precise
measurement of the nEDM has focused on three general (and overlapping)
themes: (1) new sources of CP violation beyond the standard
model~\cite{bib:cirigliano,bib:crivellin}, (2) baryogenesis scenarios,
especially new physics contributions to electroweak
baryogenesis~\cite{bib:bell} and (3) the strong CP problem, which is
in turn related to the existence of
axions~\cite{bib:carena,bib:mimura}.  The relationship of the nEDM to
the quark (chromo)EDM has also been elucidated~\cite{bib:mereghetti},
honing the relationship to the standard model effective field theory
and the low-energy effective field theory~\cite{bib:stoffer}.

Since this workshop is about feebly interacting particles, it is worth
mentioning the special relationship of the nEDM to a well-known feebly
interacting particle: the axion.  The CP-violating $\bar{\theta}$ term
in the QCD lagrangian can in principle lead to a large nEDM.  The
measured smallness of the nEDM and hence $\bar{\theta}$ provided
strong evidence Peccei-Quinn symmetry~\cite{bib:pq1,bib:pq2} which can
solve the strong CP problem.  The axion is the new boson arising from
this symmetry.  The QCD axion possesses a relationship between its
coupling and mass.  Searches for axionlike particles make no
assumption about this relationship, and the experiments may probe the
regions in coupling and mass space.  More recently, it has been
realized that an oscillating background axion field could give rise to
time-varying (oscillating) EDM's of
nucleons~\cite{bib:graham,bib:flambaum}.  This effect has been used to
place constraints on axionlike particles by searching for time
variations in the neutron EDM~\cite{bib:psiaxion}.  It has been noted
that the local axion density should not saturate the entire local dark
matter density, weakening these limits somewhat~\cite{bib:centers}.

\subsubsection{Experimental Technique}

The experimental technique used to determine the nEDM is to measure
the neutron's spin-precession frequency $\nu$ when placed in parallel
($+$) and antiparallel dc ($-$) magnetic ($B$) and electric fields
($E$)
\begin{equation}
h\nu_{\pm}=2\mu_nB\pm 2d_nE
\end{equation}
where $\mu_n$ is the neutron magnetic moment and $d_n$ its EDM.  Since
the second term reverses sign with the relative direction of the
applied fields, subtraction of the measured frequencies allows a
direct measurement of $d_n$.  Key experimental parameters in nEDM
experiments are the number of neutrons sampled, the strength of the
electric field that can be achieved, and the coherence time of the
precessing neutron spins, which should all be maximized for the best
sensitivity.  The most precise measurement to date used ultracold
neutrons (UCN).  UCN are valuable experimentally because they can be
stored in material traps.

The most recent nEDM experiment~\cite{bib:psi2020} was statistically
limited.  The leading systematic uncertainties were associated with
effects caused by magnetic field inhomogeneity.  An impressive
assortment of magnetic field
diagnostics~\cite{bib:trilogy1,bib:trilogy2,bib:trilogy3} led to a
final systematic uncertainty of $1.8\times 10^{-27}~e$cm.  This gives
confidence that the next generation of nEDM experiments should be able
to reach the $10^{-27}~e$cm level if they can meet their stated
statistical goals, with small improvements to the techniques used to
address systematic uncertainties.  New UCN source technologies are
being developed to improve the statistical error, which could lead to
a breakthrough in precision for the nEDM.  The new sources rely on
superthermal production of UCN, where excitations in materials carry
away the momentum and energy of entering neutrons.  Two leading
materials for UCN converters are superfluid $^4$He (He-II), and solid
ortho-deuterium (so-D$_2$).

\subsubsection{Ongoing and Future Experiments}

The current experimental situation and the recent theoretical work
strongly motivate a new, more precise measurement of the nEDM.
Several groups are pursuing nEDM measurements worldwide
(Table~\ref{tab:expt}).

The experiments pursue a broad variety of different experimental
techniques.  Most of the nearer term experiments pursue a similar
technique to the previous best experiment, which used ultracold
neutrons (UCN) stored in a room-temperature nEDM spectrometer.  These
experiments are grouped together in the upper portion of
Table~\ref{tab:expt}.  These experiments are motivated by access to
new superthermal sources of UCN.  Progress on UCN sources can
drastically improve the statistical precision of the experiment.

\begin{table}
\caption{\label{tab:expt}Ongoing and future nEDM experiments, a few of
  their principal features, and status.  The three sections of the
  table are meant to indicate the different experimental techniques
  being employed.  In the upper portion, ultracold neutron experiments
  performed using room-temperature nEDM spectrometers are grouped
  together.  The SNS experiment uses a polarized cold neutron beam to
  create ultracold neutrons and the EDM measurement is done entirely
  in He-II.  The lower section of the table groups together two
  neutron beam experiments.  The various experiments are discussed
  further in the text.}
\begin{center}
\begin{tabular}{lll}\hline
Experiment & Features & Status \\\hline
n2EDM (PSI)	&	spallation so-D$_2$, MSR& 2022-23 UCN comm.\\
PanEDM (ILL)    &	reactor He-II, MSR	& commissioning ongoing\\
LANL		&	spallation so-D$_2$, MSR& starting eng. run 2022\\
TUCAN (TRIUMF)  &	spallation He-II, MSR	& upgrading, first UCN 2024\\
PNPI            &	dual cell, previous meas't	& upgrading UCN source\\\hline
nEDM \@ SNS	&	cryogenic source \& expt.& 2027-\\\hline
Beam EDM	&	intense pulsed neutron beam	&	R\&D, 2025-\\
J-PARC crystal	&	high $E$ in crystal     & R\&D\\\hline
\end{tabular}
\end{center}
\end{table}

The UCN groups aim at achieving a significant improvement in the
statistical uncertainty to the $\delta d_n\sim 10^{-27}~e$cm level
with experiments beginning soon.

The n2EDM experiment will couple a completely new room-temperature EDM
spectrometer to the cryogenic solid ortho-deuterium UCN source at
PSI~\cite{bib:n2EDM,bib:thorne}.  The experimental plan embodies the
features of most of the other upcoming nEDM experiments.  The
superthermal UCN source will deliver neutrons to a superconducting
magnet through which high-field seeking UCN will pass.  The polarized
UCN will enter a magnetically shielded room (MSR) containing two
measurement cells above and below a central high-voltage electrode.  A
very homogeneous vertical magnetic field will be created within the
MSR using a system of coils.  The neutrons will be stored in the
measurement cells where an ac field will excite a $\pi/2$ spin
reorientation.  The UCN spins will then be allowed to precess about
the static field for a period of 180 s.  After this time, a second
$\pi/2$ pulse, in phase with the first, will be applied and the UCN's
will be drained from the cell to spin-analyzing detectors that will
sense the accrued phase of the UCN spins relative to the ac field's
clock.  In this way, the spin precession frequency of the UCN is
deduced.  Periodically, the polarity of the electric field will be
reversed so that each measurement cell will be able to determine the
nEDM independently.  An optically probed $^{199}$Hg comagnetometer
will be used in each cell to monitor the magnetic field at the same
time as the nEDM measurement is being conducted.  Furthermore, an
array of Cs atom magnetometers external to the measurement cell will
monitor the magnetic field and be used to understand its spatial
distribution.  Mapping of the magnetic field will be done in offline
experiments to understand the spatial distribution even more
precisely.  These techniques were used to great success in the
previous nEDM experiment at PSI.  The experiment is projected to reach
a statistical sensitivity of $1\times 10^{-27}~e$cm in 500 data days
with the demonstrated performance of the PSI UCN source.

The PanEDM experiment is undergoing commissioning at ILL.  The
experiment also features an MSR, dual measurement cells, and advanced
optical magnetometry techniques.  The UCN source for the experiment,
SuperSUN, is simultaneously being prepared for first UCN production.
The PanEDM experiment will then be coupled to the Phase I SuperSUN
source~\cite{bib:jentschel}.

A room-temperature nEDM spectrometer is being developed at Los Alamos
National Laboratory (LANL), to use a recently upgraded
spallation-driven solid ortho-deuterium UCN source.  The source has
already been demonstrated to deliver the required densities to a test
experiment~\cite{bib:itoprc}, and based on those results a statistical
uncertainty on the nEDM of $2\times 10^{-27}~e$cm will be achieved in
one live-year of running.  The MSR for the experiment has been
installed and characterized.  Currently the precession chambers and
UCN valves are being assembled.  An engineering run is planned for
2022 \cite{bib:wong}.

The double-chamber technique was pioneered by the Petersburg Nuclear
Physics Institute (PNPI) group working at ILL, who obtained a
competitive nEDM measurement~\cite{bib:serebrov2}.  The PNPI group
plans to use the same apparatus with a future He-II UCN source at the
WWR-M reactor in Gatchina.

The TUCAN EDM experiment will be discussed in more detail in
Section~\ref{sec:3}.  The experiment features a spallation-driven
He-II source of UCN coupled to a room temperature nEDM experiment.

The SNS EDM experiment will use a fully cryogenic experiment with UCN
being produced and interrogated within He-II.  The He-II will also
serve as an insulator thereby allowing a large $E$ field to be
obtained.  The experiment will use of a small amount of polarized
$^3$He within the He-II as both a comagnetometer and sensor for the
neutron spins via the strongly spin-dependent neutron capture
cross-section.  The experiment aims for first data taking in late
2027~\cite{bib:plaster}.  It is anticipated that the new techniques
developed could lead to the breakthrough to the next order of
magnitude in precision beyond that achievable in room-temperature nEDM
experiments.  Using a new dressed-spin technique to analyze the
neutron and $^3$He spins, the experiment will reach a projected 90\%
C.L. sensitivity of $2.9\times 10^{-28}~e$cm.

Finally, Table~\ref{tab:expt} lists two experiments which will use
beams of cold neutrons to measure the nEDM.

The BeamEDM experiment will be conducted at ILL and in the future at
the European Spallation Source (ESS).  It will use the method employed
prior to the UCN-based EDM measurements.  In this method, cold neutron
beams are passed through a long set of parallel plates producing a
high electric field.  Since many more neutrons can be sensed, and a
higher electric field produced, this technique can be competitive with
the UCN experiments.  A disadvantage of the technique is that
systematic shifts in the neutron spin-precession frequency arising due
to $\vec{v}\times\vec{E}$ do not average to zero as for the UCN
experiments, giving rise to a false EDM signal.  This problem can be
mitigated using pulsed neutron beams.  A prototype apparatus was
operated recently at ILL in 2020.  The experiment was used to place
limits on axionlike particles in a region of parameter space
previously inaccessible to laboratory-based
experiments~\cite{bib:schulthess,bib:beamedmprl}.  The full BeamEDM
apparatus would be developed for ESS.

An experiment conducted at JPARC aims to use diffraction of neutrons
passing through a single crystal to determine the nEDM.  The advantage
is that the crystal can be selected to have a very large internal
electric field, thereby enhancing the sensitivity to the nEDM.  The
experiment has produced initial results on diffraction
techniques~\cite{bib:itohpub}.

\subsubsection{The TUCAN EDM Experiment\label{sec:3}}

\begin{figure}[ht]
\begin{center}
\includegraphics[width=\textwidth]{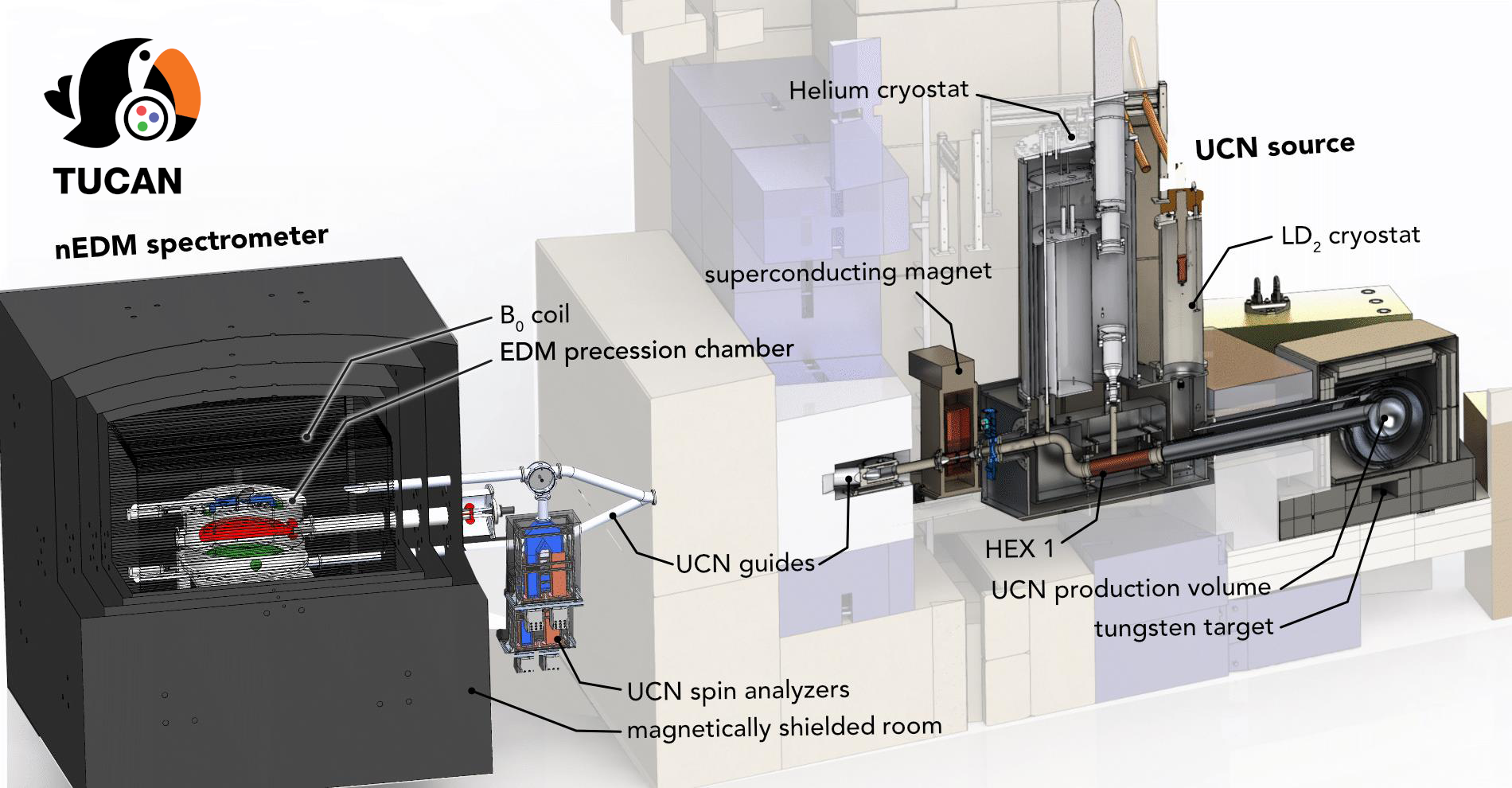}
\caption{Schematic diagram of the TUCAN EDM experiment.  The major
  portion of the biological shielding is not shown.  Protons strike a
  tungsten spallation target. Neutrons are moderated in a volume of
  LD$_2$ cooled by the LD$_2$ cryostat.  They become UCN in a UCN
  production volume containing He-II, which is cooled by a helium
  cryostat.  The UCN exit the superfluid volume and are transported
  through the superconducting magnet and UCN guides to reach the nEDM
  spectrometer located within a magnetically shielded room (MSR).  UCN
  spin analyzers detect the UCN at the end of each EDM experimental
  cycle.  For scale, the innermost layer of the MSR is a 2.4~m
  side-length cube.}
\label{fig:MesonCAD}
\end{center}
\end{figure}

The TUCAN (TRIUMF Ultra-Cold Advanced Neutron) EDM experiment aims to
measure the nEDM to a statistical precision of $d_n=1\times 10^{-27}~e$cm
in 400 days.  A layout of the planned experiment is presented in
Fig.~\ref{fig:MesonCAD}.

The facility at TRIUMF (Vancouver, Canada) previously operated a
prototype ``vertical'' UCN source that was developed in Japan, in
experiments conducted from 2017-2019.  The first results from its
operation at TRIUMF enabled characterization of the UCN source,
especially the temperature dependence of the UCN losses from the
superfluid helium (He-II) converter~\cite{bib:physrevc}.  It also
resulted in the largest UCN production results from this source, which
had been previously operated using a lower intensity spallation
facility at RCNP Osaka.  These and other results from measurements
conducted in subsequent runs were used to benchmark simulations for
the ongoing UCN source upgrade.

The base infrastructure which drives the UCN source will remain the
same for the upgrade.  This includes a fast kicker magnet which drives
a neutron spallation target, and proton beamline with appropriate
diagnostics, ending in a 20~kW tungsten spallation
target~\cite{bib:kicker,bib:bl1u}.  The entire beamline been developed
with the design beam current of 40~$\mu$A in mind.

The UCN source upgrade~\cite{bib:higuchi} will substantially increase
the UCN output compared to the vertical source.  Since the neutron
optical potential of the He-II is a mere 18 neV, the new source will
use near-horizontal extraction rather than the vertical extraction of
the prototype source (gravity represents a 100 neV/m barrier to UCN).
The cold moderator for the source will be upgraded from D$_2$O to
LD$_2$ with considerably higher production efficiency and smaller
uncertainty in the behavior of the cold neutrons (no scattering kernel
exists for D$_2$O ice).  The beam power will be increased from 1
$\mu$A to 40 $\mu$A necessitating a refrigerator upgrade from 300 mW
to 10 W at the operating temperature near 1~K.  The increased heat
flux also necessitates a new large-area heat exchanger that is
compatible with UCN.  The He-II production volume itself will also be
enlarged from 8~L to 33~L.

The helium cryostat (Fig.~\ref{fig:MesonCAD}) has been built and
tested cryogenically in Japan in 2019-2021.  It was then shipped to
TRIUMF and it is being prepared for installation.  One of the most
challenging aspects of the new cryostat is the main heat exchanger
(HEX 1) between the $^3$He refrigerant and the isotopcially pure
superfluid $^4$He used in the UCN production volume.  The UCN
production volume for the source has been fabricated, coated
internally with nickel plating, and tested with UCN at LANL.  It is
now being integrated into a cryostat that will enable it to be filled
with superfluid $^4$He.  The cryo-connection box which links these two
elements is also in preparation at the vendor site, for installation
at TRIUMF in early 2023.

The nEDM spectrometer being developed for TUCAN possesses a few
features that are unique relative to the previous generation of nEDM
experiments.  Aside from the usual features of the next generation of
EDM experiments (such as an MSR and dual measurement cells), it will
feature a self-shielded main precession field ($B_0$) coil, and Cs
magnetometers based on non-linear magneto-optical rotation (the
magnetometers also pursued by the PanEDM
collaboration~\cite{bib:rosner}).

The major subsystems of the spectrometer have been developed and the
experiment is entering the design and construction phase.  The
magnetically shielded room which will house the experiment began
installation in October 2022, and will complete installation in late
summer 2023.

The plan for the project calls for UCN production with the new TUCAN
source in 2024.  By 2025, the both the source and EDM experiment will
be ready for first data-taking.  The initial goal of the experiment is
to demonstrate the capability to reach $10^{-27}~e$cm precision.

\subsubsection{Conclusions}

Precise measurements of the nEDM carry a strong physics interest
because they place a tight constraint on CP violation.  The nEDM
addresses the strong CP problem and axions, new sources of CP
violation beyond the standard model at the TeV scale and beyond, and
baryogenesis scenarios such as those similar to or based on
electroweak baryogenesis.

The experimental situation is highly competitive with a large number
of experiments planned to commence in the near future that are
pursuing a variety of new techniques.  The next generation of nEDM
experiments aim to reduce the uncertainty by an order of magnitude, to
the $10^{-27}~e$cm level.  For room-temperature experiments, this
improvement is expected to come from an increase in the number of UCN
delivered from superthermal UCN sources of either solid
ortho-deuterium, or He-II.  Several experimental groups are also
developing innovative techniques which could surpass the next
generation of experiments, improving the precision to the
$10^{-28}~e$cm level.

%\subsubsection{Acknowledgments}

%The author is grateful to those who provided updates on their
%experiments, namely: S.~Degenkolb, P.~Fierlinger, T.~Ito, B.~Lauss,
%F.~Piegsa, G.~Pignol, B.~Plaster, A.~Serebrov, J.~Thorne and the
%author's collaborators on the TUCAN project, principally B.~Franke,
%T.~Higuchi, S.~Kawasaki, and R.~Picker.  This work was supported in
%part by the Natural Sciences and Engineering Research Council Canada,
%Canada Foundation for Innovation, and the Canada Research Chairs
%program.

%-------------------------------------------
\subsection{Limits on hadronic \texorpdfstring{$CP$}{CP}-violating interactions and axion dark matter from EDM experiments with paramagnetic molecules -- {\it V.~Flambaum~and~I.~B.~Samsonov}}
\label{ssec:flambaum}
{\it Author:  V. V. Flambaum and I.~B.~Samsonov, <v.flambaum@unsw.edu.au> - Joint Session with PSI 2022}
%-------------------------------------------
% \documentclass[12pt]{article}
% \usepackage{graphicx}
% \usepackage{amssymb}
% \usepackage{amsmath}
% \usepackage{gensymb}

% \topmargin=-1.5cm
% \oddsidemargin=0cm
% \textheight=23cm
% \textwidth=17cm

% \begin{document}

% \begin{center}
% {\Large\bf
% Limits on hadronic $CP$-violating interactions and axion dark matter from EDM experiments with paramagnetic molecules}

% \vspace{5mm}
% {\bf
% V.~V.~Flambaum and I.~B.~Samsonov}

% \it
% School of Physics, University of New South Wales, Sydney 2052, Australia

% \end{center}

\subsubsection{Introduction}

It is a fundamental problem in theoretical and experimental high-energy physics to determine the electric dipole moments (EDMs) of elementary particles, because they allow us to probe the limits of the Standard Model of elementary particles and search for the new physics beyond the Standard Model. Recently, such experiments have been used to search for axion dark matter which produces  EDMs and other $T,P$-odd effects \cite{Graham:2011qk,Graham:2013gfa,Budker:2013hfa,Stadnik:2013raa,Abel:2017rtm,Stadnik:2017hpa,Dzuba:2018anu,Roussy:2020ily,Aybas:2021nvn}.

In the recent years, atomic and molecular experiments demonstrated a tremendous progress in measuring these EDMs owing to ever growing precision in spectroscopy. Especially promising are the experiments dealing with paramagnetic molecules \cite{Cairncross:2017fip,Roussy:2022cmp,Roussy:2022cmp} which aim mainly at measuring the electron EDM. In this paper, we study the sensitivity of these experiments to proton and neutron EDMs, as well as to $P$- and $T$-symmetry violating nuclear interactions. As we will show, the constraints obtained from these experiments on the hadronic sources of $CP$ violation are becoming competitive with the corresponding constraints from experiments with diamagnetic atoms \cite{Graner:2016ses,Sachdeva:2019rkt,Allmendinger:2019jrk}.

In paramagnetic atoms, the atomic EDM appears, in particular, as a result of the following semileptonic interaction:
\begin{equation}
    {\cal L} = \frac{G_F}{\sqrt2}C^p_{SP} \bar e i\gamma_5 e \bar pp + \frac{G_F}{\sqrt2}C^n_{SP} \bar e i\gamma_5 e \bar nn\,,
    \label{L}
\end{equation}
where $G_F$ is the Fermi constant, $e$, $p$ and $n$ are respectively the electron, proton and neutron spinor fields. $C_{SP}^p$ and $C^n_{SP}$ are the electron couplings to the proton and neutron, respectively. In an atom with $Z$ protons and $N=A-Z$ neutrons, these couplings may be combined a single constant $C_{SP} = C_{SP}^p Z/A + C^n_{SP}N/A$. Exactly this combination of couplings is usually measured in experiments searching for EDMs with paramagnetic molecules. The most advanced limit on $C_{SP}$ is placed by the ACME collaboration \cite{Roussy:2022cmp}, which uses the molecule $^{232}$ThO and the JILA group \cite{Cairncross:2017fip,Roussy:2022cmp} which uses $^{180}$HfF$^+$ molecule:
\begin{equation}
    |C_{SP}|_{\rm ThO} < 7.3\times 10^{-10}\,, \qquad\qquad 
     |C_{SP}|_{\rm HfF^+} < 4.5\times 10^{-10}\,.
    \label{Climit}
\end{equation}

The coupling constant $C_{SP}$ in Eq.~(\ref{Climit}) receives various contributions from $CP$-violating interactions at the hadronic level. In Ref.~\cite{Flambaum:2019ejc}, the contributions to this coupling from the two-photon and $\pi,\eta$-meson exchanges between electrons and nucleons were calculated. In Ref.~\cite{Flambaum:2020gou}, additional contribution to $C_{SP}$ due to virtual nuclear transitions and nucleon EDMs. This allows us to find $C_{SP}$ as a function of proton $d_p$ and neutron $d_n$ EDMs, QCD $\theta$-angle, quark-chromo EDMs and $\pi$-meson-nucleon couplings to the leading order. Given this function, we extract the limits on these parameters from the constraint (\ref{Climit}). We show that some of these limits are comparable with the ones originating from EDM experiments with diamagnetic atoms.

If the dark matter is represented by axion particles, the QCD $\bar\theta$ angle may have small variations thus inducing oscillating EDMs of nucleons. Such oscillations were looked for in the work \cite{Roussy:2020ily} where null results were found over the axion mass range from $10^{-22}$ to $10^{-15}$ eV. 

%%%%%%%%%%%%%%%%%%%%%%%%%%%%%%%%%%%%%%%%%%%%%%%%%%

\subsubsection{Contribution from $CP$-odd nucleon polarizability}
\label{mesons}

\begin{figure}
    \centering
    \begin{tabular}{ccccc}
    \includegraphics[width=3cm]{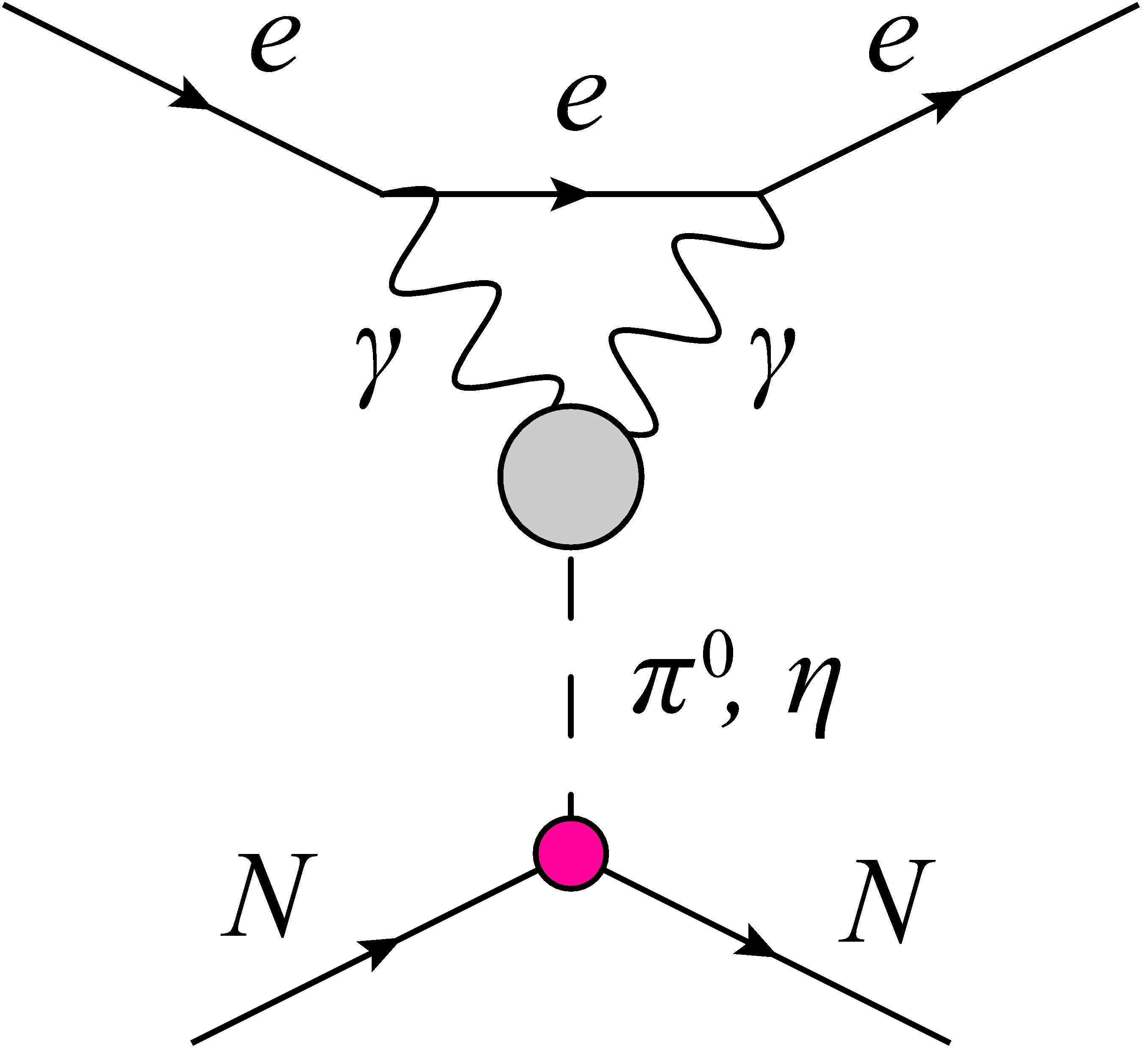} &
    \includegraphics[width=3cm]{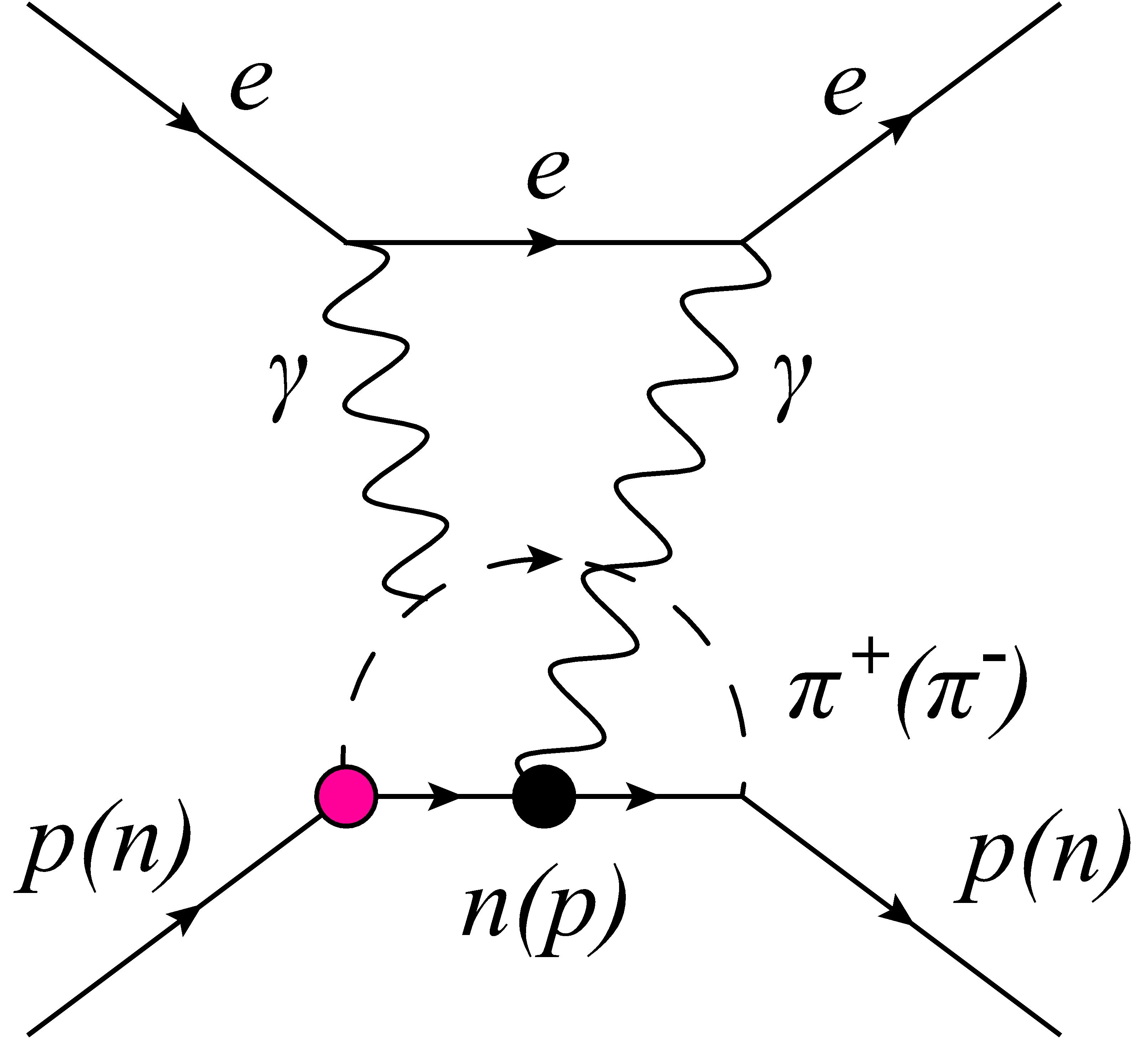} &
    \includegraphics[width=3cm]{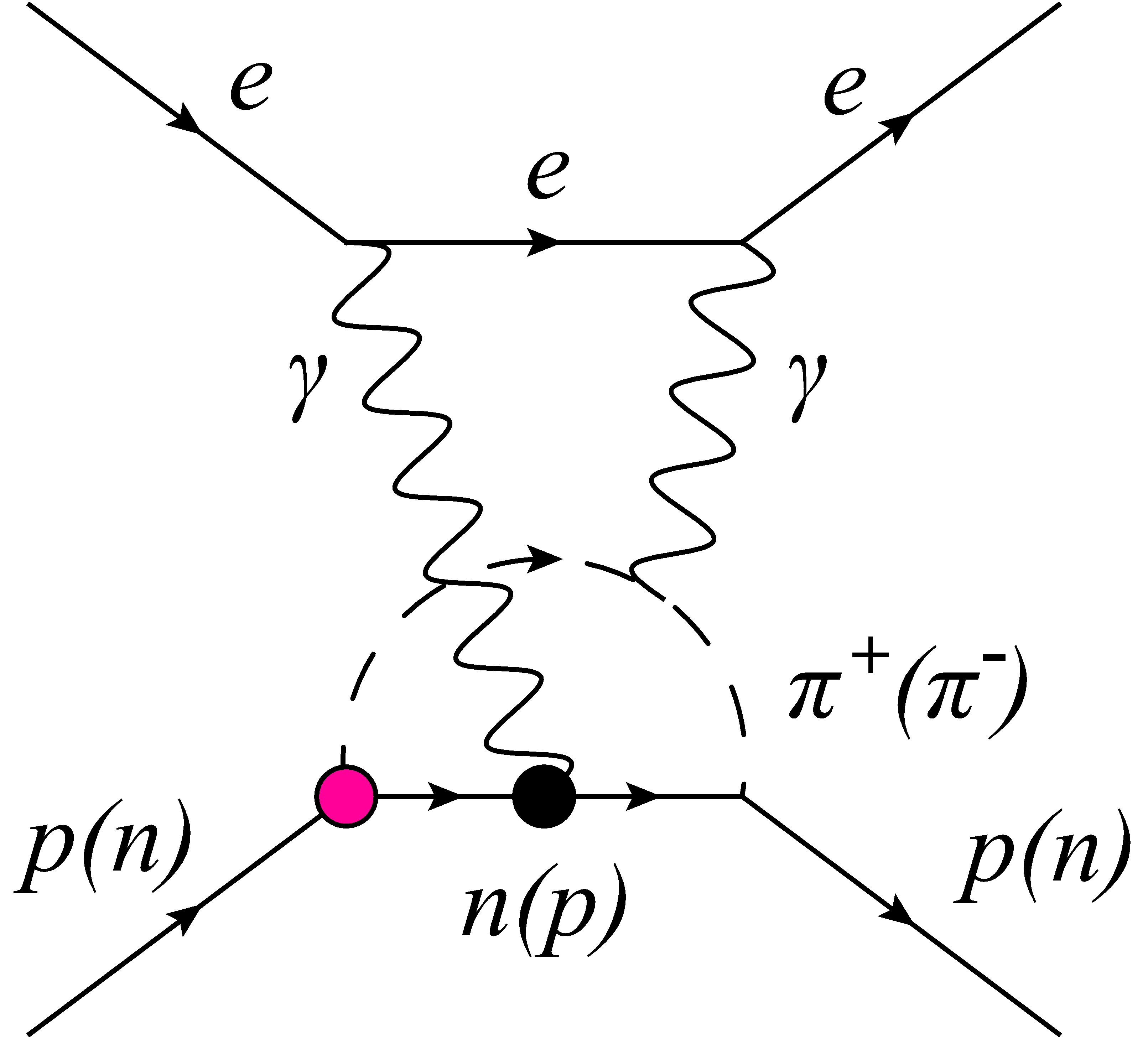} &
    \includegraphics[width=3cm]{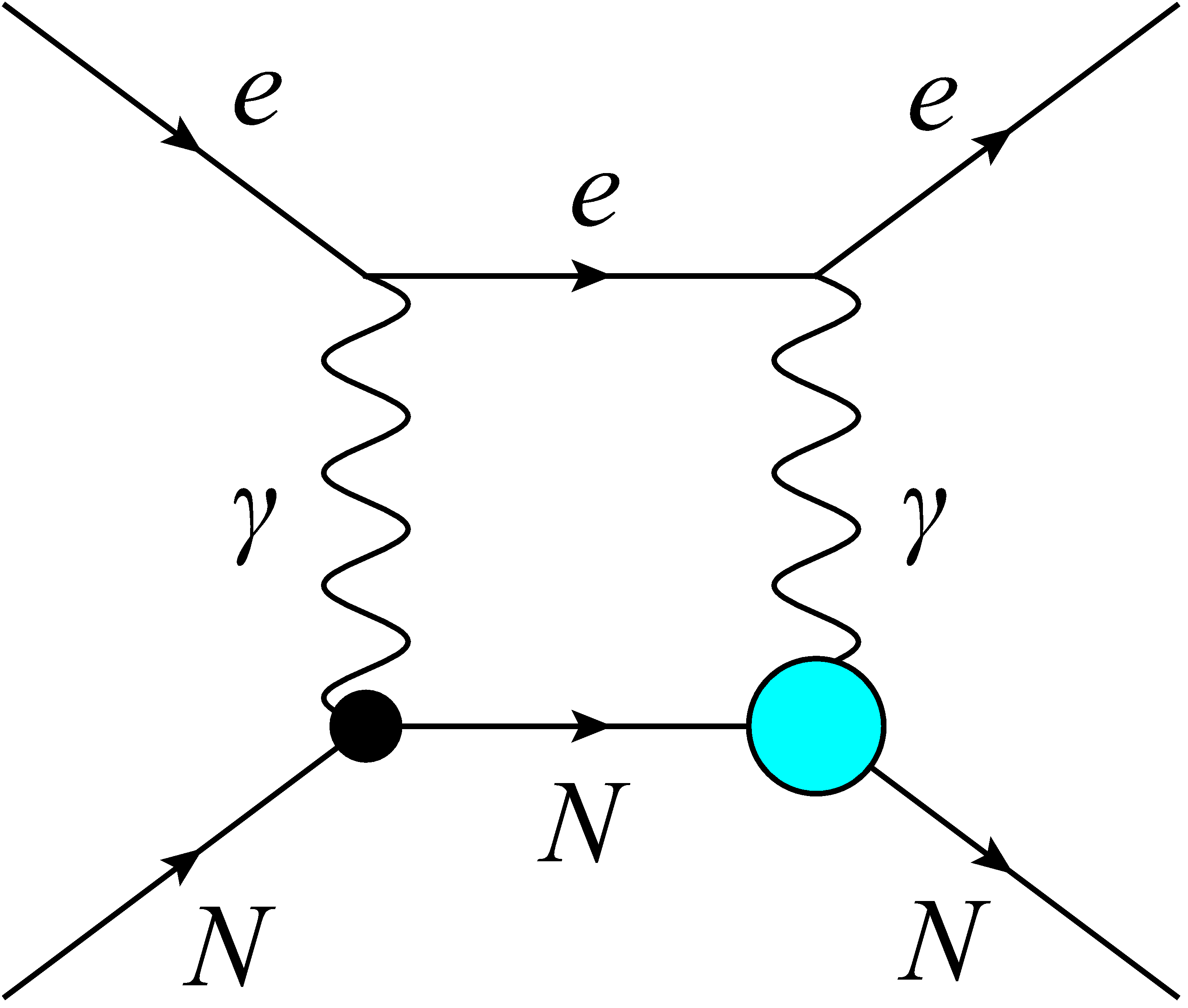} &
    \includegraphics[width=3cm]{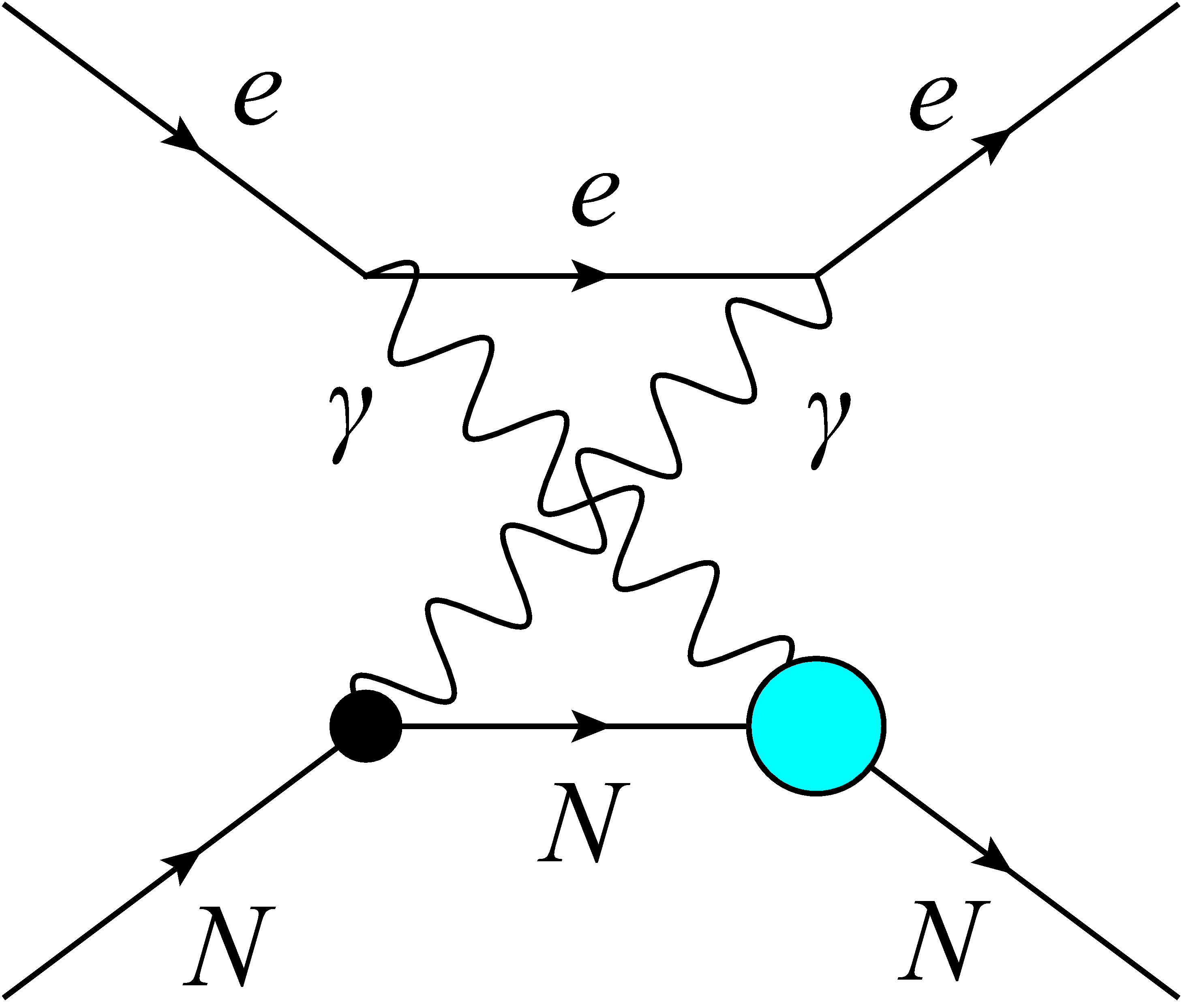} \\
    a & b & c & d &e
    \end{tabular} 
    \caption{Feynman diagrams representing different contributions to the atomic EDM due to the effective $CP$-violating  interactions (\ref{L}).}
    \label{FigDiagrams}
\end{figure}

At the hadronic level, the $CP$-violating interactions originate from nucleon EDMs $d_p$ and $d_n$, as well as from the interaction with $\pi^{0,\pm}$ pions and octet of $\eta$ mesons,
\begin{equation}
    {\cal L} = -\frac i2 F_{\mu\nu}( d_n \bar n  \sigma^{\mu\nu} \gamma_5 n + d_p \bar p  \sigma^{\mu\nu} \gamma_5 p)
    + \bar g^{(0)}_{\pi N N} \bar N \tau^a N \pi^a 
    +\bar g^{(1)}_{\pi NN} \bar N N \pi^0
    +\eta \bar N(\bar g^{(0)}_{\eta NN}+\bar g^{(1)}_{\eta NN}\tau^3)N\,,
\end{equation}
where $N = (p,n)^T$ is the nucleon doublet, $g^{(0,1)}_{\pi NN}$ are the isovector and isoscalar $CP$-odd pion-nucleon couplings, respectively. The contributions to the atomic EDM originating from the interaction (\ref{L}) are represented by Feynman diagrams in Fig.~\ref{FigDiagrams}. These diagrams involve the two-photon exchange via $CP$-odd nucleon polarizabilities $\beta_p$ and $\beta_n$,
\begin{equation}
    {\cal L}_{\rm nuc.pol.} = (\beta_p \bar p p + \beta_n \bar n n )\vec E\cdot \vec B\,,
\end{equation}
where $\vec E$ and $\vec B$ are electric and magnetic fields, respectively.
These polarizabilities were calculated in Ref.~\cite{Flambaum:2019ejc} using the chiral perturbation theory:
\begin{align}
    \beta_{p(n)}^{\rm L.O.} &= - \frac{\alpha}{\pi F_\pi m_\pi^2}
    \left[ \bar g^{(1)}_{\pi NN} +(-)\bar g^{(0)}_{\pi NN} + \frac{\bar g^{(0)}_{\eta NN}}{\sqrt3} \frac{m_\pi^2 F_\pi}{m_\eta^2 F_\eta} \right]\,,\label{betaLO}\\
    \beta_k^{\rm N.L.O.}& = \frac{\alpha g_A \bar g^{(0)}_{\pi NN}}{4 F_\pi m_N m_\pi}
    \left\{
    \begin{array}{ll}
        -\mu_n/\mu_N & \mbox{ for } k=p\\
        \mu_p/\mu_N  & \mbox{ for } k=n\,,
    \end{array}
    \right.
    \label{betaNLO}
\end{align}
where $\alpha\approx1/137$ is the fine structure constant, $F_\eta \approx F_\pi \approx 92$~MeV is the pion decay constant, $m_\pi$ and $m_\eta$ are meson masses, $g_A \approx 1.3$ is the axial triplet coupling, $m_N$ is the nucleon mass, $\mu_{n,p}$ are the nucleon magnetic moment, and $\mu_N$ is the nuclear magneton. The leading-order (L.O.) contribution (\ref{betaLO}) comes from from the diagram (a) in Fig.~\ref{FigDiagrams} which takes into account the three-level mesons exchange. The next-to-leading order (N.L.O.) contribution (\ref{betaNLO}) corresponds to one-loop charged pion exchange represented by the diagrams (b) and (c) in Fig.~\ref{FigDiagrams}.

To find the contribution to atomic EDM from the $CP$-odd nucleon polarizabilities (\ref{betaLO},\ref{betaNLO}), one has to calculate the photon loops in Feynman diagrams (a-c) in Fig.~\ref{FigDiagrams}. This calculation was performed in Ref.~\cite{Flambaum:2019ejc} with logarithmic accuracy. The corresponding contribution to the coupling constant $C_{SP}$ is
\begin{equation}
    \frac{G_F}{\sqrt2} C_{SP}^{(\beta )} = - \left( \frac{Z}{A}\beta_p + \frac{N}{A}\beta_n\right)
    \frac{3\alpha m_e}{2\pi}\ln\frac{\cal M}{m_e}\,,
    \label{Cbeta}
\end{equation}
where $\cal M$ is the renormalization scale. For the L.O.\ contributions, this scale may be taken as the $\rho$ meson mass, ${\cal M}\approx m_\rho$, while for the N.L.O.\ ones, this scale equals the pion mass, ${\cal M} = m_\pi$ \cite{Flambaum:2019ejc}. 

%%%%%%%%%%%%%%%%%%%%%%%%%%%%%%%%%%%%%%%%%%%%%%%%%%%

\subsubsection{Contribution from nucleon EDMs due to $CP$-odd nuclear transitions}
\label{NuclTransition}

In this section, we consider the contributions to the atomic EDM from the Feynman diagrams in Fig.~\ref{FigDiagrams}, which include the nucleon EDMs $d_{p,n}$ and virtual nuclear transitions. The calculation of these contributions requires accurate consideration of both $CP$-odd nuclear transitions in different nuclei and calculation of electronic matrix elements that take into account both discrete and continuum states in atoms. This calculation was performed in Ref.~\cite{Flambaum:2019ejc,Flambaum:2020gou}; here we review some features and results of this work.

{\bf Atomic EDM due to contact $CP$-odd interaction - }

We start by considering the Hamiltonian of $CP$-odd interaction between a valence electron and the nucleus \cite{Flambaum:1985gx}
\begin{equation}
    H_{\rm cont} = \frac{iG_F}{\sqrt2} A C_{SP} \gamma_0 \gamma_5 \rho(\vec R)\,,
    \label{Hcont}
\end{equation}
where $\vec R$ is the position vector of the electron and $\rho(\vec R)$ is the nuclear charge density function normalized to 1 over a spherical nucleus of radius $R_0$. In heavy atoms, the $s_{1/2}$ and $p_{1/2}$ electron wave functions have large relativistic enhancement inside the nucleus, as compared with other wave functions with higher angular momentum $l$, which have very small at the nucleus. As a result, the atomic EDM receives dominant contributions from the matrix elements of the operator (\ref{Hcont}) with the $s_{1/2}$ and $p_{1/2}$ states,
\begin{equation}
    \vec d = 2\frac{\langle s_{1/2}|e\vec R|p_{1/2}\rangle \langle p_{1/2}|H_{\rm cont}|s_{1/2}\rangle}{E_{p_{1/2}}-E_{s_{1/2}}}\,.
    \label{AtomicEDM}
\end{equation}
This matrix element may be calculated analytically using approximate wave functions at small distance \cite{Flambaum:2020gou}
\begin{equation}
    \langle p_{1/2}|H_{\rm cont}|s_{1/2}\rangle = -c_{s_{1/2}}c_{p_{1/2}} \frac{G_F C_{SP}}{10\sqrt2\pi}\frac{1+4\gamma}{\Gamma(2\gamma+1)^2}
    \frac{AZ\alpha}{R_0^2} \left(
    \frac{2ZR_0 }{a_B}
    \right)^{2\gamma} \,,
    \label{Hcontelement}
\end{equation}
where $\gamma=\sqrt{1-Z^2\alpha^2}$ is the relativistic factor. Here $c_{s_{1/2}}$ and $c_{p_{1/2}}$ are the normalization coefficients of the corresponding wave functions. Exact values of these coefficients are unimportant, as they will cancel out in the final result. 

{\bf Atomic EDM due to the nucleon permanent EDMs - }

Let $\vec d_i = d_i \vec \sigma_i$  and $\vec \mu_i = \mu_N (g_i^l \vec l_i + g_i^s \vec s_i)$ be the operators of electric and magnetic dipole moments of $i$-th nucleon in the nucleus, respectively. Here $d_i = d_p$ or $d_n$ are the proton or neutron permanent EDMs, $g_l^l = g_{p,n}^l$ and $g_i^s = g^s_{p,n}$ are the orbital and spin $g$-factors of the nucleons, $\vec l$ and $\vec s$ are the orbital momentum and spin operators, respectively. The operators $\vec d_i$ and $\vec \mu_i$ couple with electric and magnetic fields of the valence electron. The corresponding interaction Hamiltonian reads
\begin{equation}
    H = - \sum_{i=1}^A (H_i^d + H_i^\mu)\,,
    \quad
    H_i^d = \frac{e \vec d_i\cdot (\vec R - \vec r_i)}{|\vec R - \vec r_i|^3}\,,
    \quad
    H_i^\mu = \frac{e \mu_i\cdot [(\vec R - r_i)\times \vec \alpha]}{|\vec R - \vec r_i|^3}\,,
    \label{HHH}
\end{equation}
where $\vec r_i$ are the position vectors of the nucleons and $\alpha = \left( \begin{smallmatrix}  
0 &\vec \sigma \\ \vec \sigma &0
\end{smallmatrix}
\right)$ are the Dirac matrices acting on electron wave functions.

Let $m$ and $m'$ be generalized quantum numbers of atomic $|m\rangle$ and nuclear $|m'\rangle$ states, respectively. We assume that the atomic state $|mm'\rangle$ may be factorized into atomic and nuclear parts, $|mm'\rangle = |m\rangle |m'\rangle$. In this approximation, the contribution to the atomic EDM appears in the second order of perturbation theory \cite{Flambaum:2020gou},
\begin{equation}
    \vec d = -2\sum_{m,n,n'} \frac{\langle 0 |e\vec R | m\rangle \langle 0'm | H | nn'\rangle\langle n'n| H |00'\rangle}{(E_m - E_0)[\Delta E_n + {\rm sgn}(E_n)\Delta E_{n'}]}\,,
    \label{vecd}
\end{equation}
where the sum is taken over the excited states with $m\ne0$ and $nn'\ne00'$, and $\Delta E_n \equiv E_n - E_0$, $\Delta E_{n'} = E_{n'} - E_{0'}$.

It is important to note that the generalized sum in Eq.~(\ref{vecd}) involves intermediate electronic states with positive and negative states which are inherent in Dirac's theory. The negative states contribute with opposite sign of $\Delta E_{n'}$ because they may be interpreted as  blocking contributions for the electrons from the Dirac sea which cannot be excited to the occupied electronic orbitals. This account of the negative energy states is similar to the one in calculations of the atomic energy shift due to the nuclear polarizability \cite{Plunien1,Plunien:1995zz,Flambaum:2021yyz}. 

The expression for the atomic EDM (\ref{vecd}) is similar to (\ref{AtomicEDM}) and appears from it upon the substitution of the operator of the contact interaction $H_{\rm cont}$ with the following effective Hamiltonian
\begin{equation}
    H_{\rm eff} = - \sum_{nn'\ne 00'}
    \frac{|m\rangle \langle 0'm| H|nn'\rangle \langle n'n | H |00'\rangle \langle 0|}{\Delta E_n + {\rm sgn}(E_n)\Delta E_{n'}}\,.
\end{equation}
As a result, the problem is reduced to calculation of the matrix element of this effective Hamiltonian. This matrix elements may be represented in terms of the operators (\ref{HHH}) as
\begin{equation}
    \langle p_{1/2} | H_{\rm eff}| s_{1/2} \rangle = -\sum_{n'\ne0'}\sum_{i,j=1}^A 
\left[
\sum_n \frac{\langle p_{1/2} 0'|H_i^\mu|n'n\rangle\langle nn'|H_j^d|0's_{1/2}\rangle}{\Delta E_n + {\rm sgn}(E_n)\Delta E_{n'}} + (s_{1/2}
    \leftrightarrow p_{1/2})\right]\,.
    \label{calM}
\end{equation}
The two terms in the right-hand side of this equation correspond to the two Feynman diagrams (d) and (e) in Fig.~\ref{FigDiagrams}.

The matrix elements in Eq.~(\ref{calM}) were calculated in Ref.~\cite{Flambaum:2020gou}. The nuclear matrix elements in this expression were calculated separately for spherical and deformed nuclei. These matrix elements correspond to nuclear M1 spin-flip single-particle transitions. The electronic matrix elements were calculated with approximate $s_{1/2}$ and $p_{1/2}$ electron wave functions from Ref.~\cite{KhriplovichBook}, which give a good description at small distance, i.e., inside and near the nucleus. The intermediate electronic states are exact Dirac wave functions in the continuous spectrum in a Coulomb field. The integration with these wave functions was performed numerically, with numerical errors under 5\%. The results of these calculations for $^{180}$Hf and $^{232}$Th atoms are:
\begin{equation}
\begin{aligned}
    \langle p_{1/2} |H_{\rm eff} | s_{1/2} \rangle_{^{180}\rm Hf} &= -2 c_{s_{1/2}} c_{p_{1/2}} \frac{\mu_N}{a_B} (673.7 d_p - 436.2 d_n)\,,\\
    \langle p_{1/2} |H_{\rm eff} | s_{1/2} \rangle_{^{232}\rm Th} &= -2 c_{s_{1/2}} c_{p_{1/2}} \frac{\mu_N}{a_B} (1363 d_p - 1473 d_n)\,.
\end{aligned}
\label{result1}
\end{equation}
Taken into account the uncertainties of calculations of nuclear matrix elements, the total accuracy of these calculations was estimated within 50\%.

The results of the calculations (\ref{result1}) should be compared with the matrix element of contact interaction (\ref{Hcontelement}). This allows us to find the contributions to $C_{SP}$ from nucleon permanent EDMs,
\begin{equation}
    C^{(d)}_{SP} = (\lambda_1 d_p + \lambda_2 d_n) \frac{10^{13}}{e\, {\rm cm}}\,,
    \label{Cd}
\end{equation}
where the coefficients are $(\lambda_1,\lambda_2)= (12,-9)$ for $^{180}$Hf and $(\lambda_1,\lambda_2)= (6.4,-6.9)$ for $^{232}$Th. Note that, in these coefficients, we have taken into account not only the contributions from discrete nuclear transitions (\ref{result1}), but also similar contributions from nuclear excitations to continuum spectrum calculated in Ref.~\cite{Flambaum:2019ejc}. Both these contributions come with the same sign and are comparable in magnitude. In addition to contributions with $d_p$ and $d_n$ in Eq.~(\ref{Cd}), we also estimated contribution of $P,T$-odd nucleon-nucleon interaction to $C_{SP}$ \cite{Flambaum:2020xcj}. This contribution does not produce a significant change of the estimate Eq.~(\ref{Cd}).

%%%%%%%%%%%%%%%%%%%%%%%%%%%%%%%%%%%%%%%%%%%%%%%%%%%%%%

\subsubsection{Limits on $CP$-violating hadronic interactions and nucleon EDMs}

In Sect.~\ref{mesons} we calculated the contribution to $C_{SP}$ from nucleon polarizability due to $\pi$ and $\eta$ mesons exchange while in Sect.~\ref{NuclTransition} we found the contributions due to nucleon permanent EDMs. Taking into account the limits on $C_{SP}$ (\ref{Climit}) from the experiments with paramagnetic molecules \cite{Roussy:2022cmp,Roussy:2022cmp} we can find the limits on the nucleon EDMs and $\pi,\eta$-meson $CP$-violating couplings. These limits are given in Table~\ref{Limits}. 

Note that the obtained limit on the proton EDM is just 14 times weaker than the recent constraint $|d_p|< 5\times 10^{-25} e\,{\rm cm}$ \cite{Flambaum:2019kbn} from the experiment \cite{Graner:2016ses} with the $^{199}$Hg atom. Remarkably, our limit on $d_p$ are nearly 50 times more stringent than the corresponding limits from the $^{129}$Xe EDM experiment \cite{Sachdeva:2019rkt,Allmendinger:2019jrk}.

\begin{table}[tb]
\begin{center}
\begin{tabular}{|c|c|c|}
\hline
                 			  & $^{232}{\rm ThO}$                   & $^{180}{\rm HfF}^+$ \\ \hline
$|C_{SP}|$       			  & $7.3\times 10^{-10}$\ \cite{Roussy:2022cmp} & $4.5\times 10^{-10}$\ \cite{Roussy:2022cmp} \\ \hline
$|d_p|$                       & $1.1\times 10^{-23}e\cdot{\rm cm}$  & $6.8\times 10^{-24}e\cdot{\rm cm}$ \\ \hline
$|d_n|$                       & $1.0\times 10^{-23}e\cdot{\rm cm}$  & $6.2\times 10^{-24}e\cdot{\rm cm}$ \\ \hline
$|\bar{g}^{(0)}_{\pi NN}|$    & $3.1\times 10^{-10}$                & $1.9\times 10^{-10}$                 \\ \hline
$|\bar{g}^{(1)}_{\pi NN}|$    & $3.3\times 10^{-10}$                & $2.0\times 10^{-10}$                 \\ \hline
$|\tilde{d}_d|$  			  & $9.3\times 10^{-25}{\rm cm}$        & $5.7\times 10^{-25}{\rm cm}$        \\ \hline
$|\tilde{d}_u|$  			  & $1.7\times 10^{-24}{\rm cm}$        & $1.0\times 10^{-24}{\rm cm}$        \\ \hline
$|\bar{\theta}|$ 			  & $1.4\times 10^{-8}$                 & $8.6\times 10^{-9}$                 \\ \hline
\end{tabular}
\end{center}
\caption{\label{Limits} Limits on absolute values of $CP$-violating hadronic parameters.}
\end{table}

In Table~\ref{Limits}, we present also limits on quark chromo-EDMs $\tilde d_u$ and $\tilde d_d$. These limits are extracted from the obtained limits on the nucleon EDMs and $CP$-violating pion interaction constants with the use of the relations between these parameters obtained in Refs.~\cite{Pospelov:2005pr,Pospelov:1999ha,Flambaum:2014jta,Pospelov:2000bw}. These relations are explicitly presented in Ref.~\cite{Flambaum:2020gou}.

Another important result in Table~\ref{Limits} is the constraint on the QCD vacuum angle $\bar\theta$. This constraint is found from the limits on the nucleon EDMs and $CP$-odd pion couplings with the use of the following relations \cite{Pospelov:2005pr,Pospelov:1999ha,Crewther:1979pi,deVries:2015una,deVries:2015gea,Bsaisou:2014zwa,Bsaisou:2012rg}:
\begin{equation}
\label{dp-theta}
\begin{aligned}
&d_p =(2.1\pm1.2)\times10^{-16}\bar{\theta}\, e\cdot{\rm cm}\,,\qquad
d_n =-(2.7\pm1.2)\times10^{-16}\bar{\theta}\, e\cdot{\rm cm}\,,\\
&\bar{g}^{(0)}_{\pi NN}=-(15.5\pm2.5)\times 10^{-3}\,\bar{\theta}\,,\qquad
\bar{g}^{(1)}_{\pi NN}=(3.4\pm2)\times 10^{-3}\,\bar{\theta}\,.
\end{aligned}
\end{equation}
These relations may be combined with Eqs.~(\ref{Cbeta}) and (\ref{Cd}). As a result, we find 
\begin{equation}
 C_{SP}=0.067\bar{\theta} \mbox{ for $^{180}$HfF$^+$}\,,\qquad
 C_{SP}=0.051\bar{\theta} \mbox{ for $^{232}$ThO}\,.
\end{equation}
This allows us to find the limits on the QCD vacuum angle given in the last line of Table~\ref{Limits}. This limit is still nearly two orders of magnitude weaker than the currently accepted constraint from neutron and Hg atom EDM experiment $|\bar\theta|<10^{-10}$ \cite{ParticleDataGroup:2018ovx}. 

We expect that further improvement of accuracy in the experiments with paramagnetic molecules would push the limits in Table~\ref{Limits} further.

\subsubsection{Constraints on the axion dark matter}

Axion dark matter manifests itself as an oscillating QCD vacuum angle $\bar{\theta}$. JILA group \cite{Roussy:2020ily} used their electric dipole moment (EDM) measurement data and our calculations \cite{Flambaum:2019ejc,Flambaum:2020gou} of the $\bar{\theta}$ contribution to the electron-nucleus interaction constant $C_{SP}$ (\ref{Cd}) to constrain the possibility that the HfF$^+$ EDM oscillates in time due to interactions with candidate dark matter axionlike particles (ALPs). They found no evidence of an oscillating EDM over a range spanning from 27 nHz to 400 mHz, and used this result to constrain the ALP-gluon coupling over the mass range $10^{-22}-10^{-15}$ eV. This was the first laboratory constraint on the ALP-gluon coupling in the $10^{-17}-10^{-15}$  eV range, and the first laboratory constraint to properly account for the stochastic nature of the ALP field.
%%%%%%%%%%%%%%%%%%%%%%%%%%%%%%%%%%%%%%%%%%%%%%%%%%%%%%%%%%%%%%%%%%%%

% \bibliographystyle{ieeetr}
% \bibliography{biblio}

% \end{document}

%%===============
%% Joint session with PSI
%%===============

%-------------------------------------------
\subsection{Precision measurements of the fine structure constant -- {\it P.~Cladè~and~S.~Guellati}}
\label{ssec:pierre_clade}
\textit{Author: Pierre Cladé, <pierre.clade@lkb.upmc.fr>}, \\
\textit{Saïda Guellati-Khelifa, <saida.guellati@lkb.upmc.fr>}
%-------------------------------------------

This section presents the state of the art of the standard model test performed by comparing the experimental value of the magnetic moment of the electron and its theoretical value derived from quantum electrodynamics (QED) calculations using the most accurate value of the fine structure constant.

\subsubsection{Electron magnetic moment}

The magnetic moment of the electron is proportional to its spin $\bm{S}$ and the Bohr magnetons $\mu_B$:
\begin{equation}
\bm{\mu} = -g_e\mu_B \frac{\bm{S}}{\hbar}.
\end{equation}
One of the triumphs of the Dirac equation was the prediction that $g_\mathrm{e}=2$. However, in 1947, the experiment of Kusch and Foley gave the first direct indication of a deviation of the electron's g-factor from Dirac's prediction\cite{Kusch1948}. This experiment was accurate enough to confirm Schwinger's calculation of this deviation, namely the anomalous magnetic moment of the electron\cite{Schwinger1948}. Since the magnetic moment of the electron has played an important role the development of quantum electrodynamics and later of the standard model (SM). 
The anomalous magnetic moment $a_e$ is defined as $a_e = (g_e-2)/2$. This quantity can be calculated precisely from the standard model. 
It includes three contributions: 
\begin{equation}
		a_e\left(\text{SM}\right) = a_e\left(\text{QED}\right) +
		a_e\left(\text{Weak}\right)  + a_e\left(\text{Hadron}\right)
		\label{eq:prediction}
\end{equation}
where $a_e\left(\text{Weak}\right) $ and $a_e\left(\text{Hadron}\right) $ correspond to the weak interaction and the hadronic interaction and  the QED contribution $a_e\left(\text{QED}\right)$, can be decomposed as: 
\begin{equation}
a_e(\text{QED})  = \sum_{n=1}^{\infty} A^{(2n)}
		\left(\frac{\alpha}{2\pi}\right)^{n} + \sum_{n=1}^{\infty}
		A_{\mu,
		\tau}^{(2n)}\left(\frac{m_e}{m_{\mu}},
		\frac{m_e}{m_{\tau}}\right)\left(\frac{\alpha}{2\pi}\right)^{n}
\end{equation}
This expression developed as a series of power of the structure constant of the fine structure constant $\alpha$ that contains all electronic contributions of QED (terms in $A^{(2n)}$ and muonic and tau contributions $A_{\mu,\tau}^{(2n)}\left(\frac{m_e}{m_{\mu}},\frac{m_e}{m_{\tau}}\right)$. More details can be found in \cite{aoyama2019}.

\begin{figure}
\includegraphics[width=.95\linewidth]{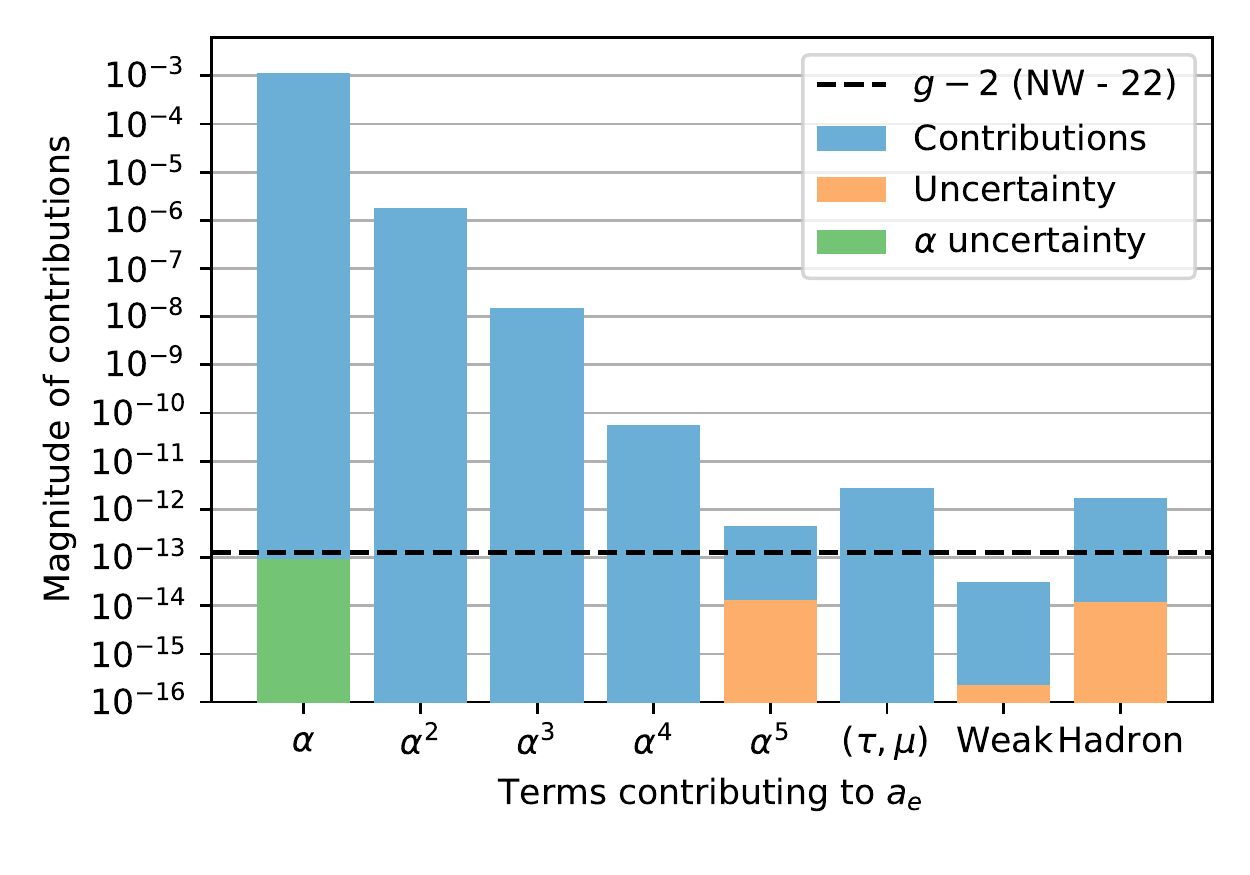}
\caption{Blue represents the magnitudes of the different contributions to $a_e$. In orange, uncertainty of the theoretical prediction (eq. \ref{eq:prediction}) and in green, uncertainty due to the fine structure constant \cite{Morel2020}. The dashed line represents the uncertainty of the direct measurement\cite{fan2022}.}
\label{fig:contributions}
\end{figure}
The magnetic moment of the electron is the most accurately determined property of an elementary particle. 
 Its experimental value is derived from the measurement of the cyclotron frequency $\nu_c = \frac{eB}{2\pi m_e}$ and the anomaly frequency $\nu_a = \nu_s - \nu_c$ (where $\nu_s=(g_e/2)\nu_c$ is the spin frequency) of a single electron in a constant magnetic field $B$.
 In 2022, G.~Gabrielse's team improved their 2008 measurement by a factor of 2.2, reaching a relative accuracy of 0.13 ppt on the value of $a_e$ \cite{fan2022}. The new value is consistent with the previous one \cite{Hanneke2008}.

\subsubsection{Fine structure constant}
Currently the accuracy of the value of $a_e$ predicted by the SM is limited by the knowledge of the fine structure constant (see Figure.~\ref{fig:contributions}).  The most precise determination of the fine structure constant relies on the determination of the ratio $h/m_\mathrm{X}$ between the Planck constant $h$ and the mass of an atom. They use the following equation: 
\begin{equation}
\alpha^2 = \frac{2R_\infty}{c} \frac{h}{m_\mathrm{e}} = \frac{2R_\infty}{c} \frac{A_\mathrm{r}(\mathrm{X})}{A_\mathrm{r}(\mathrm{e})}\frac{h}{m_\mathrm{X}} 
\end{equation}
where $R_\infty$ is the Rydberg constant, $A_\mathrm{r}(\mathrm{e})$ is the relative atomic mass of the electron and $A_\mathrm{r}(\mathrm{X})$ the relative atomic mass of the atom used. Indeed, $R_\infty$ is known at the level $\num{1.9E-12}$, $A_\mathrm{r}(\mathrm{e})$ at the level of $\num{2.9E-11}$ and, for rubidium, $A_\mathrm{r}(\mathrm{Rb})$ at the level of $\num{7E-11}$. The limiting factor is the ratio $\frac{h}{m_\mathrm{X}} $ (or simply $m_\mathrm{X}$, as $h$ is now fixed in the SI system of units).

The ratio $\frac{h}{m_\mathrm{X}} $ can be precisely determined by measuring the recoil velocity $\frac{\hbar k}{m_\mathrm{X}}$ of an atom that absorbs a photon of momentum $\hbar k$. This determination can be performed with great accuracy using a matter-wave interferometer based on cold atoms. Currently, there are two experiments that perform such a measurement with competitive uncertainty. The first one at Berkeley, uses cesium atoms, which published a determination with a relative accuracy of $\num{2.0E-10}$ in 2018 \cite{Parker2018} and the experiment we performed at Laboratoire Kastler Brossel in Paris, uses rubidium atoms, which published a determination with a relative accuracy of $\num{8E-11}$ in 2020\cite{Morel2020}. 

The principle of these experiments has been described in several articles over the years for the Cesium experiment \cite{wicht2002,muller2008,parker2016} and the Rubidium experiment \cite{Bouchendira2011,clade:052109}. In the rubidium experiment, we use the technique of Bloch oscillation to transfer to atoms 1000 recoils and measure their velocity by using a Ramsey-Bordé interferometer based on Raman transition. In 2020, we demonstrate a relative statistical uncertainty of about $\num{3E-10}$ for one hour of measurement. We were able to study several systematic effects that limit the current accuracy of the measurement. The main one is due to the knowledge of the photon momentum. Indeed, while the frequency of the laser can be measured very precisely, the momentum of a photon is well defined only for a plane wave and corrections due to the finite size of the beam \cite{weiss1994}, wavefront aberration and local intensity fluctuations \cite{bade2018} have to be taken into account.

At the moment, the values obtained by the two groups differ by $5.4\sigma$. This discrepancy remains unsolved.

\subsubsection{Discussion and perspective}

\begin{figure}
\includegraphics[width=.95\linewidth]{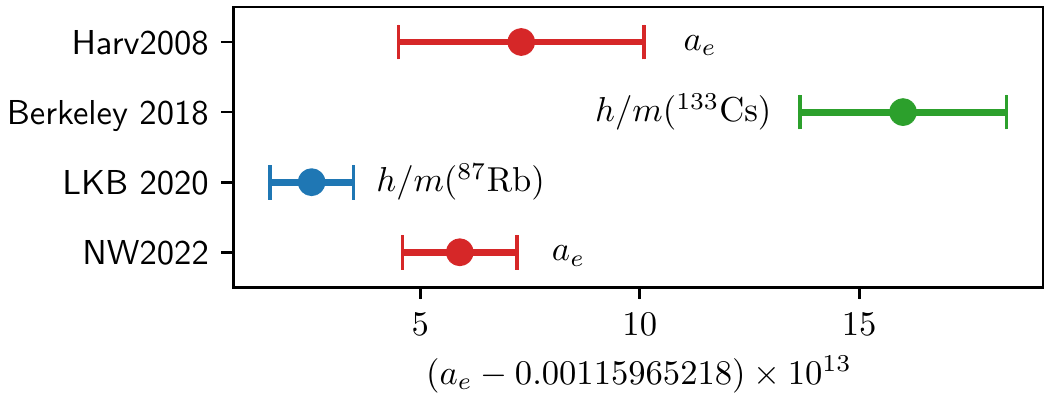}
\caption{Comparison between the determinations of $a_e$ and the prediction of the standard model using independent values of $\alpha$. Harv2008 \cite{Hanneke2008}, Berkeley 2018 \cite{Parker2018}, LKB 2020 \cite{Morel2020}, NW 2022 \cite{fan2022}.}
\label{fig:comparison_ae}
\end{figure}

 Fig.~\ref{fig:comparison_ae} shows the two most precise determinations of $a_e$ together with the prediction using the determination of $\alpha$ described in the previous section and the SM calculation \cite{aoyama2019}. Due to the large disagreement between the two values of $\alpha$, the experiment and the theory agree only at the level of $\num{1E-12}$. 

\begin{figure}
\includegraphics[width=.95\linewidth]{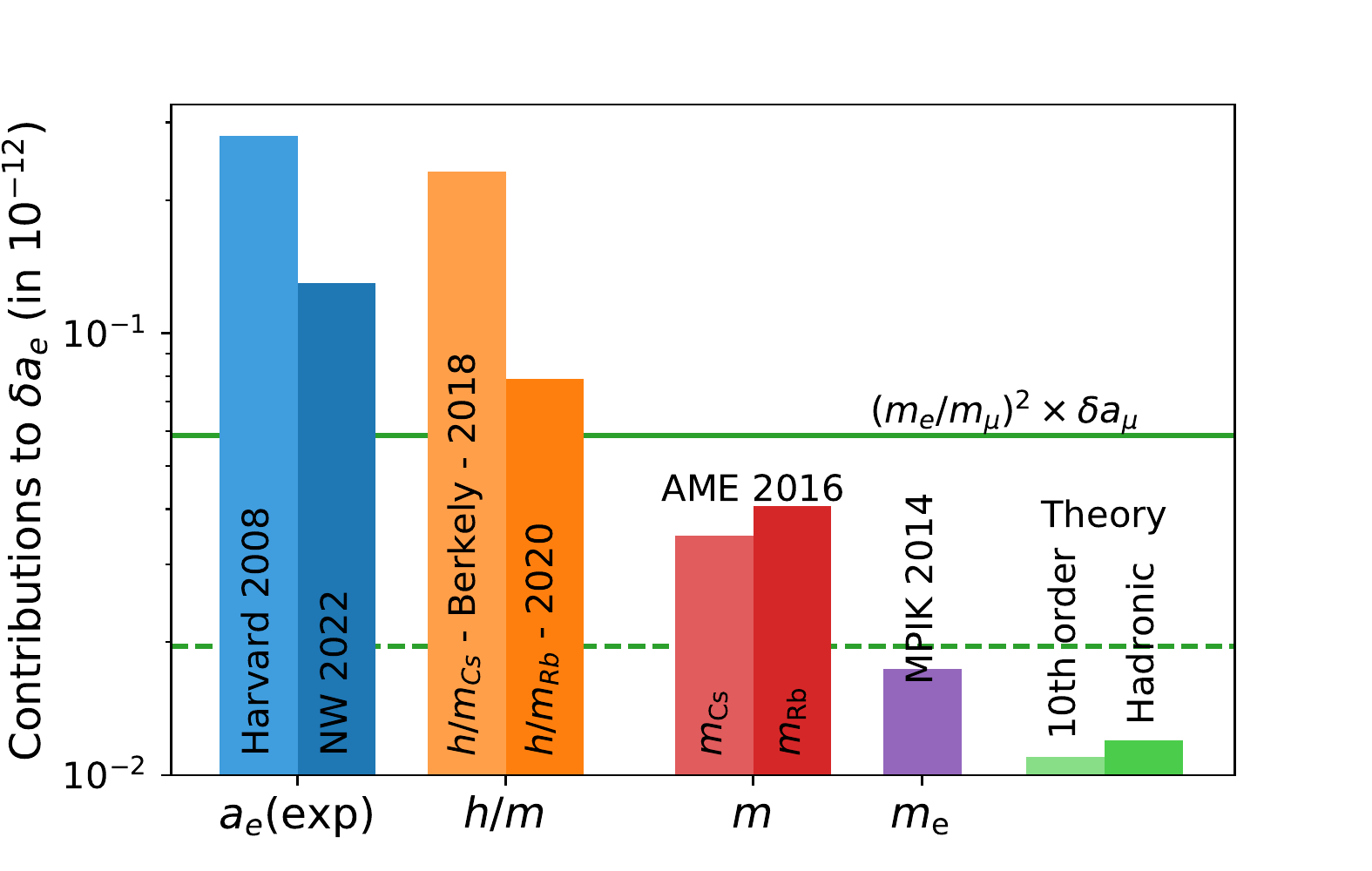}
\caption{Comparison of the different uncertainties involved in the test of the anomaly of the magnetic moment of the electron. The horizontal green line represents the current discrepancy observed for the muon scale to the electron, and the dashed line represents the uncertainty required to observe this discrepancy with $3\sigma$ accuracy. Blue : experimental determination of $a_e$ (\cite{Hanneke2008,fan2022}). Orange : determination of the ratio $h/m_\mathrm{X}$ (\cite{Parker2018, Morel2020}). Red : Atomic mass evaluation \cite{huang2021}. Purple : relative atomic mass of the electron \cite{Sturm2014}. Green : uncertainty from theory \cite{aoyama2019}.}
\label{fig:contributions_ae}
\end{figure}

The comparison between the experimental value of $a_e$ and its prediction of SM imposes constraints on the coupling with BSM particles as well as possible electron substructure. It is worth comparing the results from the electron and the muon magnetic moment. Although the precision on the measurement of $a_\mu$ is $\num{5.4E-10}$ (compared to $\num{1.3E-13}$ for the electron), it is expected that the muon is more sensitive to BSM particles: in a large class of models, new contributions to magnetic moments scale with the square of lepton masses, which gives a ratio $\num{4E4}$ between muon and electron \cite{Giudice2012,terranova2014}. Currently, there is a discrepancy $\delta a_\mu$ of the order of $\num{2.5E-9}$ between the SM prediction \cite{Aoyama2020} and the measurement of $a_\mu$ \cite{abi2021}. According to the scaling, this would lead to a discrepancy of $\num{6E-14}$ on the electron. This precision has not yet been reached but is within an order of magnitude improvement of the different experiments involved in the electron test (Fig.~\ref{fig:contributions_ae}). 

Currently, we are working on making a new determination of $\alpha$ using the same setup as in 2020, but with a Bose-Einstein condensate as a source of atoms instead of an optical molasses. This would change the contributions due to the wave front of the laser beam and make those contributions independent of the 2020 measurement. A new and experimental setup with longer interrogation area is under construction. Similar work is also in progress at Berkeley with the hope that the discrepancy between the two measurements will be understood.
figure

%-------------------------------------------
\subsection{Testing Fundamental Symmetries by Comparing the Properties of Matter/Antimatter Conjugates -- {\it S.~Ulmer}}
\label{ssec:stefan_ulmer}
\textit{Author : Stefan Ulmer, <stefan.ulmer@cern.ch>} 
%-------------------------------------------

\subsubsection{Introduction}
\begin{figure}[h!]
     % \centerline{\includegraphics[width=10 cm]{SMEClock.pdf}}
     \centerline{\includegraphics[width=12 cm]{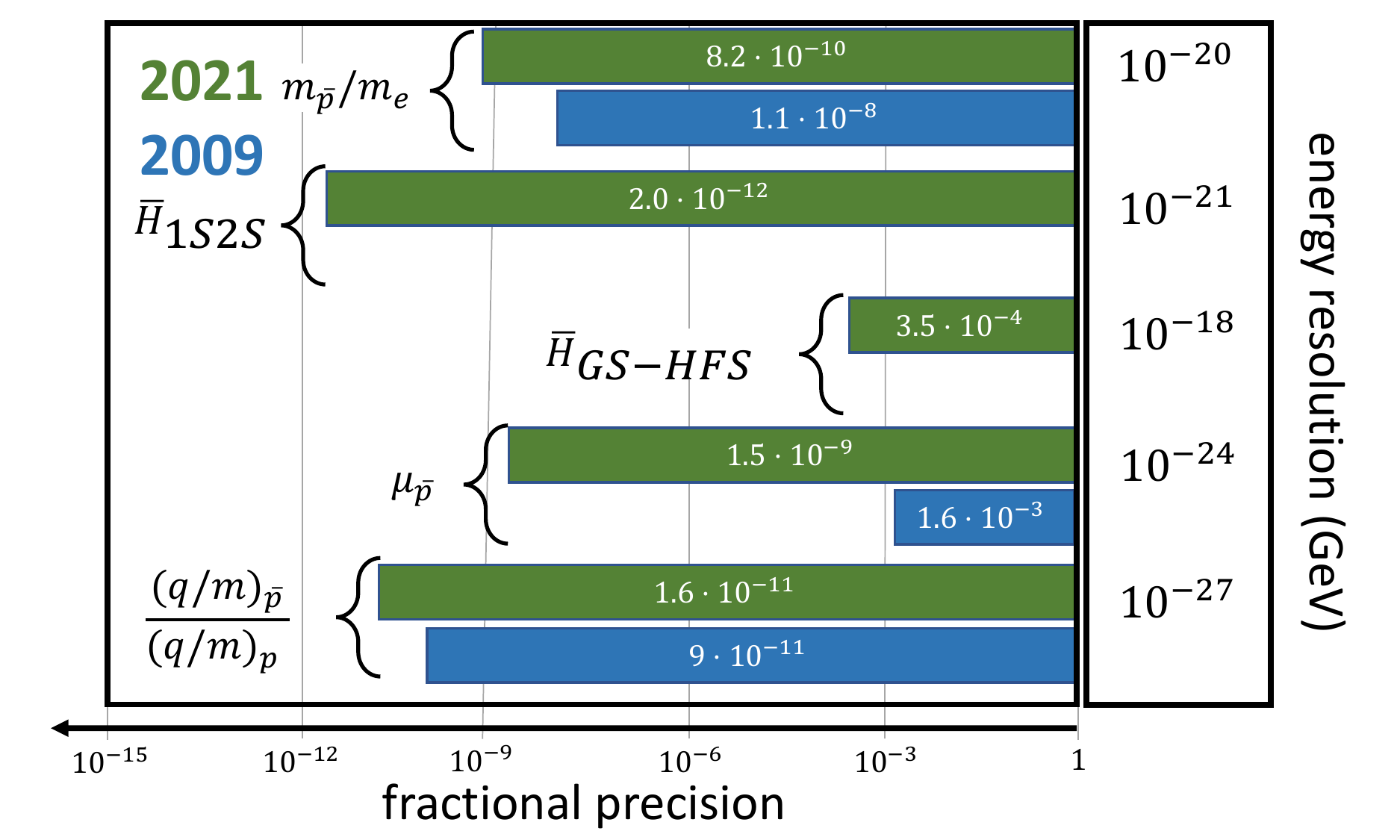}}
      \caption{Measurements of matter and antimatter fundamental properties at the AD/ELENA facility, status 2009 blue, 2021 green. The absolute energy resolution of the frequency measurements is is shown on the right.}
           \label{fig:Progress}
    \end{figure}

The experiments at the antiproton decelerator facility of CERN are motivated by the striking imbalance of matter over antimatter, that is observed on cosmological scales \cite{Dine:2003ax}. Using methods developed in atomic, molecular and optical physics - such as traps, clocks and lasers - currently five collaborations, AEgIS \cite{Amsler:2021euk}, ALPHA \cite{Ahmadi:2018eca}, ASACUSA \cite{ASACUSA:2016xeq,ASACUSA:2014mse}, BASE \cite{BASE:2015ity}, and GBAR\cite{Perez:2012cm} are comparing the fundamental properties of hydrogen/antihydrogen (H/$\bar{\text{H}}$) and protons/antiprotons (p/$\bar{\text{p}}$) with ultra-high precision.  AEgIS, GBAR, and a branch of the ALPHA collaboration, have the goal to test the weak equivalence principle by investigating the ballistic properties of $\bar{\text{H}}$ in the gravitational field of the earth. The ASACUSA collaboration is focusing on tests of CPT invariance by ground-state-hyperfine spectroscopy in a beam of polarized $\bar{\text{H}}$ atoms produced in a CUSP trap \cite{ASACUSA:2014mse}. Another effort within ASACUSA is performing high-resolution spectroscopy of antiprotonic helium \cite{ASACUSA:2016xeq}. Laser spectroscopy on this three body system gives access to the antiproton-to-electron mass ratio $m_{\bar{\text{p}}}/m_{\text{e}}$, which was determined with a fractional accuracy of $8\times10^{-10}$, consistent with recent proton-to-electron mass ratio values extracted from laser spectroscopy \cite{alighanbari2020precise} of simple molecular ions and precision Penning trap experiments \cite{Heisse:2017djn}. 
The ALPHA collaboration is performing precision measurements on the fundamental properties of $\bar{\text{H}}$ using an atom trap. The collaboration reported in 2010 the first successful demonstration of $\bar{\text{H}}$-trapping \cite{ALPHA:2010oyp}. Based on this success, ALPHA has meanwhile studied the charge neutrality of $\bar{\text{H}}$ with a record-precision at the level of $7\times10^{-10}$, and demonstrated a first low-resolution test of the free-fall weak equivalence principle. Most importantly,  the collaboration has measured the 1S/2S transition in $\bar{\text{H}}$ with a fractional resolution of 2 parts per trillion \cite{Ahmadi:2018eca}, and recently demonstrated laser cooling of antihydrogen, heralding future optical spectroscopy of $\bar{\text{H}}$ at even higher resolution.  
The BASE collaboration uses advanced Penning trap systems to compare the fundamental properties of protons and antiprotons. Using single particle nuclear magnetic resonance methods \cite{Ulmer:2011pra}, single spin quantum transition spectroscopy \cite{Smorra:2017wml}, and newly developed multi-trap techniques, they have measured the antiproton magnetic moment with a fractional accuracy of 1.5$\,$p.p.b$.$ \cite{BASE:2016yuo}, which improved the previous best measurement by more then a factor of 3000. Comparing cyclotron frequencies of antiprotons and negatively charged hydrogen ions H$^-$ \cite{BASE:2022yvh}, BASE has recently compared the proton-to-antiproton charge-to-mass ratios with a fractional accuracy of 16 parts in a trillion. This measurement also enabled a first differential test of the clock weak equivalence principle, in which a fractional accuracy of 0.03 was achieved.
The dramatic progress made by the AD-experiments in recent years is illustrated in Fig$.\,$\ref{fig:Progress}. 
\\
The exotic atom experiments conducted by ALPHA and ASACUSA, as well as the high precision comparisons of the fundamental properties of protons and antiprotons by BASE, are sensitive to exotic interactions 
\begin{eqnarray}
\Delta \mathcal{L} =\frac{\lambda}{M^k}\langle T \rangle \bar{\psi\Gamma}(i\partial)^k\psi,
\end{eqnarray}
which appear in the model-framework of the Standard Model Extension (SME) \cite{Kostelecky:2008ts}. Here, $\lambda$ is an effective coupling constant suppressed by the mass dimension $1/M^k$ of the scale at which the exotic physics occurs, $(i\partial)^k$ represents k four-derivatives acting onto the involved fermion fields, and $\Gamma$ is some gamma-matrix structure. The term $\langle T \rangle$ represents the non-zero expectation value of a function of Lorentz tensors from some higher dimension CPT breaking mechanisms \cite{Kostelecky:1988zi,Kostelecky:1991ak}. By careful consideration of different possible shapes of $\Delta \mathcal{L}$, the SME remains translationally invariant and covariant under changes of the inertial frame of the observer, but violates CPT and partially breaks covariance under particle boosts. The SME contains spontaneous CPT breaking, but features properties like microscopic causality and renormalizability.\\ 
The CPT tests produced by CERN's antimatter collaborations, some of them competitive or even at higher resolution than measurements in the matter sector, are also sensitive to oscillatory signatures induced by couplings to axion-like particles \cite{Smorra:2019qfx}. In addition, some of the measurement devices used in the community, such as for example the detection systems for non-destructive measurements of single-particle-oscillation frequencies in Penning traps, are used as haloscope-style detectors to search for ALP-induced couplings to electric $\textbf{E}$ and  magnetic $\textbf{B}$ fields, which induce the Lagrange density 
\begin{eqnarray}
   \Delta \mathcal{L}_{a\gamma\gamma}=-g_{a\gamma}a(\textbf{x})\textbf{E}(\textbf{x})\cdot\textbf{B}(\textbf{x}) \, , 
\end{eqnarray}
where $a(\textbf{x})$ is the local axion field and $g_{a\gamma}$ is the ALP-to-photon coupling constant. In \cite{Devlin:2021fpq}, the BASE collaboration has set competitive narrow-band limits on $g_{a\gamma}$ in the neV-mass range. Furthermore, the non-destructive quantum measurement methods used in ultra-low-noise Penning traps \cite{Borchert:2019ejp} allow to set ion-trap based limits on parameter ranges in which millicharged particles can exist \cite{Budker:2021quh}.

%%%%%%%%%%%%%%%%%%%%%%%%%%%%%%%%%%%%%%%%%%%%%%%%%%%
\subsubsection{SME limits from proton / antiproton comparisons}
%%%%%%%%%%%%%%%%%%%%%%%%%%%%%%%%%%%%%%%%%%%%%%%%%%%
To compare proton and antiproton charge-to-mass ratios, the cyclotron frequencies of antiprotons and negatively charged hydrogen ions H$^-$ are measured \cite{Gabrielse:1999kc}. The H$^-$ ion is a perfect proxy for the proton, with a mass
\begin{eqnarray}
m_{{\text{H}^-}}=1{.}001\,089\,218\,753\,80(3)\, m_{\text{p}},
\end{eqnarray}
a value with an uncertainty at the level of 0.03$\,$ppt \cite{BASE:2022yvh}.\\
Comparing antiproton/H$^-$ charge-to-mass ratios in a time-period of $\approx1.5\,$a, allowed the determination of the antiproton-to-proton charge-to-mass ratio      
\begin{eqnarray}
    R_{\bar{\text{p}},\text{p},\text{exp}}=-1.000\,000\,000\,003\, (16)\,.     
    \end{eqnarray}
This result has an experimental uncertainty of 16$\,$p.p.t$.$ (C.L$.$ 0.68), 
supporting CPT invariance, and provides a 4.3-fold improved limit on the coefficient $r^{H^-}$ of the minimal SME \cite{Bluhm:1997qb,Kostelecky:2008ts}, becoming $r^{H^-}<2.09\cdot10^{-27}$. In the non-minimal extension of the SME \cite{Ding:2020aew} the related charge-to-mass ratio figure of merit is 
\begin{eqnarray}
|\delta\omega_c^{\bar{p}}-R_{\bar{\text{p}},\text{p},\text{exp}}\delta\omega_c^{{p}}-2R_{\bar{\text{p}},\text{p},\text{exp}}\delta\omega_c^{{e}^-}|<1.96\times10^{-27}\,\text{GeV}, 
\end{eqnarray}
where $\frac{\delta\omega_c^{w}}{q_0 B}$ is a function of coefficients  $\tilde{b}_w$ and $\tilde{c}_w$ that characterize the strengths of feebly interacting CPT-violating background fields, coupling to particles $w$, the antiproton $\bar{\text{p}}$, the proton p, and the electron $e^-$. The measurement reported in \cite{BASE:2022yvh} sets the improved limits summarized in Table~\ref{table:SME}.
\begin{table*}[h!]
\centering
\begin{tabular}{||l | c | c | c | c ||} 
\hline
Coefficient  &Previous Limit & Improved Limit & Factor  \\ [0.5ex] 
\hline\hline
$|\tilde{c}_e^{XX}|$  & $<3.23\cdot10^{-14}$ & $<7.79\cdot10^{-15}$ & 4.14  \\
$|\tilde{c}_e^{YY}|$  & $<3.23\cdot10^{-14}$ & $<7.79\cdot10^{-15}$ & 4.14  \\
$|\tilde{c}_e^{ZZ}|$  & $<2.14\cdot10^{-14}$ & $<4.96\cdot10^{-15}$ & 4.31  \\
\hline
\hline
$|\tilde{c}_p^{XX}|, |\tilde{c}_p^{*XX}|$  & $<1.19\cdot10^{-10}$ & $<2.86\cdot10^{-11}$ & 4.14  \\
$|\tilde{c}_p^{YY}|, |\tilde{c}_p^{*YY}|$  & $<1.19\cdot10^{-10}$ & $<2.86\cdot10^{-11}$ & 4.14  \\
$|\tilde{c}_p^{ZZ}|, |\tilde{c}_p^{*ZZ}|$  & $<7.85\cdot10^{-11}$ & $<1.82\cdot10^{-11}$ & 4.31  \\
[1ex] 
\hline
\end{tabular}
\caption{Constraints on coefficients of the standard model extension. The second column describes the previous best limit based on \cite{Gabrielse:1999kc} and \cite{BASE:2015mmu},  theorized and summarized in \cite{Ding:2020aew}. The third column gives the improved limit based on the measurement presented here, the fourth column shows the ratio of the second and the third column. All entries are based on C.L$.$ 0.68. }
\label{table:SME}
\end{table*} 
\\
In addition to charge-to-mass ratio comparisons, BASE also uses elegant multi-trap-methods to measure proton and antiproton magnetic moments. The successful implementation of these techniques enabled the determination of the antiproton magnetic moment 
\begin{eqnarray}
\frac{\mu_{\bar{\text{p}}}}{\mu_\text{N}}=-2.792\,847\,3443(46),
\end{eqnarray}
as well as the  proton magnetic moment \cite{Schneider:2017lff} 
\begin{eqnarray}
\frac{\mu_{{\text{p}}}}{\mu_\text{N}}=2.792\,847\,344\,62(82).
\end{eqnarray}
These two measurements can be combined to 
\begin{eqnarray}
\left(\frac{\mu_{{\text{p}}}}{\mu_\text{N}}+\frac{\mu_{\bar{\text{p}}}}{\mu_\text{N}}\right)=0.3(8.3)\times10^{-9}.
\end{eqnarray} 
Within the uncertainty of the measurements the results test the standard model with an energy resolution of $< 8.1 \times 10^{-25}\,$GeV, and set constraints on six different exotic CPT and Lorentz violating DC coefficients of the non-minimal Standard Model Extension, with $|\tilde{b}_\text{p}^\text{Z}|<8.1\times10^{-25}\,$GeV, $|\tilde{b}_\text{F,p}^\text{XX}+\tilde{b}_\text{F,p}^\text{YY}|<4.6\times10^{-9}\,$GeV, and $|\tilde{b}_\text{F,p}^\text{ZZ}|<3.3\times10^{-9}\,$GeV for protons, as well as 
$|\tilde{b}_\text{p}^\text{*Z}|<1.5\times10^{-24}\,$GeV, $|\tilde{b}_\text{F,p}^\text{*XX}+\tilde{b}_\text{F,p}^\text{*YY}|<3.1\times10^{-9}\,$GeV, and $|\tilde{b}_\text{F,p}^\text{*ZZ}|<1.1\times10^{-9}\,$GeV for antiprotons. In addition, the measurements set limits on a possible magnetic moment splitting mediated by interactions of the form $f^0_\text{p}\bold{B}\bold{\sigma}$ which arise from ultra short distance scale physics \cite{Stadnik:2014ava}. Here $f^0_\text{p}=\mu_\text{N}(g_\text{p}-g_{\bar{\text{p}}})/4<4.5\times10^{-12}$ is obtained, which improves the previous best constraints \cite{ATRAP:2013vnt} by about three orders of magnitude.   

%%%%%%%%%%%%%%%%%%%%%%%%%%%%%%%%%%%%%%%%%%%%%%%%%%%
\subsubsection{Constraints on Antimatter / Dark Matter Interaction}
\begin{figure}[h!]
\centerline{\includegraphics[width=16.0cm,keepaspectratio]{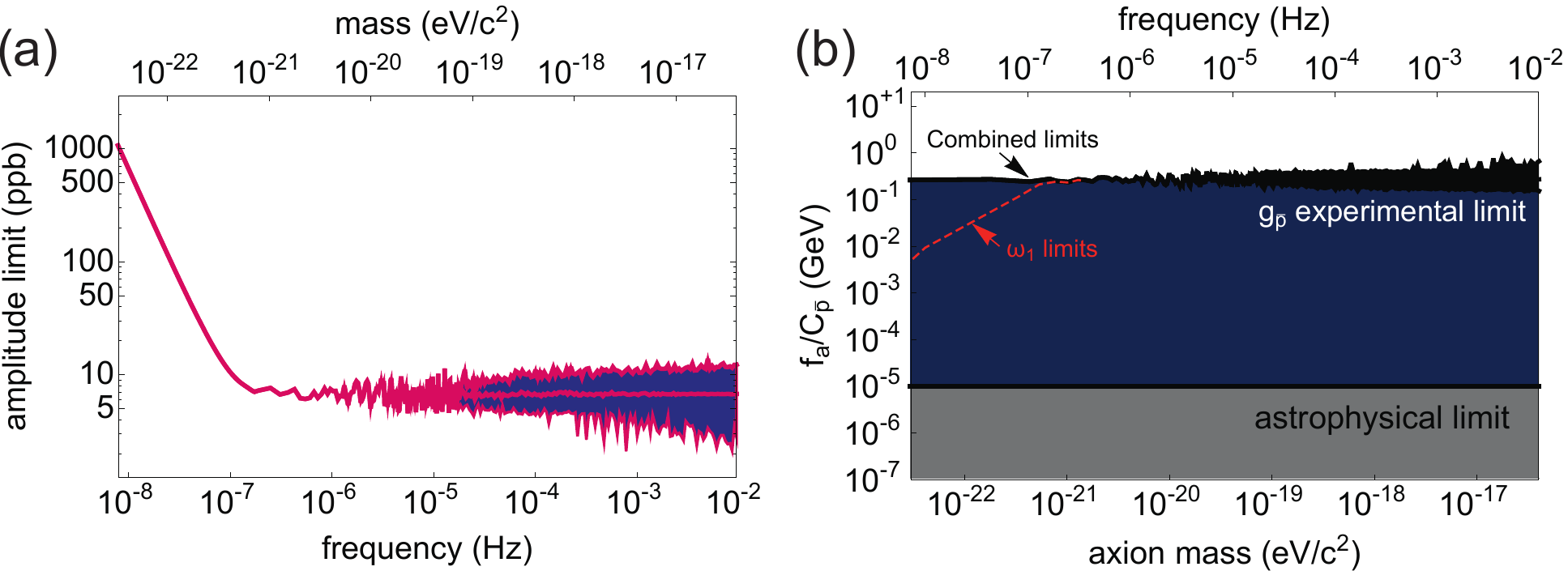}}
\caption{(a) Upper 95$\,\%$ confidence limits on the oscillation amplitude $b_\textrm{up}(\omega)$ of the antiproton Larmor frequency. (b) 95$\,\%$ confidence limits on the axion-antiproton interaction parameter as a function of the axion mass. The grey area shows the parameter space excluded by axion emission from antiprotons in SN 1987A. The dark blue area shows the parameter space excluded from our analysis of the antiproton spin-flip data. The black line shows the upper limit of the excluded area by using the most significant limit from the main and two sideband modes. The yellow dashed line shows the limits from only detecting at the main frequency $\omega_1$ for $m_a < 10^{-21} \mathrm{eV}/c^2$.}
\label{fig:Axion}
\end{figure}

%%%%%%%%%%%%%%%%%%%%%%%%%%%%%%%%%%%%%%%%%%%%%%%%%%%
In addition to constaining coefficients of the SME, the results of the BASE magnetic moment measurements allow to search for asymmetric dark matter/antimatter coupling. If ALPs exist, they may form a coherently oscillating classical field:~$a = a_0 \cos(\omega_a t)$, where the oscillation frequency is given by the compton frequency $\nu_a \approx m_a c^2 /h$. Here, $m_a$ is the axion mass, $c$ the speed of light and $h$ the reduced Planck constant. The ALPs hypothetically interact with protons and antiprotons, which would induce anomalous spin precession \cite{Stadnik:2018sas}, causing sidebands and line shape broadening effects in the sampled magnetic moment resonances. \\ 
The leading-order shift of the antiproton spin-precession frequency due to such interactions is given by: 
\begin{eqnarray}
\label{Axion_antiproton_anomalous_shift}
	\delta \omega_L^{\bar{p}}(t) &\approx \frac{C_{\bar{p}} m_a a_0 \left| \bold{v}_a \right|}{f_a} \left[ A \cos(\Omega_\textrm{sid} t + \alpha) + B \right] \sin(\omega_a t) \, ,
\end{eqnarray}
where $\left| \bold{v}_a \right| \sim 10^{-3} c$ is the average speed of the galactic axions with respect to the Solar System, $\Omega_\textrm{sid} \approx 7.29 \times 10^{-5}~\textrm{s}^{-1}$ is the sidereal angular frequency, and $\alpha \approx -25^\circ$, $A \approx 0.63$, and $B \approx -0.26$ are parameters determined by the orientation of the experiment relative to the galactic axion dark matter flux. By analyzing the time sequence data recorded in antiproton magnetic moment measurements  \cite{Smorra:2019qfx}, see Fig$.\,$\ref{fig:Axion}, 
gives limits on the asymmetric axion-antiproton coupling coefficient $f_a/C_{\overline{p}} > 0.3\,$GeV over the mass range $10^{-22}\,$eV/$c^2 < m_a < 10^{-16}\,$eV/$c^2$. 
In the considered mass range, the laboratory limits are by five orders of magnitude stronger than bounds derived from astrophysical studies. This oscillation analysis allows as well to constrain time dependent coefficients of the SME \cite{Ding:2016lwt}, here the limits $|\tilde{b}^{*X}_p| < 2.5\times 10^{-24}\,$GeV, $|\tilde{b}^{*Y}_p| < 2.5\times10^{-24}\,$GeV, $|\tilde{b}^{*XX}_p-\tilde{b}^{*YY}_p| < 1.6\times 10^{-8}\,$GeV$^{-1}$, $|\tilde{b}^{*XZ}_p| < 1.0\times 10^{-8}\,$GeV$^{-1}$, $|\tilde{b}^{*YZ}_p| < 1.0\times 10^{-8}\,$GeV$^{-1}$, and $|\tilde{b}^{*XY}_p| < 8.2\times 10^{-9}\,$GeV$^{-1}$ are obtained. 

%%%%%%%%%%%%%%%%%%%%%%%%%%%%%%%%%%%%%%%%%%%%%%%%%%
\subsubsection{Superconducting Haloscope}
\begin{figure}[h!]
\centerline{\includegraphics[width=14.0cm,keepaspectratio]{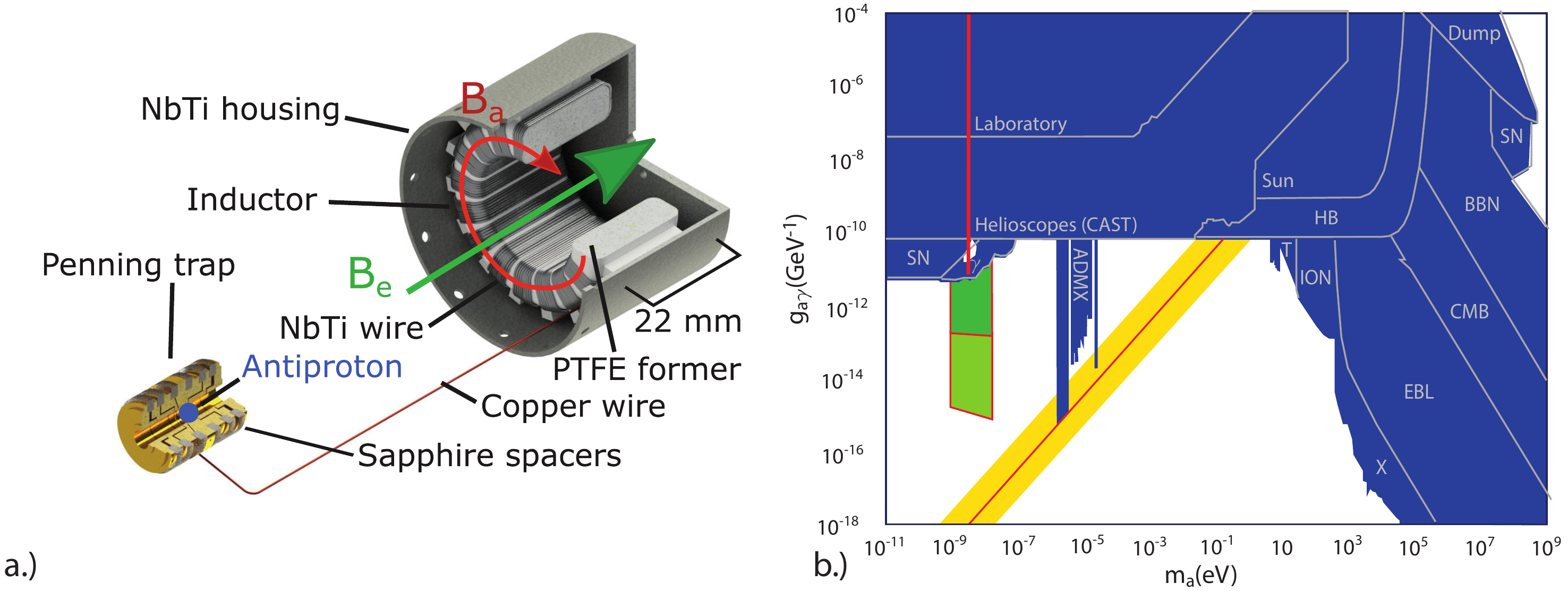}}
\caption{(a) superconducting toroidal resonator connected to a Penning trap. The green arrow indicates the external magnetic field of the superconducting magnet, the red arrow indicates the magnetic field sourced by hypothetical axions. b.) limits plot which sets the BASE measurement into context with other quests. With purpose built experiments we expect to constrain the green region.}
\label{fig:Axion2}
\end{figure}

%%%%%%%%%%%%%%%%%%%%%%%%%%%%%%%%%%%%%%%%%%%%%%%%%%
The signals sourced by axions can be picked up by sensitive resonant LC circuits in strong magnetic fields \cite{Sikivie:2013laa}. In \cite{Devlin:2021fpq} it was shown how superconducting Penning trap detectors, usually used to pick-up single particle signals  \cite{Ulmer:2013lra}, can be used as sensitive antennas for ALPs.  If ALPs oscillate through the strong magnetic field of superconducting Penning trap magnets, they source oscillating magnetic fields 
\begin{eqnarray}
\textbf{B}_a=-\frac{1}{2}g_{a\gamma}r\sqrt{\rho_a\hbar c}|\textbf{B}_e|\hat{\phi}\,,    
\end{eqnarray}
as shown in Fig$.\,$\ref{fig:Axion2}, left.
Here $\rho_a\hbar c=4\pi^2\nu_a^2|a|^2/2$ is the local axion energy density and $r$ is the radial experimental coordinate. 
The oscillating magnetic field leads to a changing flux in the inductor, which in turn produces an oscillating voltage at the input of the first cryogenic amplification stage connected to the detector\cite{Nagahama:2016lgw}. 
By calibrating the temperature of the detection system using single particle quantum-jump thermometry, and searching the detector's resonance spectra for narrow peak signatures induced by hypothetical ALPs, 
constrain the axion to photon coupling constant $g_{a\gamma} < 10^{-11}$ GeV 
in the axion mass range around 
$2.7906\,\text{neV}<m_\text{a}c^2<2.7914\,\text{neV}$. 
These constraints are more than one order of
magnitude lower than the best laboratory haloscope and 
approximately 5 times lower than the CERN axion solar telescope (CAST), see Fig$.\,$\ref{fig:Axion2}, right.  
With purpose-built experiments that cover larger magnetic volumes, that use specifically developed high-Q axion detectors and bandwidth tuners, it is anticipated to cover with such experiments the green regions shown inf Fig$.\,$\ref{fig:Axion2}, left.

%%%%%%%%%%%%%%%%%%%%%%%%%%%%%%%%%%%%%%%%%%%%%%%%%%%
\subsubsection{Summary and Conclusions}
%%%%%%%%%%%%%%%%%%%%%%%%%%%%%%%%%%%%%%%%%%%%%%%%%%%
Within this text the recent achievements made by the experiments at the AD/ELENA facility of CERN were summarized, which test CPT invariance by comparing the fundamental properties of matter/antimatter conjugates with high precision. In recent years, some of the antimatter fundamental constants measured by the community improved by several orders of magnitude. Some of the collaborations use their high-precision data, acquired by quantum-insipred measurement technology, to constrain exotic physics, and to search for asymmetric antimatter/dark matter coupling, constituting a unique window to physics beyond the standard model. By further developing measurement technologies, establishing novel cooling techniques, up-scaling and improving detection sensitivities, the experiments have potential to considerably improve the current limits in future efforts.

%%%%%%%%%%%%%%%%%%%%%%%%%%%%%%%%%
%\subsubsection*{Acknowledgements} 
%%%%%%%%%%%%%%%%%%%%%%%%%%%%%%%%%

%We acknowledge technical support by CERN, especially the Antiproton Decelerator operation group, CERN's cryolab team and engineering department, and all other CERN groups which provide support to Antiproton Decelerator experiments. 
%In addition we acknowledge financial support by RIKEN, the RIKEN EEE pioneering project funding, the RIKEN SPDR and JRA program, the Max-Planck Society, and the Max-Planck, RIKEN, PTB-Center for Time, Constants, and Fundamental Symmetries (C-TCFS). 

%-------------------------------------------
\subsection{Novel approaches to GW detection (atom interferometry) - {\it O.~Buchm\"uller}}
\label{ssec:buchmuller}
{\it Author: Oliver Buchm\"uller, <oliver.buchmuller@cern.ch> - Joint Session with PSI 2022}

%\documentclass{article}[11pt]
%\usepackage{jheppub}
%\usepackage[utf8]{inputenc}
%\usepackage{graphicx}
%\usepackage{placeins}
%\usepackage{float}
%\usepackage{subfigure}

%\title{FIPS at CERN Summary}
%\author{Oliver Buchmueller}
%\date{January 2023}

%\begin{document}

%\maketitle
%\begin{abstract}
%   We survey the prospective sensitivities of
%terrestrial and space-borne atom interferometers (AIs) to
%gravitational waves (GWs) generated by cosmological
%and astrophysical sources, and to ultralight dark matter. 
%We compare the sensitivities of LIGO and LISA with those of the 100m and 1km stages of the AION %terrestrial AI project, as well as two options for the proposed AEDGE AI space mission with cold %atom clouds either inside or outside the spacecraft. We also review the capabilities of AION and %AEDGE for detecting coherent waves of ultralight scalar dark matter.

%{\bf The text and figures for this summary are largely taken from~\cite{Badurina_2021}.}
%\end{abstract}
\subsubsection{Introduction}

Atom Interferometry (AI) is an established quantum sensor concept
based on the superposition and interference of atomic wave packets.
AI experimental designs take advantage of features used by state-of-the-art atomic clocks in combination with established techniques for building inertial sensors. 

The experimental landscape of AI projects has expanded significantly in recent years, with several terrestrial AIs based on different Cold Atom technologies currently under construction, planned or proposed. 

Four large-scale prototype projects are funded and currently under construction, namely the AION~\cite{Badurina:2019hst} in the UK, MAGIS~\cite{Abe:2021ksx} in the US, MIGA~\cite{Canuel:2017rrp} in France, and ZAIGA~\cite{Zhan:2019quq} in China. These will demonstrate the feasibility of AI at macroscopic scales, paving the way for terrestrial km-scale experiments as the next steps. There are projects to build one or several km-scale detectors, including AION-km at the STFC Boulby facility in the UK,  MAGIA-advanced and ELGAR~\cite{Canuel:2019abg} in Europe, MAGIS-km at the Sanford Underground Research facility (SURF) in the US, and advanced ZAIGA in China.  It is foreseen that by about 2035 one or more km-scale detectors will have entered operation. These km-scale experiments would not only be able to explore systematically for the first time the  mid-frequency band of gravitational waves, but would also serve as the ultimate technology readiness demonstrators for a space-based mission like AEDGE~\cite{AEDGE:2019nxb} that would reach the ultimate sensitivity to explore the fundamental physics goals outlined in this article.
     
In summary, the perspectives for large-scale atom interferometer projects are very diverse today, with a main focus on establishing the readiness of cold atom technology for use in AI experiments to explore fundamental science. 
In the following, we focus on the large-scale terrestrial project AION and the space-based AEDGE mission concept to outline the enormous science potential of AI projects.        

\subsubsection{The AION Project}
The Atom Interferometric Observatory and Network (AION) is a proposed research infrastructure allowing studies of dark matter and gravitational waves from cosmological and astrophysical sources in the theoretically relevant but currently inaccessible mid-frequency band. It will develop and demonstrate the necessary deployable and scalable quantum technology by constructing and operating 10m- and 100m-scale instruments, paving the way for a future km-scale facility and space-based experiments.

The long-term AION programme comprises:

\begin{itemize}
    \item Stage 1 (10m): Construction of a first full interferometer system, providing proof-of-principle of the basic technology, along with evidence of scalability from lab-based to purpose-built infrastructure. 
\item Stage 2 (100m): Construction will start in late 2020s, with operation foreseen for early 2030s to search for both DM and GW over an operational period of several years. 
\item Stage 3 (1km): Construction will start in the mid-2030s, with a target of reaching ultimate terrestrial sensitivity for GW / DM observation by the end of the decade. 
\item Stage 4 (1000km): A mission proposal for an Atomic Experiment for Dark Matter and Gravity Exploration in Space (AEDGE) is in preparation within the ESA Voyage 2050 programme. This would directly use AION technology and could be flying from 2045.

\end{itemize}

As for a new telescope or particle collider, the AION infrastructure will open new windows for observation of the macroscopic and microscopic structure of the universe. It will enable exploration of the properties of ULDM, and detect GWs from the very early universe and astrophysical sources in the mid-frequency band ranging from several mHz to a few Hz. The science programme spans a wide range of fundamental physics.

%\begin{figure}
%\centering 
%\includegraphics[width=6.0cm]{AION100mGGNplot.pdf}
%\includegraphics[width=6.0cm]{AIONkmGGNplot.pdf}
%\vspace{-3mm}
%\caption{\it Sensitivities of AION-100 (left panel), 
%and -km (right panel) to the density of monochromatic gravitational
%waves, $\Omega_{\rm GW} h^2$. Solid green lines assume GGN can be fully mitigated while the impact of terrestrial Gravity Gradient
%Noise (GGN) is indicated by dashed lines. The thick blue lines show the sensitivities to stochastic backgrounds with
%power-law-integrated (PI) curves while the impact of unmitigated GGN is again shown as a dashed curve.
%We also show the GGN generated by asteroids~\cite{Fedderke:2020yfy} at low frequencies, as well as the noise due to unresolved
%galactic and extragalactic binaries.
%}
%\label{fig:AIONGGN}
%\end{figure}

\subsubsection{The Space-Born Experiment AEDGE}

Atomic Experiment for Dark Matter and Gravity Exploration (AEDGE) is a proposed space experiment using cold atoms to search for ultra-light dark matter, and to detect gravitational waves in the frequency range between the most sensitive ranges of LISA and the terrestrial LIGO/Virgo/KAGRA/INDIGO experiments. This interdisciplinary experiment, will also complement other planned searches for dark matter, and exploit synergies with other gravitational wave detectors. 

The design of AEDGE requires two satellites operating along a single line-of-sight and separated by a long distance. The payload of each satellite will consist of cold atom technology as developed for state-of-the-art atom interferometry and atomic clocks.

The experimental concept is similar to the one implemented in AION, but instead using a shaft on earth it links clouds of cold atomic strontium in a pair of satellites in medium-Earth orbit via pulsed continuous-wave lasers that induce the 698\,nm atomic clock transition, and detect momentum transfers from the electromagnetic field to the strontium atoms, which act as test masses in a double atom interferometer scheme.

We consider two possible configurations for AEDGE, based on a pair of spacecraft in medium earth orbit
with a separation of 40,000~km. One is the baseline configuration considered in~\cite{AEDGE:2019nxb}, in which
the atom clouds are contained within the spacecraft and have a size $\sim 1$~m. The other configuration
assumes atom clouds with sizes $\sim 100$~m~\cite{Dimopoulos:2008sv} that are outside the spacecraft (AEDGE+).

\subsubsection{Summary of Prospective GW and ULDM Sensitivities of Atom Interferometers} 

In this section, we provide an overview of the prospective GW and ULDM sensitivities of AIs. More information is provided in the much comprehensive article~\cite{Badurina_2021}, from which the following text and figures are taken. 

\begin{figure}
\centering 
\includegraphics[width=0.47\textwidth]{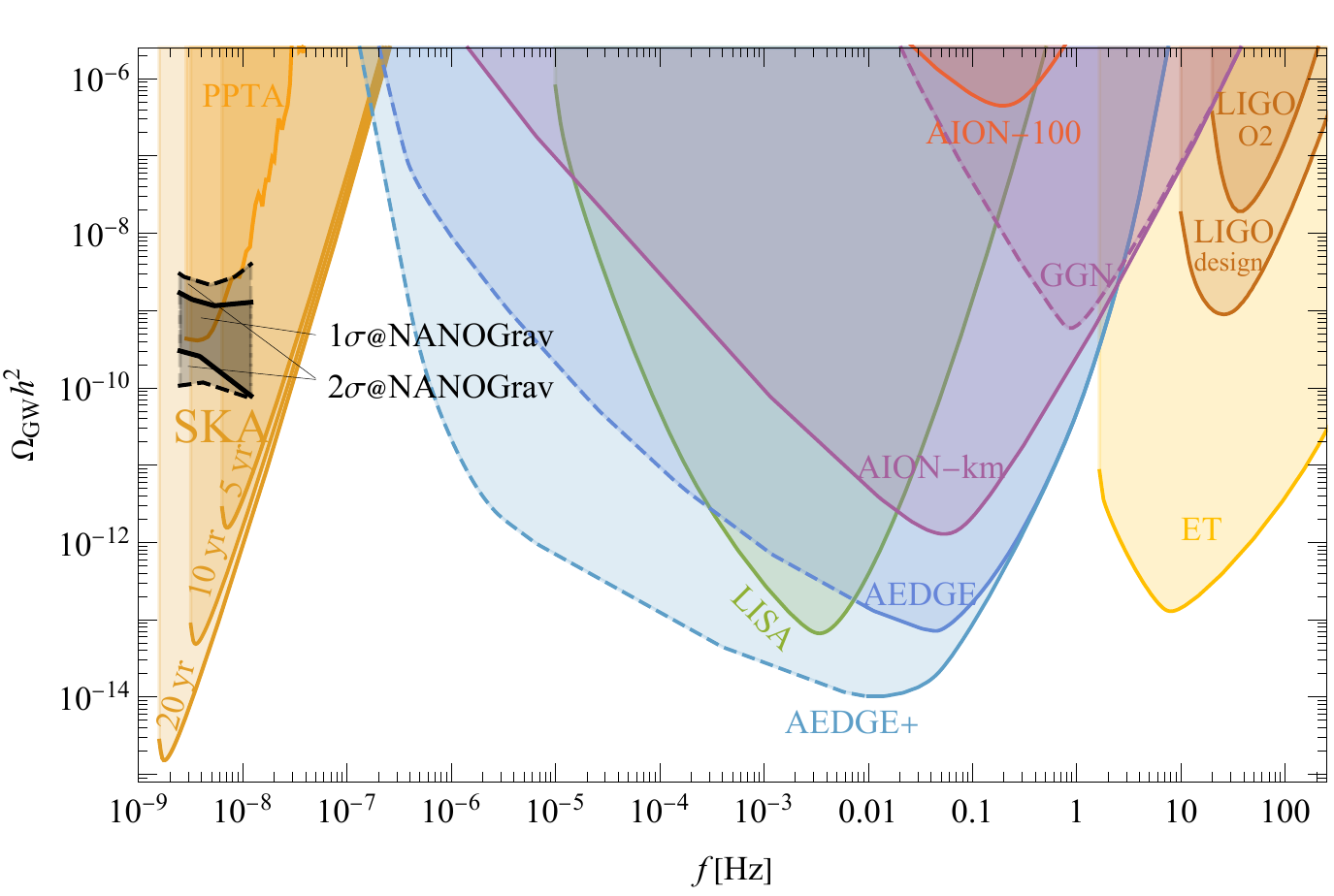}
\includegraphics[width=0.49\textwidth]{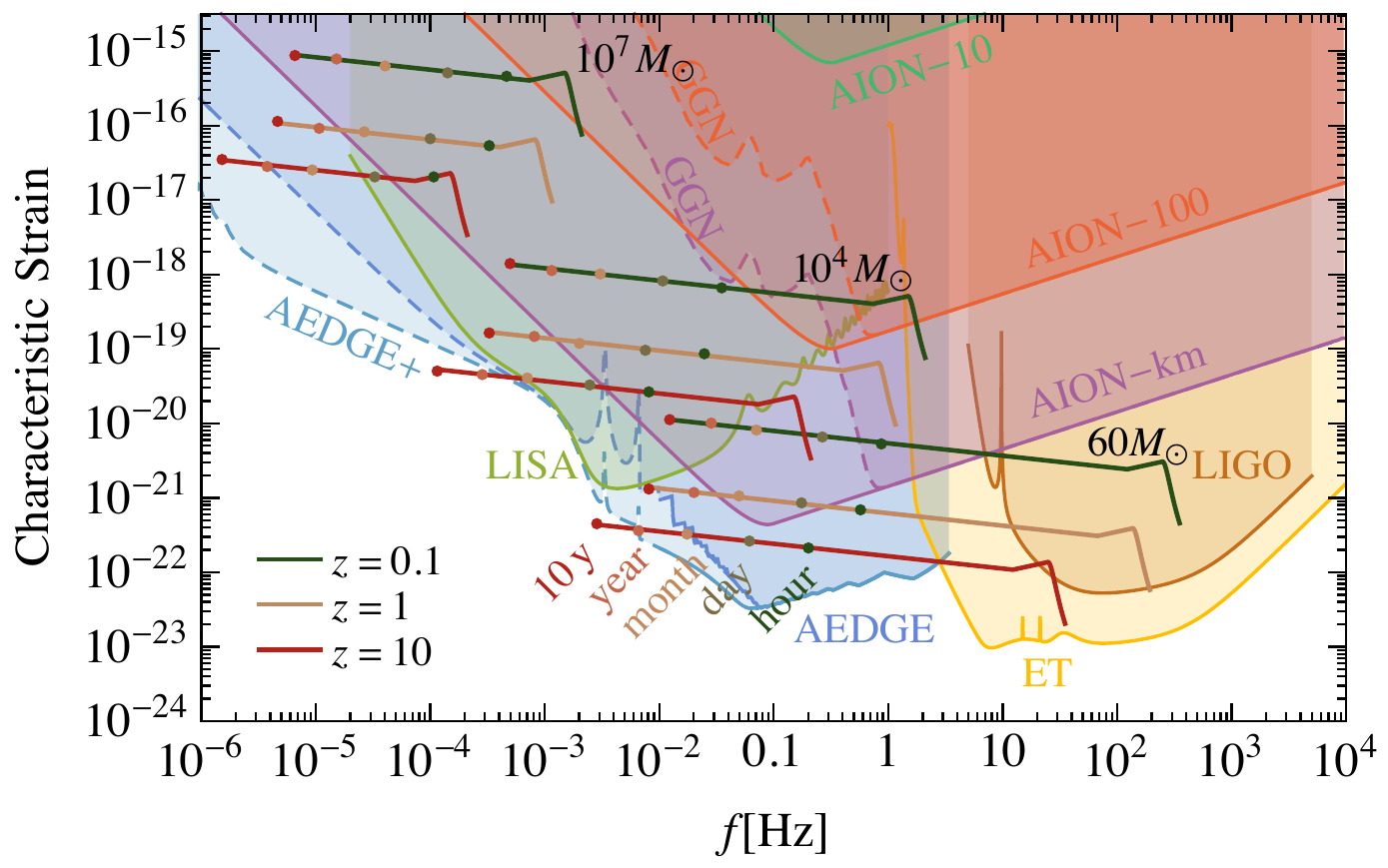}
\vspace{-3mm}
\caption{\it {\bf Left Panel:} Comparison of the $\Omega_{\rm GW} h^2$ sensitivities to PI spectra
of AION-100, AION-km, AEDGE and AEDGE+, LIGO, ET, PTAs and SKA. Also shown are power-law fits to
the NANOGrav hint of a possible GW signal at frequencies around $ 10^{-8}$~Hz. {\bf Right Panel:} Strain sensitivities of AION-10, -100 and -km, AEDGE and AEDGE+, compared
with those of LIGO, LISA and ET and the signals expected from mergers of equal-mass
binaries whose masses are $60, 10^4$ and $10^7$ solar masses.  The assumed redshifts are
$z = 0.1, 1$ and $10$, as indicated. Also shown are the remaining times during inspiral before the
final mergers.
}
\label{fig:summary}
\end{figure}

{\bf Project Sensitivity for GW}
The left panel of Fig.~\ref{fig:summary} compares the possible $\Omega_{\rm GW} h^2$ sensitivities of the
AI experiments that we consider with those of other operating,
planned and proposed experiments. At low frequencies around $ 10^{-7}$~Hz
we see the sensitivities of PTAs and SKA~\cite{Combes:2021xez}, at intermediate frequencies 
$\sim 10^{-2}$~Hz we see the expected LISA sensitivity~\cite{LISA:2017pwj}, 
and at higher frequencies around $10$~Hz
we see the sensitivity LIGO achieved during the O2 observational period and
its design goal, as well as the prospective sensitivity of the ET experiment~\cite{Maggiore:2019uih}.
These can be compared with the prospective $\Omega_{\rm GW} h^2$ sensitivities
of AION-100, AION-km, AEDGE and AEDGE+ to power-law integrated GW spectra.
For information, we also display the results from power-law fits to the NANOGrav hint~\cite{NANOGrav:2020bcs} of a possible GW signal 
at frequencies around $10^{-8}$~Hz.

The right panel of Fig.~\ref{fig:summary} displays the strain sensitivities of AION-100 and -km, AEDGE and AEDGE+, 
compared with the signals expected from mergers of equal-mass
binaries with combined masses of $60, 10^4$ and $10^7$ solar masses occurring at the redshifts
$z = 0.1, 1$ and $10$, as indicated. The dashed lines correspond to the GGN level expected in the NLNM that is
consistent with seismic measurements
made at Fermilab~\cite{Abe:2021ksx} and CERN~\cite{CERN}. Strategies to mitigate these noise levels are
under investigation, which will be increasingly challenging
at lower frequencies. The solid sensitivity curves for
AION-100 and -km assume that the GGN can be completely
mitigated. The lower AEDGE sensitivity curve is for the external cloud configuration, AEDGE+, and
shows the impacts of the extragalactic and galactic binary backgrounds at frequencies ${\cal O}(10^{-2})$
and ${\cal O}(10^{-3})$, respectively.~\footnote{We stress again that the AEDGE sensitivity curves at
low frequencies do not take into account the possible effects of instrumental noise.}
We also show for comparison the LISA sensitivity curve, which
is also impacted by the galactic binary background, as well as the sensitivity curves for LIGO and ET.
\begin{figure}
\centering 
\includegraphics[width=9cm]{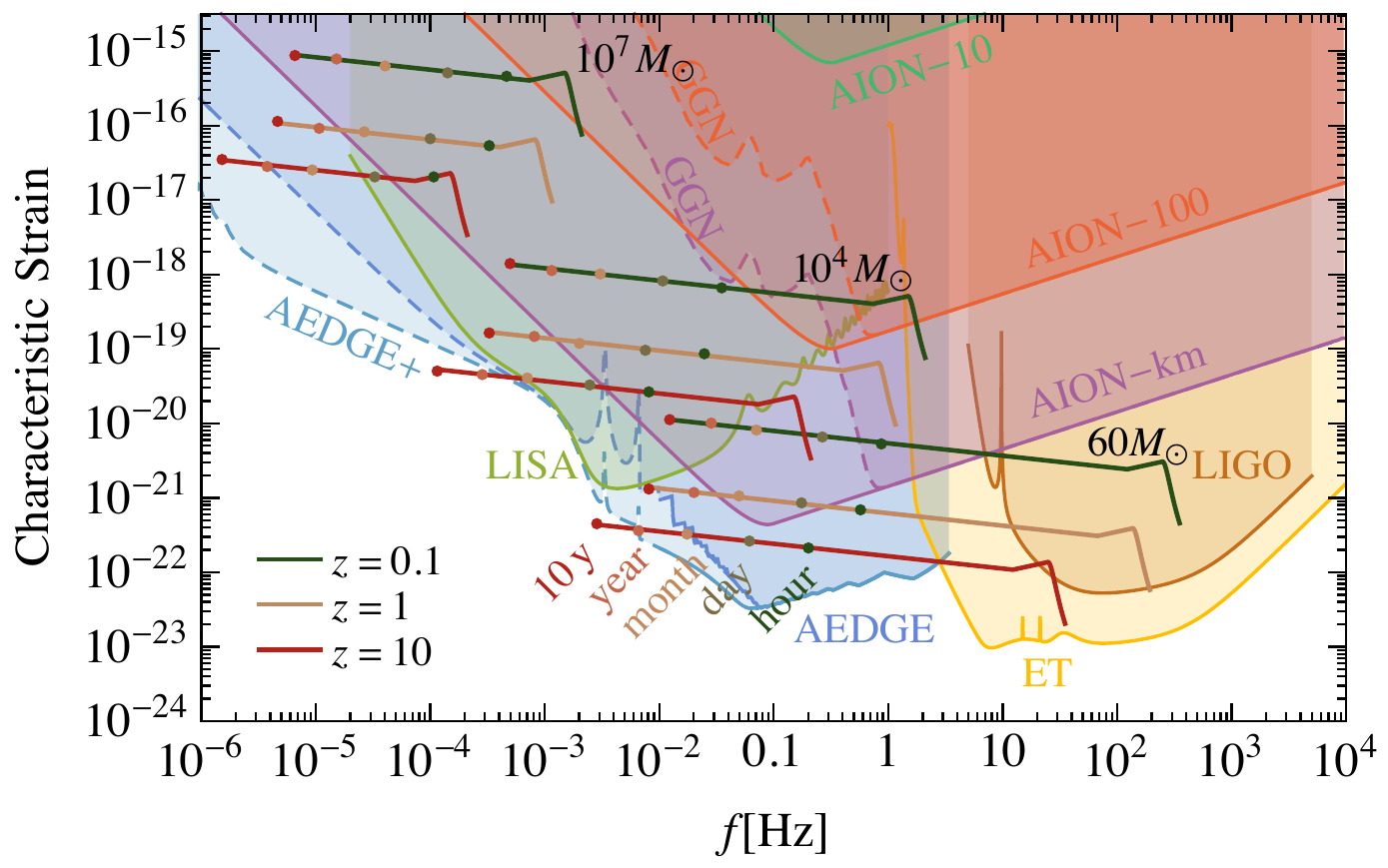}
\caption{\it Strain sensitivities of AION-10, -100 and -km, AEDGE and AEDGE+, compared
with those of LIGO, LISA and ET and the signals expected from mergers of equal-mass
binaries whose masses are $60, 10^4$ and $10^7$ solar masses.  The assumed redshifts are
$z = 0.1, 1$ and $10$, as indicated. Also shown are the remaining times during inspiral before the
final mergers.}
\label{fig:strain}
\end{figure}

\begin{figure}
\centering 
\includegraphics[width=6.0cm]{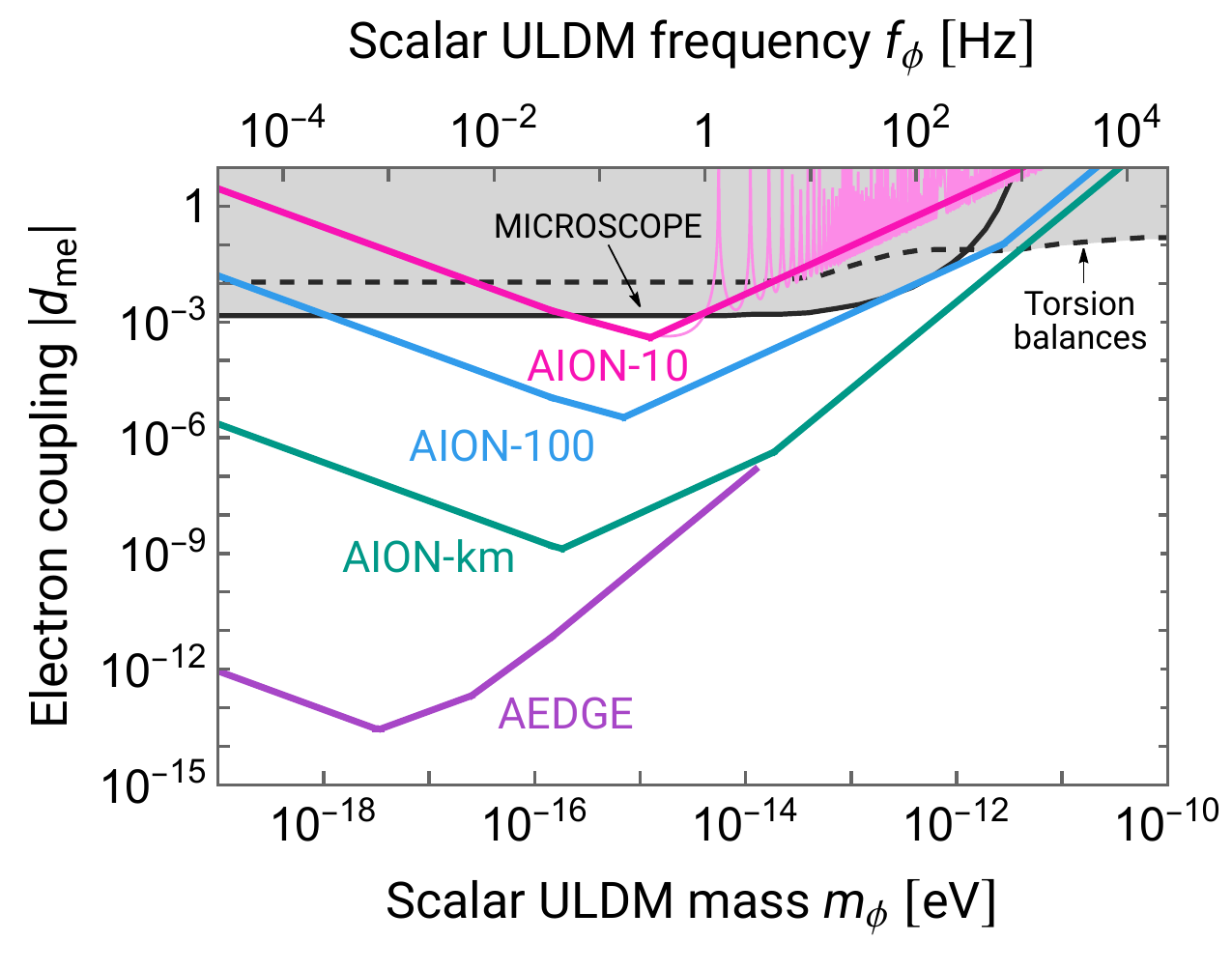}
\includegraphics[width=6.0cm]{ULF/Buchmuller/ULDM_dme.pdf}
\vspace{-3mm}
\caption{\it Sensitivity projections to ULSDM linearly coupled to electrons (left panel) and photons (right panel). The lighter-pink AION-10 curve shows the oscillatory nature of the sensitivity projections, while the darker-pink straight AION-10 curve shows the envelope of the oscillations. For clarity, for AION-100, AION-km and AEDGE, we only show the envelope. The shaded grey region shows the existing constraints from searches for violations of the equivalence principle with torsion balances~\cite{Wagner:2012ui}, atomic spectroscopy~\cite{Hees:2016gop} and the MICROSCOPE experiment~\cite{Berge:2017ovy}.
}
\label{fig:AIONULDM}
\end{figure}

{\bf Project Sensitivity for ULDM}
The left (right) panel of Fig.~\ref{fig:AIONULDM} shows sensitivity projections for ULSDM linearly coupled to electrons (photons) for a SNR = 1, using the procedure outlined in Ref.~\cite{BadurinaRefinedDM}. The AION and AEDGE sensitivity curves are compared to existing constraints, shown by the shaded grey regions. The AI sensitivity oscillates as a function of the ULSDM mass, as shown by the light-pink AION-10 curve in both panels of Fig.~\ref{fig:AIONULDM}. However, for clarity, it is often only the envelope of the oscillations that is plotted, and we have followed this procedure for the AION-100, AION-km and AEDGE projections. Also, for clarity we have plotted the AEDGE sensitivity curve only to the point where it approaches the AION-km line even though the sensitivity extends to higher frequencies: if plotted, the extension of the AEDGE sensitivity curve would lie on top of the AION-km line.
Figure~\ref{fig:AIONULDM} shows the exciting prospects for AI detectors of exploring unconstrained domains of parameter space in both the couplings to the Standard Model fields, and for ULSDM masses between $10^{-18}$~eV and $10^{-12}$~eV. AION-10 hopes to approach or even surpass existing constraints, while AION-100 and AION-km should significantly extend the reach to lower values of the coupling. These projections assume that the phase-noise is limited by atom shot-noise. Below around 0.1~Hz, it is expected that GGN will start to dominate, and we have not extrapolated the sensitivities of the terrestrial experiments below this frequency. As space-borne experiments do not have to contend with the same GGN noise, we have extended the AEDGE projections to lower frequencies, or equivalently, to lower values of the ULSDM mass that are complementary to the parameters that can be tested with terrestrial AIs.

\subsubsection{Summary}
Atom interferometry is a promising technique for many other studies in fundamental physics,
including searches for ultralight dark matter, probes of the weak equivalence
principle and tests of quantum mechanics, as well as the searches for gravitational
waves discussed here. This technique is now emerging from the laboratory to be
deployed in large-scale experiments at the 10 to 100m scale. Ideas are being
developed for possible future experiments at the km scale and in space.
In this summary article, which is based on~\cite{Badurina_2021}, we have reviewed the possible scientific capabilities of such atom
interferometer experiments, focusing on the terrestrial AION project~\cite{Badurina:2019hst} 
and its possible evolution towards a space-borne project called AEDGE~\cite{AEDGE:2019nxb}.

%\bibliographystyle{JHEP}
%\bibliography{refs}

%\end{document}

%-------------------------------------------

%-------------------------------------------
\subsection{Direct searches for Ultra-light FIPs with Gravitational-wave detectors -- {\it H.~Grote and Y.~V.~Stadnik}}
\label{ssec:grote}
{\it Authors: Hartmut Grote, <hartmut.grote@astro.cf.ac.uk> }, 
{\it Yevgeny Stadnik, <yevgenystadnik@gmail.com>} 
\subsubsection{Introduction}
\label{Sec:FIPs-LIFO_intro}

\begin{figure*}[ht]
\centering
\includegraphics[width=6.5cm]{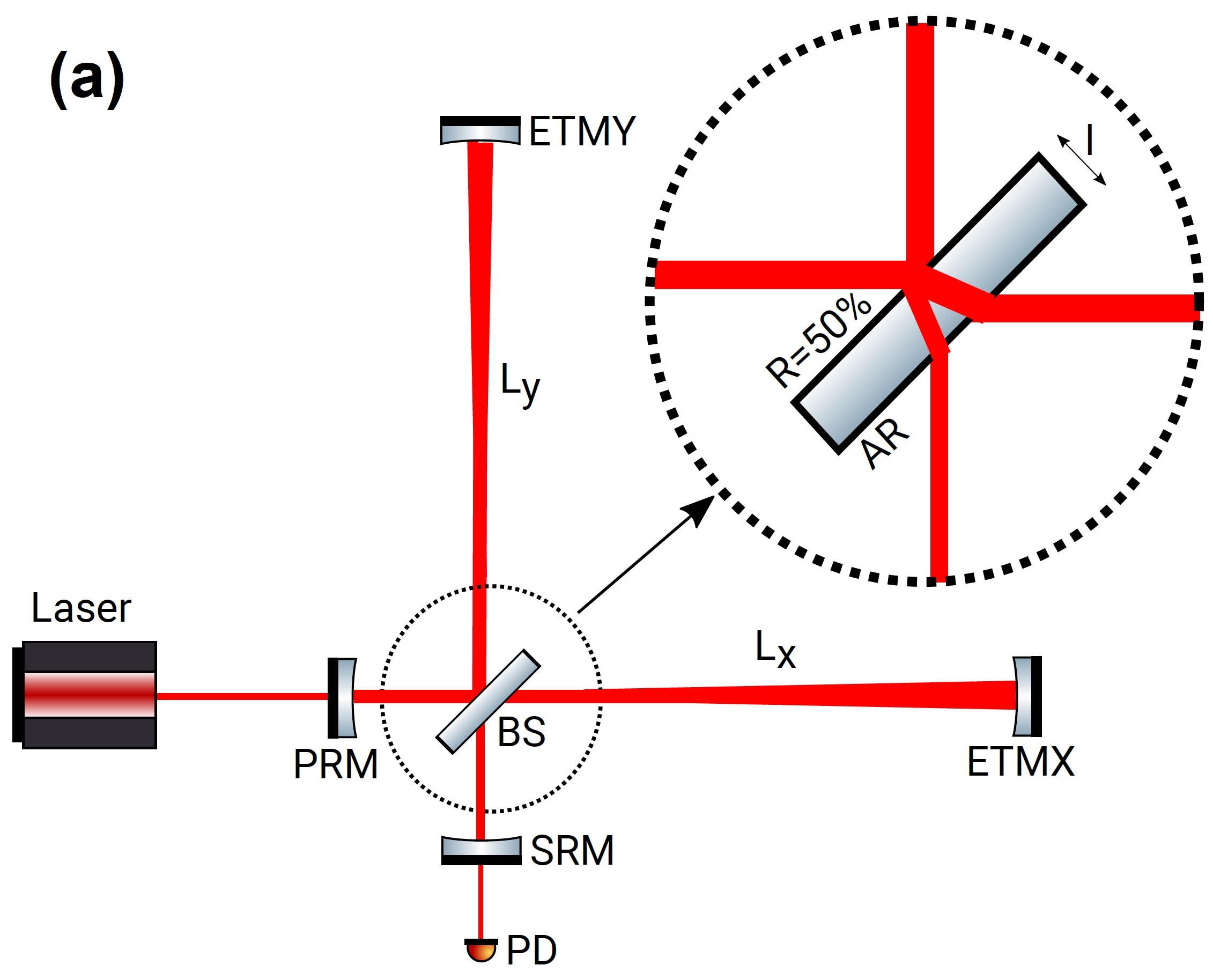}
\hspace{5mm}
%\vspace{3mm}
\includegraphics[width=6.5cm]{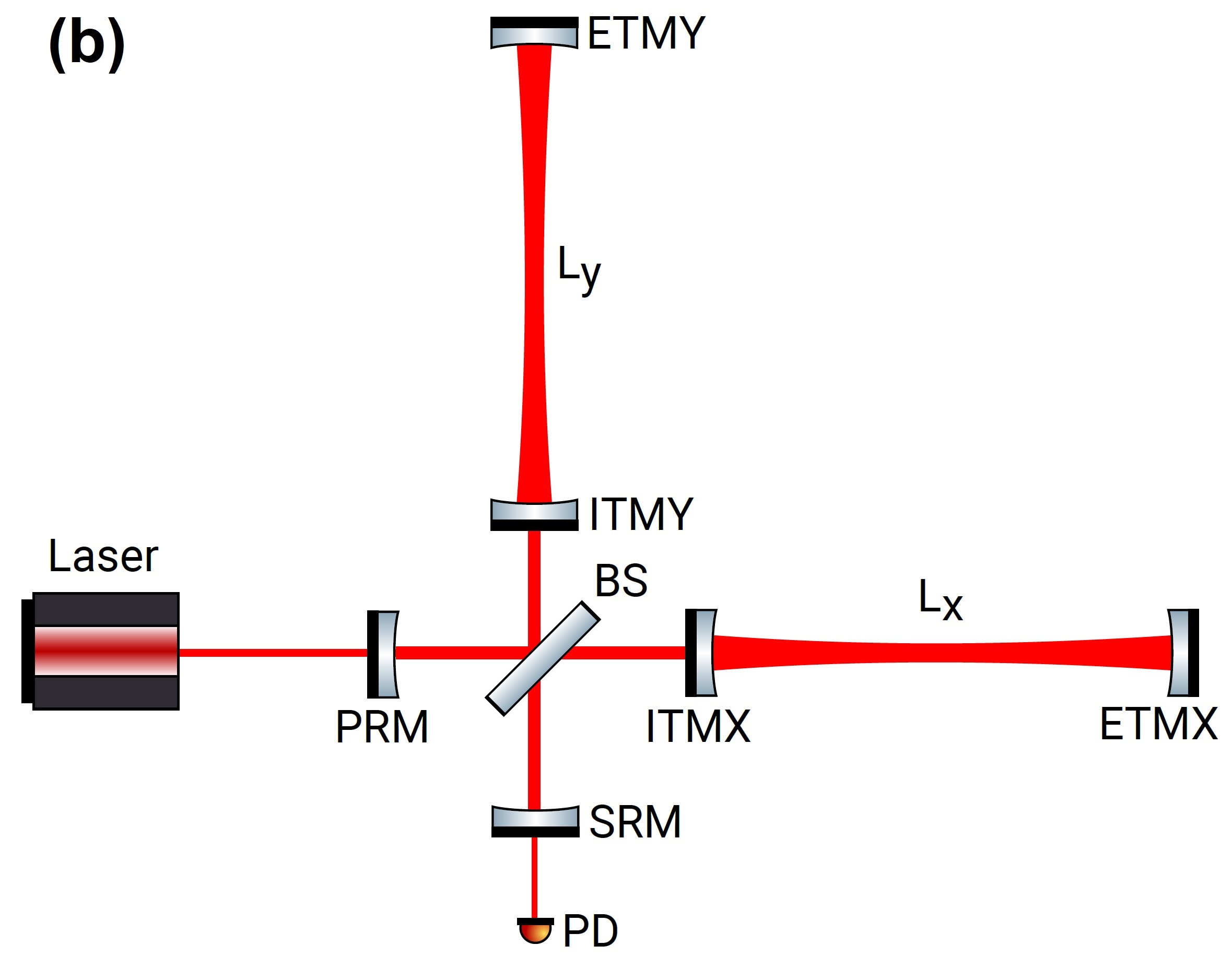}
\caption{ %\normalsize 
Simplified schematic layouts of a \textbf{(a)} dual-recycled Michelson interferometer and \textbf{(b)} dual-recycled Fabry-Perot-Michelson interferometer. 
Figures from Ref.~\cite{Grote:2019uvn}. 
}
\label{Fig:FIPs-LIFO_interferometer_schematics}
\end{figure*}

Laser interferometry has made great strides in terms of sensitivity and utility over the past century. 
The prototypical Michelson interferometry configuration used in the Michelson-Morley experiment to search for the aether \cite{Michelson:1887zz} achieved a length-equivalent accuracy of $\sim 10^{-8}~\textrm{m}$ for a 10 m arm length, corresponding to a relative accuracy of $\sim 10^{-9}$. 
A schematic of a modern Michelson interferometer is shown in Fig.~\ref{Fig:FIPs-LIFO_interferometer_schematics}(a). 
Modern gravitational-wave detectors, based on more advanced configurations that include additional optical resonators in the arms of the interferometer as shown in Fig.~\ref{Fig:FIPs-LIFO_interferometer_schematics}(b), have achieved an accuracy of $\sim 10^{-19}~\textrm{m}$ for $3-4$ km arm lengths, corresponding to a relative accuracy of $3 \times 10^{-23}$. 
This phenomenal level of precision enabled the first direct observation of gravitational waves in 2015 \cite{LIGOScientific:2016aoc}, opening up a whole new window in terms of how to view and study the Universe \cite{Bailes:2021tot}. 
This unparalleled precision motivates the use of laser interferometers as sensitive probes of certain types of new feebly-interacting particles, including particles that may contribute to the galactic dark matter.

%%%%

A variety of different approaches to search for new feebly-interacting particles using laser-interferometric gravitational-wave detectors have been explored in recent years. 
Indirect approaches have included the consideration of continuous and stochastic gravitational-wave emission from black holes surrounded by clouds of ultra-light bosons \cite{Brito:2017zvb,Tsukada:2018mbp,Palomba:2019vxe,Sun:2019mqb,KAGRA:2021tse}, as well as the effects of clouds of ultra-light bosons on binary black-hole mergers \cite{Baumann:2018vus,Choudhary:2020pxy}. 
Direct approaches have involved searches for ultra-light bosons contributing to the galactic dark matter, including scalar particles \cite{Stadnik:2014tta,Grote:2019uvn,Vermeulen:2021epa,Aiello:2021wlp}, dark photons \cite{Pierce:2018xmy,Guo:2019ker,Michimura:2020vxn,Morisaki:2020gui,LIGOScientific:2021ffg} and axion-like (pseudoscalar) particles \cite{DeRocco:2018jwe,Obata:2018vvr,Nagano:2019zit,Liu:2018icu,Martynov:2019azm}. 
Here, we focus on how laser interferometers can be used to directly search for ultra-light bosonic dark matter. 
Such searches are part of a broader recent paradigm that seeks to expand the range of tools used in dark-matter searches \cite{Bertone:2018krk}. 

Low-mass spinless bosons may be produced non-thermally in the early Universe via the ``vacuum misalignment'' mechanism \cite{Preskill:1982cy,Abbott:1982af,Dine:1982ah} and can subsequently form an oscillating classical field: 
\begin{equation}
\label{FIPs-LIFO_oscillating_scalar_field}
\phi(t) \approx \phi_0 \cos(m_\phi c^2 t / \hbar) \, , 
\end{equation}
which occurs, e.g., in the case of the harmonic potential $V(\phi) = m_\phi^2 \phi^2 / 2$ when $m_\phi \gg H(t)$, where $m_\phi$ is the boson mass and $H(t)$ is the Hubble parameter that describes the relative rate of expansion of the Universe as a function of cosmic time $t$. 
The field in Eq.~(\ref{FIPs-LIFO_oscillating_scalar_field}) carries an energy density, averaged over a period of oscillation, of $\left< \rho_\phi \right> \approx m_\phi^2 \phi_0^2 / 2$. 
Unless explicitly stated otherwise, we adopt the natural system of units $\hbar = c =1$, where $\hbar$ is the reduced Planck constant and $c$ is the speed of light in vacuum. 

\begin{figure*}[ht]
\centering
\includegraphics[width=6.5cm]{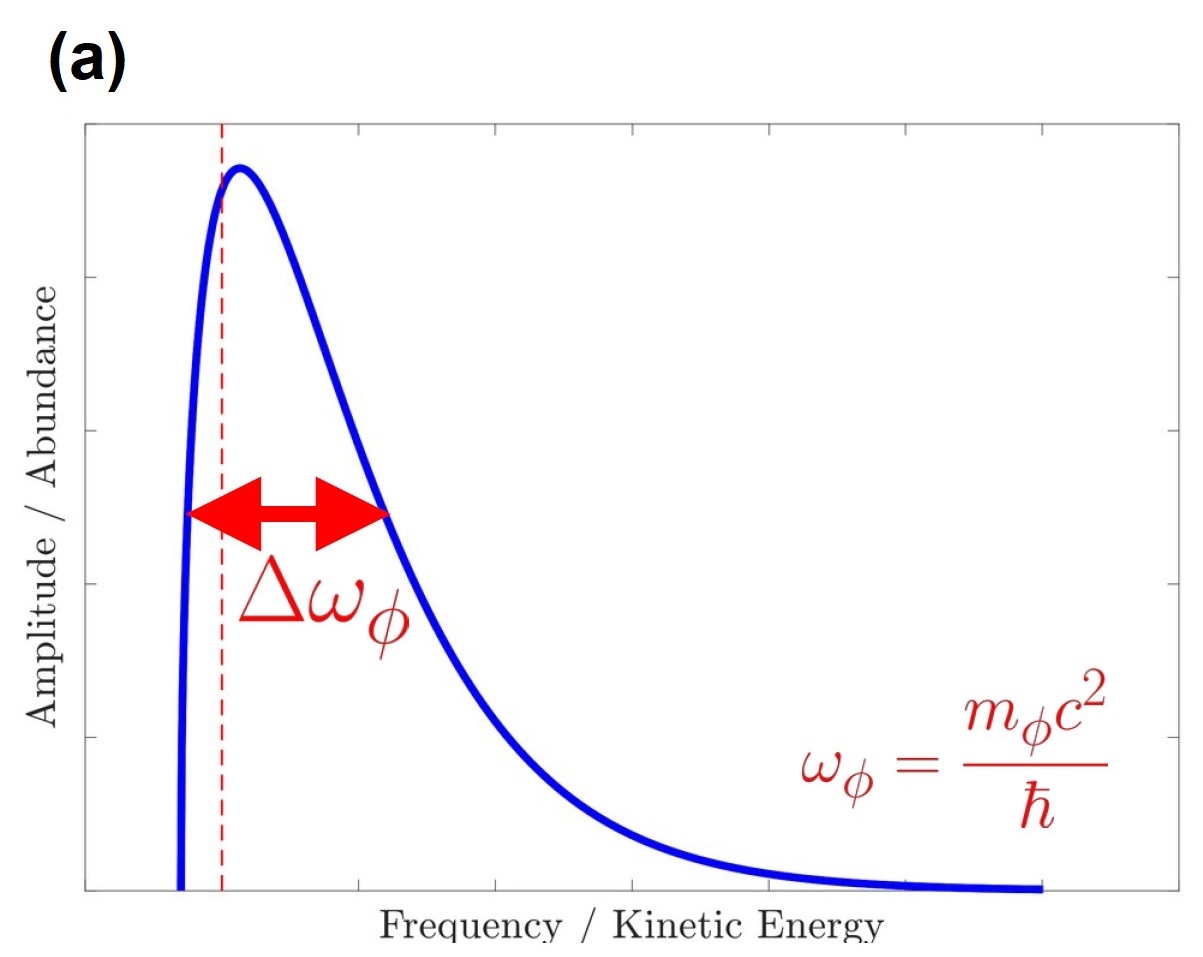}
\hspace{5mm}
%\vspace{3mm}
\includegraphics[width=6.5cm]{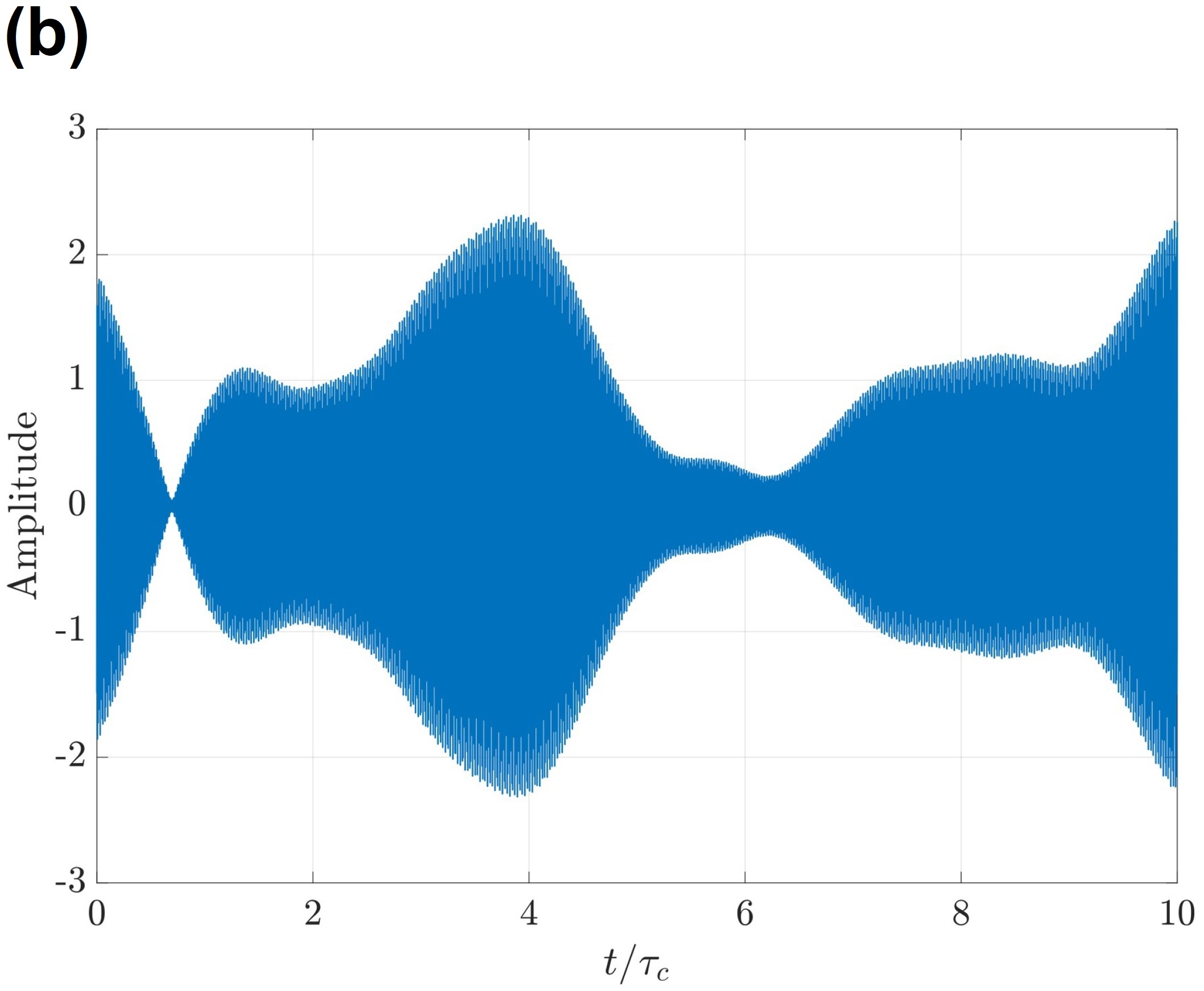}
\caption{ %\normalsize 
The virialised nature of the galactic dark matter is expected to induce \textbf{(a)} an asymmetric lineshape associated with an oscillating dark-matter field in frequency space and \textbf{(b)} a stochastically-varying amplitude of oscillation on timescales comparable to or greater than the coherence time of the oscillating dark-matter field. 
}
\label{Fig:FIPs-LIFO_stochastic_field}
\end{figure*}

The oscillations of the field in Eq.~(\ref{FIPs-LIFO_oscillating_scalar_field}) are expected to be temporally coherent on sufficiently small timescales, since the feebly-interacting bosons remain non-relativistic until the present day, which implies that all boson energies satisfy $E_\phi \approx m_\phi c^2$. 
Nowadays, the dark matter trapped in the gravitational wells of galaxies is expected to be virialised, with an estimated root-mean-square speed of $\sim 300~\textrm{km/s}$ in our local Galactic region. 
Locally, the characteristic spread in the dark-matter boson energies is hence given by $\Delta E_\phi / E_\phi \sim \left< v_\phi^2 \right> / c^2 \sim 10^{-6}$, implying a coherence time of $\tau_\textrm{coh} \sim 2 \pi / \Delta E_\phi \sim 10^6 T_\textrm{osc}$, with $T_\textrm{osc} \approx 2\pi / m_\phi$ being the oscillation period of the dark-matter field. 
In other words, the oscillations of the dark-matter field are nearly monochromatic, with an associated quality factor of $Q \sim 10^6$. 
The lineshape associated with the oscillating dark-matter field in frequency space is expected to have the asymmetric form shown in Fig.~\ref{Fig:FIPs-LIFO_stochastic_field}(a). 
On timescales exceeding the coherence time, the amplitude of oscillation $\phi_0$ fluctuates in a stochastic manner. 
Hence the signals produced by such an oscillating dark-matter field are expected to be pseudo-coherent, with an amplitude varying stochastically on timescales comparable to or greater than the coherence time, see Fig.~\ref{Fig:FIPs-LIFO_stochastic_field}(b). 
The coherence length is governed by the spatial gradients associated with the field $\phi(t,\boldsymbol{x}) \approx \phi_0 \cos(m_\phi t - m_\phi \boldsymbol{v}_\phi \cdot \boldsymbol{x})$ and is estimated to be $\lambda_\textrm{coh} \sim 2\pi / (m_\phi \sqrt{\left< v_\phi^2 \right>})$, which is $\sim 10^3$ times the Compton wavelength.

%%%%

The classical nature of the field in Eq.~(\ref{FIPs-LIFO_oscillating_scalar_field}) follows from the fitting of $\gg 1$ bosons into the reduced de Broglie volume, $n_\phi [\lambda_{\textrm{dB},\phi}/(2\pi)]^3 \ll 1$, which for the local Galactic dark-matter energy density of $\rho_\textrm{DM,local} \approx 0.4~\textrm{GeV/cm}^3$ \cite{ParticleDataGroup:2022pth} is satisfied for $m_\phi \lesssim 1~\textrm{eV}$. 
Such ultra-light dark-matter particles must necessarily be bosonic due to the Pauli exclusion principle, which prevents the haloes of galaxies and dwarf galaxies from being packed with sufficiently many ultra-light fermions. 
Ultra-light bosons would tend to suppress the formation of structures on length scales below the ground-state de Broglie wavelength of the bosons \cite{Khlopov:1985jw,Hu:2000ke}, which becomes astronomically large for sufficiently low-mass bosons. 
If ultra-light bosons account for the entirety of the observed dark matter, then their lower mass is constrained to be $m_\phi \gtrsim 10^{-21}~\textrm{eV}$ from the consideration of structures observed in Lyman-$\alpha$ forest data \cite{Irsic:2017yje,Nori:2018pka}, as well as other astrophysical observations \cite{Marsh:2018zyw,Schutz:2020jox}. 
Laser interferometers are particularly well-suited to search for the effects of oscillating dark-matter fields in the frequency range $\sim 10~\textrm{Hz} - 100~\textrm{MHz}$ corresponding to the mass range $10^{-13}~\textrm{eV} \lesssim m_\phi \lesssim 10^{-6}~\textrm{eV}$.

%%%
\subsubsection{Scalar dark matter}
\label{Sec:FIPs-LIFO-scalars}

A spinless field $\phi$ can couple to the standard-model electromagnetic field and electron via the following scalar-type interactions: 
\begin{equation}
\label{FIPs-LIFO_scalar_couplings}
\mathcal{L} = \frac{\phi}{\Lambda_\gamma} \frac{F_{\mu\nu}F^{\mu\nu}}{4} - \frac{\phi}{\Lambda_e} m_e \bar{e} e \, , 
\end{equation}
where the first term represents the coupling of $\phi$ to the electromagnetic field tensor $F$, and the second term represents the coupling of $\phi$ to the electron field $e$. 
Here $m_e$ denotes the ``standard'' mass of the electron, $\bar{e} = e^\dagger \gamma^0$ is the Dirac adjoint of the electron field, and the parameters $\Lambda_{\gamma,e}$ denote the effective new-physics energy scales of the underlying model. 
For the oscillating field in Eq.~(\ref{FIPs-LIFO_oscillating_scalar_field}), the interactions in Eq.~(\ref{FIPs-LIFO_scalar_couplings}) result in the following apparent oscillations of the electromagnetic fine-structure constant $\alpha$ and the electron mass \cite{Stadnik:2014tta}: 
\begin{equation}
\label{FIPs-LIFO_scalar_VFCs}
\frac{\delta \alpha}{\alpha} \approx \frac{\phi_0 \cos(m_\phi t)}{\Lambda_\gamma} , ~ \frac{\delta m_e}{m_e} \approx \frac{\phi_0 \cos(m_\phi t)}{\Lambda_e} \, . 
\end{equation}

Oscillations of $\alpha$ and $m_e$ would induce oscillations of solid lengths \cite{Stadnik:2014tta,Stadnik:2015xbn,Pasteka:2018tho} and refractive indices of dielectric materials \cite{Grote:2019uvn,Savalle:2019jsb}, which can be sought using laser-interferometric techniques \cite{Stadnik:2014tta,Grote:2019uvn}. 
The main experimental signatures are discussed in detail in Ref.~\cite{Grote:2019uvn}. 
For dark-matter frequencies of oscillation smaller than the fundamental vibrational frequency of the central beam-splitter, a Michelson interferometer is generally mainly sensitive to oscillations in the thickness of the beam-splitter, leading to an oscillating-in-time shift in the optical-path-length difference between the two arms of: 
\begin{equation}
\label{FIPs-LIFO_Michelson_low-freq_response}
\delta (L_x - L_y) \approx \left( \frac{1}{\Lambda_\gamma} + \frac{1}{\Lambda_e} \right) n l \phi_0 \cos(m_\phi t) \, , 
\end{equation}
where $l$ and $n$ are the thickness and refractive index, respectively, of the beam-splitter. 
On the other hand, for dark-matter frequencies of oscillation larger than the fundamental vibrational frequency of the beam-splitter, oscillations in the thickness of the beam-splitter are generally suppressed and a Michelson interferometer becomes mainly sensitive to oscillations in the refractive index of the beam-splitter. 

\begin{figure*}[ht]
\centering
\includegraphics[width=6.5cm]{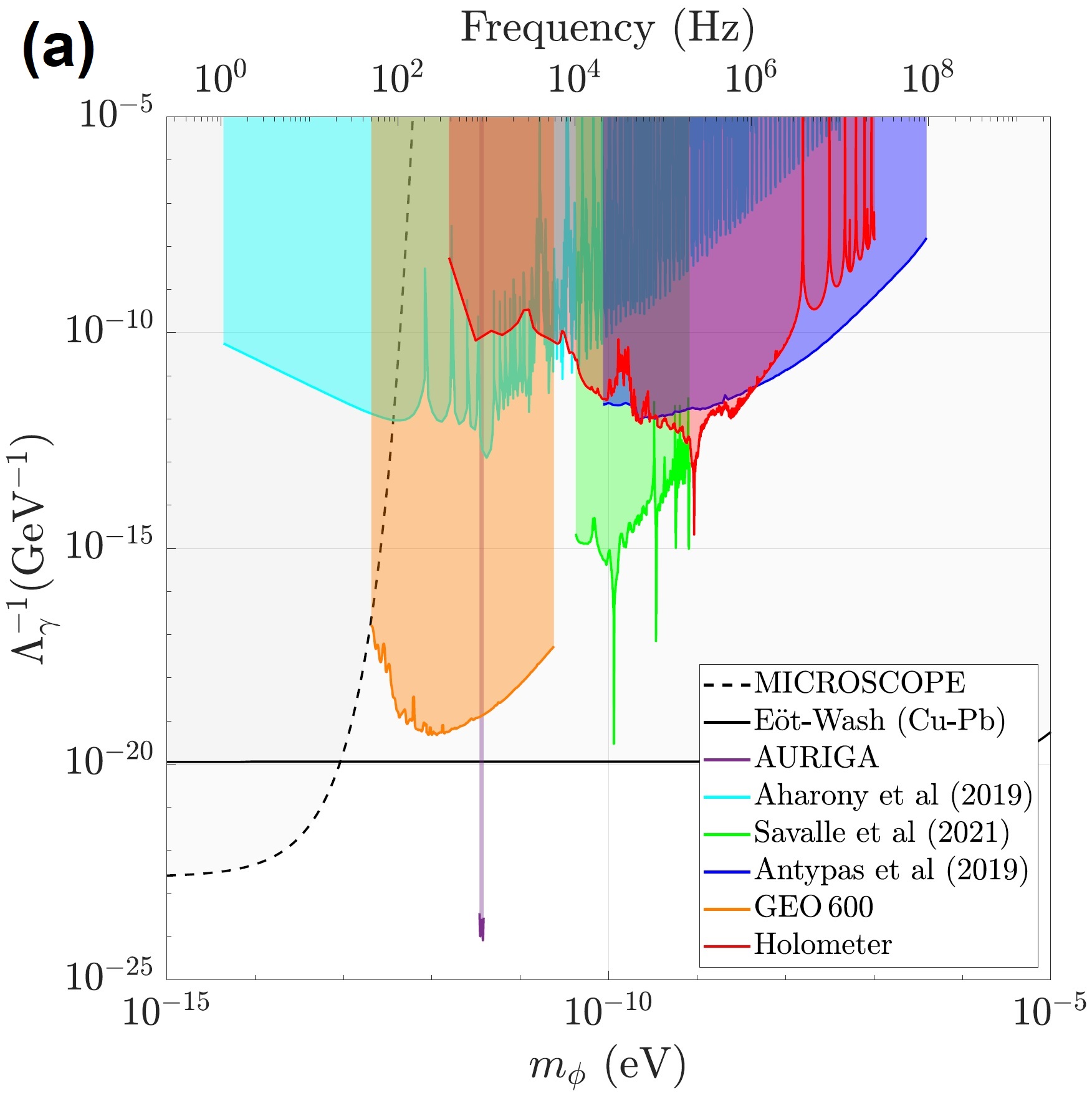}
\hspace{5mm}
%\vspace{3mm}
\includegraphics[width=6.5cm]{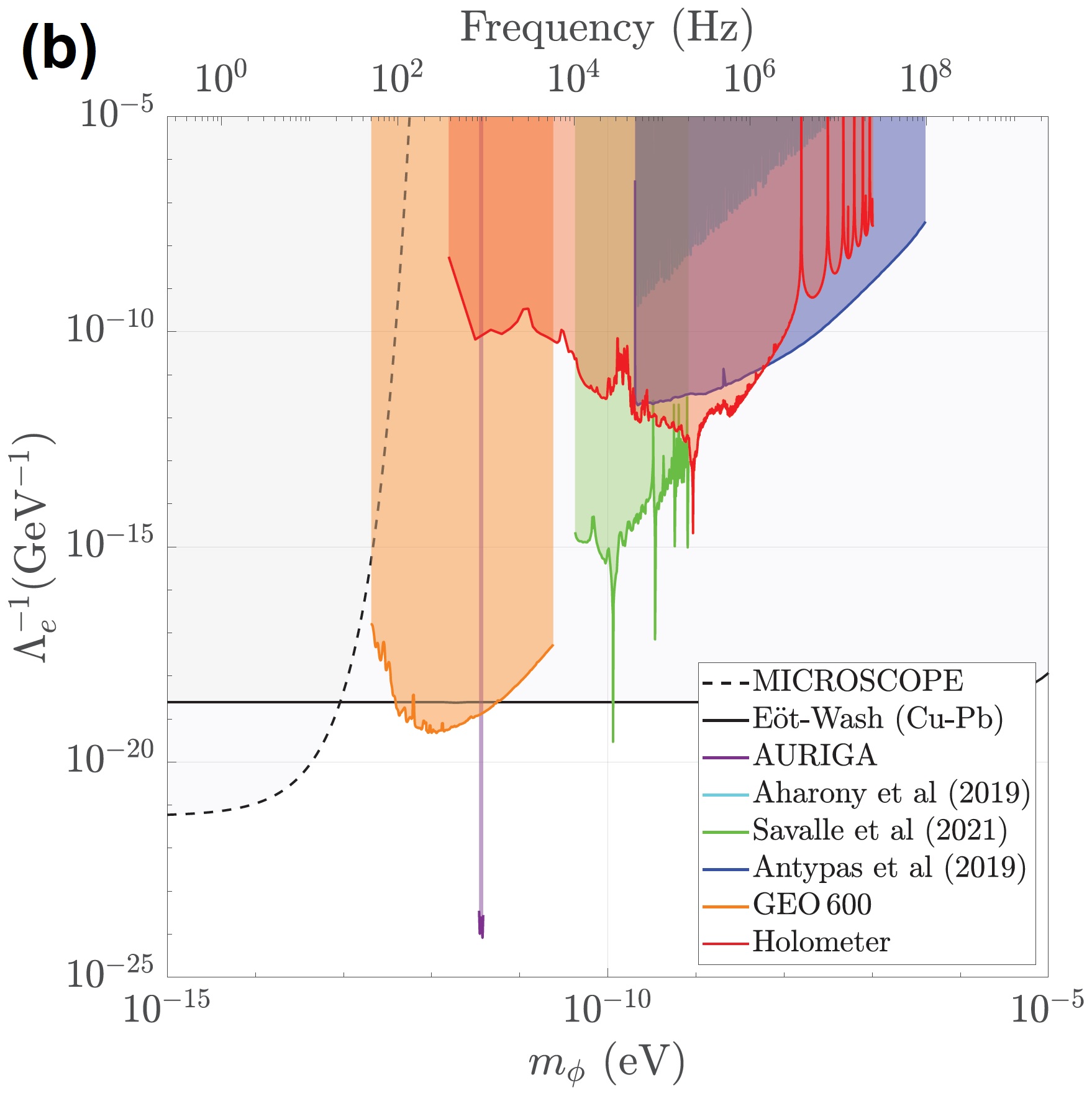}
\caption{ %\normalsize 
Constraints on the interaction parameters in Eq.~(\ref{FIPs-LIFO_scalar_couplings}) of an oscillating scalar dark-matter field coupling to \textbf{(a)} the electromagnetic field (photon) and \textbf{(b)} electron. 
Figures from Ref.~\cite{Aiello:2021wlp}. 
}
\label{Fig:FIPs-LIFO_scalar_results}
\end{figure*}

While modern gravitational-wave detectors based on the Fabry-Perot-Michelson configuration shown in Fig.~\ref{Fig:FIPs-LIFO_interferometer_schematics}(b), such as LIGO \cite{LIGOScientific:2014pky,Abbott:2016xvh}, Virgo \cite{VIRGO:2014yos} and KAGRA \cite{Aso:2013eba}, currently offer the best \textit{strain} sensitivity that is advantageous in gravitational-wave searches, interferometers based on the Michelson configuration shown in Fig.~\ref{Fig:FIPs-LIFO_interferometer_schematics}(a), such as GEO600 \cite{Grote:2013oio,Dooley:2015fpa}, can offer better \textit{phase} sensitivity that is advantageous when searching for the effects of dark matter on the central beam-splitter. 
The GEO600 detector can operate beyond the quantum shot-noise limit and has participated in joint observing runs with the Advanced LIGO detectors since 2015, with the primary aim of searching for gravitational waves. 
Members of the GEO600 collaboration recently analysed data from seven segments, each of $\sim 10^5~\textrm{s}$ duration collected in runs during 2016 and 2019, to search for a signal from an oscillating scalar dark-matter field in the frequency range $\sim 50~\textrm{Hz} - 6~\textrm{kHz}$ \cite{Vermeulen:2021epa}. 
In their analysis, they used an optimised frequency bin width for each search frequency, which yields the optimal signal-to-noise ratio for the expected signal where the linewidth is a fixed fraction of the oscillation frequency. 
No signal consistent with the coupling of an oscillating scalar dark-matter field to the interferometer was found, leading to the new bounds on the scalar-photon and scalar-electron couplings shown in Fig.~\ref{Fig:FIPs-LIFO_scalar_results}.

%%%%

The use of co-located Michelson interferometers, such as the Fermilab holometer \cite{Holometer:2015tus,Holometer:2016ipr}, can be advantageous in dark-matter searches, since the wave-like dark-matter signal can be correlated between the two interferometers, whereas the noise is uncorrelated. 
The Fermilab holometer was originally constructed to search for exotic quantum space-time correlations \cite{Holometer:2016ipr}. 
A 704-hour dataset of cross-correlation measurements has recently been analysed to search for a signal from an oscillating scalar dark-matter field in the frequency range $\sim 400~\textrm{Hz} - 25~\textrm{MHz}$ \cite{Aiello:2021wlp}. 
No signal consistent with the coupling of an oscillating scalar dark-matter field to the twin interferometers was observed, leading to new bounds on the scalar-photon and scalar-electron couplings shown in Fig.~\ref{Fig:FIPs-LIFO_scalar_results}. 
A smaller-scale experiment using a pair of co-located interferometers targetting the frequency range $\sim 1 - 100~\textrm{MHz}$ is planned in Cardiff \cite{Vermeulen:2020djm}.

%%%
\subsubsection{Axion-like dark matter}
\label{Sec:FIPs-LIFO-axions}
Besides the scalar-type interactions in Eq.~(\ref{FIPs-LIFO_scalar_couplings}) above, a spinless field $\phi$ can also couple to the standard-model electromagnetic field via the following axion-like or pseudoscalar-type interaction: 
\begin{equation}
\label{FIPs-LIFO_pseudoscalar_couplings}
\mathcal{L} = - g_{\gamma} \phi \frac{F_{\mu\nu} \tilde{F}^{\mu\nu}}{4}  \, , 
\end{equation}
where $\tilde{F}$ denotes the dual of the electromagnetic field tensor, and $g_{\gamma}$ is a coupling constant with dimensions of inverse energy. 
The canonical axion may resolve the famous strong CP problem of particle physics; see, e.g., \cite{Kim:2008hd} for a review. 
For the oscillating field in Eq.~(\ref{FIPs-LIFO_oscillating_scalar_field}), the interaction in Eq.~(\ref{FIPs-LIFO_pseudoscalar_couplings}) induces a difference in the phase velocity between left- and right-handed circularly polarised light \cite{DeRocco:2018jwe}. 
In the limit of a non-relativistic axion-like dark-matter field, the difference in phase velocities between photons of opposite polarisation states but with a common wavevector $k$ is $\Delta v_\textrm{phase} \approx \dot{\phi} g_\gamma / k$. 
Current interferometry-based gravitational-wave detector configurations are practically insensitive to such effects, but it is possible to design sensitive laser-interferometry experiments via the use of birefringent materials and birefringent cavities \cite{DeRocco:2018jwe,Obata:2018vvr,Nagano:2019zit,Liu:2018icu,Martynov:2019azm}. 
Such modified setups may even be realised on table-top scales.

%%%
\subsubsection{Dark photon dark matter}
\label{Sec:FIPs-LIFO-DP}
A dark photon is a massive vector boson associated with an additional U(1) symmetry beyond the standard model \cite{Okun:1982xi,Holdom:1985ag}. 
Low-mass dark photons may be produced via various non-thermal production mechanisms in the early Universe and subsequently form an oscillating classical field that can comprise the observed dark matter \cite{Nelson:2011sf,Arias:2012az,Graham:2015rva}. 
The wavefunction associated with a dark photon field $A_\mu$ takes the form: 
\begin{equation}
\label{FIPs-LIFO_oscillating_vector_field}
A_\mu ( t, \boldsymbol{x} ) \approx A_{\mu,0} \cos (m_A t - \boldsymbol{p}_A \cdot \boldsymbol{x}) \, , 
\end{equation}
where the local field amplitude $A_{\mu,0}$ is related to the dark photon mass $m_A$ and the local time-averaged energy density associated with the vector field according to $\left< \rho_A \right> \approx m_A^2 A_{\mu,0} A_0^\mu / 2$. 
$\boldsymbol{p}_A$ is the average momentum of a dark photon comprising the dark-matter field relative to an observer at the space-time point $x^\mu = ( t, \boldsymbol{x} )$. 

A dark photon field $A_\mu$ can couple to the standard model via direct gauge couplings of the form: 
\begin{equation}
\label{FIPs-LIFO_vector_couplings}
\mathcal{L} = -\varepsilon e A_\mu j_\textrm{SM}^\mu \, , 
\end{equation}
where $\varepsilon$ is a dimensionless coupling constant, $e$ is the electric charge quantum, and $j_\textrm{SM}^\mu$ is a 4-current (not necessarily the electromagnetic current) associated with the relevant $U(1)$ gauge symmetry. 
For instance, if one gauges the baryon number $B$ in the standard model and takes the dark photon to be the gauge boson associated with the $U(1)_B$ gauge group, then $j_\textrm{SM}^\mu$ should be identified as the baryon current. 
Another popular choice is the $B-L$ current, where $L$ is the lepton number. 
For the oscillating field in Eq.~(\ref{FIPs-LIFO_oscillating_vector_field}), the interaction in Eq.~(\ref{FIPs-LIFO_vector_couplings}) induces an oscillating-in-time acceleration on a test body carrying the relevant dark charge $Q$ \cite{Pierce:2018xmy}: 
\begin{equation}
\label{FIPs-LIFO_vector_oscillating_force}
\boldsymbol{a} (t,\boldsymbol{x}) \approx - \frac{\varepsilon e Q}{M} m_A \boldsymbol{A}_0 \sin(m_A t - \boldsymbol{p}_A \cdot \boldsymbol{x}) \, , 
\end{equation}
where $M$ is the mass of the test body. 
If the dark photon is a gauge field associated with baryon number, then $Q$ is the total baryon number associated with the test body, whereas if the dark photon gauge field is associated with $B-L$, then $Q$ is the total neutron number associated with the (electrically neutral) test body. 

\begin{figure*}[ht]
\centering
\includegraphics[width=8.5cm]{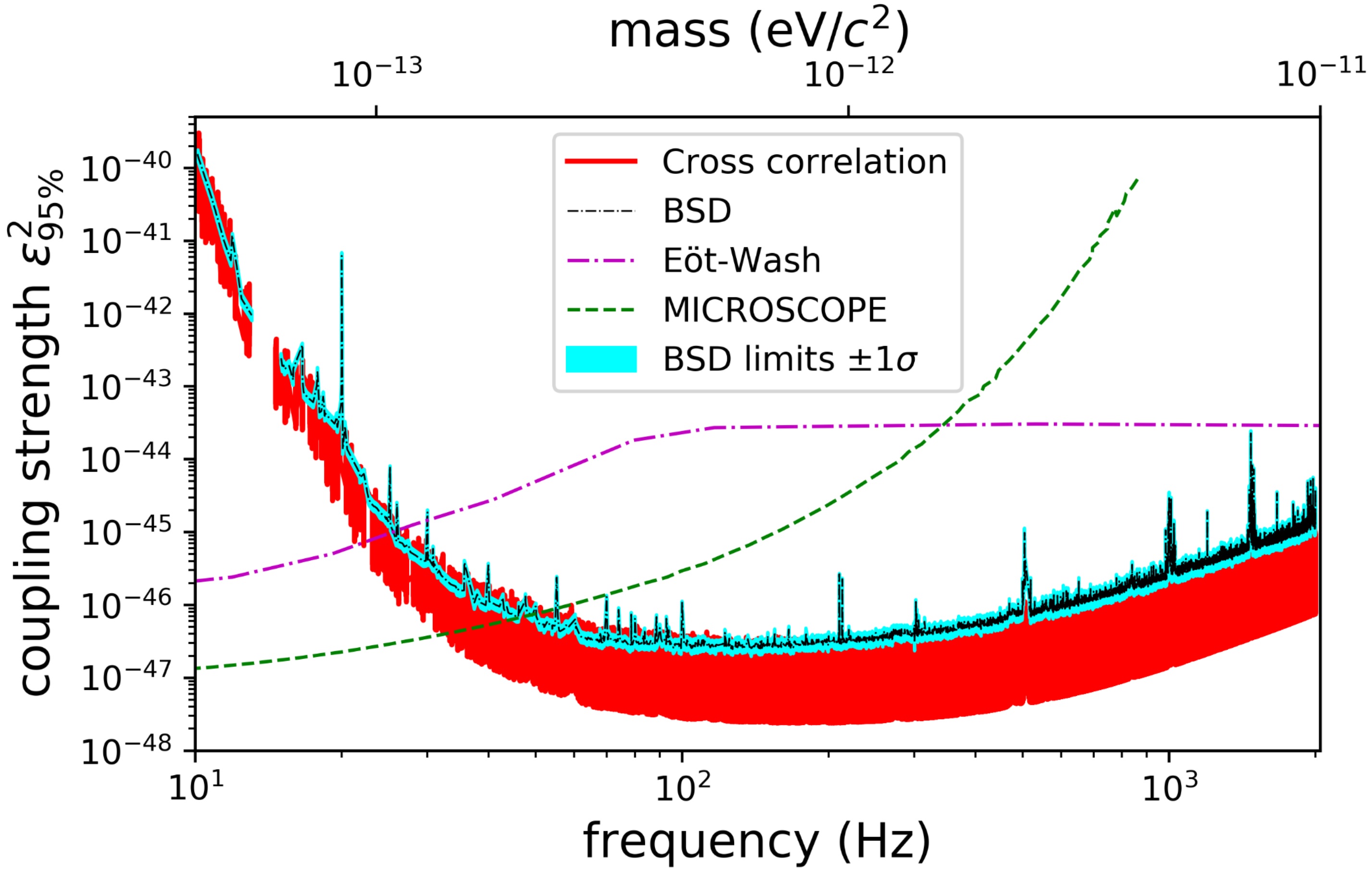}
%\hspace{5mm}
%\vspace{3mm}
%\includegraphics[width=8.5cm]{}
\caption{ %\normalsize 
Constraints on the interaction parameter in Eq.~(\ref{FIPs-LIFO_vector_couplings}) of an oscillating dark-photon dark-matter field for the baryon gauge coupling. 
Figure from Ref.~\cite{LIGOScientific:2021ffg}. 
}
\label{Fig:FIPs-LIFO_vector_results}
\end{figure*}

The acceleration in Eq.~(\ref{FIPs-LIFO_vector_oscillating_force}) is common to all optical components (mirrors and beam-splitter) of the interferometer made of the same material, up to a small difference owing to spatial gradients in the dark photon dark-matter field that lead to an oscillating-in-time shift in the optical-path-length difference between the two arms of the interferometer \cite{Pierce:2018xmy}. 
A larger signal can result if the interferometer contains optical components with different material compositions, like found in KAGRA \cite{Michimura:2020vxn}. 
A search for a signal from an oscillating dark photon dark-matter field in the frequency range $\sim 10~\textrm{Hz} - 2~\textrm{kHz}$ was performed using data from LIGO's O1 run \cite{Guo:2019ker}. 
No signal consistent with the coupling of an oscillating dark photon dark-matter field to the interferometer was found, leading to new bounds on the $B$ gauge coupling. 
It was subsequently pointed out that Refs.~\cite{Pierce:2018xmy,Guo:2019ker} had underestimated the sensitivity of these types of experimental searches due to omission of the consideration of the finite light travel time through the arms of the interferometer \cite{Morisaki:2020gui}. 
A further search in the same frequency range was performed using data from LIGO and Virgo's joint O3 run \cite{LIGOScientific:2021ffg}. 
No dark-matter signal was found, leading to the new and improved bounds on the $B$ gauge coupling shown in Fig.~\ref{Fig:FIPs-LIFO_vector_results}, as well as bounds on the $B-L$ gauge coupling. 
Further improvements in sensitivity are anticipated with new data from LIGO's upcoming O4 run, as well as future gravitational-wave detectors such as Einstein Telescope, Cosmic Explorer and LISA \cite{Morisaki:2020gui}.

%\subsubsection{Acknowledgements}
%This work has been supported by the STFC in the UK, research grants ST/T006331/1 and ST/W006456/1 and 
%The Leverhulme Trust in the UK, research grant RPG-2019-022.

%%%%

\subsection{\emph{New ideas:} Search for Dark Photon Radio Signals from White Dwarfs -- {\it N. Song}}
\label{Song}
{\it Author: Ninhqiang Song, <ningqiang.song@liverpool.ac.uk>}
%-------------------------------------------

\subsubsection{Introduction}

Dark photons being the gauge boson of an extra $U(1)$ group that couples to the Standard Model sector are well motivated in string theory compactifications~\cite{Svrcek:2006yi,Abel:2006qt,Abel:2008ai,Arvanitaki:2009fg,Goodsell:2009xc}, which may kinetically mix with Standard Model photons~\cite{Holdom:1985ag,Dienes:1996zr,Abel:2003ue}. Sufficiently long-lived dark photon with a mass $m_{A'}$ can be produced through a number of mechanisms, including inflationary fluctuations~\cite{Graham:2015rva,Ema:2019yrd,Nakai:2020cfw,Nakayama:2019rhg}, tachyonic instability~\cite{Co:2018lka,Bastero-Gil:2018uel,Agrawal:2018vin} and parametric resonance~\cite{Dror:2018pdh}, the misalignment mechanism~\cite{Arias:2012az,Alonso-Alvarez:2019ixv}, or cosmic strings~\cite{Long:2019lwl,East:2022rsi}, and constitute part of or all the cosmological dark matter. A wide variety of experiments constrain the dark photon dark matter parameter space, including light shining through wall approaches~\cite{Ehret:2010mh,Betz:2013dza}, searches for deviations from Coulomb's law~\cite{Williams:1971ms} and searches for dark photon dark matter using haloscopes~\cite{Godfrey:2021tvs,Nguyen:2019xuh,ADMX:2001nej,ADMX:2009iij,ADMX:2018gho,ADMX:2018ogs,ADMX:2019uok,Lee:2020cfj,HAYSTAC:2018rwy,HAYSTAC:2020kwv,Dixit:2020ymh,Alesini:2020vny,Cervantes:2022yzp,DOSUE-RR:2022ise,An:2022hhb,Knirck:2018ojz,Tomita:2020usq,Ramanathan:2022egk}. Additionally, conversion between dark matter dark photons and photons would cause distortion of the CMB spectrum, the non-observation of which leads to strong bounds~\cite{Jaeckel:2008fi,Mirizzi:2009iz,Arias:2012az,McDermott:2019lch}. Complementary to these terrestrial experiments and cosmological constraints,  dark photons can be searched for through their effects in astrophysical environments. For example, they contribute to the cooling of the Sun and horizontal branch stars~\cite{raffelt1996stars,An:2013yfc,An:2013yua,Hardy:2016kme,An:2020bxd}, red giant stars~\cite{Redondo:2013lna} and neutron stars~\cite{Hong:2020bxo}. 

There is a similar synergy between astrophysics and experiments in the search for dark matter axions. One promising approach consists of looking for radio waves produced by the conversion of dark matter axions to photons in the magnetosphere of neutron stars \cite{Hook:2018iia,Huang:2018lxq,Safdi:2018oeu,Witte:2021arp,Millar:2021gzs,Battye:2021xvt,Battye:2021yue,Wang:2021hfb,Foster:2022fxn}. Axions with mass $m_a\lesssim 10^{-4}$~eV can be resonantly converted to photons in regions where the resonant condition $m_a\simeq \omega_p$ is met, {\it i.e.} the axion mass is approximately equal to the plasma frequency $\omega_p=\sqrt{4\pi\alpha n_e/m_e}$, where $n_e$ is the free electron number density.
Kinetically mixed dark photons can also resonantly convert to photons in regions where $m_{A'} \simeq \omega_p$. It is suggested that such conversion in the solar corona could allow searches for dark photons with masses between $4\times 10^{-8}$~eV and $4\times 10^{-6}$~eV~\cite{An:2020jmf}. In recent work~\cite{Hardy:2022ufh}, we consider the conversion of dark photon to photon in white dwarfs in the galactic center. White dwarfs typically have a hot, dense atmosphere with an effective temperature of up to $2\times 10^5$~K, and electron densities up to $10^{17}$~cm$^{-3}$, opening up an opportunity for sensitivity to dark photons with masses as large as $10^{-3}~\rm{eV}$ when the signals are explored by radio telescopes on the Earth.  

\subsubsection{Resonant Conversion in White Dwarfs}

We consider a dark photon with the Lagrangian density
\begin{equation} \label{eq:L0}
\mathcal{L} = -\frac{1}{4} F'_{\mu\nu}F'^{\mu\nu} + \frac{1}{2} m_{A'}^2 A'_\mu A'^{\mu} + \frac{\kappa}{2} F'_{\mu\nu}F^{\mu\nu} + \mathcal{L}_{\rm SM}  ~,
\end{equation}
where $F$ ($F'$) is the SM photon (dark photon) field, $\kappa$ is the kinetic mixing, and we assume that the dynamics that give rise to the DP mass $m_{A'}$ are decoupled, {\it e.g.} that the mass comes from the Stueckelberg mechanism.  It is believed that only about 10\% of white dwarfs have a magnetic field stronger than 0.1~MG~\cite{Wang:2021hfb,Kawka:2006ub,holberg201625,Hollands_2015}, so we set $B=0$ and assume the plasma in white dwarfs to be isotropic (see~\cite{Hardy:2022ufh} however for the conversion in a general plasma where the effects of magnetic field are important, {\it e.g.} the magnetosphere of a neutron star). In such plasma, the longitudinal polarization of photon does not propagate and only the conversion of transverse dark photon modes is relevant for the signal. Assuming they propagate in the $z$ direction, the mixing of the transverse modes can be written in the symmetric form
\begin{equation}
    \left[\omega^2+\partial_z^2+\begin{pmatrix}
    -\omega_p^2&-\kappa\omega_p^2\\
    -\kappa\omega_p^2& -m_{A'}^2
    \end{pmatrix}\right]
    \begin{pmatrix}
    A_i\\
    A'_i
    \end{pmatrix}=0\,,
\end{equation}
where $i=x,y$. We can write the photon and dark photon fields in the wave form $A_i=e^{i\omega t-ikz}\tilde{A}_i(y,z)$, $A'_i=e^{i\omega t-ikz}\tilde{A}'_i(y,z)$ where $\omega^2=k^2+m_{A'}^2$. Using the WKB approximation with the assumptions $|\partial_z^2\tilde{A}_i|\,\ll k|\partial_z\tilde{A}_i|$, we obtain the first order differential equation,
\begin{equation}
    i\partial_z \tilde{A}_i=\dfrac{1}{2k}(m_{A'}^2-\omega_p^2)\tilde{A}_i-\dfrac{\kappa\omega_p^2}{2k} \tilde{A}'_i\,.
    \label{eq:dAWD}
\end{equation}
We assume the dark photon velocity has a  Maxwell-Boltzmann distribution in the galactic rest frame, $f_v(v)\simeq  (\pi v_0^2)^{-3/2} e^{-v^2/v_0^2}$. Starting from an asymptotic velocity $v_i$ far away from white dwarf of mass $M_{\rm WD}$ and radius $r_0$, the infalling dark photon accelerates to
\begin{equation}
    v\simeq \sqrt{\dfrac{2G_NM_{\rm WD}}{r_0}}= 4800~{\rm km/s}\sqrt{\dfrac{M_{\rm WD}}{M_\odot}\dfrac{0.01R_\odot}{r_0}}\gg v_i\,,
    \label{eq:vWD}
\end{equation}
near the white dwarf surface. This show that the dark photon trajectories are nearly radial with small deviations. 
 The solution of Eq.~\eqref{eq:dAWD} indicates that the conversion peaks at $m_{A'}=\omega_p$ with the probability 
$
    p_{\rm WD}\simeq \pi\kappa^2\omega_p^3/(3k\partial_r\omega_p)\,.
$
In the white dwarf atmosphere, we expect the pressure gradient $\rho^{-1}\partial P/\partial r$ to balance the gravitational potential $GM_{\rm WD}/r$, so we approximate the scale height of the atmosphere~\cite{ingham1976origin,Gill:2011yp,Wang:2021wae}
\begin{equation}
    l_a\simeq \dfrac{kT_a r_0^2}{GM_{\rm WD}\mu m_p}=0.06~{\rm km}\left(\dfrac{T_a}{10^4~\rm K}\right)\left(\dfrac{M_{\rm WD}}{M_\odot}\right)\left(\dfrac{r_0}{0.01~R_\odot}\right)^2\,,
\end{equation}
and the free electron density profile
$
    n_e(r)=n_0\exp((r-r_0)/l_a)\,,
$
where we take the mean molecular weight $\mu=0.5$ for fully ionized hydrogen plasma, and the maximum electron number density $n_0=10^{17}$~cm$^{-3}$ based on spectroscopic studies~\cite{kieu_2017,tremblay2009spectroscopic}. With this profile the resonant conversion takes place at a radius 
\begin{equation}
    r_c=r_0+l_a~\mathrm{ln}\left(\dfrac{4\pi\alpha}{m_e}\dfrac{n_0}{m_{A'}^2}\right)\simeq r_0\,,
\end{equation}
as $r_0\gg l_a$. The photon signal power produced from resonant conversion is therefore
\begin{equation}
    \dfrac{d\mathcal{P}}{d\Omega}\simeq 2p_{\rm WD}r_c^2\rho_{A'}(r_c)v_c\,,
    \label{eq:dPdOmegaWD}
\end{equation}
where the factor of 2 accounts for conversion when approaching and leaving the white dwarf, and the dark photon velocity at the resonant conversion radius $v_c\sim v$ as given in Eq.~\eqref{eq:vWD}. From Liouville's theorem, the phase space density of dark photon is conserved along the trajectories during infalling, so near $r_c$ the dark photon density 
$
\rho_{A'}(r_c)=2v_c (\sqrt{\pi} v_i)^{-1}\rho_{A'}^\infty
$,
where $\rho_{A'}^\infty$ is the energy density away from white dwarf in the dark matter halo~\cite{Millar:2021gzs,Leroy:2019ghm}. As the photons travel out of the white dwarf,  the signal might be reduced due to inverse bremsstrahlung absorption~\cite{An:2020jmf,Redondo:2008aa}.
The survival probability due to absorption is
\begin{equation}
    p^{\rm IB}_s=\exp\left(-\dfrac{2\alpha l_a m_{A'}}{27\pi}\left(\dfrac{2\pi m_e}{T_a}\right)^{1/2} \right.
    \left.\left(1-e^{-m_{A'}/T_a}\right)\left(3\log\left(\dfrac{2T_a^2}{m_{A'}^2}\right)+0.84\right)\right)\,,
\end{equation}
where we have used the approximation $\omega\simeq m_{A'}$. Explicitly, the signal power from white dwarf atmosphere
\begin{equation}
    \dfrac{d\mathcal{P}_{\rm WDa}}{d\Omega}
    =2.3\times 10^{7}~{\rm W}\ \left(\dfrac{\kappa}{10^{-8}}\right)^2
    \left(\dfrac{m_{A'}}{10^{-5}~\rm eV}\right)
    \left(\dfrac{\rho_{A'}^\infty}{0.3~{\rm GeV/cm}^3}\right)
\left(\dfrac{300~\rm km/s}{ v_0}\right)\left(\dfrac{T_a}{10^4~\rm K}\right) p_s^{\rm{IB}}\,,
\end{equation}
where we have fixed $M_{\rm WD}=M_\odot$ and $r_0=0.01R_\odot$.

\subsubsection{Detection Sensitivity}

\begin{figure}
    \centering
    \includegraphics[width=0.48\textwidth]{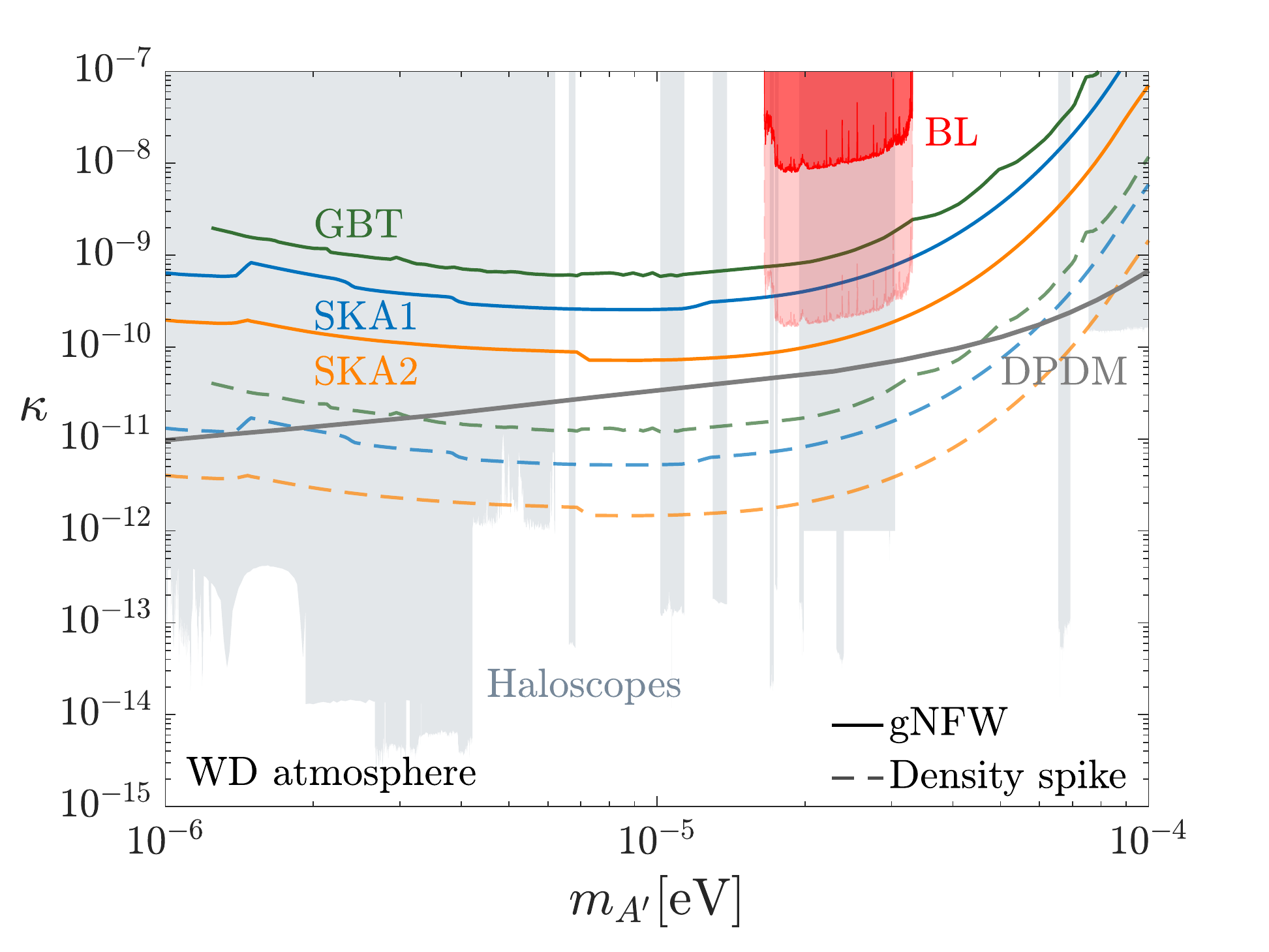}
    \includegraphics[width=0.48\textwidth]{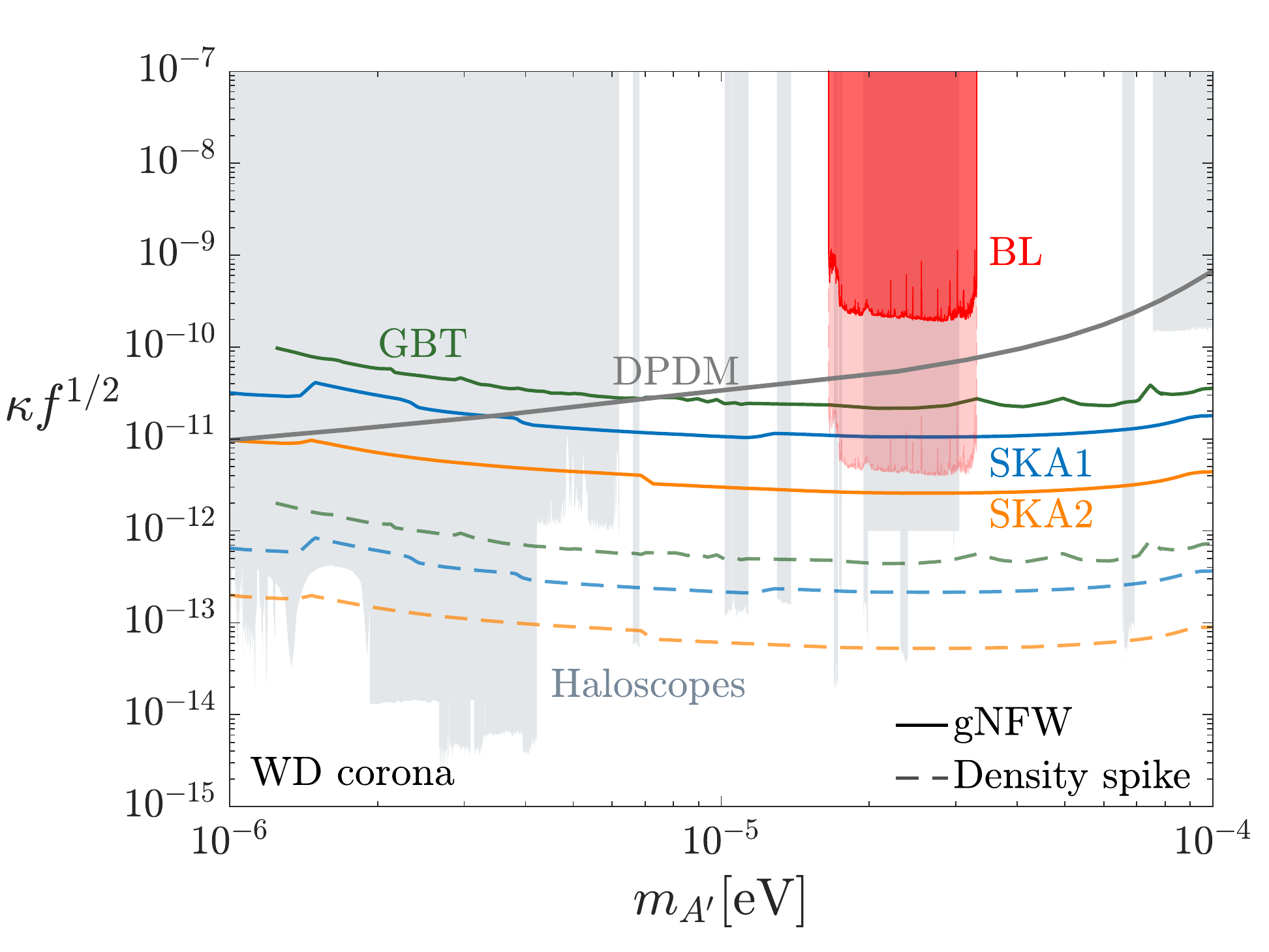}    
    \caption{Projected sensitivity of radio telescopes to dark photon dark matter from 100 hours of observation of the cumulative signals from the atmosphere (left) and corona (right) of white dwarfs within 3~pc of the galactic center. $f$ is the fraction of white dwarfs with a dense enough corona for resonant conversion, see text for details.  Colored lines show the projected sensitivity from GBT, SKA1, and SKA2 and the red shaded regions show our constraints derived from the Breakthrough Listen (BL) project~\cite{Foster:2022fxn}; for each of these we show the results both for a gNFW and a density spike profile. The grey line depicts the cosmological constraint on dark photon dark matter (DPDM) from Arias et al~\cite{Arias:2012az}. Overlaid shaded regions are limits from haloscopes~\cite{Godfrey:2021tvs,Nguyen:2019xuh,ADMX:2001nej,ADMX:2009iij,ADMX:2018gho,ADMX:2018ogs,ADMX:2019uok,Lee:2020cfj,HAYSTAC:2018rwy,HAYSTAC:2020kwv,Dixit:2020ymh,Alesini:2020vny,Cervantes:2022yzp,DOSUE-RR:2022ise,An:2022hhb,Knirck:2018ojz,Tomita:2020usq,Ramanathan:2022egk}.}    
    \label{fig:WDatm}
\end{figure}

The photon signals from dark photon conversion can be enhanced by considering a collection of stars. The signal flux density from the atmosphere of white dwarfs in the galactic center
\begin{equation}
        {S}_{\rm sig} = \dfrac{1}{\mathcal{B}d^2}\int_{R_{\min}}^{R_{\rm max}}4\pi R^2 n_{\rm WD}(R) dR
        \int f_T \dfrac{d\mathcal{P_{\rm WDa}}}{d\Omega} d\log_{10}T_a \,.
\label{eq:sigtotWDa}
\end{equation}
where $d=8$~kpc is the distance from the Earth to the galactic center, $n_{\rm WD}$ is the white dwarf number density distribution taken from~\cite{freitag2006stellar} and $f_T$ is the temperature distribution of the white dwarf atmosphere.  The signal frequencies from the individual sources are Doppler shifted differently due to the motion of the stars, leading to a total width $\mathcal{B}\simeq m_{A'} \sigma_s\sim 10^{-3}m_{A'}$, where $\sigma_s$ is the stars' velocity dispersion~\cite{Safdi:2018oeu}. The resulting sensitivity is shown in the left panel of Fig.~\ref{fig:WDatm}, where the signal flux is compared with the minimum detectable flux density of radio telescopes including SKA~\cite{dewdney2013ska1} and GBT~\cite{GBT}. We have also translated the constraint derived for axions using the flux limit data in the Breakthrough Listen (BL) project~\cite{Foster:2022fxn}. To quantify the uncertainties from the dark matter density distribution, 
we also show the sensitivities for the generalized Navarro–Frenk–White (gNFW) profile~\cite{Benito:2016kyp,Benito:2020lgu} and a density spike near the central black hole~\cite{Lacroix:2018zmg,Bertone:2002je,Lacroix:2013qka}.

Additionally, it is possible that some white dwarfs might be surrounded by a hot envelope in the outer part of their atmosphere, a so-called ``corona''~\cite{fleming1993detection,arnaud1992coronal}, which was originally proposed to account for observations of X-ray emission from such stars~\cite{musielak2003chandra}. However, these observations were later revisited and found to be either consistent with emissions
from the photosphere or with a non-detection~\cite{weisskopf2007chandra,musielak2003chandra,drake2005deposing}. Upper limits were set on the electron density in the corona, which ranges from $4.4\times 10^{11}$ to $5\times 10^{12}$~cm$^{-3}$ ~\cite{weisskopf2007chandra,zheleznyakov2004thermal}. Nevertheless, it remains possible that corona could exist in some white dwarfs, so we briefly consider the radio signals that would result. The typical suggested temperature in a corona is $10^6$~K to $\gtrsim 10^7$~K~\cite{zheleznyakov2004thermal,weisskopf2007chandra},  and the electron density profile would be similar to that in the white dwarf atmosphere. We display the sensitivities from the conversion in the possible white dwarf corona in the right panel of Fig.~\ref{fig:WDatm}. We remain agnostic about the maximum electron density in the corona, but we include a factor $f$ that quantifies the fraction of white dwarfs with a dense enough corona so that $n_0\geq \frac{m_e m_{A'}^2}{4\pi\alpha}$. Given the non-observation of corona emission, this fraction is likely to be small for $n_0\gtrsim 10^{12}~\rm{cm}^{-3}$.

Moreover, we explore in~\cite{Hardy:2022ufh} the sensitivities for dark photon conversion in the magnetosphere of neutron stars or the boundary layer of accreting white dwarfs, which could surpass the current limit by orders of magnitude in certain dark photon mass ranges. 

%-------------------------------------------
\subsection{\emph{New ideas:} A self-consistent wave description of axion miniclusters and their survival in
the galaxy --  {\it V. Dandoy}}
\label{Dandoy}
{\it Author: Virgile Dandoy, <Virgile.Dandoy@kit.edu>}
%-------------------------------------------

\subsubsection{Introduction}
The QCD axion constitutes one the best motivated extension to the Standard Model of particle physics \cite{Weinberg:1977ma, Wilczek:1977pj,Peccei:1977ur, Peccei:1977hh}. Indeed, in addition to solve the strong CP problem, it gives a generic way to produce a relic abundance of dark matter in the early Universe. If the axion is produced after inflation - so called post inflation scenario - it leads to a particularly rich phenomenology. In particular, a generic feature of this scenario is the appearance of axion miniclusters. Being gravitationaly bound structures of axions, their radius and mass is found to be of the order $R\approx 1 \text{AU} $ and $M \approx 10^{-13} \text{M}_{\odot} $ \cite{Hogan:1988mp}. This work has for purpose to calculate the abundance and survival probability of
axion miniclusters in the galaxy up to today. Indeed, assuming that a significant fraction of dark matter is bound inside such structures would have
dramatic consequences for direct axion dark matter searches (see e.g.,
\cite{ADMX:2019uok,ADMX:2021nhd,MADMAX:2019pub}). Moreover, it is well know that they may offer
additional signatures, such as radio signals
\cite{Tkachev:2014dpa,Hook:2018iia,Edwards:2020afl} or gravitational lensing
\cite{Kolb:1995bu,Fairbairn:2017dmf,Fairbairn:2017sil,Katz:2018zrn,Dai:2019lud,Ellis:2022grh}.\\
So far, the evolution of the miniclusters since their formation in the early Universe until redshift $z\approx 100$ has been conducted numerically. It is of course crucial to extend the evolution up to today and to address the question of the interactions the miniclusters would get in our galaxy.  For this reason, we calculate how an initial population of axion miniclusters in the Milky Way would have survived through the gravitational interactions with the stars in the galaxy.
This has been achieved in three steps:\\
\begin{itemize}
    \item \underline{Construction of the miniclusters}:\\
    In order to understand how the miniclusters survived in the galaxy, we first need to mathematically describe them. To this point, we will use the Schr\"odinger-Poisson system. Indeed, as the escape velocity of the axions is low enough, we could use the non relativistic approximation of the Klein-Gordon equation to describe them. Moreover, since the occupation number inside the minicluster is very high, the classical field approximation could be taken. Here the gravitational potential entering inside the Schr\"odinger equation is created by the axion field itself via the Poisson equation. Finding the wave function of this system is therefore the way to mathematically characterize the minicluster.\\

    \item \underline{Single Interaction}:\\
    With the wave function of the minicluster, one could then understand how a single star would affect a single minicluster. Since we have used the Schr\"odinger equation to define the system, we will see that the way the minicluster would react could be fully described using time dependent perturbation theory. The goal of this part is to obtain its mass and radius variation as a function of the impact parameter, the velocity and the mass of the star.\\

    \item \underline{Simulations }:\\
    Once we know how a single star would affect a single minicluster, we are going to run a Monte Carlo simulation to perturb a sample of miniclusters over their lifetime. This will extract a survival probabily as a function of the mass, radius and location.
\end{itemize}

\subsubsection{Construction of the minicluster}

Because of high occupation number and non relativistic velocity, the axion minicluster will be described through the classical field $\psi$, solution of the Schr\"odinger-Poisson system:
\begin{equation}
 \begin{split}
     &i\partial_t\psi(r) = \left(-\frac{\nabla^2}{2m_a} + m_a\phi(r)\right)\psi(r) ,\\
 & \nabla^2 \phi(r)= 4\pi G m_a|\psi|^2.
 \end{split}   
\end{equation}
In order to solve this system, we define beforehand the profile of the minicluster we would like the function $\psi(r)$ to represent. It particular, we choose and fix the potential $\phi(r)$, the density $\rho(r)$ and the distribution function $f(E)$ of the minicluster such that these last functions are self-consistently related as
\begin{eqnarray}\label{eq:rho}
\rho(r) = 4\pi m_a^2\int_{m_a\phi(r)}^0 dE\, f(E)\, \sqrt{2m_a(E - m_a\phi(r))}\enspace.\label{density_dE}
\end{eqnarray}
We then solve the Schr\"odinger equation with the fixed potential using WKB approximation such that the general solution reads
\begin{equation}\label{eq:psi_expansion}
\psi(\vec{r},t) = \sum_{nlm} C_{nlm} R_{nl}(r)Y_{lm}(\theta,\varphi) e^{-iE_n t}\enspace,
\end{equation}
where $nlm$ represent the quantum numbers, $R_{nl}$ is the radial function and $Y_{ml}$ are the spherical harmonics. All the information about the potential $\phi(r)$ is therefore encoded inside the radial function as
\begin{equation}\label{eq:R}
R_{nl}(r) = \frac{1}{\sqrt{\mathcal{N}_{nl}}}\,\frac{1}{r}\,\frac{1}{\left[2m_a (E_n - V_l(r))\right]^{1/4}}\ 
\sin\left(\int^r dr'  \sqrt{2m_a (E_n - V_l(r'))} + \pi /4\right)\enspace,
\end{equation}
with the effective potential $V_l(r) = \frac{l^2}{2m_a r^2}+m_a \phi(r)$ and $\mathcal{N}_{nl}$ the normalization constant.\\
Finally, the coefficients $C_{nlm}$ have to be found in order to fulfil the Poisson equation with the same fixed potential. Using Eq.\eqref{eq:rho}, we get 
\begin{equation}
%C_{nlm} = 4\pi \sqrt{m_a} \sqrt{f(E_n)\,\N_{nl}\,dE_n}\ e^{i\phi_{nlm}}\enspace.\label{eq:coeff}
C_{nlm} = \sqrt{(2\pi)^3 f(E_n)\, g_l(E_n) \,dE_n}\ e^{i\phi_{nlm}}\enspace,\label{eq:coeff}
\end{equation}
where $f(E)$ is the distribution function of the chosen profile, $g_l(E)$ the density of states for a given angular momentum $l$ and $\phi_{nlm}$ is a random phase uniformly distributed and uncorrelated between the different modes.\\
Such solution $\psi$ reproduces on average over the random phases a self consistent object with density $\rho(r)$, distribution function $ f(E)$ and potential $\phi(r)$.

\subsubsection{Single Encounter}
\begin{figure}[t]
\centering
  \includegraphics[scale=0.35]{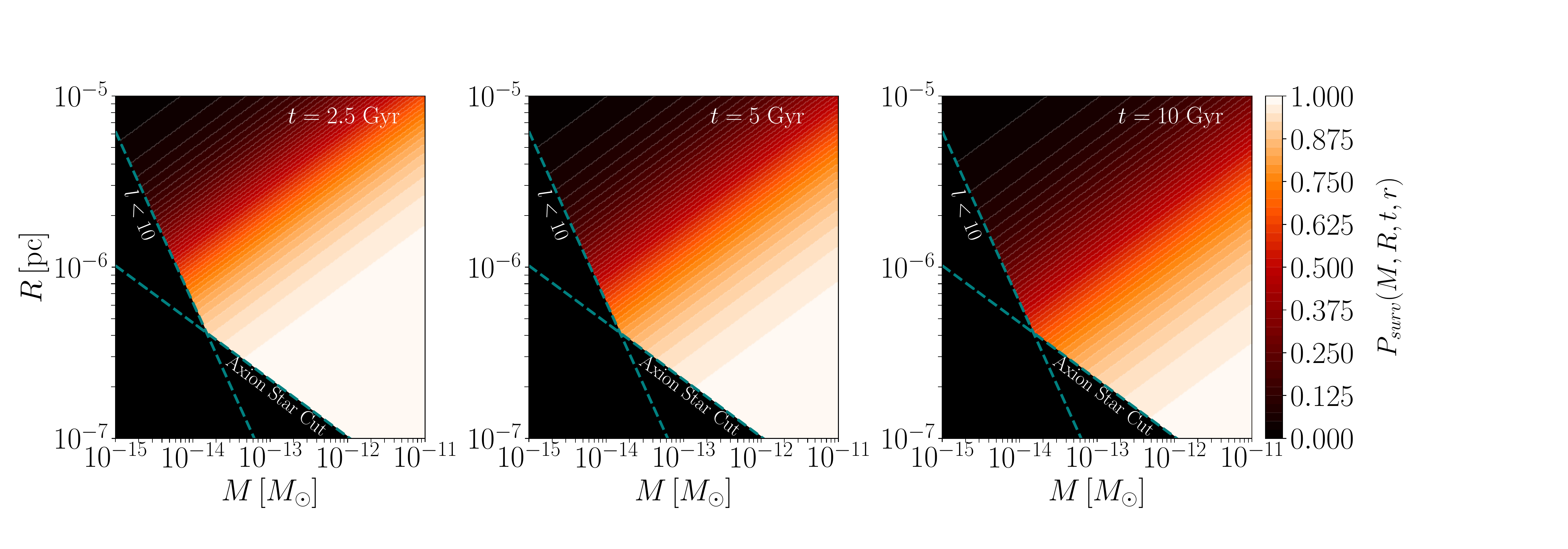}
   \includegraphics[scale=0.35]{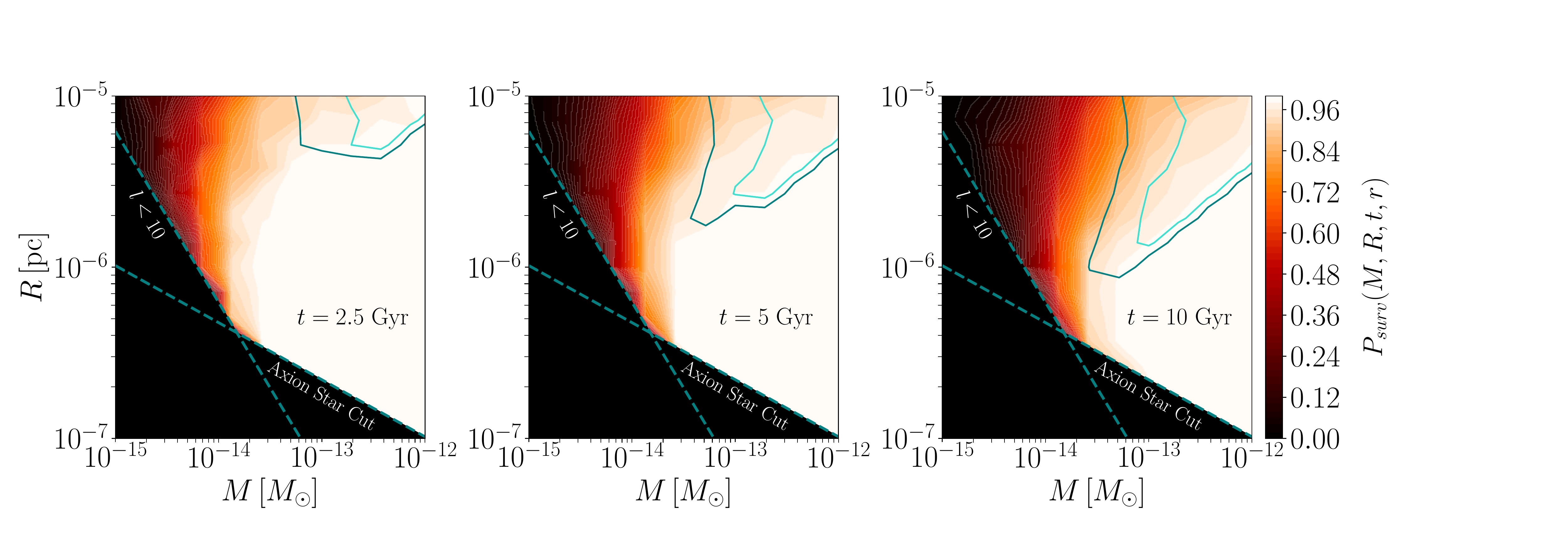}
 \caption{Time evolution of the survival probability $P_{\rm surv}(M,R,t,r)$ for the Lane-Emden \cite{Galactic} (top) and Hernquist \cite{Hernquist:1990be} (bottom) miniclusters  at the sun location $r_\odot\approx 8$ kpc. The survival probability is shown for $t=2.5$~Gyr, 5~Gyr and 10~Gyr (from left to right). The axion star line shows the limit for which the expected soliton inside the minicluster becomes bigger than the cluster itself. If the radius of the minicluster decreases below this line we assume that it becomes a bare axion star. The $l=10$ line shows the limits of our WKB approximation. For the Hernquist miniclusters we have included two contours showing the regions where 40\% (teal) and 80\% (turquoise) of the clusters that did not survive were actually destroyed instead of becoming an axion star.  }
\label{fig:Survival_MR}
\end{figure}
Once a star flies by the cluster, it creates a time dependent tidal distortion that could be expressed as a perturbation on top of the unperturbed Schr\"odinger equation
\begin{equation}
    H_1(\vec{r}, t)= m_a\phi_\mathrm{tidal} = -\frac{GM_* m_a r^2}{(b^2+v^2t^2)^{3/2}} P_2(\cos\gamma(t)),
\end{equation}
with $M_*$ the star mass, $b$ the impact parameter, $v$ the velocity of the star and $\gamma$ the angle between the location inside the cluster $\vec{r}$ and the location of the star $\vec{r}_*$.\\
If the minicluster is initially characterized by the wave function $\psi$ defined as a sum of energy levels as above, this time dependent perturbation will shift all of them by a quantity found to be 
\begin{equation}
    \delta E (E,l)=  \left(\frac{2GM_*}{b^2v}\right)^2 \, \frac{m_a}{4}<{nl}r^2{nl}>.
\end{equation}
Initially all the energy levels have negative energy. Therefore the ones that become positive after the energy shift created by the star will be considered as ejected out of the object. Hence, the variation of the density profile will be given by 
\begin{equation}
\Delta \rho(r) = 4\pi \frac{m_a^2}{r^2}\int^{l_{\rm max}(r)}_0 dl\,l  \int^0_{-\Tilde{E}(l)} dE\, \frac{f(E)}{\sqrt{2m_a\left(E-V_l(r)\right)}} \,,
\end{equation}
where $\Tilde{E}(l)$ is the minimum ejected energy level for a given angular momentum. The corresponding mass variation is given by the integral over the radial coordinate of this last expression. \\
Right after the interaction, the cluster is no longer in virial equilibrium, we therefore assume here that it would relax its radius (either by increasing it or by decreasing it) to reach a new virial configuration (see more details in Ref.\cite{Dandoy:2022prp}).
\subsubsection{Simulations}
Now that we could calculate how a star affect the mass and the radius of a single minicluster, we study their local survival in the galaxy as a function of time. To do so, we populate the galaxy with miniclusters assuming that they are the only dark matter components (this assumption could be easily relaxed to any fraction of dark matter). Hence, the orbital parameters of the clusters are distributed in such a way that they reproduced the dark matter halo of the Milky Way, which is taken as a NFW profile. We then perform a Monte Carlo simulation of the star interactions the clusters would get as a function of time. All the details on the simulation procedure are derived in Ref.\cite{Dandoy:2022prp}. The results are exposed in Fig.\ref{fig:Survival_MR} for two different minicluster profiles and as a function of time. As expected the high density miniclusters are stronger against tidal perturbations and we obtain that an important fraction of miniclusters would survive.\\
Important conclusions have to be addressed following those results. It seems that the density profile of the minicluster matters a lot on the way they would survive to tidal interactions. Intensive studies on the true profile they would get are therefore crucial to understand their survival. Finally, the question of the relaxation of the minicluster after the interaction remains an open question and could potentially only be addressed numerically.

%-------------------------------------------
\subsection{\emph{New ideas:}  Consistent Kinetic Mixing -- {\it P. Foldenauer}}
\label{Foldenauer}
{\it Author: Patrick Foldenauer, <patrick.foldenauer@csic.es>}
%-------------------------------------------

\subsubsection{Introduction}

We have ample evidence for the existence of physics beyond the Standard Model (SM) like the observation of neutrino oscillations and thus the existence of neutrino masses, or the indirect observation of dark matter (DM). These hints have motivated physicists over the past decades to construct theories completing the SM in the ultraviolet (UV) and predicting new physics. Among these are supersymmetric theories, models of grand unification or string theory. A common feature of such UV completions is the prediction of new heavy states with rather sizeable couplings to the SM sector. These new states can generically be tested by going to ever higher energies at particle physics experiments like colliders. However, the experimental landscape of particle physics is much more diverse with a myriad of experiments testing physics at low energies, but ever increasing intensities, like meson factories, beam dump experiments, or astrophysical and cosmological probes. Quite generically, extra Abelian gauge bosons are well-motivated candidates of new particles that have ever smaller couplings with decreasing mass, i.e.~that naturally live at the experimental \textit{sensitivity frontier}.

In the minimal vector portal scenario the associated gauge boson of a new $U(1)_{X}$ is kinetically mixed with the photon of QED via the operator~\cite{Okun:1982xi,Holdom:1985ag}
\begin{equation}\label{eq:a_mix}
    \mathcal{L} \supset - \frac{\epsilon_A}{2} F_{\mu\nu} X^{\mu\nu}\,,
\end{equation}
where $F_{\mu\nu}$ and $X_{\mu\nu}$ denote the QED and $U(1)_{X}$ field strength tensors, respectively. The mixing parameter $\epsilon_A$ is in principle a free parameter of the theory since the above mixing term is a gauge-invariant, renormalisable operator. In many hidden photon models, however, $\epsilon_A$ arises at the loop level from vacuum polarisation diagrams like the one in the right panel of~\ref{fig:b_mixing}, where fermions charged under both $U(1)$ symmetries run in the loop. It is hence justified to assume that in such models the kinetic mixing induced by loops scales as $\epsilon_A\propto g_x/16\pi^2$, where $g_x$ is the $U(1)_{X}$ gauge coupling constant.

The above kinetic mixing term can be diagonalised by a field redefinition of the form
\begin{align}
    A^\mu \to A^\mu - \epsilon_A \, X^\mu \qquad \Rightarrow \qquad  e\,  A_\mu\,  j^\mu_{\rm em} \to e\,  A_\mu\,  j^\mu_{\rm em} - \epsilon_A e\,  X_\mu\,  j^\mu_{\rm em}\,. 
\end{align}
As we see this leads to an interaction of the new $X$ boson with the electromagnetic current $j^\mu_\mathrm{em}$ suppressed by $\epsilon_A$, motivating the name \textit{hidden photon} for the $X$ boson.

The hidden photon can in general also acquire mass. In the most straight-forward scenario the new $U(1)_X$ is Higgsed and thus broken by the vacuum expectation value $f$ of a new scalar $S$,
\begin{align}
    \mathcal{L} = (D_\mu S)^\dagger D^\mu S \supset \frac{g_x^2 f^2}{2} \, X_\mu X^\mu \,.
\end{align}
Thus, the mass of the dark photon, $m_{A'}\propto g_x\, f$, scales proportional to the gauge coupling $g_x$. Hence, the smaller the coupling $g_x$ (and thus the weaker the hidden photon interactions)  the smaller the mass of the boson. Therefore, hidden photons can naturally hide along the sensitivity frontier.

\subsubsection{A closer look at kinetic mixing}

\begin{figure}
    \centering
    \includegraphics[width=.3\textwidth] {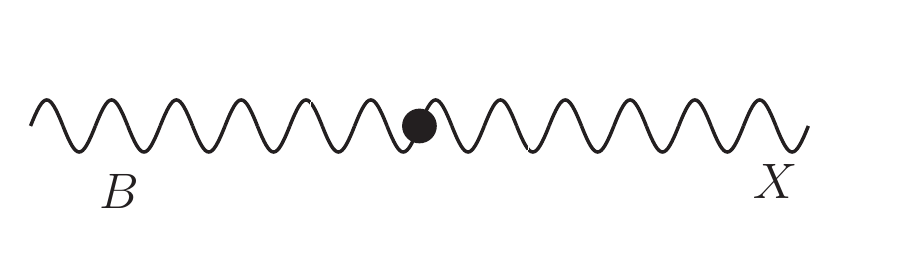}%
    \includegraphics[width=.3\textwidth]{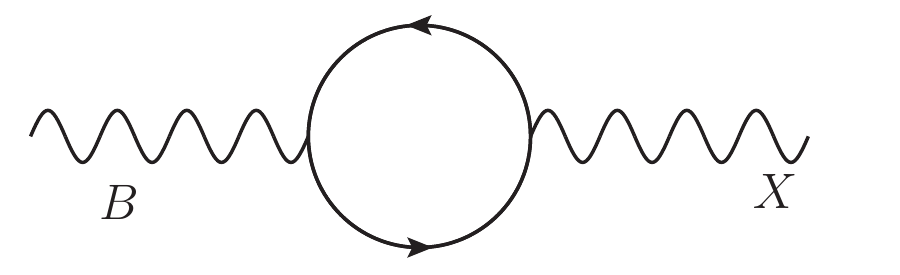}
    \caption{Tree-level (left) and loop-induced (right) mixing diagrams between the hypercharge boson $B_\mu$ and the associated gauge boson of a new $U(1)_{X}$.}
    \label{fig:b_mixing}
\end{figure}

\paragraph{Naive picture}
The kinetic mixing in~\ref{eq:a_mix} of the hidden photon with the SM photon cannot be fundamental since the $U(1)_\mathrm{em}$ only arises after electroweak symmetry breaking (EWSB). In the literature it is hence often assumed that in the UV the hidden photon fundamentally mixes with the field strength tensor $B_{\mu\nu}$ of the hypercharge boson,
\begin{equation}\label{eq:b_mix}
    \mathcal{L} \supset - \frac{\epsilon_B}{2} B_{\mu\nu} X^{\mu\nu}\,.
\end{equation}
This mixing can be either fundamental or arise at the loop-level from fermions charged under both $U(1)_Y$ and $U(1)_{X}$ as illustrated by the diagrams in~\ref{fig:b_mixing}.
Expanding the hypercharge boson into the neutral boson mass eigenstates, $B_\mu = c_w A_\mu - s_w Z_\mu$, with $s_w$ and $c_w$ being the sine and cosine of the Weinberg angle $\theta_W$, the mixing term~\ref{eq:b_mix} decomposes as
\begin{equation}\label{eq:b_dec_mix}
    \mathcal{L} \supset - c_w\frac{\epsilon_B}{2} F_{\mu\nu} X^{\mu\nu}  + s_w\frac{\epsilon_B}{2} Z_{\mu\nu} X^{\mu\nu} \,.
\end{equation}
By comparing~\ref{eq:a_mix} and~\ref{eq:b_dec_mix}, we find the matching relation between the fundamental mixing with the hypercharge boson $\epsilon_B$ and
the mixing in the broken phase with the photon $\epsilon_A$ simply to be
\begin{equation}\label{eq:naive_mix}
    \epsilon_A = c_w\, \epsilon_B\,.
\end{equation}

\paragraph{The full picture}
Treating things more carefully, it turns out that the aforementioned matching between the hypercharge and the photon mixing is not the full picture.
There exists a dimension-six operator that  introduces mixing between the hidden photon and the weak bosons~\cite{Bauer:2022nwt},
\begin{equation}\label{eq:d6_op}
    \mathcal{O}_{WX} = \frac{c_{WX}}{\Lambda^2}\, H^\dagger \sigma^i H \, W^i_{\mu\nu} X^{\mu\nu}\,.
\end{equation}
Here $H$ is the Higgs doublet, $W^i_{\mu\nu}$ the $SU(2)$ field strength tensor, and $\Lambda$ the scale of new physics responsible for generating this operator.
After EWSB this operator induces a mixing term of the form
\begin{equation}
    \mathcal{O}_{WX} \supset - \frac{\epsilon_W}{2} \, W^3_{\mu\nu} X^{\mu\nu}\,,
\end{equation}
where $\epsilon_W = c_{WX}\, v^2/\Lambda^2$.
If we again expand the neutral weak boson into the neutral mass eigenstates as before, $W^3_\mu = s_w A_\mu + c_w Z_\mu$, this translates to a mixing term of,
\begin{equation}
    \mathcal{O}_{WX} \supset - s_w \frac{\epsilon_W}{2} \, F_{\mu\nu} X^{\mu\nu} - c_w \frac{\epsilon_W}{2} \, Z_{\mu\nu} X^{\mu\nu}\,.
\end{equation}
Hence, we have to modify our matching relation to also incorporate the weak mixing contribution,
\begin{equation}\label{eq:full_mix}
    \epsilon_A = c_w\, \epsilon_B + s_w \, \epsilon_W\,.
\end{equation}

In theories where there are $SU(2)$ multiplets charged under the novel $U(1)_{X}$, this operator is necessarily generated at the one-loop level. In this case the loop contribution can be calculated analogously to the standard Abelian mixing case~\cite{Bauer:2022nwt}. We can identify the mixing as the transverse component of the vacuum polarisation amplitude,
\begin{align}
     \Pi^{\mu\nu}_{WX} & = \Pi_{WX}\, [g^{\mu\nu} p_1\cdot p_2  - p_1^\mu p_2^\nu] + \Delta_{WX} \,  g^{\mu\nu} \,,
\end{align}
and compute it as 
\begin{equation}
    {\Pi_{WX} \!=\! - \frac{ g\,g_x}{8\pi^2}\,{\sum_f} \int^1_0\!\! dx \,x(1-x) \,{T^f_3} \,{\big(v^f_{X} + a^f_{X}\big) }\,  {\log\left(\frac{\mu^2}{m_f^2-x(1-x)q^2}\right)}} \,.
\end{equation}
Here the sum runs over all $SU(2)$ degrees of freedom $f$ with $SU(2)$ charge $T^f_3$ and $U(1)_{X}$ axial and vectorial charge $a^f_X$ and $v^f_X$, respectively.

\subsubsection{A concrete example: kinetic mixing in $U(1)_{L_\mu-L_\tau}$}

In general $U(1)_{X}$ extensions of the SM the hidden photon can also interact with SM degrees of freedom via gauge couplings through a current interaction,
\begin{equation}
    \mathcal{L}_\mathrm{int} = - g_x \, j_X^\mu X_\mu\,,
\end{equation}
where the current $j_X^\mu = \sum_\psi q_\psi\, \bar \psi \gamma^\mu \psi$ 
\textit{a priori} can include all SM matter fields, in particular also the $SU(2)$ quark and lepton doublets $\psi= Q, L$. 

If we allow the current $j_X^\mu$ to only include SM fields, the minimally anomaly-free models are composed of $U(1)_{B-L}, U(1)_{L_\mu-L_e}, U(1)_{L_e-L_\tau}, U(1)_{L_\mu-L_\tau}$ and combinations thereof. In all of these models at least (some of) the lepton doublets $L_i$  are charged under the new symmetry and hence $\mathcal{O}_{WX}$ is induced via loops at the renormalizable level ($\Lambda = v$). In order to see how this affects mixing before and after EWSB we will compute the different contributions in the concrete example of $U(1)_{L_\mu-L_\tau}$.

In the broken phase the usual mixing computation in the infrared (IR) at zero momentum transfer, $q=0$, yields 
\begin{align}\label{eq:photon_mix}
    \epsilon_A = \frac{e g_{\mu\tau}}{6 \pi^2} \, \log\left(\frac{m_\mu}{m_\tau}\right)\,, &&  \qquad \qquad \qquad \qquad
    \vcenter{\hbox{\includegraphics[width=.3\textwidth]{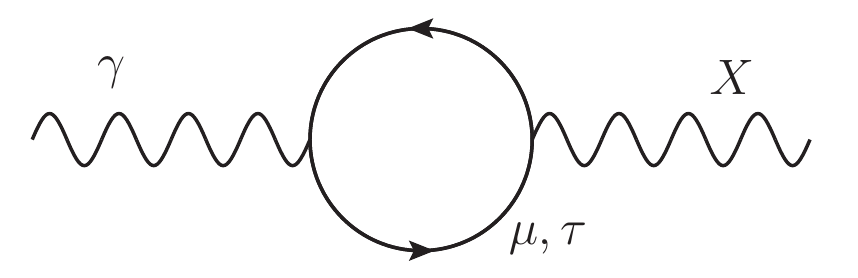}}} \,.
\end{align}

On the other hand, the naive UV computation, where we only take into account mixing with the hypercharge boson, yields a mixing of 
\begin{align}\label{eq:hyper_mix}
    \epsilon_B = \frac{g' g_{\mu\tau}}{24 \pi^2} \, \left[3\log\left(\frac{m_\mu}{m_\tau}\right) + \log\left(\frac{m_{\nu_\mu}}{m_{\nu_\tau}}\right)\right]\,, &&
    \vcenter{\hbox{\includegraphics[width=.3\textwidth]{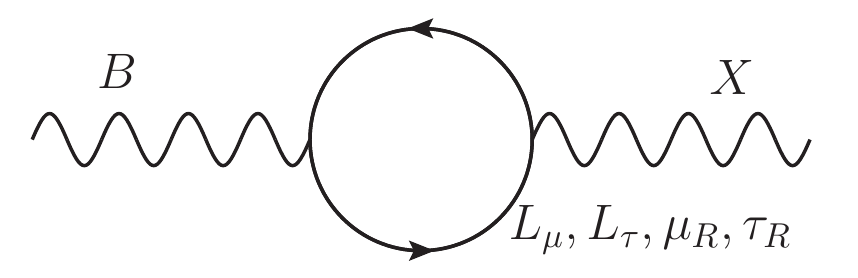}}} \,.
\end{align}
Comparing the two results in~\ref{eq:photon_mix} and ~\ref{eq:hyper_mix}, we see that the naive matching relation~\ref{eq:naive_mix} is manifestly violated and hence cannot be the correct prescription. However, we already know from the previous section how to obtain the correct matching between the mixing in the broken and unbroken phase. We have to take into account also the mixing that is generated between the hidden photon and the neutral $SU(2)$ boson.

Since in $U(1)_{L_\mu-L_\tau}$ the second and third generation lepton doublets are charged under the new symmetry the mixing with the weak boson is computed to be
\begin{align}
    \epsilon_W = \frac{g g_{\mu\tau}}{24 \pi^2} \, \left[\log\left(\frac{m_\mu}{m_\tau}\right) -\log\left(\frac{m_{\nu_\mu}}{m_{\nu_\tau}}\right)\right]\,, &&
    \vcenter{\hbox{\includegraphics[width=.3\textwidth]{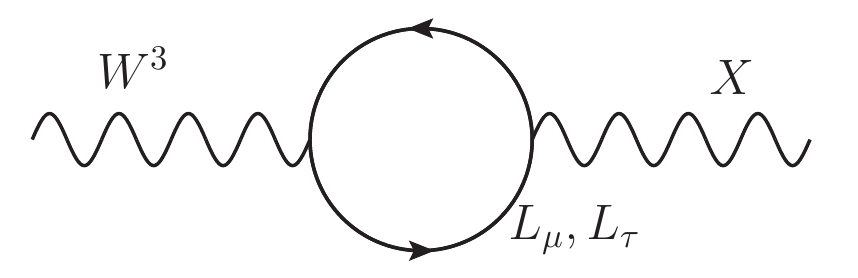}}} \,.
\end{align}
This contributes exactly the missing piece to recover the IR mixing of~\ref{eq:photon_mix} according to the matching prescription in~\ref{eq:full_mix}. In particular, the contributions from the neutrinos to $\epsilon_B$ and $\epsilon_W$ cancel off as they should since neutrinos are electrically neutral and should never contribute to the mixing with the SM QED photon.

\subsubsection{Conclusion}

The hidden photon, or more generally extra Abelian gauge bosons, are well-motivated candidates that could hide along the experimental \textit{sensitivity frontier}. 
The minimal way for these particles to interact with the SM is via kinetic mixing with the photon. In the electroweak symmetric phase this mixing has to proceed with the hypercharge boson $B$. However, there is a dimension-six operator, which can arise at the renormalisable level, that also induces mixing of the hidden photon with the neutral weak boson $W^3$. In models where there are $SU(2)$ multiplets charged under the new $U(1)_{X}$ symmetry this new type of mixing unavoidably arises at the one-loop level and crucially has to be taken into account to obtain the correct matching onto the IR effective theory. More precisely, the mixing with the photon $\epsilon_A$ is obtained from the hypercharge and weak mixing, $\epsilon_B$ and $\epsilon_W$, according to~\ref{eq:full_mix},
\begin{equation*}
    \epsilon_A = c_w\, \epsilon_B + s_w \, \epsilon_W\,.
\end{equation*}
As we have argued, this weak mixing contribution is unavoidable in the phenomenologically interesting class of hidden photon models like $U(1)_{B-L}, U(1)_{L_\mu-L_e}, U(1)_{L_e-L_\tau}, U(1)_{L_\mu-L_\tau}$ and combinations thereof.

%\subsubsection{Acknowledgements}
%PF acknowledges support by the Spanish Agencia Estatal de Investigaci\'on through the grants PID2021-125331NB-I00 and CEX2020-001007-S, funded by MCIN/AEI/10.13039/501100011033.

%-------------------------------------------
\subsection{\emph{New ideas}: Inelastic Dirac Dark Matter -- {\it S. Junius}}
\label{Junius}
{\it Author: Sam Junius, <Sam.Junius@vub.be>}
%-------------------------------------------

\subsubsection{Introduction}
Direct detection has put a lot of pressure on thermal WIMP Dark Matter (DM). Current direct detection searches are extremely sensitive and null results from these experiments hence constrain the Dark Sector (DS) - Standard Model (SM) interactions heavily. An elegant solution to explain these null results are inelastic DM models, where two DS particles couple mainly off-diagonal to the SM sector, preventing elastic scattering processes giving rise to signals in direct detection experiments. In a minimal realisation, dubbed iDM~\cite{Tucker-Smith:2001myb}, the DS consists of a pseudo-Dirac state with two majorana fermions $\chi_1$ and $\chi_2$, interacting with quarks and leptons via an abelian dark force by exchanging a dark photon $A'$ which kinetically mixes with the hypercharge gauge boson. This interaction drives the dark matter freeze-out in the early universe and the relic abundance is set through the co-annihilation process $\chi_1 \chi_2 \rightarrow f f$ into SM fermions $f$. In order to reproduce the DM abundance today, the dark fermion masses have to lie in the MeV-GeV mass range with a relative mass splitting of the order of 10\% or less. The phenomenology of this model has already been studied extensively in the literature, and has become a target model for many particle physics searches. \\
Here, we consider a novel model in the MeV-GeV range, dubbed \emph{inelastic Dirac Dark Matter} (i2DM)~\cite{Filimonova:2022pkj}, where the pseudo-Dirac states in the iDM model are promoted to Dirac states, one being charged and the other one uncharged under a dark $U(1)_D$. The symmetry is spontaneously broken by a Higgs-like mechanism, which causes the two dark fermions to mix.\footnote{Throughout our analysis, we assume that the dark scalar that is responsible for the mixing is much heavier than the other dark particles and does not affect the observables we consider.} The resulting Lagrangian reads
\begin{equation}\label{eq:lagr}
	\mathcal{L} \supset \ e \epsilon A'_{\mu} \sum_f Q_f \bar{f} \gamma^\mu f  - g_D \Big(A'_{\mu} + \epsilon\,\frac{s_W}{c_W} Z_\mu\Big) \Big(\sin^2\theta  J^\mu_1 - \sin\theta \cos\theta J^\mu_{12} + \cos^2\theta J^\mu_2\Big),
\end{equation}
where $f$ denotes the SM fermions with electric charge $Q_f$ in units of the electromagnetic coupling $e$; $g_D$ is the coupling constant of the $U (1)_D$ symmetry; $s_W$ , $c_W$ are the sine and cosine of the weak mixing angle; $\theta$ is the mixing angle between the dark fermions; and $\epsilon$ parametrizes the kinetic mixing.\footnote{In Eq.~\ref{eq:lagr}, we only show the leading terms in $\epsilon$. For higher order terms, see~\cite{Filimonova:2022pkj}.} The dark fermion currents are
\begin{equation}
J_1^\mu  = \bar{\chi}_1 \gamma^\mu \chi_1,\quad J_2^\mu = \bar{\chi}_2 \gamma^\mu \chi_2,\quad J^\mu_{12} = \bar{\chi}_1 \gamma^\mu \chi_2 + h.c.
\end{equation}
As a result, the dark matter candidate $\chi_1$ interacts only feebly if the mixing angle $\theta$  is small, while the coupling of the dark partner $\chi_2$ through the dark gauge force is unsuppressed. This allows for more processes to set the relic abundance compared to iDM, and hence opens up the viable parameter space. Furthermore, the small mass splitting, together with a small mixing angle, will result in a long lifetime of the $\chi_2$ decay into $\chi_1$ and a pair of electrons through an off-shell dark photon. Searches for such long-lived particles have already been performed at collider and fixed-target/beam-dump experiments. We estimated the sensitivity of these searches on i2DM, as well as the potential constraining power of future experiments.\\
The phenomenology of i2DM is described by six independent parameters $\{ m_1, \Delta,m_{A'},\alpha_D,\epsilon,\theta$\}, with $\alpha_D = g_D^2/(4\pi) $. The relative mass splitting between the dark fermions is defined as $\Delta = (m_2 - m_1)/m_1$. Throughout this work we focus on the mass hierarchy $m_{A'} > 2 m_2$, so that decay of the dark photon into dark fermions is kinematically allowed. Within this considered mass hierarchy region, the annihilations of the dark fermions into SM particles through an off-shell dark photon define the freeze-out dynamics of the DM candidate $\chi_1$. As a consequence, for $m_{A'} \gg m_{1,2}$, the scattering rates and decays of dark fermions scale as $y = \epsilon^2\,\alpha_D \left(m_1/m_{A'}\right)^4$. Hence, we will often parameterize the DS-SM interactions by $y$.

\begin{figure}[!t]
    \centering
    {\includegraphics[width=0.4\textwidth]{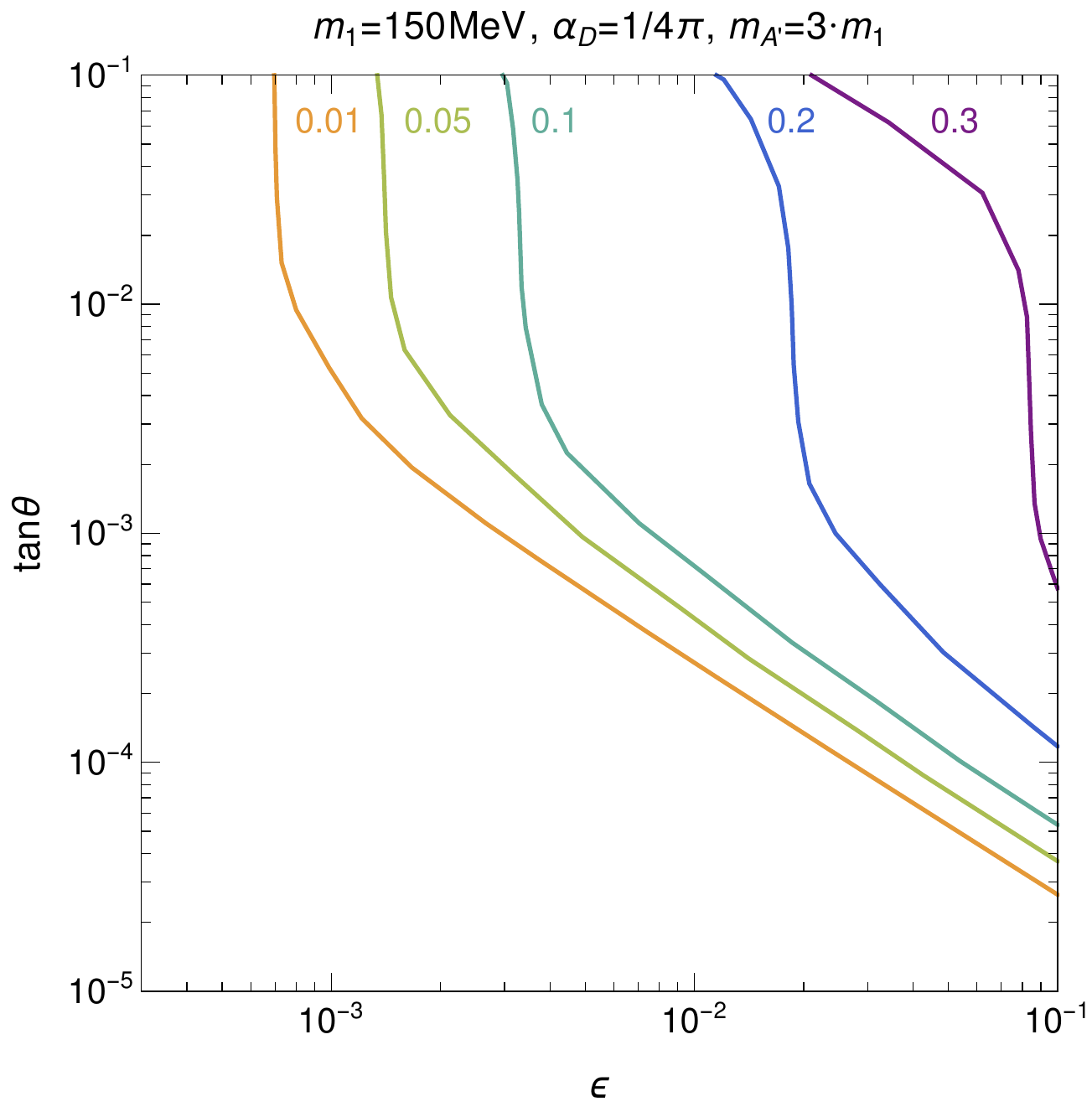}}
    \hspace{0.05\textwidth}
    {\includegraphics[width=0.4\textwidth]{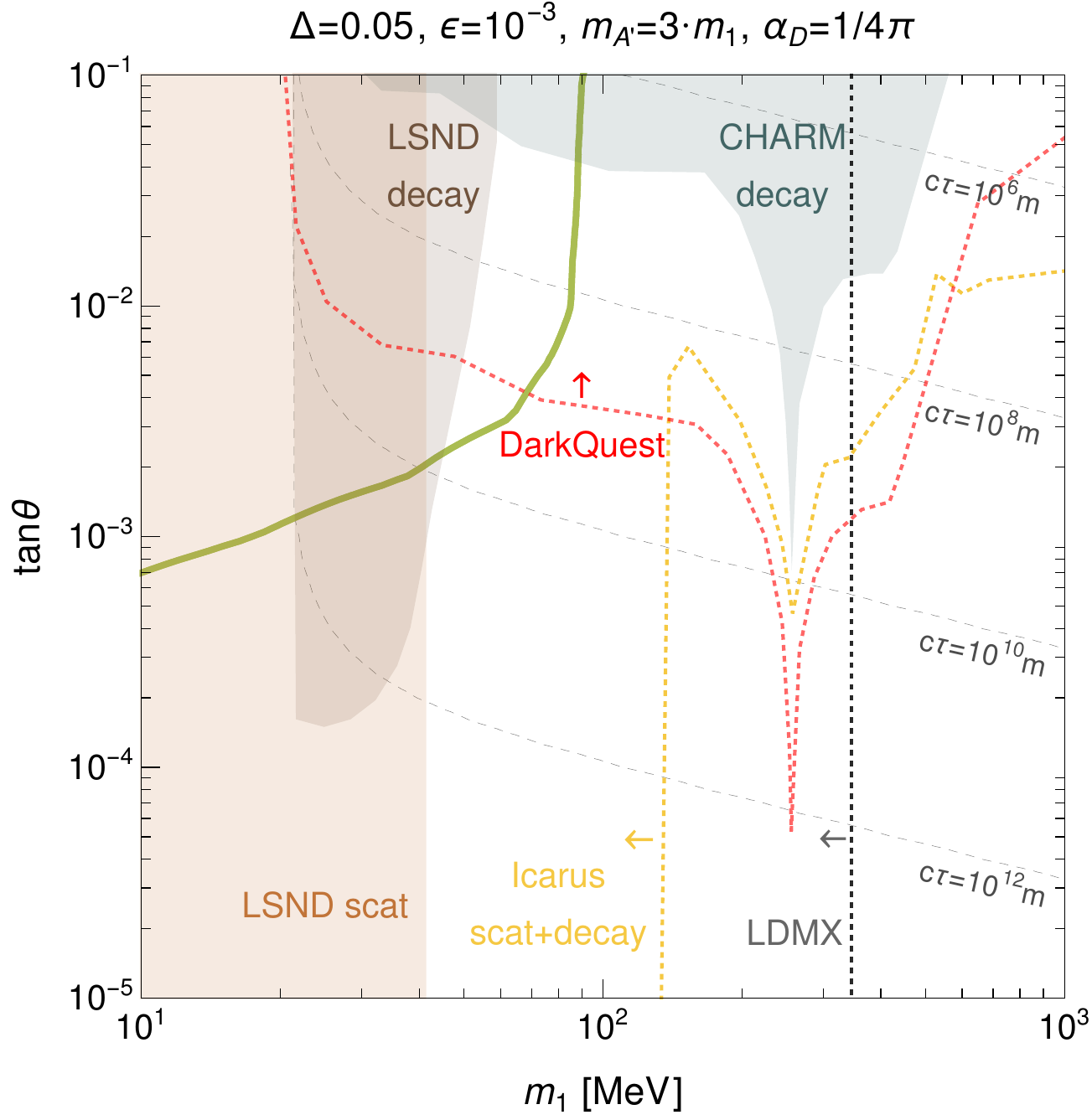}}
    \caption{LEFT: Correct DM relic abundance for i2DM as a function of the dark interaction strength, $y$, and the dark fermion mixing, $\tan \theta$, for a fixed value of the DM mass,  $m_1=150\,$MeV, and varying values of the mass splitting $\Delta = 0.01 \dots 0.3$. RIGHT: Bounds on i2DM from fixed-target experiments as a function of the dark fermion mixing $\theta$ and the dark matter mass $m_1$ for a fixed relative mass splitting $\Delta=0.05$, dark photon mixing $\epsilon = 10^{-3}$, dark photon mass $m_{A'} = 3m_1$ and dark coupling strength $\alpha_D = 1/(4\pi)$. The observed relic abundance is obtained along the green colored contour. The dashed contours indicate the proper decay length of the dark partner. For $m_1 < 2m_e/\Delta$, the dark partner can only decay into photons or neutrinos, resulting in a very long lifetime. We show existing bounds from CHARM (grey areas) and LSND (red/brown areas). The projected sensitivity of future experiments ICARUS (yellow), SeaQuest (red) and LDMX (grey) is illustrated by dotted lines; the arrow indicates the direction in parameter space that will be probed.}
    \label{fig:DMrelic}
\end{figure}

\subsubsection{Dark matter production in the early universe}
Dark matter relics in the MeV-GeV range must be feebly coupled to the thermal bath in order to account for the observed DM abundance, $\Omega h^2 = 0.12$. Moreover, viable scenarios of inelastic dark matter typically require a compressed spectrum of DS particles. Hence, for i2DM, we will always assume $\epsilon, \theta, \Delta  \ll 1$. When the spectrum is compressed, the conversion processes such as $\chi_1 f \rightarrow \chi_2 f$ are efficient in large regions of parameter space so that the dark partner can also play a role in setting the DM relic abundance through co-annihilation processes $\chi_i \chi_j \rightarrow f f$. From Eq.~\ref{eq:lagr}, it is clear that there exists a hierarchy between these processes, depending on the exact value of the mixing angle $\theta$. Hence, the relic abundance can be set by various mechanisms. We identify three different phases of freeze-out, distinguished by the processes that set the dark matter relic abundance:
\begin{itemize}[topsep=2pt,itemsep=0pt]
\item\label{item:coan} {\bf co-annihilation phase}: $\Omega_\chi h^2$ set by $ \chi_1 \chi_2 \leftrightarrow f\bar{f}$ and $\chi_2 \chi_2 \leftrightarrow ff$,
\item\label{item:partn} {\bf partner annihilation phase}: $\Omega_\chi h^2$ set by $ \chi_2 \chi_2 \leftrightarrow ff$,
\item\label{item:conv} {\bf conversion phase}: $\Omega_\chi h^2$ set by $\chi_1 f \leftrightarrow \chi_2 f$ and/or $\chi_2 \leftrightarrow \chi_1 ff$.
\end{itemize}
These three freeze-out phases can be identified in the left panel of Fig.~\ref{fig:DMrelic}, in the $\tan \theta$ vs $\epsilon$ plane. For a fixed DM mass of $m_1=150\,$MeV, we show contours where the observed DM relic abundance can be reproduced for different values of the mass splitting $\Delta$. For the largest values of $\tan \theta$ and $\Delta$ shown in the left panel of Fig.~\ref{fig:DMrelic}, we see that the relic abundance curves bend from a diagonal into a vertical line. The diagonal line can be identified with the co-annihilation phase, while for smaller values of $\tan \theta$, we enter the partner annihilation phase. In this phase, we clearly see that the relic abundance is independent of $\tan \theta$, since the partner annihilation cross section scales with $\cos^4 \theta$, which roughly equals to unity for small values of $\tan \theta$. For even smaller values of the mixing angle, the relic density contours start to deviate again from a vertical line. At this point, conversion processes are not efficient enough around the chemical freeze-out time $\chi_2$. Hence, the evolution of the $\chi_1$ abundance is not tightly tied anymore to the one of $\chi_2$ when chemical equilibrium between the two species is lost, so that its abundance will not be set anymore by $\chi_2$ annihilation, but rather by the conversion processes themselves. Since these processes depend again on $\tan \theta$, the vertical line in the left panel of Fig.~\ref{fig:DMrelic} starts to bend again in the lower regions of the plot.\footnote{In the left panel of Fig.~\ref{fig:DMrelic}, it has been assumed that both dark fermions are kept in kinetic equilibrium. While this is true for $\chi_2$, since $\chi_2$ scattering of SM fermions does not depend on $\tan \theta$, it is not always the case for $\chi_1$. In particular in the conversion phase, the conversion processes are not always efficient in keeping $\chi_1$ in kinetic equilibrium up till freeze-out. However, this does not affect qualitatively the results in Fig.~\ref{fig:DMrelic}, see~\cite{Filimonova:2022pkj}.}

\subsubsection{Probing i2DM at accelorator facilities}
Due to the small mixing angle and small kinetic mixing parameter considered in this work, direct and indirect detection experiments will fall short in probing i2DM. However, even with a relatively small kinetic mixing parameter assumed here, the dark photon can still be copiously produced in collider and fixed-target experiments. The decay of the dark photon to dark fermions is kinematically allowed, making it the main decay channel since the decay to SM fermions is suppressed by $\epsilon^2$. The dark photon will mainly decay to two $\chi_2$ particles as the other decay modes are suppressed by the small mixing angle $\theta$. The presence of $\chi_2$ can then be detected in three main ways. First, if $\chi_2$ decays after traversing the whole detector, it can be observed as missing energy. At BaBar, a search for missing energy in association with a hard initial state photon has been performed~\cite{BaBar:2017tiz}. As $e^+ e^-$ colliders such as BaBar provide a clean environment, it has a great sensitivity to long-lived neutral particles such as $\chi_2$ in the i2DM model (see the right panel of Fig.~\ref{fig:DMrelic} for characteristic lifetimes considered here). Hence, the missing energy search is able to constrain the kinetic mixing parameter governing the production of $\chi_2$ in the $e^+ e^-$ collision: $\epsilon \lesssim 10^{-3}$ for $m_{A'} \lesssim 5$~GeV. In the right panel of Fig.~\ref{fig:DMrelic}, we fix $\epsilon=10^{-3}$ to evade this bound. It has also been shown that Belle II can further improve on these bounds by roughly an order of magnitude~\cite{Duerr:2019dmv}.\\
In contrast, in fixed-target experiments, long-lived neutral particles can be probed using detectors placed several hundreds of meters away from the interaction point, hence being sensitive to long lifetimes for $\chi_2$. The two ways these detectors can detect the presence of $\chi_2$ is through scattering processes of detector material or through its decay. Neutrino experiments such as LSND~\cite{Izaguirre:2017bqb} and CHARM~\cite{Tsai:2019buq} have been searching for these scattering events, and hence place constraints on the i2DM parameter space, see the right panel of Fig.~\ref{fig:DMrelic}. In Ref.~\cite{Izaguirre:2017bqb}, this search for scattering events has been reinterpreted to also probe the decay signature in the iDM model, as both electrons from the decay of $\chi_2$ might not be resolved and give the same response in the detector as a recoiled electron from a scattering process. Similar constraints on i2DM are also shown in the right panel of Fig.~\ref{fig:DMrelic}. As these searches do not exclude the complete viable i2DM parameter space, we also show sensitivity curves from near-future fixed-target experiments such as DarkQuest~\cite{Apyan:2022tsd}, ICARUS~\cite{Batell:2021ooj} and LDMX~\cite{Berlin:2018bsc}. Searches for both scattering, decay and missing energy signals at these experiments can conclusively probe i2DM.

\subsubsection{Conclusion}
We have introduced a novel model for a MeV-GeV scale DS, dubbed \emph{inelastic Dirac Dark Matter} (i2DM). Compared to the widely studied model of inelastic Dark Matter (iDM) with Majorana fermions, i2DM has a different thermal history which we have discussed briefly. In order to reproduce the observed relic abundance, we typically need small values of the mass splitting $\Delta$, the kinetic mixing parameter $\epsilon$ and the mixing angle $\theta$. In this region of parameter space, collider and fixed-target experiments do a very good job in constraining this region. Existing limits from BaBar, LSND and CHARM already put stringent bounds on the i2DM parameters. We further showed that proposed searches/experiments from Belle II, SBN, SeaQuest and LDMX will be able to fully probe the parameter space of i2DM. Hence, fixed-target experiments have been proved to be a very powerful probe for MeV-GeV scale DS models.

%-------------------------------------------
\subsection{\emph{New ideas}: Large Energy Singles as Probe of Dark Matter  -- {\it B. Chauhan}}
\label{Chauhan}
{\it Author: Bhavesh Chauhan, <bhavesh-chauhan@uiowa.edu>}
%-------------------------------------------

\subsubsection{Overview}

The low- and medium-energy neutrinos are usually detected via their charged-current interactions, 
where the final state charged-lepton yields a visible signal. The strong correlation allows for a very 
good reconstruction of incident neutrino energy and direction. The prompt signal is followed 
by a delayed neutron capture which often allows tagging and mitigating certain backgrounds. The 
neutral-current interactions are marred by low detectable-energies, lack of directional information, and 
result in prompt-only signals, i.e., \emph{singles}. However, an advantage of the neutral-current channel is that 
it is sensitive to all flavors of neutrinos; it does not distinguish between flavors. A large 
volume scintillation detector can detect singles from 
atmospheric neutrinos as well as  new physics scenarios. In Ref.~\cite{Chauhan:2021fzu}, we have 
predicted the singles spectrum in 
JUNO~\cite{JUNO:2015zny} and note that for visible energy above 15 MeV, the events would be dominated by neutral 
current interactions of atmospheric neutrinos. If JUNO maintains a database of such Large Energy 
Singles (LES), then one can also look for signatures of new physics scenarios such as boosted dark 
sector particles. 

\subsubsection{Singles in JUNO}

In the absence of new-physics, the LES events in JUNO arise mostly from interactions of atmospheric 
neutrinos, which sets the sensitivity to new physics. The two dominant channels are: elastic scattering with protons ($\nu p$ 
ES) and quasi-elastic-like scattering with carbon ($\nu \rm C$ QEL). The scattering cross section with 
electrons is relatively smaller, and we can safely ignore these interactions. To estimate the event rates, we use 
the flux calculated by Honda et al.~\cite{Honda:2015fha} for $E_\nu>100$\,MeV, and appropriately scaled 
FLUKA flux for $E_\nu<100$\,MeV\,\cite{Battistoni:2005pd}. As the scintillation from protons is nearly 
isotropic, the angular distribution of these events is not measurable and therefore, we only use the 
angle-averaged flux and cross sections. The scintillation signal from elastic scattering and 
``proton-only" knockouts from carbon cannot be distinguished, and only the aggregate is measurable. 

To estimate the contribution of protons to LES, we use the $\nu p$ elastic scattering cross section using the 
Llewellyn-Smith form \cite{LlewellynSmith:1971uhs}. We conservatively assume that the cross section 
uncertainties (mostly arising from 
axial mass parameter and proton strangeness) are $\sim10\%$. We also account for the photosaturation 
losses, i.e., quenching, and provide the event spectrum for visible energy ($E_{\rm vis}$) which is different 
from the kinetic energy of the recoiling proton. 

To estimate the contribution of $\nu \rm C$ QEL to LES, we rely on Ref.\,\cite{Cheng:2020aaw} which reports 
the event rates for various nucleon knockout channels ($1p,~ 1n, ~1p1n,$ $~2p,~ 2n,$ ...), as well as 
the 
recoil 
proton spectrum from the sum of these channels. The event rate in JUNO with at least one proton in the final 
state is found to be $\sim$600 events for 20 kton-yr exposure. To isolate the singles events, we estimate the 
fraction of ``proton-only" knockouts that lacks a final state neutron and find that 
\begin{equation}
	\frac{N_{1p} + N_{2p} + N_{3p} + ... }{ N_{1p} + N_{1p1n} + N_{2p} + N_{1p2n} + N_{2p1n} + ... } 
	\approx 
	0.52,
\end{equation}
which implies that roughly half of the protons do not have a delayed neutron-capture signal. The singles 
spectrum from $\nu \rm C$ QEL interaction, therefore, is approximated by scaling the proton spectrum 
given in Ref. \cite{Cheng:2020aaw} by 0.52. The $E_{\rm vis}$ distribution is obtained by applying the 
effects of quenching. The results depend on choice of Monte Carlo generator and properties of the nuclear 
structure. We use the results from GENIE~\cite{Andreopoulos:2009rq}, and assume a 10\% uncertainty.  

The dominant singles background in JUNO would arise from the cosmogenic isotopes that decay via 
$\beta^\pm$, $\beta^\pm \gamma$, $\beta^\pm p$, and $\beta^\pm \alpha$  channels. We 
scale the experimentally measured yields by KamLAND \cite{KamLAND:2009zwo} where available, and 
for other cosmogenic isotopes, 
we use simulation yields from Ref.\,\cite{JUNO:2015zny}. The cosmogenic isotopes that decay within a few 
seconds of the muon passage can be tagged, and the events can be removed by imposing appropriate spatial 
and temporal cuts \cite{ Li:2015lxa}. To get rid of cosmogenic isotope decays, 
we propose a similar veto --- 
a cylindrical volume around the entire track of the cosmic muon with radius $R_{\rm veto}\sim 3$m for a time 
$\Delta t_{\rm veto}\sim$1.2 s. The fraction of cosmogenic isotopes that decay outside the $\Delta t_{\rm 
veto}$ window cannot be tagged, and constitute the irreducible background. This results in a wall-like 
background below 15 MeV. The other known singles backgrounds include intrinsic radioactivity and reactor 
neutrinos which are negligible for $E_{\rm vis}\geq$10 MeV. Incomplete reconstruction of events, e.g., 
missing one or more final state particles, can also lead to spurious LES events. However, we estimate their 
contribution to be negligibly small. 

\begin{figure}[t]
	\centering
	\includegraphics[width=0.47\textwidth]{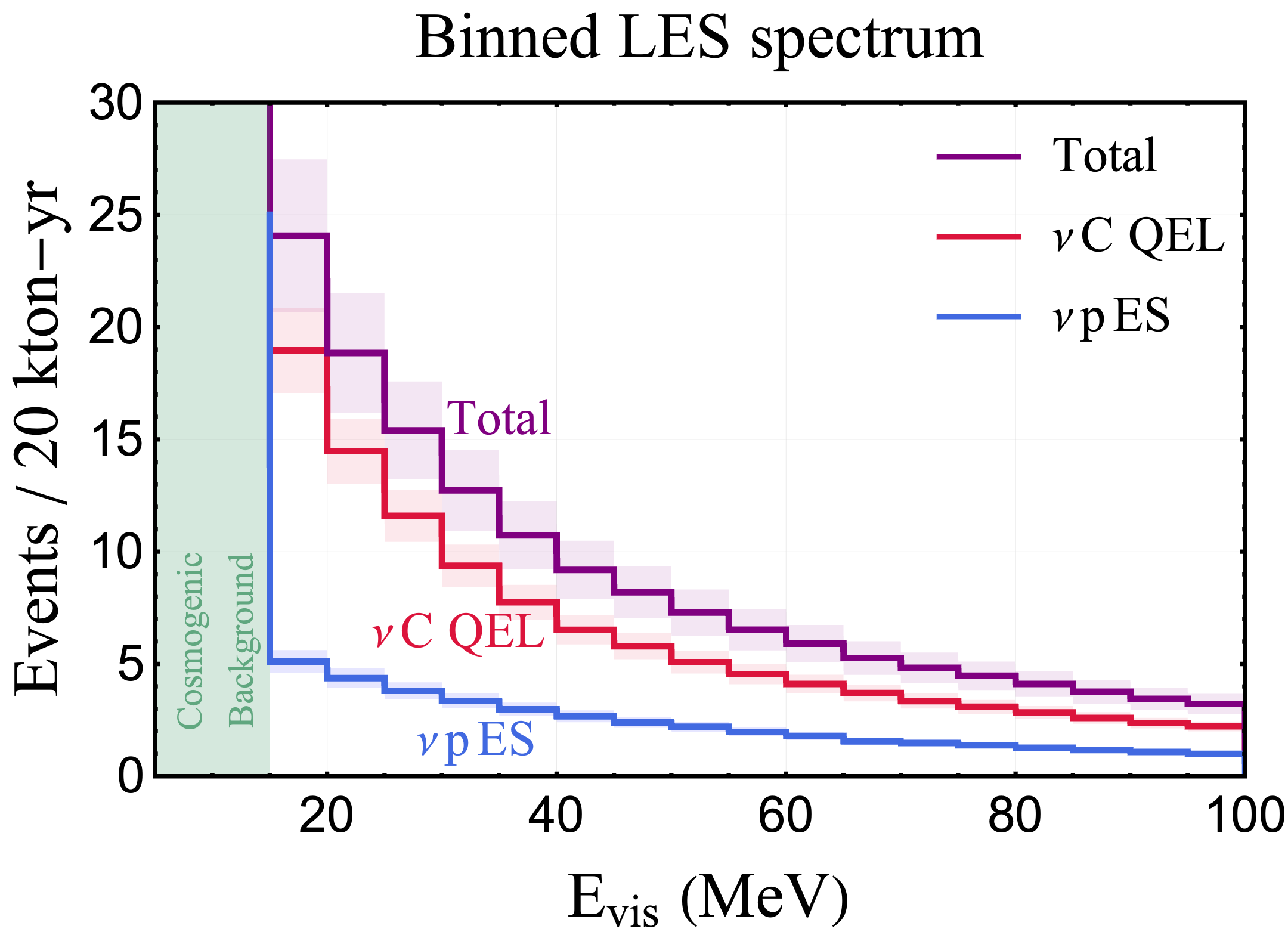}
	\caption{\label{fig:main} The binned singles spectrum in JUNO is shown with contributions 
		from atmospheric neutrino interactions with protons and carbon. The shaded region below 15 MeV 
		indicates wall-like background from decay of cosmogenic isotopes.}
\end{figure}

The binned LES event spectrum for JUNO is presented in Fig. \ref{fig:main}. Above 15 MeV visible 
energy and 20\,kton-yr exposure, we expect $\sim$40 events from $\nu p$\,ES, and $\sim$108 events from 
$\nu {\rm C}$ QEL. This \mbox{JUNO-LES}\,sample, if measured, will provide evidence of neutral-current 
interactions of atmospheric neutrinos. We estimate that JUNO can discover $\nu p $ ES at $3\, \sigma$ ($5\, 
\sigma$) with 12 (34) kton-yr exposure. Measurement of \mbox{JUNO-LES}\,sample will also open the 
window to testing many new physics scenarios. We show two examples in the next section.

\subsubsection{Dark Matter Sensitivity}

In general, an LES event may also arise from particles beyond standard model (e.g., sterile neutrino or 
dark 
matter) that scatter with protons in the detector. Model-independent limits can be obtained on the effective 
parameters, which can be easily translated depending on the new physics scenario. Here, we show two 
examples and compare the sensitivity with results from KamLAND \cite{KamLAND:2011fld}. The projected 
sensitivity of \mbox{JUNO-LES} is $\sim 100$ times more than KamLAND due to larger exposure and a wider 
$E_{\rm vis}$ range of the {\sc LES} sample. 

\begin{figure}[t]
	\centering
	\includegraphics[width=0.45\textwidth]{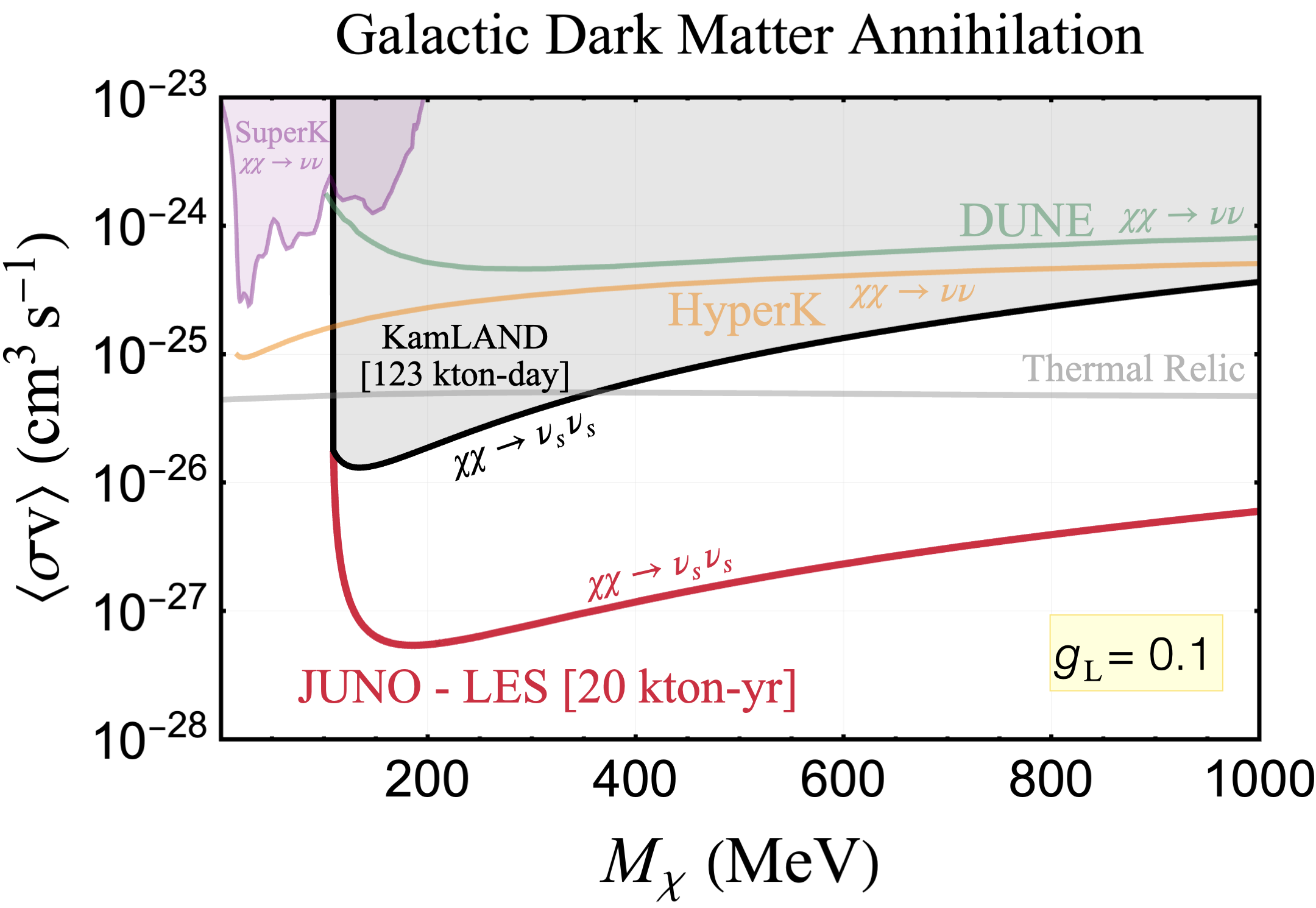}\quad
	\includegraphics[width=0.45\textwidth]{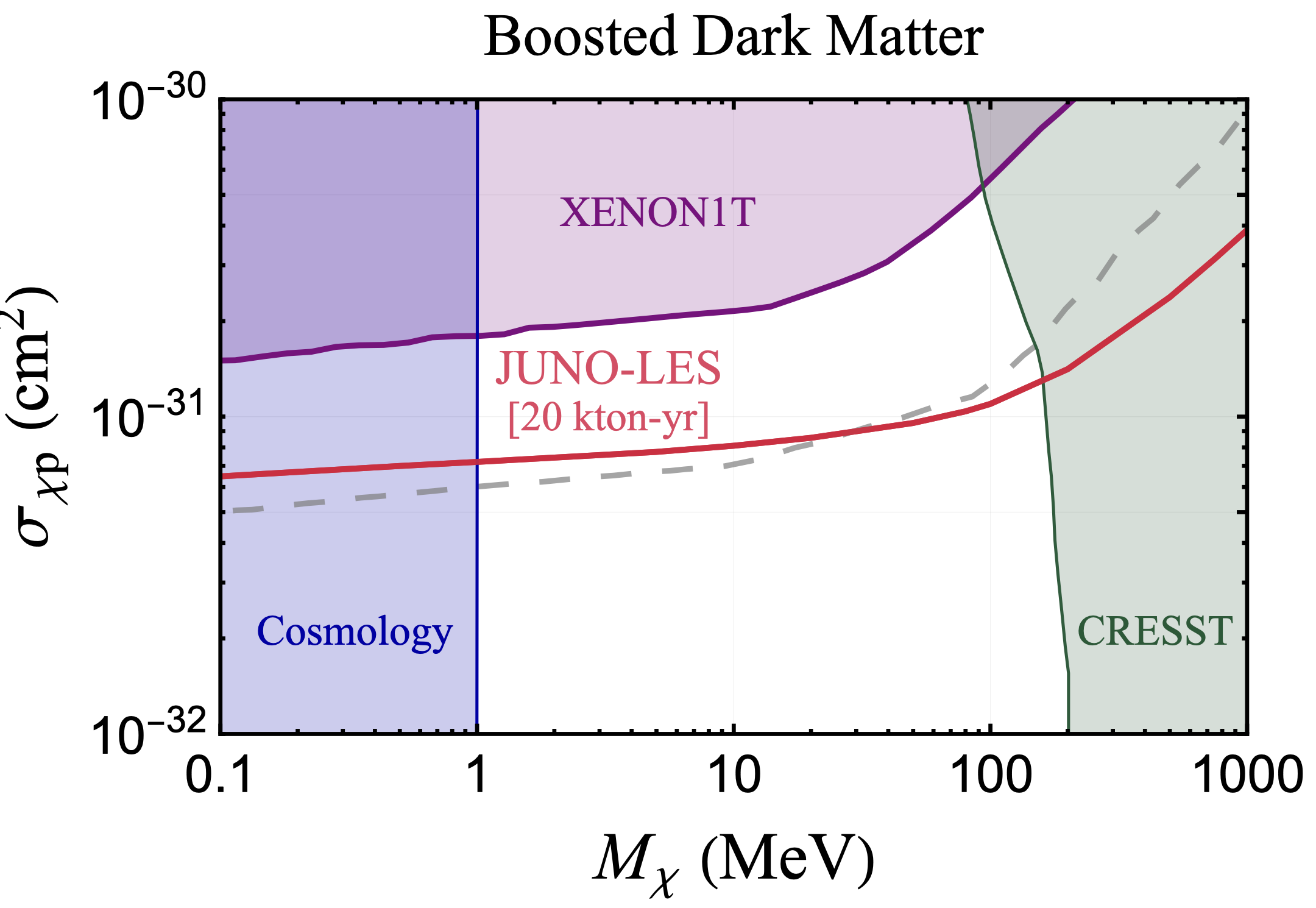}
	\caption{\label{fig:dm} The 90\% C.L. projected sensitivity of JUNO-LES to galactic dark matter 
		annihilation to sterile neutrinos (left) and cosmic ray boosted dark matter (right) is shown along 
		with 
		existing limits from other experiments.}
\end{figure}

Dark matter annihilation to sterile neutrinos is usually untestable if the mixing angle between sterile and 
active 
neutrinos is small. The \mbox{JUNO-LES} sample would be sensitive to the annihilation of galactic dark matter 
to sterile neutrinos through $\nu_s + p \rightarrow \nu_s + p$ (mediated by a new gauge boson with coupling 
strength 0.1). We assume dark matter to be Majorana fermion and ignore the extra-galactic contribution. The 
parameter space excluded by KamLAND data, and the 90\% C.L. discovery sensitivity of the 
\mbox{JUNO-LES} sample with 20 kton-yr exposure are shown in Fig.\,\ref{fig:dm}. We also 
show the thermal averaged cross section for obtaining the correct relic abundance \cite{Steigman:2012nb}, 
and find that \mbox{JUNO-LES} can probe this model for dark matter mass in the range 100 MeV to a few 
GeV. 

The interaction of cosmic rays can \emph{boost} the dark matter particles to higher velocities, 
which allows for larger momentum transfers in detectors 
\cite{Agashe:2014yua,Bringmann:2018cvk,Cappiello:2018hsu}. The 
elastic scattering between boosted dark matter and proton ($\chi + p \rightarrow \chi +p$) in JUNO  will lead 
to a singles event. We follow the method of \cite{Bringmann:2018cvk} to estimate the flux of boosted dark 
matter. We compute the total number of events with $E_{\rm vis}\in$(15, 100) MeV, and obtain the 90\% C.L. 
discovery sensitivity of JUNO by comparing with the total events in the {\sc LES} sample from atmospheric 
neutrinos. The results are shown in Fig.\,\ref{fig:dm} along with other relevant limits. Our projected sensitivity 
is consistent with the ones reported in Ref. \cite{Cappiello:2019qsw} (which approximates the sensitivity by scaling 
the KamLAND background).

\subsubsection{Summary}

We have estimated the visible energy distribution of prompt-only (i.e., \emph{singles}) at JUNO due to 
atmospheric neutrino interactions. We have determined that the background due to cosmogenic 
isotope decay would dominate for $E_{\rm vis} \leq 16.5$ MeV, and the threshold may be reduced to 
$E_{\rm vis} \sim 15$ MeV using a veto around muon track. We propose that 
JUNO can maintain a Large Energy Singles ({\sc LES}) database (i.e., $E_{\rm vis} \geq 15$\,MeV and 
no delayed neutron capture signal) wherein the neutral-current interactions of atmospheric neutrinos 
can be detected.  The {\sc LES} database can also probe new physics scenarios, which admit 
neutral-current-like interactions, for example, in the case of boosted dark sector particles. We have 
estimated the discovery sensitivity for two well-motivated new physics scenarios -- dark matter 
annihilation to sterile neutrinos, and boosted dark matter.

%-------------------------------------------
\subsection{\emph{New ideas}: Probing Axions with Novel Photon Propagation Effects at kpc, Mpc, and Gpc  -- {\it C. Sun}}
\label{ChenSun}
{\it Author: Chen Sun, <chensun@lanl.gov>}
%-------------------------------------------

Axions, as periodic scalar fields, arise ubiquitously in both low-energy phenomenological models~\cite{Peccei:1977hh, Peccei:1977ur, Weinberg:1977ma, Wilczek:1977pj, Kim:1979if, Shifman:1979if, Zhitnitsky:1980tq, Dine:1981rt} and quantum gravity theories~\cite{Svrcek:2006yi}. They serve as an important benchmark of feebly-coupled light particles beyond the Standard Model (SM). In particular, one of the most active experimental and observational targets is the coupling of an axion, $a$, to photons, which takes the form $- \frac{g_{a\gamma\gamma}}{4} a F_{\mu\nu}\tilde{F}^{\mu\nu}$.

In this note, we point out a few novel effects in photon propagation that arise from the axion-photon coupling, 
which can be utilized to probe the axion-photon coupling strength. This includes the photon disappearance due to conversion into axions~\cite{Buen-Abad:2020zbd} and photon en route production by the axions due to their stimulated decay~\cite{Buen-Abad:2021qvj}.

\subsubsection{Photon Disappearance at Gpc and Mpc Scales}
\label{sec:photon-disappearance}

In the presence of external magnetic fields, the axion-photon coupling implies that the propagation eigenstates of the photon-axion system are mixtures of axion and photon states. As a result, there is a non-zero probability $P_0$ that a photon oscillates and converts into an axion while traveling through the magnetic field, effectively resulting in photon number violation. When birefringence and Faraday rotation effects are small, as is the case with propagation in the IGM \cite{Mirizzi:2006zy}, the axion mixes only with the photon polarization parallel to the component of the magnetic field $\mathbf{B}_T$, which is transversal to the direction of motion. In the simple case of photons with energy $\omega$ propagating in a constant and homogeneous magnetic field with $B = \vert \mathbf{B}_T \vert$, the axion-photon conversion probability is given by the well-known formula \cite{Georgi:1983sy,Sikivie:1983ip,Raffelt:1987im,Csaki:2001yk} $P_0 = \frac{(2\Delta_{a\gamma})^2}{k^2} \sin^2 \left  ( \frac{k x}{2} \right ) $, 
where $x$ is the distance traveled by the photon, and
$    k  \equiv  \sqrt{ (2\Delta_{a\gamma})^2 + \left( \Delta_a - \Delta_\gamma \right)^2 }$,  
    $\Delta_{a\gamma}  \equiv  \frac{g_{a\gamma\gamma} B}{2}$,  $\Delta_a \equiv \frac{m_a^2}{2 \omega}$, $\Delta_\gamma \equiv \frac{m_\gamma^2}{2 \omega} \ ,$
in which $m_\gamma^2 \equiv \frac{4 \pi \alpha n_e}{m_e}$ is the effective photon mass squared in the presence of an ionized plasma with an electron number density $n_e$.

When the photons propagate through the intergalactic medium (IGM) or the intracluster medium (ICM) they traverse a large number of magnetic domains. 
We adopt the simple \textit{cell magnetic field} model, first introduced in \cite{Csaki:2001yk} and further developed in \cite{Grossman:2002by,Avgoustidis:2010ju}. In this model the magnetic field is assumed to be split into domains (cells) in which it can be taken to be homogeneous. 
Each {\it i}-th domain has a {\it physical} size $L_i$ and a randomly oriented magnetic field of strength $B_i$~\cite{Grossman:2002by}, whose component perpendicular to the photon's path is the same in each domain. With these simplifications, the resulting net probability of photon-axion conversion over many domains is then given by \cite{Avgoustidis:2010ju} $P_{a\gamma}(y) = (1-A) (  1 - \prod\limits( 1 - \frac{3}{2}P_{0,i} ) ) \ ,$
where $A \equiv \frac{2}{3} \left( 1 + \frac{I_a^0}{I_\gamma^0} \right)$ depends on the ratio of the initial intensities of axions and photons coming from the source, denoted by $I_a^0$ and $I_\gamma^0$ respectively; and $P_{0,i}$ is the conversion probability in the {\it i}-th magnetic domain, which can be obtained from  $P_{0}$ for $x = L_i$. 

\paragraph{Intergalactic medium propagation}
\label{subsec:igm}

At the moment, there is no direct evidence of the IGM magnetic field. Instead there are observational  lower bounds ($\sim 10^{-16}\,\mathrm{G}$ a coherent length above Mpc, more stringent at smaller coherent lengths) and upper bounds ($\sim 10^{-9} \,\mathrm{G}$) on the amplitude of the magnetic field in IGM from CMB anisotropies~\cite{Trivedi_2010, Ade:2015cva, Zucca:2016iur,Paoletti:2018uic}, the non-observation of Faraday rotation of the polarization plane of radio emission from distant quasars~\cite{Durrer:2013pga,Paoletti:2018uic}, and  the non-observation of very high energy $\gamma$-ray cascade emission~\cite{Durrer:2013pga, 2017ARA&A..55..111H, Vachaspati:2020blt}.  We take 1~nG as a convenient benchmark value for the component of the comoving magnetic field perpendicular to the line of sight. We also test the dependence on the comoving coherent length ($s_\mathrm{IGM}$) being $0.1~\mathrm{Mpc}$, $1~\mathrm{Mpc}$, and $10~\mathrm{Mpc}$ in Ref.~\cite{Buen-Abad:2020zbd}.  The bounds we derive on axion coupling from photon propagation in IGM should be understood as an upper bound on $g_{a\gamma\gamma} \times \frac{B_{\rm IGM}}{1\,{\rm nG}}$ for a fixed coherent length. 

Another important quantity of IGM that matters in our analysis is the electron density $n_e$, which determines the plasma photon mass. At low redshifts, most of the baryons are in photoionized diffuse intergalactic gas (Lyman-$\alpha$ forest,) which takes $\gtrsim 90\%$ of the total volume and $28\pm11$\% of the total mass (at $z<0.5$)~\cite{Martizzi:2018iik}.  The average electron density of Lyman-$\alpha$ forest is about $6.5\times 10^{-8}$cm$^{-3}$, assuming its mass fraction to be the central value 28\%. Furthermore, recent simulations show that for diffuse gas, most of the volume is occupied by cosmic voids and sheets, which constitute approximately $\sim 30\%$ and $\sim 40\%$ of the entire volume at $z\lesssim 2$ with mass fractions of $8\%$ and $20\%$, respectively \cite{Martizzi:2018iik}. Based on this, the electron density of the sheet component is about $3\times 10^{-8}$cm$^{-3}$; while that of the void is about $1.6 \times 10^{-8}$cm$^{-3}$ at $z=0$. We take both value and find that they lead to negligible changes in the final results.  The flux $F$ from a source of luminosity $L$ located at redshift $z$ is given by $    F(z) = P_{\gamma\gamma}(z) \frac{L}{4\pi D_L^2(z)} \ ,$ where $P_{\gamma\gamma}$ accounts for the survival probability of photon flux between the observer and the source, and the luminosity distance $D_L$ is  $    D_L(z) = (1+z) \int\limits_0^z \mathrm{d} z' ~ \frac{1}{H(z')} \ .$
The effective apparent magnitude of the SNIa located at redshift $z$ is
\begin{align}
\label{eq:meff}
    m^\mathrm{eff}(z; \boldsymbol{\theta}, M) &  =  M + 25 + 5 \log_{10} \left  ( D_L^\mathrm{eff}(z; \boldsymbol{\theta})/\mathrm{Mpc} \right ) \ , \\ 
    D_L^\mathrm{eff}(z; \boldsymbol{\theta})  & =  D_L(z; \Omega_\Lambda, H_0)/\sqrt{P_{\gamma\gamma}(z; \boldsymbol{\theta})} \ ,
\end{align}
where $\mathbf{\theta}$ is the fitting parameters $\boldsymbol{\theta} = \{ \Omega_\Lambda, H_0, m_a, g_{a\gamma\gamma} \}$. We will take $A$ 
to be 2/3 since the initial axion flux from SNIa is negligible~\cite{Grossman:2002by}. We note that the energy dependence of the photons places negligible constraints in this context.

\paragraph{Intracluster medium propagation}
\label{subsec:icm}

The angular diameter distances to galaxy clusters rely on measurements of the brightness of cluster X-ray, which travel first through the ICM and then the IGM to reach the detector.
Faraday rotation measurements in long wavelengths have shown \cite{deBruyn:2005ze,Taylor:2006ta,Bonafede:2010xg,Feretti:2012vk} that ICM has magnetic fields with a strength of order $\mathcal{O}(\mu\mathrm{G})$. Therefore, a fraction of the X-ray photons could convert into axions inside ICM~\cite{Wouters:2013hua, Berg:2016ese, Marsh:2017yvc,Conlon:2017qcw, Reynolds:2019uqt}. 

We model $n_{e,\mathrm{ICM}}$ with the double-$\beta$ profile \cite{Mohr:1999ya,Bonamente:2005ct}, 
and assume the magnetic field follows a power law on the number density~\cite{Bonafede:2010xg,Feretti:2012vk,Angus:2013sua,Reynolds:2019uqt}.
We use $r_\mathrm{ref}$ for the reference radius from the cluster's center, $B_\mathrm{ref}$ the magnetic field value at that point, and $\eta$ some power, $B_\mathrm{ICM}(r) = B_\mathrm{ref} \left( \frac{n_e(r)}{n_e(r_\mathrm{ref})} \right)^\eta \ .$ We will take the two models of the ICM magnetic field of the Perseus cluster found in \cite{Reynolds:2019uqt} and the one for the magnetic field of the Coma cluster in \cite{Bonafede:2010xg} as benchmarks for our analysis of the ICM effect, $(r_{\rm ref}/\mathrm{kpc}, B_{\rm eff}/\mu G, \eta) \in \{(0, 25, 0.7), (25, 7.5, 0.5), (0, 4.7, 0.5)\},$ denoted Model A, B, and C. 
We exclude  small radii in Model A, $r < 10\;\mathrm{kpc}$
~\cite{Reynolds:2019uqt}.
We take $L_\mathrm{ICM}=6.08~\mathrm{kpc}$ to be the (uniform) size of the magnetic domains, which is the mean of the $L^{-1.2}$ distribution between $3.5-10~\mathrm{kpc}$ proposed in \cite{Reynolds:2019uqt}.
We take the virial radius of the cluster to be $R_\mathrm{vir} = 1.8~\mathrm{Mpc}$, that of the Perseus cluster. 
In treating the orientation of the magnetic field, we assume $B_{\rm ref}$ to be the magnetic field value in the transverse direction, perpendicular to the photon's propagation direction. 
To be extra conservative, we also include a test where we set ICM magnetic fields to zero in regard to potential uncertainties in the ICM magnetic fields~\cite{Reynolds:2019uqt,Libanov:2019fzq}.
Put together, the angular diameter distance, $D_A(z) \propto \frac{\Delta T_\mathrm{SZ}^2}{S_\mathrm{X} (1+z)^4},$ inferred from galaxy clusters X-ray surface brightness and Sunyaev-Zeldovich effect (SZE) will be altered by axion as
\begin{align}
\label{eq:DAeff}
    D_A^\mathrm{eff}(z; \boldsymbol{\theta}) = D_A(z; \Omega_\Lambda, H_0) ~ \frac{P_{\gamma\gamma}^{\rm IGM}(z; \boldsymbol{\theta}, \omega_\mathrm{CMB})^2}{P_{\gamma\gamma}^{\rm IGM}(z; \boldsymbol{\theta}, \omega_\mathrm{X}, A_\mathrm{X}){\langle P_{\gamma\gamma}^\mathrm{ICM} (m_a, g_{a\gamma\gamma}) \rangle}} \ ,
\end{align}
with $D_A(z; \Omega_\Lambda, H_0)$ given by the standard cosmology formula, 
$\left <P_{\gamma\gamma}^{\rm ICM}\right > $ is the line-of-sight averaged photon survival probability weighted by electron number density squared.

\paragraph{Results}
\label{sec:results}

\begin{figure}[t]
  \centering
  \includegraphics[width=0.45\textwidth]{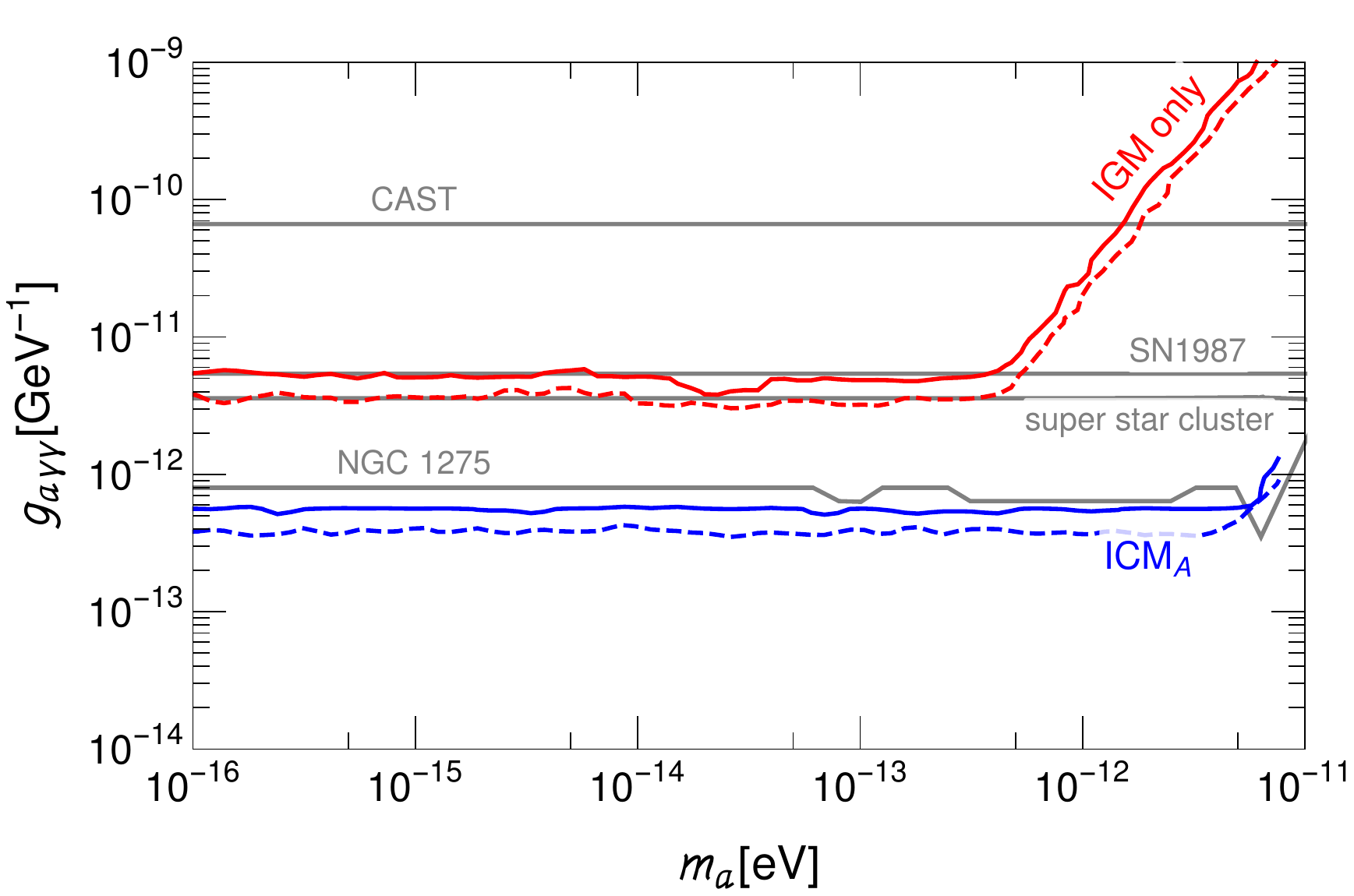}
  \caption{95\% C.L. upper limits on $g_{a\gamma\gamma}$ as a function of $m_a$. The solid curves are from "late data set" while the dashed curves are from "early data set", assuming $B_\mathrm{IGM}=1~\mathrm{nG}$ and $s_\mathrm{IGM} = 1~\mathrm{Mpc}$. To avoid clumsiness, we only show the upper limits from either assuming no ICM conversion effects on the galaxy cluster data (top red curves) or assuming model A of the ICM magnetic profile for the effect (lower blue curves). The upper limits for model B and C are in between them. We also show several existing bounds (grey lines) for comparison: CAST~\cite{Anastassopoulos:2017ftl}; SN1987a~\cite{Payez:2014xsa}; X-ray searches from super star cluster~\cite{Dessert:2020lil} and X-ray spectroscopy from AGN NGC 1275~\cite{Reynolds:2019uqt}.
  \label{fig:photobound}
}
\end{figure}
We constrain the modification of the cosmic distance $D_L$ and $D_A$ using the following combination of cosmological data sets: Pantheon~\cite{Scolnic:2017caz} + Galaxy Clusters~\cite{Bonamente:2005ct} + BAO\cite{Beutler:2011hx,Ross:2014qpa,Alam:2016hwk} + Planck~\cite{Aghanim:2018eyx} prior on the sound horizon baryon drag (denoted "early data set",) and Pantheon + Galaxy Clusters + 19 SNIa used to anchor SH0ES~\cite{Riess:2016jrr} + BAO + TDCOSMO\cite{Birrer:2020tax}, (denoted "late data set".) We find negligible difference between these two combinations, which reflects that the constraining power comes from the cosmic distance's functional shape in the redshift dependence instead its normalization. Therefore, the current ``Hubble tension'' does not affect the constraints. The results of two benchmark models are shown in Fig.~\ref{fig:photobound}.

\subsubsection{Photon Appearance at the kpc Scale}
\label{sec:photon-appearance}

\begin{figure}[h]
  \centering
  \includegraphics[trim={1mm 0 0 1mm},clip,width=0.49\textwidth]{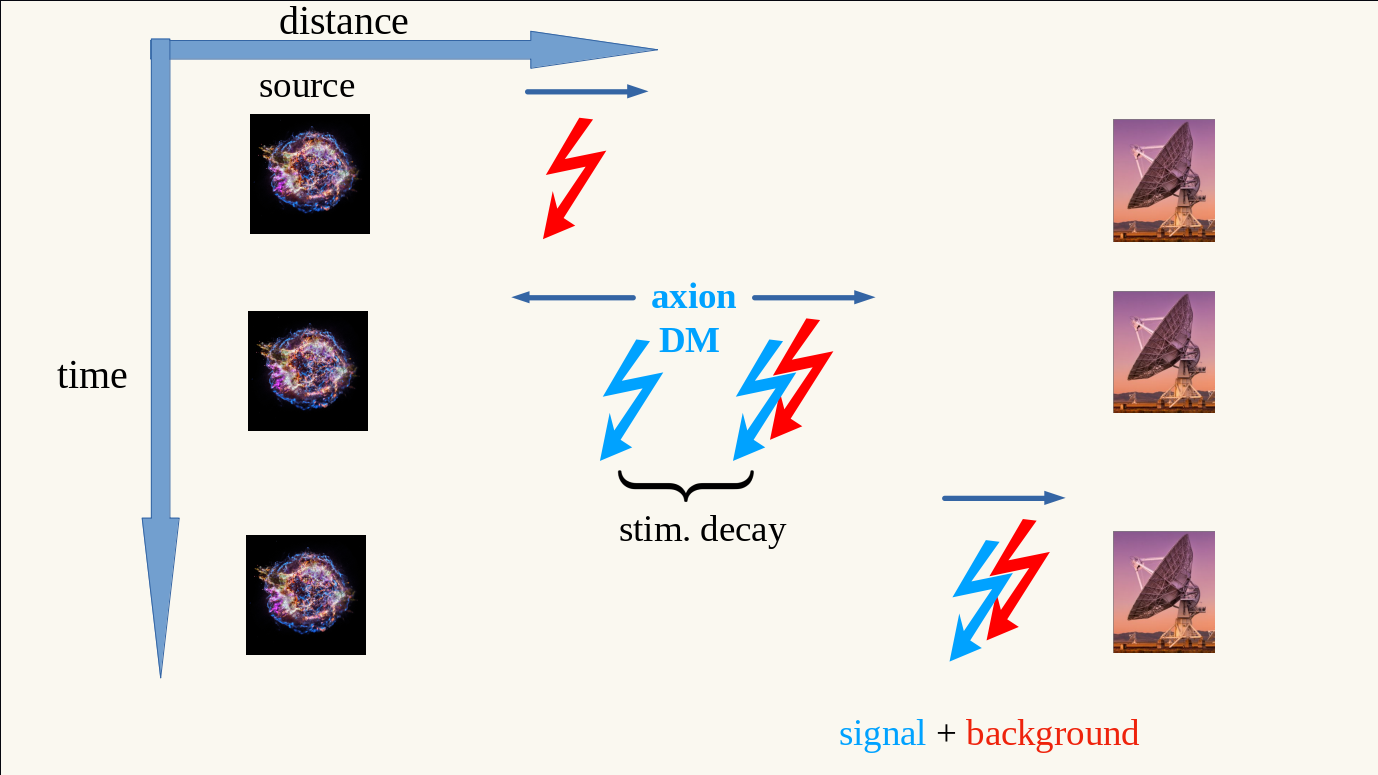}
  \includegraphics[trim={1mm 0 0 1mm},clip,width=0.49\textwidth]{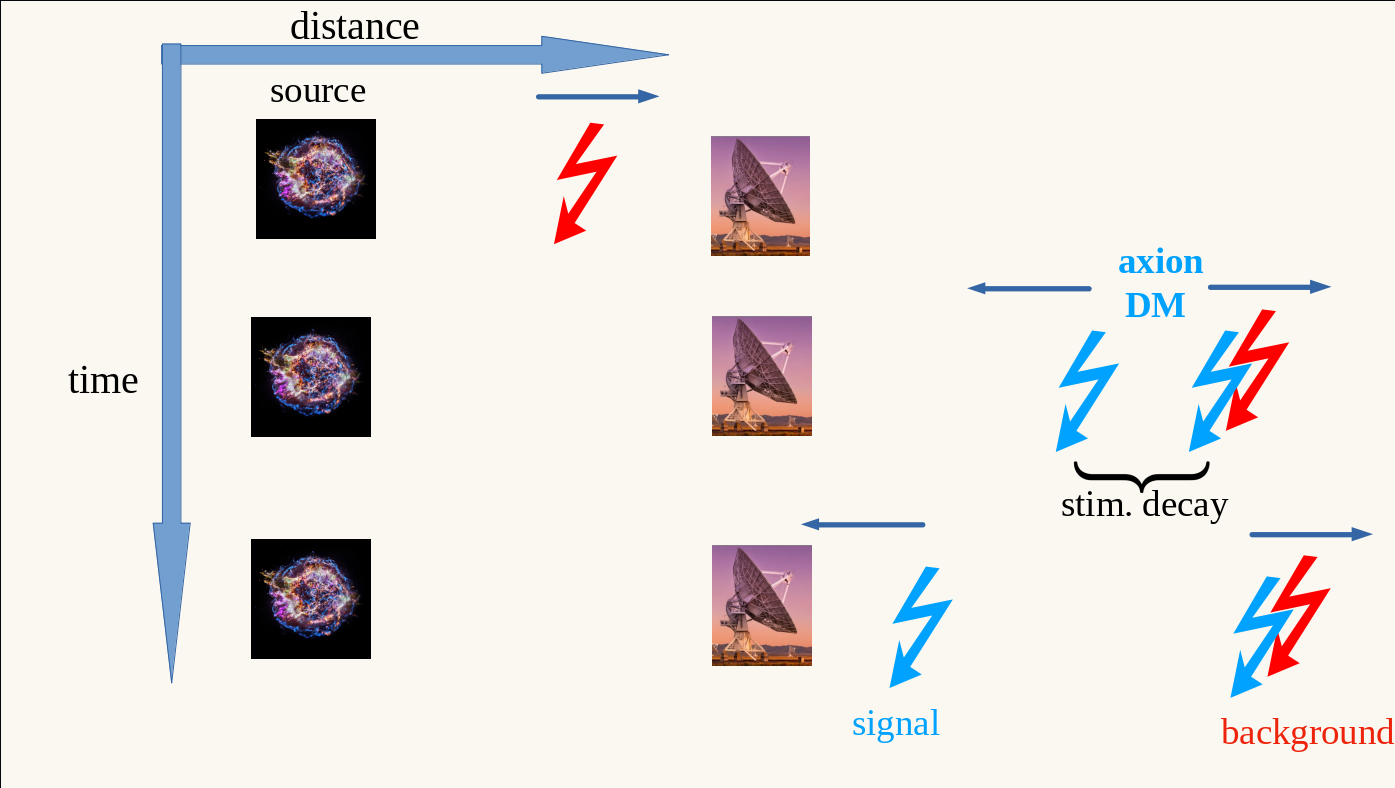}          
  \caption{Illustration of the forward-going signal (left) and the echo signal (right). The signal from axion stimulated decay is shown in blue and the continuum photon source in red.}
  
  \label{fig:illustration}
\end{figure}

In this section we will outline a novel technique to probe the coupling of axion dark matter to photons, which has been underexplored. This section will be mostly based on~\cite{Buen-Abad:2021qvj}. Similar study can be found in~\cite{Sun:2021oqp}. This is the search for an {\it "echo"}\footnote{This has also been called "axion \textit{Gegenschein}" \cite{Ghosh:2020hgd}.} signal from stimulated axion dark matter decays.\footnote{The axion dark matter's decay lifetime is still much longer than the age of the universe so it could still be the dominant component of dark matter.} The basic idea is illustrated in Fig.~\ref{fig:illustration}: 
 photons from a source traverse the axion dark matter halo. If the photon energy matches half of the axion mass, it induces stimulated decays of axion dark matter into two photons. Due to momentum conservation, these two photons travel in opposite directions with energies equal to half of the axion mass, $m_a/2$. An observer could receive two fluxes of photons from opposite directions. One flux is along the line of sight from the source to the observer, which is a superposition of the continuum emission of the source (red) and the line emission from axion decays with a frequency set by $m_a/2$ (blue). Since the line emission from the stimulated decays is a much fainter signal, it is challenging to isolate it from the bright source continuum background. The other flux is the so-called "echo" signal also from stimulated axion decays, which is along the continuation of the line of sight to the direction {\it opposite to the source}. This one could potentially be a clean signal if there is no bright source in the opposite direction.

The echo technique was first applied and developed in Refs.~\cite{Arza:2019nta, Arza:2021nec}, with the source being a powerful radio beam shooting from the Earth. Given the experimental challenges, it will be helpful to identify natural sources in the universe. So far the only astrophysical source that has been examined in the literature to trigger an echo signal is Cygnus A (Cyg~A), an extragalactic active galactic nucleus~\cite{Ghosh:2020hgd}. In~\cite{Buen-Abad:2021qvj} it is pointed out for the first time that the echo signal relies on the \textit{historical luminosity} of the photon source, unlike that the forward-going signals studied in Ref.~\cite{Caputo:2018vmy,Ayad:2020fzc,Chan:2021gjl} require the brightest radio sources {\it today}. This opens up the possibility to probe axion DM with {\it time-varying} radio sources that may be dim today but were once very bright, such as supernova remnants (SNRs).

An SNR, as a time-dependent source, typically starts with a huge photon flux from the explosion of a supernova and undergoes several different phases before eventually merging with the interstellar medium. 
In this case, the echo signal could be induced by the flux from the SNR at time much earlier than that at which the observation is made. 
In other words, the signal is an integration {\it over the history of the SNR}. It turns out that even a dim source today could still lead to an observable echo signal, since it was brighter earlier.

\paragraph{General formalism}
\label{sec:general-formalism}

The flux density of the echo signal $S_{\nu_a,\mathrm{e}}$ observable today is related to the source's flux density $S_{\nu_a,\mathrm{s}}(t)$, which is a function of time, as follows.
\begin{align}
\label{eq:delta_snu_echo}
    S_{\nu_a, \mathrm{e}}(E_\gamma, t) = \frac{\pi^2 \Gamma_a}{E_a^3} \Delta(E_\gamma - E_a) \int\limits_{0}^{t_{\rm age}/2} \! \mathrm{d} x \ \rho_a(x, -\hat{\mathbf{n}}_*) \ S_{\nu_a, \mathrm{s}}(t_{\rm age} - 2x) \ .
\end{align}
Here $\Gamma_a$ is the axion spontaneous decay width, $\nu_a = E_a / (2\pi) = m_a / (4\pi)$, $\rho_a$ the dark matter energy density $\rho_a =n_a m_a$, and unit vector $\hat{\mathbf{n}}_*$ goes from the observer to the source. The function $\Delta$ reduces to the Dirac delta function in the limit of zero signal line width and takes into account the signal width in what follows. The retarded time argument in the source's flux density $S_{\nu_a,\mathrm{s}}(t)$ reflects the fact that the portion of the echo emitted by axion dark matter at a distance $x$ along the line of sight (and being observed today, when the source has an age $t_{\rm age}$) was produced by light that first passed the observer's location a time $2x$ ago. Note also that the axion dark matter density is evaluated along the line of sight in the direction ($-\hat{\mathbf{n}}_*$) opposite to the source ($\hat{\mathbf{n}}_*$), hence the name of \textit{``echo''}. 
We assume that the dark matter is distributed according to the Navarro-Frenk-White profile \cite{Navarro:1995iw}, with a scale radius of $20~\mathrm{kpc}$ \cite{Schaller:2015mua,Calore:2015oya} and a local density of $0.4~\mathrm{GeV} \, \mathrm{cm}^{-3}$ in the solar neighborhood \cite{Sivertsson:2017rkp,Buch:2018qdr}.

In the analysis we relax the Dirac delta function $\delta(E_\gamma - E_a)$ to $\Delta(E_\gamma - E_a)$ by taking into account the finite width of order $\sim \sigma_v E_a$ caused by dark matter velocity dispersion $\sigma_v \approx 5\times 10^{-4} c$~\cite{Freese:2012xd}, as well as the approximately Gaussian shape of the signal with standard deviation $\sigma_v E_a$. 
We take the optimal $\Delta \nu$  that maximizes the ratio of the signal to the noise, with the latter scaling as $\propto \sqrt{\Delta \nu}$. Therefore, approximating the signal distribution as a Gaussian function, we find that the optimal numbers are \cite{Ghosh:2020hgd} $\Delta \nu  \approx 2.8 \, \nu_a \sigma_v$ with 84\% of the Gaussian volume within $\Delta \nu$.  We show in~\cite{Buen-Abad:2021qvj} that the 
the optical depth is negligible for the entire range of frequency we investigate, except for the echo signal coming straight from the galactic center, which we do not consider due to the background. 

We also point out that the motion of the SNR source and dark matter leads to a negligible signal reduction (unless the SNR is older than $\sim 10^4$ years.) In addition, the motion of the Earth with respect to the dark matter rest frame renders the impact of the shadow of the Earth negligible.

\paragraph{Modeling of the Supernovae Remnants}
\label{sec:modeling-snr}
In the time domain, the SNR experiences four phases: free expansion~\cite{1986ApJ...301..790W,Bietenholz:2020yvw,1999ApJS..120..299T,Draine2011jt,2013tra..book.....W}, adiabatic expansion (or Sedov-Taylor)~\cite{2013tra..book.....W,Draine2011jt}, snow plough (or radiative phase)~\cite{Pavlovic:2012az,Draine2011jt}, and dispersion phase. We only study the signal contributed by the first two phases of SNR. We model the free phase luminosity as $L_{\nu, \mathrm{free}}(t) \equiv L_{\nu, \mathrm{pk}} ~ e^{\frac{3}{2}\left(1- {t_\mathrm{pk}}/{t}\right)} \left( \frac{t}{t_\mathrm{pk}} \right)^{-1.5}$~\cite{Bietenholz:2020yvw}, and adiabatic phase $L_{\nu, \mathrm{ad}}(t) \equiv L_{\nu, {\mathrm{tran}}} ~ \left( \frac{t}{t_{\mathrm{tran}}} \right)^{-\gamma}$. The two phases are matched with each other at $t_{\rm tran}$. In Ref.~\cite{Bietenholz:2020yvw} a log normal distribution of $L_{\nu,\rm pk}$ and $t_{\rm pk}$ is derived from about 300 recent supernovae in the radio frequency 2-10\,\textrm{GHz}.
Thus, the light curve of an SNR during its first two phases can be described entirely by anchoring the light curve with early luminosity 
$\{ L_{\nu, \mathrm{pk}}; t_{\mathrm{pk}}; t_{{\mathrm{tran}}}; t_{\rm age}; \gamma(\alpha) \}$, anchoring with late luminosity, $\{L_{\nu, 0}; t_{\mathrm{pk}}; t_{\mathrm{tran}}; t_{\rm age}; \gamma(\alpha) \}$, or any other combination such as luminosity at early and late times with the age being inferred, $\{ L_{\nu, \mathrm{pk}}; t_{\mathrm{pk}}; t_{{\mathrm{tran}}}; L_{\nu, 0}; \gamma(\alpha) \}$.

In the frequency domain, the photon spectrum of a SNR follows a simple power law $  S_\nu  \propto E_\gamma^{-\alpha}$,
where $\alpha$ is the spectral index (related to $\gamma$ by $\gamma  = \frac{4}{5} (2\alpha+1) > 0 $,) and $E_\gamma$ is the photon energy. We further choose the reference (or ``pivot'') frequency to be $1\;\mathrm{GHz}$, following the convention in the literature. The spectral luminosity is then given by $  L_\nu =  S_\nu \; (4\pi D^2)
  = L_{1 {\mathrm{GHz}}} \left( \frac{\nu}{1~{\mathrm{GHz}}} \right)^{-\alpha} $, where $D$ is the distance of the source, and spherical symmetry of the source is assumed. From the Green catalog of SNRs~\cite{Green:2005yt,Green:2014cea,Green:2015isa,Green:2019mta}, one could see that spectral indices are centered around $\alpha \approx 0.5$ with considerable scattering.

\paragraph{Sensitivity from Square Kilometer Array}
\label{sec:results-from-ska}

\begin{figure}[th]
  \centering
\raisebox{0.04\height}{\includegraphics[width=0.41\textwidth]{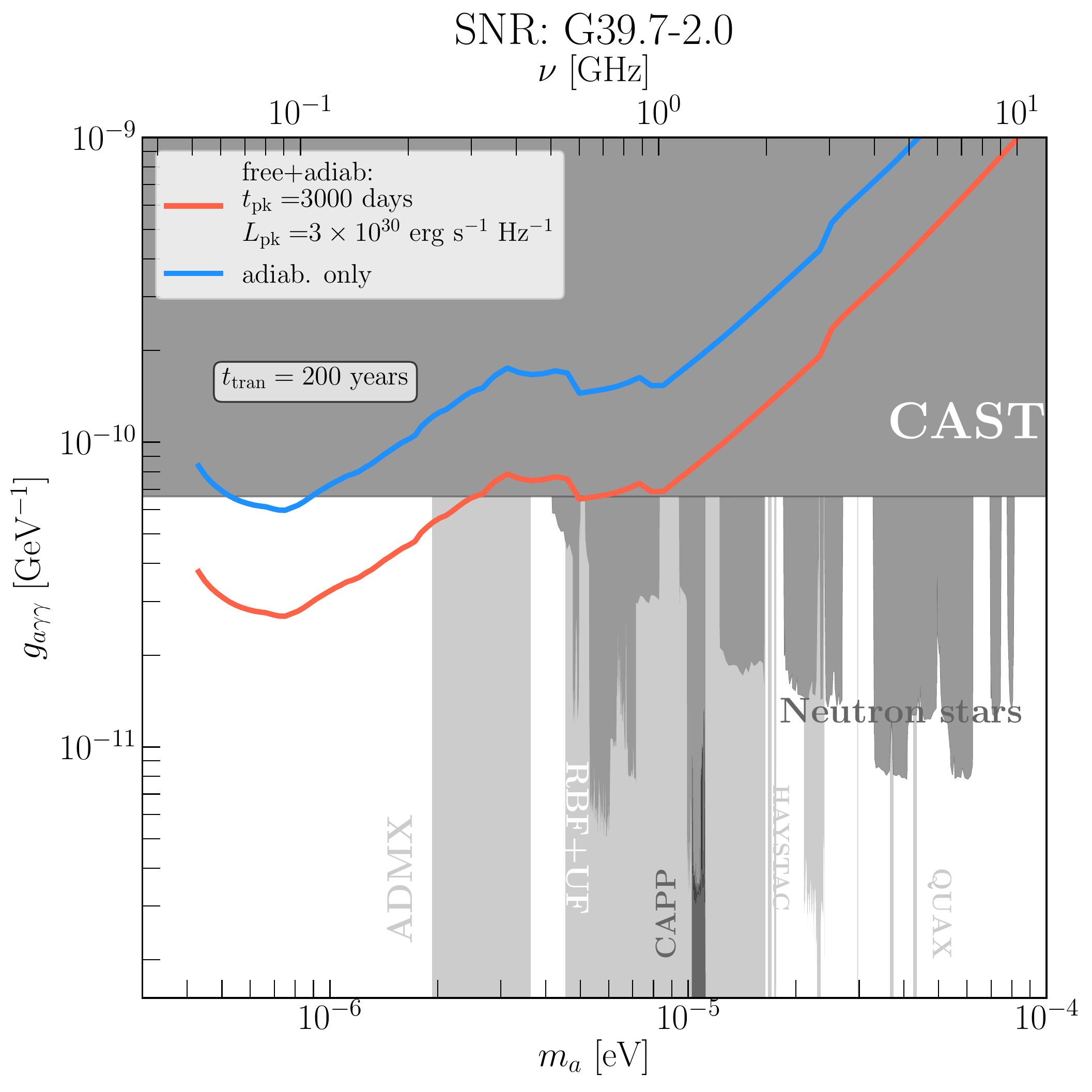}}
  \includegraphics[width=.44\textwidth]{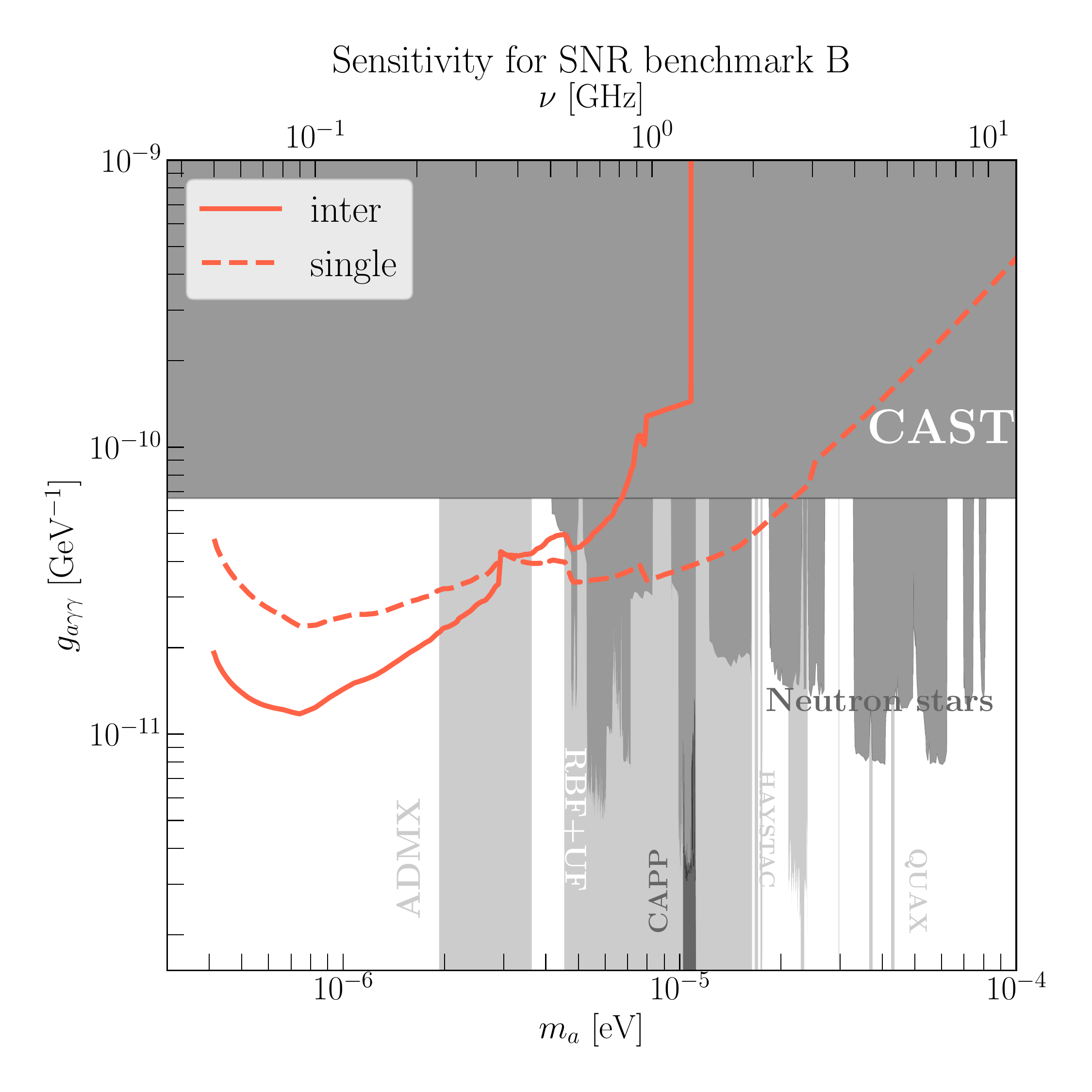}  
  \caption{\textbf{Left}: the sensitivity reach for axion dark matter coupling $g_{a\gamma\gamma}$ of SKA1, for the echo produced by SNR G39.7-2.0. We take a signal-to-noise ratio of $\mathrm{s/n}=1$. SKA1-low and SKA1-mid, in both their single dish and interferometer modalities, are combined into a single curve. We assume a typical value of $t_{\rm tran} = 200$ years. The reach for the conservative ``adiabatic-only'' case is shown in blue. The reach for the ``free+adiabatic'' case is plotted in red. For the latter case, we further assume $t_{\rm pk} = 3,000~\mathrm{days}$, which yields $L_{\rm 1GHz , \mathrm{pk}} = 3\times10^{30}~\mathrm{erg \ s^{-1} \ Hz^{-1}}$.
    \textbf{Right}: The echo signal sensitivity from an unobserved SNR that has the same luminosity today as SNR G6.4-0.1 in GC~\cite{Green:2014cea,Green:2019mta} but in a different sky location where the echo signal can be observed by SKA1.
The grey regions are existing bounds from previous literature: CAST \cite{CAST:2007jps,CAST:2017uph}, ADMX \cite{ADMX:2009iij,ADMX:2018gho,ADMX:2019uok,ADMX:2021nhd,ADMX:2018ogs,Bartram:2021ysp,Crisosto:2019fcj}, RBF+UF \cite{PhysRevLett.59.839,PhysRevD.42.1297}, CAPP \cite{Lee:2020cfj,Jeong:2020cwz,CAPP:2020utb}, HAYSTAC \cite{HAYSTAC:2020kwv}, QUAX \cite{Alesini:2019ajt,Alesini:2020vny}, and neutron stars \cite{Foster:2020pgt,Darling:2020uyo,Battye:2021yue}. These bounds are taken from \cite{AxionLimits}.}

  \label{fig:gc_reach}
\end{figure}

After properly modeling the Square Kilometer Array phase 1 (SKA1) with the aforementioned frequency cut and a benchmark observation time of 100 hours,\footnote{We note that the signal-to-noise ratio scales as $t_{\rm obs}^{-1/2}$ with $t_{\rm obs}$ the observation time. Therefore, the bounds on $g_{a\gamma}$ scales as $t_{\rm obs}^{-1/4}$.} we derive the following sensitivity shown in Fig.~\ref{fig:gc_reach}. It is observed that at low frequency the interferometer mode leads to sensitivity of parameter space not covered by the existing experiments. This serves as a proof of concept on novel radio signals induced by axion dark matter, which can potentially probe interesting unexplored parameter space.

%\subsubsection{Acknowledgements}
%CS is supported by LDRD at Los Alamos National Laboratory. The work is published under report number LA-UR-23-21432.

%-------------------------------------------
\subsection{Conclusions}
\label{ssec:ULF_conclusions}
%-------------------------------------------

% !TEX root = /Users/mauriziogiannotti/Downloads/FIPs2022_report-2/main.tex

\begin{figure*}[ht]
\centering
\includegraphics[width=\linewidth]{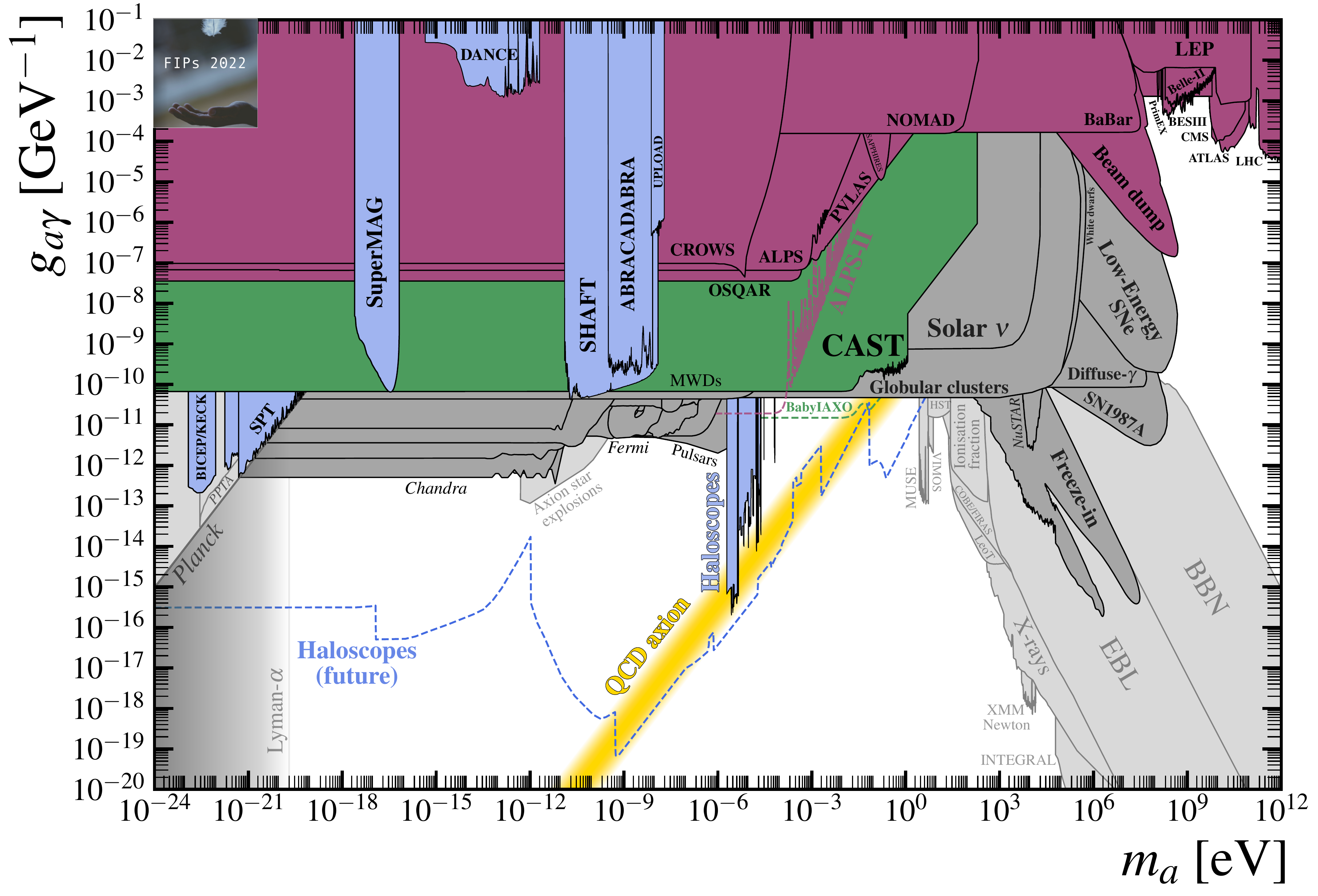}
\caption{ %\normalsize 
Summary of current bounds and prospects of ongoing and future searches for the axion-photon coupling. 
Regions shaded in colours correspond to experiments: mauve for laboratory~\cite{Ehret:2010mh,Betz:2013dza,OSQAR:2015qdv,DellaValle:2015xxa,SAPPHIRES:2021vkz,SAPPHIRES:2022bqg,Ishibashi:2023bae} and collider~\cite{NOMAD:2000usb,ATLAS:2020hii,Dolan:2017osp,CHARM:1985anb,Riordan:1987aw,Dolan:2017osp,Blumlein:1990ay,NA64:2020qwq,Belle-II:2020jti,CMS:2018erd,Jaeckel:2015jla,Knapen:2016moh,Knapen:2016moh,PrimEx:2010fvg,Aloni:2019ruo} searches for axions, green for searches for solar axions~\cite{CAST:2007jps,CAST:2017uph}, and blue for direct dark matter detection experiments~\cite{Ouellet:2018beu,Salemi:2021gck,ADMX:2018gho,ADMX:2019uok,ADMX:2021nhd,Crisosto:2019fcj,Lee:2020cfj,Jeong:2020cwz,CAPP:2020utb,Lee:2022mnc,Kim:2022hmg,Yi:2022fmn,Adair:2022rtw,Oshima:2023csb,Devlin:2021fpq,Grenet:2021vbb,HAYSTAC:2018rwy,HAYSTAC:2020kwv,HAYSTAC:2023cam,McAllister:2017lkb,Quiskamp:2022pks,Alesini:2019ajt,Alesini:2020vny,Alesini:2022lnp,CAST:2020rlf,Gramolin:2020ict,TASEH:2022vvu,Arza:2021ekq,Thomson:2019aht,Thomson:2023moc}. All astrophysical bounds are shown in grey, with searches that do not depend on the axions being dark matter shown in dark grey~\cite{Xiao:2020pra,Wouters:2013hua,Marsh:2017yvc,Reynolds:2019uqt,Reynes:2021bpe,Caputo:2022mah,Calore:2021hhn,Fermi-LAT:2016nkz,Meyer:2020vzy,Davies:2022wvj,Dolan:2022kul,Jacobsen:2022swa,HESS:2013udx,Calore:2022pks,Dessert:2021bkv,Li:2020pcn,Li:2021gxs,Noordhuis:2022ljw,Jaeckel:2017tud,Hoof:2022xbe,Payez:2014xsa,Caputo:2021rux,DeRocco:2022jyq,Dessert:2020lil,Dolan:2021rya,Langhoff:2022bij}, and those that do in a lighter grey~\cite{Foster:2022fxn,Escudero:2023vgv,BICEPKeck:2021sbt,Bolliet:2020ofj,Bernal:2022xyi,Wadekar:2021qae,Ivanov:2018byi,Fedderke:2019ajk,Castillo:2022zfl,SPT-3G:2022ods,Regis:2020fhw,Grin:2006aw,Nakayama:2022jza,Carenza:2023qxh,Foster:2021ngm,Cadamuro:2011fd,Depta:2020wmr}. Dashed lines show the prospects for ALPS-II~\cite{Ortiz:2020tgs}, BabyIAXO~\cite{IAXO:2019mpb}, and a summary of all proposed future haloscopes (see next figure for a closeup). Figure produced using resources and data found in Ref.~\cite{AxionLimits}. 
}
\label{Fig:ULF_Axion-Photon_Summary_Plot}
\end{figure*}

\begin{figure*}[ht]
\centering
\includegraphics[width=\linewidth]{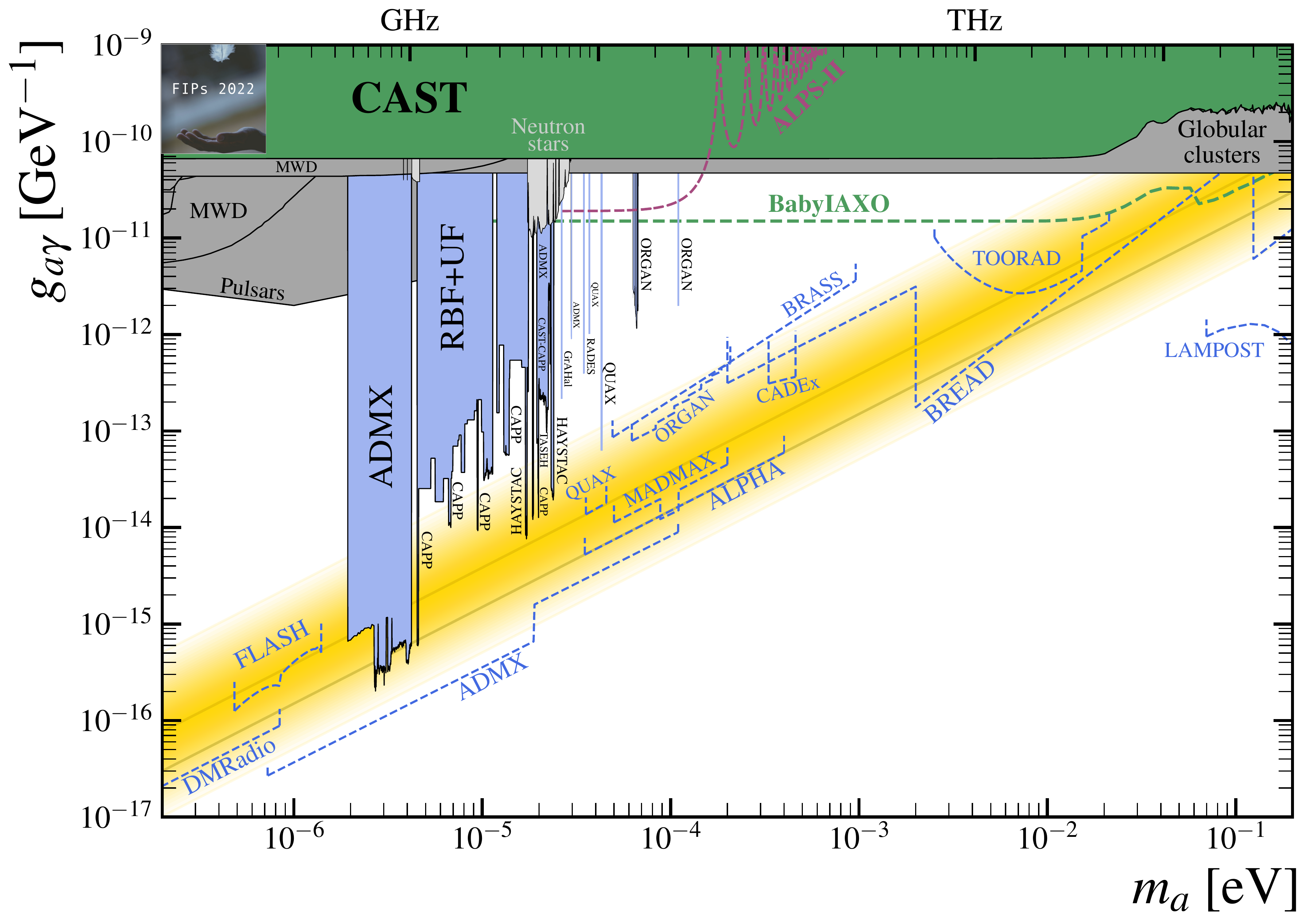}
\caption{ %\normalsize 
Closeup of the summary plot of current bounds and prospects of ongoing and future searches for the axion-photon coupling. Projected bounds from experiments include: DMRadio~\cite{DMRadio:2022pkf}, FLASH, ADMX~\cite{Stern:2016bbw}, QUAX, ORGAN~\cite{McAllister:2017lkb}, MADMAX~\cite{Beurthey:2020yuq}, ALPHA~\cite{Lawson:2019brd,Millar:2022peq}, BREAD~\cite{BREAD:2021tpx}, BRASS, CADEx~\cite{Aja:2022csb}, TOORAD~\cite{Schutte-Engel:2021bqm} and LAMPOST~\cite{Baryakhtar:2018doz}. Figure produced using resources and data found in Ref.~\cite{AxionLimits}. 
}
\label{Fig:ULF_Axion-Photon_Closeup_Plot}
\end{figure*}

\begin{figure*}[ht]
\centering
\includegraphics[width=\linewidth]{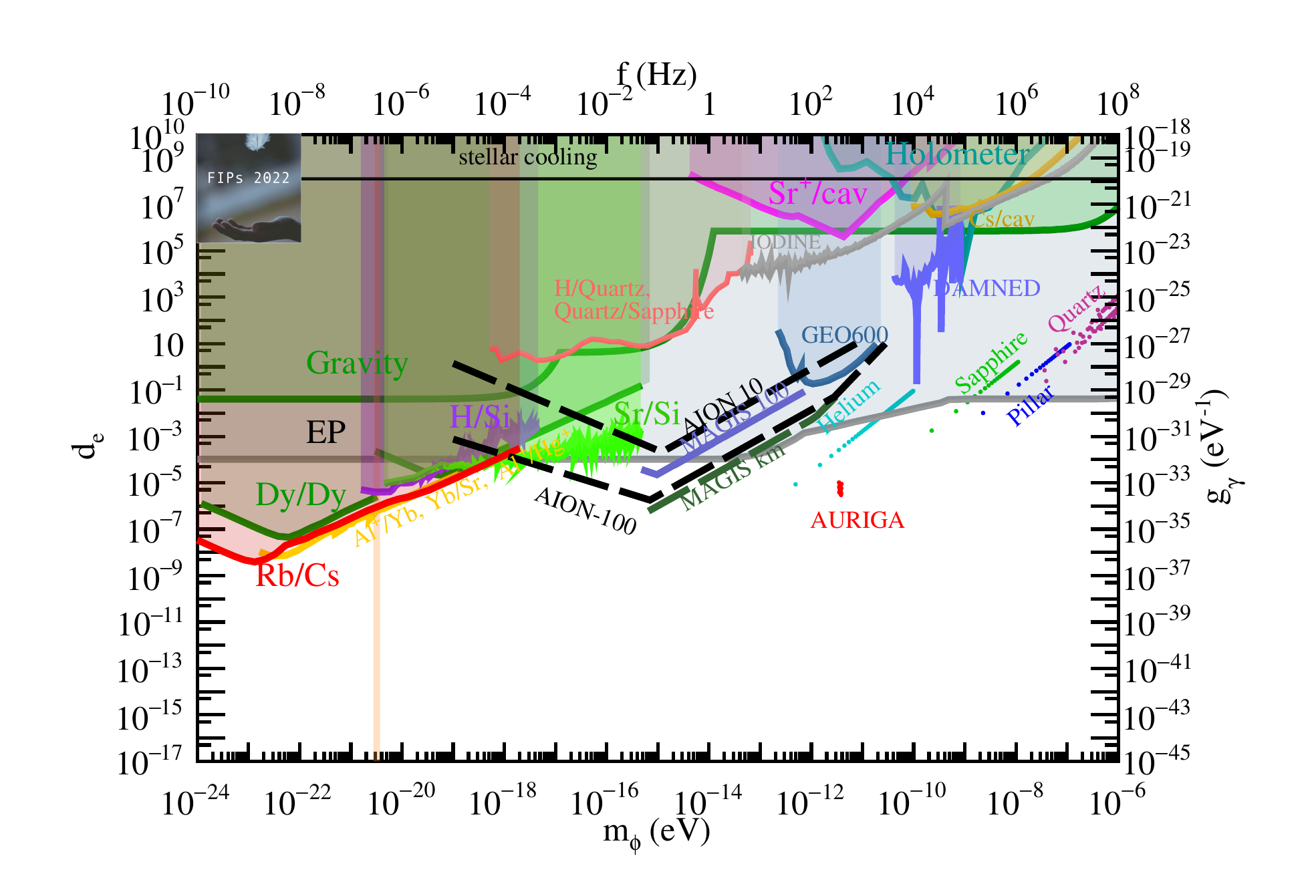}
\caption{ %\normalsize 
Summary of current bounds and prospects of ongoing and future searches for the scalar-photon coupling. 
Shaded regions above solid lines indicate regions of parameter space that have already been excluded:  AURIGA~\cite{Branca:2016rez}, H/Quartz, Quartz/Sapphire ~\cite{Campbell:2020fvq}; EP tests ~\cite{Hees:2018fpg,Berge:2017ovy, Schlamminger:2007ht,Wagner:2012ui,Touboul:2017grn, Smith:1999cr}; Sr$^+$/Cavity ~\cite{Aharony:2019iad}; Dy/Dy~\cite{VanTilburg:2015oza}; GEO600 ~\cite{Vermeulen:2021epa}; Holometer~\cite{Aiello:2021wlp}; Al$^+$/Yb, Yb/Sr, Al$^+$/Hg$^+$ \cite{boulder2021frequency}; H/Si \cite{Kennedy:2020bac}; Sr/Si \cite{Kennedy:2020bac}; Iodine~\cite{Oswald:2021vtc}; Rb/Cs~\cite{Hees:2016gop}; DAMNED~\cite{Savalle:2020vgz}; Cs/cavity \cite{Antypas:2019qji}; Stellar cooling~\cite{Raffelt:1996wa}; 
Gravity tests~\cite{Leefer:2016xfu}.
Dashed lines indicate projected sensitivities: 
MAGIS-100 and MAGIS-km~\cite{MAGIS-100:2021etm}; Super-fluid Helium (He-4), Pillar, Quartz and Sapphire~\cite{Manley:2019vxy}; AION-10 and AION-100~\cite{Badurina:2021rgt}. 
Figure produced in part using data provided by Abhishek Banerjee and Sumita Ghosh. 
}
\label{Fig:ULF_Scalar-Photon_Summary_Plot}
\end{figure*}

\begin{figure*}[ht]
\centering
\includegraphics[width=\linewidth]{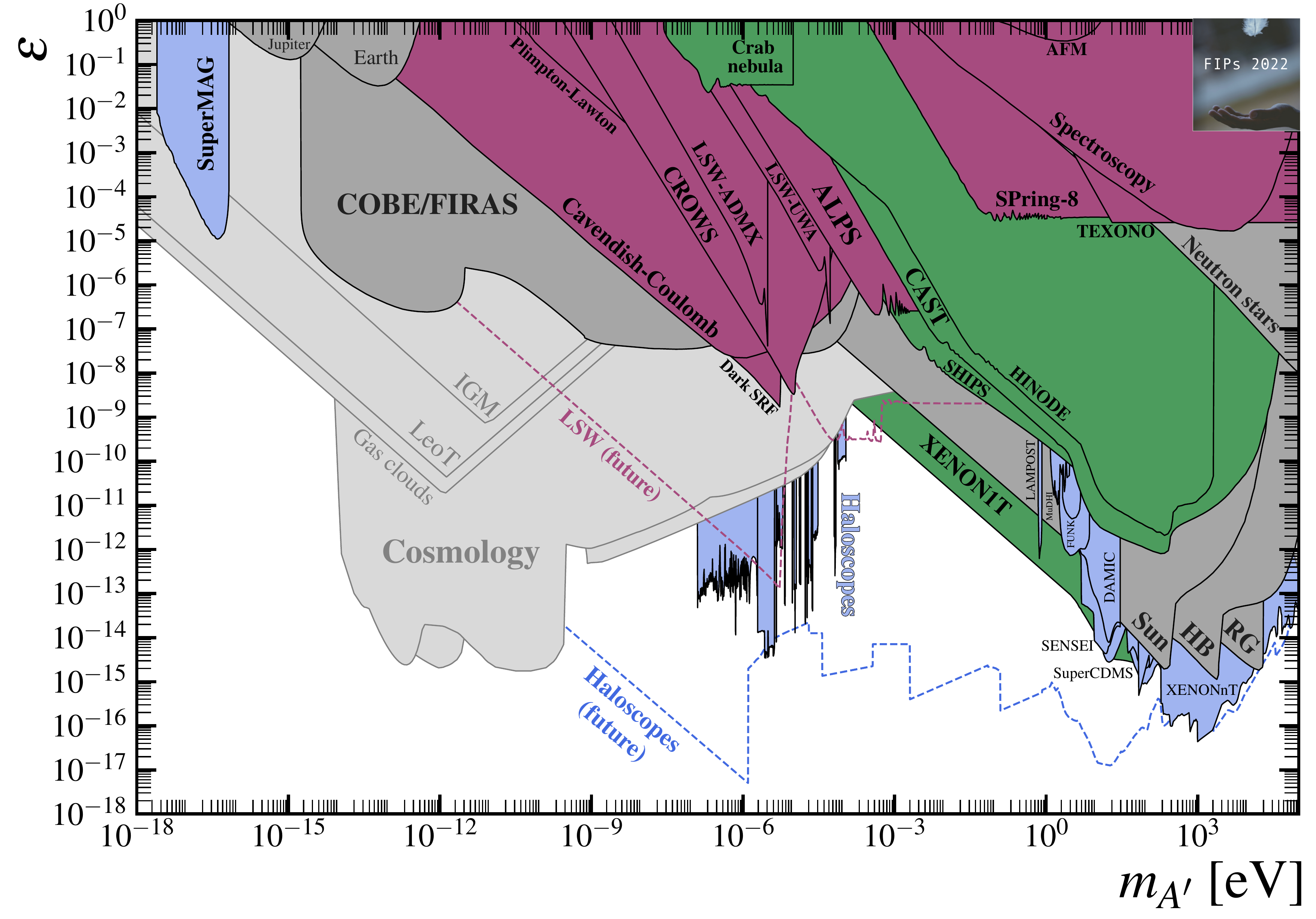}
\caption{ %\normalsize 
Summary of current bounds and prospects of (near) future searches for the dark photon kinetic mixing parameter. The colour scheme is the same as Fig.~\ref{Fig:ULF_Axion-Photon_Summary_Plot}.
Regions shaded in colours correspond to experiments: mauve for laboratory~\cite{Goldhaber:2008xy,Williams:1971ms,Bartlett:1988yy,Tu:2005ge,Kroff:2020zhp,Jaeckel:2010xx,Kroff:2020zhp,Ehret:2010mh,Inada:2013tx,Povey:2010hs,Parker:2013fxa,Wagner:2010mi,Betz:2013dza,Romanenko:2023irv,Danilov:2018bks} searches for dark photons, green for direct searches for solar emission~\cite{Redondo:2008aa,Schwarz:2015lqa,Frerick:2022mjg,XENON:2021myl}, and blue for direct dark matter detection experiments~\cite{Caputo:2021eaa,Aguilar-Arevalo:2019wdi,Godfrey:2021tvs,Phipps:2019cqy,DOSUE-RR:2022ise,An:2022hhb,Andrianavalomahefa:2020ucg,Chiles:2021gxk,An:2023wij,Manenti:2021whp,McAllister:2022ibe,Cervantes:2022yzp,Ramanathan:2022egk,Fan:2022uwu,Barak:2020fql,Brun:2019kak,Aralis:2019nfa,Cervantes:2022gtv,Fedderke:2021rrm,Fedderke:2021aqo,Dixit:2020ymh,Suzuki:2015sza,Knirck:2018ojz,Tomita:2020usq,Nguyen:2019xuh,Bloch:2016sjj,Aprile:2019xxb,Aprile:2020tmw, Bloch:2020uzh,XENON:2021myl,An:2020bxd}. All astrophysical bounds are shown in grey, with searches that do not depend on the axions being dark matter shown in dark grey~\cite{Goldhaber:1971mr,Fischbach:1994ir,Marocco:2021dku,Davis:1975mn,Marocco:2021dku,Zechlin:2008tj,Caputo:2020bdy,Hong:2020bxo,Vinyoles:2015aba}, and those that do in a lighter grey~\cite{Arias:2012az,McDermott:2019lch,Witte:2020rvb,Caputo:2020rnx,Caputo:2020bdy,Dubovsky:2015cca,Wadekar:2019xnf,Bhoonah:2019eyo}. Dashed lines show the prospects for future LSW experiments (ALPS-II~\cite{Ortiz:2020tgs}, STAX~\cite{Miyazaki:2022kxl} and DarkSRF~\cite{Berlin:2022hfx}), and a summary of all proposed future haloscopes (see next figure for a closeup). Figure produced using resources and data found in Ref.~\cite{AxionLimits}.}
\label{Fig:ULF_Dark_Photon_Kinetic_Mixing_Summary_Plot}
\end{figure*}

\begin{figure*}[ht]
\centering
\includegraphics[width=\linewidth]{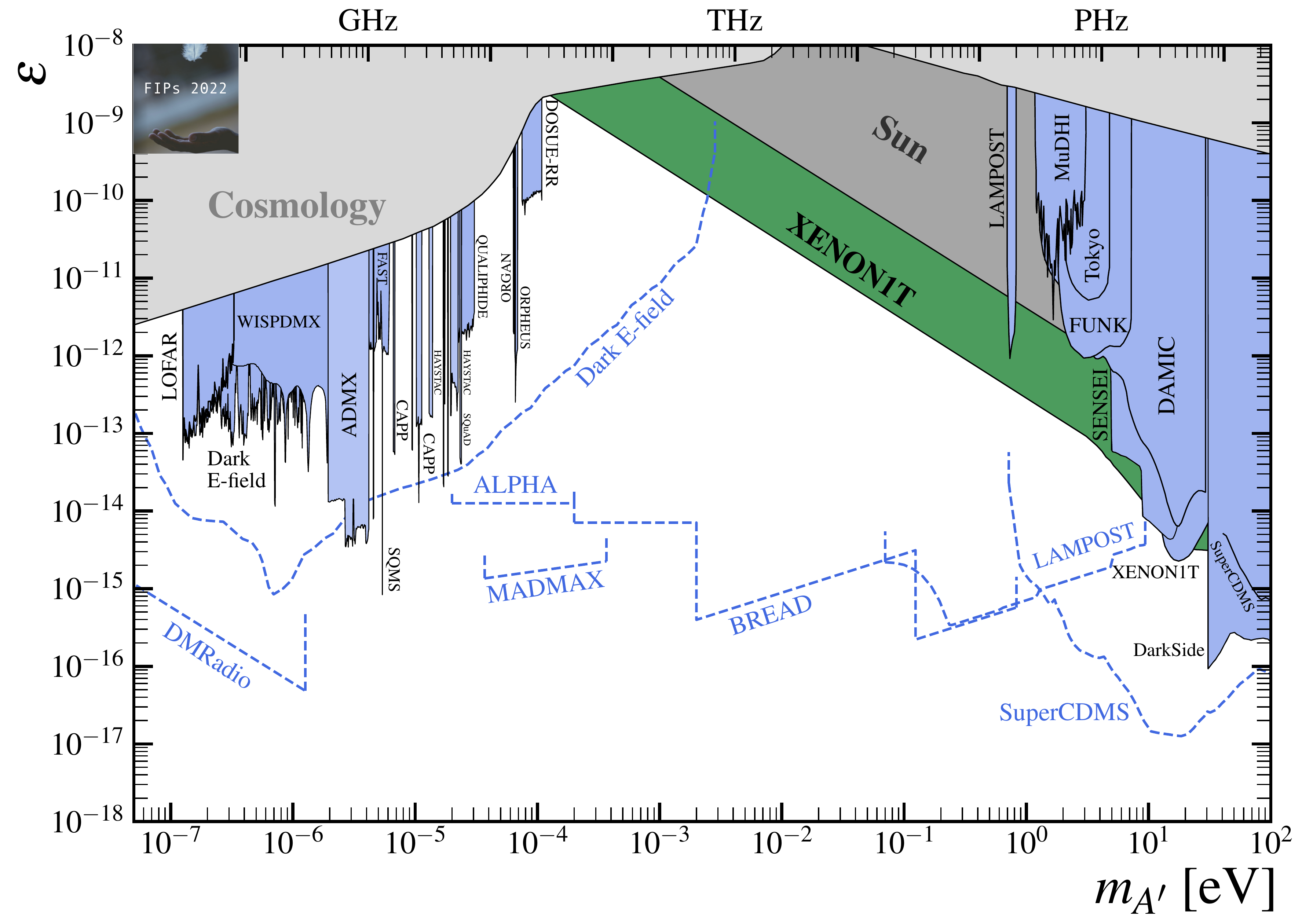}
\caption{ %\normalsize 
Closeup of the summary plot of current bounds and prospects of (near) future searches for the dark photon kinetic mixing parameter. Projected bounds from experiments include Dark E-field radio~\cite{Godfrey:2021tvs}, DMRadio~\cite{Phipps:2019cqy}, BREAD~\cite{BREAD:2021tpx}, ALPHA~\cite{Gelmini:2020kcu}, MADMAX~\cite{Beurthey:2020yuq} and SuperCDMS~\cite{Bloch:2016sjj} and LAMPOST~\cite{Baryakhtar:2018doz}.}
\label{Fig:ULF_Dark_Photon_Kinetic_Mixing_Closeup_Plot}
\end{figure*}

This chapter presented the recent developments in the field of ultralight FIPs. 
As shown in the different contributions, there is a broad range of probes available for studying new regions of as-yet-unconstrained parameter space. 
To conclude, here we present updated tables with current and proposed experiments and a set of summary plots which summarize the status of current bounds and prospects of (near) future searches in the case of the axion-photon coupling, Figs.~\ref{Fig:ULF_Axion-Photon_Summary_Plot} and~\ref{Fig:ULF_Axion-Photon_Closeup_Plot}, the scalar-photon coupling, Fig.~\ref{Fig:ULF_Scalar-Photon_Summary_Plot}; and dark photon kinetic mixing parameter, Figs.~\ref{Fig:ULF_Dark_Photon_Kinetic_Mixing_Summary_Plot} and~\ref{Fig:ULF_Dark_Photon_Kinetic_Mixing_Closeup_Plot}.  

% {\bf NEW TABLE} 

% \usepackage{longtable}
% Note: It may be necessary to compile the document several times to get a multi-page table to line up properly
\begin{longtable}{llllcc}
\caption{Running and proposed experiments dedicated to ultralight FIPs. 
Experiments labelled with an asterisk (*) have no official name.
The name appearing in the table is just a descriptive name.
For consistency with the summary tables in the other sections of this report, we use the following notation for the portals: (1) Vector, (2) Scalar, (3) Pseudoscalar.  
There is no fermion portal (number 4 in the MeV-GeV summary tables).}
\label{tab:my-table}\\
\multicolumn{6}{l}{\textbf{Running}} \\ \hline
\endfirsthead
\multicolumn{6}{c}%
{{\bfseries Table \thetable\ continued from previous page}} \\
\multicolumn{6}{l}{\textbf{Running}} \\ \hline
\endhead
\multicolumn{1}{|l|}{\textbf{Experiment}} &
  \multicolumn{1}{l|}{\textbf{Laboratory}} &
  \multicolumn{1}{l|}{\textbf{Source}} &
  \multicolumn{1}{l|}{\textbf{Method/technique}} &
  \multicolumn{1}{c|}{\textbf{Portals}} &
  \multicolumn{1}{c|}{\textbf{Ref.}} \\ \hline
\multicolumn{1}{|l|}{ABRACADABRA} &
  \multicolumn{1}{l|}{MIT} &
  \multicolumn{1}{l|}{Dark matter} &
  \multicolumn{1}{l|}{Lumped element} &
  \multicolumn{1}{c|}{(3)} &
  \multicolumn{1}{c|}{\cite{Salemi:2021gck}} \\ \hline
\multicolumn{1}{|l|}{ADMX} &
  \multicolumn{1}{l|}{UW Seattle} &
  \multicolumn{1}{l|}{Dark matter} &
  \multicolumn{1}{l|}{Microwave cavity} &
  \multicolumn{1}{c|}{(1,2,3)} &
  \multicolumn{1}{c|}{\cite{ADMX:2018gho,ADMX:2019uok,Crisosto:2019fcj}} \\ \hline
\multicolumn{1}{|l|}{ALPS} &
  \multicolumn{1}{l|}{DESY} &
  \multicolumn{1}{l|}{Laboratory} &
  \multicolumn{1}{l|}{\begin{tabular}[c]{@{}l@{}}Light-shining-\\ through-a-wall\end{tabular}} &
  \multicolumn{1}{c|}{(1,2,3)} &
  \multicolumn{1}{c|}{\cite{Ehret:2010mh, Ortiz:2020tgs}} \\ \hline
\multicolumn{1}{|l|}{\begin{tabular}[c]{@{}l@{}}Atomic \\ dysprosium \\ spectroscopy$^*$\end{tabular}} &
  \multicolumn{1}{l|}{Mainz} &
  \multicolumn{1}{l|}{\begin{tabular}[c]{@{}l@{}}Dark matter\\ Laboratory\end{tabular}} &
  \multicolumn{1}{l|}{\begin{tabular}[c]{@{}l@{}}Atomic \\ spectroscopy\end{tabular}} &
  \multicolumn{1}{c|}{(2)} &
  \multicolumn{1}{c|}{\cite{VanTilburg:2015oza,Leefer:2016xfu}} \\ \hline
\multicolumn{1}{|l|}{AURIGA} &
  \multicolumn{1}{l|}{INFN Legnaro} &
  \multicolumn{1}{l|}{Dark matter} &
  \multicolumn{1}{l|}{\begin{tabular}[c]{@{}l@{}}Resonant-bar \\ detector\end{tabular}} &
  \multicolumn{1}{c|}{(2)} &
  \multicolumn{1}{c|}{\cite{Branca:2016rez}} \\ \hline
\multicolumn{1}{|l|}{BACON} &
  \multicolumn{1}{l|}{\begin{tabular}[c]{@{}l@{}}NIST and \\ JILA Boulder\end{tabular}} &
  \multicolumn{1}{l|}{Dark matter} &
  \multicolumn{1}{l|}{\begin{tabular}[c]{@{}l@{}}Optical clock \\ comparisons\end{tabular}} &
  \multicolumn{1}{c|}{(2)} &
  \multicolumn{1}{c|}{\cite{boulder2021frequency}} \\ \hline
\multicolumn{1}{|l|}{BASE} &
  \multicolumn{1}{l|}{CERN} &
  \multicolumn{1}{l|}{Dark matter} &
  \multicolumn{1}{l|}{\begin{tabular}[c]{@{}l@{}}Antiproton trap/\\ lumped element\end{tabular}} &
  \multicolumn{1}{c|}{(3)} &
  \multicolumn{1}{c|}{\cite{Devlin:2021fpq}} \\ \hline
\multicolumn{1}{|l|}{CAPP} &
  \multicolumn{1}{l|}{IBS Daejeon} &
  \multicolumn{1}{l|}{Dark matter} &
  \multicolumn{1}{l|}{Microwave cavity} &
  \multicolumn{1}{c|}{(1,3)} &
  \multicolumn{1}{c|}{\cite{Semertzidis:2019gkj}} \\ \hline
\multicolumn{1}{|l|}{CASPEr-gradient} &
  \multicolumn{1}{l|}{Mainz} &
  \multicolumn{1}{l|}{Dark matter} &
  \multicolumn{1}{l|}{\begin{tabular}[c]{@{}l@{}}Nuclear magnetic \\ resonance\\ (liquid detector)\end{tabular}} &
  \multicolumn{1}{c|}{(3)} &
  \multicolumn{1}{c|}{\cite{JacksonKimball:2017elr, Aybas:2021nvn}} \\ \hline
\multicolumn{1}{|l|}{CASPEr-electric} &
  \multicolumn{1}{l|}{Boston} &
  \multicolumn{1}{l|}{Dark matter} &
  \multicolumn{1}{l|}{\begin{tabular}[c]{@{}l@{}}Nuclear magnetic \\ resonance\\ (solid-state detector)\end{tabular}} &
  \multicolumn{1}{c|}{(3)} &
  \multicolumn{1}{c|}{\cite{JacksonKimball:2017elr, Aybas:2021nvn}} \\ \hline
\multicolumn{1}{|l|}{CASPEr-ZULF} &
  \multicolumn{1}{l|}{Mainz} &
  \multicolumn{1}{l|}{Dark matter} &
  \multicolumn{1}{l|}{Magnetometry} &
  \multicolumn{1}{c|}{(3)} &
  \multicolumn{1}{c|}{\cite{Wu:2019exd, Garcon:2019inh}} \\ \hline
\multicolumn{1}{|l|}{CAST} &
  \multicolumn{1}{l|}{CERN} &
  \multicolumn{1}{l|}{Solar} &
  \multicolumn{1}{l|}{Helioscope} &
  \multicolumn{1}{c|}{(1,3)} &
  \multicolumn{1}{c|}{\cite{CAST:2017uph}} \\ \hline
\multicolumn{1}{|l|}{CAST-CAPP} &
  \multicolumn{1}{l|}{CERN} &
  \multicolumn{1}{l|}{Dark matter} &
  \multicolumn{1}{l|}{Microwave cavity} &
  \multicolumn{1}{c|}{(1,3)} &
  \multicolumn{1}{c|}{\cite{Adair:2022rtw}} \\ \hline
\multicolumn{1}{|l|}{CAST-RADES} &
  \multicolumn{1}{l|}{CERN} &
  \multicolumn{1}{l|}{Dark matter} &
  \multicolumn{1}{l|}{\begin{tabular}[c]{@{}l@{}}Radio-frequency \\ cavity\end{tabular}} &
  \multicolumn{1}{c|}{(3)} &
  \multicolumn{1}{c|}{\cite{CAST:2020rlf}} \\ \hline
\multicolumn{1}{|l|}{CROWS} &
  \multicolumn{1}{l|}{CERN} &
  \multicolumn{1}{l|}{Laboratory} &
  \multicolumn{1}{l|}{\begin{tabular}[c]{@{}l@{}}Light-shining-\\ through-a-wall\end{tabular}} &
  \multicolumn{1}{c|}{(2,3)} &
  \multicolumn{1}{c|}{\cite{Betz:2013dza}} \\ \hline
\multicolumn{1}{|l|}{\begin{tabular}[c]{@{}l@{}}Cs/cavity \\ comparisons$^*$\end{tabular}} &
  \multicolumn{1}{l|}{Mainz} &
  \multicolumn{1}{l|}{Dark matter} &
  \multicolumn{1}{l|}{\begin{tabular}[c]{@{}l@{}}Clock-cavity \\ comparison\end{tabular}} &
  \multicolumn{1}{c|}{(2,3)} &
  \multicolumn{1}{c|}{\cite{Antypas:2019qji, Tretiak:2022ndx}} \\ \hline
\multicolumn{1}{|l|}{DAMNED} &
  \multicolumn{1}{l|}{SYRTE Paris} &
  \multicolumn{1}{l|}{Dark matter} &
  \multicolumn{1}{l|}{\begin{tabular}[c]{@{}l@{}}Time-delay \\ interferometry\end{tabular}} &
  \multicolumn{1}{c|}{(2,3)} &
  \multicolumn{1}{c|}{\cite{Savalle:2020vgz}} \\ \hline
\multicolumn{1}{|l|}{DANCE} &
  \multicolumn{1}{l|}{U. Tokyo} &
  \multicolumn{1}{l|}{Dark matter} &
  \multicolumn{1}{l|}{optical ring cavity} &
  \multicolumn{1}{c|}{(3)} &
  \multicolumn{1}{c|}{\cite{Michimura:2019qxr}} \\ \hline
\multicolumn{1}{|l|}{Dark E-field} &
  \multicolumn{1}{l|}{UC Davis} &
  \multicolumn{1}{l|}{Dark matter} &
  \multicolumn{1}{l|}{Antenna} &
  \multicolumn{1}{c|}{(1)} &
  \multicolumn{1}{c|}{\cite{Godfrey:2021tvs}} \\ \hline
\multicolumn{1}{|l|}{DarkSRF} &
  \multicolumn{1}{l|}{Fermilab} &
  \multicolumn{1}{l|}{Laboratory} &
  \multicolumn{1}{l|}{\begin{tabular}[c]{@{}l@{}}Light-shining-\\ through-a-wall/\\ Superconducting \\ RF cavity\end{tabular}} &
  \multicolumn{1}{c|}{(1)} &
  \multicolumn{1}{c|}{\cite{Romanenko:2023irv}} \\ \hline
\multicolumn{1}{|l|}{DOSUE-RR} &
  \multicolumn{1}{l|}{U. Kyoto} &
  \multicolumn{1}{l|}{Dark matter} &
  \multicolumn{1}{l|}{\begin{tabular}[c]{@{}l@{}}cryogenic \\ millimeter-wave \\ receiver\end{tabular}} &
  \multicolumn{1}{c|}{(1)} &
  \multicolumn{1}{c|}{\cite{DOSUE-RR:2022ise}} \\ \hline
\multicolumn{1}{|l|}{Eot-Wash} &
  \multicolumn{1}{l|}{UW Seattle} &
  \multicolumn{1}{l|}{Laboratory} &
  \multicolumn{1}{l|}{Torsion pendulum} &
  \multicolumn{1}{c|}{(1,2,3)} &
  \multicolumn{1}{c|}{\cite{Schlamminger:2007ht,Wagner:2012ui,Terrano:2015sna}} \\ \hline
\multicolumn{1}{|l|}{Eot-Wash} &
  \multicolumn{1}{l|}{UW Seattle} &
  \multicolumn{1}{l|}{Dark matter} &
  \multicolumn{1}{l|}{Torsion pendulum} &
  \multicolumn{1}{c|}{(1,3)} &
  \multicolumn{1}{c|}{\cite{Terrano:2019clh,Shaw:2021gnp}} \\ \hline
\multicolumn{1}{|l|}{\begin{tabular}[c]{@{}l@{}}Ferromagnetic \\ haloscope\end{tabular}} &
  \multicolumn{1}{l|}{UWA Perth} &
  \multicolumn{1}{l|}{Dark matter} &
  \multicolumn{1}{l|}{\begin{tabular}[c]{@{}l@{}}Ferromagnetic \\ haloscope\end{tabular}} &
  \multicolumn{1}{c|}{(3)} &
  \multicolumn{1}{c|}{\cite{Flower:2018qgb}} \\ \hline
\multicolumn{1}{|l|}{FUNK} &
  \multicolumn{1}{l|}{KIT} &
  \multicolumn{1}{l|}{Dark matter} &
  \multicolumn{1}{l|}{Dish antenna} &
  \multicolumn{1}{c|}{(1)} &
  \multicolumn{1}{c|}{\cite{FUNKExperiment:2020ofv}} \\ \hline
\multicolumn{1}{|l|}{GEO600} &
  \multicolumn{1}{l|}{Hannover} &
  \multicolumn{1}{l|}{Dark matter} &
  \multicolumn{1}{l|}{\begin{tabular}[c]{@{}l@{}}Optical \\ interferometry\end{tabular}} &
  \multicolumn{1}{c|}{(1,2,3)} &
  \multicolumn{1}{c|}{\cite{Vermeulen:2021epa}} \\ \hline
\multicolumn{1}{|l|}{GrAHal} &
  \multicolumn{1}{l|}{Grenoble} &
  \multicolumn{1}{l|}{Dark matter} &
  \multicolumn{1}{l|}{Microwave cavity} &
  \multicolumn{1}{c|}{(3)} &
  \multicolumn{1}{c|}{\cite{Grenet:2021vbb}} \\ \hline
\multicolumn{1}{|l|}{\begin{tabular}[c]{@{}l@{}}H/Quartz and \\ Quartz/Sapphire \\ comparisons$^*$\end{tabular}} &
  \multicolumn{1}{l|}{UWA Perth} &
  \multicolumn{1}{l|}{Dark matter} &
  \multicolumn{1}{l|}{\begin{tabular}[c]{@{}l@{}}Clock-cavity and \\ cavity-cavity \\ comparisons\end{tabular}} &
  \multicolumn{1}{c|}{(2)} &
  \multicolumn{1}{c|}{\cite{Campbell:2020fvq}} \\ \hline
\multicolumn{1}{|l|}{HAYSTAC} &
  \multicolumn{1}{l|}{Yale} &
  \multicolumn{1}{l|}{Dark matter} &
  \multicolumn{1}{l|}{Microwave cavity} &
  \multicolumn{1}{c|}{(1,3)} &
  \multicolumn{1}{c|}{\cite{HAYSTAC:2023cam}} \\ \hline
\multicolumn{1}{|l|}{\begin{tabular}[c]{@{}l@{}}HfF$^{+}$ \\ molecular \\ EDM$^*$\end{tabular}} &
  \multicolumn{1}{l|}{JILA Boulder} &
  \multicolumn{1}{l|}{Dark matter} &
  \multicolumn{1}{l|}{Molecular EDM} &
  \multicolumn{1}{c|}{(3)} &
  \multicolumn{1}{c|}{\cite{Roussy:2020ily}} \\ \hline
\multicolumn{1}{|l|}{Holometer} &
  \multicolumn{1}{l|}{Fermilab} &
  \multicolumn{1}{l|}{Dark matter} &
  \multicolumn{1}{l|}{\begin{tabular}[c]{@{}l@{}}Optical \\ interferometry\end{tabular}} &
  \multicolumn{1}{c|}{(1,2,3)} &
  \multicolumn{1}{c|}{\cite{Aiello:2021wlp}} \\ \hline
\multicolumn{1}{|l|}{JEDI} &
  \multicolumn{1}{l|}{Jülich} &
  \multicolumn{1}{l|}{Dark matter} &
  \multicolumn{1}{l|}{Storage ring} &
  \multicolumn{1}{c|}{(3)} &
  \multicolumn{1}{c|}{\cite{JEDI:2022hxa}} \\ \hline
\multicolumn{1}{|l|}{\begin{tabular}[c]{@{}l@{}}K/He \\ co-magnetometry\end{tabular}} &
  \multicolumn{1}{l|}{Princeton} &
  \multicolumn{1}{l|}{Dark matter} &
  \multicolumn{1}{l|}{\begin{tabular}[c]{@{}l@{}}Magnetometry\\ Laboratory\end{tabular}} &
  \multicolumn{1}{c|}{(3)} &
  \multicolumn{1}{c|}{\cite{Lee:2022vvb,Vasilakis:2008yn}} \\ \hline
\multicolumn{1}{|l|}{LAMPOST} &
  \multicolumn{1}{l|}{MIT} &
  \multicolumn{1}{l|}{Dark matter} &
  \multicolumn{1}{l|}{\begin{tabular}[c]{@{}l@{}}Multilayer \\ optical dielectric \\ haloscope\end{tabular}} &
  \multicolumn{1}{c|}{(1)} &
  \multicolumn{1}{c|}{\cite{Baryakhtar:2018doz}} \\ \hline
\multicolumn{1}{|l|}{LIGO} &
  \multicolumn{1}{l|}{\begin{tabular}[c]{@{}l@{}}Hanford and \\ Livingston\end{tabular}} &
  \multicolumn{1}{l|}{Dark matter} &
  \multicolumn{1}{l|}{\begin{tabular}[c]{@{}l@{}}Optical \\ interferometry\end{tabular}} &
  \multicolumn{1}{c|}{(1,2,3)} &
  \multicolumn{1}{c|}{\cite{Guo:2019ker, LIGOScientific:2021ffg}} \\ \hline
\multicolumn{1}{|l|}{MICROSCOPE} &
  \multicolumn{1}{l|}{\begin{tabular}[c]{@{}l@{}}CNES (space \\ mission)\end{tabular}} &
  \multicolumn{1}{l|}{Laboratory} &
  \multicolumn{1}{l|}{Torsion pendulum} &
  \multicolumn{1}{c|}{(1,2)} &
  \multicolumn{1}{c|}{\cite{Touboul:2017grn, MICROSCOPE:2022doy}} \\ \hline
\multicolumn{1}{|l|}{\begin{tabular}[c]{@{}l@{}}Molecular iodine \\ spectroscopy$^*$\end{tabular}} &
  \multicolumn{1}{l|}{Duesseldorf} &
  \multicolumn{1}{l|}{Dark matter} &
  \multicolumn{1}{l|}{\begin{tabular}[c]{@{}l@{}}Molecular \\ spectroscopy\end{tabular}} &
  \multicolumn{1}{c|}{(2,3)} &
  \multicolumn{1}{c|}{\cite{Oswald:2021vtc}} \\ \hline
\multicolumn{1}{|l|}{MuDHI} &
  \multicolumn{1}{l|}{\begin{tabular}[c]{@{}l@{}}NYU, \\ Abu Dhabi\end{tabular}} &
  \multicolumn{1}{l|}{Dark matter} &
  \multicolumn{1}{l|}{\begin{tabular}[c]{@{}l@{}}Multilayer \\ optical dielectric \\ haloscope\end{tabular}} &
  \multicolumn{1}{c|}{(1)} &
  \multicolumn{1}{c|}{\cite{Manenti:2021whp}} \\ \hline
\multicolumn{1}{|l|}{NASDUCK} &
  \multicolumn{1}{l|}{Weizmann} &
  \multicolumn{1}{l|}{Dark matter} &
  \multicolumn{1}{l|}{Magnetometry} &
  \multicolumn{1}{c|}{(3)} &
  \multicolumn{1}{c|}{\cite{Bloch:2021vnn, Bloch:2022kjm}} \\ \hline
\multicolumn{1}{|l|}{nEDM} &
  \multicolumn{1}{l|}{PSI} &
  \multicolumn{1}{l|}{Dark matter} &
  \multicolumn{1}{l|}{Magnetometry} &
  \multicolumn{1}{c|}{(3)} &
  \multicolumn{1}{c|}{\cite{Abel:2017rtm, Abel:2022vfg}} \\ \hline
\multicolumn{1}{|l|}{\begin{tabular}[c]{@{}l@{}}Neutron beam \\ EDM$^*$\end{tabular}} &
  \multicolumn{1}{l|}{ILL Grenoble} &
  \multicolumn{1}{l|}{Dark matter} &
  \multicolumn{1}{l|}{Neutron EDM} &
  \multicolumn{1}{c|}{(3)} &
  \multicolumn{1}{c|}{\cite{Schulthess:2022pbp}} \\ \hline
\multicolumn{1}{|l|}{NOMAD} &
  \multicolumn{1}{l|}{CERN} &
  \multicolumn{1}{l|}{Laboratory} &
  \multicolumn{1}{l|}{\begin{tabular}[c]{@{}l@{}}Light-shining-\\ through-a-wall\end{tabular}} &
  \multicolumn{1}{c|}{(3)} &
  \multicolumn{1}{c|}{\cite{NOMAD:2000usb}} \\ \hline
\multicolumn{1}{|l|}{ORGAN} &
  \multicolumn{1}{l|}{UWA Perth} &
  \multicolumn{1}{l|}{Dark matter} &
  \multicolumn{1}{l|}{Microwave cavity} &
  \multicolumn{1}{c|}{(1,2,3)} &
  \multicolumn{1}{c|}{\cite{McAllister:2017lkb, McAllister:2022ibe}} \\ \hline
\multicolumn{1}{|l|}{ORPHEUS} &
  \multicolumn{1}{l|}{U. Washington} &
  \multicolumn{1}{l|}{Dark matter} &
  \multicolumn{1}{l|}{\begin{tabular}[c]{@{}l@{}}Dielectric loaded \\ cavity\end{tabular}} &
  \multicolumn{1}{c|}{(1,3)} &
  \multicolumn{1}{c|}{\cite{Cervantes:2022yzp}} \\ \hline
\multicolumn{1}{|l|}{OSQAR} &
  \multicolumn{1}{l|}{CERN} &
  \multicolumn{1}{l|}{Laboratory} &
  \multicolumn{1}{l|}{\begin{tabular}[c]{@{}l@{}}Light-shining-\\ through-a-wall\end{tabular}} &
  \multicolumn{1}{c|}{(2,3)} &
  \multicolumn{1}{c|}{\cite{OSQAR:2015qdv}} \\ \hline
\multicolumn{1}{|l|}{PVLAS} &
  \multicolumn{1}{l|}{INFN Ferrara} &
  \multicolumn{1}{l|}{Laboratory} &
  \multicolumn{1}{l|}{\begin{tabular}[c]{@{}l@{}}Vacuum \\ birefringence \\ and dichroism\end{tabular}} &
  \multicolumn{1}{c|}{(2,3)} &
  \multicolumn{1}{c|}{\cite{DellaValle:2015xxa}} \\ \hline
\multicolumn{1}{|l|}{QUALIPHIDE} &
  \multicolumn{1}{l|}{JPL, Caltech} &
  \multicolumn{1}{l|}{Dark matter} &
  \multicolumn{1}{l|}{Dish antenna} &
  \multicolumn{1}{c|}{(1)} &
  \multicolumn{1}{c|}{\cite{Ramanathan:2022egk}} \\ \hline
\multicolumn{1}{|l|}{QUAX-${a\gamma}$} &
  \multicolumn{1}{l|}{INFN Legnaro} &
  \multicolumn{1}{l|}{Dark matter} &
  \multicolumn{1}{l|}{\begin{tabular}[c]{@{}l@{}}Microwave \\ cavity\end{tabular}} &
  \multicolumn{1}{c|}{(1,3)} &
  \multicolumn{1}{c|}{\cite{Alesini:2019ajt,Alesini:2020vny,Alesini:2022lnp}} \\ \hline
\multicolumn{1}{|l|}{QUAX-${ae}$} &
  \multicolumn{1}{l|}{INFN Legnaro} &
  \multicolumn{1}{l|}{Dark matter} &
  \multicolumn{1}{l|}{\begin{tabular}[c]{@{}l@{}}Ferromagnetic \\ haloscope\end{tabular}} &
  \multicolumn{1}{c|}{(1,3)} &
  \multicolumn{1}{c|}{\cite{Crescini:2018qrz, QUAX:2020adt}} \\ \hline
\multicolumn{1}{|l|}{QUAX-${g_p g_s}$} &
  \multicolumn{1}{l|}{INFN Legnaro} &
  \multicolumn{1}{l|}{Laboratory} &
  \multicolumn{1}{l|}{\begin{tabular}[c]{@{}l@{}}Long-range \\ interactions\end{tabular}} &
  \multicolumn{1}{c|}{($2 \times 3$)} &
  \multicolumn{1}{c|}{\cite{Crescini:2017uxs}} \\ \hline
\multicolumn{1}{|l|}{\begin{tabular}[c]{@{}l@{}}Rb/Cs clock \\ comparison$^*$\end{tabular}} &
  \multicolumn{1}{l|}{SYRTE Paris} &
  \multicolumn{1}{l|}{Dark matter} &
  \multicolumn{1}{l|}{\begin{tabular}[c]{@{}l@{}}Microwave clock \\ comparisons\end{tabular}} &
  \multicolumn{1}{c|}{(2,3)} &
  \multicolumn{1}{c|}{\cite{Hees:2016gop}} \\ \hline
\multicolumn{1}{|l|}{\begin{tabular}[c]{@{}l@{}}Rb/Quartz \\ comparison$^*$\end{tabular}} &
  \multicolumn{1}{l|}{Mainz} &
  \multicolumn{1}{l|}{Dark matter} &
  \multicolumn{1}{l|}{\begin{tabular}[c]{@{}l@{}}Clock-cavity \\ comparison\end{tabular}} &
  \multicolumn{1}{c|}{(2)} &
  \multicolumn{1}{c|}{\cite{Zhang:2022ewz}} \\ \hline
\multicolumn{1}{|l|}{SAPPHIRES} &
  \multicolumn{1}{l|}{Hiroshima} &
  \multicolumn{1}{l|}{Laboratory} &
  \multicolumn{1}{l|}{\begin{tabular}[c]{@{}l@{}}Photon-photon \\ collider\end{tabular}} &
  \multicolumn{1}{c|}{(3)} &
  \multicolumn{1}{c|}{\cite{SAPPHIRES:2021vkz,SAPPHIRES:2022bqg, Ishibashi:2023bae}} \\ \hline
\multicolumn{1}{|l|}{SHAFT} &
  \multicolumn{1}{l|}{Boston} &
  \multicolumn{1}{l|}{Dark matter} &
  \multicolumn{1}{l|}{Lumped element} &
  \multicolumn{1}{c|}{(3)} &
  \multicolumn{1}{c|}{\cite{Gramolin:2020ict}} \\ \hline
\multicolumn{1}{|l|}{SHIPS} &
  \multicolumn{1}{l|}{\begin{tabular}[c]{@{}l@{}}Hamburg-\\ Bergedorf\end{tabular}} &
  \multicolumn{1}{l|}{Solar} &
  \multicolumn{1}{l|}{Helioscope} &
  \multicolumn{1}{c|}{(1)} &
  \multicolumn{1}{c|}{\cite{Schwarz:2015lqa}} \\ \hline
\multicolumn{1}{|l|}{SHUKET} &
  \multicolumn{1}{l|}{\begin{tabular}[c]{@{}l@{}}CEA Paris-\\ Saclay\end{tabular}} &
  \multicolumn{1}{l|}{Dark matter} &
  \multicolumn{1}{l|}{Dish antenna} &
  \multicolumn{1}{c|}{(1)} &
  \multicolumn{1}{c|}{\cite{Brun:2019kak}} \\ \hline
\multicolumn{1}{|l|}{Spring-8-LSW} &
  \multicolumn{1}{l|}{\begin{tabular}[c]{@{}l@{}}Super Photon \\ ring-8 GeV, \\ Japan\end{tabular}} &
  \multicolumn{1}{l|}{Laboratory} &
  \multicolumn{1}{l|}{\begin{tabular}[c]{@{}l@{}}Light-shining-\\ through-a-wall\end{tabular}} &
  \multicolumn{1}{c|}{(1)} &
  \multicolumn{1}{c|}{\cite{Inada:2013tx}} \\ \hline
\multicolumn{1}{|l|}{SQMS} &
  \multicolumn{1}{l|}{Fermilab} &
  \multicolumn{1}{l|}{Dark matter} &
  \multicolumn{1}{l|}{SRF Cavity} &
  \multicolumn{1}{c|}{(1)} &
  \multicolumn{1}{c|}{\cite{Cervantes:2022gtv}} \\ \hline
\multicolumn{1}{|l|}{SQuAD} &
  \multicolumn{1}{l|}{U. Chicago} &
  \multicolumn{1}{l|}{Dark matter} &
  \multicolumn{1}{l|}{\begin{tabular}[c]{@{}l@{}}Cavity + \\ Superconducting \\ Qubit\end{tabular}} &
  \multicolumn{1}{c|}{(1)} &
  \multicolumn{1}{c|}{\cite{Dixit:2020ymh}} \\ \hline
\multicolumn{1}{|l|}{\begin{tabular}[c]{@{}l@{}}Sr/Si and H/Si \\ comparisons$^*$\end{tabular}} &
  \multicolumn{1}{l|}{JILA Boulder} &
  \multicolumn{1}{l|}{Dark matter} &
  \multicolumn{1}{l|}{\begin{tabular}[c]{@{}l@{}}Clock-cavity \\ comparisons\end{tabular}} &
  \multicolumn{1}{c|}{(2,3)} &
  \multicolumn{1}{c|}{\cite{Kennedy:2020bac}} \\ \hline
\multicolumn{1}{|l|}{\begin{tabular}[c]{@{}l@{}}Sr$^{+}$/cavity \\ comparison$^*$\end{tabular}} &
  \multicolumn{1}{l|}{Weizmann} &
  \multicolumn{1}{l|}{Dark matter} &
  \multicolumn{1}{l|}{\begin{tabular}[c]{@{}l@{}}Clock-cavity\\ comparison\end{tabular}} &
  \multicolumn{1}{c|}{(2)} &
  \multicolumn{1}{c|}{\cite{Aharony:2019iad}} \\ \hline
\multicolumn{1}{|l|}{TASEH} &
  \multicolumn{1}{l|}{Taiwan} &
  \multicolumn{1}{l|}{Dark matter} &
  \multicolumn{1}{l|}{Microwave cavity} &
  \multicolumn{1}{c|}{(3)} &
  \multicolumn{1}{c|}{\cite{TASEH:2022vvu}} \\ \hline
\multicolumn{1}{|l|}{\begin{tabular}[c]{@{}l@{}}Tokyo dish \\ antennae\end{tabular}} &
  \multicolumn{1}{l|}{U. Tokyo} &
  \multicolumn{1}{l|}{Dark matter} &
  \multicolumn{1}{l|}{Dish antenna} &
  \multicolumn{1}{c|}{(1)} &
  \multicolumn{1}{c|}{\cite{Suzuki:2015sza,Knirck:2018ojz,Tomita:2020usq}} \\ \hline
\multicolumn{1}{|l|}{UPLOAD} &
  \multicolumn{1}{l|}{UWA Perth} &
  \multicolumn{1}{l|}{Dark matter} &
  \multicolumn{1}{l|}{Microwave cavity} &
  \multicolumn{1}{c|}{(3)} &
  \multicolumn{1}{c|}{\cite{Thomson:2019aht}} \\ \hline
\multicolumn{1}{|l|}{UWA-LSW} &
  \multicolumn{1}{l|}{UWA Perth} &
  \multicolumn{1}{l|}{Laboratory} &
  \multicolumn{1}{l|}{\begin{tabular}[c]{@{}l@{}}Light-shining-\\ through-a-wall\end{tabular}} &
  \multicolumn{1}{c|}{(1)} &
  \multicolumn{1}{c|}{\cite{Povey:2010hs,Parker:2013fxa}} \\ \hline
\multicolumn{1}{|l|}{WISPDMX} &
  \multicolumn{1}{l|}{U. Hamburg} &
  \multicolumn{1}{l|}{Dark matter} &
  \multicolumn{1}{l|}{RF cavity} &
  \multicolumn{1}{c|}{(1)} &
  \multicolumn{1}{c|}{\cite{Nguyen:2019xuh}} \\ \hline
\multicolumn{1}{|l|}{\begin{tabular}[c]{@{}l@{}}Yb$^+$ (E3)/\\ Yb$^+$ (E2) \\ and Yb$^+$ (E3)/\\ Sr clock \\ comparisons$^*$\end{tabular}} &
  \multicolumn{1}{l|}{\begin{tabular}[c]{@{}l@{}}PTB \\ Braunschweig\end{tabular}} &
  \multicolumn{1}{l|}{Dark matter} &
  \multicolumn{1}{l|}{\begin{tabular}[c]{@{}l@{}}Optical clock\\ comparisons\end{tabular}} &
  \multicolumn{1}{c|}{(2)} &
  \multicolumn{1}{c|}{\cite{Filzinger:2023zrs}} \\ \hline
 &
   &
   &
   &
   &
   \\
\textbf{Proposed} &
   &
   &
   &
   &
   \\ \hline
\multicolumn{1}{|l|}{\textbf{Experiment}} &
  \multicolumn{1}{l|}{\textbf{Laboratory}} &
  \multicolumn{1}{l|}{\textbf{Source}} &
  \multicolumn{1}{l|}{\textbf{Method/technique}} &
  \multicolumn{1}{c|}{\textbf{Portals}} &
  \multicolumn{1}{c|}{\textbf{Ref.}} \\ \hline
\multicolumn{1}{|l|}{AION} &
  \multicolumn{1}{l|}{\begin{tabular}[c]{@{}l@{}}Oxford (10m \\ scale), \\ CERN (100m \\ scale, TBC)\end{tabular}} &
  \multicolumn{1}{l|}{Dark matter} &
  \multicolumn{1}{l|}{\begin{tabular}[c]{@{}l@{}}Atom \\ interferometry\end{tabular}} &
  \multicolumn{1}{c|}{(2)} &
  \multicolumn{1}{c|}{\cite{Badurina:2021rgt}} \\ \hline
\multicolumn{1}{|l|}{ALPS-II} &
  \multicolumn{1}{l|}{DESY} &
  \multicolumn{1}{l|}{Laboratory} &
  \multicolumn{1}{l|}{\begin{tabular}[c]{@{}l@{}}Light-shining-\\ through-a-wall\end{tabular}} &
  \multicolumn{1}{c|}{(1,2,3)} &
  \multicolumn{1}{c|}{\cite{Ortiz:2020tgs}} \\ \hline
\multicolumn{1}{|l|}{ARIADNE} &
  \multicolumn{1}{l|}{\begin{tabular}[c]{@{}l@{}}IBS Daejeon \\ (TBC)\end{tabular}} &
  \multicolumn{1}{l|}{Laboratory} &
  \multicolumn{1}{l|}{\begin{tabular}[c]{@{}l@{}}Nuclear magnetic \\ resonance\end{tabular}} &
  \multicolumn{1}{c|}{($2\times 3,3$)} &
  \multicolumn{1}{c|}{\cite{Arvanitaki:2014dfa}} \\ \hline
\multicolumn{1}{|l|}{BabyIAXO} &
  \multicolumn{1}{l|}{DESY} &
  \multicolumn{1}{l|}{Solar} &
  \multicolumn{1}{l|}{Helioscope} &
  \multicolumn{1}{c|}{(1,3)} &
  \multicolumn{1}{c|}{\cite{IAXO:2020wwp, IAXO:2019mpb}} \\ \hline
\multicolumn{1}{|l|}{ALPHA} &
  \multicolumn{1}{l|}{\begin{tabular}[c]{@{}l@{}}Oakridge \\ National \\ Laboratory\end{tabular}} &
  \multicolumn{1}{l|}{Dark matter} &
  \multicolumn{1}{l|}{\begin{tabular}[c]{@{}l@{}}Plasma \\ haloscope\end{tabular}} &
  \multicolumn{1}{c|}{(1,3)} &
  \multicolumn{1}{c|}{\cite{Lawson:2019brd,Millar:2022peq}} \\ \hline
\multicolumn{1}{|l|}{BREAD} &
  \multicolumn{1}{l|}{Fermilab} &
  \multicolumn{1}{l|}{Dark matter} &
  \multicolumn{1}{l|}{\begin{tabular}[c]{@{}l@{}}Dish-antenna \\ haloscope\end{tabular}} &
  \multicolumn{1}{c|}{(1,3)} &
  \multicolumn{1}{c|}{\cite{BREAD:2021tpx}} \\ \hline
\multicolumn{1}{|l|}{CADEx} &
  \multicolumn{1}{l|}{Canfranc} &
  \multicolumn{1}{l|}{Dark matter} &
  \multicolumn{1}{l|}{\begin{tabular}[c]{@{}l@{}}Microwave \\ cavity\end{tabular}} &
  \multicolumn{1}{c|}{(1,2,3)} &
  \multicolumn{1}{c|}{\cite{Aja:2022csb}} \\ \hline
\multicolumn{1}{|l|}{DM-Radio} &
  \multicolumn{1}{l|}{Stanford} &
  \multicolumn{1}{l|}{Dark matter} &
  \multicolumn{1}{l|}{\begin{tabular}[c]{@{}l@{}}Lumped \\ element\end{tabular}} &
  \multicolumn{1}{c|}{(1,3)} &
  \multicolumn{1}{c|}{\cite{DMRadio:2022pkf}} \\ \hline
\multicolumn{1}{|l|}{FLASH} &
  \multicolumn{1}{l|}{INFN Frascati} &
  \multicolumn{1}{l|}{Dark matter} &
  \multicolumn{1}{l|}{\begin{tabular}[c]{@{}l@{}}Microwave \\ cavity\end{tabular}} &
  \multicolumn{1}{c|}{(1,3)} &
  \multicolumn{1}{c|}{\cite{Alesini:2019nzq}} \\ \hline
\multicolumn{1}{|l|}{IAXO} &
  \multicolumn{1}{l|}{DESY (TBC)} &
  \multicolumn{1}{l|}{Solar} &
  \multicolumn{1}{l|}{Helioscope} &
  \multicolumn{1}{c|}{(1,3)} &
  \multicolumn{1}{c|}{\cite{IAXO:2019mpb}} \\ \hline
\multicolumn{1}{|l|}{MADMAX} &
  \multicolumn{1}{l|}{DESY} &
  \multicolumn{1}{l|}{Dark matter} &
  \multicolumn{1}{l|}{\begin{tabular}[c]{@{}l@{}}Dielectric \\ haloscope\end{tabular}} &
  \multicolumn{1}{c|}{(1,3)} &
  \multicolumn{1}{c|}{\cite{Beurthey:2020yuq}} \\ \hline
\multicolumn{1}{|l|}{MAGIS} &
  \multicolumn{1}{l|}{\begin{tabular}[c]{@{}l@{}}Fermilab \\ (100m scale)\end{tabular}} &
  \multicolumn{1}{l|}{Dark matter} &
  \multicolumn{1}{l|}{\begin{tabular}[c]{@{}l@{}}Atom \\ interferometry\end{tabular}} &
  \multicolumn{1}{c|}{(2)} &
  \multicolumn{1}{c|}{\cite{MAGIS-100:2021etm}} \\ \hline
\multicolumn{1}{|l|}{STAX} &
  \multicolumn{1}{l|}{\begin{tabular}[c]{@{}l@{}}CERN/\\ Unknown\end{tabular}} &
  \multicolumn{1}{l|}{Laboratory} &
  \multicolumn{1}{l|}{\begin{tabular}[c]{@{}l@{}}Light-shining-\\ through-a-wall\end{tabular}} &
  \multicolumn{1}{c|}{(1)} &
  \multicolumn{1}{c|}{\cite{Miyazaki:2022kxl}} \\ \hline
\multicolumn{1}{|l|}{WISPLC} &
  \multicolumn{1}{l|}{U. Hamburg} &
  \multicolumn{1}{l|}{Dark matter} &
  \multicolumn{1}{l|}{Lumped element} &
  \multicolumn{1}{c|}{(3)} &
  \multicolumn{1}{c|}{\cite{Zhang:2021bpa}} \\ \hline
% \caption{Running and proposed experiments dedicated to ultralight FIPs. 
% Experiments labelled with an asterisk (*) have no official name.
% The name appearing in the table is just a descriptive name.
% For consistency with the summary tables in the other sections of this report, we use the following notation for the portals: (1) Vector, (2) Scalar, (3) Pseudoscalar.  
% There is no fermion portal (number 4 in the MeV-GeV summary tables).
% % {\color{red}--- Add hyperlink(s) to the summary tables when ready}).
% }
% \label{tab:my-table}\\
\end{longtable}

%%%%

%%%%

%%%%

%%===============
%% Light DM
%%===============

\clearpage
\section{Light (MeV-GeV) dark matter: theory and experiments}
\label{sec:light_DM} 
%========================================================

\subsection{Introduction}
\label{ssec:LDM_intro}

Sub-GeV dark sectors arise in many well-motivated extensions of the Standard Model (SM) addressing long-standing problems in fundamental physics.  
One of the sharpest motivations for dark sectors in this mass range is the need to explain the observed dark matter (DM) abundance of our universe.   In the last decade, the long-standing paradigm of Weakly Interacting Massive Particle (WIMP) DM at the GeV-TeV scale has been severely challenged by null results in direct and indirect DM experiments as well as at the LHC.  Thermal relic DM with lower masses in the MeV-GeV range, similar to that of ordinary matter, is so far much less constrained.  Unlike heavier WIMPs, however, light DM interactions with SM fields must proceed via new light mediators in order to reproduce the observed DM relic abundance, and thus result in a dark sector at (sub)-GeV scales. These light DM models present several attractive features, including sharp and testable milestones for theories where DM annihilates directly into SM particles.

Beyond DM, light dark sectors are motivated by several other open problems in fundamental physics. For example, dark sectors containing sterile neutrinos can explain the lightness of SM neutrinos; richer dark-sector models can generate the baryon-antibaryon asymmetry of the universe and potentially address the hierarchy problem; and dark sectors can address the strong-CP problem via axions or axion-like particles (ALPs).

The FIPS approach to sub-GeV dark sectors builds on the framework of Physics Beyond Colliders (PBC), which relies on a categorization of the dimension-four and -five portal interactions through which a total SM singlet can interact with SM fields. The limited number of  portals greatly simplifies both model classification and the characterization of the related experimental signatures.  Of particular interest are the renormalizable {\em vector portals} mediated by a dark vector boson; {\em scalar portals} mediated by a new scalar mixing with the SM Higgs boson; and {\em neutrino portal} operators, mediated by a heavy neutral lepton, as well as dimension-five {\em axion portal} interactions.  Vector portal and scalar portal benchmarks are in particular motivated by MeV-GeV DM models, where the new vector or scalar serves as the particle mediating DM freezeout. 

Thermal relic DM, proceeding in direct thermal contact with the SM, provides one of the simplest and most predictive scenarios for the origin of DM.  FIPs vector and scalar portal benchmarks are motivated by minimal extensions of the SM that can give rise to such thermal relic DM at the  MeV-GeV scale without violating existing cosmological, astrophysical, or  terrestrial bounds.   These minimal extensions consist of a DM particle and a mediator particle, and give rise to different experimental signatures and targets depending on whether DM freezes out directly to SM final states (``direct freezeout'') or into  mediator particles (``secluded freezeout'').  In the case of direct freezeout, the mediator can typically decay into DM, which gives rise to missing energy signatures and sharply-characterized production targets at terrestrial experiments.  For secluded freezeout, the mediator is lighter than DM, motivating searches for visibly-decaying mediators within a broad range of visible parameter space consistent with the observed DM abundance.

The {\em vector portal} benchmark scenarios describe the possible renormalizable interactions of a new vector boson with the SM.    PBC defined three benchmarks in this category, all based on a dark photon kinetically mixing with SM hypercharge.  
Benchmark BC1 covers the minimal visible dark photon model, and BC2 the invisible dark photon decaying to light dark matter.  BC3 covers millicharged particles, which arise in the limit where the dark photon mass goes to zero. 

The FIPs vector portal benchmarks have been expanded beyond the PBC dark photon benchmarks to cover two new directions.  First, we  consider models where anomaly-free global symmetries of the SM are gauged.  In these models, SM particles are directly charged under a new fifth force, and the new $U(1)$ gauge boson couples directly to the corresponding symmetry current.  The benchmark BC$_{B-L}$ introduces a new gauge boson coupled to $U(1)_{B-L}$.  Gauging $B-L$ requires the introduction of right-handed neutrinos, which offers opportunities to connect this benchmark to neutrino benchmarks.  Future directions of interest include  gauge bosons coupled to the lepton-flavor-violating anomaly-free symmetries $U(1)_{B-3L_i}$ and $U(1)_{L_i-L_j}$.  Such lepton-flavor-violating symmetries offer connections to the long-standing tensions in measurements of the muon anomalous magnetic moment, or can yield hadrophilic signatures that substantially change optimal search strategies.     
%{\bf see Foldenauer section?} 
 %
Second, BC$_{iDM}$ expands BC2 to the case of  {\em inelastic} DM, with the complex DM of BC2 now split into two non-degenerate real states. The mass splitting suppresses both direct and indirect detection signals and alters the thermal relic targets for terrestrial searches.  The off-diagonal coupling of the dark photon to the two DM states motivates a semi-visible search strategy. 
%{\bf see Junius section?} 
  
The {\em scalar portal} benchmark scenarios, from PBC, cover a new scalar mixing with the SM Higgs boson.  In BC4 the couplings of the new scalar mediator depend only on mass and mixing angle, while BC5 covers cases where the mixed quartic coupling gives rise to additional pair production in various channels. FIPS work in progress here includes development of a consistent recommendation for scalar branching ratios to SM particles.

{\em Neutrino portal} benchmarks BC6, BC7, and BC8 from PBC each introduce one new heavy neutral lepton $N$ with coupling to a single SM lepton flavor, respectively electron, muon, and tau.  FIPS is in the process of generalizing these benchmarks to include two additional scenarios with richer flavor structures that are compatible with active neutrino oscillation data.

Finally, the PBC {\em ALP portal} benchmark scenarios BC9, BC10, and BC11 cover ALP couplings to photons, SM fermions (with some flavor assumptions), and gluons respectively.  FIPS is actively developing new benchmarks here that (i) include couplings to $W$ bosons, and (ii) accurately account for the combinations of operators that arise from well-defined UV completions followed by renormalization group running.  The interplay of different combinations of operators is a generic prediction of consistent theories, includes flavor-off-diagonal axion couplings, and can have a substantial impact on the experimental signatures of ALPs.

%Both ALPs and dark sectors yielding light thermal relic DM represent well-defined and categorizable classes of minimal models and benchmarks, and provide useful and well-defined targets for experiments probing this parameter space, .  The FIPs approach to sub-GeV dark sectors is

%These interactions determine the  new boson's SM decay rates as a function of its mass.

These benchmarks serve as sharply-defined and well-motivated targets to guide exploration of the parameter space for MeV-GeV dark sectors.
Complementing direct and indirect searches for DM, a large program of accelerator-based experiments has been proposed to explore the FIP parameter space, including generic searches for dark sector mediators, exotic Higgs decays, long-lived particles, and milli-charged particles. 
This program comprises efforts at various degrees of execution, from conceptual designs to active experiments leveraging a wide range of production mechanisms and detection techniques. At colliders, FIPs can be produced either directly in electron-positron or proton-proton collisions, or through the decay of another (SM) particle. In hadronic environments, SM backgrounds are overwhelming in many cases. Dedicated detectors located next to ATLAS, CMS and LHCb have been proposed to increase the sensitivity to long-lived and milli-charged particles (neutrino physics is also accessible at some proposals). Electron-positron colliders provide complementary sensitivity to semi-visible and invisible final states. FIP searches are also performed at extracted beams with fixed target and beam dump setups, as well as meson factories and neutrino experiments. They provide unique sensitivity to (very) small dark sector-SM couplings, ideally complementing the reach of colliders. The large variety of beams and techniques used can probe many couplings and final states. For example, protons are uniquely suited to probe hadronic couplings, while muon beams open a unique window on the second generation of leptons.

This section presents an overview of current theoretical investigations and experimental concepts to search for sub-GeV dark sector particles. In Sec.~\ref{ssec:toro}, we begin the discussion with a theoretical introduction to sub-GeV dark matter, 
followed by a survey of cosmological and astrophysical probes (Sec.~\ref{ssec:boehm}), and an overview of indirect detection searches (Sec.~\ref{ssec:calore}). We then discuss the Snowmass effort in the context of FIPs at high-intensity experiments (Sec.~\ref{gori}). Secs.~\ref{ssec:harris}-\ref{ssec:kling} are dedicated to CERN experiments searching for dark particles with LHC experiments and ancillary detectors and facilities, followed by efforts with extracted beams in Sec.~\ref{ssec:crivelli}-\ref{ssec:golutvin}. The situation at DESY (Sec.~\ref{trevisani}), Frascati (Sec.~\ref{raggi}), KEK (Sec.~\ref{ssec:hearty}), JLAB (Sec.~\ref{ssec:battaglieri}), FNAL (Sec.~\ref{ssec:tran}), and SLAC (Sec.~\ref{ssec:pottgen}) is then presented. The section concludes with a collection of new ideas for searches for MeV-GeV dark sector particles at accelerator experiments.

\subsection{Light (MeV-GeV) DM and related mediators: Theory overview -- {\it N.~Toro}}
\label{ssec:toro}
{\it Author: Natalia Toro,  <ntoro@slac.stanford.edu> } 

% \documentclass[prd]{revtex4}
% %\documentclass[a4paper,11pt]{article}
% \pdfoutput=1
% %\usepackage{jheppub}

% % \usepackage[utf8]{inputenc}
% %         \usepackage[T1]{fontenc}
% % \usepackage{lmodern}
% % \usepackage{textcomp}
% % \usepackage{microtype}

% % \usepackage[english]{babel}
% % \usepackage[autostyle, english = american]{csquotes}
% % \MakeOuterQuote{"}

% % \usepackage{amsmath}
% % \usepackage{amssymb}
% % \usepackage{mathtools}
% % \usepackage{upgreek}
% % \usepackage{units}
% % \usepackage{slashed}
% % \usepackage{physics}
% % \usepackage{authblk}
% % \usepackage{booktabs}
% % \usepackage{multirow}
% \usepackage{graphicx}
% % \usepackage{adjustbox}
% % \usepackage{xcolor}
% % \usepackage{wrapfig}
% % \usepackage{xspace}

% % \usepackage{tcolorbox}
% % \usepackage{rotating}
% % \usepackage{lineno}
% % \setcounter{secnumdepth}{4}

% % \usepackage[colorlinks=true,linkcolor=black,citecolor=blue,urlcolor=blue,bookmarksopen]{hyperref}

% \newcommand{\todonote}[1]{{{\textbf{!!#1!!}}}}

% \begin{document}

% \title{MeV-to-GeV Dark Matter and Mediators: Theory Overview}
% \author{Natalia Toro}
% \affiliation{SLAC National Accelerator Laboratory, Menlo Park CA 94025, USA}

% \begin{abstract}

\subsubsection{abstract}
This contribution summarizes models of MeV-to-GeV mass dark matter, light mediators related to dark matter, and the observable signatures of each.  We focus on motivating benchmarks frequently used for accelerator-based searches aand providing a perspective on the complementarity of different experimental approaches. 
% \end{abstract}

% \maketitle

\subsubsection{Introduction and Framework}
%\subsection{The Dark Sector Picture} %[0.7p]
The nature of dark matter is among the foremost open problems facing fundamental physics today. Thinking broadly about what dark matter could be, where it came from, and how it interacts with itself and with familiar matter has been a driving force in motivating and guiding dark sector physics.  

\begin{figure*}
\includegraphics[width=0.95 \textwidth]{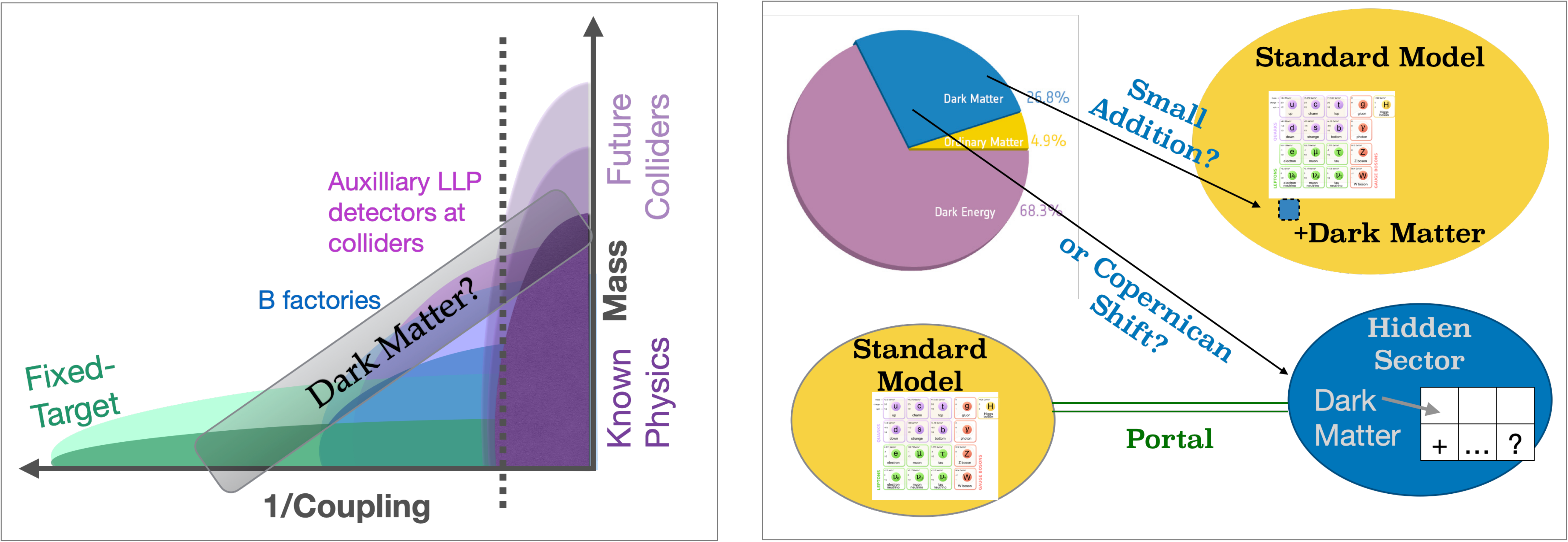}
\caption{Left: In many long-standing models, dark matter emerges as an byproduct of models that address deficiencies in the Standard Model (SM) by extend its matter and/or symmetries.  The dark sector program, by contrast, explores the possibility that dark matter is our first glimpse of physics unrelated to the SM and hence neutral under its symmetries. Dark sectors are naturally classified according to the ``portal'' interactions, restricted by symmetries, through which they couple to the SM.  Right: Schematic illustation of the mass vs. coupling space for dark matter and the complementarity between LHC experiments and other approaches probing lower energies but improving small-coupling sensitiivty through intensity and sensitivity to rare reactions and/or long lifetimes.}
\label{fig:energy-coupling-cartoon}\label{fig:darkSectorShift}
\end{figure*}

The idea that dark matter constituents could have similar mass scales to electrons and protons is appealingly simple. Its realization requires only a modest enlargement of the Standard Model, but simultaneously implies a Copernican-level shift in perspective. 
A few decades ago, conventional wisdom held that dark matter's constituents should either be much heavier than those of familiar matter --- weakly interacting massive particles, or WIMPs, with TeV-scale masses --- or exponentially lighter, as in the case of QCD axions. Both of these DM candidates revolve around the Standard Model (SM), with dark matter emerging as a byproduct of models that correct the SM's deficiencies by enlarging its matter sector and symmetries (in WIMP models, the dark matter itself is SM-charged matter; in axion models, new charged matter is added at high scales to realize the Peccei-Quinn mechanism \cite{Shifman:1979if,Kim:1979if,Zhitnitsky:1980tq,Dine:1981rt}). By contrast, DM at the GeV mass scale or below \emph{must} be neutral under Standard Model forces to be consistent with collider searches. The expectation that DM should have some non-gravitational interactions --- both to explain its production and as a prerequisite for detection --- implies that DM interacts under \emph{new} forces that operate at well-explored energy scales but do not couple appreciably to SM matter.  This is the Copernican shift in perspective alluded to earlier: the idea that the gauge symmetry structure of the SM may describe only a (cosmologically minute) fraction of the Universe, while a \emph{dark sector} with distinct but analogous structure accounts for the remainder (See Figure \ref{fig:darkSectorShift}(left)). 

A dark sector's neutrality under SM gauge symmetries makes it challenging to detect, but also provides a guide to its allowed interactions with matter.  The relevant interactions of SM-neutral matter with SM matter permitted by SM gauge symmetries and Lorentz invariance, are few in number: a dark-sector $U(1)$ gauge boson (or ``dark photon'') 
$A^\prime_\mu$, scalar $\phi$, 
or neutral fermion $\psi$ can mix with the photon, Higgs boson, or neutrinos respectively through the operators 
$\frac{\epsilon_Y}{2} F^\prime_{\mu\nu} B^{\mu\nu}$, 
$\epsilon_h |h|^2 |\phi|^2$ (and/or $A_h |h|^2 \phi$ for $\phi$ neutral), or 
$\epsilon_\nu (hL) \psi$ 
respectively, as reviewed in e.g.~\cite{Lanfranchi:2020crw, Gori:2022vri,Krnjaic:2022ozp}. These interactions are often referred to as \emph{portals} and the dark-sector particles mixing with SM fields as \emph{mediators}. 
Portal couplings are generically small, with a characteristic scale of $10^{-6}$ to $10^{-2}$ if generated from one- or two-loop radiative effects (and even smaller if generated non-perturbatively). 
Through mixing, the mediator acquires an effective coupling to SM matter proportional to its charge (dark photon portal) or Yukawa coupling (scalar portal).  The smallness of these couplings can generate small mass-scales for dark sector particles \cite{Morrissey:2009ur}, or small mass scales can arise from dimensional transmutation \cite{Strassler:2006im,Feng:2008ya}, much as the electron and proton masses are generated in the Standard Model. 

The small scales typical of dark sector masses and couplings inform searches for hidden-sector dark matter and mediators, which are limited not so much by \emph{energy} requirements (as is the case at the LHC) but by the intensity, sensitivity, and/or lifetime required to explore small couplings.  As a result, fixed-target experiments, searches at B-factories,  auxiliary LHC detectors all have substantial sensitivity and mutual complementarity (See Figure \ref{fig:energy-coupling-cartoon}(right)).  In addition, searches for dark matter and mediators at accelerators, direct and indirect detection, and cosmic probes of DM annihilation and self-interactions all have rich connections among them and opportunities for complementarity, which we emphasize throughout this contribution. 

{\bf Following the Trail of the Dark Matter Abundance - } %[0.5p]
It is encouraging that the dark sector paradigm simply explains the most salient qualitative properties of dark matter: the weakness of DM-SM interactions is explained by the SM-neutrality of DM, as discussed above.  Its cosmological stability is naturally explained if the DM is the lightest particle charged under a new dark-sector gauge symmetry. However, the only measured \emph{quantitative} property of DM is its mass abundance --- and portal interactions at the scale motivated above will necessarily thermalize DM with SM matter. Their strength and those of other DM interactions control the freeze-out of the DM from chemical equilibrium at temperatures comparable to its mass, and therefore its abundance at late times (as is also the case for thermal WIMP dark matter). Thus, the abundance of DM is a powerful hint to refine and focus the basic framework introduced above. 

Scenarios where DM was once in thermal equilibrium with ordinary matter can be governed by either \emph{direct} or \emph{indirect} freeze-out.  Direct freeze-out models --- those where DM annihilates into SM final states through its portal interactions --- define sharp milestones in parameter space within reach of upcoming experiments.  These have become important benchmarks for the accelerator and direct detection communities, with some scenarios also accessible in future indirect detection. In indirect freeze-out models, DM equilibrates with another state in dark-sector, whose decays and/or annihilations with SM particles also play a role in determining the DM abundance. Though somewhat less  predictive, these models broaden the motivation for these experiments as well as for long-lived particle decays that explore the structure of the dark sector.  We will treat each case in turn. 

\subsubsection{WIMP-Like (Direct) Freeze-Out: Sharp Predictions and Complementarity}  %[0.7p]
\label{sec:directthermal}
In the simplest thermal freeze-out scenario, the DM abundance today is determined by the interplay of the DM annihilation cross-section into SM particles --- which maintains chemical equilibrium between the SM and DM populations as the Universe cools below the DM mass --- and Hubble expansion --- which dilutes the DM abundance after it can no longer efficiently annihilate.  From the observed DM mass density, one can infer an annihilation cross-section (roughly independent of the DM mass) of $\langle \sigma v\rangle \approx 2-3\,10^{-26} \rm{cm}^3/\rm{s} \sim (20\,\rm{TeV})^{-2}$ at temperatures of $\sim 1/10$ the DM mass.  Larger cross-sections would lead to \emph{excess} DM annihilation, and hence depletion of DM to less than the observed DM abundance, while smaller cross-sections would lead to an excess of DM relative to observations unless some other depletion mechanism is present. 
The ``WIMP miracle'' refers to the proximity of this inferred cross-section scale to the $(\alpha_2/\rm{TeV})^2$ scale expected for TeV-scale particles interacting through $SU(2)_L$.  Of course, the same cross-section scale can also result from interactions with a weaker coupling and lower mass scale, as noted in \cite{Feng:2008ya}. 

This motivates examining the simplest possible thermal freeze-out scenarios in a dark sector: DM annihilating through a portal coupling into SM matter.  This scenario is readily realized in vector-portal models, where \textbf{the correct DM relic abundance is obtained by dark-sector thermal freeze-out for the perturbative mixing range motivated above, MeV-to-GeV DM masses, and modest ratios between DM and mediator masses}. By contrast, scalar portal direct freeze-out is excluded by meson decay constraints \cite{Krnjaic:2015mbs}, further motivating our focus below on the vector portal.  We also note that neutrino-portal counterparts of direct freeze-out can be realized with additional matter \cite{Batell:2017cmf,Escudero:2018fwn,Blennow:2019fhy,Ballett:2019pyw}. 

Direct vector-portal freeze-out has distinct phenomenology depending on the spin of the DM, which is usually taken to be a fermion or scalar. % (See also spin-1 DM models in \cite{}). 
Additionally, the theory admits DM mass terms that preserve the $U(1)$ gauged by the dark photon $A^\prime$, as well as DM mass terms that break this symmetry (the latter are compatible with all symmetries of the theory, since this $U(1)$  is already broken by the $A^\prime$ mass, and are readily generated by the Higgs mechanism). Depending on the relative sizes of these mass terms, fermionic DM can be Dirac, pseudo-Dirac (inelastic, i.e. DM is split into two mass eigenstates), or Majorana while the scalar theory can have elastic or inelastic interactions.  

The above-mentioned cases have surprisingly different phenomenology given their common theoretical structure. The differences arise from two effects: the velocity-scaling of annihilation and scattering cross-sections, and the depletion in inelastic models of the heavier DM state's cosmological abundance in inelastic models. Because the dark photon couples off-diagonally, tree-level interactions of the surviving light DM state are kinematically forbidden at low velocities.  These suppressions, and the kinematic conditions relevant for thermal freeze-out, CMB reionization by DM annihilation, indirect detection and direct detection are summarized in Table \ref{tab:vscaling} and elaborated below. 

\begin{table}
\begin{minipage}[ct]{0.63\textwidth}
\begin{tabular}{lcccc}
\multicolumn{2}{c}{Spin \& Mass Structure} & Annihilation & Scattering  & Prospects\\
\hline
Scalar Elastic &($m_{\slash{\!\!\!S}}=0$) & $v^2$ & $1$ & Acc, DD, ID\\
Scalar Inelastic &($m_{\slash{\!\!\!S}} \neq 0$) & $v^2 f_h$  & $f_h$ & Acc\\
\hline
Dirac& ($m_{\slash{\!\!\!S}}=0$) & $1$ & $1$ & $\times$ \tiny{excl. by CMB}\\
Pseudo-Dirac& ($m_{\slash{\!\!\!S}}\ll m_S$) & $f_h$ & $f_h$ & Acc\\
Majorana& ($m_{\slash{\!\!\!S}}\gtrsim m_S$) & $v^2$ & $v^2$ & Acc, ID, \tiny{DD}\\
\hline
\end{tabular}
\end{minipage}
\vskip 2mm
\begin{minipage}[ct]{0.24\textwidth}
\begin{tabular}{lcc}
Probe (reaction) & velocity $v$ & $f_h$\\
\hline
\textbf{Acc}elerator (prod) & $\sim c$ & N/A\\
Freeze-Out (ann) &  $\sim 1/3 c$ & $\sim 1$\\
\textbf{CMB} Reionization (ann) & $\lesssim 10^{-6}$ & $\ll 1$\\
\textbf{I}ndirect \textbf{D}etection (ann) & $\sim 10^{-3}$ & $\ll 1$\\
\textbf{D}irect \textbf{D}etection (scat) & $\sim 10^{-3}$ & $\ll 1$\\ 
\hline
\end{tabular}
\end{minipage}
\caption{\label{tab:vscaling} Left: Scaling of tree-level annihilation and scattering signals with CM-frame velocity $v$ and, in inelastic models, with the fraction $f_h$ of DM in the heavier mass eigenstate. $m_{\slash{\!\!\!S}}$ and $m_S$ refer to masses breaking and respecting the $U(1)_D$ symmetry broken by the dark photon mass.  Right: Characteristic velocity scale for DM in the epochs and conditions relevant to different experimental probes. The fractional abundance $f_h$ during the recombination epoch and in the present era depends on the precise mass splitting, but vanishes if the heavy state is unstable to decays and is exponentially suppressed but potentially phenomenologically relevant if its abundance is depopulated only through scattering (see e.g. \cite{Baryakhtar:2020rwy,CarrilloGonzalez:2021lxm}).  As a result of these suppressions, all cases considered except the Dirac fermion are presently viable; the prospects for near-term detection are indicated in the last column.}
\end{table}

{\bf Lessons from CMB and Prospects for Indirect Detection - } %[1p]
The CMB power spectrum sensitively probes DM annihilation near the time of recombination, which deposits electromagnetic energy. This energy partially reionize Hydrogen, ``smearing'' the surface of last scattering and altering the CMB power spectrum \cite{Padmanabhan:2005es,Slatyer:2009yq,Slatyer:2015jla}. Planck \cite{Planck:2018vyg} constraints $\langle \sigma v \rangle_{\mbox{eV}} \lesssim 3\cdot 10^{-26} \rm{cm}^3/\rm{s} \left(20 \rm{GeV}/m_{DM}\right)$ for annihilation to electrons (with only mild modifications to this bound different annihilation products, excepting neutrinos which are unconstrained).  This constraint robustly excludes sub-GeV thermal DM \emph{if it has a temperature-independent annihilation rate, i.e. elastic $s$-wave annihilation} into non-neutrino SM final states. From Table \ref{tab:vscaling}(left), we see that Dirac DM is the \emph{only} case of vector-portal DM satisfying this assumption.  For other DM types, low-temperature annihilations are suppressed by $v^2$ due to $p$-wave annihilation and/or by the abundance $f_h$ of the heavier component in inelastic DM, which is exponentially suppressed by collisions \cite{CarrilloGonzalez:2021lxm} and may be fully eliminated if the heavy DM state decays into the lighter one. \textbf{Hence, thermal vector-portal DM is generically consistent with CMB constraints on annihilation.}

Relative to the CMB bounds, indirect detection of light DM faces several challenges. It is subject to uncertainty in the DM halo profile, and sensitivity depends on the spectrum of gamma rays and therefore on the precise DM annihilation products.  Nonetheless, next-generation proposals for MeV gamma-ray detection, which will improve sensitivity to DM annihilation in the Milky Way by up to four orders of magnitude (see e.g. \cite{Coogan:2021sjs}), present a great opportunity.  Because the velocity dispersion of DM in the Milky Way is much larger than at recombination, models with $p$-wave annihilation predict larger cross-sections in indirect detection than at recombination. With these gains (and assuming an Einasto halo profile), proposed detectors such as AMEGO, GECCO, and MAST will probe the $10^{-31} \rm{cm}^3/\rm{s}$ annihilation cross-sections expected for thermal DM with $p$-wave annihilation.  This is one possible route to discovery of models such as Majorana and scalar elastic DM. Inelastic models may also be detectable in this way, in the subset of models where a small population of the heavy DM state survives to the present day.

\begin{figure*}
\includegraphics[width=0.95\textwidth]{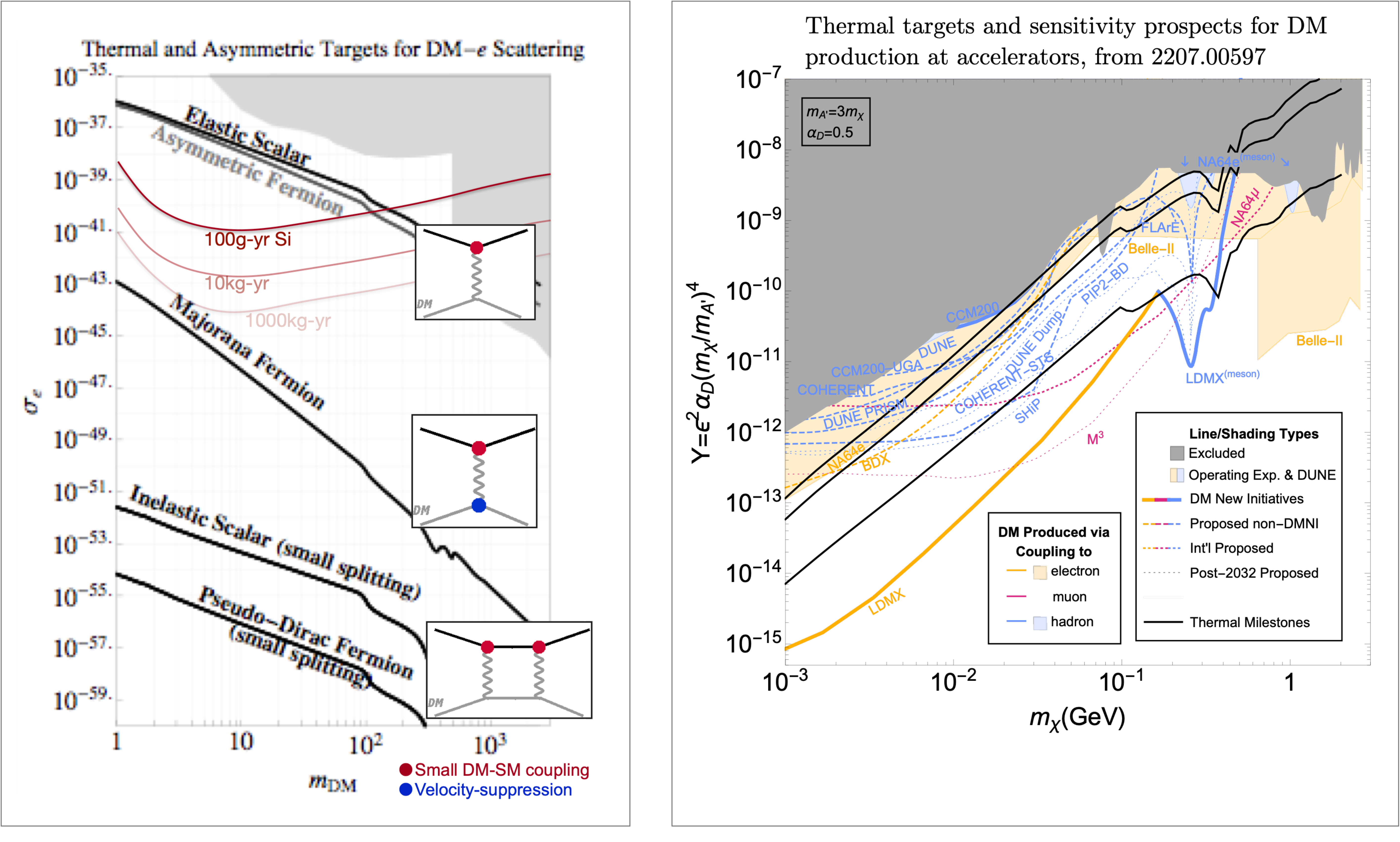}
\caption{Left: Predicted cross-sections for DM-electron scattering based on thermal freeze-out (black lines), accounting for the suppressions indicated in Table \ref{tab:vscaling}(left) (1-loop scattering dominates the predicted signal for inelastic models), along with current constraints (gray regions) and projections from \cite{Essig:2018tss} (curves in shades of red) for 100 kg to tonne-scale silicon detectors' potential sensitivity limited by yield and/or solar neutrino backgrounds. Right (copied from \cite{Krnjaic:2022ozp}): Predicted interaction strength $y \equiv \epsilon^2\alpha_D (m_\chi/m_{A'})^4$ based on thermal freeze-out (black lines), accelerator-based constraints (gray regions) and sensitivity prospects for ongoing/funded (shaded) and proposed (solid, dashed, and dotted curves) accelerator-based experiments. Plotting sensitivity in $y$ at the chosen parameter values $\alpha_D=0.5$, $m_{A^\prime}/m_{DM}=3$ is  conservative in that, as discussed in \cite{Izaguirre:2015yja,Berlin:2020uwy}, varying away from these parameters within the range of applicability of direct freeze-out generally \emph{improves} experiments' sensitivity in $y$ (except near resonance \cite{Feng:2017drg,Berlin:2020uwy} $m_{A^\prime}/m_{DM}\approx2$).}
\label{fig:accelDDprospects}\label{fig:thermalPlot}
\end{figure*}

{\bf The Need for a Multi-Pronged Program - } %[0.7p]
More broadly, an important take-away from the above discussion is that \textbf{drawing correct conclusions from experiments relying on cosmological DM requires careful consideration of interactions' dependence on both the DM velocity and, in inelastic models, the reduced abundance of the heavier mass eigenstate.} 
This is a challenge from the point of view of broadly exploring the parameter space of light DM.  From this perspective, it underscores the importance of an accelerator-based search program.
Figure \ref{fig:accelDDprospects}(left) illustrates the impact of non-relativistic suppressions on direct detection prospects (left), where the scattering rates for models in Table \ref{tab:vscaling}(left) span 20 orders of magnitude, and the relatively compact and accessible parameter space for accelerator production of light DM (right), where the (semi)relativistic production kinematics renders these effects minor.  

At the same time, the range of possibilities within this simple class of models also \textbf{underscores the tremendous value of a program that simultaneously pursues multiple avenues to DM detection}.
For example, in the ``optimistic'' elastic scalar case, near-future direct detection experiments could detect electron recoils, while ongoing or next-generation accelerator based experiments could also detect its production.  Combining the information from these two probes would allow direct confirmation of the DM's cosmological abundance and spin, independent measurements of the DM and mediator masses, measurement of the mediator coupling to SM matter, and inference of its coupling to DM itself!  In addition, future indirect detection experiments could directly probe its annihilation properties.  

Even in more pessimistic scenarios there is interesting interplay between approaches.  For example, Majorana DM can be readily produced at accelerators. Though direct and indirect detection signals are both velocity-suppressed, superfluid helium detectors can access Majorana DM scattering off nucleons for $m_{DM}\gtrsim 100$ MeV, while $p$-wave annihilation reaction could be detected in future indirect detection probes.  Inelastic scenarios offer particularly challenging phenomenology, but could be identified through long-lived decays of the heavy state at accelerators (see e.g.~\cite{Izaguirre:2015zva,Berlin:2018jbm}), boosted DM signals, or,  if the heavy state is cosmologically stable, down-scattering of this state in DM detectors which leads to a monochromatic energy deposition signal\cite{Baryakhtar:2020rwy,CarrilloGonzalez:2021lxm}.  In each case, a discovery in one probe would provide ample motivation to push the technology in others, enabling rich characterization of the physics of the dark sector!

{\bf Combination with Cosmic Probes of Light DM - }
There are several promising avenues for constraining or corroborating light hidden-sector DM using cosmological data. Two that are particularly relevant in our context are measures of the effective number of neutrino species degrees $N_{eff}$ and of DM self-interactions (SI) through its impact on cosmological structure.  Both have the greatest impact at low DM masses $\lesssim 10$ MeV, as elaborated below.  While $N_{eff}$ disfavors DM below 4--9 MeV altogether, SI bounds imply a mass-dependent upper limit on the DM-mediator coupling $\alpha_D$, which can be combined with the accelerator sensitivities shown in Fig.~\ref{fig:thermalPlot}(right) to strengthen these constraints at DM masses below 10 (40) MeV for inelastic (elastic scalar) DM.

%The former disfavors the DM models discussed in Sec.~\ref{sec:directthermal} below 4--9 MeV (depending on DM spin, etc.) and should be similar for the broader range of models discussed in Sec.~\ref{sec:generalizedthermal}.  The latter imply mass-dependent the DM-mediator coupling $\alpha_D$, which can be combined with the accelerator sensitivities shown in Fig.~\ref{fig:thermalPlot}(right) to strengthen these constraints at DM masses below 10 (40) MeV for inelastic (elastic scalar) DM.

The $N_{eff}$ limits and combination with SI bounds are subject to very different systematic uncertainties than other searches we have discussed.  It is generally agreed that direct tests of thermal models, in a terrestrial experiment, are  worthwhile even in the parameter regions most strongly disfavored by cosmology data, given the potential systematics affecting the latter.  For this reason, summary figures like Fig.~\ref{fig:thermalPlot}(right) do not generally show the combined impact of cosmology data. Nonetheless, these bounds offer important context. They shed light on the most promising places to look for light DM and could play vital roles in corroborating future signals and providing complementary insights into both the interactions and the cosmological history of light DM. 

\paragraph{Effective number of light degrees of freedom}
In \cite{Ho:2012ug,Boehm:2013jpa}, it was noted that thermal DM at few-MeV masses would decrease the apparent number $N_{eff}$ of light neutrinos from SM neutrinos by heating the electron-photon plasma relative to neutrinos.  While recent work using improved data over the last decade has strengthened this constraint (see e.g.~%~ $m_{DM}>5-10$ MeV depending on DM spin from 
\cite{Sabti:2019mhn}), it can be somewhat compensated if new dark-sector radiation degrees of freedom  \emph{increase} $N_{eff}$ to compensate. A joint CMB-BBN analysis to constrain models with dark radiation was undertaken by \cite{Giovanetti:2021izc}, finding lower bounds on DM mass of 4--8 MeV in the models discussed in \ref{sec:directthermal} depending on the DM spin.  

While the bounds from \cite{Giovanetti:2021izc} are robust to the inclusion of dark radiation, the presence of other discrepancies between CMB data and local cosmology (in particular, the $H_0$ and $\sigma_8$ tension) does raise the broader question of whether some other feature is missing from $\Lambda$CDM cosmology, the inclusion of which might alter the conclusions of these analyses.  It is in this sense that we describe the low-mass region as ``disfavored'' rather than ``excluded'' above. 

\paragraph{Self-Interactions}Another relevant probe that is advancing  rapidly is tests of DM self-interaction (SI). Though we cannot observe DM SI directly, their impact on structures at Galactic, dwarf, and cluster scales has been studied through the comparison of simulations to observations (see e.g. \cite{Bechtol:2022koa,Adhikari:2022sbh} for recent reviews). Intriguingly, some weak evidence has been found at multiple velocity scales for DM SI at the level of $\sigma/m \sim 0.1-1 \rm{cm}^2/\rm{g}$, though comparable constraints have also been obtained by other analyses.  Importantly, the SI cross-section does \emph{not} depend on the kinetic mixing $\epsilon$, but only on the couplings and masses within the dark sector (and the spin and mass structure of the DM).  It is therefore \emph{highly complementary} to both direct detection (with a signal proportional to $\epsilon^2 \alpha_D$) and accelerator probes (with signals generally scaling as $\epsilon^2$ or $\epsilon^4 \alpha_D$ in missing-energy/momentum and beam dump experiments  respectively).  

Many analyses take as an approximate constraint that the SI cross-section should be $<1 \rm{cm}^2/\rm{g}$\footnote{We note that tighter constraints have been claimed at cluster  scales (e.g. $0.13 \rm{cm}^2/\rm{g}$ from strong lensing \cite{Andrade:2020lqq} and $\sim 0.4 \rm{cm}^2/\rm{g}$ from both cluster mergers and the locations of bright central galaxies (BCGs) \cite{Harvey:2015hha,Harvey:2018uwf}, but analyses of strong lensing and mergers with different assumptions have found substantially weaker bounds \cite{Sagunski:2020spe,Wittman:2017gxn} and the BCG method is also subject to uncertainties in both theoretical and observational modeling. Given the simultaneous hints for self-interaction at the $0.1-0.5\rm{cm}^2/\rm{g}$ level, treating $1 \rm{cm}^2/\rm{g}$ as the approximate scale at which tension is generated with SI constraints seems appropriate. %velocity dependence \cite{Kaplinghat:2015aga}...
}. Using the commonly assumed mass ratio $m_{A'}/m_{DM}=3$ as a reference, this limit on SI implies
 %limitcross-section for elastic scalar DM $\sigma/m_\chi \approx \left(\frac{\alpha_D}{0.5}\right)^2 \left(\frac{10 MeV}{m_\chi}\right)^3 \cdot 72 \rm{cm}^2/\rm{g}$, while in inelastic models it is $\sim \left(\frac{\alpha_D}{0.5}\right)^4 \left(\frac{10 MeV}{m_\chi}\right)^3 \cdot 1 \rm{cm}^2/\rm{g}$ \cite{Fitzpatrick:2020vba,Fitzpatrick:2021cij}.  Thus, for inelastic (elastic scalar) models the $\alpha_D=0.5$ shown in Fig.\ref{fig:thermalPlot}(right) is viable for $m_{\chi}>10$ (60) MeV, while for lighter DM self-interaction bounds motivate focusing on 
 $\alpha_D < 0.5 \left(\frac{m_\chi}{10\,\rm{MeV}}\right)^{3/4}$ for inelastic models ($\alpha_D < 0.5 \left(\frac{m_\chi}{40\,\rm{MeV}}\right)^{3/2}$ for elastic scalar DM), using cross-sections from  \cite{Fitzpatrick:2020vba,Fitzpatrick:2021cij}.  Thus, below 10 (40) MeV, the sensitivity in the interaction strength $y$ obtained by  combining accelerator probes of $\epsilon$ with SI probes of $\alpha_D$ is stronger than the constraint from accelerators in Fig.~\ref{fig:thermalPlot}, which assumes the theoretically conservative upper bound $\alpha_D=0.5$.  The improvement is most dramatic at low masses. For example, at 1 MeV missing energy/momentum bounds on inelastic models improve by a factor of 5 (and 270 for the elastic scalar), while beam dump sensitivities improve by the square roots of these factors.  
 
\subsubsection{Generalized (Indirect) Freeze-Out: Examples and Broad Lessons} %[1.5p]
\label{sec:generalizedthermal}
Above, we started from a minimal and restrictive premise --- freeze-out of DM into SM particles controlled by the portal interaction --- and found a rich space of observational possibilities simply by altering the spin of the DM.  However, relaxing this assumption is very reasonable. The dark sector hypothesis, and viability of sub-GeV dark matter more generally, \emph{requires} new light degrees of freedom so it is natural to consider scenarios where these species also plays an important role in DM freeze-out.  \textbf{A wide variety of such models motivate a common suite of experimental probes (those presented above, plus visible decays of long-lived dark sector particles), with the complementarity across probes working somewhat differently for each model}.  This is illustrated by a few examples, which we summarize below as well as pointing out key points about their phenomenology. 

\begin{itemize}
\item \textbf{Secluded Dark Matter}, where DM annihilates into two lighter mediators, whose decays into SM matter maintain equilibrium between the dark and SM sectors.  In contrast to the direct annihilation case, secluded DM is most readily realized in scalar-portal models (where the secluded annihilation is $p$-wave), and is excluded by CMB energy injection bounds for the vector portal (where the secluded annihilation reaction is $s$-wave).  The DM annihilation cross-section is controlled by the Yukawa coupling of the DM to the dark scalar, which for sub-GeV DM must be fairly small, but thermal freeze-out yields no prediction for the mixing parameter $\sin\theta\sim A_h v_{SM}/m_h^2$.  Upcoming searches in direct detection and in long-lived meson decays will probe new parameter space in $\sin\theta$, but the most decisive prospect for testing this scenario is indirect detection with $p$-wave sensitivity using MeV gamma-ray telescopes \cite{Coogan:2021sjs}. 
\item \textbf{Strongly Interacting Dark Matter} refers to models where the DM is part of a confined dark sector (in many models, it is taken to be a stable dark-pion).  The strong dynamics of this sector can affect freeze-out in multiple ways, including 3-pion-to-2-pion annihilation through Wess-Zumino-Witten diagrams \cite{Hochberg:2014dra,Hochberg:2014kqa} and annihilation into somewhat heavier dark vector mesons \cite{Berlin:2018tvf}. In each case, the interplay of these processes' dark sector rates with the rates of scattering and/or decay reactions that maintain dark sector/SM \emph{kinetic} equilibrium \cite{Kuflik:2015isi,Kuflik:2017iqs} leads to a rich phase diagram of masses and mixings for which the observed relic abundance is realized, generally \emph{below} the thermal relic lines discussed above.  Because these mechanisms rely on strong dynamics, the presence of unstable resonances only somewhat heavier than the DM is generic.  These typically have long-lived decays to SM particles (or in some cases to a combination of SM and dark-sector particles), and offer a promising target for long-lived particle searches.  In addition,  much of this parameter space can be explored by both low-threshold direct detection and accelerator-based production experiments. In this parameter space, 2-to-2 annihilation is both $p$-wave and below the cross-section expected for thermal freeze-out, so that indirect detection is extremely challenging. 
\item \textbf{Forbidden and Not-Forbidden DM, and their variations} arise in minimal vector portal models where the DM and mediator masses are quite close ($m_\chi < m_{A'} < 2\cdot m_\chi$) so that the secluded annihilation reaction is kinematically forbidden for typical DM particles but allowed on the high-energy tail of the DM Boltzmann distribution (forbidden DM \cite{Griest:1990kh,DAgnolo:2015ujb}).  Processes with 3-particle initial states can also lead to kinematically allowed depletion of DM (not-forbidden DM \cite{Cline:2017tka}), but pay an additional Boltzmann penalty due to the involvement of 3 non-relativistic initial state particles.  Like SIMPs, there is a rich coupling-mass diagram in which the interplay of these annihilation rates with those that maintain DM-SM kinetic equilibrium leads to the observed DM abundance  \cite{Fitzpatrick:2020vba,Fitzpatrick:2021cij}.  In this mass range, the portal mediator decays visibly (since decays to DM are kinematically forbidden) through the portal coupling, which again lead to lifetimes observable in the long-lived particle  regime. Because 2-to-2 annihilation is kinematically forbidden for low DM velocities, indirect detection is impossible. Scattering in direct detection experiments is challenging to detect due to the low couplings expected in much of these models' parameter space. 
\end{itemize}

\subsubsection{Summary} %[0.75p]
The conclusions of this contribution can be summarized as follows:
\begin{itemize}
\item Sub-GeV dark matter implies a hidden sector, which motivates a powerful bottom-up classification based on the dominant ``portal'' interaction mediating dark sector interactions with the SM.
\item The new portal force must be stronger (as an effective operator) than the weak interactions, and it is reasonable to assume that the portal mediator (and possibly other dark sector particles are comparably light to the DM. This motivates looking for multi-particle dark sectors; additional effects of these particles, such as DM self-interactions can be large enough to observably affect cosmological structures, particularly in the 1-40 MeV mass range.
\item The abundance of DM is a strong and powerful hint.  Following it leads to a rich landscape of experimental opportunities. DM production at accelerators is the most robust single probe of the simplest and most WIMP-like thermal production scenarios, but direct and indirect detection and probes of DM self-interaction have promising and highly complementary prospects as well.  Examining broader models strengthens the case for pursuing all three of these classes of experiment, as well as a distinct class of accelerator-based experiments searching for long-lived dark sector particles. 
\end{itemize}
%\section{A Bit of Freeze-In} %[1p]

%\bibliographystyle{natbib}
% \bibliography{Toro}

% \end{document}
%-------------------------------------------

%-------------------------------------------
\subsection{Light Dark Matter in the MeV-GeV range: what we know from astroparticle and cosmology -- {\it C.~Boehm}}
\label{ssec:boehm}
{\it Author: Celine Boehm, <celine.boehm@sydney.edu.au> } 

\subsubsection{Introduction}
Particle Dark Matter has been the object of many studies since the late 1970s, specifically  after Hut \cite{Hut:1977zn}, Lee and Weinberg \cite{Lee:1977ua} demonstrated that thermal particles heavier than 2 GeV could explain the observed dark matter relic density  if their pair annihilation cross section was of similar strength as that expected for weak interactions. Neglecting these interactions (owing to their supposedly weak strength), cosmologists were also able to set a constraint on the dark matter mass by demonstrating that structure formation alone (independently of the relic density argument) required the dark matter particles to be heavier than a few keVs in order to explain the observed number of small cosmological structures  \cite{Bode:2000gq,Irsic:2017ixq}. This result  eventually ruled out Standard Model neutrinos and led to the birth of the Cold Dark Matter scenario, along with the Weakly Interacting Massive Particles hypothesis, which in turn has  led to a plethora of dark matter experiments and particle physics searches.

These different arguments seem to point towards heavy dark matter particles. Yet thermal dark matter particles can actually be light. In fact we showed in \cite{Boehm:2002yz} that dark matter particles could be considerably lighter than a proton providing certain conditions which are the object of this proceeding. The take away message from this research is that thermal sub-GeV dark matter particles are a possible dark matter  scenario  providing that their annihilation cross section be s-wave suppressed or a pure p-wave -- a condition which can be achieved for both fermionic or scalar dark matter particles but requires the exchange of dark gauge boson \cite{Boehm:2002yz,Boehm:2003hm} (assuming one can  evade the Gunn-Tremaine bound in the case of  fermionic dark matter \cite{Tremaine:1979we}). These studies also demonstrate that scalar dark matter particles can interact with the Standard Model if the mediator of their interactions is  fermionic. Indeed, in this very case, the dark matter pair annihilation cross section is independent of the dark matter mass. As a result, requesting that the thermal pair-annihilation cross-section be equal to a specific value to explain the observed relic density  does not constrain the dark matter mass. Instead this constrains the properties of the mediator and its couplings to the dark matter. Hence, in this specific case, dark matter can be as light as the Standard Model fermions it annihilates into. However the corresponding  annihilation cross section (whis contains a s-wave term) needs to be suppressed by about 5 order of magnitudes to avoid over producing the gamma-rays in the Milky Way and in clusters of galaxies (at least if the particles in the final state are charged).

After summarising the cosmological reasons which motivated studies of light thermal dark matter scenarios, I will summarise the Astrophysical constraints that apply on these scenarios along with the electron $g-2$ constraint which is critical for model building. I will conclude by revisiting the impact that light dark matter particles can have on cosmology and how such scenarios can be constrained with the help of modern observations of galaxy formation and large-scale-structure formation.

\subsubsection{Why studying light thermal dark matter? }
One paradox between N-body simulations and relic density/particle physics studies of dark matter is that the former usually assumes that the dark matter has no interaction with the Standard Model while the latter heavily relies on the existence of dark matter interactions in order to be able to discover it. In reality,  both should account for  interactions with the Standard Model and use the data at  disposal to determine their strength. Ideally the picture that emerges from these two different technique should be in agreement, bearing in mind  that N-body simulations probe the dark matter elastic scattering interactions with SM particles while the relic density criterion probes the annihilation cross section. Of course the result of these analyses should be consistent with that from other techniques such as dark matter direct detection (which relies on the elastic scattering cross section).    

The way to go is therefore to introduce the dark matter interactions in N-body simulations. However a first step is to estimate the damping of the dark matter fluctuations in the linear regime that stems from the dark matter interactions since the distribution of these fluctuations provide the initial conditions for structure formation. Estimating the damping is usually done by modifying the Boltzmann equations for the matter and radiation components in a CMB code to account for the interactions (see \cite{Boehm:2001hm} for a historical reference). However one can easily estimate the effect using an analytical expression and obtain a good intuition of the magnitude of the effect, as was done initially in \cite{Boehm:2000gq,Boehm:2004th}. In fact this approach  let me discovered  a new effect called "mixed" damping which eventually led me to find a mechanism to evade the Hut-Lee-Weinberg limit.

\paragraph{Generalising the Silk Damping \label{silkdamping}}

Interacting dark matter scenarios assume that dark matter can have interactions with any SM particles. As such, the dark matter primordial fluctuations are expected to be washed out below a certain scale due to dissipation. Naively the larger the interactions, the larger the damping scale (i.e.  fluctuations of larger size are being erased) and the more difficult it become to reproduce the structures that we observe in the Universe. This dissipative phenomena is called collisional damping and is similar to the so-called Silk damping for ordinary matter. 

Indeed, in the early Universe, most baryonic matter is interacting with photons. The interaction strength is given by the Compton interactions (Thomson at low energy) and is so large that the photons decouple from the ordinary matter at a very late time, around $z \simeq 3000$. Ordinary matter stops interacting with the photons a bit after,  owing to the fact that the energy density of the photons is larger than that of the baryons and therefore the baryons stay coupled to the photons later. The coupling generates the so-called Silk damping which is responsible for the tail of the CMB (i.e. the damping of the 4th to higher peaks) and also in the dark accoustic oscillations. The Silk damping length can be estimated as follows\footnote{See \cite{Boehm:2004th} for details on the derivation of this expression.}: 

$$ l_{Silk}^2 \simeq \int^{t_{dec(b-\gamma)}} \frac{\rho_\gamma c^2}{\rho_{tot} a^2 \Gamma_\gamma}  dt$$ 
where $\Gamma_\gamma$ is the photon interaction rate, $t_{dec(b-\gamma)}$ the  time when the baryons decouple from the photons. In reality this expression needs to be altered as the baryons stay coupled  to the photons after the photons have decoupled kinetically but we will not give further explanations here.  One can generalise this expression to  dark matter interactions with any Standard Model particle $i$. This leads to the collisional damping length expression: 

$$ l_{cd}^2 = \int^{t_{dec(DM-i)}} \frac{\rho_i v_i^2}{\rho_{tot} a^2 \Gamma_i} \ dt.$$

Since both photons and neutrinos have the largest velocity and densities, it will come at no surprise that the collisional damping associated with both Dark Matter-$\gamma$ and Dark Matter-$\nu$ interactions is the largest.  To prevent washing out critical scales, the damping created by these interactions must stop early and so one expects a tight constraint on the dark matter-$\gamma,\nu$ interactions from the collisional damping effect. In contrast the constraint on the dark matter-baryons and dark matter self-interactions are expected to be less stringent since both baryonic matter and dark matter become non relativistic early and have small number densities in the early Universe. There is an exception however when dark matter interacts with  baryons or itself via dark Coulomb interactions (i.e.when the cross section is inversely proportional to $v^{4}$, which can happen if the mediator of the interactions is a gauge boson/dark photon or Z'). 

There are many subsequent studies which have confirmed the conclusions of \cite{Boehm:2000gq,Boehm:2004th} and have improved the limits on these various interactions. However the DM-$\nu$ remains the most fascinating one for the following reasons. 

When dark matter interacts with neutrinos, there are four  possible situations: 
\begin{itemize} 
\item[1]  the dark matter decouples from the neutrinos before the neutrinos kinetically decouple from electrons, $\Gamma_{\nu-e}$ being the last neutrino interactions, 
\item[2] the dark matter decouples from the neutrinos before the neutrinos kinetically decouple from electrons but $\Gamma_{\nu-e}$ is not the last neutrino interactions, 
\item[3]  the dark matter decouples from the neutrinos after the neutrinos kinetically decouple from electrons, $\Gamma_{\nu-e}$ being the last neutrino interactions, 
\item[4] the dark matter decouples from the neutrinos after the neutrinos kinetically decouple from electrons but $\Gamma_{\nu-e}$ is not the last neutrino interactions, 
\end{itemize}

The maximal dissipation effect occurs for case 3, and in particular when the dark matter is as light as an electrons (i.e. the DM has a mass in the MeV range). In this case, the physics of neutrinos is unaltered so they kinetically decouple from the electrons   around a few MeVs (as in the standard model of Cosmology) and start to free-streaming immediately. However since the dark matter stays coupled to them, the neutrino free-streaming is transferred to the dark matter. This "mixed" damping effect (i.e. a mixture of collisional damping and free-streaming) is very efficient in erasing the dark matter primordial fluctuations and should result in a very strong constraint on light dark matter coupled to neutrinos. However  one   question immediately arises: if dark matter is thermal (which was the main paradigm at the time), can it be as light as a few MeVs then since the Hut-Lee-Weinberg limit should apply. In other words: how to evade the Hut-Lee-Weinberg limit?

\subsubsection{Evading the Hut-Lee-Weinberg limit}

The Hut-Lee-Weinberg limit is obtained by solving a Boltzmann equation in an expanding Universe. As such it does not require to specify any particle properties for the dark matter other than assuming that it does not annihilate nor decay (two assumptions which of course could be questioned and abandoned as many works have shown, see for example e.g.\cite{Hall:2009bx}). An additional assumption is that the DM number density is the same as that for anti-dark matter  particles (see for example \cite{Kolb:1990vq,Kaplan:2009ag}). With this in mind, one obtains a simple relationship between the dark matter relic density and the annihilation cross section which is independent of the dark matter  mass, namely $\langle \sigma v \rangle \propto 3\, (\Omega_{DM} h^2)\, 10^{-27} \rm{cm^3 s^{-1}}$. To get more out of this relationship one then needs to specify a model. Most fermionic DM models predict a cross section that depends on the DM mass. As a result, requiring that fermionic dark matter particles explain the entirety of the observed dark matter relic density  leads to a constraint on the mass of the dark matter candidate. This is true whatever mediators, including if it is a  dark gauge boson or a scalar mediator, as shown in \cite{Boehm:2003hm}. However when the mediator is a new gauge boson (or a neutral fermion), it is in fact possible to consider light dark matter scenarios providing that the mediator is also light  in order to  make the couplings relatively small (at least below the unitary limit) and maintain a cross section of weak strength, as required by the relic density criterion. 

The scalar dark matter particle scenario  leads to a very different situation. In particular, if the mediator of the dark matter-SM particles is a fermion,  the cross section is given by 

$$ \sigma v = \frac{1}{m_F^4} \left( (C_l^2 )  m_f \ + \ 2 C_l C_r m_F\right)^2$$

which is essentially 
$$ \sigma v \propto \frac{(C_l C_r)^2}{m_F^2}$$ if dark matter annihilates into light SM particles such as electrons or neutrinos. Here  $m_F$ is the mass of the mediator, $C_l, C_r$ are the left and right couplings of the DM to left and right SM fermions through the $F$ mediator (according to 
${\cal{L}} \subset \phi_{DM} \psi (C_l P_l + C_r P_r) \psi$ with $P_{l,r}$ the projectors). Hence, in this particular case, the annihilation cross section is independent of the dark matter mass. As a result imposing the relic density criterion does not constrain the DM mass to be in a specific range. It only constrains the mass of the mediator and its couplings. Therefore this case evades the Hut-Lee-Weinberg limit and the dark matter  can be as light as an electron or even lighter if it annihilates into neutrinos. However, as we will explain in the next section, dark matter cannot be arbitrarily too light if it annihilates into electrons.
 
 In summary, one can evade the Hut-Lee-Weinberg limit using two types of mediators. One is a new gauge boson called $Z'$ or dark photon (dubbed $U$ boson at the time) and the other one is a vector fermion $F$. Using simplified models \cite{Boehm:2003hm}, one can readily 
see that thermal dark matter can be very light indeed. However if the mediator of the interactions is a gauge boson then it needs to be light as well (independently of the fermionic/scalar nature of dark matter).  If the mediator is a fermion (for scalar dark matter), it must have non chiral couplings to both the left and right handed Standard Model fermions. If the mediator is a scalar (for fermionic dark matter), then it also needs to be relatively light to enable the sub-GeV dark matter mass range. However this is very strongly constrained by collider data, unless the annihilation is into neutrinos and the mediator is neutral.     

There is an interesting coincidence for the latter case. Such a dark matter-neutrino coupling can actually give rise to a neutrino self mass term (assuming Majorana neutrinos) 

$$ m_{\nu_l} = \sqrt{\frac{\langle \sigma v_r \rangle }{128 \pi^3}} m_N^2 \, (1+ \frac{m_{DM}^2}{m_N}^2) \, \ln (\frac{\Lambda^2}{m_N^2})$$

which leads to neutrino masses in the observed range \cite{Boehm:2006mi}. UV completion of this model is however very challenging \cite{Farzan:2010mr,Arhrib:2015dez}. 

%
%\begin{array}{c|c|c}
%   \textrm{ DM}  & \textrm{Z'/dark } $\gamma$ & \textrm{spin-0 or 1/2}  \hline \\
%  \textrm{ Fermionic DM } & yes & \textrm{yes but challenging } \hline  \\
%   \textrm{ Scalar DM}  & yes  & yes \hline
%\end{array}

\subsubsection{ Constraints on light dark matter scenarios}
Evading the Hut-Lee-Weinberg limit enables to consider sub-GeV DM scenarios but there are immediate issues, as discussed below.

\paragraph{Astrophysical constraints} 
DM annihilations into $e^+ e^-$ or any other charged particles eventually produce some Bremsstrahlung emission which eventually lead to the production of a  continuum of gamma-ray photons (for dark matter heavier than a few MeVs). Depending on the strength of the annihilation cross section, this emission can exceed the background and  be used to set a limit on the dark matter mass and annihilation cross section. In \cite{Boehm:2002tm}, it was shown that the gamma-ray signal from the Bremsstrahlung emission accompanying the annihilation of thermal dark matter particles into a pair of electron-positron $ DM DM \rightarrow e^+ e^- \gamma$ could exceed the background emission in galaxies and clusters of galaxies by at least 5 order of magnitudes if the dark matter is as light as a few MeVs and is more or less compatible with observations for GeV dark matter particles. \textbf{This constraint needs to be put in tension with the CMB constraint which was derived several years later using the Planck data \cite{Galli:2009zc,Galli:2011rz,Galli:2013dna}. }

Based on the constraint above, we concluded in  \cite{Boehm:2002tm} that thermal MeV DM candidates could be viable if their annihilation cross section was p-wave ($\sigma v \propto v^2$) or eventually s-wave ($\sigma v \propto cst$) but the s-wave term needs to be suppressed by at least 5 order of magnitude with respect to the relic density value. In other words,  light thermal dark matter is possible. However the gamma-ray constraint requires that the relic density be achieved through a mediator that leads to a dominant p-wave term. This means that the best mediator  for the light thermal dark matter is a new light gauge boson. This does not exclude the hypothesis that some of the dark matter annihilations proceed via a fermionic mediator but this cannot be a dominant channel in the early Universe. It could however explain  the morphology and magnitude of the 511 keV line observed in the centre of the galaxy \cite{Boehm:2003bt}, although the latter is  likely to have an astrophysical origin.

\paragraph{$g-2$ constraints and 511 keV line}

Although thermal MeV DM could explain the 511 keV, the measured value of the electron $g-2$ is severely constraining this possibility \cite{Boehm:2004gt,Ascasibar:2005rw,Boehm:2007na,Hanneke:2008tm}. Indeed a fermion mediator is needed to explain the 511 keV line morphology but the latter would actually increase the discrepancy between the predicted value of the electron $g-2$ and the SM value unless the dark matter is very light\footnote{The fermion mediator contribution to the electron $g-2$ is proportional to the dark matter mass.}. 
However a too light dark matter candidate would lead to a $N_{eff}$ value in disagreement with the BBN and CMB measurements \cite{Serpico:2004nm,Boehm:2013jpa,Boehm:2012gr}. Said differently, on one hand, dark matter would need to be ligher than 10 MeV to explain the 511 keV line given the measured value of the electron $g-2$ and on the other hand the [1-10] MeV range could be excluded by the measurement of the number of degree of freedom in the early Universe.  Based on these results, it is very unlikely that  scalar dark matter particles coupled to electrons through a fermionic mediator could be a solution to the 511 keV emission. 

A dark gauge boson mediator on the contrary is not needed to explain the 511 keV morphology but it is required to explain the dark matter relic density if dark matter is light and thermal. The latter would  increase the electron $g-2$ discrepancy between the predicted and measured values as    $$\delta a_e = 10^{-11} \, \left( \frac{z_e}{7 \ 10^{-5}} \right)^2 \, \left(\frac{m_{Z'}} {\textrm{MeV}} \right)^{-2}$$ and so would either need to be weakly coupled or much heavier than a few MeVs \cite{Boehm:2007na}. This conclusion holds in fact indepedently of the 511 keV line since the gauge mediator is only required to explain the dark matter relic density. A summary of the constraints for these different scenarios can be found in \cite{Boehm:2020wbt}.    

\subsubsection{ Cosmological implications}
One can now investigate the cosmological implications of the light thermal DM scenarios which are not excluded yet. As mentioned in Section.\ref{silkdamping}, these scenarios may eventually lead  to some collisional damping if their elastic scattering cross section is large enough.  Eventually this can change the CMB angular power spectrum, the way structures form as well as their internal dynamics (see for example \cite{Boehm:2014vja,Schewtschenko:2014fca}. 

From the particle physics point of view, a large damping effect could occur if the interactions are mediated by a dark boson as this could lead to some  Coulomb-like interactions. These scenarios could also lead to cosmological signals such as the 21cm line or non standard DM halos \cite{Fialkov:2018xre}. In addition they may be probed through direct detection experiments \cite{XENON:2019zpr}, collider physics and indirect detection (despite the p-wave cross section). As such they warrant an holistic approach to determine the part of the parameter space that is worth exploring.

In summary, DM-$\nu$ interactions do not require large interactions to lead to a different pattern of structure formation and reduce the number of companion satellites in the Milky Way \cite{Boehm:2014vja,Schewtschenko:2014fca}. They are easier to realise phenomenologically, including trough the exchange of a dark gauge boson and there are puzzling implications for neutrino physics \cite{Boehm:2006mi}.  DM annihilations into neutrinos are constrained by both CMB and BBN observations \cite{Serpico:2004nm,Boehm:2012gr} but these scenarios are in principle viable if one does not require that light thermal dark matter also explains the 511 keV line.     

A dark matter-neutrino coupling would lead to a cut-off scale (and some oscillations) in the linear matter power spectrum, which can be probed with observations at high redshift and in particular the SKA \cite{Mosbech:2022uud}. Furthermore, as pointed out in \cite{Mosbech:2022nkk}, their impact on the linear matter power spectrum changes the number of binary black hole mergers in the Universe and therefore the number of gravitational wave events to be detected with redshift. While this novel technique can actually be applied to other physical situations, it will help in particular to disentangle the dark matter microphysics properties and inform whether dark matter is made of sub-GeV particles.   

\subsubsection{Conclusion} 

Thermal dark Matter particles can be light. 
They can be in principle either fermionic or scalar. However light fermionic candidates must evade the Gunn$\&$ Tremaine bound. All (fermions or scalars) are likely to have interactions mediated through a light dark gauge boson if they are thermal (and as light as a few MeVs) as this enables to both explain the observed relic density while also be compatible with gamma-ray observations in the Milky Way and clusters of galaxies. The properties of this gauge mediator is constrained by the electron $g-2$ though these constraints would be alleviated if dark matter mostly annihilate into neutrinos in the early Universe.  

%-------------------------------------------

%-------------------------------------------
\subsection{Light Dark Matter in the MeV-GeV range: overview of indirect detection searches -- {\it F.~Calore}}
\label{ssec:calore}
{\it Author: Francesca Calore, <francesca.calore@lapth.cnrs.fr> } 

% \documentclass[aps, onecolumn,nofootinbib,notitlepage]{revtex4-1}
% \usepackage{graphicx}

% \begin{document}
% \title{Light MeV-GeV dark matter: Indirect detection searches}
% \date{\today}

% \author{Francesca Calore}\email{calore@lapth.cnrs.fr}
% \affiliation{LAPTh, USMB, CNRS,  F-74940 Annecy, France}

% \maketitle
% %we are expecting about 6 pages for a 30 minute talk, 5 pages for a 20 minute talk, 4 pages for a 20 minute talk, ad 3 pages for a 15 minute talk. You should send the .tex file directly to us. We will organize all the material in the final document. 

\subsubsection{Introduction}
Dark matter (DM) indirect detection leverages on signatures that DM particle and non-particle 
candidates can leave after interacting with the environment in various ways. 
In the more traditional approach, DM self-annihilation or decay injects in the
environment (the Galaxy or the early Universe for example) final stable products that can 
be looked for in the cosmic fluxes of photons (at multiple wavelengths, from radio to gamma rays)
and charged cosmic rays~\cite{Gaskins:2016cha}, or that can alter the recombination history 
of the Universe and leave an imprint in the CMB power spectrum, see e.g.~\cite{Slatyer:2016qyl}.
Gravitational interactions of DM with astrophysical systems lead to phenomena such as 
gravitational lensing that can be used to constrain the nature of DM interactions, see e.g.~\cite{Vegetti:2018dly}.
Finally, DM can be captured by, accreted onto or scattered off celestial bodies of different nature, from 
stars, to planets to very compact objects such as neutron stars and black holes~\cite{Baryakhtar:2022hbu}. 

A general discussion of astroparticle observables for DM can be found in the 2021 EuCAPT White Paper~\cite{AlvesBatista:2021gzc} (Fig.~10).
In general, different astroparticle observables can cover a large portion of the DM parameter space and 
probe different candidates, from fuzzy DM to primodial black holes (PBH).
In this contribution, I will focus on how one can probe light DM (in the MeV -- GeV mass range) with 
cosmic messengers. 

\subsubsection{Light DM decay and annihilation into cosmic messengers}

% particle DM signal   
Depending on their nature, particle DM candidates produce final stable particles 
(photons, electrons and positrons, etc.) through 
self-annihilation or decay which occur in the halo of the Milky Way and of external galaxies.
Today, these processes happen at rest, since DM particles are non-relativistic. 
This implies that the center of mass (CM) energy of the annihilation (decay) process is directly 
related to the mass scale of the DM particle, being $E_{\rm CM} = N m_{\rm DM}$ with $N =$ 2 (1).
We therefore expect the energy of the emitted stable particles (photons or other charged particles) 
to match the CM energy. 

If the CM energy is below the threshold for electron-positron pair production, the only allowed 
decay/annihilation channel is into two photons which are emitted with an energy $E_\gamma = N m_{\rm DM}/2$, back-to-back 
in the DM rest frame. This results in a sharp line-like signal, broadened only by the DM velocity 
dispersion in the halo and the energy resolution of the gamma-ray telescope (typically dominant).

For heavier DM masses, other final states open up progressively starting from electron-positron pairs.
Eventually, for any DM particle model one can predict the branching ratio (BR) for annihilation/decay into
specific final states. 
This results in a broader spectral energy distribution (i.e.~the number of particles produced per unit energy 
as a function of energy). In the case of photons, different terms contribute to the final spectrum: 
The prompt gamma-ray emission from decay and hadronization of
final states, gamma-ray lines, and the photons produced by higher-order corrections -- beyond tree-level processes --
from, e.g.,~leptonic final states, see~\cite{Bringmann:2012ez} for a general discussion, and 
Fig.~1 therein.

As an example, the expected flux of gamma rays from the line of sight (l.o.s.) direction 
identified by the Galactic coordinates ($\ell, b$) from DM annihilation or decay can generally 
be written as:
\begin{equation}
    \frac{{\rm d}\Phi_\gamma}{{\rm d}E}(\ell, b)= \mathcal{A}(\theta_{\rm DM}) \times \frac{{\rm d}N_\gamma}{{\rm d}E} \times \int_{\rm l.o.s.} \rho^N_{\rm DM}(s, \ell, b) ds \, ,
\end{equation}
where $\theta_{\rm DM} = \{\Gamma_\gamma, m_{\rm DM}\} $ for decay, and
$\theta_{\rm DM} = \{\langle \sigma v \rangle, m^2_{\rm DM}\}$ for annihilation.

The integral along the l.o.s.~is a geometric term which depends on the DM spatial density
distribution in the target of interested, e.g.~the halo of galaxies. 
In the Milky Way, the DM spatial density is constrained at larger radii by the rotation curve,
however to infer the DM distribution to distances closer to the Galactic center
we need to rely on semi-analytical models or fully numerical simulations of galaxy formation. 
Different parametrizations of the DM spatial distibution are commonly adopted for the Milky
Way, such e.g.~the well-known Navarro-Frenk-White profile, with a theoretical uncertainty which 
can span more than three o.d.m.~at the Galactic center. 
The uncertainty on the DM profile propagates quite severely in the 
calculation of cosmic photons and cosmic-ray fluxes from DM annihilation, given 
the dependence of the flux on the DM density squared.

Focusing on high-energy photons, these can also be produced by DM leptonic final products, 
when they interact with the interstellar medium. In particular, DM produced electrons and positrons 
can up-scatter low-energy ambient photons of the interstellar radiation field at higher energies
through the so-called inverse Compton scattering (ICS). 
The energy of the out-going photons is related to the energy of ambient photons ($E_i$) and leptons ($E_e$) 
by the equation:
\begin{equation}
    E^{\rm ICS}_f \sim 30 \left( \frac{E_i}{\rm eV} \right) \left( \frac{E_e}{\rm GeV} \right)^2 \rm MeV \, .
\end{equation}
As can be seen in Fig.~1 (left) of~\cite{Cirelli:2020bpc}, such a ``secondary'' emission process allows one to use
photons at a given energy to probe not only DM at the same energy scale but also at higher masses.
For example, sub-GeV DM can be probed by X-ray data at keV energies.
In general, lines of sight that avoid the Galactic plane (i.e.~at high latitude) are preferred by ICS emission.

% sub-GeV dark matter
Let's focus now, specifically, on DM masses from keV up to GeV. 
In this mass range, DM can produce only prompt line-like photon signals, when possible, electron-positron pairs,
and secondary photons from final state radiation and/or inverse Compton. 

{\bf Data and observables - }
What data and observables can we use to detect or constrain these candidates? 
First, one can rely on the measurement of the continuum diffuse emission of the Milky Way in hard X rays and soft gamma rays, 
mostly caused by the interactions of cosmic rays (CR, protons, electrons) with the interstellar medium. 
This emission, measured in the Galactic plane by the Compton Telescope (COMPTEL) onboard the Compton Gamma Ray Observatory~\cite{1999ApL&C..39..209S}, and by the Spectrometer (SPI) aboard the International Gamma-Ray Astrophysics Laboratory INTEGRAL~\cite{2011ApJ...739...29B,2022A&A...660A.130S}
is dominated above about 200 keV by the ICS astrophysical emission from CR electrons. 
Recently, a new analysis of 16-yr data from SPI has been performed, 
providing a new measurement of the Galactic diffuse emission up to 8 MeV, 
and superseding the 20-yr old measurement from COMPTEL in the same energy range~\cite{2022A&A...660A.130S}.
This new measurement allows to set constraints on CR transport at MeV energy,
but also on exotic emission mechanisms from particle, as we will see below, and non-particle DM.
This is currently the best sensitivity one can achieve. 
In absence of any dedicated new instrument that will cover the so-called MeV energy sensitivity gap, this will also be the best sensitivity we can hope for in the near future above a few MeV.

Besides the Galactic plane, one can also collect photons from single, promising DM targets, e.g.~dwarf  spheroidal galaxies. 
An example with SPI is presented in~\cite{Siegert:2021upf}, where 1.5 Ms of data were taken from the direction of 
the Reticulum II dwarf spheroidal galaxy. 

Light DM models can also be constrained by the observation of the 
511 keV electron-positron annihilation line, which should originate from non-thermal positrons
injected into the interstellar medium and annihilating at rest with free electrons after thermalisation in the 
interstellar medium.
Observed and characterised by SPI, the origin of the 511 keV line remains a mystery of current high-energy astrophysics, see e.g.~\cite{Prantzos:2010wi} for a review.
Light DM has been invoked in the past to explain this signal~\cite{Prantzos:2010wi}. 
More recently, the 511 keV line signal has been used to set bounds on light particles such as axion-like particles and 
sterile neutrinos~\cite{Calore:2021klc, Calore:2021lih}.

Finally, also the flux of cosmic electrons and positrons can be compared 
with light DM signal predictions. In particular, the Voyager I crossed the heliopause in 2012 and 
measured, for the very first time, the cosmic flux of interstellar electrons and positrons, without the need
of adding and modelling corrections induced by solar modulation. \footnote{\url{https://voyager.jpl.nasa.gov/mission/interstellar-mission/}}

\begin{figure}
\includegraphics[width=0.49\textwidth]{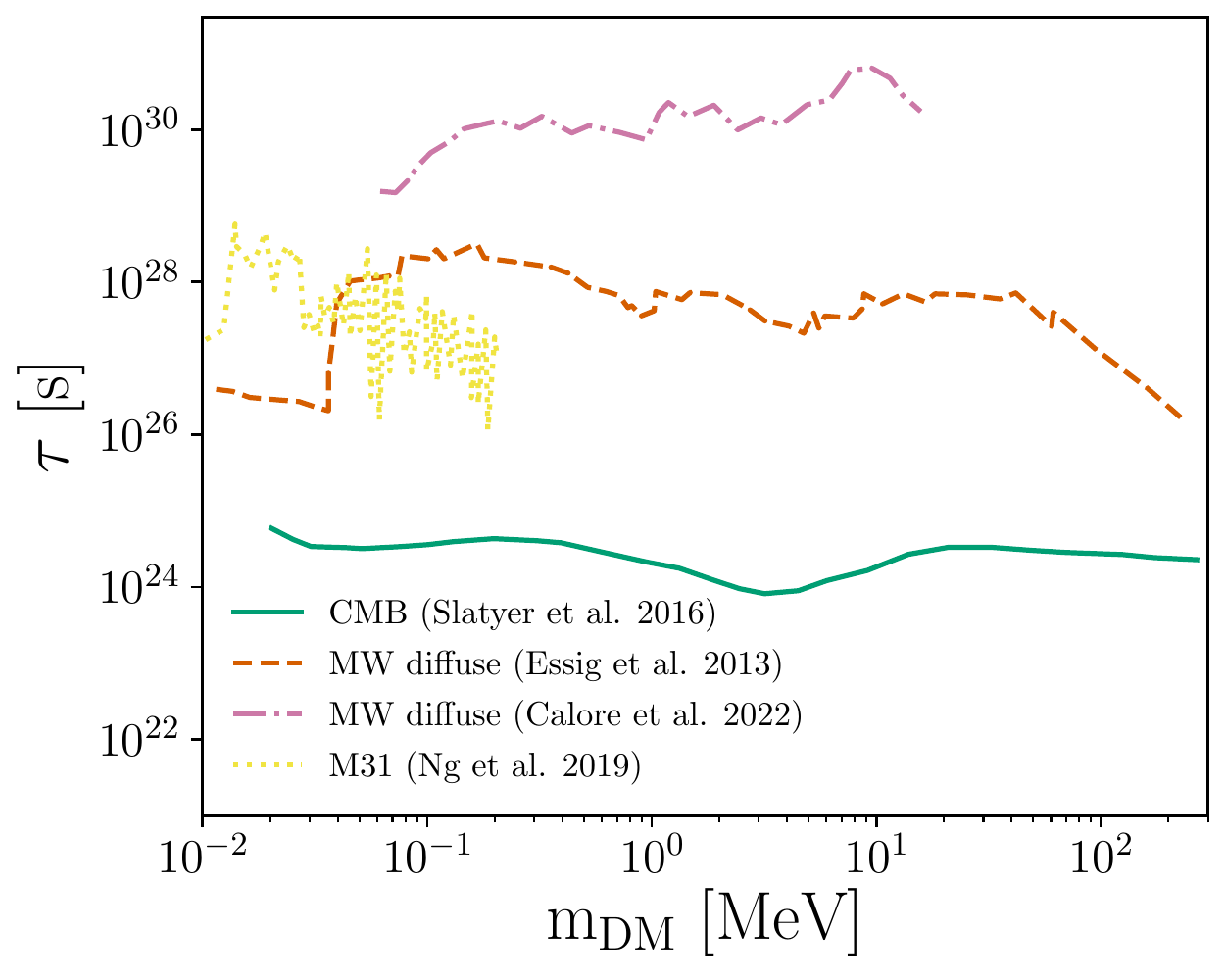}
\includegraphics[width=0.49\textwidth]{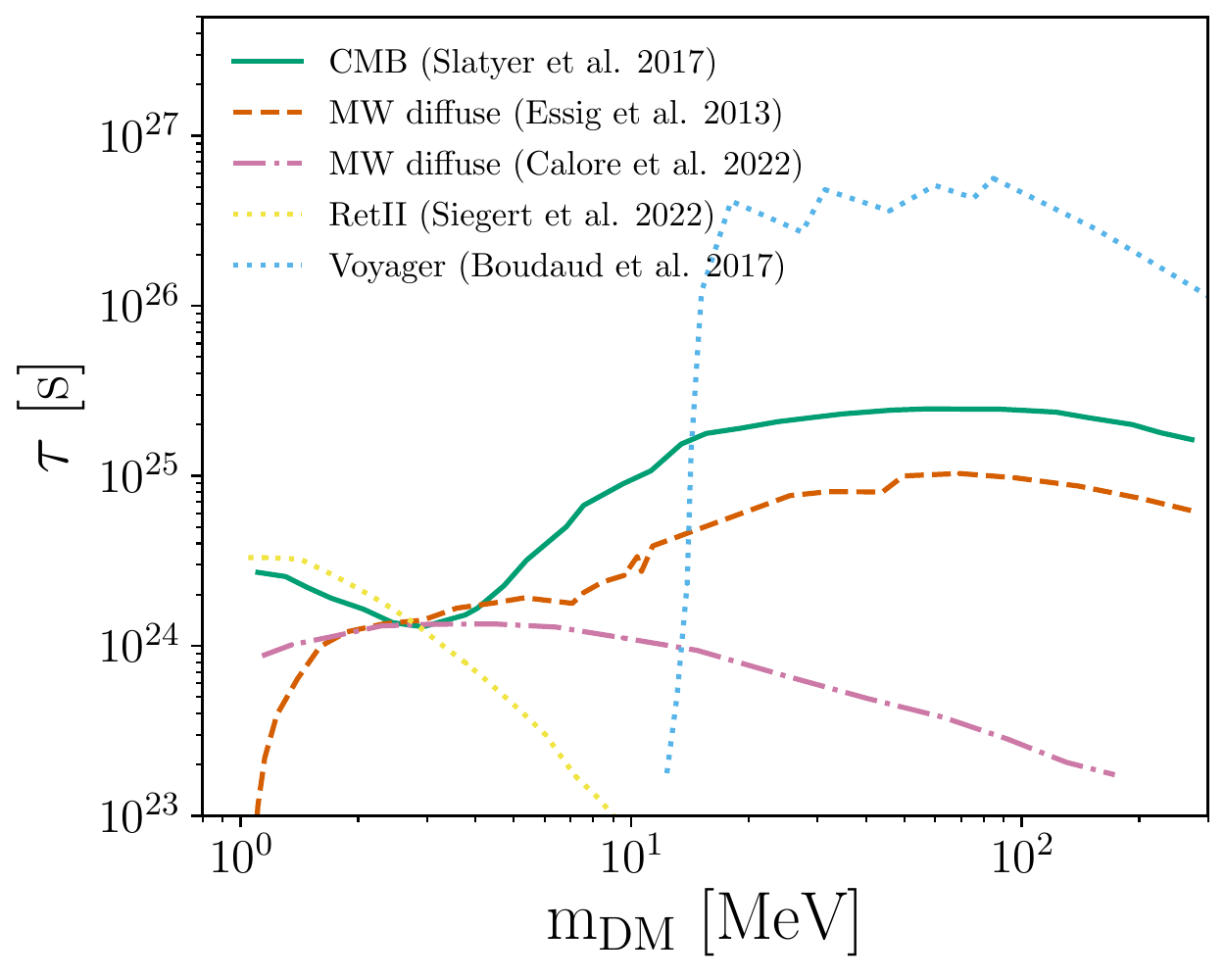}
\caption{Current constraints on the light DM lifetime for decay into two photons (left) and electron-positron
pairs (right).}
\label{fig:decay}
\end{figure}

{\bf Light DM decay: current constraints - }
I here summarise the state-of-art of indirect detection constraints on light DM 
decaying directly into photons  (line-like signal) or after final state radiation of electron-positron final pairs (FSR signal).
From the summary plots in Fig.~\ref{fig:decay} (left: line-like signal; right: FSR signal), we can notice that 
the Milky Way diffuse emission measurement from SPI and COMPTEL provide the most constraining limits for sharp line-like signals
(Fig.~\ref{fig:decay}, left), probing lifetimes of about 10$^{30}$ s
for DM masses from about 100 keV up to, at least, 10 MeV. 
The MeV diffuse emission observation supersedes in this case CMB limits.
Also single targets diffuse emission, such as from M31~\cite{Ng:2019gch} and Reticulum II~\cite{Siegert:2021upf} (not shown) result in
competitive bounds for comparatively smaller exposure times.
FSR signals from light decaying DM (Fig.~\ref{fig:decay}, right), instead, are typically less constrained by MeV diffuse observations, 
and stronger constraints are obtained from CMB observations~\cite{Slatyer:2016qyl} and Voyager I analysis~\cite{Boudaud:2016mos}.

The currently most constraining results for light decaying DM and other FIPs 
come from the new 16-yr analysis of SPI Galactic diffuse data~\cite{Calore:2022pks}.
For the first time, a spatial component corresponding to decaying DM has been 
included in the data extraction procedure of SPI data, together with other known astrophysical
components. 
No signal associated to the DM ``template'' being significantly detected, 
95\% upper limits on the decaying DM flux have been set. 
Such flux limits were then translated onto the general parameter space of light DM decay (decay rate vs mass),
by performing a multi-component spectral fit to the extracted SPI spectrum and considering 
the specific line-like and FSR spectra.
This analysis currently provides the strongest constraints on light decaying DM in the 0.1 -- 10 MeV.
\footnote{Analogously, strong constraints were derived for PBH DM which shows the same spatial 
distribution as the decaying signal~\cite{Berteaud:2022tws}.}

As a word of caution (also valid for DM annihilation processes below), most of the constraints presented in indirect detection papers are
derived under the assumption of 100\% BR into a given final state (line-like or FSR in this case).
In general, these constraints can be recasted with some care in the framework of a specific DM model.
See for example~\cite{Essig:2013goa} for a recasting within FIMP DM models, and~\cite{Calore:2022pks}
for translating the new SPI Galactic diffuse constraints into the 
parameter space of 100 keV axion-like particles and sterile neutrino DM. 

\begin{figure}
\includegraphics[width=0.49\textwidth]{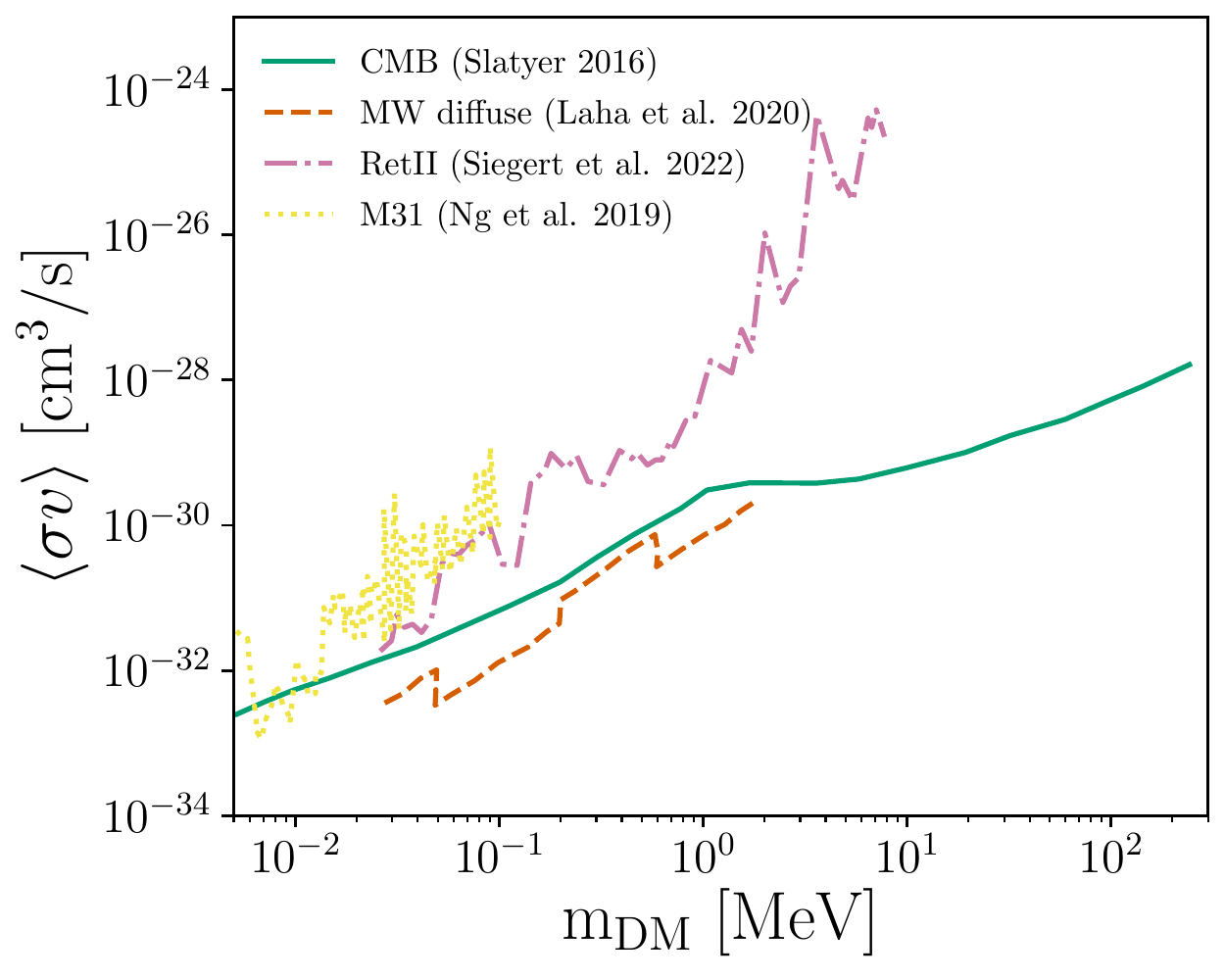}
\includegraphics[width=0.49\textwidth]{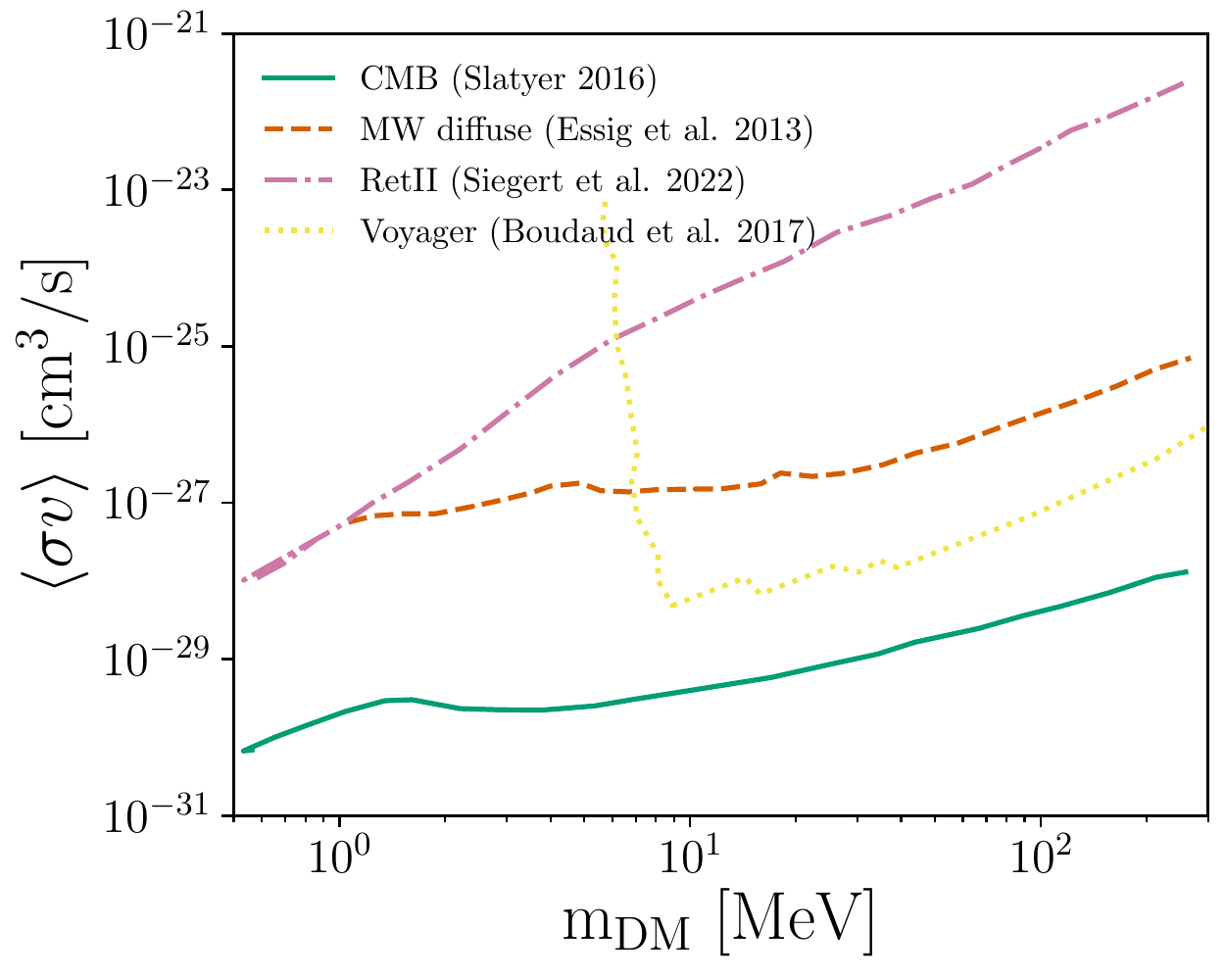}
\caption{Current constraints on the light DM velocity-averaged annihilation cross section (s-wave) for annihilation into two photons (left) and electron-positron pairs (right).}
\label{fig:ann}
\end{figure}

{\bf Light DM annihilation: current constraints - } 
Analogously to DM decay, the DM annihilation cross section as a function of DM mass can be constrained considering DM annihilation into photons directly (line-like signal) or after final state radiation of electron-positron final pairs (FSR signal).
In this case, we can refer to the summary plots in Fig.~\ref{fig:ann} (left: line-like signal; right: FSR signal).
CMB limits on s-wave (velocity independent) annihilation cross section are quite relevant, especially for the electron-positron signal~\cite{Slatyer:2015jla}. 
However, one can see that also in this case the MeV diffuse emission from SPI leads to the strongest bounds
for a DM mass range from 200  keV up to a few MeV.
Less constraining, yet competitive, are the bounds from Voyager I electrons and positrons~\cite{Boudaud:2016mos}, 
and from INTEGRAL using the ICS signal from DM annihilation~\cite{Cirelli:2020bpc}.
In this latter case, as mentioned above, the MeV diffuse emission can constrain DM masses up to a few GeV. 

In case one considers p-wave cross sections (velocity dependence as $v^2$), CMB limits 
are strongly relaxed by the small relative velocity of DM particles at recombination ($v_{\rm CMB} \lesssim 10^{-5}$ c)~\cite{Liu:2016cnk}. 
Instead,
limits from cosmic photons and electrons are not suppressed, because the annihilation process occurs 
today at rest in the halo of the Galaxy, $v_{0} \lesssim 10^{-3}$ c. This implies that, for p-wave cross section, limits from the
continuum MeV diffuse emission and Voyager I electron-positron flux are the most constraining ones, 
$\langle \sigma v \rangle_0 \sim 10^{-28}\, \rm cm^3/s$ at 10 MeV DM masses~\cite{Boudaud:2018oya}.
In contrast, for p-wave annihilation, the value of the velocity-averaged cross section (at the time of freeze out, where $v_{\rm f.~o.} \sim 0.15$ c) required to satisfy the relic density is about $10^{-26} \, \rm cm^3/s$~\cite{Diamanti:2013bia}.
The expected vanilla thermal cross section today is therefore 
$\langle \sigma v \rangle_0^{\rm f.~o.} \sim 10^{-32}\, \rm cm^3/s$~\cite{Bartels:2017dpb}.
Indirect detection probes are still far from this benchmark.

{\bf Light DM: future prospects - } 
As mentioned above, the MeV sensitivity gap currently makes the reach of current instruments limited for
discovering/constraining light DM. 
The last major experiment in the few MeV -- 100 MeV gamma-ray band was COMPTEL, which operated from 1991 to 2000. In the near term, 
the Compton Spectrometer and Imager (COSI) is planned to be launched in space in 2025, after several balloons tests.
However, it will cover an energy range from 0.2 up to 5 MeV. Yet, it will significantly improve current constraints on decaying or annihilating light DM~\cite{Aramaki:2022zpw, Caputo:2022dkz}. 
Interesting performances are planned for the Galactic
Explorer with a Coded Aperture Mask Compton Telescope (GECCO), covering up to 10 MeV~\cite{Coogan:2021sjs}.
To be able to probe higher DM masses, we will need instruments up to 100 MeV, 
such as GRAMS, APT, AMEGO-X and all-sky-ASTROGAM.
For a more general discussion about future instruments in the MeV range please see~\cite{Aramaki:2022zpw}.
At least a factor of 10 improvement is expected on the bounds on light decaying and annihilating DM, 
see for example~\cite{Cooley:2022ufh} (Fig.~4 for decay).

\subsubsection{DM capture in celestial bodies}
The sub-GeV DM parameter space can also be constrained by exploiting mechanisms of capture/scattering of DM 
particles onto celestial objects of different size and composition.
For a summary, we refer the reader to~\cite{Baryakhtar:2022hbu}, notably to Fig.~1 and table 1.
In this contribution, we focus on DM scattering and annihilation within celestial objects, 
leaving signatures in cosmic photons and neutrino fluxes. 
We will not discuss DM as an additional source of heating in planets and stars, for a discussion see for example~\cite{Leane:2020wob}.

The scattering of DM particles in galactic halos with the nuclei (or electrons) of celestial objects eventually bring
these particles into close orbits around these objects. Gravitational capture within the objects may occur with 
subsequent thermalisation of DM particles in the object core. The temporal evolution of the total number of DM particles, $\mathcal{N}$, is 
described by the following equation: 
\begin{equation}
    \frac{{\rm d}\mathcal{N}(t)}{{\rm d}t} = \mathcal{C} - \mathcal{A} \mathcal{N}^2(t) - \mathcal{E} \mathcal{N}(t) \, .
    \label{eq:nt_capture}
\end{equation}
The capture rate, $\mathcal{C}$, depends on the scattering cross section of
DM particles with nucleons, $\sigma_{\chi N}$, and controls the gravitational capture and thermalisation.
The annihilation rate, $\mathcal{A}$, depends on the DM self-annihilation cross section. The annihilation rate 
counter-balances the capture by reducing the number of DM particles in the object.
Finally, DM particles within the inner parts of the object are subject to evaporation, controlled 
by the evaporation rate, $\mathcal{E}$. Indeed, 
the finite temperature of the medium sets a minimum mass, the evaporation mass, that DM particles must have in order to remain trapped. If DM particles are too light, the velocity acquired while thermalising may 
be larger than the escape velocity of the object and make the DM particles not any longer gravitationally bounded. 
More generally, the evaporation mass sets a lower limit on the DM mass above which these type of bounds 
from celestial object capture are valid.
For a recent discussion see~\cite{Garani:2021feo}.

DM annihilation may produce different final states which are injected in the object core. 
If only feebly interacting, they can eventually escape the object and be released into the 
interstellar medium.
This is the case for neutrinos and for light, feebly interacting, mediators which DM annihilate into.
As for the case of annihilation into light, feebly interacting, mediators, these can, in turn, 
annihilate into photons in the interstellar medium, and leave a signature in high-energy gamma-ray fluxes. 
These DM models are rather commonly examined in current searches for light DM particles, within, e.g.~minimal 
and/or extended dark sectors.

If evaporation is not efficient and the object is old enough, equilibrium may be reached and the solution to 
eq.~\ref{eq:nt_capture} reads as:
\begin{equation}
    \mathcal{N} \simeq \mathcal{C} \tau_{\rm eq}  =\sqrt{ \frac{\mathcal{C}}{\mathcal{A}}} \, ,
\end{equation}
where $\tau_{\rm eq}$ is the equilibration time, to be compared to the age of the object.

Optimal targets for this type of studies are objects with large radii and high densities 
so to make the capture process easier and more efficient, as well as objects
with low core temperatures. Indeed, cold cores limit the kinetic energy that DM particles acquire 
and therefore make the evaporation mass smaller. 

In this respect, brown dwarfs (BDs) have been recently identified as promising celestial objects where DM capture and subsequent 
annihilation may leave observable features. 
Indeed, a large number of nearby BDs (more than 800 within 100 pc from Earth) is known to date and many more are expected to be discovered with 
current and future instruments (e.g.~JWST). As an example, more than $10^9$ objects are expected 
to be present in the vicinity of the Galactic center. 
BDs signatures from DM capture and annihilation were studied in~\cite{Leane:2020wob,Leane:2021ihh}.
Other celestial objects have been explored such as the Sun~\cite{Leane:2017vag, HAWC:2018szf} and Jupyter~\cite{Leane:2021tjj}.
The DM parameter space each type of objects is sensitive to is slightly different,
especially for what concerns the mediator lifetime.  

A recent analysis~\cite{Bhattacharjee:2022lts} performed a model-independent search of gamma-ray signals from 
the direction of known BDs with {\it Fermi}-LAT data. 
The selected sample of BDs consists of 9 nearby ($<$ 10pc), massive, cold BDs, which all possess
an age estimate (2 -- 10 Gyr). 
No significant gamma-ray excess emission being found towards the 9 selected BDs, 
95\% flux upper limits were set on the gamma-ray flux and recasted in the framework 
of DM capture and annihilation into gamma rays via light, long-lived mediators.
The equilibrium hypothesis being valid for the selected BDs,
one can generally write the expected DM flux from this kind of processes as:
\begin{equation}
    E^{2}\frac{{\rm d}\Phi}{{\rm d}E} \propto \frac{\mathcal{C}}{4 \pi d_\star^{2}} \times E^{2}\frac{{\rm d}N}{{\rm d}E}
\end{equation}
where $\mathcal{C}$ depends on characteristic parameters of both the BD ($M_\star, R_\star, d$) and the
DM particle model ($ \rho_{\rm DM}, \sigma_{\chi N}, m_\chi$). The spectral energy distribution ${\rm d}N/{\rm d}E$, instead, is
a box-shaped spectrum resulting from the decay of the long-lived mediators into photons.

The final constraints on DM particles (valid above 700 MeV because of the evaporation mass lower limit)
can be seen in Fig.~5 of \cite{Bhattacharjee:2022lts}.
These bounds do not suffer from large astrophysical uncertainties on, e.g.,~the 
BD distribution in the Galactic center or the DM spatial density profiles, and they 
apply to a much broader range of lifetime with respect to constraints 
from the Sun and Jupyter. 
Finally, celestial objects provide comparable bounds to DM direct detection experiments
and have the unique advantage
to extend the limits to masses lower than a few GeV with sensitivity
reaching cross section values of at least $10^{-38}$ cm$^2$.

In conclusion, celestial body capture may offer a unique way to probe DM in sub-GeV mass range. 
Lower evaporation masses can be probed by considering neutron stars and white dwarfs with a proper 
treatment of the, so far, very large astrophysical uncertainties.

\subsubsection{Conclusions}
Indirect searches for DM successfully test different DM (and FIP) models at the MeV -- GeV scale, 
probing a large portion of their parameter space.
A diversified program is in place to tackle DM over a wide spectrum of models and signatures, 
exploring also new avenues for DM capture and annihilation in celestial bodies.
More generally, lighter FIPs (ALPs, sterile neutrinos) can also be looked for with indirect detection probes, from radio wavelengths to very high-energy gamma rays.
Presently, the most urgent (experimental) need is the exploration of the MeV gap with future instruments, which can provide access to yet uncharted portions of the DM parameter space and new windows of opportunity for DM detection!

%\subsubsection*{Acknowledgements}
%F.C.~acknowledges funding by the ``Agence Nationale de la Recherche”, grant n. ANR-19-CE31-0005-01. 

% \bibliography{biblio_FIP_Calore.bib}

% \end{document}

%-------------------------------------------

%-------------------------------------------
\subsection{Light (MeV-GeV) dark matter: the Snowmass approach -- {\it S.~Gori} }
\label{gori}
{\it Author: Stefania Gori, <sgori@ucsc.edu>}

\subsubsection{Introduction}\label{Sec:intro}
The 2021-22 High-Energy Physics Community Planning Exercise (Snowmass 2021) was concluded in the summer 2022 and identified the nature of Dark Matter (DM) as one of the most important scientific questions in particle physics for the coming 
decade \cite{Butler:2023eah}. 

The Snowmass work was organized into ten "Frontiers"\footnote{These are the Accelerator (AF), Community
Engagement (CEF), Computational (CompF), Cosmic (CF), Energy (EF), Instrumentation (IF), Neutrinos
(NF), Rare Processes and Precision Measurements (RPF), Theory (TF), and Underground Facilities and
Infrastructure (UF) frontiers.} and each Frontier divided its work into several Topical Groups. The fundamental nature of Dark Matter was a central theme of the Snowmass process across several frontiers and topical groups. In a cross-cutting contribution to Snowmass, participants across frontiers collaborated to outline a road map for dark matter discovery \cite{Boveia:2022syt,Boveia:2022adi}. The absence of a clear discovery of WIMP DM has led the field to diversify into a broad program of DM searches in a range of masses spanning 90 orders
of magnitude. This was reflected in a large number of topical groups studying complementary experimental techniques to discover the nature of DM. In particular:
\begin{itemize}
\item CF1 studied particle-like DM in a wide range of masses at direct and indirect detection experiments \cite{Cooley:2022ufh}.
\item CF2 studied wave-like DM with a mass less than 1 eV (e.g. the QCD axion). Quantum measurement techniques have become crucial in this area \cite{Jaeckel:2022kwg}.
\item CF3 focused on cosmological and astrophysical probes of DM \cite{Drlica-Wagner:2022lbd}.
\item RF6 investigated the prospects of testing DM below the GeV scale using high intensity experiments \cite{Gori:2022vri}.
\item EF10 studied the production of DM at high energy colliders \cite{Bose:2022obr}.
\item NF3 explored DM production at neutrino experiments \cite{Coloma:2022dng}.
\end{itemize}
Theoretical work is inherently complementary to any of these experimental frontiers searching for DM, as it motivates specific experimental directions and sharpens the connection and complementarity between the several experimental probes \cite{Craig:2022cef}. The instrumentation, computing, underground facilities, and accelerator frontiers are also crucial for advancing our knowledge of the nature of DM. 

In the rest of these proceedings, I will briefly summarize the studies of RF6 \cite{Gori:2022vri}, highlighting the role of present and future high intensity experiments in unraveling the nature of sub-GeV DM.

%%%%%%%%%%%%%%%%%%%%%%%%%%%%%%%%%%%%%%%%%%%%%%%%%%

\subsubsection{Dark sectors at high intensity experiments}
\label{}

Sub-GeV dark sectors naturally arise in many well-motivated extensions of the Standard Model (SM).  DM can belong to a dark sector of particles similar in structure and complexity to that of ordinary matter. Generically, a dark sector is needed for MeV-GeV DM to be thermal with the SM in the early universe and undergo the freeze-out process. Beyond DM, dark sectors are motivated by several other open problems in particle physics and cosmology. For example,
dark sectors that contain sterile neutrinos can explain the lightness of SM neutrinos; richer dark-sector models can generate the baryon-antibaryon asymmetry of the universe and potentially address the hierarchy problem (e.g., in relaxion models \cite{Graham:2015cka} or in extended SUSY models like the NMSSM); dark sectors can address the strong-CP problem via axion or axion-like particles. 

Dark sectors can be systematically classified according to the "portal" that mediates their interaction with the SM. The lowest dimensional portals are the vector portal, $(\varepsilon/2)F^{\mu\nu} F'_{\mu\nu}$, the Higgs portal, $ \lambda S^2 H^\dag H$, the neutrino portal, $yNH L$, and the axion portal, $aF_{\mu\nu}\tilde F^{\mu\nu}/f_a$. In each portal, a "mediator" interacts with one or more SM particles.
Since dark sectors are generically only weakly coupled to the SM, the most powerful means to explore them is through high intensity experiments. 

In the last decade, a large experimental and theoretical effort has led to much progress in the search for the production and detection of MeV-to-GeV mass dark-sector particles. This effort started with the 
re-analyses of data from past experiments not necessarily designed for dark sector searches. Then it continued with searches at large multi-purpose experiments and at specialized smaller-scale experiments. This endeavor resulted in several community reports \cite{Battaglieri:2017aum,Alemany:2019vsk,Agrawal:2021dbo} and has recently received support from the DOE-supported DMNI program \cite{BRN}. This program selected two intensity frontier projects, CCM200 and LDMX, to explore low-mass thermal DM scenarios. 

\begin{figure}[ht]
  \centering  
  \includegraphics[width=0.8\textwidth]{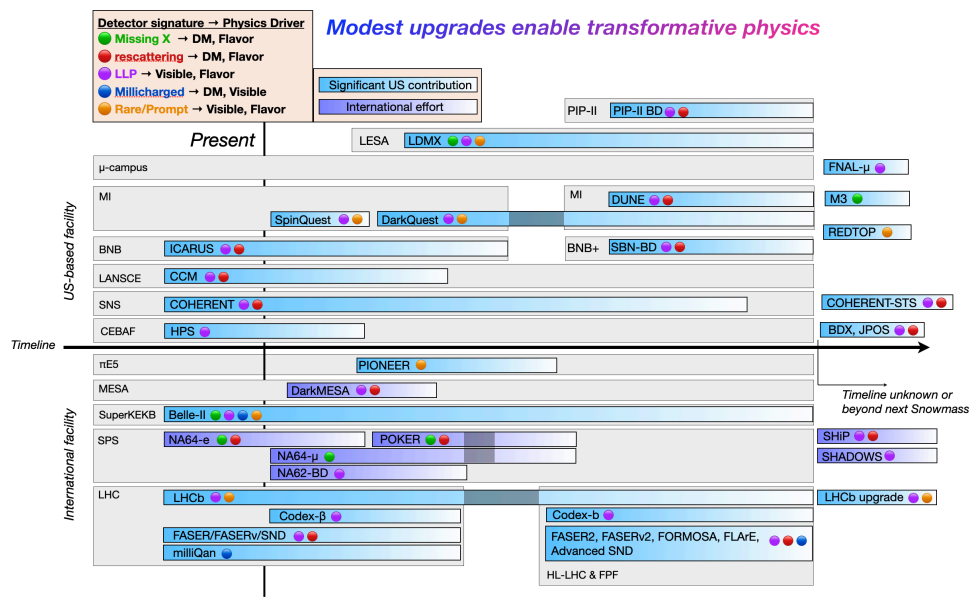}
  \vspace{-3em}
  \caption{Summary of accelerator facilities, experiments, and detector signatures (from \cite{Ilten:2022lfq}). \label{rf6-fig:exp-sum}}
\end{figure}

The work of the RF6 topical group was organized around three ``big ideas" and highlighted a vision for the next decade to achieve new milestones in the search for dark sectors. The experiments and facilities discussed by the topical group are reported in Fig. \ref{rf6-fig:exp-sum}. They are divided according to the dark sector signature that they can look for (see the different colors in the figure), a rough timeline, and their location (US-based or non-US-based).

 \vspace{0.3cm}
%%%%%%%%%%%%%
Big idea 1: ``Dark matter production at intensity-frontier experiments"\\
\noindent DM can be produced at high intensity experiments thanks to its interaction with the SM through the several portals. Once produced, it can be detected using three different search strategies \cite{Krnjaic:2022ozp}: (1) inferring missing energy, momentum, or mass; (2) detecting re-scattering of DM particles in downstream detectors; (3) observing semi-visible signatures of metastable dark-sector particles. The latter method is relevant to explore non-minimal DM models that contain additional unstable dark sector states besides the DM and the mediator (see also big idea 3). Different search strategies will have complementary strengths in the exploration of thermal freeze-out DM models and will also be complementary to low-threshold direct detection experiments \cite{Essig:2022dfa}.

\begin{figure}[p!]
\begin{center}
\includegraphics[width=0.48\textwidth]{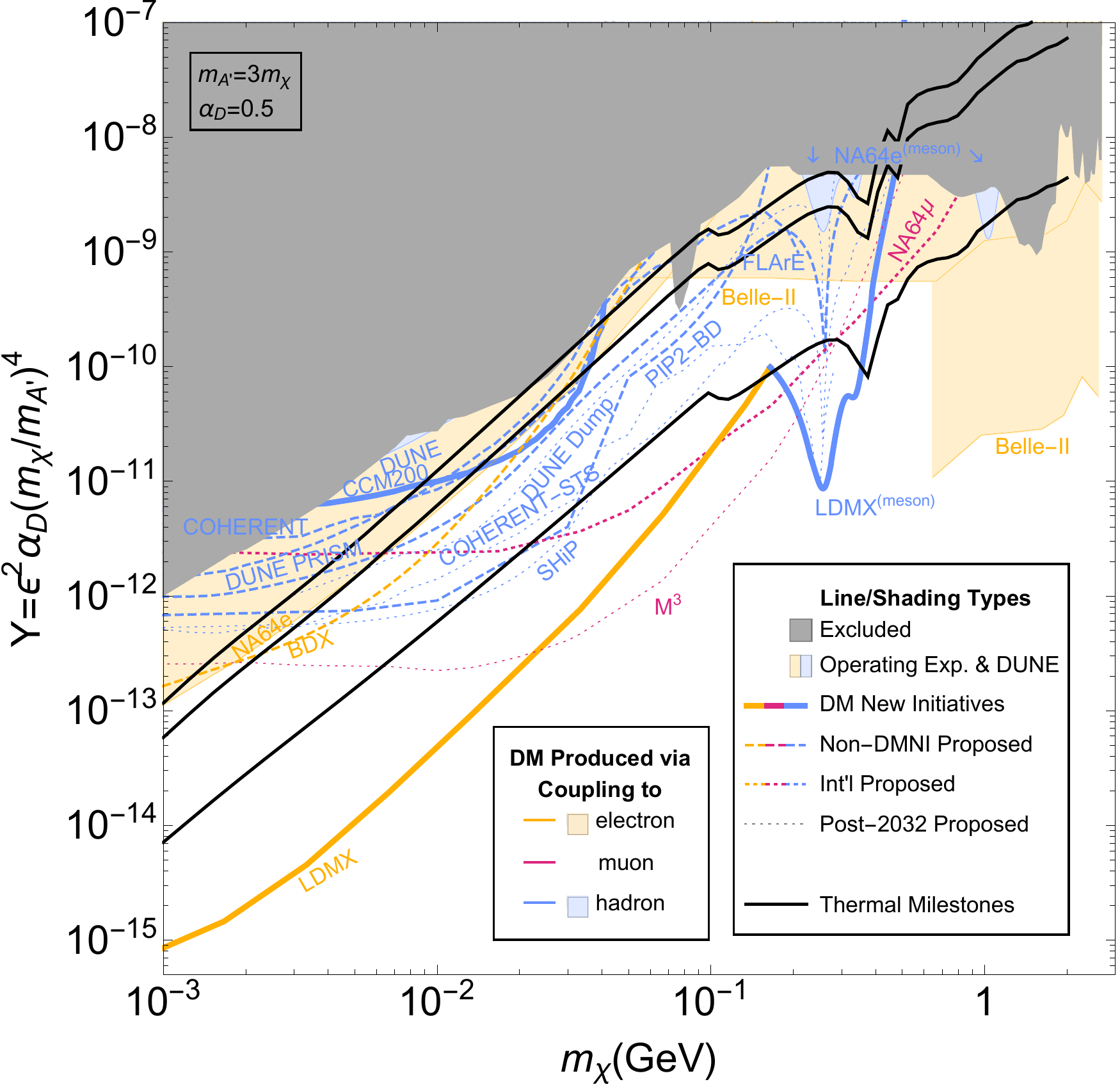}~~~
\includegraphics[width=0.46\textwidth]{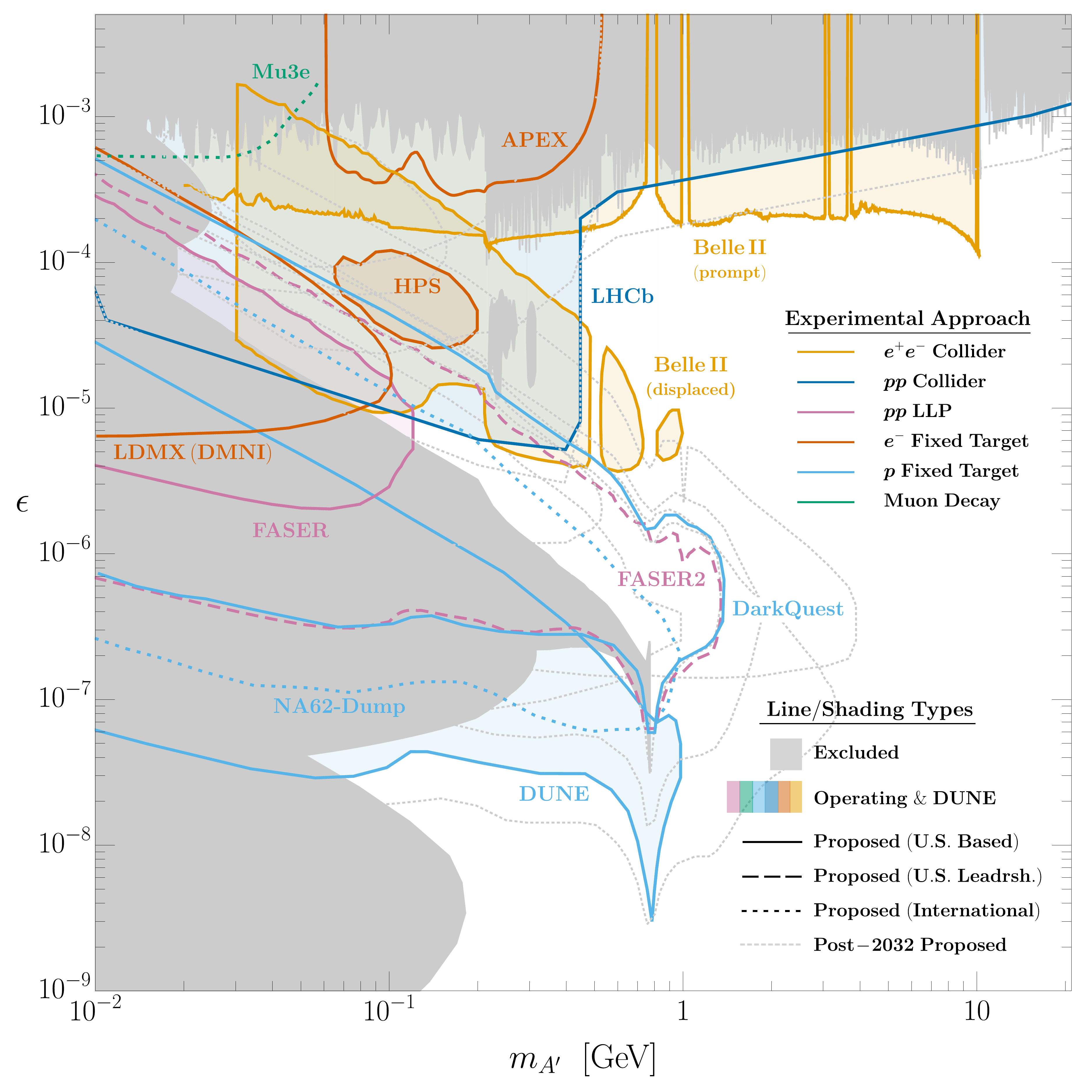}\\\vspace{1cm}
\includegraphics[width=0.46\textwidth]{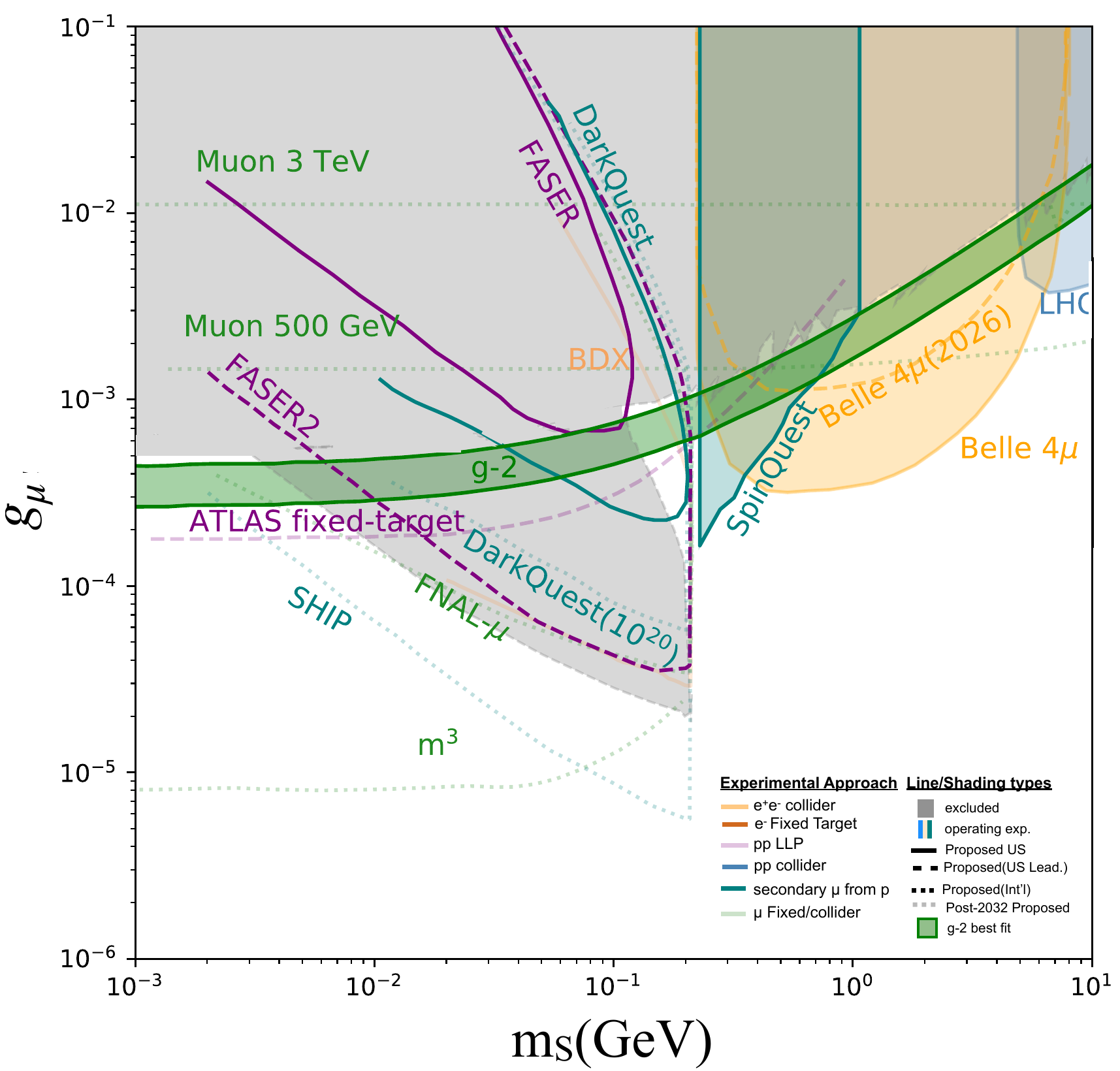}
\end{center}
\caption{
{\bf{Left:}} Reach of past, present, and future experiments on dark photon-mediated dark matter production~\cite{Krnjaic:2022ozp}. Thermal milestones are shown as black solid lines. 
{\bf{Right:}} Reach of past, present, and future experiments on the minimal vector portal \cite{Batell:2022dpx}.
{\bf{Bottom:}} Reach of past, present, and future experiments on a muon-philic scalar that can address the $(g-2)_\mu$ anomaly (green shaded region) \cite{Harris:2022vnx}.
In all panels, past experiments probed the gray-shaded regions, and future projects that are operating or have
secured full funding will probe the colored-shaded regions.
\label{rf6Figures}}
\end{figure}

A particularly interesting example are models in which the DM state interacts with the SM through the vector portal. Such simple scenarios relate the cosmological abundance of thermal DM to the signals expected at high intensity experiments, defining a sharp milestone in DM interaction strength as a function of its mass. This is shown in the left panel of Fig. \ref{rf6Figures}, where the black lines are the regions in which DM has the measured relic abundance, in the case of a scalar, Majorana, or Dirac inelastic DM state (from top to bottom). The three thermal milestones are fully achievable by near-future experiments. Particularly, Belle II and LDMX will be able to thoroughly probe the high and low DM mass region, respectively. While for this specific goal many future experiments look redundant, the use of multiple complementary techniques is important to e.g. probe a broader class of thermal freeze-out models, such as those where a mediator does not couple to electrons, or to test models where unstable particles in the dark sector play important roles in the DM cosmology. Several of these scenarios were studied during the Snowmass process \cite{Krnjaic:2022ozp}: vector ($L_\mu-L_\tau$, $B-3L_\tau$, $B$) mediated DM models; scalar (Higgs-mixed, muon-philic, neutrino-philic) mediated DM models; sterile neutrino mediated DM models; millicharged particles; inelastic DM (IDM) models as well as strongly interacting massive particle (SIMP) DM.

  \vspace{0.3cm}
%%%%%%%%%%%%%
Big idea 2: "Exploring dark-sector portals with high-intensity experiments"\\
\noindent High intensity experiments offer a unique opportunity also to explore the physics of the dark sector mediator. 
Mediators can be produced thanks to their portal interactions, and, if they are the lightest state of the dark sector, they decay back to SM particles with a lifetime proportional to the inverse of the square of the portal interaction strength. This is realized e.g. in models of secluded \cite{Pospelov:2007mp} or forbidden DM \cite{DAgnolo:2015ujb}, as well as in non-minimal models containing axion-like particles that address the strong CP problem (see e.g. \cite{Agrawal:2017ksf}). Interestingly, these models often give a lower bound on the strength of the portal coupling, thus defining a target region for high intensity experiments. This is the case of secluded DM models, in which the condition of thermalization of the dark sector with the SM does not allow too small couplings \cite{Evans:2017kti}.

The Snowmass contribution \cite{Batell:2022dpx} studied the prospects for testing minimal extensions of the SM featuring a single new light mediator coupled through one of the portal interactions at present and future high intensity experiments. This includes electron and proton beam fixed target and beam dump experiments, medium energy $e^+e^-$ colliders / meson factories, and auxiliary LHC detectors. As illustrative example, the right panel of Fig. \ref{rf6Figures} presents the near-term and future opportunities to probe the minimal vector portal model. A combination of operating, fully or partially funded and proposed near-term and future experiments will be able to search for dark photons over a broad range of currently unconstrained parameter space both in mass and lifetime. Particularly, Belle-II, LHCb, and HPS will cover regions of short dark photon lifetimes, DarkQuest, FASER2, and NA62-dump will cover intermediate values $\tau_{A^\prime}\sim\mathcal O(1$m$)$, and DUNE will reach longer lifetimes of $\mathcal O(100$m$)$.

 \vspace{0.3cm}

%%%%%%%%%%%%%
Big idea 3: "New flavors and rich structures in dark sectors"\\
\noindent Experimental work on dark sectors has been primarily focused on minimal scenarios (see big idea 1 and 2).  However, dark sectors may have a non-minimal structure, either in couplings to the SM, or in the spectra of dark sector states. For example, models that address anomalies in data like $(g-2)_\mu$ often feature a non-minimal flavor structure (i.e. dark sector particles that couple only to one generation fermions). Analogously, extended DM models such as IDM or SIMP models predict the existence of metastable DM excited states that decay partially visibly, $\chi_2\to\chi_1\ell^+\ell^-$, where $\chi_1$ is the DM state and $\ell^\pm$ are SM leptons. 

Several well-motivated non-minimal dark sector scenarios were investigated in the Snowmass contribution \cite{Harris:2022vnx}: $L_\mu-L_\tau$ visible gauge bosons, IDM and SIMP models, flavor violating QCD axions, axions coupled to gluons, $SU(2)$ gauge bosons, up or down quarks. The bottom panel of Fig. \ref{rf6Figures} shows the reach of past, present, and future high intensity experiments on a muon-philic scalar that can address the $(g-2)_\mu$ anomaly. This is one of the very few available minimal models with new particles with a mass below a few GeV that can address the anomaly. If the scalar has a mass above few MeV, this model can be fully tested in the coming years by a combination of DarkQuest, Spinquest, Belle II, and LHC (see also \cite{Forbes:2022bvo} for a more recent analysis of the parameter space with the scalar mass $m_S\gtrsim 2m_\mu$).

%%%%%%%%%%%%%
\subsubsection{Concluding remarks}
This was the conclusion of the RF6 Snowmass study: Dark sectors are a compelling possibility for physics beyond the Standard Model, to which high intensity experiments offer unique and unprecedented access. 
Maximizing the possibility of discovering a dark sector requires a four-pronged approach:
(1) support for dark-sector analyses at multi-purpose experiments; (2) the DMNI program; (3) an expansion of DMNI to include a focus on complementary signatures such as those containing visible final states (especially long-lived particles); (4) dark-sector theory. This approach will enable a robust and broad exploration of the dark sector in the coming decade and beyond.

%%%%%%%%%%%%%%%%%%%%%%%%%%%%%%%%%%%%%%%%%%%%%%%%%%%

%%%%%%%%%%%%%%%%%%%%%%%%%%%%%%%%%%%%%%%%%%%%%%%%%%%%%%%%%%%%%%%%%%%%

%\bibliographystyle{ieeetr}
%\bibliography{biblio}

%\end{document}
\afterpage{\clearpage}
%-------------------------------------------

%-------------------------------------------
\subsection{Light DM in the MeV-GeV range: results and prospects at ATLAS, CMS, LHCb -- {\it P.~Harris}}
\label{ssec:harris}
{\it Author: Philip Harris, <pcharris@mit.edu>}

\subsubsection{Introduction}
With a center of mass energy of greater than 13~TeV. The Large Hadron Collider (LHC) at CERN is the only source of laboratory collisions with a center of mass above 10~GeV on Earth. As a result, it can produce the heaviest possible particles in laboratory collision, with no other collider beyond a center of mass energy of 10~GeV expected in the near future. The LHC will remain the definitive instrument to search for heavy new particles in the controlled collider environment. 

Searches for dark sector physics in the LHC have evolved to have four characteristic signatures that drive the design and strategy of the search. These characteristic features include 
\begin{itemize}
\item {\bf Searches for Invisible Signatures:} This signature is defined as the production of a dark matter particle either through the decay of an intermediate portal particle or through the decay of an existing heavy particle, such as the Higgs boson\cite{Boveia:2016mrp,Abercrombie:2015wmb,Harris:2014hga,Buckley:2014fba,Abdallah:2015ter}. 
\item {\bf Searches for Visible Signatures:} The production of a new portal mediator that connects a dark sector with the visible sector. The observation of this particle is performed directly through the decay of this portal into standard model particles\cite{Albert:2017onk,Ilten:2015hya,Ilten:2016tkc,Ilten:2018crw}. 
\item {\bf Searches for a Long-Lived Signatures:} This signature constitutes of the production of an unstable portal or another dark sector particle that eventually decays, at least partly, to standard model particles after existing for a significant amount of time for the signatures to be displaced from the production point\cite{Alimena:2019zri}. 
\item {\bf Searches for special decays of the Higgs or Z boson:} Extended dark sector models often have additional Higgs and Z boson couplings that can enhance their production provided a Higgs or Z boson is produced, and that boson subsequently decays to the new dark sector particle\cite{Curtin:2014cca}. 
\end{itemize} 

While these signatures are characteristic of a broad range of dark sector searches, there are apparent differences when using the LHC instead of other collider experiments to search for these particles. Unlike beam dumps, the LHC is limited in total luminosity; this limits the  ability to probe the smallest possible couplings\cite{Apollinari:2116337}. Despite that, particle detectors at the LHC are some of the most well-instrumented devices in the world. This means that there is the possibility to search for distinct signatures in all or nearly every collection of collisions that occur\cite{LHCb:2018zdd,CERN-LHCC-2020-004}. Finally, the LHC is capable of producing heavy particles, including the Z boson and the Higgs boson, which can subsequently decay to dark sector particles through enhanced couplings that are present in many dark sector models, including axion-like portal, extensions of the dark photon, and mu-philic dark sector models. With a total luminosity of 100s of $fb^{-1}$, the LHC has produced orders of magnitude  more Higgs bosons, Z bosons, b mesons, and other heavy particles than any other previous experiment.  This large particle production makes it a unique tool for looking at particle decays for new dark sector particles. 

While the LHC has been performing dark sector searches throughout its existence, the results have often been presented in terms of heavy mediator searches that use different notation and model assumptions that are conventionally used in dark sector physics searches. Despite that, there has been a concerted effort to align these models to make it clear that the LHC can play a critical, complementary role to the many other lower energy dark sector experiments that are actively underway or being proposed\cite{Boveia:2022syt,Boveia:2022jox}.  

In the following conference report, we will review the various search strategies exploited at the LHC and highlight where the LHC can play a critical role in the search for dark sector physics. A reoccurring theme results from the high center of mass energy, which enables the direct production of new particles or the indirect production through the decay of the Higgs and Z-boson. 

\subsubsection{Direct Production of Invisible Particles and long-lived particles}
Invisible dark matter searches are performed at the LHC by searching for an imbalance of transverse energy when summing all visible particles. The missing transverse energy(MET) variable indicates the amount of unmeasured energy found in a collision. To measure the MET, we require the production of a mediator that decays invisibly and a recoiling visible object against that mediator to indicate a large amount of MET. This has conventionally been done through a broad range of searches known as the mono-X searches, which include the mono-jet, mono-Z boson, mono-Higgs boson, and many more\cite{ATLAS:2020uiq,ATLAS:2021hza,ATLAS:2021kxv,ATLAS:2021shl,ATLAS:2022yvh,ATLAS:2021gcn,ATLAS:2022bzt,CMS:2018zjv,CMS:2019zzl,CMS:2019ykj,CMS:2020ulv,CMS:2021far}. 

The production of invisible particles largely implies the direct production of dark matter particles. As a result, we often directly use our knowledge of the dark matter density as a driver for characterizing indivisible dark matter searches.  With both light and heavy dark sector mediators, we can characterize this by the relic density constraint present when considering the thermalization of light and dark sectors in the early universe. This, we can write as 
\begin{equation}
\langle\sigma v\rangle \propto  \frac{g^{2}_{\rm SM}g^{2}_{\rm DM}}{M^{2}_{\rm Portal}}
\end{equation}
Where $g_{\rm SM}$ is the Standard model particle coupling of the portal to the mediator. For Dark Photons, we can write the mixing parameter $\epsilon$ in terms of the above standard model coupling $g_{\rm SM}$ as: 
\begin{equation}
\epsilon=g_{\rm SM}\frac{2\left(\frac{M_{\rm Portal}}{M_{Z}} - 1\right)}{e\cos \theta_{W}}
\end{equation}
where $\theta_{W}$ is the weak mixing angle, $M_{Z}$ is the mass of the Z-boson, and $e$ is the electric charge. More generally, the above formula states that for a fixed and known dark matter relic density $\langle\sigma v\rangle_{\rm DM}$ the minimum standard model coupling gets larger, linearly with that portal mass, as one searches for heavy portal mediators\cite{Boveia:2022jox,Albert:2022xla,Krnjaic:2022ozp,Albert:2017onk,Nollett:2013pwa,Planck:2018vyg}.  

\begin{figure}[tbh!]
\centering
\includegraphics[width=0.75\textwidth]{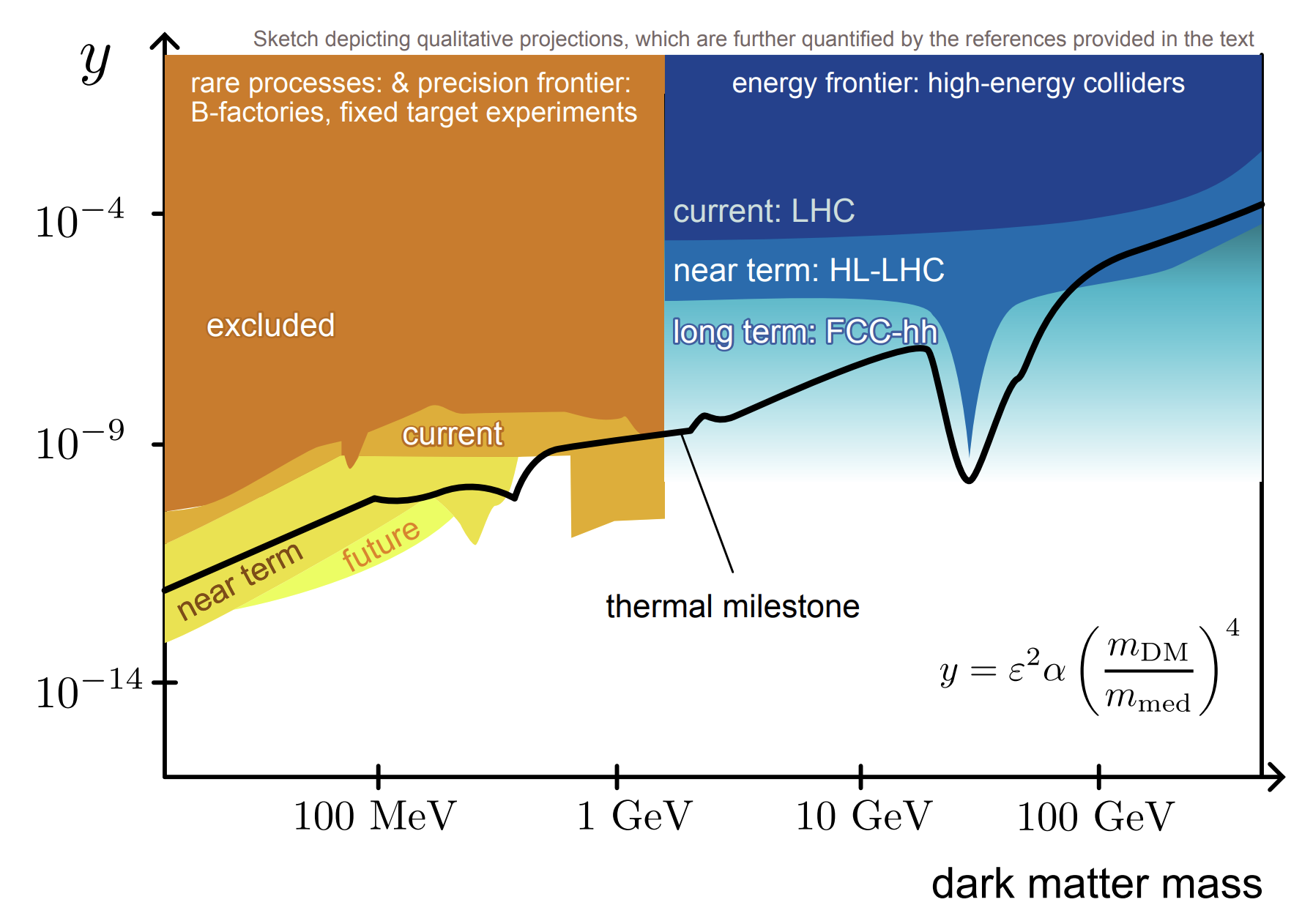}
\caption{Sketch of the dark photon bounds for dark photon decaying invisibly as a function of dark photon mass(x-axis) and dark photon mixing parameter $\epsilon$~\cite{Boveia:2022adi}, while bounds are qualitative, they illustrate the critical regions where LHC and other experiments in the next decade can probe the relic density. 
}
\label{fig:DMspin1}
\end{figure}

Consequently, for spin-1 mediators with a mass roughly greater than the $Z$ boson the standard model coupling required to attain the right relic density is large. We find that the LHC is capable of probing, or has already probed, the majority of allowed parameter space that can explain the dark matter relic density. This is indicated in figure~\ref{fig:DMspin1}\cite{Boveia:2022syt,Boveia:2022adi}. 

Likewise, a similar study can be made comparing LHC invisible searches to dark sector searches using direct detection. When the dark matter portal mass 10~GeV$< m_{\rm portal}<$2~TeV, LHC searches give complementary bounds to existing direct detection searches, which are of comparable sensitivity to current direct detection experiments, and exceed planned spin-dependent direct detection experiments\cite{Boveia:2022syt,Boveia:2022adi,Boveia:2016mrp,Abercrombie:2015wmb,Harris:2014hga,Buckley:2014fba,Abdallah:2015ter}. 

For Scalar dark sector portals, strong bounds are present from the LHC Higgs to invisible search, which relies on the distinctive vector boson fusion Higgs signature to look for invisible decays of the Higgs boson. These bounds are currently the dominant bounds for the hidden scalar portal for masses beyond a few hundred MeV\cite{Krnjaic:2015mbs}. Likewise, for Axion-like portals (ALP) with a gluon coupling, the distinctive monojet signature can probe invisible decays of the ALP portal\cite{Boveia:2022syt,Boveia:2022adi,CMS:2021far,SHROCK1982250,Argyropoulos:2021sav,Djouadi:2011aa,Baek:2012se,Djouadi:2012zc,Beniwal:2015sdl,CMS:2022qva,ATLAS:2022yvh}. 

\subsubsection{Direct Production of Visible Particles}
Direct production of visible particles is often the most striking way to observe dark sector signatures at the LHC. Visible decays of the portal mediators can lead to a set of particles that can be reconstructed, yielding a clear signature in the form of a mass resonance. When the dark sector is heavier than half $m_{\rm portal}$, standard model decays are the only signatures of the portal, making the portal a potential window into dark sector physics. However, since dark matter is not directly produced, visible searches do not directly probe dark matter but still help to explore the vast possibilities present in dark sector searches. 

Most portal mediators decay to a pair of standard model particles that allow for the reconstruction of a mass resonance. The LHC searches for resonances that decay to a pair of quarks, muons, electrons, and photons\cite{ATLAS:2018tfk,ATLAS:2019itm,ATLAS:2019erb,ATLAS:2019fgd,ATLAS:2020zms,CMS:2019emo,CMS:2019buh,CMS:2021dzg,CMS:2019gwf,CMS:2018ucw,CMS:2021ctt,CMS:2022zoc,LHCb:2019vmc,Craik:2022riw}. The sensitivity of the quark decays yields searches in the dijet or single, 2-pronged jet final state; these searches are sensitive only at large couplings to $\epsilon$ values of roughly 1 for the dark photon portal; far worse the searches with lepton final states. As a result, di-jet searches are only relevant as an experimental bound when lepton couplings are not present or heavily suppressed.  

The di-muon search has been the flagship analysis at the LHC for dark photon searches. At LHCb, the di-muon search has produced leading dark photon bounds for portal masses $ m_{A} > 220$~MeV\cite{LHCb:2019vmc}. Due to the development of the ability to save objects that operate in the trigger known as data scouting 
 or trigger level analysis, the CMS experiment has produced comparable bounds to LHCb for dark photon masses $m_{A} > 10 GeV$\cite{CMS:2019buh}. These bounds will continue to improve over the next decade. Additionally, when the portal coupling $\epsilon$ is tiny, the dark photon has a substantially large lifetime to distinguish it from the original production collision. This lifetime renders the dark photon background free, yielding an even more sensitive result.

\begin{figure}[tbh!]
\centering
\includegraphics[width=0.45\textwidth]{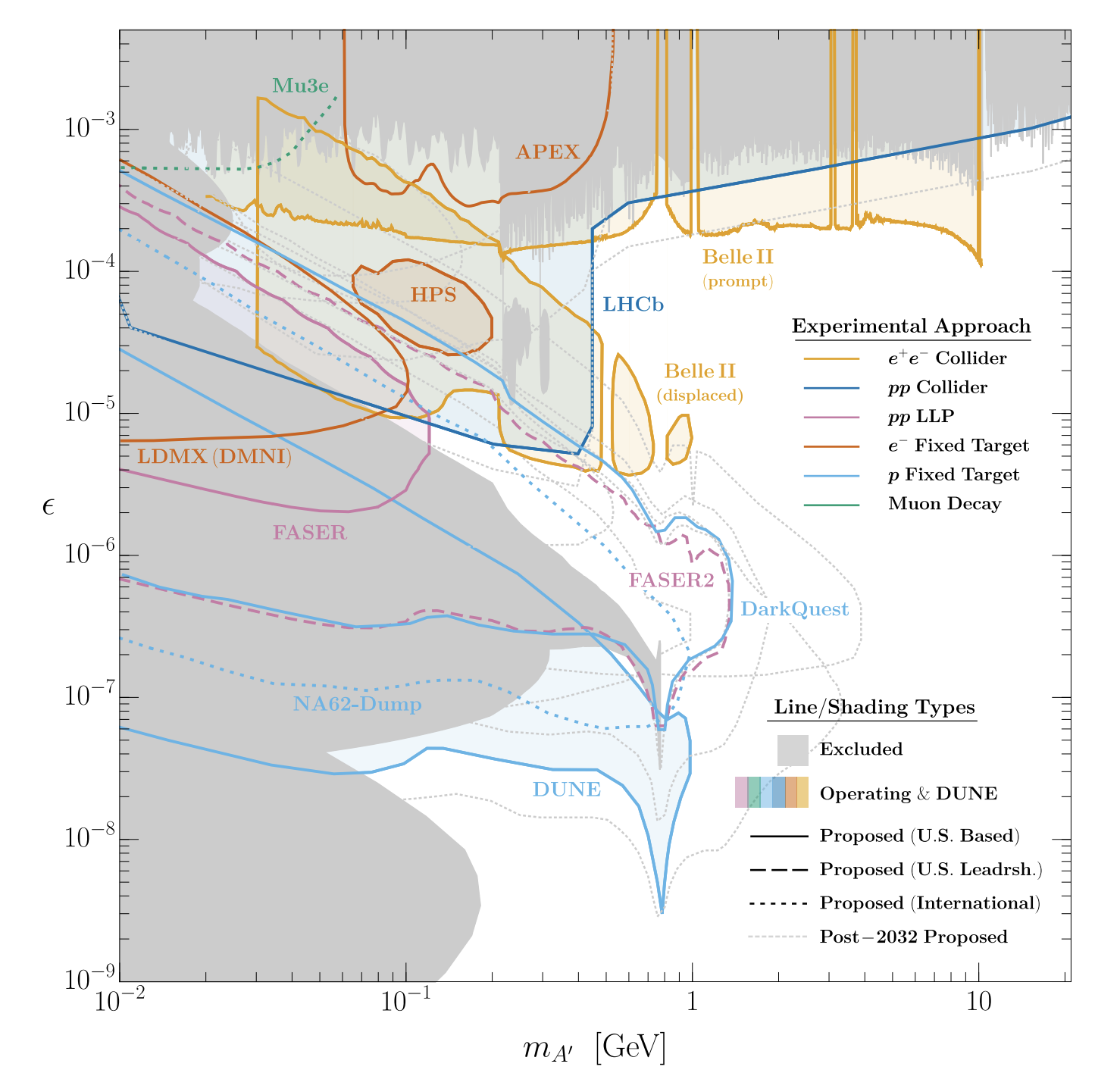}
\includegraphics[width=0.45\textwidth]{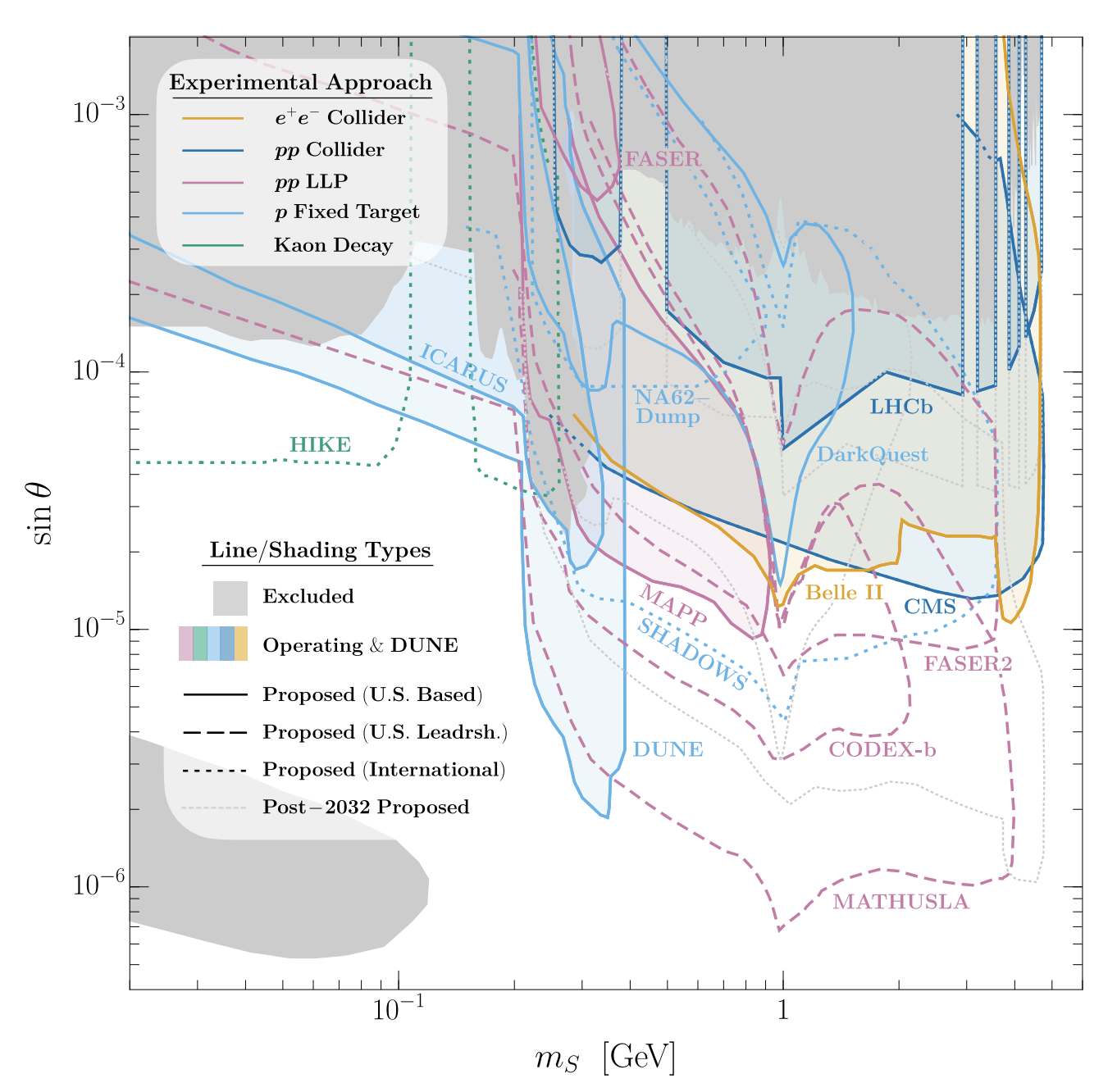}
\caption{ (left) Visible decays of the dark photon bounds on the mixing parameter $\epsilon$(y-axis) as a function of the mass of the dark photon mediator\cite{Batell:2022dpx}; (right) Visible light dark scalar bounds on the mixing parameter $\sin\theta$ (y-axis), as a function of the mediator mass\cite{Batell:2022dpx,Gori:2022vri}. The blue bounds indicate the bounds from LHC experiments.
}
\label{fig:DMvis}
\end{figure}

Similarly, when the dark photon is light, Dark photon production can occur through the decay of excited D mesons, yielding a di-electron signature resulting from an excited D-meson. This signature is distinct and produces a displaced di-electron. As a result, projections using the LHCb experiment (shown in figure~\ref{fig:DMvis}) make it such that LHCb can probe the critical $\epsilon\approx10^{-4}$ region for $10 < m_{A} < 220$~MeV\cite{Ilten:2015hya}. 

Finally, for the ALP portal, the LHC has a sensitivity to heavy axions with only a photon coupling through exclusive photon production of ALPs both in proton-proton and in heavy ion collisions\cite{Bauer:2017ris,TOTEMCollaboration:2021xam,CMS:2018erd,ATLAS:2020hii,Jaeckel:2012yz,Knapen:2016moh}. The large charge of the Pb ions substantially enhances ALP production. These searches are the most sensitive for ALPs with $5~GeV < m_{\rm portal} $. Additionally, the gluon coupling of the ALPs can be probed through $\eta\pi^{+}\pi^{-}$ decays, and $\pi^{+}\pi^{-}\gamma$ decays with the LHCb experiment giving the best sensitivity for $0.5 GeV < m_{\rm portal} < 3 GeV$\cite{Aloni:2018vki}.

\subsubsection{Direct Production of Long-lived Particles}
There are limited standard model particles with long lifetimes and distinctive signatures. As a result, there are scant long-lived standard model backgrounds when searching for particles with long lifetimes. Additionally, the prevalence of small standard model couplings among dark sector models makes long-lived signatures particularly compelling to probe dark sector physics at the LHC since small standard model couplings tend to make standard model decays of a dark sector particle long-lived. 

In inelastic dark matter models, the dark sector is extended to have an unstable intermediary particle that will then decay partly to standard model particles. Constraints from Big Bang Nucleosynthesis allow a broad range of lifetimes at values less $c\tau < 10^{4}km$~\cite{Alimena:2019zri}. The signature for inelastic dark matter can vary, but it typically involves a displaced object with limited identification. The production mode for inelastic dark matter can change. Still, often its viewed to be the same as the invisible searches with now an intermediate decay of a displaced object that enables the substantial reduction of the overall background~\cite{Harris:2022vnx,Izaguirre:2015zva,Berlin:2018jbm,Bai:2011jg,Bertuzzo:2022ozu,Curtin:2018mvb}. 

A more specific long lived signature can come from the production of a weakly coupled mediator. Light scalar mediators,$\phi$, can be produced in the decay chain of b quarks ($b\rightarrow X\phi$). The decay of these scalars can yield displaced dimuons leaving a clean, almost background free, signature. CMS has the potential to significantly take advantage of these signatures due to future track trigger capabilities that allow for the possibility of triggering on displaced tracks within the level one trigger. Bounds for this final state are shown in figure~\ref{fig:DMvis}\cite{Batell:2022dpx,Gori:2022vri,Craik:2022riw,Evans:2020aqs}. 

Figure~\ref{fig:DMllg2} shows the sensitivity of long-lived particles in the inelastic dark matter model for various experiments. For masses above a few GeV, where a large center of mass energy is required, the LHC has the largest sensitivity~\cite{Krnjaic:2022ozp,Harris:2022vnx,Berlin:2018jbm}. 

\subsubsection{Production of Extended Dark Sectors}
Finally, a unique aspect of the LHC is the substantial amount of Z and Higgs bosons produced at the LHC. Extended dark sector models, which have Higgs or Dark sector couplings, benefit substantially from the LHC. This is often quite common in dark sector models since one way to give mass to the dark sector is through a dark Higgs boson that mixes with the Higgs boson~\cite{Curtin:2014cca,ATLAS:2018coo}. 

If a Higgs boson coupling to dark photons is present, it is possible to look for the dark photon in the 4 (displaced) lepton final state. This can be a clean signature with the added mass constraint and a scheme to reduce the Z boson decays( $h\rightarrow ZZ$ background). Provided a sufficiently large Higgs to dark photon coupling such that the Br($h\rightarrow Z_{D}Z_{D}$) >
$10^{-5}$~\cite{Curtin:2014cca}. 

Similar arguments are present in the case of light scalars and ALPs. Higgs decays directly into these can yield a large production of these new particles, which can, in turn, be observed through long-lived decays. Due to the distinct lack of background branching ratios of Br$(h\rightarrow ss) > 10^{-3})$ can be probed with the existing LHC detectors and dedicated long-lived particle searches\cite{ATLAS:2018jnf,ATLAS:2020pcy,CMS:2018nsh,CMS:2019spf,CMS:2020ffa,CMS:2021pcy,CMS:2022xxa,CMS:2022fyt,ATLAS:2021jig,CMS:2021juv}. 

\begin{figure}[tbh!]
\centering
\includegraphics[width=0.45\textwidth]{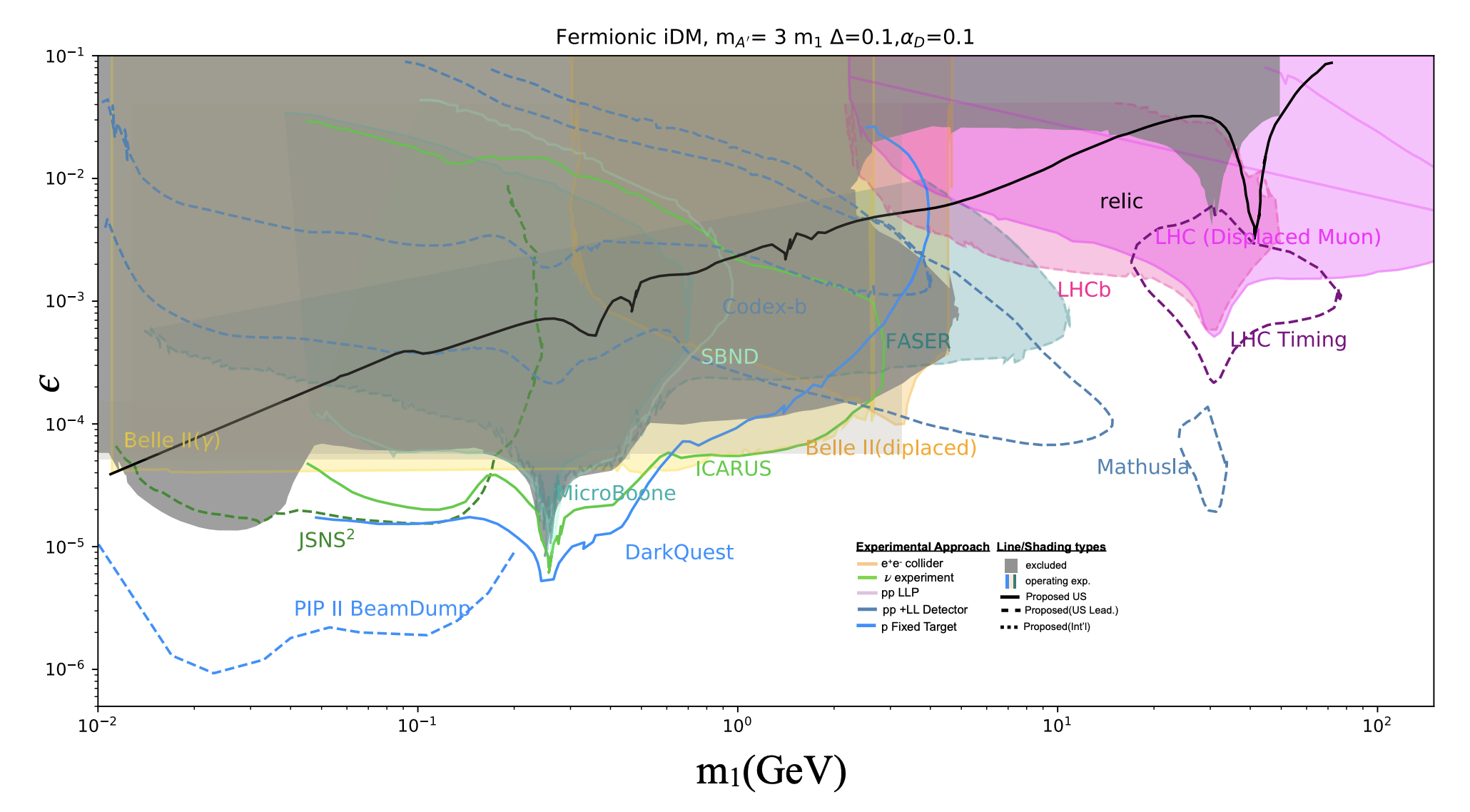}
\includegraphics[width=0.45\textwidth]{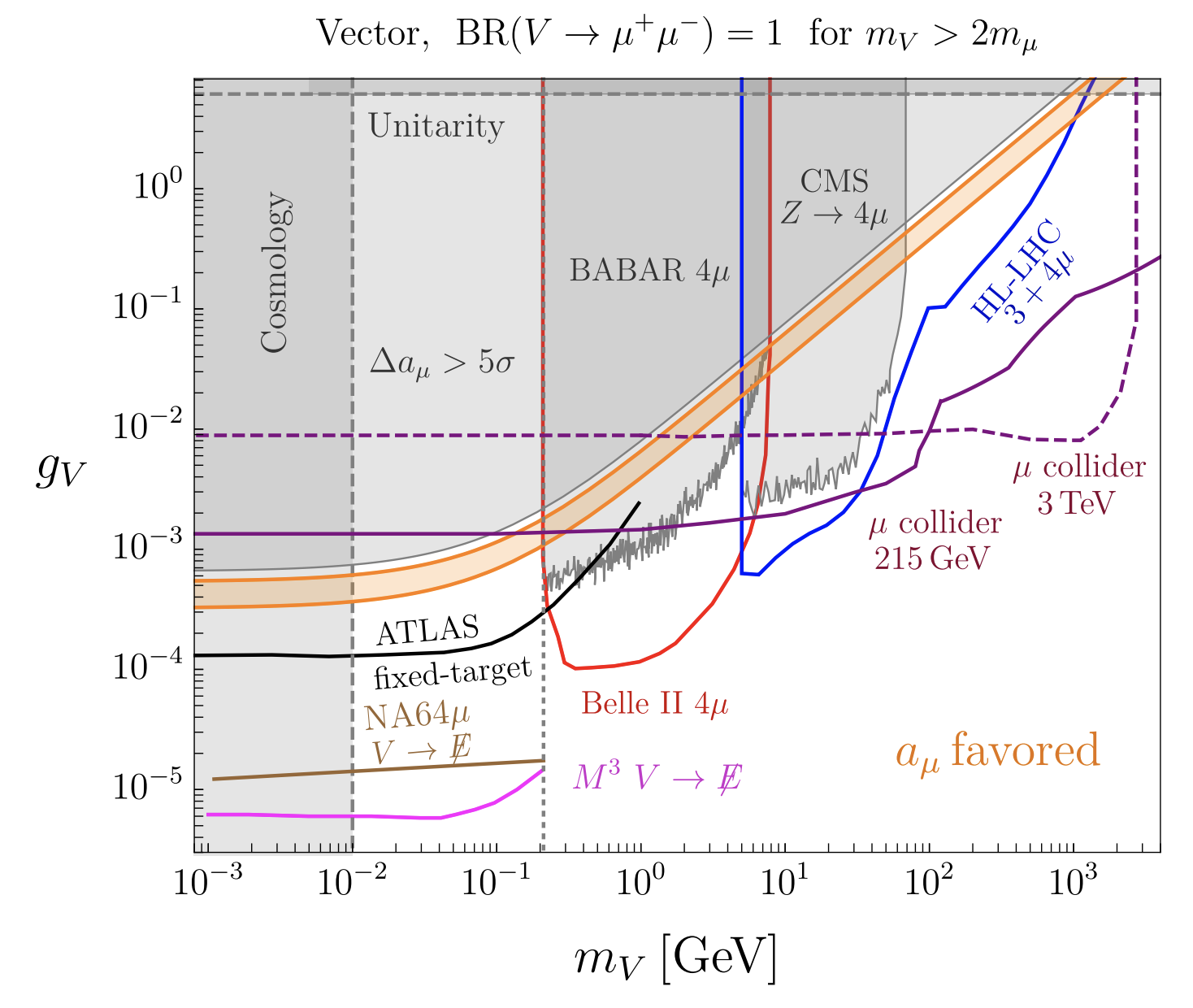}
\caption{ (Left) Bounds for the observation of a long-lived decay in the inelastic dark matter model where the dark photon decays to an unstable particle with a mass splitting between the excited particle($m_{2}$) and dark matter particle($m_{1}$) given by $\Delta=\frac{m_{2}-m_{1}}{2(m_{2}+m_{1})}=0.1$ and a dark matter coupling given by $\alpha_{D}=0.1$. In all cases, the dark photon mass is taken to be three times dark matter mass $m_{A}=3m_{1}$\cite{Harris:2022vnx}.   (Right) Comparison of coupling exclusion(y-axis) on a spin-1 vector boson that couples exclusively to muons as a function of the mediator mass (x-axis). The LHC bounds and projected bounds are the leading bounds for a mediator mass above 5~GeV\cite{Capdevilla:2021kcf}.
}
\label{fig:DMllg2}
\end{figure}

Finally, a model motivated by the $(g-2)_{\mu}$ anomaly observed at Fermilab would be explained by a portal mediator that exclusively couples to muons. This portal mediator can be produced through a radiative portal produced off a muon in a Z boson decay. This is a similar process to  the radiative 4 muon peak that is present in Z decays, except now one of the dimuon pairs will be at the mass of the portal mediator. The current and future LHC bounds can allow for the search of such a mediator that would explain the $(g-2)_{\mu}$ anomaly for masses ranging from 2~GeV to 500~GeV\cite{Capdevilla:2021kcf,CMS:2018yxg}. The bounds for this model are shown in figure~\ref{fig:DMllg2}.

\subsubsection{Conclusions}
As the only machine capable of producing a center of mass collisions above  10~GeV in the laboratory environment, the LHC will have a critical role in dark-sector searches over the next decade. These cover 4 distinct types of dark sector signatures: invisible particle searches, visible particle searches, long-lived searches, and heavy particle decays. For the first three final states(invisible, visible, long-lived), the LHC dominates the sensitivity above portal interactions at the scale of 10~GeV.  For lighter portal mediators, displaced and visible decays make the LHC the most sensitive future experiment for masses above a few hundred MeV. Finally, the LHC has produced more Higgs and Z bosons than another collider experiment and, consequently, can be sensitive to extended dark sector models that result in dark sector decays from the Higgs and Z boson. This unique element of the LHC is critical for future exploration. A summary of the sensitivity is present in table~\ref{tab:darksector}. 

%Make a table
\begin{table}
\begin{center}
\begin{tabular}{ |p{2cm}|p{4cm}|p{2cm}|p{5cm}| } 
 \hline
 Signature & Models & LHC Analysis & Critical Sensitivity \\ 
 \hline\hline
 Invisible & Dark Photon or Dark Higgs with Light Dark Matter  & Mono-X & Probes relic above 100 GeV \\ 
 \hline
 Visible & Dark Photon or Alp with Heavy/No Dark Matter  & di-lepton, di-photon & Probes masses above 250~MeV with $\epsilon~10^{-4}$ \\ 
 \hline
 Long-lived & Light Dark Higgs, inelastic Dark Matter  & long-lived particle(s) & (Scalar) Probes mixing of $10^{-4}$, (Inelastic DM) dominates above 5~GeV.  \\ 
 \hline
Extended Dark Sector &  Dark Higgs Mixing, Muphilic Mediator & Higgs and Z decays & (Dark Higgs) probes small dark photon mixing when lar mixing large, (Muphilic) dominates above 5~GeV.  \\ 
 \hline
\end{tabular}
\caption{Table describing model and ranges of sensitivity for LHC searches}
\label{tab:darksector}
\end{center}
\end{table}

When considering dark sectors in the global context of all experiments, it is essential to realize that the LHC plays a distinctive role in eliminating many possible dark sector models and extensions of models. However, when considering future projections of the LHC, there are a few critical holes to keep in mind. In particular, invisible decays with mediator masses above 10~GeV to 100~GeV will be hard to search for at the LHC to reach the desired relic density benchmark. Moreover, it is crucial to remember that the LHC long-lived searches can substantially be enhanced by next-generation experiments that allow us to increase the overall sensitivity. There is a lot of work going towards the construction and design of this signatures~\cite{Alimena:2019zri,Curtin:2018mvb,Aielli:2022awh,Bauer:2019vqk}. With an order-of-magnitude increase in data and a large variety of upgrades underway, the next decade stands to be a pivotal moment for dark-sector searches at the LHC and beyond.

\afterpage{\clearpage}
%-------------------------------------------

%-------------------------------------------
\subsection{Prospects at CERN: MATHUSLA, CODEX-b,ANUBIS -- {\it B.~Dey} }
\label{ssec:dey}
{\it Main Author: Biplab Dey, <biplab.dey@cern.ch>,\\
Contributors: Cristiano Alpigiani, Oleg Brandt, Jon Burr, David Curtin, Erez Etzion, Philip Ilten, Simon Knapen, Henry Lubatti, Steven Robertson, Toby Satterthwaite, Charles Young}

\subsubsection{Introduction}
The LHC is scheduled for ongoing and upcoming upgrades and data collection until at least 2038. A central component of this program will be searches for dark or hidden sectors beyond the standard model (BSM), in particular displaced decays-in-flight of exotic long-lived particles (LLPs). They are a compelling signature of such sectors and generically occur in theories containing a hierarchy of scales and/or small parameters. Both cases already appear in the Standard Model (SM), in which many decay widths are suppressed by the $m_W\gg \Lambda_{QCD}$ hierarchy, loop and phase-space suppressions, and/or the smallness of one or more CKM matrix elements. The $K^0_L$, $\pi^\pm$, neutron and muon are the most spectacular examples. Such LLPs also make frequent appearances in BSM scenarios featuring \emph{e.g.}~dark matter, baryogenesis, supersymmetry or neutral naturalness.

\begin{figure}\centering
    \includegraphics[width=0.6\textwidth]{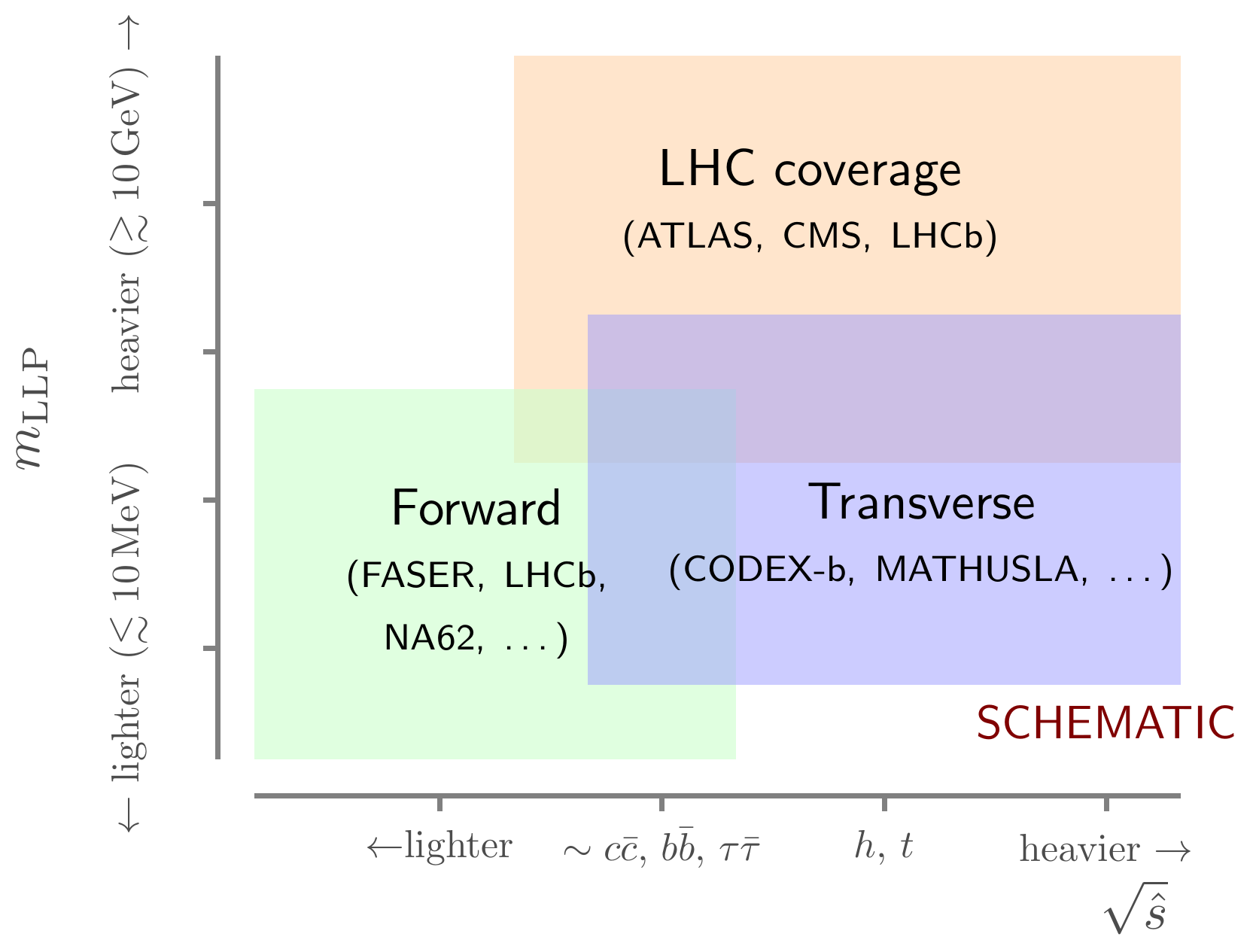} 
    \caption{Complementarity of different experiments searching for LLPs~\cite{Aielli:2019ivi}.\label{fig:schematic}}
\end{figure}

The reach of both ATLAS and CMS to the decay-in-flight of LLPs is best when the LLPs are relatively heavy (\mbox{$m\gtrsim 10\,\gev$}), though there are some important exceptions (\textit{e.g.}~\cite{CMS:2021sch,CMS:2021juv}). This is because the backgrounds and trigger challenges can strongly limit the reach for light LLPs in the complicated environment inherent to a high-energy, high-intensity hadron collider. These difficulties are reduced to a large extent by LHCb and FASER, due to a high-precision VErtex LOcator (VELO) and substantial shielding, respectively. Because of their locations and geometry, the sensitivity of LHCb and FASER is limited to relatively short lifetimes and production at low center-of-mass energies. As a result, their sensitivity to LLPs produced in, \textit{e.g.}, exotic Higgs or $B$ decays can be quite limited, especially for $c\tau \gtrsim 1\,\mathrm{m}$. To achieve comprehensive coverage of the full LLP parametric landscape, one or more high volume, \emph{transverse} LLP detectors are therefore needed (see \ref{fig:schematic}). The most viable options for such a detector at the LHC are ANUBIS~\cite{Bauer:2019vqk,ANUBIS:Satterthwaite:2839063}, \mbox{CODEX-b} \cite{Gligorov:2017nwh,Aielli:2019ivi}, and MATHUSLA~\cite{Alpigiani:2018fgd, MATHUSLA:2020uve}, as described in the remainder of these proceedings.

\subsubsection{ANUBIS}

\begin{figure}\centering
    \includegraphics[width=0.8\textwidth]{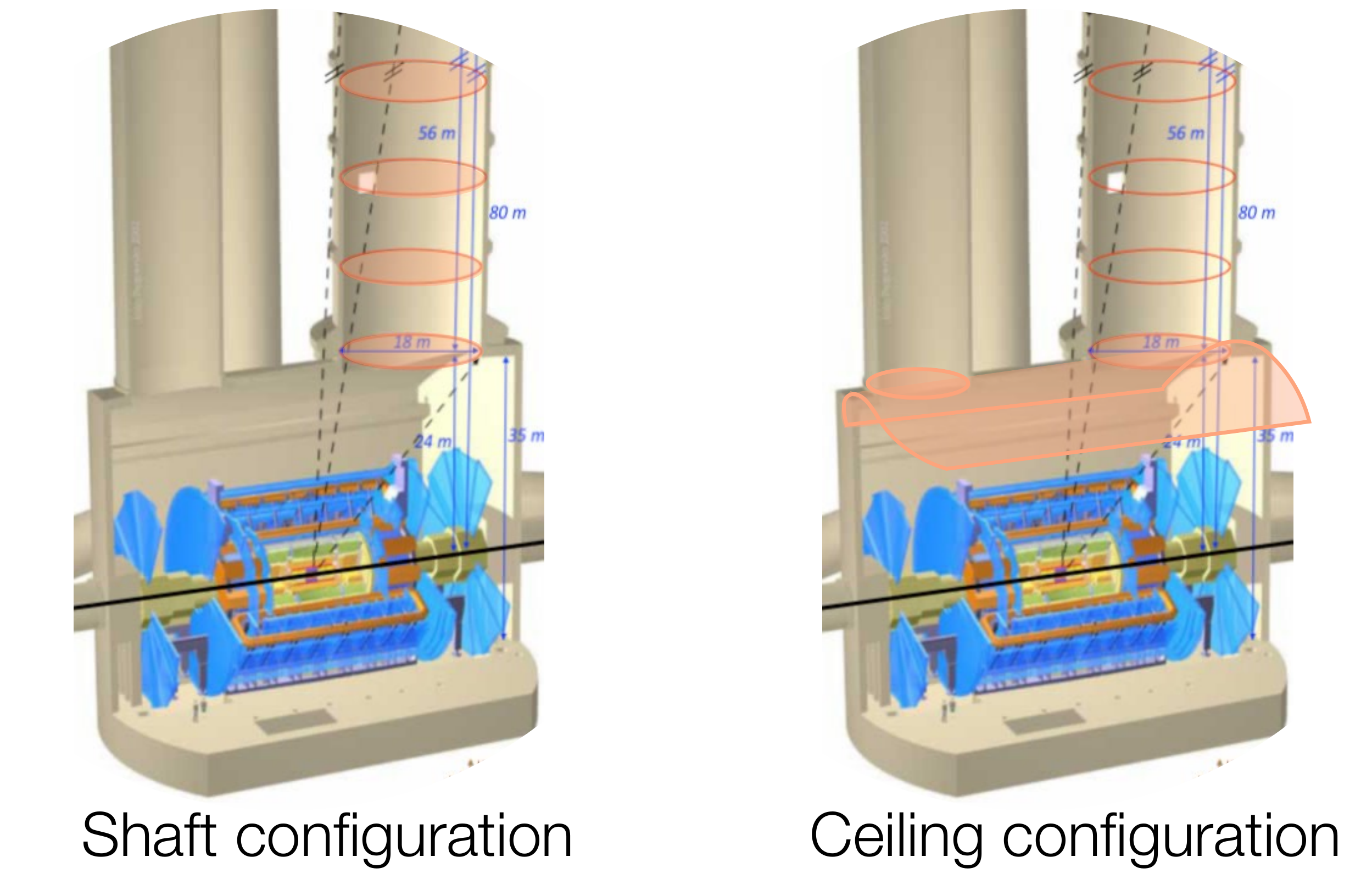} 
    \caption{The two ANUBIS configurations considered in the sensitivity study: the original shaft configuration where four tracking stations are placed in the main PX14 shaft of ATLAS, and the ceiling configuration, where a tracking station is placed on the ceiling of the US15 ATLAS cavern and one at the bottom of each of the service shafts.\label{fig:anubis_configurations}}
\end{figure}

\begin{figure}\centering
    \captionsetup[subfigure]{justification=centering}
    \centering
    \begin{subfigure}[b]{0.49\textwidth}
        \centering
        \includegraphics[width=\textwidth]{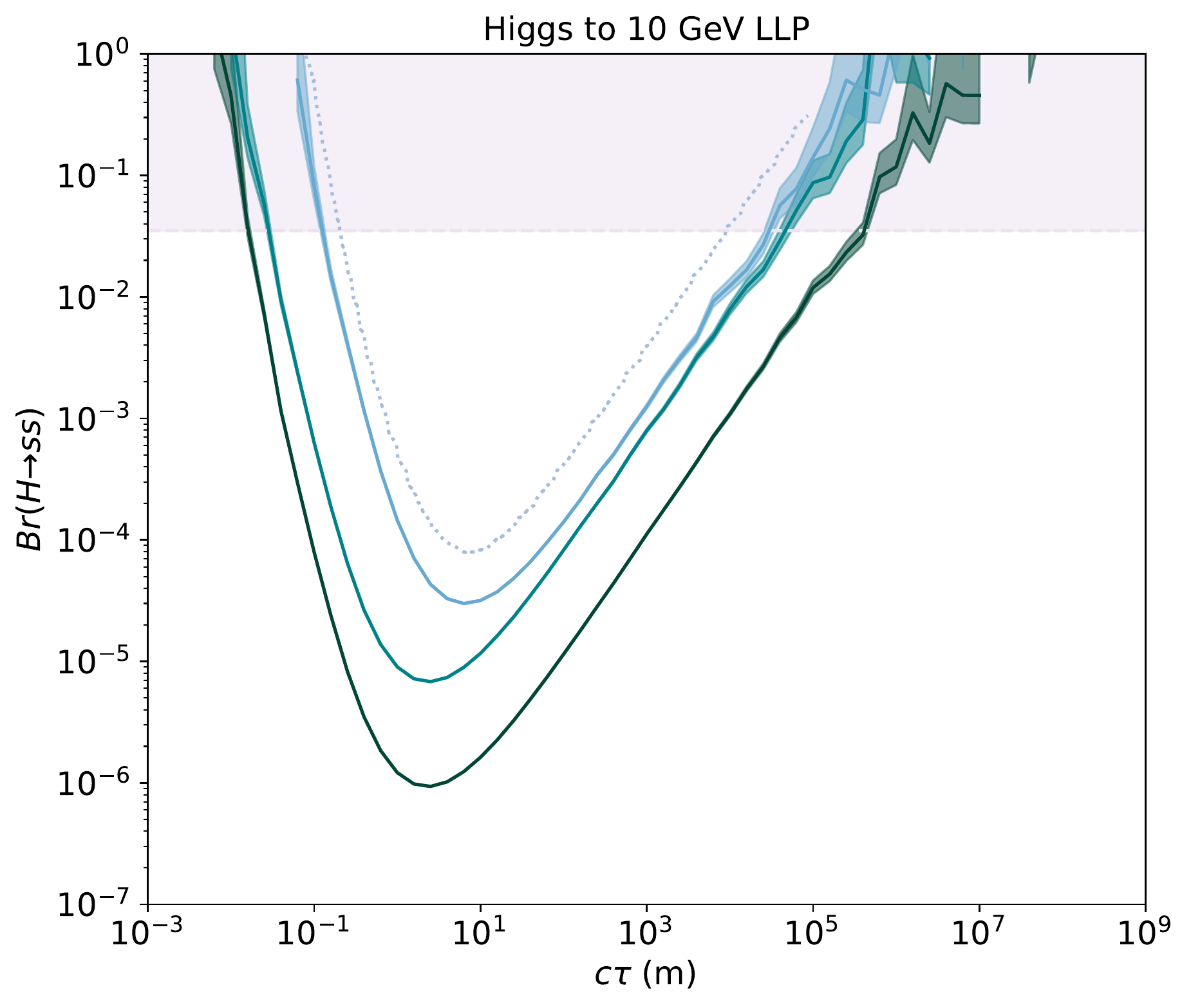}
        \caption{}
        \label{fig:br-10GeV-4evts}
    \end{subfigure}
    \hfill
    \begin{subfigure}[b]{0.49\textwidth}
        \centering
        \includegraphics[width=\textwidth]{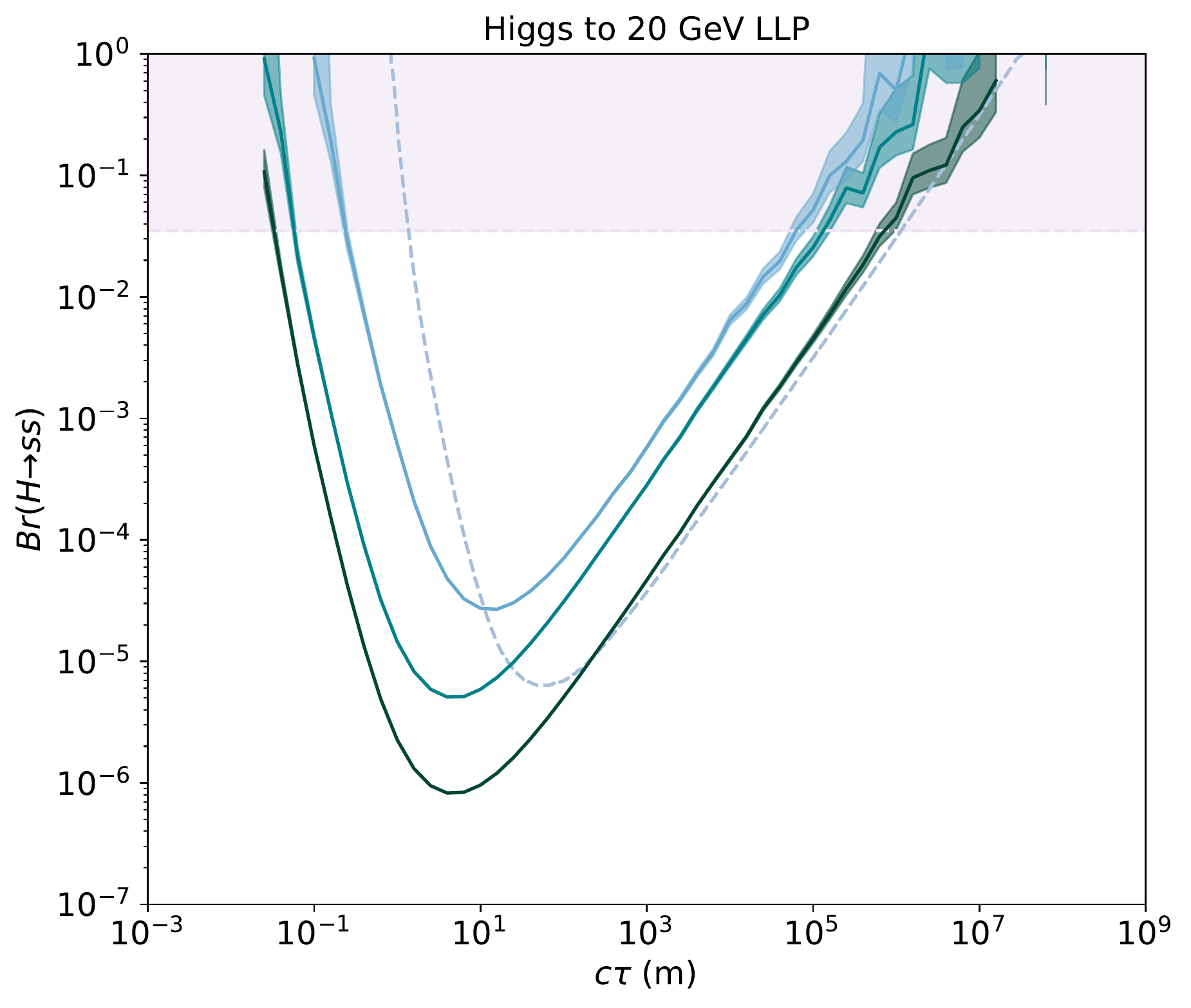}
        \caption{}
        \label{fig:br-20GeV-4evts}
    \end{subfigure}
    \hfill
    \begin{subfigure}[b]{0.49\textwidth}
        \centering
        \includegraphics[width=\textwidth]{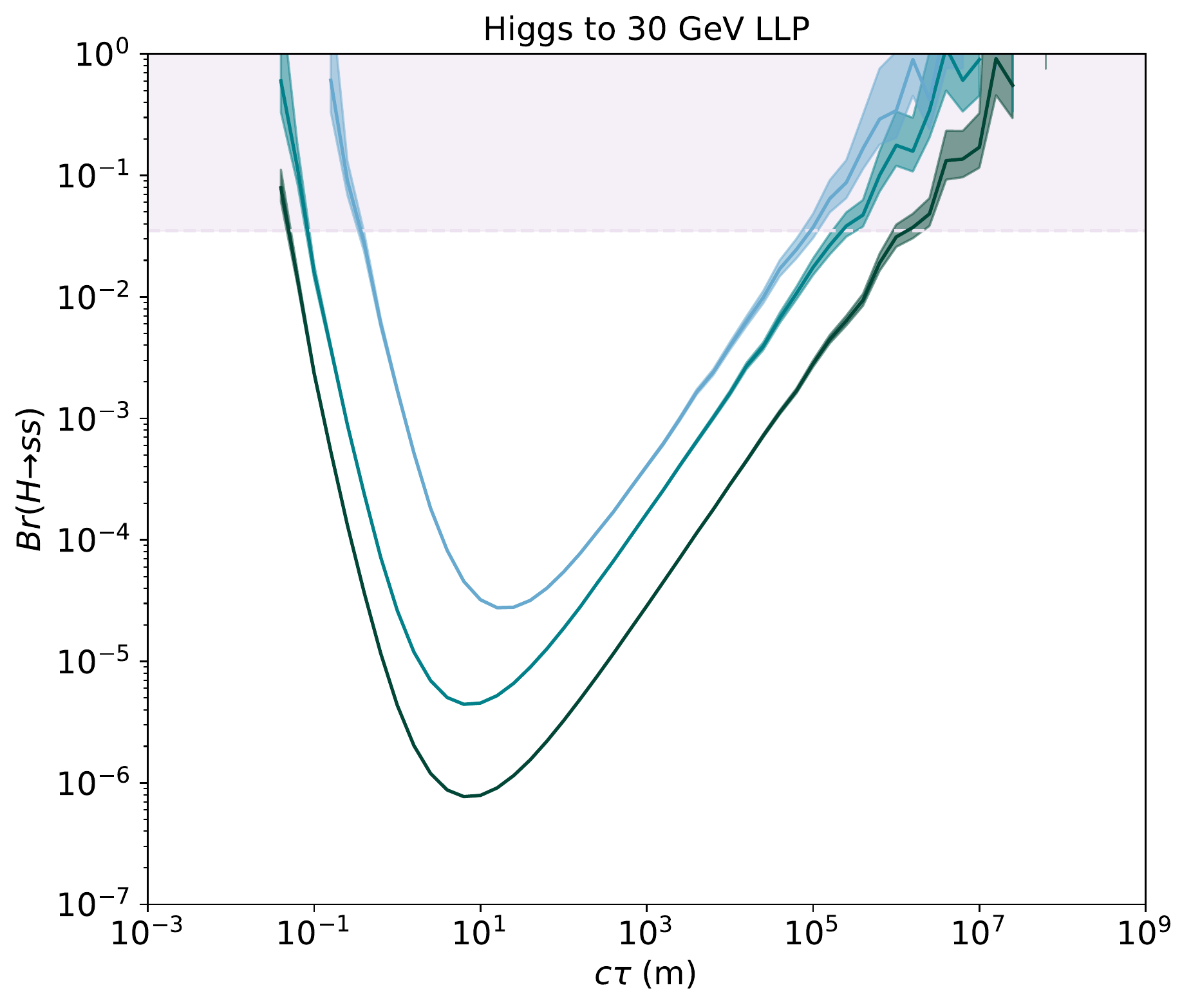}
        \caption{}
        \label{fig:br-30GeV-4evts}
    \end{subfigure}
    \hfill
    \begin{subfigure}[b]{0.49\textwidth}
        \centering
        \includegraphics[width=\textwidth]{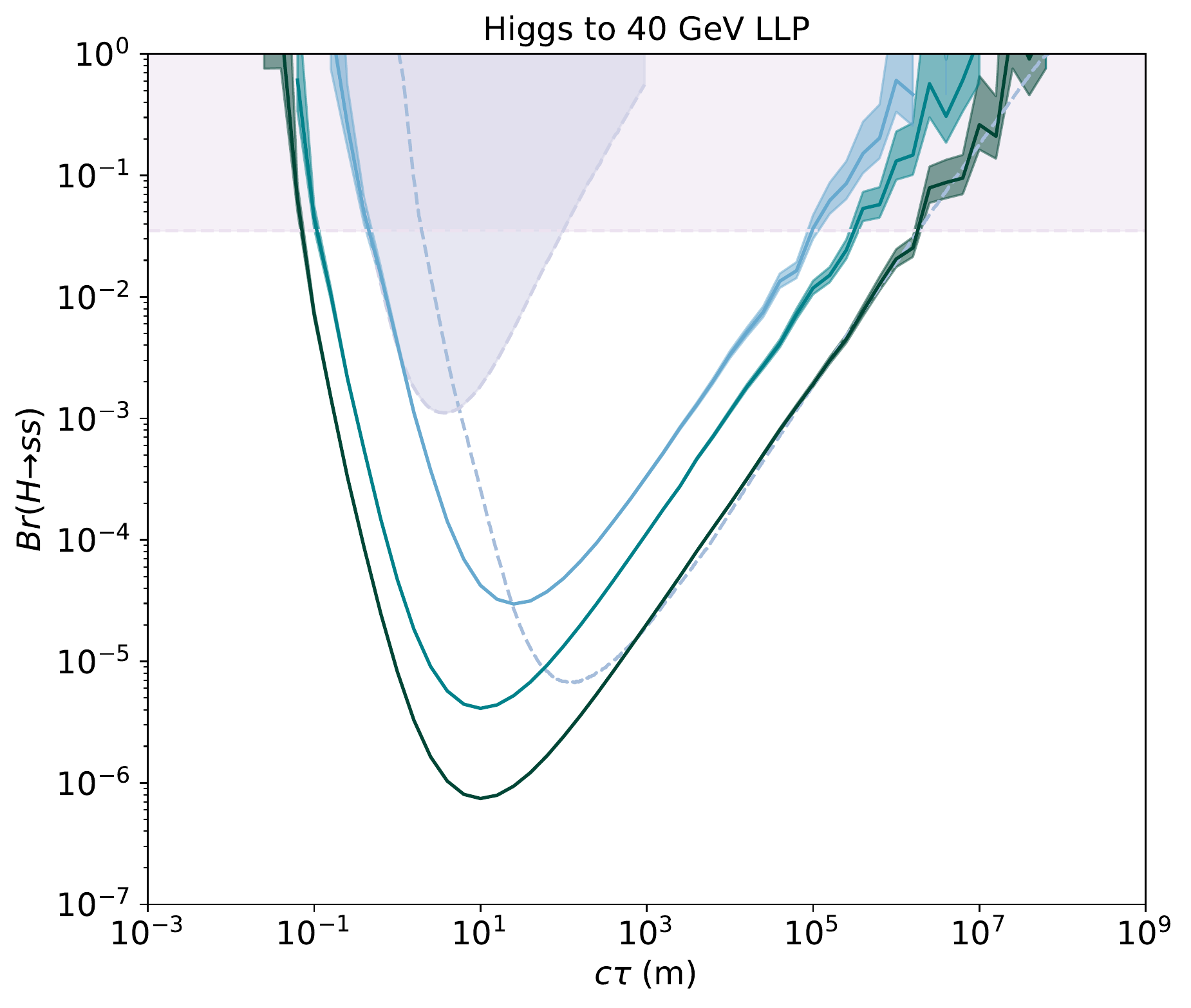}
        \caption{}
        \label{fig:br-40GeV-4evts}
    \end{subfigure}
    \hfill
    \begin{subfigure}[b]{\textwidth}
        \centering
        \includegraphics[width=\textwidth]{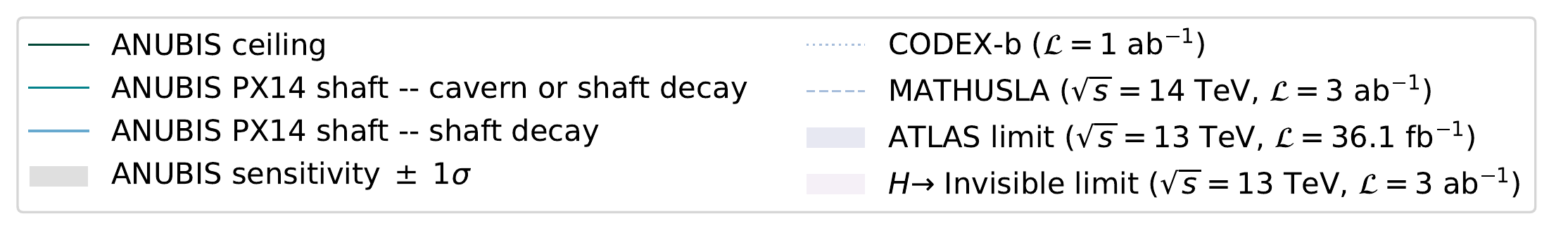}
    \end{subfigure}
    \caption{Branching ratio sensitivity of the ANUBIS cavern and shaft configurations to decays of Higgs-produced LLPs with masses of (a) 10, (b) 20, (c) 30, and (d) 40\,\gev assuming HL-LHC conditions ($\sqrt{s}=14\,\tev$, $\mathcal{L}=3\,\iab$). Uncertainties are given by Monte Carlo statistics. These are compared to ATLAS results for 40\,\gev BSM LLPs, CODEX-b projections for 10\,\gev BSM LLPs, and MATHUSLA projections for 20 and 40\,\gev BSM LLPs, as well as the projected limit for decays of the Higgs boson into invisible particles at the HL-LHC \cite{ATLAS:2018tup}\cite{Gligorov:2017nwh}\cite{Chou:2016lxi}\cite{ANUBIS:InvisibleHiggs}. For further details, see \cite{ANUBIS:Satterthwaite:2839063}.\label{fig:ANUBISSensitivity}}
\end{figure}

The ANUBIS project proposes installing a set of novel detectors as part of the ATLAS experiment in order to extend its sensitivity to neutral long-lived particles (LLPs) in the era of the HL-LHC by extending the geometrical acceptance in the transverse direction relative to the beamline~\cite{Bauer:2019vqk}. Two potential configurations are considered for the ANUBIS experiment: a ``shaft" configuration which would comprise four sets of bi-layered tracking detectors (tracking stations) installed at various heights along the main ATLAS service shaft PX14, and a ``cavern" configuration  where a tracking station is placed on the ceiling of the US15 ATLAS cavern and one at the bottom of each of the service shafts. Both of these configurations would provide tracking capabilities at distances which are significantly further from the ATLAS interaction point in the transverse direction than existing detectors. This would increase the geometrical acceptance to LLPs with decay lengths of $\mathcal{O}(10~\text{m})$ and beyond, thereby dramatically extending its sensitivity to neutral LLPs produced at the electroweak scale or above. In the case of the cavern configuration, the new detectors would be sensitive to the products of neutral LLP decays which occur between the ATLAS muon spectrometer and the cavern ceiling, while in the case of the shaft configuration, the new detectors could be sensitive to decays which occur within the PX14 service shaft. A schematic of these detector layouts and the decays to which they would be sensitive is shown in \ref{fig:anubis_configurations}.

Recent studies show that both configurations of ANUBIS would have sensitivity to LLPs produced at the electroweak scale using as a benchmark electrically neutral LLPs produced from the decays of the 125~GeV Higgs boson~\cite{ANUBIS:Satterthwaite:2839063}. These studies consider gluon-fusion and vector-boson-fusion Higgs boson production, simulated at next-to-leading order with \textsc{PowhegBox}~\cite{Oleari:2010nx}, where the Higgs boson decays into LLPs with masses of $10, 20, 30,$ and $ 40\,\gev$ with a subsequent decay into $b\overline{b}$, which are then showered and hadronized using \textsc{Pythia}~8~\cite{Sjostrand:2014zea}. A Monte Carlo method was used to investigate what fraction of these decays would occur between the existing ATLAS detectors and the proposed ANUBIS detector, and what fraction of these events would produce at least two final-state, charged-jet particles with trajectories that would allow them to be detected by ANUBIS. The sensitivity of ANUBIS to the branching ratios of four LLP masses with variable $c\tau$, assuming the full $3\,\iab$ dataset of the HL-LHC, are shown in \ref{fig:ANUBISSensitivity}. The ceiling configuration of ANUBIS would be sensitive to branching ratios of LLPs with masses of $10, 20, 30,$ and $ 40\,\gev$ reaching a limit of $\mathcal{O}(10^{-6})$ for particles with $c\tau$ of $3, 4, 6,$ or $ 10\,\text{m}$, respectively. In the shaft configuration, ANUBIS would be sensitive to decays which occur within the ATLAS cavern or the PX14 service shaft with branching ratios reaching $\mathcal{O}(10^{-5})$, and would be sensitive to decays occurring within the PX14 service shaft only with branching ratios reaching $\mathcal{O}(10^{-4})$. Each of these scenarios shows a substantial improvement to the existing capabilities of the ATLAS experiment.

A prototype detector called proANUBIS that corresponds to one element of a future tracking of the full ANUBIS detector has been constructed, and is scheduled to take data in 2023 at a representative location inside the ATLAS cavern US15. The location of the demonstrator is close to the ceiling of the ATLAS cavern, about 25\,m away from the interaction point. Most importantly, the space between proANUBIS and ATLAS is an air-filled volume, providing an ideal test bed to confirm the rate of hadronic interactions and decays of $K_L$ and $n$ predicted in Monte Carlo simulations. 

\subsubsection{CODEX-b}

\begin{figure}\centering
  	\includegraphics[width = 0.8\linewidth]{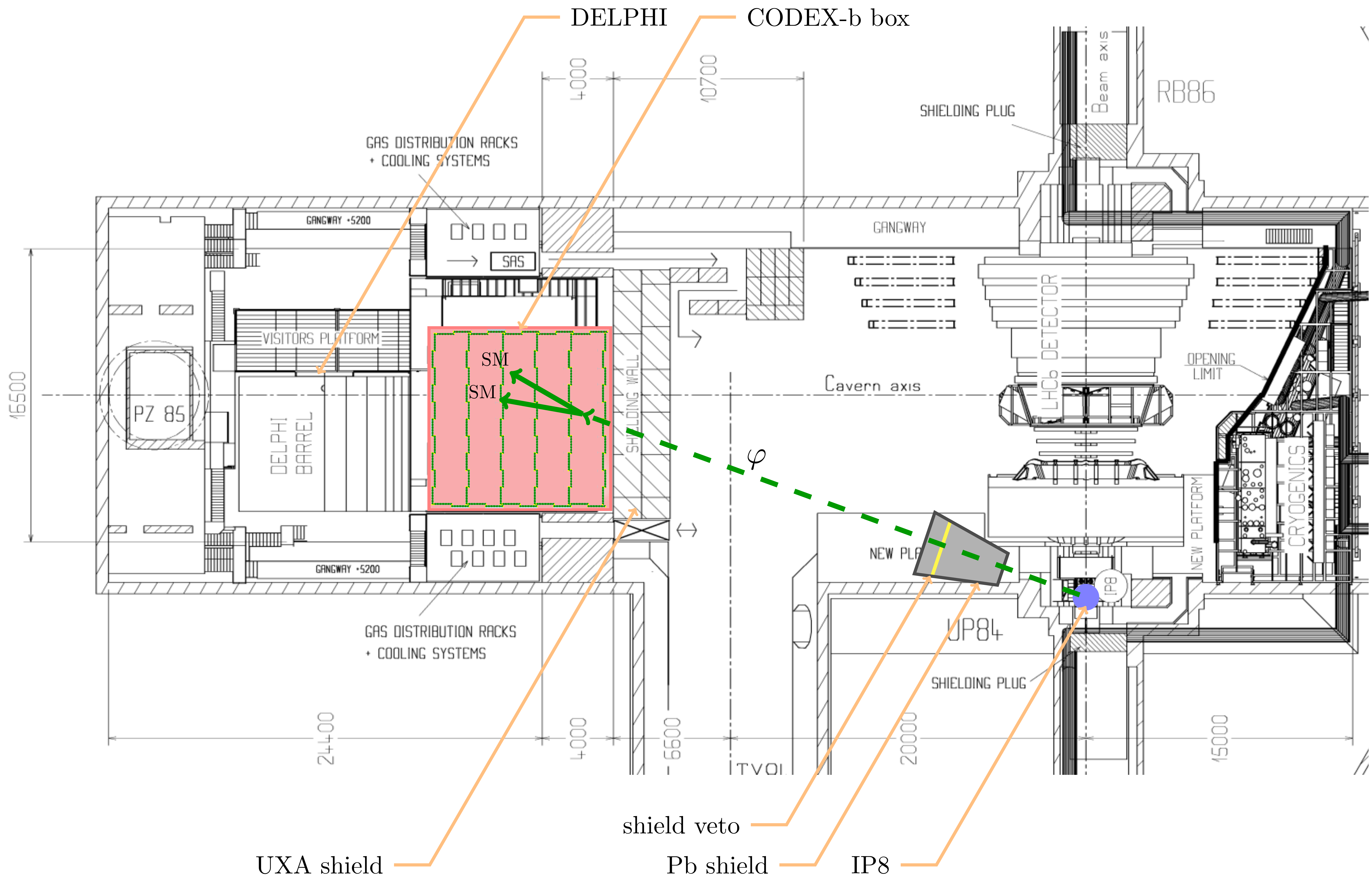}
	\caption{Top view of the LHCb cavern UX85 at point 8 of the LHC, overlaid with a top view schematic of the CODEX-b detector. Adapted from Ref.~\cite{Gligorov:2017nwh}.\label{fig:LHCbCav}}
\end{figure}

The CODEX-b experiment is proposed to be installed near the LHCb interaction point (see \ref{fig:LHCbCav}) to search for displaced decays-in-flight of exotic LLPs~\cite{Gligorov:2017nwh,Aielli:2019ivi,Dey:2019vyo}.
A recent expression of interest (EoI)~\cite{Aielli:2019ivi} and Snowmass study~\cite{Aielli:2022awh} presented the physics case and extensive experimental and simulation studies. The CODEX-b baseline configuration of Refs~\cite{Gligorov:2017nwh, Aielli:2019ivi} 
consists of a sextet RPC panels on the six outer faces of a $10\times10\times10$\,m cubic detector volume,
along with five uniformly-spaced internal stations along the $x$ axis (in beamline coordinates) containing RPC triplet. The signal reconstruction efficiency for the benchmarks presented in the figures here were found to be between 80\% and 100\%, depending the details of the model.

In a recent study, an algorithm was developed to systematically optimize the acceptance of transverse LLP detectors given a particular budget constraint \cite{Gorordo:2022rro}. This identified several alternative configurations with similar acceptance but a factor of two smaller instrumentation and correspondingly lower cost. For example, relative to the baseline configuration, it has been shown that configurations exist with half the instrumented area, which still maintain roughly 80\%  reconstruction efficiency. The process of determining which of these new configurations is most viable from an engineering point of view is currently under way.

The proposed tracking technology for CODEX-b follows the ATLAS phase-II RPC design~\cite{Collaboration:2285580}, 
so that tracking stations will consist of pairs of $1.88\times 1.03$\,m$^2$ triplet RPC panels---\textit{i.e.}~the 
fundamental array element is approximately a $2\times2$\,m$^2$ RPC triplet panel---supported 
by a structural steel frame. 

The LHCb interaction point produces a large flux of background primary hadrons and leptons, which must be mitigated for CODEX-b to be a low background experiment.  In particular, primary neutral long-lived particles---\textit{e.g.}~(anti)neutrons and $K_L^0$'s---can enter the detector and decay or scatter into tracks resembling a signal decay. These primary hadron fluxes can be suppressed with a sufficient amount of passive shielding material: for a shield of thickness $L$, the background flux suppression $\sim e^{-L/\lambda}$ where $\lambda$ is the material nuclear interaction length. For CODEX-b, the 3\,m of concrete in the UXA radiation wall, corresponding to $7\lambda$ of shielding, would need to be supplemented with an additional $4.5$\,m of Pb shield, corresponding to an additional $25\lambda$. Such a large amount of shielding material will act in turn as a source of neutral LLP secondaries, produced by muons that stream through the shielding material and scatter.

The most concerning neutral secondaries are produced $<1$\,m from the back of the shield by muons that slow down and stop before reaching the detector. For this reason, an active muon veto embedded in the shield will be needed. All these backgrounds have been modeled and quantified with extensive GEANT4 simulations; we refer to~\cite{Aielli:2019ivi} for the details.  In addition, soft cavern backgrounds have been modeled with FLUKA simulations and cross checked with with \textit{in situ} measurements \cite{Dey:2019vyo,Aielli:2022awh}. If the design criteria of the shield laid out in \cite{Aielli:2019ivi} can be met, the simulations show that CODEX-b will be able to operate with negligible backgrounds. 

In the near future, the CODEX-b collaboration proposes to install a demonstrator detector in the same location as where the final detector could be installed. This CODEX-$\beta$ detector is a small-scale version ($2\,\text{m}\times2\,\text{m}\times 2\,\text{m}$) of the full-scale CODEX-b detector. Its primary goal is to validate background estimates for the UX85 cavern, as well as demonstrate the event reconstruction capabilities for this detector concept. The latter will be done by measuring and reconstructing the $K_L$ flux passing through the detector. The CODEX-$\beta$ frame will be highly modular, such that it can be assembled with only fastening hardware and with no welding required. The modules for CODEX-$\beta$ are currently being produced, with a proposed installation in early 2024. We refer to \cite{Aielli:2022awh} for more details on its design and capabilities.

\begin{figure}\centering
	\includegraphics[width=0.7\textwidth]{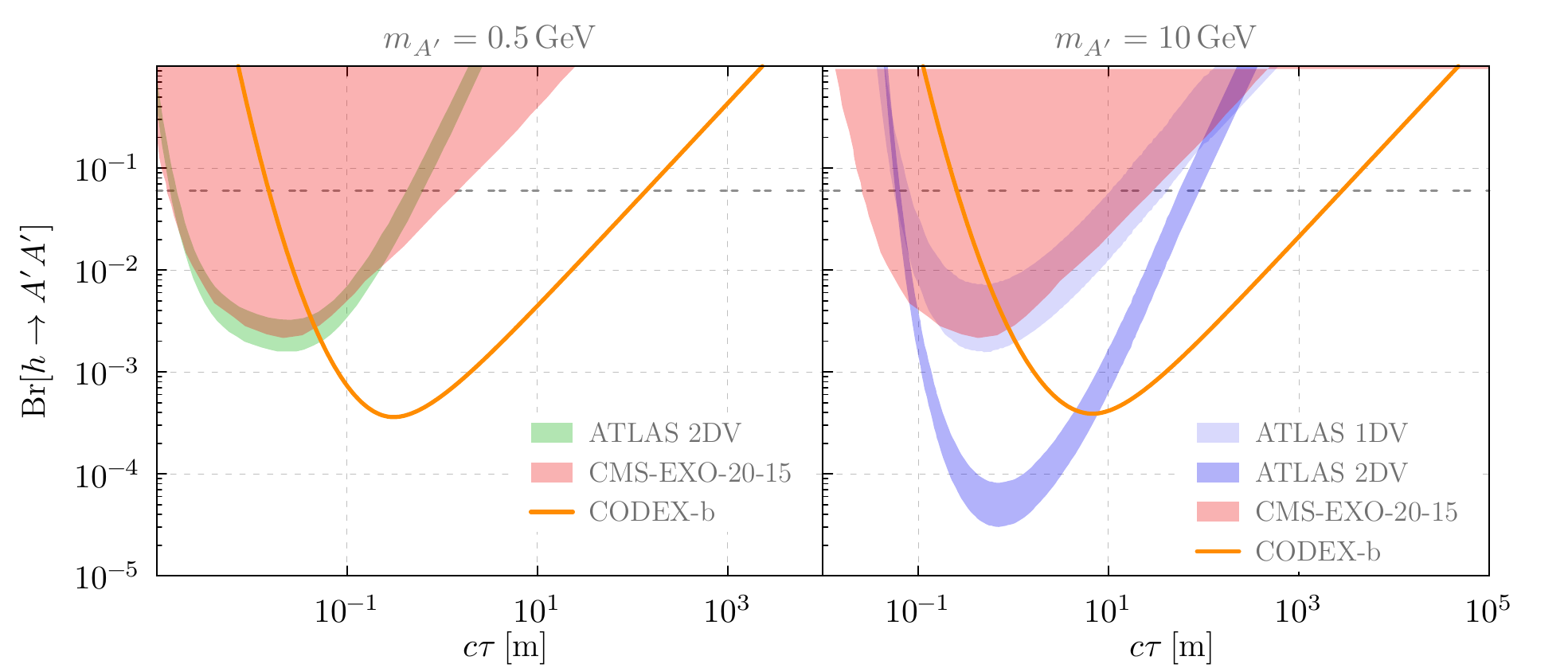}\\
  	\includegraphics[width = 0.42\textwidth]{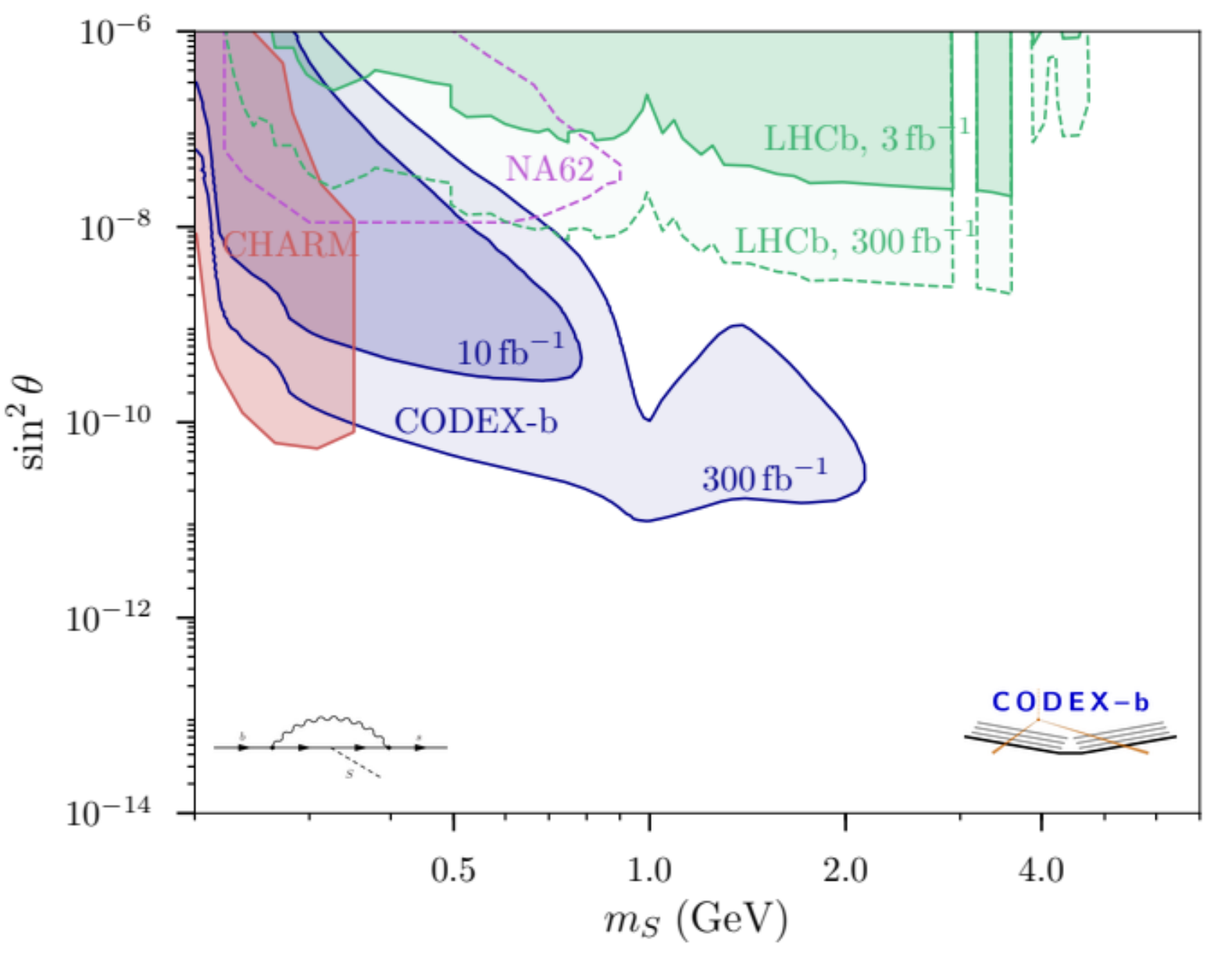}\hspace{15pt}
  	\includegraphics[width = 0.42\textwidth]{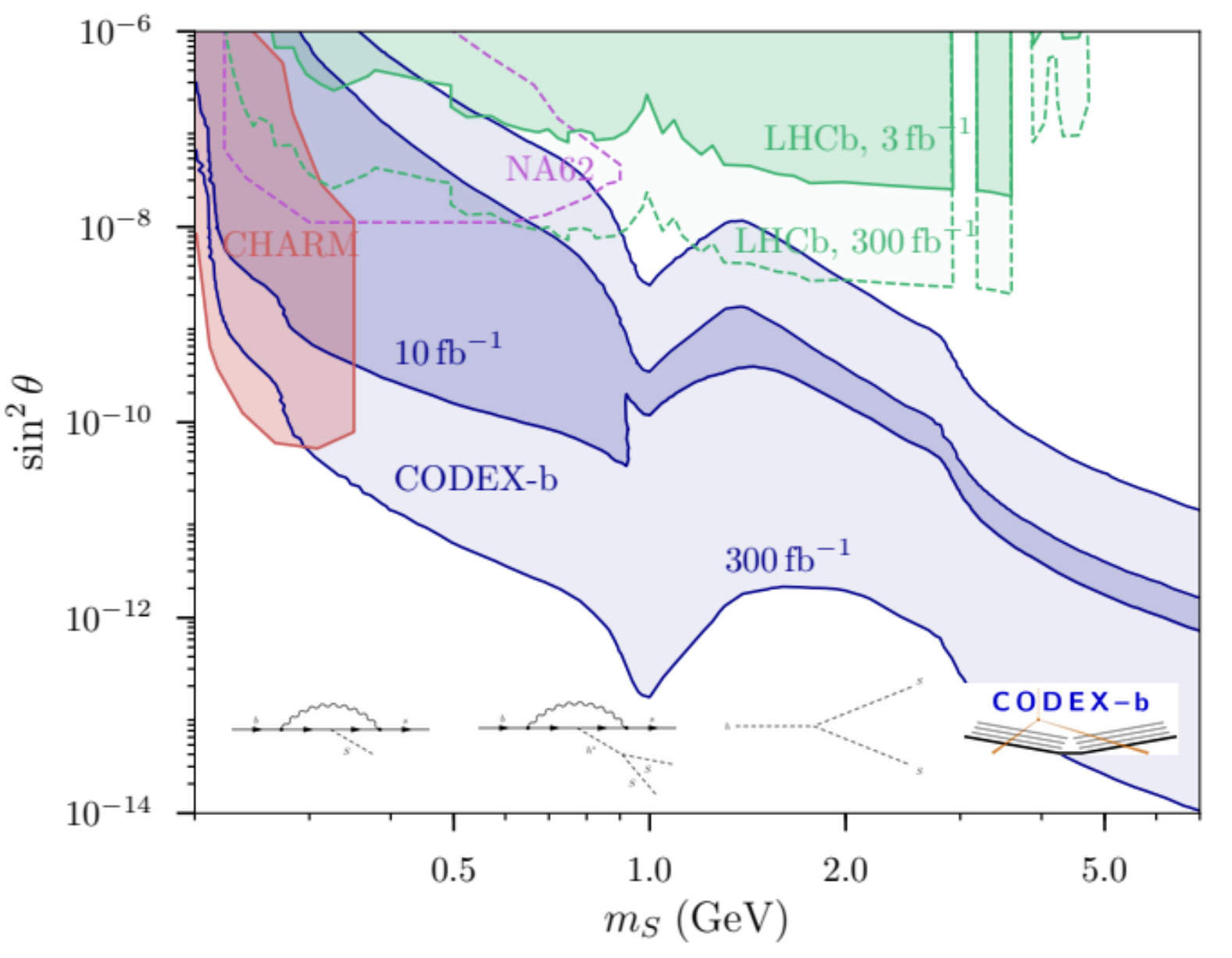}
	\caption{(top) Reach of CODEX-b for $h\to A'A'$ for two different values of the $A'$ mass, along with the (red) exclusion limit in Ref.~\cite{CMS:2021juv}; the blue and green shaded bands are expected limits for searches with the ATLAS muon systems, extrapolated to the HL-LHC~\cite{Gligorov:2017nwh}. (bottom) Projected exclusion power in the dark Higgs simplified model, for the nominal CODEX-b volume with $300\,\ifb$. The mixed quartic with the SM Higgs was chosen such that $\text{BR}[h\to SS]=0$ ($\text{Br}[h\to SS]=0.01$) in the left (right) panel~\cite{Aielli:2019ivi}. See~\cite{Aielli:2019ivi} for more details.\label{fig:portals}}
\end{figure}

\subsubsection{MATHUSLA}

MATHUSLA is a proposed auxiliary LLP detector for the HL-LHC, situated on the surface above the CMS interaction point. Its large decay volume and shielding from LHC collision backgrounds creates a low-background environment with large LLP decay acceptance that allows for the detection of many LLP types with orders of magnitude lower production rate or longer lifetime than possible for the HL-LHC main detectors, as discussed in the letters of intent~\cite{Alpigiani:2018fgd, MATHUSLA:2020uve}, the MATHUSLA physics case~\cite{Curtin:2018mvb} and the recent U.S. Snowmass study~\cite{MATHUSLA:2022sze}. In this contribution we summarize MATHUSLA's reach for several LLP benchmark models, reviewing results from~\cite{MATHUSLA:2020uve} that are based on full simulation of LLP production and kinematics but assuming zero background and 100\% reconstruction efficiency, and report briefly on progress within the collaboration. This includes an update to the default detector geometry with a front wall veto, as well as the imminent publication of a conceptual design report with updated simulation and detector design studies. 

\begin{figure}\centering
    \includegraphics[width=0.8\textwidth]{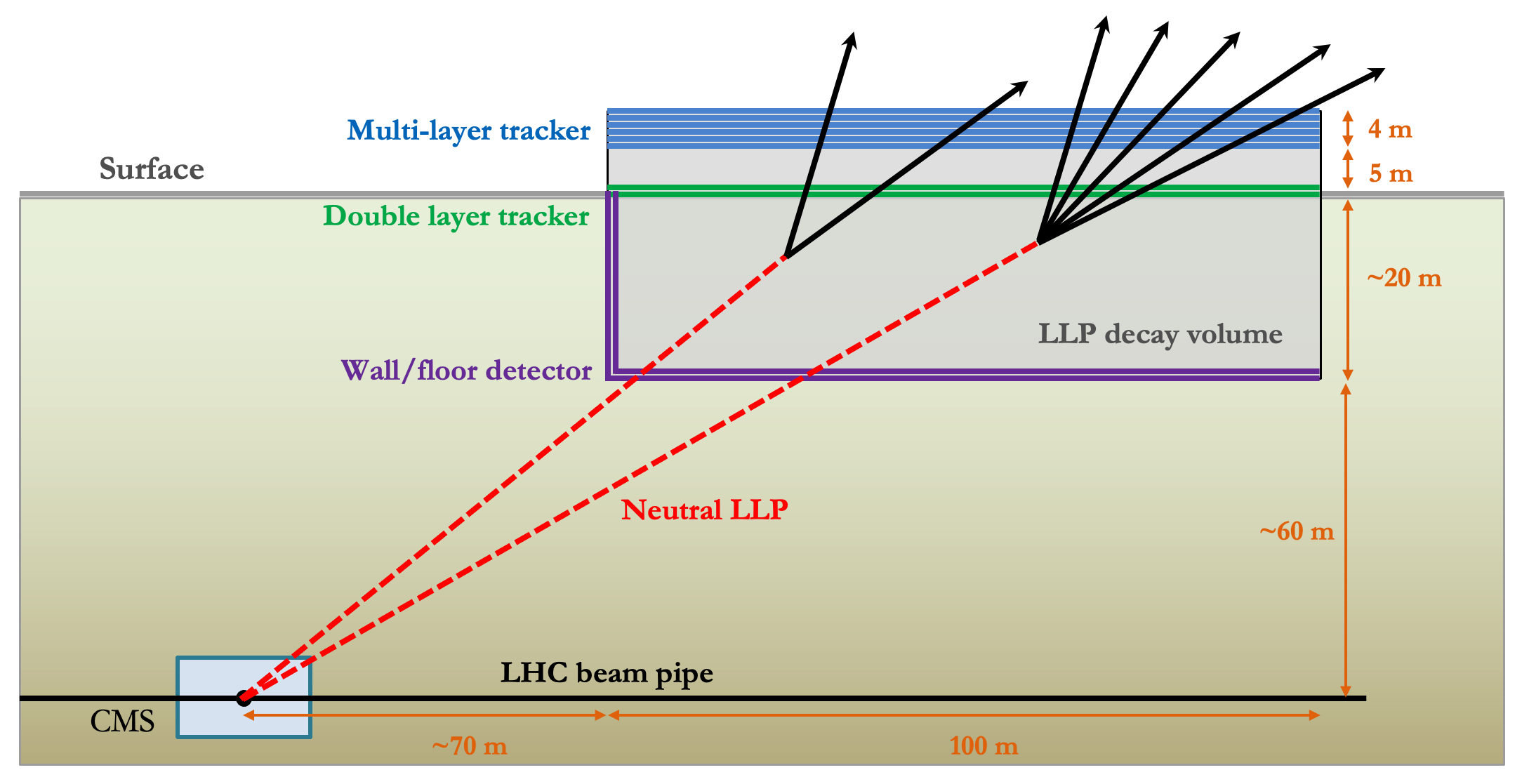}
    \caption{MATHUSLA geometry relative to the CMS collision point, illustrating how LLPs can decay in the detector and be reconstructed as displaced vertices by the trackers. For clarity, the modular structure of the MATHUSLA detector is not shown.\label{fig:mathuslacms}}
\end{figure}

MATHUSLA's basic layout is shown in \ref{fig:mathuslacms}. The decay volume of $100\,\text{m} \times 100\,\text{m } \times 25\,\text{m}$ close to CMS gives similar acceptance for LLP decays in the long lifetime limit ($c\tau \gtrsim 100\,\text{m}$) as the HL-LHC main detectors, but without the background- and trigger-limitations of the LHC collisional environment. 

The only significant background from the LHC collision is energetic muons with $E_\mu \gtrsim 40\,\gev$ that can reach the surface. To control this background, the default detector geometry includes a veto detector layer in the floor and the front-facing wall. Previous versions of the benchmark geometry only included the floor detector.

A MATHUSLA prototype is currently under construction at the University of Victoria. This unit consists of four scintillator layers with expected performance similar to current MATHUSLA design specification. Layers are separated vertically by $80\,\text{cm}$, with scintillator bars alternating in $x$ and $y$. Scintillator bars, using FNAL extrusion, are $1\,\text{cm}$ thick by $4\,\text{cm}$ wide with a single co-extruded hole for WLSF insertion. The module dimensions are approximately $80\,\text{cm}\times100\,\text{cm}$, comprised of assemblies of 20 and 24 bars, depending on bar orientation. These modules are threaded with $6\,\text{m}$ long $1.5\,\text{mm}$ WLSF (Saint Gobain BCF-92) and the fiber ends are brought to a common readout via a 64 channel SiPM array (Hamamatsu S13161-3050-AE-08). Amplitude and timing measurements are performed using a CAEN DT5202 FERS unit which is used for triggering and absolute and differential timing of signals from both ends of each WLSF. Although the scintillator bars are somewhat shorter than the nominal MATHSULA design ($\sim 2.5\,\text{m}$), the WLSFs are similar in length to the MATHSULA design. Consequently similar timing and triggering performance  is expected in this prototype to a full MATHUSLA module. This prototype will be operated as a cosmic ray detector, to test design and assembly  aspects, as well as test trigger and tracking performance
in a realistic detector.

The collaboration is currently in the final stages of completing the MATHUSLA Conceptual Design Report (CDR), which includes updates to the LLP signal simulation, reports on fully realistic background simulations in progress, and includes significant details on detector design, R\&D, DAQ and other aspects of the full-size detector. This will include new reach estimates for LLP benchmark models that improve upon existing estimates that assume 100\% reconstruction efficiency and zero backgrounds. 
For the present contribution, we simply summarize some of the results presented in~\cite{MATHUSLA:2020uve} to demonstrate MATHUSLA's physics potential for probing feebly interacting particles via LLP searches, including general hidden sectors and various dark matter candidates. 

\begin{figure}\centering
    \includegraphics[width=0.6\textwidth]{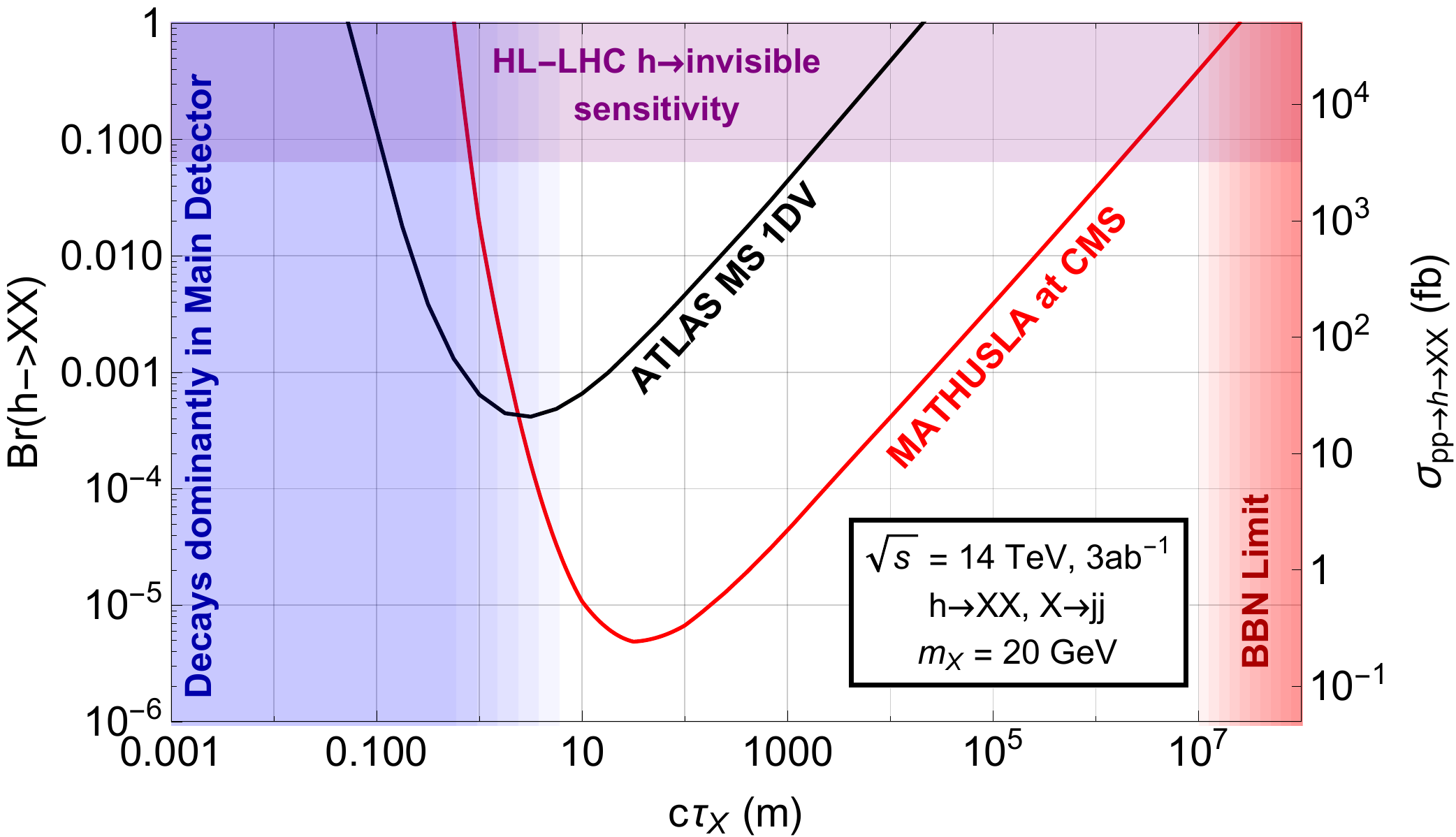}
    \caption{(red curve) MATHUSLA@CMS sensitivity, 4 observed events, for LLPs of mass $m_X = 20,\gev$ produced in exotic Higgs decays. (black curve) Reach of ATLAS search for a single hadronic LLP decay in the muon system at the HL-LHC~\cite{Coccaro:2016lnz}.\label{fig:sensitivity_higgs}}
\end{figure}

Figure~\ref{fig:sensitivity_higgs} shows MATHUSLA's sensitivity to LLPs decaying hadronically, produced in exotic Higgs decays. This type of LLP is the main physics target of MATHUSLA, with excellent and highly robust sensitivity compared to main detector estimates due to the absence of any backgrounds that can mimic a displaced vertex with $\mathcal{O}(10)$ upwards going tracks. 

The theoretical motivation for probing such LLPs is also very urgent, since a large variety of FIPs, hidden sectors, dark matter candidates and other models produce LLP signals through the higgs portal or at a similar energy scale~\cite{Curtin:2018mvb}. 

\begin{figure}\centering
    \hspace*{-0.1\textwidth}
    \begin{tabular}{cc}
    \includegraphics[width=0.5\textwidth]{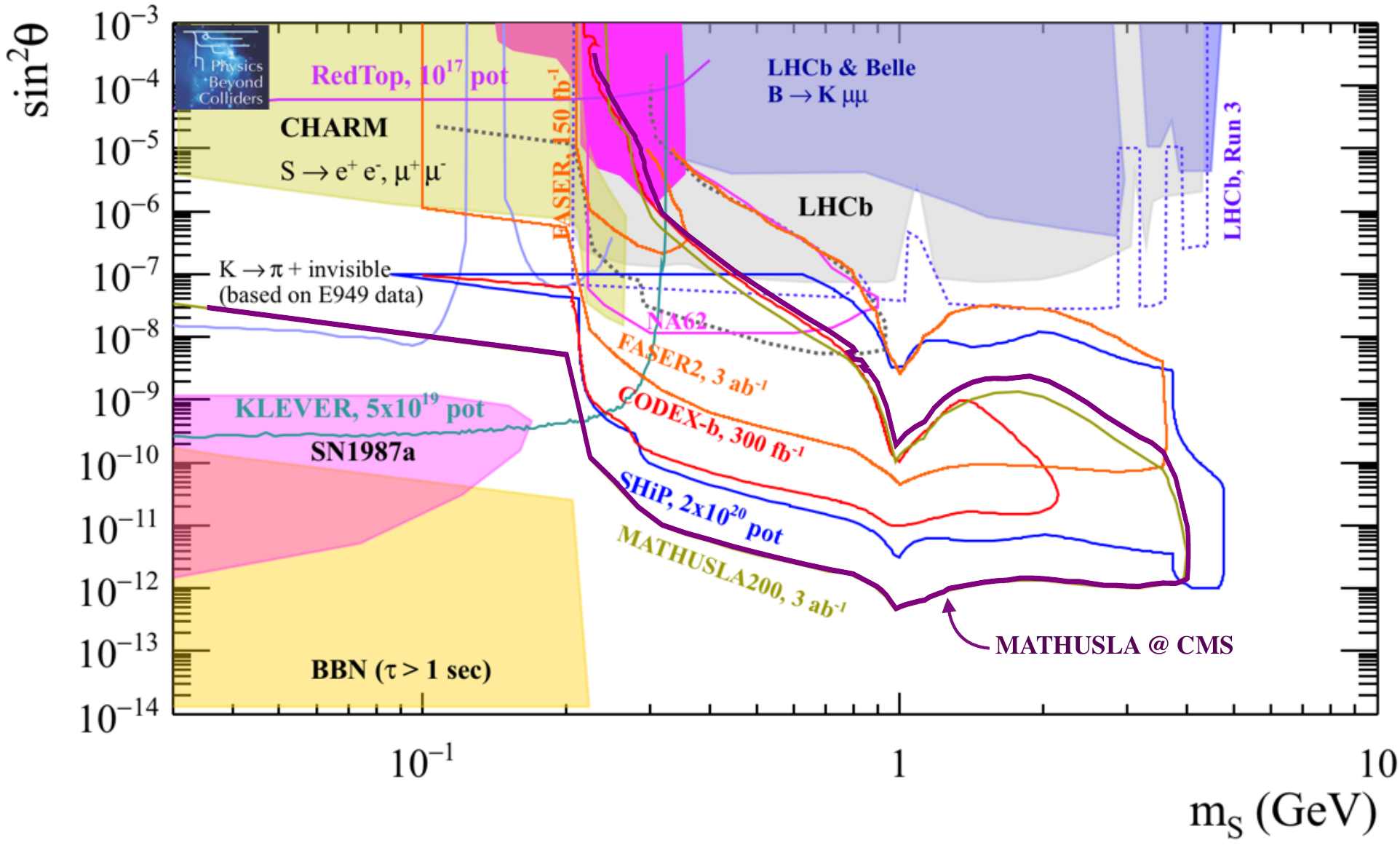} 
    &
    \includegraphics[width=0.5\textwidth]{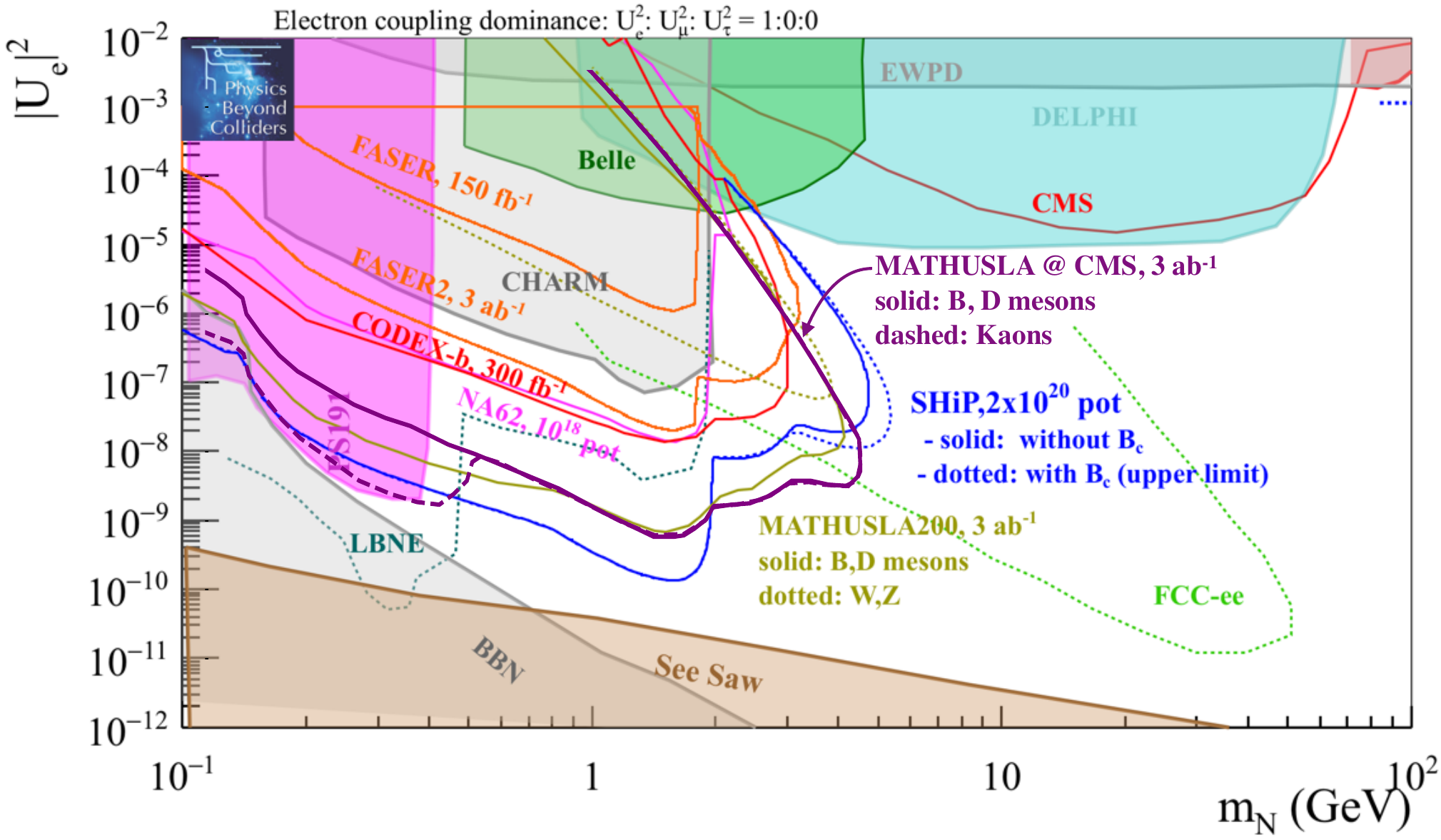}
    \\
    \includegraphics[width=0.5\textwidth]{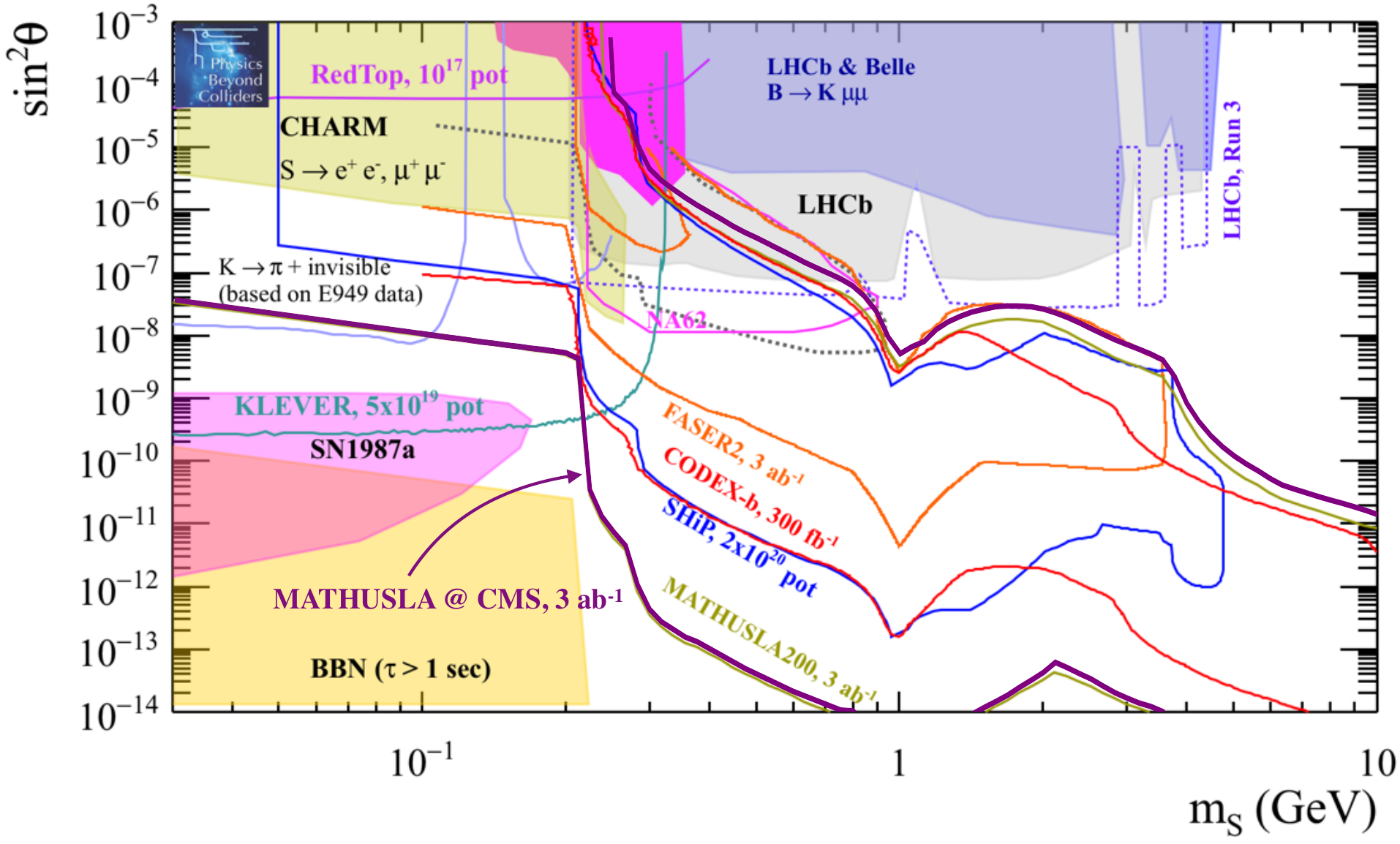}  
    &
    \includegraphics[width=0.5\textwidth]{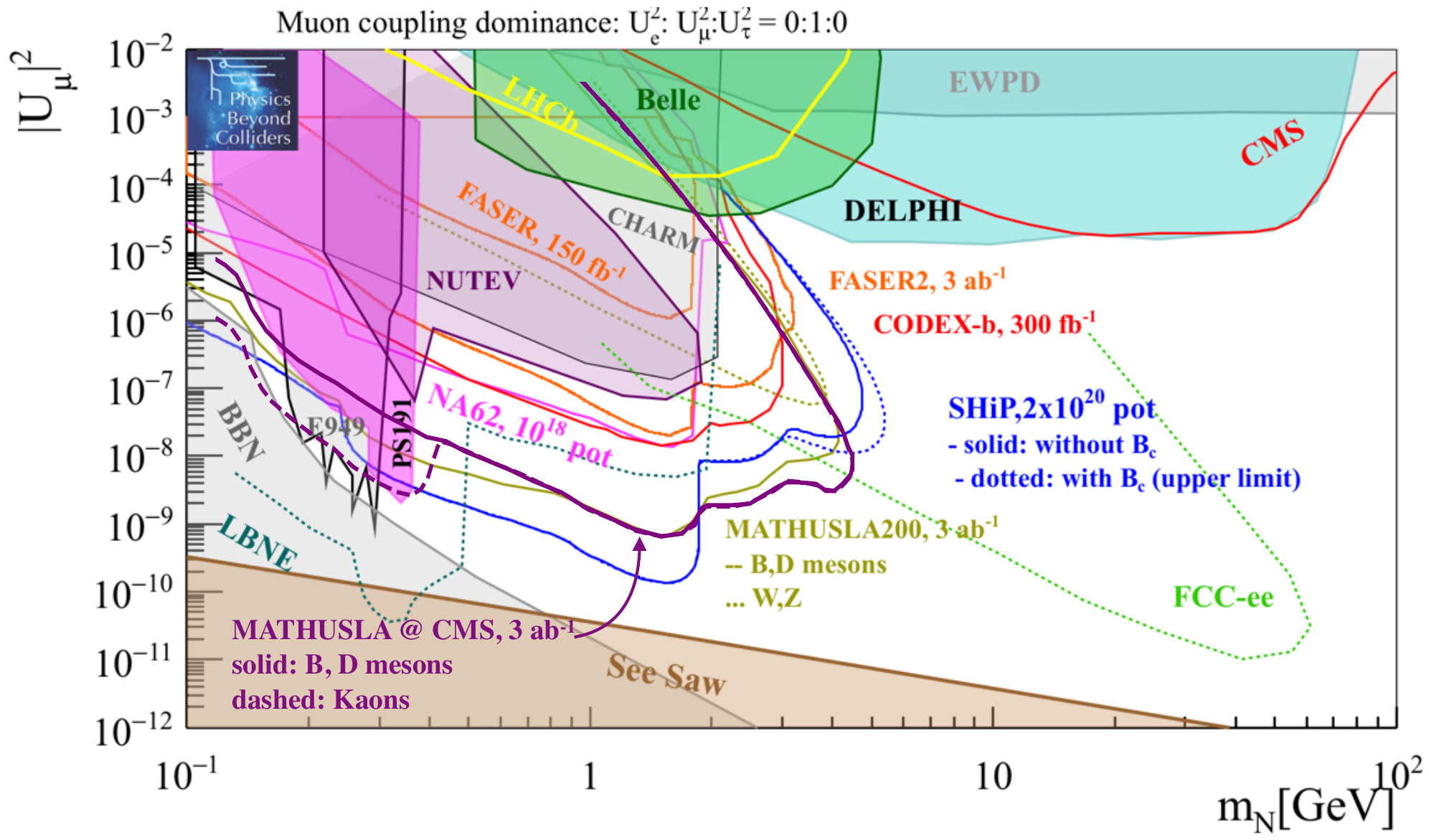} 
    \\
    \includegraphics[width=0.5\textwidth]{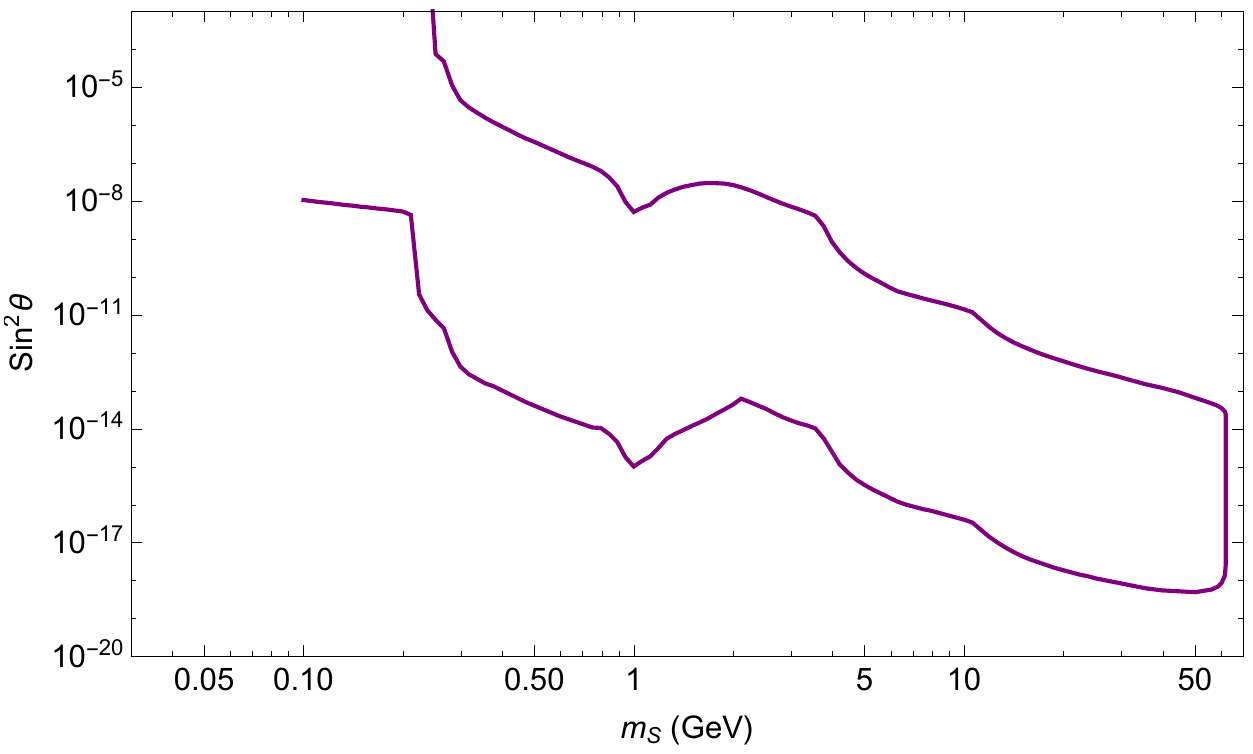} 
    &
    \includegraphics[width=0.5\textwidth]{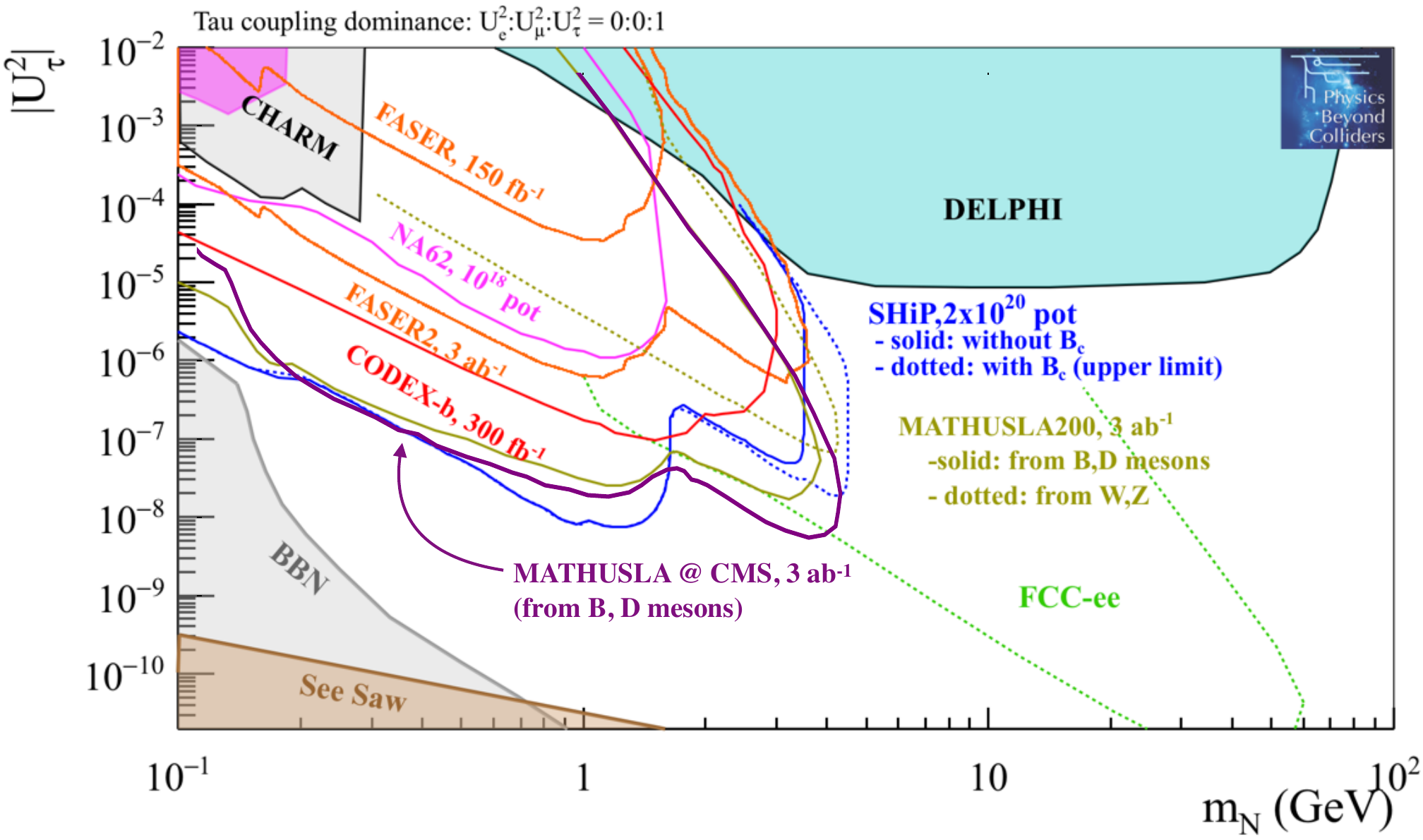} 
    \end{tabular}
    \caption{(left) Purple curves show the sensitivity of MATHUSLA@CMS for a singlet scalar LLP $s$ mixing with Higgs mixing angle $\theta$. (top) Assuming production in exotic $B$, $D$, $K$ meson decays only. (middle) Assuming additional production in exotic Higgs decays with $\mathrm{Br}(h\to ss) = 0.01$. (bottom) Same scenario as middle but showing the entire MATHUSLA sensitivity due to $h\to ss$ decays. (right) Purple curves show the sensitivity of MATHUSLA@CMS for a Heavy Neutral Lepton LLPs produced in $B, D$ (solid) and $K$ decays (dashed) for HNLs that mix predominantly with (top) electron, (middle) muon, (bottom) or tau active neutrinos. Figures from~\cite{MATHUSLA:2020uve}, augmenting plots originally from~\cite{Beacham:2019nyx}. MATHUSLA sensitivities assume 100\% LLP reconstruction efficiency and zero background.\label{fig:MATHUSLAPBCplots}}
\end{figure}

Figure~\ref{fig:MATHUSLAPBCplots} shows MATHUSLA's sensitivity to several PBC benchmark models, focusing on SM+S and heavy neutral leptons. The reach covers significant regions of new parameter space and is complementary to other transverse detectors. LLPs of this type constitute the secondary physics target of MATHUSLA, and studies to determine the robustness of this reach with respect to reconstruction efficiency and backgrounds are in progress. 

\begin{figure}
    \centering
    \begin{tabular}{cc}
    \includegraphics[width=0.5\textwidth]{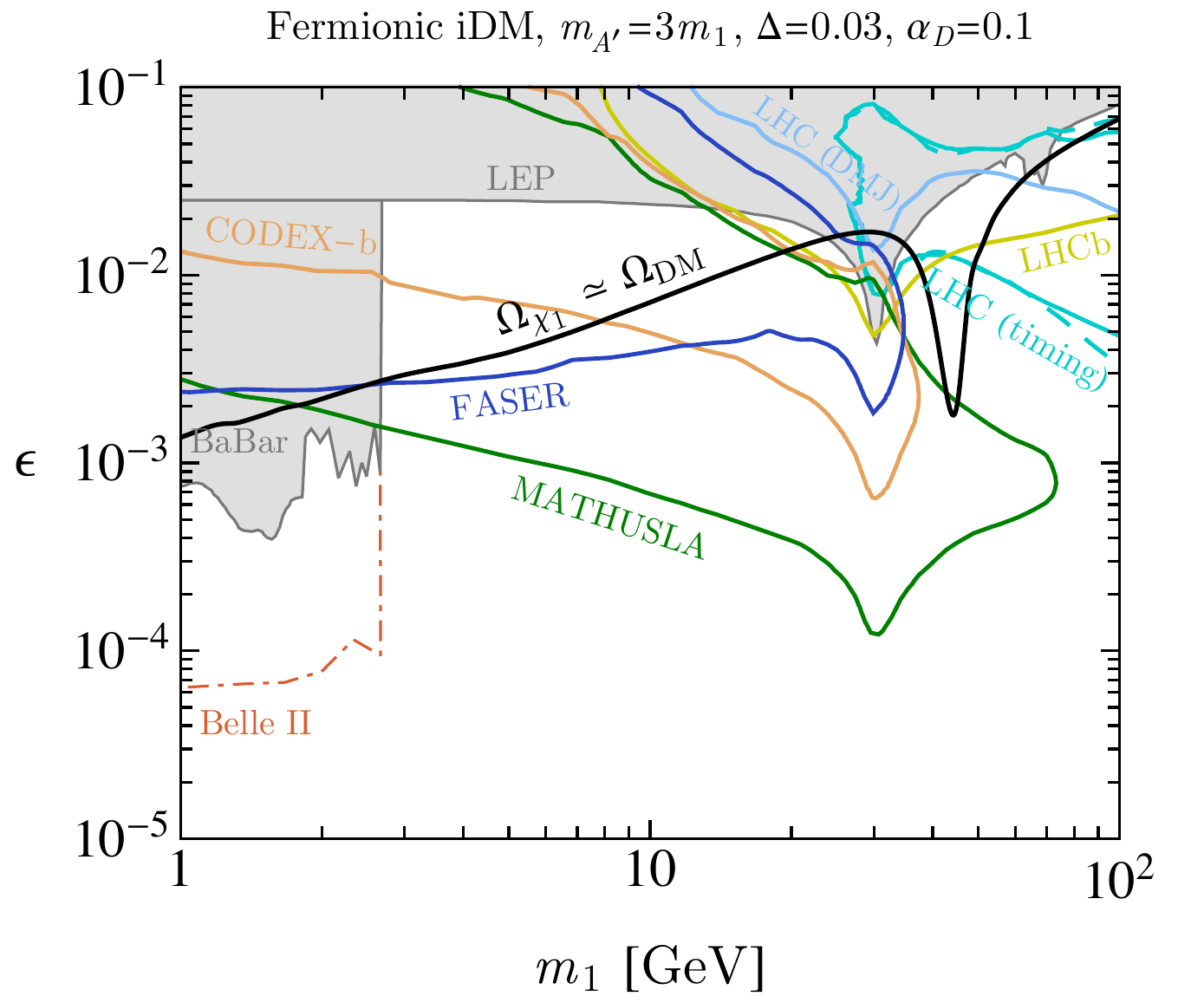}
    &
    \includegraphics[width=0.5\textwidth]{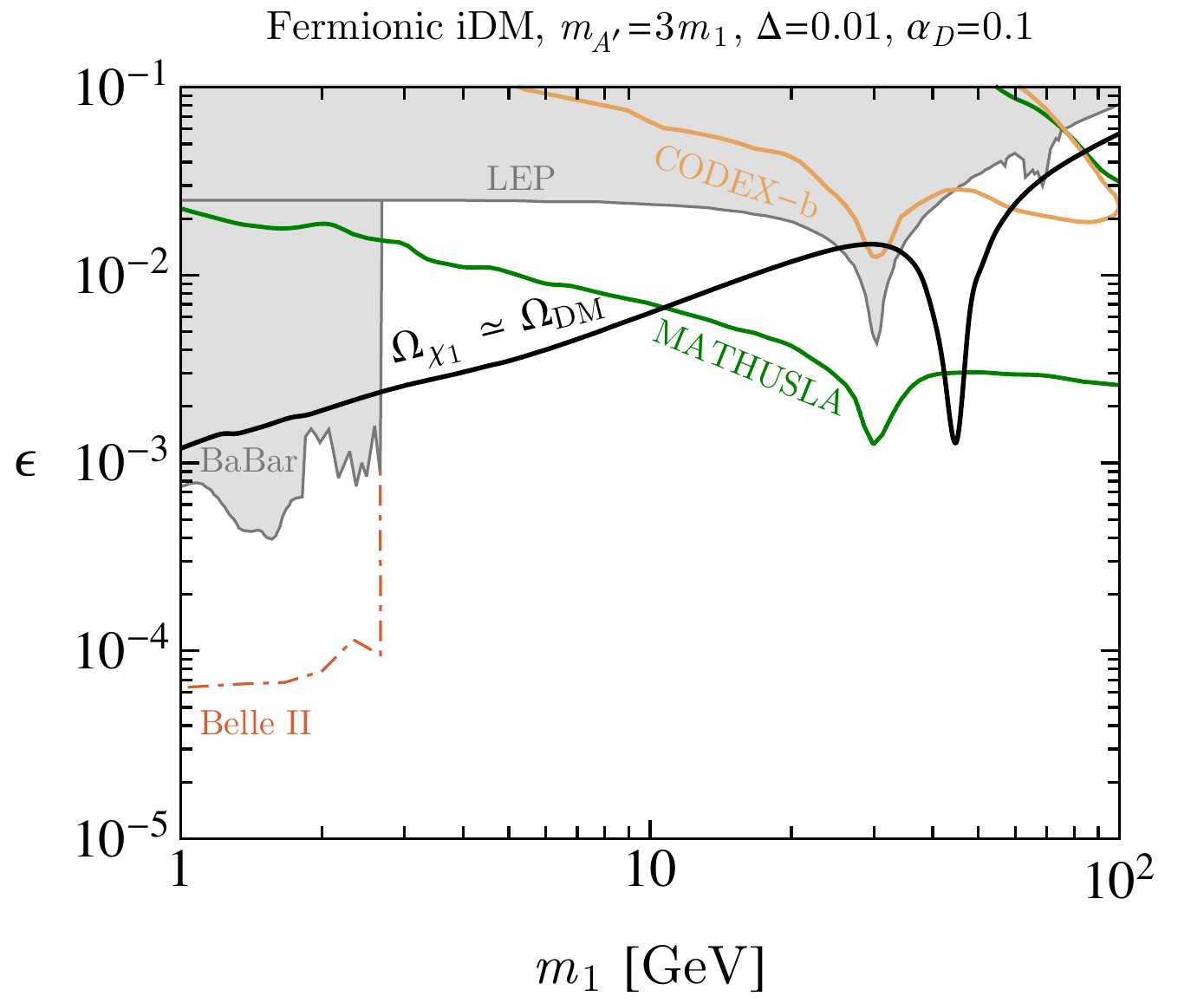}
    \end{tabular}
    \caption{Reach of MATHUSLA and other LHC experiments and searches for inelastic Dark Matter (iDM) with a dark photon  of mass $m_{A'}$ that has kinetic mixing $\epsilon$ with the SM photon, and mass splittings $\Delta$ in the percent range. The black curve indicates where thermal co-annihilations $\chi_2 \chi_1 \to A' \to f \bar f$ to SM fermions give the observed DM relic density. Figure taken from~\cite{Berlin:2018jbm}.\label{fig:iDM}}
\end{figure}

Finally, to give an illustrative example of MATHUSLA's ability to probe various dark matter models, \ref{fig:iDM} shows MATHUSLA's sensitivity to an inelastic dark matter model resulting in a meta-stable dark sector state showing up as a decaying LLP due to a dark-photon mixing with the SM photon (see~\cite{MATHUSLA:2020uve} and~\cite{Berlin:2018jbm} for more details). There are many dark matter scenarios which can only be probed via LLP decay searches, making the maximization of the LHC's reach with transverse detectors essential. 

\subsubsection{Conclusions}

To achieve the most comprehensive coverage of LLP's at the HL LHC, dedicated transverse LLP detectors are an essential piece of the puzzle. The importance of such auxiliary detectors at the LHC was stressed in particular in the recent Snowmass energy frontier summary report~\cite{Narain:2022qud}. MATHUSLA, CODEX-b and ANUBIS represent three different locations and detector concepts, covering a range of costs and providing sensitivities that are complementary to existing or approved detectors. All three collaborations have been making steady progress with their detector design, background calibrations and are planning or have already installed a demonstrator detector.

\afterpage{\clearpage}
%-------------------------------------------

%-------------------------------------------
\subsection{Prospects at CERN: FACET -- {\it C.~Zorbilmez}}
\label{ssec:zorbilmez}
{\it Author: Caglar Zorbilmez , <Caglar.Zorbilmez@cern.ch>}

\subsubsection{Introduction}
  The proposed new subsystem of CMS (Compact Muon Solenoid) called FACET for \textbf{F}orward-\textbf{A}perture \textbf{C}MS \textbf{E}x\textbf{T}ension will be sensitive to particles produced with polar angels $1<\theta<4 $ mrad (equivalently). FACET has an 18 m long section called decay volume between $z=101$  and $119$ m on the side of the IP5 interaction point, followed by an 8 m long region instrumented with various particle detectors. 
  %The lifetimes range between $c\tau \sim 0.1$ and $100$ m is covered by FACET. The Lorentz factor $\gamma$ in forward region appear to be quite high. A unique feature among the LHC experiments is the high vacuum of the decay volume, which ensure that any backgrounds from particle interactions inside a $\sim 14$ $m^3$ fiducial region is eliminated \cite{Cerci:2021nlb}.
  
  \begin{figure}[ht!]
  	  \includegraphics[width=1\textwidth]{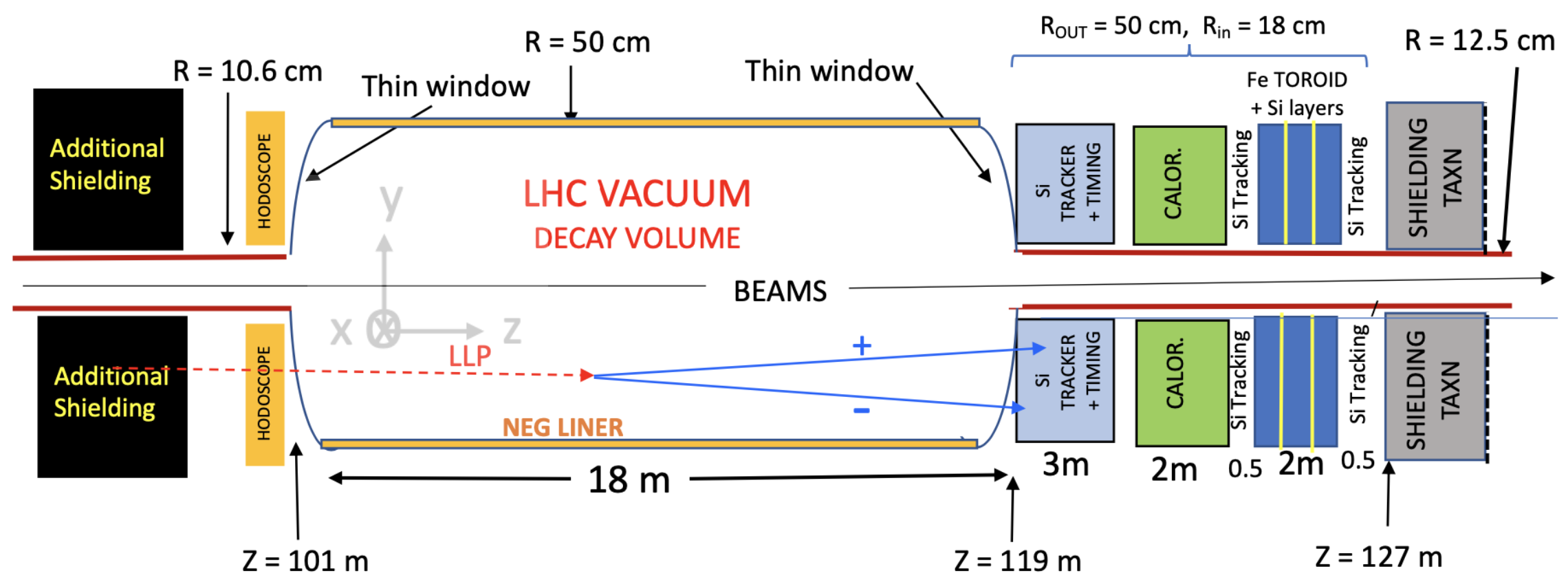}
  \caption{FACET Layout}
  \label{fig:1} 
 \end{figure}

  The detector, a schematic view of which is shown in Figure~\ref{fig:1}, requires an 18 m long section of the LHC beam pipe will be replaced with the circular pipe of a $50$ cm radius. In the FACET design, additional shielding will be placed upstream of the first detector which is hodoscope has very high efficiency to tag charged particle from interaction in the upstream shielding, consisting of a two-layer counter made of radiation-resistant quartz pads. Behind the decay volume, there is a section of precision tracking and fast timing detectors to measure the time-of-flight of charged particles, determine vertex positioning in 4 D, and measure the tracks of charged particles. After the tracking and timing system, electromagnetic and hadronic calorimeter will be placed to measure individual showers above a threshold energy (e.g., 10 GeV, but tunable) and their directions in the presence of many  low-energy showers. Behind the calorimeter, there is an iron toroid instrumented with silicon tracking to measure the charge of muons. In addition, muon detection will be made through active layers of the calorimeter.
  
  Level-1 (L1) and High-Level Trigger (HLT)  systems will be upgraded as part of CMS phase II upgrade at HL-LHC. In this context, FACET will provide an additional external trigger to the CMS Global L1 Trigger. Events triggered by FACET will include all CMS data, while events triggered by CMS will include FACET information less than \%1 of all CMS data. The FACET trigger will only be able to operate in a standalone mode where FACET information is recorded \cite{Cerci:2021nlb}.

\subsubsection{Sensitivity to Long-Lived Particles}\label{sec:LLPs}

With data of a total integrated luminosity of $ab^{-1}$ of pp collisions at a $\sqrt{s}=14$ TeV, LLP parameter space for dark photons, heavy neutral leptons, axion-like particles, and dark Higgs bosons is studied by selecting either 3 or 5 candidate events, assuming no background and that FACET can detect all penetrating neutral particle decays to $>2$ charged particles or photons occurring between $101<z<119$ m with the decay products within $18 <R<50$ cm at $z=120$ m.

%Models beyond the SM (BSM) typically predict new particles with a variety of lifetimes. In particular, new weak-scale particles can easily have long lifetimes for several reasons, including approximate symmetries that stabilize the long-lived particle (LLP), small couplings between the LLP and lighter states, and suppressed phase space available for decays. 

%Because the long-lived particles of the SM have masses $<\sim 5$ GeV and have well-understood experimental signatures, the unusual signatures of BSM LLPs offer excellent prospects for the discovery of new physics at particle colliders.

\begin{figure}[ht]
  \centering
  	  \includegraphics[width=0.4\textwidth]{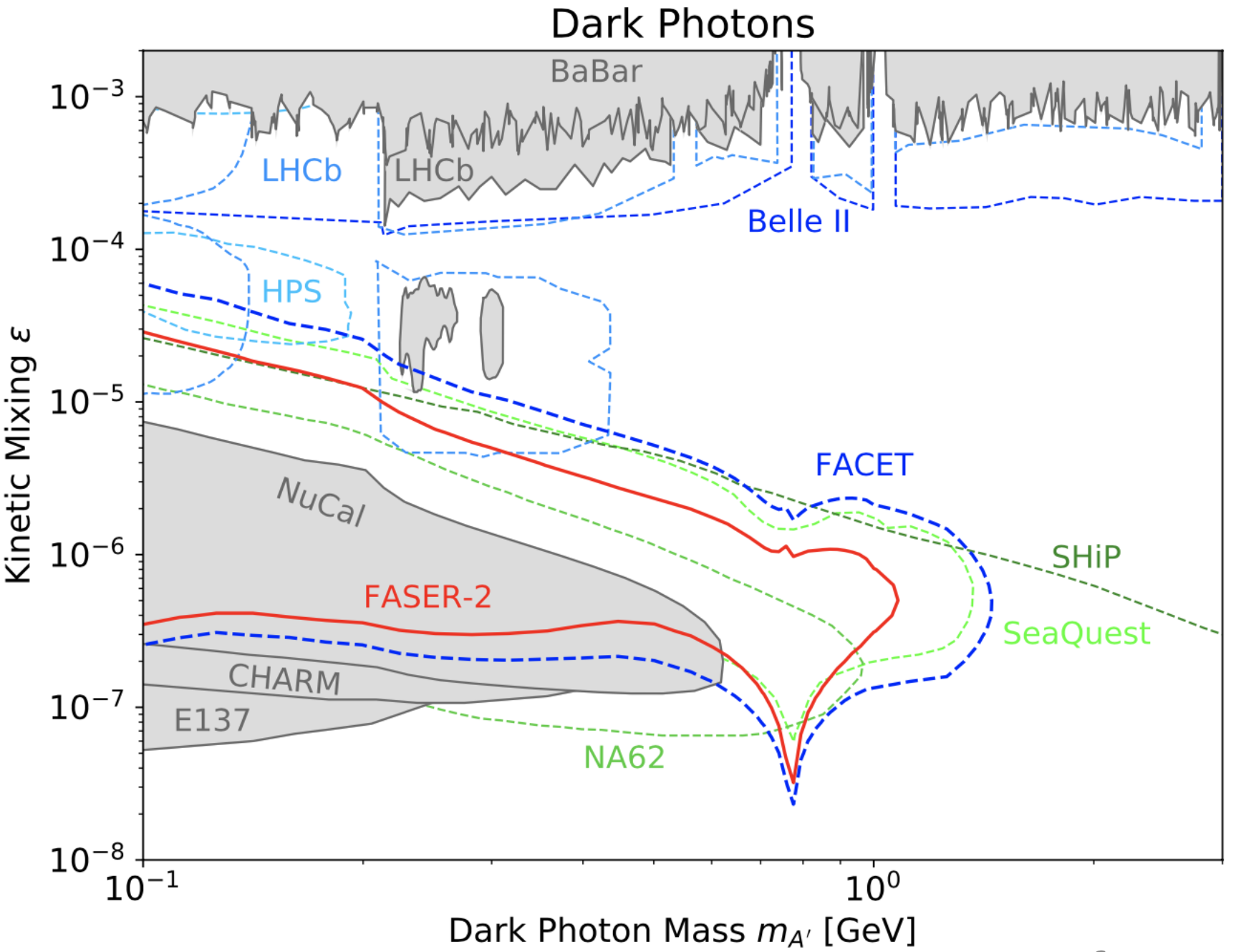}
  \caption{FACET reach for dark photons (3 event contours) in a generic model with no BSM sources, as calculated with Foresee}
  \label{fig:2} 
  \begin{subfigure}[t]{0.49\textwidth}
        \centering
        \includegraphics[width=.8\textwidth]{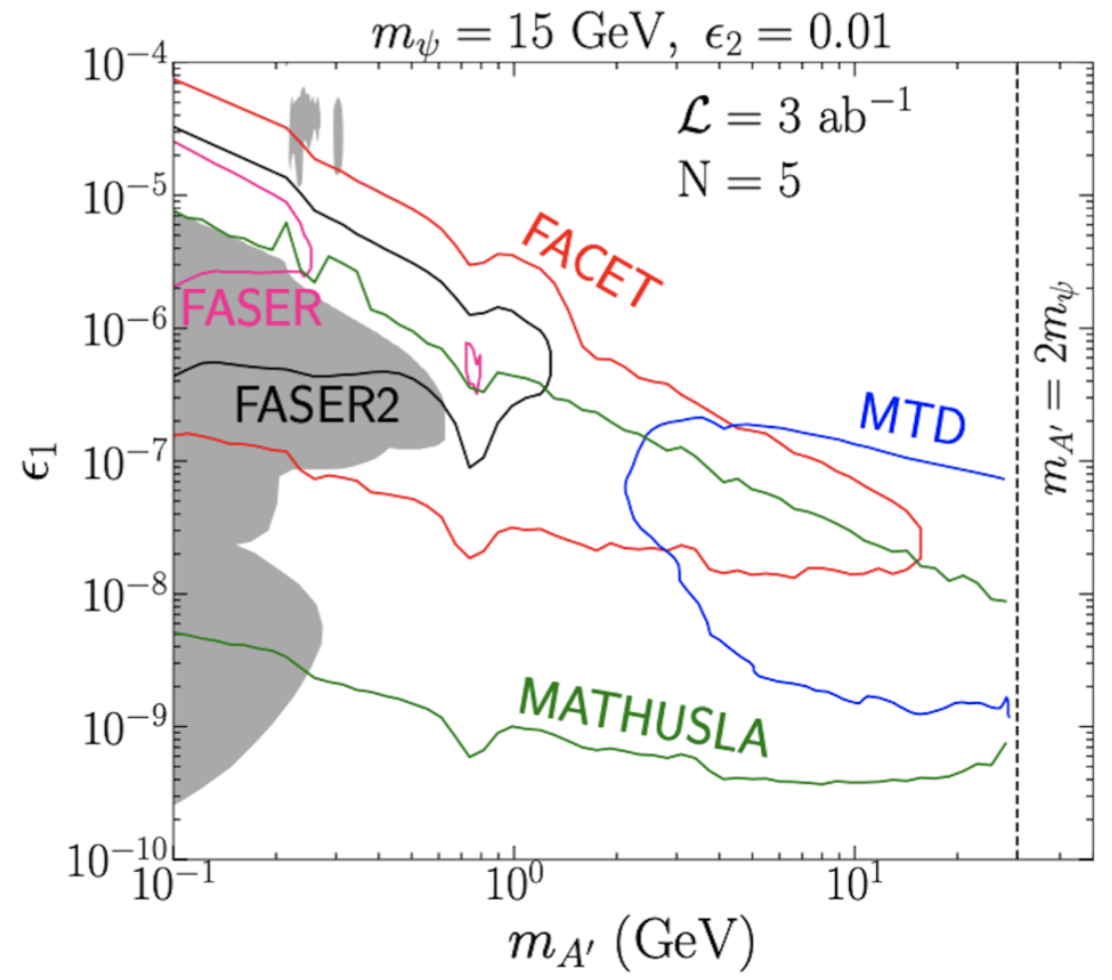}
        \caption{The model parameters corresponding to the reach}
        \label{fig:3-a}
    \end{subfigure}
    \hfill
    %\nextfloat
    \begin{subfigure}[t]{0.49\textwidth}
        \centering
        \includegraphics[width=.8\textwidth]{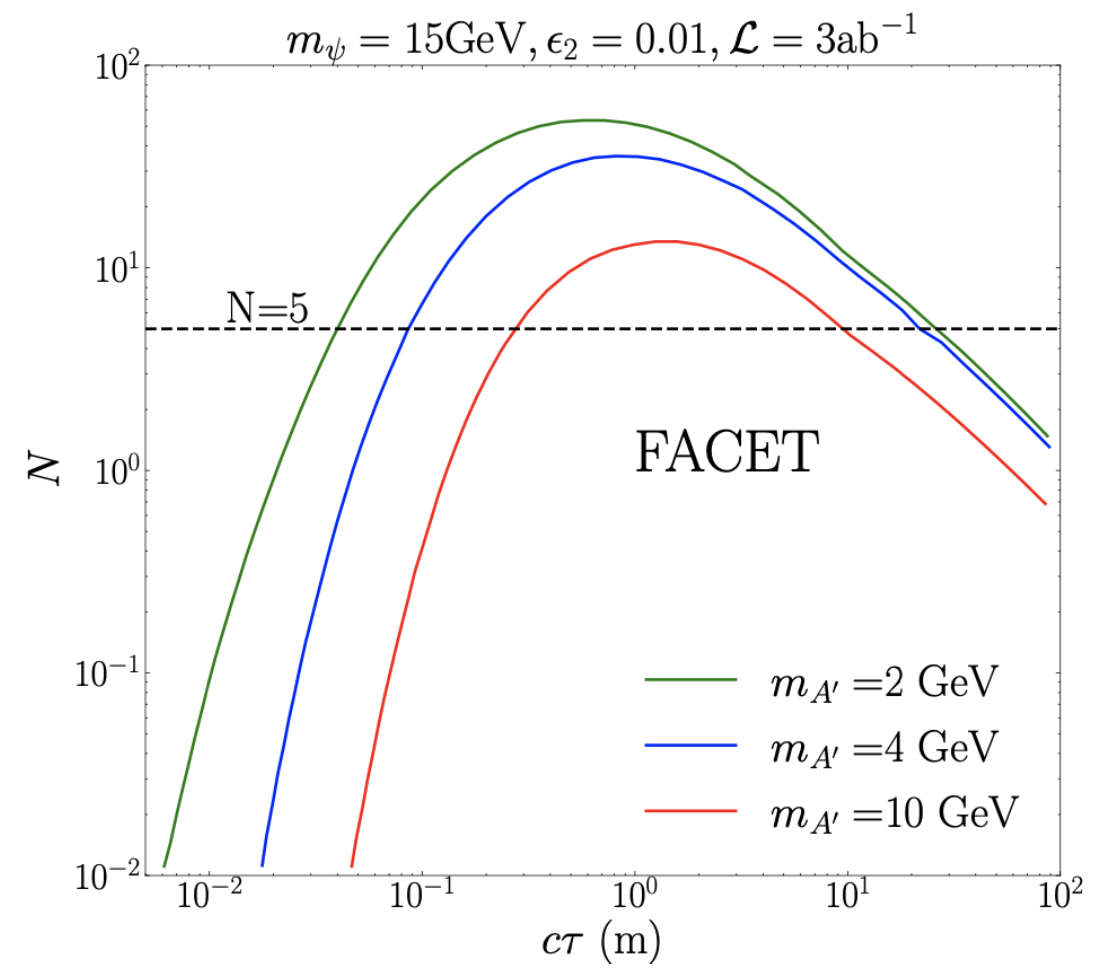}
        \caption{The number of event as a function of $c\tau$ for three $A^{'}$ masses}
        \label{fig:3-b}
    \end{subfigure}
    \caption{A comparison of FACET and other experiment dark photon reach for all final state}
    \label{fig:3}

 \end{figure}

{\bf Dark Photons - }\label{sec:DP}
Any process in hadron-hadron collision has some probability of mixing with dark photon governed by the kinetic mixing parameter ($\epsilon$) \cite{Caputo:2021eaa}. If mass of dark photon smaller than 1 GeV it assumed that the most prolific source is meson decay of $\pi^0 $, $\eta$ and $\eta^{'}$. In Figure~\ref{fig:2}, FACET reach for dark photons in generic model with no BSM source.

%\begin{figure}[h]
%    \centering
%    
%    \begin{subfigure}[t]{0.49\textwidth}
%        \centering
%        \includegraphics[width=1\textwidth]{Figs/DP1}
%        \caption{The model parameters corresponding to the reach}
%        \label{fig:3-a}
%    \end{subfigure}
%    \hfill
%    \begin{subfigure}[t]{0.49\textwidth}
%        \centering
%        \includegraphics[width=1\textwidth]{Figs/DP2}
%        \caption{The number of event as a function of $c\tau$ for three $A^{'}$ masses}
%        \label{fig:3-b}
%    \end{subfigure}
%    \caption{A comparison of FACET and other experiment dark photon reach for all final state}
%    \label{fig:3}
%\end{figure}

If mass of dark photon bigger than 1 GeV there are some main production such as Drell-Yan, Bremsstrahlung and Heavy quark decay. A comparison of the reach of FACET and other experiments is given in Figure~\ref{fig:3-a}. Figure~\ref{fig:3-b} shows the number of events as a function of lifetime for three dark photon masses.
%\newpage

\begin{figure}[ht]
  \centering
  	  \includegraphics[width=0.4\textwidth]{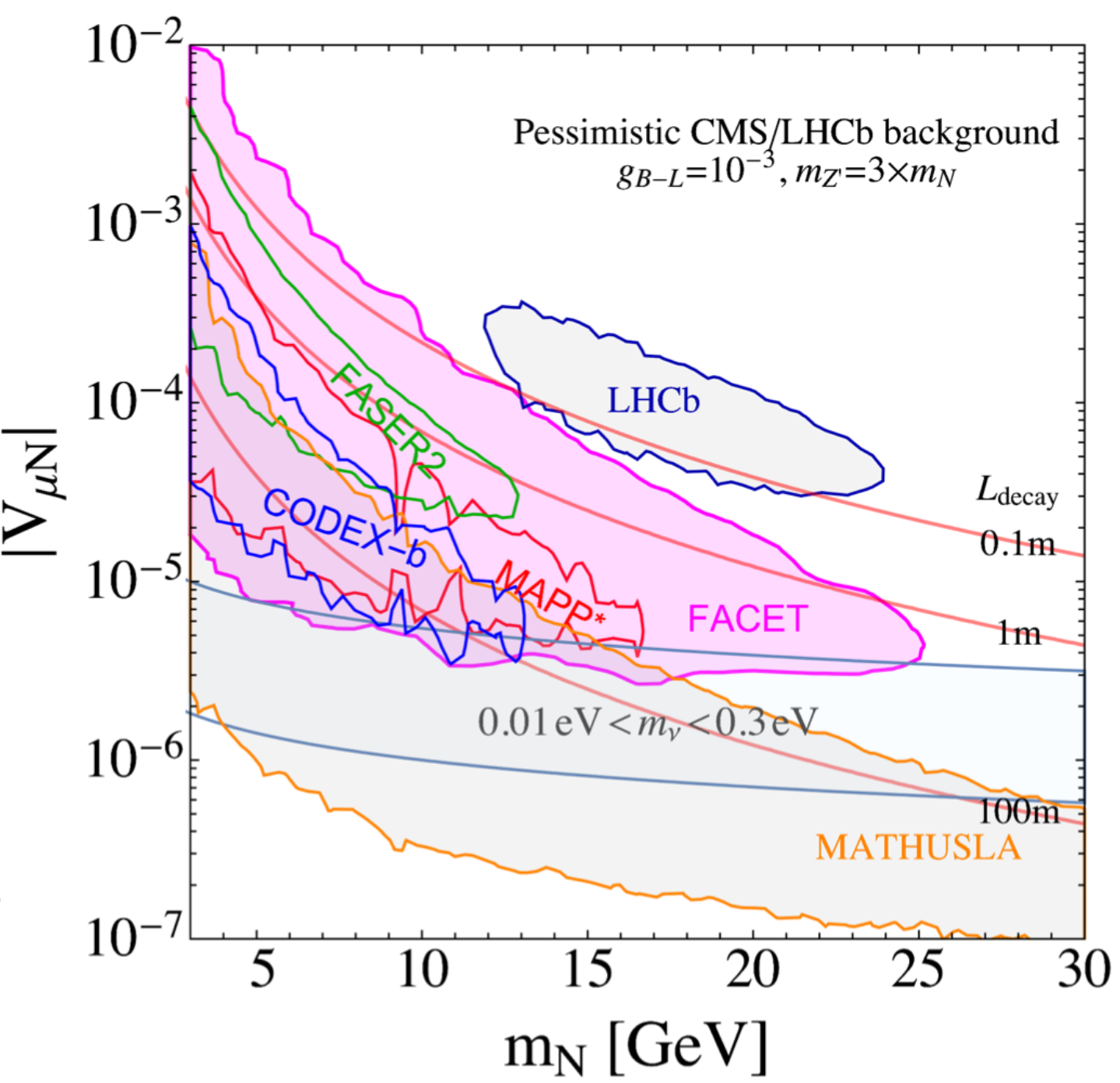}
  \caption{FACET reach in the mixing parameter vs. mass plane for a heavy neutral lepton (3 event contours), along with projections for other proposed experiments, as well as for MAPP and the upgraded LHCb detectors}
  \label{fig:4} 
 \end{figure}

{\bf Heavy Neutral Leptons - }
Heavy neutral leptons are hypothetical particles predicted by many extension of the SM. These particles can explain the origin of neutrino masses generate the observed matter-antimatter asymmetry in the universe and provide dark matter candidate. We consider a specific extension of SM with a Z prime boson and three heavy right-hand Majorana neutrinos (Ni) \cite{SHiP:2018xqw}. Figure~\ref{fig:4} shows the coverage in the mixing parameter versus mass of the Majorana neutrinos

\begin{figure}[ht]
    \centering
      
    \begin{subfigure}[t]{0.49\textwidth}
        \centering
        \includegraphics[width=.8\textwidth]{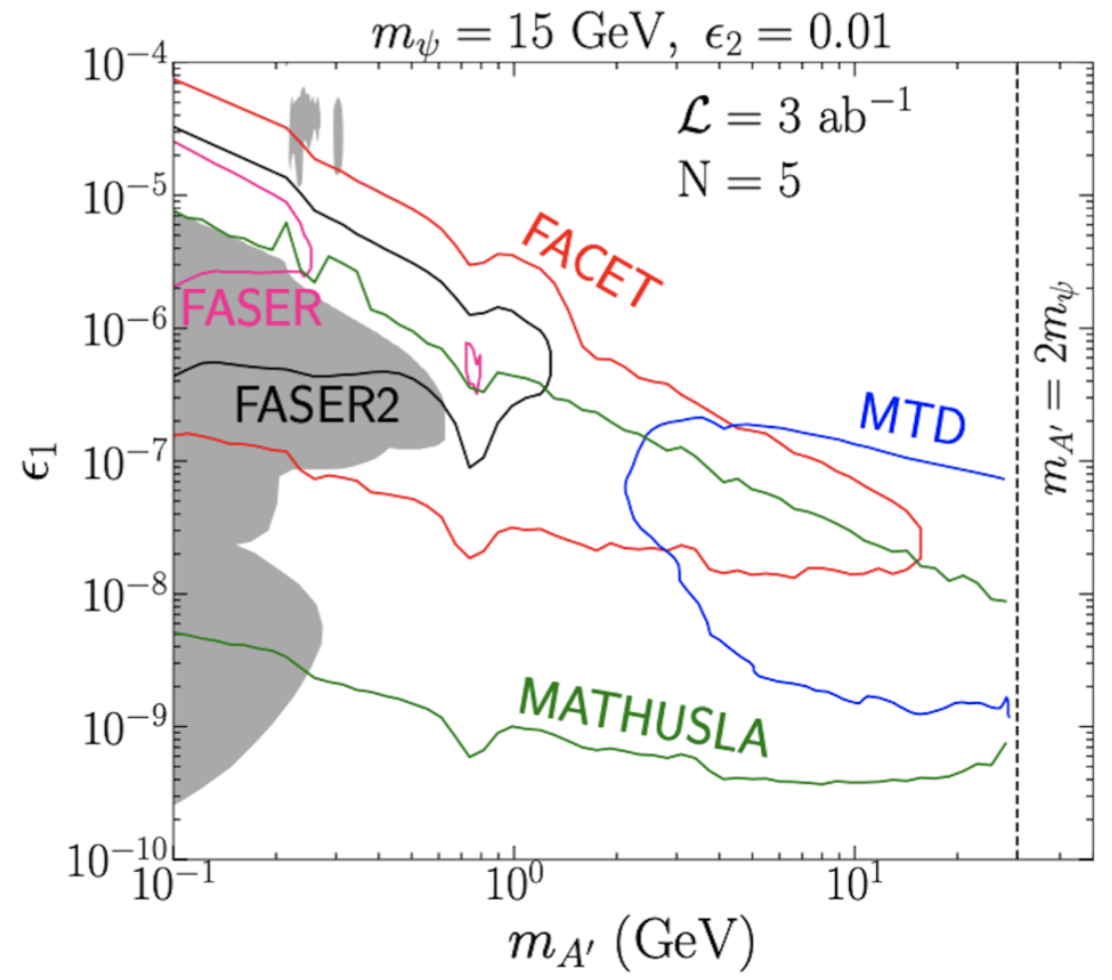}
        \caption{H(125) $ \rightarrow \phi \phi$ decay }
        \label{fig:5-a}
    \end{subfigure}
    \hfill
    \begin{subfigure}[t]{0.49\textwidth}
        \centering
        \includegraphics[width=.8\textwidth]{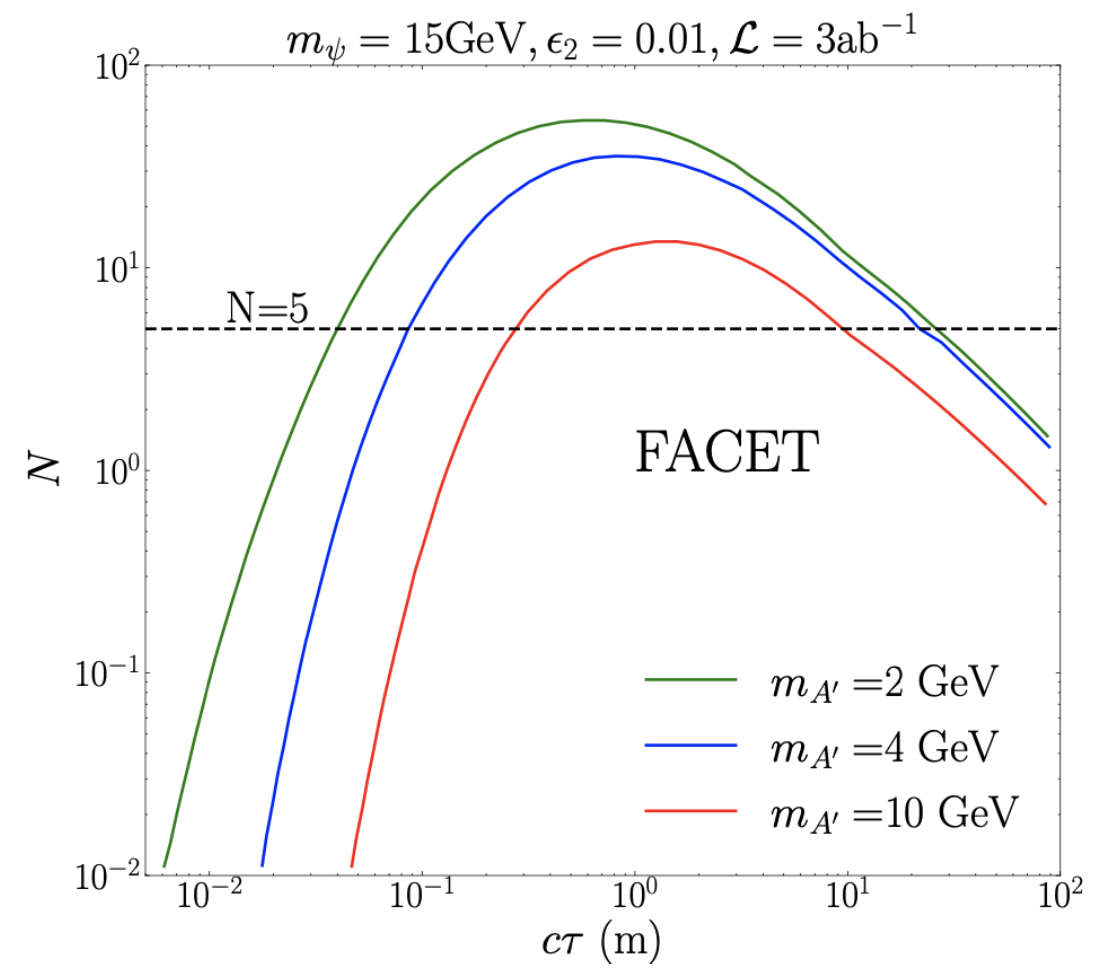}
        \caption{B meson Decay}
        \label{fig:5-b}
    \end{subfigure}
    \caption{Reach of FACET and other existing and proposed experiments for a dark Higgs boson $\phi$ (3 event contours) with the assumption of either $\%0$ (right) or $\%2.5$ (left) branching fraction for the $H(125)\rightarrow \phi \phi$.}
    \label{fig:5}
\end{figure}

{\bf Dark Higgs Bosons - }\label{sec:DHB}
Dark higgs field provides a simple mechanism to give mass to dark photon $A^{'}$. The dark higgs boson can be very long-lived due to its suppressed couplings to the accessible light SM states.The reach of FACET for the dark higgs boson decaying to a detectable final state is given in ~Figure \ref{fig:5}. FACET offers a unique sensitivity for the dark higgs boson masses.

%\newpage
\subsubsection{Summary}

 FACET will make an inclusive search for some portals with a sensitivity defined by their masses and couplings to SM particles. The searches will be background-free in many channels for long live particles. FACET will explore a unique area in parameter space of mass and couplings, largely complementary to other existing and proposed searches.

%\begin{equation}\label{eq:area}
%  S = \pi r^2
%\end{equation}
%One can refer to equations like this: see equation (\ref{eq:area}). One can also
%refer to sections in the same way: see section \ref{sec:LLPs}. Or
%to the bibliography like this: \cite{Cd95}.

% Bibliography
%-----------------------------------------------------------------
% \begin{thebibliography}{99}

% \bibitem{Cerci:2021nlb} S. Cerci et all, \emph{FACET:A new long-lived particle detector in the very forward region of the CMS experiment }, EPJ-C, (2022)
% \bibitem{Caputo:2021eaa} A. Caputo, A. J. Millar, C. A. J. O’Hare et al., \emph{Dark photon limits: A handbook}, Phys. Rev. D, (2021)
% \bibitem{Cd03} SHiP Collaboration., \emph{Sensitivity of the SHiP experiment to Heavy Neutral Leptons}, JHEP, (2019)

% \end{thebibliography}

% \end{document}

\afterpage{\clearpage}
%-------------------------------------------

%-------------------------------------------
\subsection{Status and prospects at CERN: FASER -- {\it C.~Antel} }
\label{ssec:antel}
{\it Author: Claire Antel, <claire.antel@cern.ch>}

%\documentclass{article}
%\usepackage[a4paper, total={6in, 8in}]{geometry}
%\usepackage{hyperref,graphicx}
%\usepackage[sorting=none]{biblatex}
%\addbibresource{bib.bib}

%\begin{document}
%\title{Status and Prospects at CERN: The FASER Experiment}
%\author{Claire Antel}
%\maketitle

\textbf{Introduction}\\
The FASER experiment~\cite{FASERDetector2022} is a new LHC experiment for Run 3 offering a complimentary physics programme to the other major experiments at the LHC.
While for instance ATLAS and CMS typically search for new signals that are centrally and promptly produced with high transverse momenta, FASER aims to detect light long-lived particles and neutrinos produced in the far forward region, using the LHC as a high intensity beam of light hadrons and neutrinos. The detector is located 480 meters downstream of the ATLAS proton interaction point (IP1) in a formerly decommissioned tunnel (TI12) next to the LHC main tunnel. It is used both to search for new long lived-lived particles such as dark photons, axion-like particles and heavy neutral leptons, as well as make Standard Model (SM) measurements of collider neutrino interactions.\\

%\section{Physics targets}
\textbf{Physics targets}\\
FASER searches for new particles that may be produced primarily in rare decays of SM mesons that are emitted at small angles to the beam axis, and decay within the FASER detector decay volume 480 m downstream of IP1. Due to the faraway location of FASER, the new particles are necesssarily feebly coupled to the Standard Model to be sufficiently long-lived. More importantly, the small angle of production means new particles are produced at TeV energy scales at the LHC, providing them the boost to reach FASER. The dominant meson production at the LHC are neutral pions - on the order of $10^{16}$ neutral pions will be produced at IP1 in Run 3. Thus, FASER is especially sensitive to new particles, such as a Dark Photon, in the 10-100 MeV mass range from rare neutral pion decays, but also has interesting sensitivity to new particles produced in heavy meson decays such as heavy neutral leptons and axion-like particles. A full study of FASER's physics reach is covered in \cite{FASERPhysics}.

In addition, FASER aims to make SM cross-section measurements of collider neutrinos produced at IP1 and interacting in the FASER detector, providing differential cross-section measurements at formerly uncovered energies for all three neutrino flavours\cite{FASERnuPhysics}.\\

%\section{Detector}
\textbf{Detector}\\
\begin{figure}[ht]
\centering
\begin{minipage}[c]{\textwidth}
	\centering
  \includegraphics[scale=0.3]{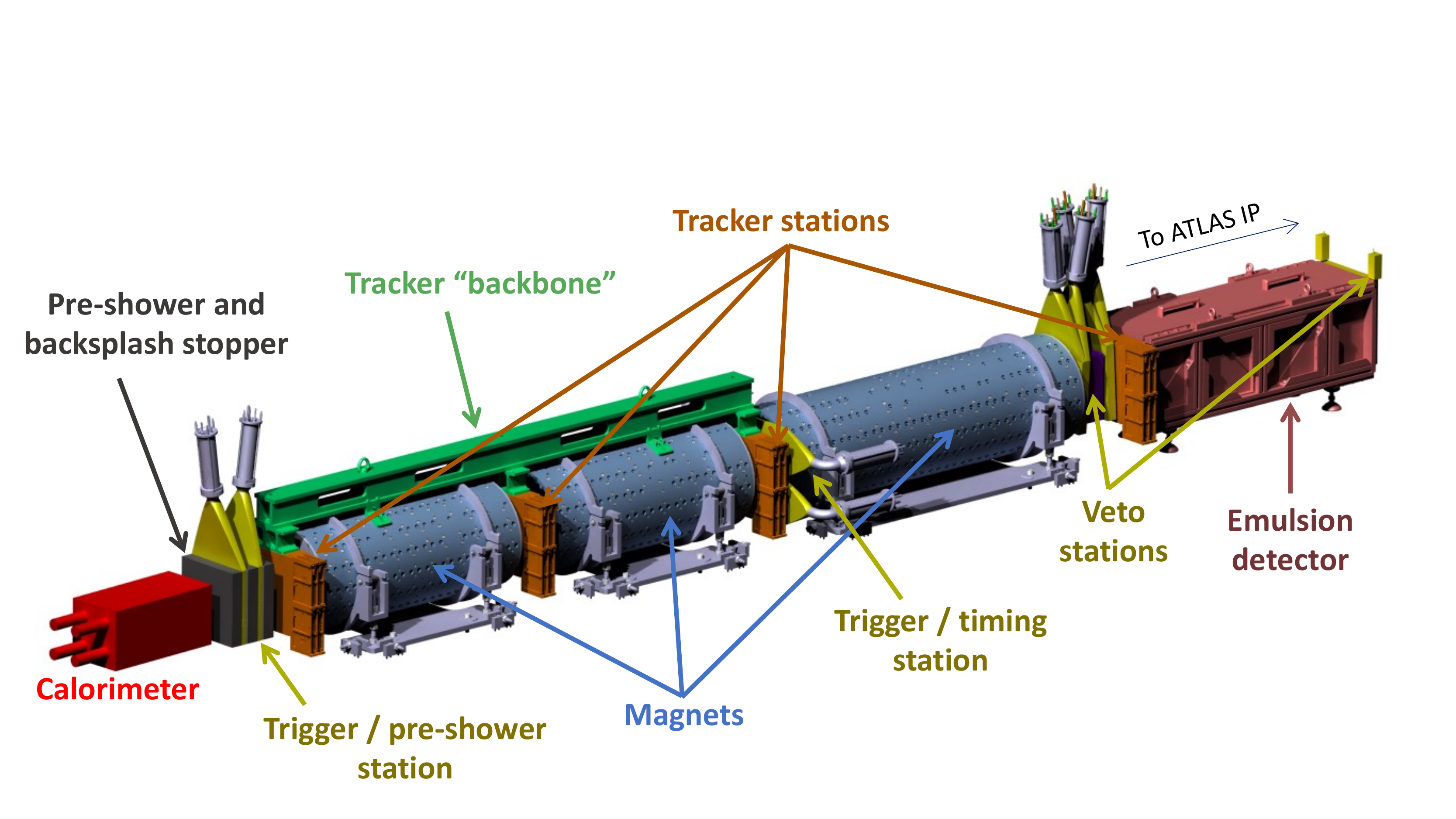}
	\caption{A labelled schematic of the FASER detector.}
	\label{fig:faser_detector}
\end{minipage}
\end{figure}
The FASER detector design, shown in Figure~\ref{fig:faser_detector}, is described in detail in \cite{FASERDetector2022}. The experiment consists of two separate detectors: The FASER spectrometer dedicated to LLP searches, and FASER$\nu$, a neutrino emulsion detector\cite{FASERnuTP}.
The FASER spectrometer is 5.5~m long and 20~cm in aperture. It consists of three permanent 0.57~T dipole magnets, three tracking stations of three tracker planes each, scintillator stations and a calorimeter. The first of the three magnets has a length of 1.5 m and defines the experiment’s decay volume. FASER$\nu$ is a 1.1~tonne emulsion film detector with interleaving tungsten plates. A tracking station interfaces it to the rest of FASER.
The important performance requirements for FASER is to tag incoming muons, the main source of background for FASER LLP searches, with extremely high efficiency using several layers of veto scintillators, trigger on and track the SM decay products of rare LLP decays in its decay volume efficiently, and capture up to TeV scale energetic showers in the calorimeter with good (1\%) energy resolution.\\
%The scintillator stations situated before the decay volume are paramount in tagging incoming muons, which are the main source of background for FASER. Scintillator stations after the decay volume are important for triggering on Standard Model decay products of the LLP in signal events. The calorimeter

%There are four tracking stations of three silicon layers each. Each silicon layer is composed of 8 double-sided silicon strip modules, which are spare ATLAS SCT modules. Four scintillator stations provide trigger signals and are connected to photomultiplier tubes (PMTs). The
%Triggering capability is also provided by the calorimeter, which is otherwise used to identify electrons and photons and to measure their energies. Four spare LHCb outer electromagnetic calorimeter [9] modules are used. They provide energy resolution of about 1\% for 100 GeV to TeV energy deposits. They have a total depth of 25 radiation lengths, but with no longitudinal shower information. They are connected to PMTs, which are read out via the same digitizer board used for the scintillators.

\begin{figure}[!h]
\centering
\begin{minipage}[c]{\textwidth}
	\centering
  \includegraphics[scale=0.1]{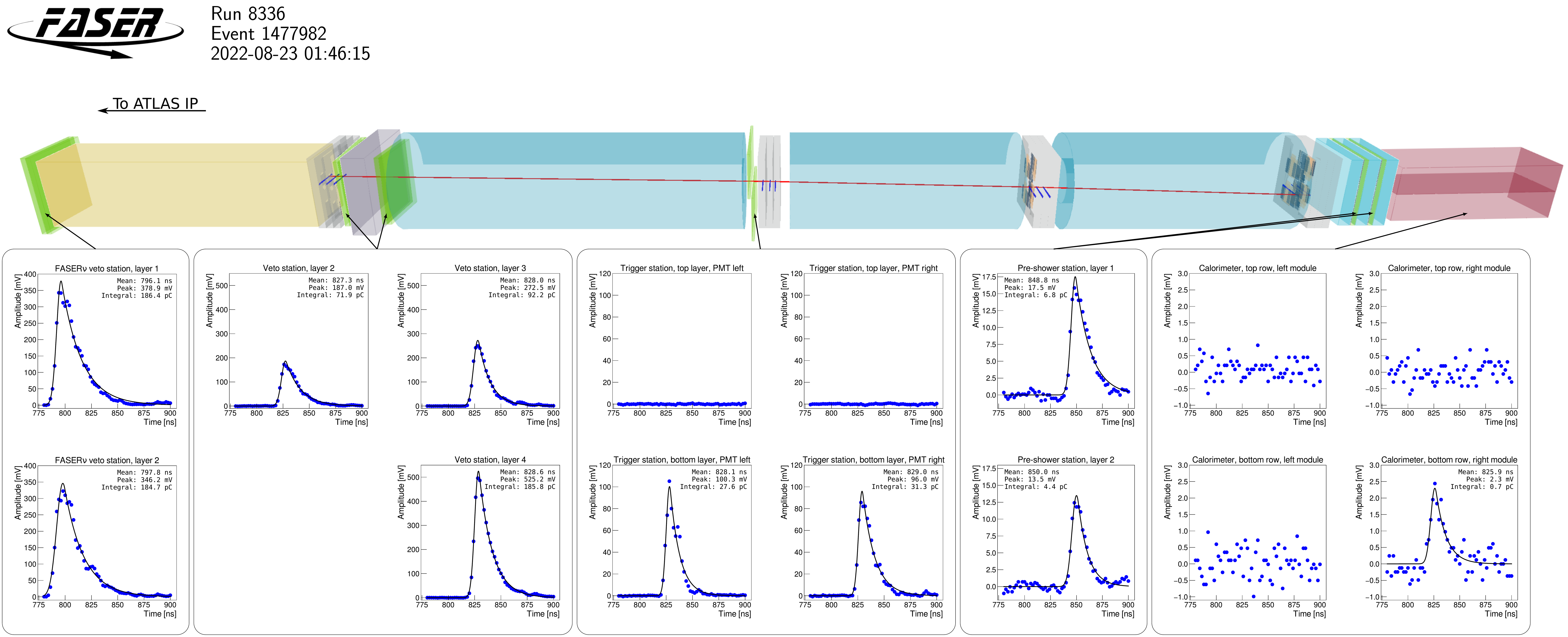}
	\caption{FASER Event Display of a muon travelling through the length of the FASER detector during 2022 proton-proton collisions. The waveforms below the display are the reconstructed PMT ADC pulses of the muon MIP signal fitted with a crystal ball function. The blue strips in the tracking layers are tracker hits while the red line is the reconstructed track.}
	\label{fig:event_display}
\end{minipage}
\end{figure}

%\section{2022 data taking experience and performance}
\textbf{2022 data taking experience and performance}\\
The FASER experiment successfully acquired 30 $fb^{-1}$ of 13.6 TeV proton-proton collision data appropriate for physics analysis during 2022 with the start of Run 3 of the LHC. Figure~\ref{fig:event_display} is an event display of an energetic muon travelling through the length of the FASER detector from the direction of IP1. The red line indicates the reconstructed track using the hits (blue strips) in each tracking layer. The bottom half shows the scintillator and calorimeter ADC waveforms, which are all consistent with a MIP signal. The triggering and tracking of an incoming particle with a collision signal time-of-flight demonstrates the detector to be successfully time tuned for data taking.\\

\begin{figure}[ht]
\centering
\begin{minipage}[c]{\textwidth}
	\centering
  \includegraphics[scale=0.2]{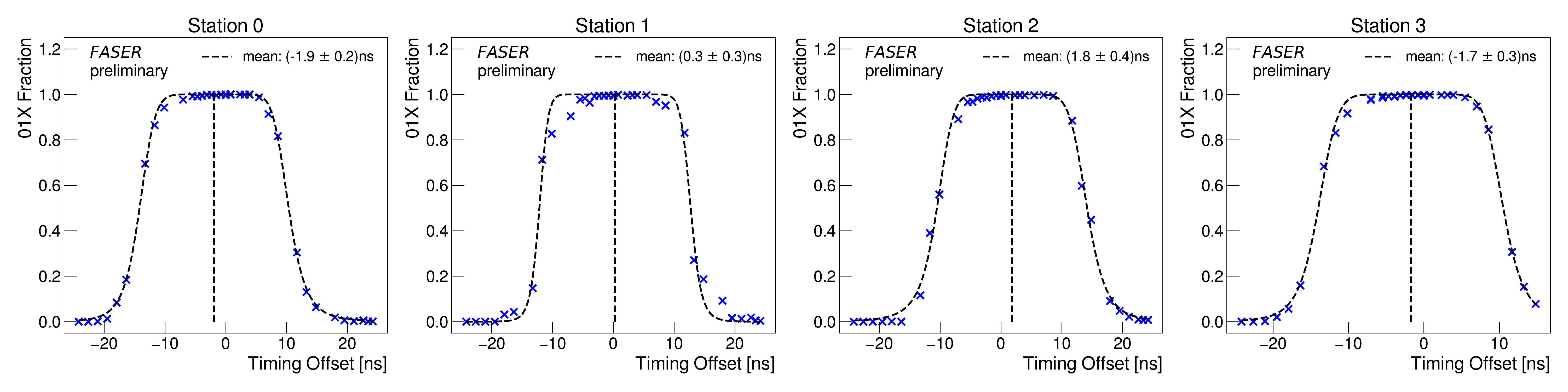}
	\caption{A fine time scan of the SCT tracker modules during early proton-proton data taking. The fraction of tracker hits with good timing as a function of the fine time delay applied in the tracker readout (in ns) is shown for each of the four tracker station. The fine time delay used in operations is shown as a dashed line, tuned to the maximum "good timing" fraction. These plots use data taken in dedicated FASER runs, where the tracker timing settings were changed, during LHC collisions in July 2022. }
	\label{fig:trk_fine_time_scan}
\end{minipage}
\end{figure}

\begin{figure}[ht]
\centering
\begin{minipage}[c]{0.48\linewidth}
	\centering
  \includegraphics[scale=0.4]{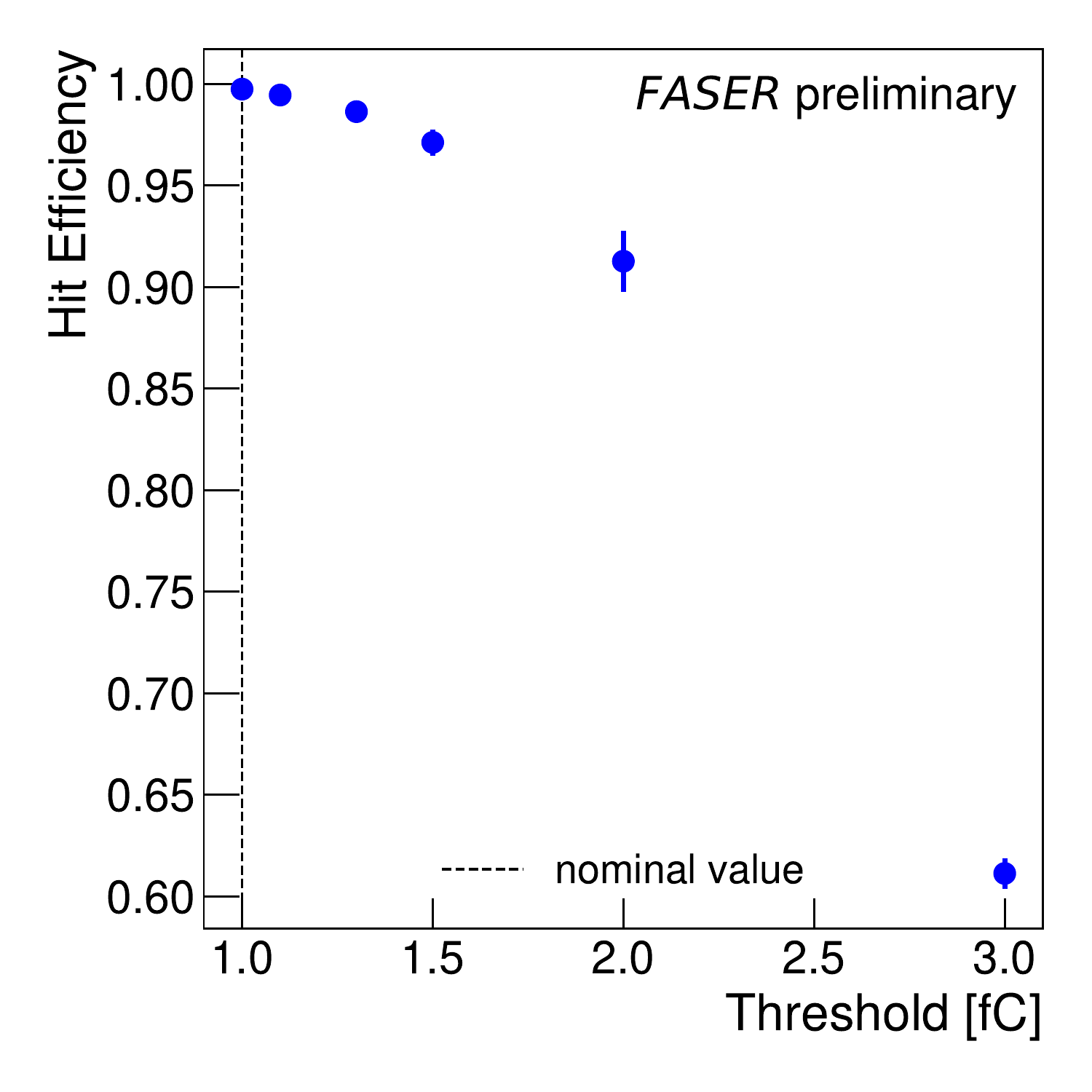}
	\caption{The SCT hit efficiency versus the applied hit threshold (in fC) is shown. The nominal threshold setting of 1~fC is indicated by a dashed line, and yields an average hit efficiency across the full tracker of $99.64\pm0.10 \%$. This plot used data taken in dedicated FASER runs, where the tracker settings were changed from nominal, during LHC collisions in July 2022. }
	\label{fig:trk_thresh_scan}
\end{minipage}%
\hspace{1em}
\begin{minipage}[c]{0.48\linewidth}
	\centering
  \includegraphics[scale=0.4]{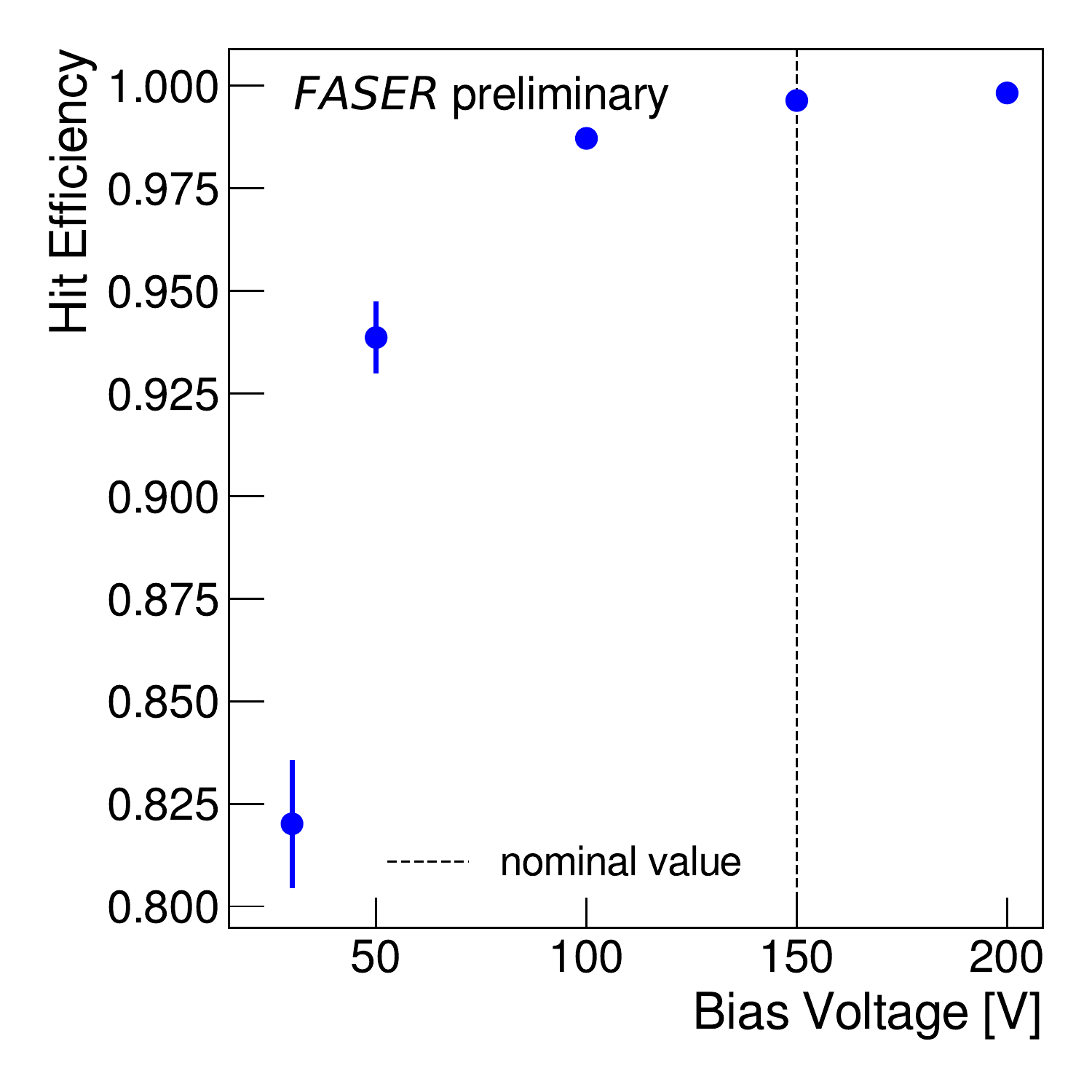}
	\caption{The SCT hit efficiency versus the applied bias voltage (in volts) is shown. The nominal setting of 150 V is indicated by a dashed line, and yields an average hit efficiency across the full tracker of $99.64\pm0.10 \%$. These plots used data taken in dedicated FASER runs, where the tracker settings were changed from nominal, during LHC collisions in July 2022. }
	\label{fig:trk_hv_scan}
\end{minipage}
\end{figure}

Early collision runs were dedicated to performance scans of the detector in running conditions, where detector settings were varied to tune operational settings for optimal performance. Figure~\ref{fig:trk_fine_time_scan} presents a fine time scan of the tracker SCT modules of each tracker station: The fine time delay was swept over a 50~ns window, and the fraction of tracker hits on reconstructed tracks well-centered in the read-out window was measured at each delay. The dashed line indicates the delay now used in operations, tuned to a value where $\sim100 \%$ of hits are perfectly centered in the readout window.
Figures~\ref{fig:trk_thresh_scan} and \ref{fig:trk_hv_scan} are scans of the applied SCT hit threshold and the bias voltage, respectively. Again, the dashed lines indicate the settings used in 2022 operations. An average hit efficiency of $99.64\pm0.10 \%$ across the full tracker for particles hitting the active part of the tracker modules is achieved for optimised threshold and voltage settings. The efficiency falls off at higher applied thresholds and lower bias voltage, as expected.\\

\begin{figure}[b]
\centering
\begin{minipage}[c]{\textwidth}
	\centering
  \includegraphics[scale=0.18]{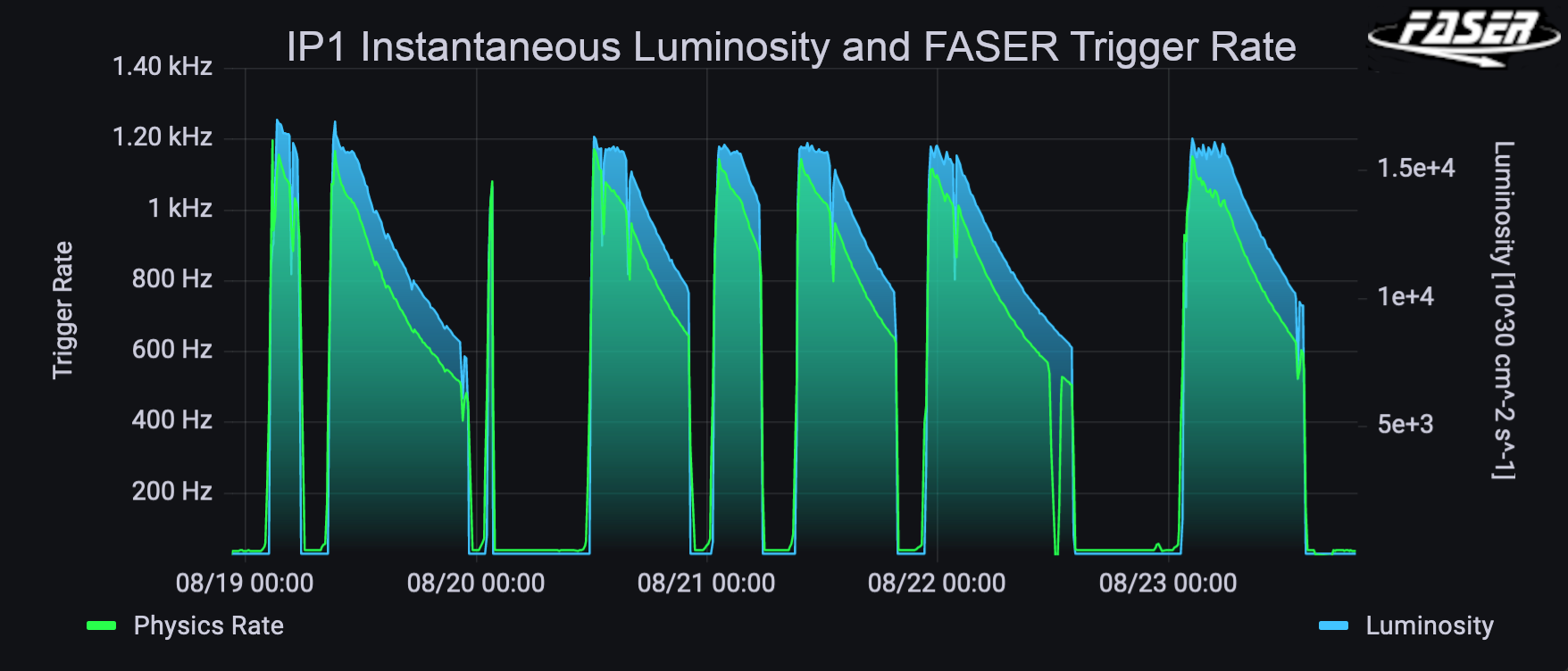}
	\caption{The instantaneous luminosity measured at Interaction Point 1 (blue) and the FASER total trigger recorded rate (green) are shown for 8 LHC fills between August 19th to 23rd, 2022. The trigger rate trend generally follows the luminosity trend but it is evident that the trigger rate falls off more strongly at the beginning of fills than the luminosity. This is suspected to be due to higher beam losses at the beginning of fills. The dip in trigger rate towards the end of the fill on the 22nd August was due to an issue with the digitizer board which halted data taking for 1 hour. Other dips in rate and luminosity are due to emission scans.} %The instantaneous luminosity as provided by ATLAS, shown in blue, is given in units of inverse microbarns per second (10^30 cm-2s-1). The trigger rate is shown in green. The trigger rate trend generally follows the luminosity trend but it is evident that the trigger rate falls off more strongly at the beginning of fills than the luminosity. This is due to higher beam losses at the beginning of fills. The dip in trigger rate towards the end of the fill on the 22nd August was due to an issue with the digitizer board which halted data taking for 1 hour and had been the only issue experienced since start of data taking. Other dips in rate and luminosity are due to emission scans.}
	\label{fig:grafana_rate_and_lumi}
\end{minipage}
\end{figure}%

\begin{figure}[b]
\centering
\begin{minipage}[c]{\textwidth}
	\centering
  \includegraphics[scale=0.38]{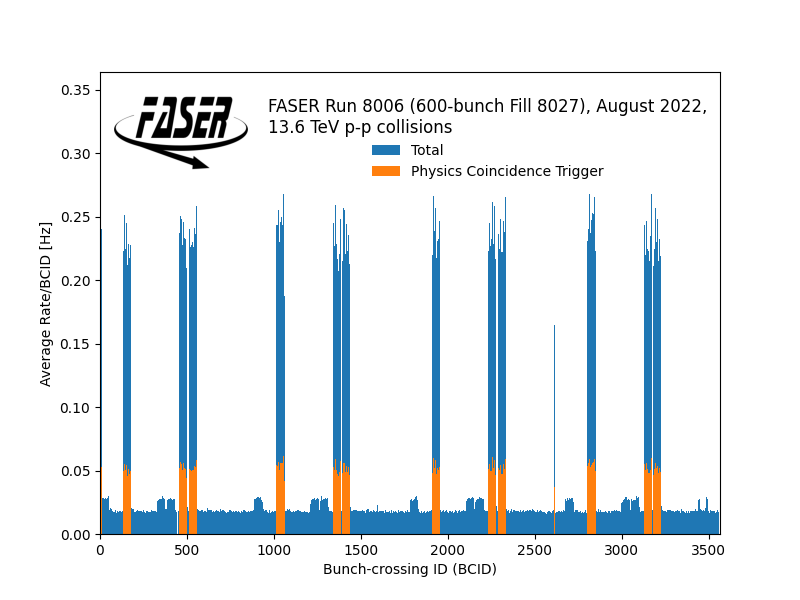}
	\caption{The FASER average trigger rate per bunch-crossng ID is shown for a particular LHC Fill with a low number of colliding bunches ($\sim600$ bunches). The blue bars represent the total recorded trigger rate, and the orange bars represent the recorded trigger rate for the physics coincidence trigger - a coincident trigger between a scintillator at the front and back of the FASER detector, signifying the likely passage of an energetic muon from the direction of IP1. The rate of the physics coincidence trigger is zero for all but colliding bunches.}
	\label{fig:rate_per_BCID}
\end{minipage}
\end{figure}

Once the LHC reached peak instantaneous luminosities during stable running in 2022, the FASER reached a peak event recording rate of 1.25~kHz, which was in fact a factor 2 higher than predicted from FLUKA simulations and in-situ measurements performed in 2018.  Figure~\ref{fig:grafana_rate_and_lumi} shows the online monitoring of the the FASER total trigger rate (green) and the instantaneous luminosity measured at IP1 (blue) for 8 LHC fills between August 19th to 23rd, 2022.
%FASER reached a peak event recording rate of
While the trigger rate shows a clear and expected correlation with luminosity at IP1, it is visibly higher at the beginning of fills relative to the luminosity before levelling off to a constant to-luminosity ratio later in the fill. The unexpected high rates is suspected to be due to (neutron-induced) cavern background in the tunnel initiated by beam losses which are prominent at the start of fills. This is supported by the measured bunch-crossing independent background rate, shown in Figure~\ref{fig:rate_per_BCID}. While a per bunch-crossing ID (BCID) total rate (blue) peaks for colliding bunches, there is a considerable level of triggered events in non-colliding bunches, indicating background sources of particles not related to proton collisions. The smaller rate bumps before each colliding bunch rate peaks are caused by secondary particles as Beam 1 interacts with a quadrupole magnet close to FASER as it passes by towards IP1. The per-BCID rate indicated in orange is the rate of a "Physics Coincidence" trigger requiring a coincidence between a front and back scintillator thereby very likely signifying an incoming energetic muon from IP1 travelling through FASER. This trigger has a non-zero rate only for colliding bunches. Overall, the additional rate has not been a hinderance for the FASER physics programme. The TDAQ system is capable of recording data with the additional rate, the deadtime remains $<2\%$ and the background events can be effectively removed using timing cuts and other event topologies that are consistent with a particle source of origin at IP1.\\

%\section{Conclusion and outlook}
\textbf{Conclusion and outlook}\\
The FASER experiment has had a successfull year of data taking in 2022, collecting $30~fb^{-1}$ of 13.6 TeV proton-proton collisions at the LHC for analysis. The data will be used to search for light long-lived particles such as dark photons, axion-like particles and heavy neutral leptons, as well as make SM measurements of collider neutrino interactions. The detector performance and overall data acquisiton operations went smoothly, despite the experiment experiencing double the trigger rate to what was predicted. Currently the collaboration is preparing to analyse the data for first results in 2023 - deriving calorimeter calibration, particle identification and tracker alignment for analysis.

%\printbibliography

%\end{document}

\afterpage{\clearpage}
%-------------------------------------------

%-----------------------------------------------
\subsection{Prospects at CERN: the Forward Physics Facility -- {\it F.~Kling}}
\label{ssec:kling}
{\it Author: Felix Kling, <felix.kling@desy.de>} 

%******************
\noindent \textbf{Introduction and Facility -} The LHC is the most energetic particle accelerator built thus far. Its primary objective is the study of the Higgs bosons and the search for new particles at the TeV scale. For this reason, the large multi-purpose detectors that surround the LHC's collision points are optimized for rare but spectacular events containing high transverse momentum particles, as expected from the decay of such heavy states. However, the vast majority of collisions at the LHC are soft, meaning that the momentum exchange between the colliding protons is small. Produced in such collisions are a large number of hadrons in the forward direction, outside the coverage of the main detectors, which can carry a sizable fraction of the beam energy. Some of these hadrons will decay into neutrinos and hence produce an intense and strongly collimated beam of highly energetic neutrinos of all three flavors along the beam direction. Similarly, these hadrons could also decay into so far undiscovered light and weakly interacting particles, as predicted by various models of new physics, and which would also form a strongly collimated beam in the forward direction~\cite{Feng:2017uoz}. This realization is the basis for the forward search and neutrino program at the LHC. 

\begin{figure*}[b]
    \includegraphics[width=0.99\textwidth]{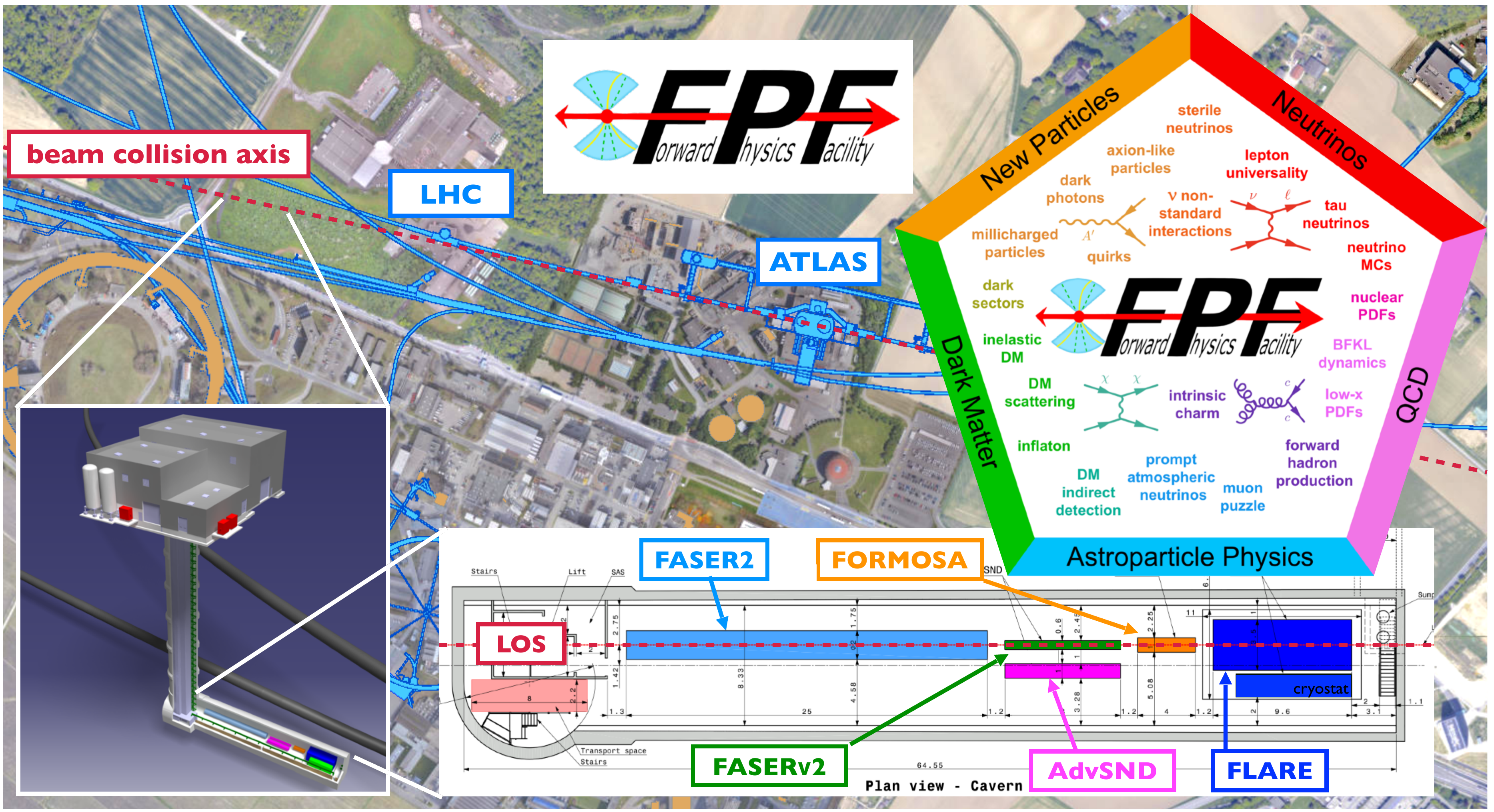} 
    \caption{Location and layout of the Forward Physics Facility. The FPF will be located along the beam collision axis about 620~m west of the ATLAS interaction point and is envisioned to be 65~m long and 8.5~m wide. It will house a diverse set of experiments to explore the many physics opportunities in the forward region. The pentagon summarizes the broad physics potential.}
    \label{fig:fpf_facility}
\end{figure*}

There are currently two experiments operating in the forward direction of the LHC that take advantage of this opportunity: FASER and SND@LHC. They are located roughly 480~m downstream of the ATLAS interaction point in the previously unused side tunnels TI12 and TI18. FASER is primarily designed to search for long-lived particles in the forward direction~\cite{FASER:2018ceo, FASER:2018eoc, FASER:2018bac, FASER:2019aik, FASER:2021cpr, FASER:2021ljd, Boyd:2803084}. In addition, the experiment contains a dedicated emulsion neutrino detector, called FASER$\nu$, at its front end~\cite{FASER:2019dxq, FASER:2020gpr}. SND@LHC is also a dedicated neutrino experiment consisting of both emulsion and electron components~\cite{SHiP:2020sos, Ahdida:2750060}. Both experiments have started data taking in summer 2022 and will continue to operate during Run~3 of the LHC.

In the last few years, it has become very clear that there is a diverse physics program that is largely unexplored in the forward region at the LHC. Since this cannot be fully exploited by the two recently installed experiments alone, several proposals for larger scale upgrades or additional detectors have been put forward~\cite{FASER:2020SnowmassLOI1, FASER:2020SnowmassLOI2, FASER:2020SnowmassLOI3, Foroughi-Abari:2020qar}. At the same time, the TI12 and TI18 tunnels, which currently house the FASER and SND@LHC experiments, will not be able to accommodate larger detectors or additional experiments. They are highly constrained by the 1980’s infrastructure that was never intended to house experiments or provide the necessary services, and also do not allow access while the LHC is running. 

To address the above-mentioned issues, the construction of a dedicated Forward Physics Facility (FPF) housing a suite of experiments has been proposed~\cite{MammenAbraham:2020hex, Anchordoqui:2021ghd, Feng:2022inv}. Several possible sites have been investigated by the CERN civil engineering team. A preferred location was identified and a baseline design was developed, as shown in Fig.~\ref{fig:fpf_facility}. The FPF will be placed along the beam collision axis about 620~m west of ATLAS interaction point. The facility will consist of a new 65m long and 8.5m wide cavern as well as a new shaft to the surface. 

Several experiments have been proposed to be housed in the FPF. Placed in the front is \textit{FLArE}, which is a liquid argon time projection chamber with a $\sim 10$~tonne target mass to detect neutrinos and other scattering signatures. \textit{FORMOSA} is a MilliQan-like experiment~\cite{Haas:2014dda, Ball:2016zrp, Ball:2020dnx, milliQan:2021lne} consisting of a plastic scintillator array to search for milli-charged particles. \textit{FASER$\nu$2} is an emulsion neutrino detector. It has a $\sim 20$~tonne target mass and the capability to see tau neutrinos. \textit{AdvSND} is an upgraded version of the SND@LHC experiment for neutrino physics. It is placed slightly off-axis and consists of electronic detector components. Placed at the end is \textit{FASER2}, which is a larger scale version of the FASER detector. It consists of $\sim 10~\text{m}^3$ decay volume followed by a magnetic spectrometer and calorimeter, designed for long-lived particle searches. It also acts as a downstream muon spectrometer for the neutrino experiments. Together, these experiments cover a broad physics program, as illustrated by the pentagon in the upper right part of Fig.~\ref{fig:fpf_facility}. This includes the search for dark matter and other new particles as well as the detection of more than a million neutrinos at TeV energies, with implications for neutrino physics, QCD and astro-particle physics. \medskip 

%******************
\noindent \textbf{Searches for BSM Physics -} 
While traditional searches for physics beyond the SM at the LHC focus on heavy and relatively strongly interacting states, new particles might also be light but very weakly interacting. Such light feebly interacting particles indeed naturally appear in many models of new physics designed to address the most significant outstanding questions in particle physics: the nature of dark, the electroweak hierarchy problem, the matter-antimatter asymmetry of the universe, the origin of neutrino masses, inflation, as well as the strong CP problem. At the LHC, such particles may be produced abundantly in the forward direction where they form a collimated beam directed at the FPF.  As illustrated in Fig.~\ref{fig:fpf_bsm}, the FPF experiments will be able to probe broad range of these models using a variety of signatures:

\begin{figure*}[t]
    \includegraphics[width=0.99\textwidth]{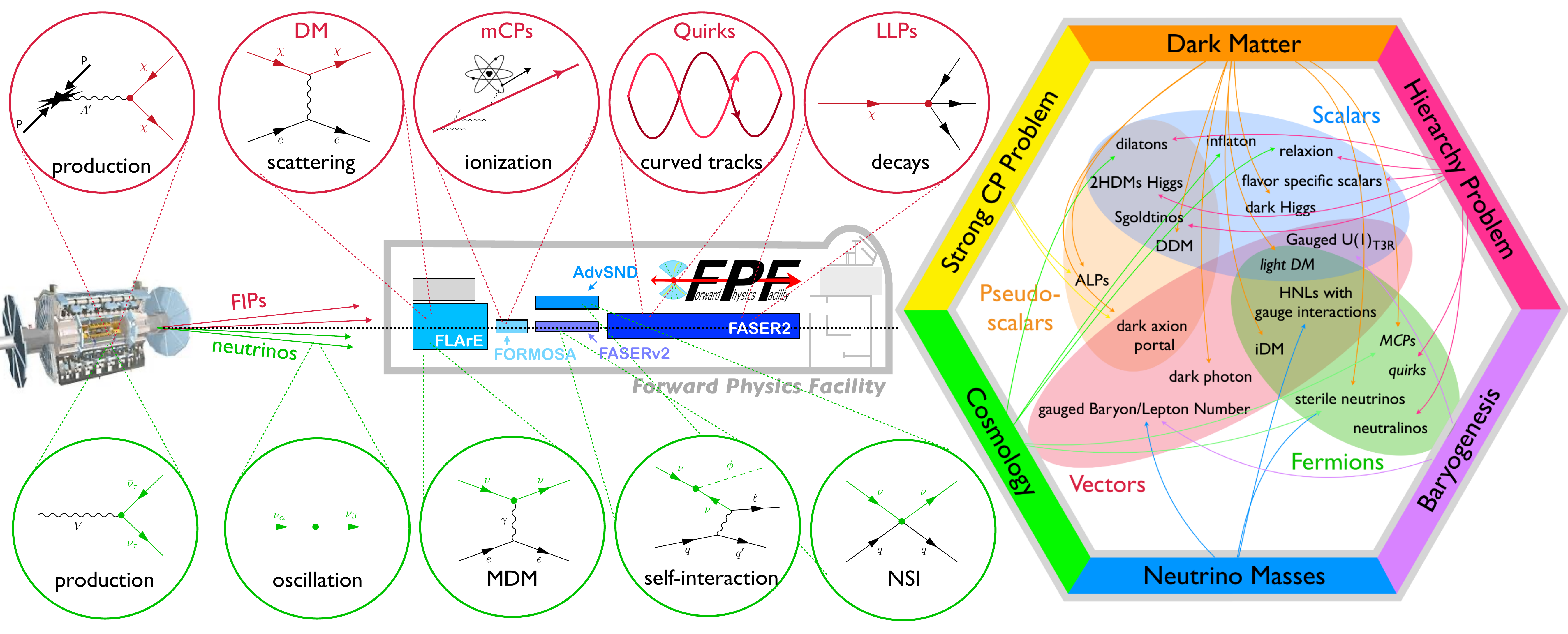} 
    \caption{New physics overview. We schematically illustrate the most important new physics signatures associated with dark sector particles and high energy neutrinos that can be tested at the FPF. The hexagon on the rights shows various new physics models that have been studied in the context of the FPF and how they are related to the outstanding questions in particle physics.}
    \label{fig:fpf_bsm}
\end{figure*}

\begin{description}
% [leftmargin=0.5cm,noitemsep,topsep=1mm,font=\normalfont\itshape]
%
\item [Long-Lived Particles] Due to their feeble interaction to the SM, new particles may have macroscopic lifetimes and decay in far away from where they are produced. The FPF is ideally located to search for such decays, for example in the FASER2 detector. Dedicated studies have shown that the FPF has the potential to discover a large variety of such long-lived particles, including the dark photon~\cite{Feng:2017uoz} and other light gauge bosons~\cite{Bauer:2018onh, Bauer:2020itv, Foguel:2022ppx}; the dark Higgs~\cite{Feng:2017vli} and other light scalars~\cite{Batell:2017kty, Boiarska:2019vid, Okada:2019opp, Kling:2020mch, Csaki:2020zqz, Batell:2021xsi, Demidov:2022ijc}; heavy neutral leptons~\cite{Kling:2018wct, Helo:2018qej} and other light fermions~\cite{Dercks:2018eua, DeVries:2020jbs, Dreiner:2022swd}; axion-like particles~\cite{Feng:2018noy, Carmona:2021seb}; inelastic dark matter~\cite{Berlin:2018jbm}; and many other non-minimal scenarios~\cite{Araki:2020wkq, Li:2021rzt, Jodlowski:2019ycu, deNiverville:2019xsx}. 
\item [DM Scattering] New light particles may also be stable on collider time scales or even cosmological time scales, in which case they would be a viable dark matter candidate. At the FPF, one can search for the scattering of these particles in the neutrino detectors. Physics sensitivity studies have been performed for a variety of models including DM interacting via a dark photon portal~\cite{Batell:2021blf, Batell:2021aja}, DM with hadrophilic mediators~\cite{Batell:2021snh}, or dark sector states with electromagnetic properties~\cite{Kling:2022ykt}. They found that the FPF experiments are often able to probe the interesting relic target region in these models.
\item [Anomalous Ionization] Another example of feebly interacting particles that can be probed at the FPF are millicharged particles, which, for example, arise in models with a massless dark photon~\cite{Holdom:1985ag}. Experimentally, these particles act like minimum ionizing particles with anomalously small energy deposits $\langle dE/dx\rangle$ which can be seen in the FORMOSA detector. Dedicated studies have shown that the FPF is the world’s most sensitive location to search for millicharged particles
in the 10 MeV to 100 GeV mass range~\cite{Foroughi-Abari:2020qar}. 
\item [Other Dark Sector Signatures] The FPF may also detect other new physics signatures. The authors of Ref.~\cite{Kling:2022ehv} pointed out that the large flux of high-energy photons at the LHC in  combination with the strong forward LHC magnets and the magnetic fields of the FASER2 naturally form a light-shining-through-walls experiment that can be used for light axion-like particle searches. They found that this setup would provide the most sensitive purely laboratory search for axion-like particles with masses between 10~meV and 10~keV. Another example are quirks~\cite{Li:2021tsy}. These are new heavy particles which are charged under both electromagnetism as well as a new confining gauge group. If the corresponding confinement scale is significantly smaller than the mass of the particle, pair produced quirks will be connected by a macroscopic flux tube. This causes them to move along helical trajectories which can be seen in the FPF detectors. Quirks therefore provide an example of heavy physics that can be probed at the FPF.  
\item [BSM Neutrino Signatures] In addition to searching for new particles, the FPF may also probe new physics phenomena in the neutrino sector using the LHC's unique beam of high energy neutrinos. A variety of such signatures are illustrated in the bottom left of Fig.~\ref{fig:fpf_bsm}. This includes searches for new tau-neutrino philic particles whose decay could enhance the tau neutrino flux~\cite{Kling:2020iar, Ansarifard:2021dju, Bahraminasr:2020ssz}; searches for sterile neutrinos induced oscillations~\cite{FASER:2019dxq, Bai:2020ukz, Aloni:2022ebm}; searches for anomalous electromagnetic properties of neutrinos such as their dipole moment~\cite{AbariTsaiAbraham, Jodlowski:2020vhr, Ismail:2021dyp}; the search for neutrino-philic particles produced in neutrino interactions~\cite{Kelly:2021mcd}; as well as the search for non-standard neutrino interactions~\cite{Ismail:2020yqc, Falkowski:2021bkq}. 
\end{description}
Together, these searches are able to probe a variety of models of new physics that have been proposed as solutions to the outstanding problems of particle physics. This is illustrated by the hexagon on the right of Fig.~\ref{fig:fpf_bsm}. \medskip 

\begin{figure*}[t]
    \includegraphics[width=0.99\textwidth]{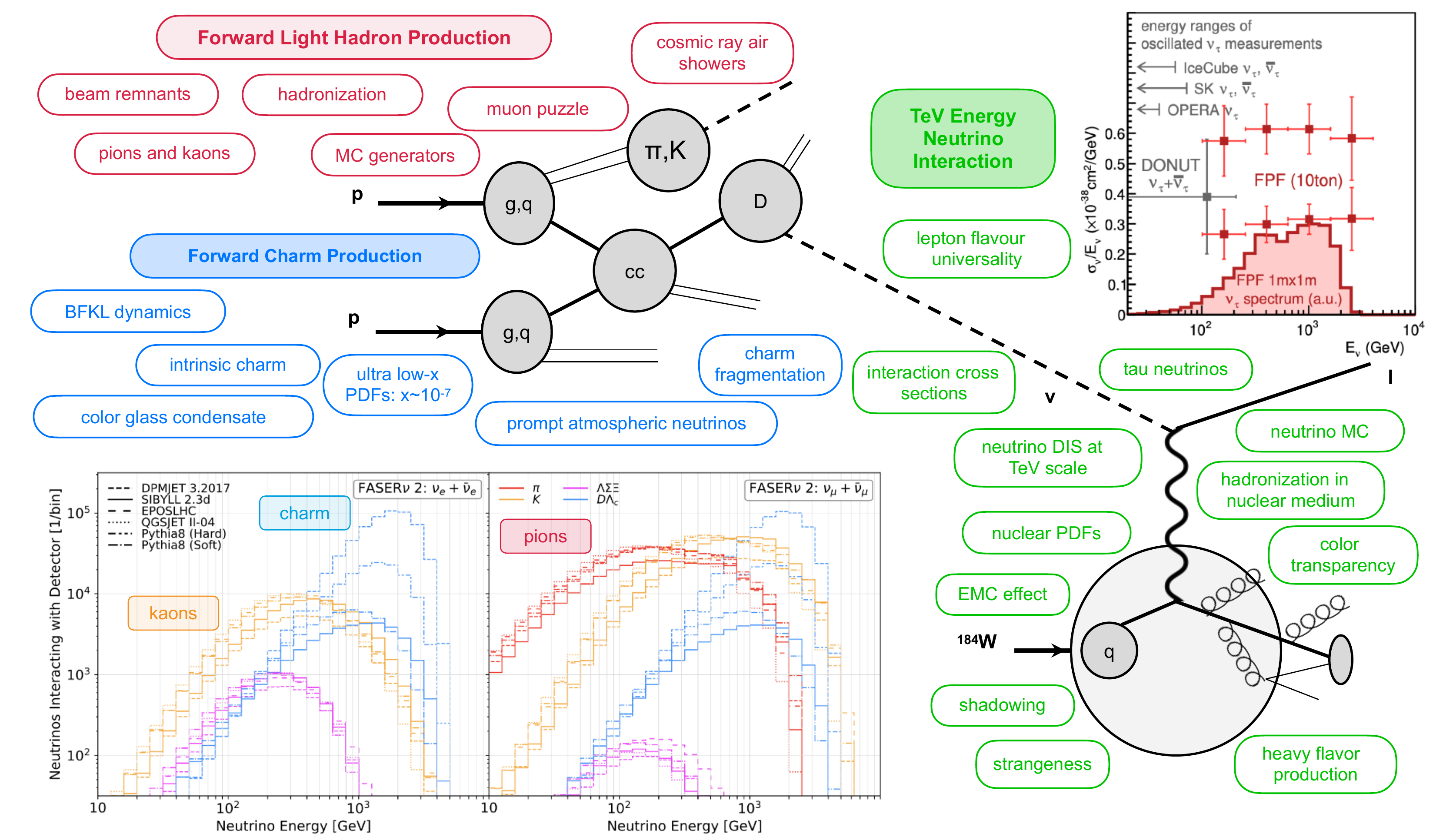} 
    \caption{SM physics overview. Shown in gray is the production of neutrinos at ATLAS collision point and subsequent interaction at the FPF detectors. Placed around are a variety of physics phenomena that could be tested using forward neutrino measurements. In the lower left we show the electron and muon neutrino energy spectrum, as obtained using different generators, separated by the production channel. In the upper right we show the spectrum of tau neutrinos and how it can be used to constrain their interaction cross section in the unconstrained TeV energy range.}
    \label{fig:fpf_sm}
\end{figure*}

%******************
\noindent \textbf{SM Measurements -}
The LHC produces an intense beam of high-energy neutrinos in the far forward direction. While this fact was known since the 80s~\cite{DeRujula:1984ns, Vannucci:253670, DeRujula:1992sn, Park:2011gh, Buontempo:2018gta, XSEN:2019bel, Foldenauer:2021gkm, Arakawa:2022rmp}, the first handful of neutrino interaction candidates have only been recorded recently by a small emulsion detector which was placed in the TI18 tunnel by the FASER collaboration in 2018~\cite{FASER:2021mtu}. The FPF experiments will be able to detect more than a million neutrinos during the HL-LHC era. This will allow them to fully utilize the opportunities offered by the LHC neutrino beam and significantly expands the LHC's physics portfolio, as illustrated in Fig.~\ref{fig:fpf_sm}. 

Shown in the lower left part are the expected number of neutrino interactions for electron and muon neutrinos in the 20~tonne FASER$\nu$2 detector as a function of the neutrino energy~\cite{Kling:2021gos}. We can see that the average neutrino energy is around a TeV. The differently colored lines correspond to the different neutrino production modes: mainly pion, kaon and charm decays. The different line styles correspond to different models for the hadronic interaction~\cite{Pierog:2013ria, Riehn:2019jet, Ostapchenko:2010vb, Fedynitch:2015kcn, Sjostrand:2014zea}. We can see that there are sizable uncertainties, especially for the modelling of forward charm production. The measurement of the neutrino fluxes at FPF experiments will provide novel input to constrain forward particle production and the underlying physics.  

Muon neutrinos and electron neutrinos with energy below about 1~TeV are mainly produced in pion and kaon decays. The production of these light hadrons is typically modeled by hadronic interaction models, of which many were originally designed for the simulation of high energy cosmic ray collisions. The neutrino flux measurement will provide crucial input to improve these models and hence reduce the uncertainties of air shower measurements. In particular, it will help to understand the observed excess of muons in cosmic-ray air showers~\cite{Albrecht:2021cxw, Anchordoqui:2022fpn, Anchordoqui:2022ivb}. High energy electron neutrinos and tau neutrinos are predominantly produced in the decay of charm hadrons. The measurement of these fluxes opens a window to the probe QCD in an otherwise experimentally inaccessible kinematic regime and for example allow to validate the predictions of collinear factorization and BFKL-based approaches; to constrain gluon PDFs at very low $x\sim10^{-7}$; to probe gluon saturation effects and non-linear dynamics; or to test models of intrinsic charm~\cite{Jeong:2020idg, Jeong:2021vqp, Bai:2021ira, Bai:2022jcs, Maciula:2020dxv, Maciula:2022lzk}. First experimental constraints on forward charm production would also constrain prompt atmospheric neutrino fluxes and hence provide valuable input for neutrino telescopes in their search for extragalactic neutrinos. 

Neutrino experiments at the FPF can also study the interactions of TeV energy neutrinos of all three flavors and, for the first time, measure their interaction cross section in this energy range. At such high energies, the interaction are well described by deep-inelastic scattering, which makes them a great laboratory to study parton distribution functions, especially for the strange quark PDF through the charm associated neutrino interactions $\nu s \to \ell c$ similar to CHORUS and NuTeV; probe nuclear effect such as shadowing, anti-shadowing and the EMC effect~\cite{Eskola:2016oht, Kovarik:2015cma, AbdulKhalek:2020yuc}; study the dynamics of the hadronic final state allowing to test hadronization inside nuclear matter, the formation zone prescription, color transparency at the highest energies, and final state interaction effects~\cite{Mosel:2022tqc}; and provide valuable input to validate and tune simulation tools for neutrino interactions~\cite{Battistoni:2009zza, Hayato:2021heg, Golan:2012rfa, Buss:2011mx, Andreopoulos:2009rq}. 

Finally, it is worth noting that the FPF experiments will detect thousands of tau neutrino interactions, allowing them to study the tau neutrino in great precision~\cite{MammenAbraham:2022xoc}. In particular, the magnetic field in the FPF neutrino detectors allows them to differentiate tau neutrinos and tau anti-neutrinos for the first time.

\afterpage{\clearpage}
%-----------------------------------------------

%-------------------------------------------
\subsection{Search for light DM at NA64 with high intensity electron and muon beams: status \& prospects at CERN -- {\it P.~Crivelli}}
\label{ssec:crivelli}
{\it Author: Paolo Crivelli, <Paolo.Crivelli@cern.ch>}

\subsubsection{The NA64 experiment}
%%%%%%%%%%%%%%%%%%%%%%%%%%%%%%%%%%%%%%%%%%%%%%%%%%%

Our proposal (P348) to search for Dark Sectors at the CERN Super Proton Synchrotron (SPS) \cite{Andreas:2013lya} was positively received by the SPS committee (SPSC) in April 2014. We were granted a test beam run in 2015 for feasibility study, and we were finally approved as the 64th CERN experiment in the North Area (NA64) in March 2016.
NA64 is designed as a hermetic general purpose detector to search for Dark Sector (DS) physics in missing energy events from electron/positron, hadron, and  muon scattering off nuclei. 
The main focus of the NA64 is Light thermal Dark Matter (LDM) interacting with the Standard Model (SM) via vector (or other) portal, such e.g. as dark photons ($A'$). 
%Because of the high energy of the incident beam, the centre-of-mass system is boosted relative to the laboratory system. Taking this boost into account resultsin enhanced  hermeticity of the detector  providing a nearly full solid angle coverage.
The experiment, in electron mode (NA64e), employs the optimized  100 GeV electron beam from the H4 beam-line at the North Area. The beam was designed to transport the electrons with the maximal intensity up to few $\simeq 10^7$ per SPS spill of 4.8 s in the momentum range between 50 and 150 GeV/c.
%The electrons produced in a converter are transported to the NA64 detector inside an evacuated beam-line tuned to an adjustable  beam momentum.
The hadron contamination in the electron beam was measured to be at a level of $\pi/e^- \lesssim 2\%$ and $K/e^- \lesssim 0.3\%$. \\
%The beam  has a transverse size  at the detector position on the order of 5 mm$^2$ and a halo with intensity  $\lesssim $ 0.5 \%.
The NA64 experiment run from 2016 until 2018, and in 2021 after the CERN long shutdown (LS2), it resumed data taking in a new permanent location at H4 CERN prepared for us. 
Despite the experiment being quite new, very interesting results were rapidly achieved \cite{Banerjee:2016tad,Banerjee:2017hhz,Banerjee:2018vgk}. In this contribution, we review the main results accomplished so far subdividing those into the $A'$ decay modes being explored. 

%\begin{figure}[!htb]
%\centering
%\includegraphics[scale=0.8]{setup-invisible-2018.png}
%\includegraphics[scale=0.5]{setup-2018-2-rotate.pdf}
%\includegraphics[scale=0.5]{NA64_setup_2018_visible.pdf}
%\includegraphics[scale=0.4]{NA64_setup_artistic.png}
%\caption{Current NA64 setup for visible mode channel $\aee$.}
%\label{fig:NA64_setups}
%\end{figure}

\begin{figure}[!htb]
\centering
\includegraphics[width=0.6\columnwidth]{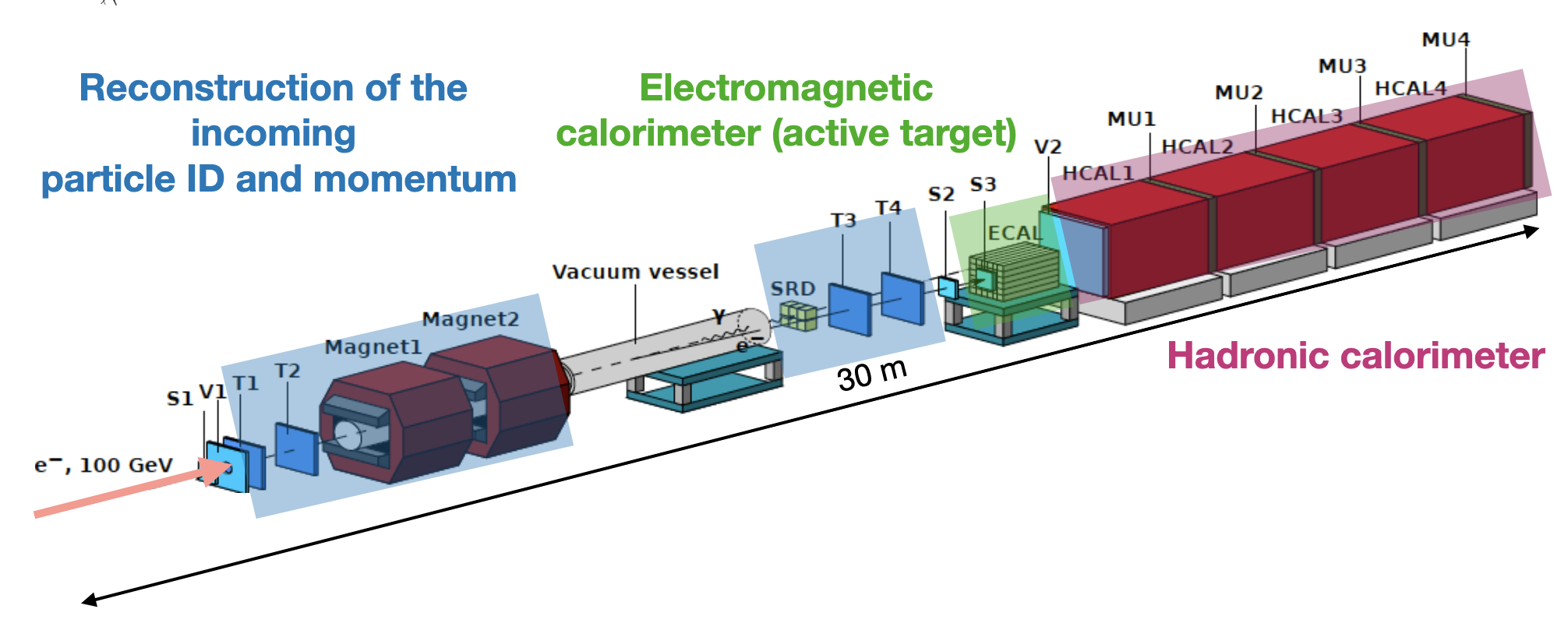}
\caption{NA64 setup and working principle for the search of dark photons through missing energy in the active target (ECAL).}
\label{fig:NA64_sketch}
\end{figure}

\textbf{Invisible mode:}
NA64 pioneered the active beam dump technique combined with the missing energy measurement to search for invisible decays of massive $A'$, produced in the ECAL target (the electromagnetic (em) calorimeter) by the dark Bremsstrahlung reaction $e^-Z \rightarrow e^-ZA'$, where electrons scatter off a nuclei of charge $Z$.  After its production, the $A'$ would promptly decay into a pair LDM candidate particles, $A'\rightarrow \chi\chi$, which would escape the setup undetected leaving missing energy as signature. For this reason we call these searches {\it invisible}. The parameter space characterized by mixing strengths 10$^{-6} < \epsilon <$ 10$^{-3}$ and masses $m_{A'} $ in the sub-GeV range is the NA64 target: a region where the DM origin can be explained as a thermal freeze-out relic. Missing energy experiments, such as NA64, require a precise knowledge of the incoming beam (momentum and particle ID) and an accurate measurement of the deposited energy from the incoming beam's interaction.
%The signal event recognition in NA64 must rely only on the detection  of the incoming and outgoing electron, since the decay product for  the $\ainv$ decay are undetectable.
  %A fraction $f$ of the  primary beam energy  $E_{A'} = f E_0$ is carried away which penetrate the detector without interactions  resulting in an event with zero-energy deposition. While the remaining part $E_e=(1-f)E_0$ is deposited in the target  by the  scattered electron. Thus, the  occurrence of $A'$  produced would appear as an excess of events whose signature is  a single e-m shower in the target with energy $E_e$ accompanied by a significant missing energy $E_{miss}=E_{A'}= E_0 - E_e$  above those expected from backgrounds. 
A signal event is defined as a single electromagnetic shower in the ECAL with an energy $E_{ECAL}$ below given threshold\footnote{The value is chosen to maximize the sensitivity of the experiment, in NA64 for 100 GeV incoming beam energy this $E_{ECAL}<50$ GeV.} accompanied by a significant missing energy $E_{miss}=E_{A'}=E_{initial}-E_{ECAL}$.  The occurrence of the $A'$ production is inferred in case these events show an excess above those expected from backgrounds. In Fig. \ref{fig:NA64_sketch}, we present a sketch of the setup and a summary of the NA64 working principle.

The signal yield for an active beam approach is proportional to $\epsilon^2$, thus, enhancing the sensitivity for NA64 with respect to the yield  $\propto \alpha_D\epsilon^4$ in traditional beam-dump approach where the A’ decay is measured in a detector further away from its production point in the dump.

The main results achieved so far in the invisible electron mode are:
\begin{enumerate}
\item 2016 July run:  $2.7\times 10^{9}$ electrons on target (EOT) have been collected and the results published in Phys. Rev. Lett. \cite{Banerjee:2016tad}. No event compatible with signal was found, thus, excluding most of the favoured region of parameter space compatible with the muon $g-2$ anomaly depicted as a band in the left plot of Fig.\ref{fig:NA64_invis}. After our results were published, BABAR completely closed the remaining region of parameter space which could provide an explanation as the Dark photon contributing to the g-2 muon \cite{Lees:2017lec}.
\item 2016 October run: the results corresponding to $4.3\times 10^{10}$ EOT have been published in Phys. Rev. D. \cite{Banerjee:2017hhz}. At this level, NA64 starts to become sensitive to light dark matter models. No event compatible with a signal was found and so a new exclusion limit could be set.
\item 2017 run: $\simeq 5.5\cdot 10^{10}$ EOT were accumulated and the combined data sample from 2016 and 2017 reaches the milestone of $\simeq 10^{11}$ EOT. 

\item 2018 run: $ 2\cdot 10^{11}$ EOT were collected. No signal-like event was detected. However, the results of the combined analysis from 2016-2018 data, illustrated in Fig. \ref{fig:NA64_invis}, set the most stringent limit for LDM below 0.1 GeV for the canonical benchmark parameters $\alpha_D=0.1$ and $m_{A'}=3m_{\chi}$, thus, NA64 became the leading beam-dump experiment in this region. These results  were selected as PRL editor's suggestion \cite{Banerjee:2019pds}.

In addition to the Bremsstrahlung reaction, the resonant $A'$ production channel through the e$^-$ annihilation with the positrons present in the electromagnetic shower has also been considered. The 90$\%$ C.L. exclusion limits from the combined analysis are shown in Fig. \ref{fig:NA64_invis}. The inclusion of the resonant process in the data analysis allows to enhance the NA64 sensitivity for a given dark photon mass resulting in the peak around 200 MeV. The addition of this process improves the NA64 sensitivity in the high mass region, where the Dark photons yield is suppressed due to the $1/m^2_{A'}$ dependency of the Bremsstrahlung cross-section (see \cite{Banerjee:2019pds}). Using positrons as a primary beam instead of electrons would increase by another order of magnitude the sensitivity of NA64 at a given mass depending on the beam energy. By scanning the positron beam energy the mass range probed by this mode can be further expanded. The drawback is that one has to deal with about an order of magnitude more hadron contamination in the beam since the secondary particles are created by the primary 400 GeV SPS protons and thus positively charged hadrons are more abundant than their negative counterpart. To study the impact of the increased hadron contamination and the possible resulting background, a first test beam with 100 GeV positron was taken during the 2022 run (see below).  
Electron/positron beam-dump experiments allow to explore alternative scenarios to the dark photon hypothesis. NA64 has already proven its potential to search for light-scalar and pseudo-scalar axion-like particles (ALPs) produced through the Primakoff reaction \cite{NA64:2020qwq}. The current NA64 coverage in these searches closes part of the gap between beam-dump and LEP bounds and it is shown in the right plot of Fig. \ref{fig:NA64_all}. A search for a generic X-boson coupling to electrons could also be performed. We were positively surprised that the NA64 sensitivity was an order of magnitude more stringent than precision experiments \cite{NA64:2021xzo}. However, one should note that in NA64 we assume the X-boson to decay invisibly while the electron g-2 \cite{Fan:2022eto} and the fine structure measurements \cite{Parker:2018vye,Morel:2020dww} are model independent. 

\begin{figure}[!htb]
\centering
\includegraphics[width=0.8\columnwidth]{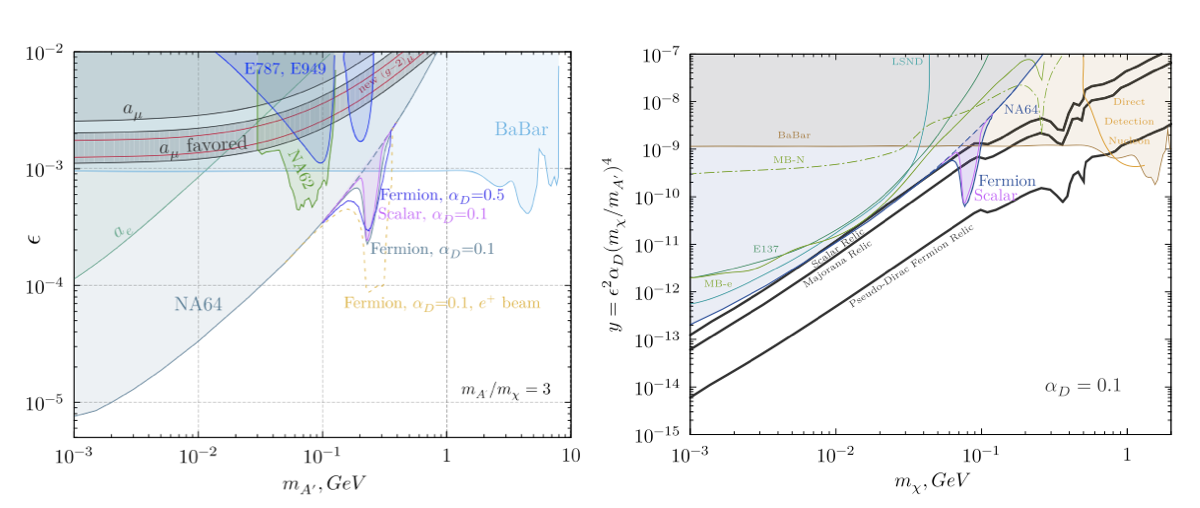}

\caption{Current status of NA64 experiment 90\% C.L. exclusion limits on $A'$ invisible decays including both the Bremsstrahlung and the resonant $A'$ production channels (Left). LDM searches (Right) \cite{Banerjee:2019pds}}.
\label{fig:NA64_invis}
\end{figure}

\end{enumerate}

\textbf{Visible mode:}
 The method for the search of $\aee$ ($\xee$) decays is described in \cite{Andreas:2013lya,Gninenko:2013rka}. 
 %Its  application to the case of the  $\xee$  decay is straightforward. 
In this case, the setup is slightly modified to include an additional compact calorimeter upstream with respect to the ECAL. 
%The high-energy electron beam can also be sent into the NA64 electromagnetic calorimeter that serves this time as an active beam dump.Typically the beam electron loses all its  shower energy in the dump. 
If the $A'$ exists, due to the $A'(X) - e^-$ coupling it would occasionally be produced by a shower electron (or positron) in its scattering off a nucleus in the dump: $e^- + Z \to e^- + Z + A'(X)  ;~ A'(X)\to \ee$. Since the $A'$ is penetrating, it would escape the beam dump and
subsequently decay into an $\ee$ pair in a downstream set of detectors. The pair energy would be equal to the energy missing from the target. 
%The apparatus is designed to identify and measure the energy of the $\ee$ pair in another calorimeter (ECAL).
Thus, the signature of the  $A'(X) \to \ee$ decay  is an event with two em-like showers
in the detector: one shower in the dump, and another one in the ECAL, located downstream in this case,  with the sum energy being equal to the beam energy.

\begin{enumerate}
\item 2017 run:  $\simeq 5.5\cdot 10^{10}$ EOT were accumulated. No candidates were found in the signal box. 
%The conclusion that the background is small is confirmed by the data. 
The combined 90\% confidence level (C.L.) upper limits for the mixing strength $\epsilon$ were obtained from the corresponding limit for the expected number of signal events. These results set the first limits on the $X-e^-$ coupling in the range $ 1.3\times 10^{-4}\lesssim \epsilon_e \lesssim 4.2\times 10^{-4}$ excluding part of the parameter space. In addition, new bounds are set on the mixing strength of photons with dark photons ($A'$) from non-observation of the decay  $A'\to$ e$^+$e$^-$ of the Bremsstrahlung $A'$ with a mass $\lesssim 23$ MeV. The corresponding paper was highlighted as an editor's suggestion in Phys. Rev. Lett.  5\cite{Banerjee:2018vgk}.
%, $N_{A'}^{90\%}$.
%The $A'$ efficiency  and its  systematic error  were determined to stem from
%the overall normalization, $A'$ yield and decay probability, which were the $A'$ mass dependent, and  also  from  
%efficiencies and their  uncertainties 
%in the primary $e^-(0.85\pm0.02)$, WCAL($0.93\pm0.05$), $V_2(0.96\pm0.03)$, ECAL($0.93\pm0.05$), $V_3(0.95\pm0.04)$, and HCAL($0.98\pm0.02$)  event detection. The later,  shown as an example values for  the 40 $X_0$ run, were determined from measurements with $e^-$ beam  cross-checked with simulations.
%To summarise the 2017 results were: 
%\begin{enumerate}
%\item The first limits on the $X-e^-$ coupling in the range $ 1.3\times 10^{-4}\lesssim \epsilon_e \lesssim 4.2\times 10^{-4}$ excluding part of the parameter space. 
%\item New bounds are set on the mixing strength of photons with dark photons ($A'$) from non-observation of
%the decay  $A'\to$ e$^+$e$^-$ of the bremsstrahlung $A'$ with a mass $\lesssim 23$ MeV.
%\item The corresponding paper was highlighted as an editor's suggestion in Phys. Rev. Lett.  5\cite{Banerjee:2018vgk}. 
%\item Promising results from the feasibility study of ALP decays search are also obtained \cite{Gninenko:2320630}.
%\end{enumerate}

\item 2018 run: about $5\times10^{10}$ EOT were collected at an energy of 150 GeV to boost the putative X bosons outside the calorimeter before it decays in order to improve the sensitivity to higher $X-e^-$ couplings. The results extend the limits to $1.2\times10^{-4}\leq\epsilon\leq6.8\times10^{-4}$ for the vector-like benchmark model and leave only a small region open to fully cover the parameter space compatible with the beryllium anomaly (see the central panel of Fig. \ref{fig:NA64_all}). The paper was published in Phys. Rev. D Rapid\cite{Banerjee:2019hmi}. Recently, these searches have been extended also to a pseudo-scalar particle decaying visibly into a lepton pair and the result has been published in Phys. Rev. D \cite{NA64:2021aiq}.
\end{enumerate}. To completely cover the remaining region of parameter space a new shorter optimized WCAL and a new spectrometer with the possibility to reconstruct the X17 invariant mass should be used as proposed in \cite{NA64:2020xxh}. Everything has been prepared and is ready for installation. However, since it cannot run in parallel with the invisible mode setup we decided to post pone this search. About 30 days of beamtime would be required to solidly probe the remaining X17 parameter space, therefore if the results from PADME currently taking data \cite{Darme:2022zfw}, would confirm this anomaly we would be able to cross check this in the 2024 run. 

\textbf{Semi-visible mode}: 
Alternative extended scenarios envisioning two DM species split in mass could result in a signature which is a combination of the two signatures described above. A very intriguing feature of this channel is related to the possibility to recovery both the DM thermal freeze-out and the $(g-2)_{\mu}$ anomaly explanations, by evading the existing experimental constraints on pure visible and invisible modes \cite{Mohlabeng:2019}. This type of models are known as inelastic DM and we refer to their signatures as semi-visible channel. An analysis based on a recast of the results from the combined 2016-2018 data \cite{NA64:2021acr} (see left plot of Fig. \ref{fig:NA64_all}) has already demonstrated the potential of NA64 to study these models. The reach of NA64 to explore in a model independent way a the broad class of parameter space is currently under study (to appear on the arxiv in January, ref will be updated).

%The status of NA64 Dark Sectors coverage sensitivities is shown in figure \ref{fig:NA64_status}.
\begin{figure}[!htb]
\centering
\includegraphics[width=\columnwidth]{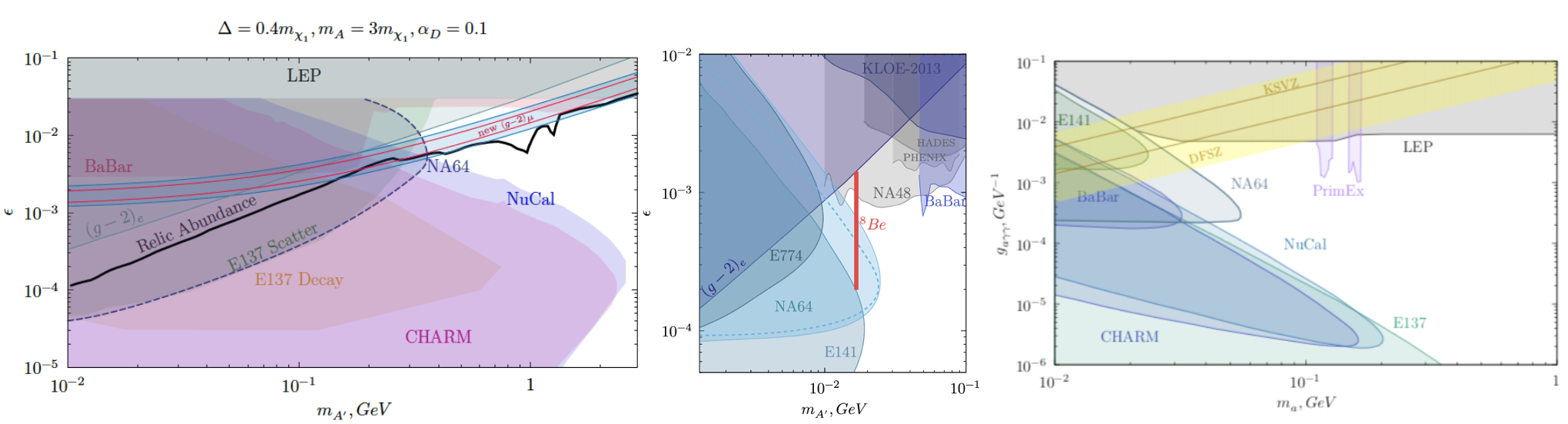}

\caption{Current status of the NA64 experiment 90\% C.L. exclusion limits on semi-visible $A'$ decays  \cite{NA64:2021acr} (Left),  $A'(X)$ visible decays (center) \cite{Banerjee:2019hmi} and NA64 coverage in ALPs searches \cite{Gninenko:2320630} (Right).}
\label{fig:NA64_all}
\end{figure}

\subsubsection{Current status and prospects of NA64} 
The very dense NA64 program restarted after LS2 in August 2021 with the installation of the setup in the new permanent experimental area in the H4 SPS beamline. The setup was ready for the 6 weeks beam-time in 2021 and it is currently taking data during the 10 weeks allocated for the 2022 run. The goal is to continue our DS exploration until LS3 and collect around $5\cdot10^{12}$ EOT in order to probe the parameter space for light DM models suggested by the observed relic density. Depending on the results of PADME, in 2024 the upgraded visible setup could be installed to probe the full parameter space of the hypothetical X17 boson which could provide an explanation of the so called beryllium anomaly.

\begin{figure}[!htb]
\centering
\includegraphics[width=0.3\columnwidth]{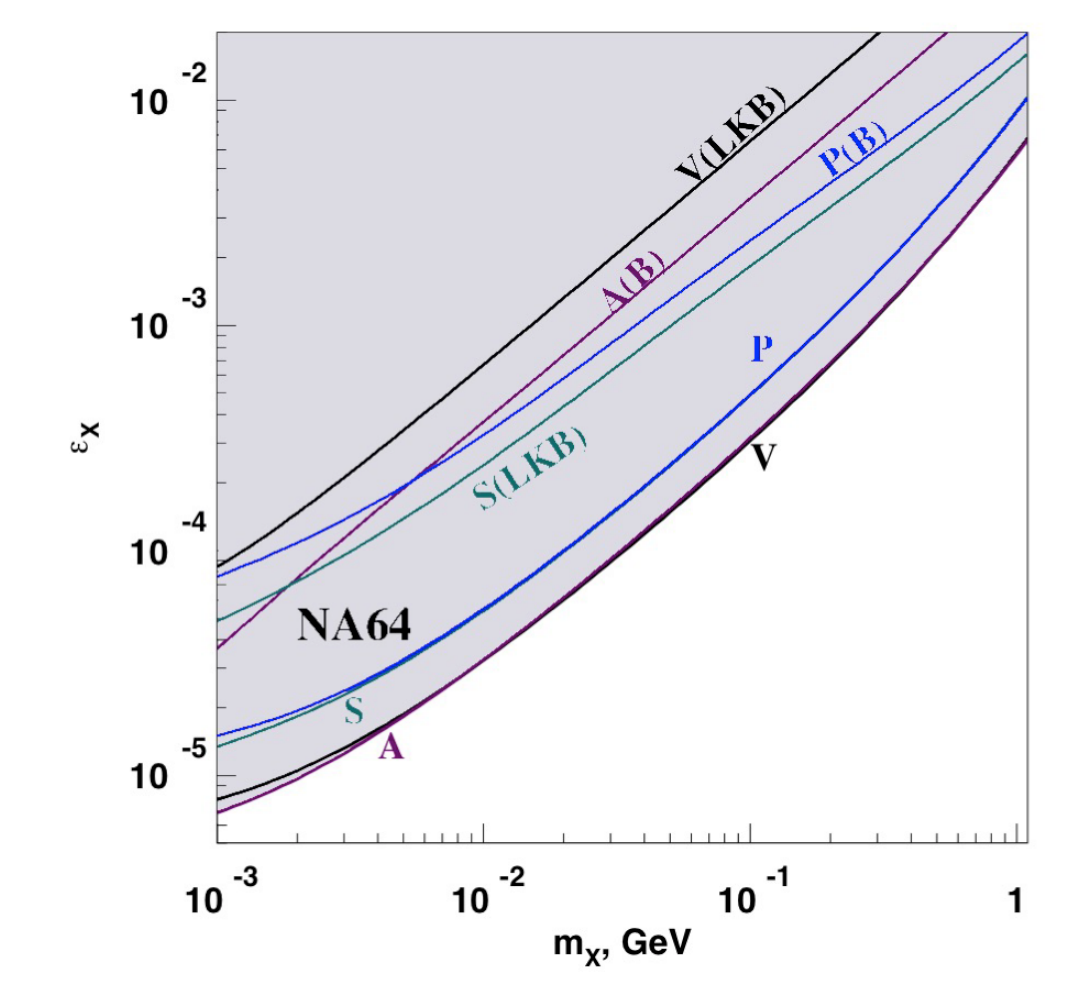}
\includegraphics[width=0.33\columnwidth]{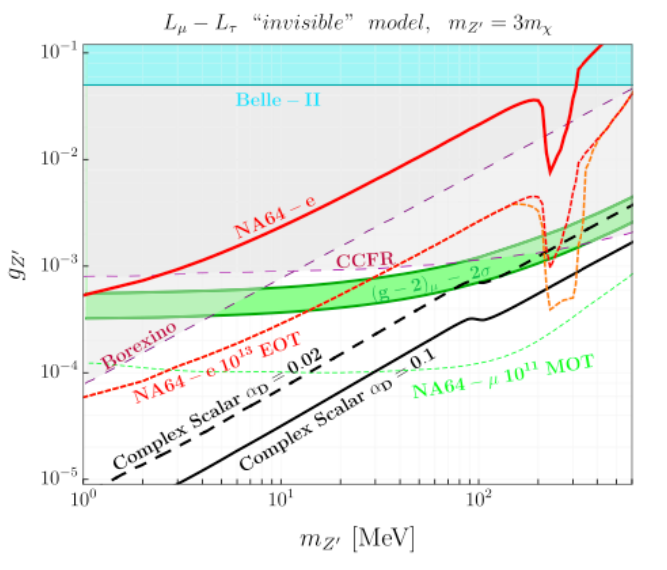}
\includegraphics[width=0.28\columnwidth]{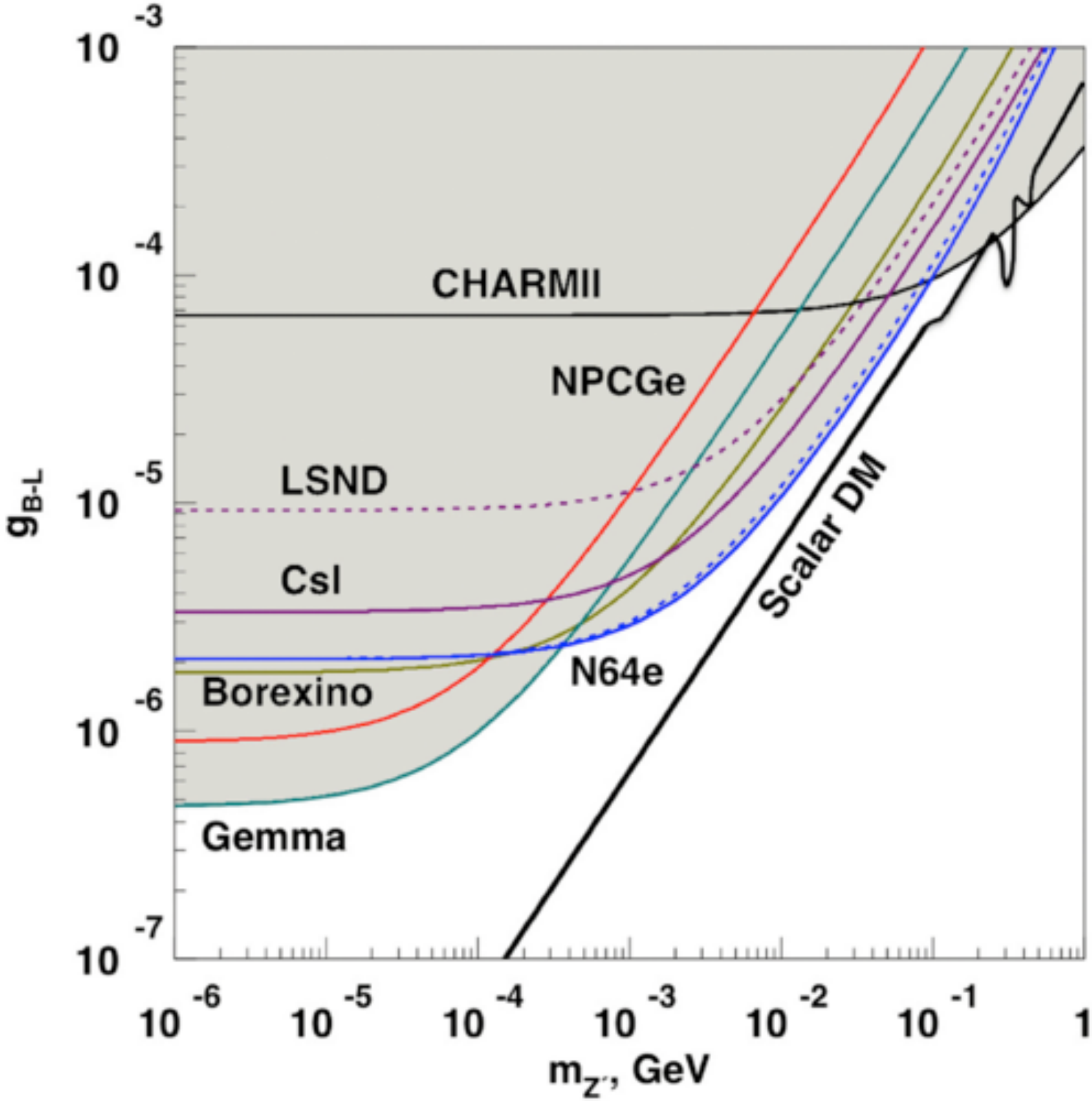}

\caption{Left: NA64 limits for a generic X boson, for the scalar (S), vector (V), pseudo-scalar (P), and axial vector (A) cases \cite{NA64:2021xzo}. Middle: NA64e exclusion limit for the  L$_\mu$-L$_\tau$ (red line) obtained with the 2016-2018 statistics for models where $Z'$ can decay to DM particles \cite{NA64:2022rme}. The red (orange) dashed curves represent the sensitivity projections for a future high-statistics NA64e run with an electron (positron) beam, for a total a of $10^{13}$ EOT, while the green dashed curve is the sensitivity projection of NA64$_\mu$. Right: NA64 exclusion limits for a new B-L Z' boson \cite{NA64:2022yly}.
}
\label{fig:NA64_2022}
\end{figure}

Combining the 2021 data with the one collected before LS2 (total statistic of $3.2\times 10^{11}$ electrons on target), we carried out for the first time using the missing-energy technique a search for a new $Z'$ gauge boson associated with (un)broken B-L symmetry in the keV-GeV mass range \cite{NA64:2022yly}. No signal events were found, thus,  new constraints on the $Z'$-e coupling strength, which, for the mass range $0.3<m_{Z'}<100$ MeV are more stringent compared to those obtained from the neutrino-electron scattering data (see Fig. \ref{fig:NA64_2022}. The data also indicate that NA64 is background free at a level of $1\times 10^{12}$.
Another possibility which is currently under investigation is based on the existence of a light $Z'$ boson resulting from gauging the difference of the lepton number between the muon and tau flavour. This hypothetical boson can couple via QED vertex corrections to the electron and its existence could explain both the muon g-2 anomaly and the DM relic composition. Moreover, this $Z'$ can be produced again through the dark Bremsstrahlung process, $e^- N \rightarrow e^- N Z'$, but also via the resonant annihilation with secondary positrons from the shower. With the 2016-2018 statistics, NA64 was able to probe in this scenario the region suggested by the $(g-2)_{\mu}$ anomaly up to $m_{Z'}\sim 1$ MeV \cite{NA64:2022rme}.  Such a light $Z'$ can additionally couple directly to muons and its search is therefore also one of the physics goals of NA64$_{\mu}$, the NA64 extension using a high energetic muon beam \cite{Gninenko:2014pea, Sieber:2021fue}.

In the beamtime 2022, 2 days were also dedicated to accumulate $\sim 10^{10}$ positrons on target to study the impact of the larger hadron contamination than in electron mode (expected) under study.

% add B-L
%UPDATE
Additionally, after LS2 NA64 started the muon program, NA64$_{\mu}$, with two pilot runs at the M2 beamline using the unique 160 GeV/c muon beam.  This will probe DS in a complementary way to the H4 measurements with electrons and will address the g-2 muon anomaly \cite{Sieber:2021fue}.  The main difference between the experimental technique used in NA64$_{\mu}$ compared to NA64e is that in this case one has to rely solely on momentum reconstruction to measure the missing energy carried away from a possible $Z'$ or $A'$ decay. This makes NA64$_{\mu}$ much more challenging than NA64e where one employs calorimeters for this purpose. During the pilot runs in 2021 and 2022, a total of $4\times10^{10}$ MOT were collected. The analysis is still ongoing but the preliminary results already hint to the fact that an additional spectrometer should be added upstream the ECAL since an accurate determination of the incoming momentum is crucial for the experiment. This will be tested in 2023 and the first physics runs are expected for 2024-2025.  

%The upgrade of the setup required to reach our physics goals is ongoing, our contribution was secured via the SNSF FLARE 186181 grant. 
It is worth mentioning, that for the new-physics process simulations and for detailed comparison between data and Monte Carlo, a new Geant4 based package called DMG4 \cite{Bondi:2021nfp} was developed by NA64 members, which has been well accepted by the community, see, e.g \cite{Eichlersmith:2022bit}.

\subsubsection{Outlook and conclusions} 
NA64 just reached a major milestone of accumulating $\sim 10^{12}$ EOT which allows one to start probing very interesting LDM benchmark models. The analysis is ongoing with the increased statistic we expect to improve the sensitivity of our searches for ALPs, L$_\mu$-L$_\tau$ and B-L $Z'$ bosons, inelastic DM. The plan until LS3 is to accumulate as much as possible electron on target as possible and if the background will be under control also use the positron mode to enhance the sensitivity in the higher $A'$ mass region. 

NA64 also started its program at the M2 beamline providing unique high intensity 160 GeV muons to explore dark sectors weakly coupled to muon.  The results of the pilot runs show that With an optimized setup, one can collect  $>10^{11}$ MOT before LS3 in order to check if an L$_\mu$-L$_\tau$ $Z'$ boson could be an explanation of the g-2 muon anomaly. After LS3 the experiment would then continue data taking to accumulate  $\sim10^{13}$ MOT to explore the $A'$ higher mass region and $\mu \to \tau$ and $\mu \to e$ LFV processes \cite{Gninenko:2022ttd}. 

In the 2022 beamtime, we accumulated $\sim 2\times 10^{9}$ pions on target (1 day) to understand potential of NA64 to explore dark sectors coupled predominantly to quarks using the missing energy technique \cite{Gninenko:2014sxa,Gninenko:2015mea}. This will further investigate and if the feasibility is demonstrated a dedicated search will be performed after LS3.  
To conclude the exploration of the NA64 physics potential has just begun. Our proposed searches with leptonic and hadronic beams provide unique sensitivities highly complementary to similar projects.

%\section*{Acknowledgments}
%I would like to acknowledge the NA64 collaboration and in particular S. Gninenko, L. Molina-Bueno and V. Poliakov. Special thanks also to the past and current ETHZ members, notably E. Depero, H. Sieber, B. Banto-Oberhauser, M. Mongillo, A. Ponten. A big thank you also to the CERN beam department: D. Banerjee, J. Bernhard, N. Charitonidis,  M. Brugger for their continuous support. My gratitude to the CERN SPSC members and NA64 referees who were fundamental for the success of NA64: particularly C. Vallé, G. Lanfranchi, G. Salam, L. Gattignon, M. Wing and G. Schnell.
%My work is supported by ETHZ and the Swiss National Science Foundation (SNSF) under the Grants No. 169133 and 186158. 

%%%%%%%%%%%%%%%%%%%%%%%%%%%%%%%%%%%%%%%%%%%%%%%%%%%%%%%%%
%%%%%%%%%%%%%%%%%%%%%%%%%%%%%%%%%%%%%%%%%%%%%%%%%%%%%%%%%
%%%%%%%%%%%%%%%%%%%%%%%%%%%%%%%%%%%%%%%%%%%%%%%%%%%%%%%%%
%%%%%%%%%%%%%%%%%%%%%%%%%%%%%%%%%%%%%%%%%%%%%%%%%%%%%%%%%

%%\input{bibliography.tex}

\afterpage{\clearpage}
%-------------------------------------------

%-------------------------------------------
\subsection{FIP results at NA62 and prospects with HIKE -- {\it E.~Goudzovski}}
\label{goudzovski}
{\it Author:Evgueni Goudzovski, <Evgueni.Goudzovski@cern.ch>}

%******************
%\subsubsection{Abstract}
%Searches for FIP production in $K^+$ decays at the NA62 experiment at CERN are reviewed. Long-term prospects for searches for FIP production in $K^+$ and $K_L$ decays within the CERN kaon decay experimental programme are discussed.

\subsubsection{Introduction}
Kaon decays represent sensitive probes of light hidden sectors, thanks to the small total kaon decay width and the availability of large datasets. The possible FIP search strategies in kaon decays are reviewed in Ref.~\cite{Goudzovski:2022vbt}, and the following are found to be most promising. Searches for the $K\to\pi X_{\rm inv}$ decay by extension of the $K\to\pi\nu\bar\nu$ measurements, where $X_{\rm inv}$ is an invisible particle, are a unique probe of the dark-scalar and ALP parameter space. Searches for heavy neutral lepton ($N$) production in $K^+\to\ell^+N$ decays are soon expected to provide sensitivity to seesaw neutrino mass models with ${\cal O}(100~{\rm MeV})$ sterile neutrinos~\cite{Abdullahi:2022jlv}. Searches for resonances in the $K\to\pi\ell^+\ell^-$ and $K\to\pi\gamma\gamma$ decay spectra are complementary to searches at beam-dump experiments for a significant ALP mass range. 
Searches for a leptonic force mediator ($X$) in $K^+\to\mu^+\nu X$ decays can probe a region of parameter space that could provide an explanation for the muon $g-2$ anomaly~\cite{Krnjaic:2019rsv}.

\subsubsection{Kaon experiments at CERN}
The NA62 experiment at CERN~\cite{NA62:2017rwk} aims to measure the $K^+\to\pi^+\nu\bar\nu$ decay rate to a 10\% precision, and pursues a wide rare $K^+$ decay programme. The experiment is served by the P42/K12 beamline from the SPS. Slowly extracted 400~GeV protons, delivered in 4.8-second spills at a nominal intensity of $3.3\times 10^{12}$ protons per pulse (ppp), impinge on a beryllium target to obtain a monochromatic (75~GeV/$c$) mixed $K^+$, $\pi^+$ and $p$ beam. The setup includes trackers and Cherenkov detectors for the incoming kaons and the final-state particles, calorimeters, and a hermetic photon veto system. The key detectors deliver a timimg precision better than 100~ps, thereby providing clean event reconstruction and manageable random veto effects at the nominal beam kaon rate of 45~MHz. The NA62 Run~1 dataset collected in 2016--2018, corresponding to $2.2\times 10^{18}$ protons on target (pot) and $6\times 10^{12}$ useful $K^+$ decays in the 60~m long fiducial region, has led to the first observation of the ultra-rare $K^+\to\pi^+\nu\bar\nu$ decay~\cite{NA62:2021zjw}.
%The dataset has also been used to establish the analysis methods for searches for FIP production in $K^+$ decays, producing world-leading limits on the dark-scalar coupling~\cite{NA62:2021zjw,NA62:2020xlg,NA62:2020pwi} and HNL mixing parameters~\cite{NA62:2020mcv,NA62:2021bji} below the kaon mass.
NA62 Run~2 started in 2021 following detector and trigger upgrades, and is approved until the Long Shutdown~3 (LS3). It should be noted that NA62 collected a dedicated sample of $1.4\times 10^{17}$ pot in beam-dump mode in 2021, and collection of $10^{18}$ pot in beam-dump mode is expected by LS3.

Beyond NA62, a comprehensive long-term kaon decay programme at CERN, High Intensity Kaon Experiments (HIKE), has been proposed~\cite{CortinaGil:2022vli}. The experiments would be based in the NA62 experimental hall, operating at a primary beam intensity of $2\times 10^{13}$~ppp in order to collect $1.2\times 10^{19}$~pot per year of operation. A staged approach is foreseen, including a multi-purpose $K^+$ decay experiment focused on the $K^+\to\pi^+\nu\bar\nu$ measurement (Phase~1), a multi-purpose $K_L$ decay experiment focused on $K_L\to\pi^0\ell^+\ell^-$ measurements (Phase~2), and an experiment dedicated to the $K_L\to\pi^0\nu\bar\nu$ measurement (Phase~3, also known as KLEVER). Collection of $5\times 10^{19}$ pot in the beam-dump mode at an intensity of $4\times 10^{13}$~ppp is also expected during dedicated periods of operation.

Results on FIP production searches in $K^+$ decays obtained from NA62 Run~1 data, along with the future NA62 and HIKE projections, are discussed below.

%%%%%%%%%%%%%%%%

% \boldmath
\subsubsection{FIP production in $K\to\pi X_{\rm inv}$ decays}
\begin{figure}[t]
\begin{center}
\resizebox{0.5\textwidth}{!}{\includegraphics{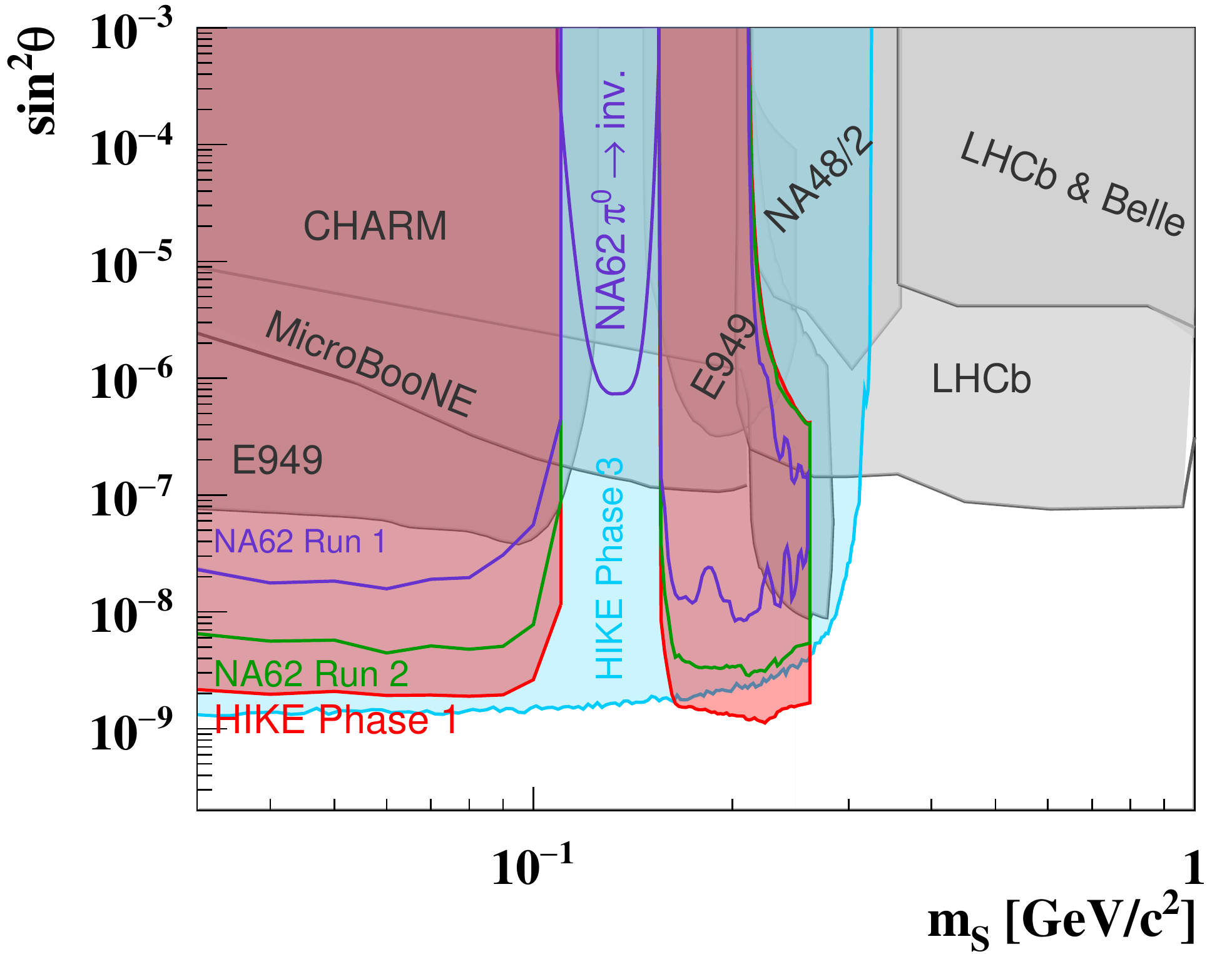}}%
\resizebox{0.5\textwidth}{!}{\includegraphics{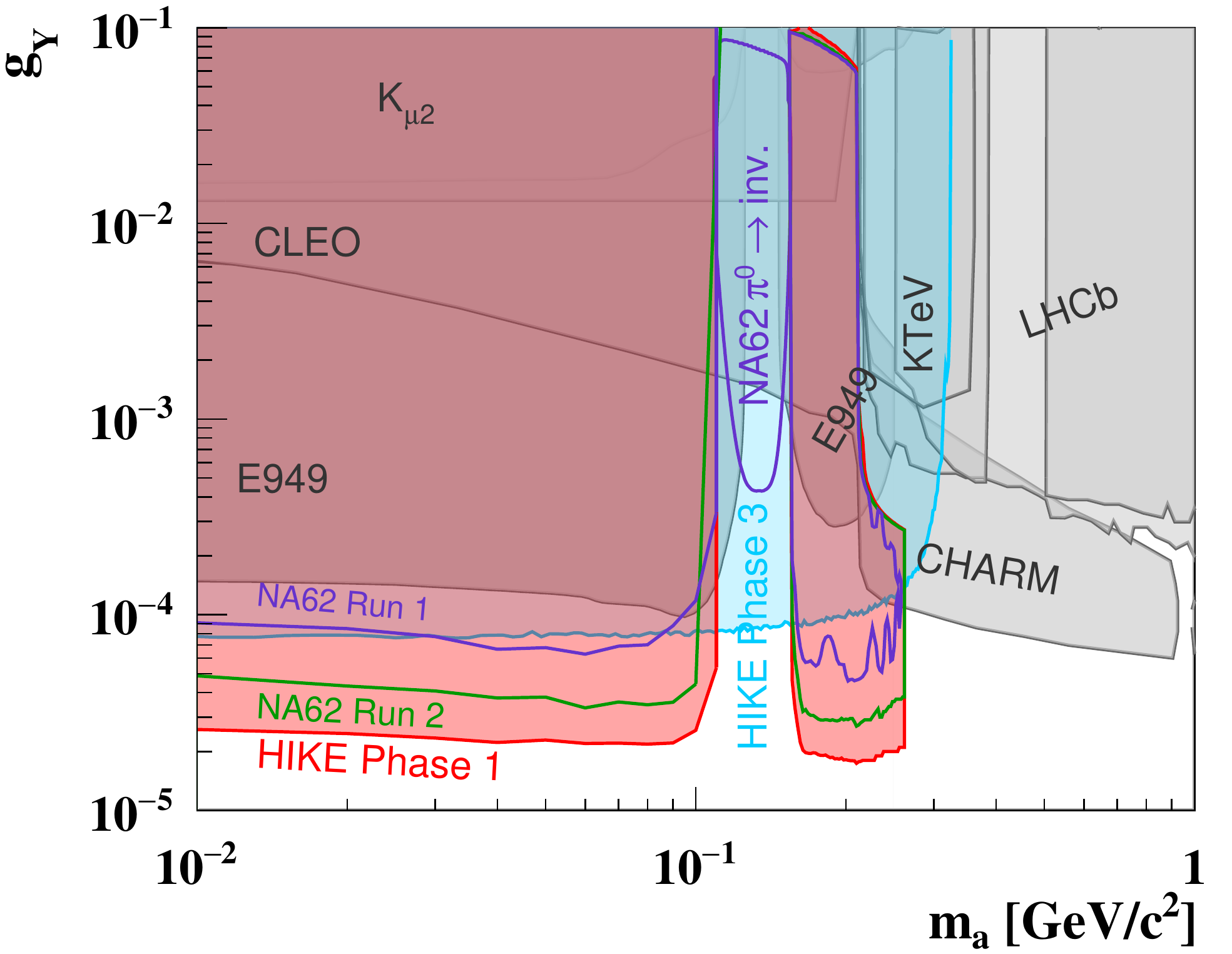}}
% \vspace{-17mm}
\end{center}
\caption{Left: excluded regions at 90\% CL of the $(m_S,\sin^2\theta)$ parameter space for the dark scalar of the BC4 model. Right: excluded regions of the parameter space $(m_{a},g_{Y})$ for the ALP of the BC10 model. Exclusion bounds from NA62 Run~1 data~\cite{NA62:2021zjw,NA62:2020xlg,NA62:2020pwi} and projected sensitivities for NA62 Run~2, HIKE Phase~1 and Phase~3 (KLEVER) are shown by coloured lines and areas; other existing bounds are shown by grey areas.}
\label{fig:KpiX}
\end{figure}

% \unboldmath
The NA62 Run~1 dataset has been used to establish upper limits on ${\mathcal B}(K^+\to\pi^+X_{\rm inv})$, where $X_{\rm inv}$ is an invisible particle, at the level of ${\cal O}(10^{-11})$ in the mass ranges 0--110~MeV/$c^2$ and 154--260~MeV/$c^2$, by extension of the $K^+\to\pi^+\nu\bar\nu$ measurement~\cite{NA62:2021zjw,NA62:2020xlg}.
%The largest background is due to the $K^+\to\pi^+\nu\bar\nu$ decay;
The two-body $K^+\to\pi^+X_{\rm inv}$ decay is characterised by a peak in the reconstructed missing mass ($m^2_{\textrm{miss}}$) distribution, with a width of ${\cal O}(10^{-3})~{\rm GeV}^2/c^4$ determined by the experimental resolution, on top of the continuous $K^+\to\pi^+\nu\bar\nu$ spectrum. The resulting constraints on the dark scalar (scenario BC4 of Ref.~\cite{Beacham:2019nyx}) and the ALP with fermion coupling (scenario BC10 of Ref.~\cite{Beacham:2019nyx}) are shown in Fig.~\ref{fig:KpiX}. Sensitivity projections for the NA62 Run~2 dataset (to be collected by LS3) and the proposed HIKE Phase~1 experiment, obtained from a dedicated study, taking into account the dominant $K^+\to\pi^+\nu\bar\nu$ background, are also shown in Fig.~\ref{fig:KpiX}. Above the di-muon threshold, large values of the coupling parameter are not excluded as the dark scalar and ALP decay length becomes comparable with the length of the experimental setup. 

The NA62 experiment has also established upper limits on ${\cal B}(K^+\to\pi^+X_{\rm inv})$ at the ${\cal O}(10^{-9})$ level in the 110--155~MeV/$c^2$ mass range, in the vicinity of the $\pi^0$ mass, using $10\%$ of the Run~1 minimum bias dataset~\cite{NA62:2020pwi}. The sensitivity is limited by the $K^+\to\pi^+\pi^0$ background with two undetected photons from the $\pi^0\to\gamma\gamma$ decay. To improve the sensitivity, the original $K^+\to\pi^+\nu\bar\nu$ event selection has been optimised for background suppression. The search has resulted in an observation of 12~candidate events, consistent with the expected background of $10^{+22}_{-8}$ events.  The resulting constraints on the FIP phase space are shown in Fig.~\ref{fig:KpiX}.

A search for the $K^+\to\pi^+X_{\rm inv}$ decay in the 260--350~MeV/$c^2$ mass range is in principle possible with the NA62 and HIKE Phase~1 data, however the sensitivity is limited by the $K^+\to\pi^+\pi^+\pi^-$ background with two undetected pions.
%Nevertheless, an ${\cal O}(10^{-9})$ sensitivity to ${\cal B}(K^+\to\pi^+X_{\rm inv})$ can be reasonably expected.
The range of mixing parameter values probed would shrink towards larger $m_X$ values (a trend already seen below 260~MeV/$c^2$) as the FIP decay length and therefore the acceptance of the event selection decreases as a function of $m_X$. Constraining the $K^+\to\pi^+X_{\rm inv}$ transition over the entire $m_X$ range is nevertheless a crucial probe for dark sectors as the invisible decay of the mediators might be the dominant one.

Searches for the $K_L\to\pi^0\nu\bar\nu$ decay naturally provide limits on $\mathcal{B}(K_L\to\pi^0X_{\rm inv})$~\cite{KOTO:2018dsc}. Since no missing mass reconstruction is performed, there is no significant reduction in the acceptance for $m_X$ in the $\pi^0$ mass region. The limits obtained from $K_L$ experiments are therefore complementary to those from the $K^+$ experiments. The expected HIKE Phase~3 sensitivity in the two scenarios considered is reported in Fig.~\ref{fig:KpiX}.

%A sensitivity of $\mathcal{O}(10^{-13})$ to ${\cal B}(K\to\pi X_{\rm inv})$ would cover the entire dark-scalar parameter space below 200~MeV, down to the Big Bang Nucleosynthesis bound~\cite{Goudzovski:2022vbt}. However this sensitivity is out of reach of the present and planned experiments.

%%%%%%%%%%%%%%%%%%%%%%%%%

%%%%%%%%%%%%%%%%%

% \boldmath
\subsubsection{Heavy neutral lepton production in $K^+\to\ell^+N$ decays}
\begin{figure}[t]
\begin{center}
\resizebox{0.6\textwidth}{!}{\includegraphics{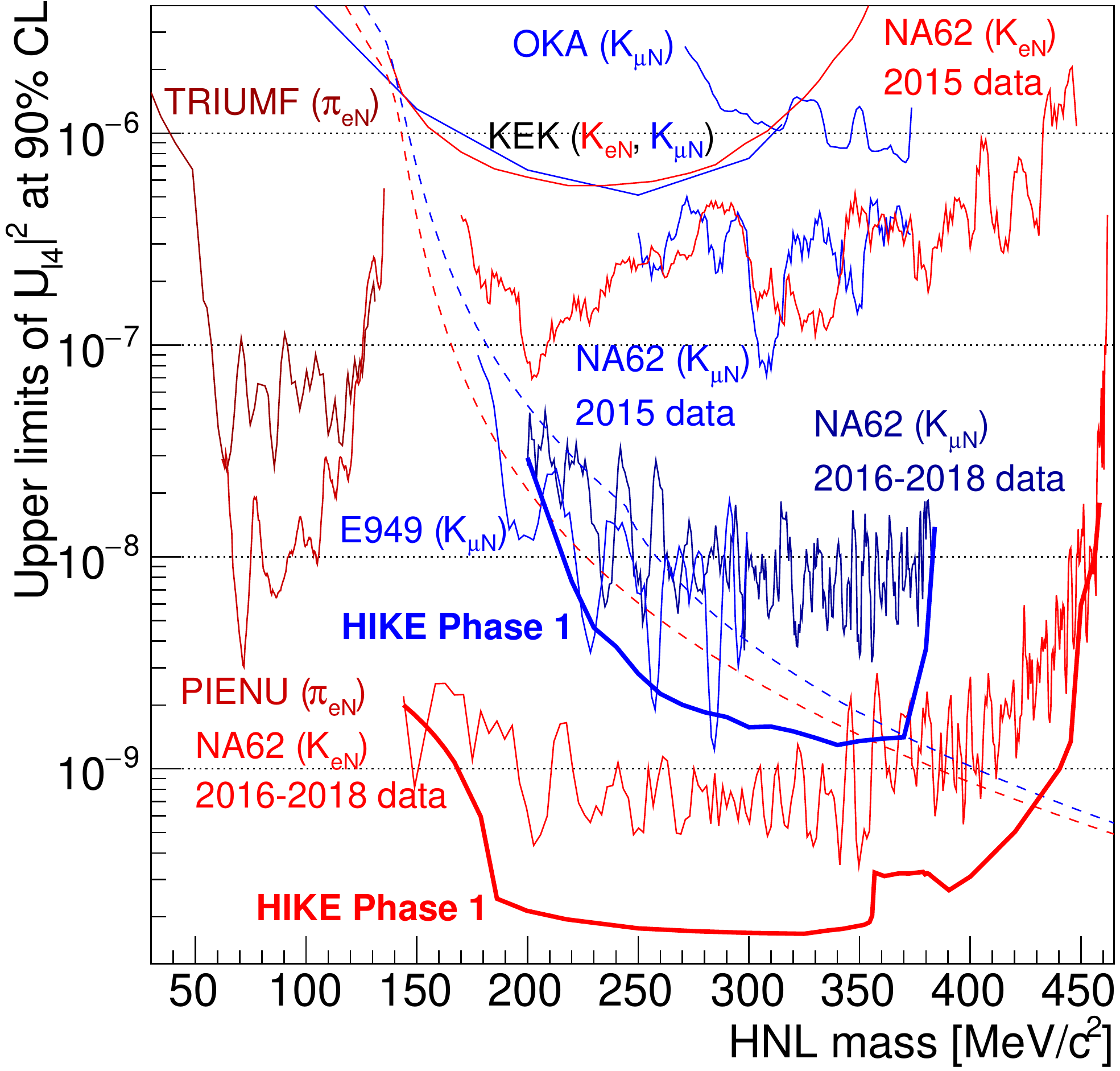}}
% \vspace{-13mm}
\end{center}
\caption{Summary of the upper limits at 90\% CL of the HNL mixing parameters $|U_{e4}|^2$ and $|U_{\mu4}|^2$ obtained from production searches, and HIKE Phase~1 sensitivity projections with $K^+\to\ell^+N$ decays. For a comparison of these limits and projections with those obtained and expected from HNL decay searches, see Ref.~\cite{Abdullahi:2022jlv}.}
\label{fig:HNL}
\end{figure}

% \unboldmath
The NA62 experiment has performed a search for HNL production in $K^+\to e^+N$ and $K^+\to\mu^+N$ decays (scenarios BC6 and BC7 of Ref.~\cite{Beacham:2019nyx}) using the Run~1 dataset, thereby establishing the analysis technique~\cite{NA62:2020mcv,NA62:2021bji}. The $K^+\to e^+N$ analysis is based on the main trigger line designed for the $K^+\to\pi^+\nu\bar\nu$ measurement, while the $K^+\to\mu^+N$ analysis uses a control trigger line downscaled by a factor of 400. Stringent exclusion bounds of the HNL mixing parameters $|U_{\ell 4}|^2$ ($\ell=e;\mu$) in HNL mass ranges of 144--462 (200--384)~MeV/$c^2$ have been established in the electron (muon) case, improving on the state-of-the art (Fig.~\ref{fig:HNL}). Both searches are limited by background. In particular, $K^+\to\mu^+\nu$ decays followed by $\mu^+\to e^+\nu\bar\nu$ decays in flight, and $\pi^+\to e^+\nu$ decays of the pions in the unseparated beam, represent irreducible backgrounds to the $K^+\to e^+N$ process. The peaking nature of the HNL production signals in terms of the reconstructed missing mass allows for data-driven background evaluation, reducing the systematic uncertainties in the background estimates.

HIKE Phase~1 sensitivity projections for HNL production searches in $K^+\to\ell^+N$ decays, obtained assuming NA62-like trigger chains and detector performance, are displayed in Fig.~\ref{fig:HNL}. The sensitivity to the HNL mixing parameters $|U_{\ell4}|^2$ scales approximately as the inverse square root of the integrated kaon flux. HIKE Phase~1 offers world-leading sensitivity to $|U_{e4}|^2$ in the region $m_N>140~{\rm MeV}/c^2$, approaching the seesaw bound and complementing future long-baseline neutrino experiments~\cite{Abdullahi:2022jlv}, and is expected to improve on the state-of-the-art for lower HNL masses via searches for the $K^+\to\pi^0e^+N$~\cite{Tastet:2020tzh} and $\pi^+\to e^+N$ decays (which is not shown in the plot). HIKE Phase~1 also provides competitive sensitivity to $|U_{\mu4}|^2$. The projection for $|U_{\mu4}|^2$ assumes data collection with a highly-downscaled control trigger, and may improve by an order of magnitude if a software trigger based on the streaming readout is employed.

%%%%%%%%%%%%%%%%

%%%%%%%%%%%%%%%%%

\subsubsection{Other NA62 results}
Beyond the searches directly related to the benchmark scenarios~\cite{Beacham:2019nyx}, NA62 Run~1 dataset has been used to establish upper limits on the production of an invisible scalar of vector particle in the $K^+\to\mu^+\nu X$ decay~\cite{NA62:2021bji} and invisible dark photon in the $\pi^0\to\gamma A^\prime$ decay~\cite{NA62:2019meo}. The stringent limits obtained by NA62 on a number of $K^+$ and $\pi^0$ decays violating lepton flavour and lepton number conservation~\cite{NA62:2019eax,NA62:2021zxl,NA62:2022tte,NA62:2022exp} can be interpreted in terms of models involving Majorana neutrinos or flavour violating ALPs and $Z^\prime$ particles. Recent measurements of the $K^+\to\pi^+\mu^+\mu^-$ decay~\cite{NA62:2022qes} and the $K^+\to\pi^+\gamma\gamma$ decay (first presented in September 2022) at a new level of precision with the NA62 Run~1 dataset are to be followed by dedicated peak searches in the decay spectra.

%NA62 has also reported the first results in the beam dump mode [REF].

%\subsubsection{Conclusions}
%The NA62 kaon decay experiment at CERN is pursuing a comprehensive FIP search programme, and has recently reported a series of FIP production searches using the Run~1 dataset, providing new constraints on a number of PBC benchmark scenarios~\cite{Beacham:2019nyx}. The NA62 Run~2 dataset (which is currently being collected) and the proposed HIKE programme at CERN are expected to provide significant improvements in sensitivity to many new physics scenarios involving dark sectors.

%%%%%%%%%%%%%%%%

% \bibliographystyle{JHEP}
% \bibliography{goudzovski_fips22_bib}

\afterpage{\clearpage}
%-------------------------------------------

%-------------------------------------------
\subsection{SHADOWS project at CERN: status and prospects -- {\it A.~Paoloni}}
\label{paoloni}
{\it Author: Alessandro Paoloni, <Alessandro.Paoloni@cern.ch>} 

% \documentclass{paper}
% \usepackage{graphicx}

% \begin{document}
% \section{Shadows project at CERN: status and perspectives.}
SHADOWS \cite{Baldini:2021hfw} \cite{Alviggi:2839484} is a newly proposed proton beam dump experiment in the CERN North Area, exploiting the expected intensity upgrade of the 400 GeV primary proton beam up to $2 \times 10^{13}$ pot on 4.8 s spills.

\begin{figure}[ht]
\begin{center}
\includegraphics[width=\textwidth]{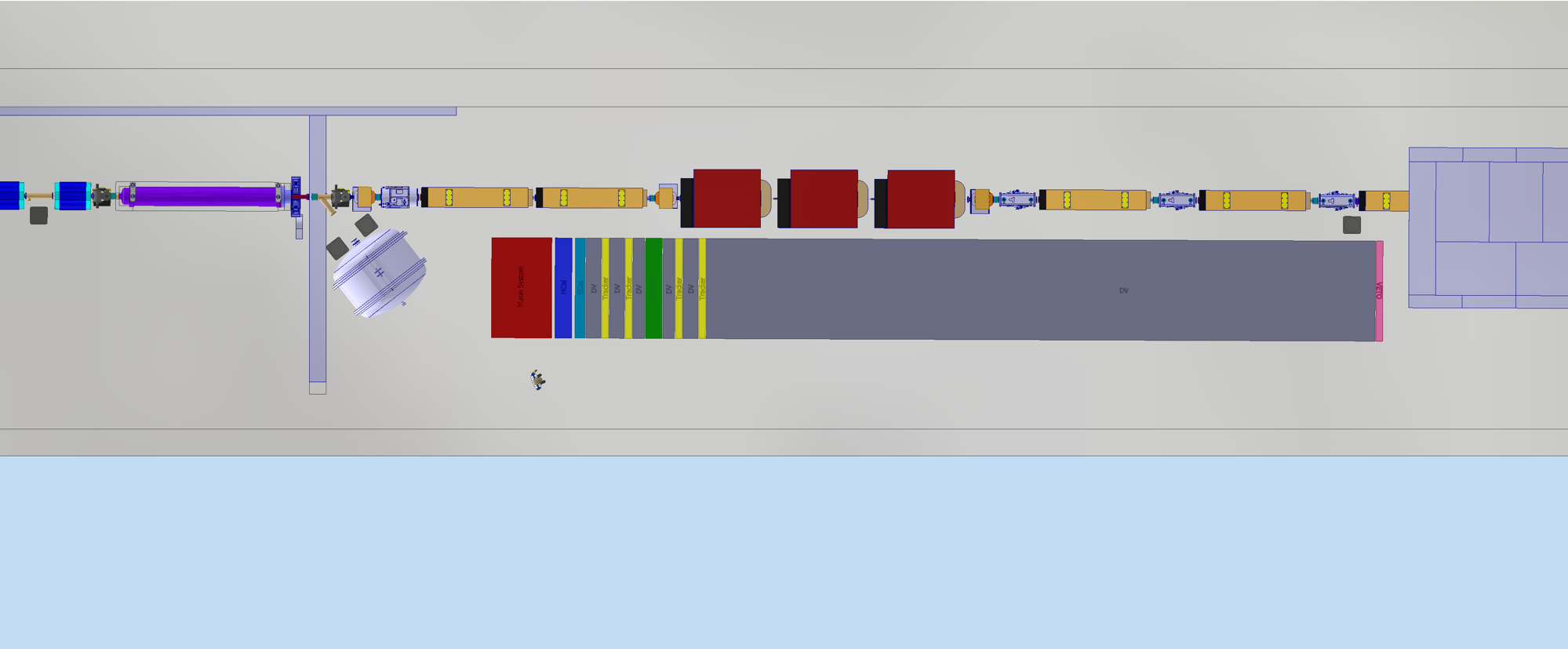}
\end{center}
\caption{\small Top view of the SHADOWS detector in the experimental hall. The beam is entering from right to left. The shielded area containing the TAXes dump is visible at right. SHADOWS detector is sketched at the bottom of the figure. The dimensions of the detector and the beamline elements are to scale.}
\label{fig:shadows_layout}
\end{figure}

The experiment is conceived to detect Feebly Interacting Particles from the decay of Charm and Beauty hadrons produced in the interaction of the protons on target.
As shown in figure \ref{fig:shadows_layout}, it is designed to be located in the TCC8/ECN3 hall, where the K12 beam for NA62 is produced, running concurrently with HIKE \cite{HIKE:2022qra} when the facility is operated in dump mode (i.e. when the primary protons are dumped on TAXes, Target Attenuator for eXperimental areas).

The experiment is conceived as a 20 m long decay volume, large $2.5 \times 2.5$ m$^2$, followed by detectors for the reconstruction of the final state.
It is placed 10 m downstream of the beam dump, to collect sufficient statistics, and 1 m off-axis from the beam to suppress background.
Indeed the principal background source is particle forward production in the primary proton interactions, especially muons, while Charm and Beauty hadrons decay products have a higher p$_T$. 

\begin{figure}[ht]
\begin{center}
\includegraphics[trim={0 12cm 0 0}, clip, width=1.0\textwidth]{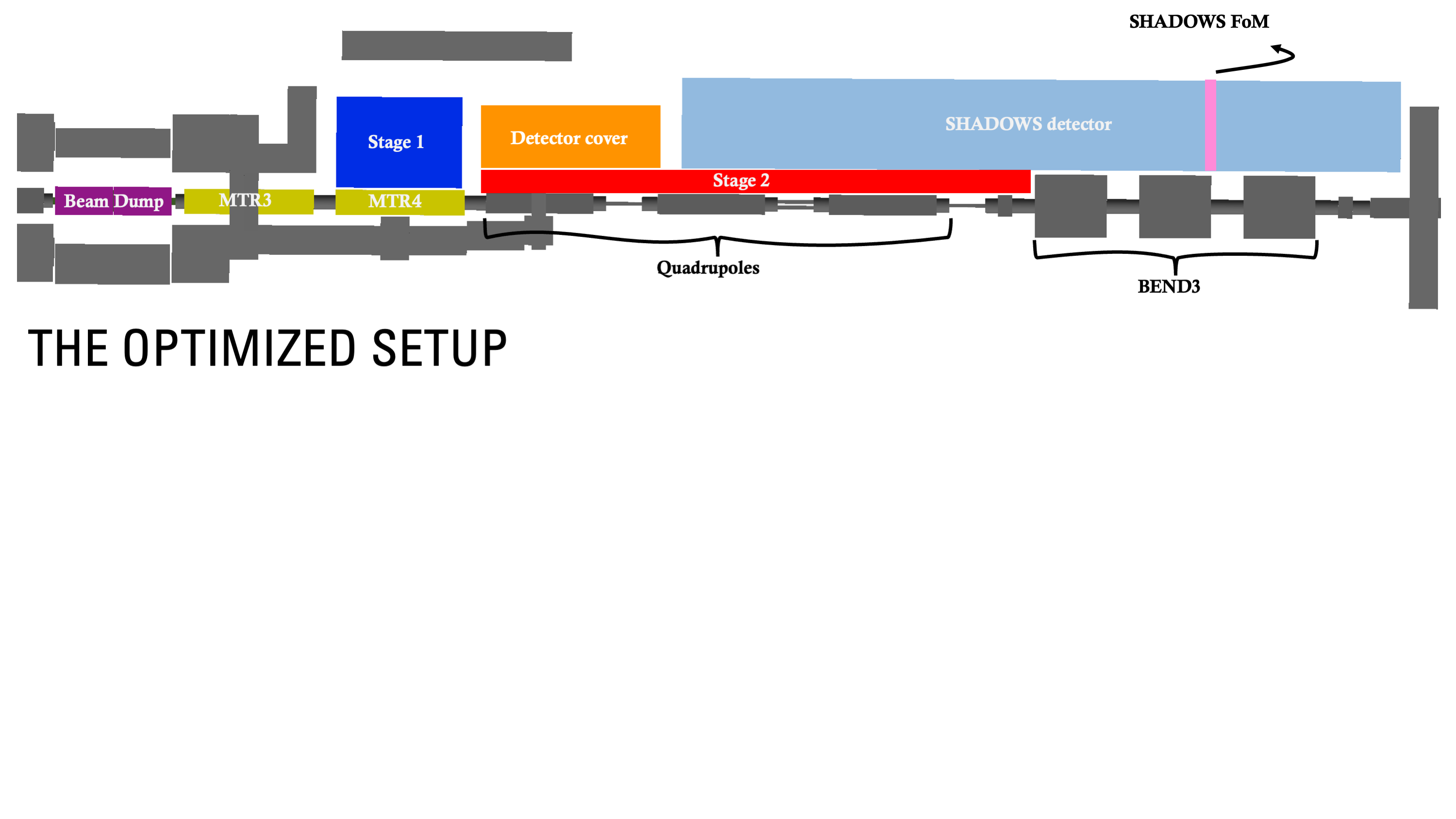}
\end{center}
\caption{\small The optimized SHADOWS setup including the MIB system. The muons created in the beam dump (purple) are pushed off-axis by the MTR3 (B1) and MTR4 (B2) magnet (yellow) to reduce the muon background for HIKE in beam dump mode. The 3.5~m long Stage 1 magnet (blue) is placed right after MTR3, Stage 2 (red) is placed directly after Stage 1 to move the separated muons off-axis right away.}
\label{fig:TheOptimizedSetup}
\end{figure}

In order to further reduce beam background, the CERN BE EA LE group studied a MIB (Magnetized Iron Block) system stopping and deflecting the muons produced in the primary proton beam interactions, that inside SHADOWS acceptance have a momentum spectrum peaked at few GeV and ending at 30 GeV.
A scheme of the MIB system is shown in figure \ref{fig:TheOptimizedSetup}. 
Its effect on the muon flux inside SHADOWS acceptance is reported in table \ref{tab:muon_MIB}.
For an accurate description the reader is referred to \cite{Alviggi:2839484}.

\begin{table}[htbp]
\caption{Muon flux rate in SHADOWS acceptance without and with the MIB for the two muon charges together and for each charge separately.}
\label{tab:muon_MIB}
\vspace{.1cm}
\begin{center}
\begin{small}
\begin{tabular}{llll}
\hline \hline
  & $\mu^+ + \mu^-$ &  $\mu^+$ & $\mu^-$ \\ \hline
  rate without MIB & 100 MHz  & 50 MHz  & 50 MHz \\
  MIB reduction factor & $\sim 120$ & $\sim 110$ & $\sim 150$ \\
  rate with MIB & 0.8 MHz &  0.5 MHz & 0.3 MHz \\
\hline \hline
\end{tabular}
\end{small}
\end{center}
\end{table}

The strong reduction of the muon flux is fundamental in suppressing the most dangerous background, represented by random coincidences of opposite charge muons entering the decay vessel and mimicking a decay vertex in the fiducial volume.
Other sub-dominant backgrounds are neutrino inelastic interactions with the air of the decay volume (mitigated with low pressure) and muon inelastic interactions upstream or inside decay vessel producing a V$_0$ mimicking the signal.
For a more detailed description of the backgrounds the reader is referred to \cite{Alviggi:2839484}.

The SHADOWS detector is designed to reconstruct and identify most of the visible final states of FIPs decays.
Its sketch is shown in figure \ref{fig:shadows_layout}. 
Starting from the beam dump the experiment is composed by:

\begin{itemize}
\item An upstream VETO detector with an efficiency of 99.5\%, a time resolution of few ns and a moderate $\sim$ 1 cm space resolution. A double MicroMegas layer \cite{Alviggi:2018yko} with pad read-out will be realized in order to stand a rate of several kHz/cm$^2$.
\item A 20 m long decay volume with 1 mbar pressure to minimize interactions.
\item A tracker system made of 4 tracking layer followed by drift volumes with a warm dipole magnet in the middle for charge measurement. Two different options, based on straw tubes \cite{NA62:2017rwk} and fiber trackers \cite{LHCb:2014uqj}, are being considered. Their space resolution must ensure a cm resolution on the reconstructed vertex inside the decay volume.
\item A timing layer with a 99.5\% efficiency and a time resolution of 100 ps, necessary for mitigating the background due to random coincidences. A rate capability of 100 Hz/cm$^2$ is also required. Different techniques (scintillating bars \cite{SHiP:2019hap} or pads\cite{Balla:2021njd}, glass RPCs \cite{Williams:2001ms}) are under investigation.
\item An electromagnetic calorimeter of few cm granularity for particle identification with a 10\% energy resolution up to 100 GeV. Also in this case different techniques are being considered (Shashlik \cite{Atoian:2007up}, PbWO$_4$ crystals \cite{CMS:1997ysd} and a Pointing Calorimeter \cite{Chau:2022lcd}).
\item A muon detector consisting of four stations alternated to iron absorbers. A global efficiency of 99\% is required with 150 ps time resolution. The stations will be composed by overlapping scintillating tiles of $15 \times 15$ cm$^2$ area with direct SiPM read-out, whose optimization is described in \cite{Balla:2021njd}.
\end{itemize}

All the considered techniques are sufficiently mature and no heavy R\&D activity is needed.
More details are given in \cite{Alviggi:2839484}.

SHADOWS is expected to start data taking after the Long Shutdown 3 in 2028.
The physics reach of the experiment with $5 \times 10^{19}$ integrated pot is presented in \cite{Alviggi:2839484} and elsewhere in this paper, where exclusion plots for light dark scalars, Axion-like particles at QCD scale and Heavy Neutral Leptons are compared to present experiments and future projects.
SHADOWS has similar or better sensitivities than CODEX-b \cite{Aielli:2022awh} (with 300 fb$^{-1}$) and FASER2 \cite{FASER:2022hcn} (with 3 ab$^{-1}$) for Feebly Interacting Particles produced in the decay of Charm and Beauty hadrons.
It naturally complements HIKE-dump, that is mostly sensitive to very forward objects, and HIKE-K, that is mostly sensitive to FIPs below the K-mass.

\afterpage{\clearpage}
%-------------------------------------------

%-------------------------------------------
\subsection{SHiP project at the Beam Dump Facility: status and prospects -- {\it A.~Golutvin}}
\label{ssec:golutvin}
{\it Author:Andrey Golutvin, <Andrey.Goloutvin@cern.ch>}

%\documentclass[a4paper,11pt]{article}
%\usepackage[a4paper, total={15cm, 23cm}]{geometry}
%\usepackage[colorinlistoftodos]{todonotes}
%\usepackage{placeins}
%\usepackage{siunitx}
%\usepackage{enumerate}
%\usepackage{lscape}
%\usepackage{rotating}
%\usepackage{float}
%\usepackage{caption}
%\usepackage{url}
%\usepackage[switch]{lineno}
%\usepackage[sectionbib]{chapterbib}
%\usepackage{xpatch}

%\newboolean{uprightparticles} 
%\setboolean{uprightparticles}{false} %Set true for upright particle symbols 
%\input{lhcb-symbols-def.tex}

%\pdfoutput=1

%\date{\today}

%\title{BDF/SHiP at the ECN3 high-intensity beam facility} 

%\author{SHiP Collaboration \\
%Presented by Andrey Golutvin (Imperial College London) }

%\newcounter{mybibstartvalue}

%\xpatchcmd{\thebibliography}{%
%  \usecounter{enumiv}%
%}{%
%  \usecounter{enumiv}%
%  \setcounter{enumiv}{\value{mybibstartvalue}}%
%}{}{}

%\begin{document}

%\clearpage\thispagestyle{empty}
%\maketitle
%\thispagestyle{empty}
%###############################
%\begingroup\baselineskip.99\baselineskip
%\endgroup
%#########################

%\setcounter{page}{1}

%\section{Introduction}
%\label{sec:introduction}

BDF/SHiP is a state-of-the-art experimental setup designed to perform a generic and exhaustive search for feebly interacting particles (FIPs) in a region of mass and coupling that is only accessible with a dedicated beam-dump configuration. The proposal aims at taking full advantage of the opportunities offered by the available but unused $4\times10^{19}$ protons at 400\,GeV at the CERN SPS accelerator. The physics programme includes a search for New Physics through both decay and scattering signatures. The details of the development of the project starting from the proposal in 2013, including references to all related documents, can be found in the recently published BDF/SHiP LoI~\cite{Aberle:2839677}. 
%The reference section at the end of this LoI also includes the complete lists of reports submitted on the facility~[15-40] and detector developments~[41-76], physics studies~[77-87], development in theory~[88-107], and PhD theses~[108-127] in the context of BDF/SHiP. 

The Deliberation Document of the 2020 Update of the European Strategy for Particle Physics~[128] (ESPPU2020) recognised the BDF/SHiP proposal as one of the front-runners among the new facilities. To respond to the financial constraints that prevented considering the project for approval in 2020, a continued programme of R\&D was launched as part of the CERN Medium Term Plan 2021-2025 to review the design of the facility, aiming for an alternative implementation in an existing beam facility around the SPS in order to significantly reduce the cost with respect to the initial proposal while preserving the original physics scope and reach of the facility.
This effort has been accompanied by a revision of the detector layout with the goal to fit in existing underground areas. The result of this location and layout optimisation study~[1] identified ECN3 as the most suitable and cost-effective option.
%The decision of the CERN management to review the post-LS3\footnote{Long Shutdown 3, currently scheduled over 2026 - 2027 for the CERN injectors} physics programme in ECN3~[135,136] prompted the BDF/SHiP collaboration to pursue the studies of the facility and of the SHiP detector aimed at ECN3, and to verify, by full simulation, the physics performance. 
The use of ECN3 for BDF/SHiP entails a major cost-saving when compared to the original proposal. 
%This document reports on the results of the studies performed for ECN3 and formally express intent to construct BDF/SHiP in ECN3.}

At the SPS, the optimal experimental conditions for BDF/SHiP are obtained with a proton beam energy of $400$ GeV and slow extraction of the proton spills over one second. At a nominal spill intensity of $4\times 10^{13}$ protons, and $10^6$ spills per year, up to $2\times 10^{20}$ protons on target could be delivered to BDF in about five years of nominal operation. The recent upgrades of the SPS may lead to the capability of delivering more than $4\times10^{13}$ protons per spill in the future, potentially allowing BDF to receive up to $6\times10^{19}$ protons on target per year.

\begin{figure}[hbt]
    \centering
    \includegraphics[width=0.7\linewidth]{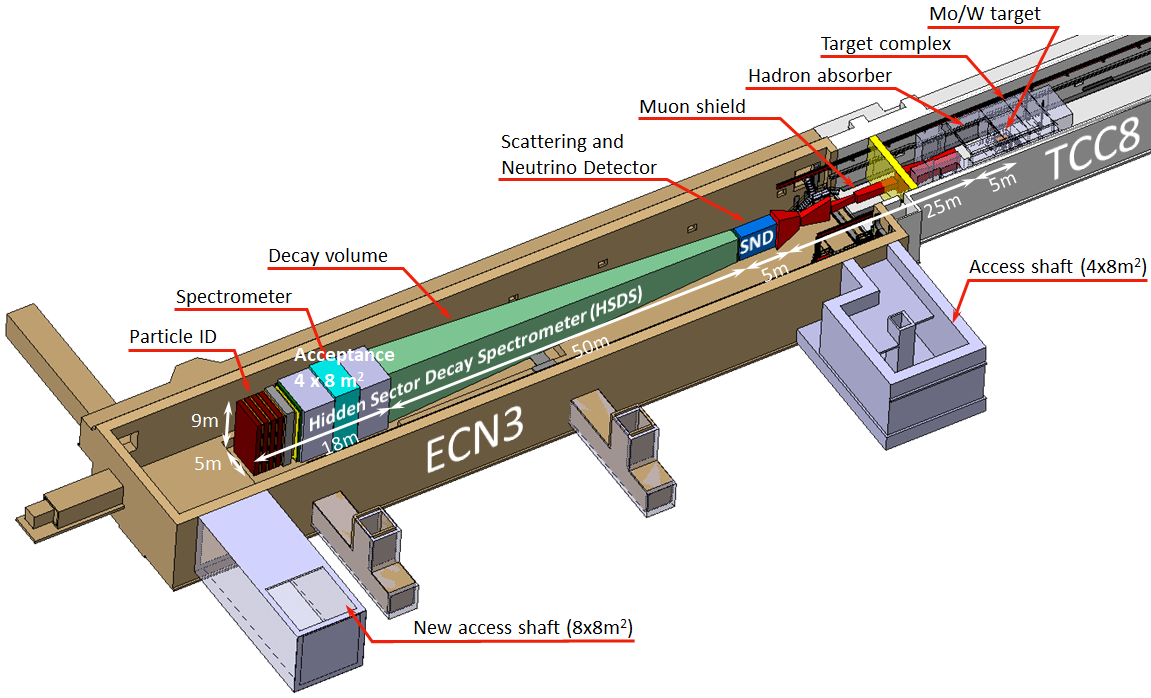}
    \caption{Overview of the BDF/SHiP experimental setup in the SPS ECN3 beam facility.}
    \label{fig:SHiP}
\end{figure}

The layout of BDF/SHiP at the end of TCC8 and throughout ECN3 is shown in Fig.~\ref{fig:SHiP}. 

The setup consists of the high-density proton target, effectively acting as a beam dump and absorber, followed by a hadron absorber and a magnetic muon shield immediately downstream. The shield deflects the muons produced in the beam dump in order to reduce the flux in the detector acceptance to an acceptable level. The hadron absorber is an integral part of the overall shielding that is completely surrounding the target system. 

The SHiP experiment is composed of a dual system of complementary apparatuses. The upstream system, the Scattering and Neutrino Detector (SND), is designed to search for LDM scattering and perform neutrino physics. %It also provides normalisation of the yield for the FIP search. 
The downstream system, the Hidden Sector Decay Search (HSDS) detector is designed to reconstruct the decay vertices of FIPs, measuring invariant mass and providing particle identification of the decay products in a nearly zero background-level environment. 

The suppression of physics background in SHiP is based on a combination of minimising the beam-induced particle rates in the detector acceptance by the highly optimised target design, hadron absorber and muon shield, and rejecting residual backgrounds by applying veto taggers, temporal, spatial and kinematics cuts, and particle identification in the detector. 

%For the optimisation in the CDS phase, a working point was chosen that put emphasis on the muon shield, while relying on only loose selection criteria based on the detectors, hence leaving a high level of redundancy in the overall background suppression. 
The revision of SHiP for the ECN3 experimental hall has started from a simple shortening of the muon shield in the first iteration, shifting the working point for the combined background suppression towards a slightly higher reliance on the detector, in particular on the veto systems. As a result of bringing the experiment closer to the proton beam dump, the detector can be reduced in lateral size while the signal acceptance is preserved for all physics modes, production and scattering/decay kinematics convolved together. Thus far, in order to remain conservative, the adaptation for ECN3 has made no assumptions about using magnet technologies for the muon shield that allow higher field gradients than conventional warm magnets.

The BDF/SHiP physics performance is anchored in a highly efficient background suppression, provided by the design of the target, hadron absorber and the muon shield. The background suppression %in the FIP decay search 
is further guaranteed by the detector systems and, in the case of the search for FIP decays, also by maintaining the decay volume under vacuum. 
All physics sensitivities below are based on acquiring $2\times 10^{20}$ protons on target. 
%which is achievable in five years of nominal operation at ECN3.

The background at ECN3 has been studied with the full simulation and the whole detector and underground geometry implemented. The original SHiP experimental setup was designed with the concept of redundancy built in to the combined performance of the suppression of beam-induced particles rates and the detector.%, as well as in the selection criteria. 
The adaptation to ECN3 and the results of the background studies bear witness of this strategy. The summary of the expected background levels is shown in  Table~\ref{Tab:bkgs}, and does not differ significantly from the original proposal given to the redundancy of the selection criteria.
%would also allow determining the background directly from experimental data and, in case of signal evidence, to perform cross checks that minimise the probability of false positives. 

\begin{table}
\centering
 \begin{tabular}{l r r} 
  Background source & Expected events  \\  
  \hline\hline
  Neutrino DIS & $<0.1$ (fully) / $<0.3$ (partially)\\ 
  Muon DIS (factorisation) & $<\,10^{-4}$ (fully) / $<\,10^{-2}$ (partially)\\
  Muon combinatorial & $ 2.1\times 10^{-3}$\\
  \hline
     \end{tabular} 
     \caption{Expected background in ECN3 in the search for FIP decays at 90\%\,CL for $2\times 10^{20}$ protons on target after applying the pre-selection, the timing, and the UBT and SBT veto. The neutrino- and muon-induced backgrounds are given separately for the set of criteria corresponding to the fully and partially reconstructed signal modes.}
\label{Tab:bkgs}
\end{table} 

\begin{figure}[pt]
\centering
\includegraphics[width=0.8\columnwidth]{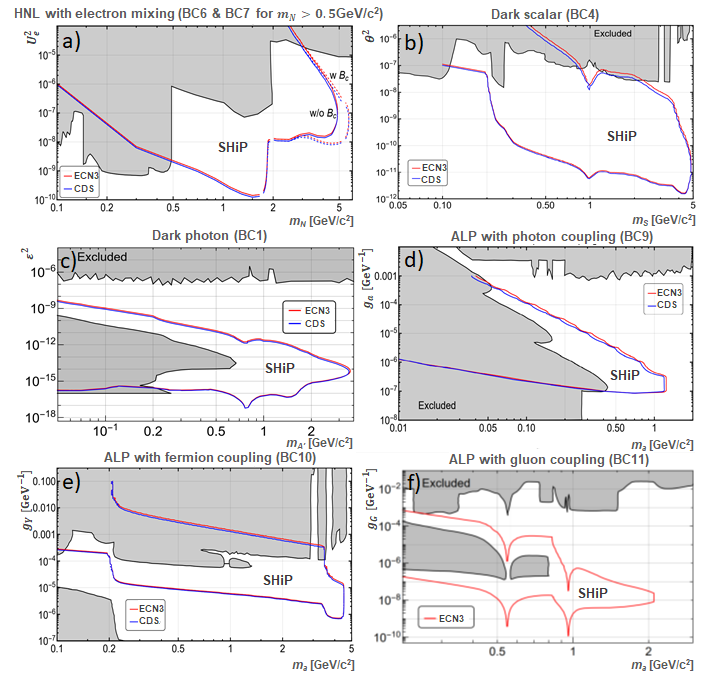}
\caption{a) SHiP's sensitivity in ECN3 compared to CDS for HNLs, b) dark scalars for which only the contribution from $B$ mesons is taken into account, c) dark photons from adding up the three production mechanisms described in the text, and axion-like particles coupled to d) photons, e) fermions, and f) gluons. All plots are based on $2\times 10^{20}$ protons on target and limits correspond to 90\% CL. The benchmark models (BC) used are those in~\cite{Beacham:2019nyx}. Regions shaded in grey are excluded by past and current experiments. %References to the current exclusion curves shown and more details can be found in~\cite{Beacham:2019nyx}.
}
\label{fig:HSsensitivity1}
\end{figure}

The BDF/SHiP sensitivities in ECN3 and in the original proposal are practically identical as shown in Fig.~\ref{fig:HSsensitivity1}. 
In most parts of the sensitive parameter space where at least a handful of events would be measured, BDF/SHiP is able to discriminate between the various models.

In conclusion, the SPS is currently under-exploited and could provide a yield of protons to ECN3 that puts BDF/SHiP in an outstanding position world-wide to make a break-through in the range of FIP masses and couplings that is not accessible to the energy and precision frontier experiments. ECN3 makes it possible to implement BDF at a fraction of the original cost without compromising on the physics scope and the physics reach. SHiP has demonstrated the feasibility to construct a large-scale, versatile detector capable of coping with 4$\times 10^{19}$ protons per year at 400\,GeV/c and ensure an extremely low background environment. With the feasibility of the facility and the detector proven, the BDF WG and the SHiP collaboration are ready to proceed with the TDR studies and commence implementation in CERN's Long Shutdown 3.
 
%\addcontentsline{toc}{section}{References}
%\bibliographystyle{JHEPx} %
%\bibliography{references}%

%\newpage

%\end{document}

\afterpage{\clearpage}
%-------------------------------------------

%-------------------------------------------
\subsection{Search for FIPs at DESY: status and prospects with LUXE -- {\it N.~Trevisani}}
\label{trevisani}
{\it Author: Nicolo Trevisani, <nicolo.trevisani@kit.edu>} 

% \documentclass{article}
% \usepackage[utf8]{inputenc}

% Line numbers
% \usepackage{lineno}
% \linenumbers

% Additional packages
% \usepackage{mathrsfs}
% \usepackage{amssymb}
% \usepackage{amsmath}

% \usepackage{authblk}

% Sub-figures packages
% \usepackage{graphicx}
% \usepackage{caption}
% \usepackage{subcaption}

% \title{Search for Feebly Interacting Particles with LUXE}
% \author{Nicol\`o Trevisani, on behalf of the LUXE Collaboration}
% \affil{KIT, Karlsruhe Institute of Technology, Kaiserstraße 12, 76131 Karlsruhe, Germany}
% \date{November 2022}

% \begin{document}

% \maketitle

% \begin{abstract}
%     The LUXE experiment will produce high-intensity electron-laser interactions to study the QED in the non-perturbative regime. 
%     These interactions have as a secondary product a flux of hard photons with energy up to a few GeV. 
%     The photons are directed onto a physical dump to produce axion-like particles (ALPs), which can decay into pair of photons, detected by an electromagnetic calorimeter.
%      We present an overview of the experimental apparatus and the primary signal production mode. Then, we discuss the elements affecting signal production and acceptance and the corresponding expected results.
%     Assuming that no background reaches the calorimeter, we show that LUXE has the potential to inspect an uninvestigated ALPs phase space, consisting of light ALPs (m$_{a} \sim \mathcal{O}(100)$~MeV) with large couplings (g$_{a \gamma \gamma} \sim 10^{-3} - 10^{-4}$~GeV$^{-1}$).
% \end{abstract}

% \clearpage

\subsubsection{Abstract}
    The LUXE experiment will produce high-intensity electron-laser interactions to study the QED in the non-perturbative regime. 
    These interactions have as a secondary product a flux of hard photons with energy up to a few GeV. 
    The photons are directed onto a physical dump to produce axion-like particles (ALPs), which can decay into pair of photons, detected by an electromagnetic calorimeter.
     We present an overview of the experimental apparatus and the primary signal production mode. Then, we discuss the elements affecting signal production and acceptance and the corresponding expected results.
    Assuming that no background reaches the calorimeter, we show that LUXE has the potential to inspect an uninvestigated ALPs phase space, consisting of light ALPs (m$_{a} \sim \mathcal{O}(100)$~MeV) with large couplings (g$_{a \gamma \gamma} \sim 10^{-3} - 10^{-4}$~GeV$^{-1}$).
    
\subsubsection{Introduction}

\begin{figure}[htbp]
    \centering
    \begin{subfigure}[b]{0.45\textwidth}
        \centering
        \includegraphics[width=\textwidth]{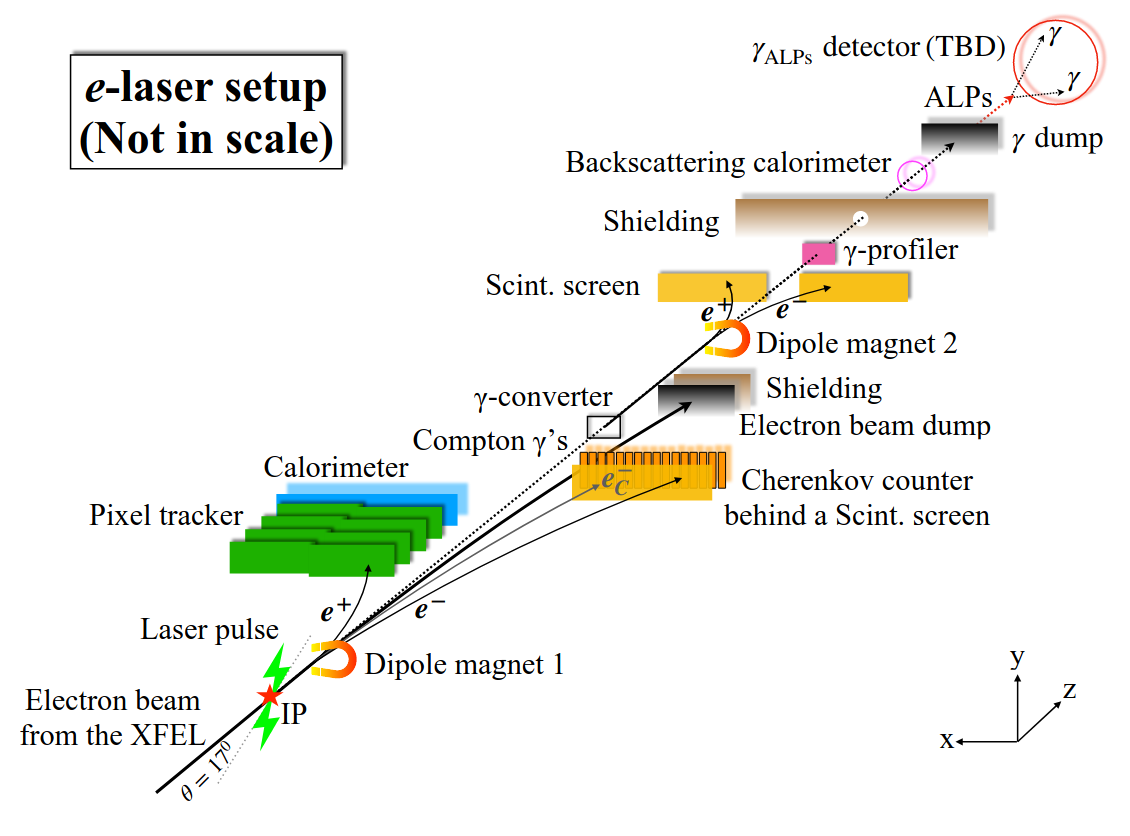}
        \caption{}
        \label{subfig:e_laser}
    \end{subfigure}
    \hspace{0.5cm}    
    \begin{subfigure}[b]{0.45\textwidth}
         \centering
         \includegraphics[width=\textwidth]{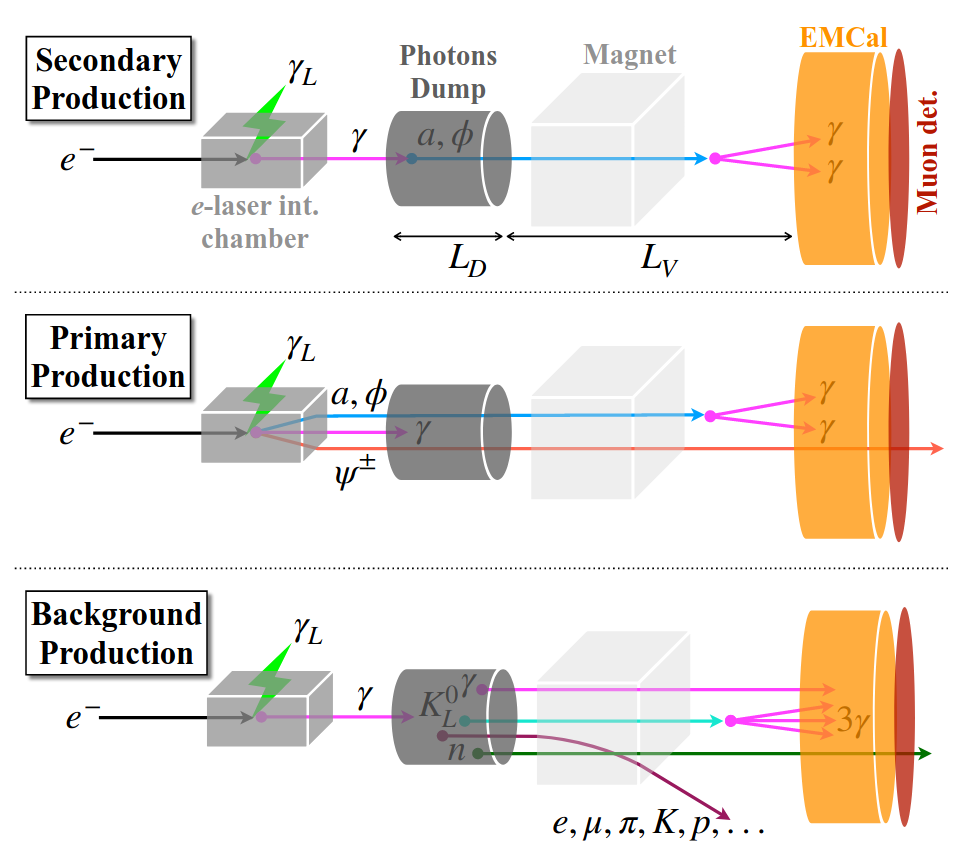}
         \caption{}
         \label{subfig:dump_prod}
    \end{subfigure}
    \caption{e-laser experimental setup (a) and signal and backgrounds productions in LUXE-NPOD (b). The secondary signal production happens with the interaction of the photons in the physical dump material, with ALPs masses up to $\mathcal{O}(1)$~GeV. 
    %and is the main focus of LUXE-NPOD. 
    In the primary production, the ALPs are produced at the electron-laser interaction, with masses limited to $\mathcal{O}(100)$~keV~\cite{NPOD}.}
    \label{fig:apparatus}
\end{figure}

LUXE~\cite{LUXE} (Laser Und XFEL Experiment) is a future experiment 
% that will be 
located at DESY.
In its design, the XFEL accelerator provides a 16.5~GeV electron beam - or \emph{bremsstrahlung} photons obtained using a converter target - that interacts with a powerful laser to obtain high-intensity interactions.
The data-taking is expected to start in 2026 and will be split into two phases: in the first period, a 40~TW laser will be used and then replaced by a 350~TW one in the second stage.
The physics goal of the LUXE experiment is twofold:
\begin{itemize}
    \item Compare the predictions of the non-perturbative QED in the Schwinger limit with the experimental results. 
    The Schwinger limit is a scale above which the electromagnetic field becomes non-linear:
    \begin{equation}
        \mathscr{E} = \frac{m_e^2 c^3}{e \hslash} = 1.32 \cdot 10^{18}~\mathrm{V/m}.
    \end{equation}
    \item Thanks to the specific laser properties, the electrons from the XFEL see the laser as a solid dump, producing \emph{bremsstrahlung} photons with energies up to 15~GeV. These photons travel until a \emph{physical} dump located at the end of the experimental apparatus, where they can produce axion-like particles (ALPs) interacting with the material of the dump~\cite{NPOD}. The ALPs then decay into pairs of photons. If they do so in the decay volume after the dump, the photons are detected by an electromagnetic calorimeter.
\end{itemize}
Figure~\ref{subfig:e_laser} shows the LUXE setup in the e-laser interaction mode, while in Figure~\ref{subfig:dump_prod}, the signal and backgrounds production in the \emph{new physics searches with an optical dump} (NPOD) layout are sketched.

\subsubsection{Current studies}

\begin{figure}[htbp]
    \centering
    \begin{subfigure}[b]{0.51\textwidth}
        \centering
        \includegraphics[width=\textwidth]{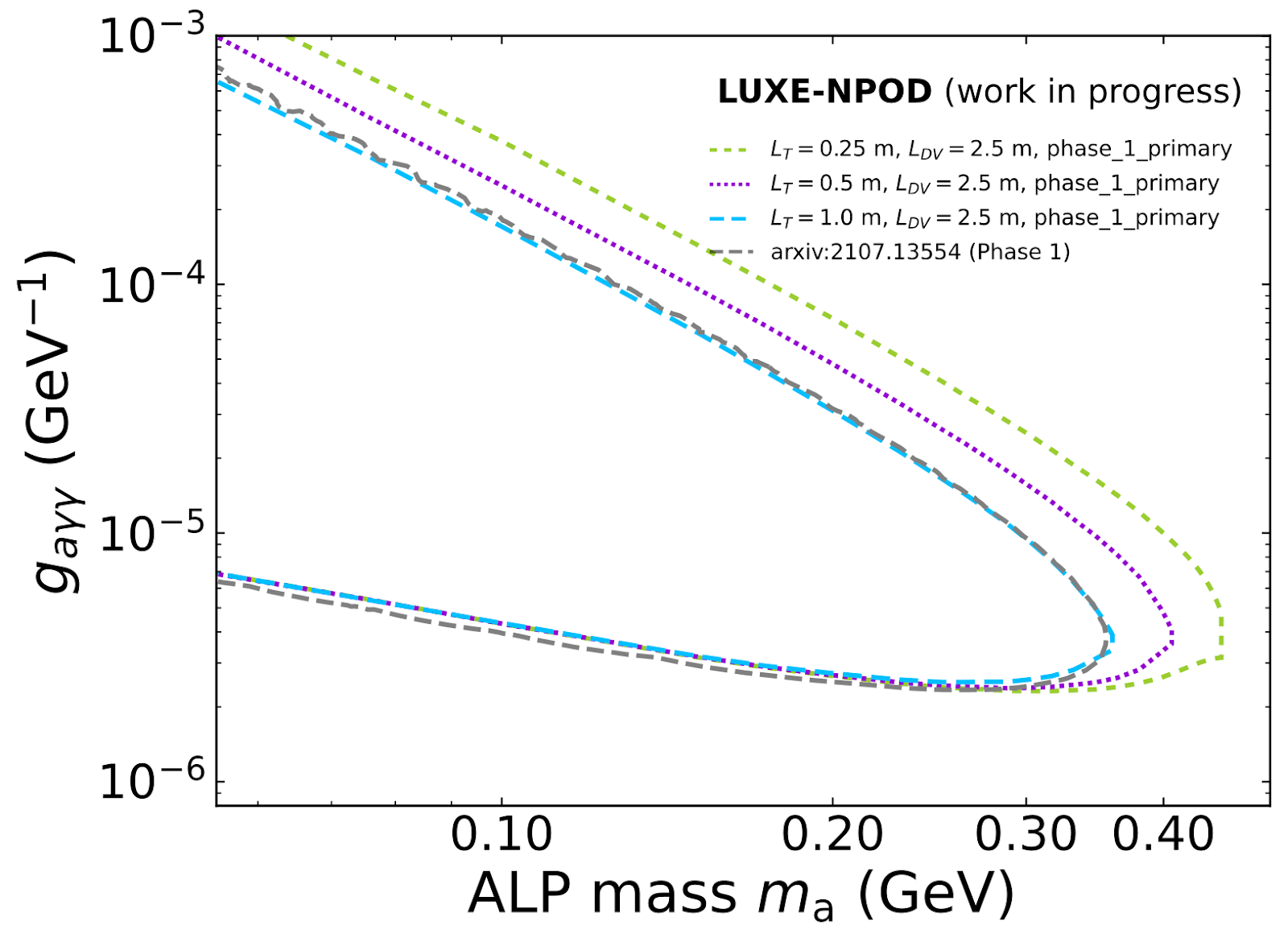}
        \caption{}
        \label{subfig:sensitivity_dump}
    \end{subfigure}
    \hspace{0.5cm}    
    \begin{subfigure}[b]{0.39\textwidth}
         \centering
         \includegraphics[width=\textwidth]{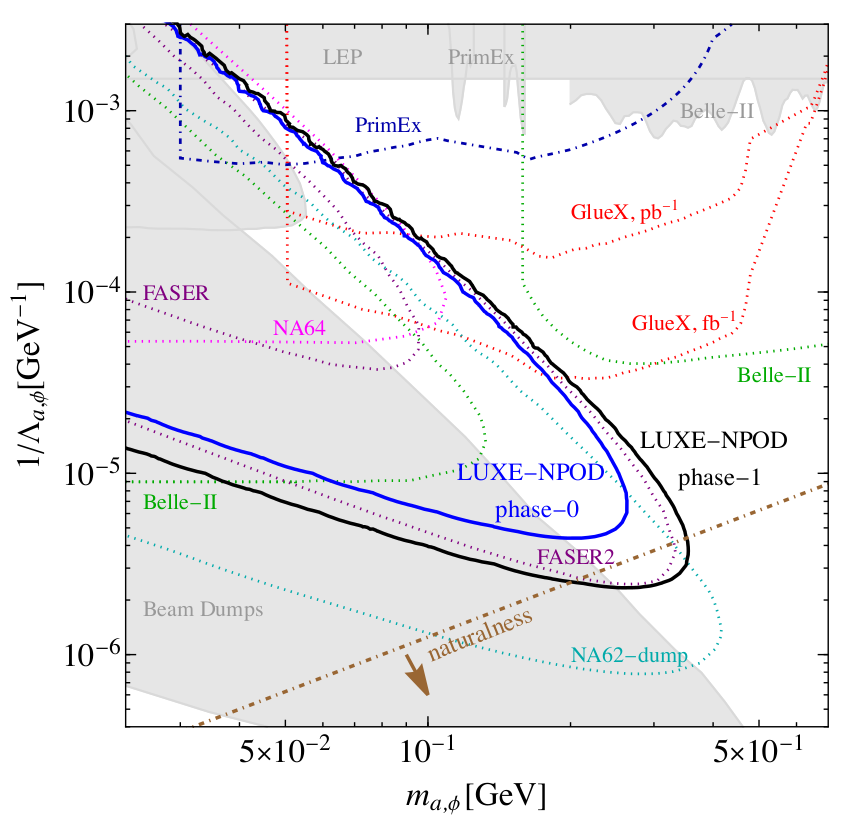}
         \caption{}
         \label{subfig:sensitivity_npod}
    \end{subfigure}
    \caption{Expected sensitivity to ALP mass and coupling to photons as a function of the physical dump length (a) and comparison of results with other experiments. In grey, the phase space excluded by previous experiments~\cite{NPOD} (b).}
    \label{fig:results}
\end{figure}

In the LUXE-NPOD apparatus, three elements can 
%be adjusted to optimize 
affect the sensitivity of the experiment: the physical dump (length and material), the decay volume length, and the detector.
%
% \begin{itemize}
%     \item The physical dump (length and material)
%     \item The decay volume length
%     \item The detector
% \end{itemize}
%
The ALPs production in the physical dump happens via the Primakoff mechanism, i.e., the photon interaction with the magnetic field of the dump material nuclei. More in general, the expected number of ALPs detected at the calorimeter is approximately:
\begin{equation}
    N_{a} \approx \mathcal{L}_{eff} \int dE_{\gamma} \frac{dN_{\gamma}}{dE_{\gamma}} \sigma_{a}(E_{\gamma})\left(\exp^{-\frac{L_{D}}{L_{a}}} - \exp^{-\frac{L_{V} + L_{D}}{L_{a}}} \right) \mathcal{A},
\end{equation}
where $\mathcal{L}_{eff} = N_{e} N_{p} \dfrac{9\rho_{N} X_{0}}{7 A_{N} m_{0}}$ is the \emph{effective luminosity}, and depends on:
\begin{itemize}
    \item the number of electrons per bunch, $N_{e} = 1.5 \cdot 10^9$;
    \item the number of bunches per year, $N_{p} \sim 10^7$;
    \item the properties of the material of the dump, currently tungsten: the density $\rho_N$, the mass number $A_N$, and the radiation length $X_0$;
    \item the nucleon mass, $m_0 = 1.66 \cdot 10^{-24}$~g ($\sim$ 930~MeV).
\end{itemize}
$L_{V}$ is the decay volume length, $L_{D}$ is the dump length, and $\mathcal{A}$ is the detector acceptance.
%
%the number of electrons per bunch ($N_{e} = 1.5 \cdot 10^9$), number of bunches per year ($N_{p} \sim 10^7$), and properties of the material of the dump, currently tungsten; $L_{V}$ is the decay volume length; $L_{D}$ is the dump length; $\mathcal{A}$ is the detector acceptance.
%
Starting from published results~\cite{NPOD} that considered $L_{V} = 2.5$~m and $L_{D} = 1$~m, a methodical scan on these two parameters has been carried out
to estimate 
the signal acceptance and the sensitivity to the ALPs mass and coupling to photons (g$_{a \gamma \gamma}$).
Assuming zero background events reach the calorimeter, shorter dumps enhance the sensitivity to more massive ALPs (Figure~\ref{subfig:sensitivity_dump}), while longer decay volumes allow probing smaller couplings.
These are, however, already excluded by previous beam dump experiments (Figure~\ref{subfig:sensitivity_npod}).
The assumption of zero background has been tested on simulation~\cite{NPOD} but would benefit from updated studies and a larger simulated dataset.
In particular, a non-negligible number of neutrons may reach the calorimeter. The detector design thus requires high signal efficiency but also background suppression capability:
\begin{itemize}
    \item signal efficiency: photons shower separation ($\sim 2$~cm);
    \item precise reconstruction of ALP invariant mass: good resolution of photons direction and energy;
    \item background suppression:
    \begin{itemize}
        \item vertex resolution (rejection of non-resonant photons pairs);
        \item shower shape determination (neutrons identification);
        \item good time resolution ($< 1$~ns) (neutrons rejection).
    \end{itemize}
\end{itemize}
All these requirements are fulfilled by a \emph{tracking calorimeter} (e.g., high-granularity calorimeter, HGCal) followed by an existing crystal or Spaghetti-Calorimeter (SpaCal) to get the total photon energy. Cheaper or existing options are also considered, such as a scintillator/absorber detector or the H1 lead/scintillating-fiber calorimeter~\cite{Glazov:2010zza}.

\subsubsection{Conclusions}

The LUXE-NPOD project has the potential to inspect an uninvestigated ALPs phase space, in particular, light ALPs (m$_{a} \sim \mathcal{O}(100)$~MeV) with large couplings (g$_{a \gamma \gamma} \sim 10^{-3} - 10^{-4}$~GeV$^{-1}$).
The results presented assume zero background.
Preliminary studies~\cite{NPOD} need to be reviewed and expanded with more accurate simulations to overcome the lack of simulated events in the relevant phase space.
An experimental setup to achieve high signal efficiency and suppress the residual background is essential.
For that, the geometry of the apparatus, the detector technology, and the analysis techniques will be optimized.

% \clearpage

% \bibliographystyle{plain} % {unsrt}
% \bibliography{main}

% \end{document}

\afterpage{\clearpage}
%-------------------------------------------

%-------------------------------------------
\subsection{Search for light DM and mediators at LNF: PADME -- {\it M.~Raggi}}
\label{raggi}
{\it Author: Mauro Raggi, <mauro.raggi@roma1.infn.it>} 

PADME is a fixed-target missing-mass experiment, at the Laboratori Nazionali di Frascati near Rome, searching for the dark photon and other dark sector candidates using a beam of positrons with energy $<$500 MeV . PAMDE has already collected a first set of physics-grade data over the last few years. We discuss the first physics results of PADME on Run II data set, the most precise measurement of the total cross-section of electron-positron annihilation into photons below 1 GeV. We also discuss expected sensitivity of the PADME Run III data set, collected at $\sqrt{s}\sim$17 MeV, to the production of on-shell X17. PADME is likely capable of providing independent confirmation of the excesses observed in the ATOMKI spectroscopic measurements with Beryllium and Helium.

\subsubsection{Introduction}
\begin{figure}[th]
\centering
%\begin{center}
\includegraphics[scale=0.65]{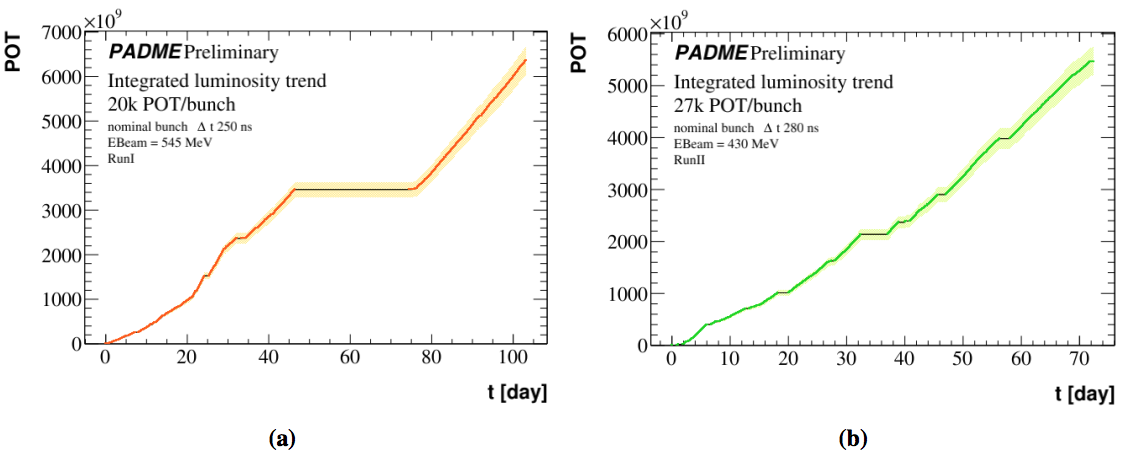}
\caption{PADME collected luminosity during Run I (a) and Run II (b).}
\label{fig:padmedata}% Give a unique label
\end{figure}

The Positron Annihilation into Dark Matter Experiment (PADME) was designed to search for invisible dark photon decays produced via the annihilation process $e^+e^-\to\gamma A^{\prime} $\cite{Raggi:2014zpa}.
The experiment relies on the missing mass technique, thus a peak in the $M^2$ miss distribution will be the signature of the existence of an invisibly decaying dark sector candidate. In more recent times, the experiment is exploiting the charged particle detectors to extend
its sensitivity to dark sector candidate with visible decays.
The PADME detector main components, Fig.\ref{fig:padmegg} (a), are: 
a $100 \mu m$ thick diamond target, a magnetic dipole bending the beam
outside the experimental acceptance, a high resolution electromagnetic calorimeter (ECAL), a small angle electromagnetic calorimeter (SAC) capable to sustain a high rate, and an in vacuum charged particle veto system for positrons(PVeto) and electrons(EVeto) detection.
Details on the detector and its performance can be found in \cite{Albicocco:2022ukx}.
The PADME first data taking period Run I, took place from November 2018 till March 2019 with a beam energy of 490 MeV. Fig. \ref{fig:padmedata} (a) shows Run I collected luminosity.
During Run I data analysis an important source of background coming from the beam halo interactions with the beam line materials was identified \cite{PADME:2022ysa}. 
%Monte Carlo studies allowed to design an optimized beam configuration.   
The Run II started in September and ended in December 2020 with a beam energy of 432.5 MeV. Data were collected with an improved beam line configuration, designed following the outcome of a detailed Monte Carlo simulation, producing a much lower background in the detectors.
Fig. \ref{fig:padmedata} (b) shows the integrated luminosity collected in Run II $\sim6\times{10^{12}}$ Positrons on Target (POT). 
After the angular anomaly observed in the Internal Pair Creation on $^8Be$\cite{kr16} nuclei has been confirmed by measurement on $^4He$\cite{kr21}, PADME has developed a strategy to constrain the existence of the postulated new particle X17. A dedicated data taking campaign called Run III has been planned for the end of 2022. 

\subsubsection{Measurement of the cross-section of electron-positron annihilation into photons}
\begin{figure}[th]
\centering
%\begin{center}
\includegraphics[scale=0.35]{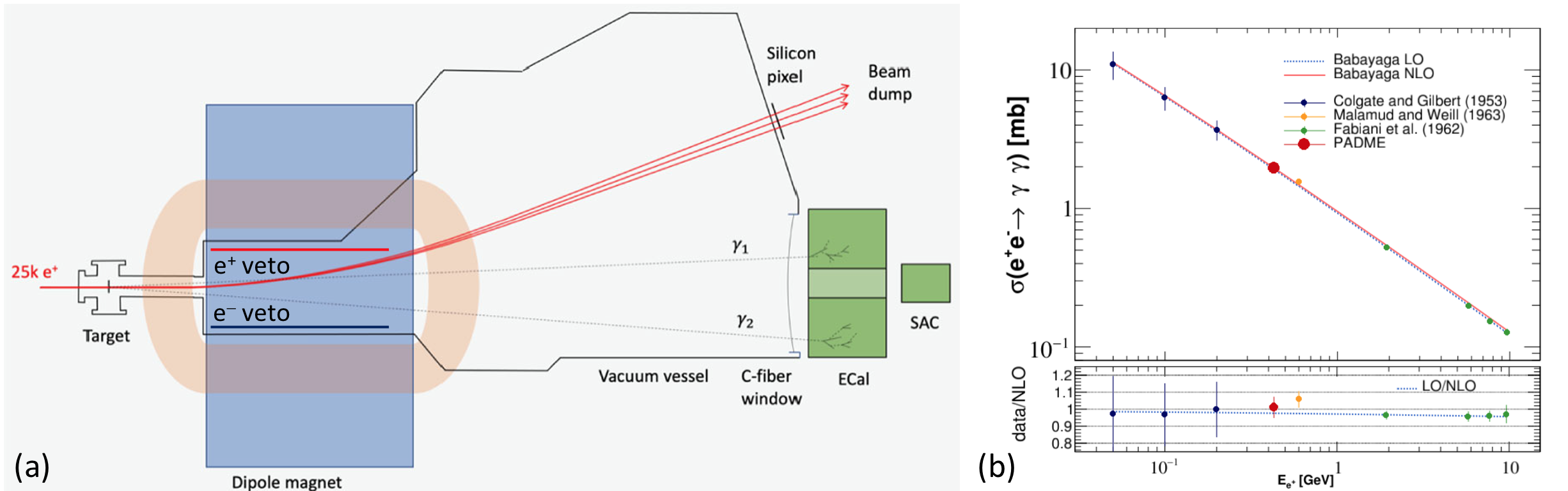}
\caption{(a) The PADME detector layout. The gamma gamma detection technique is also highlighted. (b) Status of $e^+e^- \to \gamma\gamma$ cross-section measurements.}
\label{fig:padmegg}% Give a unique label
\end{figure}

The $e^+e^-\to\gamma \gamma$ is a very important SM ``candle'' process for the PADME experiment. It allows monitoring the calorimeter energy scale, the  collected number of positrons on target, the  
detector geometry and the beam characteristics.
The experimental signature consists in two in time photons detected by the electromagnetic calorimeter with a total energy and energy to angle correlation consistent with the process kinematic, see Fig. \ref{fig:padmegg} (a). The PADME cross-section measurement is obtained on a subset of the PADME Run II data set of $~4\times 10^{11}$ POT collected with a positron energy of 432.5 MeV, corresponding to $\sqrt{s}$=21 MeV.

To obtain a pure data driven result, a tag-and-probe technique is applied to measure the
efficiency due to detector defects, asymmetries in the acceptance of the apparatus, reconstruction efficiency and pile up caused by beam background. The PADME final result is\cite{PADME:2022tqr}:
\begin{equation}
 \sigma_{e^+e^-\to \gamma\gamma} = (1.977 \pm 0.018_{stat} \pm 0.119_{syst}) \hspace{0.5cm} \mathrm{mb}    
\end{equation}
in good agreement with NLO QED predictions $(1.948 \pm0.002$) mb obtained using Babayaga generator\cite{Balossini:2008xr}.
Fig. \ref{fig:padmegg} b) the cross-section value obtained by PADME, red dot, is compared to previous results.
The systematic error is dominated by the determination of the number of collisions which accounts for target effective thickness and target calibration, leaving open the possibility of further improvements.   
The PADME measurement exploits the reconstruction of the photon pair for the first time at beam energies below 1 GeV. 
Previous results \cite{Malamud}\cite{Colgate} were based on positron disappearance rate, which might be affected by beyond-the-Standard-Model contributions leading to undetected final states. The PADME measurement, on the contrary, by explicitly detecting the photon pairs would suffer from this limitation.

\subsubsection{Searching for X17 at PADME}

An experimental group of the ATOMKI Institute for Nuclear Research
in Debrecen recently reported an anomaly in
the angular correlation spectra of e+e- pairs emitted in $^8$Be and $^4$He Internal Pair Creation nuclear transitions\cite{kr16}\cite{kr21}. The excesses observed in both spectra at different opening angles 
are compatible with the production and subsequent decay into
an electron positron pair of a new boson, that was named after the
fitted value of its mass X17. Attempts of constraining the X17 
parameter space, both in the vector and scalar hypothesis, have provided 
significant but not yet final results. 
The most stringent bound on a new particle decaying into electron positron 
pairs arise from the process $\pi^0\to \gamma X17 \to \gamma e^+e^-$, which has been investigated by the NA48/2 collaboration\cite{NA482:2015wmo}.
As pointed out for the first time by J. Feng et al., in \cite{Feng:2016jff} the X17 needs to have small coupling to quarks to evade the NA48/2 constraints.
The dominant lower bound to the X17 coupling to electrons comes from the NA64 experiment 
\cite{NA64:2019auh}\cite{NA64:2021aiq}, while the visible search from the 
KLOE experiment\cite{Anastasi:2015qla} provides an upper limit at the per-mil level.
Exploring the remaining allowed parameter space for X17 is of utmost importance to provide a definite answer on the possible new physics origin of the anomaly. Experimentally, this is very hard to achieve: beam dump experiments, like NA64, are limited by the very short lifetime of X17 while the ``bump hunt'' techniques are limited by the overwhelming SM background. In this context a new highly efficient production mechanism for X17 is needed to improve the signal to background ratio. As pointed out for the first time in \cite{Nardi:2018cxi}, the resonant production has a rate which exceed by more than 3 orders of magnitude the radiative production rates commonly used in ``bump hunt'' experiments. The drawback of such a production mechanism is the tiny width of the dark sector particles resonances, $<eV$, which drastically reduces the accessible mass region. In the peculiar case of X17, in which the mass is known to the level of few hundreds KeV, the resonant production can play a very significant role as pointed out by Darmé et al. \cite{Darme:2022zfw}.  
Profiting by the unique possibility of having positrons in the energy range 270-300 MeV, the PADME experiment is the ideal positions to perform a resonance search in the X17 mass region. 

Phenomenological studies have been performed to establish the PADME Run III
sensitivity based on two different scenarios for the total number of collected PoT per point and beam energy resolution:
- Conservative: 12 points summing up to $2\times10^{11}$ total PoT, with a 0.5\% beam energy spread, in the energy range [265, 297] MeV\\
- Aggressive: 14 points summing up to $4\times10^{11}$ total PoT, with  0.25\% beam spread, in the narrower energy range [273, 291] MeV.

\begin{figure}[th]
\centering
%\begin{center}
\includegraphics[scale=0.35]{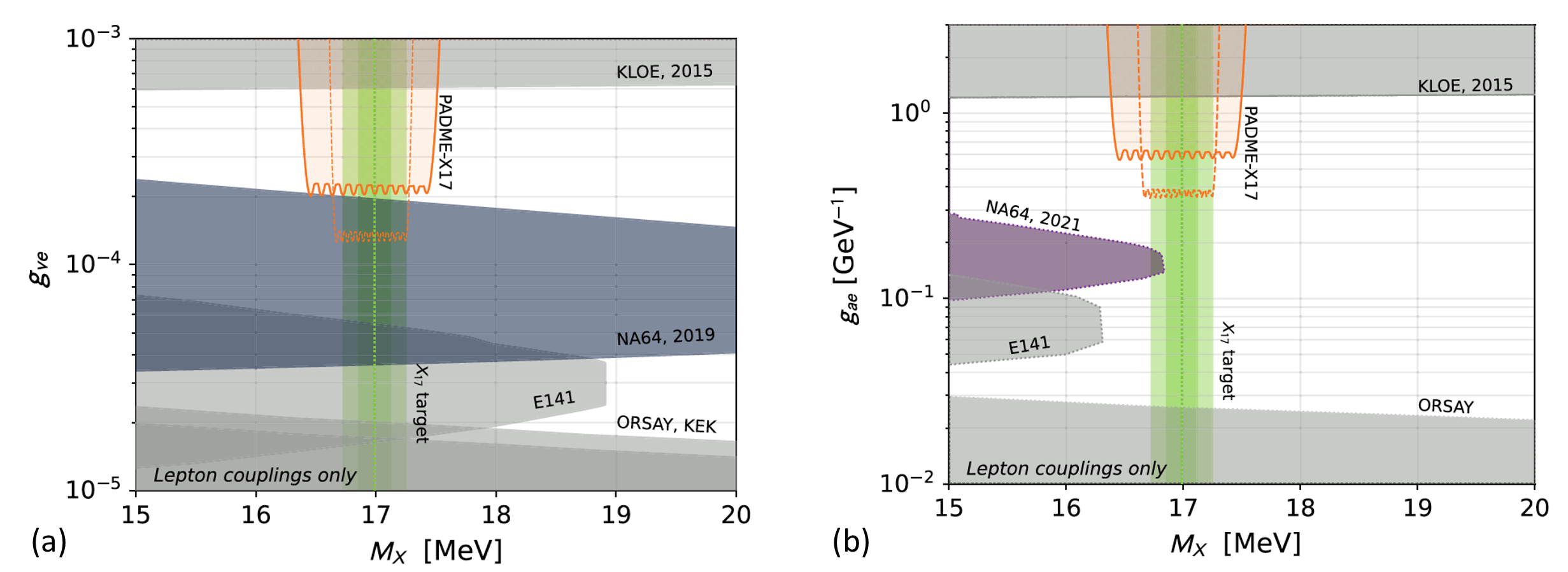}
\caption{(a) PADME expected sensitivity to vector X17. (b) PADME expected sensitivity to pseudo-scalar X17 \cite{Darme:2022zfw}.}
\label{fig:padmex17}% Give a unique label
\end{figure}

Fig \ref{fig:padmex17} shows the projected $90 \%$ C.L. sensitivity of PADME Run-III on the  $g_{ve}$ and $g_{ae}$ couplings of a X17 boson for the conservative (solid orange line) and aggressive (dashed orange line) setups.

The strategy adopted during PADME Run III is to perform a scan in steps of 
$\sim$ 0.7 MeV collecting $1\times10^{10}$ POT per point in the positron 
energy region starting at 265 MeV up to 295 MeV. The scan will consist of more than 40 points for a total of $>4\times10^{11}$ PoT. 
Using a beam intensity of $\sim 1.2\times 10^5$ PoT/s PADME is able to collect a point of the scan with $1\times10^{10}$ PoT per each day of data taking. 
The data set collected should allow PADME to graze the optimistic scenario limit shown in Fig. \ref{fig:padmex17} over a mass range of ~1 MeV around the X17 mass.
The higher granularity of the scan, with respect to the study in \cite{Darme:2022zfw}, should in addition allow to obtain a completely flat limit over the entire mass range explored. Thanks to the very good performance of the LNF LINAC the entire scan has been completed during PADME Run III from October to December 2022. Data are currently being analysed.

\subsubsection{Conclusion}

The PADME experiment was designed to search for invisible decaying dark sector candidates. During Run I and Run II PADME collected $5\times10^{12}$ PoT at energies of 490 MeV and 430 MeV respectively. The experiment has recently obtained the best measurement of the two photons annihilation cross-section at $\sqrt{s}=$21 MeV:
 $\sigma_{e^+e^-\to \gamma\gamma} = (1.977 \pm 0.018_{stat} \pm 0.119_{syst}) \hspace{0.3cm} \mathrm{mb}$\\
 During last part of 2022 PADME has collected during Run III a data set of $4\times 10^{11}$ PoT the $\sqrt{s}\sim$ 17 MeV.  
The experiment is currently analyzing Run III data set to test the existence of the X17 boson exploiting the resonant production technique.

\afterpage{\clearpage}
%-------------------------------------------
%-------------------------------------------
\subsection{Status and prospects for light DM and mediator searches at Belle II -- {\it C.~Hearty}}
\label{ssec:hearty}
{\it Author: Christopher Hearty,  <hearty@physics.ubc.ca>} \\

% \documentclass[11pt, oneside]{article}   	% use "amsart" instead of "article" for AMSLaTeX format
% \usepackage{geometry}                		% See geometry.pdf to learn the layout options. There are lots.
% \geometry{letterpaper}                   		% ... or a4paper or a5paper or ... 
% %%\geometry{landscape}                		% Activate for for rotated page geometry
% %%\usepackage[parfill]{parskip}    		% Activate to begin paragraphs with an empty line rather than an indent
% \usepackage{graphicx}				% Use pdf, png, jpg, or eps§ with pdflatex; use eps in DVI mode
% 								% TeX will automatically convert eps --> pdf in pdflatex		
% \usepackage{amssymb}

% \title{Status and prospects for dark sector searches at Belle II}
% \author{Christopher_Hearty, on behalf of the Belle II collaboration\\
% Department of Physics and Astronomy\\
% University of British Columbia\\
% Vancouver, BC, Canada\\
% {~~and}\\
% Institute of Particle Physics\\
% Victoria, BC, Canada
% }
% %%\date{}							% Activate to display a given date or no date

% \begin{document}
% \maketitle
% %%\section{}
% %%\paragraph{}

\subsubsection{Belle II and SuperKEKB}
Belle~II is an almost complete upgrade of the original Belle experiment, with better performance and higher rate capabilities \cite{Belle-II:2010dht}. It is located at the SuperKEKB $e^+e^-$ collider at the KEK laboratory in Tsukuba, Japan. In addition to dark sector searches, its physics program includes rare and forbidden $B$ meson decays, lepton flavour and CP asymmetries, and charm and tau physics \cite{Belle-II:2022cgf}. Belle~II has recorded 428\,fb$^{-1}$ of data since March 2019. It is currently in long shutdown 1 (July 2022 -- September 2023) to install a new two-layer pixel vertex detector. 

SuperKEKB is the world's highest instantaneous luminosity collider, reaching a peak of $4.7 \times 10^{34}$\,cm$^{-2}$s$^{-1}$. To reach the target of $6 \times 10^{35}$\,cm$^{-2}$s$^{-1}$, accelerator developments will focus on increasing current while reducing injection backgrounds; reducing catastrophic beam loss events; and controlling emittance blowup and beam instability. An international task force is providing input, including on a possible upgrade of the final focus or other hardware upgrades for long shutdown 2 in 2028. 

\subsubsection{$Z^\prime$ and leptophilic dark scalars}

$L_\mu$ -- $L_\tau$ models seek to explain possible Standard Model anomalies in muon $(g-2)$ and in $B$ decays to leptons. These models include a vector gauge boson $Z^\prime$ that couples to only the second and third generations leptons of the Standard Model, thereby evading strong limits from processes involving electron production or decay. Existing limits from BaBar \cite{BaBar:2016sci}, CMS \cite{CMS:2018yxg}, and Belle \cite{Belle:2021feg} strongly the constrain the parameter space relevant for $(g-2)_\mu$ for $Z^\prime$ masses above $2 m_\mu$ through searches for resonances in four muon final states. If, however, the anomaly is related to a scalar that couples only to muons, the existing constraints do not cover the relevant parameter space.  A  Belle~II search in the four muon final state will provide powerful constraints with a few ab$^{-1}$ of data \cite{Harris:2022vnx}. 

\paragraph{Search for the invisible decay of the $Z^\prime$:}

Below the $2m_\mu$ threshold, the $L_\mu$ -- $L_\tau$ $Z^\prime$ would decay exclusively to $\nu_\mu$ or $\nu_\tau$. It is also possible that the $Z^\prime$ is the mediator between dark matter ($\chi$) and the Standard Model. In this case, the decay $Z^\prime \to \chi \chi$---which also produces no particles detectable by Belle~II---would be dominant even above $2m_\mu$ threshold. 

\begin{figure}[htbp]
   \centering
   \includegraphics[width=0.49\textwidth]{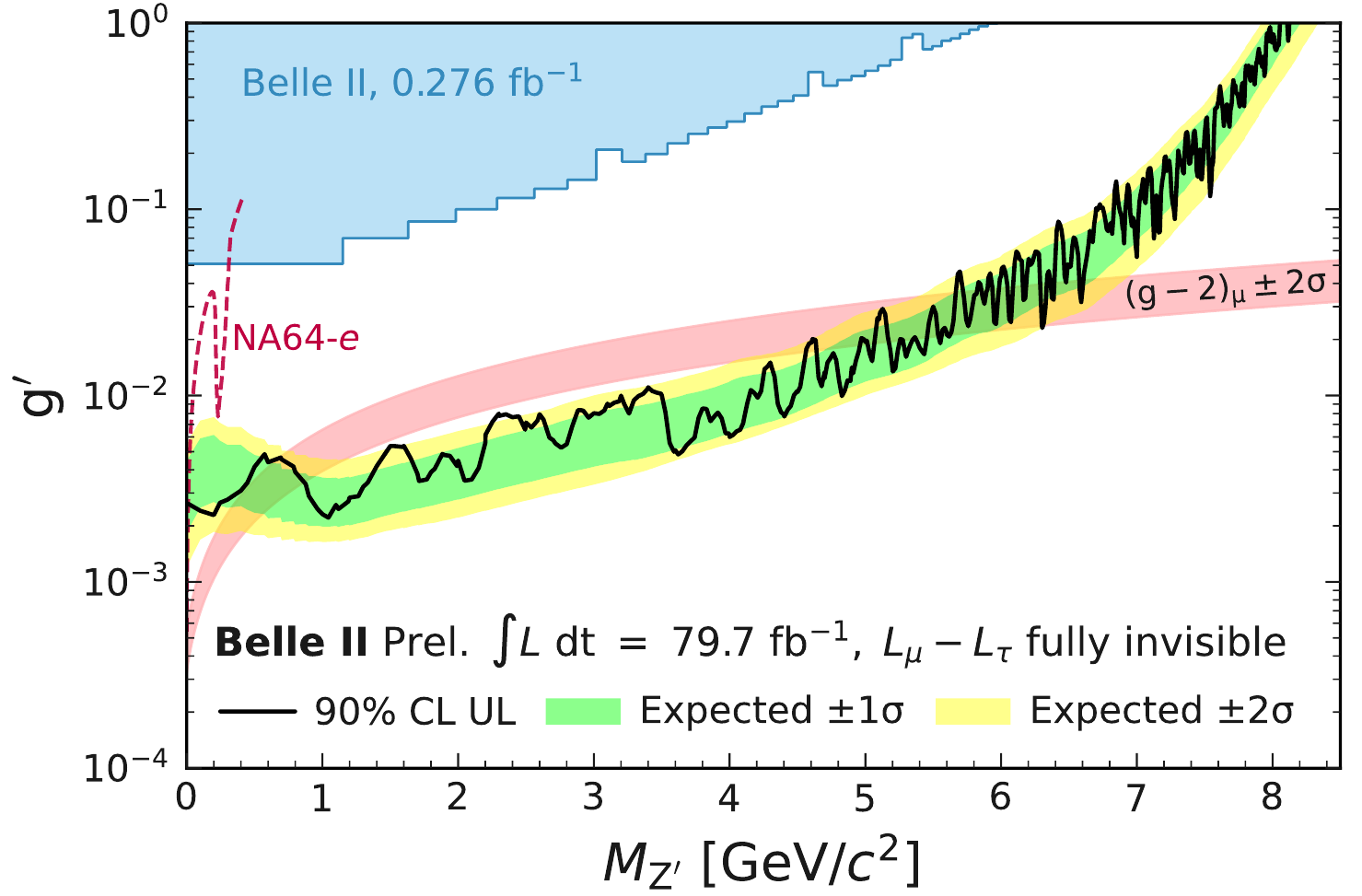} 
   \caption{Belle~II limits on the coupling $g^\prime$ between the $Z^\prime$ and the muon  as a function of $Z^\prime$ mass, under the hypothesis that the $Z^\prime$ decays only invisibly \cite{Belle-II:2022yaw}. Also shown are limits from NA-64 \cite{NA64:2022rme}.}
   \label{fig:zToInvisible}
\end{figure}

Limits on these invisible $Z^\prime$ decays have been set by an early Belle~II analysis searching for $e^+e^- \to \mu^+ \mu^- Z^\prime$, where the mass of the $Z^\prime$ is deduced from the missing mass recoiling against the muon pair \cite{Belle-II:2019qfb}. Belle~II has recently updated this analysis, with a dramatic increase in sensitivity \cite{Belle-II:2022yaw}. This is due in part to a $300\times$ increase in integrated luminosity. The new analysis also significantly improves the suppression of Standard Model backgrounds, which are primarily muon pairs accompanied by one or more undetected photons, tau pairs decaying to muons, and two-photon-fusion production of muon pairs. This is accomplished by a boosted decision tree that exploits the kinematics resulting from the $Z^\prime$ being produced in final-state radiation.  The resulting limits, under the assumption that $\mathcal{B}(Z^\prime)\to \mathrm{invisible} = 1$, exclude the parameter space that would explain $(g-2)_\mu$ for $0.8 < m_{Z^\prime} < 5.0$\,GeV/c$^2$ (Fig.~\ref{fig:zToInvisible}). 

Under the hypothesis that the $Z^\prime$ decays only to Standard Model particles, the new paper improves the existing Belle~II limits below the $2m_\mu$ threshold by more than an order of magnitude. Current sensitivity for these masses does not reach the $(g-2)_\mu$ parameter space, but future luminosity increases and planned analysis improvements will provide significant improvements.

\paragraph{Search for a $\tau^+\tau^-$ resonance in $e^+e^- \to \tau^+\tau^-$:}

A $L_\mu$ -- $L_\tau$ $Z^\prime$ boson produced via final-state radiation from a muon pair could also decay to a tau pair, if it is massive enough. Due to the neutrinos produced by the tau decays, the sensitivity is poor compared to the muon pair final state. However, if the new particle is a leptophilic scalar with mass-dependent couplings, the loss of kinematic constraints is more than compensated by the increased branching fraction. The signature is four tracks including at least two muons, with missing energy, consistent with a particle produced as final state radiation recoiling against a muon pair.  A preliminary Belle~II result sets the first leptophilic scalar results above 6.5\,GeV/c$^2$. 

\subsubsection{Dark photons}
Dark photons are produced via initial state radiation, and then decay either invisibly or to a pair of leptons or other standard model particles. Partially visible decays are also possible in slightly more complicated cases, such as indirect dark matter \cite{Duerr:2019dmv}, or dark showers \cite{Bernreuther:2022jlj}. Both of these involve displaced vertices, which significantly reduces Standard Model backgrounds.  Analyses have so far focused on the case where the dark photon mass is less than the center of mass energy, resulting in on-shell production. 

\paragraph{Invisible dark photon decays:}

\begin{figure}[htbp]
   \centering
   \includegraphics[width=0.49\textwidth]{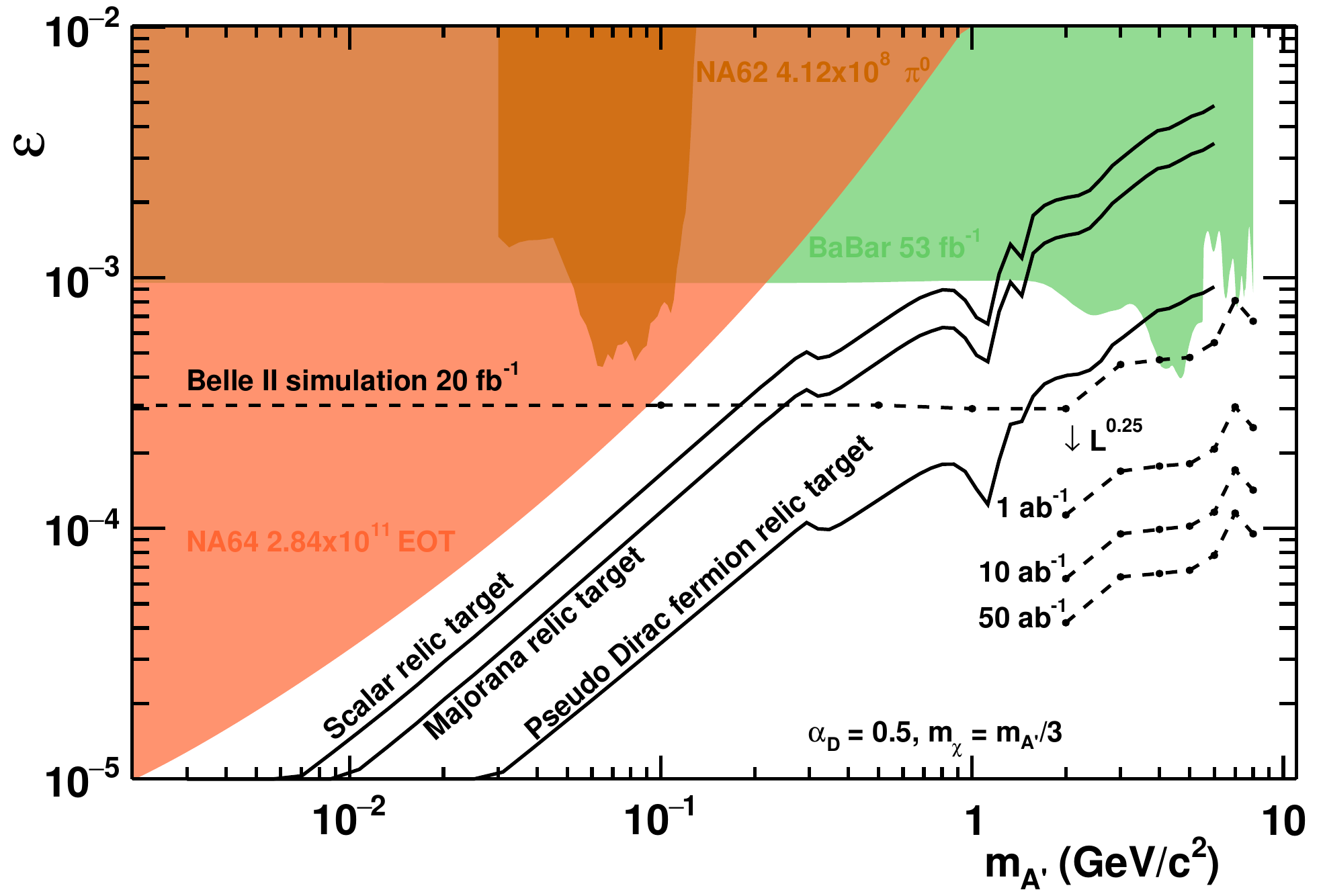} 
   \includegraphics[width=0.49\textwidth]{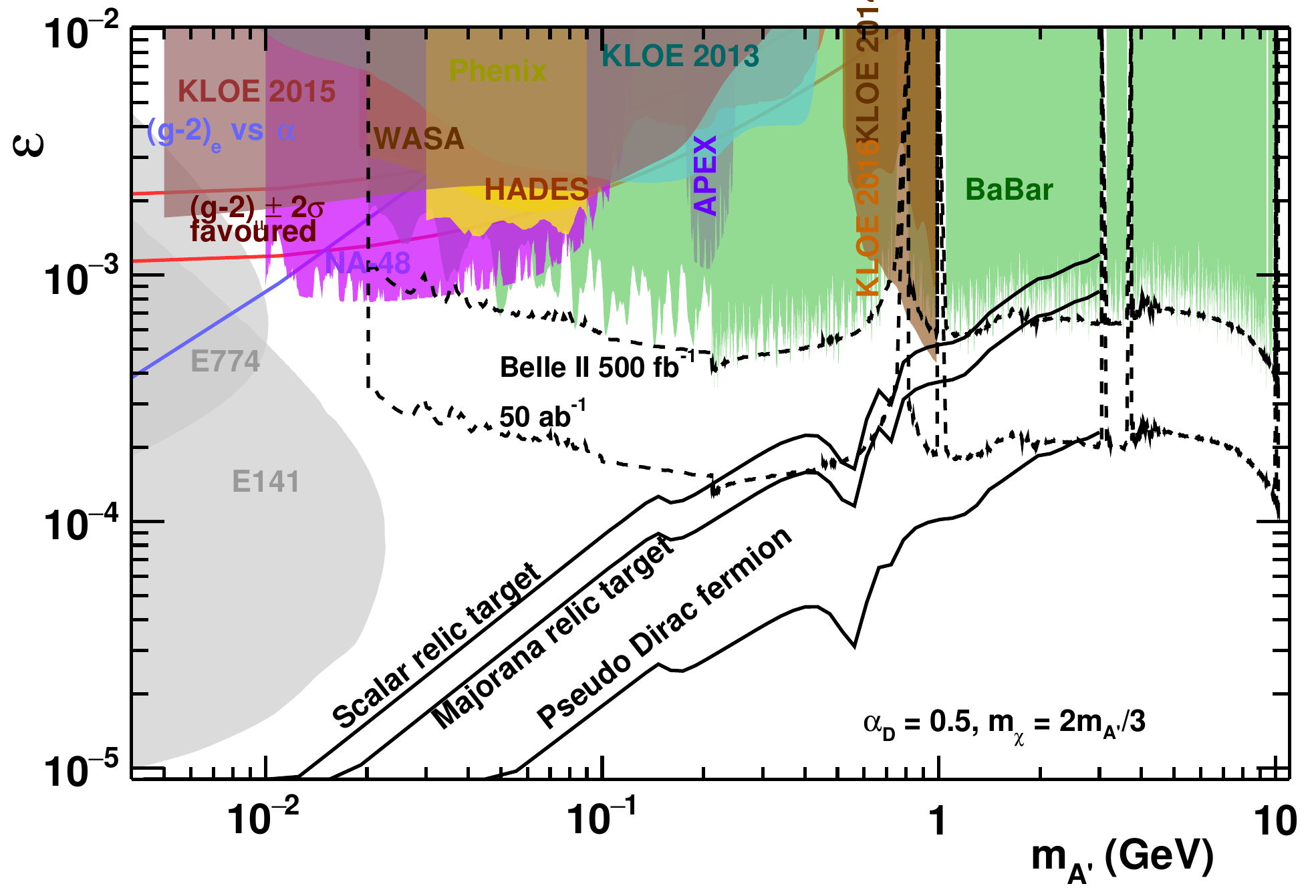} 
   \caption{Projected Belle~II sensitivity for (a) invisible dark photon decay and (b) visible dark photon searches in the kinetic mixing parameter $\varepsilon$ versus dark photon mass plane. Adapted from \cite{Belle-II:2022cgf}. Also shown are the dark matter relic density targets.}
   \label{fig:belle2visible}
\end{figure}

The visible final state in this case is a single photon. Although its decay products are not reconstructed, the mass of the dark photon $m_{A^\prime}$ is directly related to the center-of-mass energy of the photon under the assumption that there are no additional initial-state radiation photons in the event. Backgrounds include $e^+e^- \to \gamma \gamma (\gamma)$, where only one photon is detected; $e^+e^- \to e^+e^- \gamma$, with both final state electrons out of the detector acceptance; cosmic rays; and beam backgrounds. The analysis relies on quantifying each of these backgrounds in the missing mass squared vs.\ polar angle plane. Belle~II, in its initial analysis, will have sensitivity to regions of the parameter space that would correspond to that expected for the observed astronomical dark matter (Fig.~\ref{fig:belle2visible}a).

\paragraph{Visible decays of the dark photon:}

For visible decays of the dark photon, the final state consists of a photon and a pair of leptons.  The initial search at Belle~II will reconstruct all three particles. The signal consists of a narrow resonance in the $\ell^+\ell^-$ pair on top of a large but generally smooth standard model background. The $J/\psi$ and similar mass regions are excluded from the search. 

Belle~II has a considerably larger drift chamber than BaBar, which will give better mass resolution. The Belle~II projection \cite{Belle-II:2022cgf}, shown in Fig.~\ref{fig:belle2visible}b, is derived from the BaBar limits \cite{BaBar:2014zli}, assuming a factor of two improvement in resolution. 

Belle~II is also studying the statistically independent sample in which the initial-state-radiation photon is at low angles, and only the lepton pair is reconstructed.  A larger data set will enable a displaced vertex search, with significant reach \cite{Ferber:2022ewf}. The BaBar analysis did not set limits below $20\,\mathrm{MeV/c^2}$. Given the importance of investigating the ATOMKI anomaly \cite{Krasznahorkay:2015iga}, Belle~II is undertaking an analysis focused on this mass region. 

\paragraph{Dark photon with invisible dark Higgs:}

A natural dark sector extension is to include a dark Higgs ($h^\prime$), which can be produced in association with a dark photon. This model has two additional parameters, the dark Higgs mass $m_{h^\prime}$ and the coupling between the dark Higgs and the dark photon, $\alpha_D$. Belle~II has studied the case where the dark photon decays visibly ($m_{A^\prime}<2m_\chi$), and the dark Higgs is long lived ($m_{h^\prime}>m_{A^\prime}$), leaving no visible signal in the detector. KLOE has previous studied this configuration, at much lower masses \cite{KLOE-2:2015nli}; BaBar \cite{BaBar:2012bkw} and Belle \cite{Jaegle:2015fme} have published searches for the case where the dark Higgs decays to a pair of dark photons. 

Belle~II searches for dark photons decaying to a muon pairs. The final state is a pair of muons and missing momentum, in synergy with the invisible $Z^\prime$ analysis. There are additional kinematic constraint: the invariant mass of the muon pair is equal to the dark photon mass, while the missing mass is equal to the dark Higgs mass. 

\begin{figure}[htbp]
   \centering
   \includegraphics[width=0.49\textwidth]{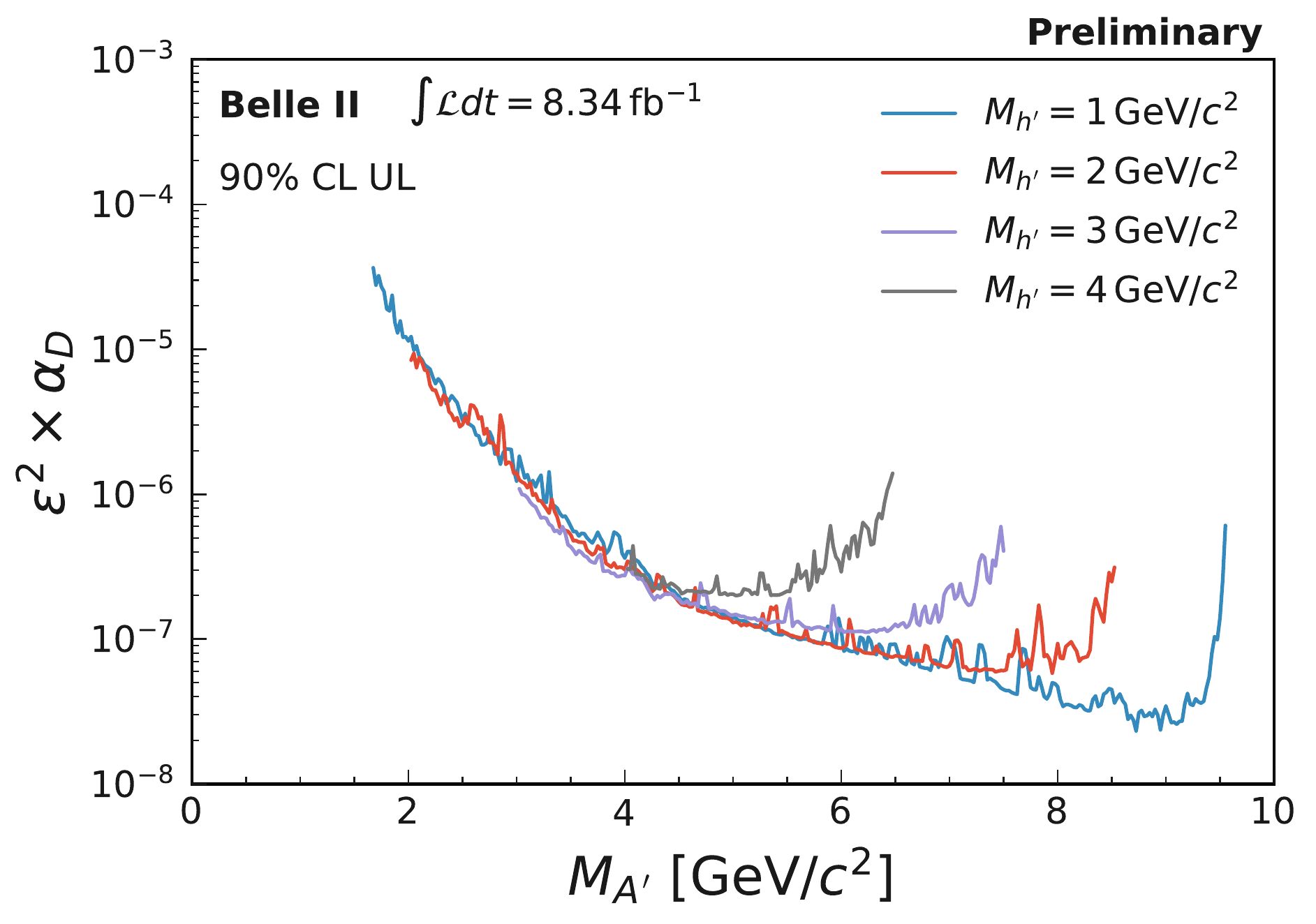} 
      \caption{Belle~II upper limits on  $\alpha_D \varepsilon^2$ as a function of dark photon mass, for four different dark Higgs masses \cite{Belle-II:2022jyy}.}
   \label{fig:DarkHiggsLimits}
\end{figure}

A signal would be a concentration of events in the $m^2_\mathrm{recoil}$ vs.\ $m_{\mu^+\mu^-}$ plane. The largest background is from $e^+e^- \to \mu^+\mu^- \gamma (\gamma) $ events with undetected photons.
The resulting limits, in the $\alpha_D \varepsilon^2$ vs.\ $m_{h^\prime}$ or $m_{A^\prime}$ planes, are the first in this mass range \cite{Belle-II:2022jyy}. For $\alpha_D = 1$, the limits are stronger than those from existing visible dark photon decay analyses (Fig.~\ref{fig:DarkHiggsLimits}).

\subsubsection{Summary}

Belle~II has accumulated a near-BaBar sized data set, which has been used to produce several world-leading dark sector searches. The current data will be used for several other high-profile analyses, including searches for invisible and visible dark photon decays, and leptophilic / muonphilic scalars decaying to muon pairs. Over the next decade, increases in the SuperKEKB luminosity, including an upgrade of the final focus, will increase the data size by a factor of 100. This data set, the clean $e^+e^-$ environment, and inclusive triggers, will give Belle~II unique sensitivity to dark sector physics.

% \bibliography{BelleIIDarkSector}
% \bibliographystyle{plain}
% \end{document}

\afterpage{\clearpage}
%-------------------------------------------

%-------------------------------------------
\subsection{Search for light DM with extracted beams: prospects at JLAB -- {\it M.~Battaglieri} }
\label{ssec:battaglieri}
{\it Author: Marco Battaglieri,  <battaglieri@ge.infn.it>} \\

\subsubsection{Introduction} 
The traditional approach to explain cosmic anomalies considers dark matter (DM) made by heavy particles (M >> 10GeV) interacting at the weekly scale (weekly interacting massive particles or WIMPs). Despite the enormous current and planned word-wide effort, the lack of evidence calls for an extension of the hunting territory to include unexplored mass ranges.
Light dark matter (M $\sim$ 1 - 10 GeV), or LDM, is a well-theoretically motivated region to explore. If coupled with the DM relic origin hypothesis (aka, the current DM to baryonic matter ratio is the same as the one at the time of the primordial Universe expansion), the LDM interaction is naturally explained by a new vector gauge boson, mediator of a new force (the heavy or dark photon $A'$). This leads to predict a feeble interaction of LDM  with Standard Model particles that could be studied in a current accelerator-based experiment where moderate beam energy ($\sim 10$ GeV) and high intensity (I$\sim$ 100 $\mu$A) are available. In the dark-photon mediator scenario, lepton beams are particularly suited to study LDM since, via kinetic-mixing, ordinary gammas couple to the $A'$ via the electromagnetic charge suppressed by a small coupling constant $\epsilon$ ($O<<1$). In the so-called Vector-Portal scenario, the LDM interaction with SM particles, mediated by the  $A'$,  can produce a visible decay (with SM particles in the final state) or invisible (when the $A'$ decays in DM particles) depending on the relative $A'$ vs. LDM mass hierarchy. The two options lead to different experimental techniques to search for LDM. Missing momentum/energy experiments look for a relevant (over a certain threshold) momentum/energy disappearance after that the primary beam interacts with the target. Beam dump experiments expect to produce LDM in the interaction of the primary beam with a thick target. A downstream detector, shielded by the copious SM radiation produced therein, will measure the LDM via the energy deposited by scattered electrons and protons. Other experiments aims at the direct detection of the mediator $A'$ assuming its visible decay in lepton (e$^+$e$^-$, $\mu^+ \mu^-$),  or hadron ($\pi^+ \pi^-$, K$^+$K$^-$) pairs.
The $A'$ can be produced by a lepton (beam) via different mechanisms.
The $A'-strhlung$ describes an incoming electron/positron radiating an $A'$ when in the target material's electric field, similar to the regular $bremsstrahlung$. Only for positrons, annihilation (resonant and non-resonant) on the target's electrons in a gamma and an $A'$ provides a supplemental production mechanism. Each of those has a different  ($\epsilon \times \alpha^n$) dependence and a characteristic $A'$ kinematics. In particular, positron-electron annihilation is the production mechanism used to study visible and invisible $A'$ decay at colliders in experiments such as BaBar, Belle-II, and KLOE.

\subsubsection{Jefferson Lab: the intensity frontier}
Jefferson Lab, located in Newport News (VA-US), is recognized worldwide as one of the major nuclear physics fixed target facilities. JLab hosts the CEBAF (Continuous Electron Beam Facility) accelerator able to deliver up to 12 GeV, CW, electron beam simultaneously to four experimental halls (A, B, C, and D). The 100$\%$  duty factor, coupled with an intense electron beam current (up to 100 $\mu$A) and a high beam polarization (up to 80$\%$) provide a unique high-intensity high-precision beam to study QCD in fixed target experiments. The four halls host different detectors ranging from the small acceptance/high-precision magnetic spectrometers (Hall-A and -C) to $4\pi$-acceptance detectors such as CLAS12 (Hall-B) and GLUEX (Hall-D). The main lab scientific goals include: identifying the role of the gluonic excitation in the spectroscopy of light mesons, understanding the spin of the nucleon and the role of the quark orbital momentum, revealing a novel landscape of nucleon sub-structure through 3D imaging at the femtometer scale, studying the short-range N-N correlations and the partonic structure of nuclei to clarify the nature of nuclear force.
The 12 GeV experimental program is in full swing (33 experiments were completed out of 91 approved) and, assuming 30 weeks/year operations, (at least) another decade of physics is expected. Other opportunities including CEBAF 20+ GeV energy upgrade as well as running a 12 GeV highly polarized positron beam are under study as well as pushing the current luminosity to unprecedented level (e.g. the SOLID experiment is expected to run at 10$^{38-39}$ cm$^{-2}$ s$^{-1}$). \\
Leveraging the electron beam quality and the apparata installed in the experimental halls, in the last years the lab extended its program to include experiments aiming to discover evidence for physics beyond the Standard Model of particle physics. Four experiments cover the Dark Sector searches: APEX, HPS, BDX, and X-17. In the following,  I will briefly describe the main features of each experiment reporting the current status and perspectives.

\paragraph{APEX}
The A-Prime EXperiment (APEX) is a traditional fixed target experiment installed in Hall-A searching for the $A'$ by its visible decay in (e$^+$ e$^-$) pairs.
The experimental technique used in APEX is the so-called 'bump-hunting' where the $A'$ is searched for as a narrow resonance in M$_{e^+ e-}$ spectrum dominated by the smooth QED Bethe-Heitler background. The key feature of the experiment is the use 
of the two-arm High-Resolution Spectrometer (HRS) to detect the lepton pair in coincidence. A similar detector stack instruments the two arms: scintillators for timing, a vertex detector for tracking, and Cherenkov and electromagnetic calorimeters for particle identification. The HRS resolution ($\delta M$/$ M\sim$ 1 MeV) allowed to set stringent exclusion limits on $A'$ masses between 80 MeV and 200 MeV with a sensitivity down to $\epsilon^2<10^{-7}$. APEX published results from the engineering run in 2010~\cite{Buncic:2015ari}. The analysis of data collected in the 2019 run is currently undergoing. Results are expected to be published soon.

\paragraph{HPS}
The Heavy Photon Search Experiment (HPS) is an experiment running in Hall-B. Similarly to APEX, HPS searches for the $A'$ by its visible decay in (e$^+$ e$^-$) pairs. In addition to the 'bump hunting' technique, the decay vertex measurement greatly extends the reach in a region of the parameter space difficult to access  ($A'$ mass between 20 MeV to 220 MeV and $\epsilon^2 \sim 10^{-8} - 10^{-10}$). The enhanced experimental reach is possible thanks to a sophisticated high-resolution ($\delta x \sim 1mm$) tracking system located very close to the center of the beam (less than an mm) sensitive to vertex positions in cm-range from the target. The technique considers that Bethe-Heitler QCD background is produced inside the target while a feeble coupling $A'$ (small $\epsilon$) decays outside the target where QCD radiation ranges out. A good spacial resolution is the key of the experiment.
\begin{figure}[ht!]
 \begin{center}
 \centering
    \includegraphics[width=0.45\textwidth]{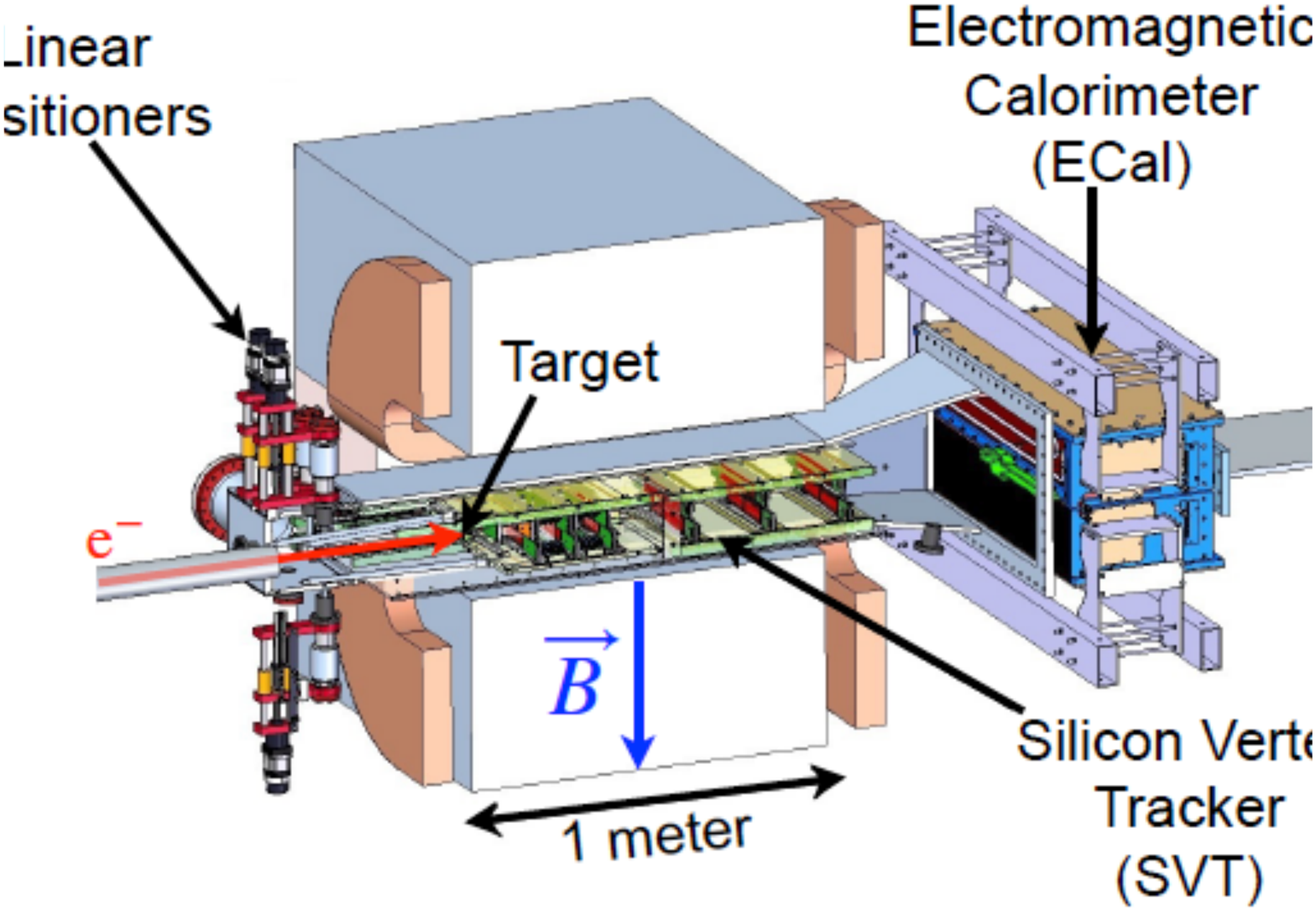}
 \end{center}
  \caption{The HPS detector.}
 \label{figure-hps-detector}
\end{figure}

\begin{figure}[ht!]
 \begin{center}
 \centering
    \includegraphics[ width=0.60\textwidth]{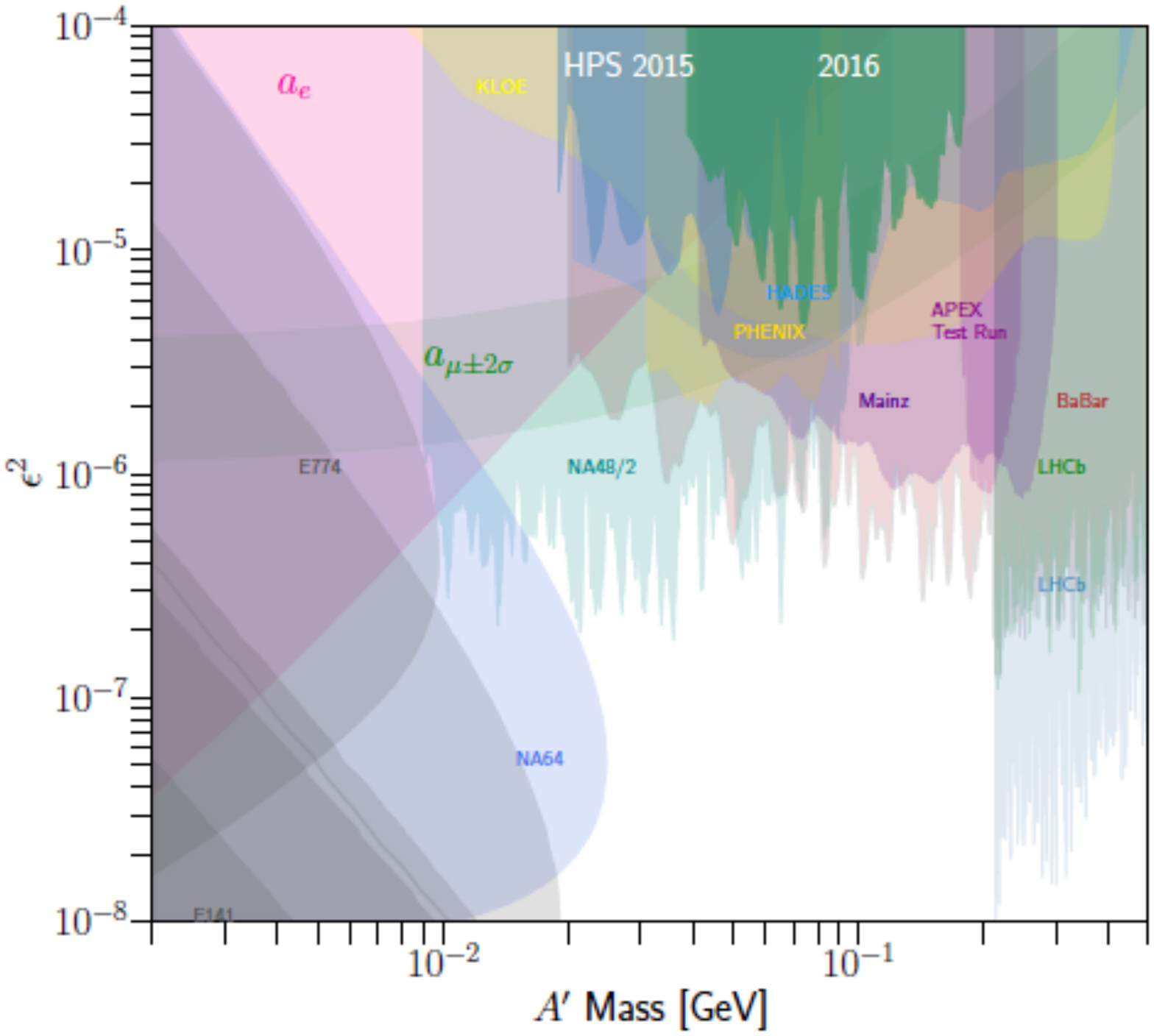}
 \end{center}
  \caption{The physics reach of the HPS experiment.}
 \label{figure-hps-reach}
\end{figure}

HPS makes use of the Hall-B  I$_e$ = 200 nA electron beam impinging on a thin (4 $\mu$m) tungsten target. For the above-mentioned reasons, the target vertex has to be as defined as possible. The beam energy changed in different  runs (E$_e$ = 1.1 GeV, 2.3 GeV, 4.4 GeV) to cover a larger area in the parameter space (the e$^+$ e$^-$  invariant mass within the acceptance of the detector increases by increasing the beam energy). The HPS apparatus is composed of a silicon vertex tracker (SVT) inserted in 0.24 T dipole magnetic field for precise electron/positron  momentum and vertex determination, and a PBWO$_4$ calorimeter used to trigger on the electromagnetic shower produced by leptons pairs. A plastic scintillator hodoscope helps in identifying photons from charged lepton increasing the trigger efficiency. The HPS detector is shown in Fig.~\ref{figure-hps-detector}.
The experiment published the 2015 engineering run 'bump hunting’ results showing the potentiality of the proposed technique ~\cite{PhysRevD.98.091101}. Preliminary results from the 2016 run (including vertexing) have been submitted for publication on PRD~\cite{https://doi.org/10.48550/arxiv.2212.10629}.
Figure~\ref{figure-hps-reach} shows the experimental reach obtained in a few days of engineering runs.
Data collected in 2019 and 2020 are currently being analyzed.

\begin{figure}[!ht]
 \begin{center}
 \centering
    \includegraphics[ width=0.60\textwidth]{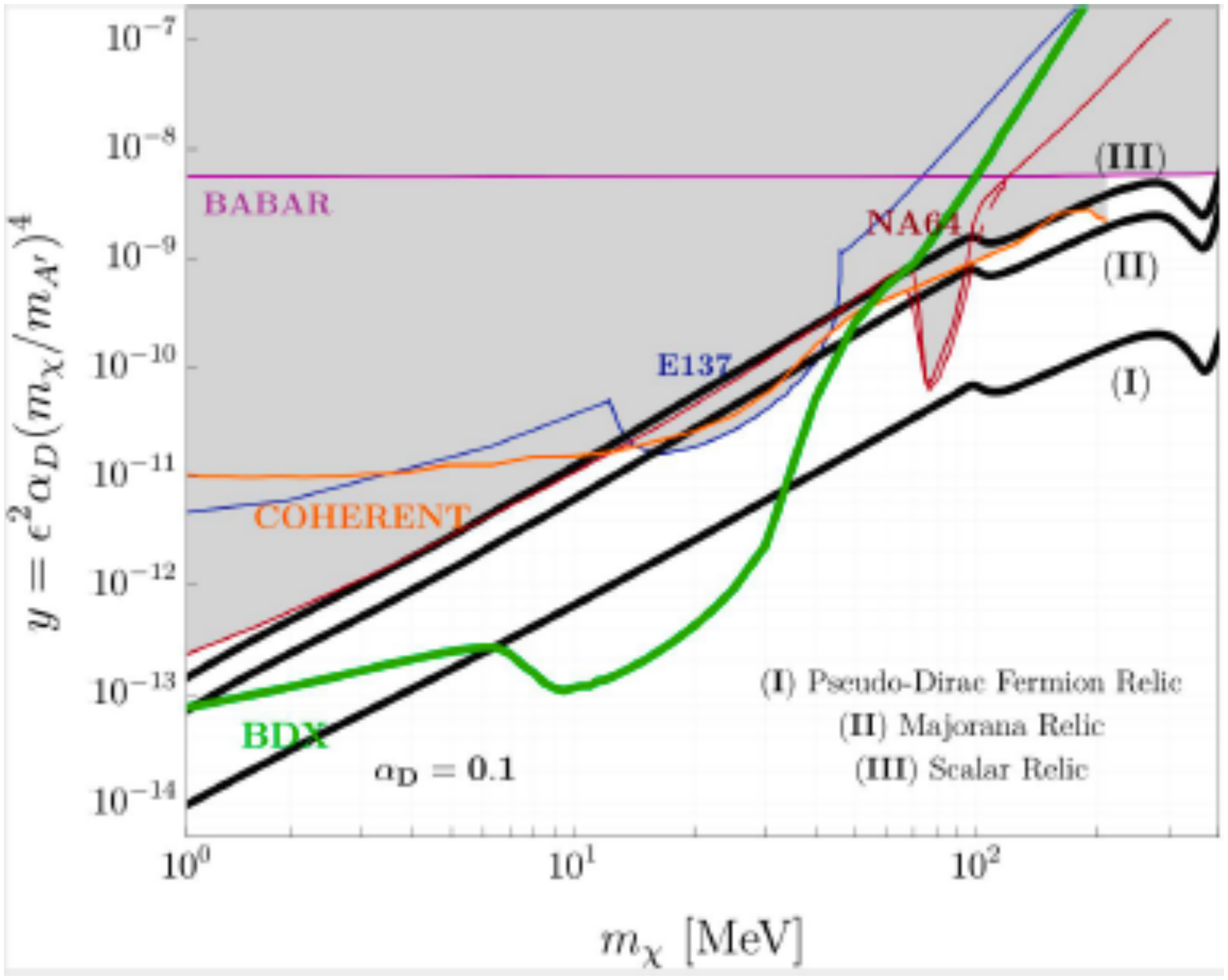}
 \end{center}
  \caption{The expected physics reach of the BDX experiment.}
 \label{figure-bdx-reach}
\end{figure}

\paragraph{BDX}
The Beam-Dump eXperiment (BDX) is a JLab-approved experiment expected to be scheduled in the near future.
The intense 11 GeV electron beam directed in the experimental Hall-A, after interacting with the an hadronic, is dumped in a cooled thick block of aluminum (several radiation lengths).  The beam produces an electromagnetic shower, releasing the initial energy in form of a copious number of standard particles (muons, neutrons, gamma, neutrinos, ...) and, if they exist, DM particles. While the SM particles are ranged out by an appropriate shielding (all but neutrinos),  the DM beam propagates downstream up to a detector where LDM could interact with nuclei and electrons. Interactions with detector's electrons
may produce a significant recoil energy detected via $O( GeV)$ EM shower that, in turn, provides a clear experimental signature. In a BDX-like experiment, DM needs to be produced and then detected requiring an intense primary beam or, equivalently, a large charge (the expected number of detectable events scale as the fourth power of the feeble coupling $\epsilon$). The combination of the CEBAF 11 GeV beam and the high current operation of Hall-A (I$_e\sim$ 60 -70 $\mu$A  corresponding to $\sim$10$^{22}$ electron on target - EOT - in a year time), makes JLab the ideal place to run BDX. \\
The BDX detector is made by an electromagnetic calorimeter surrounded by a multi-layers hermetic active and passive veto. The calorimeter will re-use $\sim$1000  CsI(Tl) crystals formerly installed in BaBar EMCal with updated photo-sensors (SiPMs) and a modern DAQ (digitizers). The Inner and Outer vetos will be made by plastic scintillator paddles read out by SiPMs via wavelength shifter fibers.  A layer of lead interleaved between the two layers of active vetos  prevents low-energy radiation to penetrate into the calorimeter's fiducial volume. Figure~\ref{figure-bdx-reach} shows the expected BDX reach in a 1-year measurement. As shown in the graph, BDX sensitivity is 10-100 times better than existing limits on LDM. 
The experiment was proposed to run in parallel (and fully parasitically) with respect to the Moeller experiment scheduled in Hall-A. BDX Collaboration is currently waiting for the lab to provide the necessary infrastructure to run the experiment. This includes a new underground experimental hall located about twenty meters downstream of the Hall-A beam dump to deploy the detector and a sizable iron shielding between the current concrete beam-dump vault and the new hall to range out SM background (cosmic background will be rejected by the active veto system).

\paragraph{BDX-MINI} 
While waiting for BDX to run, the BDX Collaboration installed and ran a reduced version of the experiment called BDX-MINI~\cite{PhysRevD.106.072011}. BDX-MINI took advantage of the 2020 Hall-A low-energy (E$_e$=2.17 GeV) run by installing a small detector in a well dug in the future BDX  location to make a precise assessment of the expected beam-related background~\cite{Battaglieri:2019ciw}. The low beam energy did not require additional shielding making effective the 26 m of dirt present between the beam-dump and the detector.
The BDX-MINI detector~\cite{Battaglieri:2020lds} is made by  a PbWO$_4$ calorimeter, surrounded by two layers of active and W passive vetos. While the detector concept is identical to BDX, the limited space  required a denser material in calorimeter and passive vetos.
The experiment ran smoothly for six months collecting a total charge of 2.6 10$^{21}$ EOT (25$\%$ of the charge expected to be collected by BDX).
\begin{figure}[!ht]
 \begin{center}
 \centering
    \includegraphics[ width=0.5\textwidth]{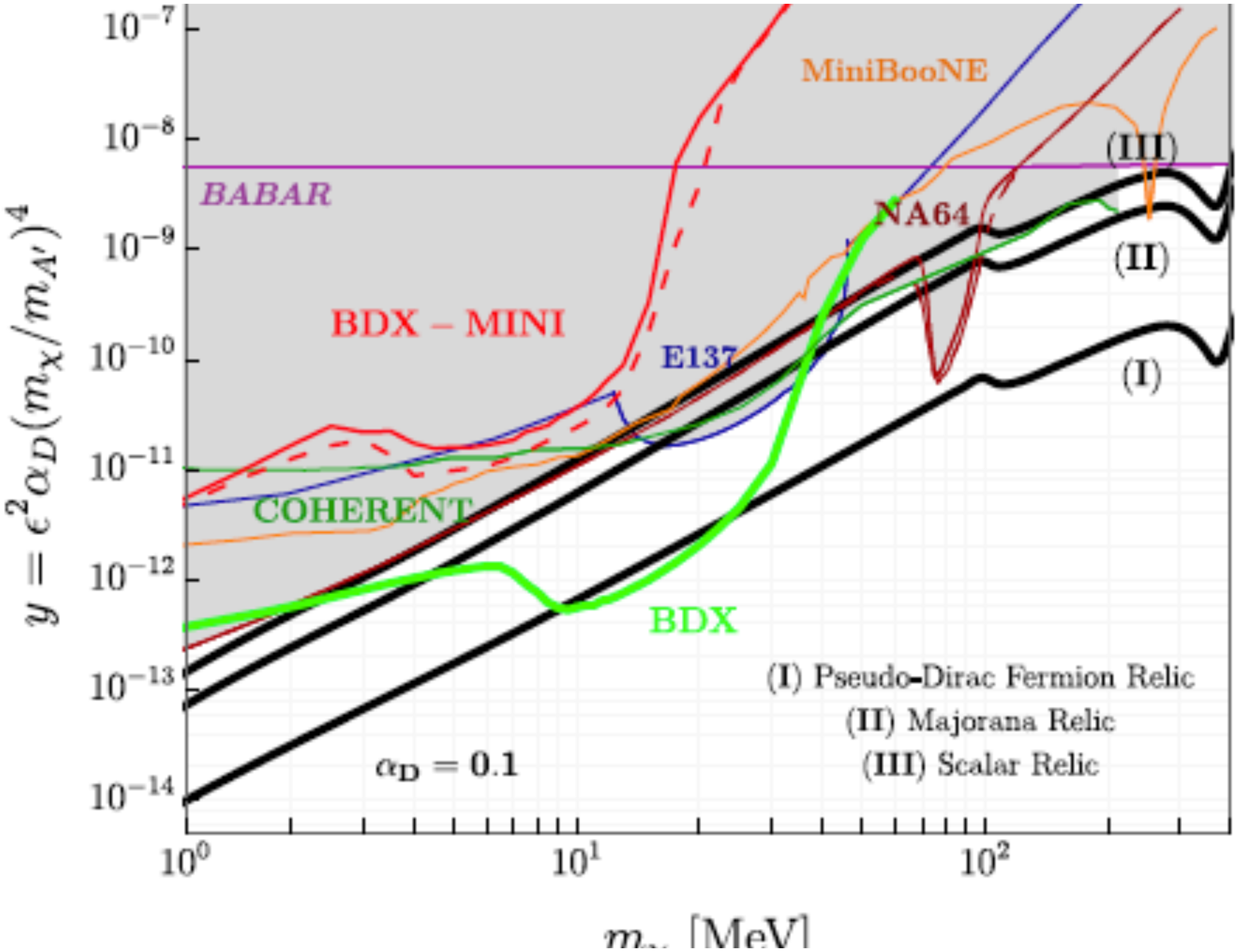}
 \end{center}
  \caption{The physics reach of the BDX-MINI experiment.}
 \label{figure-bdx-mini-reach}
\end{figure}
Figure~\ref{figure-bdx-reach} shows the exclusion limit set by the BDX-MINI experiment. Despite the limited active volume (a few percent of BDX), BDX-MINI provided exclusion limits similar to the best existing experiments demonstrating the potential of the full experiment that will run at JLab in the future.

\begin{figure}[ht!]
 \begin{center}
 \centering
    \includegraphics[ width=0.60\textwidth]{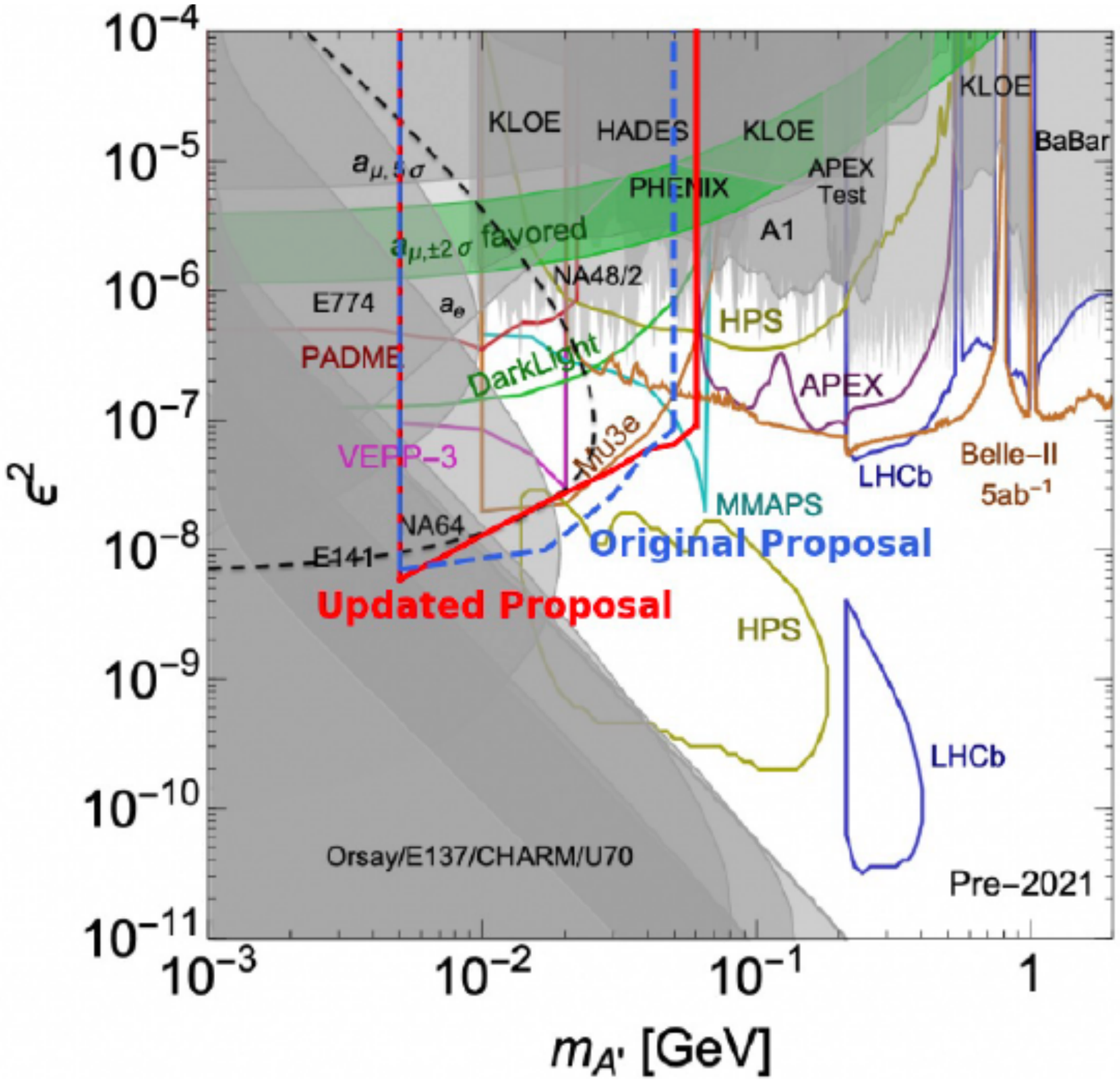}
 \end{center}
  \caption{The expected physics reach of the X-17 experiment.}
 \label{figure-x-17-reach}
\end{figure}

\paragraph{X-17} 
The X-17 experiment was approved to run in Hall-B using a dedicated apparatus. X-17 will  search for new hidden sector particles in the 3 - 60 MeV mass range 
in forward electroproduction scattering from a heavy nuclear target (Ta). The experimental technique, similar to the one used by APEX and HPS, searches for a bump in  $A'$ visible decay (e$^+$e$^-$) pairs detected in a PbWO$_4$ calorimeter. The expected reach for 120 days of running, is shown in Fig.~\ref{figure-x-17-reach}.  X-17  will cover the interesting mass range where the 'Atomki anomaly', an excess of   (e$^+$e$^-$) pairs in the decay of excited $^8$Be nuclei),  was reported.

\paragraph{Further opportunities} 
Jefferson Lab is currently exploring new directions to extend the 12 GeV physics program.\\
High-intensity secondary beams of neutrons, muons, and neutrinos (besides the LDM, if existing) are produced by the interaction of the 12 GeV electron beam with the dump. Simulations show that muons are produced at a rate of $10^{-6}$ per EOT by Bethe-Heitler radiation with a Bremsstrahlung-like energy spectrum more focused than in proton-produced beams where the main production mechanism is  meson decays. The resulting high intensity (up to $10^8$ $\mu$/s) would make the up-to 6 GeV JLab muon 'beam' well suited for a missing momentum experiment {\it \`a la M3} or a beam dump experiment  {\it \`a la BDX}. An evaluation of the experimental reach is currently undergoing.
Besides muons, neutrinos are also copiously produced in the dump. Simulations show that a large off-beam-axis $\nu$ flux  (up to $10^{18}$ $\nu$/y/m$^2$) is produced  with a Decay-At-Rest energy spectrum. Preliminary estimates showed that on a m$^3$ active volume of LAr or CsI ({\it \`a la COHERENT}), $O(5000)$ CE$\nu$NS interactions are expected in one-year exposure. This would represent 50 times the entire world statistics available for this process. \\
Studies are underway to deploy a positron beam of high energy (up to 11 GeV), high current (I$_{e+}\sim$ 0.5-1 $\mu$A), and high polarization (P$_{e+}\sim$ 60$\%$). The idea is to inject $\sim$100 MeV positron in CEBAF and deliver the positron beam to the existing experimental halls. From a first assessment, the existing detectors should be adequate to run a reach physics program with a positron beam. As far as the Dark Sector searches are concerned, experiments on thin and thick targets were considered. In the first case, the positron beam annihilation on target atomic electrons  will provide an excellent way to produce the $A'$ in the $e^+ e^- \to \gamma A'$ reaction. Measuring the resulting photon, the $A'$ mass will be detected by identifying a peak in the missing mass spectrum. This technique was successfully tested in the PADME experiment at LNF. The   JLab positron beam's high energy and large intensity will significantly extend the sensitivity of this experiment to cover  M$_{A'}$ up to 100 MeV. 
Similarly,
a missing energy experiment, ({\it \'a la NA64}) will dump the positron beam on a thick active target. Positron annihilation would enhance the resonant $A'$ production providing a unique experimental signature. The reach of the experiment could be extended to a larger range of masses by varying the positron energy.\\
Last, but not least, the lab is studying the possibility of upgrading CEBAF in order to reach an electron beam energy of up to 20 GeV. The energy upgrade will significantly benefit each of the above-mentioned experiments and will open up even further opportunities in the future.

\subsubsection{Conclusions}
Jefferson Lab hosts one of the best medium energy (up to 10 GeV) and high-intensity (up to 0100 $\mu$A) electron beam available in the world. These conditions are ideal to explore the intensity frontier searching for new physics in the 1 MeV - 1 GeV range. A significant experimental program exploring the Dark Sector is ongoing. Experiments such as APEX, HPS, and BDX-MINI collected data in the past years and more is expected in the near future. Approved experiment, such as BDX and X-17, covering unexplored regions of the parameter space, have sensitivity for new physics or, in case of a null result, will extend the current exclusion limits by orders of magnitude.  Further experiments, leveraging the available high-intensity electron beam to produce secondary $mu$, $\nu$, and neutron beams, are under study. In the future, a unique positron beam and the energy upgrade to 20 GeV will open new opportunities to explore BSM physics at the intensity frontier.

\subsubsection{Acknowledgements}
This material is based upon work supported by the U.S. Department of Energy, Office of Science, Office of Nuclear Physics under contract DE-AC05-06OR23177.
\afterpage{\clearpage}
%-------------------------------------------

%-------------------------------------------
\subsection{Search for light DM and mediators : results and prospects at FNAL -- {\it N.~Tran}}
\label{ssec:tran}
{\it Author:  Nhan Tran, <ntran@fnal.gov>} \\

In these proceedings for the FIPS 2022 workshop, we describe results and prospects for experiments at Fermilab.  We describe the status and results of planned experiments ArgoNeuT, SBND, SpinQuest.  We also discuss future proposed experiments at Fermilab including PIP-II Beam Dump, M$^3$, 3\,GeV muon beam dump, and REDTOP.  Other dark sector searches at experiments at Fermilab, MicroBooNE and DUNE, are covered elsewhere in these proceedings.

\subsubsection{Introduction}

Searches for dark sector particles in the GeV mass range and below at relativistic particle accelerators are a highly-motivated physics opportunity in the next decade.
The physics drivers are categorized described in more detail in~\cite{bi1,bi2,bi3}. We summarize the motivating scenarios briefly here as:
\begin{itemize}
    \item \textbf{thermal dark matter}: We denote this as ``DM'' below. This physics driver focuses on searches for dark matter produced in accelerator based-experiments. These types of searches have sharp milestones to reach in mass-coupling space under the assumption that dark matter was in thermal equilibrium with the SM in the early universe. 
    \item \textbf{visible dark portals}: We denote this as ``Visible'' below. This physics driver focuses on searches for dark mediator particles that decay back to SM particles. In certain mass hierarchies of the dark sector, we are most sensitive to such scenarios by looking for visible SM signatures. 
    \item \textbf{new flavors and rich dark sectors}: We denote this as ``Flavor'' below. This physics driver has focuses on two classes of models: dark sectors with rich structure, as we have in the SM, that may result in both invisible and visible particle final states; and dark sectors motivated by current anomalies from existing experimental results such as $g-2$~\cite{Muong-2:2001kxu,Muong-2:2015xgu}, the MiniBooNE excess~\cite{MiniBooNE:2018esg}, and flavor anomalies~\cite{Belle:2009zue,BaBar:2012mrf,BaBar:2013mob,LHCb:2015gmp,LHCb:2017avl,Belle:2017oht,LHCb:2017vlu,LHCb:2019hip,LHCb:2020lmf}.
\end{itemize}

\begin{figure}[tbh]
    \centering
    \includegraphics[width=0.8\textwidth]{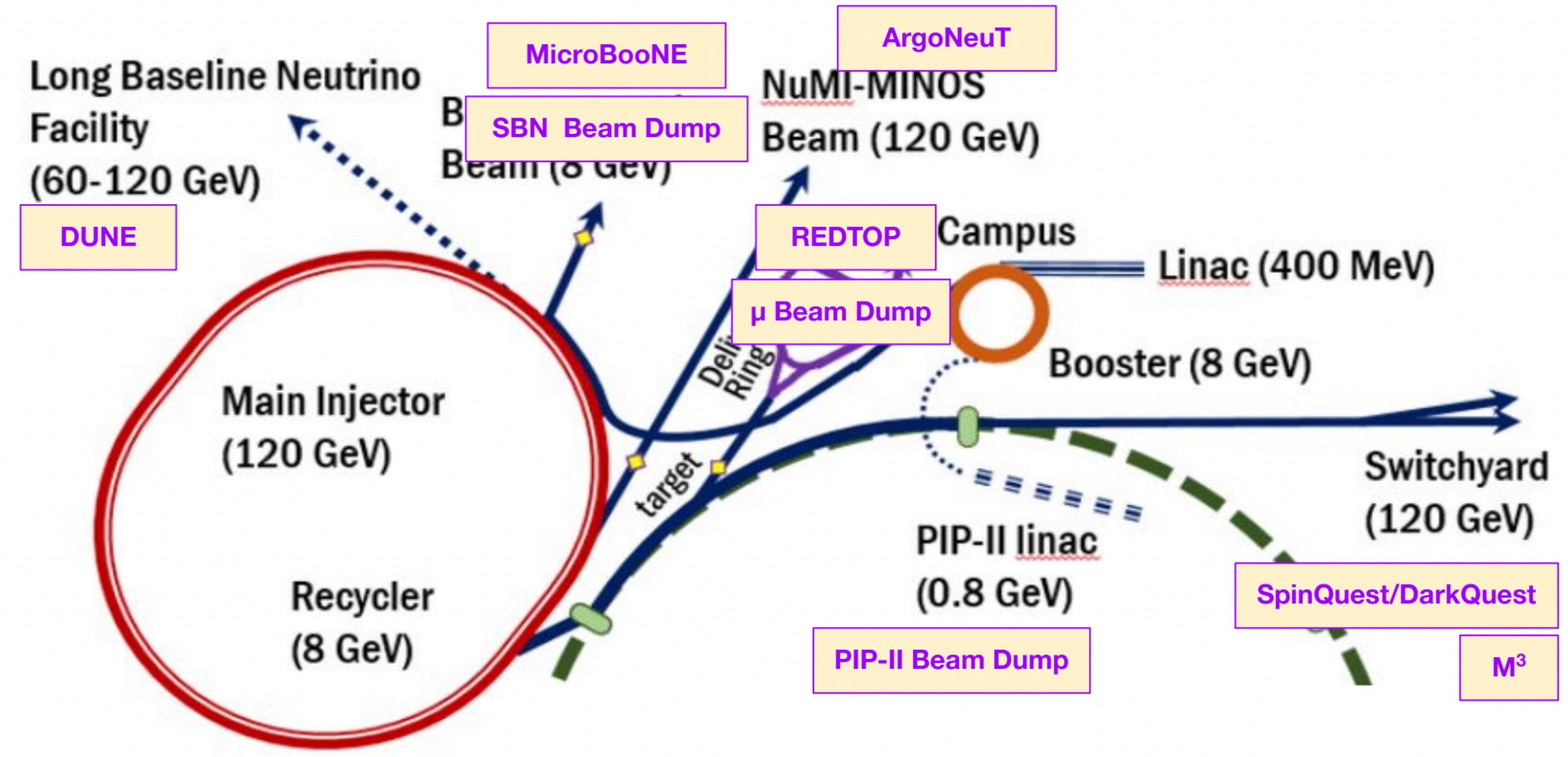}
    \caption{Fermilab accelerator complex and locations of current, planned, and proposed dark sector experiments}
    \label{fig:my_label}
\end{figure}

This proceedings summarizes the current experimental program for sub-GeV dark sector physics at Fermilab.  We describe the status and results of planned experiments ArgoNeuT, SBND, SpinQuest/DarkQuest.  We also discuss future potential experiments at Fermilab including PIP-II Beam Dump, M$^3$, 3\,GeV muon beam dump, and REDTOP.  Other dark sector searches at experiments at Fermilab, MicroBooNE and DUNE, are covered elsewhere in these proceedings. In Fig.~\ref{fig:my_label}, we should the Fermilab accelerator complex and on which beam lines these experiments are or would be located.  Some of the experimental descriptions are from~\cite{Ilten:2022lfq}

\subsubsection{Results and status of planned experiments}

\paragraph{ArgoNeuT}

ArgoNeuT was a 0.24 ton LArTPC placed in the NuMI
beam line at Fermilab for five months in 2009-2010. The
TPC is 47[w] × 40[h] × 90[l] cm$^3$, with two instrumented
wire planes, each containing 240\,wires angled at 60\,deg
to the horizontal and spaced at 4\,mm. ArgoNeuT was
placed 100 m underground in the MINOS Near Detector
hall. A detailed description of the ArgoNeuT detector and
its operations is given in Ref.~\cite{Anderson_2012}.

A search for millicharged particles~\cite{PhysRevLett.124.131801}, a simple extension of the standard model, has been performed with the ArgoNeuT detector exposed to the Neutrinos at the Main Injector beam at Fermilab. The ArgoNeuT liquid argon time projection chamber detector enables a search for millicharged particles through the detection of visible electron recoils. A search for an event signature with two soft hits (MeV-scale energy depositions) aligned with the upstream target. For an exposure of the detector of $1.0\times10^{20}$ protons on target, one candidate event has been observed, compatible with the expected background. This search is sensitive to millicharged particles with charges between $10^{-3}e$ and $10^{-1}e$ and with masses in the range from 0.1 to 3 GeV. This measurement provides leading constraints on millicharged particles in this large unexplored parameter space region.

A similar search is planned at the SBND experiment which will improve an the ArgoNeuT limits in the mass range from approximately 100--500\,MeV.

\paragraph{SpinQuest/DarkQuest} %The SpinQuest/DarkQuest experimental concept is a high sensitivity, near-term, modest-cost opportunity to explore new parameter space in dark-sector physics scenarios. 
SpinQuest/DarkQuest is a proton fixed-target beam-dump spectrometer experiment on the neutrino-muon beamline of the Fermilab accelerator complex, where it would receive a high-intensity beam of $120$\,GeV protons from the main injector. It takes advantage of the long history of investment in the existing E906/E1039 SeaQuest/SpinQuest spectrometer experiments at Fermilab, which focus on proton parton-distribution-function measurements in Drell-Yan events.
The DarkQuest detector concept proposes to add an electromagnetic calorimeter (EMCal) detector to the SpinQuest spectrometer that will open up two additional orders of magnitude in mass parameter space. An additional tracking layer is also proposed to extend the acceptance of the experiment and enable it to withstand higher instantaneous luminosity. This will allow DarkQuest to explore dark sector signatures from $\mathcal{O}$(MeV) to $\mathcal{O}$(GeV) in a variety of new final states, thus enabling DarkQuest to be a high impact dark sector experiment on the world stage.

DarkQuest can reach interesting parameter space in several dark sector scenarios. The purely standard model signals are studied in the context of dark photon, sterile neutrinos, and axion-like particle models (ALPs), while the dark matter (DM) and rich dark sectors are captured by models of inelastic dark matter  and  strongly interacting massive particles (SIMPs). The latter models offer the possibility of explaining the dark matter of the universe in a predictive framework with concrete experimental targets. Many more details on the physics sensitivity of the experiment are given in~\cite{Berlin:2018pwi,Batell:2020vqn,Blinov:2021say}. DarkQuest can also play an important part of a potential program at Fermilab to explore how light new physics could contribute to $g-2$~\cite{g2workshop,Forbes:2022bvo}.
The physics case and more detailed \textsc{Geant}-based~\cite{GEANT4:2002zbu} simulations to understand detector performance for dark sector signatures are described in more detail in~\cite{Apyan:2022tsd}.

Further upgrades beyond DarkQuest are also envisioned. In the LongQuest proposal~\cite{Tsai:2019mtm}, three potential types of installations were proposed. They are: installing long-baseline detectors behind the iron block in the backroom to search for long-lived and millicharged particles; installing a ring-imaging Cherenkov detector or a hadron blind detector to improve particle identification and background reduction; and adding a new front-dump and fast-tracking detector to search for promptly decaying particles. More details are discussed in~\cite{Tsai:2019mtm,Apyan:2022tsd}. 

\subsubsection{Future prospective experiments}

\paragraph{PIP-II Beam Dump} Using the PIP-II proton fixed-target facility, PIP2-BD envisions a $100$~ton scale liquid argon single-phase scintillation-only detector located $18$~m from a carbon target~\cite{Toups:2022yxs}. An $\mathcal{O}$(1~GeV) proton beam colliding with the fixed target produces charged mesons. Dark matter models predict dark matter production via neutral mesons such as pions and $\eta$ mesons, also produced by the proton collisions with the fixed target.

PIP2-BD has the capability to be a world-leading probe of both the vector portal kinetic mixing and leptophobic models assuming a 5-year run of the C-PAR scenario. A different phenomenology arises if there are two new particles $\chi_1$ and $\chi_2$, where $\Delta=(m_{\chi_2} - m_{\chi_1})/m_{\chi_1}>0$. The $\chi_2$ travels to the detector and decays into a $\chi_1$ and an $e^+e^-$ pair if the $\chi_2$ is sufficiently long lived. If the $\chi_2$ decay is not kinematically allowed, the dark matter signal is detectable through up- or down-scattering off of electrons. A recent study of the detection of inelastic dark matter at JSNS$^2$~\cite{Jordan:2018gcd} showed the possibility of signal/background separation in a scintillation-only detector. 

\paragraph{FNAL-$\mu$ beam dump} FNAL-$\mu$ has been proposed to search for light dark-sector particles that dominantly couple to muons that can explain the observed muon $g-2$ anomaly~\cite{Chen:2017awl}. FNAL-$\mu$ is a muon beam-dump experiment at the muon campus of Fermilab using the existing Fermilab muon beam source with the anomalous energy deposition downstream from the dump. The proposed incident muon beam energy is around 3\,GeV, as the accelerator complex is already tuned to this energy for the Muon $g-2$ experiment. Such a beam will be completely stopped within a $1.5$\,m-thick tungsten target. Dark sector particles that are produced through muon-nucleon bremsstrahlung interactions can then visibly decay inside a $3$\,m detector equipped with an electron or photon tracker/calorimeter.

FNAL-$\mu$ has a simple and compact design that could be facilitated at the $g-2$ hall of Fermilab. It could run in parallel with the on-going Muon $g-2$ experiment. With a beam intensity of $10^7$ muons per second, a one-month run corresponding to $2.5\times 10^{13}$ muons on target is expected to reach a sensitivity of $3\times10^{-4}$ for the muonic dark scalar couplings. Such a sensitivity completely explores the parameter space for light muonic dark scalars ($m_S< 2 m_\mu$) that explains the muon $g-2$ anomaly. A one-year run or $3\times 10^{14}$ muons on target could reach a sensitivity of $\mathcal{O}(10^{-5})$ for the muonic dark scalar coupling. Combined with the E137 experiment that probed the muonic dark sector through the secondary muons~\cite{Marsicano:2018vin}, such a sensitivity could completely explored the parameter space for light muonic dark scalars scalars ($m_S< 2 m_\mu$) above the supernova 1987A cooling limit.

\paragraph{M$^3$} M$^3$ is a muon beam missing momentum experiment which expects measures $\mathcal{O}$(15\,GeV) muons on a thin active target. The secondary muon beam is produced from the 120\,GeV Main Injector proton beam at the Fermilab accelerator complex. As a muon beam experiment, M$^3$ is uniquely sensitive new physics coupled to muons, particularly light new physics related to the g-2 anomaly or muon-philic dark matter.  In this setup, a relativistic muon beam impinges on a thick active target.  The signal consists of events in which a muon loses a large fraction of its incident momentum inside the target without initiating any detectable electromagnetic or hadronic activity in downstream veto
systems. 
A two-phase experiment is envisioned based at Fermilab. 
Phase 1 with 10$^{10}$ muons on target can test the remaining parameter space for which light invisibly decaying particles can resolve the (g-2)$_\mu$ anomaly, 
while Phase 2 with 10$^{13}$ muons on target can test much of the predictive parameter space over which sub-GeV dark matter achieves freeze-out via muon-philic forces, including gauged U(1)$_{L\mu-L\tau}$.

\paragraph{REDTOP}  The Rare $\eta$/$\eta'$ Decays To Probe New Physics experiment (REDTOP) is a fixed-target meson factory~\cite{REDTOP:2022slw} searching for new physics in flavor-conserving rare decays of the $\eta$ and $\eta'$ mesons. Such particles are almost unique in the particle universe, carrying the same quantum number as the Higgs boson except for parity, and no standard model (SM) charges, the dynamics of their decays is highly constrained. Conservation of their quantum numbers impose that all electromagnetic and strong decays are forbidden at the tree-level. Rare decays, are, therefore, enhanced compared to the remaining, flavor-carrying mesons. REDTOP aims at producing $10^{14}$($10^{12}$) $\eta(\eta')$ mesons. Such a sample can probe many recent theoretical models, providing enough sensitivity to explore all four portals connecting the dark sector with the SM while also probing conservation laws.
%The $\eta$/$\eta'$ hadro-production mechanism is based on a proton beam with energy equal to $1.8~\GeV$ ($3.6~\GeV$ for the $\eta'$). Several intra-nuclear baryonic resonances ($\Delta$s, N(1440), N(1535), \etc) are created in nucleon-nucleon collisions, whose decay produce an $\eta$ or $\eta'$ meson. The inclusive $\eta/\eta'$ production cross-section has been calculated for $p$-Li scattering, using several theoretical models. The average value is $1.2\times10^{-23}~\si{centi\meter}^{-2}$ for the $\eta$ meson, and $6.8\times10^{-20}~\si{centi\meter}^{-2}$ for the $\eta$' meson. Assuming $10^{18}$ protons on target (POT) per year, the desired yield can be achieved in a few years of running. A CW proton beam delivering $1\times10{}^{11}$ POT$/\second$ would generate a rate of inelastic interactions of about $0.7~\GHz$ and an $\eta$-meson yield of $4\times10{}^{6}~\eta/\second$, corresponding to $4\times10{}^{13}~\eta/\mathrm{year}$. The beam power corresponding to the above parameters is approximately $30~\watt$ ($60~\watt$ for the $\eta$'), of which less than 1\% (or $300~\si{\milli\watt}$) is absorbed in the target systems. 
At the beam energies considered above, the $\eta$ and $\eta$' mesons are produced almost at rest in the lab frame, receiving only a small boost in the direction of the incoming beam. Consequently, a hermetic, collider-style, detector covering most of the solid angle, is one of the requirements for REDTOP. 
%A schematics of REDTOP detector is shown in \cref{fig:REDTOP}.

% The REDTOP target system is split in 10 foils, each $\sim0.7~\si{milli\meter}$ thick, of a low-$Z$ material (Li or Be), spaced $10~\si{\centi\meter}$ apart. This improves the measurement of the $z$ coordinate of the $\eta$ production vertex, while minimizing the multiple scattering affecting the $\eta$ decay products when they escape from the target. It also allows any eventual BSM, long-lived particle, and its decay products, to leave the target undisturbed. The vertex detector is the innermost active component of REDTOP. It has four main tasks: identifying events with a detached vertex; participating in the reconstruction of charged tracks originating in the target region; rejecting photon converting in the target materials; and reconstructing tracks with very low transverse momentum. An event with a detached vertex found not belonging to the very few produced $K_{short}$ will be a clear indication that new physics is being observed.

The REDTOP sensitivity studies performed probe the conservation of discrete symmetries of the universe. $CP$ symmetry can be explored at REDTOP in a number of ways, thanks to the large statistics available for certain channels and a well known background. In all cases, $CP$-violation is observed via asymmetries of the $\eta$ decays. An almost unique technique proposed by REDTOP is based on the measurement of the polarization of the muon, copiously produced in some decays of the $\eta$ mesons and tagged by fully reconstructing the decay of the latter. This could be achieved either in the high-granularity, polarization-conserving, calorimeter or in a dedicated polarimeter.
These sensitivity studies performed for REDPTOP also include the search for new particles produced in rare decays of the $\eta$ and $\eta'$ mesons. Most of these particles could be visibly observed in the detector, assuming that they decay within about $50$\,cm from the $\eta$ or $\eta'$ meson production point. With the proposed statistics, REDTOP has excellent sensitivity to all four portals connecting the standard model with the dark sector.

%REDTOP is especially well suited to appreciably improve the precision of lepton universality measurements, thanks to the very large number of semi-leptonic decays of the $\eta$ meson which can be fully reconstructed, as well as a well known background. Several sensitivity studies on lepton universality have been performed, and indicate that the experiment is able to decisively probe such symmetry with unprecedented precision. Despite being more challenging at an $\eta$/$\eta'$ factory, the violation of leptonic flavor could also be probed. REDTOP could not compete on this measurement with dedicated experiments based on intense muon beams. Still, the different mechanism of production of the muons would make this exploration interesting and worth pursuing. 

The beam requirements for REDTOP are modest, and several existing laboratories worldwide have the capability to host the experiment. The REDTOP collaboration, consisting of more than one hundred scientists from fifty-three institutions, has engaged in a broad exploration, and identified BNL, Fermilab, GSI, and HIAF as possible candidates to host REDTOP.

\afterpage{\clearpage}
%-------------------------------------------

%-------------------------------------------
\subsection{Search for light DM with primary electron beams: prospects at SLAC -- {\it R.~Pottgen} }
\label{ssec:pottgen}
{\it Author: Ruth Pottgen, <ruth.pottgen@hep.lu.se>} \\

% \documentclass[12p]{article}
% \usepackage[utf8]{inputenc}
% \usepackage{graphicx}
% \usepackage{lineno}
% \usepackage{geometry}
% \geometry{
% a4paper,
% %total={170mm,257mm},
% left=25mm,
% right = 25mm,
% top=25mm,
% bottom = 25mm
% }
 
% \title{Search for light DM with primary electron beams: Prospects at SLAC}
% \author{Ruth Poettgen}
% \date{December 2022}

% \begin{document}

% \maketitle

\subsubsection{Introduction}
\begin{figure}
\centering
\includegraphics[width=0.7\textwidth]{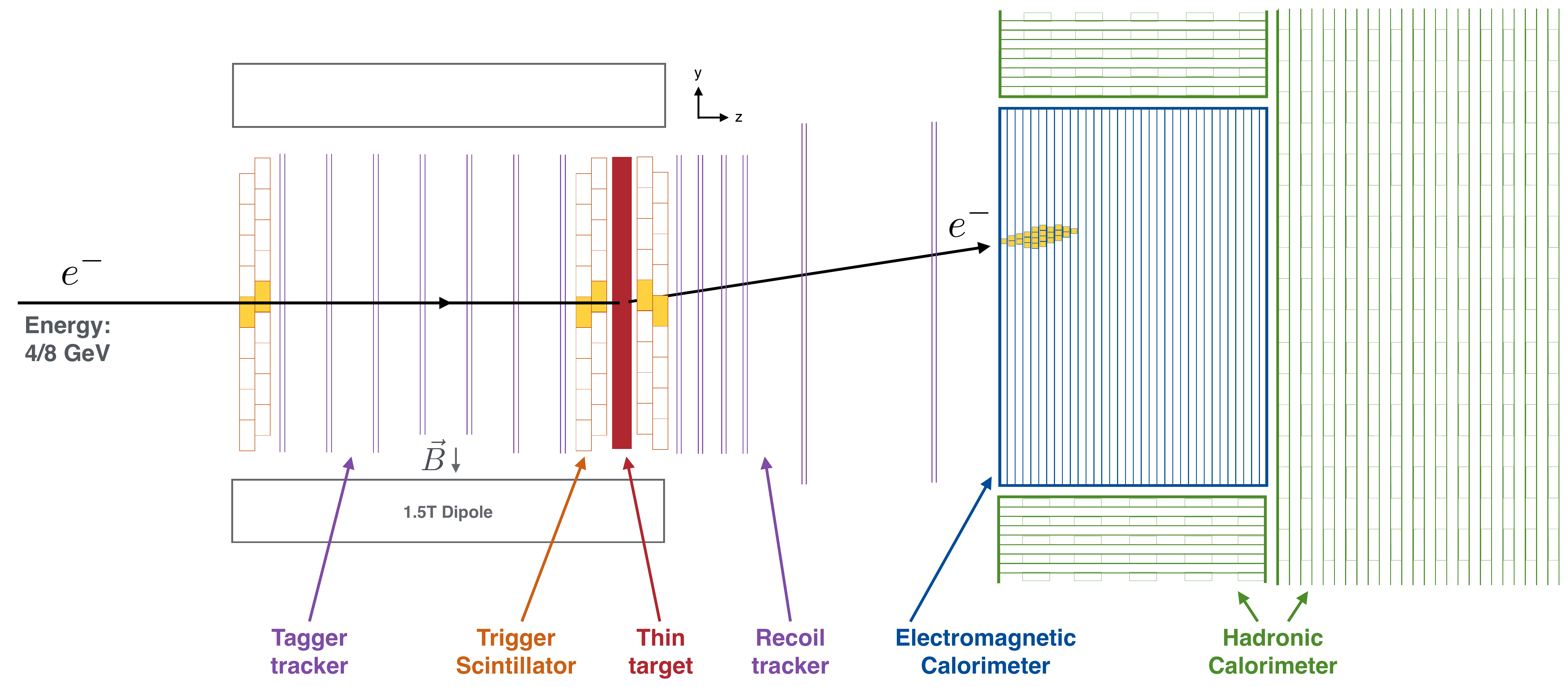}
\caption{A sketch of the conceptual layout of the LDMX detector intended to measure the missing energy and missing transverse momentum of scattered electrons.}
\label{fig:DetSketch}
\end{figure}

At SLAC, a primary, few-GeV electron beam is available via the accelerator of the Linac Coherent Light Source II (LCLS-II)~\cite{osti_1029479}. The Light Dark Matter eXperiment (LDMX) aims to make use of this opportunity to search for sub-GeV dark matter particles. The layout of the experiment is shown in Fig.~\ref{fig:DetSketch}: The incoming electrons scatter on a thin (tungsten) target, and in the interactions with the nuclei dark matter could be produced for example via bremsstrahlung of a dark photon ($A'$) and its subsequent decay into dark matter. This would lead to a significant energy loss of the electron, and this missing energy is determined from a measurement of the remaining electron energy in an electromagnetic calorimeter (ECal). 
Momentum conservation in the emission of a heavy particle leads to a sizable transverse momentum of the outgoing electron, or, equivalently, missing transverse momentum. 
Measuring this momentum forms the second pillar of the new-physics discovery strategy. 
The measurement is provided by tracking detectors in a magnetic field before and after the target. 
The main physics trigger will rely on the (missing) energy measurement in the ECal in combination with a count of the number of incoming electrons provided by scintillator arrays upstream of the tracking and close to the target. 
The design is completed by a large hadronic calorimeter (HCal) needed to reject backgrounds that look like signal events in the other detector systems. 

The dark matter signal signature in this setup would be a \emph{scattered electron that has lost most of its energy and obtained significant transverse momentum}, and no other activity anywhere in the detector. 
The goal of LDMX is to \emph{individually} measure up to $10^{16}$ electrons on target (EoT), which places a number of requirements on the beam, as well be discussed in Sec.~\ref{sec:Beam}. Before going there, I note that the setup described above offers sensitivity not only to sub-GeV dark matter, but in fact to a rich spectrum of physics beyond the Standard Model~\cite{Berlin:2018bsc,Akesson:2022vza}, both invisible and visible signatures, and in addition will allow first-time measurements of photo- and electronuclear processes in hitherto little explored phase-space regions~\cite{Ankowski:2019mfd}.

\subsubsection{The Beam} 
\label{sec:Beam}
The ambition to measure up to $10^{16}$ electrons individually defines the desired properties of the beam. 
Resolving individual electrons requires them to be spread out in time and space, which translates into a low intensity beam (meaning a small number of electrons per bunch) in combination with a large beam spot. To reach the envisaged total numbers of electrons, a high duty cycle is then needed. Favourable signal-to-background ratios for the sub-GeV dark matter search are achievable for beam energies in the range of 4\,GeV to 20\,GeV. 
At SLAC, it is possible to deliver exactly such a beam using the LCLS-II accelerator without interfering with its core science programme. 
LCLS-II will first operate at 4\,GeV, but an upgrade is planned in 2026/27 to increase the energy to 8\,GeV. 
The bunch frequency is 186\,MHz, but LCLS-II populates only one out of 2000 bunches with electrons for the photon science. The idea for LDMX is to use a fraction of the remaining bunches to provide a small number ($<10$) of electrons to the experiment with a frequency of about 40\,MHz. 
This requires a dedicated beamline (LESA -- Linac to End Station A~\cite{Markiewicz:2022xis}) to transport the electrons to the experimental hall that will host the detector. 
The first part of LESA is currently under construction as an S30 Accelerator Improvement Project and includes the kicker magnet as well as about 100\,m of beam line. LESA is expected to deliver beam to the experimental hall in autumn 2023.

\subsubsection{The LDMX Detector}
\begin{figure}[!htb]
\centering
    \includegraphics[width=0.49\textwidth]{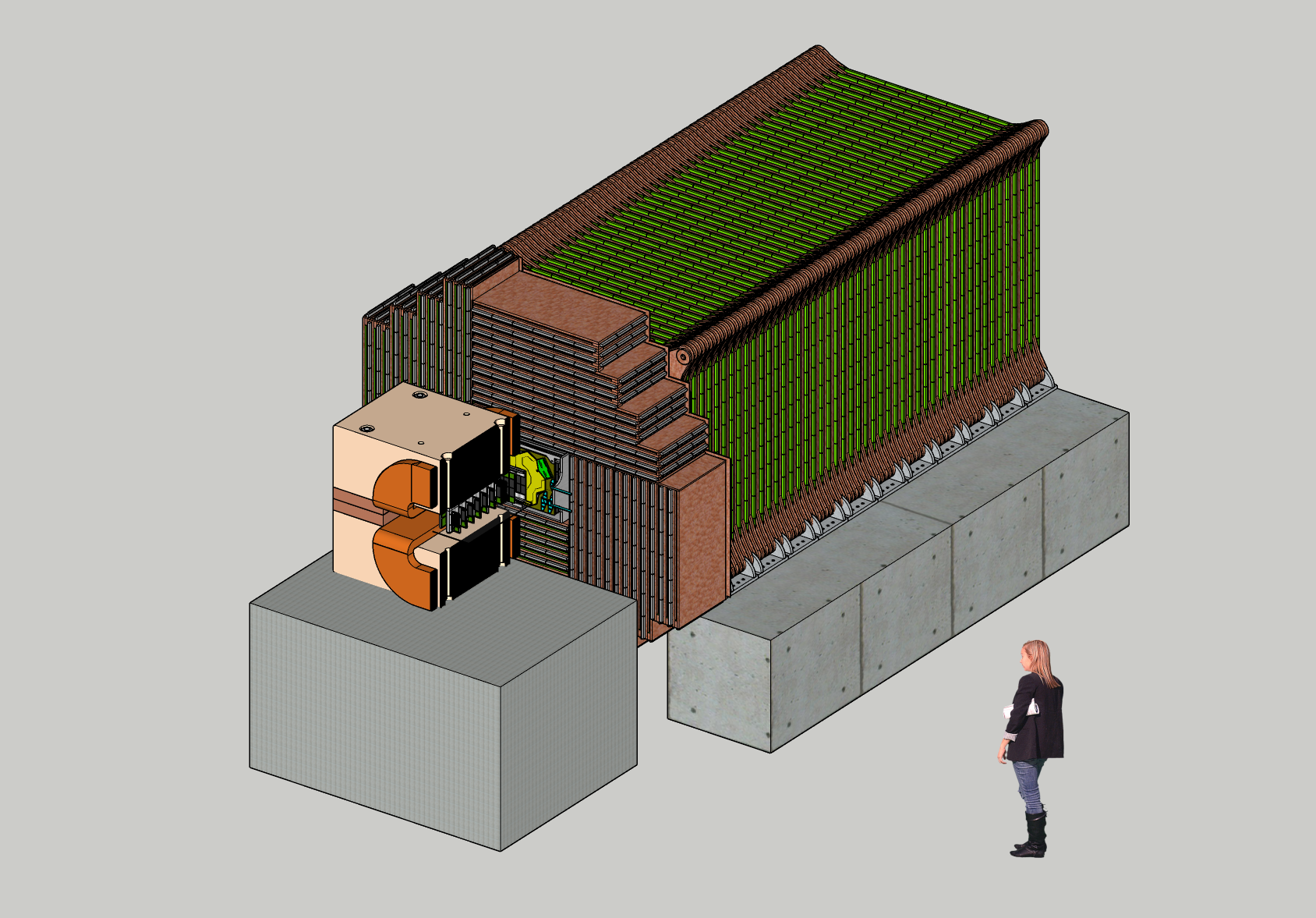}
    \includegraphics[width=0.49\textwidth]{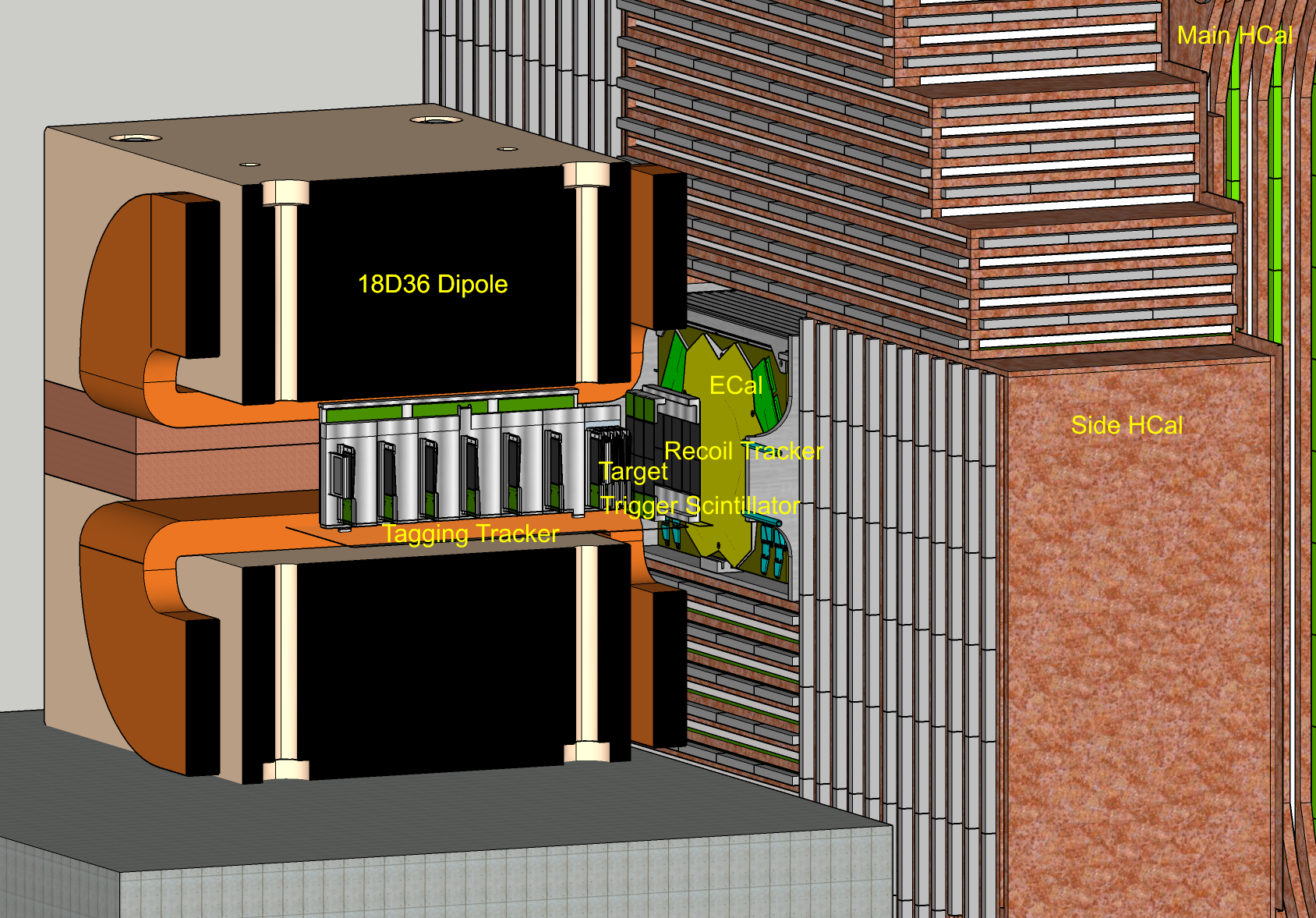}
    \caption{Left: An overview of the LDMX detector solid model. Right: A cutaway overview of the LDMX detector showing, from left to right, the trackers, trigger scintillator and target inside the spectrometer dipole, the ECal, and the Side and Main HCal.}
    \label{fig:LDMX_overview}
\end{figure}

At the beam energies available at SLAC there are no irreducible physics backgrounds to the dark matter signature, since all SM background processes produce additional particles that are in principle detectable. 
The sensitivity of the experiment thus becomes mainly a question of how efficiently these additional particles can be detected. Figure~\ref{fig:LDMX_overview} shows an overview of the planned detector design. The three main detector systems all draw on solutions developed for existing experiments, giving confidence that the design goals can be reached~\cite{LDMX:2018cma}. The tracking system is a simplified version of the Silicon Vertex Tracker of the HPS experiment~\cite{HanssonAdrian:2015cqx}, the electromagnetic SiW-sampling calorimeter uses the same modules and electronics as the CMS High-Granularity Calorimeter upgrade~\cite{CMS:2012tda}, while the hadronic veto system, a sampling calorimeter with steel absorber and plastic scintillator bars as active material, was inspired by the Cosmic Ray Veto of the Mu2e experiment~\cite{Artikov_2018,Mu2e:2014fns}. 

The main backgrounds arise from bremsstrahlung of a hard photon in the target. Particularly challenging are events in which the photon subsequently undergoes a photo-nuclear (PN) reaction in the ECal that produces only neutral particles in the final state, for example a single neutron or $K^0_L$. Although such processes occur at rate of about $10^{-9}$ relative to the incoming electron rate, this is still several orders of magnitude above the expected signal rate, meaning that these backgrounds have to be vetoed with a corresponding efficiency. This is the main task and design driver for the hadronic calorimeter. Detailed simulation studies show that the combination of detector systems as outlined above is capable of reducing photon-induced backgrounds to less than one event for $4\times10^{14}$ EoT~\cite{LDMX:2019gvz}, the statistics planned to be gathered during the first year of operation. Importantly, this is achieved \emph{without} making use of the transverse momentum information, leaving this as a powerful handle in case of unexpectedly high background levels and for signal characterisation, as discussed in the next section.

\subsubsection{Sensitivities to sub-GeV dark matter}
\begin{figure}
    \centering
    \includegraphics[width=0.6\textwidth]{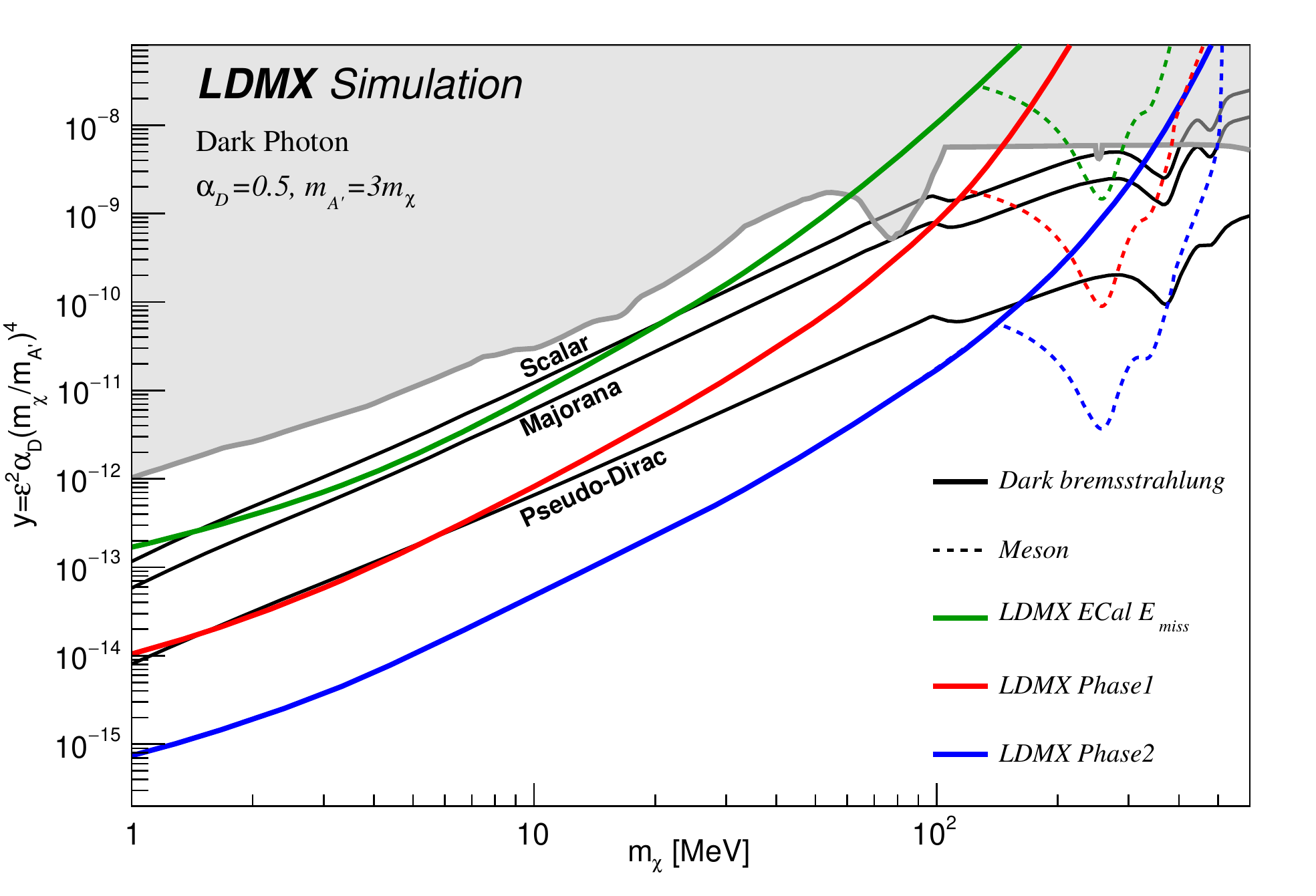}
    \caption{Projected sensitivity in the $y$ vs $m_{\chi}$ plane for a baseline LDMX run with $4\times10^{14}$~EoT with a 4~GeV beam energy (Phase 1) and $10^{16}$~EoT (Phase 2) with an 8~GeV beam energy. Projections for an early missing-energy analysis of $10^{13}$~EoT using the ECal as a target are also shown. 
    Backgrounds are assumed to be at the level of $<1$ event.
    Benchmark thermal relic targets are shown as black lines. 
    The grey region represents constraints from previous experiments. Figure from Ref.~\cite{Akesson:2022vza}.
    }
    \label{fig:reach}
\end{figure}

The LDMX sensitivity to sub-GeV dark matter is illustrated in Fig.~\ref{fig:reach} in terms of the excluded interaction strength $y$ and dark matter mass $m_{\chi}$ in the framework of a dark photon model. 
Already the first data at 4\,GeV, either with a thin target (red line) or using the ECal as an active target (green line) will provide sensitivity to several thermal targets. The second run at 8\,GeV and collecting larger statistics (blue line) allows to explore the phase space well beyond the thermal targets for masses up to O(100)\,MeV. 

\begin{figure}
\centering
\includegraphics[width=0.6\textwidth]{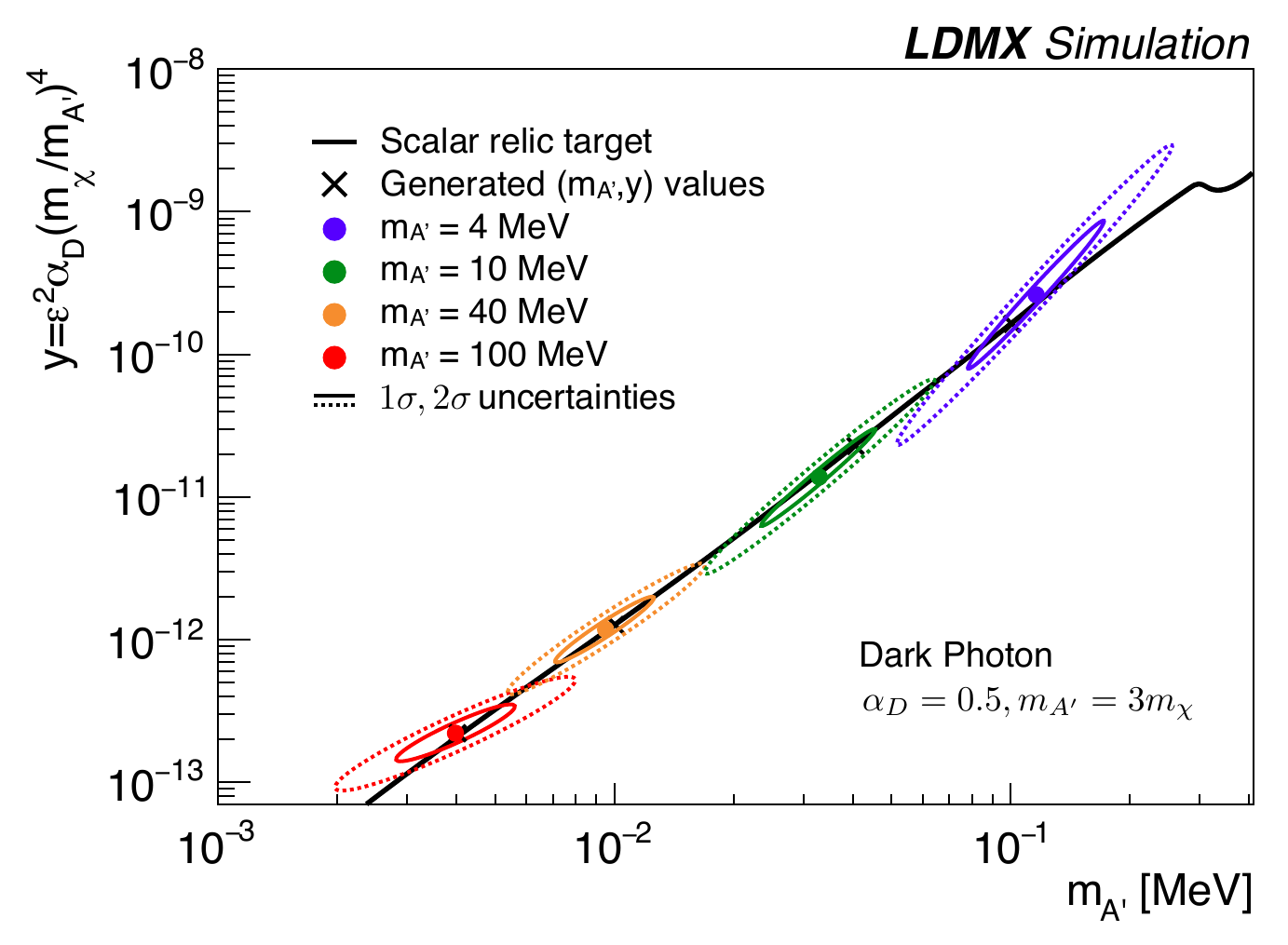}
\caption{Reconstructed mediator mass and couplings are shown for four test samples at $4\times10^{14}$~EoT.  The values of $m_{A'}$ and $y$ used to produce each sample are indicated by black crosses, while $1\sigma$ and $2\sigma$ error ellipses for the fit are denoted with solid and dashed lines, respectively.}
\label{fig:massReco}
\end{figure}

As highlighted above, these exclusion projections do not yet use the transverse momentum information of the outgoing electron. This is a strong discriminator between a dark matter signal and the SM backgrounds, and in fact can be used to constrain the dark photon mass in case an excess is observed. The potential for this mass reconstruction is shown in Fig.~\ref{fig:massReco}, demonstrating that LDMX would be able to provide an estimate of the dark photon mass as an input to further searches and characterisation of a potential signal.

\subsubsection{2022 Test Beam}
Key to achieving the sensitivities presented above is a detailed understanding of all detector and readout components. A crucial milestone for this was a successfully completed test beam campaign in April 2022. Prototypes for two detector systems were tested, the HCal and the trigger scintillators, with the aim of testing the mechanical and electronics designs as well as the readout chain and data acquisition, and eventually validating simulations against the obtained data. 
The HCal prototype comprised 19 steel+scintillator layers. The scintillator bars and their readout via Si-Photomultipliers (SiPMs) were the same as the ones to be used in the final experiment. For the trigger scintillator different prototypes with either plastic scintillator bars or a combination of plastic and LYSO bars, in each case read out via SiPMs, were tested. 

\begin{figure}[!htb]
\centering
    \includegraphics[width=0.43\textwidth]{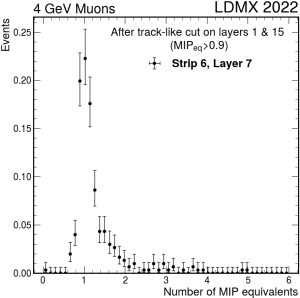}
   \includegraphics[width=0.55\textwidth]{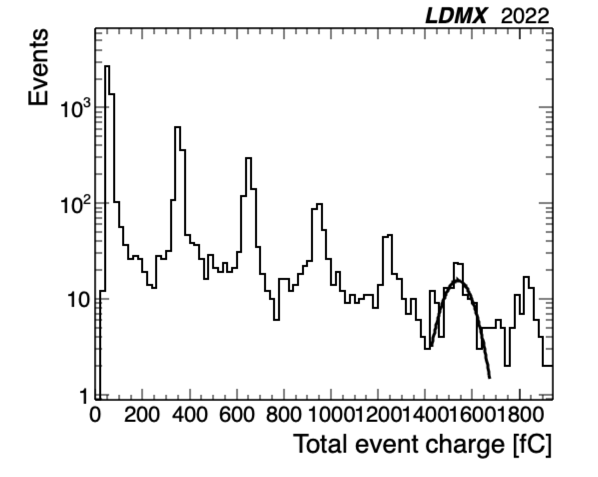}
    \caption{Left: Distribution of the sum of ADC counts for layer 7, bar 6 of the HCal prototype for 4 GeV muons, normalized by the measured values corresponding to one minimum ionizing particle, referred to as MIP equivalents (MIPeq). 
    Right: Distribution of the integrated charge recorded per event for one channel of the trigger scintillator prototype. }
    \label{fig:TBhcalMIPe}
\end{figure}

%\begin{figure}
%\begin{minipage}[r]{0.4\textwidth}
%    \includegraphics[width=\textwidth]{Figures/HCalProto.png}
%  \end{minipage}\hfill
%  \begin{minipage}[l]{0.5\textwidth}
%    \caption{
%       The assembled HCal prototype in the test beam area. The gantry on which the trigger scintillator prototype is mounted is also visible on the left, in front of the first absorber layer.
%    } \label{fig:protos}
%  \end{minipage}
%\end{figure}
%Figure~\ref{fig:protos} shows the assembled prototypes in the beam area of the T9 beam line in the East Hall at CERN. 
During two weeks, data was collected with electron, muon and hadron beams at energies between 0.1 and 8\,GeV.
The left plot in Fig.~\ref{fig:TBhcalMIPe} shows the distribution of ADC counts (normalised to the equivalent count for a minimally ionising particle, MIP) in one bar of the HCal prototype for a run with 4\,GeV muons. 
To ensure a high likelihood of a muon traversing the respective bar, only events in which the corresponding bars in a front layer and a layer further in the back measure more than 0.9 MIP equivalents are shown.

The right of Fig.~\ref{fig:TBhcalMIPe} shows the total charge measured per event in one channel of the trigger scintillator prototype with only plastic bars. This is zoomed in on the low-charge region, such that the first peak represents the pedestal of the system, the others are an integer number of 1 to 6 SiPM pixels firing. The sixth peak shows an example fit used to extract the central value of each peak for calibrating the SiPM gain and pedestal. This demonstrates that each channel can be calibrated based on the first O($10^4$) events per run.
While the analysis of the test beam data is still ongoing, these early plots showcase that good quality data was obtained and understanding of this data is developing.

%\begin{figure}[!htb]
%\centering
 %  \includegraphics[width=0.55\textwidth]{Figures/totalEventCharge_mar15_1708_chan4.png}
  %  \caption{Distribution of the integrated charge recorded per event for one channel of the trigger scintillator prototype. }
   % \label{fig:TBtsPEs}
%\end{figure}

\subsubsection{Conclusion and Outlook}
The primary electron beam becoming available at SLAC via the LESA beamline currently under construction opens exciting physics opportunities, the main motivation being the search for sub-GeV dark matter and other new physics in forward electron scattering. Using this beam, the LDMX experiment will be able to achieve outstanding sensitivity over the course of a few years. The collaboration has achieved several milestones in detector development and physics studies, and is on track to start construction in late 2023. 

%\subsubsection{Acknowledgements}
%R. Pottgen gratefully acknowledges the support through the Swedish Research Council, grant nr 2019-03436.

% \bibliography{references}
% \bibliographystyle{ieeetr}

% \end{document}

\afterpage{\clearpage}
%-------------------------------------------

%-------------------------------------------
\subsection{{\emph New ideas}: Status of the SND@LHC experiment -- {\it O. Lantwin}}
\label{Lantwin}
{\it Author: Oliver Lantwin, <oliver.lantwin@cern.ch >}

\subsubsection{Introduction}

The SND@LHC experiment\cite{SNDLHC:2022ihg} is a new experiment at the LHC situated 480 m away from the ATLAS interaction point (IP),
in the TI18 former service tunnel.
This location is symmetric to the TI12 tunnel where FASER is located.
At this location,
the experiment is shielded from the IP by 100 m of rock,
with charged particles also deflected by the LHC magnets.
The experiment covers a pseudorapidity range of $7.2<\eta<8.4$.
This coverage is chosen for the study of neutrinos originating in heavy-flavour decays \cite{Buontempo:2018gta,Beni:2019gxv,Beni:2020yfy}

\begin{figure}
  \centering
  \includegraphics[width=0.7\textwidth]{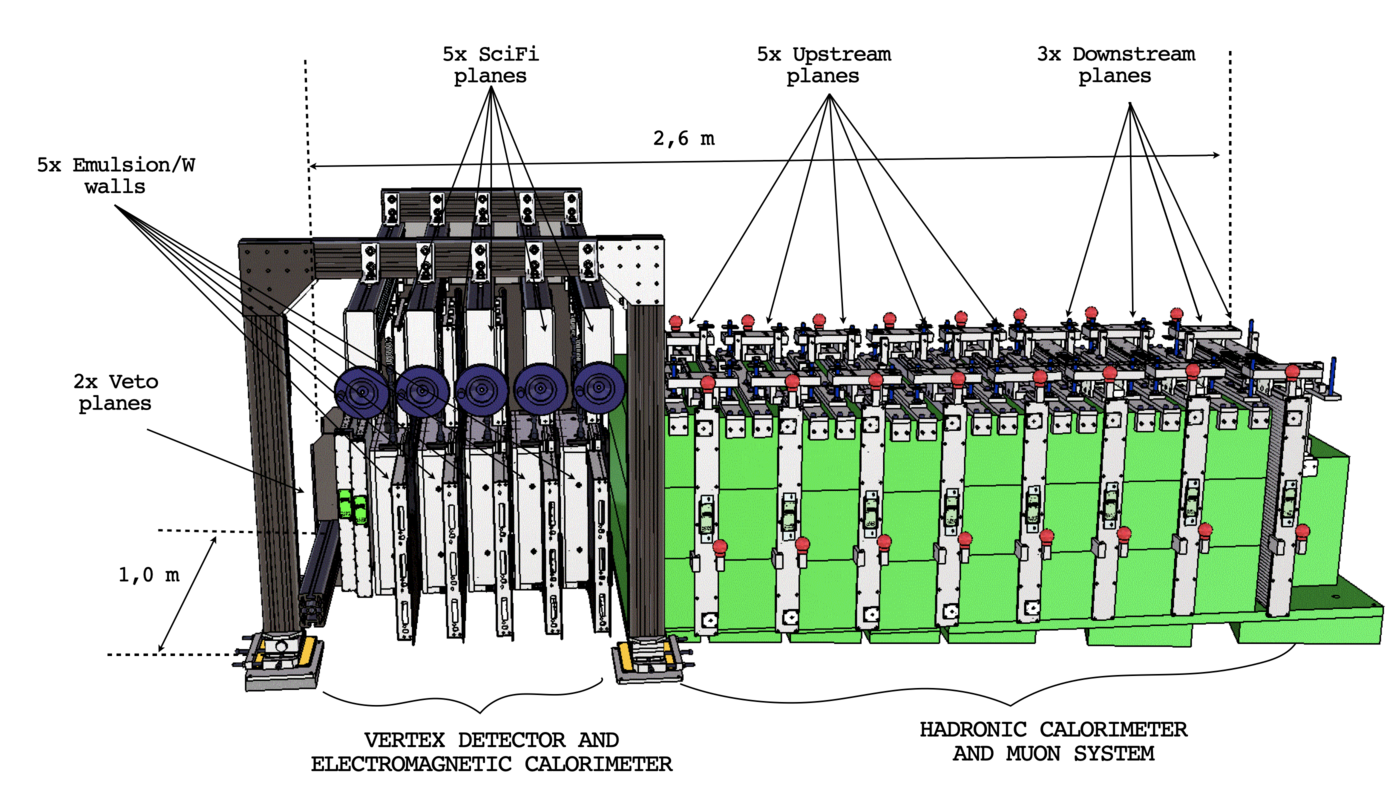}
  \caption{Overview of the SND@LHC experiment. Taken from Ref.~\cite{SNDLHC:2022ihg}.}
  \label{fig:exp}
\end{figure}

The experiment itself,
which is shown in Fig.~\ref{fig:exp},
consists of an instrumented scattering target
which fulfils the role of vertex detector and electromagnetic calorimeter,
with emulsion-cloud chambers for tracking, and scintillating fibres for timing and the energy measurement of showers,
and a muon system made up of iron walls and scintillating bars,
which also serves as a hadronic calorimeter.
In its first phase in Run 3, the experiment is expected to collect $250\;\mathrm{fb}^{-1}$.

The instrumented target offers $830\;\mathrm{kg}$ of target mass,
with a cross-section of $390\times 390\;\mathrm{mm}^{2}$.
The nuclear emulsions are replaced every $25\;\mathrm{fb}^{-1}$ or about three times per year.
About 2000 neutrino events of all species are expected in Run 3.

The experiment has now seen its first data,
with first performance studies in electronic and emulsion data currently underway,
with preliminary results looking very promising.

\subsubsection{Feebly interacting particles at SND@LHC}

For feebly interacting particles (FIPs),
SND@LHC is sensitive to both scattering and decay signatures.
A first study of sensitivities has been performed in Ref.~\cite{Boyarsky:2021moj}.
While this study is theoretical,
and several experimental assumptions need to be checked,
it gives us a first idea of SND@LHC's capabilities for feebly interacting particle searches.

\subsubsection{Scattering signatures}

\begin{figure}
  \centering
  \includegraphics[width=0.6\textwidth]{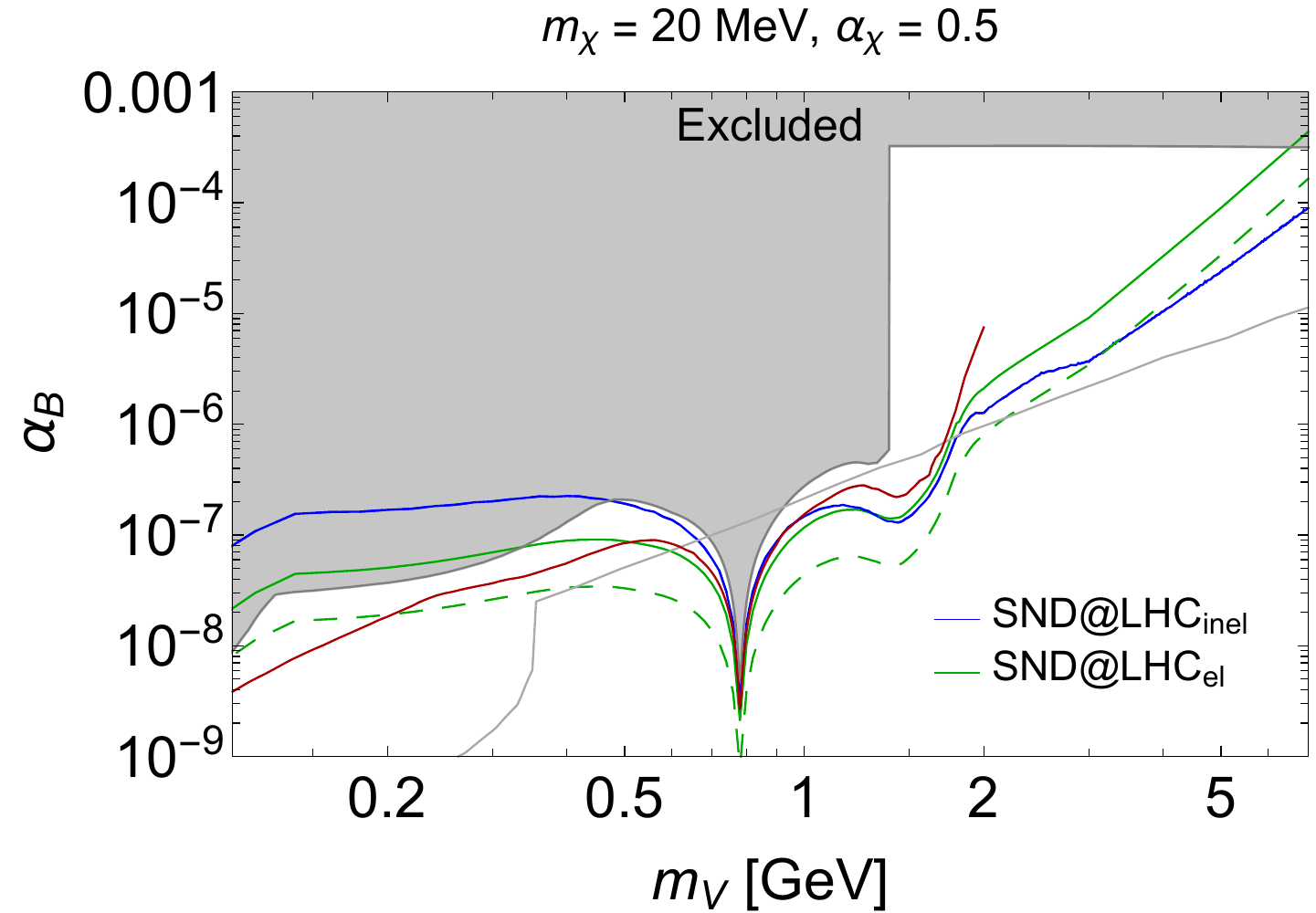}
  \caption{Sensitivity to light dark matter coupled to a leptophobic mediator. Taken from Ref.~\cite{Boyarsky:2021moj}.}
  \label{fig:ldm}
\end{figure}

The sensitivity for scattering is model-dependent.
Consider as an example light dark matter (LDM) coupled to the Standard Model particles via a leptophobic portal,
which could escape detection at many experiments looking for LDM using dark photons created in electron-positron annihilation.

At LHC energies,
elastic scattering of neutrinos is strongly suppressed,
such that elastic scattering of FIPs is essentially free from irreducible background.
While this signature is experimentally challenging,
and requires a dedicated emulsion-only analysis,
it allows covering a large range of parameter space,
especially at low masses.

In inelastic scattering,
there is a significant irreducible neutrino background,
but the ratio between neutral-current and charged-current events,
which is well predicted for neutrinos,
could allow searching for deviations from this ratio.
However,
it will need to be studied whether this ratio can be used experimentally as it is also an important handle for the understanding of the detector performance.

Preliminary sensitivities for leptophobic light dark matter are shown in Fig.~\ref{fig:ldm}.

\paragraph{Decay signatures}

\begin{figure}
  \centering
  \includegraphics[width=0.49\textwidth]{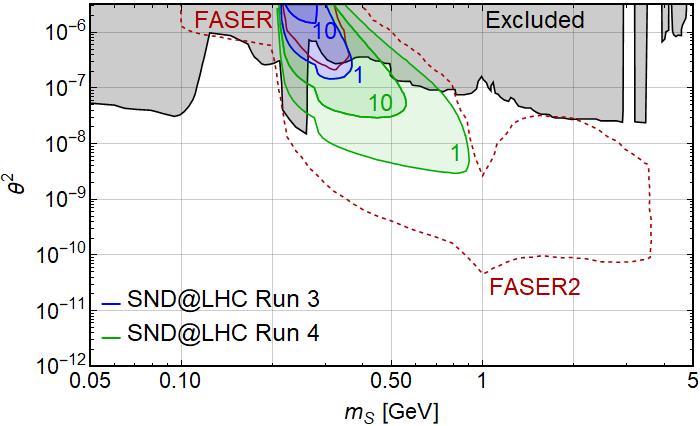}
  \includegraphics[width=0.49\textwidth]{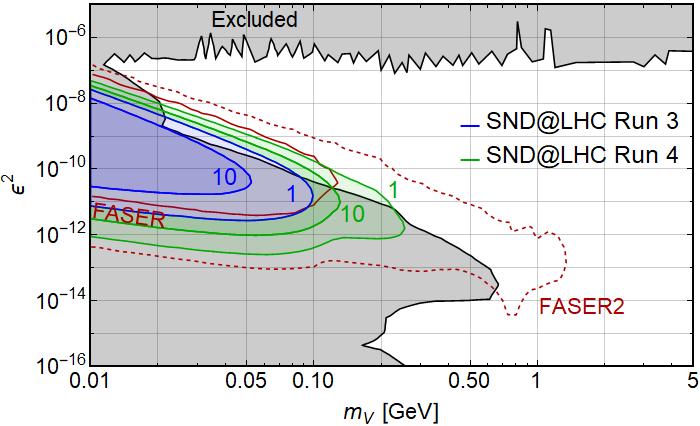}
  \includegraphics[width=0.49\textwidth]{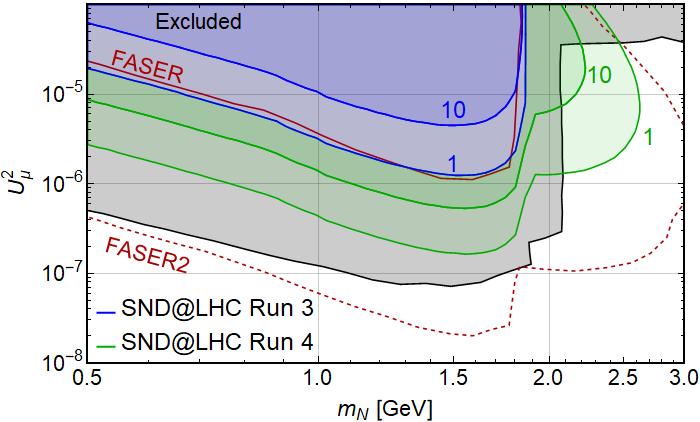}
  \caption{Sensitivity of SND@LHC and AdvSND-far (SND@LHC Run 4) for scalar, vector and heavy-neutral lepton mediators, taken from Ref.~\cite{Boyarsky:2021moj}.}
  \label{fig:fipdecay}
\end{figure}

For decays of FIPs,
the signature generally consists of a vertex with two charged tracks pointing back to the IP.
While several final states are possible,
the muonic channel is likely the experimentally favourable one.
Full studies of the backgrounds for these signatures will be needed,
with muon-induced deep inelastic scattering,
neutrino deep inelastic scattering and combinatorial backgrounds likely being the dominant backgrounds.

A preliminary estimation of the sensitivities for three common benchmark models,
as defined in Ref.~\cite{Beacham:2019nyx},
is shown in Fig.~\ref{fig:fipdecay}.

In Run 3,
SND@LHC is not expected to be competitive for decay signatures,
so that this run should be considered a proof-of-concept for AdvSND-far in Run~4,
which becomes competitive for several models.
The AdvSND-near detector,
which has a larger acceptance due to the proximity to the IP,
is not yet taken into account for the Run 4 sensitivity.
Due to the off-axis location,
AdvSND-near could benefit from reduced backgrounds.

\subsubsection{AdvSND and outlook}

The AdvSND-far and AdvSND-near detectors are two upgraded detectors planned for Run 4.
Due to the increased intensity at the HL-LHC,
SNDL@HC will need to switch from nuclear emulsions to an electronic vertex detector.
Additionally,
it is foreseen to have a magnetic spectrometer as part of AdvSND-far to allow charge-determination and momentum measurement of muons.
An overview of the key numbers is given in Tab.~\ref{tab:advsnd}.

While AdvSND-far could be situated in SND@LHC's current location in TI18,
it could also be adapted for the proposed Forward Physics Facility\cite{Feng:2022inv}.

In addition to the far detector,
a near detector called AdvSND-near,
could provide normalisation and significantly reduce systematic uncertainties,
in part due to an overlap in pseudorapidity with LHCb.

For FIP searches,
the near detector could significantly improve the acceptance of AdvSND in Run 4.
For both detectors the optimisation and choice of technologies are underway.

\begin{table}
\centering
  \begin{tabular}{r|c|c|c|l}
    & $\eta$ & mass [t] & surface [$\mathrm{cm}^{2}$] & distance [m] \\
    \toprule
    SND@LHC & $[7.2,8.4]$ & 0.83 & $39\times 39$ & 480  \\
    AdvSND-near & $[4.0,5.0]$ & 5 & $120\times 120$ & 55  \\
    AdvSND-far & $[7.2,8.4]$ & 5 & $100\times 55$ & 480 (FPF:630) \\
  \end{tabular}
  \caption{Overview of the main properties of SND@LHC and the AdvSND detectors. AdvSND-far will also have a magnet to allow for charge separation.}
  \label{tab:advsnd}
\end{table}

\subsubsection{Conclusion}

SND@LHC is a recently approved experiment for neutrino physics and feebly interacting particle searches at the LHC,
which has started taking data in April 2022
and is set to collect $250\;\mathrm{fb}^{-1}$ in Run 3.
In the scattering channels,
SND@LHC allows probing FIPs
and it is particularly powerful for the study of leptophobic light dark matter.
With the AdvSND upgrade for Run 4 of the LHC,
SND@LHC would also become competitive in searches for the decay of FIPs.
First results from performance studies are expected soon.

%-------------------------------------------

%-------------------------------------------
\subsection{\emph{New ideas:} Resonant dark sector searches at positron beam dumps -- {\it L.~Darm\'e}}
\label{darme}
{\it Author: Luc Darme, <l.darme@ip2i.in2p3.fr>}

\subsubsection{Introduction}
Despite strong improvements in both detector sensitivities and experimental luminosities in the next generation of accelerator-based FIP searches, the fundamental strategies for FIP production have not much evolved since the early days of beam dump experiments. Focusing for instance on the case of lepton-based experiment, the most recent limits  on a vector FIP $X$ from the NA64 experiment~\cite{NA64:2019auh,NA64:2021aiq} relies still on the bremsstrahlung production off a nucleus $N$, $e N \to e N X$,
as this is the dominant channel for $e^-$ based experiments. This production mechanism was  used for example in the  recasting ~\cite{Bjorken:2009mm,Andreas:2012mt} of analysis from historical beam dump experiments such as E141~\cite{Riordan:1987aw}, KEK~\cite{Konaka:1986cb} and Orsay~\cite{Davier:1989wz}. It however suffers from a strong $\alpha_{em}^2$ suppression. 
In order to unlock more effective FIP  production mechanisms in lepton-based experiment one can instead  rely on a positron beams. Indeed, while producing and accelerating positrons typically implies losses in intensity and energy, the production rate can increase dramatically if the Center-of-Mass (CoM) energy of the $e^+ e^-$ system matches precisely the FIP  mass $M_{X}$ and allows the process $e^+ e^- \to X$. 
For ultra-relativistic positrons impinging on target electrons assumed at rest,
the resonant condition is achieved for a beam energy $E_{\rm  res}$ given by
\begin{equation}
\label{eq_ResonantProd_ResMass}
E_{\rm  res}  = \frac{ M_{X}^2}{2 m_e} \ .
\end{equation}
Fulfilling this condition is the key requirement of FIP resonant production and determine to a large extent the experimental strategies. For FIPs in dark matter-motivated models  with relatively large width $\Gamma_X$, the relation in Eq.~\eqref{eq_ResonantProd_ResMass} is relaxed up to energy variations of relative order $\Gamma_X / M_X$. 

Two strategies can be considered to obtained positrons of the required energy:
\begin{itemize}
    \item Use the energy loss and secondary $e^+ e^-$ production from  electromagnetic showers in the target to ``scan''  various positron energies~\cite{Nardi:2018cxi,Marsicano:2018krp,Marsicano:2018glj,Celentano:2020vtu,Battaglieri:2021rwp,NA64:2022rme}. This  requires a target with thickness of the order of the radiation length. This strategy is particularly well-suited to study missing energy or displaced vertices signatures when the FIP or its decay product escape the target. On the other hand, the background from the electromagnetic shower typically swamp a visible $X$ prompt decay signature. 
    \item Directly scan with the beam energy around the resonant energy and search for a ``bump'' signature~\cite{Darme:2022zfw}. Since 
    for resonant production the rate is strongly enhanced, thin targets can  
    be used to avoid degrading the signal.
\end{itemize}

We will briefly present the underlying physics for the two above approaches in turn, then study an explicit application of a ``scan'' strategy motivated by the ATOMKI anomaly~\cite{Krasznahorkay:2015iga,Krasznahorkay:2021joi,Krasznahorkay:2022pxs}.

\subsubsection{Two strategies for resonant production}
\label{sec:DARME2}

\paragraph{Thick target strategy}

As they propagate through a thick target, the energy of the positrons is significantly degraded. The estimation of the proper production rate must thus account for the varying energy distribution. As long as the precise production point in the target is not required, this can be naturally described in terms of the differential track-length $\frac{dT_\pm(E)}{dE}$ which sums up the total length of target material traversed by positrons or electrons of a given energy. 

Neglecting the beam spread compared to the energy loss which occurs in the target, we can obtain the number of FIPs produced per positron-on-target from a process with cross-section $ \sigma (E)$  as:
\begin{align}
\label{eq_ResonantProd_Nperpot}
\mathcal{N} = \frac{\mathcal{N}_A X_0 \rho}{A} \int_0^{E_{+}}  dE~\frac{dT_\pm(E)}{dE}~  \sigma (E) = \mathcal{N}_A X_0 \frac{ \rho \, Z}{A} \frac{dT_\pm(E_{\rm res})}{dE}~   \frac{g_e^2}{2 m_e}   \  \ ,
\end{align}
where  $X_0$ is the radiation length of the material, $A$ its atomic mass with $\mathcal{N}_A  = 6.022 \times 10^{23}$, $\rho$ its mass density, and with $E_{+}$ the intial energy of the positron beam in the lab frame. In the second part of this expression, we used the resonant cross-section for a FIP with $e^\pm$ coupling $g_e$ in the narrow-width approximation given by:
\begin{equation}
\label{eq_ResonantProd_resdelta}
\sigma_{\rm res} (E) = \frac{g_e^2}{2 m_e}  \delta (E_ - E_{\rm  res} ) \ .
\end{equation}

The track length distribution is the most computationally challenging part of Eq.~\eqref{eq_ResonantProd_Nperpot} as it depends on the target material, lengths and shape, and must be typically estimated via MC tools such as {\sc GEANT4}~\cite{Allison:2016lfl} or  {\sc FLUKA}~\cite{Bohlen:2014buj}. It can be also derived once-and-for-all for each experimental analysis, allowing for a ``database'' approach as argued in~\cite{Darme:2021sqm,Celentano:2020vtu}. Note however that in the case of lepton beams, the resulting shower is mostly electromagnetic and can thus be described to a relatively good accuracy by established analytical formula~\cite{Tsai:1966js}. 

The analysis strategy then typically relies on an active target and searching for events with significant missing energy (see e.g.~\cite{Andreev:2021fzd,NA64:2022rme}).

\paragraph{Energy scan strategy}

In this configuration, the target is typically thin enough that the beam energy will not be degraded. The track length from Eq.~\eqref{eq_ResonantProd_Nperpot} is then replaced by the beam energy distribution itself.
We will assume that the particle width, although  small, is not too narrow so that the  energy distribution of the positrons in the beam can still be considered as 
continuous. In the CoM of the collision, this corresponds to the 
requirement
\begin{align}
    \frac{\Gamma_X M_{X}}{2 m_e} \frac{1}{\delta E} N_{\rm tot} (E) \sim \left( \frac{N_{\rm tot} (E)}{4 \cdot 10^4} \right) \left( \frac{g_e}{2\cdot 10^{-4}}\right)^2 \, \left( \frac{0.7 \,\textrm{MeV}}{\delta E}\right) \, \left( \frac{M_X}{20 \,\textrm{MeV}}\right)\gg  1 \ ,
    \label{eq_ResonantProd_continuouscondition}
\end{align}
where $N_{\rm tot} (E)$ is the total number of positron accumulated on target with a beam with energy $E$ and beam spread $\delta E$.   
The total number of produced $X$ for $N_{\rm tot}$ per positrons-on-target is given by
 \begin{align}
 \label{eq_ResonantProd_Ntot}
\mathcal{N}_{X} =N_{\rm tot} \frac{\mathcal{N}_A Z \rho}{A} L_{\rm tar}  \frac{g_e^2}{2 m_e} \frac{d f_{\textrm{beam}}}{dE}(E_{\rm res})  \ ,
\end{align}
where $L_{\rm tar}$ and  $\rho$ are respectively the target thickness and density, and 
we have assumed that the target is sufficiently thin that 
the  differential energy distribution  $\frac{d f_{\rm beam}}{dE}$ is not significantly 
modified along the positrons path through it. 
It is instructive to estimate the peak cross-section for an incoming $e^+$ with precisely the $X$ resonant energy $E_+ = E_{\rm res}$.
We obtain $ \sigma_{\rm peak} \sim 50 \, \textrm{b} \times \left(\frac{17 \, {\textrm{ MeV}}}{M_{X}} \right)$, which translates into a mean free path in a carbon target of around $200 \mu m$.  
 Thus, each positron with precisely the resonant energy will produce a $X$ with near certainty,  even in sub-millimetric targets. The $g_e^2$ suppression from Eq.~\eqref{eq_ResonantProd_Ntot} in fact arises because such ``peak'' positrons are exceedingly rare due to the narrow $X$ width, highlighting the importance of the ``continuous'' condition Eq.~\eqref{eq_ResonantProd_continuouscondition}. 

\subsubsection{Application to the ATOMKI anomaly}
\label{intro}
The anomaly observed by the  ATOMKI collaboration in the $e^+e^-$ 
angular correlation spectra in $^8$Be and $^4$He nuclear transitions~\cite{Krasznahorkay:2015iga,Krasznahorkay:2021joi,Krasznahorkay:2022pxs}, can be 
interpreted in terms of a new boson $X_{17}$ produced on shell and promptly decaying 
into an electron-positron pair. Nuclear decays of both isotopes point 
with remarkable precision to the same mass window:
\begin{eqnarray}
    M_{{X_{17}}} = \begin{cases}
        16.70 \pm 0.35 ~(\rm stat)  \pm 0.50 ~(\rm syst) \ \textrm{MeV}   \qquad   ({}^8\textrm{Be~\cite{Krasznahorkay:2015iga}}) \\[0.3em]
        16.94 \pm 0.12 ~(\rm stat) \pm 0.21 ~(\rm syst) \  \textrm{MeV} \qquad   ({}^4\textrm{He~\cite{Krasznahorkay:2021joi}}) \, \\[0.3em]
           17.03 \pm 0.11 ~(\rm stat)  \pm 0.20 ~(\rm syst) \ \textrm{MeV}  \qquad   ({}^{12}\textrm{C~\cite{Krasznahorkay:2022pxs}}) \, . 
\end{cases}
    \label{eq_ResonantProd_m17}
\end{eqnarray}
Such a small mass implies that this hypothetical  particle is potentially accessible in accelerator 
experiments with low CoM energy % $\sqrt{s}$ 
such as $e^+$ and $e^-$ beam dump experiments, 
as well as in the decays of light mesons.

The nuclear data are not currently fully sufficient to determine the 
spin/parity quantum numbers  % precise nature 
of $X_{17}$ beyond its bosonic nature (see however the discussions~\cite{Zhang:2020ukq,Feng:2020mbt,Viviani:2021stx,Barducci:2022lqd})

\begin{figure}[!t]
\begin{center}
\resizebox{0.55\textwidth}{!}{%
  \includegraphics{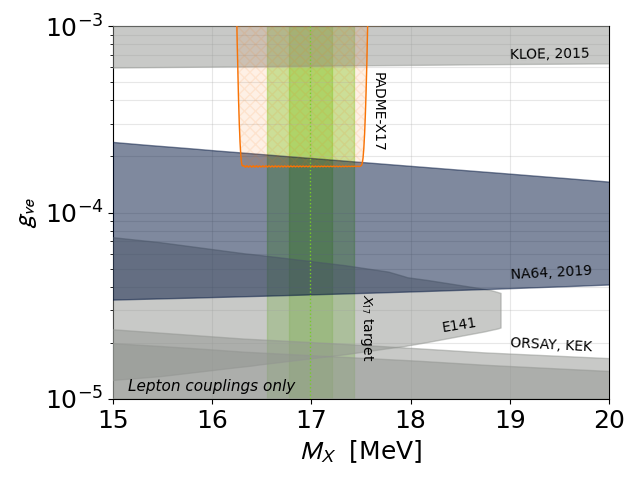}
}
\caption{Projected $90 \%$ C.L. sensitivity of PADME Run-III on the  $g_{e}$  coupling of a $X_{17}$  vector boson. 
 Lepton-based experimental limits % only are shown 
 from the KLOE~\cite{Anastasi:2015qla}, NA64~\cite{NA64:2021aiq},
 E141~\cite{Riordan:1987aw}, KEK and Orsay~\cite{Konaka:1986cb,Davier:1989wz} experiments are also shown. The dark (light) green band represents the $1\sigma$  ($2\sigma$) $X_{17}$ mass target from a naive combination of the ${}^4\textrm{He}$, ${}^8\textrm{Be}$ and  ${}^{12}\textrm{C}$ ATOMKI  results~\cite{Krasznahorkay:2018snd,Krasznahorkay:2021joi,Krasznahorkay:2022pxs}.}
\label{fig_ResonantProd_Limits}
\end{center}
\end{figure}

The $X_{17}$ must couple  to quarks in order to be produced and with electrons at a non-negligible level to allow for $X_{17}$ leptonic decays within the ATOMKI apparatus. Coupled with the small predicted mass and  given the existing limits summarised in Fig~\ref{fig_ResonantProd_Limits}, this makes a resonant search an attractive strategy to test the new physics origin of this anomaly without relying on nuclear data.
The DA$\Phi$NE beam facility at Laboratori Nazionali di Frascati (LNF) can provide a positron beam and  vary its energy in the required range of $\sim [270-290]$ MeV. Assuming a carbon target (with electron density of $10^{24} \, \textrm{cm}^{-3}$) of $100\, \mu m$ such as the one actually in use in the Positron Annihilation into Dark Matter Experiment (PADME)~\cite{Raggi:2014zpa,Raggi:2015gza}, the expected  $X_{17}$ production rate is approximately given by~\cite{Darme:2022zfw}:
\begin{equation}
% \mathcal{N}^{\rm per \ poT}_e 
\mathcal{N}^{\rm per \ poT}_{X_{17}}
\simeq 3.8 \cdot 10^{-7}\ \times \left( \frac{g_{e}}{3 \cdot 10^{-4}}\right)^2  \ \left( \frac{1 \ {\rm MeV} }{\delta E }\right) \ ,
\end{equation}    
where we have assumed a beam energy centered on $E_{\rm res}$.
Including the dominant Bhabha scattering background, we illustrate the experimental prospects following~\cite{Darme:2022zfw} in Fig.~\ref{fig_ResonantProd_Limits} with $4 \cdot 10^{11}$ total positrons-on-target, a $0.25\%$ beam spread and an energy range between $[260,300]$ MeV, probed  with $40$ energy bins.

\subsubsection{Conclusion}

Positron-based facilities allow to leverage resonant production to significantly increase  FIP signal rates. This new production mechanism can be translated into two different experimental strategies: either using a thick active target and searching for missing  energy (which works both for dark matter models and for neutrino-coupled FIPs), or using a thin target with visible decays and rely on a scan on the beam energy. The lightness of the $X_{}$ hypothetical NP particle hinted by the ATOMKI data makes it a perfect target for a demonstrator of the second case while providing an independent cross-check to this nuclear physics anomaly. In particular, we have shown that it offers a way of testing the existence of the $X_{17}$ spin-1 new state  via its  coupling to $e^\pm$, which are unavoidably required for a new physics interpretation of the ATOMKI anomaly. 

%\subsubsection{Acknowledgments}
%L.D. is  supported by the European Union’s Horizon 2020 research and innovation programme under the Marie Skłodowska-Curie grant agreement No 101028626.

\afterpage{\clearpage}
%-------------------------------------------

%-------------------------------------------
\subsection{\emph{New ideas:} Techniques for model-independent interpretations of hidden particle searches -- {\it P.~Klose}}
\label{klose}
{\it Author: Philipp Klose, <pklose@itp.unibe.ch>}

\subsubsection*{Introduction}
The standard model (SM) of particle physics is known to be incomplete.
Hidden particles can help explain many important hints for new physics, but the large variety of viable hidden sector models poses a challenge for the model-independent interpretation of hidden particle searches.

In recent years, effective field theories such as standard model effective theory (SMEFT) \cite{Buchmuller:1985jz, Grzadkowski:2010es} or Higgs effective theory (HEFT) \cite{Feruglio:1992wf,Burgess:1999ha,Barbieri:2007bh,Grinstein:2007iv}
have become a standard tools for deriving largely model-independent constraints on heavy new particles \cite{Brivio:2017vri,Ellis:2018gqa, AguilarSaavedra:2018nen, Slade:2019bjo,Dong:2022mcv,DeAngelis:2022qco}.
There has also been significant effort to develop techniques to model-independently constrain \emph{light} new particles.
One approach is to construct EFTs that encompass all possible interactions between the SM and specific candidate particles,
such as axion-like particles (ALPs) \cite{Brivio:2017ije},
heavy neutral leptons (HNLs) \cite{Liao:2016qyd,Li:2021tsq}, or non-relativistic dark matter candidates \cite{Fitzpatrick:2012ix,Cirigliano:2012pq,DelNobile:2013sia,Hoferichter:2015ipa,Hoferichter:2016nvd,Bishara:2016hek,Bishara:2017pfq,Hoferichter:2018acd,Criado:2021trs}.
Another approach is to construct ``simplified models'' that are designed to capture certain features of realistic SM extentions in a minimalist,
and therefore potentially generic, setup \cite{Alwall:2008ag,LHCNewPhysicsWorkingGroup:2011mji, Abdallah:2015ter,DeSimone:2016fbz}.
In either case, the hidden sector is typically supposed to be minimal, with only one or two hidden fields, and it is often not straightforward to translate the resulting constraints onto realistic hidden sectors.
In the following, we present two techniques that can be used to help fill this gap.

\subsubsection*{Factorizing hidden particle production rates}

Inclusive hidden particle production rates generically factorize into i) model-independent reduced matrix elements $M_d$, which depend only on SM physics,
and ii) observable-independent hidden currents $J^d$, which depend only hidden sector physics \cite{Klose:2022vro}.
To be more explicit, consider the Lagrangian of a generic hidden sector couplings to the SM,
\begin{align}
\mathcal L &= \mathcal L_\text{SM} + \mathcal L_\text{hidden} + \epsilon \sum_{d} \mathcal A_d \mathcal B^d \ .
\end{align}
The two sectors are linked by a set of portal interactions, where the parameter $\epsilon$ measures the smallness of the portal interaction,
$\mathcal A_d$ is a local operator constructed from SM fields,
and $\mathcal B^d$ is a local operator constructed from hidden fields.
Unlike the complete operator $\mathcal A_d \mathcal B^d$, $\mathcal A_d$ and $B^d$ individually do not have to be Lorentz scalars.
%For instance, if a standard model fermion $\psi$ couples to a hidden vector field $v^\mu$, one has $\mathcal A_\mu = \overline \psi \gamma_\mu \psi$ and $\mathcal B^\mu = v^\mu$.
\begin{figure}
\centering
\begin{subfigure}[b]{0.45\textwidth}
\begin{equation*}
M_d (\mathcal K \to \mathcal P) = \quad \mathcal K \ \{ \raisebox{-0.25\height}{\includegraphics[width=.5\textwidth, trim = 20 20 20 20, clip]{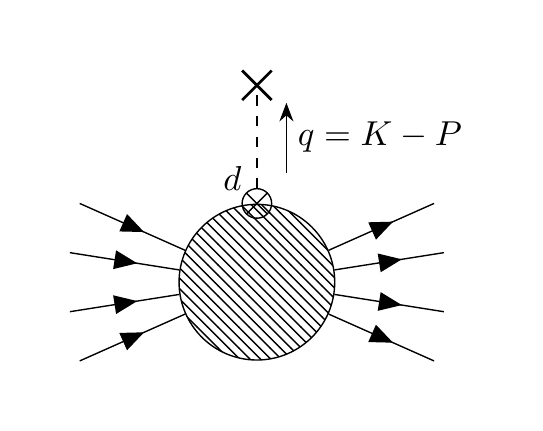}} \} \ \mathcal P
\end{equation*}
\caption{\label{subfig:reduced matrix elements}Reduced matrix elements}
\end{subfigure}
\hfill
\begin{subfigure}[b]{0.45\textwidth}
\begin{equation*}
J_d (\mathcal Q) = \quad \vcenter{\hbox{\includegraphics[width=.5\textwidth, trim = 20 20 20 20, clip]{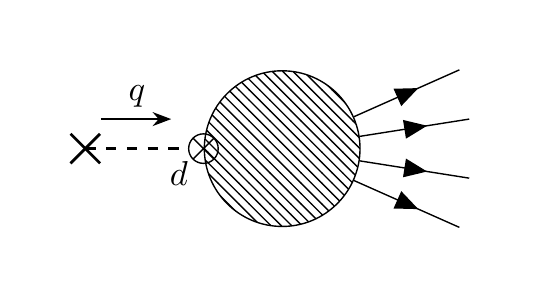}}} \} \ \mathcal Q
\end{equation*}
\caption{\label{subfig:hidden currents}Hidden currents}
\end{subfigure}
\caption{\label{fig:diagrams} Diagrammatic expressions of the reduced matrix elements $M_d$ and the hidden currents $J_d$.
Both are sums of all available connected and amputated Feynman diagrams with the appropriate number and kind of initial and final state particles.
The crossed dot denotes the required portal vertex, and the dashed line denotes the relevant missing momentum in- and outflow.
Aside from this portal vertex, diagrams for $M_d$ diagrams only contain SM vertices and propagators,
while diagrams for $J_d$ only contain hidden sector vertices and propagators. 
}
\end{figure}
We focus on the inclusive rates

\[
{\mathcal M}^2 = \sum_{k=1}^{\infty} {M_k}^2 \ \, 
{M_k}^2 = \int \text{d}^3_k \mathcal Q (2\pi)^4 \delta(K - P - Q) \mathcal M(\mathcal K \to \mathcal P; \mathcal Q)^2 \ ,
\]

where $k$ is the number of produced hidden particles,
and the matrix element $\mathcal M = \mathcal M(\mathcal K \to \mathcal P; \mathcal Q)$ measures the amplitude of transitioning from an initial state composed of $n$ standard model particles with momenta $\mathcal K = (k_1, \dots, k_n)$
into a final state composed of $m$ standard model particles with momenta $\mathcal P = (p_1, \dots, p_j)$ as well as $k$ hidden particles with momenta $\mathcal Q = (q_1, \dots, q_k)$.%
\footnote{
Here $K = \sum_i k_i$, $P = \sum_j p_j$, and $Q = \sum_l q_l$ denote the total four-momenta of the incoming and outgoing standard model and hidden particles.}
The integration measure
\begin{align}
\int \text{d}^3_k \mathcal Q &= \prod_{l=1}^k \sum_{r_l} \frac{\text{d}^3 \boldmath q_l}{(2\pi)^3} {1 \over {2\omega_l} }\ , &
q_l &= (\omega_l, \boldmath q_l) \ ,
\end{align}
encompasses a phase-space integration for each produced hidden particle as well as a sum over all available hidden particle types and helicities, collectively denoted by the indices $r_l$.
The inclusive rates factorize according to the relation
%\begin{align}
\[
{\mathcal M}^2 = \epsilon^2 M_d M_e J^{de} + {\epsilon^3}, \, \, \,
J^{de}(K-P) = \sum_{k=1}^\infty \int \text{d}^3_k \mathcal Q (2\pi)^4 \delta^4 (K-P-Q) J^e J^d \ ,
\]
%\end{align}
where the $M_d = M_d(\mathcal K, \mathcal P)$ are the reduced matrix elements, and the $J^d (\mathcal Q)$ are the hidden currents.
Both can be computed using an infinite series of Feynman diagrams as depicted in Figure \ref{fig:diagrams}.
The hidden currents $J^d$ receive contributions from all available final states, and each final state corresponds to a separate sum of Feynman diagrams constructed as depicted in Figure~\ref{subfig:hidden currents}.

This factorization can be used to streamline the adaptation of existing rates to new models and observables:
When investigating a new observable, it is sufficient to re-compute the reduced matrix elements, while the hidden currents remain the same.
Likewise, when investigating a new model, it is sufficient to re-compute the hidden currents, while the reduced matrix elements remain the same. 
The reduced matrix elements can also be used by themselves to model-independently constrain the hidden currents and with them generic hidden sectors.

\subsubsection*{Portal effective theories}

\begin{figure}
\centering
\begin{subfigure}[t]{.48\textwidth}
\includegraphics[width = \textwidth]{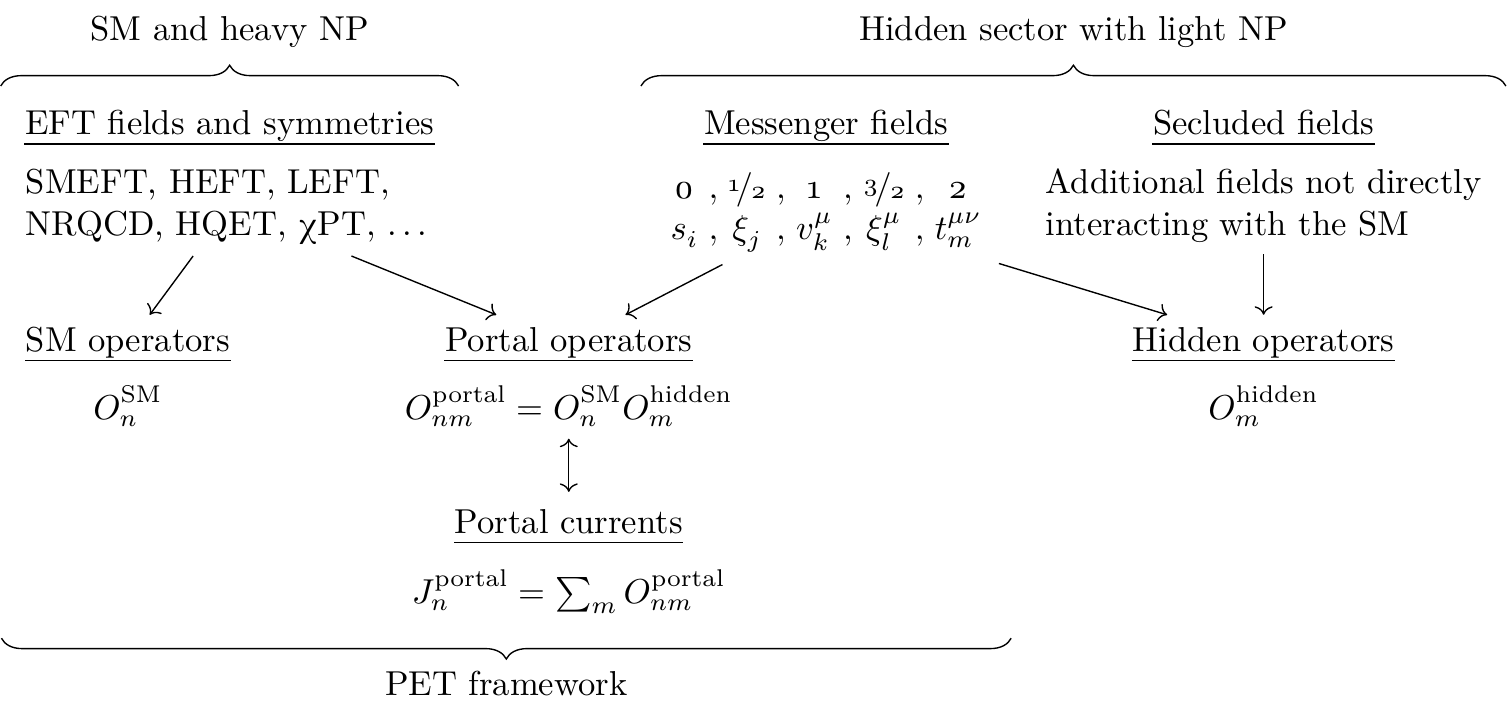}
\caption[The $\PET$ framework]{
Overview of the PET framework. We extend EFTs that capture specific SM dynamics by adding portal operators that couple them to generic messenger fields that are dynamic at the relevant energy scale.
Heavy new particles are integrated out, and the hidden sector is otherwise treated as a black box, allowing for any number of secluded fields with arbitrary couplings.
} \label{subfig:PET framework}
\end{subfigure}
\hspace{10pt}
\begin{subfigure}[t]{.48\textwidth}
\includegraphics[width = \textwidth]{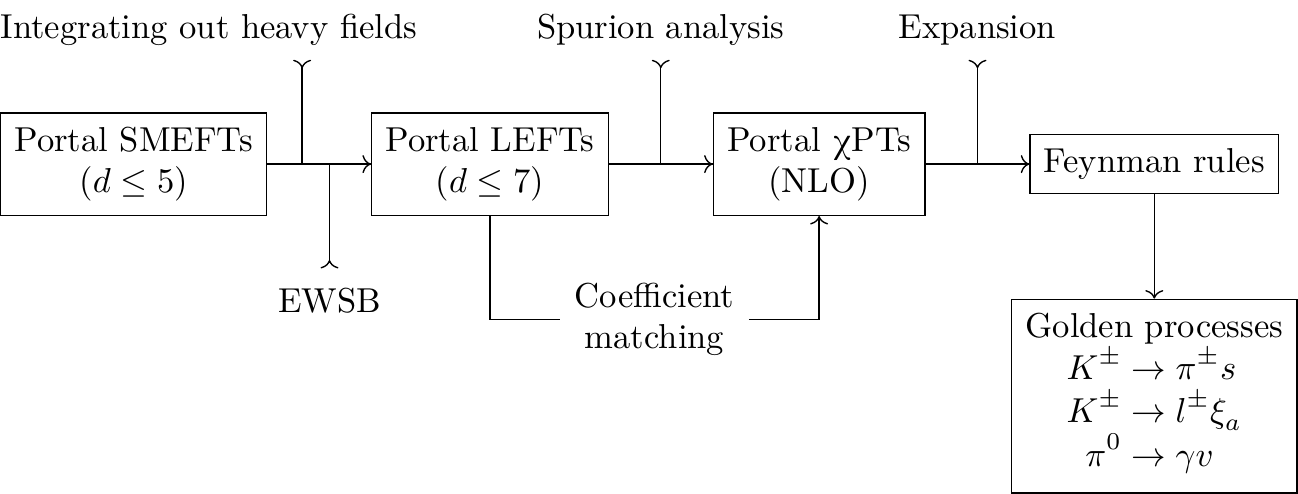}
\caption[Procedure to derive the \PET Lagrangian coupling mesons to messengers]{
Overall strategy for constructing PETs that couple SMEFT, LEFT, and ChPT to spin 0, $1/2$, and 1 messengers.
We apply the final portal ChPT Feynman rules to compute universal amplitudes for the three mentioned golden processes.
} \label{subfig:flowchart}
\end{subfigure}
\end{figure}

To make full use of the above factorization, one has to supply a comprehensive list of portal operators.
We developed a framework for constructing portal effective theories (PETs) that extend EFTs which either encompass the SM or specific parts of it by coupling them to generic new particles \cite{Arina:2021nqi}.
Figure~\ref{subfig:PET framework} depicts its overall structure.
Beside Lorentz invariance, we do not impose simplifying symmetries such as parity (P) or charge parity (CP) conservation on to the portal Lagrangian.
In contrast to other approaches, the hidden sector is treated largely as a black box, allowing for any number of secluded particles with arbitrary masses and interactions.
Heavy new particles are integrated out, and their impact is captured by including higher dimensional portal and hidden sector operators.

Following the procedure depicted in Figure~\ref{subfig:flowchart}, we construct PETs that couple SMEFT,
which encompasses both renormalizeable and higher dimensional operators constructed from SM fields,
low energy effective theory (LEFT), which captures the dynamics of the SM particles at the strong scale,
and chiral perturbation theory (ChPT), which captures the light pseudoscalar mesons,
to a single generic messenger fields of spin 0, $1/2$, or 1.
This messenger can then couple to an arbitrary number of secluded fields.

The resulting portal SMEFTs contain all portal operators of dimension $d\leq5$, while the portal LEFTs additionally contain all leading quark-flavour violating operators of dimension $d=6,7$.
These $d=6,7$ operators can contribute to quark-flavour violating transitions at the same order as $d\leq5$ portal operators because they receive contributions from virtual $W^\pm$ boson exchanges.
We further discriminate between the portal LEFT operators by using naive dimensional analysis (NDA) power-counting rules to guess the loop order of the leading matching contribution to higher dimensional operators.
Ten portal currents can then be used to parametrize the coupling of QCD to generic new particles at the strong scale.
We employ a spurion approach to determine how these currents couple to chiral perturbation theory,
and the resulting leading order portal ChPT action contains 27 new low-energy constants (LECs), 21 of which are not fixed by SM observations.
We determine some of these LECs by using large $n_c$ techniques and by matching portal ChPT to hadronic matrix elements computed from the lattice as well as the trace of the QCD stress-energy tensor.
To demonstrate the utility of portal ChPT, we finally compute generic decay rates for three golden channels:
i) Charged kaon decays into a charged pion and any new spin 0 particle,
ii) charged kaon decays into a charged lepton and any new spin $1/2$ particle,
iii) neutral pion decays into a photon and any new spin 1 particle.

\subsubsection*{Summary}

The factorization approach helps streamline the computation of hidden particle production rates.
When combined with a comprehensive list of portal operators, constructed using e.g. the PET framework, it can be used to derive model-independent constraints on realistic light hidden sectors.
We constructed portal SMEFTs, LEFTs, and ChPTs, and computed generic light meson decay rates.
Given the importance of new physics constraints from $B$ and $D$ meson decays, one promising line of future research is to also construct portal heavy quark effective theories (pHQETs)
and portal soft-collinear effective theories (pSCETs) and to then compute generic $B$ and $D$ meson decay amplitudes.

%-------------------------------------------

%-------------------------------------------
\subsection{\emph{New ideas}: Jupiter missions as probes of dark matter and long-lived mediators  -- {\it L. Li}}
\label{Li}
{\it Author: Lingfeng Li, <lingfeng\_li@brown.edu>}

%\subsubsection*{Jupiter missions as probes of dark matter and long-lived mediators}
Since 1973, Jupiter has been visited by nine spacecraft. Amongst them, the two dedicated Jupiter missions, namely the Galileo mission~\cite{1992SSRv...60....3J} and the Juno mission~\cite{2017SSRv..213....5B} orbit around the planet for years, collecting precious \textit{in situ} measurement data about the planet and the magnetosphere around it. In the past decades, the data from these missions greatly deepened our knowledge of Jupiter and supported planetary science.

\begin{figure}[h!]
\centering
\includegraphics[width=8.2 cm]{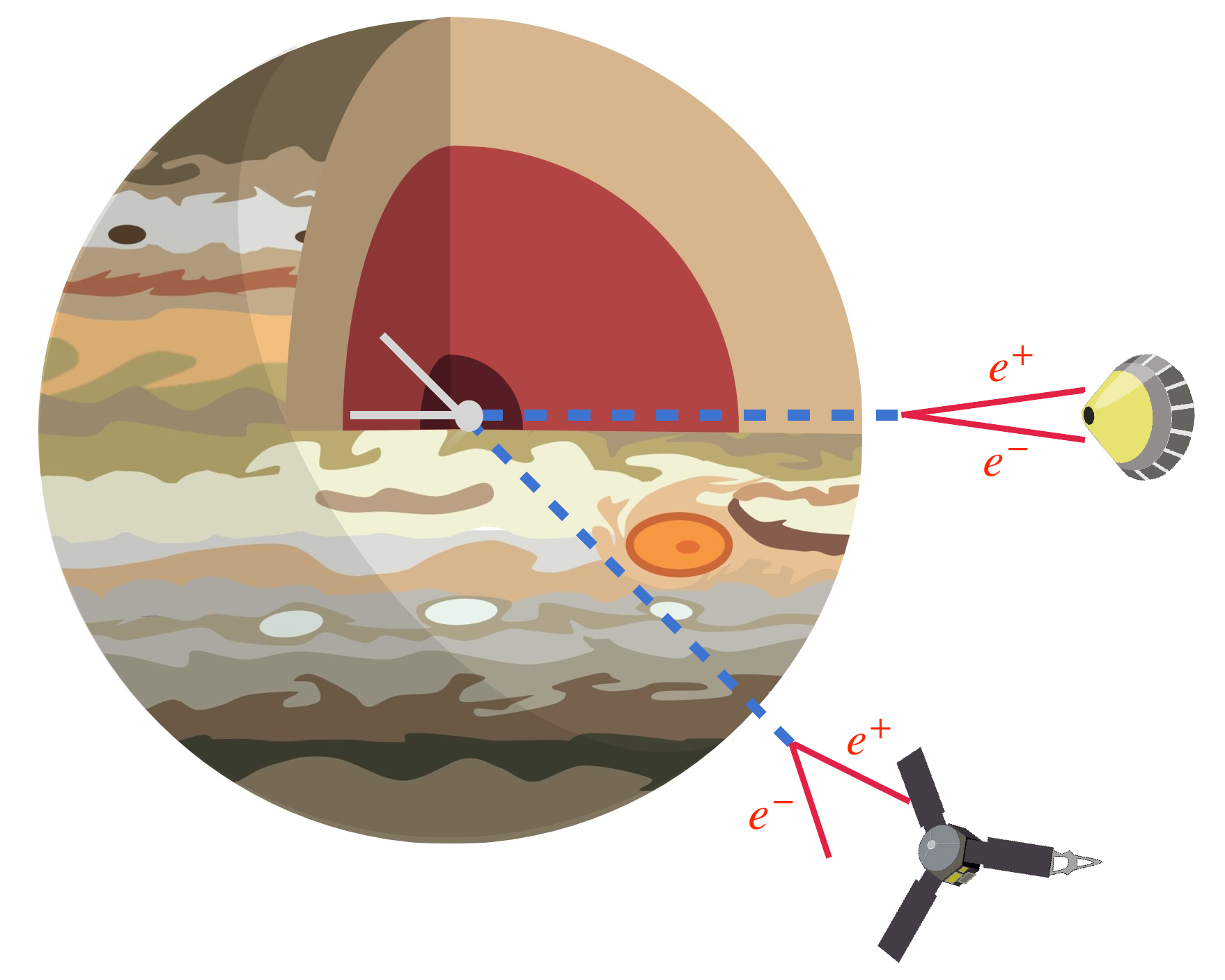}
\includegraphics[width=7.4 cm]{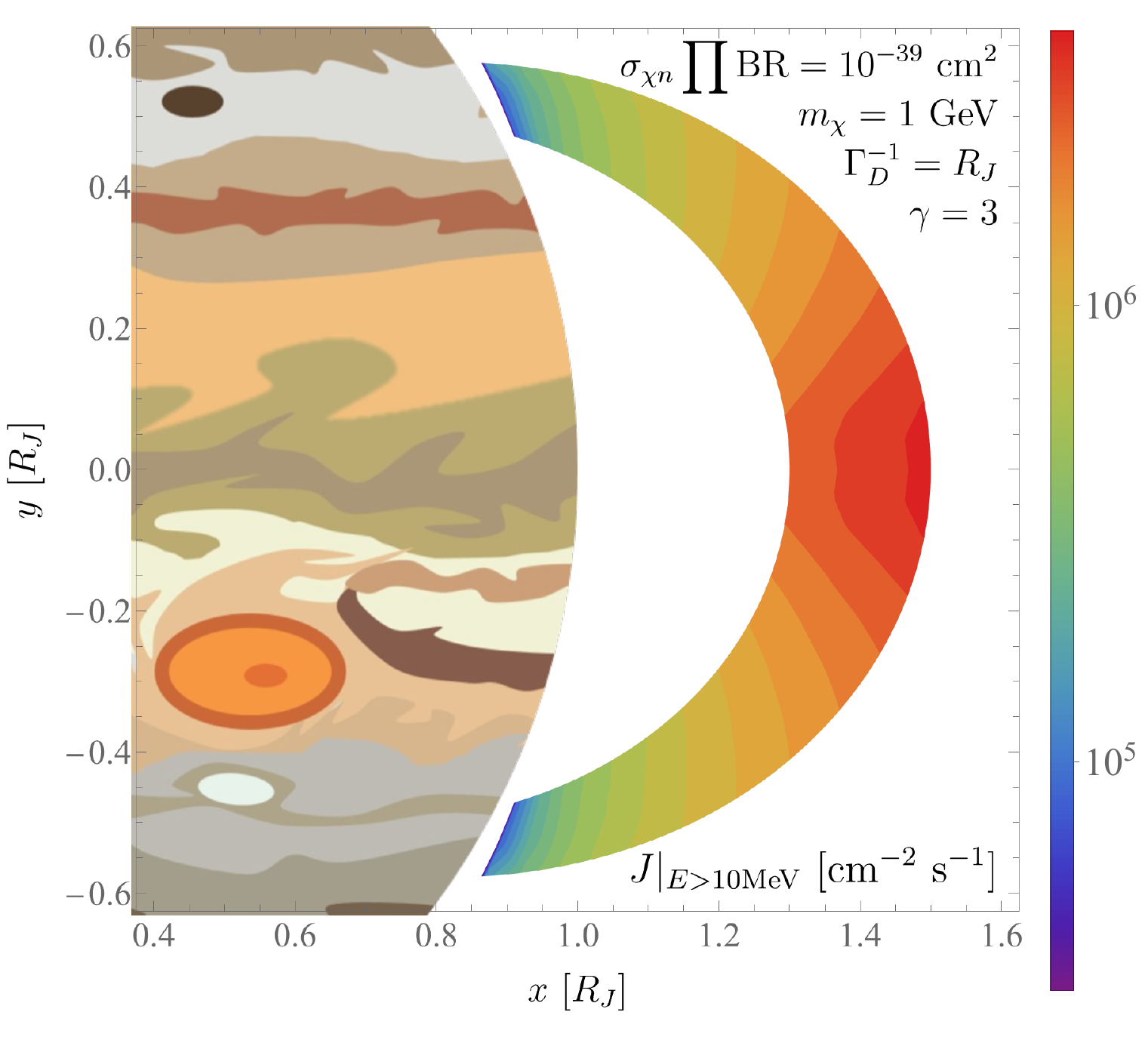}
\caption{\textbf{LEFT: }A schematic illustration of the process. Captured DM particles (solid white) annihilate into a pair of dark mediators (dashed blue), which decay into $e^+e^-$ (red solid) outside Jupiter. The energetic electrons or positrons could be measured by the Jupiter missions. \text{RIGHT: }The spatial distribution of the omnidirectional $e^\pm$ flux $J$ above 10~MeV for a benchmark model in the fully-trapped region. See~\cite{Li:2022wix} for more details.}
\label{fig:schemejupiter}
\end{figure}

A natural question arises if such a large amount of \textit{in situ} measurement data could contribute to high energy physics. More specifically, if it can help the hunt for the dark matter (DM) or similarly feebly interacting dark sector particles. One may recognize Jupiter with a deep gravity potential and large geometrical size as a giant DM collector and a potential detector in the outer space~\cite{Kawasaki:1991eu, Adler:2008ky, Leane:2021ihh, Leane:2021tjj}. Here a new search targeting DM and a long-lived mediator was proposed using the local measurements from both Jupiter missions aforementioned~\cite{Li:2022wix}. The left panel of Fig.~\ref{fig:schemejupiter} depicts the physics of this search, including the steps follow: 

\paragraph{DM stopped and captured by Jupiter}
If the DM particle has a non-vanishing elastic scatter amplitude with the SM baryons, it could lose most of its kinetic energy after the scattering. Once its mass drops below the escape velocity of Jupiter $\sim 60$~km/s,  it becomes gravitationally bound to Jupiter. Jupiter's DM capture rate as a function of DM mass and its elastic scattering cross section with nucleons $\sigma_{\chi n }$ is calculated with the method in~\cite{Gould:1987ir,Gould:1987ww}. For the benchmark of a 1 GeV DM with $\sigma_{\chi n}=10^{-38}$~cm$^{-2}$, it reaches $\mathcal{O}(10^{21})$ DM particles per second.

\paragraph{DM annihilation in pairs of long-lived mediators inside Jupiter}
DM particles trapped in the planet lose their kinetic energies and accumulate around the center of Jupiter, greatly enhancing the annihilation rates. Given the upper limit on the annihilation cross section estimated from CMB measurements~\cite{Leane:2018kjk}, the time scale for the equilibrium between DM capture and annihilation for a 1 GeV DM with $\sigma_{\chi n}\sim 10^{-38}$~cm$^{-2}$ is much shorter than the lifetime of Jupiter. In this case, the DM capture and annihilation rates are related by a factor of two.

\paragraph{Mediators decay, injecting hard $e^+e^-$ pairs into the magnetosphere}
If DM annihilates into dark mediators, they could decay into $e^+e^-$ pairs outside Jupiter as long as the mediator is sufficiently secluded from the SM sector. This could happen when the decay length of the mediator is comparable to the Jupiter radius, a sub-GeV scale dark photon with kinetic mixing parameter $\epsilon \sim 10^{-10}$ which is elusive for lab experiments. Such a long-lived mediator is often needed to produce the correct DM relic abundance~\cite{Cheng:2018vaj}. Note that $e^\pm$ final state is common when the mediator that decays to charged particles since most charged particles in the SM will eventually end up with at least one $e^\pm$ at the large length scale concerned.

\paragraph{The magnetic field of Jupiter traps $e^\pm$, increasing the flux}
If the $e^\pm$ are released in the proper position and momentum, it will be trapped in the strong magnetic field for a long time. The long time scale of the trapped electrons compensates for the low $\rho_D$ when the DM-nucleon scattering cross section $\sigma_{\chi n}$ is small, extending the reach of sensitivity. The $e^\pm$ flux distribution is achieved by solving the diffusion equation of the phase space numerically~\cite{schulz2012particle,nenon2018rings}. Two major regions of interest are found, namely the fully-trapped and quasi-trapped regions. In these cases, the relevant timescales for ultrarelativistic $e^\pm$ flux are determined by synchrotron radiation energy loss and azimuthal drift period, respectively.

\paragraph{Jupiter missions measuring the local flux intensity}
Both Jupiter missions can profile and measure relativistic $e^\pm$ fluxes in the Jovian magnetosphere. For the Galileo mission, the data comes from the Energetic Particles Investigation (EPI) carried by a smaller detector, the Galileo probe, which was released from the orbiter and dived into the atmosphere~\cite{fischer1996high}. For the Juno mission, the data comes from the Radiation Monitoring (RM) investigation that utilizes the radiation noise signatures in images of several cameras and science instruments~\cite{becker2017juno}. %Though accurate particle identification and spectroscopy are difficult in the outer solar system, all the above measurements are more sensitive to $e^\pm$ harder than $\mathcal{O}(10)$~MeV since their penetration efficiency through the radiation shielding is higher. 
By comparing the observed fluxes and the prediction of DM models, one can set conservative upper bounds on the maximum $e^\pm$ from dark mediator decays and thus the less explored region in the parameter space.
  
\begin{figure}[h!]
\centering
\includegraphics[width=6.8 cm]{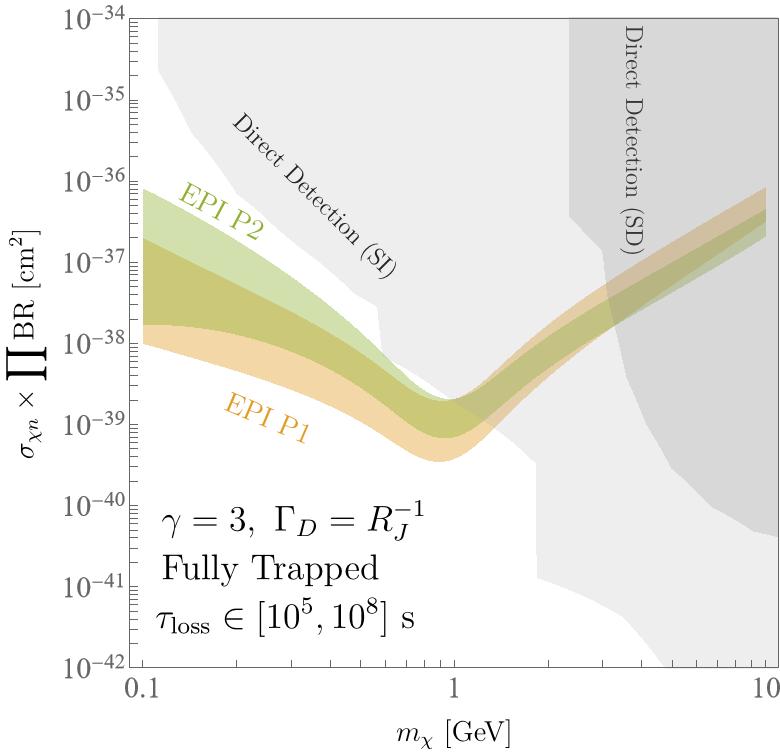}
\includegraphics[width=6.8 cm]{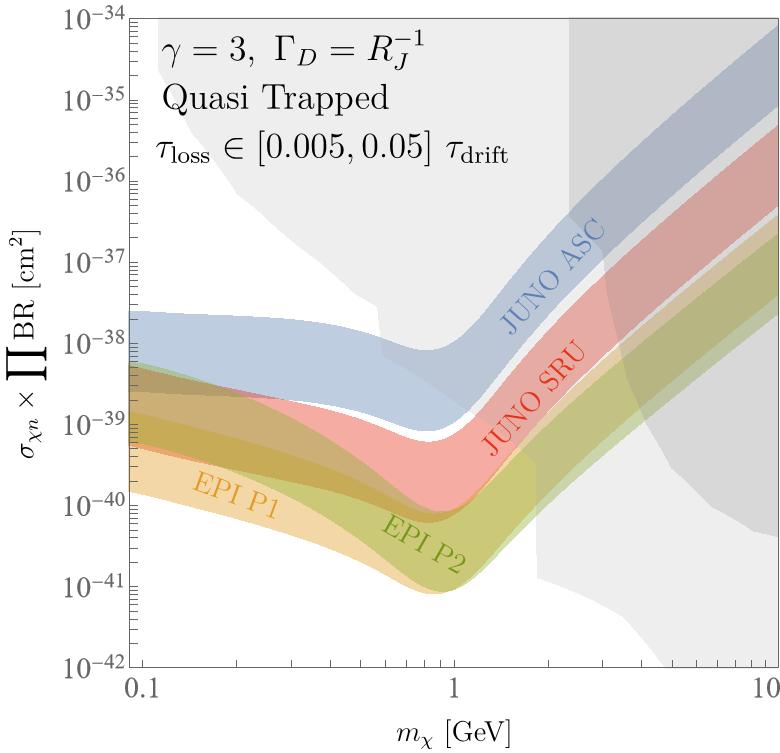}
\caption{Observed upper bounds on the corrected DM-nucleon scattering cross section due to the relativistic $e^\pm$ flux from dark mediator decays in different trapping scenarios~\cite{Li:2022wix}. Models above the bands are excluded, while the width of bands are determined by the uncertainty of environment of Jupiter. \textbf{LEFT:} limits from the Galileo probe EPI channels in the fully trapped scenario. \textbf{RIGHT:} limits from the two Galileo EPI channels as well as Juno measurements in the quasi-trapped scenario. The lighter (darker) grey regions are constraints on $\sigma_{\chi n}$ from direct detection experiments, assuming spin-(in)dependent scattering~\cite{XENON:2018voc,CDEX:2019hzn,XENON:2019zpr, PandaX-4T:2021bab, LUX-ZEPLIN:2022qhg, Behnke:2016lsk,SuperCDMS:2017nns, PICO:2019vsc}. }
\label{fig:limitsjupiter}
\end{figure}

In Fig.~\ref{fig:limitsjupiter}, the data collected by the Galileo probe and Juno is applied. Constraints are set on the product of $\sigma_{\chi n}$ and $\prod$ Br, the total branching fraction of DM annihilations ending in $e^\pm$ final state. The limit is the strongest for DM with a mass $\sim$1 GeV, reaching $10^{-41}$~cm$^2$ from the data collected by Galileo EPI probes in the quasi-trapped regime. However, the large systematic uncertainties from the magnetic field and material modeling leave the bounds to be fully verified in future studies. A weaker but potentially more reliable set of constraints comes from the $e^\pm$ flux in the fully-trapped scenario, where the systematic effects are subdominant. The upper limit on $\sigma_{\chi n} \times \prod$ Br in this case is of $\mathcal{O}^{-39}$ cm$^2$ for DM at 1 GeV. These bounds could be comparable to or stronger than the current GeV-scale DM direct detection searches, especially when the DM-nucleon scattering is spin-dependent.

The study serves as a proof of concept and an initial step to explore the full potential of the large datasets from the Jupiter missions to search for new physics that only feebly interacts with the SM. The more precise magnetosphere and particle source modeling could improve the analysis and strengthen the bounds set on DM models. Future Jupiter missions with advanced particle identification and spectroscopy and better position coverage will further enhance the sensitivity~\cite{2021ExA...tmp..136R}. We also hope the \textit{in situ} data of Jupiter or other planets beyond the magnetic field and $e^\pm$ flux measurements to be utilized to study the beyond Standard Model physics. 

\paragraph{Acknowledgment} LL is supported by the DOE grant DE-SC-0010010, the NASA grant 80NSSC18K1010 and 80NSSC22K081.

%-------------------------------------------

%-------------------------------------------
\subsection{\emph{New ideas:} New bounds on axion-like particles from neutrino experiments -- {\it S.~Urrea} }
\label{urrea}
{\it Author: Salvador Urrea, <salvador.urrea@ific.uv.es>}

%\begin{abstract}
%Neutrino experiments lie at the edge of the intensity frontier and therefore can be exploited to search for new light particles weakly coupled to the visible sector. In this work we derive new constraints on axion-like particles (ALPs) recasting the results from a search for $e^+ e^-$ pairs at MicroBooNE pointing in the direction of the NuMI absorber. In particular, we consider the addition of higher-dimensional effective operators coupling the ALP to the electroweak gauge bosons. These would induce $K\to \pi a$ from kaon decay at rest in the NuMI absorber, as well as ALP decays into pairs of leptons or photons.
%\end{abstract}

\subsubsection{Introduction}
In this proceeding we summarize the results of \cite{Coloma_2022}. In neutrino experiments using a conventional beam, a fraction of the accelerated protons (which can be as large as 10-15\%) does not get stopped by the target and ends up hitting the absorber at the end of the decay pipe. The kaons produced in such collisions, after losing energy as they interact with the medium, eventually decay at rest within the absorber. In Ref.~\cite{MicroBooNE:2021usw}, the MicroBooNE collaboration searched for monoenergetic scalars decaying into $e^{-}e^{+}$ coming from the direction of the NuMI hadron absorber, located at a distance of only 100 m from the detector. We use this analysis to set a bound on ALPs decaying to $e^{+}e^{-}$, and we also derive sentivitity regions for possible searches for pairs of muons or photons which are also predicted in this scenario.

%%%%%%%%%%%%%%%%%%%%%%%
\subsubsection{Theoretical framework and notation}
%%%%%%%%%%%%%%%%%%%%%%%

We consider the following set of effective operators describing ALP interactions with EW bosons at some high scale $\Lambda$ (which we take as  $\Lambda = f_a$)
\begin{equation}
\label{eq:Lag}
\delta\mathcal{L}_{\rm EW} = ic_{\phi} \frac{\partial^\mu a }{f_a}\phi^\dagger \overleftrightarrow{D}_\mu \phi  - c_{B}\frac{a}{f_a}B_{\mu\nu}\widetilde{B}_{\mu\nu} - c_W \frac{a}{f_a}W^I_{\mu\nu}\widetilde{W}^I_{\mu\nu} \, ,
\end{equation}
Here, $\phi$ is the Higgs doublet while $B$ and $W^I$ stand for the EW vector bosons, and $a$ is the ALP field. The dual field strengths are defined as $\widetilde{X}^{\mu\nu} \equiv \frac{1}{2}\epsilon^{\mu\nu\rho\sigma}X_{\rho\sigma}$, with $\epsilon^{0123} = 1$, and $\phi^\dagger \overleftrightarrow{D}_\mu \phi \equiv \phi^\dagger \big{(}D_\mu\phi\big)-\big{(}D_\mu \phi\big{)}^\dagger \phi {}$. 

We are interested in these hadronic processes at energies below the EW scale, in particular we set the low scale at an energy of 2 GeV. The relevant couplings are the induced ALP couplings to the light quark currents:

\begin{equation}
\frac{\partial_\mu a(x)}{f_a} \left(\sum_{q}~\bar{q}_{R} ~k_{q}  \gamma^\mu  q_{R} + \sum_{Q}  \bar{Q}_L ~k_Q\gamma^\mu Q_L\right) \, ,
\end{equation}

where following the notation of Ref.~\cite{Bauer:2021wjo}, lower and upper case for the quark field refers to the right-handed/left-handed quarks respectively, and $k_q$ and $k_Q$ are $3\times 3$ matrices with indices $(u,d,s)$. 

The standard approach to incorporate these non-standard interactions in hadronic physics is to match the theory to Chiral Perturbation Theory ($\chi$PT) ~\cite{Bauer:2021wjo,Georgi:1986df}. Setting $\Lambda = 1~\rm{TeV}$, the only flavour-changing coupling induced by solving the RG equations down to 2 GeV is given by \cite{Bauer_2021}: 
\begin{align}
\label{eq:keff-value}
\frac{[k_Q(2\text{GeV})]_{ds}}{V_{td}^*V_{ts}}\bigg|_{\Lambda=\rm{1 TeV}} \simeq   -9.7\times 10^{-3}c_W(\Lambda)  + 8.2\times 10^{-3}c_{\phi}(\Lambda) - 3.5\times 10^{-5}c_{B}(\Lambda) \, .
\end{align}

The decay width of $K^+\to \pi^+ a$ is governed by this flavour-changing coupling and takes the following form:
\begin{align}
\label{eq:BKpia}
\Gamma(K^+\to \pi^+ a) &=\dfrac{m_{K}^3 \big|[k_Q(2\text{GeV})]_{sd}\big|^2}{64\pi} \lambda_{\pi a}^{1/2}\left(1-\dfrac{m_{\pi}^2}{m_{K}^2}\right)^2\,,
\end{align}
with the definitions 
\begin{align}
\lambda_{\pi a}  \equiv \lambda(1, m_a^2/m_K^2, m_\pi^2/m_K^2) , \hspace{1cm} \lambda(a,b,c)  = a^2 + b^2 + c^2 - 2ab - 2ac - 2bc \, .
\end{align}

\paragraph{ALP decay channels}
\label{sec:ALPdecay}

For ALP masses below 400~MeV, the decay channels that are kinematically open are $a\to \gamma\gamma$, $a\to e^+e^-$ and $a\to\mu^+ \mu^-$. The decay width into leptons can be written as:
\begin{equation}
\Gamma (a\to \ell^+\ell^-) =  
|c_{\ell\ell}|^2\dfrac{ m_a m_\ell^2 }{8\pi f_a^2} \sqrt{1-\dfrac{4 m_\ell^2}{m_a^2}} \, ,
\label{eq:GammaLepton}
\end{equation}
where at low energies ($\mu \sim 2~\mathrm{GeV}$) $c_{\ell\ell}$ is given at one loop by~\cite{Gavela:2019wzg,Bauer:2017ris}
\begin{align}
\begin{split}
c_{\ell\ell} = c_{\phi}&+\frac{3\,\alpha}{4\pi} \left(\frac{3\,c_W }{s_w^2}+ \frac{5\,c_B}{c_w^2 } \right) \log \dfrac{f_a}{m_W} 
+ \dfrac{6\, \alpha}{\pi}\left(c_B \, c_w^2+c_W\, s_w^2\right) \log \dfrac{m_W}{m_\ell} \,,
\end{split}
\label{eq:cll}
\end{align}
and to simplify the notation we have written $c_i \equiv c_i(\Lambda)$. 
Similarly, the decay width into two photons reads 
 \begin{align}
\label{eq:GammaPhoton}
\Gamma (a\to \gamma\gamma)& =  |c_{\gamma\gamma}|^2 \dfrac{m_a^3 }{4\pi f_a^2}\,,
\end{align}
where the effective coupling at low energies is given at one loop by~\cite{Gavela:2019wzg,Bauer:2017ris}
\begin{align}
\begin{split}
c_{\gamma \gamma } =  &c_W\,\Big[s_w^2\,+\frac{2\,\alpha}{\pi} B_2(\tau_W)\Big] +c_B\,c_w^2 
- c_{\phi} \,\frac{\alpha}{4\pi}\,\bigg( B_0 
- \frac{m_a^2}{m_{\pi}^2-m_a^2}\bigg)\, .
\end{split}
\label{eq:cgg}
\end{align}
Here, $B_0$ and $B_2$ are loop functions (which can be found in \cite{Coloma_2022}), and $\tau_W = 4m_W^2/m_a^2$.

%%%%%%%%%%%%%%%%%%%%%%%
\subsubsection{Results}
%%%%%%%%%%%%%%%%%%%%%%%

\begin{figure}[ht!]
\begin{center}
  \includegraphics[width=0.85\textwidth]{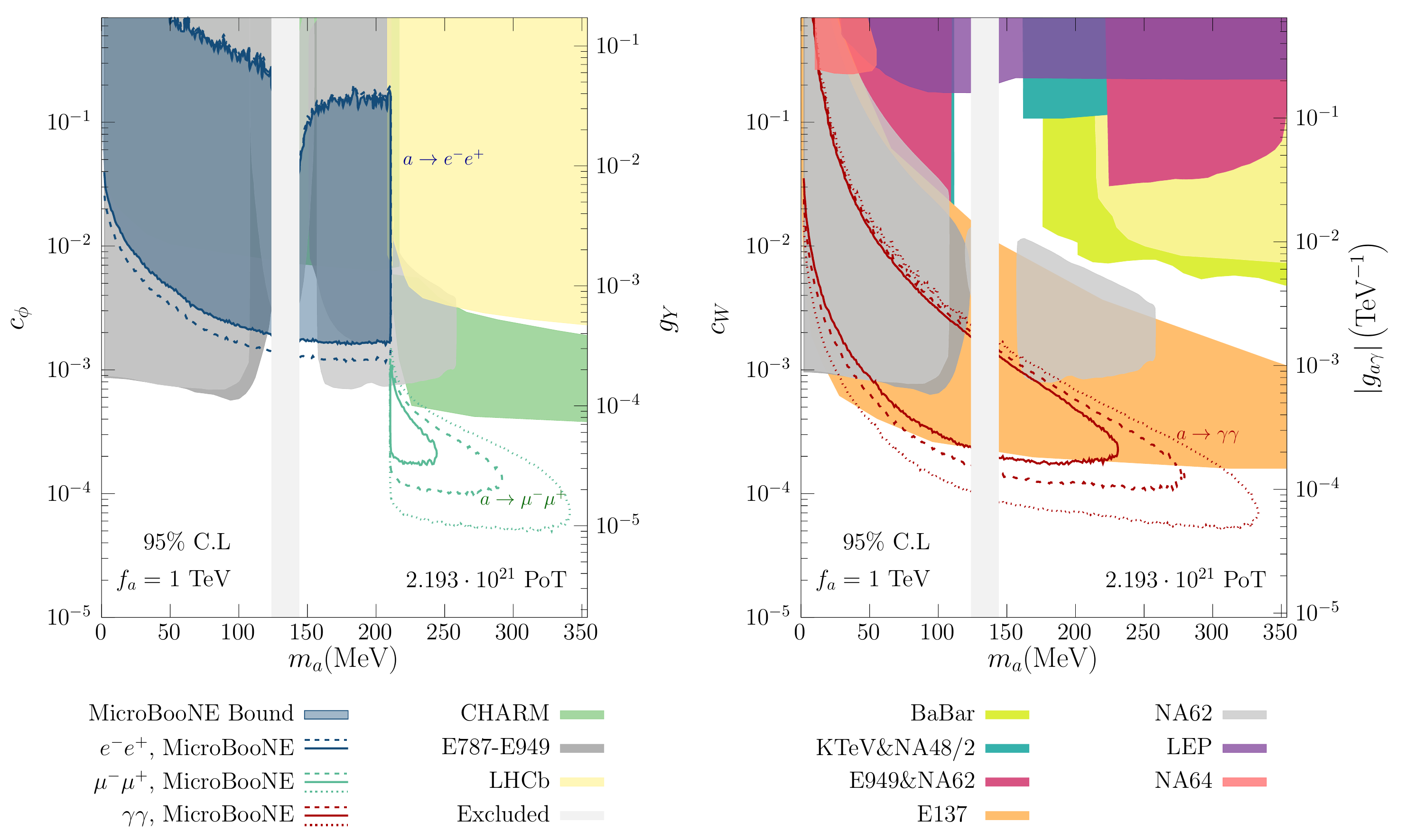}
\end{center}
\caption{\label{fig:micro-sens} In shaded blue we present our MicroBooNE bound for the $e^{-}e^{+}$ channel obtained from the data extracted from Ref.\cite{MicroBooNE:2021usw} and $N_\text{PoT}=1.93\times 10^{20}$. We also give the MicroBooNE sensitivity projections, for $c_{\phi}$ (left panel) and $c_{W}$ (right panel) as a function of $m_a$, assuming $f_a=1\;\text{TeV}$, for $N_\text{PoT}=2.2\times 10^{21}$. The regions enclosed by the colored lines satisfy $\Delta \chi^2 > 3.84$, corresponding to 95\% CL for 1 degree of freedom (d.o.f.). Solid lines assume assumed the systematics used in ~\cite{MicroBooNE:2021usw} for the $e^{+}e^{-}$ search; dashed lines assume that those systematics can be reduced to 20$\%$; and dotted lines indicate the no-background limit, taking 20$\%$ of systematics (see \cite{Coloma_2022} for details). The shaded areas show current bounds from BaBar~\cite{BaBar:2021ich}, E137~\cite{Bjorken:1988as,Dolan:2017osp}, NA62~\cite{NA62:2021zjw,NA62:2020xlg}, E787 \& E949~\cite{BNL-E949:2009dza}, LHCb~\cite{LHCb:2015nkv,LHCb:2016awg,Gavela:2019wzg}, CHARM~\cite{CHARM:1985anb,Essig:2010gu}, NA64~\cite{NA64:2020qwq} and LEP~\cite{Jaeckel:2015jla}. We also show bounds from measurements of $K\to\pi\gamma\gamma$ at E949~\cite{E949:2005qiy}, NA48/2~\cite{NA48:2002xke}, NA62~\cite{NA62:2014ybm} and KTeV\cite{KTeV:2008nqz} (taken from Ref.~\cite{Goudzovski:2022vbt}). The shaded vertical band is disfavored due to the large $a-\pi^0$ mixing, and is taken from Ref.~\cite{Gavela:2019wzg}. The right axes show the corresponding limits on the effective couplings for the so-called photon-dominance and fermion-dominance scenarios, see \cite{Beacham_2019}. }
\end{figure}

For a total number of $N_K$ kaon decays, the event rate expected from ALP decays into $\text{XX}$ pairs inside the MicroBooNE detector can be computed as
\begin{equation}
N_{events} = \frac{N_K \times \text{BR}(K \to \pi a)}{4\pi} \; \text{BR} (a \to \text{XX}) \; \epsilon_{eff}\int_{\Delta\Omega_\text{det}} d\Omega~ P_\text{decay} (\Omega) \, ,
\end{equation}
where the integral runs over all trajectories with solid angle $\Omega$ intersecting the detector, and $\Delta\Omega_\text{det}$ is the solid angle of the detector as seen from the absorber. Here $\epsilon_{eff}$ stands for the detection efficiency. Finally, $P_\text{decay}$ represents the probability of an ALP to decay inside the detector: 
\begin{equation}
\label{eq:Pdecay}
P_\text{decay} = e^{-\frac{\ell_\text{det} }{L_a }} \left[ 1 - e^{-\frac{\Delta \ell_\text{det}}{L_a}} \right] \, ,
\end{equation}
where $\ell_\text{det}$ is the distance traveled before it reaches the detector, and $\Delta \ell_\text{det}$ is the length of the ALP trajectory intersecting the detector. In practice, both $\ell_\text{det}$ and $\Delta \ell_\text{det}$ depend on the angle of the ALP trajectory. Our numerical results are shown in Fig \ref{fig:micro-sens}. 

%%%%%%%%%%%%%%%%%%%%%
%%%%%%%%%%%%%%%%%%%%%

\subsubsection{Conclusions}
In summary, this work stands out as a clear example of the multiple capabilities of neutrino experiments to search for new physics, not only in the neutrino sector but in other sectors as well, obtaining competitive constraints. 

%\textit{Acknowledgments}: I acknowledge support
%from Generalitat Valenciana through the plan GenT program (CIDEGENT/2018/019).

\afterpage{\clearpage}
%-------------------------------------------

%-------------------------------------------
% mocwd to Light DM as requested by author
%\subsection{\emph{New ideas}: Jupiter missions as probes of dark matter and long-lived mediators  - {\it L. Li}}
%\label{Li}
%{\it Author: Lingfeng Li, <lingfeng\_li@brown.edu>}
%-------------------------------------------

%-------------------------------------------
\subsection{Conclusions}
\label{ssec:LDM_conclusions}
In summary, a synergistic approach including accelerator-based experiments, direct detection experiments and astrophysical probes is essential to robustly explore the vast FIP parameter space, and establish the new particle properties in the case of an observation. A vast program is underway to explore FIPs at the MeV-GeV scale, with a particular focus on hidden sector DM. Direct freeze-out defines sharp targets for the coupling between the dark sector mediator and  SM fields, which will be probed by dedicated experiments in the near future. Considerable experimental activities are also devoted to searches for generic dark sector mediator signatures, in particular visible and invisible dark photon decays. Indirect searches offer complementary avenues to search for light dark matter, although current constraints are still far above the levels required to probe thermal freeze-out. A next-generation gamma-ray instrument is especially needed to further explore the MeV region and close the so-called MeV gap.

\begin{figure}[ht!]
\centering
\includegraphics[width=\linewidth]{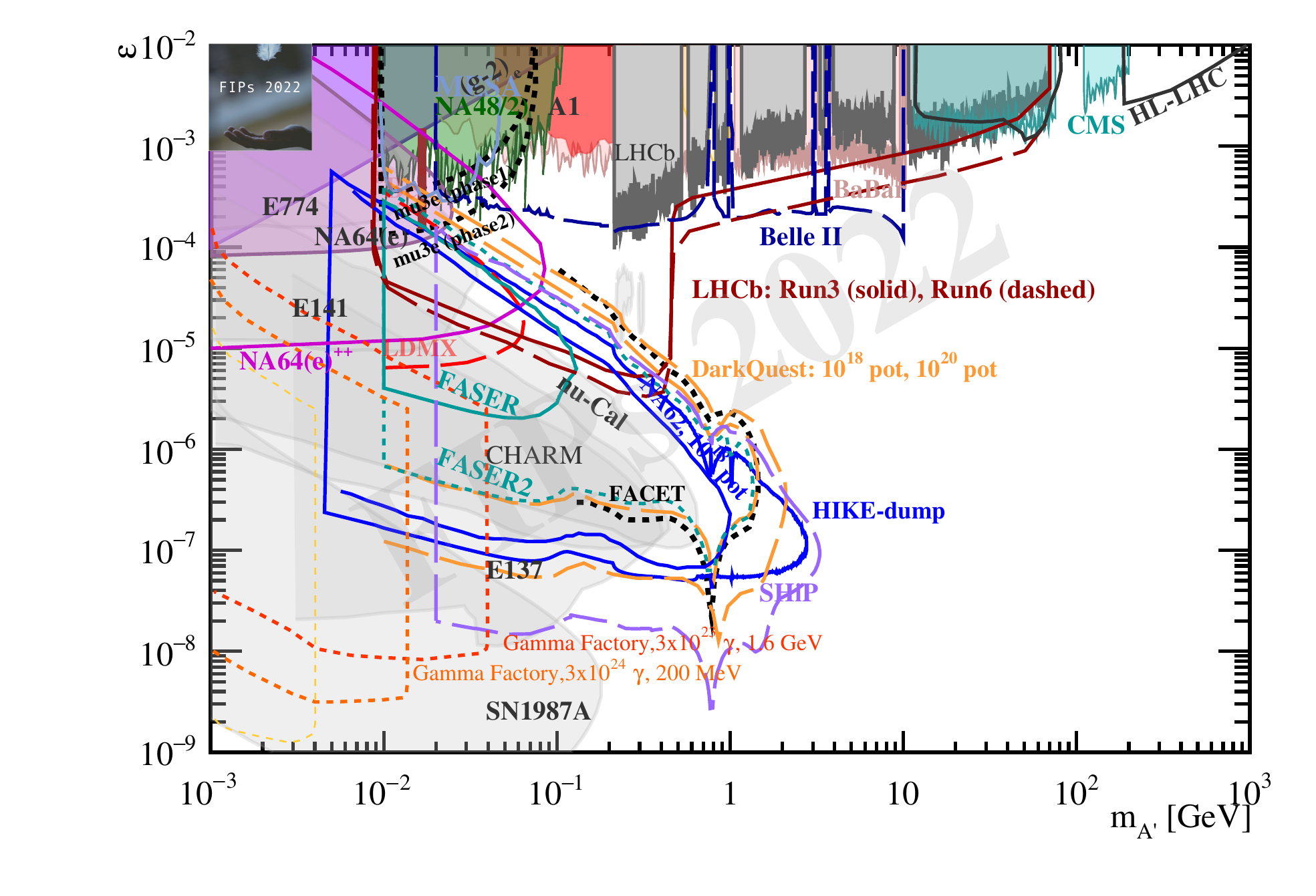}
\caption{ {\bf Dark photon into visible final states (BC1):} $\varepsilon$ versus $m_{A'}$.  Current bounds and future projections for 90\% CL exclusion limits. {\bf Legend:} filled gray areas are bounds coming from interpretation of old data sets or astrophysical data; filled coloured areas are bounds set by experimental collaborations; Solid coloured lines are projections based on existing data sets; 
Dashed coloured lines are projections based on full Monte Carlo simulations; Dotted coloured lines are projections based on toy Monte Carlo simulations. Filled areas are existing limits from searches at experiments at collider/fixed target (A1~\cite{Merkel:2014avp}, LHCb~\cite{LHCb:2019vmc},CMS~\cite{CMS:2019kiy},BaBar~\cite{BaBar:2014zli}, KLOE~\cite{KLOE-2:2011hhj,KLOE-2:2012lii,KLOE-2:2014qxg,KLOE-2:2016ydq}, NA64(e)~\cite{Andreev:2021fzd} and NA48/2~\cite{NA482:2015wmo}) and old  beam dump:  E774~\cite{Bross:1989mp}, E141~\cite{Riordan:1987aw}, E137~\cite{Bjorken:1988as,Batell:2014mga,Marsicano:2018krp}), $\nu$-Cal~\cite{Blumlein:2011mv,Blumlein:2013cua}, CHARM (from~\cite{Gninenko:2012eq}), and BEBC (from~\cite{Marocco:2020dqu}). Bounds from supernovae~\cite{Chang:2016ntp} and $(g-2)_e$~\cite{Pospelov:2008zw} are also included. Coloured curves are projections for existing and proposed experiments: Belle II~\cite{Belle-II:2018jsg}; LHCb upgrade~\cite{LHCb:upgrade}; NA62 in dump mode with $10^{18}$~\cite{NA62:dump} and HIKE-dump with $5 \times 10^{19}$ pot~\cite{HIKE-LoI}; NA64(e)~\cite{Gninenko:2013rka, Andreas:2013lya}; FASER~\cite{Ariga:2018uku} and FASER2~\cite{Anchordoqui:2021ghd, Feng:2022inv};  
FACET~\cite{Cerci:2021nlb};
DarkQUEST~\cite{Apyan:2022tsd}; 
LDMX~\cite{Berlin:2018bsc};
%HPS~\cite{HPS:2018xkw};  
DarkMESA~\cite{Doria:2019sux}; Mu3e~\cite{Echenard:2014lma};  
HL-LHC~\cite{Curtin:2014cca};
Gamma Factory~\cite{Chakraborti:2021hfm}.
}
\label{fig:BC1}
\end{figure}

\begin{figure}[ht!]
\centering
\includegraphics[width = 0.9\linewidth] {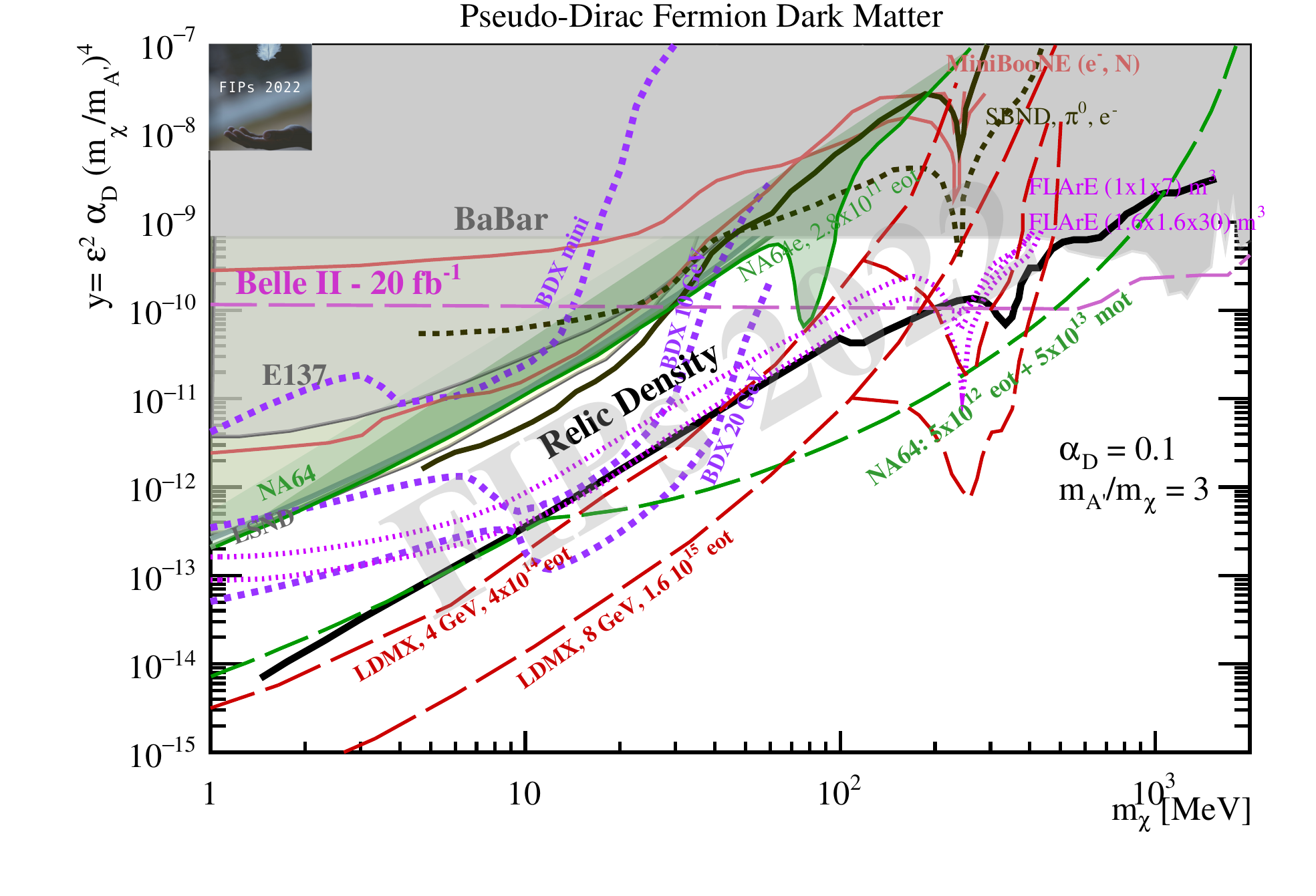}
\includegraphics[width = 0.9\linewidth] {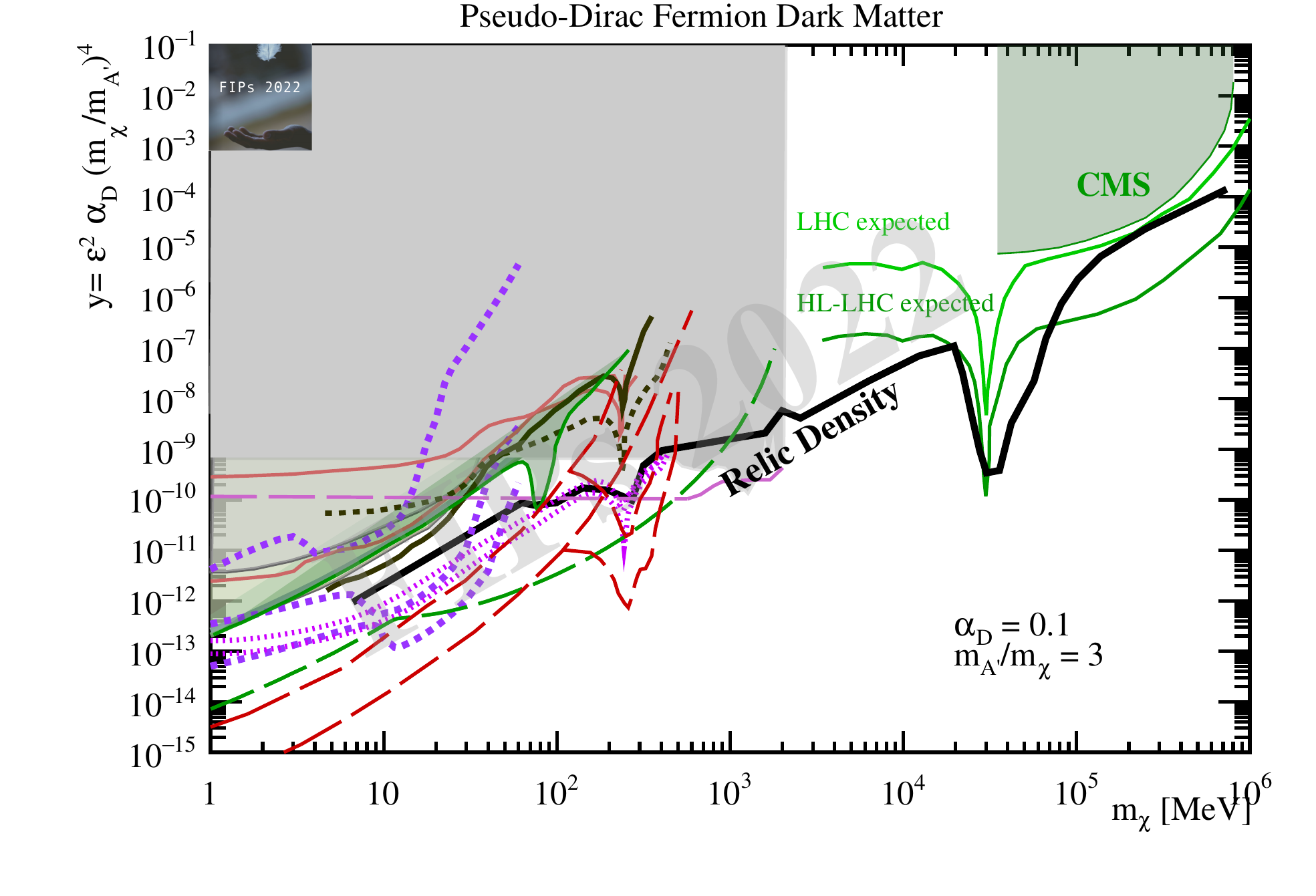}
\caption{\small {\bf Dark Photon into invisible final states (BC2)}.
  Current bounds and future projections for 90\% CL exclusion limits
 for light dark matter production through a dark photon in the plane defined by the "yield" variable $y$ as a function of DM mass $m_{\chi}$ for a specific choice of $\alpha_D = 0.1$ and $m_{A'}/m_{\chi} = 3$. The DM candidate is assumed to be a pseudo-Dirac fermion.  Top plot shows the DM mass range up to a few GeV, bottom plot up to 1 TeV. 
{\bf Legend:} filled gray areas are bounds coming from interpretation of old data sets or astrophysical data; filled coloured areas are bounds set by experimental collaborations; Solid coloured lines are projections based on existing data sets; 
Dashed coloured lines are projections based on full Monte Carlo simulations; Dotted coloured lines are projections based on toy Monte Carlo simulations.
  Current limits shown as filled areas come from: BaBar~\cite{BaBar:2017tiz}; CMS~\cite{CMS:2021far}; NA64$_e$~\cite{Andreev:2021fzd}; reinterpretation of the data from E137~\cite{Batell:2014mga} and LSND~\cite{deNiverville:2011it};  result from MiniBooNE~\cite{MiniBooNEDM:2018cxm}. The projected sensitivities, shown as solid, dashed, or dotted lines, come from: %SHiP~\cite{SHiP-ECN3-LoI}, 
  BDX-mini~\cite{Battaglieri:2020lds};
  %BDX~\cite{Battaglieri:2016ggd},
  SBND~\cite{MicroBooNE:2015bmn},
  NA64~\cite{Gninenko:2019qiv}; FLArE~\cite{Anchordoqui:2021ghd},
LDMX~\cite{LDMX:2018cma,Akesson:2022vza},%
  Belle-II~\cite{Belle-II:2018jsg}. The "LHC expected" and "HL-LHC expected" sensitivities come from \cite{Boveia:2022adi}. 
  }
\label{fig:BC2}
\end{figure}

\begin{figure}[t]
\centering
\includegraphics[width=\linewidth]{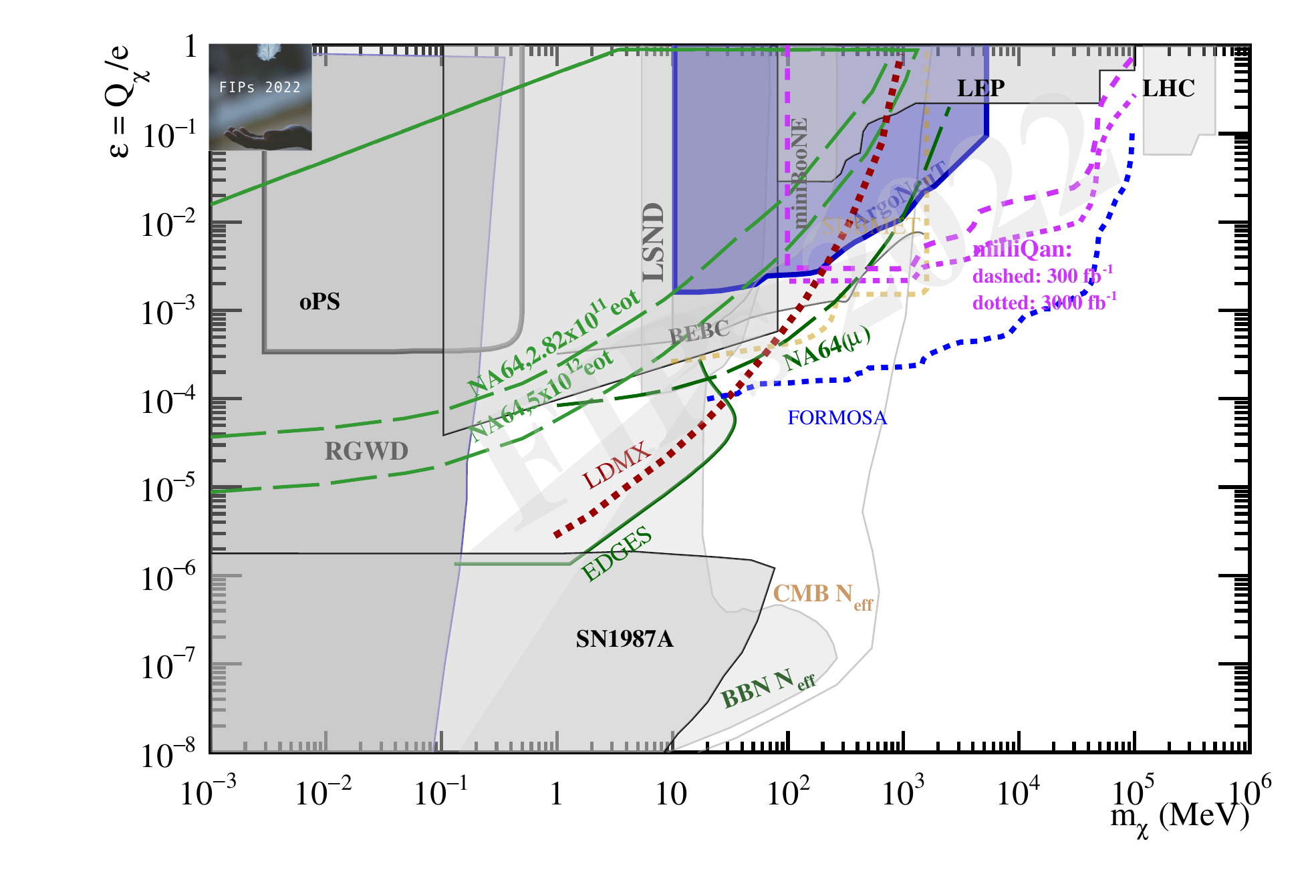}

\caption{\small  \label{fig:milli} {\bf Dark Photon milli-charged particles (BC3)}.  
Current bounds and future projections for 90\% CL exclusion limits. 
{\bf Legend:} filled gray areas are bounds coming from interpretation of old data sets or astrophysical data; filled coloured areas are bounds set by experimental collaborations; Solid coloured lines are projections based on existing data sets; 
Dashed coloured lines are projections based on full Monte Carlo simulations; Dotted coloured lines are projections based on toy Monte Carlo simulations.
Existing limits: interpretation of stellar evolution (
%RGWD~\cite{Vogel:2013raa} and 
SN1987~\cite{Chang:2018rso});
$N_{eff}$ during BBN and CMB~\cite{Vogel:2013raa};
%invisible decays of ortho-positronium (oPS)~\cite{Badertscher:2006fm};
%|SLAC milliQ experiment~\cite{Prinz:1998ua};
reinterpretation of data from LSND and MiniBooNE~\cite{Magill:2018tbb};
interpretation of BEBC data \cite{Marocco:2020dqu};
interpretation of the anomalous 21~cm hydrogen absorption signal by EDGES~\cite{Kovetz:2018zan};
interpretation of searches at LEP~\cite{Davidson:2000hf} and LHC~\cite{Jaeckel:2012yz}. Bounds from ArgoNeuT~\cite{ArgoNeuT:2019ckq}
Future sensitivities:
 NA64(e)~\cite{Arefyeva:2022eba}; NA64($\mu$)~\cite{NA64:2018iqr}; %FerMINI~\cite{Kelly:2018brz};
 milliQAN~\cite{Ball:2016zrp}; FORMOSA~\cite{Feng:2022inv}; LDMX~\cite{Berlin:2018bsc}. %LDMX~\cite{Akesson:2018vlm}. 
 %See text for details.  %Figure revised from Ref.~\cite{Fabbrichesi:2020wbt}.
 }
\label{fig:BC3}
\end{figure}

\begin{figure}[h]
\begin{center}
\includegraphics[width=\textwidth]{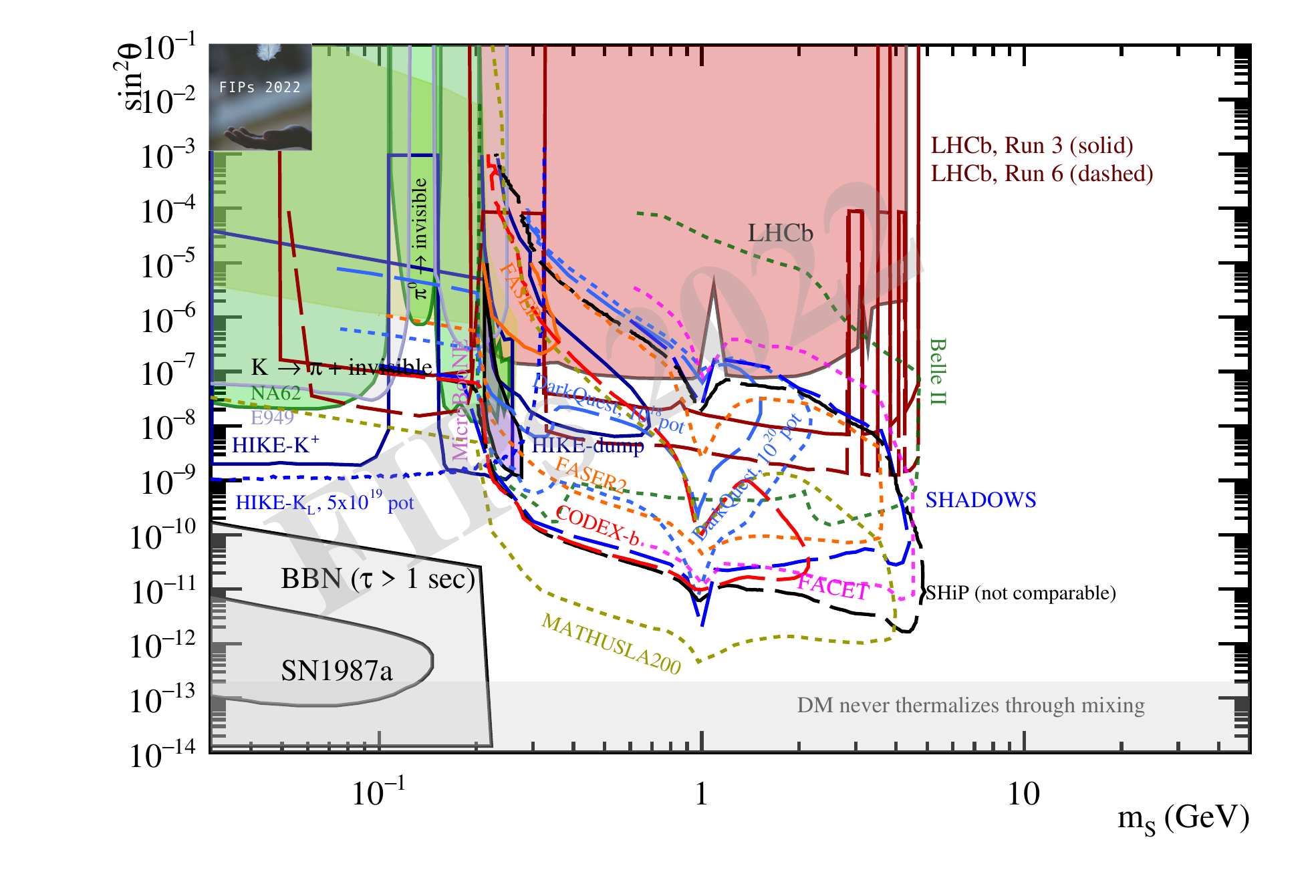}
\end{center}
\caption{{\bf Sensitivity to light dark scalar (BC4).} Current bounds and future projections for 90\% CL exclusion limits. 
{\bf Legend:} filled gray areas are bounds coming from interpretation of old data sets or astrophysical data; filled coloured areas are bounds set by experimental collaborations; Solid coloured lines are projections based on existing data sets; 
Dashed coloured lines are projections based on full Monte Carlo simulations; Dotted coloured lines are projections based on toy Monte Carlo simulations.
Filled areas come from: reinterpretation~\cite{Winkler:2018qyg} of results from CHARM experiment~\cite{Bergsma:1985qz};
NA62~\cite{CortinaGil:2020fcx, NA62:2021zjw, NA62:2020pwi};
E949~\cite{Artamonov:2008qb,Dev:2019hho}; 
MicroBooNE from NuMI data~\cite{MicroBooNE:2022ctm}; LHCb~\cite{Aaij:2016qsm, Aaij:2015tna} and Belle~\cite{Wei:2009zv}.
Coloured lines are projections of existing or proposed experiments: SHiP~\cite{SHiP-ECN3-LoI} (the legend "not comparable" means that SHiP has used the exclusive scalar production processes,
while all the others have used the inclusive one);
HIKE-K$^+$, HIKE-dump, and  HIKE-$K_L$~\cite{HIKE-LoI}; SHADOWS, slightly revised from~\cite{SHADOWS-LoI}; DarkQuest~\cite{Batell:2020vqn, Apyan:2022tsd}, Belle II~\cite{Filimonova:2019tuy}, LHCb Run 3 and Run 6 ~\cite{Craik:2022riw}, FASER2~\cite{Feng:2022inv}, CODEX-b~\cite{Aielli:2019ivi,Aielli:2022awh},  MATHUSLA~\cite{Alpigiani:2020tva}, and 
FACET~\cite{Cerci:2021nlb}.
BBN and SN1987A are from ~\cite{Fradette:2017sdd} and ~\cite{Dev:2020eam}. 
}
\label{fig:BC4}
\end{figure}

\begin{figure}[h]
\begin{center}
\includegraphics[width=\textwidth]{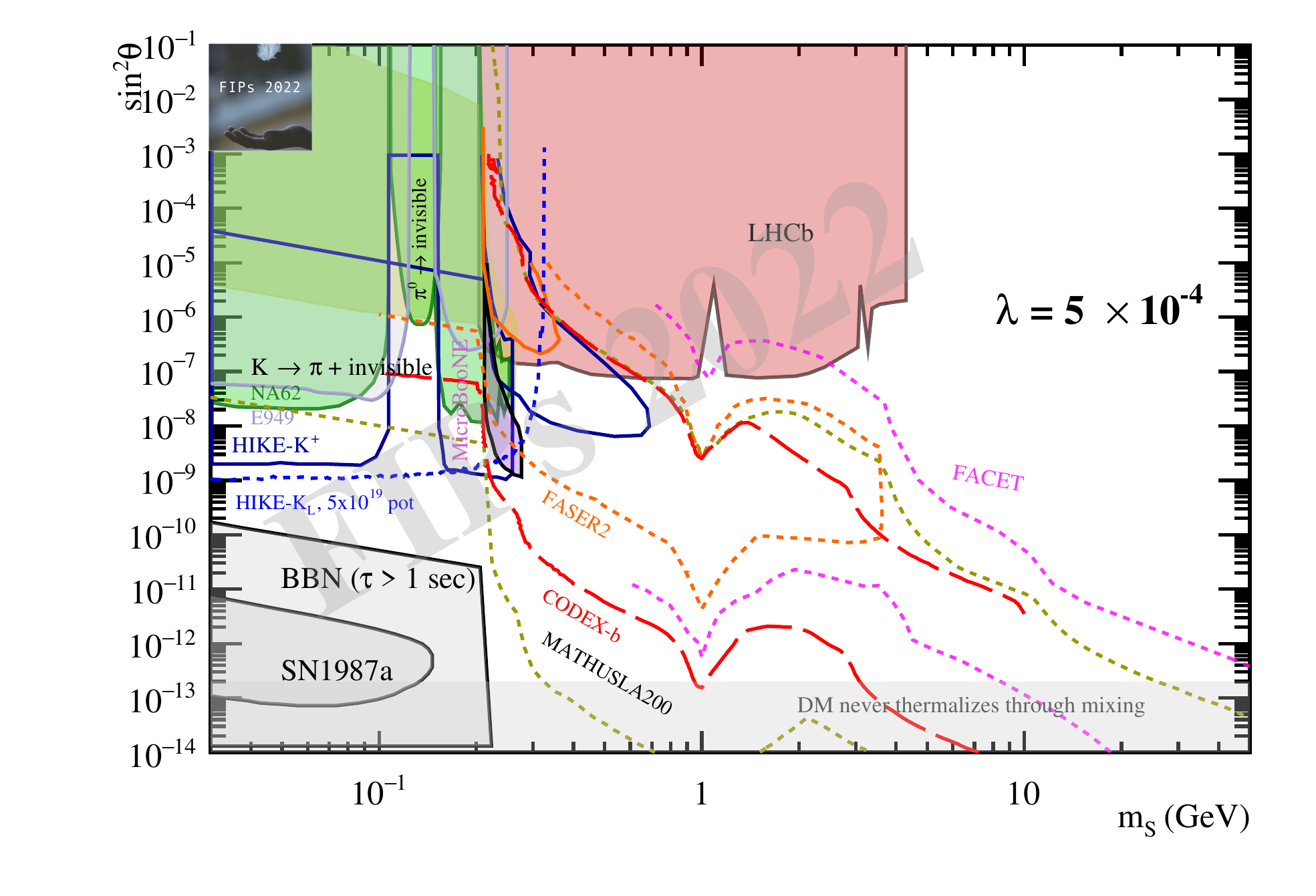}
\end{center}
\caption{{\bf Sensitivity to light dark scalar (BC5).} Current bounds and future projections for 90\% CL exclusion limits. 
{\bf Legend:} filled gray areas are bounds coming from interpretation of old data sets or astrophysical data; filled coloured areas are bounds set by experimental collaborations; Solid coloured lines are projections based on existing data sets; 
Dashed coloured lines are projections based on full Monte Carlo simulations; Dotted coloured lines are projections based on toy Monte Carlo simulations.
Filled areas come from: reinterpretation~\cite{Winkler:2018qyg} of results from CHARM experiment~\cite{Bergsma:1985qz};
NA62~\cite{CortinaGil:2020fcx, NA62:2021zjw, NA62:2020pwi};
E949~\cite{Artamonov:2008qb,Dev:2019hho}; MicroBooNE from NuMI data~\cite{MicroBooNE:2022ctm};
LHCb~\cite{Aaij:2016qsm, Aaij:2015tna} and Belle~\cite{Wei:2009zv}.
Coloured lines are projections of existing or proposed experiments: SHiP~\cite{SHiP-ECN3-LoI};
HIKE-K$^+$~\cite{HIKE-LoI, HIKE:2022qra};  HIKE-$K_L$~\cite{HIKE-LoI, HIKE:2022qra}; SHADOWS~\cite{SHADOWS-LoI};  FASER2~\cite{Feng:2022inv}, CODEX-b~\cite{Aielli:2019ivi,Aielli:2022awh},  MATHUSLA~\cite{Alpigiani:2020tva}, and FACET~\cite{Cerci:2021nlb}.  BBN and SN 1987A are from ~\cite{Fradette:2017sdd} and ~\cite{Dev:2020eam}. 
}
\label{fig:BC5}
\end{figure}

\begin{figure}[ht!]
\centering
\includegraphics[width=\linewidth]{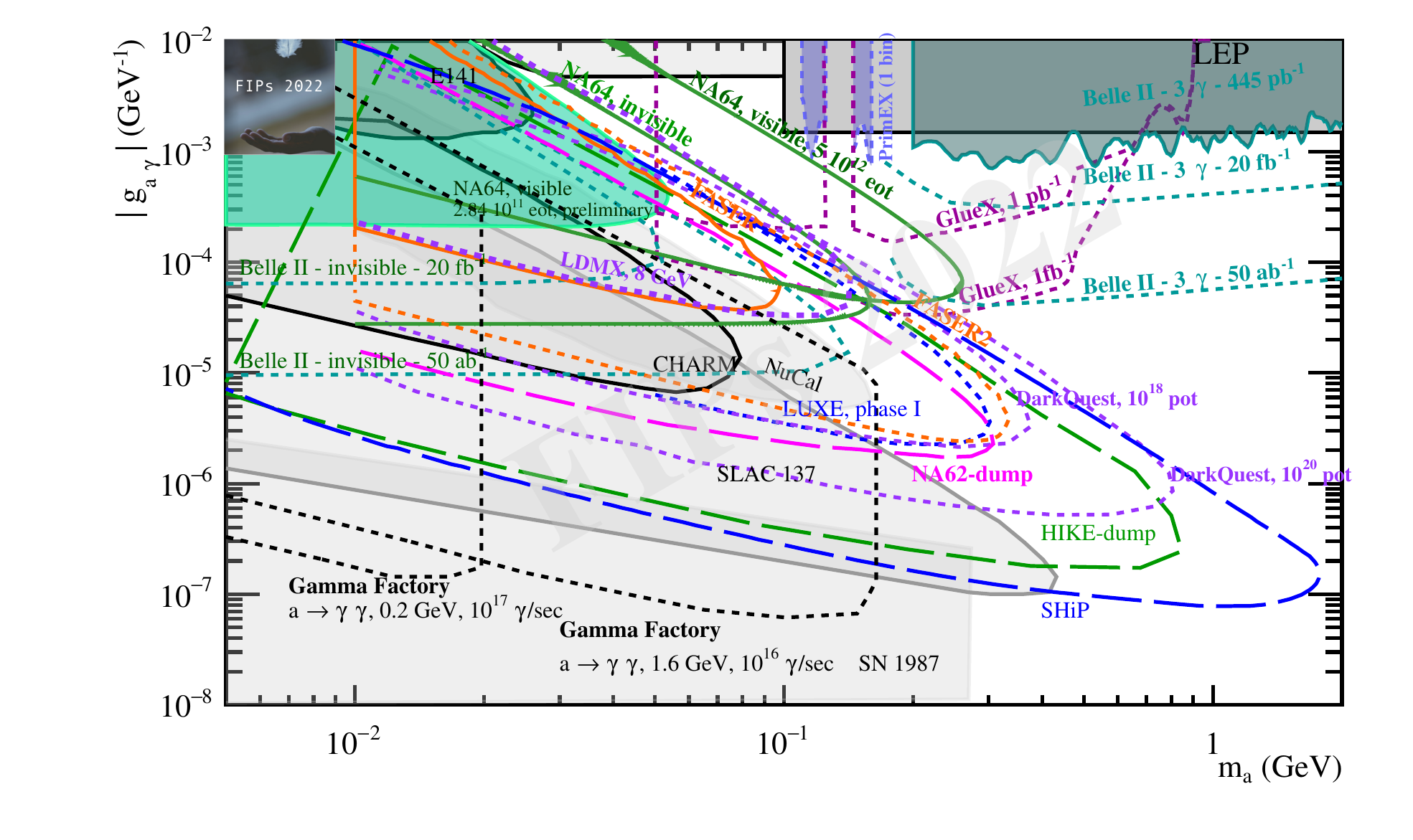}
\caption{{\bf Axions/ALPs with photon coupling (BC9).}  Region of interest for accelerator-based experiments up to a few GeV. 
Current bounds and future projections for 90\% CL exclusion limits.
{\bf Legend:} filled gray areas are bounds coming from interpretation of old data sets or astrophysical data; filled coloured areas are bounds set by experimental collaborations; Solid coloured lines are projections based on existing data sets; 
Dashed coloured lines are projections based on full Monte Carlo simulations; Dotted coloured lines are projections based on toy Monte Carlo simulations.
Shaded areas are excluded regions from:
LEP (data:~\cite{L3:1994shn, DELPHI:1991emv, DELPHI:1994mra, L3:1995nbq}; interpretation:~\cite{Knapen:2016moh} above 100 MeV and ~\cite{Jaeckel:2015jla} below 100 MeV. Caveat: the LEP line above 100 MeV is likely extendable also in the region below 100 MeV, down to the current bound from NA64);
Belle II~\cite{Belle-II:2020jti};
E137~\cite{Bjorken:1988as}; 
NA64~\cite{NA64:2020qwq};
CHARM~\cite{Gninenko:2012eq}; 
NuCal~\cite{Blumlein:1990ay}.
%PrimEx~\cite{Aloni:2019ruo} based on~\cite{Larin:2010kq}.
Curves are projections from: NA62-dump~\cite{NA62_Addendum};
LDMX with 8 GeV and $10^{16}$~eot~\cite{Berlin:2018bsc};
Belle II~\cite{Dolan:2017osp}
for 20~fb$^{-1}$ and 50~ab$^{-1}$;
SHiP~\cite{SHiP-ECN3-LoI};
FASER~\cite{Ariga:2018uku} and FASER2~\cite{Feng:2022inv};
NA64$_e^{++}$~\cite{NA64:eplus} in visible and invisible modes; LUXE-phase 1~\cite{Bai:2021gbm}; HIKE-dump~\cite{HIKE-LoI}; Gamma Factory~\cite{Balkin:2021jdr}.
interpretation of the physics reach~\cite{Aloni:2019ruo}
of PrimEx~\cite{PrimEx:2010fvg} and GlueX experiments at JLab.
}
\label{fig:BC9}
\end{figure}

\begin{figure}[h]
\begin{center}
\includegraphics[width=\textwidth]{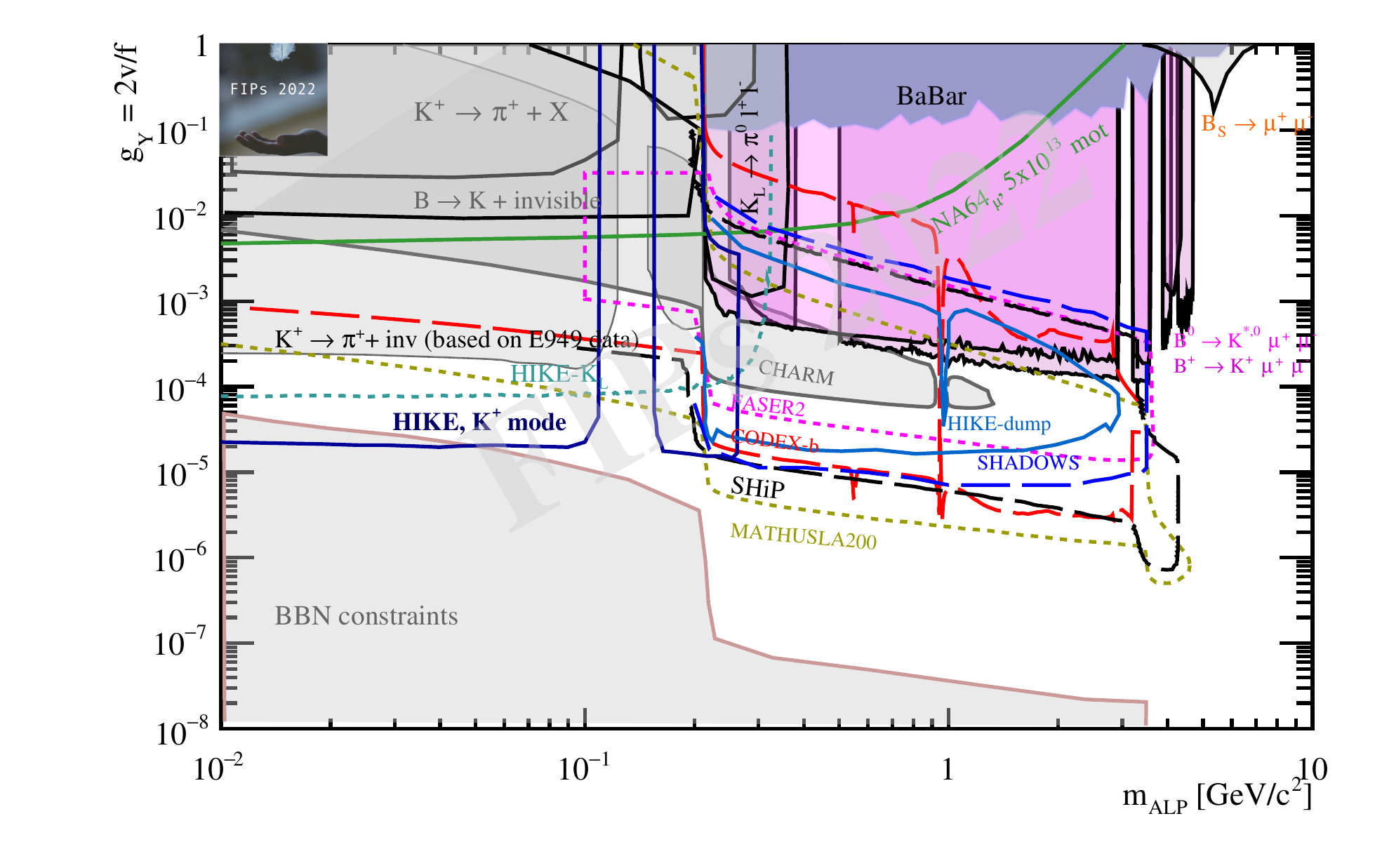}
\end{center}
\caption{{\bf Sensitivity to ALPs with fermion couplings (BC10).} Current bounds and future projections for 90\% CL exclusion limits. {\bf Legend:} filled gray areas are bounds coming from interpretation of old data sets or astrophysical data; filled coloured areas are bounds set by experimental collaborations; Solid coloured lines are projections based on existing data sets; 
Dashed coloured lines are projections based on full Monte Carlo simulations; Dotted coloured lines are projections based on toy Monte Carlo simulations. Current bounds and prospects from FASER2~\cite{Ariga:2018uku}, CODEX-b~\cite{Aielli:2019ivi}, MATHUSLA~\cite{Beacham:2019nyx}, HIKE-$K^+$, HIKE-$K_L$, and HIKE-dump ~\cite{HIKE-LoI}, SHADOWS~\cite{SHADOWS-LoI}, and SHiP~\cite{SHiP-ECN3-LoI}.
CHARM and LHCb filled areas have been adapted
by F.~Kahlhoefer, following Ref.~\cite{Dobrich:2018jyi}. 
}
\label{fig:BC10}
\end{figure}

\begin{figure}[h]
\begin{center}
\includegraphics[width=\textwidth]
{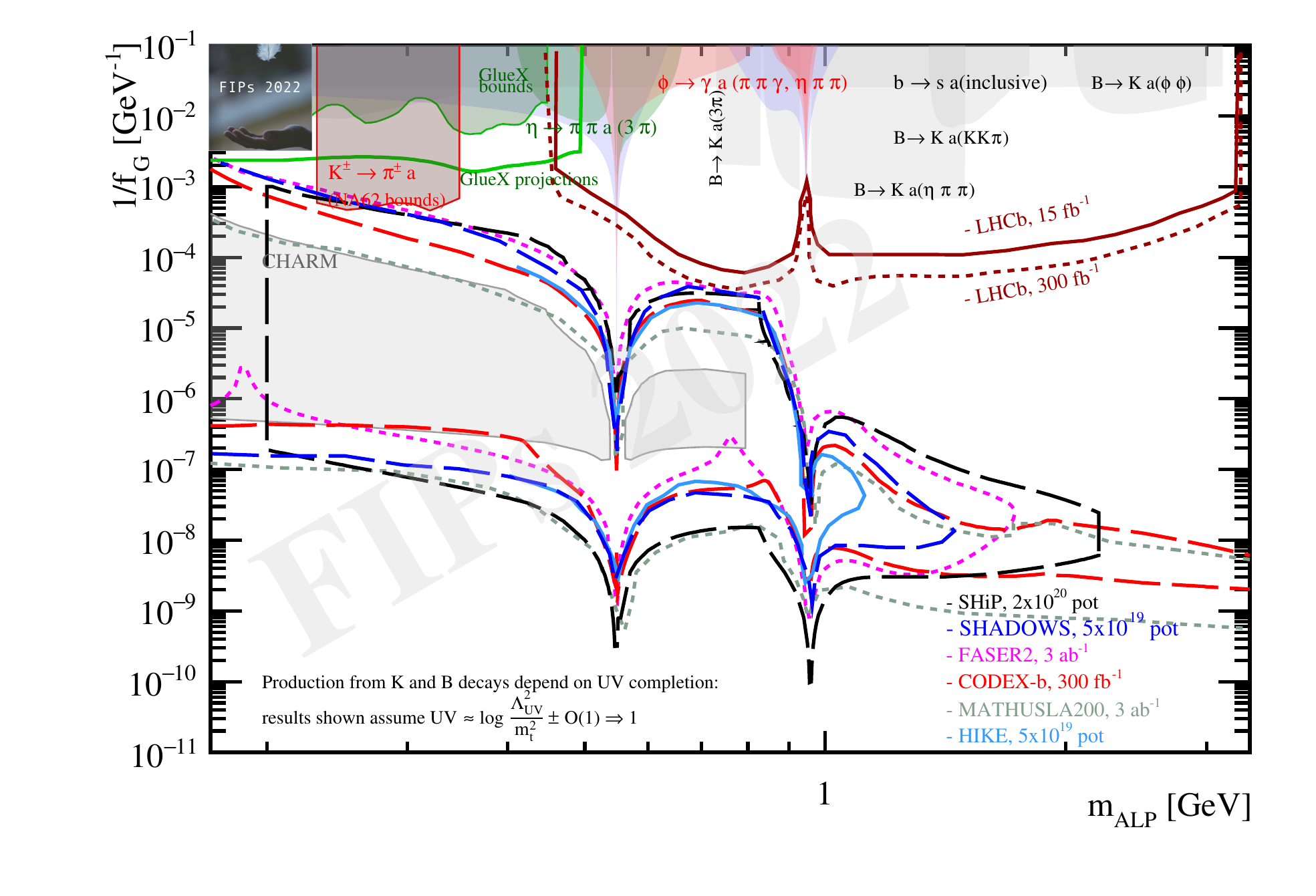}
\end{center}
\caption{{\bf Sensitivity to ALPs with gluon coupling (BC11).} 
Current bounds and future projections for 90\% CL exclusion limits. 
{\bf Legend:} filled gray areas are bounds coming from interpretation of old data sets or astrophysical data; filled coloured areas are bounds set by experimental collaborations; Solid coloured lines are projections based on existing data sets; 
Dashed coloured lines are projections based on full Monte Carlo simulations; Dotted coloured lines are projections based on toy Monte Carlo simulations.
{\bf Current bounds:} CHARM gray filled area has been computed by F.~Kling, recasting the search for long-lived particles decaying to two photons performed at CHARM~\cite{Bergsma:1985qz}. Other coloured filled areas are kindly provided by Mike Williams and revisited from Ref.~\cite{Aloni:2018vki}.
The gray areas depend on UV completion and the results shown assume $\approx [{\rm log \Lambda^2_{UV}}/m^2_t \pm {\mathcal{O}}(1)] \Rightarrow 1$. {\bf Projections:} LHCb with 15 fb$^{-1}$ and 300 fb$^{-1}$ \cite{Craik:2022riw}; CODEX-b with 300 fb$^{-1}$ \cite{Aielli:2022awh}); MATHUSLA with 3 ab$^{-1}$ (estimate from \cite{Aielli:2022awh}); FASER2 with  3 ab$^{-1}$ ~\cite{Feng:2022inv}; SHiP with $2 \times 10^{20}$~pot \cite{SHiP-ECN3-LoI}; SHADOWS~\cite{SHADOWS-LoI} and HIKE-dump~\cite{HIKE-LoI} with $5 \times 10^{19}$~pot each.
SHiP, HIKE-dump, and SHADOWS projections are based on Ref.~\cite{Jerhot:2022chi} which resums the logarithmic dependence on $\Lambda_{UV}$ through RGE evolution.
}
\label{fig:BC11}
\end{figure}

Tables~\ref{tab:exps1} and Table~\ref{tab:exps2}  show past, extant and future (proposed or approved) experiments running at accelerators that have sensitivity to search FIPs in a mass range between $\sim$~MeV and 100~GeV. Figures~\ref{fig:BC1}-\ref{fig:BC11} show the state of the art of existing limits and future projections (both at  90 \% CL) for all the accelerator-based experiments worldwide for the eight PBC benchmarks~\cite{Beacham:2019nyx} that might be connected with the existence of a light Dark Matter particle.
 
 \begin{itemize}

\item {\em BC1, Minimal dark photon model}: the SM is augmented by a single new state $A'$. DM is assumed to be either heavy or contained in a different sector. In that case, once produced, the dark photon decays back to SM states. The parameter space of this model is $\{ m_{A'}, \epsilon \}$.

\item{\em BC2, Light dark matter coupled to dark photon}: model of minimal dark freezeout. The benchmark values of dark coupling constant $\alpha_{D} = g_D^2/(4\pi)$ are such that the decay of $A'$ occurs predominantly into the invisible $\chi\chi^*$ final state. The parameter space for this model is $\{ m_{A'}, \epsilon, m_\chi, \alpha_D \}$ with further model-dependence associated with properties of $\chi$ (boson or fermion). The suggested choices for the parameter space are (i) $\epsilon$ vs $m_{A'}$ with $\alpha_D \gg \epsilon^2 \alpha$ and $2m_\chi <m_{A'}$, (ii) $y$ vs. $m_\chi$ plot where the {\it yield} variable $y$,  $y = \alpha_D \epsilon^2 (m_\chi/m_{A'})^4$, contains a combination of parameters relevant for the freeze-out and DM-SM particles scattering cross section. One possible choice is $\alpha_D = 0.1$ and $m_{A'}/m_\chi = 3$. 

\item{\em BC3, Millicharged particles (mCP):}
This benchmark can be seen as a specific limit of the vector portal theory when $m_{A'}$ goes to zero and the parameter space simplifies to the mass ($m_\chi$) and effective charge ($|Q| = |\epsilon g_D e|$) of millicharged particles. The suggested choice of parameter space is $(m_\chi, \epsilon = Q_\chi/e)$ and $\chi$ can be taken to be a fermion. 

\item{\em BC4, Higgs-mixed scalar: minimal scenario}
 This benchmark considers a dark scalar $S$ mixing with the Higgs, taking the quadratic coupling constant $\lambda = 0$. All production and decay are controlled by the mixing angle $\theta$ of the singlet scalar with the SM Higgs boson. The parameter space for this model is ${\theta, m_S}$.

\item {\em BC5} Higgs-mixed scalar with large pair-production channel:  In this model, by contrast, the quartic coupling constant $\lambda$ is  assumed to dominate the production via {\em e.g.} $h\to SS$, $B \to K^{(*)}SS$, $B^0 \to SS$ etc., and the parameter space is accordingly $\{\lambda, \theta, m_S\}$. In the sensitivity plots a value  of the branching fraction $BR({h \to SS})$ close to $10^{-2}$ is assumed in order to be complementary to the LHC searches for the Higgs to invisible channels.

\item  {\em BC9, photon dominance:} Assuming a single ALP state $a$, with a predominant coupling to photons, all phenomenology (production, decay, oscillation in the magnetic field) can be determined as functions of the $\{m_a, g_{a\gamma} \}$ parameter space. 

\item {\em BC10 fermion dominance:} Assuming a single ALP state $a$, with a predominant coupling to fermions, all phenomenology (production and decay) can be determined as functions of  $\{m_a, f^{-1}_l, f^{-1}_q \}$. For the sake of simplicity, we take $f_q=f_l$.  

 \item  {\em BC11, gluon dominance:} This case assumes an ALP coupled to gluons. The parameter space is $\{m_a, f^{-1}_G \}$. In this case, the limit of $m_a < m_{a,QCD}|_{f_a=f_G}$ is unnatural as it requires fine-tuning and therefore is less motivated.

\end{itemize} 

All the plots have been made using the following graphical conventions: filled coloured areas indicate existing bounds, while filled shaded gray areas are interpretations of astrophysical/cosmological measurements and/or reinterpretation of old data sets performed by people not belonging to the original collaboration. Dotted lines are projections obtained using a toy Monte Carlo; dashed lines are projections obtained using a full Monte Carlo with background simulated (at different levels); solid lines are extrapolations from existing data sets.

%====================
\clearpage
%====================

%=============================================

\begin{table}[h]
\tabcolsep7.5pt
\caption{Main past accelerator-based experiments sensitive to FIPs searches. Legend for portals: 1: Vector; 2: Scalar; 3: Pseudo-scalar; 4: Fermion. The techniques used are: i) visible decays; ii) invisible decays; $e^-$ or nucleon recoil; missing mass $\cancel{\it{M}}$, missing momentum $\cancel{\it{p}}$ and missing energy $\cancel{\it{E}}$. }
\label{tab:exps1}
\begin{center}
\begin{tabular}{|l|l|c|c|c|c|}
\hline 
Experiment & lab  & beam & particle yield/$\mathcal{L}$ & technique & portals \\
\hline
%=========================================
{\bf past} &     & & &  &   \\ \hline
%=========================================

{\scriptsize ArgoNeuT~\cite{Anderson:2012vc}} & {\scriptsize FNAL}   & {\scriptsize $p$, 120~GeV} & {\scriptsize $1.25 \times 10^{20}$} & {\scriptsize visible}  & {\scriptsize (1,4) } \\ 

{\scriptsize BaBar~\cite{BaBar:2001yhh}} & {\scriptsize SLAC}   & {\scriptsize $e^+ e^-$, 10.58~GeV} & {\scriptsize 514~fb$^{-1}$} & {\scriptsize visible, invis.}  & {\scriptsize (1) } \\ 

{\scriptsize BEBC-WA66~\cite{1986253}}    & {\scriptsize CERN}    & {\scriptsize $p$, 400~GeV} & {\scriptsize } &{\scriptsize visible} &{\scriptsize (1,4) }  \\ 

{\scriptsize Belle~\cite{Belle:2000cnh}}    & {\scriptsize KEK}    & {\scriptsize $e^+ e^-$, 10.58~GeV} & {\scriptsize 0.6-0.8~fb$^{-1}$} &{\scriptsize visible} &{\scriptsize (1,2,4) }  \\ 

{\scriptsize CHARM~\cite{Winter:1982sk}}    & {\scriptsize CERN}  &  {\scriptsize $p$, 400~GeV} & {\scriptsize $2.4~\cdot~10^{18}$} & {\scriptsize visible} & {\scriptsize (1,2,3,4)}  \\ 

{\scriptsize E137~\cite{Bjorken:1988as}}     & {\scriptsize SLAC}   & {\scriptsize $e^-$, 20~GeV} & {\scriptsize $2\cdot 10^{20}$ (30~C)} & {\scriptsize visible} & {\scriptsize (1,3)}  \\ 

{\scriptsize E141~\cite{Riordan:1987aw}}     & {\scriptsize SLAC}  &  {\scriptsize $e^-$, 9~GeV} & {\scriptsize $2 \cdot 10 yes^{15}$} & {\scriptsize visible} & {\scriptsize (1,3)}  \\ 

{\scriptsize E774~\cite{Bross:1989mp}}     & {\scriptsize FNAL}  &  {\scriptsize $e^-$, 275~GeV} & {\scriptsize $2\cdot 10^{15} $ } & {\scriptsize visible} & {\scriptsize (1)}  \\

{\scriptsize KLOE~\cite{Adinolfi:2002uk,Adinolfi:2002zx}}     & {\scriptsize LNF}   & {\scriptsize $e^+ e^-$, 1~GeV} & {\scriptsize up to 1.7~fb$^{-1}$} &{\scriptsize visible, inv.} & {\scriptsize (1)} \\ 

{\scriptsize LSND~\cite{deNiverville:2011it}}     & {\scriptsize LANL}   &  {\scriptsize $p$, 800~MeV} & {\scriptsize $10^{23}$~pot} & {\scriptsize $e^-$ recoil} & {\scriptsize (1)} \\

{\scriptsize MiniBooNE~\cite{MiniBooNE:2008paa}} & {\scriptsize FNAL}  & {\scriptsize $p$, 8~GeV} & {\scriptsize $1.9 \cdot 10^{20}$} & {\scriptsize recoil $e,N$} &  {\scriptsize (1)} \\

{\scriptsize NA48/2~\cite{NA482:2015wmo} }  & {\scriptsize CERN}   & {\scriptsize $K^{\pm}$, 60~GeV} & {\scriptsize $1.6 \cdot 10^{11} \; K$-decays} & {\scriptsize visible} & {\scriptsize (1)}  \\ 

{\scriptsize NuCAL~\cite{Blumlein:2011mv,Blumlein:2013cua}}    & {\scriptsize Serpukhov}     & {\scriptsize $p$, 70~GeV} & {\scriptsize $1.7 \cdot 10^{18} $} & {\scriptsize visible} & {\scriptsize (1,3)} \\

{\scriptsize PIENU~\cite{Malbrunot:2011zz}}  & {\scriptsize TRIUMF}   &{\scriptsize $\pi^+$, 75~MeV}   & {\scriptsize $10^7$}  & {\scriptsize $\cancel{\it{M}}$} & {\scriptsize (4)} \\ \hline

\hline
\end{tabular}
\end{center}
\end{table}

\clearpage
\begin{table}[h]
\tabcolsep7.5pt
\caption{Main current, and future (proposed or approved) accelerator-based experiments sensitive to FIPs searches. Legend for portals: 1: Vector; 2: Scalar; 3: Pseudo-scalar; 4: Fermion.}
\label{tab:exps2}
\begin{center}
\begin{tabular}{|l|l|c|c|c|c|c|}
\hline 
Experiment & lab  & beam & particle yield/$\mathcal{L}$ & technique & portals & timescale \\
\hline
%=========================================
{\bf current}  &       & & &  &  & \\ \hline
%=========================================

%{\scriptsize APEX~\cite{APEX:2011dww}}  & {\scriptsize JLAB} & {\scriptsize $e^+$, 2.2~GeV} & {\scriptsize up to 150~$\mu$A} & {\scriptsize visible} & {\scriptsize (1) } & {\scriptsize unknown} \\

{\scriptsize ATLAS~\cite{ATLAS:2008xda}}  & {\scriptsize CERN}  &  {\scriptsize $pp$, 13-14~TeV} & {\scriptsize up to 3~ab$^{-1}$} & {\scriptsize visible, invis.} & {\scriptsize (1,2,3,4) } & {\scriptsize 2042}\\

{\scriptsize Belle II~\cite{Belle-II:2010dht}} & {\scriptsize KEK} & {\scriptsize $e^+ e^-$, 11~GeV} & {\scriptsize up to 50~ab$^{-1}$ } & {\scriptsize visible, invis.} & {\scriptsize (1,2,3,4)} &  {\scriptsize 2035} \\

{\scriptsize CMS~\cite{CMS:2008xjf}}   & {\scriptsize CERN} & {\scriptsize $pp$, 13-14~TeV} & {\scriptsize up to 3~ab$^{-1}$} & {\scriptsize visible, invis.} & {\scriptsize (1,2,3,4)} & {\scriptsize 2042} \\

{\scriptsize Dark(Spin)Quest~\cite{Berlin:2018pwi}} & {\scriptsize FNAL}   & {\scriptsize $p$, 120~GeV} & {\scriptsize $10^{18} \to 10^{20}$} & {\scriptsize visible }
& {\scriptsize (1,2,3,4)}  & {\scriptsize 2024}\\

{\scriptsize FASER~\cite{FASER:2018eoc}}  & {\scriptsize CERN}  & {\scriptsize $pp$, 14~TeV} & {\scriptsize 150~fb$^{-1} $} & {\scriptsize visible}  & {\scriptsize (1,2,3,4)} & {\scriptsize 2025} \\

%{\scriptsize HPS~\cite{Celentano:2014wya} } &    {\scriptsize JLAB}    & {\scriptsize $e^-$, 2-6~GeV} & {\scriptsize $\sim 10^{20}$~eot} & {\scriptsize visible} & {\scriptsize (1,3)} & {\scriptsize unknown} \\

{\scriptsize LHCb~\cite{LHCb:2008vvz}}  & {\scriptsize LHC} & {\scriptsize $pp$, 13-14~TeV}  & {\scriptsize up to 300~fb$^{-1}$} & {\scriptsize visible}  & {\scriptsize (1,2,3,4)} & {\scriptsize 2042} \\

{\scriptsize MicroBooNE~\cite{MicroBooNE:2016pwy}}  & {\scriptsize FNAL}   & {\scriptsize $p$, 120~GeV (NuMi)} & {\scriptsize $\sim 7 \times 10^{20}$~pot}  & {\scriptsize visible} &  {\scriptsize(2,4)} & {\scriptsize 2015-2021} \\

%{\scriptsize BES III~\cite{BESIII:2020nme} } & {\scriptsize BEPCII }  & {\scriptsize $e^+ e^-$, 3.7~GeV} & {\scriptsize up to 40~fb$^{-1}$} & {\scriptsize invis.} &  {\scriptsize (1)} &  \\

{\scriptsize NA62~\cite{NA62:2017rwk}}   & {\scriptsize CERN}   &{\scriptsize $K^+$, 75~GeV} & {\scriptsize a few $10^{13}$ K decays} &  {\scriptsize visible, invis.} & {\scriptsize (1,2,3,4)} & {\scriptsize 2025} \\

{\scriptsize NA62-dump~\cite{NA62_Addendum}}  & {\scriptsize CERN}   & {\scriptsize $p$, 400~GeV}  & {\scriptsize $\sim 10^{18}$~pot} & {\scriptsize visible}   & {\scriptsize (1,2,3,4)}  &  {\scriptsize 2025} \\

{\scriptsize NA64$_{e}$~\cite{NA64:eplus}} & {\scriptsize CERN }   & {\scriptsize $e^-$/$e^+$, 100~GeV} & {\scriptsize up to $1\cdot~10^{13}$~$e^-/e^+$} & {\scriptsize $\cancel{\it{E}}$, visible} & {\scriptsize (1,3)} & {\scriptsize $< 2032$} \\

{\scriptsize PADME~\cite{Raggi:2015gza}}  & {\scriptsize LNF} & {\scriptsize $e^+$, 550~MeV} & {\scriptsize $5 \cdot 10^{12}$ $e^+$ot} & {\scriptsize missing mass}  & {\scriptsize (1)} & {\scriptsize $<2023$} \\ 

{\scriptsize T2K-ND280~\cite{T2K:2019bbb}} & {\scriptsize JPARC} &  {\scriptsize $p$, 30~GeV} &   {\scriptsize $10^{21}$~pot}  &  {\scriptsize visible}  &  {\scriptsize (4)} & {\scriptsize running} \\
 \hline

%=========================================
{\bf proposed} &            & & & & & \\ \hline
%=========================================

{\scriptsize BDX~\cite{BDX:2014pkr}}  &   {\scriptsize JLAB}   & {\scriptsize $e^-$, 11~GeV} & {\scriptsize $\sim 10^{22}$ eot/year} & {\scriptsize recoil $e$} 
& {\scriptsize (1,3)} & {\scriptsize 2024-2025}\\

{\scriptsize CODEX-b~\cite{Aielli:2019ivi}} & {\scriptsize CERN}     &{\scriptsize $pp$, 14~TeV}   & {\scriptsize 300~fb$^{-1}$ }       & {\scriptsize visible}     & 
{\scriptsize (1,2,3,4)} & {\scriptsize 2042} \\ 

{\scriptsize Dark MESA~\cite{Christmann:2020qav}}  & {\scriptsize Mainz}     & {\scriptsize $e^-$, 155~MeV} & {\scriptsize 150~$\mu$A}  & {\scriptsize visible} & {\scriptsize (1)} & {\scriptsize $<2030$} \\

{\scriptsize FASER2~\cite{Feng:2022inv}}  & {\scriptsize CERN}  & {\scriptsize $pp$, 14~TeV} & {\scriptsize 3~ab$^{-1}$} & {\scriptsize visible}  & {\scriptsize (1,2,3,4)} & {\scriptsize 2042} \\ 

{\scriptsize FLaRE~\cite{Feng:2022inv}}  & {\scriptsize CERN}  & {\scriptsize $pp$, 14~TeV} & {\scriptsize 3~ab$^{-1}$} & {\scriptsize visible, recoil}  & {\scriptsize (1)} & {\scriptsize 2042} \\ 

{\scriptsize FORMOSA~\cite{Feng:2022inv}}  & {\scriptsize CERN}  & {\scriptsize $pp$, 14~TeV} & {\scriptsize 3~ab$^{-1}$} & {\scriptsize visible}  & {\scriptsize (1)} & {\scriptsize 2042} \\ 

{\scriptsize Gamma Factory}~\cite{Krasny:2015ffb}  & {\scriptsize CERN}  & {\scriptsize photons}  & {\scriptsize up to $10^{25}\; \gamma$/year} & {\scriptsize visible}  & {\scriptsize (1,3)} & {\scriptsize 2035-2038?} \\ 

{\scriptsize HIKE-dump~\cite{HIKE-LoI, HIKE:2022qra}}   & {\scriptsize CERN}   & {\scriptsize $p, 400$~GeV}  & {\scriptsize 5 $\cdot 10^{19}$~pot} & {\scriptsize visible}   & {\scriptsize (1,2,3,4)} & {\scriptsize $<$2038}\\ 

{\scriptsize HIKE-K$^+$~\cite{HIKE-LoI, HIKE:2022qra}}   & {\scriptsize CERN}   & {\scriptsize $K^+, 75$~GeV}  & {\scriptsize $10^{14}$ K decays} & {\scriptsize visible, inv.}   & {\scriptsize (1,2,3,4)} & {\scriptsize $<$2038}\\ 

{\scriptsize HIKE-K$_L$~\cite{HIKE-LoI, HIKE:2022qra}}   & {\scriptsize CERN}   & {\scriptsize $K_L, 40$~GeV}  & {\scriptsize $10^{14}$ K decays} & {\scriptsize visible, inv.}   & {\scriptsize (1,2,3,4)} & {\scriptsize $<$2042}\\ 

{\scriptsize LBND (DUNE)~\cite{Berryman:2019dme}} &  {\scriptsize FNAL}     & {\scriptsize $p$, 120~GeV}   & {\scriptsize $\sim 10^{21}$~pot}    & {\scriptsize recoil $e,N$} 
& {\scriptsize (1,2,3,4)} & {\scriptsize  $<$ 2040}\\

{\scriptsize LDMX~\cite{LDMX:2018cma}}  & {\scriptsize SLAC}        & {\scriptsize $e^-$, 4,8~GeV}  & {\scriptsize $2 \cdot 10^{16}$~eot} & {\scriptsize $\cancel{\it{p}}$, visible} & {\scriptsize (1)} & {\scriptsize $< 2030$} \\ 

{\scriptsize M$^3$~\cite{Kahn:2018cqs}} & {\scriptsize FNAL} & {\scriptsize $\mu$, 15~GeV} & {\scriptsize $10^{10}$ ($10^{13})$~mot} & $\cancel{\it{p}}$ & {\scriptsize(1)} & {\scriptsize proposed}\\

{\scriptsize MATHUSLA~\cite{MATHUSLA:2018bqv}} & {\scriptsize CERN }  & {\scriptsize $pp$, 14~TeV}   & {\scriptsize 3~ab$^{-1}$ } & {\scriptsize visible}    & 
{\scriptsize (1,2,3,4)} & {\scriptsize 2042} \\ 

{\scriptsize milliQan~\cite{Ball:2016zrp} } & {\scriptsize CERN} &   {\scriptsize $pp$, 14~TeV}  & {\scriptsize 0.3-3~ab$^{-1}$} & {\scriptsize visible} & {\scriptsize (1)} & {\scriptsize $< 2032$}\\

{\scriptsize MoeDAL/MAPP~\cite{Frank:2019pgk}}  & {\scriptsize CERN} & {\scriptsize $pp$, 14~TeV} & 
 {\scriptsize 30~fb$^{-1}$} & {\scriptsize visible} & {\scriptsize (4)} & {\scriptsize $< 2032$}\\

{\scriptsize Mu3e~\cite{Mu3e:2020gyw}}   & {\scriptsize PSI}  &  {\scriptsize 29~MeV}  & {\scriptsize  $10^{8} \to 10^{10} \mu$/s} & {\scriptsize visible} & {\scriptsize (1)} & {\scriptsize $< 2038 ?$} \\

{\scriptsize NA64$_{\mu}$~\cite{NA64-mu}} & {\scriptsize CERN}  & {\scriptsize $\mu$, 160~GeV} & {\scriptsize up to $2 \times 10^{13}$~mot} & {\scriptsize $\cancel{\it{p}}$} & {\scriptsize (1)} & {\scriptsize $< 2032$}\\

{\scriptsize PIONEER~\cite{PIONEER:2022alm}}   & {\scriptsize PSI}  &  {\scriptsize 55-70~MeV}, $\pi^+$  & {\scriptsize  $0.3\cdot 10^{6} \pi$/s} & {\scriptsize visible} & {\scriptsize (4)} & {\scriptsize phase I approved}\\

{\scriptsize SBND~\cite{McConkey:2017dsv}}  & {\scriptsize FNAL}   &{\scriptsize $p$, 8~GeV}     & {\scriptsize $6 \cdot 10^{20}$~pot} & {\scriptsize recoil Ar}  & {\scriptsize (1) } & {\scriptsize $< 2030$}  \\

{\scriptsize SHADOWS~\cite{SHADOWS-LoI}}   & {\scriptsize CERN}   & {\scriptsize $p, 400$~GeV}  & {\scriptsize $5\cdot 10^{19}$~pot} & {\scriptsize visible}   & {\scriptsize (2,3,4)} & {\scriptsize $<$2038}\\  

{\scriptsize  SHiP~\cite{SHiP-ECN3-LoI}}   & {\scriptsize CERN}   & {\scriptsize $p, 400$~GeV}  & {\scriptsize $2\cdot 10^{20}$~pot} & {\scriptsize visible, recoil}   & {\scriptsize (1,2,3,4)} & {\scriptsize $<$2038}\\ 

\hline
\end{tabular}
\end{center}
\end{table}
%=========================================

%=====================
%  BC1
%=====================

\clearpage

\clearpage
%=====================
%  BC2
%=====================

%--------------------------------------------

\clearpage
%=====================
%  BC3
%=====================

%-------------------------------------------------

\clearpage
%=====================
%  BC4
%=====================
%------------------------------------------

\clearpage
%=====================
%  BC5
%=====================
%------------------------------------------
%------------------------------------------

\clearpage

%=====================
%  BC9
%=====================
%--------------------------------------------------------

\clearpage
%--------------------------------------------------
%--------------------------------------------------

%--------------------------------------------------
%--------------------------------------------------

%-------------------------------------------

\clearpage
%=============================================
\section{Heavy Neutral Leptons and possible connections with active neutrino physics}
\label{sec:HNL}
%=============================================

%-------------------------------------------
\subsection{Introduction}
\label{ssec:HNL_Intro}
%-------------------------------------------

Heavy neutral leptons (HNL) are mainly-singlet fermions with masses $\gg$~eV. They emerge in extensions of the Standard Model (SM), in which fermionic gauge singlets of the SM (named ``sterile neutrinos"), $N$, are included. Indeed, this constitutes a very minimal (if not the most minimal) extension of the SM, that can also account for neutrino masses and the baryon asymmetry of the Universe.  

Barrying new symmetries that distinguish these new fields $N$ from the SM ones, Yukawa interactions between $N$, the leptonic doublet $L\equiv (\nu_L^T, \ell^T)^T$ and the Higgs doublet $H$ are allowed by gauge interactions, 
$    - \mathcal{L}_y = i \overline{L} \cdot \sigma_2 H^\ast~ N + \mathrm{h.c.}$. After electroweak symmetry breaking, a Dirac mass $m_D$ arises, proportional to the Yukawa coupling and the vacuum expectation value $v$ of the Higgs field. 

This renormalisable term is dubbed the ``neutrino portal" and is one of key bridges between the SM and hypothetical dark sectors (DS), as the sterile neutrinos can further couple to other fields in the DS. As neutrinos are the least known of the SM fermions, they provide a privileged window on the physics beyond the SM (BSM) and could hide a non-standard behaviour still to be uncovered, e.g. see Sec.~\ref{hostert}.

The Yukawa coupling and the resulting Dirac mass conserve lepton number. The latter is an accidental global symmetry of the SM and it does not need to be conserved at a fundamental level. If this is the case, a Majorana mass $M_R$ for the sterile neutrinos should be included in the theory as it is allowed by the SM gauge interactions. The magnitude of the scale of $M_R$ is not known. Though studying the properties of light neutrinos imposes constraints on the HNLs if they are involved in the generation of the light neutrino masses (cf.~Secs.~\ref{klaric} and \ref{Gonzalez-Garcia}), light neutrino oscillation data alone is not sufficient to pin down $M_R$ because it is only sensitive to a specific combination of $M_R$ and the HNL couplings. In the type-I seesaw this is, e.g., given by 
\eqref{eq:mlseesaw} in the notation of \eqref{eq:numassgi}.
On the other hand, different values of $M_R$ can have very different impact on particle physics and cosmology, cf.~e.g.~\cite{Drewes:2013gca,Abdullahi:2022jlv}. 
Eigenvalues of $M_R$ below the TeV scale are particularly interesting because the HNLs are accessible to searches at accelerator-based experiments (cf.~e.g.~\cite{Atre:2009rg,Deppisch:2015qwa,Cai:2017mow,Agrawal:2021dbo,Abdullahi:2022jlv}) 
while being able to address cosmological problems 
(cf.~e.g.~\cite{Boyarsky:2009ix,Drewes:2013gca,Drewes:2016upu,Agrawal:2021dbo,Abdullahi:2022jlv}).
Theoretical motivation for a low scale of $M_R$ can be drawn from different ideas, including symmetry considerations, technical naturalness and minimality. A partial summary can be found in Sec.~5.1 of \cite{Agrawal:2021dbo}.

In presence of both Dirac and Majorana mass terms, the light neutrinos acquire Majorana masses. For an introduction see Sec.s~\ref{Gonzalez-Garcia} and \ref{klaric} and a detailed discussion is given in Ref.~\cite{Agrawal:2021dbo}. This provides very strong theoretical motivation for HNLs. With the discovery of neutrino masses thanks to neutrino oscillations, their origin is one of the most compelling questions in particle physics and, so far, the only particle physics evidence of physics BSM. They are not only non-zero but also much smaller than those of all other SM fermions, with a mild hierarchy differently than the charged leptons and the quarks. Moreover, leptonic mixing is large contrary to that of the quarks. All these indications point towards a different origin of neutrino masses. HNLs offer a natural avenue to implement it, the simplest of which is the see-saw mechanism. For $m_D\ll M_R$, neutrino masses are suppressed by the scale $M_R$ and are of Majorana type, providing a target for neutrinoless double beta decay experiments, see Sec.~\ref{deppisch}, and offering a way to explain the different mass hierarchy and leptonic mixing structure from the other SM fermions.

The heavy states remain heavy and constitute the HNLs discussed in this section. They remain mainly in the sterile neutrino direction but have a small mixing with the active neutrinos. Via this small mixing they participate in the charged and neutral current SM gauge interactions, albeit with a much suppressed coupling controlled by the mixing angle. For a detailed explanation of the connection between the HNL mixing and neutrino masses see Sec.~\ref{klaric} and also Ref.~\cite{Agrawal:2021dbo}. Typically, the resulting HNL mixing angles are small as they are given by $m_D/ M_R$ and therefore are controlled by neutrino masses. In extensions of the see-saw mechanism, e.g. in inverse, extended, linear see-saw models, the connection with neutrino masses can be broken and much larger mixing angles can be allowed. For instance, in inverse see-saw models, a global lepton number symmetry is quasi-conserved and light neutrino masses are controlled by the small explicit lepton number breaking terms. Large Yukawa coupling, and consequently large mixing angles, are allowed.

It is possible to include these models in further extensions of the SM. For instance, right-handed neutrinos are required in left-right models, $SU(2)_L\times SU(2)_R$, that restore the parity between the left- and right-chiral fields, and in larger grand unified theories (GUTs) in which such group can be embedded, e.g. $SO(10)$. 

The HNLs introduced to explain neutrino masses can also generate the baryon asymmetry of the Universe, via leptogenesis, that provides an additional strong theoretical motivation for their existence, see Sec.s~\ref{petcov} and \ref{sandner}. In the simplest picture, in the early Universe the HNLs were part of the thermal plasma subsequently getting out of equilibrium when the temperature drops below their mass. Their decays can generate a lepton asymmetry if the decay channel into leptons and Higgs and its conjugate proceed at different rates due to CP-violation in the Yukawa couplings. This lepton asymmetry is converted into a baryon asymmetry via SM non-perturbative effects, named sphalerons. Variations of this simplest scenario have been proposed and leptogenesis can be successful from GUT-inspired scales down to the GeV one. As discussed in Sec.s~\ref{klaric} and \ref{sandner}, the same mixing angles that controls the HNL phenomenology can be linked to the observed baryon asymmetry.

	HNLs can have different experimental signatures, opening up a wealth of opportunities for discovery.
	As particles that mix with the light neutrinos, HNLs can appear in any process that produces the light neutrinos, as long as such processes are kinematically allowed.
	In particular, light HNLs can be produced in the decays of light mesons.
	One class of experiments relying on this process are the \emph{peak searches}, where light mesons (e.g. Kaons or pions) decay into a lepton and an HNL.
	By measuring the momenta of the decaying meson and the produced lepton, one can reconstruct the missing mass of the HNL.
	Alternatively, one can instead also look for the decay products of the long-lived HNLs in \emph{beam-dump experiments}, where a large number of mesons is produced by the interaction of a proton beam with a target or a dump.
	Depending on the energy of the proton beam, the HNLs could also be produced in the decays of heavier mesons, allowing HNL searches above the Kaon mass.
	HNLs can also be produced directly in collider experiments. If the HNLs are sufficiently long lived, they can lead to a distinct experimental signature in the form of a \emph{displaced vertex},
	where the HNL travels a macroscopic distance before decaying.
	As Majorana particles, HNLs can lead to another experimental signature that is absent in the SM - they could violate lepton number.

In this section an overview of current theoretical and experimental investigations on the motivation and search for HNL particles is presented. We start in Sec~\ref{klaric} with the proposal of additional benchmarks mainly motivated by the connection to neutino mass and baryon asymmetry generation mechanisms, following the FIP Physics Centre approach to clasiffy and study HNL models. The connection between HNL physics and neutrino oscillation data is further studied in Sec~\ref{Gonzalez-Garcia}, while Sec.~\ref{Wong} focuses on the sensitivity of cosmological observables to neutrino masses. A theoretical overview of the role of HNLs in neutrinoless double beta decay processes is then presented in Sec~\ref{deppisch}. The following three sections are dedicated to current and future sensitivity studies of several neutrino experiments as MicroBooNE (Sec.~\ref{soldner-rembold}), DUNE and HyperKamiokande (Sec.~\ref{Joachim_Kopp}), and neutrino telescopes (Sec.~\ref{coloma}).  Sec.~\ref{Bob_Velghe} presents the sensitivity of the PIONEER experiment to the HNL mixing couplings $|U_e|$ and $|U_\mu|$ via pion decay measurements. Collider searches are discussed in the next three sections. Sec.~\ref{middleton} presents a new analysis of BaBar data setting limits on the HNL mixing with the $\tau$ leptons $|U_\tau|$. Sec.~\ref{Joscha_Knolle} summarizes the current status of HNL searches performed by ATLAS and CMS experiments at LHC, while in Sec.~\ref{ripellino} the focus is on the study of long-lived signatures at the future FCC-ee with particular attention to HNL searches. Sec~\ref{milstead} is devoted to the HIBEAM/NNBAR project analyzing the sensitivity to baryon number violating processes in which only B is violated. 
The following sections present several new theoretical and phenomenological ideas: the correlation between the baryon asymmetry generation via leptogenesis and experimental oservables (Secs.~\ref{petcov} and \ref{sandner}), HNLs as a portal between the SM and DS (Sec.~\ref{hostert}), the reinterpretation of FIP searches (Sec.~\ref{tastet}), heavy neutrino-antineutrino oscillations (Sec. \ref{hajer}) and bounds on HNLs from a phenomenological reanalysis of BEBC WA66 data (Sec.~\ref{marocco}). Finaly, in Sec.~\ref{ssec:HNL_conclusions} we briefly summarise and discuss the outlook of the field.

%--------------------------------
\subsection{Heavy Neutral Leptons: the FIP physics centre approach {\it J.~Klarić}}
\label{klaric}
{\it Author: Juraj Klarić <juraj.klaric@uclouvain.be>}
%--------------------------------
\subsubsection{Introduction}

Extending the Standard Model (SM) with Heavy Neutral Leptons (HNLs) can simultaneously resolve several shortcomings of the standard model (SM):
1) they can explain the origin of the neutrino masses~\cite{Minkowski:1977sc,Glashow:1979nm,Gell-Mann:1979vob,Mohapatra:1979ia,Yanagida:1980xy,Schechter:1980gr}, 2) generate the baryon asymmetry of the Universe through leptogenesis~\cite{Fukugita:1986hr} and 3) serve as a DM candidate~\cite{Dodelson:1993je}, all while being within reach of existing and near future experimental searches.
To explore the allowed parameter space of HNLs, it is crucial to map the realistic models involving HNLs onto simple phenomenological benchmarks.
Phenomenological studies usually assume only one HNL, that interacts with the SM via the Lagrangian:
\begin{equation}
	\mathcal L
	\supset
	- \frac{m_W}{v} \overline N \theta^*_\alpha \gamma^\mu e_{L \alpha} W^+_\mu
	- \frac{m_Z}{\sqrt 2 v} \overline N \theta^*_\alpha \gamma^\mu \nu_{L \alpha} Z_\mu
	- \frac{M}{\sqrt 2 v} \theta_\alpha h \overline{\nu_L}_\alpha N
	+ \text{h.c.}
	\,,\label{PhenoModelLagrangian}
\end{equation}
where $v\simeq 174$ GeV is the Higgs vacuum expectation value (VEV), $m_W$ and $m_Z$ are the W and Z boson masses.
The fields $e_{L_\alpha}$ and $\nu_{L_\alpha}$ correspond to the charged left-handed leptons and left-handed neutrinos, whereas $N$ is the HNL field with a mass $M$.
The \emph{mixing angles} $\theta_\alpha$ quantify the coupling of the HNLs to the SM generations with $\alpha=e\,, \mu\,, \tau$.
Up to kinematic factors, the cross-sections for the production and decay of HNLs in association with a lepton flavour $\alpha$ is determined by the magnitudes $U_\alpha^2 = |\theta_\alpha|^2$.
To quantify the overall coupling between the HNL and the SM, it is useful to introduce the flavour-independent mixing angle $U^2 = \sum_\alpha U_\alpha^2$.
Finally, it is important to note that in the phenomenological Lagrangian~\eqref{PhenoModelLagrangian}, the HNL field $N$ can be either a Dirac or a Majorana field.
These two cases can have very different phenomenological implications.
In the case of a Dirac HNL, lepton number is conserved in the decays of HNLs, whereas in the Majorana case it is violated.
The amount of lepton number violation is given by the ratio of lepton number violating (LNV) and lepton number conserving (LNC) decays, $R_{\ell \ell}$, which takes the values $R_{\ell \ell} = 0$ in the case of Dirac HNLs (LNC), and $R_{\ell \ell} = 1$ in the case of Majorana HNLs (equal amount of LNV and LNC decays).
In models with more than one HNL, this ratio is not necessarily constrained to the two limiting values,
but can instead take any value in between $R_{\ell \ell} \in [0,1]$~\cite{Deppisch:2015qwa,Anamiati:2016uxp}.
All together, this leads to very interesting phenomenology~\cite{Atre:2009rg,Drewes:2013gca,Deppisch:2015qwa,Antusch:2016ejd,Chun:2017spz,Cai:2017mow,Abdullahi:2022jlv}.

While the models described by \ref{PhenoModelLagrangian} are a good starting point for HNL searches, they cannot be used to explain the observed light neutrino masses.
In realistic neutrino mass models based on the type-I seesaw mechanism, the number of HNL species is greater or equal to the number of massive light neutrinos.
Therefore, in the case of $m_\text{lightest}=0$, where only two of the light neutrinos are massive and $n \gtrsim 2$, whereas if we ever measure $m_\mathrm{lightest}>0$ at least $n=3$ HNLs will be necessary.
Compared to the phenomenological model with one HNL, realistic models typically have many more parameters ($7n-3$ for $n$ HNLs), only $6$ of which can be constrained by the neutrino oscillation data
(the two light neutrino mass differences, three angles  and one CP violating phase of the Pontecorvo–Maki–Nakagawa–Sakata matrix $V_\nu$).
These constrain the properties of HNLs in realistic models, for instance by limiting the range of allowed flavour ratios $U_e^2: U_\mu^2: U_\tau^2$.
In contrast, the phenomenological Lagrangian which is effectively described with only five free parameters: $(M, U_e^2, U_\mu^2, U_\tau^2, R_{\ell \ell})$,
is completely unconstrained by the light neutrino properties.

Although the parameter space of the simplified phenomenological Lagrangian \ref{PhenoModelLagrangian} is much smaller than that of realistic type-I seesaw models,
it still has too many free parameters to be fully explored in experimental searches.
In particular, for a fixed value of $U^2$, the sensitivity of different experiments is highly sensitive to the flavour ratios $U_e^2: U_\mu^2: U_\tau^2$~\cite{SHiP:2018xqw,Drewes:2018gkc,Tastet:2021vwp}.
Currently, most searches consider a (Dirac or Majorana) HNL coupled to a single SM flavour, which corresponds to the three benchmarks BC6, BC7 and BC8 defined in~\cite{Beacham:2019nyx},
\begin{subequations}
	\label{OldBenchmarks}
	\begin{align}
		U_e^2 : U_\mu^2 : U_\tau^2 = 1:0:0 &&  \text{BC6,}\\
		U_e^2 : U_\mu^2 : U_\tau^2 = 0:1:0 &&  \text{BC7,}\\
		U_e^2 : U_\mu^2 : U_\tau^2 = 0:0:1 &&  \text{BC8,}
	\end{align}
\end{subequations}
with $R_{\ell \ell}=0$ for Dirac or $R_{\ell \ell}=1$ for Majorana HNLs.
However, such single flavour benchmarks are not typically realised in realistic models of neutrino masses.
Past studies have shown that even a percent-level deviation from the single flavour benchmarks can significantly affect the experimental sensitivities,
particularly in the case of a pure-$\tau$ mixing as in BC8~\cite{Drewes:2018gkc}.
As a part of the efforts of the FIPs physics centre HNL working group, in~\cite{Drewes:2022akb} we proposed two additional benchmarks that are compatible with realistic neutrino mass models:
\begin{subequations}
	\label{NewBenchmarks}
	\begin{align}
		\label{NObenchmark}
		U_e^2 : U_\mu^2 : U_\tau^2 &= 0:1:1\,,\\
		\label{IObenchmark}
		U_e^2 : U_\mu^2 : U_\tau^2 &= 1:1:1\,,
	\end{align}
\end{subequations}
with the choices $R_{\ell \ell}=0$ and $R_{\ell \ell}=1$. The three old benchmarks in \ref{OldBenchmarks} and the two new benchmarks in \ref{NewBenchmarks} are shown in \ref{fig:ternary}.
Together, the new \eqref{NewBenchmarks} and old \eqref{OldBenchmarks} benchmarks can be used to effectively approximate the phenomenology of realistic models at accelerator-based experiments.

\subsubsection{Criteria for benchmark selection}

When deciding on new benchmark values for the ratios of the flavoured mixing angles $U_e^2:U_\mu^2:U_\tau^2$, our goal is to capture the physics of realistic neutrino mass models as accurately as possible.
In doing so, we consider the following criteria:
\begin{enumerate}
	\item \textbf{Consistency with neutrino oscillation data.} \label{crit1}
		Explaining the light neutrino masses and their properties is one of the main motivations for the existence of HNLs.
		When we map realistic type-I seesaw models onto the Lagrangian \ref{PhenoModelLagrangian},
		the allowed range of the ratios $U_e^2:U_\mu^2:U_\tau^2$, is closely connected to the properties of the light neutrinos.
		Compared to previous studies~\cite{SHiP:2018xqw, Drewes:2018gkc,Tastet:2021vwp}, we use an updated global fit to light neutrino oscillation data~\cite{Esteban:2020cvm}.
		Furthermore, we estimate how the measurements of the light neutrino properties at future facilities - (specifically the determination of the phase $\delta_{CP}$) affects the allowed range of flavour ratios as shown in \ref{fig:ternary}.
	\item \textbf{Added value.} \label{crit2}
		New benchmark points can only be justified if they can lead to significantly different predictions compared to those of the existing single-flavour benchmarks in \ref{OldBenchmarks}.
		This favours benchmark points with a sizeable mixture of more than one lepton flavour.
		Percent-level changes of the ratios $U_e^2:U_\mu^2:U_\tau^2$ near the single-flavour benchmarks can significantly affect the expected experimental sensitivity, especially so in the case of BC8~\cite{Drewes:2018gkc}.
		On the other hand, the sensitivities near benchmarks with more than one lepton flavour can be quite robust to percent-level changes.
		Additionally, such benchmarks can also lead to lepton-flavor violating signatures at accelerator-based experiments, which are not possible in the case of the single-flavour benchmarks.
	\item \textbf{Symmetry considerations.}
		Since our approach to benchmark selection is motivated in a ``bottom-up'' way, the criteria \ref{crit1} and \ref{crit2} are the most important.
		Nonetheless, model building constraints can be taken into account when there are no conclusive choices based on \ref{crit1} and \ref{crit2}.
		In particular, models based on discrete symmetries of the fermion mixing matrices~(cf.~\cite{Xing:2015fdg,King:2017guk,Xing:2020ijf}) can be used to further motivate certain flavour ratios.
	\item \textbf{Simplicity.}
		Ratios of $U_e^2:U_\mu^2:U_\tau^2$ that are simple and can easily be communicated with the community are preferred if they are equally motivated based on the other criteria.
	\item \textbf{Leptogenesis.}
		Baryogenesis through leptogenesis is one of the main motivations for the existence of HNLs. In particular, for light HNL masses this corresponds to the low-scale leptogenesis mechanisms~\cite{Akhmedov:1998qx,Pilaftsis:2003gt,Asaka:2005pn}. Scans of the parameter space indicate that a significant part of this parameter space is already within reach of existing experiments (cf.~\cite{Klaric:2020phc,Klaric:2021cpi,Hernandez:2022ivz})
		Flavour ratios that can allow for leptogenesis within the experimentally testable part of the parameter space are favoured.
\end{enumerate}

In the following sections, we assess how the two newly proposed benchmarks satisfy these criteria when viewed from the minimal type-I seesaw model in \ref{sec:2HNLs} and in the non-minimal model with three HNLs \ref{sec:3HNLs}.

\subsubsection{The minimal seesaw model}
\label{sec:2HNLs}

The minimal type-I seesaw model that can reproduce the observed light neutrino masses and mixings is the model with two HNLs $N_I$ (with $I=1, 2$), with masses $M_I$ and mixing angles $\theta_{\alpha I}$.
These masses and mixing angles correspond to only 11 physical parameters.
The requirement to reproduce the properties of the light neutrinos leaves only four of these parameters completely free,
while the others can be indirectly determined by the light neutrino masses $m_i$ and by the angles and phases of the PMNS mixing matrix $V_\nu$.
This model can accommodate a wide range of masses and mixing angles for the HNLs.
From the point of view of the accelerator based searches,
the accessible regime corresponds to HNL masses below the TeV scale, and mixing angles much larger than what is implied by the naive seesaw formula
\begin{align}
	\label{LargeMixings}
	|\theta_{\alpha I}|^2 \gg \frac{\sum_i m_i}{M_I}\,.
\end{align}
This limit corresponds to a class of \emph{symmetry protected scenarios}, where the two HNLs approximately respect a generalisation of the global $U(1)_{B-L}$ symmetry of the SM~\cite{Shaposhnikov:2006nn,Kersten:2007vk}.
Such an approximate symmetry is realised in a number of low-scale seesaw models, such as the \emph{linear}~\cite{Akhmedov:1995ip,Akhmedov:1995vm,Gavela:2009cd} and \emph{inverse}~\cite{Wyler:1982dd,Mohapatra:1986aw,Mohapatra:1986bd,Bernabeu:1987gr,Branco:1988ex} seesaws, as well as in the $\nu$MSM~\cite{Asaka:2005pn,Asaka:2005an}.
In this class of models, the light neutrino masses are not suppressed by the large HNL masses, but instead by the small size of the generalised $B-L$ breaking parameters.
This has the advantage that the mixing angles do not need to be suppressed and can be as large as $\mathcal{O}(1)$ while keeping the light neutrino masses stable under radiative corrections.
The approximate $B-L$ symmetry also imposes strong relations between the HNL parameters, namely the two HNLs have quasi-degenerate masses $M_1 \approx M_2 \equiv M$,
and the relative phase between their couplings is maximal $\theta_{\alpha 2} \approx i \theta_{\alpha 1}$.
After applying these constraints, the only free parameters are effectively the mass scale $M$, and the mixing angle $U^2$, while the flavour ratios $U_\alpha^2/U^2$
are completely determined by the light neutrino parameters in $V_\nu$ and $m_i$.

\begin{figure}
	\centering
	\includegraphics[height=0.25\textheight]{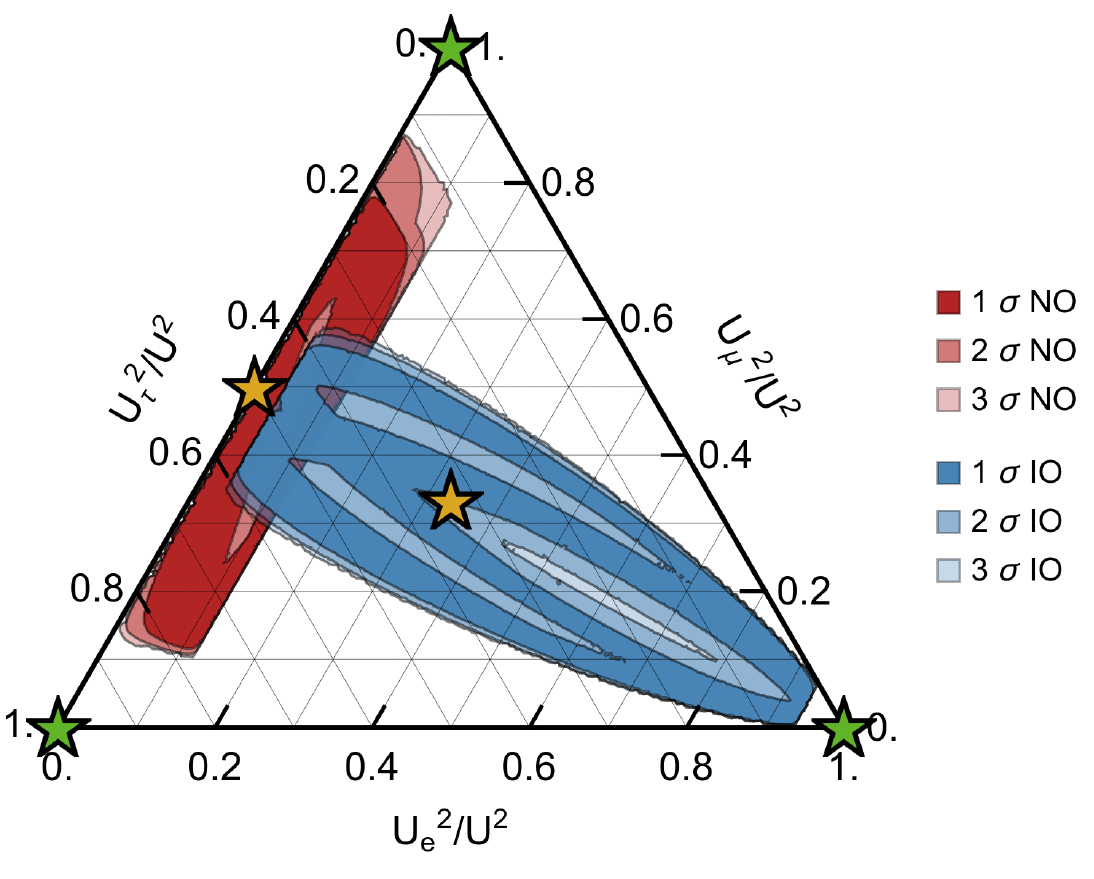}
	\includegraphics[height=0.25\textheight]{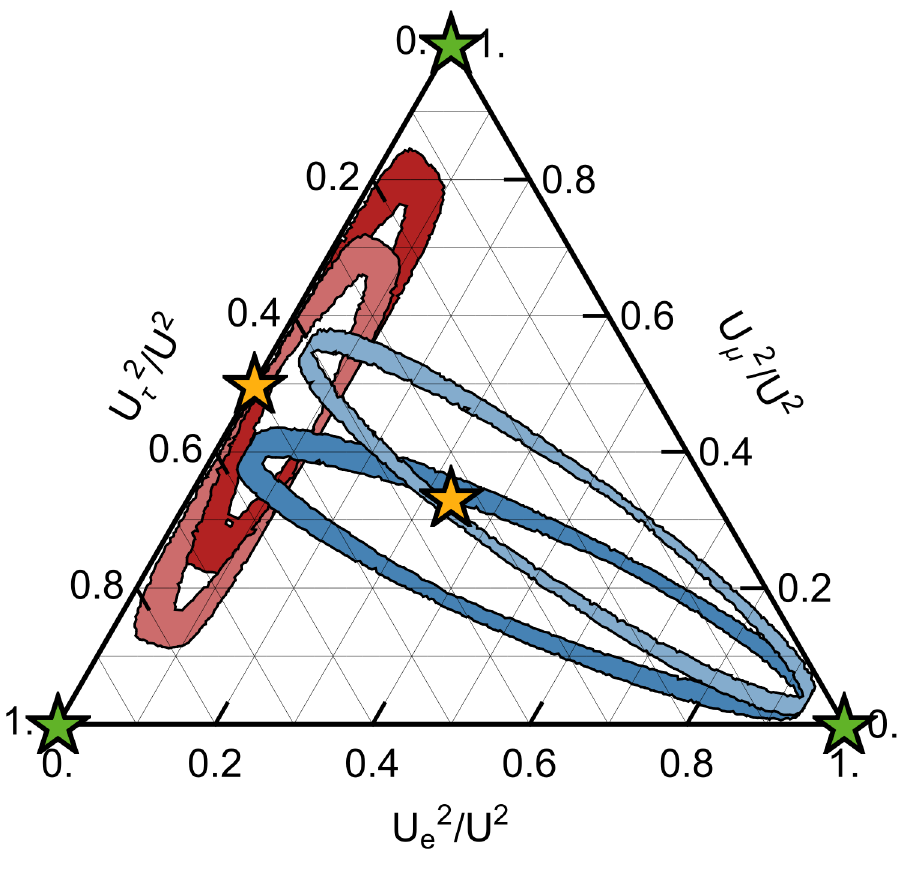}
	\caption{
		The allowed range of mixing ratios $U_e^2 : U_\mu^2 : U_\tau^2$ in the minimal seeseaw model for NO (red) and IO (blue),
		compared with the new benchmarks \ref{NewBenchmarks} (yellow) and the old benchmarks \ref{OldBenchmarks} (green).
		\textbf{Left:} Contours indicating the range allowed by the current neutrino oscillation data~\cite{Esteban:2020cvm}.
		\textbf{Right:} For comparison we also include the future projection based of 15 years of data taking at DUNE~\cite{DUNE:2020jqi}, assuming the true value of $\delta=-\pi/2$, and two benchmark values of the light neutrino mixing angle $s_{23}^2\equiv \sin^2\theta_{23}=0.58$ (darker region) and $s_{23}^2=0.42$ (lighter regions).
		Comparable sensitivity can be expected at Hyper-K~\cite{Hyper-Kamiokande:2018ofw,Ballett:2016daj}. Figures taken from~\cite{Drewes:2022akb}.
}
	\label{fig:ternary}
\end{figure}

Only five of the seven light neutrino parameters have been measured in the minimal model: two mass splittings and three mixing angles.
Since the lightest neutrino mass is exactly zero in the model with two HNLs,
the two mass splittings are sufficient to determine the $m_i$ up to the \emph{ordering} of neutrino masses, which may be \emph{normal} (NO) or \emph{inverted} (IO).
The remaining two unknown parameters are the CP violating phases: the \emph{Dirac} phase $\delta_{CP}$
which is already being probed in neutrino oscillation experiments~\cite{Esteban:2020cvm}, and a \emph{Majorana} phase which can be probed in rare processes such as neutrinoless double beta $(0\nu\beta\beta)$ decay.
Due to these two undetermined phases, the flavour ratios $U_e^2 : U_\mu^2 : U_\tau^2$ are not completely fixed,
but are instead limited to a specific range depending on the light neutrino mass ordering as shown in the ternary diagram Figure \ref{fig:ternary}.
To generate this diagram, we vary the neutrino oscillation parameters within the allowed $3 \sigma$ region.
Besides the unknown Majorana phase, the two least determined parameters are the CP phase $\delta_{CP}$ and one of the mixing angles $\theta_{23}$.
Since the combination of the latter two parameters has a large deviation from Gaussianity, to generate these figures, we use the 2-d projection of $\chi^2$ from~\cite{Esteban:2020cvm}.

Since consistency with neutrino oscillation data is one of the main criteria for benchmark selection,
it is important to confront the allowed range of mixing ratios from \ref{fig:ternary} with the proposed benchmarks.
While certain choices of parameters can lead to a mixing pattern dominated by one of the flavours,
strictly speaking neither of the three single-flavour benchmarks BC6, BC7 and BC8 can be realised within the minimal seesaw model.
As the expected sensitivity of the different experiments
can change by orders of magnitude even for a small deviation from the single-flavour mixing limit,
the existing data highly motivate the two new benchmarks in \ref{NewBenchmarks}.
The benchmark \ref{IObenchmark} is only realised in the case of an inverted neutrino mass ordering.
Nonetheless, it can serve as proxy for the NO scenarios with a large electron mixing where $U_e^2/U^2 \sim 0.1$,
as the experimental sensitivities do not change significantly when changing the flavoured mixings within one order of magnitude.
The new benchmark \ref{NObenchmark} is not strictly realised in either of the two neutrino mass orderings,
it is a good approximation for the NO case with a minimal electron mixing where $U_e^2/U^2 \sim 10^{-3}$.

\begin{figure}
	\centering
	\includegraphics[height=0.3\textheight]{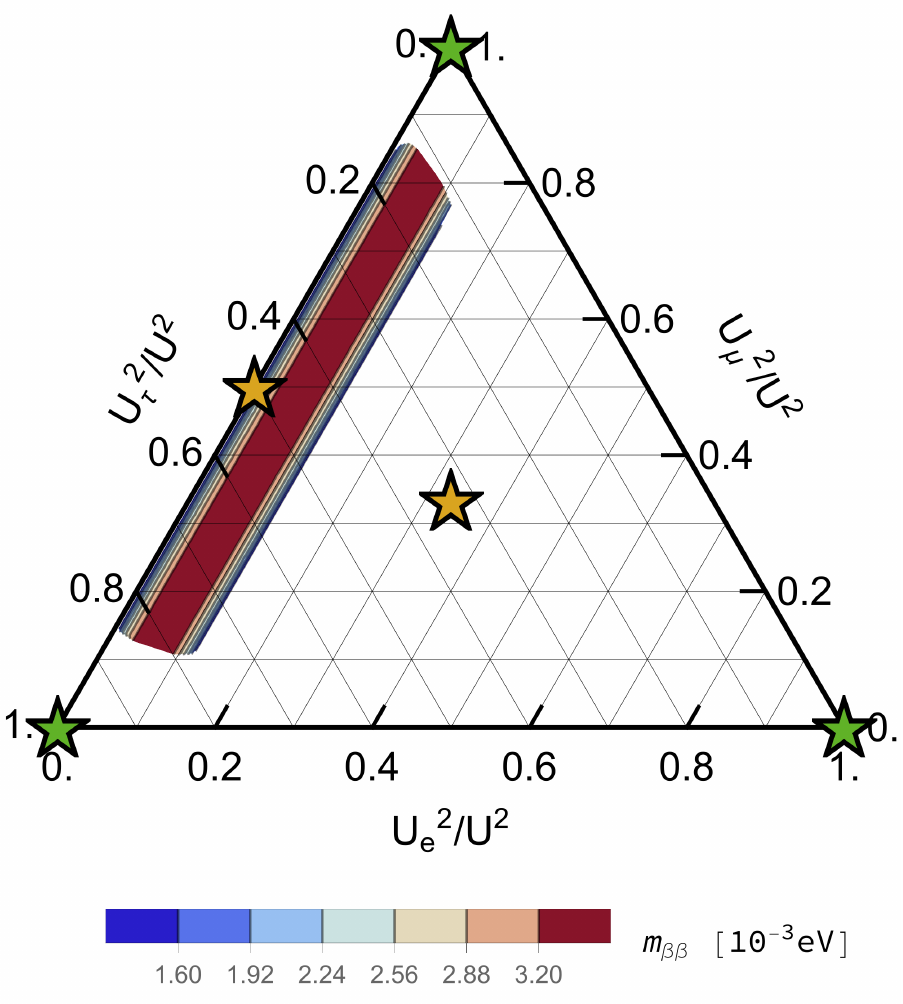} \includegraphics[height=0.3\textheight]{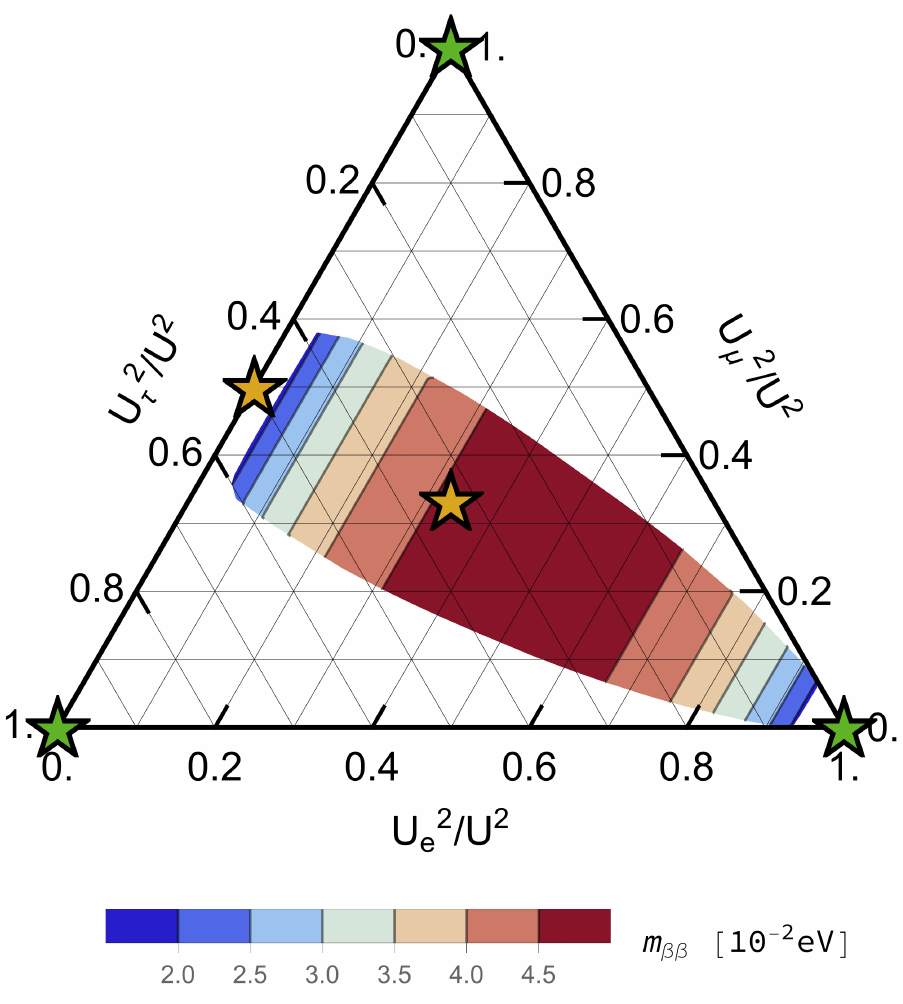}
	\caption{
		The parameter $m_{\beta\beta}$ indicating the rate of neutrinoless double beta decay in the minimal model as a function of the mixing ratios for NO (left) and IO (right).
		We assume a negligible contribution from the HNL exchange, as well as the large mixing limit \ref{LargeMixings}.
		Unfortunately, due to the theoretical uncertainty associated with the calculation of the neutrinoless double beta decay rate, it cannot be used to indirectly measure the flavour ratios,
		however, it may be used to constrain the allowed range of values of $U_e^2/U^2$. Figures taken from~\cite{Drewes:2022akb}.
	}
	\label{fig:0nubb}
\end{figure}

\textbf{Neutrinoless double beta decay -}
The range of allowed mixing angles also depends on the Majorana phase, which is a parameter that is essential for neutrinoless double beta decay.
The relation between this parameter and the electron mixing takes a rather simple form in the minimal model,
allowing us to correlate rate of the $0\nu\beta\beta$ decay with the mixing pattern as shown in \ref{fig:0nubb}.
This indicates that a non-observation of $0\nu\beta\beta$ decay may exclude certain flavour mixing patterns,
however the possibility of an indirect determination of $U_e^2/U^2$ remains slim due to the large theoretical uncertainties in the computed $0\nu\beta\beta$ decay rate.
It is important to note that the simple relation shown in \ref{fig:0nubb} only holds for $M \gg 100$ MeV, where we can neglect the HNL contribution to this process.

\textbf{Leptogenesis -}
Another advantage of the minimal scenario is that it can be used to explain the observed BAU via leptogenesis in a wide range of masses and mixing angles. %\cref{fig:leptoMU2}.
Low-scale leptogenesis is consistent with the full range of mixing patterns that are allowed by the seesaw mechanism.
Nonetheless, specific flavour patterns may be more favourable and allow for a larger range of mixing angles or mass splittings between the HNLs.
This typically happens in the case of a \emph{flavour asymmetric washout}, where one of the lepton flavours approximately decouples,
thereby hiding the lepton asymmetry from being erased in the early Universe.
In particular, in the case of NO this further motivates benchmark choice \ref{NObenchmark}, where the electron coupling is suppressed.

\textbf{Dirac or Majorana HNLs -} When mapping the minimal type-I seesaw model onto the phenomenological Lagrangian \ref{PhenoModelLagrangian},
an important consideration is whether one should treat the HNLs as Dirac ($R_{\ell \ell}=0$) or Majorana ($R_{\ell \ell}=1$) particles.
While each of the two HNLs in the minimal seesaw model is individually a Majorana state,
in the symmetry protected scenarios the two Majorana HNLs can be combined into an pseudo-Dirac state which preserves lepton number.
In practice, this means that depending on the HNL parameters, all values of $R_{\ell \ell}$ between $0$ and $1$ are possible.

\subsubsection{The model with three HNLs}
\label{sec:3HNLs}

The scenario with two HNLs can only give masses to two of the light neutrinos.
The minimal type-I seesaw model that can simultaneously explain massess of three light neutrinos is the model with three HNLs.
Compared to the model with two HNLs, this model is far less predictive, with 18 free parameters.
In this model the flavour ratios $U_\alpha^2/U^2$ do not only depend on the light neutrino parameters, but also on additional HNL phases and parameters.
The limit $m_\mathrm{lightest} \rightarrow 0$ restricts the allowed range of mixing ratios $U_\alpha^2/U^2$ to a range similar to what is shown in \ref{fig:ternary}.
However, even a value of $m_\mathrm{lightest} \gtrsim 10^{-5}$ eV can cause a significant deviation,
with $m_\mathrm{lightest} \sim 10^{-2}$ eV allowing the full range of flavour ratios $U_\alpha^2/U^2$~\cite{Chrzaszcz:2019inj}.
%sentence about m_{lightest} in cosmology
An advantage of the model with three HNLs is that it allows for leptogenesis with significantly larger $U^2$ than in the minimal model.
Due to the additional dynamical effects, such large mixing angles no longer depend on flavour hierarchical couplings,
which means that the full range of flavour ratios can be consistent with leptogenesis~\cite{Abada:2018oly,Drewes:2021nqr}.
Taking these considerations into account, the model with three HNLs does not seem to motivate any additional benchmarks,
but it can serve as an additional motivation for the single flavour benchmarks BC-6-BC-8, which cannot be realised within the minimal model.

\subsubsection{Summary}

The interpretation of results of experimental searches for HNLs fundamentally depends on the assumptions about their properties.
Since even the simplest phenomenological model of HNLs \ref{PhenoModelLagrangian} comes with as many new parameters,
experimental searches can be greatly simplified if we assume a certain relation between the HNL parameters.
The most commonly used benchmarks BC6-BC8 for HNL searches assume that the HNLs only couple to a single active lepton flavour~\eqref{OldBenchmarks}.
Such a single-flavour mixing assumption is typically not realised in realistic models of neutrino masses based on the type-I seesaw mechanism.
To resolve this shortcoming we propose  two additional benchmark points for HNL searches~\eqref{NewBenchmarks} and summarize the main motivation behind them.
Together, the new~\eqref{NewBenchmarks} and old~\eqref{OldBenchmarks} benchmarks can cover a wide range of phenomena that can be realised in realistic models of neutrino masses.

%-------------------------------------------
\subsection{NHL's and light oscillating neutrinos in 2022 -- {\it M.~C.~Gonzalez-Garcia}}
\label{Gonzalez-Garcia}
{\it Author: Maria Concepcion Gonzalez-Garcia}
%-------------------------------------------

% \documentclass[preprintnumbers,nofootinbib,noshowpacs,eqsecnum,prd,superscriptaddress,letterpaper]{revtex4}

% \usepackage{amsmath}
% \usepackage{amssymb}
% \usepackage{multirow}
% \usepackage{rotating}

% %%%%%%%%%%%%%%%%%%%%%%%%%%%%%%%%%%%%%%%%%%%%%%%%%%
% %                                                %
% %    BEGINNING OF TEXT                           %
% %                                                %
% %%%%%%%%%%%%%%%%%%%%%%%%%%%%%%%%%%%%%%%%%%%%%%%%%%

% \begin{document}
% \title{Neutral Heavy Leptons and Light Oscillating Neutrinos in 2022}
% \author{M.C. Gonzalez-Garcia}
% \affiliation{Departament  de  Fisica  Quantica  i  Astrofisica
%  and  Institut  de  Ciencies  del  Cosmos,  Universitat
%  de Barcelona, Diagonal 647, E-08028 Barcelona, Spain\\
% Instituci\'o Catalana de Recerca i Estudis Avancats,
% Pg. Lluis  Companys  23,  08010 Barcelona, Spain.\\
% C.N. Yang Institute for Theoretical Physics, Stony Brook University,
% Stony Brook NY11794-3849,  USA}
% \maketitle

\subsubsection{Introduction: Neutral Heavy Leptons and Active Neutrino Oscillations}
I was asked by the organizers to discuss the possible connection between
neutral heavy leptons (NHL) and the light active neutrino physics. The most
model-independent  connection appears in the context of the extensions of
the Standard Model (SM)  required to explain  the well-stablished observation
that:

$\bullet$ Atmospheric $\nu_\mu$  and
$\bar\nu_\mu$ disappear most likely converting to $\nu_\tau$
and $\bar\nu_\tau$.

$\bullet$ Accelerator $\nu_\mu$  and
$\bar\nu_\mu$ disappear over long baseline (LBL)
distances of $\sim$ 200 to 700 Km. 

$\bullet$  Solar $\nu_e$ convert to $\nu_{\mu}$ and/or $\nu_\tau$.

$\bullet$ Reactor $\bar\nu_e$ disappear over distances of $\sim$ 200 Km
and $\sim$ 1.5 km with different probabilities.

$\bullet$ Accelerator $\nu_\mu$ and $\bar\nu_\mu$  appear as $\nu_e$
and $\bar\nu_e$ at LBL distances  $\sim$ 200 to 700 Km. 

All these results imply that lepton flavours are not conserved in
neutrino propagation and that requires physics beyond the SM.  The
logic behind this statement is that a fermion mass term couples
right-handed and left-handed fermions. But the SM, a gauge theory
based on the gauge symmetry $SU(3)_{\rm C}\times SU(2)_{\rm L}\times
U(1)_{\rm Y}$ -- spontaneously broken to $SU(3)_{\rm C}\times
U(1)_{\rm EM}$ by the the vacuum expectation value of a Higgs doublet
field $\phi$ --, contains three fermion generations which reside in
the chiral representations of the gauge group required to describe
their interactions.  As such, right-handed fields are included for
charged fermions since they are needed to build the electromagnetic
and strong currents.  But no right-handed neutrino is included in the
model because neutrinos are neutral and colourless and therefore the
right-handed neutrinos are singlets of the SM group (hence
unrequired).  This implies that each of the lepton flavour is
conserved and therefore total lepton number ($L$) is a global a
symmetry of the model. A symmetry which is part of $B-L$
($B$ being the total baryon number) which is
non-anomalous.  As a consequence of this symmetry, within the
framework of the SM no mass term can be built for the neutrinos at any
order in perturbation theory and it cannot be generated by
non-perturbative effects either.  This is, the SM predicts that
neutrinos are {\sl strictly} massless and that there is no 
flavour mixing nor CP violation in the leptonic sector. Clearly this
is in contradiction with the neutrino data as summarized above.

The simplest extension to introduce lepton flavour violation in the
model is to enlarge the SM Lagrangian with mass terms for the neutrino.
There are two type of neutrino mass operators which can be built
depending on the fields introduced and symmetries assumed:
\begin{itemize}
\item One can introduce 3 electroweak (EW) singlets right-handed neutrinos,
  $\nu_R$, and impose $L$ conservation. With this particle contens it is
  possible
  to build a Yukawa operator in the Lagrangian involving the
  left-handed lepton doublets, the Higgs doublet,  and the $\nu_R$'s,
  with couplings $\lambda^\nu$,
  in total analogy with the corresponding operators generating the mass
  for charged fermions.
  After spontaneous EW symmetry 
  breaking this leads to 
\begin{equation}
  {\cal L}_D={\cal L}_{SM}-M_\nu \bar \nu_L \nu_R +h.c. \;\;\;
  {\rm with}\;\;\;  M_\nu=\frac{v}{\sqrt{2}} \lambda^\nu\,
\label{eq:dirac}
\end{equation}
with $v$ being the Higgs vev. In this case there are 3 mass eigenstate
neutrinos which are Dirac fermions, 
ie $\nu^C\neq \nu$.
\item One can construct a mass term only with the same chirality neutrino
  fields 
  (either with the $\nu_L$'s from the SM or with additional $\nu_R$'s)
by allowing $L$ violation 
\begin{equation}
  {\cal L}_M={\cal L}_{SM}-\frac{1}{2}M_\nu \bar \nu \nu^c +h.c.
  \label{eq:maj}
\end{equation}
In this case the mass eigenstates are Majorana fermions, $\nu^C=\nu$.
Notice that the Majorana mass term for $\nu_L$ above  breaks EW
gauge invariance while for $\nu_R$ it is gauge-invarant.
\end{itemize}
Altogether if  one adds some number $S$ of EW singlet
neutrinos and allow for both type of mass terms
the most general gauge-invariant form of the neutrino mass can be written as
\begin{equation}
    \label{eq:mnu}
    -\mathcal{L}_{M_\nu}
    = \frac{1}{2} \overline{\vec\nu^c} M_\nu \vec\nu + {\rm h.c.} \,,
\end{equation}
where 
\begin{equation}
    M_\nu = 
    \begin{pmatrix}
	0 & D \\
	D  & M
    \end{pmatrix}, 
\label{eq:numassgi}
\end{equation}
and $\vec\nu = (\vec \nu_{L}, \, \vec{\nu^c_{R}} )^T$ is a 
$(3+S)$-dimensional vector.  $D$ is a $3\times 3$ Dirac-mass term
\eqref{eq:dirac} while $M$ is a $S\times S$ Majorana-mass term \eqref{eq:maj}
for the right-handed neutrinos.  

Once the neutrino mass term is included in the Lagrangian,
leptonic flavours are
mixed in the CC interactions of the leptons, and a leptonic mixing matrix
appears in similarity to the CKM matrix for the quarks.
The discussion of mixing in the leptonic sector, however, is more general
than in the quark sector because the number of  massive neutrinos can be
larger than three, since
there are no constraints on the number of EW-singlet neutrinos to be included
in the model.
In particular, if we denote the neutrino mass eigenstates by $\nu_i$,
$i=1,2,\ldots,n$, ($n=3+S$) and the charged lepton mass eigenstates 
by $l_i=(e,\mu,\tau)$, in the mass basis, leptonic CC interactions 
are given by
\begin{equation}
-{\cal L}_{\rm CC}={\frac{g}{2}}\, \overline{{l_i}_L} 
\, \gamma^\mu \, U_{\rm LEP}^{ij}\,  \nu_j \; W_\mu^+ +{\rm h.c.}.
\label{CClepmas}  
\end{equation} 
Here $U_{\rm LEP}$ is a $3\times n$ matrix.

A consequence of the presence of the leptonic mixing is the
possibility of flavour oscillations of the neutrinos \cite{Pontecorvo:1967fh}
in their propagation which provides the explanation to  the experimental
observations. 
Neutrino oscillations appear because of the misalignment between 
the interaction neutrino eigenstates and the propagation eigenstates (
which  for propagation in vacuum  are the mass eigenstates). 
Thus a neutrino of energy $E$ produced in a CC
interaction with a charged lepton $l_\alpha$ can be detected via a CC
interaction with a charged lepton $l_\beta$ with a probability which
presents an oscillatory behaviour, with oscillation lengths
given by the phase difference between the different propagation eigenstates
-- which in the ultrarelativistic limit is 
$L_{0,ij}^{\rm osc}=\frac{4 \pi E}{\Delta m_{ij}^2}$ -- and amplitude
that is proportional to elements in the mixing matrix.
Thus neutrino oscillations are only sensitive to mass squared
differences and they do not give us information on the absolute value
of the masses. 
Furthermore the presence of matter in the neutrino propagation alters both the
oscillation frequencies and the amplitudes \cite{Mikheev:1986gs}.

In the absence of any other form of new physics affecting the charged leptons
one can choose without loss of generality the flavour and mass states of
the charged leptons to coincide. In that basis the $3\times (3+S)$
leptonic mixing matrix above coincides with first three rows of the unitary
$V^\nu$, which is a  $(3+S)\times (3+S)$ matrix
relating the neutrino weak eigenstates to the
neutrino mass eigenstates and which can write in block terms as
$${ V^\nu}
=\begin{array}{c|ll}
{\rm dim}   &\;\; 3   &  \;\;S \\[-0.1cm]\hline 
 3 & \left({ K_l}\right. & \left.{ K_h}\right)\equiv { U_{\rm LEP}}\\
 S & \tilde K_h &\tilde K_H
\end{array}$$
In this notation the interactions of the 3 lightest {$\nu_l$}
and the $S$ heaviest {$N$} (from here on refer to as NHL's) states
are \cite{Schechter:1980gr}
\begin{eqnarray}
{\cal L}_{CC}&=&-{\frac{g}{2}}\left( {\overline{\ell}}
  \gamma^\mu { K_l\, \nu_l}\,
  + {\overline{\ell}}
  \gamma^\mu { K_h\,  N}\,\right)  W^+_\mu \,+{\rm h.c.}\\
{\cal L}_{NC}&=&-{\frac{g}{2}}\left(
    { \overline{\nu_l}}
  \gamma^\mu { K_l^\dagger K_l}\, { \nu_l}\,+
    { \overline{\nu_l}}
  \gamma^\mu { K_l^\dagger}{ K_h}\, { N}\, +
    { \overline{N}}
  \gamma^\mu { K_h^\dagger}{ K_l}\, { \nu_l}\, +
    { \overline{N}}
    \gamma^\mu { K_h^\dagger K_h}\, { N}\,\right) Z_\mu
\end{eqnarray}  
Both  $K_{l}$ (which is $3\times 3$) and ${ K_{h}}$ ($3\times S$)
are non-unitary
matrices but they are related by the unitarity of the full $V^\nu$ matrix
which implies that
\begin{equation}
{ V^\nu}{ {V^\nu}^\dagger}=\left(\begin{array}{cc}
    \!\!{ K_l K_l^\dagger}+
{ K_h K_h^\dagger} & \dots\\
\dots & \dots\end{array}\!\!\right)=
  \left(\begin{array}{cc}  
 \!\!\!   I_{3\times 3} & 0 \\0 &\!\!\! I_{S\times S}\end{array}\right)
\Rightarrow \hspace*{0.1cm}
{ U_{\rm LEP}}{ U^\dagger_{\rm LEP}}
={ K_l K_l^\dagger}+
{ K_h K_h^\dagger}
=I_{3\times 3}\;.\end{equation}
The relation above implies a generic, model-independent, connection between
the NHL couplings and the violation of unitarity in the $\nu_l$ oscillations.

A convenient parametrization to characterize and quantify the presently allowed
violation of unitarity in the $\nu_l$ sector is \cite{Xing:2011ur,Escrihuela:2015wra,Blennow:2016jkn}
\begin{equation}
{ K_l}=\left[I_{3\times 3}-
      {\left(\begin{array}{ccc}\alpha_{ee}&0&0\\
    \alpha_{e\mu}&\alpha_{\mu\mu}&0\\
    \alpha_{e\tau}&\alpha_{\mu\tau}&\alpha_{\tau\tau}\end{array}\right)}
      \right]{U_{\rm 3\times 3}}
\end{equation}
with
\begin{equation}
  U_{\rm 3\times 3}
  =\left(\begin{array}{ccc}
	1 & 0 & 0 \\
	0 & c_{23}  & {s_{23}} \\
	0 & -s_{23} & {c_{23}}
    \end{array}\right)
    \cdot
    \left(\begin{array}{ccc}
	c_{13} & 0 & s_{13} e^{-i\delta_{\rm CP}} \\
	0 & 1 & 0 \\
	-s_{13} e^{i\delta_{\rm CP}} & 0 & c_{13}
    \end{array}\right)
    \cdot
    \left(\begin{array}{ccc}
	c_{21} & s_{12} & 0 \\
	-s_{12} & c_{12} & 0 \\
	0 & 0 & 1
    \end{array}\right)
    \cdot
    \left(\begin{array}{ccc}
	e^{i \eta_1} & 0 & 0 \\
	0 & e^{i \eta_2} & 0 \\
	0 & 0 & 1
    \end{array}\right),
    \label{eq:U3m}
\end{equation}
where $c_{ij} \equiv \cos\theta_{ij}$ and $s_{ij} \equiv\sin\theta_{ij}$.
For vanishing $\alpha$'s one then recuperates the
parametrization of the mixing matrix usually employed in the global
analysis of oscillation data in the context of $3\nu$-mixing.

Let us first discuss the expectations in  the canonical type-I see saw
\cite{Minkowski:1977sc,Ramond:1979py,GellMann:1980vs,Yanagida:1979as,Mohapatra:1979ia}.
This corresponds to the general neutrino mass matrix \eqref{eq:numassgi}
with $M\gg D$. The diagonalization of $M_\nu$ leads to three light,
$\nu_l$, and $S$ NHL's, $N$: 
\begin{equation}
    -\mathcal{L}_{M_\nu}
    = \frac{1}{2}\bar{\nu}_{l}  M_{\nu_l} \nu_{l} +
    \frac{1}{2}\bar{N} M_N {N} 
\end{equation}
with 
\begin{equation}
    \label{eq:mlseesaw}
    M_{\nu_l}\simeq -V_l^T D^T M^{-1} D V_l,
    \qquad M_N \simeq V_h^T M V_h
\end{equation}
and
\begin{equation}
    \label{eq:Useesaw}
    V^\nu \simeq
    \begin{bmatrix}
	\left(1 - \frac{1}{2}D^\dagger {M^*}^{-1} M^{-1} D
	\right) V_l & D^\dagger {M^*}^{-1} V_h
	\\
	-M^{-1} D V_l
	& \left(1 - \frac{1}{2}{M}^{-1} D D^\dagger
	{M^*}^{-1} \right) V_h
    \end{bmatrix}
\end{equation}
where $V_l$ and $V_h$ are $3\times 3$ and $S\times S$ unitary matrices
respectively. In this scenario the lightness of the light neutrino mass
is explained as it is suppressed by the heaviness of the NHL mass.
For $D$ with characteristic size of the charged fermion masses, the current
bounds on the light neutrino mass implies that $M_N\gg$ TeV.

From the expressions above we find  that
\begin{equation}
      {\left(\begin{array}{ccc}2\alpha_{ee}&\alpha_{e\mu}^*& \alpha_{e\tau}^*\\
    \alpha_{e\mu}&2\alpha_{\mu\mu}&\alpha_{\mu\tau}^*\\
    \alpha_{e\tau}&\alpha_{\mu\tau}&2\alpha_{\tau\tau}\end{array}\right)}
      \simeq K_h\,K_h^\dagger=D^\dagger {M^*}^{-1} M^{-1} D\sim
      \frac{M_{\nu_l}}{M_N}\ll 10^{-9}\;.
\end{equation}
Thus in the canonical type-I see-saw scenario one expects the violation of
unitarity in the light neutrino sector (and hence the NHL couplings)
to be too small to be experimentally  accessible.

In more general see-saw type-I scenarios the bounds on the $\alpha$'s
depend on the NHL mass range. A detailed description and quantification
can be found in Ref.~\cite{Blennow:2016jkn}. Summarizing, their results
show that
\begin{itemize}
\item for NHL's with masses below of ${\cal O}(10)$ eV one has direct
 effects of the NHL's states in some of the neutrino oscillation experiments.
 This is the type of scenarios employed to explain some of the ``so-called''
 short baseline anomalies (see for example Ref.~\cite{Dentler:2018sju}).
 Current fits to oscillation data constraint $\alpha$'s$\lesssim 0.1$-$0.01$
\item for NHL's with masses above $\sim$ 10 eV but still well below the
  EW scale, additional bounds arise from searches for the NHL's at beam-dump
  experiments.
\item for NHL's with masses around or above EW scale, the violation
  of unitarity and universality of the weak interactions of the charged
  leptons are severely constrained by EW precision data
  which imply $\alpha$'s$\lesssim 10^{-3}$-$10^{-4}$
\end{itemize}  
The outcome of this study is that in the full range of NHL masses,
unitarity violation is constrained to the level that it can be still ignored
in the present oscillation analysis performed in the 3$\nu$ scenario.
Unfortunately this means that the model-independent 
correlation between the NHL effects and the active light-neutrino oscillations
is not accessible with present data.

However, within a given model, ie, within a given form of
$M_\nu$, the entries of $K_l$ and $K_h$ are related, and therefore
model-dependent relations between the NHL couplings and the 3$\nu$ oscillation
parameters exist. In this respect, maximally testable scenarios are those
with minimal lepton flavour violation \cite{Gavela:2009cd} for which,
up to an overall normalization and some phases, the entries of $K_h$
can be determined once we know the light 3$\nu$ masses and mixing.
Consequently the flavour dependence of the NHL's signals in those
scenarios can be predicted from the results of the 3$\nu$ oscillation
analysis.

It is with this motivation in mind, that I summarize next the present
determination of the 3$\nu$ flavour parameters from the global
description of the neutrino oscillation data.

\subsubsection{Status of the 3$\nu$ Global Description}

In brief the experimental results which have establish with high precision
that neutrinos are massive are:

$\bullet$ Atmospheric $\nu_\mu$  and $\bar\nu_\mu$ disappear most likely
converting to $\nu_\tau$ and $\bar\nu_\tau$.
The results show an energy and distance dependence perfectly described by 
mass-induced oscillations. 

$\bullet$ Accelerator $\nu_\mu$  and
$\bar\nu_\mu$ disappear over distances of $\sim$ 200 to 800 km. 
The energy spectrum of the results show a clear oscillatory behaviour
also in accordance with mass-induced oscillations with wavelength in
agreement with the effect observed in atmospheric neutrinos. 

$\bullet$ Accelerator $\nu_\mu$ and $\bar\nu_\mu$  appear as $\nu_e$
and $\bar\nu_e$ at distances  $\sim$ 200 to 800 km. 

$\bullet$ Solar $\nu_e$ convert to $\nu_{\mu}$ and/or $\nu_\tau$.
The observed energy dependence of the effect is well described by  
neutrino conversion in the Sun matter according to the MSW effect
~\cite{Wolfenstein:1977ue, Mikheev:1986gs}

$\bullet$ Reactor $\bar\nu_e$ disappear over distances of $\sim$ 200 km
and $\sim$ 1.5 km with different probabilities. The observed energy spectra
show two different mass-induced oscillation wavelengths:
at short distances in agreement with the one observed in accelerator
$\nu_\mu$  disappearance, and a long distance compatible with the required
parameters for MSW conversion in the Sun.

The minimum scenario to describe these results
requires the mixing between the
three flavour neutrinos of the standard model in three distinct mass
eigenstates. In this case, neglecting the small effect of unitarity
violation discussed in the previous section, we can approximate the
leptonic mixing matrix by $U_{3\times 3}$ in Eq.~(\ref{eq:U3m}).
The angles $\theta_{ij}$ can be taken without loss
of generality to lie in the first quadrant, $\theta_{ij} \in [0,
  \pi/2]$, and the phase $\delta_{\rm CP} \in [0, 2\pi]$. Values of $\delta_{\rm CP}$
different from 0 and $\pi$ imply $CP$ violation in neutrino oscillations
in vacuum. The Majorana phases $\eta_1$ and $\eta_2$ play
no role in neutrino oscillations.

Neutrino oscillations are only sensitive to mass squared
differences and do not give us information on the absolute value
of the masses. The observed oscillation patterns 
require two distinctive  oscillation wavelengths.
There are two possible non-equivalent orderings for the mass 
eigenvalues: $m_1\ll m_2< m_3$ so $\Delta m^2_{21} \ll \Delta m^2_{32} (\simeq \Delta m^2_{31} > 0) $, ($\Delta m^2_{ij} \equiv m_i^2 - m_j^2$)
refer to as Normal ordering (NO), 
and $m_3\ll m_1< m_2$ so $\Delta m^2_{21} \ll -(\Delta m^2_{31} \simeq  \Delta m^2_{32} < 0)$ refer to as
Inverted ordering (IO).

In total the 3-$\nu$ oscillation analysis of the existing data involves six
parameters: 2 mass differences (one of which can be positive or negative), 
3 mixing angles, and the CP phase. I summarize in Table \ref{tab:expe}
the different experiments contributing dominantly to the present determination
of the different parameters.

\begin{table}[ht]
\centering
\caption{Experiments contributing to the present determination of  
the oscillation parameters.} 
\begin{tabular}{l|l|l|}
Experiment & Dominant &  Important  \\
\hline
Solar Experiments &  {$\theta_{12}$} 
&  {$\Delta m^2_{21}$}  , {$\theta_{13}$} 
\\
Reactor LBL (KamLAND)  &  
{$\Delta m^2_{21}$}  
& {$\theta_{12}$}   , {$\theta_{13}$} 
\\
Reactor MBL (Daya-Bay, Reno, D-Chooz)   
&  {$\theta_{13}$}, {$|\Delta m^2_{31,32 }|$} &
\\
Atmospheric Experiments (SK, IC-DC)  
&   &
{$\theta_{23}$},{$|\Delta m^2_{31,32}|$}, 
{$\theta_{13}$},{$\delta_{\rm CP}$}
\\
Accel LBL $\nu_\mu$,$\bar\nu_\mu$,
Disapp (K2K, MINOS, T2K, NO$\nu$A) 
&  {$|\Delta m^2_{31,32 }|$}, {$\theta_{23}$} &
\\
Accel LBL $\nu_e$,$\bar\nu_e$ App (MINOS, T2K, NO$\nu$A)  
&  {$\delta_{\rm CP}$}   &  
$\theta_{13}$  ,  {$\theta_{23}$}\\\hline  
\end{tabular}
\label{tab:expe}
\end{table}

\begin{figure}[ht]
 \centering
  \includegraphics[width=0.9\textwidth]{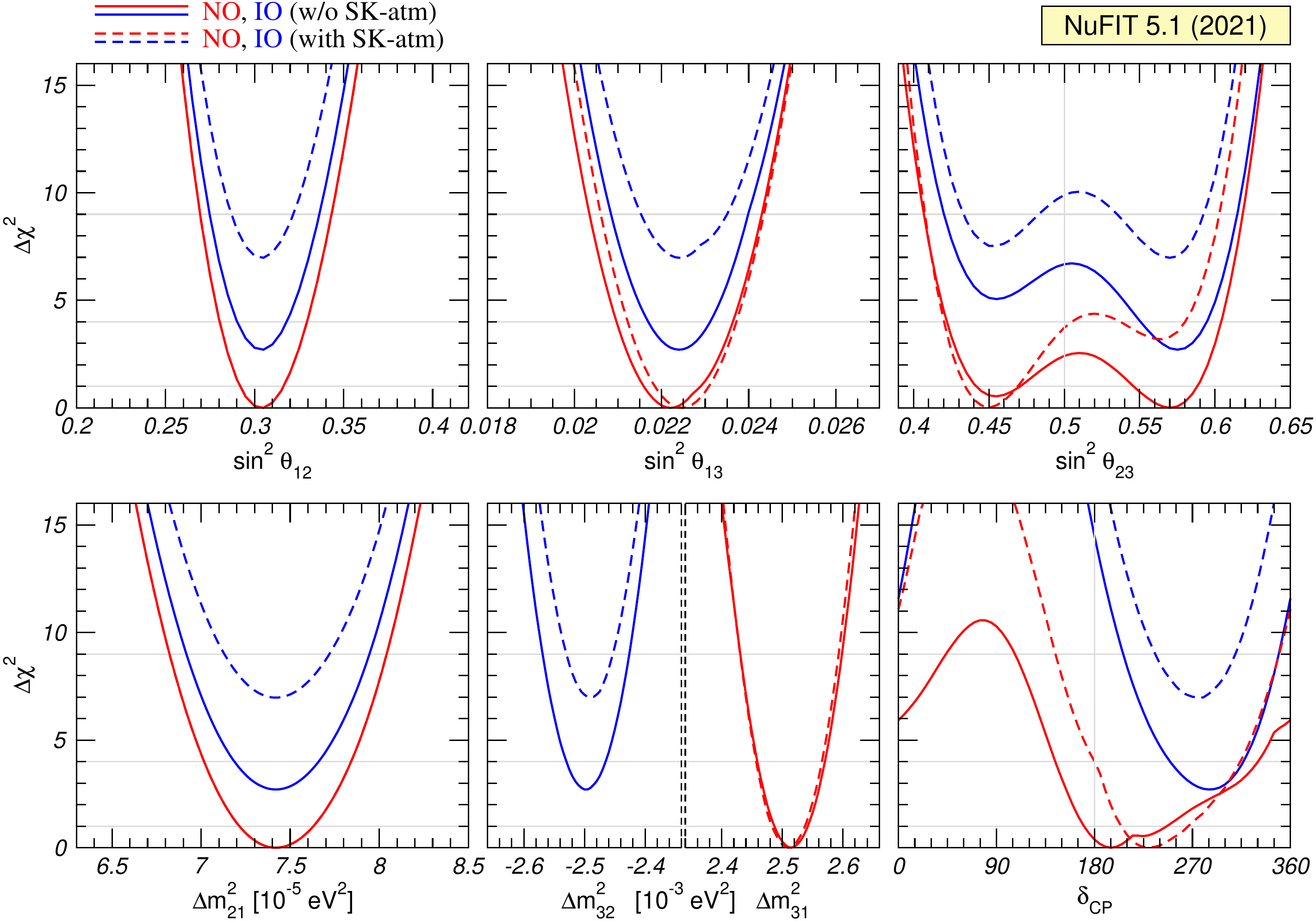}
  \caption{Global $3\nu$ oscillation analysis
    from  fit to global data NuFIT~5.1. We show $\Delta\chi^2$
    profiles minimized with respect to all undisplayed parameters. The
    red (blue) curves correspond to Normal (Inverted) Ordering. Solid
    (dashed) curves are without (with) adding the tabulated SK-atm
    $\Delta\chi^2$.  Note that as atmospheric mass-squared splitting
    we use $\Delta m^2_{31}$ for NO and $\Delta m^2_{32}$ for IO.}
  \label{fig:chisq-glob}
\end{figure}

At present the determination of the leptonic parameters requires
global analysis of the data which is in the hands of a few
phenomenological groups (see, for example,~\cite{deSalas:2020pgw,Capozzi:2021fjo}). The allowed parameter ranges obtained
by the different groups are generically in good agreement, which provides
a test of the robustness of the results. I summarize here the results
from the latest analysis of the NuFIT~\cite{nufit} Collaboration
in Refs.~\cite{Esteban:2020cvm,Gonzalez-Garcia:2021dve}.
In Fig\ref{fig:chisq-glob} I show
different projections of the allowed six-dimensional parameter
space. The best fit values and the derived ranges for the six parameters at 
the $1\sigma$ ($3\sigma$) level are given in Tab.~\ref{tab:results}. For each
parameter the ranges are obtained after marginalizing with respect to
the other parameters.  

\begin{table}[ht!]
%\centering %Please confirm if it have table header.
  \caption{Determination of  three-flavour oscillation
    parameters
    from  fit to global data
    NuFIT~5.1.  Results in the
    first and second columns correspond to  analysis performed under
    the assumption of NO and IO, respectively; therefore, they are
    confidence intervals defined relative to the respective local
    minimum.  Results shown in the upper and lower sections
    correspond to analysis performed without and with the addition of
     tabulated SK-atm $\Delta\chi^2$ data respectively.  In quoting
     values for the largest mass splitting, we  defined
     $\Delta m^2_{3\ell} \equiv \Delta m^2_{31} > 0$ for NO and
     $\Delta m^2_{3\ell}\equiv \Delta m^2_{32} < 0$ for IO.}
  \begin{small}
  \setlength{\tabcolsep}{2.35mm}
    \begin{tabular}{clcccc}
           \hline\hline
      \multirow{11}{*}{\begin{sideways}\hspace*{-7em}without SK atmospheric data\end{sideways}} &
      & \multicolumn{2}{c}{Normal Ordering (Best Fit)}
      & \multicolumn{2}{c}{Inverted Ordering ($\Delta\chi^2=2.6$)}
      \\
      \cline{3-6}
      && bfp $\pm 1\sigma$ & $3\sigma$ Range
      & bfp $\pm 1\sigma$ & $3\sigma$ Range
      \\
      \cline{2-6}
      \rule{0pt}{4mm}\ignorespaces
      & $\sin^2\theta_{12}$
      & $0.304_{-0.012}^{+0.013}$ & $0.269 \to 0.343$
      & $0.304_{-0.012}^{+0.012}$ & $0.269 \to 0.343$
      \\[1mm]
      & $\theta_{12}/^\circ$
      & $33.44_{-0.74}^{+0.77}$ & $31.27 \to 35.86$
      & $33.45_{-0.74}^{+0.77}$ & $31.27 \to 35.87$
      \\[3mm]
      & $\sin^2\theta_{23}$
      & $0.573_{-0.023}^{+0.018}$ & $0.405 \to 0.620$
      & $0.578_{-0.021}^{+0.017}$ & $0.410 \to 0.623$
      \\[1mm]
      & $\theta_{23}/^\circ$
      & $49.2_{-1.3}^{+1.0}$ & $39.5 \to 52.0$
      & $49.5_{-1.2}^{+1.0}$ & $39.8 \to 52.1$
      \\[3mm]
      & $\sin^2\theta_{13}$
      & $0.02220_{-0.00062}^{+0.00068}$ & $0.02034 \to 0.02430$
      & $0.02238_{-0.00062}^{+0.00064}$ & $0.02053 \to 0.02434$
      \\[1mm]
      & $\theta_{13}/^\circ$
      & $8.57_{-0.12}^{+0.13}$ & $8.20 \to 8.97$
      & $8.60_{-0.12}^{+0.12}$ & $8.24 \to 8.98$
      \\[3mm]
      & $\delta_{\rm CP}/^\circ$
      & $194_{-25}^{+52}$ & $105 \to 405$
      & $287_{-32}^{+27}$ & $192 \to 361$
      \\[3mm]
      & $\dfrac{\Delta m^2_{21}}{10^{-5}~{\ensuremath{{\rm eV}^2}}}$
      & $7.42_{-0.20}^{+0.21}$ & $6.82 \to 8.04$
      & $7.42_{-0.20}^{+0.21}$ & $6.82 \to 8.04$
      \\[3mm]
      & $\dfrac{\Delta m^2_{3\ell}}{10^{-3}~{\ensuremath{{\rm eV}^2}}}$
      & $+2.515_{-0.028}^{+0.028}$ & $+2.431 \to +2.599$
      & $-2.498_{-0.029}^{+0.028}$ & $-2.584 \to -2.413$
      \\[2mm]
      \hline\hline
      \multirow{11}*{\begin{sideways}\hspace*{-7em}with SK atmospheric data\end{sideways}} &
      & \multicolumn{2}{c}{Normal Ordering (Best Fit)}
      & \multicolumn{2}{c}{Inverted Ordering ($\Delta\chi^2=7.0$)}
      \\
      \cline{3-6}
      && bfp $\pm 1\sigma$ & $3\sigma$ range
      & bfp $\pm 1\sigma$ & $3\sigma$ range
      \\
      \cline{2-6}
      \rule{0pt}{4mm}\ignorespaces
      & $\sin^2\theta_{12}$
      & $0.304_{-0.012}^{+0.012}$ & $0.269 \to 0.343$
      & $0.304_{-0.012}^{+0.013}$ & $0.269 \to 0.343$
      \\[1mm]
      & $\theta_{12}/^\circ$
      & $33.45_{-0.75}^{+0.77}$ & $31.27 \to 35.87$
      & $33.45_{-0.75}^{+0.78}$ & $31.27 \to 35.87$
      \\[3mm]
      & $\sin^2\theta_{23}$
      & $0.450_{-0.016}^{+0.019}$ & $0.408 \to 0.603$
      & $0.570_{-0.022}^{+0.016}$ & $0.410 \to 0.613$
      \\[1mm]
      & $\theta_{23}/^\circ$
      & $42.1_{-0.9}^{+1.1}$ & $39.7 \to 50.9$
      & $49.0_{-1.3}^{+0.9}$ & $39.8 \to 51.6$
      \\[3mm]
      & $\sin^2\theta_{13}$
      & $0.02246_{-0.00062}^{+0.00062}$ & $0.02060 \to 0.02435$
      & $0.02241_{-0.00062}^{+0.00074}$ & $0.02055 \to 0.02457$
      \\[1mm]
      & $\theta_{13}/^\circ$
      & $8.62_{-0.12}^{+0.12}$ & $8.25 \to 8.98$
      & $8.61_{-0.12}^{+0.14}$ & $8.24 \to 9.02$
      \\[3mm]
      & $\delta_{\rm CP}/^\circ$
      & $230_{-25}^{+36}$ & $144 \to 350$
      & $278_{-30}^{+22}$ & $194 \to 345$
      \\[3mm]
      & $\dfrac{\Delta m^2_{21}}{10^{-5}~{\ensuremath{{\rm eV}^2}}}$
      & $7.42_{-0.20}^{+0.21}$ & $6.82 \to 8.04$
      & $7.42_{-0.20}^{+0.21}$ & $6.82 \to 8.04$
      \\[3mm]
      & $\dfrac{\Delta m^2_{3\ell}}{10^{-3}~{\ensuremath{{\rm eV}^2}}}$
      & $+2.510_{-0.027}^{+0.027}$ & $+2.430 \to +2.593$
      & $-2.490_{-0.028}^{+0.026}$ & $-2.574 \to -2.410$
      \\[2mm]
      \hline\hline
    \end{tabular}
  \end{small}
  \label{tab:results}
\end{table}

These results yield the present determination of the modulus of the
leptonic mixing matrix
\begin{equation}
|U_{\rm 3\times 3}|_{3\sigma} = \begin{pmatrix}
    0.801 \to 0.845 &\qquad
    0.513 \to 0.579 &\qquad
    0.143 \to 0.156
    \cr
    0.244 \to 0.499 &\qquad
    0.505 \to 0.693 &\qquad
    0.631 \to 0.768
    \cr
    0.272 \to 0.518 &\qquad
    0.471 \to 0.669 &\qquad
    0.623 \to 0.761\end{pmatrix}\;,
\end{equation}

The results obtained show that the ranges of
$\Delta m^2_{21}$, $\theta_{21}$, $|\Delta m^2_{3\ell}|$ and $\theta_{13}$
has been rather robust over the last years. In particular it had been
a result of global analyses for the last decade, that the
value of $\Delta m^2_{21}$ preferred by KamLAND was somewhat higher than the
one from solar experiments.  The tension appeared due to a combination
of two effects: the well-known fact that the $^{8}${B} measurements
performed by SNO, SK and Borexino showed no evidence of the low energy
spectrum turn-up expected in the standard
LMA-MSW solution for the
value of $\Delta m^2_{21}$ favored by KamLAND, and the observation of a
non-vanishing day-night asymmetry in SK, whose size is larger than the
one predicted for the $\Delta m^2_{21}$ value indicated by KamLAND.
Altogether this resulted in slightly over $2\sigma$ discrepancy
between the best fit $\Delta m^2_{21}$ value indicated of KamLAND and the
solar results. But with the inclusion of the latest SK4 2970-day
the tension decreased to $\sim 1.1\sigma$ 
due to both, the smaller day-night asymmetry and the slightly more pronounced
turn-up in the low energy part of the spectrum.

Conversely the determination of the mass ordering (MO), the octant of
$\theta_{23}$, and the significance of CP violation has been changing
as new data from LBL (and also SK-ATM) has been presented.
This is so because they all correspond to subdominant 3$\nu$ oscillation
effects to be observed over the dominant 2$\nu$ oscillations in
the different experiments. This is illustrated in the table below where
I present a brief summary of the present status of the ``hints'' for
these effects in the combination of different experiments\\

{\centering\begin{tabular}{c|cccccc}
& best fit MO & $\Delta\chi^2$(MO) & best fit $\delta_{\rm CP}$ &
$\Delta\chi^2$(CPC) & oct.$\theta_{23}$ & $\Delta\chi^2$(oct)\\\hline
    T2K     & NO &  2.5 & 265$^\circ$ & 4 & 2nd & 0.5\\
    NO$\nu$A& NO & 0.3 & 135$^\circ$ & 0.5& 2nd & 0.1 \\
   T2K+NO$\nu$A & IO & 1.5 & 275$^\circ$ & 2.0 & 2nd & 2.2\\
              +reactors & NO & 2.7 & 195$^\circ$& 0.4 & 2nd & 0.5\\
      + SK-Atm 328 kt-y (NuFIT 5.0) & NO & 7.1 & 197$^\circ$  & 0.5 & 2nd  & 2.5\\
   or + SK-Atm 373 kt-y  (NuFIT 5.1) & NO & 7.0 & 230$^\circ$  & 4.0 & 1st  & 3.2
\end{tabular}}\\[+0.2cm]

Definite determination of the MO, the octant of $\theta_{23}$, and of leptonic
CP violation will most likely require the results from the new generation of
neutrino experiments.
Juno \cite{JUNO:2015zny} is a new reactor neutrino experiment already in
its last phases of construction. It is expected to improve
the precision on the determination of $\Delta m^2_{21}$, $|\Delta m^2_{31}|$,
and $\sin^2\theta_{12}$ by one order of magnitude within 10 years of operation.
It is also designed to provide a determination of the MO to a confidence
level which depends on the ordering and their final performance. Being a
disappearance experiment does not have sensitivity to CP violation.
The next generation LBL experiments 
HyperKamiokande ~\cite{Hyper-Kamiokande:2018ofw}, and DUNE ~\cite{DUNE:2015lol}
are designed to provide the  definite precise determination of the MO and of
leptonic CP violation.

%\subsubsection*{Acknowledgements}
%This project has received funding/support from the European Union's Horizon
%2020 research and innovation program under the Marie Skłodowska-Curie
%grant agreement No 860881-HIDDeN, as well as from grant PID2019-105614GB-C21, 
%"Unit of Excellence Maria de Maeztu 2020-2023"
%award to the ICC-UB CEX2019-000918-M,  
%funded by MCIN/AEI/10.13039/501100011033, and by USA-NSF grant PHY-1915093.

% \bibliographystyle{JHEP}
% \bibliography{fips22_mcgg}
% \end{document} 

%-------------------------------------------
\subsection{Prospects for the measurement of the absolute neutrino masses in cosmology -- {\it Y.~Wong}}
\label{Wong}
{\it Author: Yvonne Wong, <yvonne.y.wong@unsw.edu.au>}
%-------------------------------------------

% \documentclass[11pt,a4paper]{article}
% \pdfoutput=1

% %\setlength{\arraycolsep}{1.5pt}

% \usepackage{jcappub}
% \usepackage{slashed}
% \usepackage{braket}
% \usepackage{hyperref}
% \usepackage{float}
% \usepackage{bbold}
% \usepackage{amsmath}
% \usepackage{amssymb}
% \usepackage{graphicx}
% \usepackage{xcolor}
% \usepackage{mathrsfs}
% \usepackage{comment}
% \usepackage{bm}
% \usepackage[normalem]{ulem}
% \usepackage{cancel}

% %package for CONCEPT eqs
% \usepackage[scr=boondox]{mathalfa}

% \begin{document}

% \title{Prospects for the measurement of the absolute neutrino mass in cosmology}

% \author[a]{Yvonne Y. Y. Wong}

% \affiliation[a]{Sydney Consortium for Particle Physics and Cosmology, School of Physics, The University of New South Wales,  Sydney NSW 2052, Australia} 

% \emailAdd{yvonne.y.wong@unsw.edu.au}

% \notoc

% \maketitle

% \newpage

\subsubsection{Introduction}

Cosmological bounds in the absolute neutrino mass scale go back a long way.  Already in 1972, Cowsik and McClelland argued that an upper limit of $m_\nu \lesssim  8$~eV could be set on the individual neutrino mass from cosmology, assuming three standard-model (SM) neutrino species~\cite{Cowsik:1972gh}.  Their argument was simple: since the standard hot big bang theory predicts a background of relic neutrinos---the so-called cosmic neutrino background (C$\nu$B)---whose present-day reduced energy density  is $\Omega_\nu = \sum m_\nu/(94\,  h^2~{\rm eV})$, such a large amount of energy could overclose the universe if the neutrino mass $m_\nu$  was too large.  Then, by  demanding that $\Omega_\nu \lesssim 1$, one arrives at $\sum m_\nu \lesssim 24$~eV  for a reduced Hubble parameter value  $h=0.5$.

Modern cosmological limits on $\sum m_\nu$ have evolved to be far more sophisticated than the original estimate of~\cite{Cowsik:1972gh}, and derive essentially from the gravitational effects of the C$\nu$B on the events that take place after its formation at 1s post big bang.    These events include the formation of the cosmic microwave background (CMB)  anisotropies at 400k~years, as well as the large-scale matter distribution in the present-day universe.  However, even these modern bounds are subject to what we predict to be the properties of the C$\nu$B given SM particle and gravitational physics.  Our first task, therefore, is to summarise these SM predictions.

%%%%%%%%%
%%%%%%%%%

\subsubsection{The cosmic neutrino background}
\label{sec:background}

In the standard hot big bang model SM neutrinos are held in a  state of thermodynamic equilibrium with other SM particles via the weak interaction in the first second post-big bang.  As the universe expands and cools, these interactions become less frequent. When the universe cools to a temperature of ${\cal O}(1)$~MeV, the interaction rate per neutrino drops below the Hubble expansion rate.  From this point onwards, neutrinos free-stream to infinity, forming the C$\nu$B.  Like its cousin the CMB, the C$\nu$B has a thermal spectrum well described by the relativistic Fermi-Dirac distribution $f(p) = [1+\exp(p/T_\nu)]^{-1}$, where  at $T \ll m_e$ the temperature parameter is given by $T_\nu = (4/11)^{1/3} T_\gamma$, with $T_\gamma$ the CMB temperature.

Given its momentum distribution and present-day temperature $T_{\nu,0} \simeq 2 \times 10^{-4}$~eV, we deduce the present-day C$\nu$B number density to be $n_\nu \sim 110 \, {\rm cm}^{-3}$ per family of neutrinos and anti-neutrinos.  Then, for those neutrino families with masses satisfying $T_{\nu 0}  \ll m_\nu$ and are hence non-relativistic today, it follows simply that their present-day energy density is $\rho _\nu= m_\nu n_\nu$.  Summing up all massive families and normalising to the present-day critical density, we find the reduced neutrino energy density  $\Omega_\nu = \sum m_\nu/(94\,  h^2~{\rm eV})$.

%%%%%%%%%%
%%%%%%%%%%%

\subsubsection{Signatures of neutrino masses on cosmological observables}
\label{sec:signatures}

%%%%%%%%%%%
\begin{figure}
	\centering
\includegraphics[width=0.75\textwidth]{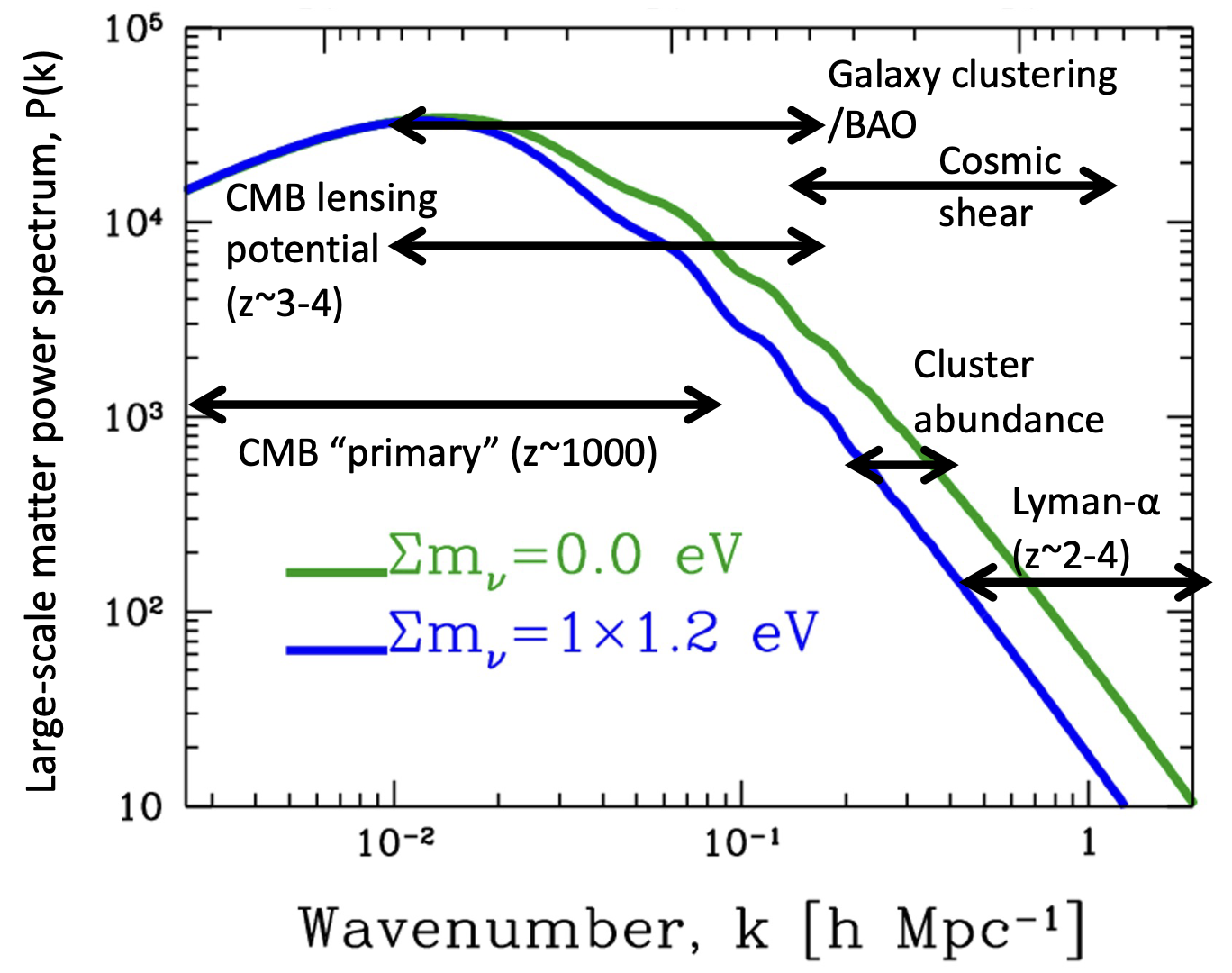}
\caption{Two predictions of the large-scale matter power spectrum $P(k)$ from linear perturbation theory.  The green line denotes the standard $\Lambda$CDM prediction, while the blue line represents the case in which some (cold) dark matter  energy density $\Omega_{\rm dm}$ has been substituted with a non-zero $\Omega_\nu$ comprising of one massive neutrino species of mass 1.2~eV.  All other cosmological parameters are held fixed between the two cases.  The $k$-regions spanned by the black arrows indicate the wave numbers probed by different classes of cosmological observables at various redshifts.
	\label{fig:powerspec}}
\end{figure}
%%%%%%%%%%%%

Signatures of neutrino masses in cosmological observables have been discussed extensively in the literature (e.g., \cite{Wong:2011ip}).  Broadly speaking, we expect a power suppression in the  matter power spectrum $P(k)$ at large wave numbers $k$ roughly proportional to the neutrino fraction $f_\nu \equiv \Omega_\nu/\Omega_m$, where $\Omega_m$ is the total matter density.  This suppression arises because  ultra-relativistic C$\nu$B neutrinos at $T_\nu \gg m_\nu$ are highly resistant to gravitational capture on small scales.  This remains true even after the neutrinos have become non-relativistic at $T_\nu \ll m_\nu$.  The net effect is, in two universes with the same $\Omega_m$, that which contains the larger $f_\nu$ will see fewer structures form on small  scales.  Figure~\ref{fig:powerspec} illustrates this effect.

Also shown  in figure~\ref{fig:powerspec} are various classes of cosmological observables, as well as the wave number $k$ and redshift $z$ ranges to which they are sensitive.  Clearly, only the CMB primary anisotropies measure $P(k)$ at large scales/small wave numbers, which serves the important purposes of ``pinning'' down the background cosmology.  For this reason, full-sky CMB data from, e.g., the Planck mission~\cite{Planck:2015fie,Planck:2018vyg}, are unavoidable  in cosmological parameter estimation.

Another point of interest is that the power spectra in figure~\ref{fig:powerspec} have been computed using linear perturbation theory.  In reality, as density perturbations grow over time, linear theory must finally break down starting with the large wave numbers.  At $z=0$, it fails at $k \gtrsim 0.1\, h{\rm Mpc}^{-1}$; at $z=3$, at $k \gtrsim 0.5\, h{\rm Mpc}^{-1}$.  Thus, except  for the CMB, all current  and upcoming measurements of $P(k)$ are sensitive to nonlinear dynamics to various extents.

However, it must be stressed that (i) not all observables suffer from the same nonlinearities, and (ii) some forms of nonlinearity are inherently much less tractable than others.  Nonlinear gravitational clustering of dark matter, for example, is common to all small-scale observables.  It is however tractable, in the sense that one could write down a Lagrangian to describe the nonlinear physics and compute the nonlinearities from first principles.  

The same cannot be said however for the baryonic astrophysics, e.g., star formation, supernova feedback, critical to the modelling of the hydrogen clouds at the heart of the Lyman-$\alpha$ forest: these come from empirical modelling.  Probes that use tracers, e.g., the galaxy power spectrum, which assumes galaxies to track dark matter, likewise rely on empirical bias relations that have no first-principles basis.  The same is true for cluster abundance observations that use the cluster X-ray temperature or richness as proxies for the cluster mass.  Some degree of prudent scepticism should be exercised when interpreting these measurements.

%%%%%%%%%%%%%
%%%%%%%%%%%%%%

\subsubsection{Current constraints and caveats}
\label{sec:current}

Table~\ref{tab:constraints} shows the 95\% upper limits on $\sum m_\nu$ derived from various data combinations.  At face value, the best number is  $\sum m_\nu\lesssim 0.12$~eV, coming from the full Planck 2018 data set (temperature, polarisation, and lensing potential) and baryon acoustic oscillations (BAO) measurements from SDSS-III BOSS DR12~\cite{BOSS:2016wmc}.
Formally, this constraint is similar to that derived previously from the Planck 2015 data plus the Lyman-$\alpha$ power spectrum~\cite{Palanque-Delabrouille:2015pga}.  However, in view of the non-trivial modelling of the latter system, the Planck 2018 CMB+BAO bound, which relies on well-scrutinised linear physics alone, is arguably the more robust of the two.

%%%%%%%%%%%%%%%%%%%%%%%
\begin{table}
	\centering
	\begin{tabular}{|l|c|c|c|c|}
		\hline\hline
		&&+CMB lens & +BAO & +CMB lens+BAO \\
		\hline
		Planck2018 TT+lowE &0.54 & 0.44 & 0.16 & {\bf 0.13} \\
		2015 numbers & 0.72 & 0.68& 0.21 & n/a \\
		\hline
		Planck2018 TT+lowE+TE+EE &0.26 &0.24& 0.13 & {\bf 0.12} \\
		CamSpec numbers & 0.38& 0.27& n/a & {\bf 0.13}\\	
		2015 numbers & 0.49 &0.59 & 0.17 & n/a \\
		\hline\hline	
	\end{tabular}		
	\caption{1D marginal 95\% upper limits on the neutrino mass sum $\sum m_\nu$ in units of eV derived from various data combinations. ``Planck2018'' numbers are taken from~\cite{Planck:2018vyg}, while the ``2015 numbers'' come from~\cite{Planck:2015fie}.  The label ``CamSpec'' refers to an alternative polarisation likelihood function employed in~\cite{Planck:2018vyg}.  All fits assume the standard $\Lambda$CDM model   extended to include a floating $\sum m_\nu$. 
		\label{tab:constraints}}	
\end{table}
%%%%%%%%%%%%%%%%%%%%%%%%

Nonetheless, as robust as any modelling may be, it must be stressed that  some assumptions underpin these constraints.  To begin with, to even constrain neutrino masses cosmologically, we must assume that a C$\nu$B exists.  Fortunately, there is no reason to think that this is not the case: current observations are consistent to high statistical significance with the presence of three non-interacting neutrino families at the time of CMB formation~\cite{Planck:2018vyg}.   Still, there are caveats and some small room for play when interpreting these bounds.

%%%%%%%%%%%%%%%%%
%%%%%%%%%%%%%%%%

\paragraph{Caveat 1: Which mass ordering? }

Most neutrino mass fits  assume three neutrinos of degenerate masses, because it saves time: specifying three different mass values in a power spectrum calculation triples the computation time relative to the single-mass case.  It is however useful to note that switching to a realistic ordering can change the $\sum m_\nu\lesssim 0.12$~eV bound by up to 40\%.  Specifically, reference~\cite{RoyChoudhury:2019hls}  reported $\sum m_\nu \lesssim 0.146$~eV and $\sum m_\nu \lesssim 0.172$~eV for  the normal and  inverted mass ordering respectively, using the
same Planck 2018 CMB+BAO data as in table~\ref{tab:constraints}.

%%%%%%%%%%%
%%%%%%%%%%%%

\paragraph{Caveat 2: Model dependence }
\label{sec:modeldependence}

The constraints in table~\ref{tab:constraints} have been derived from a 7-parameter fit.  That is,  the free variables are the standard six parameters of the flat $\Lambda$CDM model---matter density $\omega_m$, baryon density $\omega_b$, reduced Hubble rate $h$, the primordial fluctuation amplitude $A_s$ and spectral index $n_s$, and the optical depth to reionosation $\tau$---and the neutrino mass sum $\sum m_\nu$ distributed equally in three species. It is however conceivable that the underlying cosmology is more complicated and that  relaxing some of the $\Lambda$CDM assumptions might loosen the $\sum m_\nu$ bound.

Table~\ref{tab:extended} shows how the $\sum m_\nu$ bound varies with several popular extensions to the flat $\Lambda$CDM model.  Notably, introducing a dynamical dark energy equation of state parameter  $w(z) = w_0 + w_a (1-a)$ that is allowed to go ``phantom'', i.e., $w(z) < -1$,   relaxes the $\sum m_\nu$ bound by a factor of two to $\sum m_\nu\lesssim  0.249$~eV for degenerate masses.  Note however that this is about the maximum gain one could expect from playing this sort of games, as the signatures of neutrino masses are generally difficult to completely negate by physically unrelated effects.

%%%%%%%%%%%%%%
\begin{table}
	\centering
	\begin{tabular}{|l|c|c|c|}
		\hline\hline
		& Degenerate & Normal & Inverted \\
		\hline
		Baseline $\Lambda$CDM$+\sum m_\nu$ & 0.121 & 0.146 & 0.172 \\
		$+r$ & 0.115 & 0.142 & 0.167 \\
		$+w$ &0.186 & 215 & 0.230 \\
		$+w_0 w_a$ & {\bf 0.249} & {\bf 0.256} & {\bf 0.276} \\
		$+w_0 w_a, w(z) > -1$ & {\bf 0.096} & {\bf 0.129} & {\bf 0.157} \\
		$+\Omega_k$ & 0.150 & 0.173 & 0.198 \\	
		\hline\hline
	\end{tabular}
	\caption{1D marginal 95\% upper limits on the neutrino mass sum $\sum m_\nu$ in units of eV in several extended cosmologies, derived in~\cite{RoyChoudhury:2019hls} using Planck  2018 TT+TE+EE+lowE+lensing+BAO.   Extensions to the $\Lambda$CDM baseline include a nonzero primordial tensor-to-scalar ratio~$r$, a dark energy equation of state parameter differing from $w=-1$,
		and a nonzero curvature energy density $\Omega_k$.\label{tab:extended}}
\end{table}
%%%%%%%%%%%%%%%

Moreover, extending the cosmology does {\it not} always give the desired outcome. Consider  again a dynamical $w(z)$, but now restricted to $w(z) \geq -1$.  As  shown in table~\ref{tab:extended},  the degenerate-mass bound in fact winds up 20\% tighter at  $\sum m_\nu \lesssim 0.096$~eV, because of volume effects inherent in all Bayesian credible intervals.  We stress that this somewhat unintuitive result does not in any way represent a problem: volume is a feature of Bayesian statistics, not a bug.  But this example does caution against over-interpretation of marginal improvements or degradations of any credible interval;  there is often no deep meaning behind them.

%%%%%%%%%%%%%%%%%%%%%
%%%%%%%%%%%%%%%%%%%%%%%%

\paragraph{Caveat 3: More data does not equate to  improved constraints }

Adding a new data set sometimes brings new physics information and in turn improves the $\sum m_\nu$ constraint.  This is the case, for example, when CMB polarisation (i.e., TE+EE) is added  to the fit, lifting the degeneracy between $A_s$ and~$\tau$.  Referring to the second column in table~\ref{tab:constraints}, the $\sum m_\nu$ bound also improves by a factor of two as a result, from $\sum m_\nu \lesssim 0.54$~eV  to $\sum m_\nu \lesssim 0.26$~eV.   Another example is BAO, which provides a low-redshift data point on the distance-redshift ladder; going from the first to the third column of table~\ref{tab:constraints}, we see that the $\sum m_\nu$ bound tightens significantly---in some cases by more than a factor of three.

In contrast, the gain from adding the CMB lensing potential is marginal; the $\sum m_\nu$ bound improves by at most 20\%. This is  unsurprising: the CMB TT power spectrum is itself lensed by $P(k)$ at $\ell \gtrsim 500$ and hence already contains much the same information as the lensing potential.   So one might even question if the 20\% gain is  merely a side effect of the Bayesian machinery.  Indeed, previous analyses~\cite{Planck:2015fie} have shown that adding the Planck 2015 CMB lensing potential to the fit in fact {\it worsens} the $\sum m_\nu$ bound by 20\%.  The degradation can ultimately be traced to a marginal  incompatibility between the lensing potential and the TT power spectrum.  But this example serves as another cautionary note that the Bayesian machinery can easily turn a marginal incompatibility between two data sets into an equally marginal shift of the $\sum m_\nu$ bound.   Again, one should not read into  these minor changes.

%%%%%%%%%%%%%%
%%%%%%%%%%%%%%%%

\paragraph{Caveat 4: Non-standard neutrino physics  }

Standard neutrino mass bounds assume SM neutrino physics.  Non-standard neutrino physics could alter the properties of the C$\nu$B and in turn relax cosmological constraints on $\sum m_\nu$.  Relaxing the $\sum m_\nu$ bound by fiddling with the properties of the C$\nu$B usually buys more room for play than by changing the background cosmology alone.  We discuss three cases below.

\begin{itemize}
\item[-]{\bf Non-relativistic neutrino decay into dark radiation} \\
Reference~\cite{FrancoAbellan:2021hdb} proposes a scenario in which the C$\nu$B decays  non-relativistically into massless, invisible particles  with a lifetime of $\tau_\nu \sim 0.1$~Myr.  Then, cosmological data can tolerate $\sum m_\nu \lesssim 0.42$~eV, i.e., more than three times the allowed region of the standard $\Lambda$CDM benchmark $\sum m_\nu \lesssim 0.12$~eV.  Shorter lifetimes could conceivably lead to the same outcome.  However, as the modelling of neutrino decay in cosmology is generally non-trivial~\cite{Barenboim:2020vrr,Chen:2022idm} and the result of~\cite{FrancoAbellan:2021hdb} is already at the edge of validity of their non-relativistic approximation, this possibility remains unexplored.

\item[-]{\bf Neutrino spectral distortion} \\
Current cosmological measurements are not sensitive to the C$\nu$B momentum distribution, a fact that can be exploited to relax the $\sum m_\nu$ bound.  The idea is as follows.  Decay into neutrinos or neutrino interactions can cause the C$\nu$B distribution to deviate from a relativistic Fermi-Dirac form.  If the new physics  enhances the C$\nu$B average momentum, then one could maintain the early-time neutrino energy density at its standard value (i.e., $N_{\rm eff}=3.0440$~\cite{Bennett:2020zkv}), while simultaneously {\it lowering} the C$\nu$B number density $n_\nu$.  A smaller $n_\nu$ in turn means a larger $\sum m_\nu$ can be accommodated by a given~$\Omega_\nu$.

Following this line of argument, reference~\cite{Oldengott:2019lke}
	 shows that relaxing the $\sum m_\nu$ by a factor of two can be easily accomplished with a moderate spectral distortion.  In a more daring study, reference~\cite{Alvey:2021sji} finds that the bound could even relax to $\sum m_\nu \lesssim 3$~eV if a Gaussian momentum distribution was assumed (although of course it is unclear what kind of physical process would lead to a Gaussian momentum distribution).

\item[-] {\bf Late-time neutrino mass generation mechanism} \\
Yet another possibility is that neutrinos do not gain a mass until a suitably late time.  This type of scenarios has previously been discussed in the context of mass-varying neutrinos in~\cite{Fardon:2003eh}.  
More recent variants include a late-time phase transition~\cite{Dvali:2016uhn}.  But, purely from a phenomenological perspective, reference~\cite{Lorenz:2021alz} finds that while constraints on $\sum m_\nu$ are generally quite tight at $z \gtrsim 1$, a much larger mass sum, $\sum m_\nu \lesssim 1.46$~eV, can be tolerated at $z \lesssim 1$.
\end{itemize}

%%%%%%%%%%
%%%%%%%%%%

%%%%%%%%%%%%%

\subsubsection{Future probes}
\label{sec:future}

Several upcoming cosmological observations are likely to improve the $\sum m_\nu$ bound or even measure $\sum m_\nu$. The ESA Euclid mission, a space-based dedicated cosmic shear survey to launch in 2024, has a purported $1 \sigma$ sensitivity to $\sum m_\nu$ of $0.011 \to 0.02$~eV\cite{EuclidTheoryWorkingGroup:2012gxx}.  Further in the future, the ground-based, stage-4 CMB polarisation experiment, CMB-S4, also claims a similar sensitivity to $\sum m_\nu$~\cite{CMB-S4:2016ple}.  In other words, if the true neutrino mass sum was $\sum m_\nu = 0.06$~eV, then measuring it at $3 \to 5\sigma$ significance would be possible in the next decade.

But, as with the interpretation of current $\sum m_\nu$ bounds, there are assumptions behind the derivation of these forecasted sensitivities.  For example, to obtain the Euclid numbers, strong assumptions need to be made about how well we understand the redshift evolution of the galaxy bias, good modelling of the cluster mass function, etc., all of which involve highly nonlinear physics that have no fundamental description.   Thus, not only it is {\it not} entirely foolproof that $\sum m_\nu$ will finally be measured to the claimed statistical significance, the extent to which one can trust a claimed ``measurement'' of $\sum m_\nu$ is also less than clearcut.  Until two experiments measure the same $\sum m_\nu$ value, prudent scepticism should be exercised.

Finally, irrespective of the actual sensitivity finally realised, it will {\it not}  be sufficient to resolve the individual neutrino masses~\cite{Hamann:2012fe}.  For the foreseeable future, the question of the neutrino mass ordering can only be addressed by cosmological observations  if the measured $\sum m_\nu$ value should fall below the minimum for the inverted mass ordering, $\sum m_\nu\simeq 0.11$~eV.

%%%%%%%%%%%%%%
%%%%%%%%%%%%%%

\subsubsection{Final remarks}

These is no question that neutrino masses induce some non-trivial effects on cosmological observables, which can in turn be used to  measure or constrain the absolute neutrino mass scale.  While these measurements and constraints have merits,  it is important not to over-interpret them, as there are assumptions underpinning these numbers that may not be as well understood as claimed.  Nor should one completely ignore these numbers: while it is possible to evade the tightest constraints to a good extent, it is not a situation of ``anything goes''.  In our opinion, the best one could do is to treat all bounds on and forecasted sensitivities to $\sum m_\nu$  as ballpark figures. Until multiple observations have measured the same $\sum m_\nu$ value, one should take all claimed ``measurements'' of $\sum m_\nu$ {\it  cum grano salis}.

% 	\bibliographystyle{utcaps}
% \bibliography{NeutrinoBib}

% \end{document}

%-------------------------------------------
\subsection{Heavy Neutral Leptons and their Connection with Neutrinoless Double Beta Decay: Theory Overview -- {\it F.~Deppisch}}
\label{deppisch}
{\it Author: Frank F. Deppisch, <f.deppisch@ucl.ac.uk>}
%-------------------------------------------

%\documentclass[11pt,a4paper]{article}
%\usepackage{amssymb}
%\usepackage{mathtools}
%\usepackage{hyperref}
%\usepackage[margin=2cm]{geometry}

%\begin{document}
%\title{Heavy Neutral Leptons and their Connection with\\ Neutrinoless Double Beta Decay: Theory Overview}

%\author{Frank F. Deppisch\footnote{f.deppisch@ucl.ac.uk},\\ Department of Physics and Astronomy, University College London,\\Gower Street, London WC1E 6BT, United Kingdom}

%\maketitle

%\subsubsection{abstract}
%Neutrinoless double beta decay is a crucial process probing the Majorana nature of light neutrinos. In addition, it is sensitive to other sources of lepton number violation, generically arising in models of light Majorana neutrino mass generation. We here provide a brief overview of the phenomenology of one of the most popular scenarios incorporating heavy right-handed neutrinos in a seesaw type-I mechanism of neutrino mass generation. Specifically, we highlight the interplay and complementarity of direct experimental searches for such heavy neutral leptons and neutrinoless double beta decay. Using a phenomenological parametrization of the active-sterile mixing we discuss the current constraints as well as the future sensitivity to HNLs and their nature.

\subsubsection{Introduction}
The Standard Model (SM) has five distinct species of matter building blocks, i.e., Weyl fermion states, namely the left-handed lepton and quark $SU(2)$ doublets as well as the right-handed lepton, up-quark and down-quark singlets. Curiously absent are right-handed neutrino states $N_j$. While they would nicely complete the picture, their non-observation is not wholly surprising; in order to fit them into the SM, they would be \emph{sterile}, i.e., uncharged under the SM gauge interactions. They can only couple with other SM particles through the neutrino portal, i.e., the Yukawa interaction $-y_\nu^{\ell j} \bar L_\ell\cdot H N_j$ with a left-handed lepton doublet $L_\ell$ and the Higgs doublet $H$. If they exist, their observation is suppressed by the light neutrino masses $m_\nu \lesssim 0.1$~eV required due to oscillations but strongly constrained by absolute mass searches such as tritium decay and cosmological observations. 

This suppression occurs for \emph{Dirac} neutrinos, i.e., when the left-handed and right-handed neutrinos combine to form a Dirac fermion with a mass induced by the Yukawa coupling after electroweak symmetry breaking, but it also applies for \emph{Majorana} neutrinos. While the left-handed SM neutrino states cannot have a Majorana mass term $-\frac{1}{2}m_\nu^{ij} \bar\nu_{L_i}^C \nu^{\phantom{C}}_{L_j}$ because it is not allowed given that the SM is a gauge invariant and renormalizable theory, the equivalent mass term for the sterile right-handed neutrinos, $-\frac{1}{2}M^{ij} \bar N_i^C N^{\phantom{C}}_j\!\!$, would not be forbidden. This opens up the possibility of \emph{lepton number violation} as the hallmark of the Majorana nature of neutrinos.

Combining the Yukawa and heavy sterile neutrino Majorana mass terms leads to the well-known \emph{seesaw mechanism} (of type-I), with the light neutrino mass scale given by $|m_\nu| \sim |V_{\ell N}|^2 m_N$. Here, $|V_{\ell N}|$ is the active-sterile mixing strength and $m_N$ is the heavy right-handed neutrino mass. The mixing induces suppressed charged and neutral currents between the sterile states $N_j$ and a SM lepton of flavour $\ell = e, \mu, \tau$. This forms a \emph{seesaw floor}: the minimal active-sterile neutrino mixing required to explain the observed light neutrino masses, $|V_{\ell N}|^2 = m_\nu / m_N \approx 10^{-11}(10~\text{GeV}/m_N)$ with $m_\nu = 0.1$~eV as illustration.

This is not to say that the active-sterile mixing must always satisfy the seesaw relation. If smaller, there must be an additional source of neutrino mass generation, e.g., another right-handed neutrino or another mechanism such as seesaw type-II. If larger, a cancellation between two or more contributions, e.g., from two sterile states, is required. Related to this, the sterile neutrino mass scale in the above seesaw relation is not necessarily the mass of the sterile neutrino. In general, it is connected to the scale of lepton number violation. This may manifest itself in a \emph{quasi-Dirac} sterile neutrino, i.e., two Majorana sterile neutrinos that have a opposite $CP$ parities and a small mutual mass splitting $\Delta m_N$. This is realized in extended scenarios like the \emph{inverse} and \emph{linear} seesaw mechanisms.

Nevertheless, sterile neutrinos or more generally, Heavy Neutral Leptons (HNLs)\footnote{We will use the term HNL throughout, though it is more general and may include uncharged and uncoloured fermions that do not contribute to light neutrino masses or which may have interactions beyond the SM.} are well motivated feebly interacting particles. Because of the small active-sterile neutrino mixing, necessarily suppressed to explain the lightness of neutrinos, HNLs are weakly coupled to the SM and they are typically \emph{long-lived particles (LLPs)}. Their proper decay length is approximately
\begin{align}
	L_N^0 \approx 
	100~\text{m}\times \frac{10^{-11}}{|V_{\ell N}|^2} \left(\frac{10~\text{GeV}}{m_N}\right)^5 
	\approx 100~\text{m}\times \left(\frac{10~\text{GeV}}{m_N}\right)^4,
\end{align}
for $1~\text{GeV} \lesssim m_N \lesssim m_W$. The latter expression applies at the intersection with the seesaw floor for $m_\nu = 0.1$~eV. HNLs in such a mass range thus naturally lead to LLP signatures.

To indirectly probe the scenario described above, we should observe light neutrinos as Majorana fermions. The only realistic way to do this, given the strong mass constraint $m_\nu \lesssim 0.1$~eV, is neutrinoless double beta ($0\nu\beta\beta$) decay. This rare nuclear decay, energetically allowed in certain isotopes such as $^{76}\text{Ge} \to {}^{76}\text{Se} + e^- e^-$, is only possible if total lepton number is violated. Besides proving the Majorana nature of light active neutrinos, it is generally sensitive to other exotic sources of lepton number violation, at or below the $\mathcal{O}(10)$~TeV scale. We will here focus on HNL contributions to $0\nu\beta\beta$ decay in comparison with direct HNL searches.

\subsubsection{Current Direct Constraints and Future Sensitivities on HNLs}

\begin{figure}[t!]
	\centering
	\includegraphics[width=0.75\textwidth]{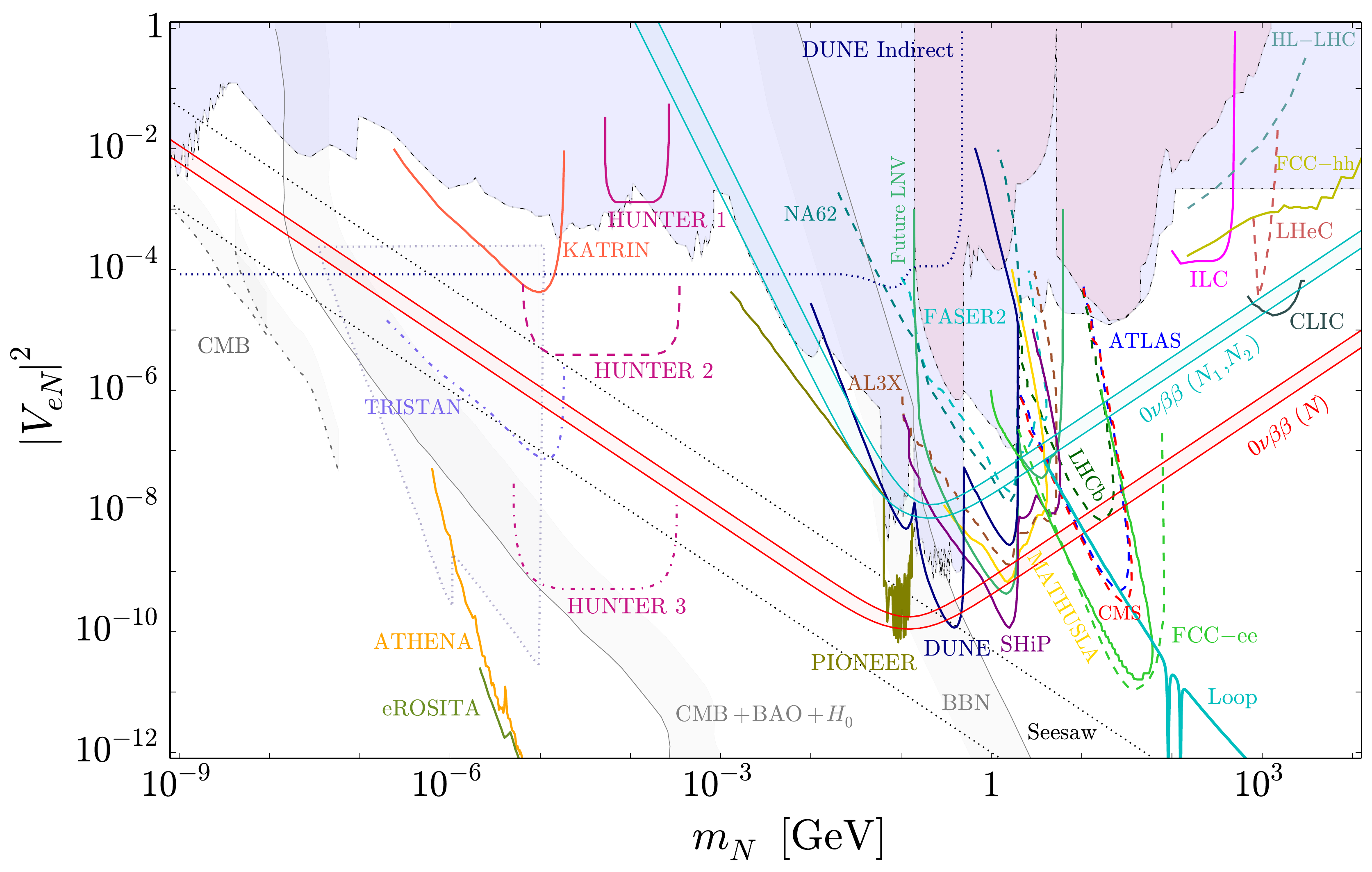}
	\caption{Current constraints (shaded regions) and projected sensitivities of future searches (open curves), including $0\nu\beta\beta$ decay, on the electron-flavour active-sterile neutrino mixing strength $|V_{eN}|^2$ as a function of the HNL mass $m_N$. The light blue (red) region is excluded for both Dirac and Majorana (Majorana only) HNLs. The $0\nu\beta\beta$ decay sensitivities are for a Majorana HNL (red band) and quasi-Dirac HNL (teal band) for the half-life $T^{0\nu}_{1/2} = 10^{28}$~yr in $^{76}$Ge with uncertainties from $0\nu\beta\beta$ nuclear matrix elements. The diagonal band labelled \emph{Seesaw} indicates the seesaw floor $m_\nu = |V_{eN}|^2 m_N$ with $0.001~\text{eV} < m_\nu < 0.05$~eV. Taken from \cite{Bolton:2022pyf}, with detailed descriptions of the various probes in \cite{Bolton:2019pcu} and data available at \url{www.sterile-neutrino.org}.}
	\label{fig:future}
\end{figure}

The LLP property of HNLs is critical in direct searches \cite{Bolton:2019pcu, Abdullahi:2022jlv}. Focussing on the electron-flavour mixing strength $|V_{eN}|$ later relevant for $0\nu\beta\beta$ decay, Fig.~\ref{fig:future} provides an overview of current constraints and future sensitivities over the broad HNL mass range $0.1~\text{eV} < m_N < 10$~TeV. The diagonal band labelled \emph{Seesaw} indicates the seesaw floor, $m_\nu = |V_{eN}|^2 m_N$ with $0.001~\text{eV} < m_\nu < 0.05$~eV. 

Direct searches can be broadly categorized as follows. For large mass, HNLs are produced efficiently in \emph{high-energy collisions} through charged and neutral currents, $pp\to W^+\to e^+ N$ and $pp\to Z\to \nu N$. \emph{Beam-dump experiments} and \emph{meson factories} generate mesons abundantly, with HNLs produced in their decay, e.g., $K^+\to e^+ N$. Sufficiently light HNLs mixing with electron-flavour emerge in \emph{nuclear beta decays} and other weak nuclear processes. Here, there is strong potential for development, e.g., the recent BeEST experiment~\cite{Friedrich:2020nze} has improved limits by almost two orders of magnitude. While there is no hint of a signal for HNLs, there are anomalies in oscillation experiments around the squared mass difference $\Delta m_{14}^2 = m_N^2 - m_\nu^2 \approx 1~\text{eV}^2$, which can be interpreted as a hint for light sterile neutrinos, although there no consistent interpretation. Regardless, \emph{neutrino oscillations} are sensitive to eV-scale sterile neutrinos but also indirectly to more massive HNLs through the resulting deficit of active neutrinos detected, if the absolute neutrino flux is sufficiently well known. Similarly, a non-vanishing active-sterile mixing means that the mixing matrix among the active neutrinos is no longer unitary, changing \emph{electroweak precision data} observables. 

Finally, for HNL masses $m_N \lesssim 1$~GeV, constraints from \emph{cosmology and astrophysics} are severe. Most relevant around this mass scale, HNLs thermalize in the early universe through scattering or oscillation, even for very small mixing, and if decaying at times later than $\sim 1$~s, they disturb big bang nucleosynthesis (BBN). Lighter and consequently longer-lived HNLs may inject light neutrino degrees of freedom or over-close the universe as Dark Matter. 

\subsubsection{HNLs in Neutrinoless Double Beta Decay}
\label{sec:0vbb}
Future neutrinoless double beta decay searches will be sensitive to light Majorana neutrinos with masses $m_\nu \approx 10^{-2}$~eV as well as New Physics of lepton number violation below $\mathcal{O}(10)$~TeV \cite{Doi:1981, Cirigliano:2017djv, Graf:2018ozy, Deppisch:2020ztt}. This includes Majorana HNLs and the $0\nu\beta\beta$ decay half-life $T^{0\nu}_{1/2}$ due to the combined effect of light active Majorana neutrinos and HNLs is
\begin{align}
\label{eq:inv_0vbb}
	\left[T^{0\nu}_{1/2}\right]^{-1} = 
	\big|m_{\beta\beta}^\text{eff}\big|^2 \big|\mathcal{M}^{0\nu}_\nu\big|^2 G^{0\nu} / m_e^2.
\end{align}
Here, $G^{0\nu} \sim 10^{-15}$~yr$^{-1}$ is the phase space factor setting the overall decay rate, $\mathcal{M}^{0\nu}_\nu$ is the dimensionless nuclear matrix element for light neutrino exchange and the electron mass $m_e$ is conventionally used for normalization. The nuclear matrix element is $|\mathcal{M}^{0\nu}_\nu| \sim \mathcal{O}(1)$ but it has a large theoretical uncertainty due to differences in nuclear structure calculations. Both $G^{0\nu}$ and $M^{0\nu}_\nu$ depend on the $0\nu\beta\beta$ decaying isotope. Existing $0\nu\beta\beta$ decay searches have not observed a signal, setting a limit $T^{0\nu}_{1/2} \gtrsim 10^{26}$~yr~\cite{PhysRevLett.117.082503, GERDA:2020xhi}. Future planned experiments such as LEGEND-1000~\cite{Zsigmond:2020bfx} have a  projected sensitivity reaching $T^{0\nu}_{1/2} \approx 10^{28}$~yr. The effective $0\nu\beta\beta$ mass in Eq.~\eqref{eq:inv_0vbb} is
\begin{align}
	\label{eq:mbb_eff_full}
	\big|m_{\beta\beta}^\text{eff}\big| = 
	\left|\sum^3_{i=1} U_{ei}^2    m_{\nu_i} 
	    + \sum^2_{a=1} V_{e N_a}^2 m_{N_a} 
	    \frac{\mathcal{F}(m_{N_a})\langle\mathbf{p}^2\rangle}{\langle\mathbf{p}^2\rangle + m_{N_a}^2}\right|,
\end{align}
summing over the three light active neutrinos $\nu_i$, with the Pontecorvo–Maki–Nakagawa–Sakata mixing matrix $U$, and two HNLs $N_a$ as illustration. The $0\nu\beta\beta$ energy scale is $\langle\mathbf{p}^2\rangle \approx (100\text{MeV})^2$. For $m_{N_a}^2 \approx \langle\mathbf{p}^2\rangle$, the momentum dependence on the nuclear matrix elements would need to be included more carefully but the above expression is a good approximation. It interpolates between the light and heavy HNL regimes \cite{Babic:2018ikc}, using the weakly varying correction factor $\mathcal{F}(m_{N_a}) \approx 1$ to account for the variation of the $0\nu\beta\beta$ energy scale in different types of nuclear matrix elements \cite{Dekens:2020ttz, Bolton:2022tds}. 

Different from direct searches, $0\nu\beta\beta$ decay depends on the coherent sum over all contributions as apparent from Eq.~\eqref{eq:mbb_eff_full}, and the rate depends on the relative $CP$ phases in the potentially complex active-active and active-sterile neutrino mixing matrix elements $U_{ei}$ and $V_{eN_a}$, respectively. Quasi-Dirac HNLs have (approximately) opposite mixing phases, $V_{eN_2}^2 \approx -V_{eN_1}^2$ and a small mass difference, $m_{N_2} - m_{N_1} = \Delta m_N \ll m_{N_a}$. Their contributions thus cancel each other, as is generally expected in any lepton number violating process such as $0\nu\beta\beta$ decay must vanish in the limit of exact Diracness. For masses, $m_N \gtrsim 100$~MeV, the contribution of a quasi-Dirac HNL to $0\nu\beta\beta$ decay is roughly 
\begin{align}
	\label{eq:0vbb:half-life-heavy-quasi-dirac}
	\frac{10^{28}~\text{yr}}{T_{1/2}^{0\nu}} 
	\approx \left(\frac{\Delta m_N/m_N}{10^{-2}}
	\cdot \frac{|V_{eN}|^2}{10^{-7}}
	\cdot \frac{1~\text{GeV}}{m_N}\right)^2,
\end{align}
for $^{76}$Ge~\cite{Deppisch:2020ztt} and normalized to the LEGEND-1000 sensitivity. This expression also covers the limiting case of a Majorana HNL with effectively $\Delta m_N/m_N = 1$. Future $0\nu\beta\beta$ decay experiments are thus sensitive to active-sterile mixing strengths down to $|V_{eN}|^2 \approx 10^{-10}$ to $10^{-9}$, in the range $10~\text{MeV}\lesssim m_N \lesssim 1$~GeV. This is also shown in Fig.~\ref{fig:future} via the red band with the width indicating nuclear matrix element uncertainties including the potential quenching of the axial nuclear coupling strength~\cite{Deppisch:2016rox}. 

As noted, the sensitivity is accordingly reduced for quasi-Dirac HNLs. This is shown by the teal band in Fig.~\ref{fig:future}, for a mass splitting $\Delta m_N / m_N = 10^{-2}$. It also includes the effect of the interference with the contribution from the light active neutrinos. On general grounds, this interference is expected to be destructive: for light HNLs, $m_N \ll 100$~MeV, Eq.~\eqref{eq:mbb_eff_full} simplifies to $\sum_i U_{ei}^2 m_{\nu_i} + \sum_a V_{e N_a}^2 m_{N_a} = 0$ due to the seesaw relation between the masses and mixing matrix elements\footnote{This assumes that there are no other sources contributing to light neutrino masses.}. This results in the steep slope of the teal band for $m_N < 100$~MeV, with the light neutrino contribution being $|m_{\beta\beta}| = 10^{-3}$~eV. 

Phenomenologically speaking, it is possible to consider HNLs with arbitrary mass, mass splitting and mixing strength. While larger masses $m_N \gtrsim 10$~GeV and splitting $\Delta m_N / m_N$ are allowed they lead to larger loop corrections to the light neutrino masses~\cite{Mitra:2011qr, Lopez-Pavon:2012yda, Bolton:2019pcu}. The above discussion only includes the seesaw-induced mass which corresponds to the tree level. In Fig.~\ref{fig:future}, the region to the right of the curve labelled \emph{Loop} is disfavoured in the quasi-Dirac case by requiring that the loop induced neutrino mass is less than 10\% of the tree-level seesaw mass.

\begin{figure}[t!]
	\centering
	\includegraphics[width=0.38\textwidth]{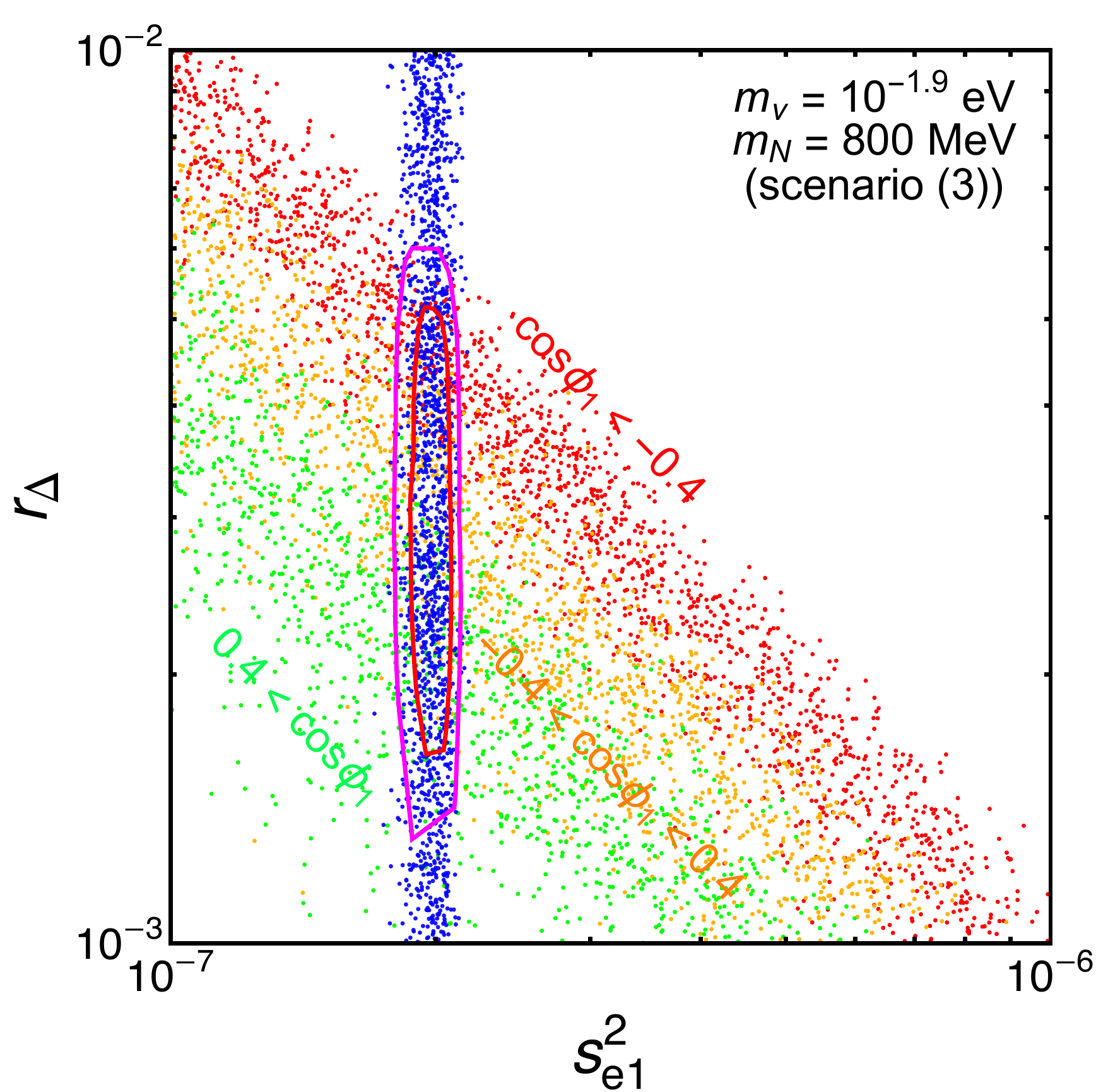}
	\includegraphics[width=0.38\textwidth]{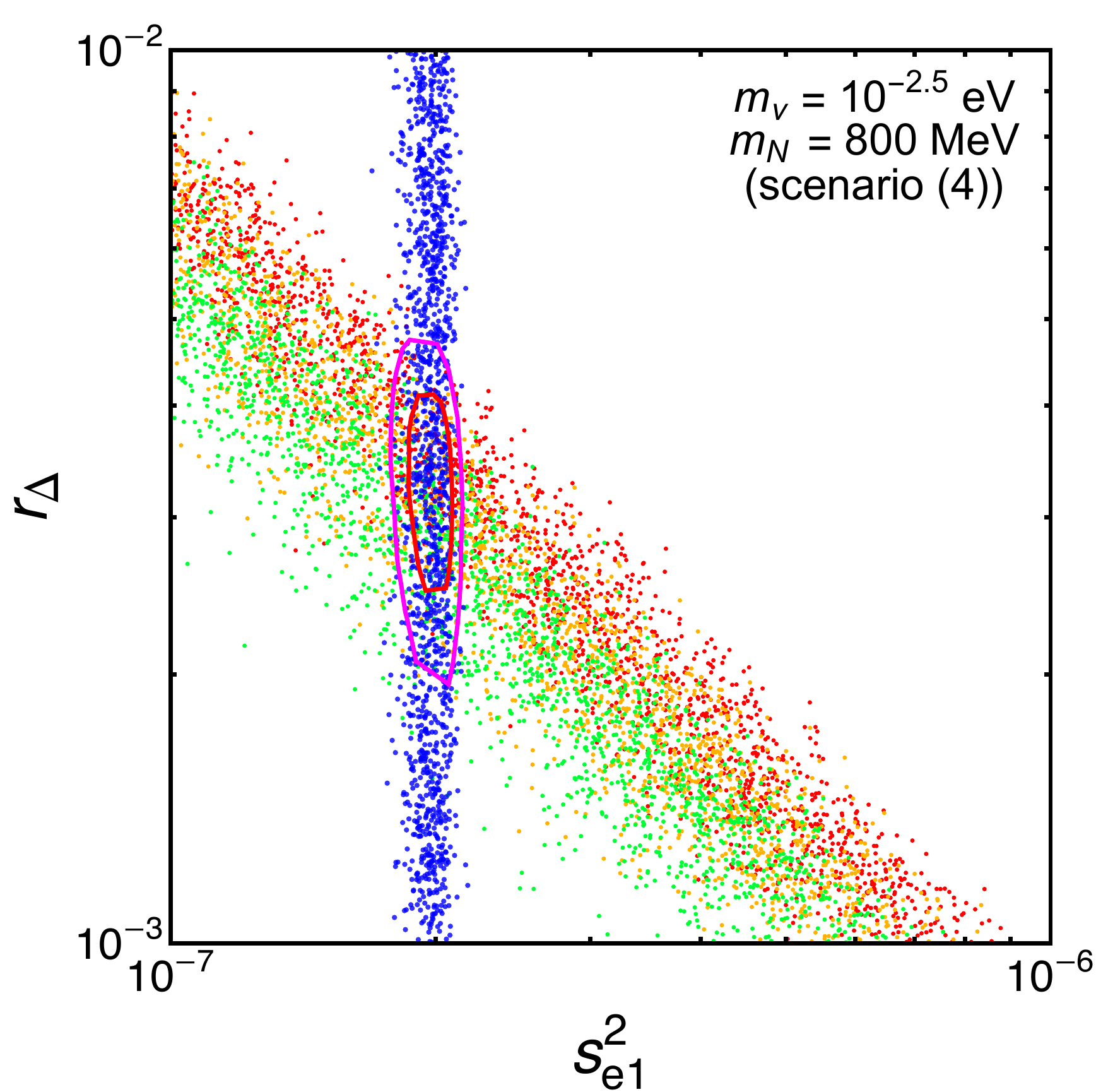}
	\caption{Active-sterile mixing strength $s_{e1}^2 \approx |V_{eN_1}|^2$ and mass splitting $r_\Delta = \Delta m_N/ m_N$ constrained by a hypothetical observation of an HNL signal at DUNE (vertical blue band) and LEGEND-1000 (diagonal band) in a scenario with a pair of quasi-Dirac HNLs with mass $m_{N_1} = 800$~MeV and a light active neutrino mass $m_\nu = 10^{-1.9}$~eV (left) and $m_\nu = 10^{-2.5}$~eV~(right). The colours on the $0\nu\beta\beta$ band indicate the $CP$ phase $\phi_1$ of HNL $N_1$ relative to that of light neutrinos. The red contours give the 68\% and 95\% credible regions in combining both observations. Taken from \cite{Bolton:2022tds}.}
	\label{fig:complementarity}
\end{figure}
It is worthwhile to note that both $0\nu\beta\beta$ decay and direct searches, e.g., at DUNE, are expected to probe a parameter space close to the seesaw floor, for $m_N \lesssim 10$~GeV. This potentially allows probing detailed properties of HNLs by combining results from direct and $0\nu\beta\beta$ searches \cite{Bolton:2022tds}. Fig.~\ref{fig:complementarity} shows the constraints on the active-sterile mixing strength $s_{e1}^2 \approx |V_{eN_1}|^2$ and the mass splitting $r_\Delta = \Delta m_N / m_N$ of a pair of quasi-Dirac HNLs expected from hypothetical observations of an HNL signal at DUNE (blue band) and LEGEND-1000 (diagonal coloured band) with $\approx 300$~events and 10~events, respectively. As a direct search, DUNE is largely insensitive to the mass splitting as well as any $CP$ phases, thus fixing $|V_{eN_1}|^2$ for the assumed HNL mass $m_{N_1} = 800$~MeV. On the other hand, $0\nu\beta\beta$ decay is suppressed by the quasi-Dirac mass splitting, resulting in a diagonal band in the parameter space shown. The left and right plot are for a light neutrino mass of $m_\nu = 10^{-1.9}$~eV and $m_{\nu} = 10^{-2.5}$~eV, respectively, additionally contributing to $0\nu\beta\beta$ decay. The former value saturates the future $0\nu\beta\beta$ decay rate on its own. The leads to the broadening of the band in the left plot but it also activates a sensitivity to the $CP$ phase $\phi_1$ of the HNL as indicated. In combining both hypothetical observations at DUNE and LEGEND-1000 would pinpoint the active-sterile mixing and the mass splitting within the 68\% and 95\% credible contours shown.

\subsubsection{Conclusion}
HNLs are well-motivated by the need to understand the lightness of active neutrinos, which also naturally makes them feebly interacting. They can be searched for in a large number of experiments and observations over a broad mass range $1~\text{eV} \lesssim m_N \lesssim 10$~TeV. Future direct searches will exploit the expected LLP signature to reach unprecedented sensitivities close to or drilling through the seesaw floor of light neutrino mass generation. This connection to the light neutrinos, specifically their mass and their possible Majorana nature, also allows HNLs to be probed in $0\nu\beta\beta$ decay where future searches such as LEGEND-1000 are expected to improve the sensitivity on the $0\nu\beta\beta$ decay half life by two orders of magnitude to $T_{1/2}^{0\nu} \approx 10^{28}$~yr. The main goal of $0\nu\beta\beta$ decay searches is to prove the Majorana nature of light neutrinos. Its discovery would provide the strongest motivation for HNLs.

In addition, direct searches and $0\nu\beta\beta$ decay are highly complementary. The former mainly rely on real HNL production and thus depend on the HNL mass and the modulus of the active-sterile mixing strength. On the other hand, $0\nu\beta\beta$ decay involves a virtual HNL exchange and is delicately sensitive to the nature of HNLs, an HNL mass splitting and the $CP$ phases in the HNL sector\footnote{The SM-allowed two-neutrino double beta decay can also be used to directly search for MeV-scale HNLs \cite{Deppisch:2020mxv, Deppisch:2020sqh, Bolton:2020ncv}.}. Observing HNLs in direct and $0\nu\beta\beta$ decay searches can thus shed light on the detailed properties of HNLs and their role in generating the light neutrino masses as well as the matter-antimatter asymmetry of the universe through \emph{leptogenesis}.

We have here focussed on the sterile nature of HNLs, only including a mass mixing with the active neutrinos and thus allowing HNLs to participate in SM charged and neutral currents, albeit strongly suppressed by the small active-sterile mixing. Going further beyond the SM, HNLs can have other interactions, with SM or exotic particles, such as transition magnetic moments~\cite{Bolton:2021pey} or new gauge forces~\cite{Liu:2022kid, Padhan:2022fak}. They may also not be immediately recognizable as sterile neutrinos, e.g., neutralinos in $R$-parity violating supersymmetry \cite{Bolton:2021hje}. 

%\subsubsection*{Acknowledgments}
%The author would like to thank Patrick Bolton, Bhupal Dev, Mudit Rai and Zhong Zhang for collaborations on which this proceedings report is mainly based on. The author acknowledges support from the UK Science and Technology Facilities Council (STFC) via the Consolidated Grant ST/T000880/1.

%\bibliographystyle{unsrt}
%\bibliography{deppisch}

%-------------------------------------------
\subsection{Search for heavy neutral leptons and Higgs portal scalars with MicroBooNE -- {\it S.~S\"oldner-Rembold}}
%-------------------------------------------
\label{soldner-rembold}
{\it Author: Stefan S\"oldner-Rembold, <Stefan.Soldner-Rembold@cern.ch>}
%-------------------------------------------

%\usepackage{xcolor}
%\usepackage{graphicx}% Include figure files
%\usepackage{dcolumn}% Align table columns on decimal point
%\usepackage{bm}% bold math
%\usepackage[pdftex]{hyperref} % Extensive support for hypertext in LATEX
%\usepackage[mathlines]{lineno}% Enable numbering of text and display math

%\usepackage{svg}

%\usepackage{adjustbox}
%\usepackage{multirow}
%\usepackage{lineno}

\newcommand{\umu}{$\lvert U_{\mu4}\rvert^2$}
\newcommand{\umusq}{\lvert U_{\mu4}\rvert^2}
\newcommand{\emix}{\lvert U_{e 4}\rvert^2}
\newcommand{\mumix}{\lvert U_{\mu 4}\rvert^2}
\newcommand{\alphamix}{\lvert U_{\alpha 4}\rvert^2}
\newcommand{\thetaeffmix}{\theta_{\textrm{eff}}^2}
\newcommand{\thetamix}{\theta^2}
\def\mhps {{\ensuremath{m_{\mathrm{HPS}}}}}
\def\mhnl {{\ensuremath{m_{\mathrm{HNL}}}}}
\def\mllp {{\ensuremath{m_{\mathrm{LLP}}}}}
\def\mllpsq {{\ensuremath{m^2_{\mathrm{LLP}}}}}
\hyphenation{SliceID}
\hyphenation{MicroBooNE}

%\begin{document}

%\widetext
%\leftline{FERMILAB-PUB-22-507-ND}

%\centerline{\em INTERNAL DOCUMENT -- NOT FOR PUBLIC DISTRIBUTION}

%\begin{center}

%{\bf Search for heavy neutral leptons and Higgs portal scalars with MicroBooNE\footnote{to be included in the proceedings for FIPS 2022, October 2022, CERN}}

%\author{Stefan S\"oldner-Rembold, on behalf of the MicroBooNE Collaboration}\affiliation{The University of Manchester}

%Stefan S\"oldner-Rembold, {\em The University of Manchester}, for the MicroBooNE Collaboration
%\bigskip
%\end{center}

%\date{\today}

%\begin{abstract}
%We present a new search for long-lived Higgs portal scalars (HPS) and heavy neutral leptons (HNL) decaying in the MicroBooNE liquid-argon time projection chamber.  The measurement is performed using data collected synchronously with the NuMI neutrino beam from Fermilab's Main Injector. The measurement represents a significant improvement in sensitivity compared to a previous HNL search performed with
%the Booster Neutrino Beam searching for HNL decays that occur after the neutrino spill reaches the detector.
%\end{abstract}

%\maketitle

%% SECTIONS

\subsubsection{Introduction}

The MicroBooNE detector \cite{MicroBooNE:2016pwy} was a liquid-argon Time Projection Chamber exposed to both the booster neutrino beam (BNB) and
the neutrino beam from the main injector (NuMI) at Fermilab. 
In this document, we present recent results on searches 
for the production and decay of long-lived heavy neutral leptons (HNL) and Higgs portal scalars (HPS). The document is an 
abbreviated presentation of the results described in Ref.~\cite{MicroBooNE:2022ctm}, which should be used as the primary reference. 

Previously, MicroBooNE published upper limits on the production of HNLs decaying to $\mu \pi$ pairs for an exposure of $2.0 \times 10^{20}$ protons on target (POT) from the BNB, using a dedicated trigger configured to detect HNL decays that occur after the neutrino spill reaches the detector. That search yielded upper limits at the $90\%$ confidence level (CL) on the element $\umusq$ of the extended PMNS mixing matrix $\umusq$ for Dirac and Majorana HNLs
in the HNL mass range $260\le \mhnl\le 385$~MeV and assuming $\lvert U_{e 4}\rvert^2 = \lvert U_{\tau 4}\rvert^2 = 0$~\cite{MicroBooNE:2019izn}.    
MicroBooNE has also published a search for HPS decaying to $e^+e^-$ pairs assuming the HPS originate from kaons decaying at rest after having been produced at the NuMI absorber~\cite{MicroBooNE:2021usw}. 
A data set corresponding to $1.93 \times 10^{20}$~POT is used to set limits on $\theta^2$ in the range $10^{-6}$--$10^{-7}$ at the $95\%$ CL for the mass range directly below the range considered here ($0 < \mhps< 211$~MeV).

The HNLs would be produced in the decays of charged kaons and pions originating from the proton interactions on the targets of the BNB or NuMI neutrino beams. If the HNL lifetime is sufficiently long to allow the HNL to reach the MicroBooNE detector, they can decay into Standard Model (SM) particles within the argon volume.  We consider the production channel $K^{+}\rightarrow \mu^{+} N$
with the decay $N \rightarrow \mu^{\mp} \pi^{\pm}$.
The HNL production rate and decay width into $\mu \pi$ are each proportional to $\lvert U_{\mu 4}\rvert^2$, and the total rate therefore to $\lvert U_{\mu 4}\rvert^4$,
assuming $\lvert U_{e 4}\rvert^2 = \lvert U_{\tau 4}\rvert^2 = 0$~\cite{Atre:2009rg}. We thus place limits exclusively on the \umu{} mixing matrix element.
The accessible HNL masses are constrained by the condition $m_K - m_{\mu}>m_{\rm HNL}>m_{\mu} + m_{\pi}$.
HNL states can include Dirac and Majorana mass terms, where
Majorana HNLs would decay in equal numbers into $\mu^+\pi^-$ and $\mu^-\pi^+$ final states. 
Dirac HNLs from $K^+$ decays could only decay to the charge combination $\mu^-\pi^+$ to conserve lepton number.

The Higgs portal model~\cite{Patt:2006fw} is an extension to the SM, where an electrically-neutral singlet scalar boson mixes with the 
Higgs boson with a mixing angle $\theta$. 
Through this mixing, this HPS boson
acquires a coupling to SM fermions via their Yukawa couplings, which is proportional to $\sin\theta$. 
The phenomenology of the Higgs portal model, including the equations describing production and decay of the scalar boson, are discussed in Ref.~\cite{Batell:2019nwo}. 
The  dominant production channel considered here is the two-body decay $K^+\rightarrow \pi^+S$ (where the HPS is denoted by $S$).  
The dominant decay mechanism is a penguin diagram with a top quark contributing in the loop. 

The partial decay width for decays to charged leptons is proportional to $m_{\ell}^2$~\cite{Batell:2019nwo}. If there are no dark sector particles with masses $<\mhps/2$,
the branching fraction into $\mu^+\mu^-$ pair is $\approx 100\%$
for $m_{\mu^+\mu^-}<\mhps<m_{\pi^0\pi^0}$.
The decays into $\pi^+\pi^-$ pairs become accessible at $\mhps>279.1$~MeV. 
The $\pi^+ \pi^-$ decay signatures would appear very similar to $\mu^+\mu^-$ decays, but
the analysis is not sensitive to HPS decays in the $\pi^+ \pi^-$ decay channel, as
the HPS would decay before the detector.
We set limits as a function of the mixing angle $\theta$ as  
HPS production and decay rate are $\propto\theta^2$. 

\begin{figure*}[ht!]
  \centering
  \includegraphics[width=0.8\textwidth]{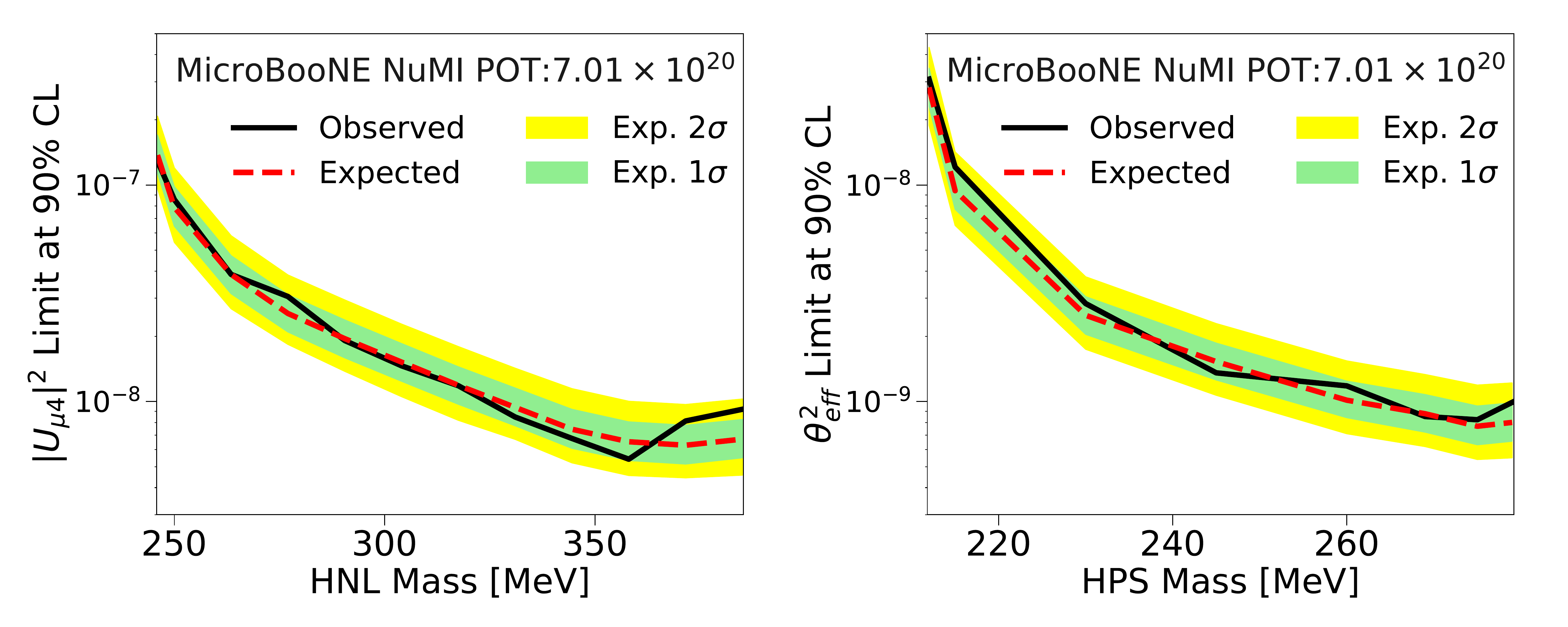}
\caption{
Limits at the $90\%$ confidence level as function of mass for (left) $\mumix$, assuming a Majorana HNL decaying into $\mu\pi$ pairs, and (right) $\thetaeffmix$ of an HPS decaying into $\mu\mu$ pairs. 
The observed limits are compared to the median expected limit.}
% with the $1$ and $2$ standard deviations ($\sigma$) bands.}
  \label{fig:brazil}
\end{figure*}

We generate the signal using the flux of charged kaons that
produce neutrinos, decaying them instead into an HNL or an HPS through the processes $K\rightarrow \mu N$ or $K\rightarrow \pi S$. The exponential decay of the HPS flux is accounted for when selecting a decay vertex. The HPS lifetime is proportional to $\theta^{-2}$~\cite{Batell:2019nwo}.
The decay length of an HPS with $m_{\rm HPS}>2m_\mu$ and a mixing angle in the region of
interest of $10^{-7}<\theta< 10^{-9}$ is similar to the distance between the absorber and the MicroBooNE detector. Therefore some HPS will decay before reaching the detector, reducing the flux in the MicroBooNE detector.
For large values of $\theta^2$, only a small fraction of the HPS reach the detector before decaying, which restricts the upper reach of exclusion contours as a function of $\theta^2$.
The exponential decay of the HNL flux is negligible for the mixing angles considered here
as the HNL lifetime is much longer than the time needed to reach the MicroBooNE detector. The number of HNL decaying before reaching the detector is therefore neglected, and the final event rate is proportional to $\lvert U_{\mu 4}\rvert^4$.

\subsubsection{Results}

We train Boosted Decision Trees (BDTs) that discriminate between the
signal and the background passing an initial selection, separately
for each $m_{\rm HNL}$ and $m_{\rm HPS}$ mass point.
The background sample contains events where hits from overlaid cosmic events are mis-reconstructed as signal candidates. 
Therefore, cosmic-ray background is also rejected by the BDT, even without training on a beam-off sample.
In total, we use 21 BDT input variables.
Uncertainty sources are considered for the background and signal samples by applying variations that modify the BDT score distributions.
For the simulated background sample describing neutrino interactions in the cryostat, we consider the impact of the flux simulation, cross-section modeling, hadron interactions with argon, and detector variations.

The BDT score distributions are used as input to a modified frequentist CL$_s$ calculation to set upper limits on the signal strength for each model and mass point. 
The observed and median expected $90\%$~CL limits on $\mumix$ are shown for each HNL mass point in Fig.~\ref{fig:brazil}. 
The $1$- and $2$-standard deviation intervals cover the range of expected limits produced by $68\%$ and $95\%$ of background prediction outcomes around the median expected value.
The observed limits are contained in the $1$-standard-deviation interval for all mass points with the exception of $\mhnl=371.5$~MeV and $385.0$~MeV, and for $\mhps=215$~MeV, where the observed limit lies within $2$ standard deviations and use a linear interpolation between the mass points when drawing contours. 
We derive the limits assuming HNLs are Majorana particles. For a Dirac HNL, only decays to the charge conjugated final state $\mu^-\pi^+$ are allowed in $K^+$ decays. The expected number of decays is therefore a factor of two smaller for the same $\mumix$ value.
The limits for Dirac HNLs are calculated from the Majorana limit by applying a factor of $\sqrt{2}$ to account for the reduced decay rate, since the difference due to angular distributions of the decay is found to be negligible.

\begin{figure*}[htbp]
  \centering
  \includegraphics[width=0.45\textwidth]
  {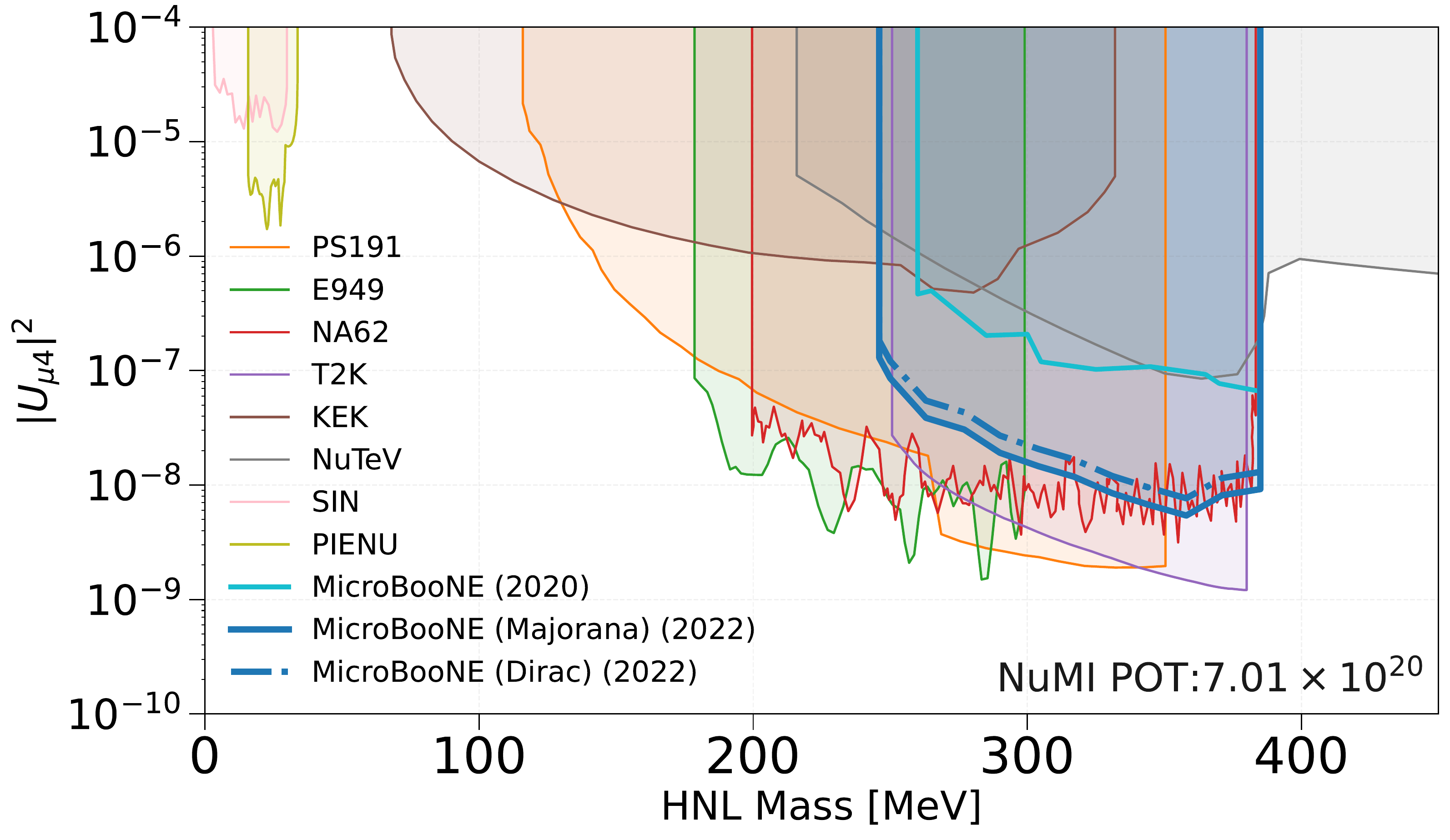}
    \includegraphics[width=0.45\textwidth]
    {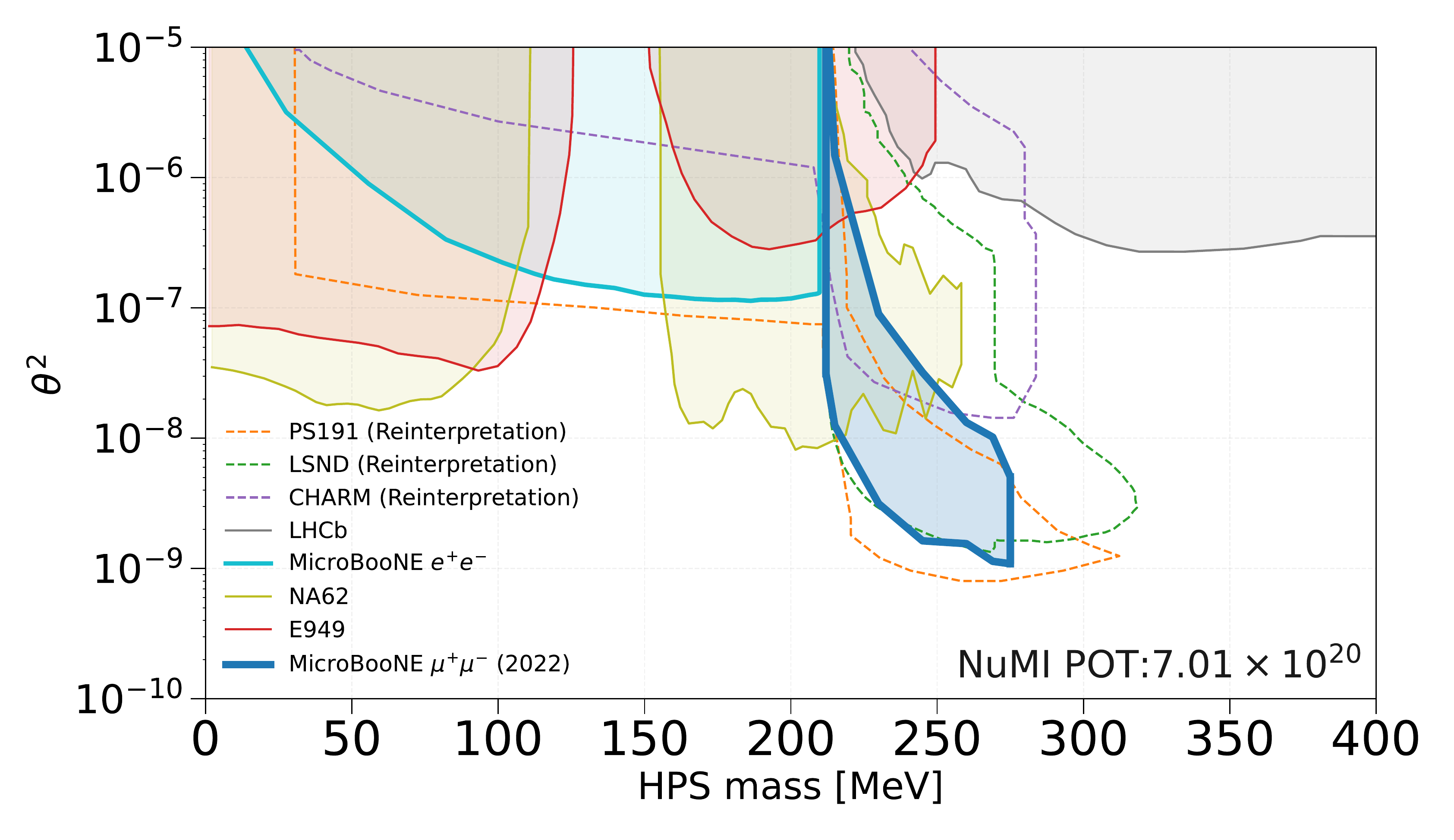}
  \caption{(a) Limits on $\mumix$ at the $90\%$ CL as function of mass for Majorana and Dirac HNL compared to the
  results of the SIN~\cite{Daum:1987bg}, PIENU~\cite{PIENU:2019usb}, KEK~\cite{Hayano:1982wu}, NA62~\protect\cite{NA62:2021bji}, E949~\cite{E949:2014gsn}, PS191~\protect\cite{Bernardi:1987ek}, T2K~\protect\cite{T2K:2019jwa} and NuTeV~\cite{NuTeV:1999kej} collaborations.
  (b) Limits at the $90\%$~CL on the scalar-Higgs mixing angle $\theta$ as a function of $\mhps$ compared to reinterpretations of 
  CHARM~\protect\cite{Winkler:2018qyg}, LSND~\protect\cite{Foroughi-Abari:2020gju}, and PS191~\protect\cite{Gorbunov:2021ccu} measurements.
  In other mass ranges, limits are from a MicroBooNE search for the $e^+e^-$ final state~\protect\cite{MicroBooNE:2021usw} (at the $95\%$~CL), and from searches by the NA62~\protect\cite{NA62:2021bji,NA62:2021zjw} and E949 collaborations~\protect\cite{BNL-E949:2009dza} for charged kaon decays to pions and an HPS.
The LHCb collaboration performed two searches for an HPS with short lifetime, which would be produced and subsequently decay within the detector~\protect\cite{LHCb:2015nkv,LHCb:2016awg}. The joint
coverage of the LHCb result is shown at the $95\%$~CL.
  }
  \label{fig:limscomphps}
\end{figure*}

In Fig.~\ref{fig:limscomphps}, we compare the observed limits to the existing experimental limits in similar regions of parameter space for both models. The results extend MicroBooNE's sensitivity to $\mumix$ by approximately an order of magnitude compared to the previous MicroBooNE HNL result~\cite{MicroBooNE:2019izn}.
The T2K~\cite{T2K:2019jwa} and NuTeV detectors~\cite{NuTeV:1999kej} were also located in a neutrino beam. The PS191 experiment~\cite{Bernardi:1985ny,Bernardi:1987ek} at CERN was specifically designed to search for massive decaying neutrinos. The NA62~\cite{NA62:2017qcd} and E949~\cite{E949:2014gsn} collaborations performed a peak search for HNLs in kaon decays. The muon spectrum measured in stopped $K^+\to\mu^+\nu$ decays $(K_{2\mu})$ has also been used to set limits on HNLs~\cite{Asano:1981he,Hayano:1982wu}.
In the mass range $300<\mhnl<385$~MeV, this search has similar sensitivity as NA62~\cite{NA62:2021bji}. 
The E949~\cite{E949:2014gsn}, PS191~\cite{Bernardi:1987ek}, and T2K~\cite{T2K:2019jwa} limits are stronger across the range $300<\mhnl<385$ MeV.
The T2K collaboration provides no limit point for masses above $380$~MeV. Here, the MicroBooNE limit is of equal or greater sensitivity than the NA62 result.

For the HPS model, we constrain a region of parameter space for $212<\mhps<275$~MeV not previously excluded by any dedicated experimental search.
The existing limits in this region are reinterpretations of decades old CHARM~\cite{Winkler:2018qyg}, LSND~\cite{Foroughi-Abari:2020gju}, and PS191~\cite{Gorbunov:2021ccu} measurements, performed by authors outside the respective collaborations
without access to the original experimental data or MC simulation.
 Reinterpretations depend on external beamline, flux, and detector simulations. 
 If the signal topology differs from the original selection criteria, the results also depend on estimated detection efficiencies. In the case of the CHARM experiment, 
 the more recent sensitivity estimate in Ref.~\cite{Winkler:2018qyg} disagrees by nearly an order of magnitude from the estimate in Ref.~\cite{Clarke:2013aya}.

In summary, we set upper limits on the mixing parameter $\mumix$ ranging from $\mumix=12.9\times 10^{-8}$ for Majorana HNLs with a mass of $\mhnl=246$~MeV to $\mumix=0.92 \times 10^{-8}$ for $\mhnl=385$~MeV, assuming $\lvert U_{e 4}\rvert^2 = \lvert U_{\tau 4}\rvert^2 = 0$ and HNL decays
into $\mu^\pm\pi^\mp$ pairs.
These limits on $\mumix$ are of similar sensitivity to those published by the NA62 collaboration~\cite{NA62:2021bji} and they represent an order of magnitude improvement in sensitivity compared to the previous MicroBooNE result~\cite{MicroBooNE:2019izn}.
We also constrain the scalar-Higgs mixing angle $\theta$ by searching for HPS decays into $\mu^+\mu^-$ final states, excluding a contour in the parameter space with lower bounds of 
$\theta^2<31.3 \times 10^{-9}$ for $\mhps=212$~GeV  
and
$\theta^2<1.09 \times 10^{-9}$ for $\mhps=275$~GeV.
These are the first constraints in this region of the $\theta^2$--$\mhps$ parameter space from a dedicated experimental search. It is also the first search in this mass range using a liquid-argon TPC. Other MicroBooNE results searching for beyond the Standard Model physics, such as HNL decays, milli-charged particles, axions, and dark matter particles are in preparation.

%-------------------------------------------
% \subsection{Prospects at DUNE near detectors and short baseline experiments in US}
\subsection{Heavy Neutral Leptons at Short-Baseline Neutrino Detectors -- {\it J.~Kopp} }
\label{Joachim_Kopp}
{\it Author: Joachim Kopp, <jkopp@cern.ch>}
%-------------------------------------------

% \documentclass[onecolumn,oneside,11pt,altaffilletter,tightenlines,showpacs,showkeys,
%                notitlepage,preprintnumbers,nofootinbib,superscriptaddress,a4paper,        floatfix]{revtex4-2}

% \usepackage{graphicx}
% \usepackage{amsmath}
% \usepackage{cleveref}
% \usepackage{siunitx}

% \begin{document}

% %---------------------------------------------------------------------------
% \title{Heavy Neutral Leptons at Short-Baseline Neutrino Detectors}
% \author{Joachim Kopp (CERN and JGU Mainz)}

% \maketitle
%---------------------------------------------------------------------------

One of the most promising ways of searching for heavy neutral leptons
(also known as sterile neutrinos, right-handed neutrinos, or singlet
fermions) is to leverage the power of existing and upcoming neutrino beams
\cite{Breitbach:2021gvv}.
Notably, long-baseline neutrino experiments are typically equipped with
a sophisticated suite of \emph{Near Detectors}, placed of order hundreds
of meters away from the target.  The main role of these Near Detectors
in oscillation physics is the precise measurement of the unoscillated
neutrino flux and spectrum, the monitoring of the beam to quickly detect
deviations from nominal parameters, and the measurement of neutrino--nucleus
interaction cross sections.  To achieve these goals, the detectors
benefit from a (for neutrino standards) incredibly large event rate
of for example tens of events per beam spill for DUNE, and from excellent
event reconstruction capabilities.  In addition, it is desirable to measure
the neutrino flux not only on the beam axis but also in off-axis locations
by either making detectors movable or by operating several of them in different
locations. This is useful because the neutrino spectrum changes
as a function of the off-axis angle, but neutrino cross obviously do not.
Therefore, off-axis measurements allow us to disentangle systematic uncertainties
associated with the beam itself from those associated with neutrino cross sections.

All of these features make near detectors of oscillation experiments -- or
short-baseline neutrino detectors more generally -- ideally suited for
heavy neutral lepton (HNL) searches. These searches, too, benefit from
extremely high beam intensity, excellent event reconstruction,
and the possibility to go off-axis. Therefore, they can be carried out
fully parasitically.

In the following, we will elaborate on the prospects of HNL searches
in such experiments, following mainly the discussion in ref.~\cite{Breitbach:2021gvv},
but also adding some new aspects.

%---------------------------------------------------------------------------
\subsubsection{Heavy Neutral Lepton Fluxes}
\label{sec:fluxes}
%---------------------------------------------------------------------------

Our starting point is the Neutrino Portal operator
\begin{align}
  \mathcal{L} \supset y \, \bar{L} (i \sigma^2 H^*) N \,,
  \label{eq:L}
\end{align}
where $y$ is a dimensionless Yukawa coupling, $L$ is a Standard Model (SM)
lepton doublet field, $H$ is the Higgs doublet field, and $N$ is the HNL field.
For clarity, we have omitted flavour indices here, which are understood to be
carried by $L$ and $N$.  We remind the reader that Eq. \ref{eq:L} is the only
renormalizable operator through which SM particles can couple to a singlet
fermion, making it particularly interesting in the context of dark sector
searches.

After the Higgs field acquires its vacuum expectation value, Eq.~\ref{eq:L} leads
to mixing between active neutrinos and $N$, and this in turn implies that any
process that in the SM produces neutrinos can now also produce HNLs, as long as
the latter are kinematically accessible.  Compared to SM neutrino production,
HNL production is of course suppressed by a mixing angle.  In an accelerator
neutrino experiment, neutrinos are chiefly produced in charged meson decays,
mostly $\pi \to \nu \mu$, but also kaon an $D$ meson decays.

To model meson decays to HNLs, we start from the results of DUNE's beamline
simulation, which have been released in the form of Monte Carlo event files
\cite{DUNEfluxes}. We extract the properties of the neutrinos' parent mesons
from these files and then re-decay the mesons into final states containing
HNLs instead of SM neutrinos.  We thus obtain accurate predictions for HNL
fluxes from pion and kaon decays, which are most relevant for HNLs whose
mass is below the pion resp.\ kaon mass (minus a charged lepton mass).
For heavier HNLs (up to the $D$ meson mass), the most important production channel
in DUNE is $D_s$ decay, which is not included in DUNE's simulations as it
is irrelevant for oscilaltion physics.  We estimate HNL fluxes from $D_s$
decay using the {\sc NuShock} code~\cite{NuShock}, which has been released
together with ref.~\cite{Ballett:2019bgd} (see ref.~\cite{Breitbach:2021gvv}
for further details).

\begin{figure}
  \centering
  \hspace*{-1.5cm} \includegraphics[width=1.2\textwidth]{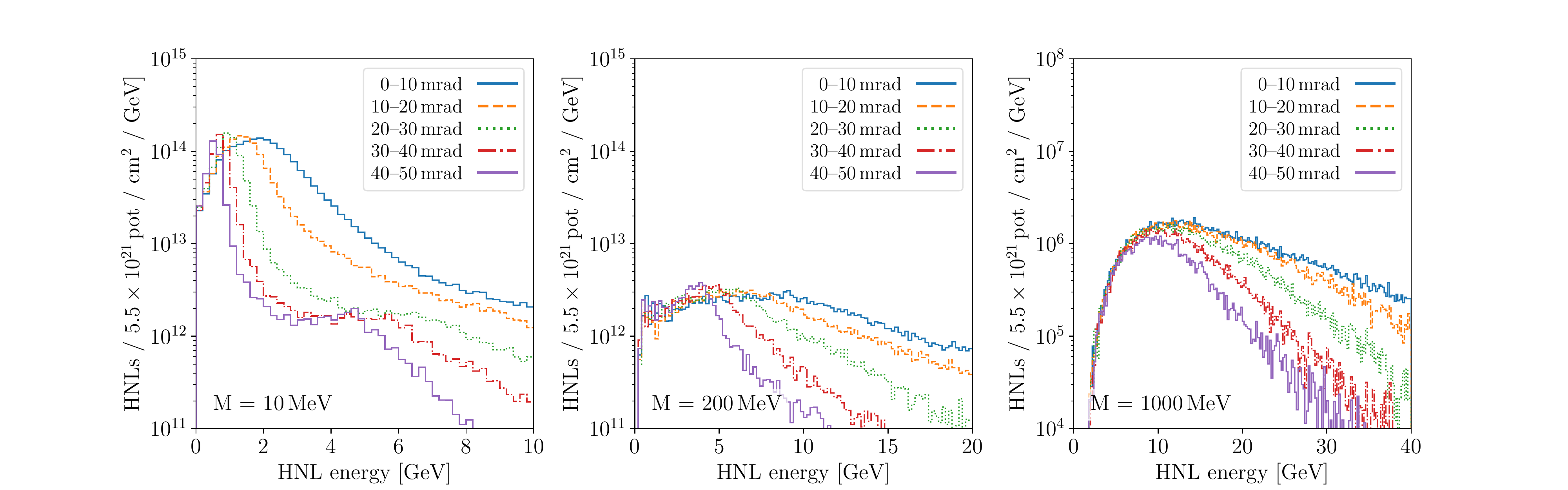}
  \caption{Predicted Heavy Neutral Lepton (HNL) fluxes at different on-axis
    and off-axis locations in the DUNE beam. Figure taken from
    ref.~\cite{Breitbach:2021gvv}.}
  \label{fig:hnl-flux}
\end{figure}

Our predictions for the HNL fluxes at DUNE are shown in Figure~\ref{fig:hnl-flux},
where the three panels correspond to different HNL masses, $M_N$: in the left
panel, $M_N \ll m_\pi$ has been assumed, so the flux is dominated by HNL
production in pion decays. In the middle panel, production in pion decays
is kinematically forbidden, but kaon decays are kinematically allowed.
The panel on the right, finally, considers a rather heavy HNL that can only
be produced in $D_s$ decays. Note the different scaling of the vertical axis
for this much reduced flux.  Comparing different off-axis locations (different
colored histograms), we observe a spectrum that gets narrower and softer
further away from the beam axis.  However, this softening is less
pronounced for heavier HNLs (note the different scale on the horizontal
axis in the three panels of Figure~\ref{fig:hnl-flux}), as can be understood
from the fact that heavier parent mesons are less boosted in the forward
direction. This suggests that looking for HNLs off-axis may be advantageous,
as the background due to SM neutrino interactions drops more rapidly
than the HNL signal. We will investigate this possibility in the following.

%---------------------------------------------------------------------------
\subsubsection{Heavy Neutral Lepton Sensitivity in DUNE}
\label{sec:dune-sensitivity}
%---------------------------------------------------------------------------

To further investigate the prospects for detecting HNLs in the DUNE
Near Detectors, we have set up a simulation that propagates the HNLs
(and SM background neutrinos) through DUNE's liquid argon and gaseous
argon near detectors and decaying them randomly.  Only fully leptonic decay
modes and decay modes involving pions are considered as these decays
have the largest branching ratios across the HNL mass spectrum
\cite{Ballett:2019bgd, Coloma:2020lgy, Breitbach:2021gvv, Abdullahi:2022jlv}.
SM backgrounds are simulated again using the {\sc NuShock} code~\cite{NuShock},
which takes as input a list of neutrino scattering events (from GENIE~v3.00.06~\cite{Andreopoulos:2009rq} in our case) and applies simple detector
simulation and event reconstruction methods to them. We conservatively
assume no charge identification capabilities.

\begin{figure}
  \centering
  \includegraphics[width=\textwidth]{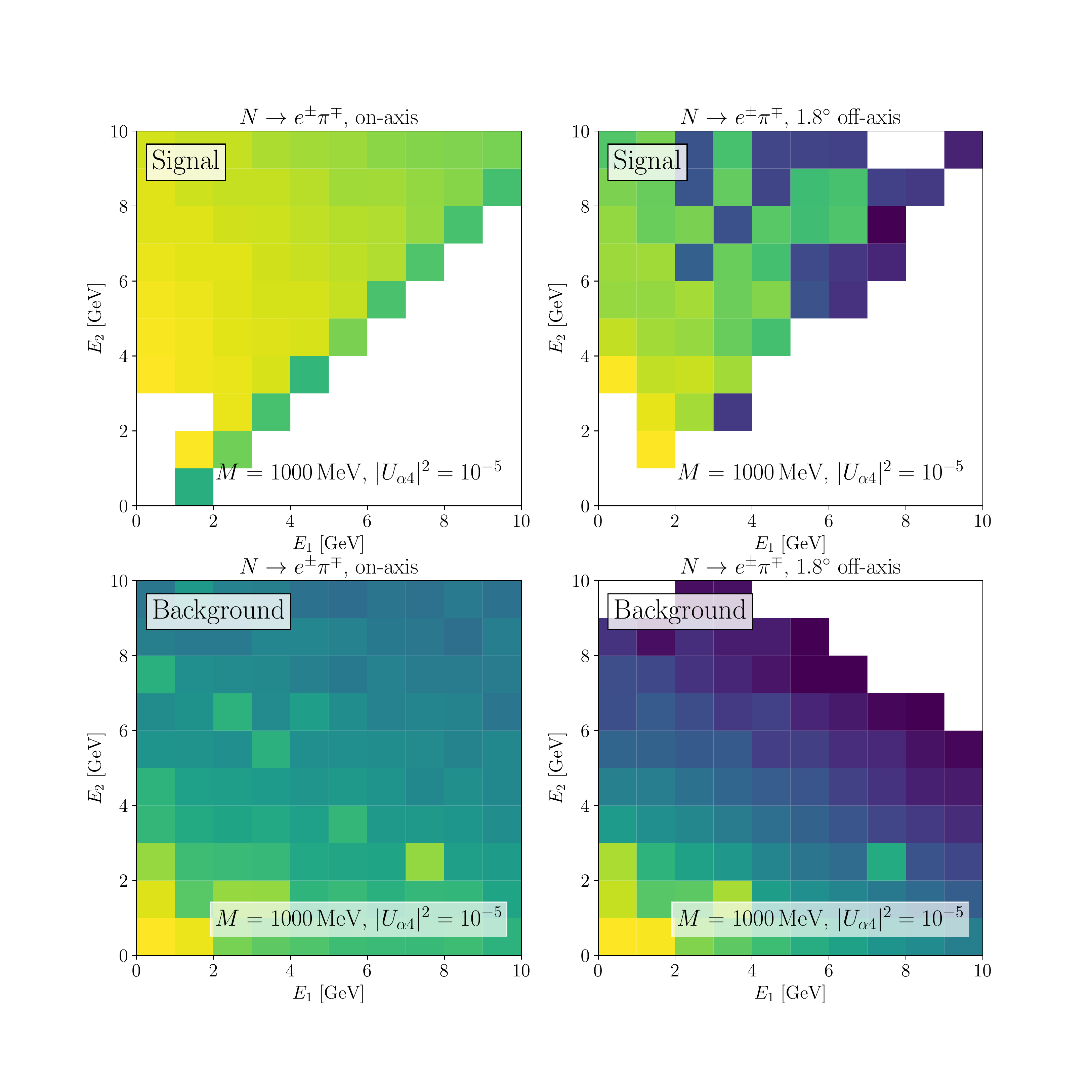}
  \caption{Two-dimensional distributions of signal events (top) and
    background events (bottom) in the plane spanned by the energies of the
    two HNL decay products. The large differences in the distributions
    help to suppress backgrounds.  Panels on the left are for an
    on-axis measurement, while those on the right are for off-axis
    operation of the detector. In the latter case, backgrounds are
    significantly reduced, but so is the signal.}
  \label{fig:hnl-2d-event-spectra}
\end{figure}

We then apply a cut on the angle $\theta$ between the mean direction of the two
HNL decay product candidates and the beam axis, requiring $\theta < M_N / (E_1
+ E_2)$, where $E_1$ and $E_2$ are the energies of the two particles
\cite{Ballett:2019bgd}.  This cut, which turns out to be quite effective in
reducing backgrounds, exploits the fact that HNL decays are typically
relatively forward, whereas neutrino interaction products are distributed more
isotropically due to the large mass of the recoil nucleus.  Events surviving
the cut are binned the ($E_1$, $E_2$) plane (see
Figure~\ref{fig:hnl-2d-event-spectra}), and a maximum likelihood analysis is carried
out on the resulting two-dimensional histograms. 

\begin{figure}
  \centering
  \includegraphics[width=\textwidth,clip=true,trim=0cm 0cm 0cm 32cm]{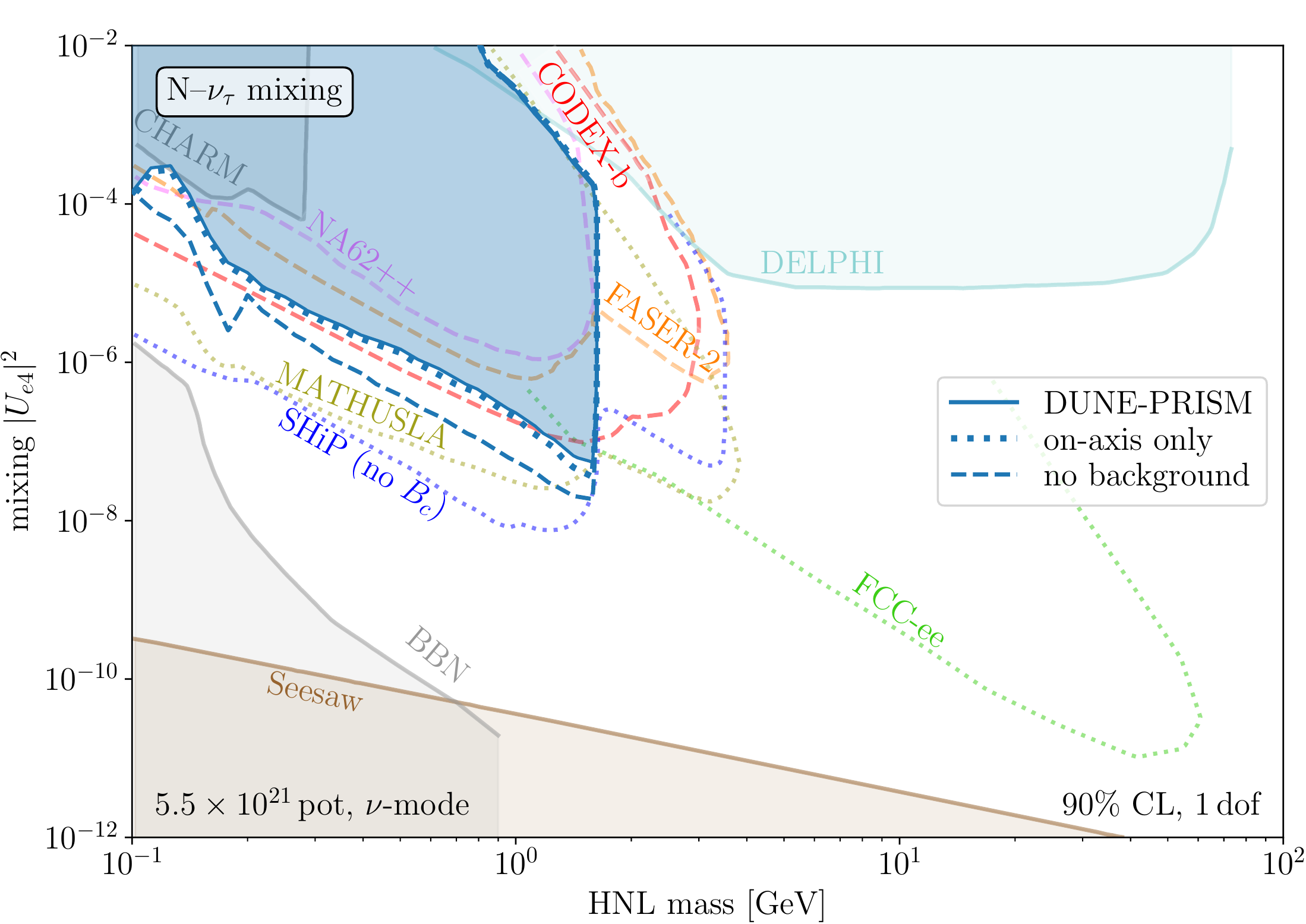}
  \caption{Existing and predicted future limits on HNLs mixing with $\nu_\tau$.
    Figure taken from ref.~\cite{Breitbach:2021gvv}.}
  \label{fig:sensitivity}
\end{figure}

This leads to sensitivity projections, some
of which we show in Figure~\ref{fig:sensitivity}.  In that plot, we focus
specifically on HNLs mixing with $\nu_\tau$, where we see that the DUNE
near detectors can cover significant unexplored parameter space.
Interestingly, we also observe that on-axis-only running (dotted blue curve) gives
sensitivities comparable to a running strategy that involves both
on-axis and off-axis running (solid blue curve).  This is because
the loss in signal and the loss in background balance each other when
going off-axis, given the efficient background suppression cuts
discussed above.

An update of these projections above with the most recent information on detector response included, and for the expected staging scenario of the DUNE near detectors was recently presented in 
\cite{Abdullahi:2022jlv}.

%---------------------------------------------------------------------------
\subsubsection{HyperKamiokande}
\label{sec:hk}
%---------------------------------------------------------------------------

Given the excellent prospects for HNL searches in the DUNE near detectors,
we can assume that also the near detectors in the current T2K and future T2HK
experiments should be very well suited for such searches. This is indeed
the case, though there are a number of differences which ultimately
put T2K and T2HK at a slight disadvantage compared to DUNE:
\begin{enumerate}
  \item Beam energy. The lower energy of the proton beam J-PARC (50\,{\rm GeV})
    compared to Fermilab's Main Injector beam (up to 120\,{\rm GeV}) implies
    that virtually no charm quarks are produced at J-PARC. As only charmed
    mesons (and bottom mesons) can decay to $\nu_\tau$, this means that
    HNL mixing with $\nu_\tau$ is inaccessible at J-PARC.

  \item No low-density detector. The current plans for the J-PARC Near Detector
    suite do not foresee a large gaseous detector like DUNE's ND-GAr. This
    means that HNL searches there cannot benefit from the better signal-to-background
    ratio offered by gaseous detectors in searches where the signal scales as
    the detector volume, whereas the background scales as the detector mass. However,
    as we have shown above, efficient background rejection can be possible in 
    HNL searches, alleviating this disadvantage.
\end{enumerate}

Initial sensitivity projections made by T2HK are presented in
\cite{Abdullahi:2022jlv}.

%---------------------------------------------------------------------------

% \bibliography{refs-jk}
% \bibliographystyle{JHEP}

% \end{document}

%-------------------------------------------
\subsection{Heavy Neutral Leptons at neutrino telescopes -- {\it P.~Coloma}}
\label{coloma}
{\it Author: Pilar Coloma, <pilar.coloma@ift.csic.es>}
%-------------------------------------------

%\usepackage{jheppub}

%\usepackage{mathrsfs}
%\usepackage{amsmath,amsthm,amssymb,slashed}
%\usepackage{color}
%\usepackage{multirow}
%\usepackage[normalem]{ulem}

%\usepackage{afterpage}

%\newcommand{\Dmq}{\Delta m^2}
%\newcommand{\be}{\begin{equation}}
%\newcommand{\ee}{\end{equation}}
%\newcommand{\bea}{\begin{eqnarray}}
%\newcommand{\eea}{\end{eqnarray}}
%\newcommand{\ba}{\begin{array}}
%\newcommand{\ea}{\end{array}}

%\allowdisplaybreaks

%\title{Heavy Neutral Leptons at neutrino telescopes}

%\author[a]{Pilar Coloma,}
%\affiliation[a]{Instituto de Física Teórica UAM/CSIC, Calle de
%  Nicolás Cabrera 13--15, Universidad Autónoma de Madrid,
%  Cantoblanco, E-28049 Madrid, Spain}
%\emailAdd{pilar.coloma@ift.csic.es}

%\subsubsection{Abstract}
%The addition of heavy neutral leptons to the Standard Model (SM) particle content offers perhaps the simplest possibility to generate the light neutrino masses, one of the main open problems in particle physics today. After being produced, such heavy neutrinos may decay back to SM particles, leaving a visible signal in neutrino detectors. This talk summarizes some of the main possibilities to search for such signals using neutrino telescopes and atmospheric neutrino detectors. 

\subsubsection{Introduction and motivation}

The addition of heavy neutrinos to the Standard Model (SM) particle content is well-motivated: they can generate the light neutrino masses through the well-known Seesaw mechanism, and open a  renormalizable portal to the hidden sector (the so-called \emph{neutrino portal}). The addition of a heavy neutrino introduces a new physics scale associated to its mass, which may even be related to the violation of the lepton number symmetry if neutrinos are Majorana fermions. In traditional type-I seesaw models~\cite{Minkowski:1977sc,Gell-Mann:1979vob,Yanagida:1979as,Mohapatra:1979ia}, the SM neutrino masses are inversely proportional to the Majorana mass, which is set to very high scales (at the GUT scale) in order to favor Yukawa couplings of order one. However, these models generate large corrections to the Higgs mass~\cite{Casas:2004gh}, and may only be tested indirectly since the new particles are too heavy to be produced in laboratory experiments. 

Alternatively, an appealing possibility is that the right-handed neutrinos are relatively light, and lie at the GeV scale or below. In the literature these neutrinos are often referred to as Heavy Neutral Leptons (HNLs). Low-scale seesaws explain the smallness of neutrino masses from the conservation of a global symmetry, lepton number~\cite{Mohapatra:1986bd,Bernabeu:1987gr,Branco:1988ex}. Besides being able to reproduce the observed pattern of light neutrino masses and mixing, these models could perhaps also solve some of the open problems of the SM such as the generation of the baryon asymmetry in the Universe or the dark matter abundance~\cite{Akhmedov:1998qx,Asaka:2005pn}. 

Thus, since a priori there is no strong argument to set the HNL mass to any given value it seems logical to try to explore the phenomenological consequences of HNL across a wide range of scales and experimental setups. For the purposes of this talk it is enough to consider one HNL with a mass in the MeV-GeV range, that mixes with the SM neutrinos. Then, a neutrino of flavor $\alpha$ can be written as a superposition of the four mass eigenstates:
\begin{equation}
 \nu_\alpha = \sum_i U_{\alpha i} \nu_i + U_{\alpha 4} N \, , \end{equation} 
where $N$ refer to the HNL, and $U$ is the full mixing matrix which parametrizes the change between the flavor and mass bases. Hereafter, in order to simplify our notation, we will use $U_\alpha \equiv U_{\alpha 4}$.

After being produced, such heavy neutrinos may decay back to SM particles, leaving a visible signal in neutrino detectors. This talk summarizes some of the main possibilities to search for such signals using neutrino telescopes and atmospheric neutrino detectors. 

\subsubsection{Searches for HNL decays in neutrino telescopes}

In the atmosphere HNLs can be directly produced from meson decays, assuming they are kinematically accessible, through their mixing with the SM neutrinos. Alternatively, they could be produced in the up-scattering of SM neutrinos as they interact with matter (for example, $\nu_\alpha \mathcal{N} \to N \mathcal{N}$ for an elastic interaction, where $\mathcal{N}$ is a nucleus present in the Earth). 

Under the assumption that the HNL interacts only through mixing with the active neutrinos, this implies that once it has been produced the HNL will eventually decay to SM particles (a combination of mesons and leptons). Also, it will be typically long-lived since the decay rates are proportional to the mixing. As an example, to a good approximation, the decay length (in the laboratory frame) for a HNL with mass $m_N = 1~\mathrm{GeV}$ reads:
\[
L_{lab, N} \simeq 30 \left(\frac{10^{-3}}{|U_{\tau }|^2} \right) \left( \frac{E_N}{10~\mathrm{GeV}}\right) ~\mathrm{m} \, ,
\]
where $E_N$ is the energy of the HNL, and we have set $U_{e } = U_{\mu }=0$ for concreteness. 

Non-minimal extensions of the SM also contemplate the possibility that the HNLs interacts with the SM particles through effective operators, or new interactions with additional mediators. A simple example consists in the addition of an effective ($d=5$) dipole operator $\mu_{\nu_\alpha} (\bar\nu_{L,\alpha} \sigma^{\mu\nu} N ) F_{\mu\nu} $,
where $\mu_{\nu_\alpha}$ is the transition dipole moment, $F_{\mu\nu}$ is the electromagnetic field strenght tensor and $\sigma^{\mu\nu} \equiv \frac{i}{2}\left[ \gamma^\mu, \gamma^\nu \right]$. 
For $m_N \sim 100~\mathrm{MeV}$, the associated decay length in this case is:
\[ L_{lab, N} \simeq 100 \left( \frac{10^{-8} \mu_B}{\mu_\nu} \right) \left( \frac{E_N}{10~\mathrm{GeV}}\right)~\mathrm{m} \, ,\] 
where $\mu_B$ is Bohr's magneton. Additional extensions considered in the literature include HNLs which feel a new interaction mediated by a relatively light $Z'$, which have been put forward in order to explain the MiniBooNE anomaly~\cite{Ballett:2018ynz,Bertuzzo:2018itn}.

It thus follows that, in both minimal and non-minimal scenarios, after being produced the HNLs may travel for tens or even hundreds of meters before decaying, leading to distinct signatures in neutrino detectors. In particular, this talk focuses on the possibilities to detect the decay products of HNLs inside atmospheric neutrino detectors and neutrino telescopes. Hereafter I will distinguish three types of searches for HNL depending on their production point, as outlined below, since they lead to different phenomenological consequences.

{\bf Production inside neutrino detectors -}

In the SM, tau neutrinos with ultra-high energies are expected to leave a characteristic signal in neutrino telescopes: a first energy deposition from a $\nu_\tau$ charged-current (CC) interaction in the detector, followed by a second energy deposition associated to the decay of the tau lepton produced~\cite{Learned:1994wg}. This signal is usually referred to as a \emph{double-bang}, a \emph{double-cascade} or a \emph{double-pulse} signal. 

As proposed in Ref.~\cite{Coloma:2017ppo}, a similar signal would be expected in certain BSM models and, in particular, in extensions of the SM with HNLs. In this case, since the HNLs are long-lived and the neutrino flux follows a steep power law with energy, the events are expected to take place with energies in the ballpark of tens of GeV (at higher energies the HNL would be too boosted, exiting the detector before decaying). Neutrino telescopes, such as Icecube, may be sensitive to such a signal provided that: (1) the first shower triggers the detector; (2) the two depositions take place sufficiently close to a photomultiplier so the light does not get absorbed by the ice before being detected; and (3) the two energy depositions are sufficiently separated so they can be distinguished from each other. 

The main advantage of searching for HNLs in this way resides in the availability of a large $\nu_\tau$ flux from atmospheric $\nu_\mu \to \nu_\tau$ oscillations, as opposed to laboratory experiments where tau neutrinos are difficult to produce. As a first estimate of the sensitivity to such a process, in Ref.~\cite{Coloma:2017ppo} we determined that at least one event would be observed at Icecube/Deepcore over a period of 6 years of data taking, for a HNL with $m_N \sim 1~\mathrm{GeV}$ and mixing $U_{\tau 4} \sim 2 \times 10^{-4}$.  In the neutrino magnetic moment scenario, under the same assumptions the corresponding sensitivity could potentially reach values of $\mu_{\nu_\tau} \sim 10^{-9}\mu_B$ for $m_N \sim 100~\mathrm{MeV}$, well below current constraints from laboratory experiments. The background from  coincidental atmospheric neutrino events was estimated to be negligible in Ref.~\cite{Coloma:2017ppo}. However, the Icecube collaboration is now studying the sensitivity to such a signal in depth, including a more detailed assessment of the expected backgrounds. 

\begin{figure}[ht!]
\begin{center}
  \includegraphics[width=0.97\textwidth]{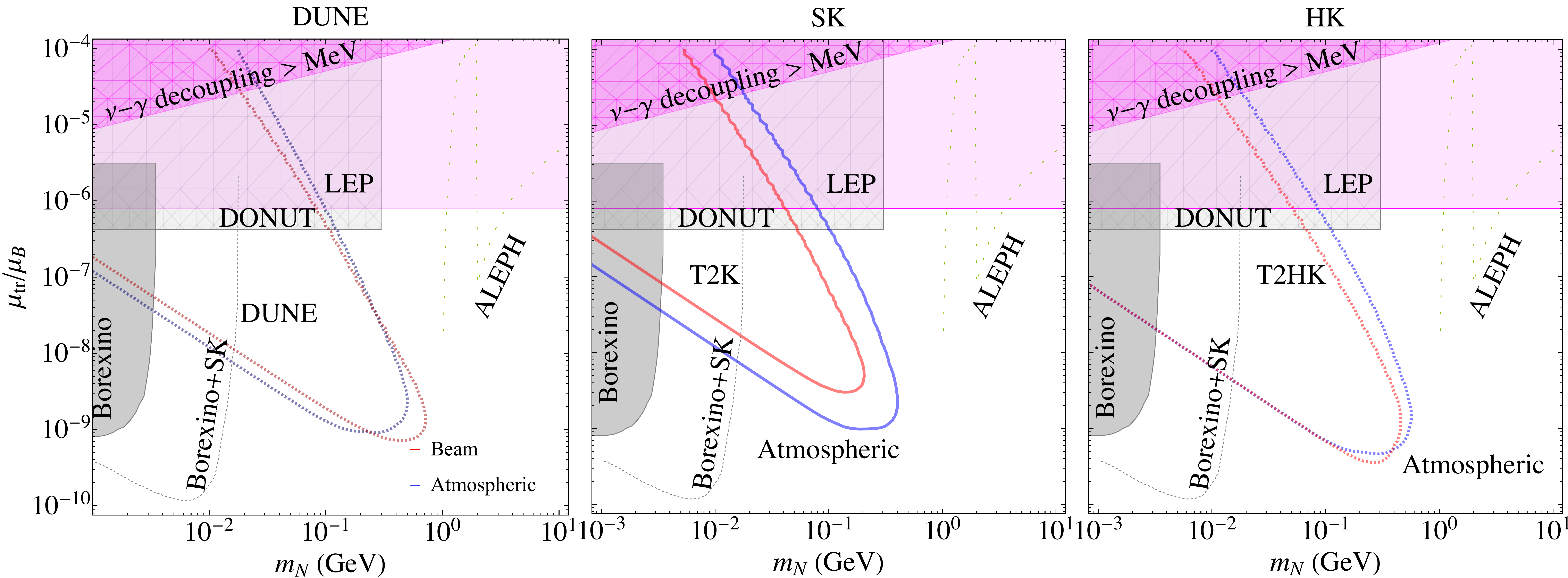}
\end{center}
\caption{\label{fig:MMsens} Sensitivity to a transition magnetic moment leading to the up-scattering $\nu_\tau \to N$ and subsequent decay of the HNL, for DUNE (left), SK (center) and HK (right), as a function of the HNL mass, $m_N$. Backgrounds are assumed to be negligible, see Ref.~\cite{Atkinson:2021rnp} for details. In each case the sensitivity achievable with atmospheric neutrino data (blue lines) and with beam data (right lines) is shown separately. Figure taken from Ref.~\cite{Atkinson:2021rnp}. }
\end{figure}

In a follow-up work~\cite{Atkinson:2021rnp} we recently considered the possibility to search for these events using atmospheric neutrino detectors, such as Super-Kamiokande (SK), its planned upgrade Hyper-Kamiokande (HK), and the DUNE experiment (which will also collect a large atmospheric neutrino sample). These detectors offer several advantages with respect to Icecube/DeepCore: (i) their much better spatial resolution allows them to observe events with smaller separation between the two cascades; (ii) they are sensitive to lower energies; (iii) they will be exposed to a long-baseline beam, which constitutes an additional (and controlled) neutrino source; and (iv) near detectors would also be available. We found that better sensitivities would be expected compared to those attainable at Icecube/DeepCore, in particular for the dipole scenario where values as low as $\mu_{\nu_\tau} \sim 5\times 10^{-10}\mu_B$ could be probed, as shown in Fig.~\ref{fig:MMsens}. The study in Ref.~\cite{Schwetz:2020xra} also found similar conclusions, although in that case only events produced by the long-baseline beam were considered.

%%%%%%%%%%%%%%%%%%%%%
%%%%%%%%%%%%%%%%%%%%%

{\bf Production through up-scattering inside the Earth -}

As the HNL becomes more long-lived it becomes difficult to observe a double-cascade signal, since one of the two showers will tend to take place outside the detector with a large probability. In this case, however, it is still possible to gain sensitivity to HNLs produced from neutrino up-scattering inside the Earth. This idea takes advantage of the large volume of the Earth surrounding the detector to enhance the production rate, while the visible signal would be the energy deposited from the decay of the HNL inside the detector. Of course, in this case the main background would come from standard neutrino interactions. However, it was shown in Refs.~\cite{Plestid:2020vqf,Plestid:2020ssy} that the event rates from solar neutrino up-scattering would be large enough to lead to new limits using past data from Borexino and SK, for masses in the range around $m_N \sim 1-10~\mathrm{MeV}$ in the dipole scenario, and for masses in the range $m_N \sim  10-20 ~\mathrm{MeV}$ in the mixing scenario. 

The same main idea has been explored further in the literature, using neutrinos from different sources. The authors of Ref.~\cite{Gustafson:2022rsz} studied the same process, but using the flux of atmospheric neutrinos instead: although it is less intense than the solar neutrino flux, it allows to produce heavier HNLs since it reaches higher neutrino energies. Similarly, the authors of Ref.~\cite{Schwetz:2020xra} also considered the same type of signal at DUNE, produced in this case by the long-baseline neutrino beam. Finally, in Ref.~\cite{Huang:2022pce} a similar idea was explored but using instead the flux of ultra-high energy astrophysical neutrinos. In this case a signal could be searched using future $\nu_\tau$ telescopes such as GRAND, POEMMA and Trinity. The authors found that new regions of parameter space could be probed for HNL masses of $\sim 30~\mathrm{TeV}$, well beyond the reach of present colliders.

{\bf Production in the upper layers of the atmosphere -}

A third possibility, studied in Refs.~\cite{Arguelles:2019ziu,Coloma:2019htx,Kusenko:2004qc,Asaka:2012hc,Masip:2014xna} is to consider the production of HNL in the upper layers of the atmosphere. The idea is to take advantage of the large center-of-mass energy available in the collision of cosmic rays, which allows to copiously produce not only light mesons such as kaons and pions but also heavier ones (such as $D$ and $D_s$) and tau leptons, which would allow to produce HNLs in the MeV-GeV range as a byproduct of their decays. If sufficiently boosted, the HNLs could then reach neutrino telescopes and decay inside, leaving a visible signal. 

In Ref.~\cite{Arguelles:2019ziu} we used Icecube~\cite{IceCube:2014rwe} and SK~\cite{Super-Kamiokande:2017yvm} publicly available data to derive new bounds on HNLs. Instead of focusing on a particular model, we produced model-independent limits on their production branching ratio (BR) as a function of their lifetime ($c\tau$), for HNLs produced from heavy meson and $\tau$ decays. Our analysis, performed with publicly available data, indicated that production branching ratios of the order of $10^{-4}$ could be probed at the 90\% CL, for values of the HNL lifetime around $c\tau \sim \mathcal{O}(0.01)~\mathrm{km}$ for Icecube and $c\tau \sim \mathcal{O}(1)~\mathrm{km}$ for SK. Unfortunately, these limits fall short to probe the minimal scenario where the HNL is produced and decays through mixing. The reason is that in the minimal scenario the two are correlated and, in order to reach the optimal values of the lifetime, the production BR would be lower than $10^{-4}$. However, our model-independent constraints may be easily recasted to other non-minimal scenarios. 

\begin{figure}[ht!]
\begin{center}
  \includegraphics[width=0.48\textwidth]{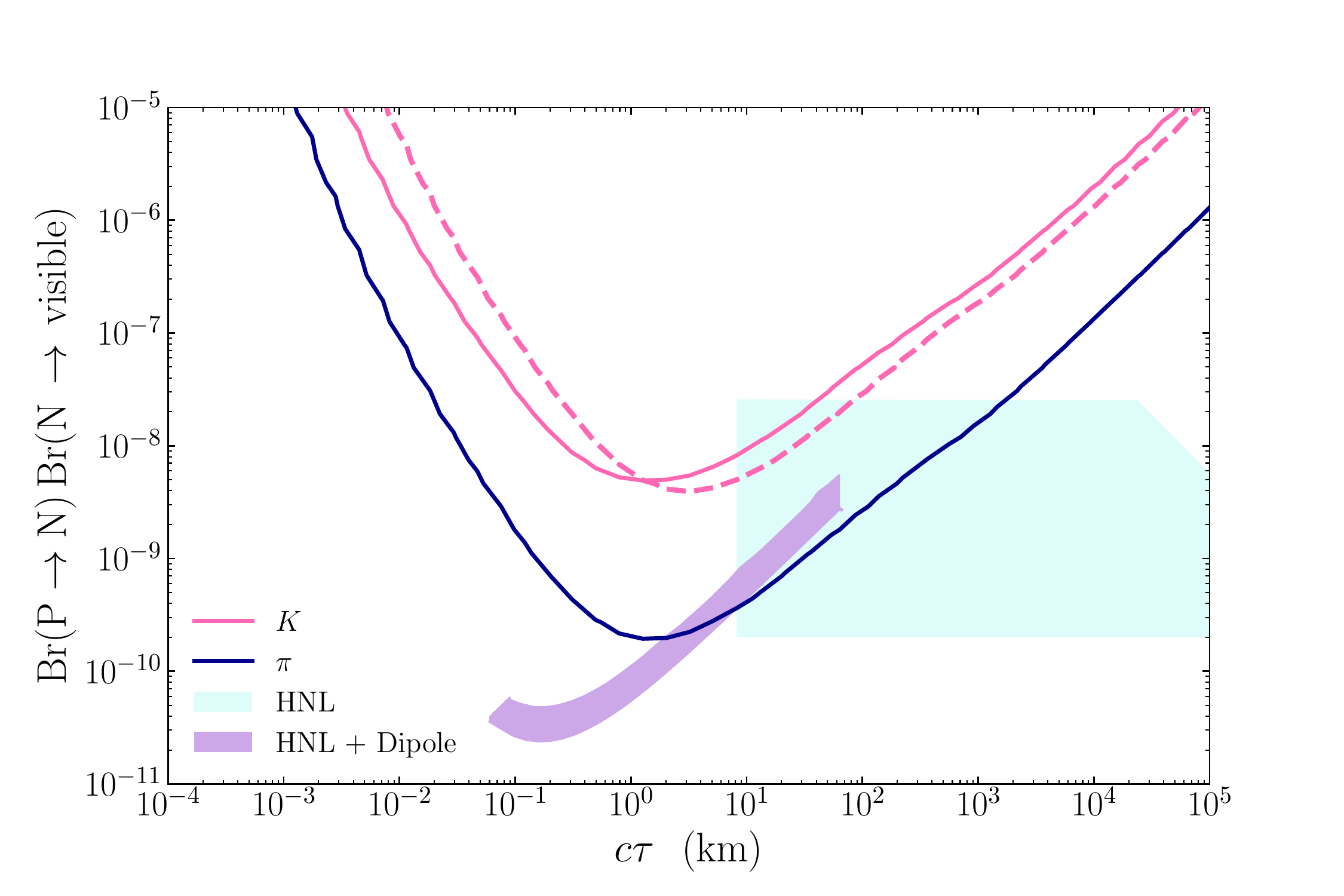}
  \includegraphics[width=0.44\textwidth]{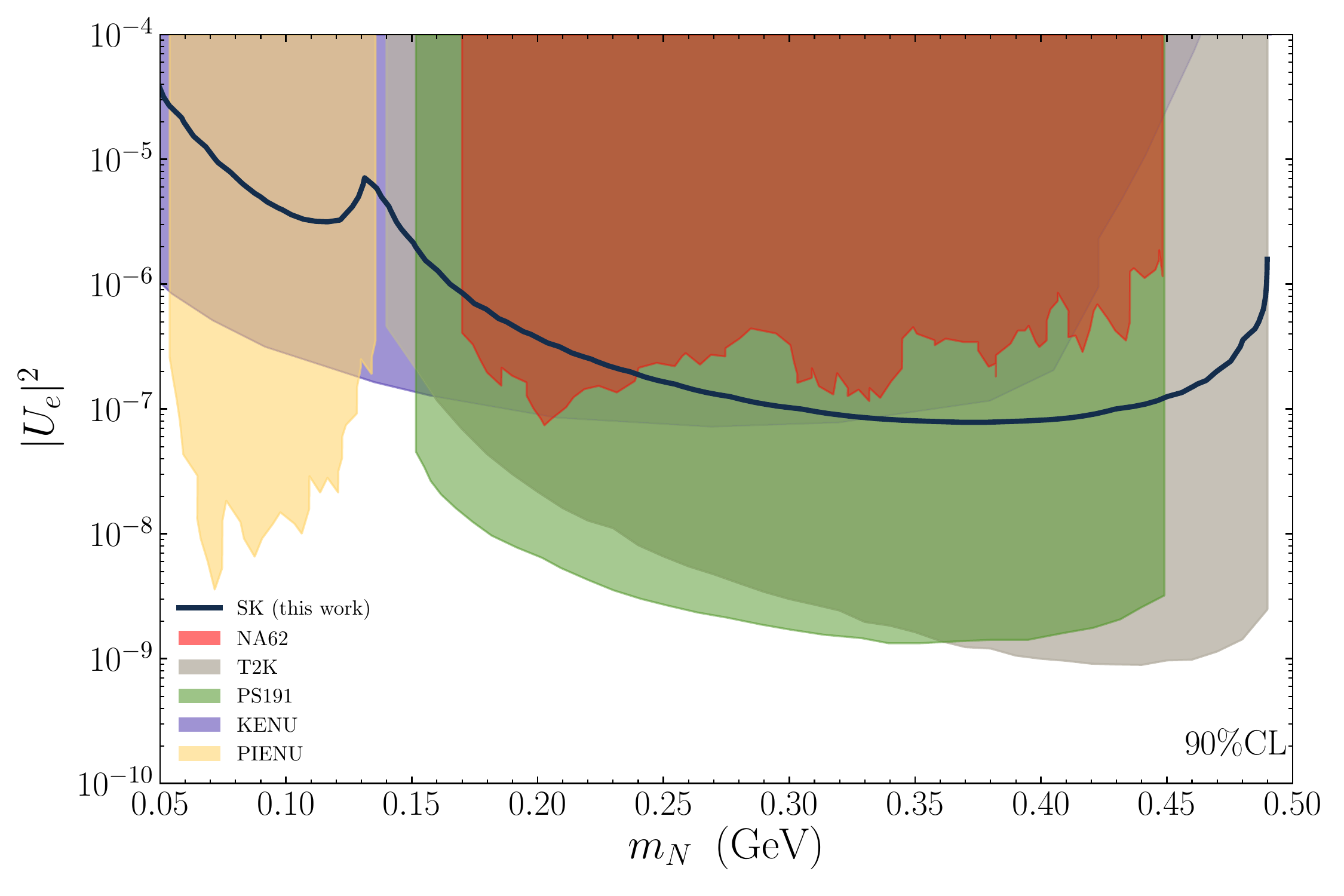}
\end{center}
\caption{\label{fig:HNLatmos} Left: our constraints on the production branching ratio from a reanalysis of SK atmospheric data, for the model-independent scenario where the production branching ratio and HNL lifetime are assumed to be uncorrelated. Each line assumes the HNL is produced from a given parent meson, as indicated. The solid lines correspond to $m_N=100~\mathrm{MeV}$ while the dashed line has been obtained for $m_N=250~\mathrm{MeV}$. Right: same limit, but derived instead for the minimal HNL scenario when both production and decay take place through its mixing with electron neutrinos ($U_{e}$). Figure adapted from Ref.~\cite{Coloma:2019htx}, see text for details.}
\end{figure}

In a follow-up work~\cite{Coloma:2019htx} we considered the production from lighter mesons instead (kaons and pions), for which the produced fluxes in the atmosphere are much higher. Using SK atmospheric neutrino data we are able to set a tight model-independent limit for the production branching ratio, shown in the left panel in Fig.~\ref{fig:HNLatmos}. In particular we obtain $\mathrm{BR} < 2\times 10^{-10}$ ($\mathrm{BR} < 4\times 10^{-9}$) at 90\% CL for $m_N = 100~\mathrm{MeV}$ ($m_N=250~\mathrm{MeV}$) and optimal values of its lifetime, $c\tau \sim \mathcal{O}(\mathrm{1-3})~\mathrm{km}$. In this case we show separately the sensitivities obtained assuming that the HNL is produced from kaon decays (pink lines) or from pion decays (blue line). The light blue shaded region shows the allowed parameter space for the minimal scenario when the HNL only interacts through mixing, for a HNL with mass $m_N = 250~\mathrm{MeV}$ (see Ref.~\cite{Coloma:2019htx} for details), while the shaded purple region shows the region where a HNL with a dipole interaction may be able to explain the MiniBooNE anomaly~\cite{Fischer:2019fbw}. The excellent sensitivity obtained in our analysis translates into very good limits also for the minimal scenario, as indicated by the partial overlap between the shaded blue region and the pink lines. In particular, we find $|U_{\alpha}|^2 < \mathcal{O}(10^{-7})$ ($\alpha=e,\mu$) for masses in the range $m_N \sim 300-400~\mathrm{MeV}$, at 90\% CL, as shown in the right panel of Fig.~\ref{fig:HNLatmos} for $\alpha = e$.

%%%%%%%%%%%%%%%%%%%%%
%%%%%%%%%%%%%%%%%%%%%

\subsubsection{Summary}

To summarize, in this talk I have given an overview of different possibilities to search for HNLs in neutrino telescopes and atmospheric neutrino detectors. For concreteness, I have focused on the case where only one HNL is produced, considering its visible decays to SM particles. I have discussed three types of searches, depending on the production point of the HNL: (i) searches for HNLs produced inside the detector, leading to double-cascade signals; (ii) searches for HNLs produced from upscattering of light neutrinos inside the Earth, which subsequently decay inside neutrino detectors; and (iii) searches for HNLs produced in the upper layers of the atmosphere, which then decay inside neutrino telescopes or atmospheric neutrino detectors. In most cases I have presented sensitivities to the minimal scenario where the HNL interacts only through mixing with the light neutrinos, as well as for the case where it interacts through an effective dipole operator. For the analyses performed in Refs.~\cite{Arguelles:2019ziu,Coloma:2019htx}, our limits are also provided as a function of the production branching ratio and the lifetime of the HNL, which are model independent and can be easily recasted to non-minimal scenarios. 

All analyses presented in this talk have been performed outside the experimental collaborations, with (in some cases rather limimted) publicly available information. Therefore, there is still plenty of room for improvement, which may lead to a better sensitivity in some of the scenarios considered. As an example, for the studies performed in Refs.~\cite{Arguelles:2019ziu,Coloma:2019htx} the inclusion of a two-dimensional binning in the data (in both zenith angle and energy) is expected to lead to significanlty better results, as explained in detail in Ref.~\cite{Arguelles:2019ziu}. Finally, it should be stressed that while I have focused on the HNL scenario, the methodology developed for these studies may also be applicable to searches for other types of long-lived particles. 

%%%%%%%%%%%%%%%%%%%%%%%%%%%%
 
%\subsubsection*{Acknowledgments} 
 
 %This work has received support from the IFT Centro de Excelencia Severo Ochoa Grant No.~CEX2020-001007-S and by Grant PID2019-108892RB-I00 (funded by \newline MCIN/AEI/10.13039/501100011033), and from Grant RYC2018-024240-I (funded by \newline MCIN/AEI/10.13039/501100011033 and by ``ESF Investing in your future''). This work has also received partial funding/support from the European Union’s Horizon 2020 research and innovation program under the Marie Sklodowska-Curie grant agreement No. 860881-HIDDeN.

%\bibliographystyle{JHEP}
%\bibliography{references}

%\end{document}

%-------------------------------------------
\subsection{PIONEER at PSI Prospects for Rare Pion Decays -- {\it B.~Velghe}}
\label{Bob_Velghe}
{\it Author: Bob Velghe, < bvelghe@triumf.ca >}
Experimental hints pointing to a possible violation of lepton flavour universality (LFU) and a discrepancy from unitarity of the quark mixing matrix (CKM) are stacking up~\cite{Fischer2022,Bryman2022}.
However, present uncertainties on key measurements are limiting the reach of related new physics searches.

Specifically, the ratio $R_{e/\mu} = \Gamma(\pi^+ \to e^+ \nu (\gamma))/\Gamma(\pi^+ \to \mu^+ \nu (\gamma))$ offers the most stringent test of electron-muon universality, probing new pseudoscalar couplings up to the PeV scale~\cite{Bryman2011}.
In the Standard Model (SM), $R_{e/\mu}$ has been computed to an exceptional precision of 0.01\%~\cite{Cirigliano2007}, contrasting with the current experimental value, $R_{e/\mu}^\mathrm{Exp.} = 1.2327(23) \times 10^{-4}$, which has relative uncertainty of 0.2\%~\cite{PDG2022}.

In its first phase of operation, the new PIONEER~\cite{PIONEER2022a} experiment aims to match the $R_{e/\mu}^\mathrm{Exp.}$ uncertainty to that of the SM prediction.
The experiment has been approved at the Paul Scherrer Institut (PSI), where low energy and high intensity pion beams are available. 
In a second phase, PIONEER will measure the $\pi^+ \to \pi^0 e^+ \nu$ decay branching fraction, which allows for the theoretically cleanest extraction of the CKM matrix element $\left|V_{ud}\right|$.
As we will outline, the setup also allows for direct and indirect searches for feebly-interacting particles (FIPs) with masses below the pion mass.

\begin{figure}
    \includegraphics[width=\textwidth]{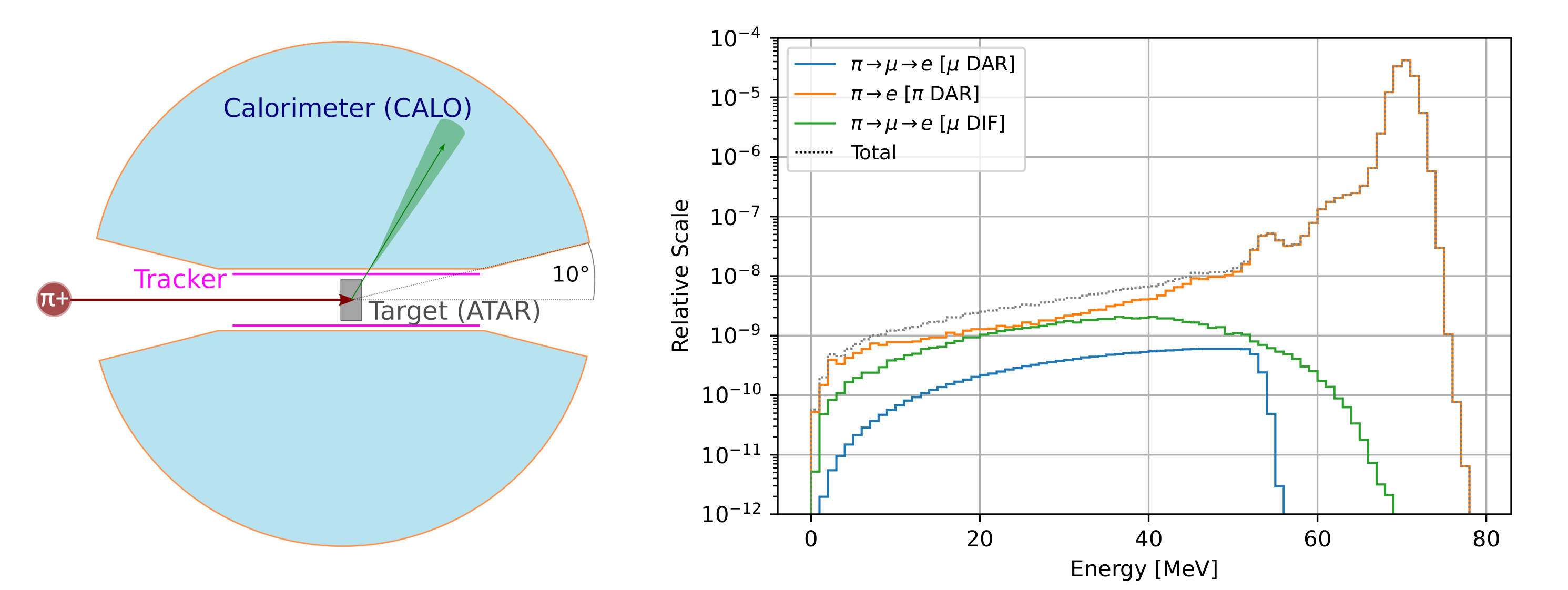}
    \caption{Sketch of the apparatus (left). Typical calorimeter energy spectrum for $\pi^+ \to e^+$ and $\pi^+ \to \mu^+ \to e^+$ processes, after background suppression. Note $\pi^+ \to e^+$ the low energy tail (right). DAR refers to decay-at-rest and DIF stands for decay-in-flight.}
    \label{fig:exp-setup}
\end{figure}
The PIONEER experiment, shown in Fig. \ref{fig:exp-setup}, will significantly improve on the techniques developed by the PIENU~\cite{AguilarArevalo2015} and PEN~\cite{Pocanic2014} collaborations. 
Incoming beam pions are stopped in a 6~mm thick segmented active target (ATAR)~\cite{Mazza2021}. Muons originating from the predominant $\pi^+ \to \mu^+ \nu$ decay have a range of about 1~mm and thus cannot escape the target volume.
The energy of the outgoing positrons is measured with an uncertainty of less than 2\% by a 25 radiation length deep calorimeter covering a $3 \pi$~sr solid angle. 
A cylindrical tracker inserted between the ATAR and the CALO helps identify events with a positron entering the CALO fiducial volume.
The simulated background-suppressed positron energy spectrum of the prompt $\pi^+ \to e^+$ and delayed $\pi^+ \to \mu^+ \to e^+$ events is illustrated in Fig. \ref{fig:exp-setup} (right).

\begin{figure}
    \includegraphics[width=\textwidth]{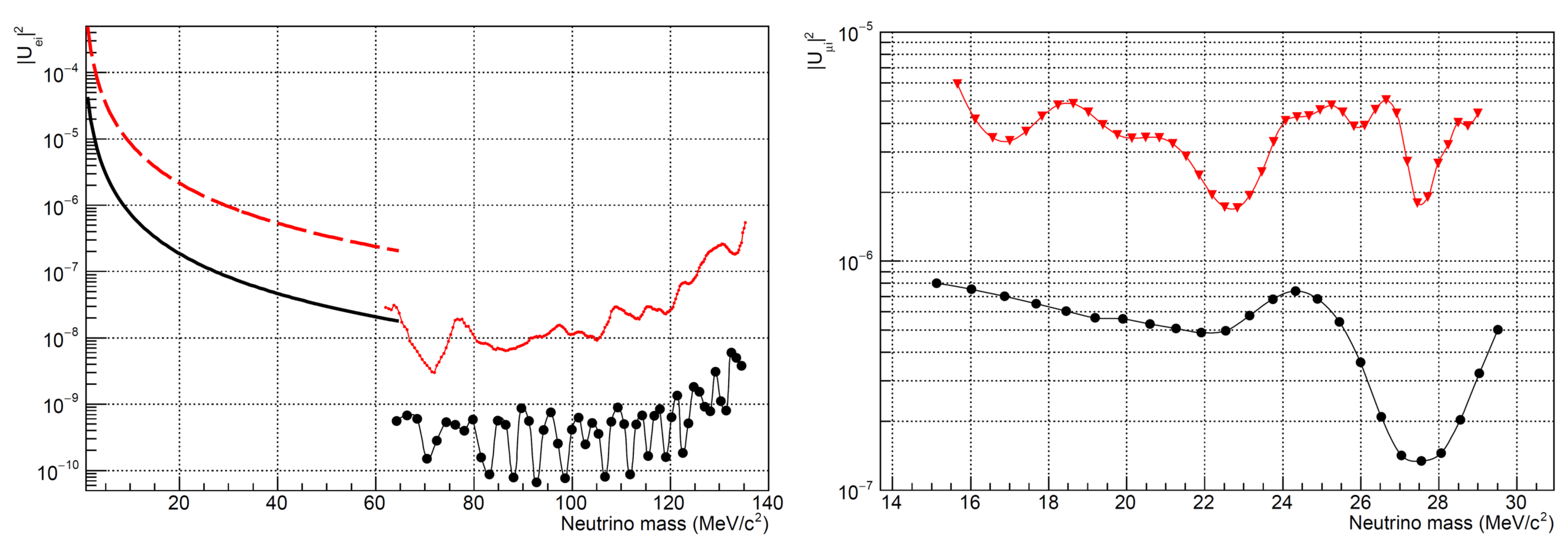}
    \caption{PIENU exclusion limits (red) and PIONEER projections (black) for the mixing matrix elements $\left|U_{e4}\right|$ (left) and $\left|U_{\mu4}\right|$ (right). The dashed bound is extracted from $R_{e/\mu}^\mathrm{Exp.}$.}
    \label{fig:pioneer_proj}
\end{figure}
PIENU searched for the  $\pi^+ \to e^+ \nu_H$ decay by looking for peaks in the background-suppressed positron energy spectrum~\cite{Aoki2011,AguilarArevalo2018}. 
This peak search method does not give access to the heavy neutral lepton (HNL) mass region below 60~MeV/$c^2$, yet, a bound can be derived from $R_{e/\mu}^\mathrm{Exp.}$~\cite{Bryman2019a,Bryman2019b}. 
The corresponding limits on $\left|U_{e4}\right|$ are shown in Fig. \ref{fig:pioneer_proj} (left)
where $\left|U_{l4}\right|$ ($l = e, \mu$) quantifies the mixing between a hypothetical HNL $\nu_H$ and the electron or muon neutrino, respectively.
A search for $\pi^+ \to \mu^+ \nu_H$ was also performed using the same technique on the energy deposited in the active target by decay-at-rest muons~\cite{AguilarArevalo2019}; the limits are reported in Fig. \ref{fig:pioneer_proj} (right).
PIONEER will improve those bounds by an order of magnitude.

Rare pion decays can also put constraints on axion-like particles~\cite{Wilczek1982,Altmannshofer2020,AguilarArevalo2020},
%Charged lepton flavour violating decay $\mu^+ \to e^+ X_H$, where $X_H$ is a massive neutral boson. 
and other exotic three body decays such as $\pi^+ \to l^+ \nu X$, where $X$ is a weakly interacting particle~\cite{AguilarArevalo2021}.
%Possible footnote:
%\footnote{
%The overall landscape of present and expected exclusion limits on HNLs couplings to the first and second lepton generation is nicely summarized in Fig. 26--27 of Ref.~\cite{Abdullahi2022}. 
%However, in those plots, the limit derived from primordial Big Bang Nucleosynthesis (BBN) constraints depend on model assumptions, notably on the HNLs lifetime~\cite{Boyarsky2021}.}

% \printbibliography

% \end{document}

%-------------------------------------------
\subsection{New results from Babar -- {\it S.~Middleton}}
\label{middleton}
{\it Author: Sophie Charlotte Middleton, <smidd@caltech.edu>}
%-------------------------------------------

%\mg{Note that I had to Change to BaBar since I was getting an error with the macro defined in the original contribution.}

%******************
\noindent \subsubsection{Abstract}
A model-independent search for a mostly sterile Heavy Neutral Lepton (HNL), capable of mixing with the Standard Model $\nu_{\tau}$ neutrino, was presented. A total of 424 $fb^{-1}$ of BaBar~ data has been analyzed. No significant signal is seen and 95 $\%$ confidence level upper limits are set on the extended Pontecorvo–Maki–Nakagawa–Sakata (PMNS) matrix element, $|U_{\tau 4}|^{2}$. The limits vary from $2.31 \times 10^{-2}$ to  $5.04 \times 10^{-6}$, with all uncertainties considered, across the mass range $100 < m_{4} < 1300$ MeV/$c^{2}$. More stringent limits are placed on higher masses. The technique employed uses only kinematic information and makes no assumptions about the model behind the origins of the HNL, its lifetime, or decay modes. This novel approach can be applied to future experimental searches.

\subsubsection{Introduction}
The latest search for a heavy neutral lepton (HNL) at BaBar~ is presented in Ref.~\cite{BaBar:2022cqj}. New limits on the square of the extended Pontecorvo Maki Nakagawa Sakata (PMNS) matrix element describing mixing between the $\tau$ sector and a hypothetical 4th neutrino mass state, $|U_{\tau4}|^{2}$, are found in the range 100 $<m_{4}<$ 1300 MeV/$c^{2}$.  The data sample used corresponds to an integrated luminosity of 424 fb$^{-1}$.  \medskip 

\noindent \textbf{Experimental Strategy -}
% \label{exp_strat}
An overview of the ~BaBar~ detector can be found in Ref.~\cite{BaBar:2013byz}. The analysis presented explores the scenario where a HNL can interact with the $\tau$ via charged-current weak interactions. The key principle is that if the decay products of the $\tau$ have recoiled against a HNL, the kinematics of the outgoing visible decay products would be modified with respect to Standard Model (SM) $\tau$ decay. In this analysis, it is assumed that the HNL does not decay within the detector.

The 3-prong, pionic, $\tau$ decay is used, giving access to the region 300$<m_{4}<$1360 MeV$/c^{2}$, which, historically, has weaker constraints. Denoting the three charged pions as a hadronic system $h^{-}$, the decay can be considered a two-bodied:

\begin{equation}  
\tau^{-} \rightarrow h^{-} (E_{h}, \vec{p}_{h}) + \nu  (E_{\nu}, \vec{p}_{\nu}), 
\end{equation}
 where $\nu$ describes the outgoing neutrino state. The allowed phase space of the reconstructed energy, $E_{h}$, and invariant mass, $m_{h}$, of the hadronic system varies as a function of the HNL mass. The heavier the HNL the smaller the fraction of the incoming $\tau$-lepton's energy measured in the visible pion system.

In the center-of-mass frame the $\tau$-lepton energy is $\sqrt{s}/2$. $E_{h}$ must then fall between two extremes:

\begin{equation} E_{\tau} - \sqrt{m^{2}_{4} + q_{+}^{2}} < E_{h} < E_{\tau} - \sqrt{m^{2}_{4} + q_{-}^{2}}, \end{equation}

where %begin{widetext}
\begin{equation}
q_{\pm} = \frac{m_{\tau}}{2} \bigg ( \frac{m_{h}^{2} - m_{\tau}^{2} - m_{4}^{2}}{m_{\tau}^{2}} \bigg ) \sqrt{\frac{E_{\tau}^{2}}{m_{\tau}^{2}} - 1} \pm \frac{E_{\tau}}{2} \sqrt{\big ( 1- \frac{(m_{h}+ m_{4})^{2}}{m_{\tau}^{2}}\big ) \big ( 1- \frac{(m_{h} - m_{4})^{2}}{m_{\tau}^{2}} \big )};
\end{equation}
%\end{widetext}
and $3m_{\pi^{\pm}} < m_{h} < m_{\tau} - m_{4}$. The analysis proceeds by constructing template histograms in the ($E_{h}$, $m_{h}$) plane for signal and all possible background processes. As the HNL mass increases, the allowed phase space of the visible system is reduced in the $E_{h}$, $m_{h}$ plane. Only channels in which the non-signal (tag) $\tau$ decays leptonically are used in this analysis since these provide a cleaner environment.  \medskip 

\noindent \textbf{Signal and Background Simulations -}
All SM background contributions are estimated from Monte Carlo (MC) simulations. Generated processes are passed through a GEANT4 model of the BaBar~ detector and the same reconstruction and digitization routines are used as with the data. 

Events emanating from $\tau$-pairs are simulated using the KK2F \cite{kk2f} generator and TAUOLA \cite{tauola} which uses the averaged experimentally measured $\tau$ branching rates \cite{PDG2020}. Several non-$\tau$ backgrounds must also be understood. These include $e^{+}e^{-} \rightarrow \Upsilon(4S) \rightarrow B^{+} B^{-}$ and $B^{0}$ $\bar{B}^{0}$)  which are simulated using EvtGen \cite{Lange:2001uf}; $e^{+}e^{-} \rightarrow  q\bar{q}$ which are simulated using JETSET \cite{Sjostrand:1985ys} \cite{Sjostrand:1986hx} and  $e^{+}e^{-} \rightarrow \mu^{+} \mu^{-} (\gamma)$ which are simulated using KK2F. In total, 26 signal samples were produced using KK2F and TAUOLA, one for each of the HNL masses across the range 100 MeV/$c^{2}$ $< m_{4} <$ 1300 MeV/$c^{2}$, at 100 MeV/$c^{2}$~ increments. For each of these HNL masses, both a $\tau^{+}$ and $\tau^{-}$ signal channel was simulated.  \medskip 

\noindent \textbf{Analysis Procedure -}
The contents of a given bin, $i,j$, in the $(m_{h},E_{h})$ data histogram are assumed to be from a Poisson distribution. Its assumed that a bin may contain events emanating from any of the SM background processes, and potentially HNL signal events.  The likelihood to observe the selected candidates in all the $(m_{h}, E_{h})$  bins is then a product of the Poisson probability to observe events in each bin:
%----er

%\begin{widetext}

\begin{align*}
\mathcal{L} =   \prod_{\text{charge}}^{+ -} \bigg ( \prod_{\text{channel}}^{e \mu } \bigg (  \prod^{ij}_{\text{bin}}  \bigg ( \frac{1}{n_{\text{obs},ij}!}\bigg[ N_{\tau,\text{gen}}\cdot |U_{\tau 4}|^{2} \cdot p_{\text{\tiny HNL}, ij}+N_{\tau,\text{gen}}\cdot (1-|U_{\tau 4}|^{2}) \cdot p_{\tau-\text{SM}, ij} + n^{\text{reco}}_{BKG,ij}\bigg]^{(n_{\text{obs}})_{ij}} \times
\end{align*}
\begin{equation}
exp\bigg[-(N_{\tau,\text{gen}}\cdot |U_{\tau 4}|^{2} \cdot p_{HNL, ij} + N_{\tau,\text{gen}}\cdot (1-|U_{\tau 4}|^{2}) \cdot p_{\tau-SM, ij}+ n^{\text{reco}}_{BKG,ij}) \bigg] \bigg )_{\text{bin}}
\times  \prod_{k} f(\theta_{k},\tilde{\theta}_{k}) \bigg )_{\text{channel}} \bigg )_{\text{charge}},
\label{eq:lik}
\end{equation}
%\end{widetext}
where $n_{\text{obs}}$ is the number of observed events in the bin $ij$, $N_{\tau,\text{gen}}$ is the number of generated $\tau$'s, $p_{HNL (\tau-SM), ij}$ is the probability of a reconstructed event being in a given bin in the HNL ($\tau-SM$) 2D template and $n^{\text{reco}}_{BKG,ij}$ is the expected number of non-$\tau$ background events. The final product is a set of Gaussian nuisance parameters. The expression involves a product over all bins, $ij$, over the two 1-prong channels, and over both $\tau$-lepton charges ($\pm$). 

A test statistic, $q$, is defined:

\begin{equation}
    q = -2 \text{ln} \bigg (  \frac{\mathcal{L}_{H_{0}}(|U_{\tau 4}|_{0}^{2};\hat{\hat{\theta}}_{0},\text{data})}{\mathcal{L}_{H_{1}}(|\hat{U}_{\tau 4}|^{2};\hat{\theta},\text{data}) } \bigg ) = -2\text{ln}(\Delta \mathcal{L}),
\end{equation}
where $\mathcal{L}$ in the numerator and denominator describes the maximized likelihood for two instances. The denominator is the maximized (unconditional) likelihood giving the maximum likelihood estimator of $|U_{\tau 4}|^{2}$ and the set of nuisance parameters ($\hat{\theta}$); $\hat{\theta}$ is a vector of nuisance parameters that maximize the likelihood. In the numerator, the nuisance parameters are maximized for a given value of $|U_{\tau 4}|^{2}$. The analysis aims to find the value of $|U_{\tau 4}|^{2} $ that minimizes this quantity at the 95 $\%$ confidence level.

\subsubsection{Uncertainties}

Gaussian nuisance parameters are used to parameterize systematic uncertainties on the yields including: luminosity (0.44 $\%$), $\sigma(ee\rightarrow\tau\tau)$ (0.31 $\%$), leptonic branching fractions ($\sim$ 0.2 $\%$), 3-prong branching fraction (0.57 $\%$), PID Efficiency ($e: 2\%$, $\mu $: 1$\%$, $\pi $:  3$\%$). Shape uncertainty due to the MC modeling must also be accounted for. For many hadronic $\tau$ decay channels the relative uncertainties from experimental results are large. A $\tau$-lepton decay to three charged pions is mediated by the  $a_{1}(1260)$ resonance which decays through the intermediate $\rho \pi$ state. In the MC samples used in this analysis the PDG \cite{PDG2020} average of $m_{a_{1}} = 1230 \pm 40$ MeV/$c^{2}$ and a Breit-Wigner averaged width of $\Gamma_{a_{1}} = 420 \pm 35 $ are used. Reference~\cite{PDG2020}  quotes the estimated width to be between 250 - 600 MeV/$c^{2}$. The uncertainty associated with the $a_{1}$ resonance represents the dominant contribution to the systematic error in the analysis. In order to understand the effects of the uncertainty on the $a_{1}$ mass on the final results in this analysis several additional MC simulations were built, in which the $m_{a_{1}}$ was varied to $\pm 1 \sigma$ of the experimental average.

\subsubsection{Results}

\begin{figure}[t]
         \centering
         \includegraphics[width=4in]{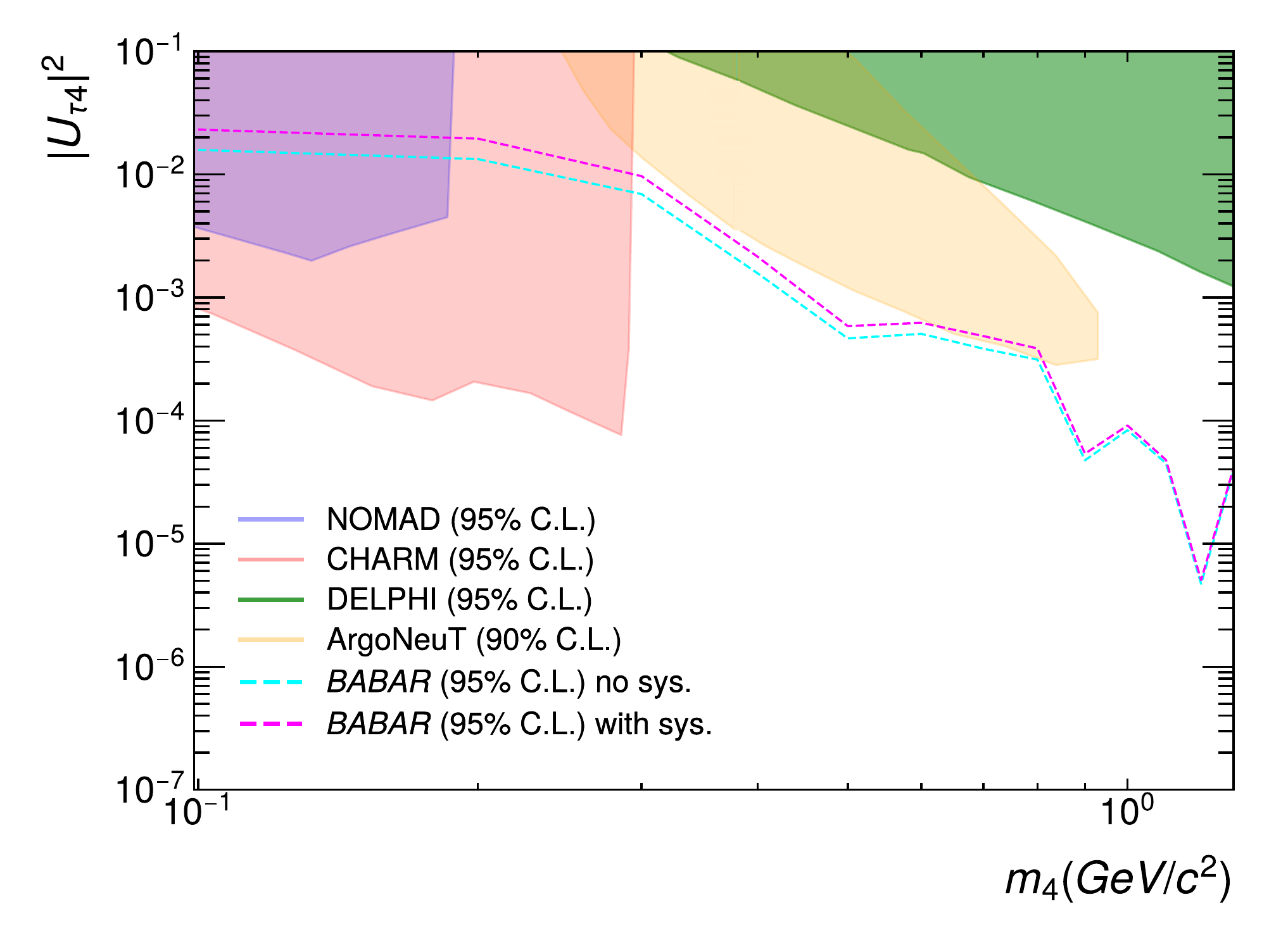}
        \caption{Upper limits at 95$\%$ confidence level (C.L.) on $|U_{\tau 4}|^{2}$. The magenta line represents the result when uncertainties are included. The magenta line is expected to be a very conservative upper limit. }
        \label{fig:limits}
\end{figure}

Figure~\ref{fig:limits} shows the upper limit at the 95$\%$ confidence level provided by this analysis using the described binned likelihood technique. The magenta line represents the upper limit when all systematic uncertainties are considered. The dominant systematic uncertainty is that due to the assumptions made within our simulation.

\subsubsection{Conclusions}
To conclude, the new analysis from BaBar~ has set limits on $|U_{\tau 4}|^{2}$ in the range $100 < m_{4} < 1300$ MeV/$c^{2}$. The technique used can be applied to future searches.

% \bibliographystyle{utcap_mod}
% \bibliography{babar_references}

%-------------------------------------------
\subsection{Searches for heavy neutral leptons at the LHC -- {\it J.~Knolle}}
\label{Joscha_Knolle}
{\it Author: Joscha Knolle, <joscha.knolle@cern.ch>}
%-------------------------------------------

% \pdfsuppresswarningpagegroup=1
% \documentclass[a4paper,12pt]{article}
% \usepackage[left=2.5cm, right=2.5cm, top=2.5cm, bottom=3.5cm]{geometry}
% \usepackage[utf8]{inputenc}
% \usepackage{textcomp}
% \usepackage{graphicx}
% \usepackage{amsmath}
% \usepackage{array}
% \usepackage{multirow}
% \usepackage{booktabs}
% \usepackage{cite}
% \usepackage[hang,flushmargin]{footmisc}
% \bibliographystyle{cms_unsrt}
% \usepackage[hidelinks]{hyperref}
% \renewcommand\thesubsubsection{\arabic{subsubsection}}
% \usepackage{etoolbox}
% \patchcmd{\thebibliography}{\section*{\refname}}{\subsubsection*{\refname}}{}{}
% \setlength\abovecaptionskip{0pt}

% \begin{document}
%{\let\thefootnote\relax
%\footnotetext{Copyright 2022 CERN for the benefit of the ATLAS and CMS collaborations. %Reproduction of this article or parts of it is allowed as specified in the CC-BY-4.0 license.}}

%\subsection*{Searches for heavy neutral leptons at the LHC}

%\textbf{Joscha Knolle} on behalf of the ATLAS \& CMS Collaborations \\
%Ghent University, Sint-Pietersnieuwstraat 33, 9000 Gent, Belgium \\
%E-mail: \url{joscha.knolle@cern.ch}

\subsubsection{Introduction}

Heavy neutral leptons (HNLs) are introduced in extensions of the standard model (SM) of particle physics to explain the nonzero neutrino masses via the seesaw mechanism.
They can also be a dark matter candidate, or provide a mechanism for the generation of the matter-antimatter asymmetry in the universe.
At the CERN Large Hadron Collider (LHC), HNLs could be produced in proton-proton (pp) collisions through various processes.
The ATLAS~\cite{ATLAS:2008xda} and CMS~\cite{CMS:2008xjf} experiments have performed HNL searches covering a wide mass range from a few GeV to several TeV, making use of the excellent reconstruction of charged leptons, jets, and missing momentum with their detectors.

Here, a selection of recent ATLAS and CMS results is presented that are based on the full pp collision data set at $\sqrt{s}=13\,\mathrm{TeV}$ recorded during Run~2 of the LHC (2015--2018), corresponding to an integrated luminosity of about 140\,fb\textsuperscript{--1}.
The searches cover a wide range of HNL phenomenologies:\
HNL production in type-I seesaw models through the Drell--Yan (DY) or a $t$-channel vector boson fusion (VBF) process~\cite{CMS:2022fut, ATLAS:2022atq, CMS:2022rqc};
pair production of heavy neutral and charged leptons in type-III seesaw models~\cite{ATLAS:2022yhd};
and HNL pair production in decays of a new heavy $\mathrm{Z}^\prime$ boson~\cite{CMS:2022irq}.

\subsubsection{Searches for long-lived HNLs in DY production}

\begin{figure}[t!]
\centering
\begin{minipage}[b]{0.42\textwidth}\raggedleft
\includegraphics[width=0.87\textwidth]{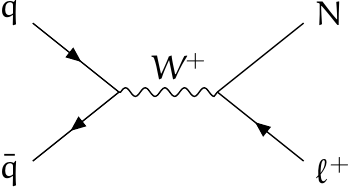} \\[1ex]
\includegraphics[width=\textwidth]{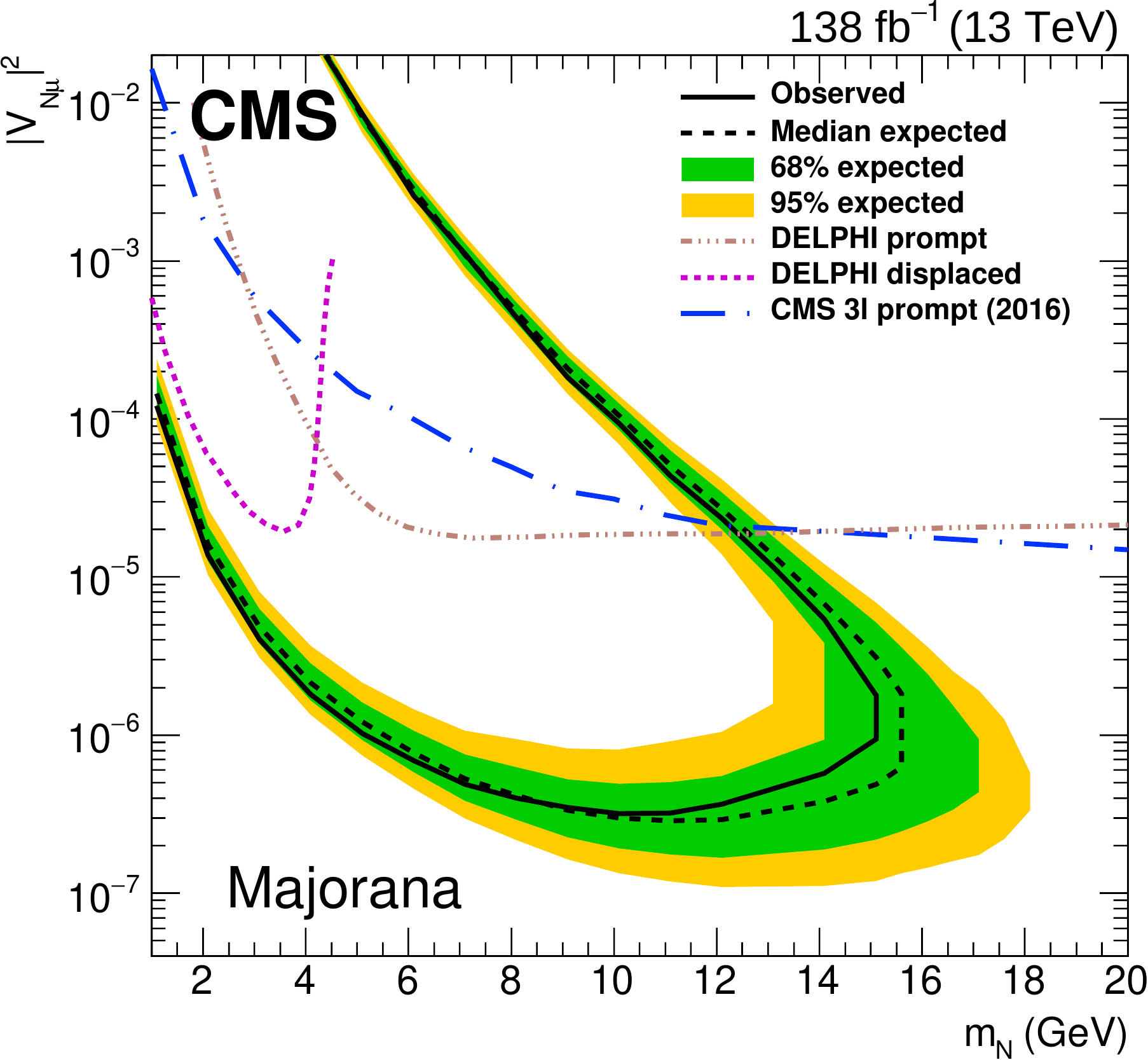}
\end{minipage}
\hspace*{0.05\textwidth}
\includegraphics[width=0.48\textwidth]{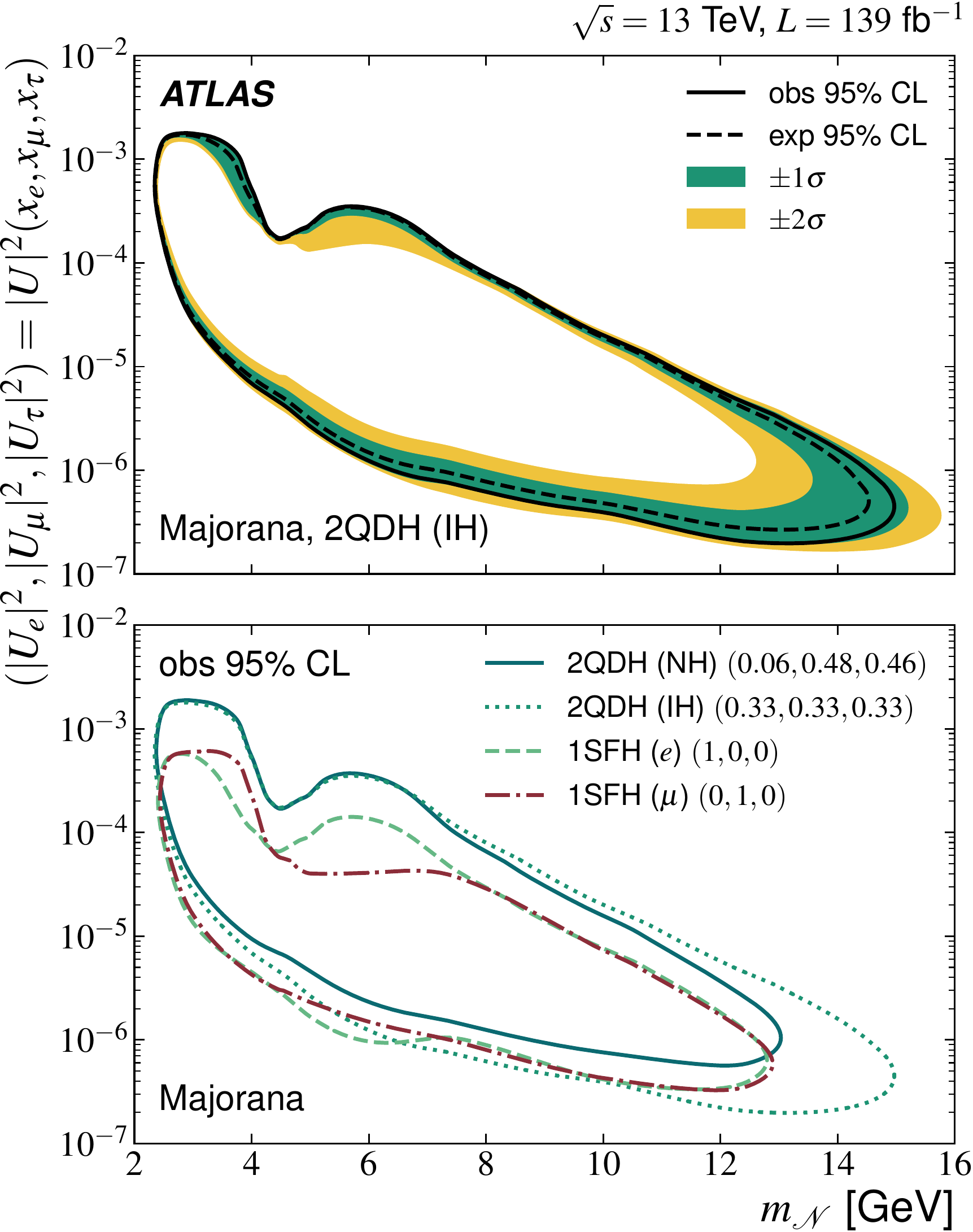}
\caption{%
    (Upper left) Example Feynman diagram for HNL production via the DY process in type-I seesaw models.
    (Lower left) Exclusion limits on the mixing parameter $|V_{\mathrm{N}\mu}|^2$ between HNL and muon in type-I seesaw models as a function of the HNL mass $m_{\mathrm{N}}$ from Ref.~\cite{CMS:2022fut}.
    (Right) Exclusion limits on the mixing parameter $|U|^2$ in type-I seesaw models with different mixing assumptions as a function of the HNL mass $m_{\mathcal{N}}$ from Ref.~\cite{ATLAS:2022atq}.
}
\label{JK:fig01}
\end{figure}

In a type-I seesaw model, HNLs of Majorana or Dirac nature can be produced in the DY process via their mixing with SM neutrinos, as illustrated in Fig.~\ref{JK:fig01} (upper left).
The HNL decay can proceed to two charged leptons and a SM neutrino (resulting in a trilepton final state), or to a charged lepton and two quarks (dilepton final state).
Searches for this production mode in both final states have been performed by the ATLAS, CMS, and LHCb experiments, covering HNL masses between 1\,GeV and 1.6\,TeV~\cite{CMS:2018iaf, CMS:2018jxx, ATLAS:2019kpx, LHCb:2020wxx}.
For HNL masses below 20\,GeV, the HNL lifetime becomes so large that the displacement of the HNL decay vertex can be resolved by the experiments, such that dedicated searches for long-lived decays become possible.

In Ref.~\cite{CMS:2022fut}, the CMS Collaboration presents a search for HNL production through the DY process in the trilepton final state with a secondary dilepton vertex.
Events are categorized by the invariant mass of the secondary dilepton system and the transverse displacement of the secondary vertex.
Background contributions arise from nonprompt leptons selected for the secondary dilepton vertex, and are estimated with a ``loose-not-tight ratio'' method from data.
Exclusion limits are derived from a template fit, considering exclusive HNL couplings to either electrons or muons, with one example shown in Fig.~\ref{JK:fig01} (lower left).

In Ref.~\cite{ATLAS:2022atq}, the ATLAS Collaboration presents a search for the same HNL process, using events with a displaced dilepton vertex.
The HNL mass is reconstructed from the selected leptons using a W~boson mass constraint and imposing the vector from the primary to the displaced vertex as the HNL flight direction, and is used to categorize the events.
Background contributions arise from random track crossings, and are estimated with an ``event shuffling'' method from data.
Exclusion limits are derived from a template fit, considering exclusive HNL couplings to either electrons or muons, as well as mixed-coupling scenarios with two quasi-degenerate HNLs.
Examples are shown in Fig.~\ref{JK:fig01} (right).

The exclusion limits of the long-lived searches significantly improve over the results from the prompt searches in the mass range 1--15\,GeV.
Differences between the results from the two experiments arise from different $p_{\mathrm{T}}$ thresholds in the single-lepton triggers, different strategies for the selection of a secondary or displaced vertex, and different requirements for the removal of background contributions from low-mass resonances.

\subsubsection{Search for HNLs in \textit{t}-channel VBF production}

\begin{figure}[t!]
\centering
\raisebox{2.5em}{\includegraphics[width=0.4\textwidth]{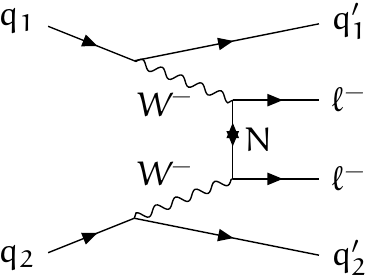}}
\hspace*{0.05\textwidth}
\includegraphics[width=0.42\textwidth]{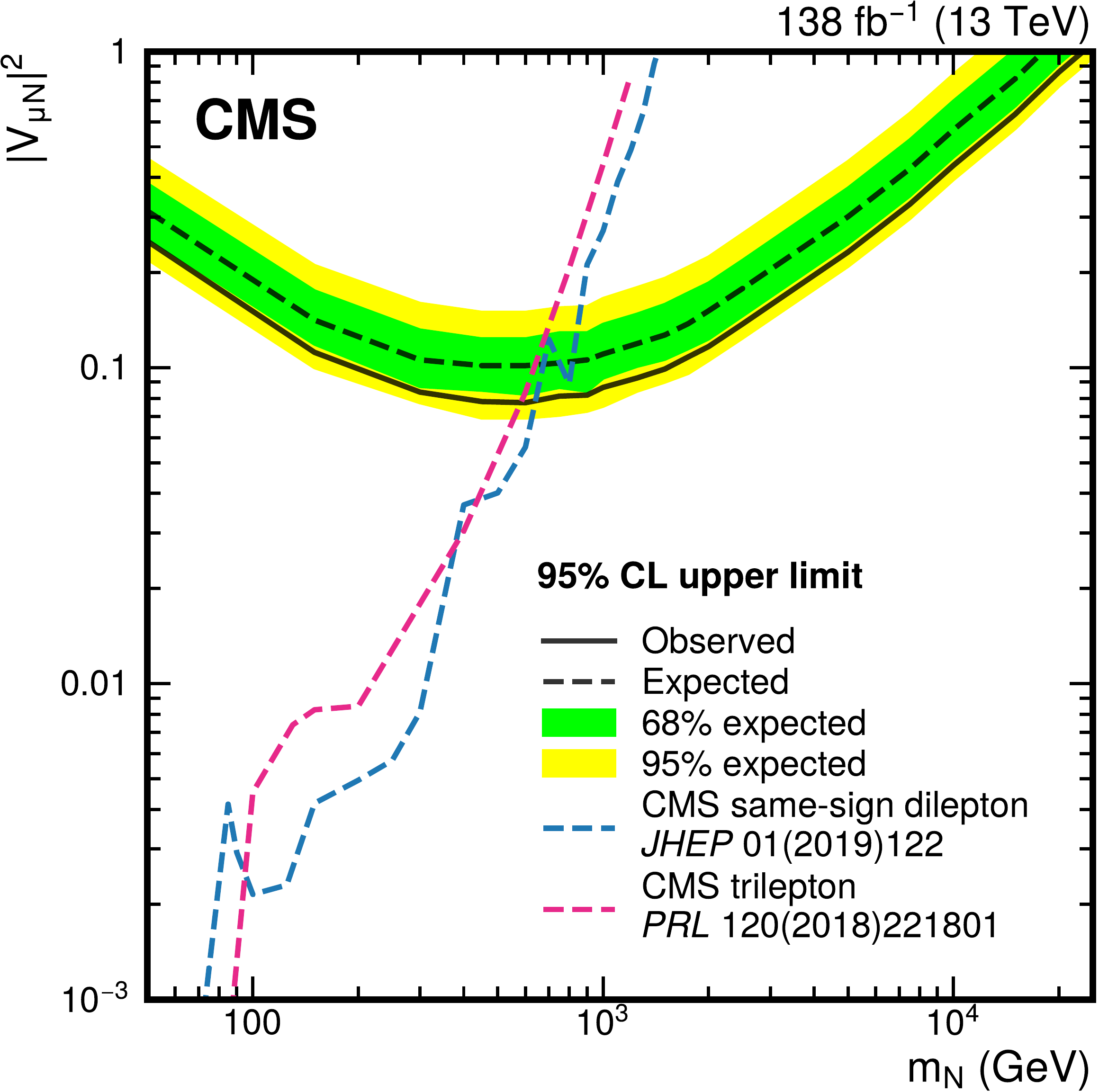}
\caption{%
    (Left) Example Feynman diagram for HNL production in a $t$-channel VBF process in type-I seesaw models.
    (Right) Exclusion limits on the mixing parameter $|V_{\mu\mathrm{N}}|^2$ between HNL and muon in type-I seesaw models as a function of the HNL mass $m_{\mathrm{N}}$ from Ref.~\cite{CMS:2022rqc}.
}
\label{JK:fig02}
\end{figure}

The production of HNLs in type-I seesaw models is also possible via VBF processes, relevant typically only at large HNL masses.
In Ref.~\cite{CMS:2022rqc}, the CMS Collaboration presents a search for HNL production in a $t$-channel VBF process, as illustrated in Fig.~\ref{JK:fig02} (left), characterized by the presence of two jets with large rapidity separation, using events with two same-sign muons.
An HNL of Majorana nature with exclusive couplings to muons is considered.

The selected events are categorized by the azimuthal angle between the two muons.
Background contributions with nonprompt muons are estimated from data, while contributions from diboson and associated top quark plus boson production are estimated from simulation.
Exclusion limits are derived from template fits to the distribution of the ratio of the scalar $p_{\mathrm{T}}$ sum of all selected jets to the leading muon $p_{\mathrm{T}}$.
The limits are shown in Fig.~\ref{JK:fig02} (right).
The search in the $t$-channel VBF production process provides better sensitivity than previous searches in the DY production process~\cite{CMS:2018iaf, CMS:2018jxx} for HNL masses above 650\,GeV.

\subsubsection{Search for heavy leptons in type-III seesaw models}

\begin{figure}[!b]
\centering
\raisebox{0.8em}{\includegraphics[width=0.4\textwidth]{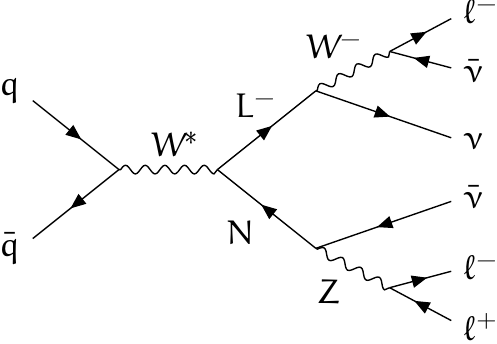}}
\hspace*{0.05\textwidth}
\includegraphics[width=0.5\textwidth]{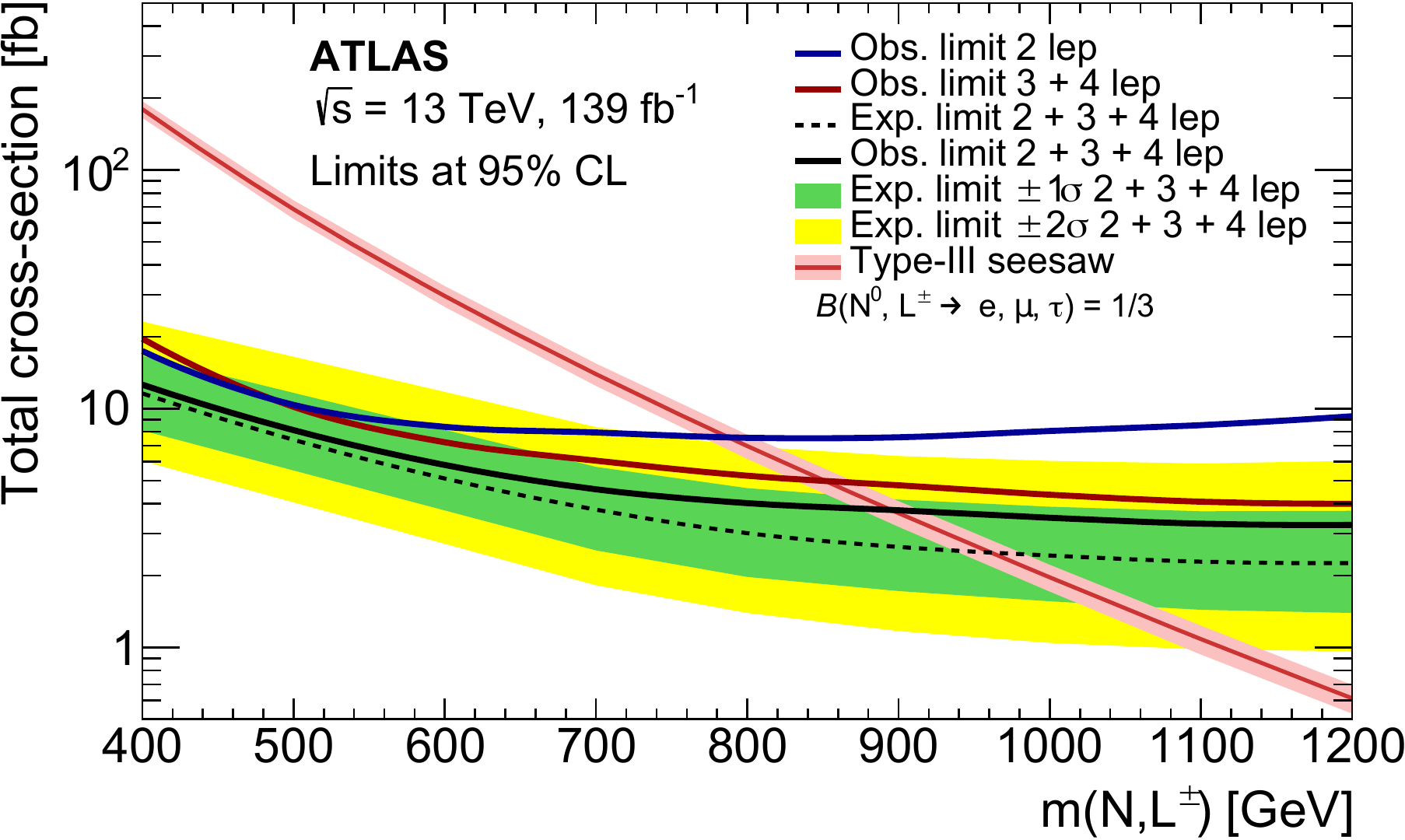}
\caption{%
    (Left) Example Feynman diagram for HNL and heavy charged lepton pair production in type-III seesaw models.
    (Right) Exclusion limits on the production cross section as a function of the degenerate HNL and heavy charged lepton mass $m(\mathrm{N},\mathrm{L}^\pm)$ for events with different number of leptons and their combination from Ref.~\cite{ATLAS:2022yhd}.
}
\label{JK:fig03}
\end{figure}

In a type-III seesaw model, the SM is extended with an additional fermionic $\mathrm{SU}(2)_L$ triplet, resulting in new heavy neutral and charged leptons.
These can be pair produced via $s$-channel production of virtual electroweak gauge bosons, and decay each to a boson (H, Z, or W) and a charged lepton or neutrino. An example is illustrated in Fig.~\ref{JK:fig03} (left).

In Ref.~\cite{ATLAS:2022yhd}, the ATLAS Collaboration presents a search for pair production of heavy neutral and charged leptons, in a mass-degenerate scenario with ``democratic'' mixing parameters, using final states with three or four charged leptons.
The selected events are categorized by the presence of Z~boson candidate (an opposite-sign same-flavour lepton pair with invariant mass close to the Z~boson mass) and the number of jets, targeting different possible final state combinations of the bosons in the decay.
Background contributions with nonprompt leptons and charge-misidentified electrons are estimated from data, while diboson and $\mathrm{t\bar{t}}$+boson contributions are estimated from simulation.

Exclusion limits are derived from template fits to the distribution of the transverse mass of the three-lepton system (the scalar $p_{\mathrm{T}}$ sum of all reconstructed charged leptons and jets and of the missing transverse momentum) for three-lepton (four-lepton) events.
The results are also combined with a similar search using dilepton events from Ref.~\cite{ATLAS:2020wop}.
The individual and combined limits are shown in Fig.~\ref{JK:fig03} (right).
Heavy neutral and charged leptons in type-III seesaw models are excluded up to 910\,GeV.

\subsubsection{Search for HNL pair production in \texorpdfstring{$\boldsymbol{\mathrm{Z}^\prime}$}{Z'} decays}

\begin{figure}[!b]
\centering
\raisebox{0.8em}{\includegraphics[width=0.45\textwidth]{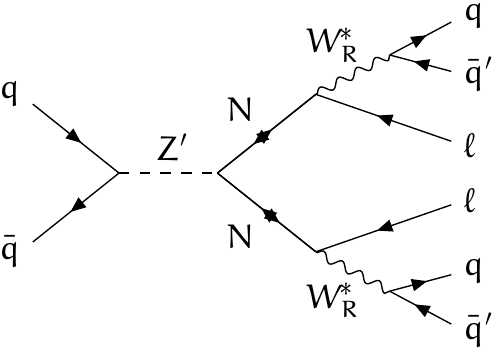}}
\hspace*{0.05\textwidth}
\includegraphics[width=0.45\textwidth]{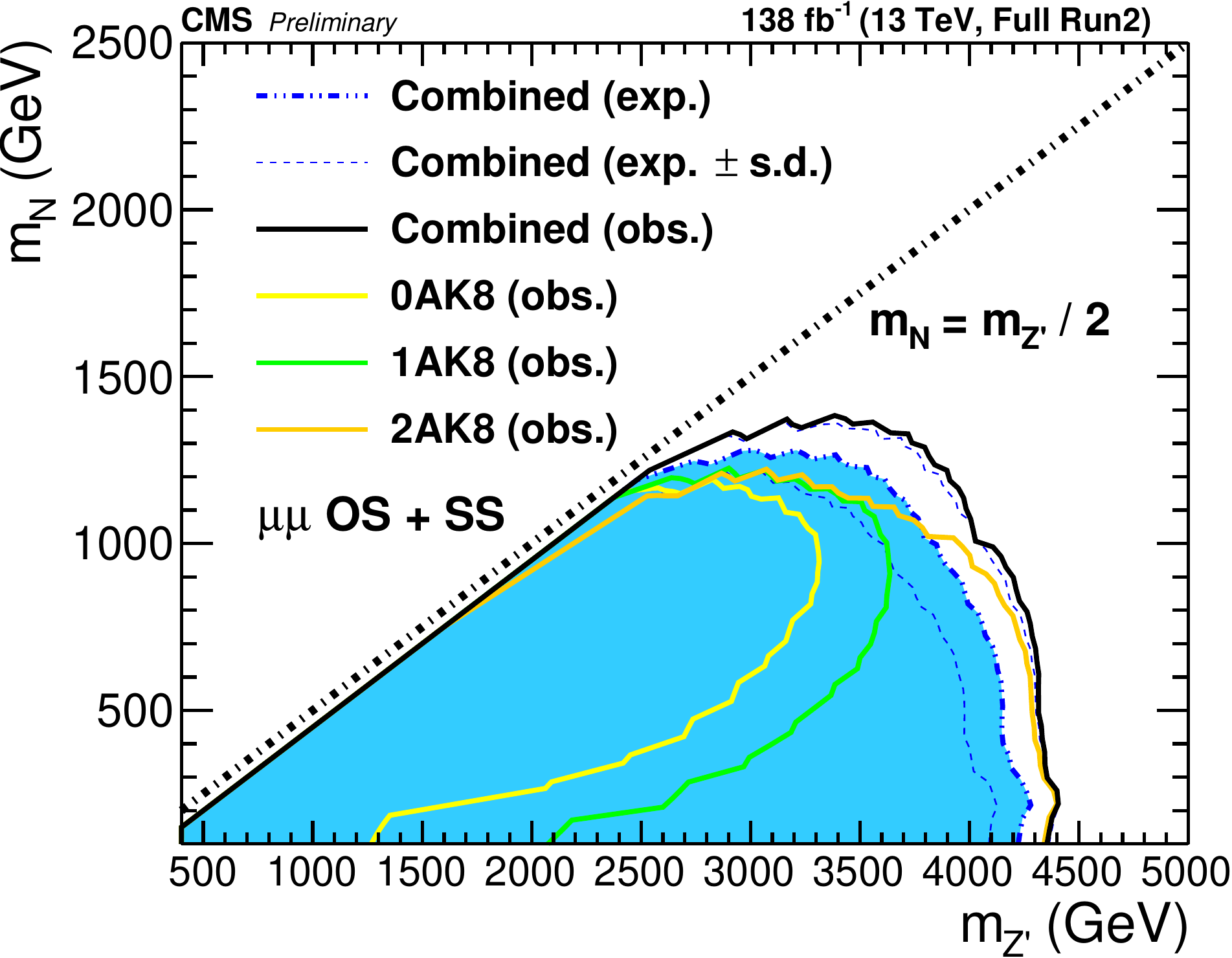}
\caption{%
    (Left) Example Feynman diagram for HNL pair production in $\mathrm{Z}^\prime$ decays.
    (Right) Exclusion limits derived from dimuon events on LRSM models as a function of HNL mass $m_{\mathrm{N}}$ and $\mathrm{Z}^\prime$ mass $m_{\mathrm{Z}^\prime}$ for events with different number of large-radius jets (``AK8'') and their combination from Ref.~\cite{CMS:2022irq}.
}
\label{JK:fig04}
\end{figure}

In a left-right symmetry model (LRSM), the SM is extended with three additional gauge bosons ($\mathrm{W}_{\mathrm{R}}^\pm$ and $\mathrm{Z}^\prime$) and three right-handed neutrinos to establish a symmetry between left and right $\mathrm{SU}(2)$ groups.
The unique LRSM signature yields events with both extra gauge bosons and right-handed neutrinos.

In Ref.~\cite{CMS:2022irq}, the CMS Collaboration presents a search for HNL pair production in $\mathrm{Z}^\prime$ decays, illustrated in Fig.~\ref{JK:fig04} (left), targeting scenarios with a large mass gap between the HNL and the $\mathrm{Z}^\prime$ such that the HNLs and their decay products are highly boosted.
The HNL decays via a virtual $\mathrm{W}_{\mathrm{R}}^\pm$ to a charged lepton and two quarks are selected either with a boosted (non-isolated charged lepton and a large-radius jet) or a resolved (isolated charged lepton and two small-radius jets) topology.
Events are categorized by the lepton flavour and the number of large-radius jets, and the $\mathrm{Z}^\prime$ is reconstructed from the selected HNL decay products.
Background contributions arise dominantly from $\mathrm{t\bar{t}}$ and $\ell^+\ell^-$ production, and are estimated from simulation.

Exclusion limits are derived from template fits to the reconstructed $\mathrm{Z}^\prime$ mass distribution, and one example is shown in Fig.~\ref{JK:fig04} (right).
For the lowest considered HNL mass of 100\,GeV, the LRSM model is excluded for a $\mathrm{Z}^\prime$ mass of up to 2.8 (4.35)\,TeV in the electron (muon) channel.
The largest excluded values of the HNL mass are 1.2 (1.4)\,TeV at a $\mathrm{Z}^\prime$ mass of about 3.0 (3.4)\,TeV in the electron (muon) channel.

\subsubsection{Outlook}

While the ATLAS and CMS collaborations have presented many results on HNL searches with the full Run~2 data set, further analysis efforts are ongoing to extend results that were so far only performed on a partial data set, to include new models and final states, and to use new and improved analysis techniques.
Additionally, the LHC has resumed data taking in 2022 at an increased energy $\sqrt{s}=13.6\,\mathrm{TeV}$, providing interesting prospects, especially for HNL models with very high masses.

% \bibliography{references}

% \end{document}

%------------------------------------------
\subsection{Opportunities for FIPs searches at FCC-ee -- {\it G.~Ripellino}}
\label{ripellino}
{\it Author: Giulia Ripellino}
%---------------------------------

%\documentclass{article}

% Language setting
% Replace `english' with e.g. `spanish' to change the document language
%\usepackage[english]{babel}

% Set page size and margins
% Replace `letterpaper' with `a4paper' for UK/EU standard size
%\usepackage[letterpaper,top=2cm,bottom=2cm,left=3cm,right=3cm,marginparwidth=1.75cm]{geometry}

% Useful packages
%\usepackage{amsmath}
%\usepackage{amssymb}
%\usepackage{graphicx}
%\usepackage{subfigure}
%\usepackage[colorlinks=true, allcolors=blue]{hyperref}

%\title{Opportunities for FIPs searches at FCC-ee}
%\author{Giulia Ripellino}

%\begin{document}
%\maketitle

\subsubsection{Introduction}
The electron-positron stage of the Future Circular Collider, FCC-ee, is a frontier factory for Higgs, top, electroweak, and flavor physics, designed to operate in a 100\,km circular tunnel built at CERN~\cite{CERN-ESU-015}. In addition to its unique program of high-precision Standard Model (SM) measurements, it will offer powerful opportunities to discover evidence of physics beyond the SM. Here, we discuss the opportunities for searches for feebly interacting particles (FIPs) at FCC-ee. 

General-purpose detectors can search for FIPs in many different ways. However, since FIPs are typically light particles with very low production cross-sections, the SM backgrounds are often overwhelming. This can be counteracted by searching for FIPs with signatures that are uncharacteristic of the SM, such as displaced vertices or delayed jets from long-lived particles (LLPs). FCC-ee offers exciting potential for the study of LLPs, where searches can be highly competitive to similar searches at collider and non-collider experiments. Three physics cases producing long-lived signatures at FCC-ee are highlighted here; heavy neutral leptons (HNLs), axion-like particles (ALPs), and exotic decays of the Higgs boson. 
% The searches motivate out-of-the-box optimization of experimental conditions and analysis techniques for the design of the future FCC-ee detectors.

\subsubsection{Heavy Neutral Leptons}
\label{sec:analysis_hnl}
\begin{figure*}
\centering
\includegraphics[width=0.7\textwidth]{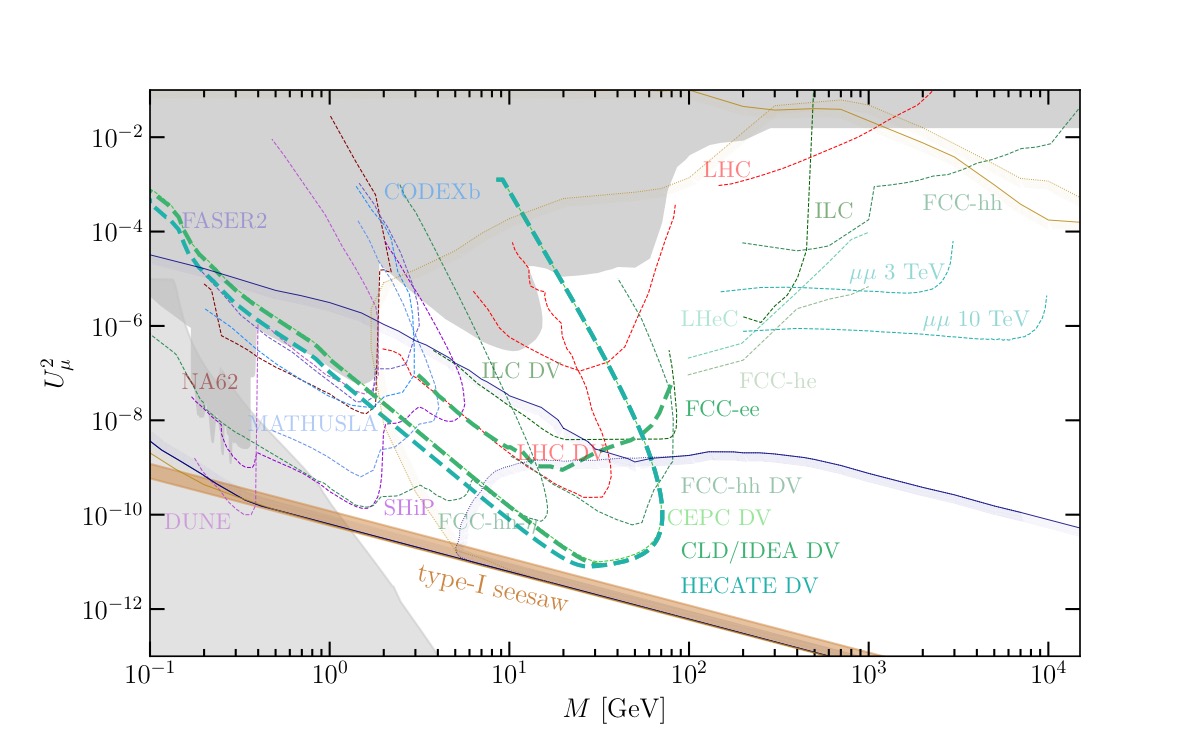}
\caption{
Constraints on the active-sterile mixing and mass of HNLs from past experiments together with projections for future experiments~\cite{Blondel:2022qqo}. The bold green line shows the sensitivity 
of displaced vertex searches at FCC-ee. 
% The parameter region inside the curves
% corresponds to more than four observed HNL decays 
% with $|V_{ \ell N }|^2=\delta_{\ell \mu}U_\mu^2$
% from 
% $5\times 10^{12}$ $Z$ bosons, assuming no background events and $95\%$ reconstructed HNL decays (i.e., all decays except the invisible decay) inside the main detectors with a displacement of over 400~$\mu$m. 
}
\label{fig:HNLsummary}
\end{figure*}

A common feature of several popular solutions to the origin of neutrino masses is the hypothetical existence of heavy, sterile neutrinos, also referred to as heavy neutral leptons (HNLs). Depending on the precise scenario, they can be Dirac or Majorana fermions, and mediate processes that violate lepton flavor symmetries. If HNLs mix with the SM neutrinos, they can participate in the SM weak interaction with couplings proportional to the active-sterile mixing matrix elements and the HNL mass. In the kinematically accessible regime, FCC-ee is an excellent machine to discover HNLs and to study their properties. 
% In particular, the large statistics FCC run around the Z pole, producing 5 $10^{12}$ Z bosons \cite{Benedikt:2651299}, is expected to be particularly powerful in this area~\cite{Blondel:2014bra,Abada:2019zxq,Klaric:2020phc,Abada:2018oly}. 
The sensitivity to active-sterile mixing is shown in Figure~\ref{fig:HNLsummary} for current and proposed detectors, including an FCC-ee displaced vertex analysis. 

Dirac and Majorana HNLs are studied for FCC-ee considering only the lightest heavy mass eigenstate, denoted by $N$, with mass $m_N$ and mixing $V_{\ell N}$. 
The  processes
\begin{subequations}
\label{eq:sim_HNL}
\begin{align}
\textbf{Majorana}~N &:
    e^+ e^- \to Z \to N \nu_e + N\overline{\nu_e}, \quad\text{with}\quad N \to e^+e^-\nu_e + e^+e^-\overline{\nu_e},
    \label{eq:sim_Majorana}
    \\
\textbf{Dirac}~N &:
    e^+ e^- \to Z \to N \overline{\nu_e} + \overline{N}\nu_e, \quad\text{with}\quad N~(\overline{N}) \to e^+e^-\nu_e~(\overline{\nu_e})\,,
    \label{eq:sim_Dirac}    
\end{align}
\end{subequations}
are simulated for $e^+e^-$ collisions at $\sqrt{s}=91$~GeV using the \texttt{HeavyN}~\cite{Alva:2014gxa,Degrande:2016aje} and \texttt{HeavyN\_Dirac}~\cite{Degrande:2016aje,Pascoli:2018heg} 
Universal \texttt{FeynRules} Object~\cite{Christensen:2008py,Degrande:2011ua,Alloul:2013bka} libraries and simulation details according to Ref.~\cite{Blondel:2022qqo}.
% in conjunction with \texttt{MadGraph5\_aMC@NLO}~\cite{Stelzer:1994ta,Alwall:2014hca}. The parton-level events are passed to \texttt{Pythia}~\cite{Sjostrand:2014zea} to simulate parton showering and hadronization and 
The detector response is simulated with \texttt{Delphes}~\cite{deFavereau:2013fsa}, using the latest Innovative Detector for Electron–positron Accelerators (IDEA) FCC-ee detector concept~\cite{Antonello:2020tzq} card. 

% The IDEA detector comprises of a silicon pixel vertex detector; a large-volume, light short-drift wire chamber surrounded by a layer of silicon micro-strip detectors; a thin, low-mass superconducting solenoid coil; a pre-shower detector; a dual-readout calorimeter; and muon chambers within the magnet return yoke.

\begin{figure}[!t]
\centering
\includegraphics[width=0.45\textwidth]{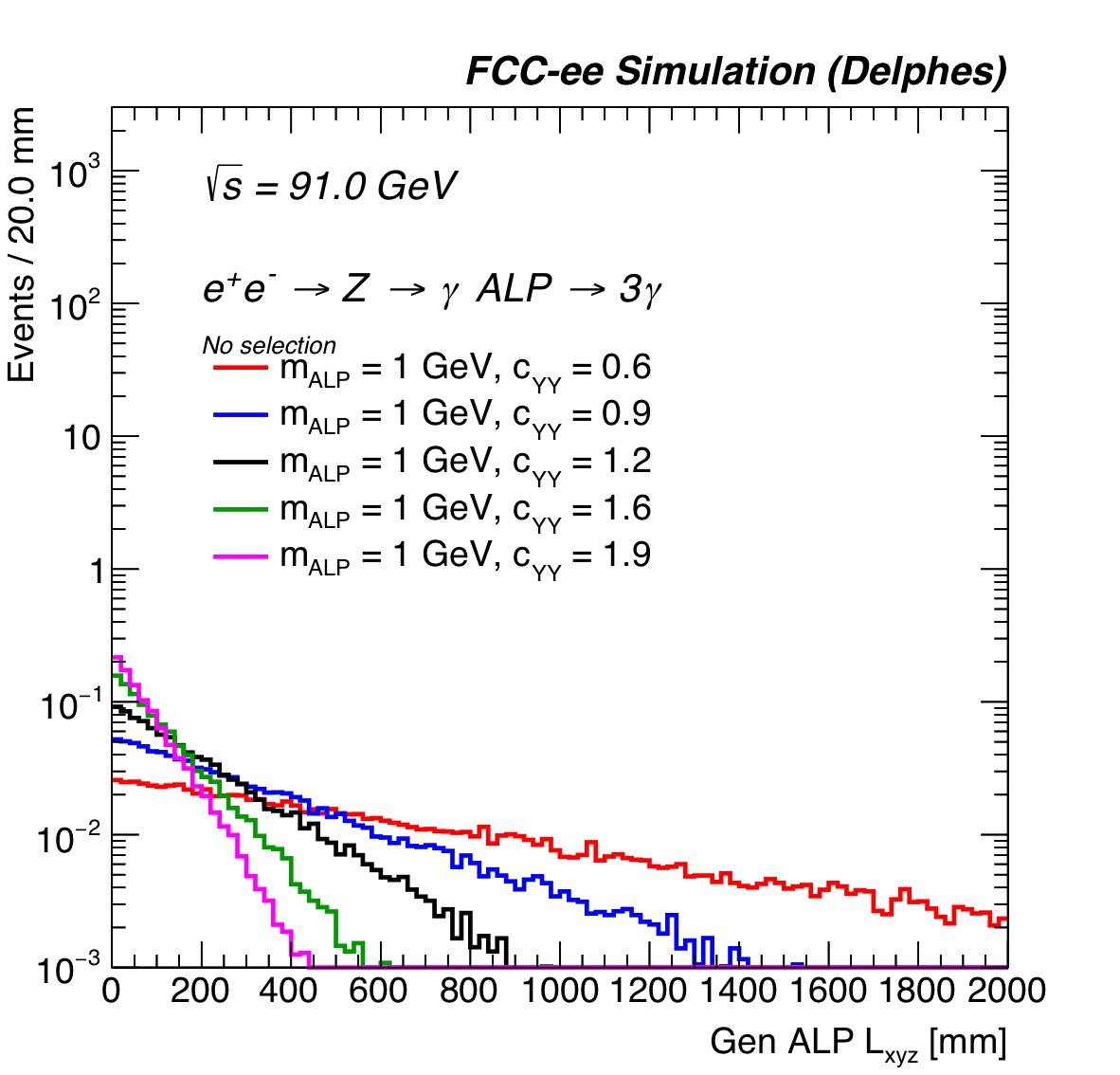} \label{fig:HNLlifetime_gen}
\includegraphics[width=0.45\textwidth]{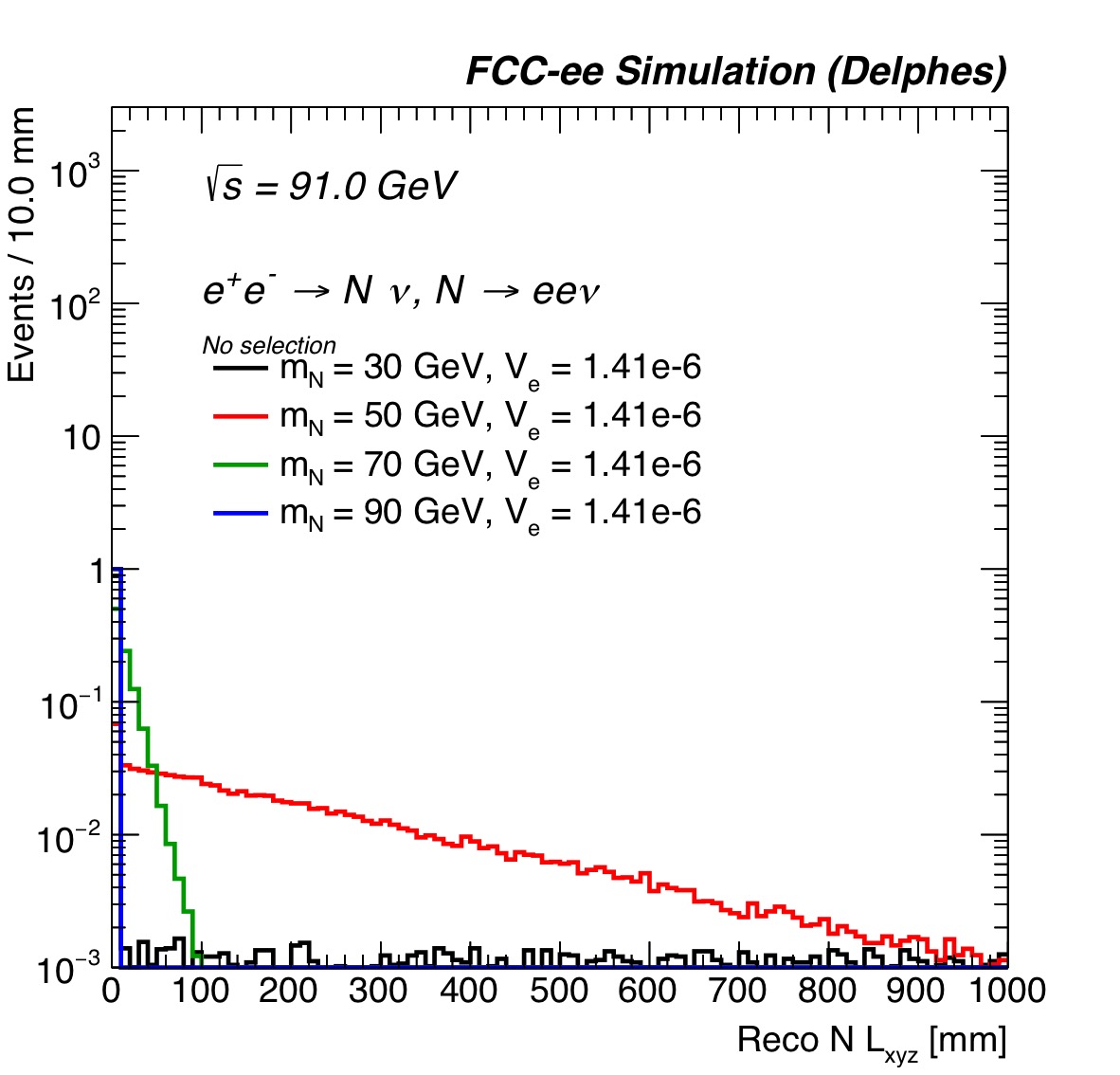} \label{fig:HNLlifetime_reco}
\caption{
Normalized distributions of the generator-level lifetime of $N$ in the lab frame (a) and
the reconstruction-level three-dimensional decay length $L_{xyz}$ (b) of the $N$ for representative HNL masses and an active-sterile mixing $\vert V_{eN}\vert = 1.41 \times 10^{-6}$~\cite{Blondel:2022qqo}. 
}
\label{fig:HNLlifetime}
\end{figure}

In the absence of additional new physics, light HNLs with active-sterile mixing much smaller than unity are generically long-lived. To explore this at FCC-ee, Figure~\ref{fig:HNLlifetime} shows the generator-level lifetime of $N$ and the reconstructed three-dimensional decay length $L_{xyz}$ of the HNL.
For a fixed width of $\vert V_{eN}\vert = 1.41\times10^{-6}$, different qualitative features can be observed for the representative $m_N$. For the smallest considered masses, characteristic decay lengths readily exceed several meters, resulting in decay vertices outside the fiducial coverage of the detectors. Such HNLs will appear as missing momentum in the $e^+e^-$ collision events.
For heavier $N$, lifetimes are drastically smaller, with decay lengths that are mostly within 100~mm, making these signals suitable e.g for displaced vertex searches.

\begin{figure}[!t]
\centering
\includegraphics[width=0.45\textwidth]
{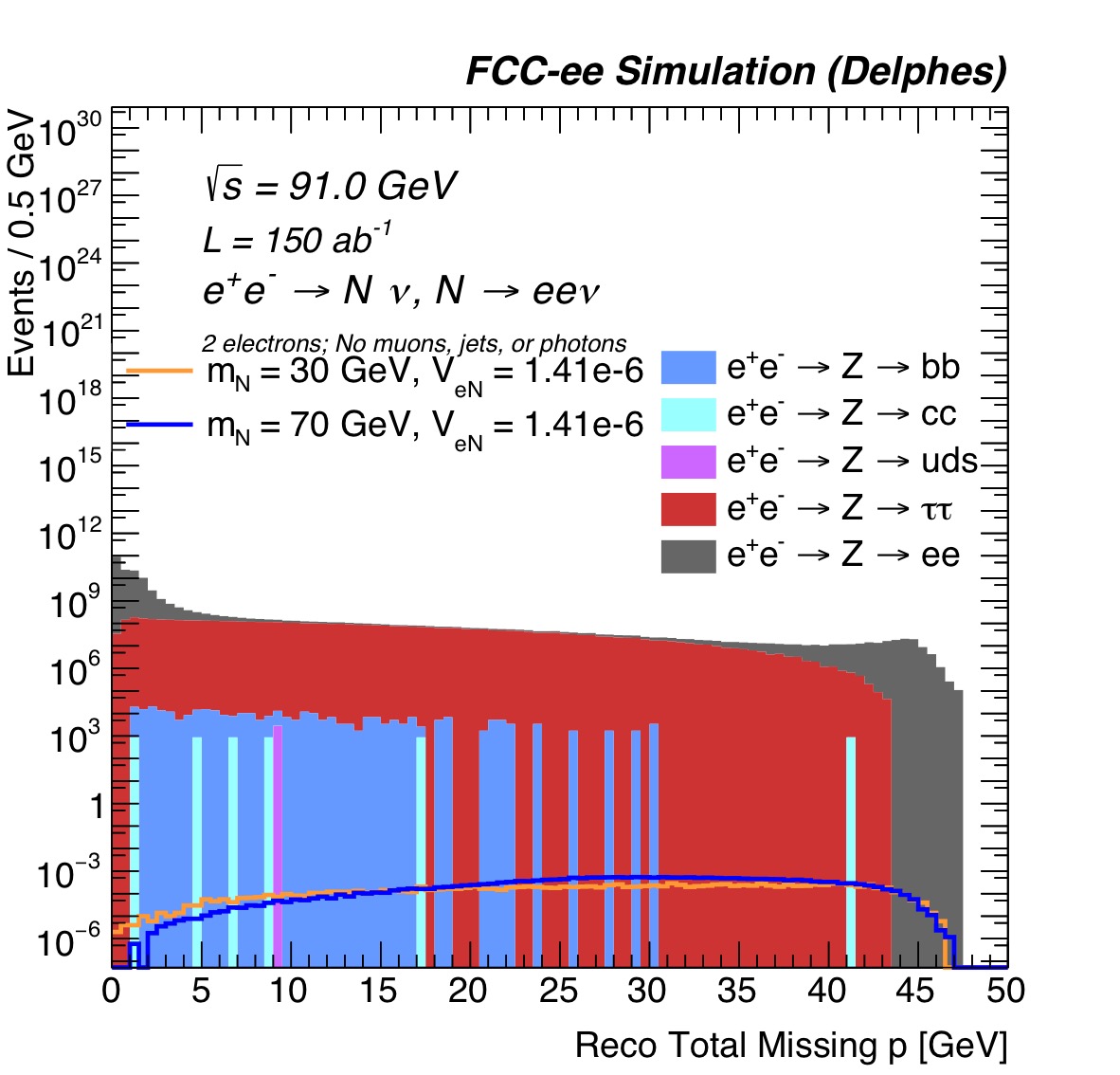}
\label{fig:HNLmissingenergy}
\includegraphics[width=0.45\textwidth]{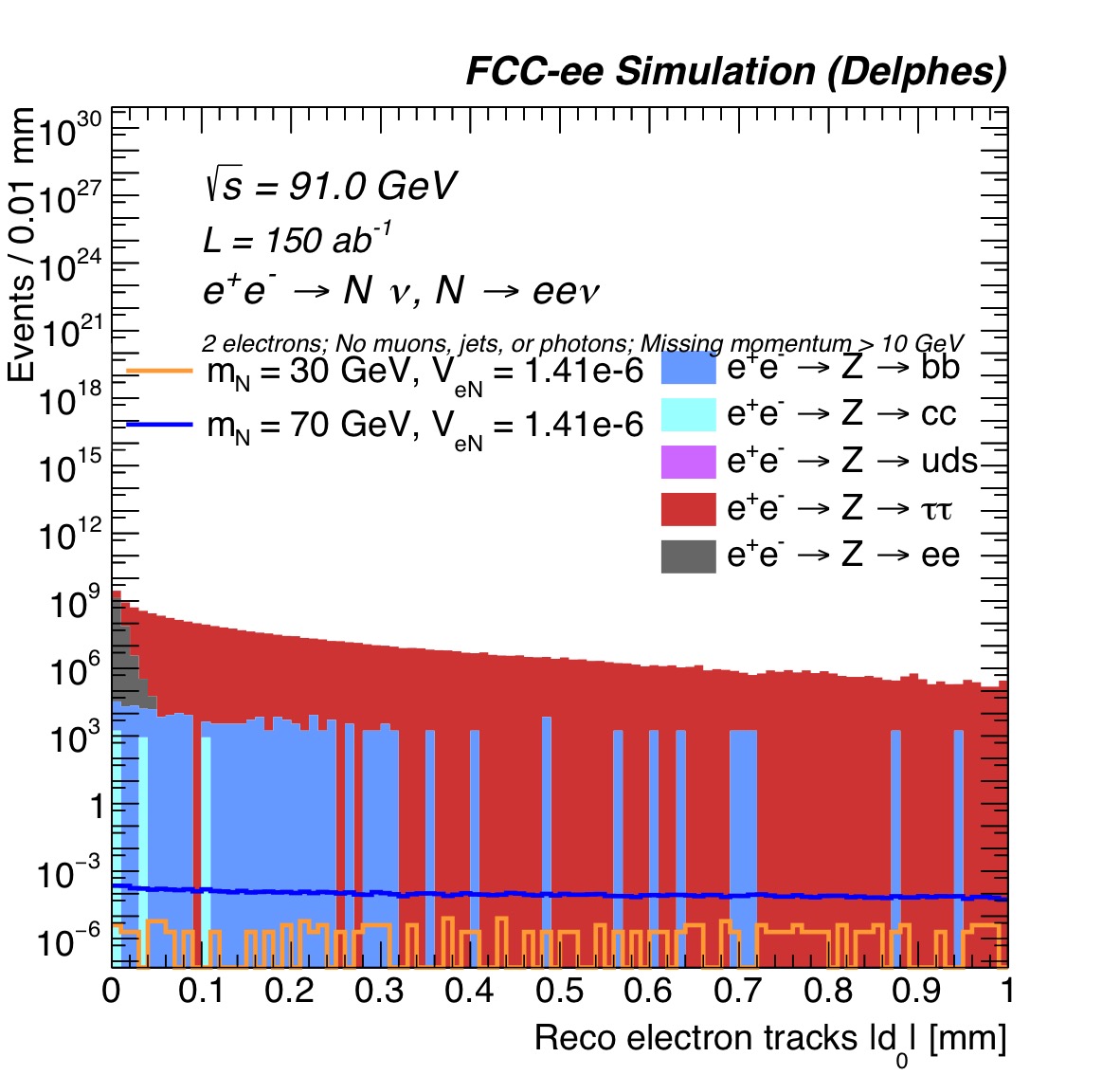}
\label{fig:HNLd0}
\caption{Normalized, reconstructed-level distributions of the total missing momentum (a) and the absolute value of the transverse impact parameter $|d_0|$ (b), for representative HNL signal benchmark mass and $\vert V_{eN}\vert$ choices, as well as background processes~\cite{Blondel:2022qqo}.
}
\label{fig:HNLexpdist}
\end{figure}

As a first step towards a sensitivity analysis for FCC-ee, several backgrounds to the HNL processes are considered; $Z$ bosons decaying to electron-positron pairs, to tau pairs, to light quarks, to charm quark pairs, and to $b$ quark pairs. These background processes are simulated with the same conditions as the signal. Figure~\ref{fig:HNLexpdist} shows distributions of the total missing momentum $\not\! p$ and the electron-track transverse impact parameter $|d_0|$. Requiring $\not\! p > 10$~GeV significantly reduces the background while maintaining a high efficiency for the HNL signal. Similarly, a requirement of $|d_0|>0.5$~mm removes the vast majority of the background. A similar event selection would therefore be suitable for an HNL search at FCC-ee.

\subsubsection{Axion-like particles}
\label{sec:analysis_alp}

\begin{figure}[hbtp]
\centering
\includegraphics[width=0.4\textwidth]
{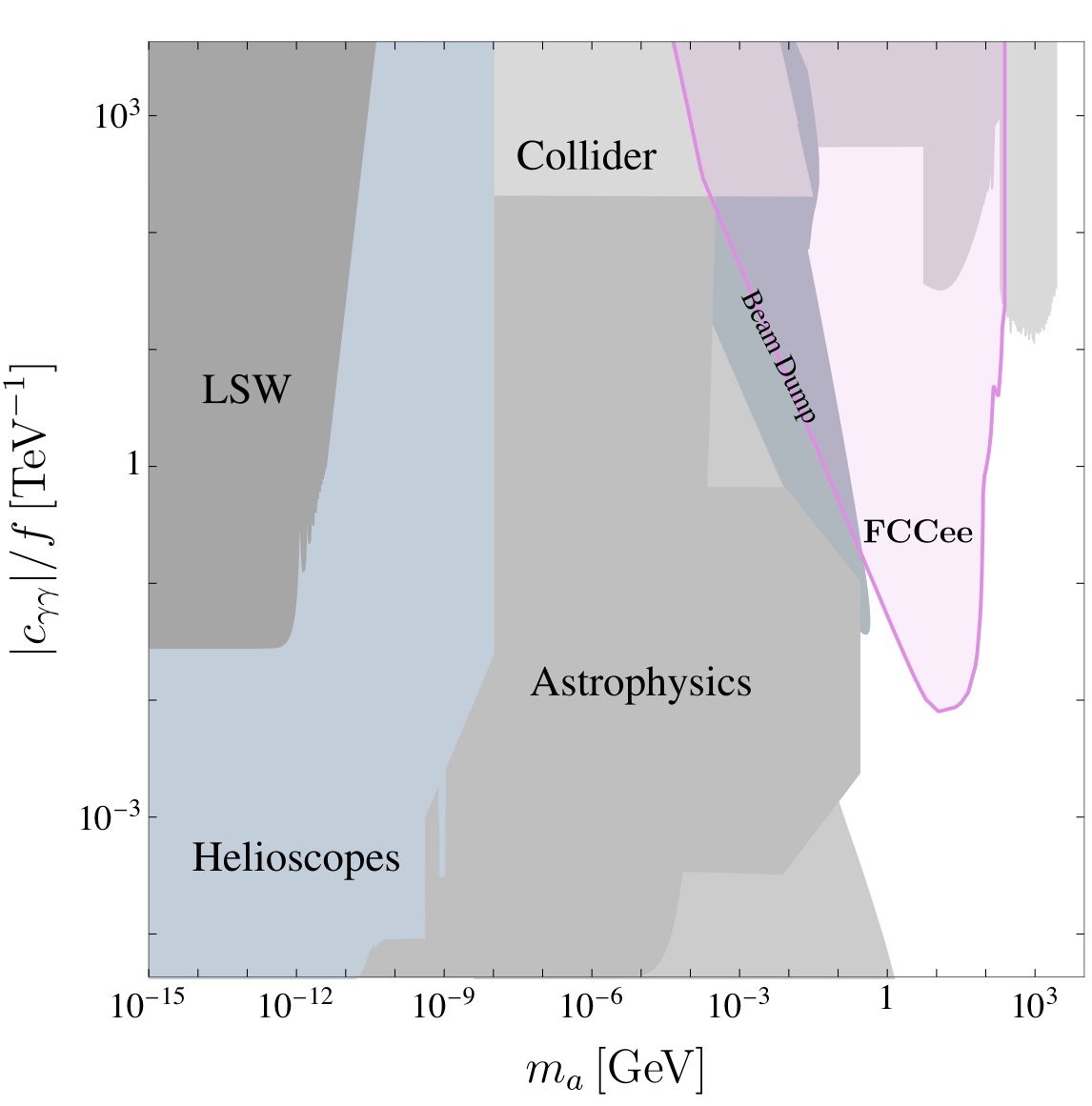}
% {GenALP_lifetime_xyz_ALP_selNone_nostack_log.jpg}}
\includegraphics[width=0.45\textwidth]{HNL/Giulia_Ripellino/GenALP_Lxyz_ALP_selNone_nostack_log.jpg}
\caption{\label{fig:ALPSensitivity}(a): Projected sensitivity of FCC-ee to the process $e^+ e^- \to \gamma a \to 3 \gamma$~\cite{Blondel:2022qqo}. (b): Normalized distribution of $L_{xyz}$ for $m_{\text{ALP}}= 1$~GeV and several benchmark choices of $c_{\gamma\gamma}$~\cite{Blondel:2022qqo}.}
\label{fig:ALPs}
\end{figure}

% Many models that address open, fundamental problems of the SM are governed by global symmetries. If an approximate global symmetry is spontaneously broken, a pseudo Nambu-Goldstone boson appears in the theory that is light compared to the symmetry breaking scale. If this pseudo Nambu-Goldstone boson is a pseudoscalar, it is often referred to as an axion-like particle or ALP.
ALPs appear in many models that address open, fundamental problems in the SM and could for instance provide the evidence for a Dark Sector. At FCC-ee, ALPs are predominantly produced in association with a photon, Z boson, or Higgs boson with subsequent decays into gauge bosons, leptons, or quarks. Here, ALPs $a$ from $Z$ decays at FCC-ee are studied for the process
\begin{equation}
    e^+e^- \to Z \to a\gamma, \quad\text{with}\quad a \to \gamma\gamma \,,
\end{equation}
which is simulated using the model libraries of Ref.~\cite{Bauer:2018uxu} and the same setup as described in Section~\ref{sec:analysis_hnl}. Figure~\ref{fig:ALPSensitivity} shows the projected sensitivity of FCC-ee to this process~\cite{Bauer:2018uxu}. 

% The corresponding ALP decay width is given by
% %
% \begin{align}
%    \Gamma(a\to\gamma\gamma)
%    &= \frac{\alpha^2\space m_a^3}{64\pi^3 f^2}  c_{\gamma\gamma}^2 \,,
% \end{align}
% where $c_{\gamma\gamma}$ is the ALP-photon coupling.

For small ALP-photon coupling $c_{\gamma\gamma}$ and light ALPs, the ALP decay vertex can be considerably displaced from the production vertex. 
Figure~\ref{fig:ALPSensitivity} shows the generated three-dimensional decay length $L_{xyz}$ for an ALP mass of 1~GeV and several benchmark choices of the coupling $c_{\gamma\gamma}$. Characteristic decay lengths are typically within the fiducial volume of the detectors for the chosen parameter values. This makes the process a suitable benchmark signal e.g for a displaced photon search. 

% This analysis assumes at least four signal events and combines the $Z$-pole run with runs at $\sqrt{s} = 2m_W$ and $\sqrt{s} = 250\,$ GeV. Further details are provided in~\cite{Bauer:2018uxu}.

% \begin{figure}[hbtp]
% \centering
% \includegraphics[width=0.45\textwidth]{ALPs_sensitivity_cyy.jpg}
% \caption{Projected sensitivity of FCC-ee in $e^+ e^- \to \gamma a \to 3 \gamma$~\cite{Blondel:2022qqo}.}
% \label{fig:ALPSensitivity}
% \end{figure}

\subsubsection{Exotic Higgs Boson Decays} \label{sec:analysis_higgs}

\begin{figure}[hbtp]
\centering
\includegraphics[width=0.45\textwidth]
{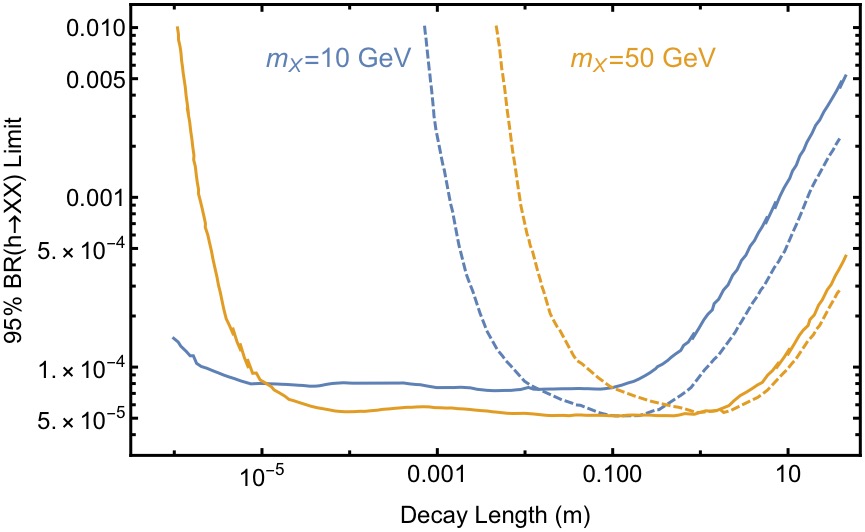}
\includegraphics[width=0.45\textwidth]{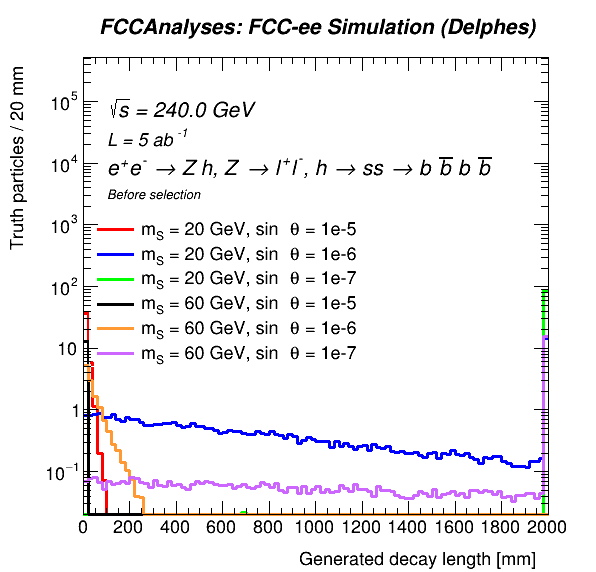}
\caption{\label{fig:ExoticHiggsLLPSensitivity}(a): Sensitivity of FCC-ee to exotic Higgs boson decays to LLPs~\cite{Alipour-Fard:2018lsf}. The solid line employs an invariant mass cut to improve sensitivity at shorter decay lengths while the dashed line relies on longer decay lengths to reduce SM backgrounds. (b): Generated dark scalar decay length $L_{xyz}$ for a dark scalar mass $m_{\text{S}}= 20$ or 60~GeV and a mixing angle $\sin{\theta}=10^{-5}, 10^{-6}$, or, $10^{-7}$.}
\label{fig:DarkScalars}
\end{figure}

Current bounds on the Higgs width leave plenty of room for decays that are not predicted by the SM. The products of these exotic Higgs boson decays can decay promptly themselves or be completely stable, each of which present their own experimental challenges and advantages. Figure~\ref{fig:ExoticHiggsLLPSensitivity} displays an illustration, taken from Ref.~\cite{Alipour-Fard:2018lsf}, of how sensitive FCC-ee can be to Higgs boson decays to long-lived scalar, pseudo-scalar, or vector particles. 

Here, we study exotic decays to long-lived scalars. Such decays may result from simple constructions, such as adding a single scalar field $S$ to the SM according to
\begin{equation}
    V_\text{scalar}=V_H+V_S+c_1S|H|^2+c_2S^2|H|^2~\,,
\end{equation}
where $H$ is the SM Higgs doublet. They may also arise in rich, hidden sectors such as Hidden Valley models~\cite{Strassler:2006im,Strassler:2006ri,Han:2007ae}. 
In many models the long-lived scalar inherits much of the Higgs' coupling structure with the actual size of the couplings reduced by a common small mixing angle $\theta$. 

% The branching fraction for the dark scalar ($s$) decays into a pair of SM particles $X$ is then given by
% \begin{equation}
%     \Gamma\left(s\to X_\text{SM}X_\text{SM}\right)\ =\ \sin^2\theta \ \Gamma\left(h(m_s)\to X_\text{SM}X_\text{SM}\right)\,,
% \end{equation}
% where $h$ is the SM Higgs.

Exotic Higgs decays to dark scalars at FCC-ee are studied for the process
\begin{equation}
    e^+ e^- \to Z h \; \textrm{with} \: Z \to e^+ e^- \: \textrm{or} \: \mu^+ \mu^- \: \textrm{and} \: h \to s s \to b \bar{b} b \bar{b}\,,
\end{equation}
using the HAHM\_MG5Model\_v3~\cite{Curtin:2013fra,Curtin:2014cca} and simulation as detailed in Section~\ref{sec:analysis_hnl}, with the exception of the $e^+e^-$ center-of-mass energy which is set to $\sqrt{s}=240$~GeV. Figure~\ref{fig:ExoticHiggsLLPSensitivity} shows the generated three-dimensional decay length $L_{xyz}$ for several benchmark choices of the mixing angle $\theta$ and the dark scalar mass. Characteristic decay lengths are typically within the fiducial volume of the detectors, making these signals suitable e.g for a displaced vertex search. 

%\bibliographystyle{ieeetr}
%\bibliography{sample}

%\end{document}

%-------------------------------------------
\subsection{The HIBEAM/NNBAR experiment for the European Spallation Source  -- {\it D.~Milstead}}
%-------------------------------------------
\label{milstead}
{\it Author: David Milstead, <milstead@fysik.su.se>}
%-------------------------------------------

%\documentclass{article}

% Language setting
% Replace `english' with e.g. `spanish' to change the document language
%\usepackage[english]{babel}

% Set page size and margins
% Replace `letterpaper' with `a4paper' for UK/EU standard size
%\usepackage[letterpaper,top=2cm,bottom=2cm,left=3cm,right=3cm,marginparwidth=1.75cm]{geometry}

% Useful packages
%\usepackage{amsmath}
%\usepackage{graphicx}
%\usepackage[colorlinks=true, allcolors=blue]{hyperref}

%\title{The HIBEAM/NNBAR experiment for the European Spallation Source}
%\author{David Milstead, Fysikum, Stockholm University}

%\begin{document}
%\maketitle

%\begin{abstract}
%The HIBEAM/NNBAR program offers a unique high precision probe with neutrons for baryon number violation and the existence of a dark sector of particles. The work exploits the world's brightest neutron source, the European Spallation Source. Neutron conversions to anti-neutrons and/or sterile neutrons are sought. The processes are rare and may occur spontaneously as well as being induced via feeble interactions with a magnetic field. An ultimate sensitivity improvement for free neutrons converting to anti-neutrons of three orders of magnitude can be achieved. Such a leap in sensitivity in the testing of a global symmetry is a rare one.

%\end{abstract}

\subsubsection{Introduction}

The conservation of baryon number ($B$) is among the most fragile of the empirically observed conservation laws. Baryon number violation (BNV) is a fundamental Sakharov condition for baryogenesis~\cite{Sakharov:1967dj}. In the Standard Model (SM), baryon number conservation corresponds to an accidental symmetry at perturbative level. Extensions of the SM, such as  supersymmetry~\cite{Calibbi:2016ukt} and extra dimensions~\cite{Nussinov:2001rb}, therefore routinely violate $B$. The non-perturbative sector of the SM itself predicts BNV via ultra-rare sphaleron interactions. It is perhaps more pertinent not to ask if it takes place but which processes and at which scales BNV takes place. A common way of searching for BNV is via single nucleon decay. However, such searches require the simultaneous violation of $B$ and lepton number $L$. Conversions of neutrons to anti-neutrons and/or sterile neutrons offer high sensitivity to BNV processes in which only $B$ is violated. Like the photon and neutrino, which may mix with axion-like particles and sterile neutrinos, respectively, neutron mixing with a putative dark sector with is well suited as a portal to physics beyond the SM~\cite{Berezhiani:2009ldq,Bringmann:2018sbs,Berezhiani:2020vbe}. The cleanest means of performing such searches is with high flux beam neutrons though progress is limited to the capabilities of neutron sources. Presently under construction, the European Spallation Source (ESS)~\cite{peggseuropean} will be the world's brightest neutron source and will open a new discovery window for BNV searches. The HIBEAM/NNBAR project~\cite{White,Yiu:2022faw,Backman:2022szk,Abele:2022iml} is a proposed multi-stage program for the ESS to take advantage of this that can achieve an ultimate search sensitivity improvement of three orders of magnitude compared to that available at other facilities.

The program's first stage is termed {\it High Intensity Baryon Extraction and Measurement} (HIBEAM). This stage probes a discovery window for sterile neutron searches and also includes a pilot experiment for free neutron-antineutron conversions - the first such experiment at a spallation source. The second stage (NNBAR) is a dedicated search for free neutrons converting to anti-neutrons and can achieve a sensitivity increase of three orders of magnitude.

 HIBEAM/NNBAR is a cross-disciplinary milieu with participants from a number of European countries and the United Status. Scientists come from a range of specialised areas including experimental particle and nuclear physics, magnetics, neutronics and vacuum design. 

\subsubsection{The European Spallation Source}
The European Spallation Source (ESS) is currently under construction in Lund, Sweden. It will eventually provide a suite of 22 neutron beam instruments. Protons from a linac collide with a rotating tungsten neutron target. Neutrons are then slowed in a moderator. The ESS accelerator provides a 2 GeV proton pulse with 2.86 ms length, at a repetition rate of 14~Hz.

HIBEAM would operate as a standard ESS instrument with a beamport of normal size. The cold neutron intensity will be around $10^{11}$ n/s and the flight path is $\sim 50$~m.

The NNBAR experiment would use the Large Beam Port (LBP). The LBP, so named given its size compared to the standard ESS beamports, is included in the ESS infrastructure to allow the extraction of a large integrated neutron flux. The LBP is as large as around three normal-sized beamports. To appropriately exploit the LBP, a liquid deuterium moderator is being designed as part of the HighNESS project~\cite{Santoro:2022tvi}. The NNBAR experiment can expect a cold neutron intensity of over $10^{13}$ n/s at the detector area which lies after a flight path of around $200$~m.

\subsubsection{HIBEAM and NNBAR}
\begin{figure}[ht!]
	\centering
	\includegraphics[width=.75\textwidth]{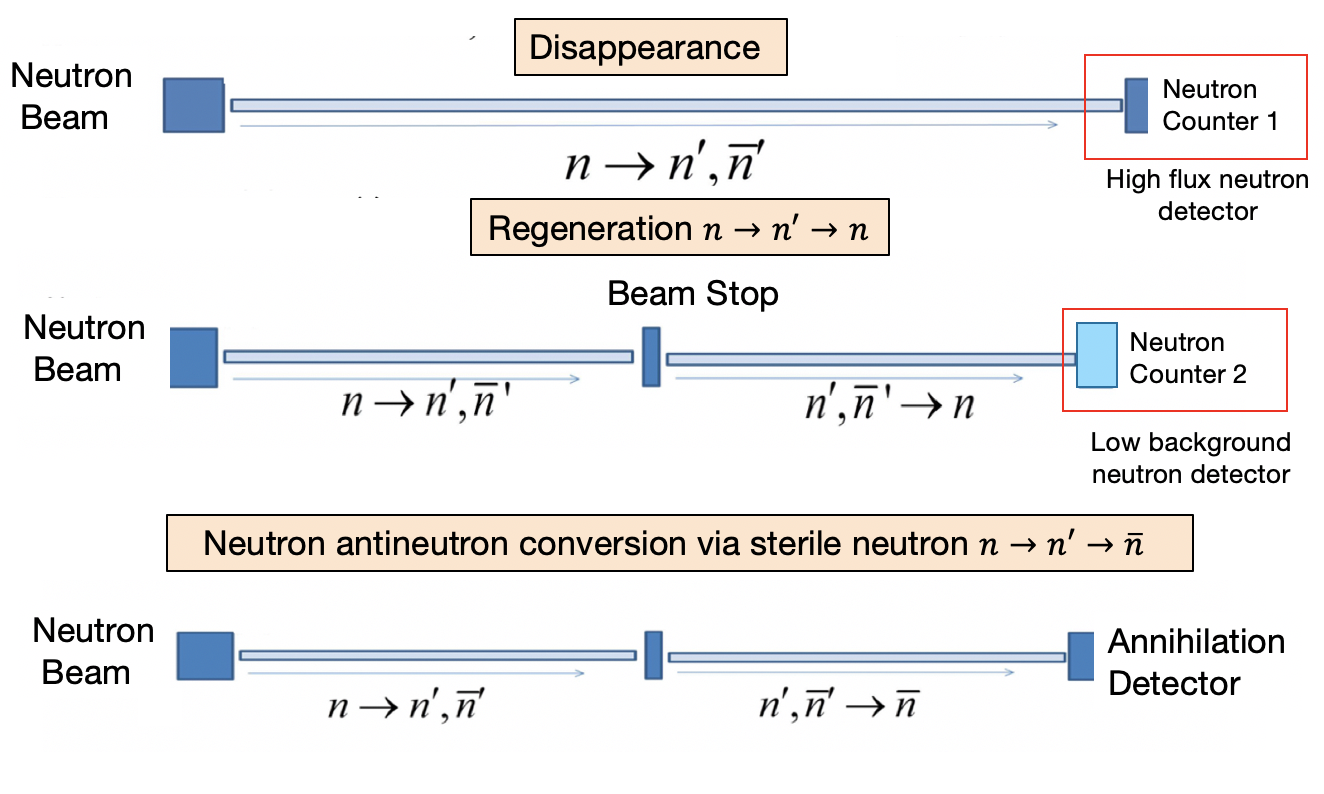}
%	\centerfloat	
	\caption{\label{fig-3chann} Illustration of the principles of searches for sterile neutrons at HIBEAM: (a) regeneration, (b) disappearance, and (c) neutron-antineutron transition via a sterile neutron state. Modes (b) and (c) require a neutron absorber to be placed at the halfway point of the beamline.}
	%form the layers (the dashed line shows one of them). The inner nested layers of a reflector starting at $z$ are constructed from the outer ones.
\end{figure}

HIBEAM consists of a number of search experiments, as illustrated in 
Figure~\ref{fig-3chann}. One such experiment looks for an anomalous loss of flux following propagation over around 50~m ("neutrons disappearance"). Neutrons can transform to sterile neutron states ($n,n'$) which escape detection. {\it Neutron regeneration} is also shown. Here, the neutron flux is absorbed in a beam stop. However, any sterile neutron states produced in front of the beam stop would propagate through and then be able to transform back to neutron states prior to detection in a neutron counter. The final mode is neutron to antineutron transformation via sterile neutron states. The antineutrons would then annihilate in a thin carbon foil enclosed by an annihilation detector, which is designed to measure the expected multi-pion state from anti-neutron-nucleon annihilation~\cite{Golubeva:2018mrz}. Different detector choices are under consideration. These include the development of a detector concept based on a TPC, hadronic range detector and lead-glass-based electromagnetic calorimeter~\cite{Yiu:2022faw}, as well as a modified version of the WASA detector~\cite{CELSIUSWASA:2008vnq}.  

The propagation region for neutrons in each of the above configurations is magnetically controlled, such that the magnetic field would match that in any dark sector, thereby ensuring degeneracy and avoiding suppression of the neutron-antineutron transition while allowing the feeble mixing to take place. The magnetic field region extends to the multi-Gauss (G) region and would be scanned in steps of several mG. The sensitivity in oscillation time of up to at least 400~s can be achieved. With the oscillation time and magnetic field sensitivities, HIBEAM can make order of magnitude improvements compared with other searches~\cite{White,Berezhiani:2017jkn,Ayres:2021zbh}.     

\begin{figure}[tbp!] 
	\centering
	\includegraphics[width=0.75\columnwidth]{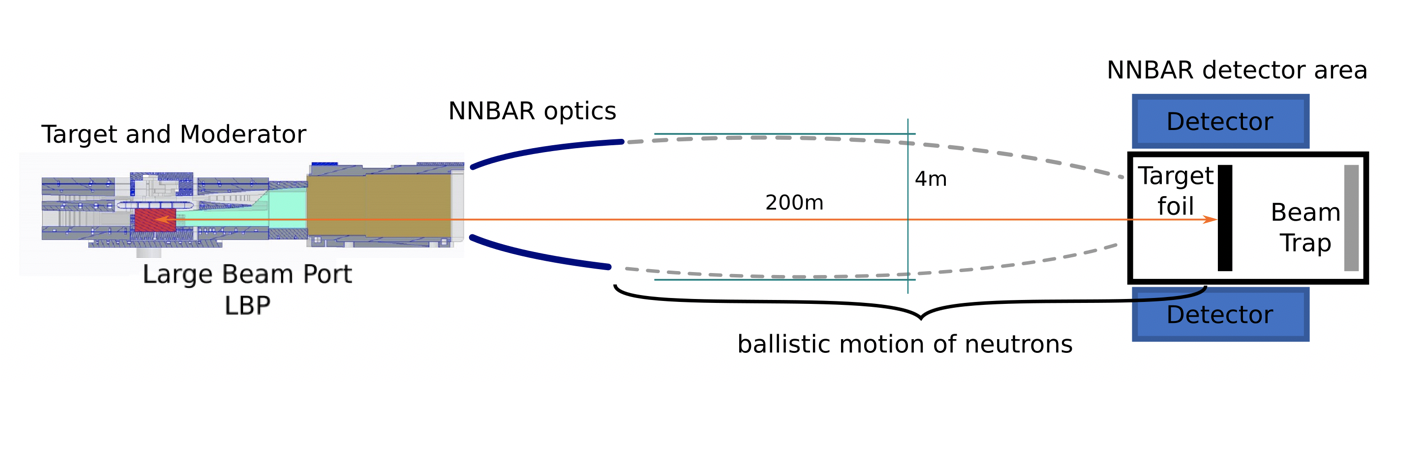}
	\caption{Schematic overview of the NNBAR experiment (not to scale).
	The moderator (source) to detector distance is 200~m. The vacuum tube and magnetic shielding are not shown.}
	\label{fig:NNBAR_schematic}
\end{figure}

In addition to the program described above, HIBEAM will also pioneer a first search for free neutrons transforming to anti-neutrons at a spallation source. Here, the propagation region is field-free ($<0.1$~mG). This pilot experiment demonstrates the ability to characterise and control spallation backgrounds, as well as allows the development and testing of neutron optics and the annihilation detector (also used in the $n \rightarrow \bar{n}$ search with sterile neutrons). The pilot would be a mini-NNBAR experiment ahead of the full NNBAR experiment which which would exploit the LBP. A schematic diagram of the NNBAR experiment is shown in Figure~\ref{fig:NNBAR_schematic}. It show the extraction of neutrons from the target and moderator complex and the focusing of neutrons over a propagation distance of 200~m to the annihilation detector area. The final sensitivity of NNBAR is expected to be around $10^3$ times that of the previous experiment~\cite{BaldoCeolin:1989qd}. The gains arise from the large flux from the LBP, the propagation length, improvements in neutronics and a three-year running period~\cite{White}. 

\begin{figure}[ht!] % replace 't' with 'b' to force it to be on the bottom
	\centering
	\includegraphics[width=0.85\columnwidth]{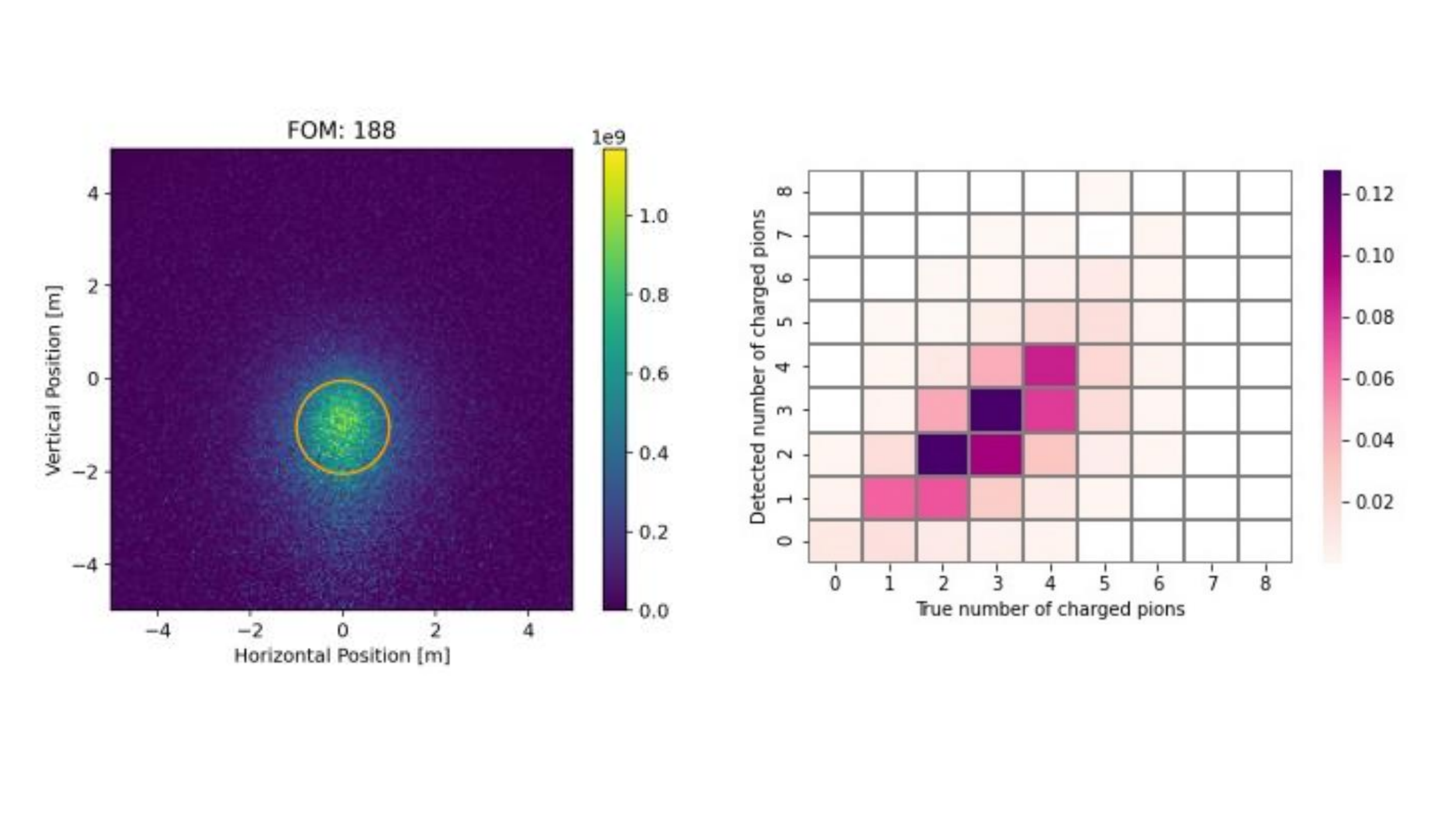}
	
 \vspace{-1.5cm}
	\caption{Left: Focused beam at the detector area in a view transverse to the beamline. Right: the reconstructed and true charged pion multiplicities.}
	\label{fig:focpi}
\end{figure}

Work towards conceptual design reports for both the HIBEAM and NNBAR stages, as well as prototype development work are ongoing. See, eg, Refs.~\cite{Backman:2022szk,Dunne:2021arq}. All aspects of the experiments, from moderator to detector design are under study. As a representative sample of the work, Figure~\ref{fig:focpi} shows the focused neutron beam over 200m from a neutron supermirror complex to the detector area and the reconstructed pion multiplicity compared with the true charged pion multiplicities. A software package incorporating the various simulation programs is has been developed~\cite{Barrow:2021deh}. 

%\subsubsection{Acknowledgements}
%The author gratefully acknowledges the Swedish Research Council for support for the design of the %HIBEAM program. 
%The work for NNBAR is part of the HighNESS project supported by the European Union Framework Programme for Research and Innovation Horizon2020 initiative for the HighNESS project under~grant agreement 951782. 

%\bibliographystyle{utphys}
%\bibliographystyle{alpha}
%\bibliography{sample}

%\end{document}

%---------------------------------
\subsection{\emph{New ideas:}Tests of Low-Scale Leptogenesis in
Charged Lepton Flavour Violation Experiments -- {\it A.~Granelli, J.~Klarić, S.~T.~Petcov}}
\label{petcov}
{\it Authors: Alessandro Granelli, Juraj Klarić, Serguey T.~Petcov <petcov@sissa.it>}
%---------------------------------

%\documentclass[11pt,a4paper]{article}
%\usepackage{dsfont}
%\usepackage[square, numbers, sort&compress]{natbib}
%\usepackage{amssymb}
%\usepackage{amsfonts}
%\usepackage{amssymb}
%\usepackage{graphicx}
%\usepackage[centertags]{amsmath}
%\usepackage[normalem]{ulem}
%\usepackage[colorlinks=true,urlcolor=black,linkcolor=black,citecolor=black]{hyperref}
%\usepackage{a4wide}
%\usepackage{colordvi}

%\begin{document}

%\noindent\textbf{Tests of Low-Scale Leptogenesis in
%Charged Lepton Flavour Violation Experiments}\\
%\noindent\textit{Authors: Alessandro Granelli$^{~a,b,c}$, Juraj Klari\'c$^{~d}$  and Serguey T.~Petcov$^{~a,b,e}$ \footnote{Also at:
%Institute of Nuclear Research and Nuclear Energy,
%Bulgarian Academy of Sciences, 1784 Sofia, Bulgaria.} \\ (speaker) $<$petcov@sissa.it$>$}\\

%\vspace{2mm}
%\noindent$^{a}$\,{\it SISSA, via Bonomea 265, 34136 Trieste, Italy.} \\
%$^{b}$\,{\it INFN, Sezione di Trieste, via Valerio 2, 34127 Trieste, Italy.} \\
%$^{c}$\,{\it IFPU, via Beirut 2, 34151 Trieste, Italy.}\\
%$^{d}$\,{\it Centre for Cosmology, Particle Physics and Phenomenology, Université Catholique de Louvain, Louvain-la-Neuve B-1348, Belgium.}\\
%$^{e}$\,{\it Kavli IPMU (WPI), UTIAS, The University of Tokyo, Kashiwa,
%Chiba 277-8583, Japan.}\\

The origin of the matter-antimatter (or baryon) asymmetry of the Universe is still a fundamental and unresolved problem in Particle Physics and Cosmology, i.e., in Astroparticle Physics. Its solution requires physics beyond that predicted by the Standard Model (SM). The mechanism of leptogenesis (LG) offers a particularly appealing solution as it relates the generation and smallness of neutrino masses to the generation of the baryon asymmetry of the Universe (BAU). In its simple realisation, a lepton charge CP-violating asymmetry is generated in the early Universe in the CP and lepton charge non-conserving decays of the heavy Majorana neutrinos of the (type-I) seesaw mechanism of neutrino mass generation. This asymmetry is converted into the BAU by the (B+L) violating but (B-L) conserving sphaleron processes, which exist in the SM and are effective at $T\sim (132-10^{12})$ GeV.

In high scale LG, the masses of the heavy Majorana neutrinos are by few to several orders of magnitude below the GUT scale $2\times 10^{16}$ GeV. This makes high scale LG practically untestable in low-energy experiments. A unique possibility to test experimentally the LG idea is provided by the low-scale scenarios based on the type-I seesaw mechanism proposed in Refs.~\cite{Pilaftsis:1997jf, Pilaftsis:1998pd, Akhmedov:1998qx, Asaka:2005pn}. In these scenarios, viable LG is possible with two or three quasi-degenerate in mass heavy Majorana neutrinos having masses even below the electroweak scale.

Of crucial importance for the low-scale LG experimental tests are also the magnitudes of the heavy Majorana neutrino couplings to the charged and neutral currents in the weak interaction Langrangian. It was shown in Ref.~\cite{Drewes:2021nqr} (see also Ref.~\cite{Abada:2018oly}) that in the case of low-scale LG based on the type-I seesaw mechanism with three heavy Majorana neutrinos $N_{1,\,2,\,3}$ having quasi-degenerate masses $M\equiv M_1\simeq M_2\simeq M_3$, the sum of their squared couplings, $\sum_{\ell, i} |(RV)_{\ell i}|^2$, $\ell = e,\,\mu,\,\tau$, $i = 1,\,2,\,3$, to the charged lepton current can be relatively large: $\sum_{\ell, i} |(RV)_{\ell i}|^2\lesssim 10^{-2}$. More specifically, in Ref.~\cite{Granelli:2022lgt}, we have derived the maximal values of the couplings of the heavy Majorana neutrinos to the electron and muon, $|\sum_i(RV)_{\mu i}^*(RV)_{e i}|$, that are compatible with successful LG in the mass range $0.1\,\text{GeV} \lesssim M\lesssim 500\,\text{TeV}$ for either vanishing or thermal initial abundances of the heavy Majorana neutrinos (denoted in what follows by VIA and TIA cases, respectively). We have shown that such couplings, which induce charged lepton flavour violating processes involving muons (i.e., $\mu$LFV processes), can be so large as to be probed in the current and upcoming experiments on $\mu \to e \gamma$ and $\mu \to eee$ decays, and $\mu - e$ conversion in nuclei. In what follows, we report briefly the results of our analysis performed in Ref.~\cite{Granelli:2022lgt}.

The MEG \cite{MEG:2016leq},
SINDRUM \cite{Bellgardt:1987du}
and SINDRUM II \cite{Dohmen:1993mp,Bertl:2006up} Collaborations reported the following experimental limits on the branching ratios of $\mu \rightarrow e\gamma$ and
$\mu \rightarrow eee$ decays,
$\textrm{BR}(\mu \rightarrow e \gamma)$ and
$\textrm{BR}(\mu \rightarrow eee)$, and on the relative
$\mu -  e$ conversion cross section in a nucleus ${}_{Z}^{A}\textrm{X}$, $\textrm{CR}(\mu\,{}^{A}_{Z}\textrm{X} \rightarrow e\,{}^{A}_{Z}\textrm{X})$ ($Z$ and $A$ being respectively the atomic and mass numbers):
$\textrm{BR}(\mu \rightarrow e\gamma) < 4.2 \times 10^{-13}$, $\textrm{BR}(\mu \rightarrow eee) < 1.0 \times 10^{-12}$, $\textrm{CR}(\mu\,{}_{22}^{48}\textrm{Ti} \rightarrow e\,{}_{22}^{48}\textrm{Ti}) < 4.3 \times 10^{-12}$, $\textrm{CR}(\mu\,{}_{~79}^{197}\textrm{Au} \rightarrow e\,{}_{~79}^{197}\textrm{Au}) < 7.0 \times 10^{-13}$ (all at $\textrm{90\% C.L.}$). The upcoming experiments MEG II on the $\mu \rightarrow e\gamma$ decay,
Mu3e on $\mu \rightarrow eee$ decay,
Mu2e and COMET on $\mu - e$ conversion in aluminium
and PRISM/PRIME on $\mu - e$ conversion in titanium plan to increase these sensitivities further, aiming to reach, respectively, $\textrm{BR}(\mu \rightarrow e\gamma) \simeq 6 \times 10^{-14}$ \cite{MEGII:2018kmf}, $\textrm{BR}(\mu \rightarrow eee) \,\sim\, 10^{-15}~(10^{-16})$ \cite{Arndt:2009}, $\textrm{CR}(\mu\,{}_{13}^{27}\textrm{Al} \rightarrow e\,{}_{13}^{27}\textrm{Al}) \sim\,
6 \times 10^{-17}$ \cite{Bartoszek:2015, Abramishvili:2020} and $\textrm{CR}(\mu\,{}_{22}^{48}\textrm{Ti} \rightarrow e\,{}_{22}^{48}\textrm{Ti}) \,\sim\, 10^{-18}$ \cite{Barlow:2011zza}.

Using the analytical expressions for $\textrm{BR}(\mu \rightarrow e \gamma)$,
$\textrm{BR}(\mu \rightarrow eee)$ and $\textrm{CR}(\mu\,{}^{A}_{Z}\textrm{X} \rightarrow e\,{}^{A}_{Z}\textrm{X})$ derived, e.g., in Ref.~\cite{Ibarra:2011xn}, which depend on the quantity $|\sum_{i=1,2,3}(RV)_{\mu i}^* (RV)_{e i}|$,
and taking into account that the mass splittings of $N_{1,\,2,\,3}$ are negligible, we have derived, as a function of the mass scale $M$, the limits on the viable LG parameter space that follow from the current limits on the $\mu$LFV \cite{MEG:2016leq, Bellgardt:1987du, Dohmen:1993mp,Bertl:2006up} and shown the potential of the upcoming experiments MEG II, Mu3e, COMET and PRISM/PRIME of testing this low-scale LG scenario. More specifically, we have shown that the indicated upcoming experiments can probe significant region of the viable LG parameter space. 

\begin{figure}
    \centering
\includegraphics[width=0.49\textwidth]{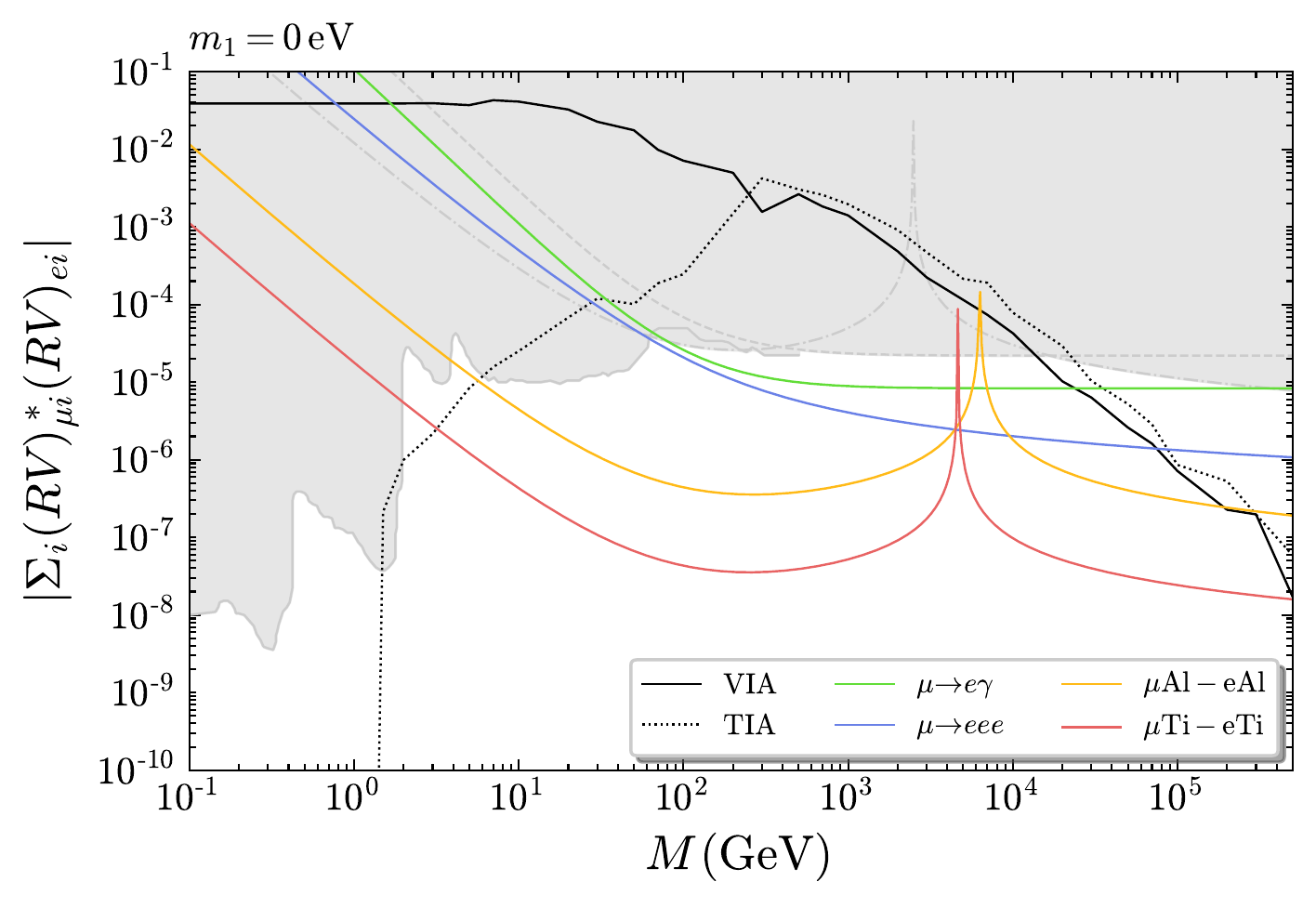}
\includegraphics[width=0.49\textwidth]{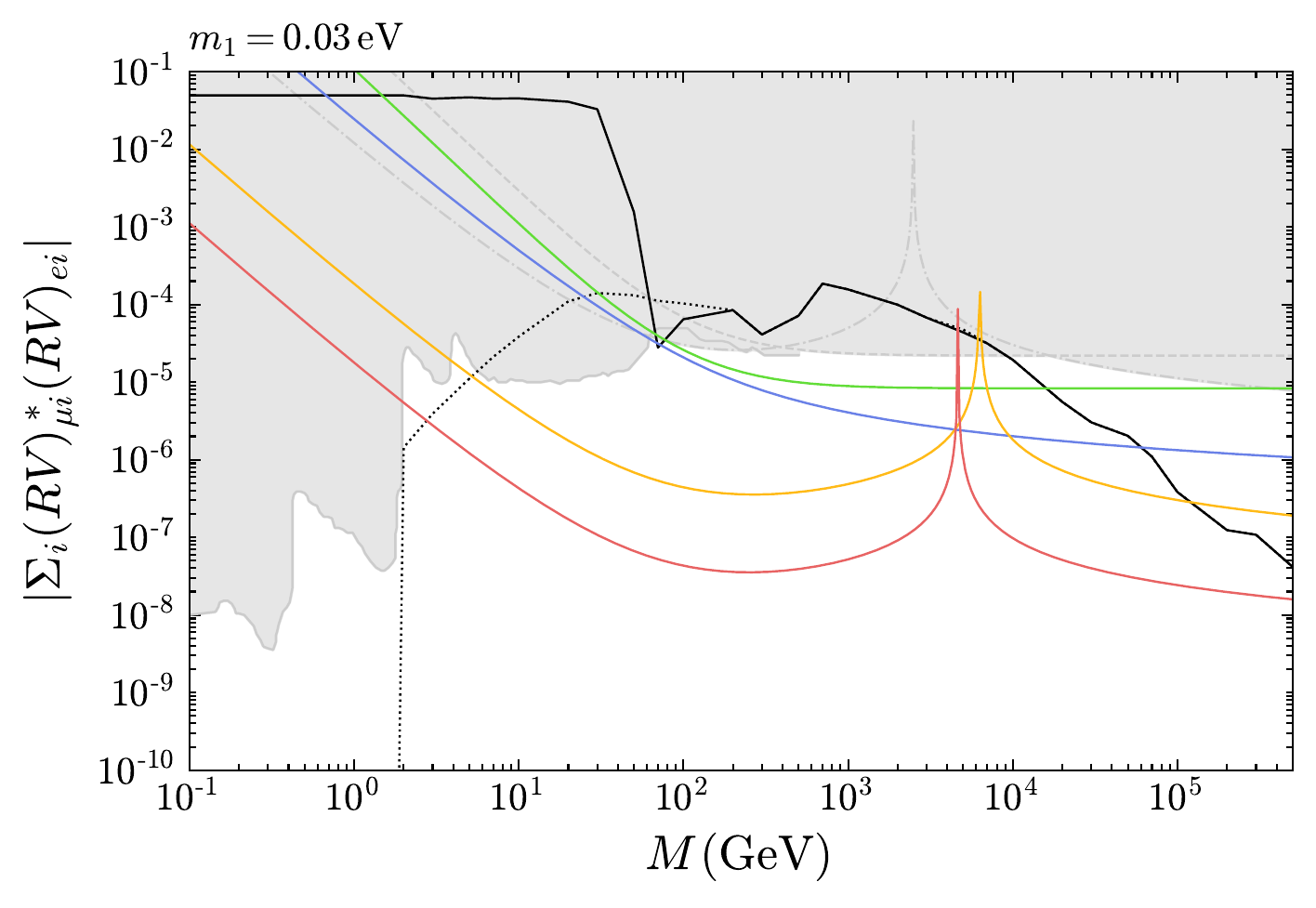}
    \caption{the region of successful low-scale LG in the
$|\sum_{i=1,2,3} (RV)^*_{\mu i} (RV)_{e i}| - M$ plane
for light neutrino mass spectrum with normal ordering with $m_1 = 0$ (left panel) and $m_1 = 0.03$ eV (right panel). The white area corresponds to the region of viable LG that is not excluded by current low-energy data \cite{Chrzaszcz_2020, MEG:2016leq, Bertl:2006up}.
%For points in the area beneath the solid (dotted) black curves, LG is viable in the VIA (TIA) case. The grey shaded region is excluded up to $M\sim 500$ GeV by low-energy experiments \cite{Chrzaszcz_2020} and above by the current upper limits $\text{BR}(\mu \rightarrow e\gamma) < 4.2 \times 10^{-13}$ \cite{MEG:2016leq} and $\text{CR}(\mu\,{}_{13}^{27}\text{Au} \rightarrow e\,{}_{13}^{27}\text{Au}) < 7 \times 10^{-13}$ \cite{Bertl:2006up}. 
The coloured lines correspond, from top to bottom, to the sensitivities of the upcoming experiments on $\mu \to e \gamma$ and $\mu \to eee$ decays, and on $\mu - e$ conversion in aluminium and titanium. See the text for further details.}
\label{fig:cLFV}
\end{figure}

Our results are shown graphically in Fig.~\ref{fig:cLFV}. In the figure, the light neutrino mass spectrum is assumed to be with normal ordering, with the lightest neutrino mass set to $m_1 = 0$ (left panel) and $m_1 = 0.03$ eV (right panel). The region of viable LG in the VIA (TIA) case is the area below the solid (dotted) black lines, while the subregion which is excluded by the current low-energy data \cite{Chrzaszcz_2020}, including the current upper
limits on $\textrm{BR}(\mu \rightarrow e \gamma)$ and on
$\textrm{CR}(\mu\,{}_{~79}^{197}\textrm{Au} \rightarrow e\,{}_{~79}^{197}\textrm{Au})$ given above, is shown in grey.
The green, blue, yellow and red lines represent, from top to bottom, the prospective sensitivities of the planned experiments on $\mu \rightarrow e\gamma$ and $\mu \rightarrow eee$ decays, as well as on $\mu - e$ conversion in aluminium and titanium. It is clear from the figure that the planned experiments on $\mu$LFV processes can probe directly significant region of the LG parameter space. 
In particular, the future MEG II and Mu3e experiments on $\mu \rightarrow e \gamma$ and $\mu \rightarrow eee$ decays will be able to probe the currently allowed LG regions, which extend respectively from $M\cong 90$ GeV to $M\cong 2\times 10^4$ GeV and from $M\cong 60$ GeV to $M\cong 7\times 10^4$ GeV in the VIA case and to slightly larger values in the TIA case; they will probe values of the parameter $|\sum_{i=1,2,3}(RV)_{\mu i}^{*}\, (RV)_{ei}|$ down to $8\times 10^{-6}$ and $1.5\times 10^{-6}$.
Except for a narrow region in the vicinity of the spike at 6.0 TeV, in the VIA (TIA) case the upcoming experiments on $\mu - e$ conversion in aluminium Mu2e~\cite{Bartoszek:2015} and COMET~\cite{Abramishvili:2020} will probe the allowed LG region within the interval $M\cong (4~(6) - 3\times 10^5)$ GeV and values of $|\sum_{i=1,2,3}(RV)_{\mu i}^{*}\, (RV)_{ei}|$ down to $2\times 10^{-7}$, while the planned experiment with higher sensitivity on  $\mu - e$ conversion in titanium PRISM/PRIME~\cite{Barlow:2011zza} will test (apart from a narrow interval around the spike at 4.5 TeV) the LG region in the range of $M\cong (2~(3) - 5\times 10^5)$ GeV and values of $|\sum_{i=1,2,3}(RV)_{\mu i}^{*}\, (RV)_{ei}|$ as small as $1.6\times 10^{-8}$.

The analysis we have performed in Ref.~\cite{Granelli:2022lgt} also revealed that, in the region of parameter space where LG is successful, the heavy Majorana neutrinos can have sizeable charged current couplings not only to the electron and muon, but to the electron, muon and tauon simultaneously. Consequently, even experiments on $\tau \to eee(\mu\mu\mu)$ and $\tau \to e(\mu)\gamma$ decays (e.g., at BELLE II \cite{BELLEIIbook}) can probe a part of the LG parameter space, although a relatively narrower one.

We can thus conclude that the upcoming and planned experiments on $\mu$LFV processes can provide unique possibility for a direct test of the low-scale LG scenario based on the type-I seesaw mechanism with three quasi-degenerate in mass heavy Majorana neutrinos, with the potential for a discovery.

\paragraph*{Acknowledgements} The work of A.G.~and S.T.P.~was supported in part by the European Union's Horizon 2020 research and innovation programme under the Marie Skłodowska-Curie grant agreement No.~860881-HIDDeN, and by the Italian INFN program on Theoretical Astroparticle Physics. S.T.P.~acknowledges partial support from the World Premier International Research Center Initiative (WPI Initiative, MEXT), Japan.
J.K.~acknowledges the support of the Fonds de la Recherche Scientifique - FNRS under Grant No.~4.4512.10.
Computational resources have been provided by the Consortium des Équipements de Calcul Intensif (CÉCI), funded by the Fonds de la Recherche Scientifique de Belgique (F.R.S.-FNRS) under Grant No.~2.5020.11 and by the Walloon Region.

%\end{document}

\clearpage
%-------------------------------------------
\subsection{\emph{New ideas:} Bounds on right handed neutrinos from observable leptogenesis -- {\it S.~Sandner}}
\label{sandner}
{\it Author: Stefan Sandner, <stefan.sandner@ific.uv.es>}

\subsubsection{Introduction}
\noindent
Leptogenesis addresses the problem of the observed baryon asymmetry of our Universe (BAU) within a framework which can simultaneously provide masses to the active neutrinos. 
The minimal type-I seesaw model in accordance with neutrino oscillations data extends the Standard Model (SM) with two Majorana singlet fermions (HNLs) that couple to the SM via the fermion portal.
It is also able to explain the BAU for heavy state masses ranging from sub-GeV to $\sim 10^{15}\,\rm{GeV}$.
An interesting scenario is that the Majorana masses are in the range of $(0.1 - 100)\,\rm{GeV}$, such that the HNLs can be produced at colliders.
In this case, the relevant process to generate the BAU is via HNL oscillations during its freeze-in~\cite{Akhmedov:1998qx, Asaka:2005pn}.
Although this model has been extensively studied in the past (see~\cite{Drewes:2017zyw} for a review), an accurate analytical understanding of the parameter space that leads to successful baryogenesis was first derived in~\cite{Hernandez:2022ivz}.
In particular, the use of parametrization-independent CP flavour invariants allows to express the analytical solutions in terms of other flavour observables.
This allows to either analytically predict the constraints on the BAU arising from putative future measurements of HNLs, CP violation in neutrino oscillations and neutrinoless double-beta decay or, alternatively, to set bounds on HNL parameters from the BAU.
In the following I review the method we derived in~\cite{Hernandez:2022ivz} to analytically solve the complete linearized set of quantum Boltzmann equations.
The solutions take into account mass effects in the interaction rates and cover all washout regimes.
In section~\ref{sec:cons} I show some of the resulting constraints from the BAU.

\subsubsection{The model and analytical approximation}
\label{sec:ana_aprox}
\noindent
The model considered is the type-I seesaw, which adds to the SM $n$ fermion singlets $N^i$. 
The Lagrangian therefore reads 
\begin{eqnarray}
\label{eq:lag}
{\cal L} = {\cal L}_{SM}- \sum_{\alpha,i} \bar L^\alpha Y^{\alpha i} \tilde\Phi N^i - \sum_{i,j=1}^n {\frac{1}{2}} \bar N^{ic} M_{Rij} N^j+ h.c.\,, \nonumber
\end{eqnarray}
where $Y$ is a $3\times n$ complex Yukawa matrix and $M_R$  is a $n\times n$ complex symmetric matrix. 
$L$ is the fermion doublet and $\tilde\Phi = i\sigma_2 \Phi^*$ is the Higgs doublet.
We consider the minimal model with $n=2$.
An approximate lepton number (LN) symmetry leads to testable mixings between the HNLs and the SM sector and, therefore, exceed the naive seesaw scaling~\cite{WYLER1983205, Gavela:2009cd} .
Assigning the LN $L(N_1) = -L(N_2) = 1$, $Y$ and $M_R$ take the following form
\begin{eqnarray}
\label{eq:textures}
Y=\begin{pmatrix}
y_{e} e^{i \beta_e} & y'_e e^{i \beta_e'}\\
y_{\mu} e^{i \beta_\mu} & y'_{\mu}  e^{i \beta_\mu'}\\\
y_{\tau} e^{i \beta_\tau} & y'_{\tau} e^{i \beta_\tau'}\ 
\end{pmatrix},\;\;
M_R=\begin{pmatrix}
\mu_1 & \Lambda \\
\Lambda  & \mu_2 
\end{pmatrix}\,.
\end{eqnarray}
Here, with $y^2 \equiv \sum_\alpha y_\alpha^2$, we have $y_\alpha'/y \ll 1$ and $\mu_i/\Lambda \ll 1$, because $y_\alpha'$ and $\mu_i$ break the LN symmetry.
In particular, this guarantees the light neutrino masses to be under perturbative control, $m_\nu = f(y'/y, \mu_i/\Lambda)$, while leading to unsuppressed HNL mixings $U^2 \simeq (yv/M)^2$, with $v = 246\,\rm{GeV}$ the Higgs vev and $M$ being the average of the physical HNL masses $\Lambda = (M_1 + M_2)/2 \equiv M$.
We will use eq.~\eqref{eq:textures} to analytically solve for the baryon asymmetry $Y_B$ by perturbing around the symmetric limit.
To do so, we make the following approximations.
We first linearize the system, assume the interaction rates to evolve only linearly with the temperature at leading order and that the lepton chemical potentials do not receive flavour cross contributions from the $B/3 - L_\alpha$ chemical potentials.
To find a closed form solution, we further need to employ an adiabatic approximation for cases in which there is a large hierarchy between the vacuum oscillation rate and thermalization rate of the right handed neutrinos, i.e. $\epsilon =\Gamma_{\rm osc}/\Gamma \ll 1$ or $\epsilon^{-1} \ll 1$.
If however $\epsilon \ll 1$ only until some temperature $ T_0 > T_{\rm EW}$ but then $\epsilon^{-1} \ll 1$ from $T_0$ down to $T_{\rm EW}$, a solution can be found via the projection of the solution found at $T_0$ onto the subsystem of the weak washout modes.
This method allows to cover the \textit{intermediate} regime which has not been considered in the literature before.
Comparing the analytical result to the full numerical solution we find an agreement within at most a factor of two.
Further imposing that the model resembles neutrino oscillations data, the parameter space gets tightly constrained and correlated.
It is described by only $6$ free parameters which are one Yukawa scale, two HNL masses and 3 phases encoding CP violation, i.e. the Dirac and Majorana PMNS phases and a high scale phase, see~\cite{Hernandez:2022ivz} for the parametrization.
Our analytical approximations for the baryon asymmetry depends on CP flavour invariants, which can be used to derive robust connections between the generation of the baryon asymmetry and other observables, which I will discuss in the next section, but more details can be found in~\cite{Hernandez:2022ivz}.

\subsubsection{Constraints from the baryon asymmetry}
\label{sec:cons}
\noindent
By employing the perturbative methods discussed in the previous section, we can derive the constraints imposed by successful baryogenesis on the masses and mixings of the HNLs as well as the CP violating phases.
Here I consider two particular examples, but see~\cite{Hernandez:2022ivz} for further details.
On the one hand, I show how to derive an absolute upper bound on the mixing of the HNLs with the active neutrinos for which leptogenesis is possible.
On the other hand, I show correlations in the flavour ratios $|U_\alpha^2|/U^2$ and implications on the amplitude of neutrinoless double-beta decay driven by $m_{\beta\beta}$.

\begin{figure}[!t]
\centering
\begin{tabular}{cc}
\hspace{-0.5cm}
\includegraphics[width=0.5\textwidth]{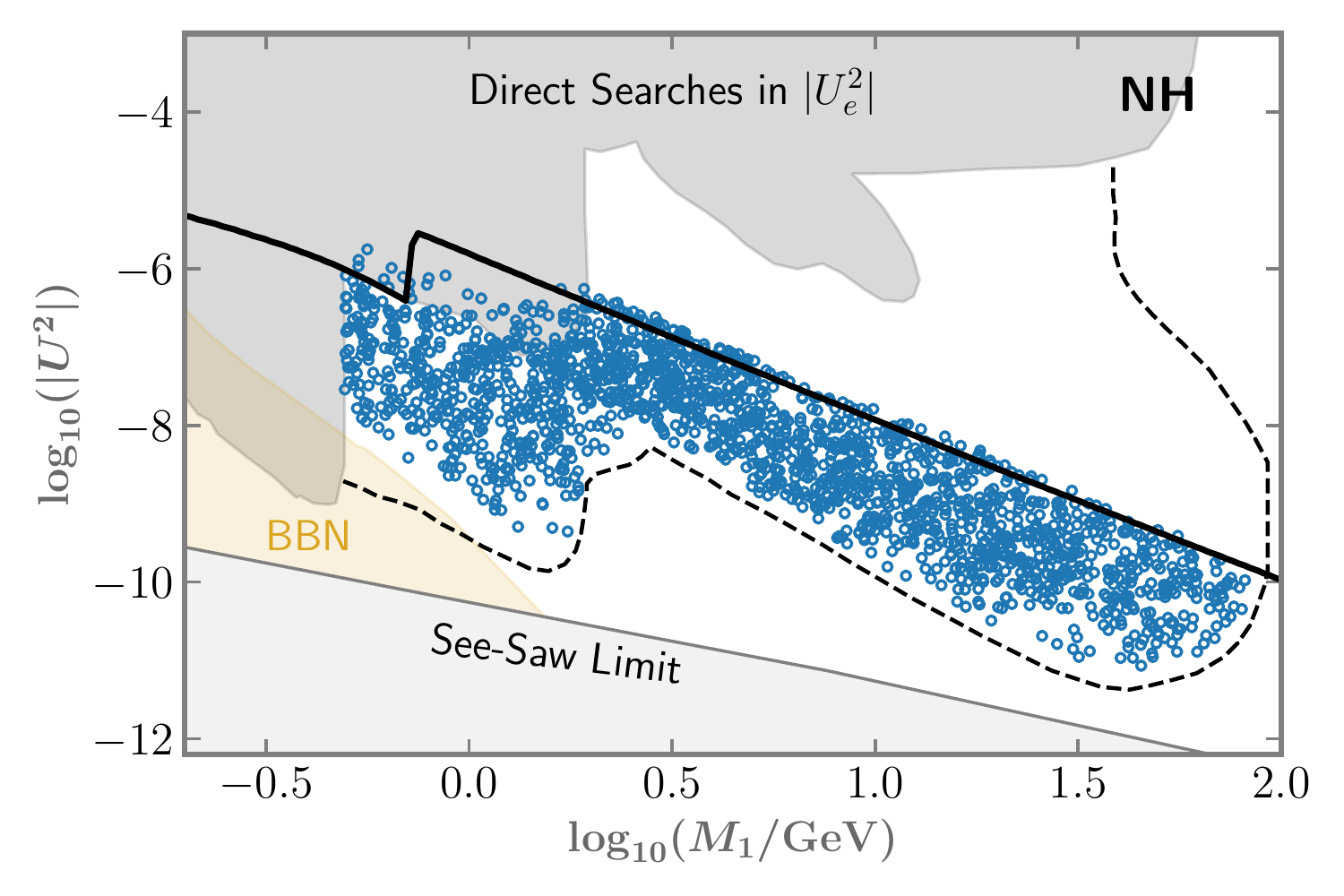} & 
\hspace{-0.55cm} 
\includegraphics[width=0.5\textwidth]{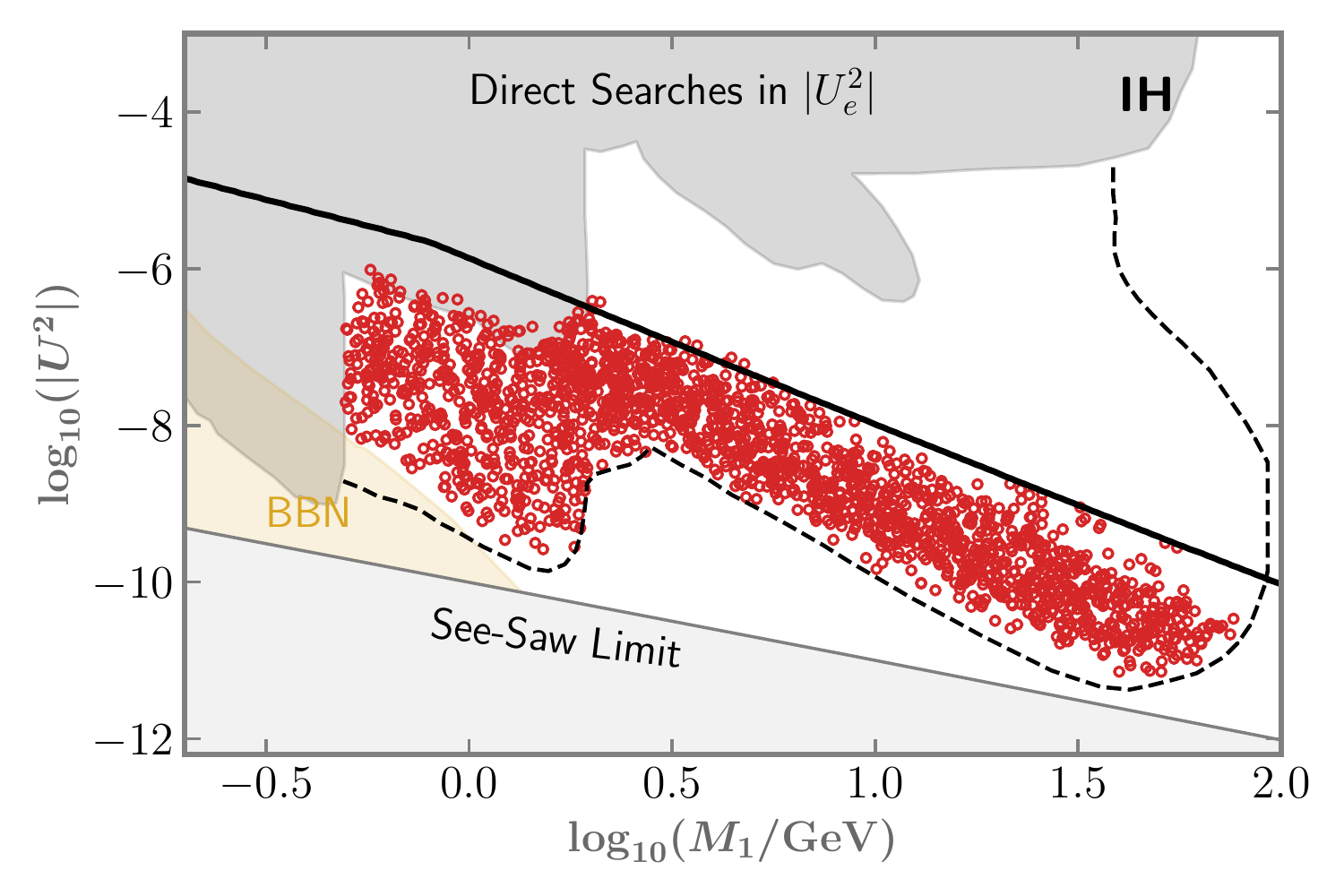}   
\end{tabular}
\vspace{-0.4cm}
\caption{ 
Numerical result of a Bayesian analysis (blue (red) points for NH (IH)) together with the analytical derived upper bound on the HNL mixing (black line).
The grey shaded region is excluded by direct searches or neutrino masses (seesaw limit), while the yellow one is excluded by big bang nucleosynthesis constraints.
}
\label{fig:scan_general}
\end{figure}

\noindent
\textbf{Analytical upper bound.}
The largest mixings of the HNLs compatible with the BAU can be achieved if one weak mode ensures the out-of-equilibrium condition~\cite{Sakharov:1991} at the electroweak phase transition.
The following physical scenario guarantees exactly this.
As long as the LN symmetry of eq.~\eqref{eq:textures} is approximately exact, the two HNLs are nearly degenerate, i.e. $\Delta M = \mu_1 + \mu_2 \ll 1$.
Furthermore, in the same basis it is evident that $N_2$ interacts with the thermal plasma only via the perturbatively small coupling $y' \ll y$.
Imposing the constraints arising from neutrino oscillations it can be shown that in fact $y' \propto y^{-1}$~\cite{Hernandez:2022ivz}.
This means that the larger the HNL mixing the farther $N_2$ is kept out of thermal equilibrium.
Such a scenario is known as the \textit{overdamped} regime, in which the vacuum oscillation length of $N_1 \to N_2$, dictated by $\Delta M$, is larger than its plasma free streaming length.
On the other hand, the analytical solution of the quantum kinetic equations reveals that the baryon asymmetry behaves as $Y_B \sim \mathcal{C}_1 y'/y^3 + \mathcal{C}_2 y'/y$ if helicity conserving interactions are weak or as $Y_B \sim \mathcal{C}_3 y'/y$ if they are strong.
This means that there is a non-trivial interplay between the generation of the light neutrino masses and the baryon asymmetry, which leads to an upper bound on the HNL mixing.
Figure~\ref{fig:scan_general} shows the upper bound analytically derived in ref.~\cite{Hernandez:2022ivz} for both, normal and inverted, hierarchies.
It is compared to the full numerical solution of a parameter space scan within the sensitivity reach of the future colliders \texttt{SHiP} and \texttt{FCC-ee}.
We can appreciate an excellent agreement.

\begin{figure}[!t]
\centering
\begin{tabular}{cc}
\hspace{-0.5cm}
\includegraphics[width=0.5\textwidth]{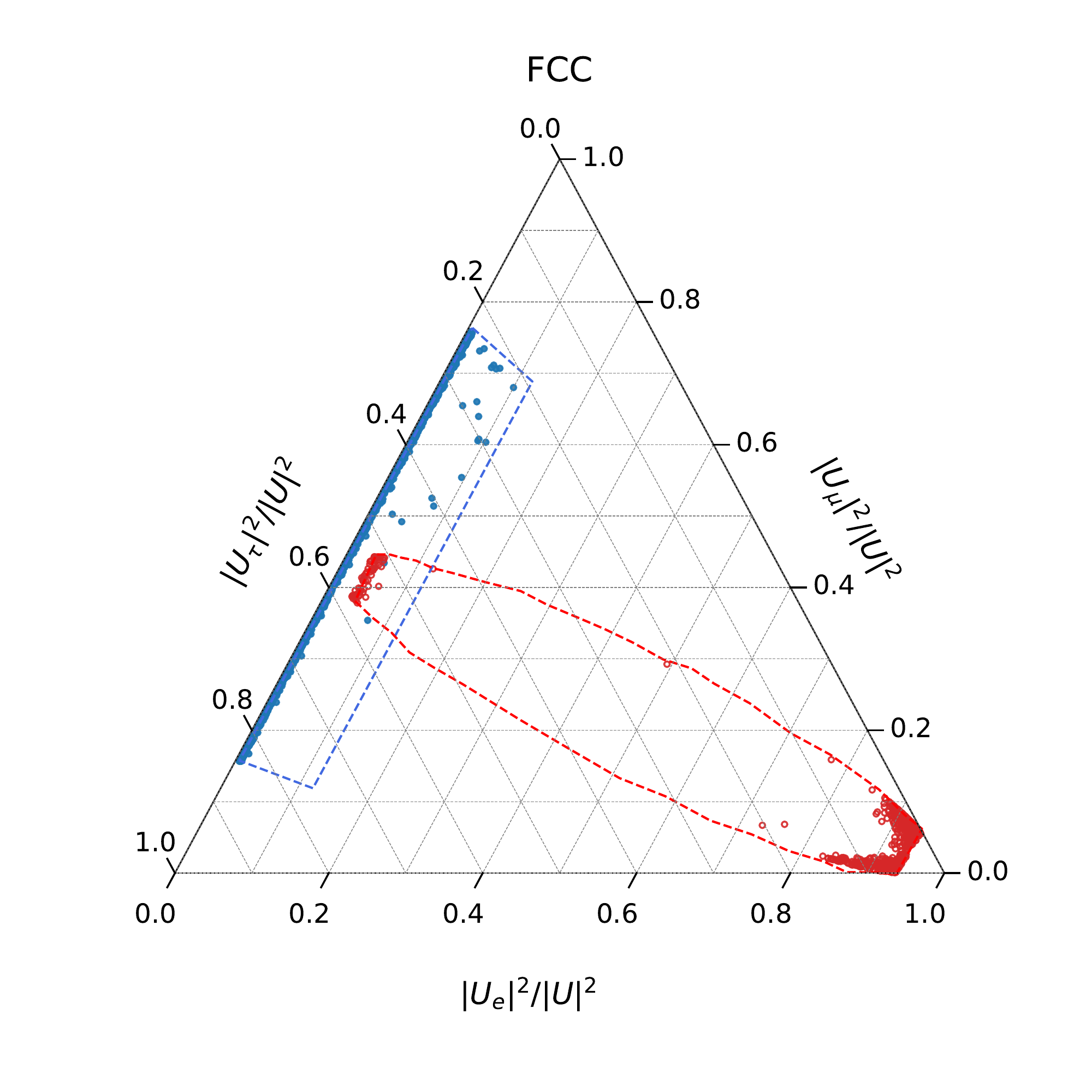} & 
\hspace{-0.55cm} 
\includegraphics[width=0.5\textwidth]{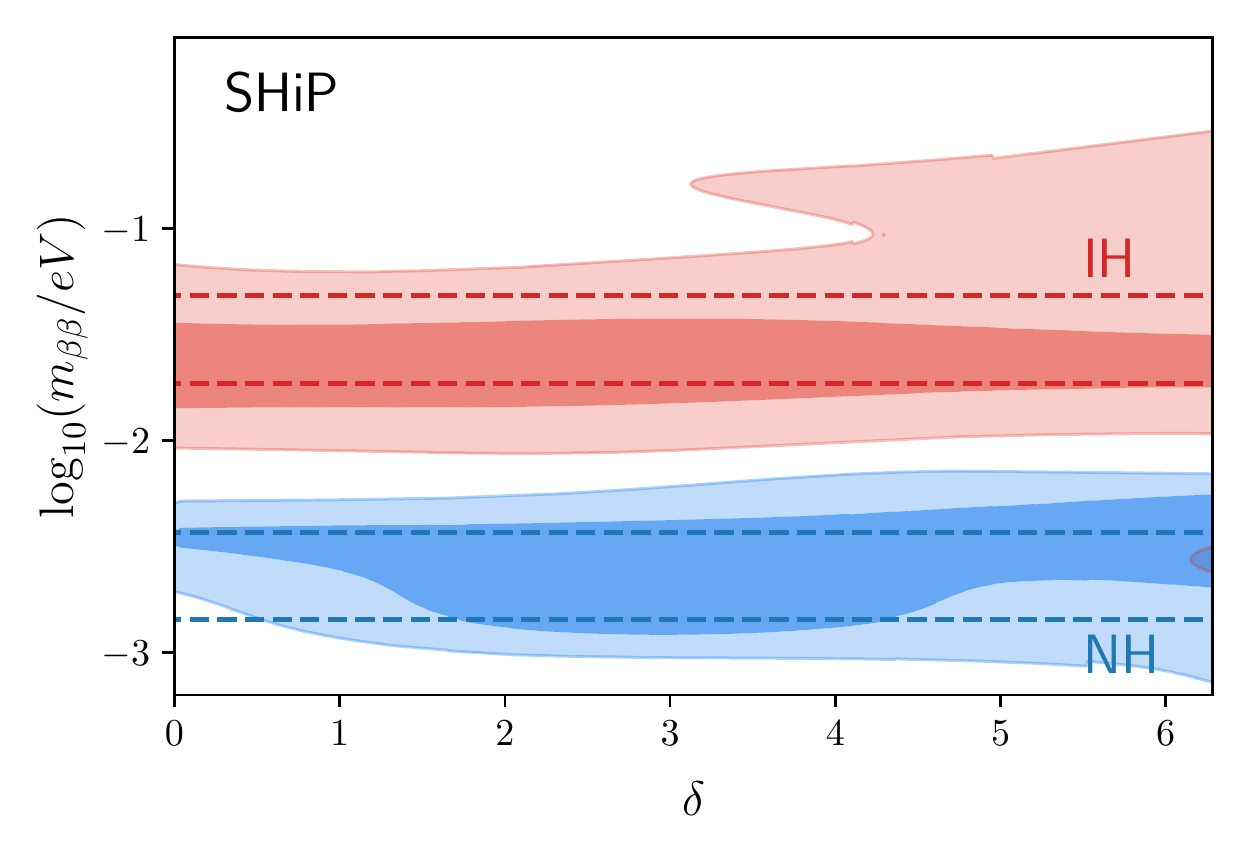}   
\end{tabular}
\vspace{-0.4cm}
\caption{ 
Solutions of a numerical scan with fixed $\Delta M/M = 10^{-2}$ for which the BAU can be explained. NH (IH) is shown in blue (red).
Left: Flavour ratio of points testable at \texttt{FCC-ee}. The dashed lines correspond to the region compatible with neutrino oscillations data.
Right: $1$ and $2\sigma$ region of points testable at \texttt{SHiP} on the plane $(\delta, m_{\beta\beta})$. The standard light neutrino contribution is contained within the dashed bands. 
}
\label{fig:constraints}
\end{figure}

\noindent
\textbf{Correlations of the BAU to other observables.}
For concreteness I will focus on the scenario in which $\epsilon^{-1} \ll 1$ at the time of the first oscillation $T_{\rm osc}$.
This is known as the \textit{fast oscillations} regime.
It requieres flavour hierarchical interactions, $\Gamma_\alpha(T_{\rm EW}) < H(T_{\rm EW})$ for some flavour $\alpha$, to achieve HNL mixings inside the sensitivity reach of \texttt{SHiP} and \texttt{FCC-ee}.
Such hierarchies are controlled by $\epsilon_\alpha = y_\alpha^2/y^2$, which is naturally expected to be $\mathcal{O}(1)$.
The farther suppressed $\epsilon_\alpha$ is compared to $\mathcal{O}(1)$ the more pronounced is the flavour selection in $|U_\alpha|^2/U^2$.
This can be seen in the left panel of fig.~\ref{fig:constraints} for an exemplary and potentially measurable relative mass splitting of $\Delta M/M = 10^{-2}$.
The right panel of figure~\ref{fig:constraints} shows the total contribution of active neutrinos and HNLs to $m_{\beta\beta}$ for HNLs with mixings to the active neutrinos which are testable at \texttt{SHiP}.
Remarkably, the presently preferred range of $\delta \geq \pi$~\cite{Esteban:2020cvm,deSalas:2020pgw} corresponds to the region where HNL effects lead to an enhancement of $m_{\beta\beta}$.

 %%%%%%%%%%%%%%%%%%%%%%%%%%%%%%%%%%%%%%%%%%%%%%%%%%%%%%%%
%%%%%%%%%%%%%%%%%%%%%%%%%%%%%%%%%%%%%%%%%%%%%%%%%%%%%%%%
%%%%%%%%%%%%%%%%%%%%%%%%%%%%%%%%%%%%%%%%%%%%%%%%%%%%%%%%

%-------------------------------------------
\subsection{\emph{New ideas:} Heavy neutrinos coupled to dark forces -- {\it M.~Hostert}}
\label{hostert}
{\it Author: Matheus Hostert, <mhostert@perimeterinstitute.ca>}

%\documentclass[aps,twocolumn,prd,showpacs,showkeys,preprintnumbers,superscriptaddress,nobibnotes,floatfix,longbibliography,notitlepage,nofootinbib]{revtex4-2}

%\pdfoutput=1
%\usepackage{amsmath}
%\usepackage{amsfonts}
%\usepackage{amssymb}
%\usepackage{mathrsfs}
%\usepackage{color}
%\usepackage{slashed}
%\usepackage{graphicx}
%\usepackage{xcolor}
%\usepackage{hyperref}
%\hypersetup{colorlinks=true,linkcolor=blue}
%\usepackage{multirow, makecell}
%\usepackage{listings}
%\usepackage{upgreek}
%\usepackage{bm}
%\usepackage[capitalise]{cleveref}
%\usepackage[normalem]{ulem}
%\usepackage{siunitx}
%\usepackage{booktabs}
%\usepackage{tabularx}
%\usepackage{physics}

%%%%%%% A few editorial macros. %%%%%%%
%\definecolor{MH}{rgb}{0.0,0.9,0}
%\newcommand{\mh}[1]{\textcolor{MH}{#1}}

%%%%%%%%%%%%%%%%%%%%%%%%%%%%%%%%%%%%%%%

%\begin{document} 

%\title{Heavy neutrinos coupled to dark forces}
%\author{Matheus Hostert}
%\email{mhostert@perimeterinstitute.ca}
%\affiliation{Perimeter Institute for Theoretical Physics, Waterloo, ON N2J 2W9, Canada}
%\affiliation{School of Physics and Astronomy, University of Minnesota, Minneapolis, MN 55455, USA}
%\affiliation{William I. Fine Theoretical Physics Institute, School of Physics and Astronomy, University of Minnesota, Minneapolis, MN 55455, USA}

%\maketitle 

The interaction of heavy neutral leptons (HNL) with the Standard Model (SM) is of sub-Weak strength, characterized by $|U_{\alpha 4}| G_F$, where $|U_{\alpha 4}|$ is the active-heavy neutrino mixing element. 
Naturally, at the scales relevant for FIPs ($\lesssim 10$~GeV), these particles have lifetimes much larger than the muon. 
In this case, the best strategies consist of looking for $N$ as a missing-energy resonance, such as in $\pi, K^+\to \ell^+ N$ decays, or for its decays in flight at high-intensity beam dump and neutrino experiments. 
An upper limit on the HNL lifetime can be derived from Big Bang Nucleosynthesis, requiring very approximately that $c\tau_N^0\lesssim 3\times 10^{4}$~km~\cite{Sarkar:1995dd,Dolgov:2000jw,Ruchayskiy:2012si,Hufnagel:2017dgo,Boyarsky:2020dzc,Sabti:2020yrt}.
We note, however, that while HNLs are not allowed to be arbitrarily long-lived, the upper bound on their lifetime is a built-in feature of the sub-Weak interactions of the minimal model.

From a phenomenological point of view, expanding our reach to FIPs will also require searching for behavior that is not described in the minimal models.
In fact, despite the impressive experimental coverage of the minimal HNL models, very few searches focus on HNLs that decay within the scale of the detector.
In addition, existing searches for prompt, lepton-number-violating HNL decays, such as $K^+\to \ell^+ N \to \ell^+ \ell^+\pi^-$, still rely on two key assumptions: that the CC branching rations of the HNL are large and that HNLs are Majorana. Neither of these assumptions may hold in non-minimal models where HNLs decay predominantly through a stronger-than-Weak force.

Furthermore, the allowed parameter space for HNL below the kaon mass is very small due to the complementarity between limits from lab-based experiments, which provide upper limits on the mixing angles, and cosmology, which provides lower limits~\cite{Bondarenko:2021cpc}.
However, as discussed in Ref.~\cite{Arguelles:2021dqn}, new forces can relax cosmological limits more than they strengthen lab-based ones, opening a larger window for HNLs below the kaon mass.

The cases of interest for non-minimal HNL models can be divided into two main groups. 
Firstly, invisibly-decaying HNLs, where the new mediators only interact with invisible particles like neutrinos and dark matter.
This happens in Majoron models, for example.
Secondly, visibly-decaying HNLs, where the mediators also talk to the SM particles, such as in a kinetically-mixed dark photon, $(B-L)$ gauge boson, or in models where the HNL has transition magnetic moments.
Recasting existing searches in models with invisible decays is usually straightforward, as many searches for HNLs do not rely on observing their decays. 
For the case of visible decays, however, the recasting will depend on the branching ratios and new production mechanisms of the HNL. 
For instance, beam dump constraints are modified due to the decoupling of the production rate and the lifetime of the HNL. 
For extremely short lifetimes, these are avoided altogether. 
Other constraints are affected due to modified branching ratios (no LNV decays, for instance) or modified detector acceptance. 
For instance, peak searches in $K\to \ell N$ are impacted, as hermetic detectors can veto the additional visible decay products of $N$.

From the theoretical point of view, it is easy to conceive of models where the SM-singlet fermions interact via additional forces that dominate over the mixing-suppressed Weak force.
For the sake of concreteness, we discuss two models where $N$ can be produced and decay through new forces but note that several examples have been discussed in the literature~\cite{Khalil:2006yi,Perez:2009mu,Okada:2014nsa,Diaz:2017edh,Nomura:2018ibs,Hagedorn:2018spx,Shakya:2018qzg,Pospelov:2011ha,Pospelov:2012gm,Pospelov:2013rha,Harnik:2012ni,McKeen:2018pbb,Cline:2020mdt,Cline:2022gcg}.

An interesting connection between HNLs coupled to dark forces can be made with experimental anomalies in the neutrino sector.
These types of models with $\mathcal{O}(100)$~MeV HNLs have been proposed as a solution of the MiniBooNE low-energy excess~\cite{Aguilar-Arevalo:2020nvw}
(for a review of the MiniBooNE and other anomalies in short-baseline neutrino experiments, see Ref.~\cite{Acero:2022wqg}).
The general idea behind these scenarios is that the new forces mediate neutrino upscattering into $N$, which can subsequently decay inside neutrino detectors into electromagnetic showers, mimicking the electron-like events observed by MiniBooNE~\cite{
Gninenko:2009ks,Gninenko:2010pr,Gninenko:2012rw,Masip:2012ke,Radionov:2013mca,Magill:2018jla,Vergani:2021tgc,Alvarez-Ruso:2021dna,Bertuzzo:2018itn,Ballett:2018ynz,Ballett:2019pyw,Abdullahi:2020nyr,Datta:2020auq,Dutta:2020scq,Abdallah:2020biq,Abdallah:2020vgg,Hammad:2021mpl,Dutta:2021cip,Fischer:2019fbw,Chang:2021myh,Abdallah:2022grs}.
Below, we will discuss new progress in constraining the models invoked in these explanations using existing accelerator neutrino data.

%%%%%%%%%%%%%%%%%%%%%%%%%%%%%%%%%%%%%%%%%%%%%%%%%%%%%%%%%%%%%%%%%%%%%%%%%%%%%%%%%%%%%%%%%%
\subsubsection{HNLs and Dark Photons}

\begin{figure}[t]
    \centering
  \includegraphics[width=0.49\textwidth]{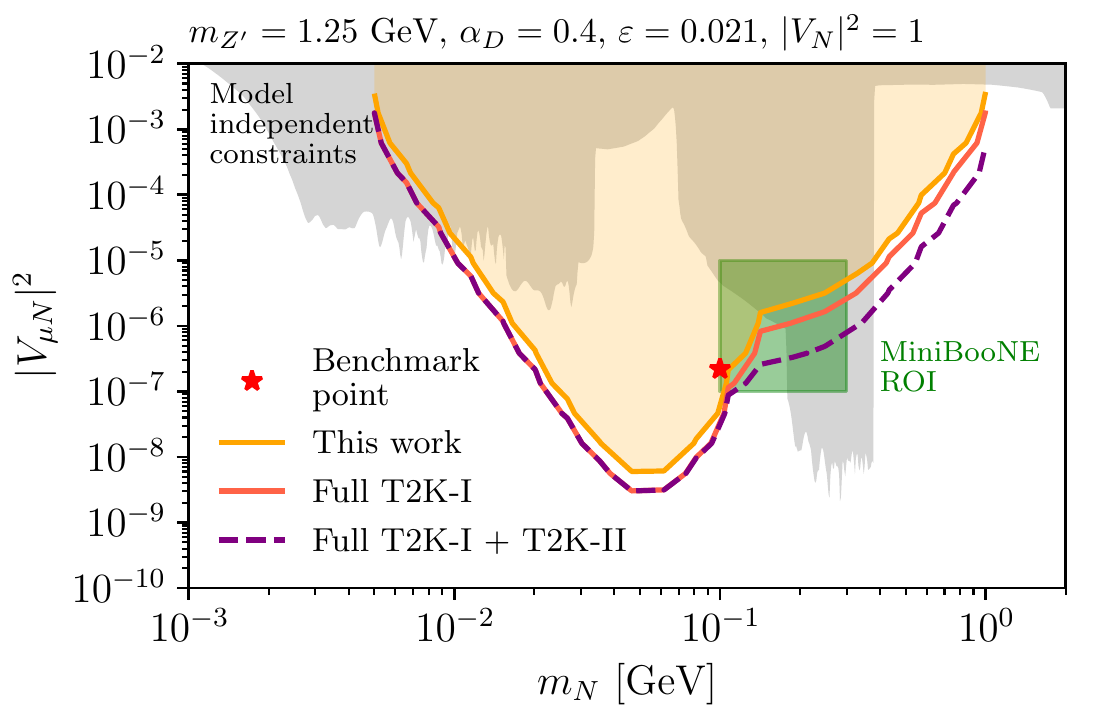}
    \caption{Parameter space of a simplified model where an HNL $N$ is produced and decay via dark photon interactions. In orange, we show new constraints derived from T2K data~\cite{Arguelles:2022lzs}.
    The signature is the upscattering of neutrinos into $N$ in the ND280 near detector, and its subsequent displaced decay into electron-positron pairs inside the Gaseous Argon TPCs, $N\to \nu e^+e^-$.
    \label{fig:HNL_darkphoton}
    }
\end{figure}

The first type of model we consider is that of HNLs in a secluded $U(1)_D$ gauge symmetry.
In this kind of dark sector, $N$ can interact with the massive dark photon mediator $Z^\prime$ as well as with the dark higgses that break the symmetry.
The mediators, in turn, can interact with the rest of the SM via kinetic or higgs mixing portals.
The interactions of $N$ with the rest of SM fermions will then be proportional to multiple portal couplings, such as the product of neutrino mixing and kinetic mixing parameter.
Specific realizations of these models have been discussed in Refs.~\cite{Bertuzzo:2018ftf,Ballett:2019pyw,Abdullahi:2020nyr}.

For the purposes of the phenomenology, aiming to cover the simplest models of HNLs coupled to dark forces, we consider a simplified model of an HNL $N$ and a kinetically-mixed dark photon.
The relevant interaction Lagrangian is given by
\begin{equation}\label{eq:HNL_darkphoton}
    \mathscr{L}_{\rm int} \supset  Z^\prime_\mu \left(  e \epsilon {J}_{\rm EM}^\mu +  g_D \sum_{i,j}^{n+3} V_{i j} \overline{\nu_i} \gamma^\mu \nu_j \right),
\end{equation}
where $\nu_i$ stands for one of the neutral lepton states, $i \in \{1,\dots, 3+n\}$, where $n$ is the number of HNLs.
The interaction vertices $V_{ij}$ are model-dependent and are proportional to the mixing between the neutral leptons. 
For simplicity, we define the interaction vertex between low-energy flavor states,$<{\hat{\nu}_{\alpha}} = \sum_{i=1,2,3} U_{\alpha i}^* <{\nu_i}$ , and the heavy neutrino $N$ as $V_{\alpha N} \simeq \sum_{i\le3} U_{\alpha i}^* V_{i N}$, which is constrained to be small.
This is the parameter that controls the interactions of $N$ with active neutrinos.
The decay of the HNL $N$ will also depend on the model considered.
If $N = \nu_5$, for example, its decay can be much faster than $\nu_4$ as the dark photon interactions between states with $i > 3$ are much stronger than those involving light neutrinos.
To account for this possibility, we define the parameter $|V_N|^2 = \sum_{i<N} |V_{iN}|^2 < 1$, which will allow us to decouple the $N$ lifetime from its production cross sections.
This neglects the mass splitting between $N$ and the daughter neutral leptons with $3 < i < N$, which can be important for the energy spectrum of the final states.
However, this is not important for the limits discussed below and allows us to consider short-lived HNLs with minimal addition to the number of parameters.

The idea behind the UV completions of Eq.\ref{eq:HNL_darkphoton} is to mix $N$ with fermions that are charged under the new gauge symmetry, $\nu_D$.
The breaking of $U(1)_D$ and the electroweak symmetry then leads to the mixing between all neutral leptons.
The mixing can be induced by two types of operators,
\begin{equation}
(\overline{L} \tilde{H}_D)\nu_D \text{ and }
(\overline{L} \tilde{H})(\Phi \nu_D).
\end{equation}
In the first case, new scalar doublets, $H_D$, charged under both the dark symmetry and  $SU(2)_L$ are required.
In the second method, two new ingredients in addition to $\nu_D$ are needed: the dark Higgs $\Phi$, which breaks the $U(1)_D$, and a new mediator particle that completes the dimension-5 operators.
For instance, a fully sterile neutrino, $\nu_s$, can realize this case by means of two neutrino portal interactions, $ \mathcal{L} \supset y_1 (\overline{L}\tilde{H})\nu_s + y_2 \overline{\nu_s}(\nu_D \Phi)$. 

The most striking signature in these models is that of coherent neutrino upscattering on nuclei $A$,
\begin{equation}
    \nu_\alpha A \to (N \to \nu \ell^+\ell^-) A,
\end{equation}
where $\ell$ can be an electron or muon.
For $e^+e^-$ pairs, the boost of $N$ can give rise to very collimated showers, faking single photon signatures in neutrino detectors.
This has been used to explain the MiniBooNE excess in Refs.~\cite{Bertuzzo:2018itn,Ballett:2018ynz,Ballett:2019pyw,Abdullahi:2020nyr}.

Strong constraints on these models can be placed by detectors with a combination of high and low-density materials, like the near detector of T2K, ND280~\cite{T2K:2011qtm}.
Because of the large-Z material of the $\pi^0$ detector, a combination of iron, lead, and water at the front of the detector, coherent-upscattering production of HNLs is enhanced.
Downstream of the detector, ND280 contains three gaseous argon (GAr) time-projection-chambers (TPCs) and three fine-grained scintillator detectors (FGDs). 
Because of their low density, the number of neutrino interactions is very small, and HNL decay signatures can be searched for in a background-free environment~\cite{Abe:2019kgx}.
Furthermore, a 0.2~T magnetic field helps to separate the collimated $e^+e^-$ pairs inside the detector.
While T2K has only searched for the decay in flight (DIF) of HNLs~\cite{Abe:2019kgx}, new constraints on the upscattering signature can be obtained by making use of the $N \to \nu e^+e^-$ DIF channel~\cite{Abe:2019kgx} as well as the photon-like sideband of the $\nu_e$CCQ measurement by ND280~\cite{T2K:2020lrr}.
The latter measures the $e^+e^-$ from photon conversions in the FGD.

In Ref.~\cite{Arguelles:2022lzs}, new constraints were found using a dedicated Monte-Carlo and a Kernel-Density-Estimator to predict the differential event rate as a function of the several different mass and coupling parameters of the theory.
The resulting constraints in a slice of parameter space are shown in Figure \ref{fig:HNL_darkphoton}.
The transition from a long-lived HNL to a short-lived HNL can be seen around $m_{N}\sim 150$~MeV.
At that point, HNLs produced in the heavy $\pi^0$ detector decay before reaching the fiducial volume of the GAr TPCs.
In that case, the dominant constraint is the one obtained from the FGD.
Also shown are the projections for the total number of POTs expected at T2K with the same ND280 design (Full T2K-I) as well as the combination with the POTs expected with a new design of ND280 (Full T2K-I + T2K-II).
In T2K-II, the $\pi^0$ detector at the front of ND280 will be removed to give space for two SuperFGDs, an improved design of the FGD detectors~\cite{Abe:2016tii}.

Other future directions to further constrain HNLs coupled to dark photons include direct searches for $e^+e^-$ pairs in liquid argon detectors of the Short-Baseline Neutrino program at Fermilab~\cite{Machado:2019oxb}. 
Kaon decays can also provide strong limits on the direct production and decay of HNLs.
NA62 could search for new resonances in the decay $K^+\to \ell^+ (N\to \nu e^+e^-)$, where in the case the dark photon is produced on-shell, both the HNL and the dark photon masses could be reconstructed~\cite{Ballett:2019pyw}.
Finally, searches for semi-visible dark photons at $e^+e^-$ colliders like Belle-II and fixed-target experiments like NA64 can constrain the production of HNLs in dark photon decays~\cite{Abdullahi:2020nyr}.

%%%%%%%%%%%%%%%%%%%%%%%%%%%%%%%%%%%%%%%%%%%%%%%%%%%%%%%%%%%%%%%%%%%%%%%%%%%
\subsubsection{HNLs with Transition Magnetic Moments}

Another possibility we discuss is that of an HNL with transition magnetic moments (TMM). 
In this case, at low energies, the only new particles to be considered is the HNL, $N$.
Its new interactions then proceed through a dimension-five TMM operator,
\begin{equation}\label{eq:dipole}
     \mathscr{L}_{\rm int} \supset  d_{\alpha N}\, \overline{\nu_\alpha} \sigma_{\mu\nu}  F^{\mu \nu} N_R  + \text{ h.c.}
\end{equation}
At the SU$(2)$-invariant level, the operator arises at dimension six and can generate transition moments via the $W$ and $Z$ boson. 
Because the photon is massless, however, the operator in Eq. \ref{eq:dipole} dominates low-energy observables.
In principle, on top of its TMM, HNL can still interact with light neutrinos via mixing.
In the TMM model, this mixing may be large, as it arises from the Dirac mass term, $m_D \overline{\nu}_\alpha N_R$, which is correlated with $d_{\alpha N}$~\cite{Voloshin:1987qy}.
Nevertheless, for simplicity, it is assumed that the effects of mixing are much smaller than those of the TMM, implying that some underlying mechanism is at play to suppress the Dirac mass term~\cite{Voloshin:1987qy,Barbieri:1988fh,Babu:1989px,Babu:1989wn,Leurer:1989hx,Lindner:2017uvt,Babu:2020ivd}.

\begin{figure}[t]
    \centering
  \includegraphics[width=0.49\textwidth]{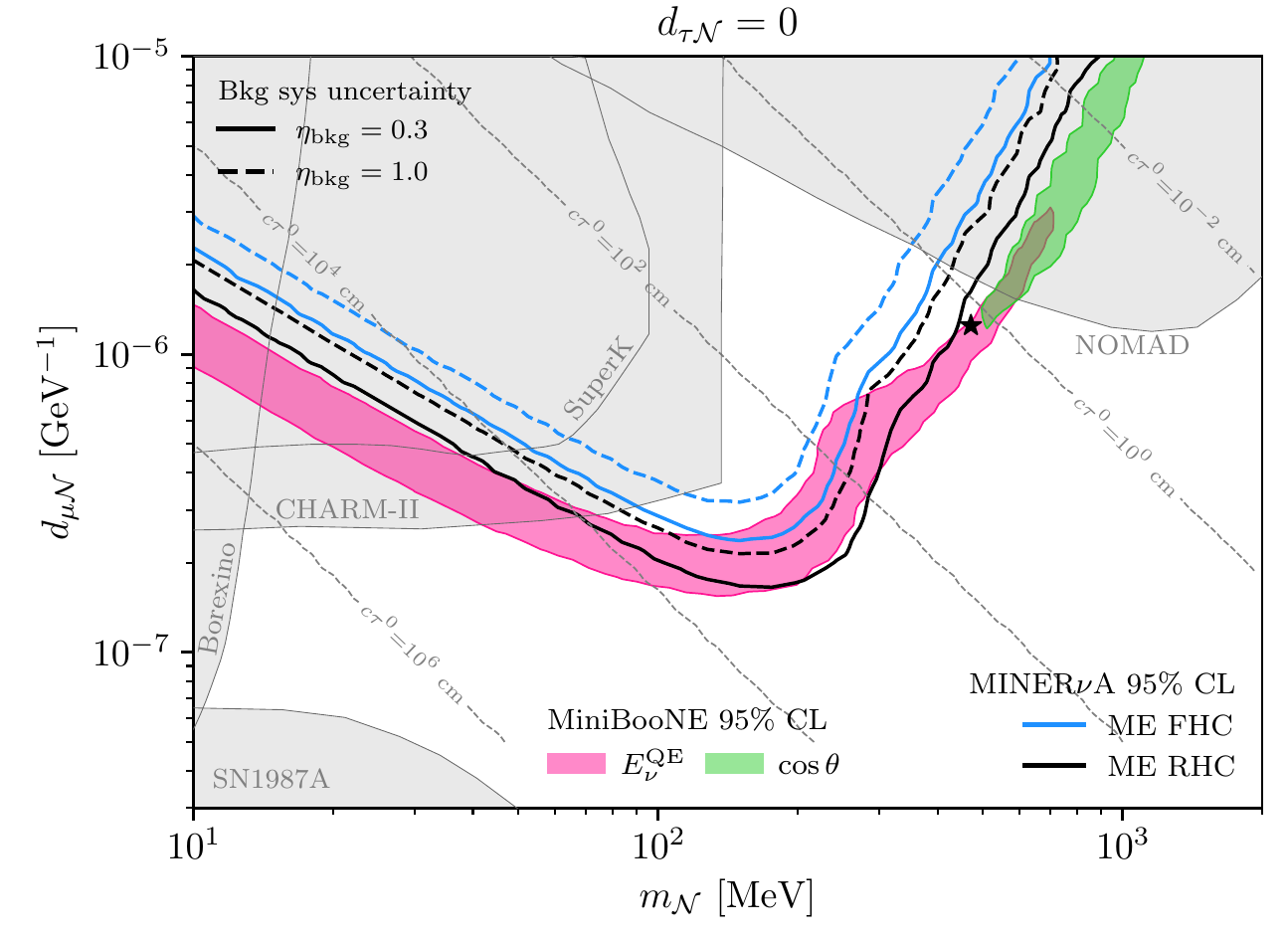}
    \caption{The TMM of an HNL $N$ as a function of its mass.
    The solid (dashed) lines show the MINERvA neutrino and antineutrino mode constraints on the dipole coupling using nominal (conservative) systematic uncertainties on the photon-like background.
    The best-fit regions to explain the MiniBooNE energy and angular spectrum are shown in pink and green, respectively.
    \label{fig:HNL_TMM}
    }
\end{figure}

Constraints on the TMM operator have been derived using a variety of low-energy data, including neutrino experiments like NOMAD~\cite{Gninenko:1998nn}, Borexino and Super-Kamiokande~\cite{Plestid:2020vqf,Gustafson:2022rsz}, and LSND and MiniBooNE~\cite{Magill:2018jla}.
A collection of constraints, including astrophysical ones, is presented in Refs.~\cite{Magill:2018jla,Brdar:2020quo}.

Similarly to the case of dark photons, active neutrinos can coherently upscatter on nuclei, $A$, via the exchange of a photon and produce $N$.
Depending on its lifetime, $N$ can then decay into a single photon inside the active volume of neutrino detectors,
\begin{equation}
    \nu A \to (N \to \nu \gamma) A.
\end{equation}
For smaller masses, the lifetime of $N$ can be significant, so its production in the dirt between the target and the detector should be considered.
This has been put forward as an explanation of the MiniBooNE and LSND puzzles~\cite{Gninenko:2009ks,Alvarez-Ruso:2021dna}.
A detailed fit to the MiniBooNE energy and angular spectrum were recently performed~\cite{Vergani:2021tgc,Kamp:2022bpt}.

In Figure \ref{fig:HNL_TMM}, we show new constraints on the parameter space of a TMM between muon-neutrinos and a HNL~\cite{Kamp:2022bpt}.
These limits were obtained using MINERvA data on elastic neutrino-electron scattering~\cite{Park:2013dax,Valencia:2019mkf,MINERvA:2022vmb}.
Two MINERvA Medium-Energy (ME) beam measurements have been considered. 
One using the beam in forward-horn-current (FHC) and another in reverse-horn-current (RHC) mode. 
These enhance the number of neutrinos and antineutrinos in the beam, respectively.
The main measurement relied on separating single electrons from photons using the showers energy deposition, $dE/dX$.
By using the photon-like $dE/dX$ sample (the sideband of the analysis), new constraints can be placed on the number of forward-going single photons in the MINERvA detector.
Because this sample is used to tune the Monte-Carlo predictions, we assign a conservative systematic uncertainty to the backgrounds.
In the nominal case, we take an overall $30\%$ and in the conservative case a $100\%$ overall normalization uncertainty.
The latter covers the worst-case scenario where the entire large $dE/dX$ sample is mis-modelled.

A similar analysis was employed in Ref.~\cite{Arguelles:2018mtc} to constrain the dark photon model above as well, exploring the misidentification of collimated $e^+e^-$ pairs as a single photon.
Future progress to constrain these models can be made at the liquid argon detectors of the SBN program, as well as future experiments like DUNE~\cite{Atkinson:2021rnp,Schwetz:2020xra}.

%\bibliographystyle{apsrev4-1}
%\bibliography{lib}{}

%\end{document}

%-------------------------------------------
\subsection{\emph{New ideas:} How to simplify the reinterpretation of FIP searches -- {\it J.-L.~Tastet}}
\label{tastet}
{\it Author: Jean-Loup Tastet, <jean-loup.tastet@uam.es>}
%-------------------------------------------

\subsubsection{Motivation}

Virtually all models of feebly-interacting particles depend on some free parameters which aren't predicted by theory.
Typical parameters include the mass of the new particle, mixing angles with Standard Model particles, a kinetic mixing coefficient, a mass scale and/or a Wilson coefficient.
Since we are considering \emph{feebly}-interacting particles, in general at least one of these parameters should be small.
The sensitivity of a given experiment to a FIP candidate largely depends on those model parameters, such that the non-observation of a signal lets us constrain the set of allowed parameters (usually by placing an upper bound on the ``small'' parameter as a function of the remaining ones).
However, different experiments or measurements may depend on model parameters differently, potentially leading to complementary constraints.
When combined, such constraints can lead to stronger limits than the ones obtained by any individual experiment.
This is especially true when considering a model that makes strong cosmological and/or astrophysical predictions (such as e.g.\ the $\nu$MSM~\cite{Asaka:2005pn,Asaka:2005an}) as shown recently in ref.~\cite{Bondarenko:2021cpc}, but this applies more generally to any global scan or Bayesian analysis.
However, performing such a combination requires knowing the limits on the set of all model parameters.
Therefore, in order to perform such a study without having to reinterpret (with varying accuracy) the experimental results, each experiment should report results for every possible combination of model parameters.

\begin{figure}
    \centering
    \includegraphics[width=\textwidth]{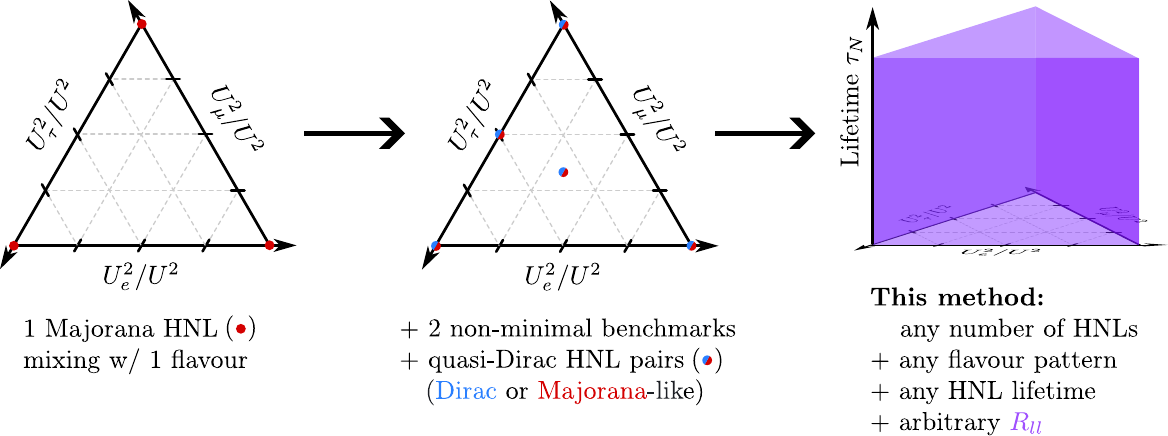}
    \caption{Evolution of the interpretation of HNL searches: from simplified benchmarks featuring a single Majorana HNL (\textbf{left}), to more realistic benchmarks featuring (in addition) quasi-Dirac HNLs (\textbf{center}), and finally to the method proposed here, which allows interpreting the results throughout the entire parameter space (\textbf{right}). $U_{\alpha}^2$ denotes $|\Theta_{\alpha}|^2$ or its sum over indistinguishable mass eigenstates, when applicable.}
    \label{fig:evolution-of-hnl-interpretation}
\end{figure}

This is usually deemed impractical, and with good reason: experiments typically perform extensive validation of their signal and background models, which may quickly become time-consuming when the number of parameter combinations to test becomes formally infinite.
Instead, they report results for a small set of so-called ``benchmark models'' which, as a whole, should be representative of the phenomenology of the model.
This may lead to problems if the benchmarks are too simple, and miss some important phenomenological aspects of the complete models.
This is the case for instance with heavy neutral leptons (HNLs).
Many experiments have reported limits on a simplified model consisting of one Majorana HNL mixing with a single neutrino flavour (see fig.~\ref{fig:evolution-of-hnl-interpretation}, left).
However, it was later noticed \cite{Drewes:2016jae,Caputo:2017pit} that such a model cannot account for the observed neutrino masses (assuming that they are produced through a low-scale seesaw mechanism).
Worse, low-scale seesaw models may lead to an approximate lepton number conservation \cite{Kersten:2007vk} (but see also \cite{Drewes:2019byd,Fernandez-Martinez:2022gsu}) that may suppress the lepton-number-violating signatures that some experiments rely on.
In this example, the simplified benchmarks were clearly inadequate.
In order to (partially) address this problem, two additional benchmark points have been proposed in the previous edition of these proceedings \cite{Agrawal:2021dbo}, and later slightly simplified in ref.~\cite{Drewes:2022akb} (see fig.~\ref{fig:evolution-of-hnl-interpretation}, center).
These benchmarks have recently been used by ATLAS to report the results of a search for displaced HNLs \cite{ATLAS:2022atq}. In this search, the limits did not vary much across benchmarks, which suggests that the parameter space was well covered, without any obvious blind spots.
However, those new, more realistic benchmarks still do not address the issue of combining constraints.

In an effort to reconcile the seemingly conflicting requirements of extensive validation and broad parameter space coverage, we propose a rather generic method that leverages the \emph{scaling properties} shared by many FIP signatures in order to express the sensitivity/limits for the entire set of model parameters (see fig.~\ref{fig:evolution-of-hnl-interpretation}, right).
Similar methods have already been discussed in refs.~\cite{SHiP:2018xqw,Abada:2018sfh,Tastet:2021vwp,Abada:2022wvh}.
We pay particular attention to the data which, if published by experiments along with their results, would enable theorists to employ this method.

\subsubsection{Signal scaling}

Consider a feebly-interacting particle that interacts with the Standard Model through a set of small couplings $\theta_1,\dots,\theta_N$ (e.g.\ $\Theta_e, \Theta_{\mu}, \Theta_{\tau}$ for HNLs).
Because of their smallness, we can safely perform a tree-level expansion in these couplings and ignore all higher-order corrections.
For the sake of simplicity, in what follows we shall focus on searches for \emph{decaying} FIPs whose width is sufficiently small that we can use the narrow-width approximation, i.e.\ the FIP is produced on-shell in the collision/decay of Standard Model particles, propagates (possibly over a macroscopic distance), and finally decays back to a final state containing Standard Model particles.
If the FIP is not observed in the experiment (e.g.\ in missing mass searches), the method remains generally applicable, but the scaling properties will be different because the parameters involved in the decay do not play a role any more.
Similarly, if the FIP re-interacts with Standard Model particles (e.g.\ in an emulsion target), the scaling properties will again be different and may e.g.\ involve the interaction length of the FIP instead of its decay width.

\begin{figure}[!ht]
    \centering
    \includegraphics[height=0.86\textheight]{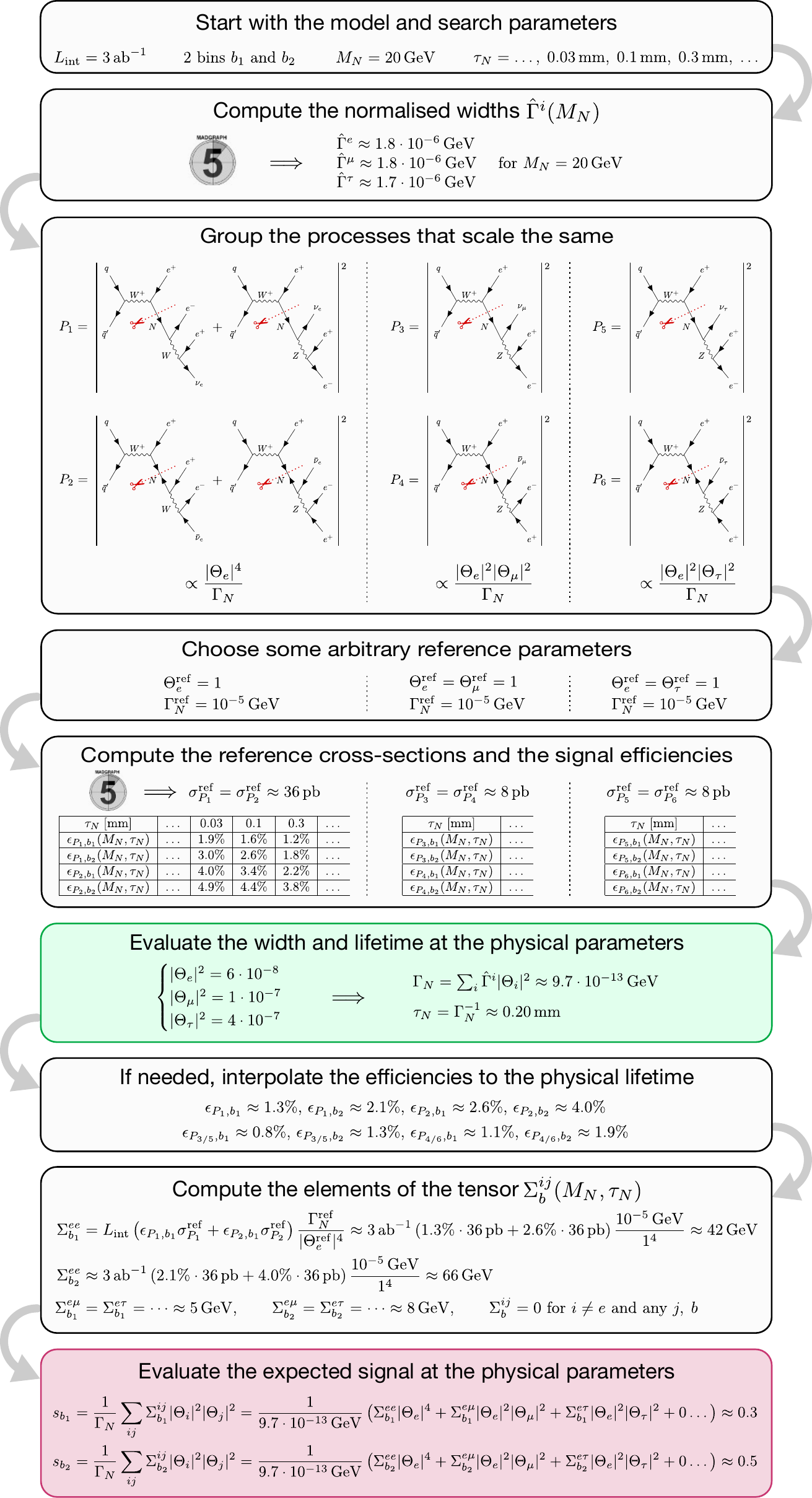}
    \caption{Worked out example of the signal rescaling method, for one Majorana HNL and the $e^+e^+e^-$ trilepton final state, using two bins and dummy (but not unrealistic) values for the efficiencies. Here the interpolation to the physical lifetime is done before computing the tensor elements. In practice it can be beneficial to swap these steps and to precompute the $\Sigma$ tensor on an $M_{\mathrm{FIP}} \times \tau_{\mathrm{FIP}}$ grid, and to only perform the interpolation when evaluating the expression for~$s_b$.}
    \label{fig:signal-rescaling-workflow}
\end{figure}

Consider now a generic Feynman diagram mediating the process described above, such as one of the diagrams shown in the third panel of fig.~\ref{fig:signal-rescaling-workflow}.
Denote by $\theta_{i_{\mathrm{prod}}}$ the small coupling involved in the FIP production vertex, by $\theta_{i_{\mathrm{decay}}}$ the one involved in the FIP decay vertex, and by $M_{\mathrm{FIP}}$ and $\Gamma_{\mathrm{FIP}}(M_{\mathrm{FIP}}, \{\theta_i\})$ the mass and width of the FIP (with the latter being a function of all parameters).
Generically, the \emph{amplitude} of such a diagram will scale as:
\begin{equation*}
    \propto \frac{\theta_{i_{\mathrm{prod}}} \theta_{i_{\mathrm{decay}}}}{(q_{\mathrm{FIP}}^2 - M_{\mathrm{FIP}}^2 + iM_{\mathrm{FIP}}\Gamma_{\mathrm{FIP}})}
\end{equation*}
In many models (such as HNLs), all the diagrams contributing to a given process\footnote{Our definition of ``process'' here matches the one used in e.g.\ \texttt{MadGraph5\_aMC@NLO}~\cite{Alwall:2014hca}.} (initial + final state) involve the same small parameters and therefore scale the same, i.e.\ there is no \emph{interference} between multiple diagrams which depend on different $\theta$'s.
In this simpler (but experimentally relevant) case, the cross-section of each process~$P$ will obey the following scaling law (where we have used the narrow-width approximation and integrated over phase space):
\begin{equation*}
    \sigma_P \propto \frac{|\theta_{i_{\mathrm{prod},P}}|^2 |\theta_{i_{\mathrm{decay},P}}|^2}{\Gamma_{\mathrm{FIP}}}
    \;\Longrightarrow\;
    \sigma_P = \sigma^{\mathrm{ref}}_P \cdot \frac{\tau_{\mathrm{FIP}} |\theta_{i_{\mathrm{prod},P}}|^2 |\theta_{i_{\mathrm{decay},P}}|^2}{\tau^{\mathrm{ref}}_{\mathrm{FIP}} |\theta^{\mathrm{ref}}_{i_{\mathrm{prod},P}}|^2 |\theta^{\mathrm{ref}}_{i_{\mathrm{decay},P}}|^2}
\end{equation*}
i.e.\ the cross-section can be computed once for a set of reference parameters $(\tau^{\mathrm{ref}}_{\mathrm{FIP}}, \{\theta^{\mathrm{ref}}_i\})$, and then rescaled exactly to a different set $(\tau_{\mathrm{FIP}}, \{\theta_i\})$ (keeping in mind that the physical lifetime\footnote{Natural units are assumed throughout this contribution.} $\tau_{\mathrm{FIP}} = \Gamma_{\mathrm{FIP}}^{-1}$ is a function of $M_{\mathrm{FIP}}$ and of the $\theta$'s).

Considering now a signal region and bin (that we will generically denote with $b$), we can obtain the scaling law for the signal count resulting from each process by multiplying the above equation by the integrated luminosity $L_{\mathrm{int}}$ and the efficiency $\epsilon_{P,b}(M_{\mathrm{FIP}},\tau_{\mathrm{FIP}})$, with an important subtlety: if the FIP is sufficiently displaced, the efficiency will depend on the $\theta$'s through its lifetime. We will clarify how to deal with this shortly. The ``prompt'' part of the efficiency otherwise does not depend on $\theta$'s\footnote{In the absence of interference involving multiple $\theta$'s.}, because they enter the amplitude multiplicatively and therefore do not affect the distribution of final-state particles.
Summing over all the processes (which generally have different scaling) that contribute to bin~$b$, we obtain the scaling law for the signal count~$s_b$:
\begin{equation*}
    s_b = L_{\mathrm{int}} \times \sum_{\mathclap{P \in \text{processes}}} \epsilon_{P,b} \sigma_P = \tau_{\mathrm{FIP}} \times \sum_P |\theta_{i_{\mathrm{prod},P}}|^2 |\theta_{i_{\mathrm{decay},P}}|^2 \frac{L_{\mathrm{int}} \epsilon_{P,b} \sigma^{\mathrm{ref}}_P}{\tau^{\mathrm{ref}}_{\mathrm{FIP}} |\theta^{\mathrm{ref}}_{i_{\mathrm{prod},P}}|^2 |\theta^{\mathrm{ref}}_{i_{\mathrm{decay},P}}|^2}
\end{equation*}
The sum can be reordered to factor out the dependence on model parameters:
\begin{equation*}
\begin{split}
    s_b &= \tau_{\mathrm{FIP}} \times \sum_{j,k} |\theta_j|^2 |\theta_k|^2 \left( \sum_P  \delta_{ji_{\mathrm{prod},P}} \delta_{ki_{\mathrm{decay},P}} \frac{L_{\mathrm{int}} \epsilon_{P,b}(M_{\mathrm{FIP}}, \tau_{\mathrm{FIP}}) \sigma^{\mathrm{ref}}_P(M_{\mathrm{FIP}})}{\tau^{\mathrm{ref}}_{\mathrm{FIP}} |\theta^{\mathrm{ref}}_{i_{\mathrm{prod},P}}|^2 |\theta^{\mathrm{ref}}_{i_{\mathrm{decay},P}}|^2} \right) \\
    &\equiv \sum_{j,k} \tau_{\mathrm{FIP}} \Sigma_b^{jk}(M_{\mathrm{FIP}}, \tau_{\mathrm{FIP}}) |\theta_j|^2 |\theta_k|^2
\end{split}
\end{equation*}
where we have defined the tensor $\Sigma_b^{jk}$ as the term in parentheses and made its dependence on $M_{\mathrm{FIP}}$, $\tau_{\mathrm{FIP}}$ and the bin~$b$ explicit. Upon closer examination of this term, we realise that it actually gives us a simple recipe for computing the $\Sigma$ tensor, and thus the signal for arbitrary $\theta$'s and $\tau_{\mathrm{FIP}}$ (the latter still being treated as an independent parameter for the moment).
Let's go through this recipe step by step:
\begin{enumerate}
    \item Identify the scaling properties of the relevant processes.
    \item Group together processes which scale similarly with respect to $\theta$'s (same $i_{\mathrm{prod}/\mathrm{decay}}$).
    \item For each process, select an arbitrary set of \emph{reference parameters}\footnote{Depending on your Monte-Carlo generator, you may want to choose them so that the width of the FIP is small enough for the narrow-width approximation to hold. E.g.\ in \texttt{MadGraph5\_aMC@NLO}, the reference width must be small to trigger the narrow-width treatment.}.
    \item Using these parameters, compute the numbers of events $(L_{\mathrm{int}} \epsilon_{P,b} \sigma^{\mathrm{ref}}_P)$ (for all bins~$b$) produced by each process and normalise them by the reference parameters present in the denominator. A scan over the mass $M_{\mathrm{FIP}}$ is usually unavoidable at this point, and possibly a scan over $\tau_{\mathrm{FIP}}$ if the FIP is significantly displaced.
    \item Sum these normalised numbers of events over all processes from a group and assign the result to the corresponding tensor element, e.g.\ for the group of processes that scale with $\theta_j$ at the production vertex and with $\theta_k$ at the decay vertex, the sum of their normalised numbers of events in bin~$b$ should be assigned to element~${}_b^{jk}$.
\end{enumerate}
With the $\Sigma$ tensor at hand, all we still need to rescale the expected signal to arbitrary parameters is to finally sort out the lifetime dependence.
We proceed similarly, by noticing that the total width $\Gamma_{\mathrm{FIP}}$ can be broken down into the partial widths mediated by each $\theta_i$, which are respectively proportional to $|\theta_i|^2$:
\begin{equation*}
    \tau_{\mathrm{FIP}}^{-1} = \Gamma_{\mathrm{FIP}}(M_{\mathrm{FIP}}, \{\theta_i\}) = \sum_i \hat{\Gamma}^i(M_{\mathrm{FIP}}) |\theta_i|^2
\end{equation*}
where the individual elements $\hat{\Gamma}^i$ can be easily computed as a function of $M_{\mathrm{FIP}}$ by setting $\theta_i = 1$ and all other $\theta$'s to zero.
The physical lifetime $\tau_{\mathrm{FIP}}$ can then be substituted in the expression for $s_b$, if needed by interpolating the efficiencies (or directly the $\Sigma$ elements) between grid points.
We now have all the ingredients we need to accurately compute the signal $s_b$ for arbitrary model parameters, with the only potential source of error being the interpolation between mass and lifetime points.
Fig.~\ref{fig:signal-rescaling-workflow} shows a worked out example in the case of HNLs.

\subsubsection{Remarks}

A few comments are in order.
First, although the $\Sigma$~tensor is not symmetric, only its symmetric part will contribute to the signal.
Second, this method can easily generate sparse tensors, in particular if the signal is insensitive to some of the $\theta$'s.
Third, often the lifetime dependence of the efficiencies can be approximated by a simple ansatz. For instance, in a prompt search, the efficiency should become constant as $\tau_{\mathrm{FIP}} \to 0$, and decrease as $1/\tau_{\mathrm{FIP}}$ when $\tau_{\mathrm{FIP}}$ is much larger than the size of the experiment, thus constraining its functional form (see e.g.\ the fit in ref.~\cite{Tastet:2021vwp}).
Fourth, discrete model parameters, such as for instance the number of HNLs or the (non-)conservation of lepton number, can easily be added to the scaling laws. For instance, in the case of a pair of quasi-Dirac HNLs, this takes the form of additional coefficients in front of the width and cross-sections, with different coefficients for lepton number conserving and violating processes that depend on the model and the value of $R_{ll}$. The interested reader is referred to ref.~\cite{Tastet:2021vwp} for more details.
Fifth, and finally, the above formalism can be generalised to models where multiple diagrams depending of different (and possibly complex) $\theta$'s interfere and contribute to the same process, at the cost of slightly more complicated computations. $\hat{\Gamma}^i$ and $\Sigma_b^{jk}$ are respectively promoted to hermitian tensors $\hat{\Gamma}^{ij}$ and $\Sigma_b^{(ij)(kl)}$, with the scaling laws:
\begin{align*}
    s_b = \sum_{i,j,k,l} \tau_{\mathrm{FIP}} \Sigma_b^{(ij)(kl)}(M_{\mathrm{FIP}}, \tau_{\mathrm{FIP}}) \theta_i^* \theta_j^* \theta_k \theta_l &
    & \tau_{\mathrm{FIP}}^{-1} = \sum_{i,j} \hat{\Gamma}^{ij}(M_{\mathrm{FIP}}) \theta_i^* \theta_j
\end{align*}
Computing the tensor elements then requires being able to explicitly compute the interference terms (and the corresponding efficiencies) for all combinations of $\theta$'s. In $\Sigma$ this corresponds to elements outside of the ${}^{(ij)(ij)}_b$ diagonal.
Although the number of elements can appear daunting, in general it is mitigated by the symmetry properties and the sparsity of the tensors.\footnote{See the \href{https://indico.cern.ch/event/1119695/contributions/5033188/attachments/2531642/4355958/slides_JLTastet_FIPs2022.pdf}{backup slides on Indico} for more details.} % <--- Remove this footnote if the link is problematic.

\subsubsection{Computing the limits}

We are now equipped with a method for accurately rescaling the signal to arbitrary model parameters, with the only potential source of errors being the interpolation in mass and lifetime.
In a background-free experiment, this would be fully sufficient for computing the sensitivity or limits.
For the projected ``discovery'' sensitivity, all one would need to do is to require $s_{\mathrm{tot}} = \sum_b s_b \geq 1$, while in the event of a null result, parameters such that $s_{\mathrm{tot}} \geq 3$ (resp.\ $2.3$) would be excluded at the $95\%$ (resp.\ $90\%$) confidence level.
This corresponds to the ``sensitivity matrix'' approach employed by the SHiP collaboration in ref.~\cite{SHiP:2018xqw}.

In the presence of background, it becomes necessary to compare the expected signal to the observed counts and to the expected background using some statistical procedure (such as e.g.\ the $\mathrm{CL}_s$ test).
In order to use our method, one must then have access to a model of the background.
This can take the form of (ideally) an open likelihood or, if the likelihood is approximately Gaussian at the $2\sigma$ level, of the covariance matrix $B_{bb'}$ between the background counts in all signal regions and bins (including both the statistical and systematic components).\footnote{Because such a method is inherently approximate, it has the added ``benefit'' that theorists cannot in good faith use it to claim a discovery.}
Because the present method deals with the expected signal, we do not have much to add here. Instead we will point the interested reader to the recommendations from the LHC Reinterpretation Forum~\cite{LHCReinterpretationForum:2020xtr} before concluding.

\subsubsection{Summary}

We have presented a method for accurately rescaling to arbitrary model parameters the signal resulting from a feebly-interacting particle.
In order to work, this method requires:
\begin{itemize}
    \item The signal efficiencies $\epsilon_{P,b}(M_{\mathrm{FIP}}, \tau_{\mathrm{FIP}})$ as a function of the FIP mass (and lifetime if displaced), \textbf{disaggregated}\footnote{If the Monte-Carlo truth is available, disaggregating the efficiencies by process should only require a negligible amount of extra computation.} by process (or group of processes that scale the same)\footnote{If multiple diagrams involving different $\theta$'s interfere, efficiencies should be reported by pair of interfering diagrams (some of which may have a negative contribution to the cross-section).} and by signal region/bin.
    \item (Optionally) The cross-sections of each process (or group of processes that scale the same). Although this can usually be recomputed easily, it provides a useful cross-check in case different computations disagree.
    \item Alternatively, experiments may choose to compute and report the $\Sigma$ and $\hat{\Gamma}$ tensors on a suitable grid of masses and lifetimes.
\end{itemize}
A specific example showing how to report these data can be found in this Zenodo dataset~\cite{zenodo5547940}.

The rescaled signal can then be used to accurately extrapolate the limits to arbitrary model parameters, with no additional requirements if the search is background-free, and otherwise if the following data are available:
\begin{itemize}
    \item The observed counts in each signal region and bin.
    \item A (possibly simplified) background model, such as e.g.\ an open likelihood or the covariance matrix of background counts (including systematics).
\end{itemize}
Providing these data would allow to probe not just a finite, zero-measure set of benchmark points, but an entire volume of parameter space, opening the door for fast and accurate global scans and Bayesian analyses.

%\bigskip
%{\itshape JLT is grateful to Oleg Ruchayskiy, Pavlo Kashko and Inar Timiryasov for stimulating discussions about various methods for simplifying the reinterpretation of HNL searches.
%JLT acknowledges partial financial support by the Spanish Research Agency (Agencia Estatal de Investigaci\'on) through the grant IFT Centro de Excelencia Severo Ochoa No CEX2020-001007-S, by the grant PID2019-108892RB-I00 funded by MCIN/AEI/ 10.13039/501100011033, by the European Union's Horizon 2020 research and innovation programme under the Marie Sk\l odowska-Curie grant agreement No 860881-HIDDeN, and by the grant Juan de la Cierva FJC2021-047666-I funded by MCIN/AEI/10.13039/501100011033 and by the European Union ``NextGenerationEU''/PRTR.}

%-------------------------------------------
\subsection{\emph{New ideas:} Simulation of heavy neutrino-antineutrino oscillations -- {\it J.~Hajer}}
\label{hajer}
{\it Author: Jan Hajer}

%\documentclass[a4paper]{article}

%
%\usepackage{amsmath}
%\usepackage{slashed}
%\usepackage{cancel}
%\usepackage{units}

%
%\usepackage[font=small]{caption}
%\usepackage[font=small]{subcaption}
%\usepackage{graphicx}
%\usepackage{booktabs}
%\renewcommand{\floatpagefraction}{.95}

%
%\usepackage{geometry}
%\usepackage[hidelinks]{hyperref}

%
%
%
%
%
%
%
%
%
%

%\author{Jan Hajer}

%\title{Simulation of heavy neutrino-antineutrino oscillations}

%\date{}

%\begin{document}

%\maketitle

Extending the Standard Model (SM) such that mass terms for the neutrinos are generated can explain the observed neutrino oscillations data.
In the framework of type~I seesaw models current observations require at least two massive sterile neutrinos.
In this case the seesaw equation relating the Dirac~$\vec m_D^{(i)}$ and Majorana~$m_M^{(i)}$ masses of the heavy neutrinos with the light neutrino mass matrix reads
\begin{equation} \label{eq:seesaw}
M_\nu = \frac{\vec m_D^{(1)} \otimes \vec m_D^{(1)}}{m_M^{(1)}} + \frac{\vec m_D^{(2)} \otimes \vec m_D^{(2)}}{m_M^{(2)}} \,.
\end{equation}
Taking this equation as a starting point, there are three ways to archive small neutrino masses via the seesaw mechanism.
Both, the high scale seesaw with very heavy Majorana mass terms as well as the small coupling seesaw with tiny Dirac masses are not observable at current collider experiment.
However, when the two terms in relation~\eqref{eq:seesaw} cancel, all parameters can have values such that detection might be possible.
In this case the two Majorana neutrinos form a pseudo-Dirac pair characterised by a tiny mass splitting.
In collider experiments one expects, therefore, to discover neither a single Majorana nor a Dirac heavy neutrino but a pseudo-Dirac heavy neutrino.

In the symmetry protected seesaw scenario (SPSS) this cancellation is ensured by an almost exact \emph{lepton number}-like symmetry~\cite{Antusch:2015mia, Antusch:2016ejd, Antusch:2022hhh}.
When the symmetry is conserved the SM Lagrangian is extended by
\begin{equation} \label{eq:symmetric Lagrangian}
\mathcal L_\text{SPSS}^L =
\overline{N_i^c} i \slashed \partial N_i^{}
- y_{1\alpha} \overline{N_1^c} \widetilde H^\dagger \ell_\alpha
- \overline{N_1^c} m_M^{} N_2^{}
+ \dots + \text{H.c.} \,,
\end{equation}
and the two sterile neutrinos form an exact Dirac pair.
Furthermore, the resulting mass matrix
\begin{equation}
M_n^L =
\begin{pmatrix}
0 & \vec m_D^{} & 0 \\
\vec m_D^T & 0 & m_M^{} \\
0 & m_M^{} & 0
\end{pmatrix}
\,,
\end{equation}
does not generate masses for the light neutrino.
Only in the presence of the small symmetry breaking terms
\begin{equation} \label{eq:broken Lagrangian}
\mathcal L_\text{SPSS}^{\cancel L} =
- y_{2\alpha} \overline{N_2^c} \widetilde H^\dagger \ell_\alpha
- \mu_M^\prime \overline{N_1^c} N_1^{}
- \mu_M^{} \overline{N_2^c} N_2^{}
+ \dots + \text{H.c.} \,,
\end{equation}
does the general mass matrix
\begin{equation} \label{eq:broken mass matrix}
 M_n^{\cancel L} =
\begin{pmatrix}
 0 & \vec m_D^{} & \vec \mu_D^{} \\
 \vec m_D^T & \mu_M^\prime & m_M^{} \\
 \vec \mu_D^T & m_M^{} & \mu_M^{}
\end{pmatrix}
\,,
\end{equation}
lead to light neutrino masses and a mass splitting $\Delta m$ between the two heavy states.
In order to ensure that the light neutrinos remain almost massless these lepton number violating (LNV) effects must be small.
In this case the two heavy Majorana neutrinos necessarily form a pseudo-Dirac pair with a similarly small mass splitting.
This setup encompasses well known models such as the linear and the inverse seesaw.

\begin{figure}
\small
\begin{subfigure}{.5\linewidth}\raggedright
\includegraphics[width=\linewidth]{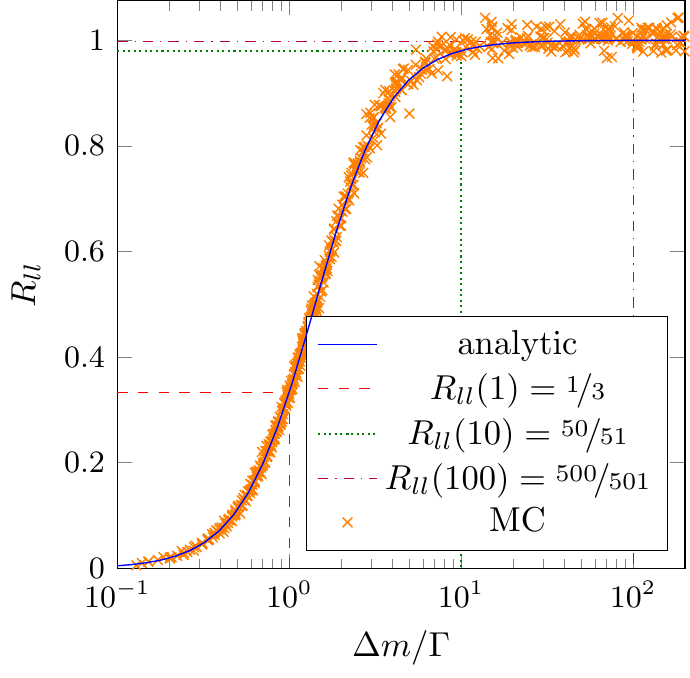}
\caption{Simulation and calculation of $R_{ll}$} \label{fig:Rll}
\end{subfigure}%
\begin{subfigure}{.5\linewidth}\raggedleft
\includegraphics[width=\linewidth]{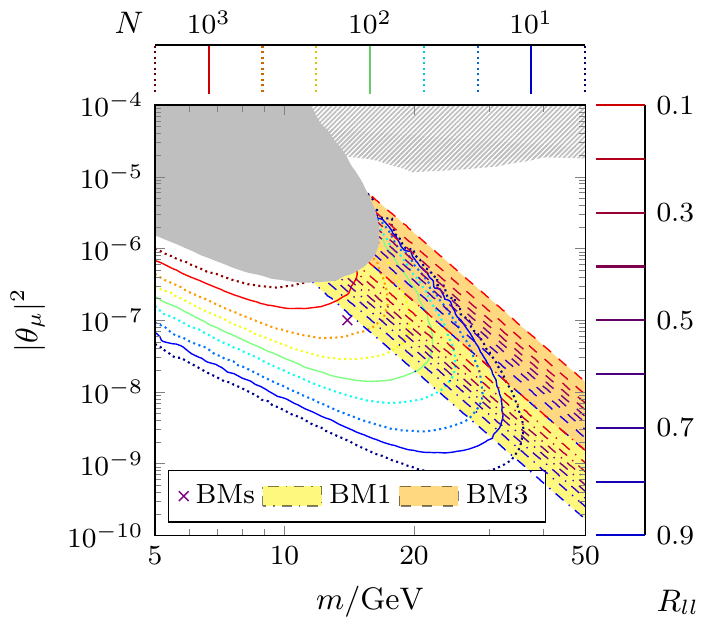}
\caption{Simulation of expected number of events} \label{fig:events}
\end{subfigure}%
\caption{%
Examples of simulations possible with the model file and patch presented here.
Panel~\subref{fig:Rll}: Comparison between the analytic calculation~\eqref{eq:Rll} and the simulation of $R_{ll}$.
Panel~\subref{fig:events}: Simulation of the number of expected events in a displaced vertex search at the HL-LHC in comparison to the current exclusion bounds given in grey \cite{ATLAS:2015gtp, CMS:2018iaf, ATLAS:2020xyo, CMS:2022fut}.
The purple cross indicates the BM points given in figure~\ref{fig:BMs} and the two bands correspond to parameter values leading to a $R_{ll}$ between $0.1$ and $0.9$.
}
\end{figure}

\begin{figure}
\small
\begin{subfigure}[b]{.5\linewidth}
\centering
\begin{tabular}{cccc} \toprule
BM & $\Delta$ m / $\mu$ eV & $c \tau_\text{osc} / {mm}$ & $R_{ll}$ \\ \midrule
1 & 82.7 & 15 & 0.9729 \\
2 & 207 & 6 & 0.9956 \\
3 & 743 & 1.67 & 0.9997 \\
\bottomrule \end{tabular}
\vspace{8ex}
\caption{Benchmark model points} \label{fig:BMs}
\end{subfigure}%
\begin{subfigure}[b]{.5\linewidth}
\includegraphics[width=\linewidth]{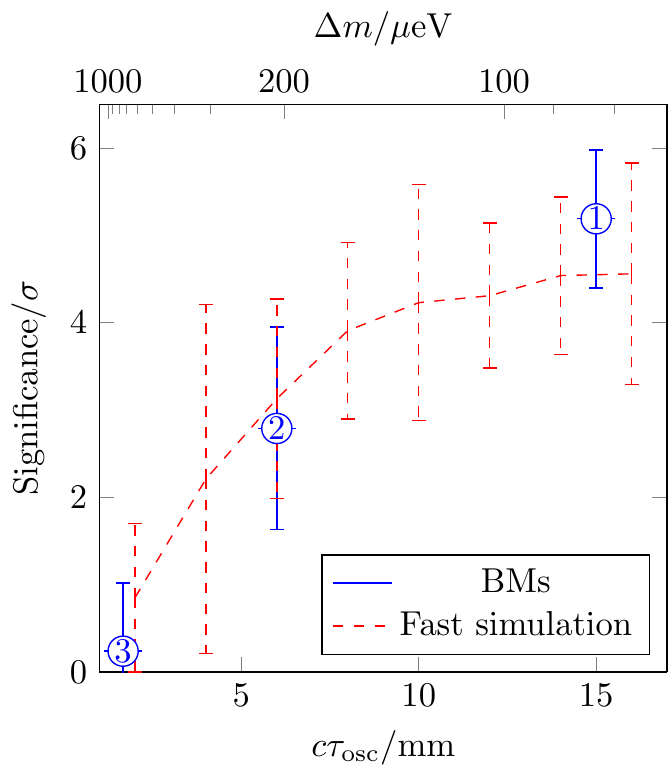}
\caption{Significance as function of oscillation period} \label{fig:sig}
\end{subfigure}%
\caption{
Discoverability of heavy neutrino-antineutrino oscillations.
As indicated in figure \ref{fig:events} all three BM points have a mass of 14 GeV and an active sterile mixing of $|\theta_\mu| = 10^{-7}$ leading to a decay width of $\Gamma = 13.8 \,{\rm \mu eV}$.
Panel~\subref{fig:BMs}: The three mass splittings $\Delta m$, corresponding oscillations periods $\tau_\text{osc}$ and resulting $R_{ll}$ are given.
Panel~\subref{fig:sig}: The significance to discover oscillations at these points with $3\,{\rm ab^{-1}}$ of HL-LHC data, corresponding to $90$ events after cuts in our analysis, has been simulated using MadGraph and Delphes.
The intermediate points have been derived using a fast simulation.
}
\end{figure}

A crucial feature of such pseudo-Dirac pairs is the ability to oscillate between the neutrino and antineutrino eigenstates \cite{Antusch:2017ebe, Antusch:2020pnn}.
When expressed as function of the proper time $\tau$ these oscillations are governed by the mass splitting
\begin{equation} \label{eq:oscillations}
P^{\frac{\text{LNC}}{\text{LNV}}}_\text{osc}(\tau) = \frac{1 \pm \cos(\Delta m \tau) \exp(-\lambda)}2 \,,
\end{equation}
where $\lambda$ is a damping parameter covering potential decoherence effects.
We have checked that it can be taken to be zero for the parameter values in which heavy neutrinos are long-lived.
Integrating over the oscillations~\eqref{eq:oscillations} of the decaying heavy neutrino allows to calculate the ratio between lepton number violating and conserving processes
\begin{equation} \label{eq:Rll}
R_{ll} = \frac{N_{ll}^\text{LNV}}{N_{ll}^\text{LNC}} = \frac{\Delta m^2}{\Delta m^2 + 2 \Gamma^2} \,.
\end{equation}
The functional dependence of this ratio is depicted in figure~\ref{fig:Rll}.

In the phenomenological symmetry protected seesaw scenario (pSPSS), instead of introducing all possible symmetry breaking terms at the Lagrangian level, the heavy neutrino mass splitting is directly introduced as a parameter, such that \cite{Antusch:2022hhh}
\begin{equation} \label{eq:Delta m mass splitting}
%m_{\nicefrac45}^{} 
m_{4/5}^{} 
= m_M^{} \left(1 + \frac12 |\vec \theta|^2 \right) \mp \frac12 \Delta m \,.
\end{equation}
This model captures all leading order LNV effects and depends in addition to the usual set of parameters consisting of the Majorana mass of the lightest pseudo-Dirac pair $m_M$ and its active-sterile mixing parameter $\vec \theta$ only on the mass splitting of the pseudo-Dirac pair $\Delta m$ and the damping parameter $\lambda$.
The pSPSS has been implemented in FeynRules and published online~\cite{pSPSS}.
In order to be able to simulate the heavy neutrino-antineutrino oscillations using this model file the Monte Carlo (MC) generator used for the analysis needs to be patched; such a patch for the MadGraph MC generator is also available~\cite{Antusch:2022hhh, pSPSS}.

This implementation can be used to study properties of the heavy neutrinos.
Figure~\ref{fig:events} shows the number of events appearing in a displaced vertex search at the HL-LHC, reproducing previous results~\cite{Drewes:2019fou}.
Additionally, we have studied the discoverability of heavy neutrino-antineutrino oscillations~\cite{Antusch:2022ceb}.
To this end we have defined three benchmark (BM) points with fixed mass and coupling.
Their varying mass splitting is given in figure~\ref{fig:BMs}.
Figure~\ref{fig:sig} shows, that a $5{\sigma}$ discovery of heavy neutrino-antineutrino oscillations is possible at the HL-LHC.

%\bibliographystyle{unsrt}
%\bibliography{bibliography}

%\end{document}

%-------------------------------------------
\subsection{\emph{New ideas:} Constraints on heavy neutral leptons from the BEBC WA66 beam dump experiment -- {\it G.~Marocco}}
\label{marocco}
{\it Author: Giacomo Marocco, <gmarocco@lbl.gov>}

%\documentclass[]{article}
%\usepackage{geometry}
%\geometry{margin=1in}
%\usepackage{amsmath,amssymb}
%\usepackage{graphicx}
%\usepackage{float}
%\usepackage{makecell}
%\usepackage{siunitx}
%\usepackage{cancel}
%\usepackage{caption}
%\usepackage{subcaption}

%\begin{document}

%\begin{center}{\Large \textbf{
%Constraints on heavy neutral leptons from the BEBC WA66 beam dump experiment
%}}\end{center}

% TODO: write the author list here. Use initials + surname format.
% Separate subsequent authors by a comma, omit comma at the end of the list.
% Mark the corresponding author with a superscript *.
%\begin{center}
%Giacomo Marocco \\
%gmarocco@lbl.gov
%\end{center}

\subsubsection{Introduction}

Neutrinos have small but non-zero masses, the origin of which is unknown. An attractive explanation involves extending the Standard Model (SM) by adding to it right-handed neutrinos, thus generating small masses for the left-handed neutrinos. The most popular is the ‘seesaw’ mechanism for Majorana masses which has many variants \cite{Mohapatra:2005wg}. Such models have open parameter space where the heavy neutral leptons (HNL) mix only with a single flavour of active neutrinos \cite{Cordero-Carrion:2019qtu}. We have carried out work studying the currently least constrained possibility of mixing between HNLs and the tau neutrino \cite{Barouki:2022bkt}. In this simple model, an HNL $N$ has a mass $m_N$ and mixes with the $\nu_\tau$ with a strength given by $U_{\tau N}$. This mixing arises from one of  the few renormalisable operators --- the  so-called neutrino portal  --- that may consistently be added to the Standard Model to couple it to a `dark sector', so is a promising target in the search for new physics beyond the electroweak scale \cite{Lanfranchi:2020crw}. Our bounds also apply to neutrino portal dark sector models where the HNL is a Dirac fermion \cite{Smirnov:2019msn}

Data from the Big European Bubble Chamber (BEBC) WA66 experiment in the 1982 CERN beam dump (400 GeV protons from the SPS) \cite{Grassler:1986vr} had been used to carry out a dedicated search for HNLs \cite{WA66:1985mfx} contemporaneously with CHARM. However that analysis focussed on HNL production (and decay) via mixing with electron and muon neutrinos; the production of HNLs in $\tau$ decays was not considered nor were decays via neutral currents taken into account. Given that BEBC continues to set world-leading bounds on other new physics such as dark photons \cite{Buonocore:2019esg}, magnetic moments, and millicharged particles \cite{Marocco:2020dqu}, we reassess its sensitivity to HNLs mixing with $\nu_\tau$, addressing the above lacunae. We also carry out a reanalysis of HNL mixing with $\nu_e$ in order to include all relevant decay modes and correct a decay rate in \cite{WA66:1985mfx} that omitted an interference contribution, thus obtaining a more restrictive bound on the mixing angle. The bounds from BEBC \cite{WA66:1985mfx} have not been noted in many otherwise comprehensive recent discussions on HNLs e.g. \cite{Beacham:2019nyx,Agrawal:2021dbo,Abdullahi:2022jlv,Batell:2022xau}.  

\subsubsection{HNL fluxes}

An important step in establishing bounds on HNLs is finding both the differential spectrum of the HNLs emerging from a beam dump, as well as their overall flux normalisation. Both of these depend on non-perturbative hadronic physics.

To eliminate systematic errors in the extraction of these quantities, associated e.g. with the adopted model of proton-nucleon interactions in the beam dump, we calibrate  this directly using the \emph{concommitant} flux of active neutrinos. This was measured at BEBC \cite{Grassler:1986vr}, and is consistent with their dominant source being the three-body prompt decays of $D^\pm$ and $D^0$ mesons \cite{WA66:1985mfx}. Hence the total number of HNLs produced, $\mathcal{N}_N$, can be directly related to the total number of ($\sim$massless) active neutrinos of a particular species $\mathcal{N}_{\nu_\ell}$ via
\begin{equation}
    \frac{\mathcal{N}_N}{\mathcal{N}_{\nu_\ell}} \simeq \frac{\sum_i \sigma(pN \rightarrow P_i + X)\text{Br}(P_i \rightarrow N + Y)}{\sigma(pN \rightarrow D^+ D^- + X)\text{Br}(D^\pm \rightarrow \ell \nu_\ell + X)+\sigma(pN \rightarrow D^0 \bar{D}^0 + X)\text{Br}(D^0 \rightarrow \ell \nu_\ell + X)},
    \label{eqn:normalisation}
\end{equation}
where we sum over all parent particles $P_i$ that produce HNLs in their decays. We take $4\sigma(pN \rightarrow D_s + X) =  2\sigma(pN \rightarrow D^+ D^- + X) = \sigma(pN \rightarrow D^0 \bar{D}^0 + X) $ in accordance with data from the Fermilab E769 experiment \cite{E769:1996xad}, so that all cross-sections in the denominator above are proportional to each other. If all the production cross sections $\sigma(pN\rightarrow X)$ in the numerator too are proportional (to be justified when we identify the $P_i$ that appear in this equation), then the hadronic dependence drops out \emph{modulo} the proportionality constants, thus simplifying the calculation considerably and yielding a robust constraint. In the WA66 experiment, it was estimated that \SI{4.1e-4}{} muon neutrinos were produced via $D$ decays per proton on target \cite{Grassler:1986vr}, which allows for direct calculation of $\mathcal{N}_{\nu_\ell}$. The above procedure minimises systematic uncertainties in the overall flux normalisation when the angular distribution is known.

The angular distribution of HNLs may in turn be calculated from first principles, given particular parent meson distributions.
The $D_s$ meson is the dominant source of $\tau$ leptons, which in turn decay to heavy HNLs. We thus need their differential distribution which is usually parameterised as \cite{Frixione:1994nb}:

\begin{equation}
\frac{\mathrm{d}^2 \sigma}{\mathrm{d}x_\mathrm{F} \mathrm{d}p_\mathrm{T}^2} \propto \mathrm{e}^{-b p_\mathrm{T}^2} (1- \lvert x_\mathrm{F} \rvert)^n,
\label{eqn:DsProd}
\end{equation}
where $x_\mathrm{F} = 2 p_L^\mathrm{CM}/\sqrt{s}$ is twice the longitudinal momentum in the cms frame (relative to the cms energy), $p_\mathrm{T}$ is the transverse momentum and the parameters $b$ and $n$ must be extracted from data. While $b$ can be considered to be independent of both the cms energy and the quark content of the charmed meson \cite{DsTau:2019wjb}, $n$ may in general depend on both of these quantities. In the absence of specific data, we take $n$ for $D_s$ production to be the same as for $D^0, \, D^\pm$ mesons. To parameterise the production in the WA66 experiment, we use the results from the WA82 experiment \cite{Adamovich:1992cv}, since both experiments used the same target material (copper), as well as similar beam energies (370 GeV for WA82 \emph{cf.} 400 GeV for WA66). We adopt $b = 0.93 \pm 0.09$ GeV$^{-2}$ and $n = 6.0 \pm 0.3$ \cite{Adamovich:1992cv}; somewhat different values for $n$ were quoted by other experiments e.g. \cite{FermilabE653:1991vmo,Ammar:1988ta}, but this is not as important for HNL production as the transverse momentum distribution which is set by $b$. 

\subsubsection{HNL detection}
In order to be detected, the HNLs produced in the beam dump must reach the detector and then decay within it. The probability for the HNL to reach the detector depends on all the possible detection channels open to it, as well as the mediating interactions. For simplicity, we consider only SM particles in the final state, i.e. HNL decay via the known electroweak bosons. Our analysis is easily generalised to decays via other mediators, see e.g. \cite{Deppisch:2019kvs,Jho:2020jfz}. 
The detection probability depends solely on the  decay channels for which a search was carried out in BEBC, and is associated with an experimental efficiency $\epsilon$. The number of observed events is related to the number of HNLs produced in the beam dump as
\begin{equation}
    \mathcal{N} = \mathcal{N}_N \,
  \Omega \, \langle P \rangle_\Omega  \sum_\alpha  \frac{\Gamma_\alpha}{\Gamma} \cdot \epsilon_\alpha,
  \label{nEvents}
\end{equation}
where $\mathcal{N}_N$ is given by Eq.\eqref{eqn:normalisation}, $\Omega$ is the geometric acceptance set by the solid angle subtended by the detector and $\langle \cdot \rangle_\Omega$ indicates an average over HNLs that lie within this acceptance, while the sum is over experiment-specific channels. 
The efficiency $\epsilon$ is a combination of factors which depends on both the detector response and the HNL decay channel. At BEBC, searches were made for $\ell^- \pi^+$/$\ell^+ \pi^-$ and $ \ell^- \ell^+ \nu$ where $\ell = e, \mu$ \cite{WA66:1985mfx}. HNL decay candidates were required to have an oppositely charged particle pair (with momentum $>1$~GeV/$c$ for scanning efficiency $>97\%$) and no associated neutral hadron interactions or neutral strange particle decays. Cuts were made on the energy and angle 
%(with respect to the beam axis) 
of the charged decay products to ensure consistency with the assumed production/decay channel. The decay distributions of the HNLs were calculated using the full differential decay rate implemented in our Monte Carlo.

Since the sensitivity to HNLs depends on the experimental cuts that were used to isolate signal events, we must take all of this into account 
%(unlike in \cite{Boiarska:2021yho}) 
to extract bounds on the HNL mixing angles. In particular, we require that HNL events pass a cut on the invariant transverse mass $M_\mathrm{T}$, defined by
\begin{equation}
    M_\mathrm{T} \equiv (p_\mathrm{T}^2 + M_I^2)^{1/2} + p_\mathrm{T} < m_D - m_\mu.
    \label{eqn:mT}
\end{equation}
We further adopt a lepton identification efficiency of 96\%. This was the  detection efficiency of the WA66 experiment for electron tracks of momentum $> 0.8$~GeV/$c$, while it was 97\% for muons of momentum $>3$~GeV/$c$ \cite{WA66:1985mfx}. (While the cuts used depended on the HNL under consideration,  no specific results for mixing with $\nu_\tau$s were given, hence we conservatively use the same cuts that were placed on HNLs mixing with $\nu_\mu$s.)

There were no surviving candidates in WA66 for the HNL decay channels $ee\nu$, $e\mu\nu$ or $\mu\mu\nu$, or for $e\pi$, and there was only 1 candidate for $\mu^+\pi^-$ (with invariant mass $\sim 1$~GeV). The background for this decay channel was estimated using data from the WA59 experiment \cite{WA59:1984wvj} in which BEBC, filled with a Ne/H$_2$ mix similar to WA66, was exposed to a conventional `wide band' beam (in which the fraction of HNLs would have been $<1\%$ of that in the beam dump beam). This background was $0.6 \pm 0.2$ events \cite{WA66:1985mfx} corresponding to an upper limit of 3.5 events @ 90\% CL with one candidate event. Since there were no candidate events in the 3-body channels available to $U_{eN}$ or $U_{\tau N}$ mixing, we have conservatively adopted an upper limit of 2.3 signal events \cite{Junk:1999kv}.

\begin{figure}
     \centering
     \begin{subfigure}{0.45\textwidth}
         \centering
         \includegraphics[width=\textwidth]{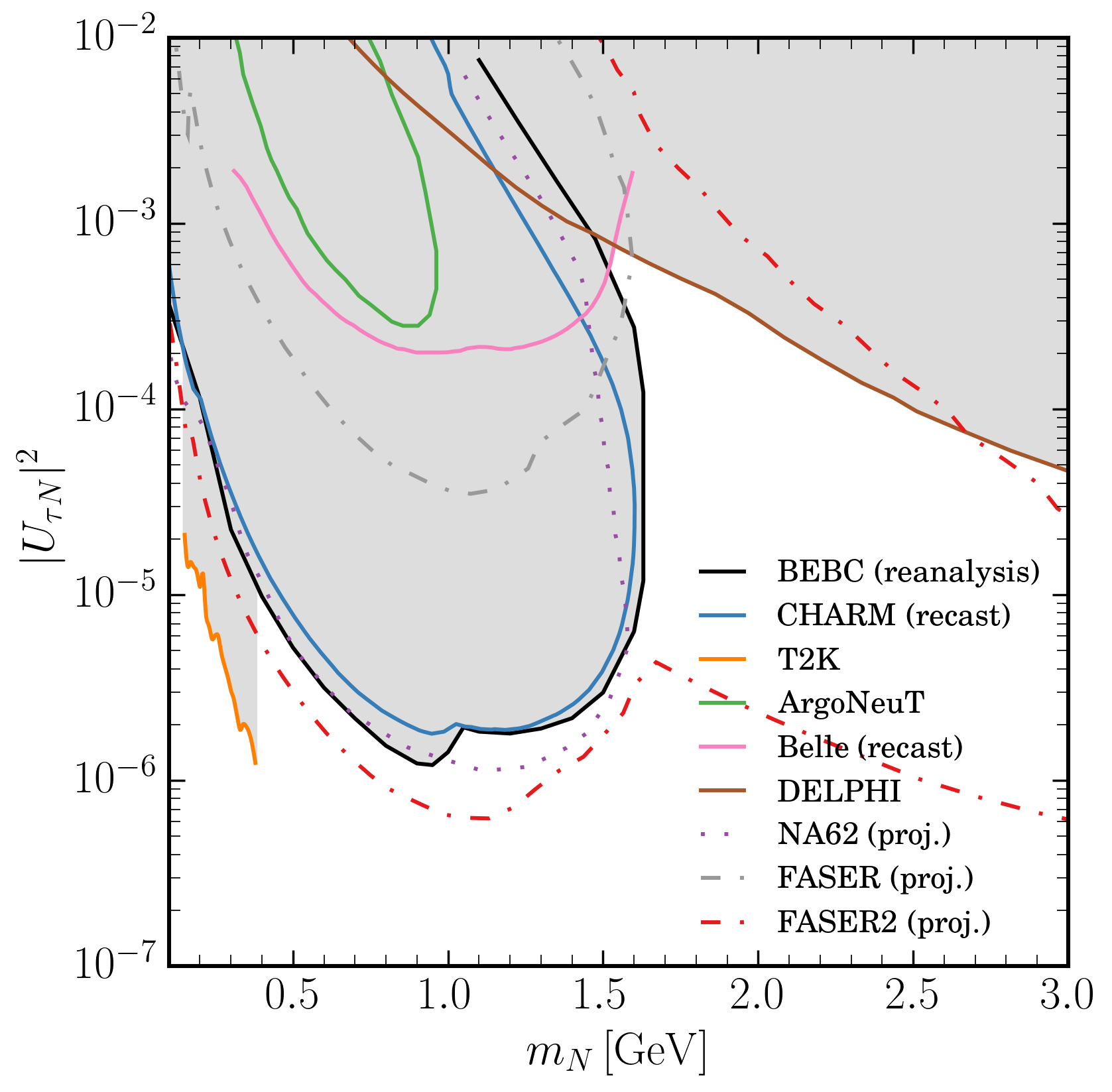}
         \label{fig:tau}
     \end{subfigure}
          \centering
     \begin{subfigure}{0.45\textwidth}
         \centering
         \includegraphics[width=\textwidth]{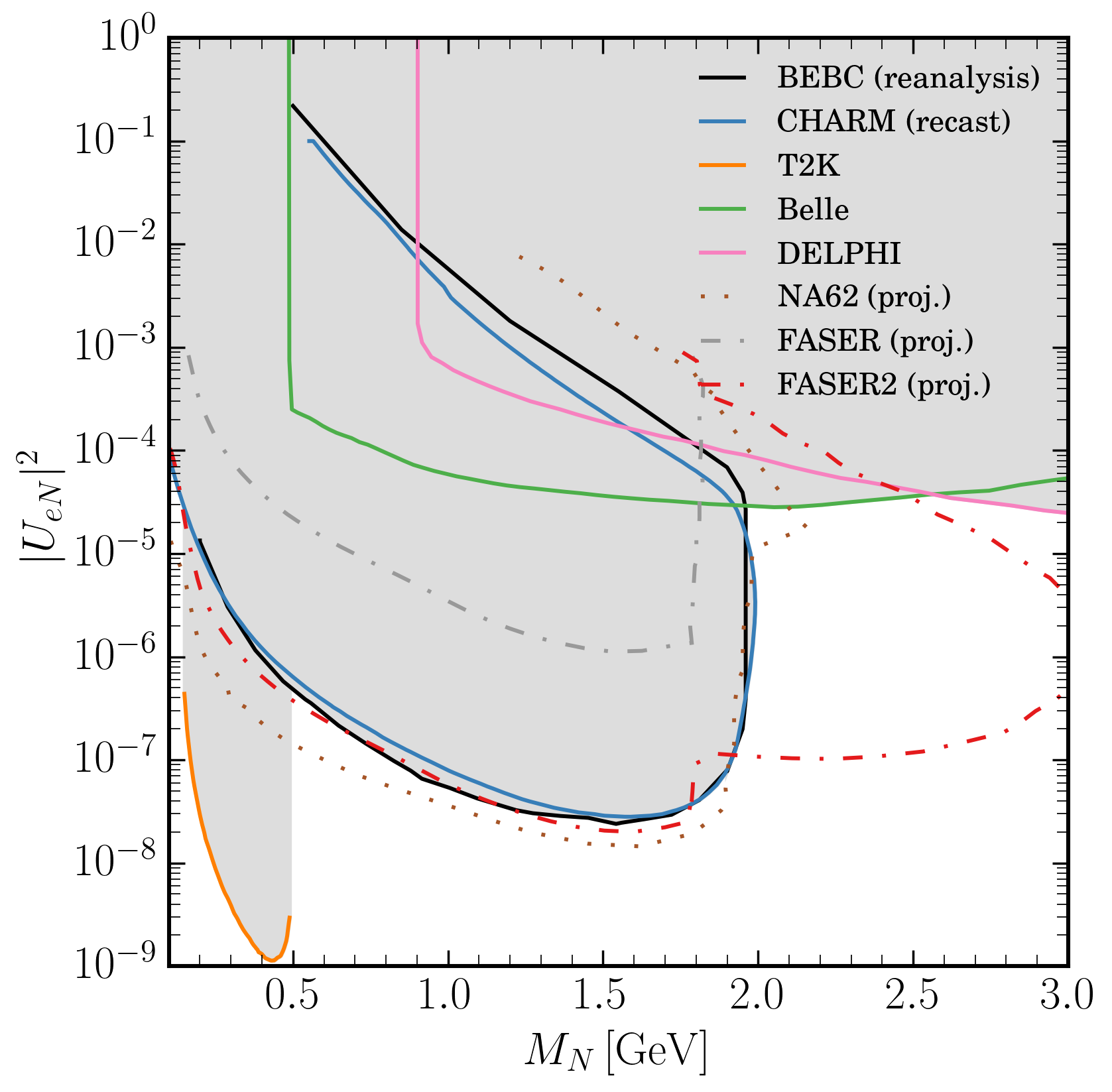}
         \label{fig:electron}
     \end{subfigure}
\caption{\textit{(Left)} The 90\% exclusion region in the HNL mass versus its mixing with $\nu_\tau$ set by reanalysis of BEBC WA66, compared to the recast \cite{Boiarska:2021yho} of the CHARM bound. Also shown are bounds from T2K \cite{T2K:2019jwa}, ArgoNeuT \cite{ArgoNeuT:2021clc}, a recast of Belle \cite{Dib:2019tuj}, and DELPHI \cite{DELPHI:1996qcc}, as well as the projected sensitivities of NA62 in beam dump mode \cite{Drewes:2018gkc} and FASER/FASER2 \cite{FASER:2018eoc}. \textit{(Right)} The 90\% exclusion region in the HNL mass versus its mixing with $\nu_e$ set by this reanalysis of BEBC WA66.
%, compared to the previously set bound \cite{WA66:1985mfx}.
The recast bound \cite{Boiarska:2021yho} from CHARM is also shown, as are bounds from T2K \cite{T2K:2019jwa}, Belle \cite{Belle:2013ytx}, and DELPHI \cite{DELPHI:1996qcc}, as well as the projected sensitivities of NA62 in beam dump mode \cite{Drewes:2018gkc} and FASER/FASER2 \cite{FASER:2018eoc}.}
\label{fig:bounds}
\end{figure}

The bounds resulting from these observations are shown in Fig. \ref{fig:bounds}. Remarkably, BEBC WA66  outperforms all other experiments, including the much bigger CHARM detector. This is primarily because its decay region was off-axis to the beam so it had a lower geometric acceptance than BEBC, as well as receiving a smaller fraction of high energy HNLs. Consequently the on-axis BEBC sets a tighter bound as the HNL mass increases and the transverse momentum gets smaller. We also show in Fig.~\ref{fig:bounds} updated bounds from BEBC WA66 \cite{WA66:1985mfx} on $U_{e N}$, the mixing with the electron neutrino. Using a corrected formula for the HNL decay probabilities, additional production channels, as well as an improved fit for the $D$ meson distribution 
%(which implies a more collimated  beam) 
results in a two-fold improvement over the bounds previously obtained. 

We have demonstrated the continued capability of the BEBC detector to place world-leading bounds on hypothetical particles of interest. This reanalysis has taken into account production and decay channels of HNLs with non-zero $\nu_\tau$ mixings that have not been much considered earlier, thus providing an up-to-date set of exclusions.
%\bibliographystyle{h-physrev}
%\bibliography{references}

%%%%%%%%%%%%%%% Appendix %%%%%%%%%%%%%%%%%%%%%%%%%%

%\appendix

%\end{document}

%-------------------------------------------

\subsection{Conclusions on HNLs}
\label{ssec:HNL_conclusions}
%-------------------------------------------

Models with Heavy Neutral Leptons constitute a minimal extension of the SM that can also explain neutrino masses, the baryon asymmetry of the Universe and even dark matter. They have a rich phenomenology in a variety of experiments, ranging from peak searches to beam dump experiments, to colliders and to lepton number violating processes. It is also possible to inscribe them into more complex scenarios, providing a connection with other dark sectors, such as dark photons and dark scalars.

Here, we summarise the current status and future sensitivity of HNL searches in the standard scenario in which only HNL are added to the SM. For reference, we consider the following simplified benchmark points %given in \eqref{OldBenchmarks} %based on~\cite{Beacham:2019nyx} 
that assume a single HNL coupled exclusively to one SM flavor
\begin{subequations}
\begin{equation}
    U_e^2 : U_\mu^2 : U_\tau^2 = 1 : 0 : 0 \quad \text{BC6}
    \end{equation}
    \begin{equation}
    U_e^2 : U_\mu^2 : U_\tau^2 = 0 : 1 : 0  \quad \text{BC7}
    \end{equation}
    \begin{equation}
      U_e^2 : U_\mu^2 : U_\tau^2 = 0 : 0 : 1 \quad \text{BC8}
      \end{equation}
\end{subequations}
Therefore, for these simplified benchmarks the HNL production and decay can be determined just as function of the parameter space $(m_N,|U_\alpha|^2)$. 
As discussed in Sec.~\ref{klaric} these benchmarks are not consistent with neutrino oscillation data in minimal seesaw scenarios. To address this shortcoming, the following additional benchmarks %\eqref{NewBenchmarks} 
have been proposed,
\begin{subequations}
	\begin{align}
		U_e^2 : U_\mu^2 : U_\tau^2 &= 0:1:1\,,\\
		U_e^2 : U_\mu^2 : U_\tau^2 &= 1:1:1\,,
	\end{align}
\end{subequations}
In addition to the flavour mixing pattern $U_e^2 : U_\mu^2 : U_\tau^2$ it is useful to fix the ratio of lepton number violating to lepton number conserving HNL decays $R_{\ell \ell}=1$ to define benchmarks that can be used to compare the sensitivity of different experiments. While these simple additions cannot describe all aspects of complete neutrino mass models (cf.~Sec.~\ref{hajer}), they represent an improved effective description of realistic models.

\begin{figure}[h]
\begin{center}
\includegraphics[width=\textwidth]{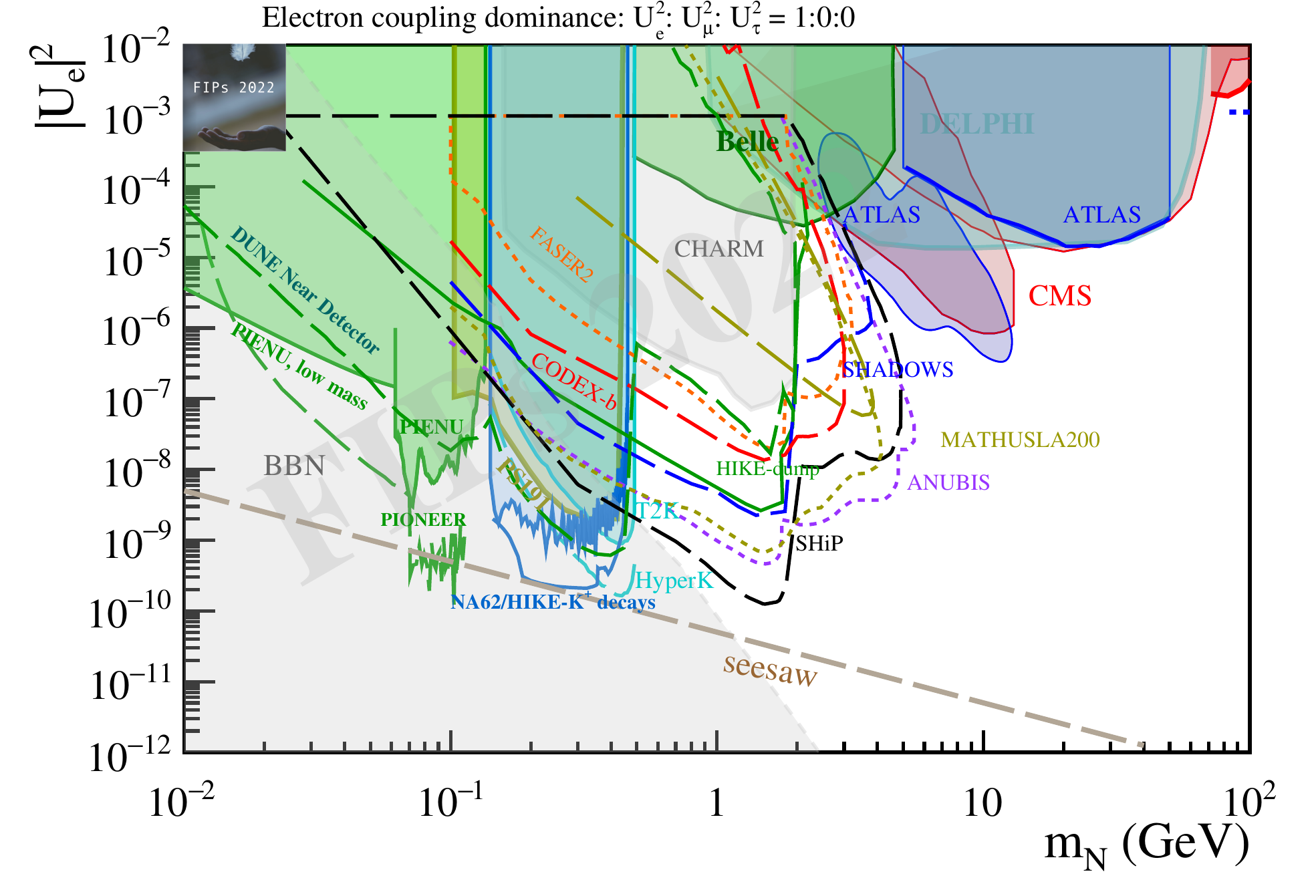}
\end{center}
\caption{{\bf Sensitivity to HNL with electron coupling (BC6).} Current bounds and future projections for 90\% CL exclusion limits. {\bf Legend:} filled gray areas are bounds coming from interpretation of old data sets or astrophysical data; filled coloured areas are bounds set by experimental collaborations; Solid coloured lines are projections based on existing data sets; 
Dashed coloured lines are projections based on full Monte Carlo simulations; Dotted coloured lines are projections based on toy Monte Carlo simulations.
Filled areas are existing bounds from:
PS191~\cite{Bernardi:1987ek}, CHARM~\cite{Bergsma:1985qz}, PIENU~\cite{Aguilar-Arevalo:2017vlf},
NA62 ($K_{eN}$)~\cite{NA62:2020xlg},
T2K~\cite{extracted_beams:T2K_HNL_results},
Belle~\cite{Liventsev:2013zz}, DELPHI~\cite{Abreu:1996pa}, ATLAS~\cite{ATLAS:2019kpx, ATLAS:2022atq}, and CMS~\cite{CMS:2018iaf, CMS:2022fut}.
Coloured curves are projections from:  PIONEER~\cite{PIONEER:2022yag}, HIKE-$K^+$~\cite{HIKE-LoI, HIKE:2022qra};
HIKE-dump~\cite{HIKE-LoI, HIKE:2022qra};
DarkQuest~\cite{Blinov:2021say},
Belle II~\cite{Dib:2019tuj},
FASER2~\cite{Ariga:2018uku};  DUNE near detector~\cite{Abdullahi:2022jlv}, Hyper-K (projections based on ~\cite{T2K:2019jwa}), CODEX-b~\cite{Aielli:2019ivi}, SHiP~\cite{SHiP-ECN3-LoI}, SHADOWS~\cite{SHADOWS-LoI} and MATHUSLA200~\cite{Alpigiani:2020tva}.
The BBN bounds are from \cite{Boyarsky:2020dzc}%\cite{Sabti:2020yrt}. 
The dashed seesaw  line is given by $|U_\alpha|^2 =\sqrt{\Delta m^2_{atm}}/m_N$ corresponding to the naive seesaw scaling and should
be considered only as indicative.
}
\label{fig:BC6}
\end{figure}

\begin{figure}[h]
\begin{center}
\includegraphics[width=\textwidth]{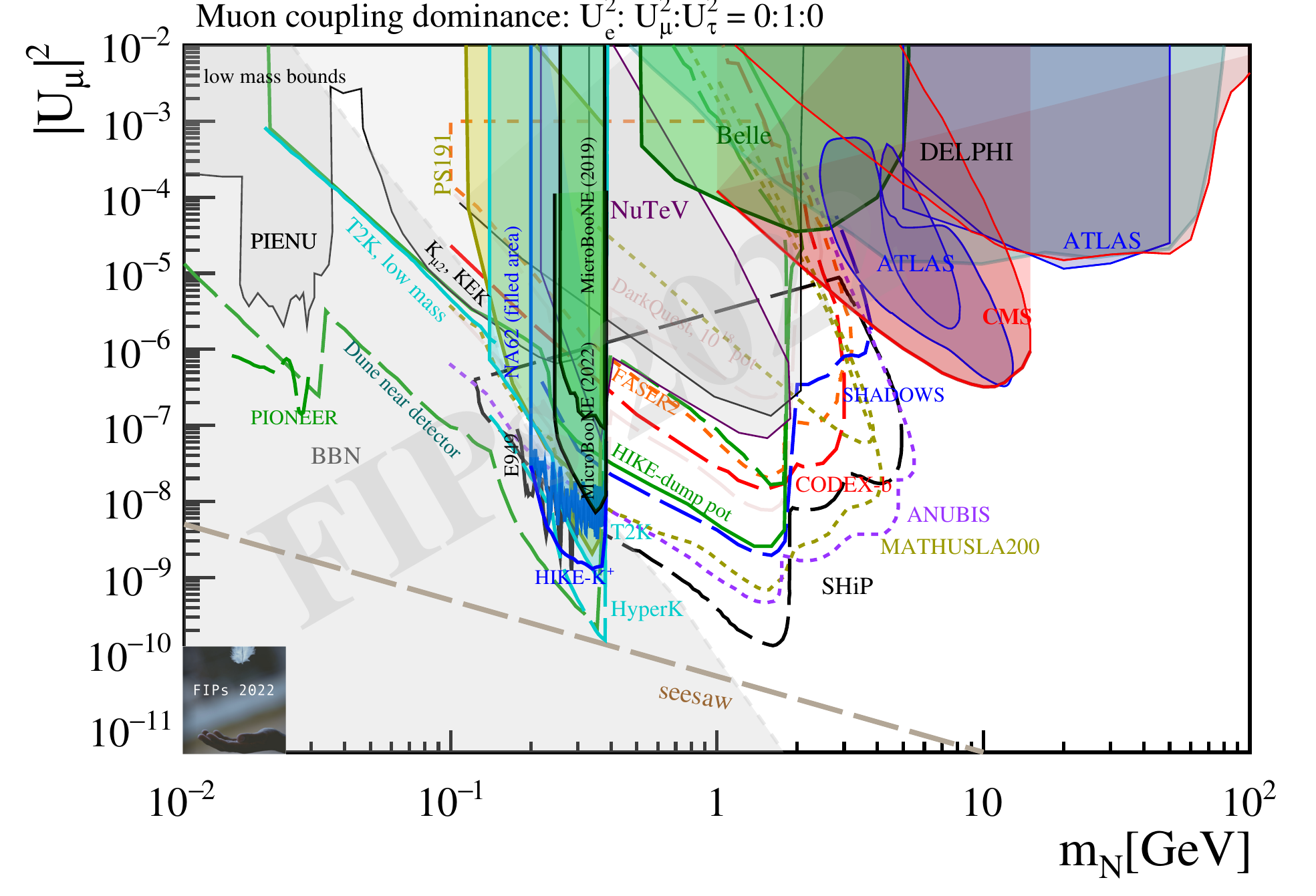}
\end{center}
\caption{{\bf Sensitivity to HNL with muon coupling (BC7).} Current bounds and future projections for 90\% CL exclusion limits. {\bf Legend:} filled gray areas are bounds coming from interpretation of old data sets or astrophysical data; filled coloured areas are bounds set by experimental collaborations; Solid coloured lines are projections based on existing data sets; 
Dashed coloured lines are projections based on full Monte Carlo simulations; Dotted coloured lines are projections based on toy Monte Carlo simulations. Filled coloured areas are existing bounds from:
PS191~\cite{Bernardi:1987ek}, CHARM~\cite{Bergsma:1985qz}, PIENU~\cite{Aguilar-Arevalo:2017vlf},
NA62 ($K_{\mu N}$)~\cite{NA62:2021bji},
T2K~\cite{extracted_beams:T2K_HNL_results},
Belle~\cite{Liventsev:2013zz}; DELPHI~\cite{Abreu:1996pa}, ATLAS~\cite{ATLAS:2019kpx, ATLAS:2022atq} and CMS~\cite{CMS:2018iaf, CMS:2022fut}; MicroBooNE  2019~\cite{MicroBooNE:2019izn} and 2022~\cite{MicroBooNE:2022ctm}
Coloured curves are projections from:  NA62-dump~\cite{NA62:dump,Beacham:2019nyx};
HIKE-$K^+$~\cite{HIKE-LoI, HIKE:2022qra};
HIKE-dump~\cite{HIKE-LoI, HIKE:2022qra};
DarkQuest~\cite{Batell:2020vqn},
Belle II~\cite{Dib:2019tuj};
FASER2~\cite{Ariga:2018uku}; DUNE near detector~\cite{Abdullahi:2022jlv}; Hyper-K (projections based on ~\cite{T2K:2019jwa}); SHiP~\cite{SHiP-ECN3-LoI}, CODEX-b~\cite{Aielli:2019ivi}, and MATHUSLA200~\cite{Alpigiani:2020tva}.
The BBN bounds are from \cite{Boyarsky:2020dzc}.
%\cite{Sabti:2020yrt}. 
The dashed seesaw  line is given by $|U_\alpha|^2 =\sqrt{\Delta m^2_{atm}}/m_N$ corresponding to the naive seesaw scaling and should
be considered only as indicative.
}
\label{fig:BC7}
\end{figure}

\begin{figure}[h]
\begin{center}
\includegraphics[width=\textwidth]{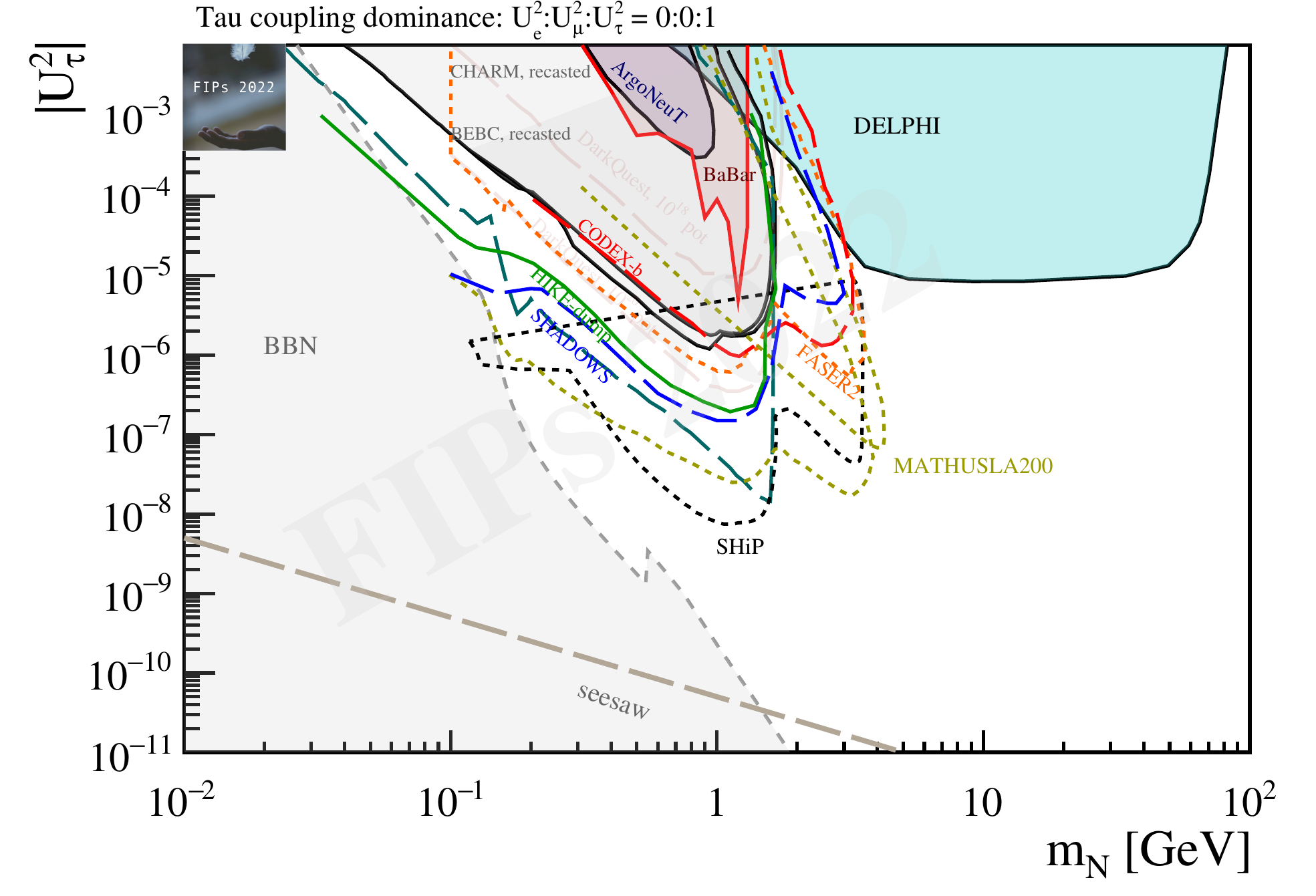}
\end{center}
\caption{{\bf Sensitivity to HNL with tau coupling (BC8).} Current bounds and future projections for 90\% CL exclusion limits. 
{\bf Legend:} filled gray areas are bounds coming from interpretation of old data sets or astrophysical data; filled coloured areas are bounds set by experimental collaborations; Solid coloured lines are projections based on existing data sets; 
Dashed coloured lines are projections based on full Monte Carlo simulations; Dotted coloured lines are projections based on toy Monte Carlo simulations.
Filled areas are existing bounds from:
CHARM (recasted)~\cite{Bergsma:1985qz} and BEBC (recasted)~\cite{Barouki:2022bkt};  %Belle~\cite{Liventsev:2013zz}; 
ArgoNeut~\cite{ArgoNeuT:2021clc};
BaBar~\cite{BaBar:2022cqj};
DELPHI~\cite{Abreu:1996pa};
T2K~\cite{extracted_beams:T2K_HNL_results}
Coloured curves are projections from:  HIKE-dump~\cite{HIKE-LoI},
SHiP~\cite{SHiP-ECN3-LoI},
DarkQuest~\cite{Batell:2020vqn},
Belle II~\cite{Dib:2019tuj},
DUNE~\cite{Abdullahi:2022jlv},
FASER2~\cite{Ariga:2018uku};  CODEX-b~\cite{Aielli:2019ivi}, and MATHUSLA200~\cite{Alpigiani:2020tva}.
The BBN bounds are from \cite{Boyarsky:2020dzc}.
%\cite{Sabti:2020yrt}. 
The dashed seesaw  line is given by $|U_\alpha|^2 =\sqrt{\Delta m^2_{atm}}/m_N$ corresponding to the naive seesaw scaling and should
be considered only as indicative.
%The seesaw bounds are computed under the hypothesis of two HNLs mixing with  active neutrinos, and should be considered only indicative.
%{\it SHADOWS improves the bound by more than one order of magnitude in coupling below the D mass with respect to the CHARM results and by more than 2 orders of magnitude between the D and the B masses improving over Belle and Delphi. In terms of overall coverage, SHADOWS is better of any other experiment running or proposed except SHiP and MATHUSLA.
}
\label{fig:BC8}
\end{figure}

Taking given values of the HNL mixing angles and masses, a contribution to neutrino masses arises in see-saw models and their extensions. In Figs.~\ref{fig:BC6},\ref{fig:BC7} and ~\ref{fig:BC8}, we report a line that corresponds to the naive values of the mixing angles required to explain neutrino masses, specifically $|U_{\alpha}|^2 =\sqrt{\Delta m^2_\mathrm{atm}}/m_N$, where $\Delta m^2_\mathrm{atm}$ is the light neutrino atmospheric mass-squared difference. We note that larger values of the mixing angles are allowed from a theoretical perspective, especially in symmetry protected realizations of the minimal type I see-saw models, e.g. inverse or linear see-saw ones, where the HNL mixing is expected to naturally be much larger than the naive seesaw scaling.

In summary, as described in Figs.~\ref{fig:BC6},\ref{fig:BC7} and ~\ref{fig:BC8}, current and future sensitivities reach very small mixing angles that, for masses $\sim 100$--$400$~MeV even reach the naive see-saw type I prediction. Future experiments will significantly improve the physics reach across all the relevant mass range, and most strongly around the GeV region.

%=====================
%  BC6
%=====================

%--------------------------------------------------

%=====================
%  BC7
%=====================
%--------------------------------------------------

%--------------------------------------------------
%--------------------------------------------------

\clearpage
%=============================================
\section{Conclusions and Outlook}
\label{sec:conclusions}
%=============================================
%Conclusions here

%\section{Conclusions and Outlook}

% The last few years have been characterized by an undisputed increment
% of interest in the physics of Feebly-interacting particles (FIPs),
% shared by large sections of the physics community.
Feebly-interacting particles (FIPs) are currently one of the most
discussed topics in fundamental physics. 
Breaking with the traditional approach, which favoured the existence
of new particles with relatively large couplings to the Standard Model
(SM) and masses commensurate to the electroweak scale, an ever
increasing effort has been devoted to the low energy frontier, at
lower masses and much more feeble interaction strengths.
% The relevance of experimental studies of FIPs has been recognized by
% the international community and discussed in recent strategy reports.
The relevance of experimental FIPs studies has
been widely recognized by the international community, as outlined in recent strategy
reports.

The interest is well motivated. FIPs may provide answers to many outstanding problems 
in particle physics, including the
nature of dark matter, the origin of the neutrino masses, the strong
CP problem, the origin of the matter-antimatter asymmetry,
or cosmological inflation.
Owing to their very nature, namely low masses and feeble interactions,
FIPs are also extremely interesting in astrophysical and cosmological contexts. 
Undeniably, the existence of FIPs would have a tremendous
impact on astrophysics, for example on stellar
evolution and on the propagation of photons on astrophysical or
cosmological distances.
The ability to detect FIPs could provide the tools to
explore regions of the universe not accessible in any other way.
Thus, within the FIP paradigm, fundamental physics questions might be
naturally intertwined and results from terrestrial based laboratory
experiments should be compared with data coming from astrophysics and
cosmology.

The purpose of this report is to present an updated guide to the enormous progress in
all of the most relevant aspects of this research.
It provides an overview of the recent experimental and theoretical
progress, as well as summaries and perspectives from major
laboratories and reviews of the latest strategy reports.
Each of the major sections, Ultralight FIPs, Light dark matter, and
Heavy Neutral Leptons, includes a set of updated summary plots and
tables with running and proposed experiments.

As evident from our report, the FIP community is active and
diversified, including theorists in particle physics, astrophysics and
cosmology, as well as experimentalists working on colliders,
fixed-target, beam-dump and direct  DM detection experiments.
The intense FIP investigation of the last few years has produced a
plethora of innovative theoretical ideas, and catalyzed significant advancements in
existing technologies and revolutionary new techniques in optics,
nuclear magnetic resonance (NMR), resonant cavities, gravitational wave
detection, and others. As illustrated in the summary figures and tables, 
% at the end of each section, 
rapid progress is expected in the coming years. 
% As evident from the tables and summary figures reported at the end of
% our main sections, the coming years should produce a wealth of
% experimental results from all major laboratories in the world. 
% (SHOULD WE BE MORE SPECIFIC HERE?)

The breadth of the open questions in particle physics and their deep
interconnection, together with the (so far) failure of previous
paradigms, make the search for FIPs a compelling and timely endeavor, and
provides hope and opportunities for exciting and profound discoveries
in the near future.

\clearpage
%=============================================
\section{Acknowledgements}
\label{sec:acknowledgements}
%=============================================
We thank CERN and the Physics Beyond Colliders study group for the support during the organization of the workshop and we acknowledge the invaluable help of the secretariat team, P.~Mage-Granados and C.~Cazenoves, that allowed us to smoothly run a workshop with hundreds of participants. We are grateful to the colleagues that helped in steering the scientific programme of the workshop, namely: T.~Ferben, P.~Hernandez,  G.~F.~Giudice, I.Irastorza, J.~Jaeckel,   J.~Monroe, P.~Schuster, and M.~Shaposhnikov for interesting discussions and exchanges. 

\noindent
{\it  C.~Sun} is supported by LDRD at Los Alamos National Laboratory (US) and this report is published under report number LA-UR-23-21432. 

\noindent
{\it G. Lanfranchi, C. Gatti} and {\it A. Paoloni} are supported by INFN, Laboratori Nazionali di Frascati (Italy) and this work is published under the report number INFN-23-14-LNF. 

\noindent
{\it M. Battaglieri}'s contribution is supported by Jefferson Laboratory (US) and this report is published with the report number JLAB-PHY-23-3789.

\noindent
{\it F.~Kling} and {\it A. Lindner} are supported by the DESY Laboratory (Germany) and this work is published under the number DESY-23-050. 

\noindent
{\it N.Tran}'s work is supported by Fermi National Laboratory (US) and this report is published under the number FERMILAB-PUB-23-149-PPD. 

\noindent
{\it J.~Kopp} is supported by CERN Theory Department and by the Mainz Excellence Centre and this report is published under CERN-TH-2023-061 and MITP-23-015 report numbers.

\noindent
{\it B.~Wallisch} acknowledges support from the Swedish Research Council under Grant~\mbox{638-2013-8993}.

\noindent
{\it C.~Gatti}'s contribution is supported by EU through FET Open SUPERGALAX project, grant agreement N.863313.

\noindent
{\it H. Grote}'s contribution has been supported by the STFC in the UK, research grants ST/T006331/1 and ST/W006456/1 and 
The Leverhulme Trust in the UK, research grant RPG-2019-022.

\noindent
{\it Y. V. Stadnik} was supported
by the Australian Research Council under the Discovery Early Career Researcher Award
DE210101593. 

\noindent 
{\it F. Kling} acknowledges support by the Deutsche Forschungsgemeinschaft under Germany's Excellence Strategy - EXC 2121 Quantum Universe - 390833306.

\noindent 
{\it V.V. Flambaum}  work was supported by the Australian Research Council Grants No. DP230101058 and DP200100150.

\noindent
{\it J. Martin} is grateful to those who provided updates on their
experiments, namely: S.~Degenkolb, P.~Fierlinger, T.~Ito, B.~Lauss,
F.~Piegsa, G.~Pignol, B.~Plaster, A.~Serebrov, J.~Thorne and the
author's collaborators on the TUCAN project, principally B.~Franke,
T.~Higuchi, S.~Kawasaki, and R.~Picker.  {\it J. Martin}'s work was supported in part by the Natural Sciences and Engineering Research Council Canada, Canada Foundation for Innovation, and the Canada Research Chairs program. {\it L. Li} is supported by the DOE grant DE-SC-0010010, the NASA grant 80NSSC18K1010 and 80NSSC22K081. 

\noindent
{\it P. Foldenauer} acknowledges support by the Spanish Agencia Estatal de Investigaci\'on through the grants PID2021-125331NB-I00 and CEX2020-001007-S, funded by MCIN/AEI/10.13039/501100011033.

\noindent
The work of {\it P.~Carenza} is supported by the European Research Council under Grant No.~742104 and by the Swedish Research Council (VR) under grants  2018-03641 and 2019-02337.  

\noindent
{\it S. Ulmer} acknowledges the technical support by CERN, especially the Antiproton Decelerator operation group, CERN's cryolab team and engineering department, and all other CERN groups which provide support to Antiproton Decelerator experiments and the financial support by RIKEN, the RIKEN EEE pioneering project funding, the RIKEN SPDR and JRA program, the Max-Planck Society, and the Max-Planck, RIKEN, PTB-Center for Time, Constants, and Fundamental Symmetries (C-TCFS). {\it F. Calore} acknowledges funding by the ``Agence Nationale de la Recherche”, grant n. ANR-19-CE31-0005-01. 

\noindent
{\it L.~Darm\'e} is  supported by the European Union’s Horizon 2020 research and innovation programme under the Marie Skłodowska-Curie grant agreement No 101028626. 

\noindent
{\it M. Battaglieri}'s contribution is based upon work supported by the U.S. Department of Energy, Office of Science, Office of Nuclear Physics under contract DE-AC05-06OR23177 and this work is published under the report number JLAB-PHY-23-3789. 

\noindent
{\it N.~Tran}'s contribution is supported by the Fermi National Laboratory and this work his published under the report number FERMILAB-PUB-23-149-PPD. 

\noindent
{\it P.~Crivelli}
would like to acknowledge the NA64 collaboration and in particular S. Gninenko, L. Molina-Bueno and V. Poliakov. Special thanks also to the past and current ETHZ members, notably E. Depero, H. Sieber, B. Banto-Oberhauser, M. Mongillo, A. Ponten. A big thank you also to the CERN beam department: D. Banerjee, J. Bernhard, N. Charitonidis,  M. Brugger for their continuous support and expresses gratitude to the CERN SPSC members and NA64 referees who were fundamental for the success of NA64: particularly C. Vallé, G. Lanfranchi, G. Salam, L. Gatignon, M. Wing and G. Schnell.
P. Crivelli's work is supported by ETHZ and the Swiss National Science Foundation (SNSF) under the Grants No. 169133 and 186158.

\noindent
{\it R. Pottgen} gratefully acknowledges the support through the Swedish Research Council, grant nr 2019-03436. 

\noindent
{\it S.~Urrea} acknowledges support from Generalitat Valenciana through the plan GenT program (CIDEGENT/2018/019). 

\noindent
{\it F.~Deppisch } would like to thank Patrick Bolton, Bhupal Dev, Mudit Rai and Zhong Zhang for collaborations on which this proceedings report is mainly based on and acknowledges support from the UK Science and Technology Facilities Council (STFC) via the Consolidated Grant ST/T000880/1.

\noindent
{\it M.~C.~Gonzales-Garcia} has received funding/support from the European Union's Horizon
2020 research and innovation program under the Marie Skłodowska-Curie
grant agreement No 860881-HIDDeN, as well as from grant PID2019-105614GB-C21, 
"Unit of Excellence Maria de Maeztu 2020-2023"
award to the ICC-UB CEX2019-000918-M,  
funded by MCIN/AEI/10.13039/501100011033, and by USA-NSF grant PHY-1915093.

\noindent
{\it J.~L.~Tastet} is grateful to Oleg Ruchayskiy, Pavlo Kashko and Inar Timiryasov for stimulating discussions about various methods for simplifying the reinterpretation of HNL searches.
He acknowledges partial financial support by the Spanish Research Agency (Agencia Estatal de Investigaci\'on) through the grant IFT Centro de Excelencia Severo Ochoa No CEX2020-001007-S, by the grant PID2019-108892RB-I00 funded by MCIN/AEI/ 10.13039/501100011033, by the European Union's Horizon 2020 research and innovation programme under the Marie Sk\l odowska-Curie grant agreement No 860881-HIDDeN, and by the grant Juan de la Cierva FJC2021-047666-I funded by MCIN/AEI/10.13039/501100011033 and by the European Union ``NextGenerationEU''/PRTR.

\noindent
{\it D.~Milstead} gratefully acknowledges the Swedish Research Council for support for the design of the HIBEAM program. 
The work for NNBAR is part of the HighNESS project supported by the European Union Framework Programme for Research and Innovation Horizon2020 initiative for the HighNESS project under~grant agreement 951782. 

\noindent
{\it P.~Coloma}'s
 work has received support from the IFT Centro de Excelencia Severo Ochoa Grant No.~CEX2020-001007-S and by Grant PID2019-108892RB-I00 (funded by \newline MCIN/AEI/10.13039/501100011033), and from Grant RYC2018-024240-I (funded by \newline MCIN/AEI/10.13039/501100011033 and by ``ESF Investing in your future''). {\it P.~Coloma}'s work has also received partial funding/support from the European Union’s Horizon 2020 research and innovation program under the Marie Sklodowska-Curie grant agreement No. 860881-HIDDeN.

 \noindent
{\it J.~Shelton}'s work was partially supported by US Department of Energy grant grant DE-SC0023365.

\noindent
{\it B. Velghe's} contribution is supported by NSERC (Natural Sciences and Engineering Research Council), funding SAPPJ-2018-0017, Canada.

\noindent
{\it J. Lopez-Pavon} acknowledges support from the European Union's Horizon 2020 research and innovation programme under the Marie Sk\l odowska-Curie grant agreement No 860881-HIDDeN and the Marie Skłodowska-Curie Staff Exchange grant agreement No 101086085 – ASYMMETRY, the Generalitat Valenciana through the plan GenT program (CIDEGENT/2018/019), and from the Spanish Ministerio de Ciencia e Innovacion through the project PID2020-113644GB-I00.

%======================
% The bibliography
%======================
\clearpage

\bibliographystyle{JHEP}
\bibliography{biblio/biblio.bib,biblio/biblio2.bib,biblio/biblio_plots}

\end{document}
#